\documentclass[11pt,a4paper]{article}
\usepackage{latexsym}
\usepackage{graphicx}
\usepackage{psfrag}
\usepackage{amsmath}
\usepackage{amssymb}
\usepackage{color}
\usepackage{mathtools}
\usepackage{rotating}
\usepackage{multirow}
\usepackage{colortbl}
\usepackage{lscape,epsfig}
\usepackage{a4wide}
\usepackage[]{xcolor}
\usepackage{authblk}
\usepackage{verbatim}
\usepackage{rotating}
\usepackage{fancyheadings}
\usepackage{minibox}
\usepackage{enumitem}
\usepackage[colorlinks=true,a4paper=true,backref=false,linktocpage=true,citecolor=blue,urlcolor=blue,linkcolor=blue,pdfpagemode=UseOutlines]{hyperref}
\usepackage[sort&compress,square,comma,numbers]{natbib}
\usepackage[nottoc]{tocbibind} 
\usepackage{lineno}
\usepackage{placeins}
\usepackage[normalem]{ulem}
\usepackage{supertabular}

\definecolor{lightgreen}{rgb}{.85,.99,.85}
\definecolor{darkgreen}{rgb}{0,.7,0}
\definecolor{orange}{rgb}{1.0,.6,0}
\newcommand{\tbr}{\hspace{0.25mm}{\color{red}$\blacksquare$}} 
\newcommand{\tbg}{{\color{green}$\bigstar$}}

\newcommand{\rC}{C}
\newcommand{\gA}{A}
\newcommand{\oP}{P}

\newcommand{\bda}{\begin{\displaymath}\begin{array}{rl}}
\newcommand{\eda}{\end{array}\end{displaymath}}
\newcommand{\be}{\begin{equation}}
\newcommand{\ee}{\end{equation}}
\newcommand{\bdm}{\begin{displaymath}}
\newcommand{\edm}{\end{displaymath}}
\newcommand{\bea}{\begin{eqnarray}}
\newcommand{\eea}{\end{eqnarray}}

\newcommand{\fs}{\,.}
\newcommand{\co}{\,,}

\newcommand{\ubar}{\overline{\rule[0.42em]{0.4em}{0em}}\hspace{-0.5em}u}
\newcommand{\dbar}{\,\overline{\rule[0.65em]{0.4em}{0em}}\hspace{-0.6em}d}
\newcommand{\sbar}{\,\overline{\rule[0.42em]{0.4em}{0em}}\hspace{-0.5em}s}
\newcommand{\lbar}{\bar{\ell}}

\newcommand{\lsim}{\,\raisebox{-0.3em}{$\stackrel{\raisebox{-0.1em}{$<$}}{\sim}$
}\,} 
\newcommand{\gsim}{\,\raisebox{-0.3em}{$\stackrel{\raisebox{-0.1em}{$>$}}{\sim}$
}\,}
\newcommand{\lvac}{\langle 0|\,}

\newcommand{\al}{&\!\!\!}

\newcommand{\Ch}{$\chi$} 
\newcommand{\mat}[1]{\begin{pmatrix}#1\end{pmatrix}}

\newcommand{\off}[1]{{}}
\newcommand{\flagold}{FLAG 13}

\newcommand{\bi}{\begin{itemize}}
\newcommand{\ei}{\end{itemize}}

\newcommand{\beq}{\begin{equation}}
\newcommand{\eeq}{\end{equation}}

\newcommand{\Mpi}{M_\pi}
\newcommand{\Fpi}{F_\pi}
\newcommand{\Mka}{M_K}
\newcommand{\Fka}{F_K}
\newcommand{\Met}{M_\et}

\newcommand{\<}{\langle}
\renewcommand{\>}{\rangle}

\newcommand{\MeV}{\,\mathrm{MeV}}
\newcommand{\Refs}{\,\mathrm{Refs.}}

\ifdefined\Ref
  \renewcommand{\Ref}{\,\mathrm{Ref.}}
\else
  \newcommand{\Ref}{\,\mathrm{Ref.}}
\fi

\newcommand{\GeV}{\,\mathrm{GeV}}
\newcommand{\fm}{\,\mathrm{fm}}

\newcommand{\ep}{\epsilon}
\newcommand{\de}{\delta}

\newcommand{\et}{\eta}

\newcommand{\rme}{{\rm e}}

\newcommand{\msbar}{{\overline{{\rm MS}}}}
\newcommand{\lms}{\Lambda_\msbar}

\def\mev{{\rm MeV}}
\def\gev{{\rm GeV}}
\def\tev{{\rm TeV}}
\def\fm{{\rm fm}}
 
\def\psibar{\bar{\psi}}

\def\ubar{\bar{u}} 

\def\csw{c_{\rm sw}}

\def\gbar{\bar{g}}

\newcommand{\bd}{\begin{displaymath}}
\newcommand{\ed}{\end{displaymath}}

\newcommand{\eq}[1]{Eq.~(\ref{#1})}
\newcommand{\fig}[1]{Fig.~\ref{#1}}
\newcommand{\tab}[1]{Tab.~\ref{#1}}
\newcommand{\sect}[1]{Sec.~\ref{#1}}

\newcommand{\figurebox}[2]{\fbox{\vbox to#2in{\hbox to #1in{\hfil}\vfil}}}
\newcommand{\bm}[1]{\mbox{\boldmath ${#1}$}}

\newcommand{\gtaeq}{\raisebox{-.6ex}{$\stackrel{\textstyle{>}}{\sim}$}}

\newcommand{\Nf}{N_{\hspace{-0.08 em} f}}
\newcommand{\Tr}{{\rm Tr}\,}

\newcommand{\fKfpicharged}{ \frac{f_{K^\pm}}{f_{\pi^\pm}}}
\newcommand{\fKfpichargedr}{ {f_{K^\pm}}/{f_{\pi^\pm}}}

\newcommand{\mpi}{m_\pi}

\newcommand{\fpi}{f_\pi}

\newcommand{\half}{\textstyle{1\over2}}

\newcommand{\abar}{\overline{a}}

\newcommand{\rb}[1]{\raisebox{1.5ex}[-1.5ex]{#1}}

\newcommand{\NLo}{\stackrel{\rule[-0.1cm]{0cm}{0cm}\mbox{\tiny NLO}}{=}}

\definecolor{Gray}{rgb}{0.5,0.5,0.5}
\definecolor{Black}{rgb}{0.0,0.0,0.0}

\def\good{\makebox[1em]{\centering{\mbox{\color{green}$\bigstar$}}}}
\def\bad{\makebox[1em]{\centering\color{red}\tiny$\blacksquare$}}
\def\soso{\makebox[1em]{\centering{\mbox{\raisebox{-0.5mm}{\color{green}\Large$\circ$}}}}}
\def\okay{\hspace{0.25mm}\raisebox{-0.2mm}{{\color{green}\large\checkmark}}}

\newcommand{\mr}{\mathrm}

\def\half{{1\over2}}
\def\bra#1{\left\langle #1\right|}
\def\ket#1{\left| #1\right\rangle}

\def\Tr{\,\mathrm{Tr}}

\def\fm{\mathrm{fm}}
\def\ev{\mathrm{e\kern-0.1em V}}
\def\kev{\mathrm{ke\kern-0.1em V}}
\def\mev{\mathrm{Me\kern-0.1em V}}
\def\gev{\mathrm{Ge\kern-0.1em V}}
\def\tev{\mathrm{Te\kern-0.1em V}}

\let\Re=\re \let\Im=\im

\def\n#1e#2n{{#1}\times 10^{#2}}

\def\bea{\begin{eqnarray}}
\def\eea{\end{eqnarray}}
\def\nn{\nonumber}

\def\oO{\mathcal{Q}}
\def\cO{\mathcal{O}}

\def\ods2{\mathcal{O}_{\Delta S=2}}
\def\zds2{Z_{\Delta S=2}}

\def\msbar{{\overline{\mathrm{MS}}}}

\def\lqcd{\Lambda_\mathrm{QCD}}
\def\lat{\mathrm{lat}}

\def\spose#1{\hbox to 0pt{#1\hss}}
\def\ltapprox{\mathrel{\spose{\lower 3pt\hbox{$\mathchar"218$}}
 \raise 2.0pt\hbox{$\mathchar"13C$}}}
\def\gtapprox{\mathrel{\spose{\lower 3pt\hbox{$\mathchar"218$}}
 \raise 2.0pt\hbox{$\mathchar"13E$}}}
\def\inapprox{\mathrel{\spose{\lower 3pt\hbox{$\mathchar"218$}}
 \raise 2.0pt\hbox{$\mathchar"232$}}}

\makeatletter
\def\slash#1{{\mathpalette\c@ncel{#1}}} 
\def\big#1{{\hbox{$\left#1\vbox to1.012\ht\strutbox{}\right.\n@space$}}}
\def\Big#1{{\hbox{$\left#1\vbox to1.369\ht\strutbox{}\right.\n@space$}}}
\def\bigg#1{{\hbox{$\left#1\vbox to1.726\ht\strutbox{}\right.\n@space$}}}
\def\Bigg#1{{\hbox{$\left#1\vbox
to2.083\ht\strutbox{}\right.\n@space$}}}
\makeatother

\newcommand{\nl}{\nonumber \\}

\newcommand{\delv}{{\bf \nabla}}
\newcommand{\delvt}{\tilde{{\bf \nabla}}}

\newcommand{\delfour}{{\Delta^{(4)}}}
\newcommand{\delsq}{\Delta^{(2)}}

\newcommand{\Ev}{\tilde{{\bf E}}}
\newcommand{\Bv}{\tilde{{\bf B}}}
\newcommand{\sigmav}{\mbox{\boldmath$\sigma$}}

\def\spose#1{\hbox to 0pt{#1\hss}}
\def\ltapprox{\mathrel{\spose{\lower 3pt\hbox{$\mathchar"218$}}
\raise 2.0pt\hbox{$\mathchar"13C$}}}
\def\gtapprox{\mathrel{\spose{\lower 3pt\hbox{$\mathchar"218$}}
\raise 2.0pt\hbox{$\mathchar"13E$}}}
\def\inapprox{\mathrel{\spose{\lower 3pt\hbox{$\mathchar"218$}}
\raise 2.0pt\hbox{$\mathchar"232$}}}

\newcommand{\cred}{\color{red}}

\newcommand{\alphah}{\alpha_\mathrm{V'}}
\newcommand{\alphav}{\alpha_\mathrm{V}}
\newcommand{\alphap}{\alpha_\mathrm{P}}

\newcommand{\SLfnalmilcBDstar}{FNAL/MILC 14}   
\newcommand{\SLfnalmilcBD}{FNAL/MILC 15C}               
\newcommand{\SLfnalmilcBpi}{FNAL/MILC 15}                              
\newcommand{\SLfnalmilcBsK}{FNAL/MILC 19}                              
\newcommand{\SLhpqcdBD}{HPQCD 15}                              
\newcommand{\SLhpqcdBsK}{HPQCD 14}                                              
\newcommand{\SLrbcukqcdBpi}{RBC/UKQCD 15}                                       
\newcommand{\SLLambdabp}{Detmold 15\\ $\Lambda_b \to p$}                                                
\newcommand{\SLLambdabc}{Detmold 15\\ $\Lambda_b \to \Lambda_c$}                                                
\newcommand{\SLhpqcdBK}{HPQCD 13E}                                              
\newcommand{\SLfnalmilcBK}{FNAL/MILC 15D}                              

\newcommand{\FLAGAVBEGIN}{}
\newcommand{\FLAGAVEND}{}

\newcommand{\scal}{\mathcal{S}}

\title{{\bf \Huge FLAG Review 2021}\\
  \vspace{0.5cm}
  {\bf \large Flavour Lattice Averaging Group (FLAG)} }

\newcounter{affilctr} 

\addtocounter{affilctr}{1}
\newcounter{RIKEN}
\author[\theRIKEN]{Y.~Aoki}
\affil[\theRIKEN]{RIKEN Center for Computational Science, Kobe 650-0047, Japan}

\addtocounter{affilctr}{1}
\newcounter{Connecticut}
\addtocounter{affilctr}{1}
\newcounter{RIKENBNL}
\author[\theConnecticut,\theRIKENBNL]{T.~Blum}
\affil[\theConnecticut]{Physics Department, University of Connecticut, Storrs, CT 06269-3046, USA}
\affil[\theRIKENBNL]{RIKEN BNL Research Center, Brookhaven National Laboratory, Upton, NY 11973, USA}

\addtocounter{affilctr}{1}
\newcounter{Bern}
\author[\theBern]{G.~Colangelo}
\affil[\theBern]{Albert Einstein Center for Fundamental Physics,
Institut f\"ur Theoretische Physik, Universit\"at Bern, Sidlerstr. 5, 3012 Bern, Switzerland}

\addtocounter{affilctr}{1}
\newcounter{Regensburg}
\author[\theRegensburg]{S.~Collins}
\affil[\theRegensburg]{Institut f\"ur Theoretische Physik, Universit\"at Regensburg, 93040 Regensburg, Germany}

\addtocounter{affilctr}{1}
\newcounter{Odense}
\author[\theOdense]{M.~Della~Morte}
\affil[\theOdense]{CP3-Origins and IMADA, University of Southern
  Denmark, Campusvej 55, DK-5230 Odense M, Denmark}

\addtocounter{affilctr}{1}
\newcounter{Parma}
\addtocounter{affilctr}{1}
\newcounter{ParmaINFN}
\author[\theParma,\theParmaINFN]{P.~Dimopoulos}
\affil[\theParma]{
  Dipartimento di Scienze Matematiche, Fisiche e Informatiche, Universit\`a di Parma, 43124 Parma, Italy}
\affil[\theParmaINFN]{INFN,
Gruppo Collegato di Parma, Parco Area delle Scienze 7/a (Campus), 43124 Parma, Italy}

\addtocounter{affilctr}{1}
\newcounter{Wuppertal}
\addtocounter{affilctr}{1}
\newcounter{Julich}
\author[\theWuppertal,\theJulich]{S.~D\"urr}
\affil[\theWuppertal]{University of Wuppertal, Gau{\ss}stra{\ss}e\,20, 42119 Wuppertal, Germany}
\affil[\theJulich]{J\"ulich Supercomputing Center, Forschungszentrum J\"ulich,
  52425 J\"ulich, Germany}

\addtocounter{affilctr}{1}
\newcounter{Peking}
\addtocounter{affilctr}{1}
\newcounter{PekingCIC}
\addtocounter{affilctr}{1}
\newcounter{PekingHEP}
\addtocounter{affilctr}{1}
\newcounter{PekingNP}
\author[\thePeking,\thePekingCIC,\thePekingHEP,\thePekingNP]{X.~Feng}
\affil[\thePeking]{School of Physics, Peking University, Beijing 100871, China}
\affil[\thePekingCIC]{Collaborative Innovation Center of Quantum Matter, Beijing 100871, China}
\affil[\thePekingHEP]{Center for High Energy Physics, Peking University, Beijing 100871, China}
\affil[\thePekingNP]{State Key Laboratory of Nuclear Physics and Technology, Peking University, Beijing 100871, China}

\addtocounter{affilctr}{1}
\newcounter{Osaka}
\author[\theOsaka]{H.~Fukaya}
\affil[\theOsaka]{Department of Physics, Osaka University, Toyonaka, Osaka 560-0043, Japan}

\addtocounter{affilctr}{1}
\newcounter{SF}
\author[\theSF]{M.~Golterman}
\affil[\theSF]{Dept. of Physics and Astronomy, San Francisco State University, San Francisco, CA 94132, USA}

\addtocounter{affilctr}{1}
\newcounter{Indiana}
\author[\theIndiana]{Steven Gottlieb}
\affil[\theIndiana]{Department of Physics, Indiana University, Bloomington, IN 47405, USA}

\addtocounter{affilctr}{1}
\newcounter{LANL}
\author[\theLANL]{R.~Gupta}
\affil[\theLANL]{Los Alamos National Laboratory, Theoretical Division T-2, Los Alamos, NM 87545, USA}

\addtocounter{affilctr}{1}
\newcounter{KEK}
\addtocounter{affilctr}{1}
\newcounter{Sokendai}
\author[\theKEK,\theSokendai]{S.~Hashimoto}
\affil[\theKEK]{High Energy Accelerator Research Organization (KEK), Tsukuba 305-0801, Japan}
\affil[\theSokendai]{School of High Energy Accelerator Science,
The Graduate University for Advanced Studies (Sokendai), Tsukuba 305-0801, Japan}

\addtocounter{affilctr}{1}
\newcounter{APS}
\author[\theAPS]{U.~M.~Heller}
\affil[\theAPS]{American Physical Society (APS), One Research Road, Ridge, NY 11961, USA}

\addtocounter{affilctr}{1}
\newcounter{Madrid}
\author[\theMadrid]{G.~Herdoiza}
\affil[\theMadrid]{Instituto de F\'{\i}sica Te\'orica UAM/CSIC and
  Departamento de F\'{\i}sica Te\'orica, Universidad Aut\'onoma de Madrid, Cantoblanco 28049 Madrid, Spain}

\addtocounter{affilctr}{1}
\newcounter{IFIC}
\author[\theIFIC]{P.~Hernandez}
\affil[\theIFIC]{IFIC (CSIC-UVEG), Parc Cient\'{\i}fic de la Universitat de Val\`encia, E-46980 Paterna, Spain}

\addtocounter{affilctr}{1}
\newcounter{Edinburgh}
\author[\theEdinburgh]{R.~Horsley}
\affil[\theEdinburgh]{Higgs Centre for Theoretical Physics, School of Physics and Astronomy, University of Edinburgh, Edinburgh EH9 3FD, UK}

\addtocounter{affilctr}{1}
\newcounter{Southampton}
\addtocounter{affilctr}{1}
\newcounter{STAG}
\addtocounter{affilctr}{1}
\newcounter{CERN}
\author[\theSouthampton,\theSTAG,\theCERN]{A.~J\"uttner}
\affil[\theSouthampton]{School of Physics \& Astronomy, University of Southampton, Southampton SO17 1BJ, UK}
\affil[\theSTAG]{STAG Research Center, University of Southampton, Highfield, Southampton SO17 1BJ, UK}
\affil[\theCERN]{CERN, Theoretical Physics Department, Geneva, Switzerland}

\author[\theKEK,\theSokendai]{T.~Kaneko}

\author[\theIndiana]{E.~Lunghi}

\addtocounter{affilctr}{1}
\newcounter{Arizona}
\author[\theArizona]{S.~Meinel}
\affil[\theArizona]{Department  of  Physics,  University  of  Arizona,  Tucson,  AZ  85721,  USA}

\addtocounter{affilctr}{1}
\newcounter{WandM}
\addtocounter{affilctr}{1}
\newcounter{JLab}
\author[\theWandM,\theJLab]{C.~Monahan}
\affil[\theWandM]{Department of Physics, The College of William \& Mary, Williamsburg, VA  23187,  USA}
\affil[\theJLab]{Theory Center, Thomas Jefferson National Accelerator Facility, Newport News, VA 23606, USA}

\addtocounter{affilctr}{1}
\newcounter{ChapelHill}
\author[\theChapelHill]{A.~Nicholson}
\affil[\theChapelHill]{Dept.~of Physics and Astronomy, University of North Carolina, Chapel Hill, NC 27516-3255, USA}

\author[\theOsaka]{T.~Onogi}

\author[\theMadrid]{C.~Pena}

\addtocounter{affilctr}{1}
\newcounter{BNL}
\author[\theBNL]{P.~Petreczky}
\affil[\theBNL]{Physics Department, Brookhaven National Laboratory, Upton, NY 11973, USA}

\author[\theEdinburgh]{A.~Portelli}

\author[\theIFIC]{A.~Ramos}

\addtocounter{affilctr}{1}
\newcounter{Seattle}
\author[\theSeattle]{S.~R.~Sharpe}
\affil[\theSeattle]{Physics Department, University of Washington, Seattle, WA 98195-1560, USA}

\addtocounter{affilctr}{1}
\newcounter{Fermilab}
\author[\theFermilab]{J.~N.~Simone}
\affil[\theFermilab]{Fermi National Accelerator Laboratory, Batavia, IL 60510 USA}

\addtocounter{affilctr}{1}
\newcounter{RomaTre}
\author[\theRomaTre]{S.~Simula}
\affil[\theRomaTre]{INFN, Sezione di Roma Tre, Via della Vasca Navale 84, 00146 Rome, Italy}

\addtocounter{affilctr}{1}
\newcounter{TrinityCollege}
\affil[\theTrinityCollege]{School of Mathematics \& Hamilton Mathematics Institute, Trinity College Dublin, Dublin 2, Ireland}
\author[\theTrinityCollege]{S.~Sint}

\addtocounter{affilctr}{1}
\newcounter{NIC}
\addtocounter{affilctr}{1}
\newcounter{Humboldt}
\author[\theNIC,\theHumboldt]{R.~Sommer}
\affil[\theNIC]{John von Neumann Institute for Computing (NIC), DESY, Platanenallee~6, 15738 Zeuthen, Germany}
\affil[\theHumboldt]{Institut f\"ur Physik, Humboldt-Universit\"at zu Berlin, Newtonstr. 15, 12489 Berlin, Germany}

\addtocounter{affilctr}{1}
\newcounter{TorVergata}
\author[\theTorVergata]{N.~Tantalo}
\affil[\theTorVergata]{INFN, Sezione di Tor Vergata, c/o Dipartimento di Fisica,
 Universit\`a di Roma Tor Vergata, Via della Ricerca Scientifica 1, 00133 Rome, Italy}

\author[\theFermilab]{R.~Van de Water}

\author[\theBern,\theCERN]{U.~Wenger}

\addtocounter{affilctr}{1}
\newcounter{Mainz}
\author[\theMainz]{H.~Wittig}
\affil[\theMainz]{PRISMA Cluster of Excellence, Institut f\"ur
Kernphysik and Helmholtz Institute Mainz, University of Mainz, 55099 Mainz,
Germany}

\date{\today}

\begin{document}
 \hfill
\begin{minipage}{0.4\textwidth}
\begin{flushright}    
 \vspace{-1cm}
 CERN-TH-2021-191 \\ 
 JLAB-THY-21-3528 \\ 
\end{flushright}
\end{minipage}
{\let\newpage\relax\maketitle} 

\abstract{ We review lattice results related to pion, kaon, $D$-meson,
  $B$-meson, and nucleon physics with the aim of making them easily
  accessible to the nuclear and particle physics communities. More
  specifically, we report on the determination of the light-quark
  masses, the form factor $f_+(0)$ arising in the semileptonic $K \to
  \pi$ transition at zero momentum transfer, as well as the decay
  constant ratio $f_K/f_\pi$ and its consequences for the CKM matrix
  elements $V_{us}$ and $V_{ud}$. Furthermore, we describe the results
  obtained on the lattice for some of the low-energy constants of
  $SU(2)_L\times SU(2)_R$ and $SU(3)_L\times SU(3)_R$ Chiral
  Perturbation Theory. We review the determination of the $B_K$
  parameter of neutral kaon mixing as well as the additional four $B$
  parameters that arise in theories of physics beyond the Standard
  Model.  For the heavy-quark sector, we provide results for $m_c$ and
  $m_b$ as well as those for the decay constants, form factors, and
  mixing parameters of charmed and bottom mesons and baryons.  These
  are the heavy-quark quantities most relevant for the determination
  of CKM matrix elements and the global CKM unitarity-triangle fit.
  We review the status of lattice determinations of the strong
  coupling constant $\alpha_s$.  We consider nucleon matrix elements,
  and review the determinations of the axial, scalar and tensor
  bilinears, both isovector and flavor diagonal.  Finally, in this
  review we have added a new section reviewing determinations of
  scale-setting quantities.  }

\newpage

\fancyhead{}
\fancypagestyle{defaultstyle}
               {
               }
\fancypagestyle{updXXX20XXstyle}
               {
                 \fancyfoot[R]{\emph{Updated XXX.~20XX}}
               }
\fancypagestyle{upddec2020style}
               {
                 \fancyfoot[R]{\emph{Updated Dec.~2020}}
               }
\fancypagestyle{draft}
               {
                 \fancyfoot[R]{\emph{Draft 09/11/2021}}
               }
\renewcommand{\headrulewidth}{0pt}
\thispagestyle{plain}

\def\reducedapptables{}


\def\noglossary{}

\tableofcontents

\clearpage
\section{Introduction}
\label{sec:introduction}

Flavour physics provides an important opportunity for exploring the
limits of the Standard Model of particle physics and for constraining
possible extensions that go beyond it. As the LHC explores 
a new energy frontier and as experiments continue to extend the precision 
frontier, the importance of flavour physics will grow,
both in terms of searches for signatures of new physics through
precision measurements and in terms of attempts to construct the
theoretical framework behind direct discoveries of new particles. 
Crucial to such searches for new physics is the ability to quantify
strong-interaction effects.
Large-scale numerical
simulations of lattice QCD allow for the computation of these effects
from first principles. 
The scope of the Flavour Lattice Averaging
Group (FLAG) is to review the current status of lattice results for a
variety of physical quantities that are important for flavour physics. Set up in
November 2007, it
comprises experts in Lattice Field Theory, Chiral Perturbation
Theory and Standard Model phenomenology. 
Our aim is to provide an answer to the frequently posed
question ``What is currently the best lattice value for a particular
quantity?" in a way that is readily accessible to those who are not
expert in lattice methods.
This is generally not an easy question to answer;
different collaborations use different lattice actions
(discretizations of QCD) with a variety of lattice spacings and
volumes, and with a range of masses for the $u$- and $d$-quarks. Not
only are the systematic errors different, but also the methodology
used to estimate these uncertainties varies between collaborations. In
the present work, we summarize the main features of each of the
calculations and provide a framework for judging and combining the
different results. Sometimes it is a single result that provides the
``best" value; more often it is a combination of results from
different collaborations. Indeed, the consistency of values obtained
using different formulations adds significantly to our confidence in
the results.

The first four editions of the FLAG review were made public in
2010~\cite{Colangelo:2010et}, 2013~\cite{Aoki:2013ldr},
2016~\cite{Aoki:2016frl}, and 2019~\cite{Aoki:2019cca}
(and will be referred to as FLAG 10, FLAG 13, FLAG 16, and FLAG 19, respectively).
The fourth edition reviewed results related to both light 
($u$-, $d$- and $s$-), and heavy ($c$- and $b$-) flavours.
The quantities related to pion and kaon physics  were 
light-quark masses, the form factor $f_+(0)$
arising in semileptonic $K \rightarrow \pi$ transitions 
(evaluated at zero momentum transfer), 
the decay constants $f_K$ and $f_\pi$, 
the $B_K$ parameter from neutral kaon mixing,
and the kaon mixing matrix elements of new operators that arise in 
theories of physics beyond the Standard Model.
Their implications for
the CKM matrix elements $V_{us}$ and $V_{ud}$ were also discussed.
Furthermore, results
were reported for some of the low-energy constants of $SU(2)_L \times
SU(2)_R$ and $SU(3)_L \times SU(3)_R$ Chiral Perturbation Theory.
The quantities related to $D$- and $B$-meson physics that were
reviewed were the masses of the charm and bottom quarks
together with the decay constants, form factors, and mixing parameters
of $B$- and $D$-mesons.
These are the heavy-light quantities most relevant
to the determination of CKM matrix elements and the global
CKM unitarity-triangle fit. 
The current status of 
lattice results on the QCD coupling  $\alpha_s$ was reviewed.
Last but not least, we reviewed calculations of nucleon matrix elements 
of flavor nonsinglet and singlet bilinear operators, including the nucleon axial charge
$g_A$ and the nucleon sigma term.
These results are relevant for constraining $V_{ud}$, for searches for new physics
in neutron decays and other processes, and for dark matter searches.

In the present paper we provide updated results for all the above-mentioned
quantities, but also extend the scope of the review by adding a section on
scale setting, Sec.~\ref{sec:scalesetting}. 
The motivation for adding this section is that uncertainties in the value of
the lattice spacing $a$ are a major source of error for the calculation of
a wide range of quantities. Thus we felt that a systematic compilation of results,
comparing the different approaches to setting the scale, and summarizing the present status,
would be a useful resource for the lattice community.
An additional update is the inclusion, in Sec.~\ref{sec:Kpipi_amplitudes}, of a brief description
of the status of lattice calculations of $K\to\pi\pi$ decay amplitudes. Although some aspects of these calculations
are not yet at the stage to be included in our averages, they are approaching this stage, and
we felt that, given their phenomenological relevance, a brief review was appropriate.

For the most precisely determined quantities, 
isospin breaking---both from the up-down quark mass difference and from QED---must be included.
A short review of methods used to include QED in lattice-QCD simulations is given
in Sec.~\ref{sec:latticeqed}.
An important issue here is that, in the context of a QED$+$QCD theory,
the separation into QED and QCD contributions to a given physical quantity
is ambiguous---there are several ways of defining such a separation.
This issue is discussed from different viewpoints
in the section on quark masses---see Sec.~\ref{sec:physical point and isospin}---and 
that on scale setting---see Sec.~\ref{sec:scalesetting}.
We stress, however, that the physical observable in QCD$+$QED is defined 
unambiguously. Any ambiguity only  arises because we are trying to separate a well-defined, physical quantity into two unphysical parts that provide useful
information for phenomenology.

Our main results are collected in Tabs.~\ref{tab:summary1}, \ref{tab:summary2}, \ref{tab:summary3}, \ref{tab:summary4} and \ref{tab:summary5}.
As is clear from the tables, for most quantities there are results from ensembles with
different values for $N_f$.  In most cases, there is reasonable agreement among
results with $N_f=2$, $2+1$, and $2+1+1$.  As precision increases, we may
some day be able to distinguish among the different values of $N_f$, in
which case, presumably $2+1+1$ would be the most realistic.  (If isospin
violation is critical, then $1+1+1$ or $1+1+1+1$ might be desired.)
At present, for some quantities the errors in the $N_f=2+1$ results are smaller
than those with $N_f=2+1+1$ (e.g., for $m_c$), while for others
the relative size of the errors is reversed. 
Our suggestion to those using the averages is to take whichever of the
$N_f=2+1$ or $N_f=2+1+1$ results has the smaller error.
We do not recommend using the $N_f=2$ results, except for studies of
the $N_f$-dependence of condensates and $\alpha_s$, as these have an uncontrolled
systematic error coming from quenching the strange quark.

Our plan is to continue providing FLAG updates, in the form of a peer
reviewed paper, roughly on a triennial basis. This effort is
supplemented by our more frequently updated
website \href{http://flag.unibe.ch}{{\tt
http://flag.unibe.ch}} \cite{FLAG:webpage}, where figures as well as pdf-files for
the individual sections can be downloaded. The papers reviewed in the
present edition have appeared before the closing date {\bf 30 April 2021}.\footnote{%
  Working groups were given the option of including papers submitted to {\tt arxiv.org}
  before the closing date but published after this date. This flexibility allows this review to
  be up-to-date at the time of submission. A single paper of this type was included.
  }

\clearpage
\begin{sidewaystable}[ph!]
\vspace{-1cm}
\centering
\begin{tabular}{|l|l||l|l||l|l||l|l||l||l|l|}
\hline
Quantity \rule[-0.2cm]{0cm}{0.6cm}    & \hspace{-1.5mm}Sec.\hspace{-2mm} &$N_f=2+1+1$ & Refs. &  $N_f=2+1$ & Refs. &$N_f=2$ &Refs. \\
\hline \hline
$ m_{ud}$[MeV]&\ref{sec:msmud}&$3.410(43)$&\cite{Bazavov:2018omf,Carrasco:2014cwa}&$3.381(40)$&\cite{Blum:2014tka,Durr:2010vn,Durr:2010aw,McNeile:2010ji,Bazavov:2010yq}&&\\[1mm]
$ m_s   $[MeV]&\ref{sec:msmud}&$93.40(57)$&\cite{Bazavov:2018omf,Lytle:2018evc,Carrasco:2014cwa,Chakraborty:2014aca}&$92.2(1.0)$&\cite{Bazavov:2009fk,Durr:2010vn,Durr:2010aw,McNeile:2010ji,Blum:2014tka}&&\\[1mm]
$ m_s / m_{ud} $&\ref{sec:msovermud}&$27.23(10)$&\cite{Bazavov:2017lyh,Carrasco:2014cwa,Bazavov:2014wgs}&$27.42(12)$&\cite{Bruno:2019vup,Blum:2014tka,Bazavov:2009fk,Durr:2010vn,Durr:2010aw}&&\\[1mm]
$ m_u $[MeV]&\ref{subsec:mumd}&$2.14(8)$&\citep{Giusti:2017dmp,Bazavov:2018omf}&$2.27(9)$&\citep{Fodor:2016bgu}&&\\[1mm]
$ m_d $[MeV]&\ref{subsec:mumd}&$4.70(5)$&\citep{Giusti:2017dmp,Bazavov:2018omf}&$4.67(9)$&\citep{Fodor:2016bgu}&&\\[1mm]
$ {m_u}/{m_d} $&\ref{subsec:mumd}&$0.465(24)$&\citep{Giusti:2017dmp,Basak:2018yzz}&$0.485(19)$&\citep{Fodor:2016bgu}&&\\[1mm]
\hline
$\overline{m}_c(\mbox{3 GeV}) $[GeV]&\ref{sec:mcnf4}&$0.988(11)$&\cite{Carrasco:2014cwa,Chakraborty:2014aca,Alexandrou:2014sha,Bazavov:2018omf,Hatton:2020qhk}&$0.992(5)$&\cite{McNeile:2010ji,Yang:2014sea,Nakayama:2016atf,Petreczky:2019ozv}&&\\[1mm]
$ m_c / m_s $&\ref{sec:mcoverms}&$11.768(34)$&\cite{Chakraborty:2014aca,Carrasco:2014cwa,Bazavov:2018omf}&$11.82(16)$&\cite{Yang:2014sea,Davies:2009ih}&&\\[1mm]
\hline
$\overline{m}_b(\overline{m}_b) $[GeV]&\ref{s:bmass}&$4.203(11)$&\cite{Hatton:2021syc,Colquhoun:2014ica,Bussone:2016iua,Gambino:2017vkx,Bazavov:2018omf}&$4.171(20)$&\cite{McNeile:2010ji}&&\\[1mm]
\hline
$ f_+(0) $&\ref{sec:Direct}&$0.9698(17)$&\cite{Carrasco:2016kpy,Bazavov:2018kjg}&$0.9677(27)$&\cite{Bazavov:2012cd,Boyle:2015hfa}&$0.9560(57)(62)$&\cite{Lubicz:2009ht}\\[1mm]
$ f_{K^\pm} / f_{\pi^\pm}  $&\ref{sec:Direct}&$1.1932(21)$&\cite{Dowdall:2013rya,Carrasco:2014poa,Bazavov:2017lyh,Miller:2020xhy}&$1.1917(37)$&\cite{Follana:2007uv,Bazavov:2010hj,Durr:2010hr,Blum:2014tka,Durr:2016ulb,Bornyakov:2016dzn}&$1.205(18)$&\cite{Blossier:2009bx}\\[1mm]
$ f_{\pi^\pm}$[MeV]&\ref{sec:fKfpi}&&&$130.2(8)$&\cite{Follana:2007uv,Bazavov:2010hj,Blum:2014tka}&&\\[1mm]
$ f_{K^\pm}  $[MeV]&\ref{sec:fKfpi}&$155.7(3)$&\cite{Dowdall:2013rya,Bazavov:2014wgs,Carrasco:2014poa}&$155.7(7)$&\cite{Follana:2007uv,Bazavov:2010hj,Blum:2014tka}&$157.5(2.4)$&\cite{Blossier:2009bx}\\[1mm]
\hline
$ \text{Re}(A_2) $[GeV]&\ref{sec:Kpipi_amplitudes}&&&$1.50(4)(14)\times10^{-8}$&\cite{Blum:2015ywa}&&\\[1mm]
$ \text{Im}(A_2) $[GeV]&\ref{sec:Kpipi_amplitudes}&&&$-8.34(1.03)\times10^{-13}$&\cite{Blum:2015ywa}&&\\[1mm]
$\hat{B}_{K} $&\ref{sec:BK lattice}&$0.717(18)(16)$&\cite{Carrasco:2015pra}&$0.7625(97)$&\cite{Durr:2011ap,Laiho:2011np,Blum:2014tka,Jang:2015sla}&$0.727(22)(12)$&\cite{Bertone:2012cu}\\[1mm]
$ B_2$&\ref{sec:Bi}&$0.46(1)(3)$&\cite{Carrasco:2015pra}&$0.502(14)$&\cite{Jang:2015sla,Garron:2016mva}&$0.47(2)(1)$&\cite{Bertone:2012cu}\\[1mm]
$ B_3$&\ref{sec:Bi}&$0.79(2)(5)$&\cite{Carrasco:2015pra}&$0.766(32)$&\cite{Jang:2015sla,Garron:2016mva}&$0.78(4)(2)$&\cite{Bertone:2012cu}\\[1mm]
$ B_4$&\ref{sec:Bi}&$0.78(2)(4)$&\cite{Carrasco:2015pra}&$0.926(19)$&\cite{Jang:2015sla,Garron:2016mva}&$0.76(2)(2)$&\cite{Bertone:2012cu}\\[1mm]
$ B_5$&\ref{sec:Bi}&$0.49(3)(3)$&\cite{Carrasco:2015pra}&$0.720(38)$&\cite{Jang:2015sla,Garron:2016mva}&$0.58(2)(2)$&\cite{Bertone:2012cu}\\[1mm]
\hline

\end{tabular}\\[0.2cm]


\caption{\label{tab:summary1} Summary of the main results of this review concerning quark 
masses, light-meson decay constants, and hadronic kaon-decay and kaon-mixing parameters.
These are grouped in terms of $\Nf$, the number of dynamical quark flavours in lattice simulations. 
Quark masses are given in the $\msbar$ scheme at running scale  $\mu=2\,\gev$ or as indicated.
BSM bag parameters $B_{2,3,4,5}$ are given in the $\msbar$ scheme at scale $\mu=3\,\gev$.
Further specifications of the quantities are given in the quoted sections.
Results for $N_f=2$ quark masses are unchanged since FLAG~16~\cite{Aoki:2016frl},
and are not included here.
For each result we list the references that enter the FLAG average or estimate,
and we stress again the importance of quoting these original works when referring to
FLAG results. From the entries in this column one
can also read off the number of results that enter our averages for each quantity. We emphasize that these numbers only give a very rough indication of how thoroughly the quantity in question has been explored on the lattice and recommend consulting the detailed tables and figures in the relevant section for more significant information and for explanations on the source of the quoted errors.}
\end{sidewaystable}

\clearpage
\begin{sidewaystable}[ph!]
\vspace{-1cm}
\centering
\begin{tabular}{|l|l||l|l||l|l||l|l||l|l||l|l||l|l|}
\hline
Quantity \rule[-0.2cm]{0cm}{0.6cm}    & \hspace{-1.5mm}Sec.\hspace{-2mm} &$N_f=2+1+1$ & Refs. &  $N_f=2+1$ & Refs. &$N_f=2$ &Refs. \\
\hline \hline
$ f_D $[MeV]&\ref{sec:fD}&$212.0(7)$&\cite{Bazavov:2017lyh,Carrasco:2014poa}&$209.0(2.4)$&\cite{Na:2012iu,Bazavov:2011aa,Boyle:2017jwu}&$208(7)$&\cite{Carrasco:2013zta}\\[1mm]
$ f_{D_s} $[MeV]&\ref{sec:fD}&$249.9(5)$&\cite{Bazavov:2017lyh,Carrasco:2014poa}&$248.0(1.6)$&\cite{Davies:2010ip,Bazavov:2011aa,Boyle:2017jwu,Yang:2014sea}&$246(4)$&\cite{Balasubramanian:2019net,Carrasco:2013zta}\\[1mm]
$ f_{D_s}\over{f_D} $&\ref{sec:fD}&$1.1783(16)$&\cite{Bazavov:2017lyh,Carrasco:2014poa}&$1.174(7)$&\cite{Na:2012iu,Bazavov:2011aa,Boyle:2017jwu}&$1.20(2)$&\cite{Carrasco:2013zta}\\[1mm]
$ f_+^{D\pi}(0)$&\ref{sec:DtoPiK}&$0.612(35)$&\cite{Lubicz:2017syv}&$0.666(29)$&\cite{Na:2011mc}&&\\[1mm]
$ f_+^{DK}(0)  $&\ref{sec:DtoPiK}&$0.7385(44)$&\cite{Lubicz:2017syv,Chakraborty:2021qav}&$0.747(19)$&\cite{Na:2010uf}&&\\[1mm]
$ f_{B} $[MeV]&\ref{sec:fB}&$190.0(1.3)$&\cite{Dowdall:2013tga,Bussone:2016iua,Hughes:2017spc,Bazavov:2017lyh}&$192.0(4.3)$&\cite{Bazavov:2011aa,McNeile:2011ng,Na:2012sp,Aoki:2014nga,Christ:2014uea}&$188(7)$&\cite{Carrasco:2013zta,Bernardoni:2014fva}\\[1mm]
$ f_{B_{s}} $[MeV]&\ref{sec:fB}&$230.3(1.3)$&\cite{Dowdall:2013tga,Bussone:2016iua,Hughes:2017spc,Bazavov:2017lyh}&$228.4(3.7)$&\cite{Bazavov:2011aa,McNeile:2011ng,Na:2012sp,Aoki:2014nga,Christ:2014uea}&$225.3(6.6)$&\cite{Carrasco:2013zta,Bernardoni:2014fva,Balasubramanian:2019net}\\[1mm]
$ f_{B_{s}}\over{f_B} $&\ref{sec:fB}&$1.209(5)$&\cite{Dowdall:2013tga,Bussone:2016iua,Hughes:2017spc,Bazavov:2017lyh}&$1.201(16)$&\cite{Bazavov:2011aa,McNeile:2011ng,Na:2012sp,Aoki:2014nga,Christ:2014uea,Boyle:2018knm}&$1.206(23)$&\cite{Carrasco:2013zta,Bernardoni:2014fva}\\[1mm]
$ f_{B_d}\sqrt{\hat{B}_{b_d}}$[MeV]&\ref{sec:BMix}&$210.6(5.5)$&\cite{Dowdall:2019bea}&$225(9)$&\cite{Gamiz:2009ku,Aoki:2014nga,Bazavov:2016nty}&$216(10)$&\cite{Carrasco:2013zta}\\[1mm]
$ f_{B_s}\sqrt{\hat{B}_{B_s}}$[MeV]&\ref{sec:BMix}&$256.1(5.7)$&\cite{Dowdall:2019bea}&$274(8)$&\cite{Gamiz:2009ku,Aoki:2014nga,Bazavov:2016nty}&$262(10)$&\cite{Carrasco:2013zta}\\[1mm]
$ \hat{B}_{B_d}$&\ref{sec:BMix}&$1.222(61)$&\cite{Dowdall:2019bea}&$1.30(10)$&\cite{Gamiz:2009ku,Aoki:2014nga,Bazavov:2016nty}&$1.30(6)$&\cite{Carrasco:2013zta}\\[1mm]
$ \hat{B}_{B_s}$&\ref{sec:BMix}&$1.232(53)$&\cite{Dowdall:2019bea}&$1.35(6)$&\cite{Gamiz:2009ku,Aoki:2014nga,Bazavov:2016nty}&$1.32(5)$&\cite{Carrasco:2013zta}\\[1mm]
$ \xi $&\ref{sec:BMix}&$1.216(16)$&\cite{Dowdall:2019bea}&$1.206(17)$&\cite{Aoki:2014nga,Bazavov:2016nty}&$1.225(31)$&\cite{Carrasco:2013zta}\\[1mm]
$ B_{B_s}/B_{B_d}  $&\ref{sec:BMix}&$1.008(25)$&\cite{Dowdall:2019bea}&$1.032(38)$&\cite{Aoki:2014nga,Bazavov:2016nty}&$1.007(21)$&\cite{Carrasco:2013zta}\\[1mm]

\hline
Quantity \rule[-0.2cm]{0cm}{0.6cm}    & \hspace{-1.5mm}Sec.\hspace{-2mm} &\multicolumn{3}{c|}{$N_f=2+1$ and $N_f=2+1+1$} & Refs. & & \\
\hline
$ \alpha_{\overline{\rm MS}}^{(5)}(M_Z) $&\ref{s:alpsumm}&\multicolumn{3}{c|}{$0.1184(8)$}&\cite{Ayala:2020odx,Bazavov:2019qoo,Cali:2020hrj,Bruno:2017gxd,Chakraborty:2014aca,McNeile:2010ji,Aoki:2009tf,Maltman:2008bx}&&\\[1mm]
$ \Lambda_{\overline{\rm MS}}^{(5)} $[MeV]&\ref{s:alpsumm}&\multicolumn{3}{c|}{$214(10)$}&\cite{Ayala:2020odx,Bazavov:2019qoo,Cali:2020hrj,Bruno:2017gxd,Chakraborty:2014aca,McNeile:2010ji,Aoki:2009tf,Maltman:2008bx}&&\\[1mm]
$ \Lambda_{\overline{\rm MS}}^{(4)} $[MeV]&\ref{s:alpsumm}&\multicolumn{3}{c|}{$297(12)$}&\cite{Ayala:2020odx,Bazavov:2019qoo,Cali:2020hrj,Bruno:2017gxd,Chakraborty:2014aca,McNeile:2010ji,Aoki:2009tf,Maltman:2008bx}&&\\[1mm]
$ \Lambda_{\overline{\rm MS}}^{(3)} $[MeV]&\ref{s:alpsumm}&\multicolumn{3}{c|}{$339(12)$}&\cite{Ayala:2020odx,Bazavov:2019qoo,Cali:2020hrj,Bruno:2017gxd,Chakraborty:2014aca,McNeile:2010ji,Aoki:2009tf,Maltman:2008bx}&&\\[1mm]
\hline

\end{tabular}
\caption{\label{tab:summary2}Summary of the main results of this review concerning heavy-light 
mesons and the strong coupling constant. These are grouped in terms of $\Nf$, the number of dynamical quark flavours in lattice simulations.   The  quantities listed are specified in the quoted sections.
For each result we list the references that enter the FLAG average or estimate,
and we stress again the importance of quoting these original works when referring to
FLAG results.
From the entries in this column one
can also read off the number of results that enter our averages for each quantity. We emphasize that these numbers only give a very rough indication of how thoroughly the quantity in question has been explored on the lattice and recommend consulting the detailed tables and figures in the relevant section for more significant information and for explanations on the source of the quoted errors. 
}
\end{sidewaystable}
\clearpage

\begin{sidewaystable}[ph!]
\vspace{-1cm}
\centering
\begin{tabular}{|l|l||l|l||l|l||l|l||l|l||l|l||l|l|}
\hline
Quantity \rule[-0.2cm]{0cm}{0.6cm}    & \hspace{-1.5mm}Sec.\hspace{-2mm} &$N_f=2+1+1$ & Refs. &  $N_f=2+1$ & Refs. &$N_f=2$ &Refs. \\
\hline \hline
$ \Sigma^{1/3}$[MeV]&\ref{sec:SU2_averages}&$286(23)$&\cite{Cichy:2013gja,Alexandrou:2017bzk}&$272(5)$&\cite{Bazavov:2010yq,Borsanyi:2012zv,Durr:2013goa,Boyle:2015exm,Cossu:2016eqs,Aoki:2017paw}&$266(10)$&\cite{Baron:2009wt,Cichy:2013gja,Brandt:2013dua,Engel:2014eea}\\[1mm]
$ {\Fpi}/{F}$&\ref{sec:SU2_averages}&$1.077(3)$&\cite{Baron:2011sf}&$1.062(7)$&\cite{Bazavov:2010hj,Beane:2011zm,Borsanyi:2012zv,Durr:2013goa,Boyle:2015exm}&$1.073(15)$&\cite{Frezzotti:2008dr,Baron:2009wt,Brandt:2013dua,Engel:2014eea}\\[1mm]
$ \lbar_3$&\ref{sec:SU2_averages}&$3.53(26)$&\cite{Baron:2011sf}&$3.07(64)$&\cite{Bazavov:2010hj,Beane:2011zm,Borsanyi:2012zv,Durr:2013goa,Boyle:2015exm}&$3.41(82)$&\cite{Frezzotti:2008dr,Baron:2009wt,Brandt:2013dua}\\[1mm]
$ \lbar_4$&\ref{sec:SU2_averages}&$4.73(10)$&\cite{Baron:2011sf}&$4.02(45)$&\cite{Bazavov:2010hj,Beane:2011zm,Borsanyi:2012zv,Durr:2013goa,Boyle:2015exm}&$4.40(28)$&\cite{Frezzotti:2008dr,Baron:2009wt,Brandt:2013dua,Gulpers:2015bba}\\[1mm]
$ \lbar_6$&\ref{sec:SU2_averages}&&&&&$15.1(1.2)$&\cite{Frezzotti:2008dr,Brandt:2013dua}\\[1mm]
$ a_0^2\Mpi$&\ref{sec:SU2_averages}&$-0.0441(4)$&\cite{,Helmes:2015gla}&&&$-0.04385(47)$&\cite{Feng:2009ij}\\[1mm]
$ \Sigma_0^{1/3}      $[MeV]&\ref{sec:SU3_averages}&&&$245(8)$&\cite{Bazavov:2009fk}&&\\[1mm]
$ \Sigma/\Sigma_0     $&\ref{sec:SU3_averages}&&&$1.48(16)$&\cite{Bazavov:2009fk}&&\\[1mm]
$ F_0                 $[MeV]&\ref{sec:SU3_averages}&&&$80.3(6.0)$&\cite{Bazavov:2010hj}&&\\[1mm]
$ F/F_0               $&\ref{sec:SU3_averages}&&&$1.104(41)$&\cite{Bazavov:2009fk}&&\\[1mm]
$ B/B_0               $&\ref{sec:SU3_averages}&&&$1.21(7)$&\cite{Bazavov:2009fk}&&\\[1mm]
$ L_4 $&\ref{sec:SU3_averages}&$+0.09(34)\times10^{-3}$&\cite{Dowdall:2013rya}&$-0.02(56)\times10^{-3}$&\cite{Bazavov:2010hj}&&\\[1mm]
$ L_5 $&\ref{sec:SU3_averages}&$+1.19(25)\times10^{-3}$&\cite{Dowdall:2013rya}&$+0.95(41)\times10^{-3}$&\cite{Bazavov:2010hj}&&\\[1mm]
$ L_6 $&\ref{sec:SU3_averages}&$+0.16(20)\times10^{-3}$&\cite{Dowdall:2013rya}&$+0.01(34)\times10^{-3}$&\cite{Bazavov:2010hj}&&\\[1mm]
$ L_8 $&\ref{sec:SU3_averages}&$+0.55(15)\times10^{-3}$&\cite{Dowdall:2013rya}&$+0.43(28)\times10^{-3}$&\cite{Bazavov:2010hj}&&\\[1mm]
$ a_0^{1/2}\mu_{\pi K}  $&\ref{sec:SU3_averages}&$0.127(2)$&\cite{Helmes:2018nug}&&&&\\[1mm]
$ a_0^{3/2}\mu_{\pi K}  $&\ref{sec:SU3_averages}&$-0.0463(17)$&\cite{Helmes:2018nug}&&&&\\[1mm]
$ a_0^1 M_K             $&\ref{sec:SU3_averages}&$-0.388(20)$&\cite{Helmes:2017smr}&&&&\\[1mm]
\hline
\hline

\end{tabular}
\caption{\label{tab:summary3}Summary of the main results of this review concerning LECs, grouped in terms of $\Nf$, the number of dynamical quark flavours in lattice simulations.   The  quantities listed are specified in the quoted sections.
For each result we list the references that enter the FLAG average or estimate,
and we stress again the importance of quoting these original works when referring to
FLAG results.
From the entries in this column one
can also read off the number of results that enter our averages for each quantity. We emphasize that these numbers only give a very rough indication of how thoroughly the quantity in question has been explored on the lattice and recommend consulting the detailed tables and figures in the relevant section for more significant information and for explanations on the source of the quoted errors. 
}

\end{sidewaystable}
\clearpage
\begin{sidewaystable}[ph!]
\vspace{-1cm}
\centering
\begin{tabular}{|l|l||l|l||l|l||l|l||l|l||l|l||l|l|}
\hline
Quantity \rule[-0.2cm]{0cm}{0.6cm}    & \hspace{-1.5mm}Sec.\hspace{-2mm} &$N_f=2+1+1$ & Refs. &  $N_f=2+1$ & Refs. &$N_f=2$ &Refs. \\
\hline \hline
$ g_A^{u-d} $&\ref{sec:gA-IV}&$1.246(28)$&\cite{Gupta:2018qil,Chang:2018uxx,Walker-Loud:2019cif}&$1.248(23)$&\cite{Liang:2018pis,Harris:2019bih}&&\\[1mm]
$ g_S^{u-d} $&\ref{sec:gS-IV}&$1.02(10)$&\cite{Gupta:2018qil}&$1.13(14)$&\cite{Harris:2019bih}&&\\[1mm]
$ g_T^{u-d} $&\ref{sec:gT-IV}&$0.989(34)$&\cite{Gupta:2018qil}&$0.965(61)$&\cite{Harris:2019bih}&&\\[1mm]
$ g_A^u  $&\ref{sec:gA-FD}&$\phantom{-}0.777(25)(30)$&\cite{Lin:2018obj}&$\phantom{-}0.847(18)(32)$&\cite{Liang:2018pis}&&\\[1mm]
$ g_A^d  $&\ref{sec:gA-FD}&$-0.438(18)(30)$&\cite{Lin:2018obj}&$-0.407(16)(18)$&\cite{Liang:2018pis}&&\\[1mm]
$ g_A^s  $&\ref{sec:gA-FD}&$-0.053(8)$&\cite{Lin:2018obj}&$-0.035(6)(7)$&\cite{Liang:2018pis}&&\\[1mm]
$ \sigma_{\pi N} $[MeV]&\ref{sec:gS-sum}&$64.9(1.5)(13.2)$&\cite{Alexandrou:2014sha}&$	39.7(3.6)$&\cite{Durr:2011mp,Durr:2015dna,Yang:2015uis}&$37(8)(6)$&\cite{Bali:2012qs}\\[1mm]
$ \sigma_{s} $[MeV]&\ref{sec:gS-sum}&$41.0(8.8)$&\cite{Freeman:2012ry}&$52.9(7.0)$&\cite{Durr:2011mp,Freeman:2012ry,Junnarkar:2013ac,Durr:2015dna,Yang:2015uis}&&\\[1mm]
$ g_T^u  $&\ref{sec:gT-FD}&$\phantom{-}0.784(28)(10)$&\cite{Gupta:2018lvp}&&&&\\[1mm]
$ g_T^d  $&\ref{sec:gT-FD}&$-0.204(11)(10)$&\cite{Gupta:2018lvp}&&&&\\[1mm]
$ g_T^s  $&\ref{sec:gT-FD}&$-0.0027(16)$&\cite{Gupta:2018lvp}&&&&\\[1mm]
\hline
\hline

\end{tabular}
\caption{\label{tab:summary4}Summary of the main results of this review concerning nuclear matrix elements, grouped in terms of $\Nf$, the number of dynamical quark flavours in lattice simulations.   The  quantities listed are specified in the quoted sections.
For each result we list the references that enter the FLAG average or estimate,
and we stress again the importance of quoting these original works when referring to
FLAG results.
From the entries in this column one
can also read off the number of results that enter our averages for each quantity. We emphasize that these numbers only give a very rough indication of how thoroughly the quantity in question has been explored on the lattice and recommend consulting the detailed tables and figures in the relevant section for more significant information and for explanations on the source of the quoted errors. 
}

\end{sidewaystable}
\clearpage
\begin{sidewaystable}[ph!]
\footnotesize
\vspace{-1cm}
\centering
\begin{tabular}{|l|l||l|l||l|l||l|l||l|l||l}
\hline
Quantity \rule[-0.2cm]{0cm}{0.6cm}    & \hspace{-1.5mm}Sec.\hspace{-2mm} &$N_f=1+1+1+1$ &Refs. &$N_f=2+1+1$ & Refs. &  $N_f=2+1$ & Refs. &$N_f>2+1$ &Refs. \\
\hline \hline
$ \sqrt{t_0} $[fm]&\ref{s:flowscales}&&&$0.14186(88)$&\cite{Miller:2020evg,Bazavov:2015yea,Dowdall:2013rya}&$0.14464(87)$&\cite{Bruno:2016plf,Blum:2014tka,Borsanyi:2012zs}&&\\[1mm]
$ w_0 $[fm]&\ref{s:flowscales}&$0.17236(70)$&\cite{Borsanyi:2020mff}&$0.17128(107)$&\cite{Miller:2020evg,Bazavov:2015yea,Dowdall:2013rya}&$0.17355(92)$&\cite{Blum:2014tka,Bazavov:2014pvz,Borsanyi:2012zs}&$0.17177(67)$&\cite{Miller:2020evg,Borsanyi:2020mff,Bazavov:2015yea,Dowdall:2013rya}\\[1mm]
$ r_0 $[fm]&\ref{s:flowscales}&&&$0.474(14)$&\cite{Carrasco:2014cwa}&$0.4701(36)$&\cite{Bazavov:2014pvz,Yang:2014sea,Aoki:2010dy,Gray:2005ur,Aubin:2004wf}&&\\[1mm]
$ r_1 $[fm]&\ref{s:flowscales}&&&$0.3112(30)$&\cite{Dowdall:2013rya}&$0.3127(30)$&\cite{Aoki:2010dy,Bazavov:2010hj,Davies:2009tsa,Gray:2005ur,Aubin:2004wf}&&\\[1mm]
\hline
\hline

\end{tabular}
\caption{\label{tab:summary5}Summary of the main results of this review concerning setting of the lattice scale, grouped in terms of $\Nf$, 
the number of dynamical quark flavours in lattice simulations.   The  quantities listed are specified in the quoted sections.
For each result we list the references that enter the FLAG average or estimate,
and we stress again the importance of quoting these original works when referring to
FLAG results.
From the entries in this column one
can also read off the number of results that enter our averages for each quantity. We emphasize that these numbers only give a very rough indication of how thoroughly the quantity in question has been explored on the lattice and recommend consulting the detailed tables and figures in the relevant section for more significant information and for explanations on the source of the quoted errors. 
}

\end{sidewaystable}
\clearpage



This review is organized as follows.  In the remainder of
Sec.~\ref{sec:introduction} we summarize the composition and rules of
FLAG and discuss general issues that arise in modern lattice
calculations.  In Sec.~\ref{sec:qualcrit}, we explain our general
methodology for evaluating the robustness of lattice results.  We also
describe the procedures followed for combining results from different
collaborations in a single average or estimate (see
Sec.~\ref{sec:averages} for our definition of these terms). The rest
of the paper consists of sections, each dedicated to a set of
closely connected physical quantities, 
or, for the final section, to the determination of the lattice scale.
Each of these
sections is accompanied by an Appendix with explicatory notes.\footnote{%
In some cases, in order to keep the length of this review within reasonable bounds,
we have dropped these notes for older data, since they can be found in 
previous FLAG reviews~\cite{Colangelo:2010et,Aoki:2013ldr,Aoki:2016frl,Aoki:2019cca} .}

In previous editions, we have provided, in an appendix, a glossary summarizing
some standard lattice terminology and
describing the most commonly used lattice techniques and methodologies.
Since no significant updates in this information have occurred since our
previous edition, we have decided, in the interests of reducing the length
of this review, to omit this glossary, 
and refer the reader to FLAG 19 for this information~\cite{Aoki:2019cca}.
This appendix also contained, in previous versions, 
a tabulation of the actions used in the papers that were reviewed.
Since this information is available in the discussions in the separate sections,
and is time-consuming to collect from the sections, we have dropped these tables.
We have, however, kept a short appendix, Appendix~\ref{sec:zparam}, describing the
parameterizations of semileptonic form factors that are used in Sec.~\ref{sec:BDecays}.
Moreover, in Appendix~\ref{app:acronyms}, we have added a summary and explanations of acronyms introduced in the manuscript. Collaborations referred to by an acronym can be identified through the corresponding bibliographic reference.

\subsection{FLAG composition, guidelines and rules}

FLAG strives to be representative of the lattice community, both in
terms of the geographical location of its members and the lattice
collaborations to which they belong. We aspire to provide the nuclear- and
particle-physics communities with a single source of reliable
information on lattice results.

In order to work reliably and efficiently, we have adopted a formal
structure and a set of rules by which all FLAG members abide.  The
collaboration presently consists of an Advisory Board (AB), an
Editorial Board (EB), and nine Working Groups (WG). The r\^{o}le of
the Advisory Board is to provide oversight of the content, procedures, schedule
and membership of FLAG, to help resolve disputes, to serve as a source
of advice to the EB and to FLAG as a whole, and to provide a
critical assessment of drafts.
They also give their approval of the final version of the preprint before
it is rendered public. The Editorial Board coordinates the activities
of FLAG, sets priorities and intermediate deadlines, organizes votes on
FLAG procedures, writes the introductory sections, and takes care of
the editorial work needed to amalgamate the sections written by the
individual working groups into a uniform and coherent review. The
working groups concentrate on writing the review of the physical
quantities for which they are responsible, which is subsequently
circulated to the whole collaboration for critical evaluation.

The current list of FLAG members and their Working Group assignments is:
\begin{itemize}
\item
Advisory Board (AB):\hfill
G.~Colangelo, M.~Golterman, P.~Hernandez, T.~Onogi, \\
\hbox{} \hfill and R.~Van de Water
\item
Editorial Board (EB):\hfill
S.~Gottlieb, A.~J\"uttner, S.~Hashimoto, S.R.~Sharpe, \\
\hbox{} \hfill and U.~Wenger
\item
Working Groups (coordinator listed first):
\begin{itemize}
\item Quark masses \hfill T.~Blum, A.~Portelli, and A.~Ramos
\item $V_{us},V_{ud}$ \hfill T.~Kaneko, J.~N.~Simone, S.~Simula, and N.~Tantalo
\item LEC \hfill S.~D\"urr, H.~Fukaya, and U.M.~Heller
\item $B_K$ \hfill P.~Dimopoulos, X.~Feng, and G.~Herdoiza
\item $f_{B_{(s)}}$, $f_{D_{(s)}}$, $B_B$ \hfill Y.~Aoki, M.~Della Morte, and C. Monahan
\item 
$b$ and $c$ semileptonic and radiative decays
\hfill E.~Lunghi, S.~Meinel, and C.~Pena
\item $\alpha_s$ \hfill S.~Sint, R.~Horsley, and P.~Petreczky
\item NME \hfill R.~Gupta, S.~Collins, A.~Nicholson, and
 H.~Wittig
 \item Scale setting \hfill R.~Sommer, N.~Tantalo, and U.~Wenger
\end{itemize}
\end{itemize}

The most important FLAG guidelines and rules are the following:
\begin{itemize}
\item
the composition of the AB reflects the main geographical areas in
which lattice collaborations are active, with members from
America, Asia/Oceania, and Europe;
\item
the mandate of regular members is not limited in time, but we expect that a
certain turnover will occur naturally;
\item
whenever a replacement becomes necessary this has to keep, and
possibly improve, the balance in FLAG, so that different collaborations, from
different geographical areas are represented;
\item
in all working groups the 
members must belong to 
different lattice collaborations;
\item
a paper is in general not reviewed (nor colour-coded, as described in
the next section) by any of its authors;
\item
lattice collaborations 
will be consulted on the colour coding
of their calculation;
\item
there are also internal rules regulating our work, such as voting procedures.
\end{itemize}
 
As for FLAG 19, for this review we sought the advice of external reviewers
once a complete draft of the review was available. For each review section, we
have asked one lattice expert (who could be a FLAG alumnus/alumna) and
one nonlattice phenomenologist for a critical assessment. 
The one exception is the scale-setting section, where only a lattice
expert has been asked to provide input.
This is similar
to the procedure followed by the Particle Data Group in the creation of the
Review of Particle Physics.  The reviewers provide comments and feedback on
scientific and stylistic matters. They are not anonymous, and enter into a discussion with
the authors of the WG. Our aim with this additional step is to make sure that a wider
array of viewpoints enter into the discussions, 
so as to make this review more useful for its intended audience.

\subsection{Citation policy}
We draw attention to this particularly important point.  As stated
above, our aim is to make lattice-QCD results easily accessible to
those without lattice expertise,
and we are well aware that it is likely that some
readers will only consult the present paper and not the original
lattice literature. It is very important that this paper not be the
only one cited when our results are quoted. We strongly suggest that
readers also cite the original sources. In order to facilitate this,
in Tabs.~\ref{tab:summary1}, \ref{tab:summary2}, \ref{tab:summary3}, \ref{tab:summary4}, and \ref{tab:summary5}, besides
summarizing the main results of the present review, we also cite the
original references from which they have been obtained. In addition,
for each figure we make a bibtex file available on our webpage
\cite{FLAG:webpage} which contains the bibtex entries of all the
calculations contributing to the FLAG average or estimate. The
bibliography at the end of this paper should also make it easy to cite
additional papers. Indeed, we hope that the bibliography will be one of
the most widely used elements of the whole paper.

\subsection{General issues}

Several general issues concerning the present review are thoroughly
discussed in Sec.~1.1 of our initial 2010 paper~\cite{Colangelo:2010et},
and we encourage the reader to consult the relevant pages. In the
remainder of the present subsection, we focus on a few important
points. Though the discussion has been duly updated, it is similar
to that of Sec.~1.2  in the previous three reviews~\cite{Aoki:2013ldr,Aoki:2016frl,Aoki:2019cca}.

The present review aims to achieve two distinct goals:
first, to provide a {\bf description} of the relevant work done on the lattice;
and, second,
to draw {\bf conclusions} on the basis of that work,  summarizing
the results obtained for the various quantities of physical interest.

The core of the information about the work done on the lattice is
presented in the form of tables, which not only list the various
results, but also describe the quality of the data that underlie
them. We consider it important that this part of the review represents
a generally accepted description of the work done. For this reason, we
explicitly specify the quality requirements
used and provide sufficient details in appendices so that the reader
can verify the information given in the tables.\footnote{%
We also use terms
like ``quality criteria", ``rating", ``colour coding", etc., when referring to
the classification of results, as described in Sec.~\ref{sec:qualcrit}.}

On the other hand, the conclusions drawn 
on the basis of the available lattice results
are the responsibility of FLAG alone. Preferring to
err on the side of caution, in several cases we draw
conclusions that are more conservative than those resulting from
a plain weighted average of the available lattice results. This cautious
approach is usually adopted when the average is
dominated by a single lattice result, or when
only one lattice result is available for a given quantity. In such
cases, one does not have the same degree of confidence in results and
errors as when there is agreement among several different
calculations using different approaches. The reader should keep
in mind that the degree of confidence cannot be quantified, and
it is not reflected in the quoted errors. 

Each discretization has its merits, but also its shortcomings. For most
topics covered in this review we
have an increasingly broad database, and for most quantities
lattice calculations based on totally different discretizations are
now available. This is illustrated by the dense population of the
tables and figures in most parts of this review. Those
calculations that do satisfy our quality criteria indeed lead, in almost all cases, to
consistent results, confirming universality within the accuracy
reached. The consistency between independent lattice
results, obtained with different discretizations, methods, and
simulation parameters, is an important test of lattice QCD, and
observing such consistency also provides further evidence that
systematic errors are fully under control.

In the sections dealing with heavy quarks and with $\alpha_s$, the
situation is not the same. Since the $b$-quark mass can barely be resolved
with current lattice spacings, most lattice methods for treating $b$
quarks use effective field theory at some level. This introduces
additional complications not present in the light-quark sector.  An
overview of the issues specific to heavy-quark quantities is given in
the introduction of Sec.~\ref{sec:BDecays}. For $B$- and $D$-meson
leptonic decay constants, there already exists a good number of
different independent calculations that use different heavy-quark
methods, but there are only a few independent calculations of
semileptonic $B$, $\Lambda_b$, and $D$ form factors and of $B-\bar B$ mixing
parameters. 
For $\alpha_s$, most lattice methods involve a range of
scales that need to be resolved and controlling the systematic error
over a large range of scales is more demanding. The issues specific to
determinations of the strong coupling are summarized in Sec.~\ref{sec:alpha_s}.
\smallskip
\\{\it Number of sea quarks in lattice simulations:}\\
\noindent
Lattice-QCD simulations currently involve two, three or four flavours of
dynamical quarks. Most simulations set
the masses of the two lightest quarks to be equal, while the
strange and charm quarks, if present, are heavier
(and tuned to lie close to their respective physical values). 
Our notation for these simulations indicates which quarks
are nondegenerate, e.g., 
$\Nf=2+1$ if $m_u=m_d < m_s$ and $\Nf =2+1+1$ if $m_u=m_d < m_s < m_c$. 
Calculations with $\Nf =2$, i.e., two degenerate dynamical
flavours, often include strange valence quarks interacting with gluons,
so that bound states with the quantum numbers of the kaons can be
studied, albeit neglecting strange sea-quark fluctuations.  The
quenched approximation ($N_f=0$), in which all sea-quark contributions 
are omitted, has uncontrolled systematic errors and
is no longer used in modern lattice simulations with relevance to phenomenology.
Accordingly, we will review results obtained with $N_f=2$, $N_f=2+1$,
and $N_f = 2+1+1$, but omit earlier results with $N_f=0$. 
The only exception concerns the QCD coupling constant $\alpha_s$.
Since this observable does not require valence light quarks,
it is theoretically well defined also in the $N_f=0$ theory,
which is simply pure gluodynamics.
The $N_f$-dependence of $\alpha_s$, 
or more precisely of the related quantity $r_0 \Lambda_\msbar$, 
is a theoretical issue of considerable interest; here $r_0$ is a quantity
with the dimension of length that sets the physical scale, as discussed in
 Sec.~\ref{sec:scalesetting}.
We stress, however, that only results with $N_f \ge 3$ 
are used to determine the physical value of $\alpha_s$ at a high scale.
\smallskip
\\{\it Lattice actions, simulation parameters, and scale setting:}\\
\noindent
The remarkable progress in the precision of lattice
calculations is due to improved algorithms, better computing resources,
and, last but not least, conceptual developments.
Examples of the latter are improved
actions that reduce lattice artifacts and actions that preserve
chiral symmetry to very good approximation.
A concise characterization of
the various discretizations that underlie the results reported in the
present review is given in 
Appendix~A.1 of FLAG 19.

Physical quantities are computed in lattice simulations in units of the
lattice spacing so that they are dimensionless.
For example, the pion decay constant that is obtained from a simulation
is $f_\pi a$, where $a$ is the spacing between two neighboring lattice sites.
(All simulations with results quoted in this review use hypercubic lattices,
i.e., with the same spacing in all four Euclidean directions.)
To convert these results to physical units requires knowledge
of the lattice spacing $a$ at the fixed values of the bare QCD parameters
(quark masses and gauge coupling) used in the simulation.
This is achieved by requiring agreement between
the lattice calculation and experimental measurement of a known
quantity, which thus ``sets the scale" of a given simulation.
Given the central importance of this procedure,
we include in this edition of FLAG a dedicated section, Sec.~\ref{sec:scalesetting},
discussing the issues and results. 
\smallskip
\\{\it Renormalization and scheme dependence:}\\
\noindent
Several of the results covered by this review, such as quark masses,
the gauge coupling, and $B$-parameters, are for quantities defined in a
given renormalization scheme and at a specific renormalization scale. 
The schemes employed (e.g., regularization-independent MOM schemes) are often
chosen because of their specific merits when combined with the lattice
regularization. For a brief discussion of their properties, see
Appendix A.3 of FLAG 19.
The conversion of the results obtained in
these so-called intermediate schemes to more familiar regularization
schemes, such as the $\msbar$-scheme, is done with the aid of
perturbation theory. It must be stressed that the renormalization
scales accessible in simulations are limited, because of the presence
of an ultraviolet (UV) cutoff of $\sim \pi/a$.
To safely match to $\msbar$, a scheme defined in perturbation theory,
Renormalization Group (RG) running to higher scales is performed,
either perturbatively or nonperturbatively (the latter using
finite-size scaling techniques).
\smallskip
\\{\it Extrapolations:}\\
\noindent
Because of limited computing resources, lattice simulations are often
performed at unphysically heavy pion masses, although results at the
physical point have become increasingly common. Further, numerical
simulations must be done at nonzero lattice spacing, and in a finite
(four-dimensional) volume.  In order to obtain physical results,
lattice data are obtained at a sequence of pion masses and a sequence
of lattice spacings, and then extrapolated to the physical pion mass
and to the continuum limit.  In principle, an extrapolation to
infinite volume is also required. However, for most quantities
discussed in this review, finite-volume effects are exponentially
small in the linear extent of the lattice in units of the pion mass,
and, in practice, one often verifies volume independence by comparing
results obtained on a few different physical volumes, holding other
parameters fixed. To control the associated systematic uncertainties,
these extrapolations are guided by effective theories.  For
light-quark actions, the lattice-spacing dependence is described by
Symanzik's effective theory~\cite{Symanzik:1983dc,Symanzik:1983gh};
for heavy quarks, this can be extended and/or supplemented by other
effective theories such as Heavy-Quark Effective Theory (HQET).  The
pion-mass dependence can be parameterized with Chiral Perturbation
Theory ($\chi$PT), which takes into account the Nambu-Goldstone nature
of the lowest excitations that occur in the presence of light
quarks. Similarly, one can use Heavy-Light Meson Chiral Perturbation
Theory (HM$\chi$PT) to extrapolate quantities involving mesons
composed of one heavy ($b$ or $c$) and one light quark.  One can
combine Symanzik's effective theory with $\chi$PT to simultaneously
extrapolate to the physical pion mass and the continuum; in this case,
the form of the effective theory depends on the discretization.  See
Appendix A.4 of FLAG 19
for a brief description of the different
variants in use and some useful references.  Finally, $\chi$PT can
also be used to estimate the size of finite-volume effects measured in
units of the inverse pion mass, thus providing information on the
systematic error due to finite-volume effects in addition to that
obtained by comparing simulations at different volumes.
\smallskip
\\{\it Excited-state contamination:}\\
\noindent
In all the hadronic matrix elements discussed in this review, the hadron in question
is the lightest state with the chosen quantum numbers. This implies that it dominates the
required correlation functions as their extent in Euclidean time is increased. Excited-state
contributions are suppressed by $e^{-\Delta E \Delta \tau}$, 
where $\Delta E$ is the gap between
the ground and excited states, and $\Delta \tau$ the relevant separation in Euclidean time. 
The size of $\Delta E$ depends on the hadron in question, and in general
is a multiple of the pion mass. In practice, as discussed at length in Sec.~\ref{sec:NME},
the contamination of signals due to  excited-state contributions is a much more
challenging problem for baryons than for the other particles discussed here.
This is in part due to the fact that the signal-to-noise ratio drops exponentially for
baryons, which reduces the values of $\Delta \tau$ that can be used.
\smallskip
\\{\it Critical slowing down:}\\
\noindent
The lattice spacings reached in recent simulations go down to 0.05 fm
or even smaller. In this regime, long autocorrelation times slow down
the sampling of the
configurations~\cite{Antonio:2008zz,Bazavov:2010xr,Schaefer:2010hu,Luscher:2010we,Schaefer:2010qh,Chowdhury:2013mea,Brower:2014bqa,Fukaya:2015ara,DelDebbio:2002xa,Bernard:2003gq}.
Many groups check for autocorrelations in a number of observables,
including the topological charge, for which a rapid growth of the
autocorrelation time is observed with decreasing lattice spacing.
This is often referred to as topological freezing. A solution to the
problem consists in using open boundary conditions in time~\cite{Luscher:2011kk}, 
instead of the more common antiperiodic ones. More recently,
two other approaches have been proposed, one based on a multiscale
thermalization algorithm \cite{Endres:2015yca,Detmold:2018zgk} and another based on
defining QCD on a nonorientable manifold \cite{Mages:2015scv}.  The
problem is also touched upon in Sec.~\ref{s:crit}, where it is
stressed that attention must be paid to this issue. While large scale
simulations with open boundary conditions are already far advanced
\cite{Bruno:2014jqa}, 
few results reviewed here have been obtained with any of the above methods.
It is usually {\it  assumed} that the continuum limit can be reached by extrapolation
from the existing simulations, and that potential systematic errors due
to the long autocorrelation times have been adequately controlled.
Partially or completely frozen topology would produce a mixture of different $\theta$ vacua, and 
the difference from the desired $\theta=0$ result
may be estimated in some cases using  
 chiral perturbation theory, which gives predictions for the $\theta$-dependence of the 
physical quantity of interest \cite{Brower:2003yx,Aoki:2007ka}. These ideas have been systematically and successfully tested in various models in \cite{Bautista:2015yza,Bietenholz:2016ymo}, and a numerical test on MILC ensembles indicates that the topology dependence 
for some of the physical quantities reviewed here is small, consistent with theoretical 
expectations~\cite{Bernard:2017npd}.
\smallskip
\\ {\it Simulation algorithms and numerical errors:}\\
\noindent
Most of the modern lattice-QCD simulations use exact algorithms such 
as those of Refs.~\cite{Duane:1987de,Clark:2006wp}, which do not produce any systematic errors when exact 
arithmetic is available. In reality, one uses numerical calculations at 
double (or in some cases even single) precision, and some errors are 
unavoidable. More importantly, the inversion of the Dirac operator is 
carried out iteratively and it is truncated once some accuracy is 
reached, which is another source of potential systematic error. In most 
cases, these errors have been confirmed to be much less than the 
statistical errors. In the following we assume that this source of error 
is negligible. 
Some of the most recent simulations use an inexact algorithm in order to 
speed up the computation, though it may produce systematic effects. 
Currently available tests indicate that errors from the use of inexact
algorithms are under control~\cite{Bazavov:2012xda}.


\section{Quality criteria, averaging and error estimation}
\label{sec:qualcrit}

The essential characteristics of our approach to the problem of rating
and averaging lattice quantities 
have been outlined in our first publication~\cite{Colangelo:2010et}. 
Our aim is to help the reader
assess the reliability of a particular lattice result without
necessarily studying the original article in depth. This is a delicate
issue, since the ratings may make things appear 
simpler than they are. Nevertheless,
it safeguards against the 
possibility
of using lattice results, and
drawing physics conclusions from them, without a critical assessment
of the quality of the various calculations. We believe that, despite
the risks, it is important to provide some compact information about
the quality of a calculation. We stress, however, the importance of the
accompanying detailed discussion of the results presented in the various
sections of the present review.
 
\subsection{Systematic errors and colour code}
\label{sec:color-code}

The major sources of systematic error are common to most lattice
calculations. These include, as discussed in detail below,
the chiral, continuum, and infinite-volume extrapolations.
To each such source of error for which
systematic improvement is possible we
assign one of three coloured symbols: green
star, unfilled green circle
(which replaced in Ref.~\cite{Aoki:2013ldr}
the amber disk used in the original FLAG review~\cite{Colangelo:2010et})
or red square.
These correspond to the following ratings: 
\begin{itemize}[noitemsep,nolistsep] 
\item[\good] the parameter values and ranges used 
to generate the data sets allow for a satisfactory control of the systematic uncertainties;
\item[\soso] the parameter values and ranges used to generate
the data sets allow for a reasonable attempt at estimating systematic uncertainties, which
however could be improved;
\item[\bad] the parameter values and ranges used to generate
the data sets are unlikely to allow for a reasonable control of systematic uncertainties.
\end{itemize}
The appearance of a red tag, even in a
single source of systematic error of a given lattice result,
disqualifies it from inclusion in the global average.

Note that in the first two editions~\cite{Colangelo:2010et,Aoki:2013ldr},
FLAG used the three symbols in order to rate the reliability of the systematic errors 
attributed to a given result by the paper's authors.
Starting with FLAG 16~\cite{Aoki:2016frl} the meaning of the 
symbols has changed slightly---they now rate the quality of a particular simulation, 
based on the values and range of the chosen parameters,
and its aptness to obtain well-controlled systematic uncertainties. 
They do not rate the quality of the analysis performed by the authors 
of the publication. The latter question is
deferred to the relevant sections of the present review, 
which contain detailed discussions of 
the results contributing (or not) to each FLAG average or estimate. 

For most quantities the colour-coding system refers to the following  
sources of systematic errors: (i) chiral extrapolation; 
(ii) continuum extrapolation; (iii) finite volume. 
As we will see below, renormalization is another source of systematic
uncertainties in several quantities. This we also classify using the 
three coloured symbols listed above, but now with
a different rationale:  they express how reliably these quantities are 
renormalized, from a field-theoretic point of view
(namely, nonperturbatively, or with 2-loop or 1-loop perturbation theory).

Given the sophisticated status that the field has attained,
several aspects, besides those rated by the coloured symbols,
need to be evaluated before one can conclude
whether a particular analysis leads to results that should be included in an
average or estimate. Some of these aspects are not so easily expressible
in terms of an adjustable parameter such as the lattice spacing, the pion mass
or the volume. As a result of such considerations,
it sometimes occurs, albeit rarely, that a given
result does not contribute to the FLAG average or estimate, 
despite not carrying any red tags.
This happens, for instance, whenever aspects of the analysis appear 
to be incomplete 
(e.g., an incomplete error budget), so that the presence
of inadequately controlled systematic effects cannot be excluded. 
This mostly refers to results with a statistical error only, or results
in which the quoted error budget obviously fails to account 
for an important contribution.

Of course, any colour coding has to be treated with caution; we emphasize
that the criteria are subjective and evolving. Sometimes, a single
source of systematic error dominates the systematic uncertainty and it
is more important to reduce this uncertainty than to aim for green
stars for other sources of error. In spite of these caveats, we hope
that our attempt to introduce quality measures for lattice simulations
will prove to be a useful guide. In addition, we would like to
stress that the agreement of lattice results obtained using
different actions and procedures provides further validation.

\subsubsection{Systematic effects and rating criteria}
\label{sec:Criteria}

The precise criteria used in determining the colour coding are
unavoidably time-dependent; as lattice calculations become more
accurate, the standards against which they are measured become
tighter. For this reason FLAG reassesses criteria with each edition and as a result
some of the quality criteria (the one on chiral extrapolation for instance) have been tightened up   
over time~\cite{Colangelo:2010et,Aoki:2013ldr,Aoki:2016frl,Aoki:2019cca}.

In the following, we present the rating criteria used in the current report. 
While these criteria apply to most quantities without modification
there are cases where they need to be amended or additional criteria need to be defined. 
For instance, when discussing
results obtained in the $\epsilon$-regime of chiral perturbation theory in Sec.~\ref{sec:LECs}
the finite volume criterion listed below for the $p$-regime is no longer appropriate.\footnote{We refer to Sec.~\ref{sec:chPT} 
for an explanation of the various regimes of chiral perturbation theory.} Similarly, the discussion
of the strong coupling constant in Sec.~\ref{sec:alpha_s} requires tailored criteria
for renormalization, perturbative behaviour, and continuum extrapolation. 
Finally, in the section on nuclear matrix elements, Sec.~\ref{sec:NME},
the chiral extrapolation criterion is made slightly stronger, and a new criterion is adopted for
excited-state contributions.
In such cases,
the modified criteria are discussed in the respective sections. Apart from only a few exceptions the 
following colour code applies in the tables:

\begin{itemize}
\item Chiral extrapolation:
\begin{itemize}[noitemsep,nolistsep] 
	\item[\good] $M_{\pi,\mathrm{min}}< 200$ MeV, with three or more pion masses used in the extrapolation \\
	\underline{or} two values of $M_\pi$ with one lying within 10 MeV of 135MeV (the physical neutral pion mass) and the other one below 200 MeV  
	\item[\soso]  200 MeV $\le M_{\pi,{\mathrm{min}}} \le$ 400 MeV, with three or more pion masses used in the extrapolation \\\underline{or} two values of $M_\pi$ with $M_{\pi,{\mathrm{min}}}<$ 200 MeV \\\underline{or} a single value of $M_\pi$, lying within 10 MeV of 135 MeV (the physical neutral pion mass)
	\item[\bad] otherwise  
	\end{itemize}
This criterion is unchanged from FLAG 19. 
In Sec.~\ref{sec:NME} the upper end of the range for $M_{\pi,{\mathrm{min}}}$
in the green circle criterion is lowered to 300 MeV, as in FLAG 19.
\item 
Continuum extrapolation:
\begin{itemize}[noitemsep,nolistsep] 
	\item[\good] at least three lattice spacings \underline{and} at least two points below 0.1 fm \underline{and} a range of lattice spacings satisfying $[a_{\mathrm{max}}/a_{\mathrm{min}}]^2 \geq 2$
	\item[\soso] at least two lattice spacings \underline{and} at least one point below 0.1 fm 
	\underline{and} a range of lattice spacings 
	satisfying $[a_{\mathrm{max}}/a_{\mathrm{min}}]^2 \geq 1.4$
	\item[\bad] otherwise
\end{itemize}
It is assumed that the lattice action is $\cO(a)$-improved (i.e., the
discretization errors vanish quadratically with the lattice spacing);
otherwise this will be explicitly mentioned. For
unimproved actions an additional lattice spacing is required.
This condition is unchanged from FLAG 19.
\item 
Finite-volume effects:\\ 
The finite-volume colour code used for a result is 
chosen to be the worse of the QCD and the QED codes, as described below. If only QCD is used the QED colour code is ignored.

\emph{-- For QCD:}
\begin{itemize}[noitemsep,nolistsep] 
	\item[\good] $[M_{\pi,\mathrm{min}} / M_{\pi,\mathrm{fid}}]^2 \exp\{4-M_{\pi,\mathrm{min}}[L(M_{\pi,\mathrm{min}})]_{\mathrm{max}}\} < 1$,
	\underline{or} at least three volumes
	\item[\soso] $[M_{\pi,\mathrm{min}} / M_{\pi,\mathrm{fid}}]^2 \exp\{3-M_{\pi,\mathrm{min}}[L(M_{\pi,\mathrm{min}})]_{\mathrm{max}}\} < 1$,
	\underline{or} at least two volumes
	\item[\bad]  otherwise 
\end{itemize}
where we have introduced $[L(M_{\pi,\mathrm{min}})]_{\mathrm{max}}$, which is the maximum box size used in 
the simulations performed at the smallest pion mass $M_{\pi,{\rm min}}$, as well as a fiducial pion mass 
$M_{\pi,{\rm fid}}$, which we set to 200
MeV (the cutoff value for a green star in the chiral extrapolation). 
It is assumed here that calculations are in the $p$-regime of chiral perturbation
theory, and that all volumes used exceed 2~fm. 
The rationale for this condition is as follows.
Finite volume effects contain the universal factor $\exp\{- L~M_\pi\}$,
and if this were the only contribution a criterion based on
the values of $M_{\pi,\textrm{min}} L$ would be appropriate. 
However, as pion masses decrease, one must also account for
the weakening of the pion couplings. In particular,
1-loop chiral perturbation theory~\cite{Colangelo:2005gd} 
reveals a behaviour proportional to
$M_\pi^2 \exp\{- L~M_\pi\}$. 
Our 
condition includes this weakening of the coupling, 
and ensures, for example, that simulations with
$M_{\pi,\mathrm{min}} = 135~{\rm MeV}$ and $L~M_{\pi,\mathrm{min}} =
3.2$ are rated equivalently to those with $M_{\pi,\mathrm{min}} = 200~{\rm MeV}$
and $L~M_{\pi,\mathrm{min}} = 4$.

\emph{-- For QED (where applicable):}
\begin{itemize}[noitemsep,nolistsep]
	\item[\good]$1/([M_{\pi,\mathrm{min}}L(M_{\pi,\mathrm{min}})]_{\mathrm{max}})^{n_{\mathrm{min}}}<0.02$,
		\underline{or} at least four volumes
	\item[\soso] $1/([M_{\pi,\mathrm{min}}L(M_{\pi,\mathrm{min}})]_{\mathrm{max}})^{n_{\mathrm{min}}}<0.04$,
		\underline{or} at least three volumes
	\item[\bad]  otherwise 
\end{itemize}
Because of the infrared-singular structure of QED, electromagnetic finite-volume effects decay only like a power of the inverse spatial extent. In several cases like mass splittings~\cite{Borsanyi:2014jba,Davoudi:2014qua} or leptonic decays~\cite{Lubicz:2016xro}, the leading corrections are known to be universal, i.e., independent of the structure of the involved hadrons. In such cases, the leading universal effects can be directly subtracted exactly from the lattice data. We denote $n_{\mathrm{min}}$ the smallest power of $\frac{1}{L}$ at which such a subtraction cannot be done. In the widely used finite-volume formulation $\mathrm{QED}_L$, one always has $n_{\mathrm{min}}\leq 3$ due to the nonlocality of the theory~\cite{Davoudi:2018qpl}.
The QED criteria are used here only in Sec.~\ref{sec:qmass}.
Both QCD and QED criteria are unchanged from FLAG 19.
\item Isospin breaking effects (where applicable):
\begin{itemize}[noitemsep,nolistsep]
	\item[\good] all leading isospin breaking effects are included in the lattice calculation
	\item[\soso] isospin breaking effects are included using the electro-quenched approximation
	\item[\bad]otherwise
\end{itemize}
This criterion is used for quantities which are breaking isospin symmetry or which can be determined at the sub-percent accuracy where isospin breaking effects, if not included, are expected to be the dominant source of uncertainty. In the current edition, this criterion is only used for the up- and down-quark masses, and related quantities ($\epsilon$, $Q^2$ and $R^2$).
The criteria for isospin breaking effects 
are unchanged from FLAG 19.
\item Renormalization (where applicable):
\begin{itemize}[noitemsep,nolistsep]
	\item[\good]  nonperturbative
	\item[\soso]  1-loop perturbation theory or higher  with a reasonable estimate of truncation errors
	\item[\bad]  otherwise 
\end{itemize}	
In Ref.~\cite{Colangelo:2010et}, we assigned a red square to all
results which were renormalized at 1-loop in perturbation theory. In 
FLAG 13~\cite{Aoki:2013ldr}, we decided that this was too restrictive, since 
the error arising from renormalization constants, calculated in perturbation theory at
1-loop, is often estimated conservatively and reliably. 
These criteria have remained unchanged since then.

\item Renormalization Group (RG) running (where applicable): \\ 
For scale-dependent quantities, such as quark masses or $B_K$, it is
essential that contact with continuum perturbation theory can be established.
Various different methods are used for this purpose
(cf.~Appendix A.3 in FLAG 19 \cite{Aoki:2019cca}): Regularization-independent Momentum
Subtraction (RI/MOM), the Schr\"odinger functional, and direct comparison with
(resummed) perturbation theory. Irrespective of the particular method used,
the uncertainty associated with the choice of intermediate
renormalization scales in the construction of physical observables
must be brought under control. This is best achieved by performing
comparisons between nonperturbative and perturbative running over a
reasonably broad range of scales. These comparisons were initially
only made in the Schr\"odinger functional approach, but are now
also being performed in RI/MOM schemes.  We mark the data for which
information about nonperturbative running checks is available and
give some details, but do not attempt to translate this into a
colour code. 
\end{itemize}

The pion mass plays an important role in the criteria relevant for
chiral extrapolation and finite volume.  For some of the
regularizations used, however, it is not a trivial matter to identify
this mass. 
In the case of twisted-mass fermions, discretization
effects give rise to a mass difference between charged and neutral
pions even when the up- and down-quark masses are equal: the charged pion
is found to be the heavier of the two for twisted-mass Wilson fermions
(cf.~Ref.~\cite{Boucaud:2007uk}).
In early works, typically
referring to $N_f=2$ simulations (e.g., Refs.~\cite{Boucaud:2007uk}
and~\cite{Baron:2009wt}), chiral extrapolations are based on chiral
perturbation theory formulae which do not take these regularization
effects into account. After the importance of accounting for isospin
breaking when doing chiral fits was shown in Ref.~\cite{Bar:2010jk},
later works, typically referring to $N_f=2+1+1$ simulations, have taken
these effects into account~\cite{Carrasco:2014cwa}.
We use $M_{\pi^\pm}$ for $M_{\pi,\mathrm{min}}$
in the chiral-extrapolation rating criterion. On the
other hand, 
we identify $M_{\pi,\mathrm{min}}$ with
the root mean square (RMS) of $M_{\pi^+}$,
$M_{\pi^-}$ and $M_{\pi^0}$ in the finite-volume rating criterion.

In the case of staggered fermions,
discretization effects give rise to several light states with the
quantum numbers of the pion.\footnote{
We refer the interested reader to a number of reviews on the
subject~\cite{Durr:2005ax,Sharpe:2006re,Kronfeld:2007ek,Golterman:2008gt,Bazavov:2009bb}.}
The mass splitting among these ``taste'' partners represents a
discretization effect of $\cO(a^2)$, which can be significant at large
lattice spacings but shrinks as the spacing is reduced. In the
discussion of the results obtained with staggered quarks given in the
following sections, we assume that these artifacts are under
control. We conservatively identify $M_{\pi,\mathrm{min}}$ with the root mean
square (RMS) average of the masses of all the taste partners, 
both for chiral-extrapolation and finite-volume criteria.

In some of the simulations, the fermion formulations employed for the
valence quarks are different from those used for the sea quarks. Even
when the fermion formulations are the same, there are cases where the
sea and valence quark masses differ. In such cases, we use the smaller
of the valence-valence and valence-sea $M_{\pi_{\rm min}}$ values in the
finite-volume criteria, since either of these channels may give the leading
contribution depending on the quantity of interest at the one-loop
level of chiral perturbation theory. For the chiral-extrapolation
criteria, on the other hand, we use the unitary point, where the sea and
valence quark masses are the same, to define $M_{\pi_{\rm min}}$.

The strong coupling $\alpha_s$ is computed in lattice QCD with methods
differing substantially
from those used in the calculations of the other quantities 
discussed in this review. Therefore, we have established separate criteria for
$\alpha_s$ results, which will be discussed in Sec.~\ref{s:crit}.

In the section on nuclear matrix elements, Sec.~\ref{sec:NME},
an additional criterion is used.
This concerns the level of control over contamination from excited states,
which is a more challenging issue for nucleons than for mesons. 
In response to an improved understanding of the impact of this
contamination, the excited-state contamination criterion has been made more stringent compared
to that in FLAG 19.

\subsubsection{Heavy-quark actions}
\label{sec:HQCriteria}

For the $b$ quark,
the discretization of the
heavy-quark action follows a very different approach from that used for light
flavours. There are several different methods for
treating heavy quarks on the lattice, each with its own issues and
considerations.  Most of these methods use
Effective Field Theory (EFT) at some point in the computation, either
via direct simulation of the EFT, or by using EFT
as a tool to estimate the size of cutoff errors, 
or by using EFT to extrapolate from the simulated
lattice quark masses up to the physical $b$-quark mass. 
Because of the use of an EFT, truncation errors must be
considered together with discretization errors. 

The charm quark lies at an intermediate point between the heavy
and light quarks. In our earlier reviews, the calculations
involving charm quarks often treated it using one of the approaches adopted
for the $b$ quark. Since FLAG 16~\cite{Aoki:2016frl}, however, 
most calculations simulate the charm quark using light-quark actions.
This has become possible thanks to the increasing availability of
dynamical gauge field ensembles with fine lattice spacings.
But clearly, when charm quarks are treated relativistically, discretization
errors are more severe than those of the corresponding light-quark quantities.

In order to address these complications, 
the heavy-quark section adds an additional, bipartite,
treatment category to the rating system. The purpose of this
criterion is to provide a guideline for the level of action and
operator improvement needed in each approach to make reliable
calculations possible, in principle. 

A description of the different approaches to treating heavy quarks on
the lattice 
can be found in Appendix A.1.3 of FLAG 19~\cite{Aoki:2019cca}.
For truncation errors we use HQET power counting throughout,
since this review is focused on heavy-quark quantities involving $B$
and $D$ mesons rather than bottomonium or charmonium quantities.  
Here we describe the criteria for how each approach
must be implemented in order to receive an acceptable rating (\okay) 
for both the heavy-quark actions and the weak operators.  Heavy-quark
implementations without the level of improvement described below are
rated not acceptable (\bad). The matching is evaluated together with
renormalization, using the renormalization criteria described in
Sec.~\ref{sec:Criteria}.  We emphasize that the heavy-quark
implementations rated as acceptable and described below have been
validated in a variety of ways, such as via phenomenological agreement
with experimental measurements, consistency between independent
lattice calculations, and numerical studies of truncation errors.
These tests are summarized in Sec.~\ref{sec:BDecays}.  \smallskip
\\ {\it Relativistic heavy-quark actions:} \\
\noindent 
\okay \hspace{0.2cm}   at least tree-level $\cO(a)$ improved action and 
weak operators  \\
This is similar to the requirements for light-quark actions. All
current implementations of relativistic heavy-quark actions satisfy
this criterion. \smallskip \\
{\it NRQCD:} \\
\noindent 
\okay \hspace{0.2cm}   tree-level matched through $\cO(1/m_h)$ 
and improved through $\cO(a^2)$ \\
The current implementations of NRQCD satisfy this criterion, and also
include tree-level corrections of $\cO(1/m_h^2)$ in the action. 
\smallskip \\
{\it HQET: }\\
\noindent 
\okay \hspace{0.2cm}  tree-level  matched through $\cO(1/m_h)$ 
with discretization errors starting at $\cO(a^2)$ \\
The current implementation of HQET by the ALPHA collaboration
satisfies this criterion, since both action and weak operators are
matched nonperturbatively through $\cO(1/m_h)$.  Calculations that
exclusively use a static-limit action do not satisfy this criterion,
since the static-limit action, by definition, does not include $1/m_h$
terms.  We therefore include static computations in our final estimates only if truncation errors (in $1/m_h$)  are discussed and included in the systematic uncertainties.\smallskip \\
{\it Light-quark actions for heavy quarks:}  \\
\noindent 
\okay \hspace{0.2cm}  discretization errors starting at $\cO(a^2)$ or higher \\
This applies to calculations that use the twisted-mass Wilson action, a
nonperturbatively improved Wilson action, domain wall fermions or the HISQ action for charm-quark 
quantities. It also applies to calculations that use these light
quark actions in the charm region and above together with either the
static limit or with an HQET-inspired extrapolation to obtain results
at the physical $b$-quark mass. In these cases, the continuum-extrapolation criteria described earlier 
must be applied to the entire range of heavy-quark masses used in 
the calculation.

\subsubsection{Conventions for the figures}
\label{sec:figurecolours}

For a coherent assessment of the present situation, the quality of the
data plays a key role, but the colour coding cannot be carried over to
the figures. On the other hand, simply showing all data on equal
footing might give the misleading impression that the overall
consistency of the information available on the lattice is
questionable. Therefore, in the figures we indicate the quality of the data
in a rudimentary way, using the following symbols:
\begin{itemize}[noitemsep,nolistsep]
	\item[\raisebox{0.3mm}{\hspace{0.65mm}{\color{darkgreen}$\blacksquare$}}] corresponds to results included in the average or estimate (i.e., results that contribute to the black square below);
	\item[\raisebox{0.3mm}{\hspace{0.65mm}{\color{lightgreen}$\blacksquare$\hspace{-0.3cm}\color{darkgreen}$\square$}}] corresponds to results that are not included in the average but pass all quality criteria;
	\item[\raisebox{0.3mm}{\hspace{0.65mm}{\color{red}$\square$}}] corresponds to all other results;
	\item[\raisebox{0.3mm}{\hspace{0.65mm}{\color{black}$\blacksquare$}}]corresponds to FLAG averages or estimates; they are also highlighted by a gray vertical band.
\end{itemize} 
The reason for not including a given result in
the average is not always the same: the result may fail one of the
quality criteria; the paper may be unpublished; 
it may be superseded by newer results;
or it may not offer a complete error budget. 

Symbols other than squares are
used to distinguish results with specific properties and are always
explained in the caption.\footnote{%
For example, for quark-mass results we
distinguish between perturbative and nonperturbative renormalization, 
for low-energy constants we distinguish between the $p$- and $\epsilon$-regimes, 
and for heavy-flavour results we distinguish between
those from leptonic and semi-leptonic decays.}

Often, nonlattice data are also shown in the figures for comparison. 
For these we use the following symbols:
\begin{itemize}[noitemsep,nolistsep]
	\item[\raisebox{0.15mm}{\hspace{0.65mm}\color{blue}\Large\textbullet}]
	corresponds to nonlattice results;
	\item[\raisebox{0.35mm}{\hspace{0.65mm}{\color{black}$\blacktriangle$}}] corresponds to Particle Data Group (PDG) results.
\end{itemize}
\subsection{Averages and estimates}\label{sec:averages}

FLAG results of a given quantity are denoted either as {\it averages} or as {\it estimates}. Here we clarify this distinction. To start with, both {\it averages} and {\it estimates} are based on results without any red tags in their colour coding. For many observables there are enough independent lattice calculations of good quality, with all sources of error (not merely those related to the colour-coded criteria), as analyzed in the original papers, appearing to be under control. In such cases, it makes sense to average these results and propose such an {\it average} as the best current lattice number. The averaging procedure applied to this data and the way the error is obtained is explained in detail in Sec.~\ref{sec:error_analysis}. In those cases where only a sole result passes our rating criteria (colour coding), we refer to it as our FLAG {\it average}, provided it also displays adequate control of all other sources of systematic uncertainty.

On the other hand, there are some cases in which this procedure leads to a result that, in our opinion, does not cover all uncertainties. Systematic  errors are by their nature often subjective and difficult to estimate, and may thus end up being underestimated in one or more results that receive green symbols for all explicitly tabulated criteria.   
Adopting a conservative policy, in these cases we opt for an {\it estimate} (or a range), which we consider as a fair assessment of the knowledge acquired on the lattice at present. This {\it estimate} is not obtained with a prescribed mathematical procedure, but reflects what we consider the best possible analysis of the available information. The hope is that this will encourage more detailed investigations by the lattice community.

There are two other important criteria that also play a role in this
respect, but that cannot be colour coded, because a systematic
improvement is not possible. These are: {\em i)} the publication
status, and {\em ii)} the number of sea-quark flavours $\Nf$. As far as the
former criterion is concerned, we adopt the following policy: we
average only results that have been published in peer-reviewed
journals, i.e., they have been endorsed by referee(s). The only
exception to this rule consists in straightforward updates of previously
published results, typically presented in conference proceedings. Such
updates, which supersede the corresponding results in the published
papers, are included in the averages. 
Note that updates of earlier results rely, at least partially, on the
same gauge-field-configuration ensembles. For this reason, we do not
average updates with earlier results. 
Nevertheless, all results are
listed in the tables,\footnote{%
Whenever figures turn out to be overcrowded,
older, superseded results are omitted. However, all the most recent results
from each collaboration are displayed.}
and their publication status is identified by the following
symbols:
\begin{itemize}
\item Publication status:\\
\gA  \hspace{0.2cm}published or plain update of published results\\
\oP  \hspace{0.2cm}preprint\\ 
\rC  \hspace{0.2cm}conference contribution
\end{itemize}
In the present edition, the
publication status on the {\bf 30th of April 2021} is relevant.
If the paper appeared in print after that date, this is accounted for in the
bibliography, but does not affect the averages.\footnote{%
As noted above in footnote 1, one exception to this deadline was made, Ref.~\cite{Chakraborty:2021qav}.
}

As noted above,
in this review we present results from simulations with $N_f=2$,
$N_f=2+1$ and $N_f=2+1+1$ (except for $ r_0 \Lambda_\msbar$ where we
also give the $N_f=0$ result). We are not aware of an {\em a priori} way
to quantitatively estimate the difference between results produced in
simulations with a different number of dynamical quarks. We therefore
average results at fixed $\Nf$ separately; averages of calculations
with different $\Nf$ are not  provided.

To date, no significant differences between results with different
values of $N_f$ have been observed in the quantities 
listed in Tabs.~\ref{tab:summary1}, \ref{tab:summary2}, \ref{tab:summary3}, \ref{tab:summary4}, and \ref{tab:summary5}.
In particular, differences between results from simulations with $\Nf = 2$ and $\Nf = 2 + 1$ 
would reflect Zweig-rule violations related to strange-quark loops. 
Although not of direct phenomenological relevance,
the size of such violations is an interesting theoretical issue {\em per se}, 
and one that can be quantitatively addressed only with lattice calculations.
It remains to be seen whether the status presented here will change in the future,
since this will require dedicated $N_f=2$ and $N_f=2+1$ calculations, which
are not a priority of present lattice work.

The question of differences between results with $\Nf=2+1$ and
$\Nf=2+1+1$ is more subtle.
The dominant effect of including the charm sea quark is to
shift the lattice scale, an effect that is accounted for by
fixing this scale nonperturbatively using physical quantities.
For most of the quantities discussed in this review, it is 
expected that residual effects are small in the continuum limit,
suppressed by $\alpha_s(m_c)$ and powers of $\Lambda^2/m_c^2$.
Here $\Lambda$ is a hadronic scale that can only be
roughly estimated and depends on the process under consideration.
Note that the $\Lambda^2/m_c^2$ effects have been addressed 
in~Refs.~\cite{Bruno:2014ufa,Knechtli:2017xgy,Athenodorou:2018wpk,Cali:2019enm,Cali:2021xwh},
and found to be small for the quantities considered.
Assuming that such effects are generically small, it might be reasonable to
average the results from $\Nf=2+1$ and $\Nf=2+1+1$ simulations,
although we do not do so here.

\subsection{Averaging procedure and error analysis}
\label{sec:error_analysis}

In the present report, we repeatedly average results
obtained by different collaborations, and estimate the error on the resulting
averages. 
Here we provide details on how averages are obtained.

\subsubsection{Averaging --- generic case}
We follow the procedure of the previous  two editions~\cite{Aoki:2013ldr,Aoki:2016frl},
which we describe here in full detail.

One of the problems arising when forming averages is that not all
of the data sets are independent.
In particular, the same gauge-field configurations,
produced with a given fermion discretization, are often used by
different research teams with different valence-quark lattice actions,
obtaining results that are not really independent.  
Our averaging procedure takes such correlations into account. 

Consider a given measurable quantity $Q$, measured by $M$ distinct,
not necessarily uncorrelated, numerical experiments (simulations). The result
of each of these measurement is expressed as
\begin{equation}
Q_i \,\, = \,\, x_i \, \pm \, \sigma^{(1)}_i \pm \, \sigma^{(2)}_i \pm \cdots
\pm \, \sigma^{(E)}_i  \,\,\, ,
\label{eq:resultQi}
\end{equation}
where $x_i$ is the value obtained by the $i^{\rm th}$ experiment
($i = 1, \cdots , M$) and $\sigma^{(\alpha)}_i$ (for $\alpha = 1, \cdots , E$) 
are the various errors.
Typically $\sigma^{(1)}_i$ stands for the statistical error 
and $\sigma^{(\alpha)}_i$ ($\alpha \ge 2$) are the different
systematic errors from various sources. 
For each individual result, we estimate the total
error $\sigma_i $ by adding statistical and systematic errors in quadrature:
\begin{eqnarray}
Q_i \,\, &=& \,\, x_i \, \pm \, \sigma_i \,\,\, ,
\nonumber \\
\sigma_i \,\, &\equiv& \,\, \sqrt{\sum_{\alpha=1}^E \Big [\sigma^{(\alpha)}_i \Big ]^2} \,\,\, .
\label{eq:av-err-Qi}
\end{eqnarray}
With the weight factor of each total error estimated in standard fashion,
\begin{equation}
\omega_i \,\, = \,\, \dfrac{\sigma_i^{-2}}{\sum_{i=1}^M \sigma_i^{-2}} \,\,\, ,
\label{eq:weighti}
\end{equation}
the central value of the average over all simulations is given by
\begin{eqnarray}
x_{\rm av} \,\, &=& \,\, \sum_{i=1}^M x_i\, \omega_i \,\, . 
\end{eqnarray}
The above central value corresponds to a $\chi_{\rm min}^2$ weighted
average, evaluated by adding statistical and systematic errors in quadrature.
If the fit is not of good quality ($\chi_{\rm min}^2/{\rm dof} > 1$),
the statistical and systematic error bars are stretched by a factor
$S = \sqrt{\chi^2/{\rm dof}}$.

Next, we examine error budgets for
individual calculations and look for potentially correlated
uncertainties. Specific problems encountered in connection with
correlations between different data sets are described in the text
that accompanies the averaging.
If there is reason to believe that a source of error is correlated
between two calculations, a $100\%$ correlation is assumed.
The correlation matrix $C_{ij}$ for the set of correlated lattice results is
estimated by a prescription due to Schmelling~\cite{Schmelling:1994pz}.
This consists in defining
\begin{equation}
\sigma_{i;j} \,\, = \,\, \sqrt{{\sum_{\alpha}}^\prime \Big[ \sigma_i^{(\alpha)} \Big]^2 } \,\,\, ,
\label{eq:sigmaij}
\end{equation}
with $\sum_{\alpha}^\prime$ running only over those errors of $x_i$ that
are correlated with the corresponding errors of the measurement $x_j$. 
This expresses the part of the uncertainty in $x_i$
that is correlated with the uncertainty in $x_j$. 
If no such correlations are known to exist, then
we take $\sigma_{i;j} =0$. 
The diagonal and off-diagonal elements of the correlation
matrix are then taken to be
\begin{eqnarray}
C_{ii} \,\,&=& \,\, \sigma_i^2 \qquad \qquad (i = 1, \cdots , M) \,\,\, ,
\nonumber \\
C_{ij} \,\,&=& \,\, \sigma_{i;j} \, \sigma_{j;i} \qquad \qquad (i \neq j) \,\,\, .
\label{eq:Ciiij}
\end{eqnarray}
Finally, the error of the average is estimated by
\begin{equation}
\sigma^2_{\rm av} \,\, = \,\, \sum_{i=1}^M \sum_{j=1}^M \omega_i \,\omega_j \,C_{ij}\,\,,
\label{eq:sigma2av}
\end{equation}
and the FLAG average is
\begin{equation}
Q_{\rm av} \,\, = \,\, x_{\rm av} \, \pm \, \sigma_{\rm av} \,\,\, .
\end{equation}
 \subsubsection{Nested averaging}
\label{sec:nested_average}

We have encountered one case
where the correlations between results are more involved,
and a nested averaging scheme is required.
This concerns the $B$-meson bag parameters discussed in Sec.~\ref{sec:BMix}.
In the following, we describe the details of the nested averaging scheme.
This is an updated version of the section added in the web update of the FLAG 16 report.

The issue arises for a quantity $Q$ that is given by a ratio, $Q=Y/Z$.
In most simulations, both $Y$ and $Z$ are calculated, and the error in $Q$ can be
obtained in each simulation in the standard way.
However, in other simulations only $Y$ is calculated,
with $Z$ taken from a global average of some type.
The issue to be addressed is that this average value $\overline{Z}$ has errors
that are correlated with those in $Q$.

In the example that arises in Sec.~\ref{sec:BMix},
$Q=B_B$,  $Y=B_B f_B^2$ and $Z=f_B^2$.
In one of the simulations that contribute to the average, 
$Z$ is replaced by $\overline{Z}$, 
the PDG average for $f_B^2$~\cite{Rosner:2015wva}
(obtained with an averaging procedure similar to that used by FLAG).
This simulation is labeled with $i=1$, so that
\begin{equation}
 Q_1 = \frac{Y_1}{\overline{Z}}.
  \label{eq:FNAL_B_PDG}
\end{equation}
The other simulations have results labeled $Q_j$, with $j\ge 2$.
In this set up, the issue is that $\overline{Z}$ is correlated with the $Q_j$, $j\ge 2$.\footnote{%
There is also a small correlation between $Y_1$ and $\overline{Z}$, but we follow the
original Ref.~\cite{Bazavov:2016nty}
 and do not take this into account. Thus, the error in $Q_1$
is obtained by simple error propagation from those in $Y_1$ and $\overline{Z}$.
Ignoring this correlation is conservative, because, as in the
calculation of $B_K$, the correlations between $B_B f_B^2$ and $f_B^2$ tend to
lead to a cancelation of errors. By ignoring this effect we are making a small overestimate
of the error in $Q_1$.}

We begin by decomposing the error in $Q_1$ in the same
schematic form as above,
\begin{equation}
 Q_1 
  = x_1 
  \pm \frac{\sigma_{Y_1}^{(1)}}{\overline{Z}}
  \pm \frac{\sigma_{Y_1}^{(2)}}{\overline{Z}} \pm\cdots
  \pm \frac{\sigma_{Y_1}^{(E)}}{\overline{Z}}
  \pm \frac{Y_1 \sigma_{\overline{Z}}}{\overline{Z}^2}.
  \label{eq:Q1nested}
\end{equation}
Here the last term represents the error propagating from that in $\overline{Z}$,
while the others arise from errors in $Y_1$.
For the remaining $Q_j$ ($j\ge 2$) the decomposition is as in Eq.~(\ref{eq:resultQi}).
The total error of $Q_1$ then reads 
\begin{equation}
 \sigma_1^2 = 
  \left(\frac{\sigma_{Y_1}^{(1)}}{\overline{Z}}\right)^2
  + \left(\frac{\sigma_{Y_1}^{(2)}}{\overline{Z}}\right)^2 +\cdots
  + \left(\frac{\sigma_{Y_1}^{(E)}}{\overline{Z}}\right)^2
  + \left(\frac{Y_1}{\overline{Z}^2}\right)^2 \sigma_{\overline{Z}}^2,
  \label{eq:sigma1}
\end{equation}
while that for the $Q_j$ ($j\ge 2$) is
\begin{equation}
 \sigma_j^2 = 
  \left(\sigma_j^{(1)}\right)^2
  + \left(\sigma_j^{(2)}\right)^2 +\cdots
  + \left(\sigma_j^{(E)}\right)^2.
  \label{eq:sigmaj}
\end{equation}
Correlations between $Q_j$ and $Q_k$ ($j,k\ge 2$) are taken care of by
Schmelling's prescription, as explained above.
What is new here is how the correlations 
between $Q_1$ and $Q_j$ ($j\ge 2$) are taken into account.

To proceed, we recall from Eq.~(\ref{eq:sigma2av}) that
$\sigma_{\overline{Z}}$ is given by
\begin{equation}
 \sigma_{\overline{Z}}^2 = \sum_{{i'},{j'}=1}^{M'} \omega[Z]_{i'}
  \omega[Z]_{j'} C[Z]_{i'j'}.
\end{equation}
Here the indices
$i'$ and $j'$ run over the $M'$ simulations that contribute to $\overline{Z}$,
which, in general, are different from those contributing to the results for $Q$.
The weights $\omega[Z]$ and correlation matrix $C[Z]$ are given an explicit
argument $Z$ to emphasize that they refer to the calculation of this quantity
and not to that of $Q$.
$C[Z]$ is calculated using the Schmelling prescription
[Eqs.~(\ref{eq:sigmaij})--(\ref{eq:sigma2av})] in terms of the errors, $\sigma[Z]_{i'}^{(\alpha)}$,
taking into account the correlations between the different calculations of $Z$.

We now generalize Schmelling's prescription for $\sigma_{i;j}$, Eq.~(\ref{eq:sigmaij}),
to that for $\sigma_{1;k}$ ($k\ge 2$), i.e., the part of the error in $Q_1$ that
is correlated with $Q_k$. We take
\begin{equation}
 \sigma_{1;k} \,\, = \,\, 
  \sqrt{
  \frac{1}{\overline{Z}^2} \sum^\prime_{(\alpha)\leftrightarrow k}
  \Big[\sigma_{Y_1}^{(\alpha)} \Big]^2 
  + \frac{Y_1^2}{\overline{Z}^4} 
  \sum_{i',j'}^{M'} \omega[Z]_{i'} \omega[Z]_{j'} C[Z]_{i'j'\leftrightarrow k}
  }
  \,\,\, .
\label{eq:sigma1k}
\end{equation}
The first term under the square root sums those sources of error in $Y_1$ that
are correlated with $Q_k$. Here we are using a more explicit notation from that
in Eq.~(\ref{eq:sigmaij}), with $(\alpha) \leftrightarrow k$ indicating that the sum
is restricted to the values of $\alpha$ for which the error $\sigma_{Y_1}^{(\alpha)}$
is correlated with $Q_k$.
The second term accounts for the correlations within $\overline{Z}$ with $Q_k$,
and is the nested part of the present scheme.
The new matrix $C[Z]_{i'j'\leftrightarrow k}$ is a restriction
of the full correlation matrix $C[Z]$, and is defined as follows.
Its diagonal elements are given by
\begin{eqnarray}
C[Z]_{i'i'\leftrightarrow k} \,\,&=& \,\, (\sigma[Z]_{i'\leftrightarrow k})^2 \qquad \qquad (i' = 1, \cdots , M') \,\,\, ,
 \\
 (\sigma[Z]_{i'\leftrightarrow k})^2 & = &
  \sum^\prime_{(\alpha)\leftrightarrow k} (\sigma[Z]_{i'}^{(\alpha)})^2,
\label{eq:sigmaZipk}
\end{eqnarray}
where the summation 
$\sum^\prime_{(\alpha)\leftrightarrow k}$ 
over $(\alpha)$ is restricted to those $\sigma[Z]_{i'}^{(\alpha)}$ that are
correlated with $Q_k$.
The off-diagonal elements are
\begin{eqnarray}
C[Z]_{i'j'\leftrightarrow k} \,\,&=& \,\, \sigma[Z]_{i';j'\leftrightarrow k} \, \sigma[Z]_{j';i'\leftrightarrow k} \qquad \qquad (i' \neq j') \,\,\, ,\\
 \sigma[Z]_{i';j'\leftrightarrow k} & = &
  \sqrt{
  \sum^\prime_{(\alpha)\leftrightarrow j'k} 
  (\sigma[Z]_{i'}^{(\alpha)})^2},
\label{eq:sigmaZipjpk}
\end{eqnarray}
where the summation 
$\sum^\prime_{(\alpha)\leftrightarrow j'k}$ 
over $(\alpha)$ is restricted to $\sigma[Z]_{i'}^{(\alpha)}$ that are
correlated with {\it both} $Z_{j'}$ and $Q_k$.

The last quantity that we need to define is $\sigma_{k;1}$.
\begin{equation}
\sigma_{k;1} \,\, = \,\, \sqrt{\sum^\prime_{(\alpha)\leftrightarrow 1} \Big[ \sigma_k^{(\alpha)} \Big]^2 } \,\,\, ,
\label{eq:sigmak1}
\end{equation}
where the summation $\sum^\prime_{(\alpha)\leftrightarrow 1}$ is
restricted to those $\sigma_k^{(\alpha)}$ that are correlated with one of
the terms in Eq.~(\ref{eq:sigma1}).

In summary, we construct the correlation matrix $C_{ij}$ using
Eq.~(\ref{eq:Ciiij}), as in the generic case, except the expressions
for $\sigma_{1;k}$ and $\sigma_{k;1}$ are now given by
Eqs.~(\ref{eq:sigma1k}) and (\ref{eq:sigmak1}), respectively. All other $\sigma_{i;j}$ are given by
the original Schmelling prescription, Eq.~(\ref{eq:sigmaij}).
In this way we extend the philosophy of Schmelling's approach while accounting
for the more involved correlations.

\clearpage
\section{Quark masses}
\label{sec:qmass}
Authors: T.~Blum, A.~Portelli, A.~Ramos\\

Quark masses are fundamental parameters of the Standard Model. An
accurate determination of these parameters is important for both
phenomenological and theoretical applications. The bottom- and charm-quark 
masses, for instance, are important sources of parametric
uncertainties in several Higgs decay modes. The up-,
down- and strange-quark masses govern the amount of explicit chiral
symmetry breaking in QCD. From a theoretical point of view, the values
of quark masses provide information about the flavour structure of
physics beyond the Standard Model. The Review of Particle Physics of
the Particle Data Group contains a review of quark masses
\cite{Zyla:2020zbs}, which covers light as well as heavy
flavours. Here we also consider light- and heavy-quark masses, but
focus on lattice results and discuss them in more detail. We do not
discuss the top quark, however, because it decays weakly before it can
hadronize, and the nonperturbative QCD dynamics described by present
day lattice simulations is not relevant. The lattice determination of
light- (up, down, strange), charm- and bottom-quark masses is
considered below in Secs.~\ref{sec:lqm}, \ref{s:cmass},
and \ref{s:bmass}, respectively.

Quark masses cannot be measured directly in experiment because
quarks cannot be isolated, as they are confined inside hadrons. From a
theoretical point of view, in QCD with $N_f$ flavours, a precise
definition of quark 
masses requires one to choose a particular renormalization
scheme. This renormalization procedure introduces a
renormalization scale $\mu$, and quark masses depend on this
renormalization scale according to the Renormalization Group (RG)
equations. In mass-independent renormalization schemes the RG equations
read
\begin{equation}
  \label{eq:qmass_tau}
  \mu \frac{{\rm d} \bar m_i(\mu)}{{\rm d}{\mu}} = \bar m_i(\mu) \tau(\bar g)\,,
\end{equation}
where the function $\tau(\bar g)$ is the anomalous
dimension, which
depends only on the value of the strong coupling $\alpha_s=\bar
g^2/(4\pi)$. Note that in QCD $\tau(\bar g)$ is the same for all quark
flavours.  The anomalous 
dimension is scheme dependent, but its 
perturbative expansion  
\begin{equation}
  \label{eq:tau_asymp}
  \tau(\bar g) \raisebox{-.1ex}{
            $\stackrel{\small{\bar g \to 0}}{\sim}$} -\bar g^2\left(
    d_0 + d_1\bar g^2 + \dots
  \right) 
\end{equation}
has a leading coefficient $d_0 = 8/(4\pi)^2$,  
which is scheme independent.\footnote{We follow the conventions of
  Gasser and Leutwyler~\cite{Gasser:1982ap}.}
   Equation~(\ref{eq:qmass_tau}), being a
first order differential equation, can be solved exactly by using
Eq.~(\ref{eq:tau_asymp}) as the boundary condition. The formal
solution of the RG equation reads
\begin{equation}
  \label{eq:qmass_rgi}
  M_i = \bar m_i(\mu)[2b_0\bar g^2(\mu)]^{-d_0/(2b_0)}
  \exp\left\{
    - \int_0^{\bar g(\mu)}{\rm d} x\, \left[
      \frac{\tau(x)}{\beta(x)} - \frac{d_0}{b_0x}
    \right]
  \right\}\,,
\end{equation}
where $b_0 = (11-2N_f/3) / (4\pi)^2$ is the universal
leading perturbative coefficient in the expansion of the
$\beta$-function, {\color{black}$\beta(\bar g)\equiv  d\bar g^2/d\log{\mu^2}$, which governs the running of the strong coupling constant near the scale $\mu$}. The renormalization group invariant
(RGI) quark masses $M_i$ are formally integration constants of the
RG Eq.~(\ref{eq:qmass_tau}). They are scale independent, and due
to the universality of the coefficient $d_0$, they are also scheme
independent. Moreover, they are nonperturbatively defined by
Eq.~(\ref{eq:qmass_rgi}). They only depend on the number
of flavours $N_f$, making them a natural candidate to quote quark
masses and compare determinations from different lattice
collaborations. Nevertheless, it is customary in the phenomenology
community to use the $\overline{\rm MS}$ scheme at a scale $\mu = 2$
GeV to compare different results for light-quark masses, and use a
scale equal to its own mass for the charm and bottom
quarks. In this review, we will quote the final averages of both
quantities.

Results for quark masses are always quoted in the four-flavour
theory. $N_{f}=2+1$ results have to be converted to the four-flavour theory. Fortunately, the charm quark is heavy $(\Lambda_{\rm
  QCD}/m_c)^2<1$, and this conversion can be performed in perturbation
theory with negligible ($\sim 0.2\%$) perturbative
uncertainties. Nonperturbative corrections in this matching are more
difficult to estimate. Since these effects are suppressed by a factor of 
$1/N_{\rm c}$, and a factor of the strong coupling at the scale of the
charm mass, naive power counting arguments would suggest that the 
effects are $\sim 1\%$. In practice, numerical nonperturbative
studies~\cite{Hollwieser:2020qri,Athenodorou:2018wpk,Bruno:2014ufa} 
have found this power counting argument to be an overestimate by one
order of magnitude in the determination of simple hadronic
quantities or the $\Lambda$-parameter. Moreover, lattice 
determinations do not show any significant deviation between the
$N_{f}=2+1$ and $N_{f}=2+1+1$ simulations. For example, the
difference in the final averages for the mass of the strange quark
$m_s$ between $N_f=2+1$ and $N_f=2+1+1$ determinations is about
{\color{black}1.3\%, or about one standard deviation}. 

We quote all final averages at $2$ GeV in the $\overline{\rm
  MS}$ scheme and also the RGI values (in the four-flavour theory). We
use the exact RG
 Eq.~(\ref{eq:qmass_rgi}). Note that to use this equation we
need the value of the strong coupling in the $\overline{\rm MS}$
scheme at a scale $\mu = 2$ GeV. All our results are obtained from the
RG equation in the $\overline{\rm MS}$ scheme and the 5-loop beta
function together with the 
value of the $\Lambda$-parameter in the four-flavour theory
$\Lambda^{(4)}_{\overline{\rm MS}} = 294(12)\, {\rm MeV}$ obtained in
this review (see Sec.~\ref{sec:alpha_s}). In the uncertainties of the RGI
masses we separate the contributions from the determination of the
quark masses and the propagation of the uncertainty of
$\Lambda^{(4)}_{\overline{\rm MS}}$. These are identified with the
subscripts $m$ and $\Lambda$, respectively. 

Conceptually, all lattice determinations of quark masses contain three
basic ingredients:
\begin{enumerate}
\item Tuning the lattice bare-quark masses to match the experimental
  values of some quantities.  Pseudo-scalar meson masses provide 
  the most common choice, since they have a strong dependence on the
  values of quark masses. In pure
  QCD with $N_f$ quark flavours these values are not
  known, since the electromagnetic interactions affect the
  experimental values of meson masses. Therefore, pure QCD
  determinations use model/lattice information to determine the
  location of the physical point. This is discussed at length in Sec.~\ref{sec:physical point and isospin}.

\item Renormalization of the bare-quark masses. Bare-quark masses
  determined with
  the above-mentioned criteria have to be renormalized. Many of the
  latest determinations use some nonperturbatively defined
  scheme. One can also use perturbation theory to connect directly the
  values of the bare-quark masses to the values in the $\overline{\rm
    MS}$ scheme at $2$ GeV. Experience shows that
  1-loop calculations are unreliable for the renormalization of
  quark masses: usually at least two loops are required to have
  trustworthy results.
   
\item If quark masses have been nonperturbatively renormalized, for
  example, to some MOM/SF scheme, the values in this scheme must be
  converted to the phenomenologically useful values in the
  $\overline{\rm MS}$ scheme (or to the scheme/scale independent RGI
  masses). Either option 
  requires the use of perturbation theory. The larger the
  energy scale of this matching with perturbation theory, the better,
  and many recent computations in MOM schemes do a nonperturbative
  running up to $3$--$4$ GeV. Computations in the SF scheme allow us to
  perform this running nonperturbatively over large energy scales and
  match with perturbation theory directly at the electro-weak scale $\sim 100$
  GeV. 
\end{enumerate}
{\color{black}Note that many lattice determinations of quark masses make use of
perturbation theory at a scale of a few GeV.}

We mention that lattice-QCD calculations of the $b$-quark mass have an
additional complication which is not present in the case of the charm
and light quarks. 
At the lattice spacings currently used in numerical simulations the
direct treatment of the $b$ quark with the fermionic actions commonly
used for light quarks is very challenging. Only two determinations of
the $b$-quark mass use this approach, reaching the physical $b$-quark
mass region at two lattice spacings with $aM\sim 1$. 
There are a few widely used approaches to treat the $b$ quark on the
lattice, which have been already discussed in the FLAG 13 review (see
Sec.~8 of Ref.~\cite{Aoki:2013ldr}). 
Those relevant for the determination of the $b$-quark mass will be
briefly described in Sec.~\ref{s:bmass}.

\medskip


\subsection{Masses of the light quarks}
\label{sec:lqm}

Light-quark masses are particularly difficult to determine because
they are very small (for the up and down quarks) or small (for
the strange quark) compared to typical hadronic scales. Thus, their impact
on typical hadronic observables is minute, and it is difficult to
isolate their contribution accurately.

Fortunately, the spontaneous breaking of $SU(3)_L\times SU(3)_R$
chiral symmetry provides observables which are particularly sensitive
to the light-quark masses: the masses of the resulting Nambu-Goldstone
bosons (NGB), i.e., pions, kaons, and eta. Indeed, the
Gell-Mann-Oakes-Renner relation~\cite{GellMann:1968rz} predicts that
the squared mass of a NGB is directly proportional to the sum of the
masses of the quark and antiquark which compose it, up to higher-order
mass corrections. Moreover, because these NGBs are light, and are
composed of only two valence particles, their masses have a
particularly clean statistical signal in lattice-QCD calculations. In
addition, the experimental uncertainties on these meson masses are
negligible. Thus, in lattice calculations, light-quark masses are
typically obtained by renormalizing the input quark mass and tuning
them to reproduce NGB masses, as described above.

\subsubsection{The physical point and isospin symmetry}\label{sec:physical point and isospin}
As mentioned in Sec.~\ref{sec:color-code}, the present review relies on the
hypothesis that, at low energies, the Lagrangian ${\cal L}_{\mbox{\tiny
QCD}}+{\cal L}_{\mbox{\tiny QED}}$ describes nature to a high degree of
precision. However, most of the results presented below are obtained in pure QCD
calculations, which do not include QED. Quite generally, when comparing QCD
calculations with experiment, radiative corrections need to be applied. In pure
QCD simulations, where the parameters are fixed in terms of the masses of some
of the hadrons, the electromagnetic contributions to these masses must be
discussed. How the matching is done is generally ambiguous because it relies on
the unphysical separation of QCD and QED contributions. In this section, and in the
following, we discuss this issue in detail. A related discussion, in the context of scale setting, is given in Sec.~\ref{sec:isobreak}.
Of course, once QED is included in
lattice calculations, the subtraction of electromagnetic contributions is no longer
necessary.

Let us start from the unambiguous case of QCD+QED. As explained in the
introduction of this section, the physical quark masses are the parameters of
the Lagrangian such that a given set of experimentally measured, dimensionful
hadronic quantities are reproduced by the theory. Many choices are possible for
these quantities, but in practice many lattice groups use pseudoscalar meson
masses, as they are easily and precisely obtained both by experiment, and through
lattice simulations. For example, in the four-flavour case, one can solve the
system
\bea
M_{\pi^+}(m_u,m_d,m_s,m_c,\alpha) &=& M_{\pi^+}^{\mathrm{exp.}}\co
\label{eq:phypt1}\\
M_{K^+}(m_u,m_d,m_s,m_c,\alpha) &=& M_{K^+}^{\mathrm{exp.}}\co
\label{eq:phypt2}\\
M_{K^0}(m_u,m_d,m_s,m_c,\alpha) &=& M_{K^0}^{\mathrm{exp.}}\co
\label{eq:phypt3}\\
M_{D^0}(m_u,m_d,m_s,m_c,\alpha) &=& M_{D^0}^{\mathrm{exp.}}\co
\label{eq:phypt4}
\eea
where we assumed that
\begin{itemize}
  \item all the equations are in the continuum and infinite-volume limits;
  \item the overall scale has been set to its physical value, generally
  through some lattice-scale setting procedure involving a fifth dimensionful
  input (see the discussion in Sec.~\ref{sec:isobreak});
  \item the quark masses $m_q$ are assumed to be renormalized from the bare,
  lattice ones in some given continuum renormalization scheme;
  \item $\alpha=\frac{e^2}{4\pi}$ is the fine-structure constant expressed as function of the positron charge $e$, generally set to the Thomson limit $\alpha=0.007297352\dots$~\cite{Zyla:2020zbs};
  \item the mass $M_{h}(m_u,m_d,m_s,m_c,\alpha)$ of the meson
  $h$ is a function of the quark masses and $\alpha$. The functional
  dependence is generally obtained by choosing an appropriate parameterization
  and performing a global fit to the lattice data;
  \item the superscript exp.~indicates that the mass is an experimental input,
  lattice groups use in general the values in the Particle Data Group
  review~\cite{Zyla:2020zbs}.
\end{itemize}

However, ambiguities arise with simulations of QCD only. In that case, there is
no experimentally measurable quantity that emerges from the strong interaction
only. The missing QED contribution is tightly related to isospin-symmetry breaking 
effects. Isospin symmetry is explicitly broken by the differences
between the up- and down-quark masses $\delta m=m_u-m_d$, and electric charges
$\delta Q=Q_u-Q_d$. These effects are, respectively, of order $\cO(\delta
m/\lqcd)$ and $\cO(\alpha)$, and are expected to be $\cO(1\%)$ of a typical
isospin-symmetric hadronic quantity. Strong and electromagnetic isospin-breaking
effects are of the same order and therefore cannot, in principle, be evaluated
separately without introducing strong ambiguities. Because these effects
are small, they can be treated as a perturbation,
\be
X(m_u,m_d,m_s,m_c,\alpha)=\bar{X}(m_{ud}, m_s, m_c)
+\delta mA_X(m_{ud}, m_s, m_c)
+\alpha B_X(m_{ud}, m_s, m_c)\co\label{eq:isoex}
\ee
for a given hadronic quantity $X$, where $m_{ud}=\frac12(m_u+m_d)$ is the average light-quark mass. There are several things to notice
here. Firstly, the neglected higher-order $\cO(\delta m^2,\alpha\delta
m,\alpha^2)$ corrections are expected to be $\cO(10^{-4})$ relatively to $X$,
which at the moment is way beyond the relative statistical accuracy that can be
delivered by a lattice calculation.  Secondly, this is not strictly speaking an
expansion around the isospin-symmetric point, the electromagnetic interaction
has also symmetric contributions. From this last expression the previous
statements about ambiguities become clearer. Indeed, the only unambiguous
prediction one can perform is to solve Eqs.~(\ref{eq:phypt1})--(\ref{eq:phypt4})
and use the resulting parameters to obtain a prediction for $X$, which is
represented by the left-hand side of~\eq{eq:isoex}. This prediction will be the
sum of the QCD isospin-symmetric part $\bar{X}$, the strong isospin-breaking effects $
X^{SU(2)}=\delta mA_X$, and the electromagnetic effects $X^{\gamma}=\alpha B_X$. 
Obtaining any of these terms individually requires extra,
unphysical conditions to perform the separation. To be consistent with previous editions of FLAG, we also define $\hat{X}=\bar{X}+X^{SU(2)}$ to be the $\alpha\to 0$ limit of $X$.

With pure QCD simulations, one typically solves
Eqs.~(\ref{eq:phypt1})--(\ref{eq:phypt4}) by equating the QCD isospin-symmetric
part of a hadron mass $\bar{M}_h$, result of the simulations, with its
experimental value $M_h^{\mathrm{exp.}}$. This will result in an~$\cO(\delta
m,\alpha)$ mis-tuning of the theory parameters which will propagate as an error
on predicted quantities. Because of this, in general, one cannot predict
hadronic quantities with a relative accuracy higher than $\cO(1\%)$ from pure
QCD simulations, independently on how the target $X$ is sensitive to isospin-breaking effects. If one performs a complete lattice prediction of the physical
value of $X$, it can be of phenomenological interest to define in some way
$\bar{X}$, $X^{SU(2)}$, and $X^{\gamma}$. If we keep $m_{ud}$, $m_s$ and $m_c$
at their physical values in physical units, for a given renormalization scheme and scale, then these three quantities can be extracted by
setting successively and simultaneously $\alpha$ and $\delta m$ to $0$. This is
where the ambiguity lies: in general the $\delta m=0$ point will depend on the
renormalization scheme used for the quark masses. In the next section, we give
more details on that particular aspect and discuss the order of scheme
ambiguities.

\subsubsection{Ambiguities in the separation of isospin-breaking contributions}
In this section, we discuss the ambiguities that arise in the individual
determination of the QED contribution $X^{\gamma}$ and the strong-isospin
correction $X^{SU(2)}$ defined in the previous section. Throughout this section, we
assume that the isospin-symmetric quark masses $m_{ud}$, $m_s$ and $m_c$ are
always kept fixed in physical units to the values they take at the QCD+QED
physical point in some given renormalization scheme. Let us assume that both up
and down masses have been renormalized in an identical mass-independent scheme
which depends on some energy scale $\mu$. We also assume that the renormalization
procedure respects chiral symmetry so that quark masses renormalize
multiplicatively. The renormalization constants of the quark masses are
identical for $\alpha=0$ and therefore the renormalized mass of a quark has the
general form
\be
  m_q(\mu)=Z_m(\mu)[1+\alpha Q_{\mathrm{tot.}}^2\delta_{Z}^{(0)}(\mu)
  +\alpha Q_{\mathrm{tot.}}Q_q\delta_{Z}^{(1)}(\mu)
  +\alpha Q_q^2\delta_{Z}^{(2)}(\mu)
  ]m_{q,0}
  \co\label{eq:mqqedren}
\ee
up to $\cO(\alpha^2)$ corrections, where $m_{q,0}$ is the bare-quark mass, {\color{black} $Q_{\mathrm{tot.}}$ and $Q_{\mathrm{tot.}}^2$ are the sum of all quark charges and squared charges, respectively, and $Q_q$ is the quark charge, all in units of in units of the positron charge $e$.} Throughout this section, a subscript $ud$
generally denotes the average between up and down quantities and $\delta$ the
difference between the up and the down quantities. The source of the ambiguities
described in the previous section is the mixing of the isospin-symmetric mass
$m_{ud}$ and the difference $\delta m$ through renormalization. Using
\eq{eq:mqqedren} one can make this mixing explicit at leading order in
$\alpha$:
\be
\mat{m_{ud}(\mu)\\\delta m(\mu)}=Z_m(\mu)[1+\alpha Q_{\mathrm{tot.}}^2\delta_{Z}^{(0)}(\mu)+\alpha M^{(1)}(\mu)+\alpha M^{(2)}(\mu)]
\mat{m_{ud,0}\\\delta m_0}
\label{eq:isomixbr}
\ee
with the mixing matrices
\be
  M^{(1)}(\mu)=\delta_Z^{(1)}(\mu)Q_{\mathrm{tot.}}\mat{
  Q_{ud} & \frac14\delta Q\\
  \delta Q &Q_{ud}
  }\qquad\text{and}\qquad
  M^{(2)}(\mu)=\delta_Z^{(2)}(\mu)\mat{
  Q_{ud}^2 & \frac14\delta Q^2\\
  \delta Q^2 & Q_{ud}^2
  }\,,
\ee
where $Q_{ud}=\frac{1}{2}(Q_u+Q_d)$ and $\delta Q=Q_u-Q_d$ are the average and
difference of the up and down charges, and similarly $Q_{ud}^2=\frac{1}{2}(Q_u^2+Q_d^2)$
and $\delta Q^2=Q_u^2-Q_d^2$ for the squared charges.
Now let us assume that for the purpose of determining the different components
in \eq{eq:isoex}, one starts by tuning the bare masses to obtain equal up and
down masses, for some small coupling $\alpha_0$ at some scale $\mu_0$,
i.e., $\delta m(\mu_0)=0$. At this specific point, one can extract the pure QCD,
and the QED corrections to a given quantity $X$ by studying the slope of
$\alpha$ in \eq{eq:isoex}. From these quantities the strong-isospin contribution
can then readily be extracted using a nonzero value of $\delta m(\mu_0)$. However,
if now the procedure is repeated at another coupling $\alpha$ and scale $\mu$ with the
same bare masses, it appears from \eq{eq:isomixbr} that $\delta m(\mu)\neq 0$.
More explicitly,
\be
\delta m(\mu)=m_{ud}(\mu_0)\frac{Z_m(\mu)}{Z_m(\mu_0)}
[\alpha\Delta_Z(\mu)
-\alpha_0\Delta_Z(\mu_0)]\co\label{eq:dmamb}
\ee
with
\be
\Delta_Z(\mu)=Q_{\mathrm{tot.}}\delta Q\delta_Z^{(1)}(\mu)+\delta Q^2\delta_Z^{(2)}(\mu)\co
\ee
up to higher-order corrections in $\alpha$ and $\alpha_0$. In other words, the
definitions of $\bar{X}$, $X^{SU(2)}$, and $X^{\gamma}$ depend on the
renormalization scale at which the separation was made. This dependence, of
course, has to cancel in the physical sum $X$. One can notice that at no point did we
mention the renormalization of $\alpha$ itself, which, in principle, introduces
similar ambiguities. However, the corrections coming from the running of
$\alpha$ are $\cO(\alpha^2)$ relatively to $X$, which, as justified above, can be
safely neglected. Finally, important information is provided by \eq{eq:dmamb}:
the scale ambiguities are $\cO(\alpha m_{ud})$. For physical quark masses, one
generally has $m_{ud}\simeq \delta m$. So by using this approximation in the
first-order expansion~\eq{eq:isoex}, it is actually possible to define
unambiguously the components of $X$ up to second-order isospin-breaking
corrections. Therefore, in the rest of this review, we will not keep track of the
ambiguities in determining pure QCD or QED quantities. However, in the context
of lattice simulations, it is crucial to notice that $m_{ud}\simeq \delta m$ is
only accurate \emph{at the physical point}. In simulations at
larger-than-physical pion masses, scheme ambiguities in the separation of QCD
and QED contributions are generally large. Once more, the argument made here
assumes that the isospin-symmetric quark masses $m_{ud}$, $m_s$, and $m_c$ are
kept fixed to their physical value in a given scheme while varying $\alpha$.
Outside of this assumption there is an additional isospin-symmetric $\cO(\alpha
m_q)$ ambiguity between $\bar{X}$ and $X^{\gamma}$.

Such separation in lattice QCD+QED simulation results appeared for the
first time in RBC 07~\cite{Blum:2007cy} and Blum 10~\cite{Blum:2010ym}, where
the scheme was implicitly defined around the $\chi$PT expansion. In that setup,
the $\delta m(\mu_0)=0$ point is defined in pure QCD, i.e., $\alpha_0=0$ in the
previous discussion. The QCD part of the kaon-mass splitting from the
first FLAG review~\citep{Colangelo:2010et} is used as an input in RM123~11~\cite{deDivitiis:2011eh}, which focuses on QCD isospin corrections only. It therefore
inherits from the convention that was chosen there, which is also to set
$\delta m(\mu_0)=0$ at zero QED coupling. The same convention was used in the
follow-up works RM123~13~\citep{deDivitiis:2013xla} and
RM123~17~\citep{Giusti:2017dmp}. The BMW collaboration was the first to
introduce a purely hadronic scheme in its electro-quenched study of the baryon
octet mass splittings~\citep{Borsanyi:2013lga}. In this work, the quark mass
difference $\delta m(\mu)$ is swapped with the mass splitting $\Delta M^2$
between the connected $\bar{u}u$ and $\bar{d}d$ pseudoscalar masses. Although
unphysical, this quantity is proportional~\citep{Bijnens:2006mk} to
$\delta m(\mu)$ up to $\cO(\alpha m_{ud})$ chiral corrections. In this scheme,
the quark masses are assumed to be equal at $\Delta M^2=0$, and the $\cO(\alpha
m_{ud})$ corrections to this statement are analogous to the scale ambiguities
mentioned previously. The same scheme was used for the
determination of light-quark masses in BMW~16~\citep{Fodor:2016bgu} and in the recent BMW prediction
of the leading hadronic contribution to the muon magnetic moment~\citep{Borsanyi:2020mff}. The BMW
collaboration used a different hadronic scheme for its determination of the
nucleon-mass splitting in BMW~14~\citep{Borsanyi:2014jba} using full QCD+QED simulations.
In this work, the $\delta m=0$ point was fixed by imposing the baryon splitting
$M_{\Sigma^+}-M_{\Sigma^-}$ to cancel. This scheme is quite different from the
other ones presented here, in the sense that its intrinsic ambiguity is not
$\cO(\alpha m_{ud})$. What motivates this choice here is that
$M_{\Sigma^+}-M_{\Sigma^-}=0$ in the limit where these baryons are point
particles, so the scheme ambiguity is suppressed by the compositeness of the
$\Sigma$ baryons. This may sound like a more difficult ambiguity to quantify, but this
scheme has the advantage of being defined purely by measurable quantities.
Moreover, it has been demonstrated numerically in~BMW~14~\citep{Borsanyi:2014jba} that,
within the uncertainties of this study, the $M_{\Sigma^+}-M_{\Sigma^-}=0$ scheme
is equivalent to the $\Delta M^2=0$ one, explicitly
$M_{\Sigma^+}-M_{\Sigma^-}=-0.18(12)(6)\MeV$ at $\Delta M^2=0$. The calculation
QCDSF/UKQCD~15~\citep{Horsley:2015vla} uses a ``Dashen scheme,'' where quark
masses are tuned such that flavour-diagonal mesons have equal masses in QCD and
QCD+QED. Although not explicitly mentioned by the authors of the paper, this
scheme is simply a reformulation of the $\Delta M^2=0$ scheme mentioned
previously. Finally, MILC~18~\citep{Basak:2018yzz} also used
the $\Delta M^2=0$ scheme and noticed its connection to the ``Dashen scheme'' from
QCDSF/UKQCD~15.

Before the previous edition of this review, the contributions $\bar{X}$,
$X^{SU(2)}$, and $X^{\gamma}$ were given for pion and kaon masses based on
phenomenological information. Considerable progress has been achieved by the
lattice community to include isospin-breaking effects in calculations, and it is
now possible to determine these quantities precisely directly from a lattice
calculation. However, these quantities generally appear as intermediate products
of a lattice analysis, and are rarely directly communicated in publications.
These quantities, although unphysical, have a phenomenological interest, and we
encourage the authors of future calculations to quote them explicitly.
\subsubsection{Inclusion of electromagnetic effects in lattice-QCD simulations}
\label{sec:latticeqed}
Electromagnetism on a lattice can be formulated using a naive discretization of
the Maxwell action $S[A_{\mu}]=\frac{1}{4}\int d^4
x\,\sum_{\mu,\nu}[\partial_{\mu}A_{\nu}(x)-\partial_{\nu}A_{\mu}(x)]^2$. Even in
its noncompact form, the action remains gauge invariant. This is not the case
for non-Abelian theories for which one uses the traditional compact Wilson gauge
action (or an improved version of it). Compact actions for QED feature spurious
photon-photon interactions which vanish only in the
continuum limit. This is one of the main reason why the noncompact action is
the most popular so far. It was used in all the calculations presented in this
review. Gauge-fixing is necessary for noncompact actions {\color{black} because of the usual infinite measure of equivalent gauge orbits which contribute to the path integral}. It was shown~\citep{Hansen:2018zre,Lucini:2015hfa} that gauge-fixing is not necessary with compact actions, including in the construction of interpolating operators for charged states. 

Although discretization is straightforward, simulating QED in a finite volume is
more challenging. Indeed, the long range nature of the interaction suggests
that important finite-size effects have to be expected. In the case of periodic
boundary conditions, the situation is even more critical: a naive implementation
of the theory features an isolated zero-mode singularity in the photon
propagator. It was first proposed in~\citep{Duncan:1996xy} to fix the global
zero-mode of the photon field $A_{\mu}(x)$ in order to remove it from the
dynamics. This modified theory is generally named $\mathrm{QED}_{\mathrm{TL}}$.
Although this procedure regularizes the theory and has the right classical
infinite-volume limit, it is nonlocal because of the zero-mode fixing. As
first discussed in~\citep{Borsanyi:2014jba}, the nonlocality in time of
$\mathrm{QED}_{\mathrm{TL}}$ prevents the existence of a transfer matrix, and
therefore a quantum-mechanical interpretation of the theory. Another
prescription named $\mathrm{QED}_{\mathrm{L}}$, proposed
in~\citep{Hayakawa:2008an}, is to remove the zero-mode of $A_{\mu}(x)$
independently for each time slice. This theory, although still
nonlocal in space, is local in time and has a well-defined transfer matrix.
Whether these nonlocalities constitute an issue to extract infinite-volume
physics from lattice-QCD+$\mathrm{QED}_{\mathrm{L}}$ simulations is, at the time
of this review, still an open question. However, it is known through analytical
calculations of electromagnetic finite-size effects at $\cO(\alpha)$ in hadron
masses~\citep{Hayakawa:2008an,deDivitiis:2013xla,Davoudi:2014qua,Borsanyi:2014jba,Fodor:2015pna,Tantalo:2016vxk,Davoudi:2018qpl},
meson leptonic decays~\citep{Tantalo:2016vxk}, and the hadronic vacuum
polarization~\citep{Bijnens:2019ejw} that $\mathrm{QED}_{\mathrm{L}}$ does not
suffer from a problematic (e.g., UV divergent) coupling of short- and
long-distance physics due to its nonlocality. Another strategy, first proposed
in~\citep{Gockeler:1989wj} and used by the QCDSF collaboration, is to bound the
zero-mode fluctuations to a finite range. Although more minimal, it is still
a nonlocal modification of the theory and so far finite-size effects for this
scheme have not been investigated. More recently, two proposals for local
formulations of finite-volume QED emerged. The first one described
in~\citep{Endres:2015gda} proposes to use massive photons to regulate zero-mode
singularities, at the price of (softly) breaking gauge invariance. The second
one presented in~\citep{Lucini:2015hfa}, based on earlier works~\cite{Wiese:1991ku,Polley:1993bn}, avoids the zero-mode issue by using
anti-periodic boundary conditions for $A_{\mu}(x)$. In this approach, gauge
invariance requires the fermion field to undergo a charge conjugation
transformation over a period, breaking electric charge conservation. These local
approaches have the potential to constitute cleaner approaches to finite-volume
QED. All the calculations presented in this
review used $\mathrm{QED}_{\mathrm{L}}$ or $\mathrm{QED}_{\mathrm{TL}}$, with the
exception of QCDSF.

Once a finite-volume theory for QED is specified, there are various ways to
compute QED effects themselves on a given hadronic quantity. The most direct
approach, first used in~\citep{Duncan:1996xy}, is to include QED directly in the
lattice simulations and assemble correlation functions from charged quark
propagators. Another approach proposed in~\citep{deDivitiis:2013xla}, is to
exploit the perturbative nature of QED, and compute the leading-order
corrections directly in pure QCD as matrix elements of the electromagnetic
current. Both approaches have their advantages and disadvantages and as shown
in~\citep{Giusti:2017dmp}, are not mutually exclusive. A critical comparative
study can be found in~\citep{Boyle:2017gzv}.

Finally, most of the calculations presented here made the choice of computing
electromagnetic corrections in the electro-quenched approximation. In this
limit, one assumes that only valence quarks are charged, which is equivalent to
neglecting QED corrections to the fermionic determinant. This approximation reduces
dramatically the cost of lattice-QCD+QED calculations since it allows the reuse of previously generated QCD configurations. If QED is introduced pertubatively through current insertions, the electro-quenched approximation avoids computing disconnected contributions coming from the electromagnetic current in the vacuum, which are generally challenging to determine precisely.
The electromagnetic contributions from sea quarks to hadron-mass splittings are known to be flavour-$SU(3)$ 
and large-$N_c$ suppressed, thus electro-quenched simulations are
expected to have an $\cO(10\%)$ accuracy for the leading electromagnetic effects.
This suppression is in principle rather weak and results obtained from
electro-quenched simulations might feature uncontrolled systematic errors. For
this reason, the use of the electro-quenched approximation constitutes the
difference between \good~and \soso~in the FLAG criterion for the inclusion of
isospin-breaking effects.

\subsubsection{Lattice determination of $m_s$ and $m_{ud}$}
\label{sec:msmud}

We now turn to a review of the lattice calculations of the light-quark
masses and begin with $m_s$, the isospin-averaged up- and down-quark
mass $m_{ud}$, and their ratio. Most groups quote only $m_{ud}$, not
the individual up- and down-quark masses. We then discuss the ratio
$m_u/m_d$ and the individual determinations of $m_u$ and $m_d$.

Quark masses have been calculated on the lattice since the
mid-nineties. However, early calculations were performed in the quenched
approximation, leading to unquantifiable systematics. Thus, in the following,
we only review modern, unquenched calculations, which include the effects of
light sea quarks.

Tables~\ref{tab:masses3}~and~\ref{tab:masses4} list
the results of $\Nf=2+1$ and $\Nf=2+1+1$ lattice calculations
of $m_s$ and $m_{ud}$. These results are given in the $\msbar$ scheme
at $2\,\gev$, which is standard nowadays, though some groups are
starting to quote results at higher scales
(e.g.,~Ref.~\cite{Arthur:2012opa}). The tables also show the colour coding
of the calculations leading to these results. As indicated earlier in
this review, we treat calculations with different numbers, $N_f$, of
dynamical quarks separately.

\bigskip
\noindent
{\em $\Nf=2+1$ lattice calculations}
\medskip

We turn now to $\Nf=2+1$ calculations. These and the corresponding
results for $m_{ud}$ and $m_s$ are summarized in
Tab.~\ref{tab:masses3}. Given the very high precision of a number of
the results, with total errors on the order of 1\%, it is important to
consider the effects neglected in these calculations.  Isospin-breaking 
and electromagnetic\ effects are small on $m_{ud}$ and $m_s$, and have
been approximately accounted for in the calculations that will be
retained for our averages. We have already commented that the effect
of the omission of the charm quark in the sea is expected to be small,
below our current precision, and we do not add any additional uncertainty due to these
effects in the final averages. 

\begin{table}[!ht]
\vspace{2mm}
{\footnotesize{
\begin{tabular*}{\textwidth}[l]{l@{\extracolsep{\fill}}rllllllll}
Collaboration & Ref. & \hspace{0.15cm}\begin{rotate}{60}{publication status}\end{rotate}\hspace{-0.15cm} &
 \hspace{0.15cm}\begin{rotate}{60}{chiral extrapolation}\end{rotate}\hspace{-0.15cm} &
 \hspace{0.15cm}\begin{rotate}{60}{continuum  extrapolation}\end{rotate}\hspace{-0.15cm}  &
 \hspace{0.15cm}\begin{rotate}{60}{finite volume}\end{rotate}\hspace{-0.15cm}  &  
 \hspace{0.15cm}\begin{rotate}{60}{renormalization}\end{rotate}\hspace{-0.15cm} &  
 \hspace{0.15cm}\begin{rotate}{60}{running}\end{rotate}\hspace{-0.15cm}  & 
\rule{0.6cm}{0cm}$m_{ud} $ & \rule{0.6cm}{0cm}$m_s $ \\
&&&&&&&&& \\[-0.1cm]
\hline
\hline
&&&&&&&&& \\[-0.1cm]
{ALPHA 19}& \cite{Bruno:2019vup} & \gA & \soso & \good & \good &
\good & $e$  & 3.54(12)(9)  & 95.7(2.5)(2.4)\\

{Maezawa 16}& \cite{Maezawa:2016vgv} & \gA & \bad & \good & \good &
\good & $d$  & --  & 92.0(1.7)\\

{RBC/UKQCD 14B$^\ominus$}& \cite{Blum:2014tka} & \gA & \good & \good & \good &
\good & $d$  & 3.31(4)(4)  & 90.3(0.9)(1.0)\\

{RBC/UKQCD 12$^\ominus$}& \cite{Arthur:2012opa} & \gA & \good & \soso & \good &
\good & $d$  &  3.37(9)(7)(1)(2) & 92.3(1.9)(0.9)(0.4)(0.8)\\

{PACS-CS 12$^\star$}& \protect{\cite{Aoki:2012st}} & \gA & \good & \bad & \bad & \good & $\,b$
&  3.12(24)(8) &  83.60(0.58)(2.23) \\

{Laiho 11} & \cite{Laiho:2011np} & \rC & \soso & \good & \good & \soso
& $-$ & 3.31(7)(20)(17)
& 94.2(1.4)(3.2)(4.7)\\

{BMW 10A, 10B$^+$} & \cite{Durr:2010vn,Durr:2010aw} & \gA & \good & \good & \good & \good &
$\,c$ & 3.469(47)(48)& 95.5(1.1)(1.5)\\

{PACS-CS 10}& \cite{Aoki:2010wm} & \gA & \good & \bad & \bad & \good & $\,b$
&  2.78(27) &  86.7(2.3) \\

{MILC 10A}& \cite{Bazavov:2010yq} & \rC & \soso  & \good & \good &
\soso  &$-$& 3.19(4)(5)(16)&\rule{0.6cm}{0cm}-- \\

{HPQCD~10$^{\ast\ast}$}&  \cite{McNeile:2010ji} &\gA & \soso & \good & \good & $-$
&$-$& 3.39(6)$ $ & 92.2(1.3) \\

{RBC/UKQCD 10A}& \cite{Aoki:2010dy} & \gA & \soso & \soso & \good &
\good & $\,a$  &  3.59(13)(14)(8) & 96.2(1.6)(0.2)(2.1)\\

{Blum~10$^\dagger$}&\cite{Blum:2010ym}& \gA & \soso & \bad & \soso & \good &
$-$ &3.44(12)(22)&97.6(2.9)(5.5)\\

{PACS-CS 09}& \cite{Aoki:2009ix}& \gA &\good   &\bad   & \bad & \good  &  $\,b$
 & 2.97(28)(3) &92.75(58)(95)\\

{HPQCD 09A$^\oplus$}&  \cite{Davies:2009ih}&\gA & \soso & \good & \good & $-$
& $-$& 3.40(7) & 92.4(1.5) \\

{MILC 09A} & \cite{Bazavov:2009fk} & \rC &  \soso & \good & \good & \soso &
$-$ & 3.25 (1)(7)(16)(0) & 89.0(0.2)(1.6)(4.5)(0.1)\\

{MILC 09} & \cite{Bazavov:2009bb} & \gA & \soso & \good & \good & \soso & $-$
& 3.2(0)(1)(2)(0) & 88(0)(3)(4)(0)\\

{PACS-CS 08} & \cite{Aoki:2008sm} &  \gA & \good & \bad & \bad  & \bad & $-$ &
2.527(47) & 72.72(78)\\

{RBC/UKQCD 08} & \cite{Allton:2008pn} & \gA & \soso & \bad & \good & \good &
$-$ &$3.72(16)(33)(18)$ & $107.3(4.4)(9.7)(4.9)$\\

\hspace{-0.2cm}{\begin{tabular}{l}CP-PACS/\\JLQCD 07\end{tabular}} 
& \cite{Ishikawa:2007nn}& \gA & \bad & \good & \good  & \bad & $-$ &
$3.55(19)(^{+56}_{-20})$ & $90.1(4.3)(^{+16.7}_{-4.3})$ \\

{HPQCD 05}
& 
\cite{Mason:2005bj}& \gA & \soso & \soso & \soso & \soso &$-$&
$3.2(0)(2)(2)(0)^\ddagger$ & $87(0)(4)(4)(0)^\ddagger$\\

\hspace{-0.2cm}{\begin{tabular}{l}MILC 04, HPQCD/\\MILC/UKQCD 04\end{tabular}} 
& \cite{Aubin:2004fs,Aubin:2004ck} & \gA & \soso & \soso & \soso & \bad & $-$ &
$2.8(0)(1)(3)(0)$ & $76(0)(3)(7)(0)$\\
&&&&&&&&& \\[-0.1cm]
\hline
\hline\\
\end{tabular*}\\[-0.2cm]
}}
\begin{minipage}{\linewidth}
{\footnotesize 
\begin{itemize}
\item[$^\ominus$] The results are given in the $\msbar$ scheme at 3
  instead of 2~GeV. We run them down to 2~GeV using numerically
  integrated 4-loop
  running~\cite{vanRitbergen:1997va,Chetyrkin:1999pq} with $N_f=3$ and
  with the values of $\alpha_s(M_Z)$, $m_b$, and $m_c$ taken
  from~Ref.~\cite{Agashe:2014kda}. The running factor is 1.106. At
  three loops it is only 0.2\% smaller, indicating that perturbative
  running uncertainties are small. We neglect them here.\\[-5mm]
\item[$^\star$] The calculation includes electromagnetic and $m_u\ne m_d$ effects
  through reweighting.\\ [-5mm]
\item[$^+$] The fermion action used is tree-level improved.\\[-5mm]
\item[$^{\ast\ast}$] $m_s$ is obtained by combining $m_c$ and
      HPQCD 09A's $m_c/m_s=11.85(16)$~\cite{Davies:2009ih}.
      Finally, $m_{ud}$
      is determined from $m_s$ with the MILC 09 result for
      $m_s/m_{ud}$. Since $m_c/m_s$ is renormalization group invariant
      in QCD, the renormalization and running of the quark masses
      enter indirectly through that of $m_c$ (see below).\\[-5mm]
\item[$^\dagger$] The calculation includes quenched electromagnetic effects.\\[-5mm]
\item[$^\oplus$] What is calculated is $m_c/m_s=11.85(16)$. $m_s$ is then obtained by combining
      this result with the determination $m_c(m_c) = 1.268(9)$~GeV
      from~Ref.~\cite{Allison:2008xk}. Finally, $m_{ud}$
      is determined from $m_s$ with the MILC 09 result for
      $m_s/m_{ud}$.\\[-5mm]
\item[$^\ddagger$] The bare numbers are those of MILC 04. The masses are simply rescaled, using the
ratio of the 2-loop to 1-loop renormalization factors.\\[-5mm]
\item[$a$] The masses are renormalized nonperturbatively at a scale of
  2~GeV in a couple of $N_f=3$ RI-SMOM schemes. A careful study of
  perturbative matching uncertainties has been performed by comparing
  results in the two schemes in the region of 2~GeV to 3~GeV~\cite{Aoki:2010dy}.\\[-5mm]
\item[$b$] The masses are renormalized and run nonperturbatively up to
  a scale of $40\,\gev$ in the $N_f=3$ SF scheme. In this scheme,
  nonperturbative and NLO running for the quark masses are shown to
  agree well from 40 GeV all the way down to 3 GeV~\cite{Aoki:2010wm}.\\[-5mm]
\item[$c$] The masses are renormalized and run nonperturbatively up to
  a scale of 4 GeV in the $N_f=3$ RI-MOM scheme.  In this scheme,
  nonperturbative and N$^3$LO running for the quark masses are shown
  to agree from 6~GeV down to 3~GeV to better than 1\%~\cite{Durr:2010aw}.  \\[-5mm]
\item[$d$] All required running is performed nonperturbatively.
\item[$e$] Running is performed nonperturbatively from 200 MeV to the
  electroweak scale $\sim 100$ GeV.
\end{itemize}
}
\end{minipage}
\caption{$\Nf=2+1$ lattice results for the masses $m_{ud}$ and $m_s$ (MeV).} 
\label{tab:masses3}
\end{table}

The only new computation since the previous FLAG edition is the
determination of light-quark masses by the ALPHA
collaboration~\cite{Bruno:2019vup}. This work uses nonperturbatively
$\mathcal O(a)$ improved Wilson 
fermions (a subset of the CLS ensembles~\cite{Bruno:2014jqa}). 
The renormalization is performed nonperturbatively in the SF scheme
from 200 MeV up to the electroweak scale $\sim 100$
GeV~\cite{Campos:2018ahf}. This nonperturbative running over such
large energy scales avoids any use of perturbation theory at
low energy scales, but adds a cost in terms of uncertainty: the
running alone propagates to $\approx 1\%$ of the error in quark
masses. This turns out to be one of the dominant pieces of uncertainty
for the case of $m_s$. 
On the other hand, for the case of $m_{ud}$, the uncertainty is
dominated by the chiral extrapolations. 
The ensembles used include four values of the lattice spacing below
$0.09$ fm, which qualifies for a $\good$ in the continuum
extrapolation, and pion masses down to 200 MeV. 
This value lies just at the boundary of the $\good$ rating, but since
the chiral extrapolation is a substantial source of systematic
uncertainty, we opted to rate the work with a $\soso$. 
In any case, this work enters in the average and their results show a
reasonable agreement with the FLAG average. 

We now comment in some detail on previous works that also contribute
to the averages. 

RBC/UKQCD~14~\cite{Blum:2014tka} significantly improves on
their RBC/UKQCD~12B~\cite{Arthur:2012opa} work by adding three new
domain wall fermion simulations to three used previously. Two of the
new simulations are performed at essentially physical pion masses
($M_\pi\simeq 139\,\mev$) on lattices of about $5.4\,\fm$ in size and
with lattice spacings of $0.114\,\fm$ and $0.084\,\fm$. It is
complemented by a third simulation with $M_\pi\simeq 371\,\mev$,
$a\simeq 0.063$ fm and a rather small $L\simeq 2.0\,\fm$. Altogether,
this gives them six simulations with six unitary ($m_{\rm sea}=m_{\rm val}$) $M_\pi$'s in the
range of $139$ to $371\,\mev$, and effectively three lattice spacings
from $0.063$ to $0.114\,\fm$. They perform a combined global continuum
and chiral fit to all of their results for the $\pi$ and $K$ masses
and decay constants, the $\Omega$ baryon mass and two Wilson-flow
parameters.  Quark masses in these fits are renormalized and run
nonperturbatively in the RI-SMOM scheme. This is done by computing the
relevant renormalization constant for a reference ensemble, and
determining those for other simulations relative to it by adding
appropriate parameters in the global fit. This calculation passes
all of our selection criteria.

$\Nf=2+1$ MILC results for light-quark masses go back to
2004~\cite{Aubin:2004fs,Aubin:2004ck}. They use rooted staggered
fermions.  By 2009 their simulations covered an impressive range of
parameter space, with lattice spacings going down to 0.045~fm, and
valence-pion masses down to approximately
180~MeV~\cite{Bazavov:2009fk}.  The most recent MILC $\Nf=2+1$
results, i.e., MILC 10A~\cite{Bazavov:2010yq} and MILC
09A~\cite{Bazavov:2009fk}, feature large statistics and 2-loop
renormalization.  Since these data sets subsume those of their
previous calculations, these latest results are the only ones that
need to be kept in any world average.

The BMW 10A, 10B~\cite{Durr:2010vn,Durr:2010aw}
calculation still satisfies our stricter selection criteria. They
reach the physical up- and down-quark mass
by {\it interpolation} instead of by extrapolation. Moreover, their
calculation was performed at five lattice spacings ranging from 0.054
to 0.116~fm, with full nonperturbative renormalization and running
and in volumes of up to (6~fm)$^3$, guaranteeing that the continuum
limit, renormalization, and infinite-volume extrapolation are
controlled. It does neglect, however, isospin-breaking effects, which
are small on the scale of their error bars.

Finally, we come to another calculation which satisfies our selection
criteria, HPQCD~10 \cite{McNeile:2010ji}. It updates the staggered-fermions 
calculation of HPQCD~09A~\cite{Davies:2009ih}. In these
papers, the renormalized mass of the strange quark is obtained by
combining the result of a precise calculation of the renormalized
charm-quark mass, $m_c$, with the result of a calculation of the
quark-mass ratio, $m_c/m_s$. As described in Ref.~\cite{Allison:2008xk} and
in Sec.~\ref{s:cmass}, HPQCD determines $m_c$ by fitting
Euclidean-time moments of the $\bar cc$ pseudoscalar density two-point
functions, obtained numerically in lattice QCD, to fourth-order,
continuum perturbative expressions. These moments are normalized and
chosen so as to require no renormalization with staggered
fermions. Since $m_c/m_s$ requires no renormalization either, HPQCD's
approach displaces the problem of lattice renormalization in the
computation of $m_s$ to one of computing continuum perturbative
expressions for the moments. To calculate $m_{ud}$
HPQCD~10~\cite{McNeile:2010ji} use the MILC 09 determination of the
quark-mass ratio $m_s/m_{ud}$~\cite{Bazavov:2009bb}.

HPQCD~09A~\cite{Davies:2009ih} obtains
$m_c/m_s=11.85(16)$~\cite{Davies:2009ih} fully nonperturbatively,
with a precision slightly larger than 1\%. HPQCD~10's determination of the
charm-quark mass, $m_c(m_c)=1.268(6)$,\footnote{To obtain this number, 
we have used the conversion from $\mu=3\,$ GeV to $m_c$ given in Ref.~\cite{Allison:2008xk}.} is even more precise, achieving an accuracy
better than 0.5\%.

This discussion leaves us with five results for our final average for
$m_s$: 
ALPHA~19~\cite{Bruno:2019vup},
MILC~09A~\cite{Bazavov:2009fk}, BMW~10A, 
10B~\cite{Durr:2010vn,Durr:2010aw}, HPQCD~10~\cite{McNeile:2010ji} and
RBC/UKQCD~14~\cite{Blum:2014tka}. Assuming that the result from HPQCD~10 is
100\% correlated with that of MILC~09A, as it is based on a subset of the MILC~09A
configurations, we find $m_s=92.2(1.1)\,\mev$ with a $\chi^2/$dof = 1.65.

For the light-quark mass $m_{ud}$, the results satisfying our criteria
are ALPHA~19, RBC/UKQCD 14B, BMW 10A, 10B, HPQCD 10, and MILC 10A. For the
error, we include the same 100\% correlation between statistical
errors for the latter two as for the strange case, resulting in
the following (at scale 2 GeV in the $\overline{\rm MS}$ scheme, and $\chi^2/$dof=1.4),
%
\begin{align}\label{eq:nf3msmud}
&& \FLAGAVBEGIN m_{ud}&= 3.381(40)\FLAGAVEND\;\mev&&\Refs~\mbox{\cite{Bruno:2019vup,Blum:2014tka,Durr:2010vn,Durr:2010aw,McNeile:2010ji,Bazavov:2010yq}},\,\nonumber \\[-3mm]
&\Nf=2+1 :&\\[-3mm]
&&\FLAGAVBEGIN m_s    &=92.2(1.0)\FLAGAVEND\;\mev&&\Refs~\mbox{\cite{Bruno:2019vup,Bazavov:2009fk,Durr:2010vn,Durr:2010aw,McNeile:2010ji,Blum:2014tka}}, \nonumber
\end{align}
%
and the RGI values
\begin{align}\label{eq:nf3msmud rgi}
&&  M_{ud}^{\rm RGI}&= 4.695(56)_{m}(54)_{\Lambda}\;\mev&&\Refs~\mbox{\cite{Bruno:2019vup,Blum:2014tka,Durr:2010vn,Durr:2010aw,McNeile:2010ji,Bazavov:2010yq}},\,\nonumber \\[-3mm]
&\Nf=2+1 :&\\[-3mm]
&& M_s^{\rm RGI}    &=128.1(1.4)_{m}(1.5)_{\Lambda}\;\mev&&\Refs~\mbox{\cite{Bruno:2019vup,Bazavov:2009fk,Durr:2010vn,Durr:2010aw,McNeile:2010ji,Blum:2014tka}}. \nonumber
\end{align}

\bigskip
\noindent
{\em $\Nf=2+1+1$ lattice calculations}
\medskip


{\color{black}
Since the previous review a new computation of $m_s, m_{ud}$ has appeared, ETM 21A~\cite{Alexandrou:2021gqw}. Using twisted-mass fermions with an added clover-term to suppress $\cO(a^2)$ effects between the neutral and charged pions, this work represents a significant improvement over ETM 14~\cite{Carrasco:2014cwa}. 
Renormalization is performed nonperturbatively in the RI-MOM scheme. 
Their ensembles comprise three lattice spacings (0.095, 0.082, and
0.069 fm), two volumes for the finest lattice spacings with
pion masses reaching down to the physical point in the two finest
lattices allowing a controlled chiral extrapolation. 
Their volumes are large, with $m_\pi L$ between four and five. These
characteristics of their ensembles pass the most stringent FLAG criteria
in all categories. 
This work extracts quark masses from two different quantities, one based on
the meson spectrum and the other based on the baryon
spectrum. Results obtained with these two methods agree within errors. The latter agrees well with the FLAG average while the former is high in comparison
(there is good agreement with their previous results, ETM 14~\cite{Carrasco:2014cwa}). Since ETM 21A was not published by the FLAG deadline, it is not included in the averages.
}

There are three other works that enter in light-quark mass averages: 
FNAL/MILC/ TUMQCD~18~\cite{Bazavov:2018omf} (which contributes both to
the average of $m_{ud}$ and $ m_s$), and the $m_{ud}$ determinations in
HPQCD 18~\cite{Lytle:2018evc} and HPQCD
14A~\cite{Chakraborty:2014aca}.  

While the results of HPQCD 14A and HPQCD 18 agree well (using
different methods), there are several tensions in the determination of $m_s$. 
The most significant discrepancy is between ETM 21A and the FLAG
average.
But also two recent and very precise determinations (HPQCD 18 and
FNAL/MILC/TUMQCD~18) show a tension. 
Overall there is a rough agreement between
the different determinations with $\chi^2/{\rm dof} = 1.7$ (that we
apply to our average according to the standard FLAG averaging procedure).
In the case of $m_{ud}$ on the other hand only two works contribute
to the average: ETM 14 and FNAL/MILC/TUMQCD~18. 
They disagree, with the FNAL/MILC/TUMQCD~18 value basically matching
the $N_{f}=2+1$ result. 
The large $\chi^2/{\rm dof} \approx 1.7$ increases significantly the
error of the average. 
These large values of the $\chi^2$ are difficult to understand in
terms of a statistical fluctuation. 
On the other hand the $N_{f}=2+1$ and $N_{f}=2+1+1$ averages
show a good agreement, which increases our confidence in the averages
quoted below. 

The $\Nf=2+1+1$ results are summarized in 
Tab.~\ref{tab:masses4}. Note that the results of
Ref.~\cite{Chakraborty:2014aca} are reported as $m_s(2\,\gev;N_f=3)$
and those of Ref.~\cite{Carrasco:2014cwa} as
$m_{ud(s)}(2\,\gev;N_f=4)$. We convert the former to $N_f=4$ and
obtain $m_s(2\,\gev;N_f=4)=93.7(8)\mev$. The average of
FNAL/MILC/TUMQCD~18, HPQCD 18, ETM 14 and HPQCD 14A
is 93.43(70)$\mev$ with $\chi^2/\mbox{dof}=1.7$. 
For the light-quark average we use ETM 14 and FNAL/MILC/TUMQCD 18
with an average 3.410(43)$\mev$ and a $\chi^2/\mbox{dof}=1.7$.
We note these $\chi^2$ values are large. For the case of the
light-quark masses this is mostly due to ETM 14 
masses lying significantly above the rest, but in the case of $m_s$
there is also some tension between the recent and very precise
results of HPQCD~18 and FNAL/MILC/TUMQCD~18. Also note that the
2+1-flavour values are consistent with the four-flavour ones, so in
all cases we have decided to simply quote averages according to FLAG
rules, including stretching factors for the errors based on $\chi^2$
values of our fits: 
%
\begin{align}\label{eq:nf4msmud}
&&\FLAGAVBEGIN m_{ud}&= 3.410(43)\FLAGAVEND\;\mev&& \Refs~\mbox{\cite{Carrasco:2014cwa,Bazavov:2018omf}},\nonumber\\[-3mm]
&\Nf=2+1+1 :& \\[-3mm]
&&\FLAGAVBEGIN m_s   &= 93.40(57)\FLAGAVEND\; \mev&& \Refs~\mbox{\cite{Carrasco:2014cwa,Bazavov:2018omf,Lytle:2018evc,Chakraborty:2014aca}},\nonumber
\end{align}
%
and the RGI values
\begin{align}\label{eq:nf4msmud rgi}
&& M_{ud}^{\rm RGI}&= 4.736(60)_{m}(55)_{\Lambda} \,\mev&& \Refs~\mbox{\cite{Bazavov:2018omf,Carrasco:2014cwa}},\nonumber\\[-3mm]
&\Nf=2+1+1 :& \\[-3mm]
&& M_s^{\rm RGI}   &=129.7(0.8)_{m}(1.5)_{\Lambda}\,\mev&& \Refs~\mbox{\cite{Bazavov:2018omf,Lytle:2018evc,Carrasco:2014cwa,Chakraborty:2014aca}}.\nonumber
\end{align}

\begin{table}[!htb]
\vspace{2.5cm}
{\footnotesize{
\begin{tabular*}{\textwidth}[l]{l@{\extracolsep{\fill}}rllllllll}
Collaboration & Ref. & \hspace{0.15cm}\begin{rotate}{60}{publication status}\end{rotate}\hspace{-0.15cm} &
 \hspace{0.15cm}\begin{rotate}{60}{chiral extrapolation}\end{rotate}\hspace{-0.15cm} &
 \hspace{0.15cm}\begin{rotate}{60}{continuum  extrapolation}\end{rotate}\hspace{-0.15cm}  &
 \hspace{0.15cm}\begin{rotate}{60}{finite volume}\end{rotate}\hspace{-0.15cm}  &  
 \hspace{0.15cm}\begin{rotate}{60}{renormalization}\end{rotate}\hspace{-0.15cm} &  
 \hspace{0.15cm}\begin{rotate}{60}{running}\end{rotate}\hspace{-0.15cm}  & 
\rule{0.6cm}{0cm}$m_{ud} $ & \rule{0.6cm}{0cm}$m_s $ \\
&&&&&&&&& \\[-0.1cm]
\hline
\hline
&&&&&&&&& \\[-0.1cm]
{ETM 21A}& \cite{Alexandrou:2021gqw} & \oP & \good & \good & \good &
\good & $-$  & $3.636(66)(^{+60}_{-57})$  & $98.7(2.4)(^{+4.0}_{-3.2})$\\

{HPQCD 18}$^\dagger$ & \cite{Lytle:2018evc} & \gA & \good & \good & \good &
$\good$ & $-$  &  & 94.49(96) \\

{FNAL/MILC/TUMQCD 18}& \cite{Bazavov:2018omf} & \gA & \good & \good & \good &
\good & $-$  & 3.404(14)(21)  & 92.52(40)(56)\\

{HPQCD 14A $^\oplus$} & \cite{Chakraborty:2014aca} & \gA & \good & \good & \good &
$-$ & $-$  &  & 93.7(8) \\
  
{ETM 14$^\oplus$}& \cite{Carrasco:2014cwa} & \gA & \soso & \good & \good &
\good & $-$  & 3.70(13)(11)  & 99.6(3.6)(2.3)\\

&&&&&&&&& \\[-0.1cm] 
\hline
\hline\\[-2mm]
\end{tabular*}
}}
\begin{minipage}{\linewidth}
{\footnotesize 
  \begin{itemize}
  \item[$^\dagger$] Bare-quark masses are renormalized nonperturbatively in
    the RI-SMOM scheme at scales $\mu\sim 2-5$ GeV for different
    lattice spacings and translated to the $\overline{\rm MS}$
    scheme. Perturbative running is then used to run all results to a
    reference scale $\mu = 3$ GeV.
\item[$^\oplus$] As explained in the text, $m_s$ is obtained by combining the
        results $m_c(5\,\gev;N_f=4)=0.8905(56)$~GeV and
        $(m_c/m_s)(N_f=4)=11.652(65)$, determined on the same data
        set. A subsequent scale and scheme conversion, performed by
        the authors, leads to the value 93.6(8). In the table, we have converted this
        to $m_s(2\,\gev;N_f=4)$, which makes a very small change. 
\end{itemize}
}
\end{minipage}

\caption{$\Nf=2+1+1$ lattice results for the masses $m_{ud}$ and $m_s$ (MeV).} 
\label{tab:masses4}
\end{table}

\begin{figure}[!htb]
\begin{center}
\includegraphics[width=11.5cm]{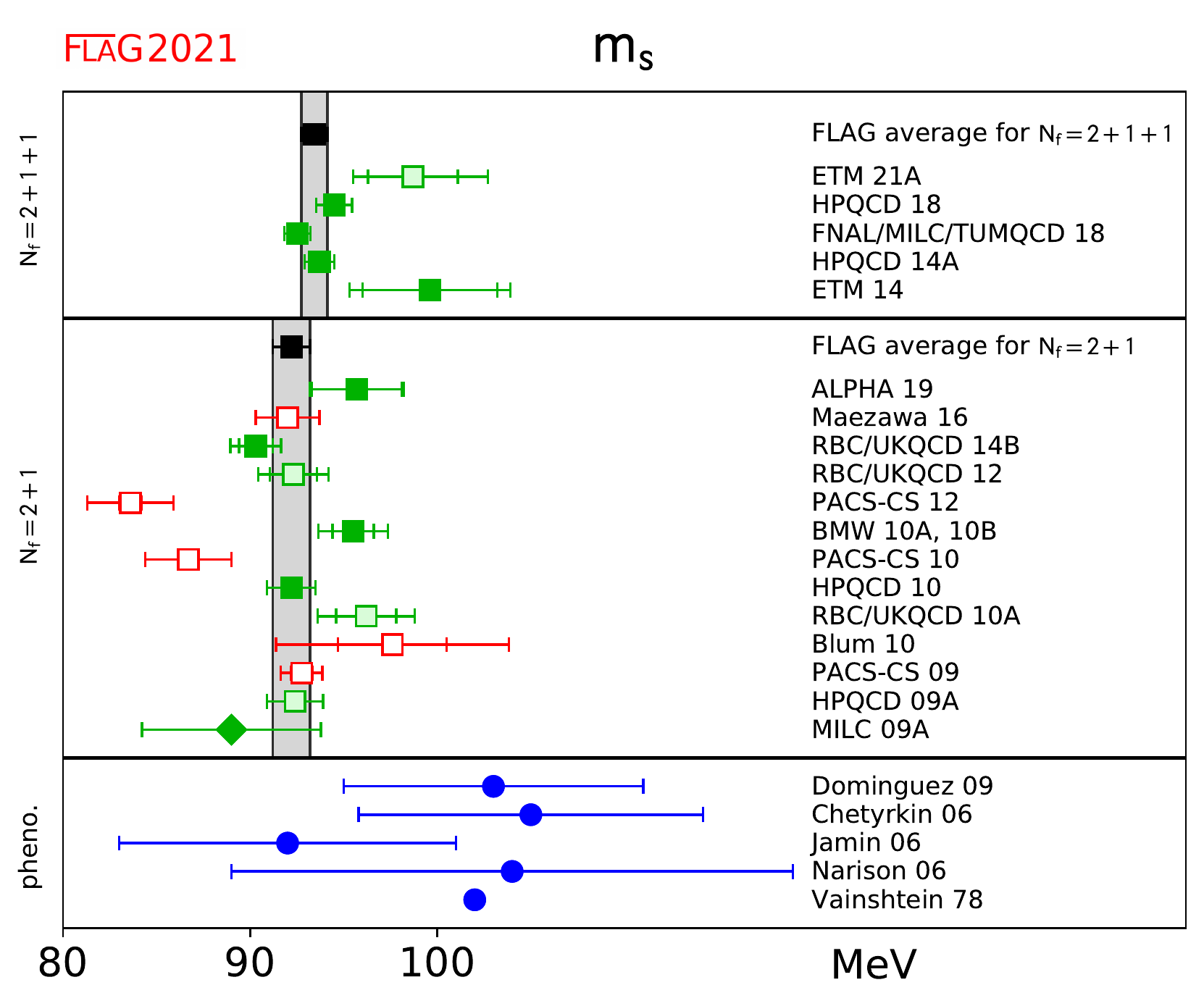}
\end{center}
\vspace{0.2cm}
\begin{center}
\caption{ \label{fig:ms} $\msbar$ mass of the strange quark (at 2 GeV scale) in MeV. 
 The upper two panels show the lattice results 
  listed in Tabs.~\ref{tab:masses3} and \ref{tab:masses4}, while 
	the bottom panel collects  sum rule 
	results~\cite{Dominguez:2008jz, Chetyrkin:2005kn,Jamin:2006tj, Narison:2005ny, Vainshtein:1978nn}.
  Diamonds and squares represent results based on perturbative and nonperturbative
  renormalization, respectively. 
 The black squares and the grey bands represent our averages (\ref{eq:nf3msmud}) and (\ref{eq:nf4msmud}). The significance of the colours is explained in Sec.~\ref{sec:qualcrit}.
}\end{center}

\end{figure}

\begin{figure}[!htb]

\begin{center}
\includegraphics[width=11.5cm]{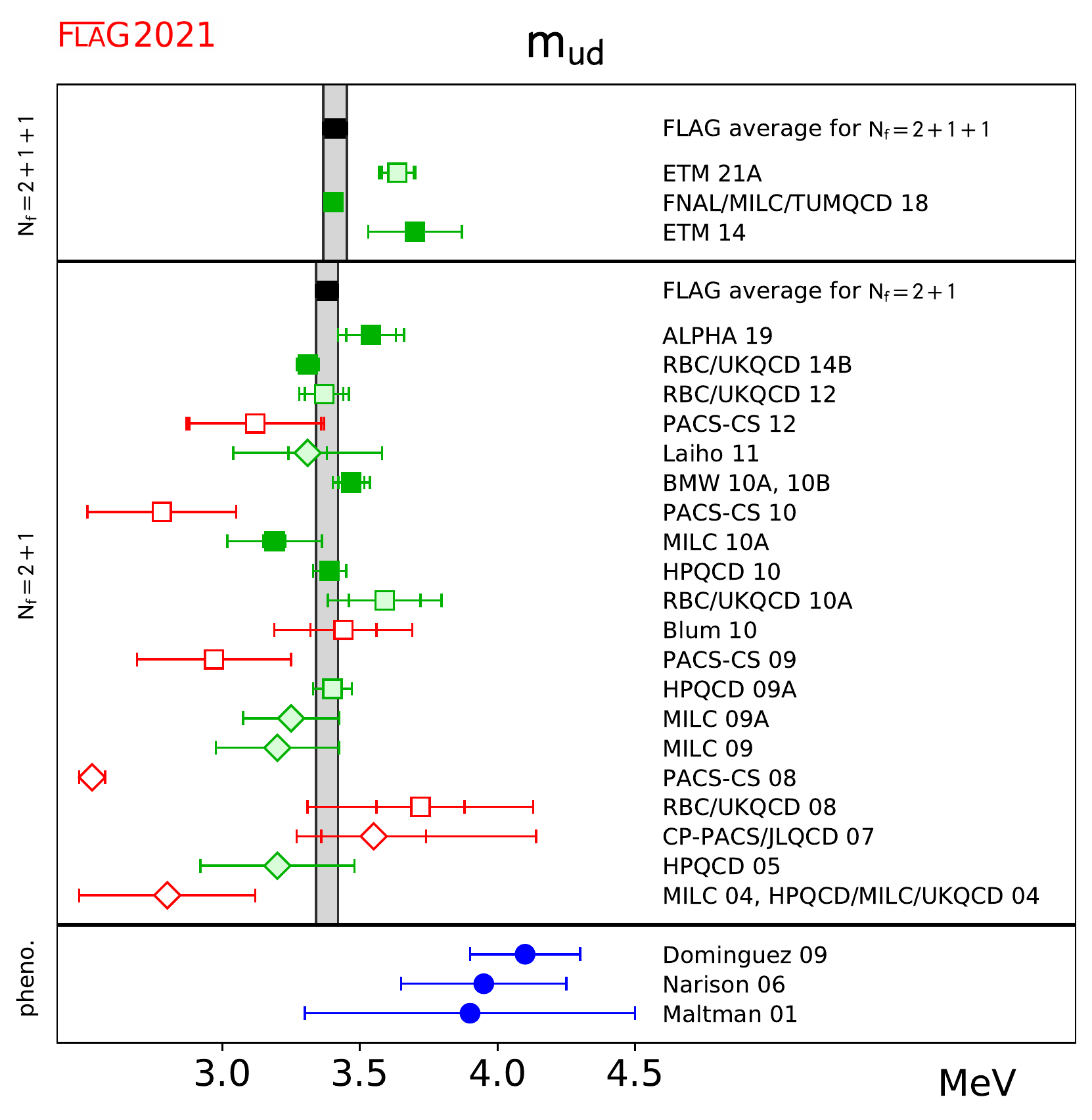}
\end{center}
\begin{center}
\caption{ \label{fig:mud} Mean mass of the two lightest quarks,
 $m_{ud}=\frac{1}{2}(m_u+m_d)$. The bottom panel shows 
 results based on sum rules~\cite{Dominguez:2008jz,Narison:2005ny,Maltman:2001nx} (for more details see Fig.~\ref{fig:ms}).}\end{center}

\end{figure}

In Figs.~\ref{fig:ms} and \ref{fig:mud} the lattice results listed in Tabs.~\ref{tab:masses3} and \ref{tab:masses4} and the FLAG averages obtained at each value of $N_f$ are presented and compared with various phenomenological results. 

\subsubsection{Lattice determinations of $m_s/m_{ud}$}
\label{sec:msovermud}

\begin{table}[!htb]
\vspace{3cm}
{\footnotesize{
\begin{tabular*}{\textwidth}[l]{l@{\extracolsep{\fill}}rllllll}
Collaboration & Ref. & $\Nf$ & \hspace{0.15cm}\begin{rotate}{60}{publication status}\end{rotate}\hspace{-0.15cm}  &
 \hspace{0.15cm}\begin{rotate}{60}{chiral extrapolation}\end{rotate}\hspace{-0.15cm} &
 \hspace{0.15cm}\begin{rotate}{60}{continuum  extrapolation}\end{rotate}\hspace{-0.15cm}  &
 \hspace{0.15cm}\begin{rotate}{60}{finite volume}\end{rotate}\hspace{-0.15cm}  & \rule{0.1cm}{0cm} 
$m_s/m_{ud}$ \\
&&&&&& \\[-0.1cm]
\hline
\hline
&&&&&& \\[-0.1cm]

{ETM 21A}& \cite{Alexandrou:2021gqw} & 2+1+1 & \oP & \good & \good & \good & 27.17(32)$^{+56}_{-38}$\\

{MILC 17 $^\ddagger$} & \cite{Bazavov:2017lyh} & 2+1+1 & \gA & \good & \good & \good & $27.178(47)^{+86}_{-57}$\\

{FNAL/MILC 14A} & \cite{Bazavov:2014wgs} & 2+1+1 & \gA & \good & \good & \good & $27.35(5)^{+10}_{-7}$\\

{ETM 14}& \cite{Carrasco:2014cwa} & 2+1+1 & \gA & \soso & \good & \soso & 26.66(32)(2)\\

&&&&&& \\[-0.1cm]
\hline
&&&&&& \\[-0.1cm]
{ALPHA 19}& \cite{Bruno:2019xed} &2+1 & \gA & \soso & \good & \good & 27.0(1.0)(0.4)\\

{RBC/UKQCD 14B}& \cite{Blum:2014tka} &2+1  & \gA & \good & \good & \good & 27.34(21)\\

{RBC/UKQCD 12$^\ominus$}& \cite{Arthur:2012opa} &2+1  & \gA & \good & \soso & \good & 27.36(39)(31)(22)\\

{PACS-CS 12$^\star$}& \cite{Aoki:2012st}       &2+1  & \gA & \good & \bad & \bad & 26.8(2.0)\\

{Laiho 11} & \cite{Laiho:2011np}              &2+1  & \rC & \soso & \good & \good & 28.4(0.5)(1.3)\\

{BMW 10A, 10B$^+$}& \cite{Durr:2010vn,Durr:2010aw} &2+1  & \gA & \good & \good & \good & 27.53(20)(8) \\

{RBC/UKQCD 10A}& \cite{Aoki:2010dy}           &2+1  & \gA & \soso & \soso & \good & 26.8(0.8)(1.1) \\

{Blum 10$^\dagger$}&\cite{Blum:2010ym}         &2+1  & \gA & \soso & \bad & \soso & 28.31(0.29)(1.77)\\

{PACS-CS 09}  & \cite{Aoki:2009ix}            &2+1  &  \gA &\good   &\bad   & \bad & 31.2(2.7)  \\

{MILC 09A}    & \cite{Bazavov:2009fk}       &2+1  & \rC & \soso & \good & \good & 27.41(5)(22)(0)(4)  \\
{MILC 09}      & \cite{Bazavov:2009bb}      &2+1  & \gA & \soso & \good & \good & 27.2(1)(3)(0)(0)  \\

{PACS-CS 08}   & \cite{Aoki:2008sm}           &2+1  & \gA & \good & \bad  & \bad & 28.8(4)\\

{RBC/UKQCD 08} & \cite{Allton:2008pn}         &2+1  & \gA & \soso & \bad  & \good & 28.8(0.4)(1.6) \\

\hspace{-0.2cm}{\begin{tabular}{l}MILC 04, HPQCD/\\MILC/UKQCD 04\end{tabular}} 
& \cite{Aubin:2004fs,Aubin:2004ck}            &2+1  & \gA & \soso & \soso & \soso & 27.4(1)(4)(0)(1)  \\
&&&&&& \\[-0.1cm]
\hline
\hline\\
\end{tabular*}\\[-0.2cm]
}}
\begin{minipage}{\linewidth}
{\footnotesize 
\begin{itemize}
\item[$^\ddagger$] The calculation includes electromagnetic effects.\\[-5mm]
\item[$^\ominus$] The errors are statistical, chiral and finite volume.\\[-5mm]
\item[$^\star$] The calculation includes electromagnetic and $m_u\ne m_d$ effects through reweighting.\\[-5mm]
\item[$^+$] The fermion action used is tree-level improved.\\[-5mm]
\item[$^\dagger$] The calculation includes quenched electromagnetic effects.
\end{itemize}
}
\end{minipage}
\caption{Lattice results for the ratio $m_s/m_{ud}$.}
\label{tab:ratio_msmud}
\end{table}

The lattice results for $m_s/m_{ud}$ are summarized in Tab.~\ref{tab:ratio_msmud}.
In the ratio $m_s/m_{ud}$, one of the sources of systematic error---the
uncertainties in the renormalization factors---drops out. 
Also other systematic effects (like the effect of the scale setting)
are reduced in these ratios.  
This might explain that despite the discrepancies that are present in
the individual quark mass determinations, the ratios show an overall
very good agreement. 

\medskip
\noindent
{\em $\Nf=2+1$ lattice calculations}
\medskip

ALPHA 19~\cite{Bruno:2019vup}, discussed already, is the
only new result for this section. 
The other works contributing to this average are
RBC/UKQCD 14B, which replaces  
RBC/UKQCD 12 (see Sec.~\ref{sec:msmud}), and the results of MILC 09A
and BMW 10A, 10B.  

The results show very good agreement with a $\chi^2/{\rm dof} = 0.14$. 
The final uncertainty ($\approx 0.5\%$) is smaller than the
ones of the quark masses themselves.
At this level of precision, the uncertainties in the electromagnetic
and strong isospin-breaking corrections might not be completely 
negligible. Nevertheless, we decided not to add any uncertainty
associated with this effect. The main reason is that most recent
determinations try to estimate this uncertainty themselves and found
an effect smaller than naive power counting estimates (see $\Nf=2+1+1$
section), 
 \be
     \label{eq:msovmud3} 
     \mbox{$N_f = 2+1$ :} \qquad \FLAGAVBEGIN{m_s}/{m_{ud}} = 27.42 ~ (12) \FLAGAVEND\qquad\Refs~\mbox{\cite{Bruno:2019vup,Blum:2014tka,Bazavov:2009fk,Durr:2010vn,Durr:2010aw}}\,.
 \ee 

\bigskip
\noindent
{\em $\Nf=2+1+1$ lattice calculations}
\medskip

For $N_f = 2+1+1$ there are three results, MILC 17~\cite{Bazavov:2017lyh}, ETM
14~\cite{Carrasco:2014cwa} and FNAL/MILC 14A~\cite{Bazavov:2014wgs},
all of which satisfy our selection criteria. 

All these works have been discussed in the previous FLAG
edition~\cite{Aoki:2019cca}, except the new result ETM 21A, that we
have already examined (and anyway does not appear in the average because it was unpublished at the deadline). 
The fit has $\chi^2/{\rm dof} \approx
2.5$, and the result shows reasonable agreement with the $N_{f}=2+1$ result.

 \be
      \label{eq:msovmud4} 
      \mbox{$N_f = 2+1+1$ :}\qquad \FLAGAVBEGIN m_s / m_{ud} = 27.23 ~ (10)\FLAGAVEND \qquad\Refs~\mbox{\cite{Bazavov:2017lyh,Carrasco:2014cwa,Bazavov:2014wgs}},
 \ee
which corresponds to an overall uncertainty equal to 0.4\%. It is
worth noting that~\cite{Bazavov:2017lyh} estimates the EM effects in
this quantity to be $\sim 0.18\%$ (or 0.049 which is less than the quoted error above).

\begin{figure}[!htb]
\begin{center}
\includegraphics[width=11cm]{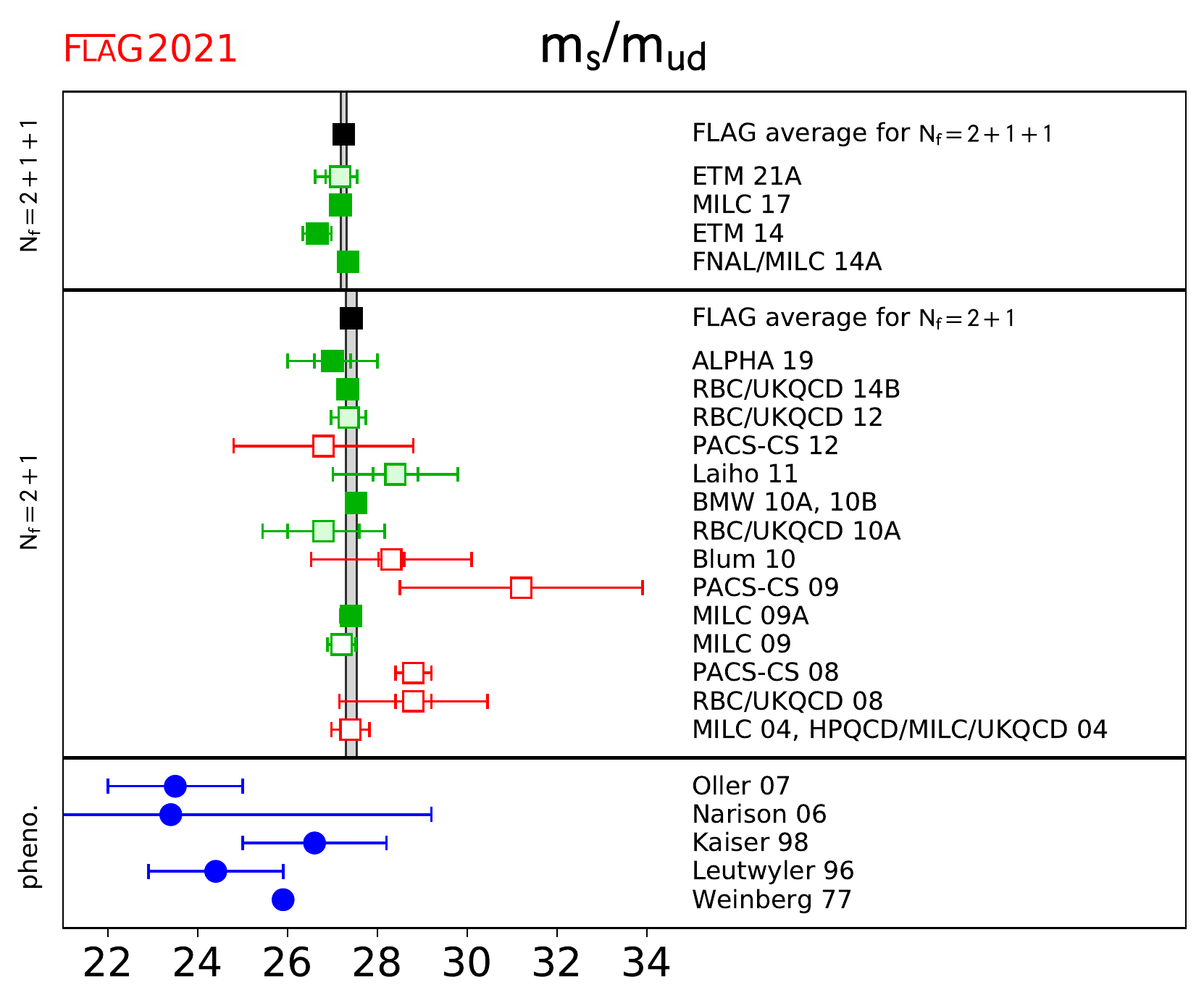}
\end{center}
%
%
%
%
%
%
\vspace{0.5cm}
\begin{center}
	\caption{ \label{fig:msovmud}Results for the ratio $m_s/m_{ud}$. The upper part indicates the lattice results listed in Tab.~\ref{tab:ratio_msmud} together with the FLAG averages for each value of $N_f$. The lower part shows results obtained from $\chi$PT and sum rules~\cite{Oller:2006xb,Narison:2005ny,Kaiser,Leutwyler:1996qg,Weinberg:1977hb}. 
}
\end{center}
\end{figure}

All the lattice results listed in Tab.~\ref{tab:ratio_msmud} as well
as the FLAG averages for each value of $N_f$ are reported in
Fig.~\ref{fig:msovmud} and compared with $\chi$PT and sum rules.

\subsubsection{Lattice determination of $m_u$ and $m_d$}
\label{subsec:mumd}
In addition to reviewing computations of individual $m_u$ and $m_d$ quark
masses, we will also determine FLAG averages for the parameter $\epsilon$ related to
the violations of Dashen's theorem
\be
\epsilon=\frac{(\Delta M_{K}^{2}-\Delta M_{\pi}^{2})^{\gamma}}
{\Delta M_{\pi}^{2}}\,,
\label{eq:epsdef}
\ee
where $\Delta M_{\pi}^{2}=M_{\pi^+}^{2}-M_{\pi^0}^{2}$ and $\Delta
M_{K}^{2}=M_{K^+}^{2}-M_{K^0}^{2}$ are the pion and kaon squared mass
splittings, respectively. The superscript $\gamma$, here and in the following, denotes corrections that arise from electromagnetic effects only. This parameter is often a crucial intermediate
quantity in the extraction of the individual light-quark masses. Indeed, it can
be shown, using the $G$-parity symmetry of the pion triplet, that $\Delta
M_{\pi}^{2}$ does not receive $\cO(\delta m)$ isospin-breaking corrections. In
other words
\be
\Delta M_{\pi}^{2}=(\Delta M_{\pi}^{2})^{\gamma}
\qquad
\text{and}
\qquad
\epsilon=\frac{(\Delta M_{K}^{2})^{\gamma}}
{\Delta M_{\pi}^{2}}-1\,,
\label{eq:epslo}
\ee
at leading-order in the isospin-breaking expansion. The difference $(\Delta
M_{\pi}^{2})^{SU(2)}$ was estimated in previous editions of FLAG through the
$\epsilon_m$ parameter. However, consistent with our leading-order truncation
of the isospin-breaking expansion, it is simpler to ignore this term. Once
known, $\epsilon$ allows one to consistently subtract the electromagnetic part of
the kaon-mass splitting to obtain the QCD splitting $(\Delta M_{K}^{2})^{SU(2)}$. In
contrast with the pion, the kaon QCD splitting is sensitive to
$\delta m$, and, in particular, proportional to it at leading order in $\chi$PT.
Therefore, the knowledge of $\epsilon$ allows for the determination of $\delta m$
from a chiral fit to lattice-QCD data. Originally introduced in another form
in~\citep{Dashen:1969eg}, $\epsilon$ vanishes in the $SU(3)$ chiral
limit, a result known as Dashen's theorem. However, in the 1990's numerous
phenomenological papers pointed out that $\epsilon$ might be an $\cO(1)$ number,
indicating a significant failure of $SU(3)$ $\chi$PT in the description of
electromagnetic effects on light-meson masses. However, the phenomenological
determinations of $\epsilon$ feature some level of controversy, leading to the
rather imprecise estimate $\epsilon=0.7(5)$ given in the first edition of FLAG.
Starting with the FLAG 19 edition of the review, we quote more precise averages for
$\epsilon$, directly obtained from lattice-QCD+QED simulations. We refer the
reader to earlier editions of FLAG and to the
review~\citep{Portelli:2015gda} for discusions of the phenomenological
determinations of $\epsilon$.

The quality criteria regarding finite-volume effects for calculations including QED are presented in Sec.~\ref{sec:Criteria}. Due to the long-distance nature of the
electromagnetic interaction, these effects are dominated by a power law in the
lattice spatial size. The coefficients of this expansion depend on the chosen
finite-volume formulation of QED. For $\mathrm{QED}_{\mathrm{L}}$, these effects
on the squared mass $M^2$ of a charged meson are given
by~\citep{Borsanyi:2014jba,Davoudi:2014qua,Davoudi:2018qpl}
\be
  \Delta_{\mathrm{FV}}M^2=
  \alpha M^2\left\{
  \frac{c_{1}}{ML}+\frac{2c_1}{(ML)^2}+
  \cO\left[\frac{1}{(ML)^3}\right]\right\}\co
\ee
with $c_1\simeq-2.83730$. It has been shown in~\citep{Borsanyi:2014jba} that the
two first orders in this expansion are exactly known for hadrons, and are equal to
the pointlike case. However, the $\cO[1/(ML)^{3}]$ term and higher orders depend
on the structure of the hadron. The universal corrections for
$\mathrm{QED}_{\mathrm{TL}}$ can also be found in \citep{Borsanyi:2014jba}. In
all this part, for all computations using such universal formulae, the QED
finite-volume quality criterion has been applied with $n_{\mathrm{min}}=3$,
otherwise $n_{\mathrm{min}}=1$ was used.

Since FLAG 19, six new results have been reported for nondegenerate light-quark
masses. In the $N_f=2+1+1$ sector, MILC~18~\citep{Basak:2018yzz} computed
$\epsilon$ using $N_f=2+1$ asqtad electro-quenched
QCD+$\mathrm{QED}_{\mathrm{TL}}$ simulations and extracted the ratio $m_u/m_d$
from a new set of $N_f=2+1+1$ HISQ QCD simulations. Although $\epsilon$ comes
from $N_f=2+1$ simulations, $(\Delta M_{K}^{2})^{SU(2)}$, which is about three
times larger than $(\Delta M_{K}^{2})^{\gamma}$, has been determined in the
$N_f=2+1+1$ theory. We therefore chose to classify this result as a four-flavour
one. This result is explicitly described by the authors as an update of
MILC~17~\citep{Bazavov:2017lyh}. In MILC~17~\citep{Bazavov:2017lyh}, $m_u/m_d$
is determined as a side-product of a global analysis of heavy-meson decay
constants, using a preliminary version of $\epsilon$ from
MILC~18~\citep{Basak:2018yzz}. In FNAL/MILC/TUMQCD~18~\citep{Bazavov:2018omf}
the ratio $m_u/m_d$ from MILC~17~\citep{Bazavov:2017lyh} is used to determine
the individual masses $m_u$ and $m_d$ from a new calculation of $m_{ud}$. The
work RM123~17~\citep{Giusti:2017dmp} is the continuation of the $N_f=2$ work
named RM123~13~\citep{deDivitiis:2013xla} in the previous edition of FLAG. This
group now uses $N_f=2+1+1$ ensembles from ETM~10~\citep{Baron:2010bv}, however,
still with a rather large minimum pion mass of $270~\mathrm{MeV}$, leading to
the \soso~rating for chiral extrapolations. In the $N_f=2+1$ sector, BMW~16~\citep{Fodor:2016bgu} reuses the data set produced from their determination of the light-baryon octet-mass splittings~\citep{Borsanyi:2013lga} using electro-quenched QCD+$\mathrm{QED}_{\mathrm{TL}}$ smeared clover fermion simulations. Finally, MILC~16~\citep{Basak:2016jnn}, which is a preliminary result for the value of $\epsilon$ published in MILC~18~\citep{Basak:2018yzz}, also provides a $N_f=2+1$ computation of the ratio $m_u/m_d$.

MILC 09A~\cite{Bazavov:2009fk} uses the mass difference
between $K^0$ and $K^+$, from which they subtract electromagnetic
effects using Dashen's theorem with corrections, as discussed in the introduction of this section.  The up  and down  
sea quarks remain degenerate in their calculation, fixed to the value of
$m_{ud}$ obtained from $M_{\pi^0}$. To determine $m_u/m_d$, BMW 10A, 10B~\cite{Durr:2010vn,Durr:2010aw}
follow a slightly different strategy. They obtain this ratio from
their result for $m_s/m_{ud}$ combined with a phenomenological
determination of the isospin-breaking quark-mass ratio $Q=22.3(8)$,
from $\eta\to3\pi$
decays~\cite{Leutwyler:2009jg} (the decay $\eta\to3\pi$ is very
sensitive to QCD isospin breaking, but fairly insensitive to QED
isospin breaking).
%
\begin{sidewaystable}[ph!]
\centering
\vspace{2.5cm}
{\footnotesize{
\begin{tabular*}{\textwidth}[l]{l@{\extracolsep{\fill}}r@{\hspace{2mm}}l@{\hspace{2mm}}l@{\hspace{1.5mm}}l@{\hspace{1.5mm}}l@{\hspace{1.5mm}}l@{\hspace{1.5mm}}l@{\hspace{1.5mm}}l@{\hspace{1.5mm}}l@{\hspace{1.5mm}}l@{\hspace{1.5mm}}l}
Collaboration \al  Ref. \al \hspace{0.15cm}\begin{rotate}{60}{publication status}\end{rotate}\hspace{-0.15cm}  \al 
\hspace{0.15cm}\begin{rotate}{60}{chiral extrapolation}\end{rotate}\hspace{-0.15cm} \al 
\hspace{0.15cm}\begin{rotate}{60}{continuum  extrapolation}\end{rotate}\hspace{-0.15cm}  \al 
\hspace{0.15cm}\begin{rotate}{60}{finite volume}\end{rotate}\hspace{-0.15cm}  \al  
\hspace{0.15cm}\begin{rotate}{60}{isospin breaking}\end{rotate}\hspace{-0.15cm} \al
\hspace{0.15cm}\begin{rotate}{60}{renormalization}\end{rotate}\hspace{-0.15cm} \al   
\hspace{0.15cm}\begin{rotate}{60}{running}\end{rotate}\hspace{-0.15cm}  \al  
\rule{0.6cm}{0cm}$m_u$\al 
\rule{0.6cm}{0cm}$m_d$ \al \rule{0.3cm}{0cm} $m_u/m_d$\\
\al \al \al \al \al \al \al \al \al \al  \\[-0.1cm]
\hline
\hline
\al \al \al \al \al \al \al \al \al \al  \\[-0.1cm]
{MILC 18} \al \cite{Basak:2018yzz} \al \gA \al \good \al \good \al  \good \al \soso
\al \good \al $-$ \al \al
\al $0.4529(48)({}^{+150}_{-67})$\\
{FNAL/MILC/TUMQCD 18$^*$} \al \cite{Bazavov:2018omf} \al \gA \al \good \al \good \al  \good \al \soso
\al \good \al $-$ \al 2.118(17)(32)(12)(03) \al 4.690(30)(36)(26)(06)
\al \\
{MILC~17$^\dagger$} \al \cite{Bazavov:2017lyh} \al \gA \al \good \al \good \al \good \al \soso 
\al \good \al $-$ \al \al \al $0.4556(55)({}^{+114}_{-67})(13)$ \\
{RM123~17} \al \cite{Giusti:2017dmp} \al \gA \al \soso \al \good \al \good
\al \soso \al \good \al $\,b$ \al 2.50(15)(8)(2) \al 4.88(18)(8)(2)
\al 0.513(18)(24)(6) \\
{ETM 14}& \cite{Carrasco:2014cwa}  \al \gA \al \good \al \good \al \good \al \bad \al \good \al 
$\,b$ \al 2.36(24) \al 5.03(26) \al 0.470(56) \\[0.5ex]
\hline
\al \al \al \al \al \al \al \al \al \al  \\[-0.2cm]
{BMW 16} \al \cite{Fodor:2016bgu} \al \gA \al \good \al \good \al \good
\al \soso \al \good \al $-$ \al 2.27(6)(5)(4) \al 4.67(6)(5)(4) \al 
0.485(11)(8)(14)\\

{MILC 16} \al \cite{Basak:2016jnn} \al \rC \al \soso \al \good \al \good 
\al \soso \al \good \al $-$ \al \al \al 
$0.4582(38)({}^{+12}_{-82})(1)(110)$ \\

{QCDSF/UKQCD 15} \al \protect{\cite{Horsley:2015eaa}} \al \gA \al \soso \al \bad \al \bad \al \good \al $-$\al $-$
\al  \al   \al 0.52(5)\\

{PACS-CS 12} \al \protect{\cite{Aoki:2012st}} \al \gA \al \good \al \bad \al \bad \al \good \al \good \al $\,a$
\al  2.57(26)(7) \al  3.68(29)(10) \al 0.698(51)\\

{Laiho 11} \al \cite{Laiho:2011np} \al \rC \al \soso \al \good \al
\good \al \bad \al \soso \al $-$ \al 1.90(8)(21)(10) \al
4.73(9)(27)(24) \al 0.401(13)(45)\\

{HPQCD~10$^\ddagger$}\al \cite{McNeile:2010ji} \al \gA \al \soso \al \good \al \good \al \bad \al \good \al
$-$ \al 2.01(14) \al 4.77(15) \al  \\

{BMW 10A, 10B$^+$}\al \cite{Durr:2010vn,Durr:2010aw} \al \gA \al \good \al \good \al \good \al \bad \al \good \al
$\,b$ \al 2.15(03)(10) \al 4.79(07)(12) \al 0.448(06)(29) \\

{Blum~10}\al\cite{Blum:2010ym}\al \gA \al \soso \al \bad \al \soso \al \soso \al \good \al $-$ \al 2.24(10)(34)\al 4.65(15)(32)\al 0.4818(96)(860)\\

{MILC 09A} \al  \cite{Bazavov:2009fk} \al  \rC \al  \soso \al \good \al \good \al \bad \al \soso \al $-$
\al 1.96(0)(6)(10)(12)
\al  4.53(1)(8)(23)(12)  \al   0.432(1)(9)(0)(39) \\

{MILC 09} \al  \cite{Bazavov:2009bb} \al  \gA \al  \soso \al  \good \al  \good \al  \bad \al \soso \al 
$-$\al  1.9(0)(1)(1)(1)
\al  4.6(0)(2)(2)(1) \al  0.42(0)(1)(0)(4) \\

\hspace{-0.2cm}{\begin{tabular}{l}MILC 04, HPQCD/\rule{0.1cm}{0cm}\\MILC/UKQCD 04\end{tabular}} \al \cite{Aubin:2004fs}\cite{Aubin:2004ck} \al  \gA \al  \soso \al  \soso \al  \soso \al \bad \al
\bad \al$-$\al  1.7(0)(1)(2)(2)
\al  3.9(0)(1)(4)(2)  \al  0.43(0)(1)(0)(8) \\

\al \al \al \al \al \al \al \al \al \al \al  \\[-0.3cm]
\hline
\hline\\
\end{tabular*}\\[-0.2cm]
}}
\begin{minipage}{\linewidth}
{\footnotesize 
\begin{itemize}
\item[$^*$] FNAL/MILC/TUMQCD~18 uses $\epsilon$ from MILC~18 to produce the individual $m_u$ and $m_d$ masses.
\item[$^\dagger$]MILC~17 additionally quotes an optional 0.0032 uncertainty on $m_u/m_d$ corresponding to QED and QCD separation scheme ambiguities. Because this variation is not per se an error on the determination of $m_u/m_d$, and because it is generally not included in other results, we choose to omit it here.
\item[$^\ddagger$]Values obtained by combining the HPQCD 10 result
  for $m_s$ with the MILC 09 results for $m_s/m_{ud}$ and
  $m_u/m_d$.\\[-5mm]
\item[$^+$] The fermion action used is tree-level improved.\\[-5mm]
\item[$a$] The masses are renormalized and run nonperturbatively up to
  a scale of $100\,\gev$ in the $N_f=2$ SF scheme. In this scheme,
  nonperturbative and NLO running for the quark masses are shown to
  agree well from 100 GeV all the way down to 2 GeV~\cite{DellaMorte:2005kg}.\\[-5mm]
\item[$b$] The masses are renormalized and run nonperturbatively up to
  a scale of 4 GeV in the $N_f=3$ RI-MOM scheme.  In
  this scheme, nonperturbative and N$^3$LO running for the quark
  masses are shown to agree from 6~GeV down to 3~GeV to
  better than 1\%~\cite{Durr:2010aw}.
\end{itemize}
}
\end{minipage}
\caption{Lattice results for $m_u$, $m_d$ (MeV) and for the ratio $m_u/m_d$. The values refer to the 
$\msbar$ scheme  at scale 2 GeV.  The top part of the table lists the results obtained with $\Nf=2+1+1$,  
while the lower part presents calculations with $N_f = 2+1$.}
\label{tab:mu_md_grading}
\end{sidewaystable}
\begin{figure}[t]
\begin{center}
\includegraphics[width=13cm]{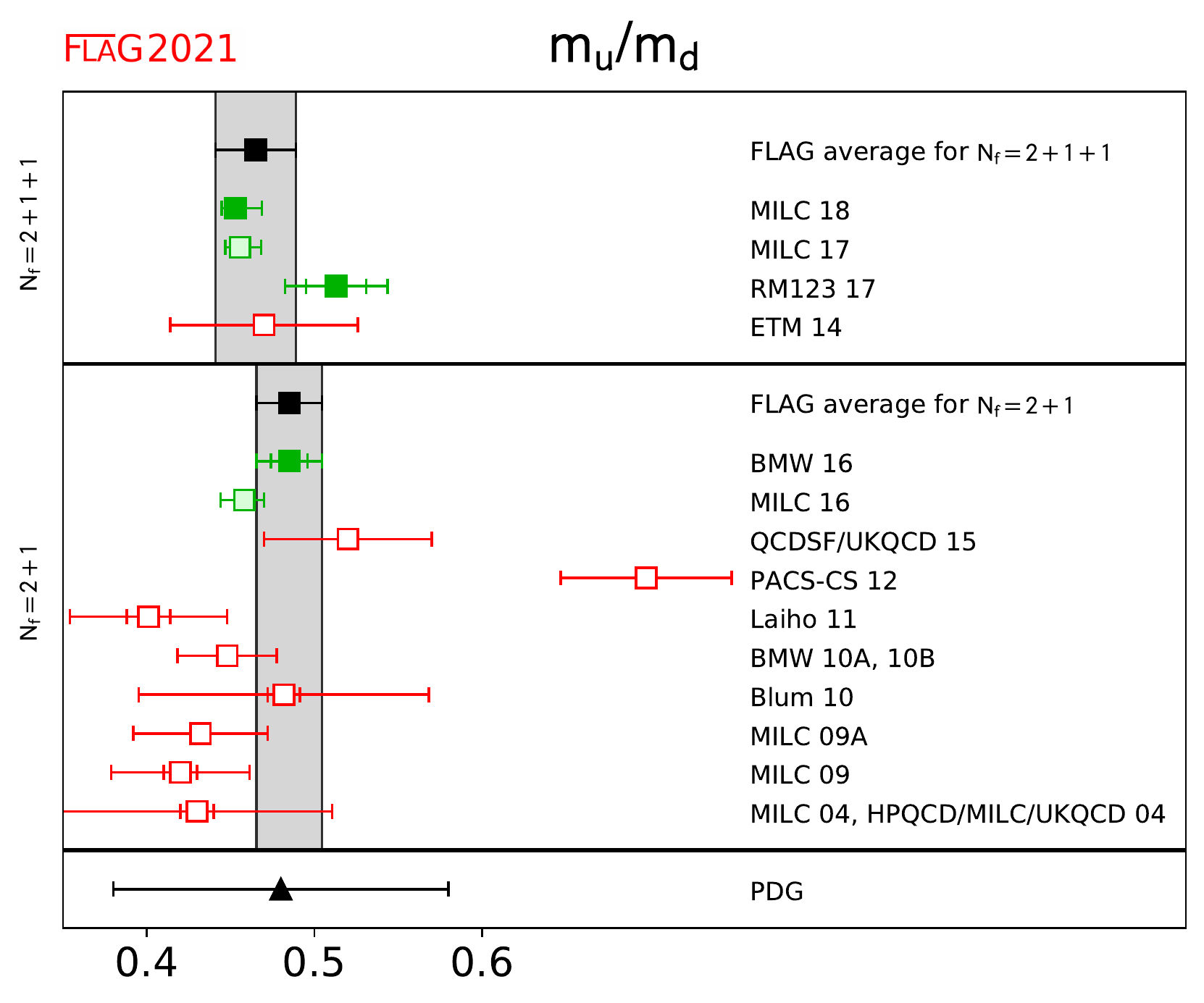}
\end{center}
\caption{\label{fig:mu over md}Lattice results and FLAG averages at $N_f = 2+1$ and $2+1+1$ for the up-down-quark masses ratio $m_u/m_d$, together with the current PDG estimate.}
\end{figure}
Instead of subtracting electromagnetic effects using phenomenology,
RBC~07~\cite{Blum:2007cy} and Blum~10~\cite{Blum:2010ym} actually include a
quenched electromagnetic field in their calculation. This means that their
results include corrections to Dashen's theorem, albeit only in the presence of
quenched electromagnetism. Since the up and down quarks in the sea are treated
as degenerate, very small isospin corrections are neglected, as in MILC's
calculation. PACS-CS 12~\cite{Aoki:2012st} takes the inclusion of
isospin-breaking effects one step further. Using reweighting techniques, it also
includes electromagnetic and $m_u-m_d$ effects in the sea. However, they  do not
correct for the large finite-volume effects coming from electromagnetism in
their $M_{\pi}L\sim 2$ simulations, but provide rough estimates for
their size, based on Ref.~\citep{Hayakawa:2008an}.
QCDSF/UKQCD~15~\cite{Horsley:2015eaa} uses QCD+QED dynamical simulations
performed at the $SU(3)$-flavour-symmetric point, but at a single lattice
spacing, so they do not enter our average. The smallest partially quenched ($m_{\rm sea}\neq m_{\rm val}$) pion
mass is greater than 200 MeV, so our chiral-extrapolation criteria require a
$\soso$ rating. Concerning finite-volume effects, this work uses three spatial
extents $L$ of $1.6~\mathrm{fm}$, $2.2~\mathrm{fm}$, and $3.3~\mathrm{fm}$.
QCDSF/UKQCD~15 claims that the volume dependence is not visible on the two
largest volumes, leading them to assume that finite-size effects are under
control. As a consequence of that, the final result for quark masses does not
feature a finite-volume extrapolation or an estimation of the finite-volume
uncertainty. However, in their work on the QED corrections to the hadron
spectrum~\citep{Horsley:2015eaa} based on the same ensembles, a volume study
shows some level of compatibility with the $\mathrm{QED}_{\mathrm{L}}$
finite-volume effects derived in~\citep{Davoudi:2014qua}. We see two issues
here. Firstly, the analytical result quoted from~\citep{Davoudi:2014qua} predicts
large, $\cO(10\%)$ finite-size effects from QED on the meson masses at the values
of $M_{\pi}L$ considered in QCDSF/UKQCD~15, which is inconsistent with the
statement made in the paper. Secondly, it is not known that the zero-mode
regularization scheme used here has the same volume scaling as
$\mathrm{QED}_{\mathrm{L}}$. We therefore chose to assign the \bad~rating for
finite volume to QCDSF/UKQCD~15. Finally, for $N_f=2+1+1$, ETM
14~\cite{Carrasco:2014cwa} uses simulations in pure QCD, but determines
$m_u-m_d$ from the slope $\partial M_K^2/\partial m_{ud}$ and the physical
value for the QCD kaon-mass splitting taken from the phenomenological
estimate in FLAG 13.

Lattice results for $m_u$, $m_d$ and $m_u/m_d$ are summarized in
Tab.~\ref{tab:mu_md_grading}. 
The colour coding is specified in detail in Sec.~\ref{sec:color-code}.
Considering the important progress in the last years on including isospin-breaking
effects in lattice simulations, we are now in a position where averages for
$m_u$ and $m_d$ can be made without the need of phenomenological inputs.
Therefore, lattice calculations of the individual quark masses using
phenomenological inputs for isospin-breaking effects will be coded \bad.

We start by recalling the $N_f=2$ FLAG average for the light-quark masses,
entirely coming from RM123 13~\cite{deDivitiis:2013xla},
\begin{align}
&& m_u &=2.40(23)   \,\mev&\Ref~\mbox{\cite{deDivitiis:2013xla}}\,,\nonumber\\
\label{eq:mumdNf2} \hspace{0cm}\Nf = 2:\hspace{0.2cm}
&& m_d &= 4.80(23)  \,\mev&\Ref~\mbox{\cite{deDivitiis:2013xla}}\,,\\
&& {m_u}/{m_d} &= 0.50(4) &\Ref~\mbox{\cite{deDivitiis:2013xla}}\,,\nonumber
\end{align}
with errors of roughly 10\%, 5\% and 8\%, respectively. In these results, the
errors are obtained by combining the lattice statistical and
systematic errors in quadrature. 
For $\Nf=2+1$, the only result, which qualifies for entering the FLAG average for quark masses, is BMW~16~\citep{Fodor:2016bgu},
%
\begin{align}
&&\FLAGAVBEGIN m_u &=2.27(9)\FLAGAVEND\,\mev&\Ref~\mbox{\cite{Fodor:2016bgu}}\,, \nonumber\\
\label{eq:mumd} \hspace{0cm}\Nf = 2+1:\hspace{0.2cm}
&&\FLAGAVBEGIN m_d &= 4.67(9) \FLAGAVEND\,\mev&\Ref~\mbox{\cite{Fodor:2016bgu}}\,,\\
&&\FLAGAVBEGIN {m_u}/{m_d} &= 0.485(19)\FLAGAVEND&\Ref~\mbox{\cite{Fodor:2016bgu}}\,,\nonumber
\end{align}
%
with errors of roughly 4\%, 2\% and 4\%, respectively. This estimate
is slightly more precise than in the previous edition of FLAG. More
importantly, it now comes entirely from a lattice-QCD+QED calculation,
whereas phenomenological input was used in previous
editions. These numbers result in the following RGI averages
\begin{align}
&& M_u^{\rm RGI} &=3.15(12)_m(4)_\Lambda \,\mev&\Ref~\mbox{\cite{Fodor:2016bgu}}\,, \nonumber\\
\label{eq:mumd rgi} \hspace{0cm}\Nf = 2+1:\hspace{0.2cm}\\[-5mm]
&& M_d^{\rm RGI} &= 6.49(12)_m(7)_\Lambda \,\mev&\Ref~\mbox{\cite{Fodor:2016bgu}}\,.\nonumber
\end{align}

Finally, for $\Nf=2+1+1$, RM123~17~\citep{Giusti:2017dmp} and FNAL/MILC/TUMQCD~18~\citep{Bazavov:2018omf} enter the average for the individual $m_u$ and $m_d$ masses, and RM123~17~\citep{Giusti:2017dmp} and MILC~18~\citep{Basak:2018yzz} enter the average for the ratio $m_u/m_d$, giving
%
\begin{align}
	&&\FLAGAVBEGIN m_u &=2.14(8)\FLAGAVEND \,\mev&\Ref~\mbox{\cite{Giusti:2017dmp,Bazavov:2018omf}}\,,\nonumber\\
\label{eq:mumd 4 flavour} \Nf = 2+1+1:\hspace{0.15cm}
        &&\FLAGAVBEGIN m_d &= 4.70(5)\FLAGAVEND \,\mev&\Ref~\mbox{\cite{Giusti:2017dmp,Bazavov:2018omf}}\,,\\
	&&\FLAGAVBEGIN {m_u}/{m_d} &= 0.465(24)\FLAGAVEND&\Ref~\mbox{\cite{Giusti:2017dmp,Basak:2018yzz}}\,.\nonumber
\end{align}
%
with errors of roughly 4\%, 1\% and 5\%, respectively. One can observe some marginal discrepancies
between results coming from the MILC collaboration and RM123~17~\citep{Giusti:2017dmp}. More specifically,
adding all sources of uncertainties in quadrature, one obtains a 1.7$\sigma$ discrepancy between RM123~17~\citep{Giusti:2017dmp} and
MILC~18~\cite{Basak:2018yzz} for $m_u/m_d$, and a 2.2$\sigma$ discrepancy between RM123~17~\citep{Giusti:2017dmp} and 
FNAL/MILC/TUMQCD~18~\citep{Bazavov:2018omf} for $m_u$. However, the values of $m_d$ and $\epsilon$ are in very good agreement between the two groups. These discrepancies are presently too weak to constitute evidence for concern, and will be monitored as more lattice groups provide results for these quantities. The RGI averages for $m_u$ and $m_d$ are

\begin{align}
	&& M_u^{\rm RGI} &=2.97(11)_m(3)_\Lambda \,\mev&\Ref~\mbox{\cite{Giusti:2017dmp,Bazavov:2018omf}}\,,\nonumber\\
\label{eq:mumd 4 flavour rgi} \Nf = 2+1+1:\hspace{0.15cm}\\[-5mm]
        && M_d^{\rm RGI} &= 6.53(7)_m(8)_\Lambda \,\mev&\Ref~\mbox{\cite{Giusti:2017dmp,Bazavov:2018omf}}\,.\nonumber
\end{align}

Every result for $m_u$ and $m_d$ used here to produce the FLAG averages relies
on electro-quenched calculations, so there is some interest to comment on the
size of quenching effects. Considering phenomenology and the lattice results
presented here, it is reasonable for a rough estimate to use the value $(\Delta
M_{K}^{2})^{\gamma}\sim 2000~\mathrm{MeV}^2$ for the QED part of the kaon-mass
splitting. Using the arguments presented in Sec.~\ref{sec:latticeqed}, one can
assume that the QED sea contribution represents $\cO(10\%)$ of $(\Delta
M_{K}^{2})^{\gamma}$. Using $SU(3)$
PQ$\chi$PT+QED~\citep{Bijnens:2006mk,Portelli:2012pn} gives a $\sim 5\%$ effect.
Keeping the more conservative $10\%$ estimate and using the experimental value
of the kaon-mass splitting, one finds that the QCD kaon-mass splitting $(\Delta
M_{K}^{2})^{SU(2)}$ suffers from a reduced $3\%$ quenching uncertainty.
Considering that this splitting is proportional to $m_u-m_d$ at
leading order in $SU(3)$ $\chi$PT, we can estimate that a similar error will
propagate to the quark masses. So the individual up and down masses look mildly
affected by QED quenching. However, one notices that $\sim 3\%$ is the level
of error in the new FLAG averages, and increasing significantly this accuracy
will require using fully unquenched calculations. 

In view of the fact that a {\it massless up quark} would solve the
strong CP problem, many authors have considered this an attractive
possibility, but the results presented above exclude this possibility:
the value of $m_u$ in Eq.~(\ref{eq:mumd}) differs from zero by $26$
standard deviations. We conclude that nature solves the strong
CP problem differently.

Finally, we conclude this section by giving the FLAG averages for $\epsilon$
defined in Eq.~(\ref{eq:epsdef}). For $\Nf=2+1+1$, we average the results of
RM123~17~\citep{Giusti:2017dmp} and MILC~18~\cite{Basak:2018yzz} with the value of $(\Delta
M_{K}^{2})^{\gamma}$ from BMW~14~\citep{Borsanyi:2014jba} combined with
Eq.~(\ref{eq:epslo}), giving
\begin{align}
\label{eq:epsilon 4 flavour} \Nf = 2+1+1:\hspace{0.15cm}\\[-5mm]
        && \epsilon &= 0.79(6) & \Ref~\mbox{\cite{Giusti:2017dmp,Basak:2018yzz,Borsanyi:2014jba}}\,.\nonumber
\end{align}

Although BMW~14~\citep{Borsanyi:2014jba} focuses on hadron masses and did not
extract the light-quark masses, they are the only fully unquenched QCD+QED
calculation to date that qualifies to enter a FLAG average. With the exception of
renormalization, which is not discussed in the paper, this work has a
\good~rating for every FLAG criterion considered for the $m_u$ and $m_d$ quark
masses. For $\Nf=2+1$ we use the results from BMW~16~\citep{Fodor:2016bgu}, 
\begin{align}
\label{eq:epsilon 3 flavour} \Nf = 2+1:\hspace{0.15cm}\\[-5mm]
        && \epsilon &= 0.73(17) & \Ref~\mbox{\cite{Fodor:2016bgu}}\,.\nonumber
\end{align}

It is important to notice that the $\epsilon$ uncertainties from BMW~16 and RM123~17
are dominated by estimates of the QED quenching effects. Indeed, in contrast
with the quark masses, $\epsilon$ is expected to be rather sensitive to the sea-quark QED contributions. Using the arguments presented in
Sec.~\ref{sec:latticeqed}, if one conservatively assumes that the QED sea
contributions represent $\cO(10\%)$ of $(\Delta M_{K}^{2})^{\gamma}$, then
Eq.~(\ref{eq:epslo}) implies that $\epsilon$ will have a quenching error of
$\sim 0.15$ for $(\Delta M_{K}^{2})^{\gamma}\sim 2000~\mathrm{MeV}^2$,
representing a large $\sim 20\%$ relative error. It is interesting to observe
that such a discrepancy does not appear between BMW~15 and RM123~17, although the
$\sim 10\%$ accuracy of both results might not be sufficient to resolve these
effects. {\color{black}On the other hand, in the context of $SU(3)$ chiral perturbation theory, Bijnens and Danielsson~\cite{Bijnens:2006mk} show that the QED quenching effects on $\epsilon$ do not depend on unknown LECs at NLO and are therefore computable at that order.  In that approach, MILC 18 finds the effect at NLO to be only 5\%.  
} To conclude, although the controversy around the value of $\epsilon$
has been significantly reduced by lattice-QCD+QED determinations, computing this
at {\color{black}few-percent accuracy} requires simulations with charged sea quarks.

\subsubsection{Estimates for $R$ and $Q$}\label{sec:RandQ}
The quark-mass ratios
\be\label{eq:Qm}
R\equiv \frac{m_s-m_{ud}}{m_d-m_u}\hspace{0.5cm} \mbox{and}\hspace{0.5cm}Q^2\equiv\frac{m_s^2-m_{ud}^2}{m_d^2-m_u^2}
\ee
compare $SU(3)$ breaking  with isospin breaking. Both numbers only depend on the ratios $m_s/m_{ud}$ and $m_u/m_d$,
\be
R=\frac{1}{2}\left(\frac{m_s}{m_{ud}}-1\right)\frac{1+\frac{m_u}{m_d}}{1-\frac{m_u}{m_d}}
\qquad\text{and}\qquad Q^2=\frac{1}{2}\left(\frac{m_s}{m_{ud}}+1\right)R\,.
\ee
The quantity $Q$ is of
particular interest because of a low-energy theorem~\cite{Gasser:1984pr},
which relates it to a ratio of meson masses,  
\begin{equation}\label{eq:QM}
 Q^2_M\equiv \frac{\hat{M}_K^2}{\hat{M}_\pi^2}\frac{\hat{M}_K^2-\hat{M}_\pi^2}{\hat{M}_{K^0}^2-
   \hat{M}_{K^+}^2}\co\hspace{1cm}\hat{M}^2_\pi\equiv\mbox{$\frac{1}{2}$}( \hat{M}^2_{\pi^+}+ \hat{M}^2_{\pi^0})
 \co\hspace{0.5cm}\hat{M}^2_K\equiv\mbox{$\frac{1}{2}$}( \hat{M}^2_{K^+}+ \hat{M}^2_{K^0})\fs\end{equation}
{\color{black}(We remind the reader that the $\,\hat{ }\,$ denotes a quantity evaluated in the $\alpha\to 0$ limit.) }
Chiral symmetry implies that the expansion of $Q_M^2$ in powers of the
quark masses (i) starts with $Q^2$ and (ii) does not receive any
contributions at NLO:
\be\label{eq:LET Q}Q_M\NLo Q \fs\ee

We recall here the $N_f=2$ estimates for $Q$ and $R$ from FLAG~16,
\be\label{eq:RQresNf2} R=40.7(3.7)(2.2)\co\hspace{2cm}Q=24.3(1.4)(0.6)\ ,\ee 
where the second error comes from the phenomenological inputs that were used.
For $\Nf=2+1$, we use Eqs.~(\ref{eq:msovmud3}) and (\ref{eq:mumd}) and obtain
\be\label{eq:RQres} R=38.1(1.5)\co\hspace{2cm}Q=23.3(0.5)\ ,\ee 
where now only lattice results have been used.
For $\Nf=2+1+1$ we obtain
\be\label{eq:RQresNf4} R=35.9(1.7)\co\hspace{2cm}Q=22.5(0.5)\ ,\ee
which are quite compatible with two- and three-flavour results. It is interesting
to notice that the most recent phenomenological determination of $R$ and $Q$
from $\eta\to 3\pi$ decay~\citep{Colangelo:2018jxw} gives the values
$R=34.4(2.1)$ and $Q=22.1(7)$, which are marginally discrepant with some of the averages
presented here. The authors of~\citep{Amoros:2001cp,Colangelo:2018jxw}
point out that this discrepancy is likely due to surprisingly large corrections to the
approximation in Eq.~(\ref{eq:LET Q}) used in the phenomenological analysis.

Our final results for the masses $m_u$, $m_d$, $m_{ud}$, $m_s$ and the mass ratios
$m_u/m_d$, $m_s/m_{ud}$, $R$, $Q$ are collected in Tabs.~\ref{tab:mudms} and
\ref{tab:mumdRQ}.

\begin{table}[!thb]\vspace{0.5cm}
{
\begin{tabular*}{\textwidth}[l]{@{\extracolsep{\fill}}cccc}
\hline\hline
$\Nf$ & $m_{ud}$ & $ m_s $ & $m_s/m_{ud}$ \\ 
&&& \\[-2ex]
\hline\rule[-0.1cm]{0cm}{0.5cm}
&&& \\[-2ex]
2+1+1 & 3.410(43) & 93.44(68) & 27.23(10)\\ 
&&& \\[-2ex]
\hline\rule[-0.1cm]{0cm}{0.5cm}
&&& \\[-2ex]
2+1 & 3.364(41) & 92.03(88) & 27.42(12)\\ 
&&& \\[-2ex]
\hline
\hline
\end{tabular*}
\caption{\label{tab:mudms} Our estimates for the strange-quark and the average
  up-down-quark masses in the $\msbar$ scheme  at running scale
  $\mu=2\,\gev$. Mass values are given in MeV. In the
  results presented here, the error is the one which we obtain
  by applying the averaging procedure of Sec.~\ref{sec:error_analysis} to the
  relevant lattice results. 
  } }
\end{table}

\begin{table}[!thb]
{
\begin{tabular*}{\textwidth}[l]{@{\extracolsep{\fill}}cccccc}
\hline\hline
$\Nf$ & $m_u  $ & $m_d $ & $m_u/m_d$ & $R$ & $Q$\\ 
&&&&& \\[-2ex]
\hline\rule[-0.1cm]{0cm}{0.5cm}
&&&&& \\[-2ex]
2+1+1 & 2.14(8) & 4.70(5) & 0.465(24) & 35.9(1.7) & 22.5(0.5) \\ 
&&&&& \\[-2ex]
\hline\rule[-0.1cm]{0cm}{0.5cm}
&&&&& \\[-2ex]
2+1 & 2.27(9) & 4.67(9) & 0.485(19) & 38.1(1.5) & 23.3(0.5) \\ 
&&&&& \\[-2ex]
\hline
\hline
\end{tabular*}
\caption{\label{tab:mumdRQ} Our estimates for the masses of the
  two lightest quarks and related, strong isospin-breaking
  ratios. Again, the masses refer to the $\msbar$ scheme  at running
  scale $\mu=2\,\gev$. Mass values are given
  in MeV.}  }
\end{table}
\clearpage


\subsection{Charm-quark mass}
\label{s:cmass}

In the following, we collect and discuss the lattice determinations of the $\overline{\rm MS}$ charm-quark mass $\overline{m}_c$.
Most of the results have been obtained by analyzing the lattice-QCD simulations of two-point heavy-light- or 
heavy-heavy-meson correlation functions, using as input the experimental values of the $D$, $D_s$, and charmonium mesons.
Some groups use the moments method.
The latter is based on the lattice calculation of the Euclidean time moments of pseudoscalar-pseudoscalar correlators for heavy-quark currents followed by an OPE expansion dominated by perturbative QCD effects, which provides the determination of both the heavy-quark mass and the strong-coupling constant $\alpha_s$.

The heavy-quark actions adopted by various lattice collaborations have been discussed in previous FLAG reviews~\cite{Aoki:2013ldr,Aoki:2016frl,Aoki:2019cca}, and their descriptions can be found in Sec.~A.1.3 of FLAG 19 \cite{Aoki:2019cca}.
While the charm mass determined with the moments method does not need any lattice evaluation of the mass-renormalization constant $Z_m$, the extraction of $\overline{m}_c$  from two-point heavy-meson correlators does require the nonperturbative calculation of $Z_m$.
The lattice scale at which $Z_m$ is obtained is usually at least of the order $2$--$3$ GeV, and therefore it is natural in this review to provide the values of $\overline{m}_c(\mu)$ at the renormalization scale $\mu = 3~\gev$.
Since the choice of a renormalization scale equal to $\overline{m}_c$ is still commonly adopted (as by the PDG~\cite{Zyla:2020zbs}), we have collected in Tab.~\ref{tab:mc} the lattice results for both $\overline{m}_c(\overline{m}_c)$ and $\overline{m}_c(\mbox{3 GeV})$, obtained  for $N_f =2+1$ and $2+1+1$. For $N_f=2$, interested readers are referred to previous reviews~\cite{Aoki:2013ldr,Aoki:2016frl}.

{\color{black}When not directly available in the published work, we apply a conversion factor using perturbative QCD evolution at five loops to run down from $\mu = 3$ GeV to the scales $\mu = \overline{m}_c$  and 2 GeV of $0.7739(60)$ and 0.9026(23), respectively, where the error comes from the uncertainty in $\Lambda_{\rm QCD}$.  We use $\Lambda_{\rm QCD} = 297(12)$ MeV for $N_f = 4$ (see Sec.~\ref{sec:alpha_s}). Perturbation theory uncertainties, estimated as the difference between results that use
  4- and 5-loop running, are significantly smaller than the
  parametric uncertainty coming from $\Lambda_{\rm QCD}$. For $\mu = \overline{m}_c$, the former is about
  about 2.5 times smaller. Given the high precision of many
  of these results, future works should take the uncertainties in $\Lambda_{\rm QCD}$
  and perturbation theory seriously.
\begin{table}[!htb]
\vspace{3cm}
{\footnotesize{
\begin{tabular*}{\textwidth}[l]{l@{\extracolsep{\fill}}rllllllll}
Collaboration & Ref. & $N_f$ & \hspace{0.15cm}\begin{rotate}{60}{publication status}\end{rotate}\hspace{-0.15cm} &
 \hspace{0.15cm}\begin{rotate}{60}{chiral extrapolation}\end{rotate}\hspace{-0.15cm} &
 \hspace{0.15cm}\begin{rotate}{60}{continuum  extrapolation}\end{rotate}\hspace{-0.15cm} &
 \hspace{0.15cm}\begin{rotate}{60}{finite volume}\end{rotate}\hspace{-0.15cm} &  
 \hspace{0.15cm}\begin{rotate}{60}{renormalization}\end{rotate}\hspace{-0.15cm} & 
  \rule{0.5cm}{0cm}$\overline{m}_c(\overline{m}_c)$ & 
  \rule{0.3cm}{0cm}$\overline{m}_c(\mbox{3 GeV})$ \\
&&&&&&&&& \\[-0.1cm]
\hline
\hline
&&&&&&&&& \\[-0.1cm]
ETM 21A & \cite{Alexandrou:2021gqw} & 2+1+1 & \oP & \good & \good & \good & \good & 1.339(22)($^{+19}_{-10}$)(10)$^\dagger$ & 1.036(17)($^{+15}_{-8}$) \\ 
HPQCD 20A & \cite{Hatton:2020qhk} & 2+1+1 & \gA & \good & \good & \good & \good & 1.2719(78) & 0.9841(51) \\ 
HPQCD 18  & \cite{Lytle:2018evc} & 2+1+1 & \gA & \good & \good & \good & \good & 1.2757(84) & 0.9896(61) \\
\hspace{-0.2cm}{\begin{tabular}{l}FNAL/MILC/\rule{0.1cm}{0cm}\\TUMQCD 18\end{tabular}}
		& \cite{Bazavov:2018omf} & 2+1+1 & \gA & \good &  \good & \good & $-$ & 1.273(4)(1)(10) & 0.9837(43)(14)(33)(5)  \\ 
HPQCD 14A  & \cite{Chakraborty:2014aca} & 2+1+1 & \gA & \good & \good & \good & $-$ & 1.2715(95) & 0.9851(63) \\ 
ETM 14A & \cite{Alexandrou:2014sha} & 2+1+1 & \gA & \soso & \good & \soso & \good & 1.3478(27)(195) & 1.0557(22)(153)$^*$ \\ 
ETM 14 & \cite{Carrasco:2014cwa} & 2+1+1 & \gA & \soso & \good & \soso & \good & 1.348(46) &1.058(35)$^*$ \\ 
&&&&&&&&& \\[-0.1cm]
\hline
&&&&&&&&& \\[-0.1cm]
ALPHA 21  & \cite{Heitger:2021apz} & 2+1 & \gA$^+$ & \good & \good & \good & \good &  1.296(19) &  1.007(16) \\
Petreczky 19  & \cite{Petreczky:2019ozv} & 2+1 & \gA$$ & \good & \good & \good & \good &  1.265(10) &  1.001(16) \\
Maezawa 16  & \cite{Maezawa:2016vgv} & 2+1 & \gA & \bad & \good & \good  & $\good$ &  1.267(12) &  \\
JLQCD 16 & \cite{Nakayama:2016atf} & 2+1 & \gA & \soso & \good & \good & $-$ & 1.2871(123) & 1.0033(96) \\
$\chi$QCD 14 & \cite{Yang:2014sea} & 2+1 & \gA& \soso & \soso & \soso & \good & 1.304(5)(20) & 1.006(5)(22) \\                  
HPQCD 10  & \cite{McNeile:2010ji} & 2+1 & \gA & \soso & \good & \soso  & $-$ & 1.273(6) & 0.986(6) \\
HPQCD 08B & \cite{Allison:2008xk} & 2+1 & \gA &  \soso & \good & \soso & $-$ & 1.268(9) & 0.986(10) \\
&&&&&&&&& \\[-0.1cm]
\hline \hline
&&&&&&&&& \\[-0.1cm]
PDG & \cite{Zyla:2020zbs} & & & & & & & 1.27(2) & \\[1.0ex]
\hline \hline
&&&&&&&&& \\
\end{tabular*}\\[-0.2cm]
}}
\begin{minipage}{\linewidth}
{\footnotesize 
\begin{itemize}
\item[$^\dagger$] We applied the running factor 0.7739(60) for $\mu=3$ GeV to $\overline{m}_c$. The errors are statistical, systematic, and the uncertainty in the running factor.\\[-5mm]
\item[$^*$] A running factor equal to 0.900 between the scales $\mu = 2$ GeV and $\mu = 3$ GeV was applied by us.  \\[-5mm]
\item[$^+$] Published after the FLAG deadline. \\[-5mm]
\end{itemize}
}
\end{minipage}
\caption{\label{tab:mc} Lattice results for the $\msbar$ charm-quark mass $\overline{m}_c(\overline{m}_c)$ and $\overline{m}_c(\mbox{3 GeV})$ in GeV, together with the colour coding of the calculations used to obtain them.}
\end{table}

In the next subsections we review separately the results for $\overline{m}_c$ with three or four flavours of quarks in the sea.

\subsubsection{$N_f = 2+1$ results}
\label{sec:mcnf3}

Since the last review~\cite{Aoki:2019cca}, there are two new results, Petreczky 19 \cite{Petreczky:2019ozv} and ALPHA 21 \cite{Heitger:2021apz}, the latter of which was not published at the FLAG deadline. 
Petreczky 19 employs the HISQ action on ten ensembles with ten lattice spacings down to 0.025 fm, physical strange-quark mass, and two light-quark masses, the lightest corresponding to 161 MeV pions. Their study incorporates lattices with 11 different sizes, ranging from 1.6 to 5.4 fm. The masses are computed from moments of pseudoscalar quarkonium correlation functions, and $\overline{\rm MS}$ masses are extracted with 4-loop continuum perturbation theory. Thus this work easily rates green stars in all categories.
ALPHA 21 uses the $\cO(a)$-improved Wilson-clover action with five lattice spacings from 0.087 to 0.039 fm, produced by the CLS collaboration. For each lattice spacing, several light sea-quark masses are used in a global chiral-continuum extrapolation (the lightest pion mass for one ensemble is 198 MeV). The authors also use nonperturbative renormalization and running through application of step-scaling and the Schr\"odinger functional scheme. Finite-volume effects are investigated at one lattice spacing and only for $\sim 400$ MeV pions on the smallest two volumes where results are compatible within statistical errors. ALPHA 21 satisfies the FLAG criteria for green-star ratings in all of the categories listed in Tab.~\ref{tab:mc}, but because it is a new result that was unpublished at the deadline, does not enter the average in this review.

Descriptions of the other works in this section can be found in the last review~\cite{Aoki:2019cca}.

According to our rules on the publication status, the FLAG average for the charm-quark mass at $N_f = 2+1$ is obtained by combining the results HPQCD 10, $\chi$QCD 14, JLQCD 16, and Petreczky 19,
\begin{align}
      \label{eq:mcmcnf3} 
&& \overline{m}_c(\overline{m}_c)         & = 1.275(5) ~ \gev          &&\Refs~\mbox{\cite{McNeile:2010ji,Yang:2014sea,Nakayama:2016atf,Petreczky:2019ozv}}\,, \\[-3mm]
&\mbox{$N_f = 2+1$:}&\nonumber\\[-3mm]
&&\FLAGAVBEGIN\overline{m}_c(\mbox{3 GeV})& = 0.992(5)\FLAGAVEND ~ \gev&&\Refs~\mbox{\cite{McNeile:2010ji,Yang:2014sea,Nakayama:2016atf,Petreczky:2019ozv}}\,,
\end{align}
where the error on $ \overline{m}_c(\overline{m}_c)$ includes a
stretching factor $\sqrt{\chi^2/\mbox{dof}} \simeq 1.16$ as discussed
in Sec.~\ref{sec:averages}. This result corresponds to the following
RGI average
\begin{align}
      \label{eq:mcmcnf3 rgi} 
&& M_c^{\rm RGI} & = 1.526(9)_m(14)_\Lambda ~ \gev&&\Refs~\mbox{\cite{McNeile:2010ji,Yang:2014sea,Nakayama:2016atf,Petreczky:2019ozv}}\,.
\end{align}

\subsubsection{$N_f = 2+1+1$ results}
\label{sec:mcnf4}

For a discussion of older results, see the previous FLAG reviews. Since FLAG 19 two groups have produced updated values with charm quarks in the sea. 

HPQCD 20A~\cite{Hatton:2020qhk} is an update of HPQCD 18, including a new finer ensemble ($a\approx 0.045$ fm) and EM corrections computed in the quenched approximation of QED for the first time. Besides these new items, the analysis is largely unchanged from HPQCD 18 except for an added $\alpha_s^3$ correction to the SMOM-to-$\overline{\rm MS}$ conversion factor and tuning the bare charm mass via the $J/\psi$ mass rather than the $\eta_c$. Their new value in pure QCD is $\overline{m}_c(3~\rm GeV)=0.9858(51)$ GeV which is quite consistent with HPQCD 18 and the FLAG 19 average. The effects of quenched QED in both the bare charm-quark mass and the renormalization constant are small. Both effects are precisely determined, and the overall effect shifts the mass down slightly to $\overline{m}_c(3~\rm GeV)=0.9841(51)$ where the uncertainty due to QED is invisible in the final error. The shift from their pure QCD value due to quenched QED is about $-0.2\%$. 

{\color{black}
ETM 21A~\cite{Alexandrou:2021gqw} is a new  work that follows a similar methodology as ETM 14, but with significant improvements. Notably, a clover-term is added to the twisted mass fermion action which suppresses $\cO(a^2)$ effects between the neutral and charged pions. Additional improvements include new ensembles lying very close to the physical mass point, better control of nonperturbative renormalization systematics, and use of both meson and baryon correlation functions to determine the quark mass. They use the RI-MOM scheme for nonperturbative renormalization. The analysis comprises ten ensembles in total with three lattice spacings (0.095, 0.082, and 0.069 fm), two volumes for the finest lattice spacings and four for the other two, and pion masses down to 134 MeV for the finest ensemble. The values of $m_\pi L$ range mostly from almost four to greater than five. According to the FLAG criteria, green stars are earned in all categories. The authors find $m_c(3~\rm GeV)=1.036(17)(^{+15}_{-8})$ GeV. In Tab.~\ref{tab:mc} we have applied a factor of 0.7739(60) to run from 3 GeV to $\overline{m}_c$. As in FLAG 19, the new value is consistent with ETM 14 and ETM 14A, but is still high compared to the FLAG average. The authors plan future improvements, including a finer lattice spacing for better control of the continuum limit and a new renormalization scheme, like RI-SMOM. This result has not been published by the deadline, so it does not yet appear in the average. 
}

Five results enter the FLAG average for $N_f = 2+1+1$ quark flavours: ETM 14, ETM 14A, HPQCD 14A, FNAL/MILC/TUMQCD 18, and HPQCD 20A. We note that while
the determinations of $\overline{m}_c$ by ETM 14 and 14A agree well with each other, they are incompatible with HPQCD 14A, FNAL/MILC/TUMQCD 18, and HPQCD 20A by several standard deviations. While the latter use the same configurations, the analyses are quite different and independent. As mentioned earlier, $m_{ud}$ and $m_s$ values by ETM are also systematically high compared to their respective averages. Combining all four results yields

 \begin{align}
      \label{eq:mcmcnf4} 
&& \overline{m}_c(\overline{m}_c)           & = 1.278(13)\gev          &&\Refs~\mbox{\cite{Carrasco:2014cwa,Chakraborty:2014aca,Alexandrou:2014sha,Bazavov:2018omf,Hatton:2020qhk}}\,, \\[-3mm]
&\mbox{$N_f = 2+1+1$:}& \nonumber\\[-3mm]
&&  \FLAGAVBEGIN\overline{m}_c(\mbox{3 GeV})& = 0.988(11)\FLAGAVEND ~ \gev&&\Refs~\mbox{\cite{Carrasco:2014cwa,Chakraborty:2014aca,Alexandrou:2014sha,Bazavov:2018omf,Hatton:2020qhk}}\,,
 \end{align}
where the errors include large stretching factors
$\sqrt{\chi^2/\mbox{dof}}\approx2.0$ and $2.5$, respectively. We have
assumed 100\% correlation for statistical errors between ETM results and the same for
HPQCD 14A, HPQCD 20A, and FNAL/MILC/TUMQCD 18. 

These are obviously poor $\chi^2$ values, and the stretching factors are quite large. While it may be prudent in such a case to quote a range of values covering the central values of all results that pass the quality criteria, we believe in this case that would obscure rather than clarify the situation.  From Fig.~\ref{fig:mc} we note that not only do ETM 21A, ETM 14A, and ETM 14 lie well above the other 2+1+1 results, but also above all of the 2+1 flavour results. A similar trend is apparent for the light-quark masses (see Figs.~\ref{fig:ms} and \ref{fig:mud}) while for mass ratios there is better agreement (Figs.~\ref{fig:msovmud},~\ref{fig:mu over md}~and~\ref{fig:mc over ms}). The latter suggests there may be underestimated systematic uncertainties associated with scale setting and/or renormalization which have not been detected. Finally we note the ETM results are significantly higher than the PDG average. For these reasons, which admittedly are not entirely satisfactory, we continue to quote an average with a stretching factor as in previous reviews.

The RGI average reads as follows,
 \begin{align}
      \label{eq:mcmcnf4 rgi} 
&&  M_c^{\rm RGI}& = 1.520(17)_m(14)_\Lambda  ~ \gev&&\Refs~\mbox{\cite{Alexandrou:2014sha,Chakraborty:2014aca,Bazavov:2018omf,Carrasco:2014cwa,Hatton:2020qhk}}\,.
 \end{align}

Figure~\ref{fig:mc} presents the values of $\overline{m}_c(\overline{m}_c)$ given in Tab.~\ref{tab:mc} along with the FLAG averages obtained for $2+1$ and $2+1+1$ flavours.
\begin{figure}[!htb]
\begin{center}
\includegraphics[width=11.5cm]{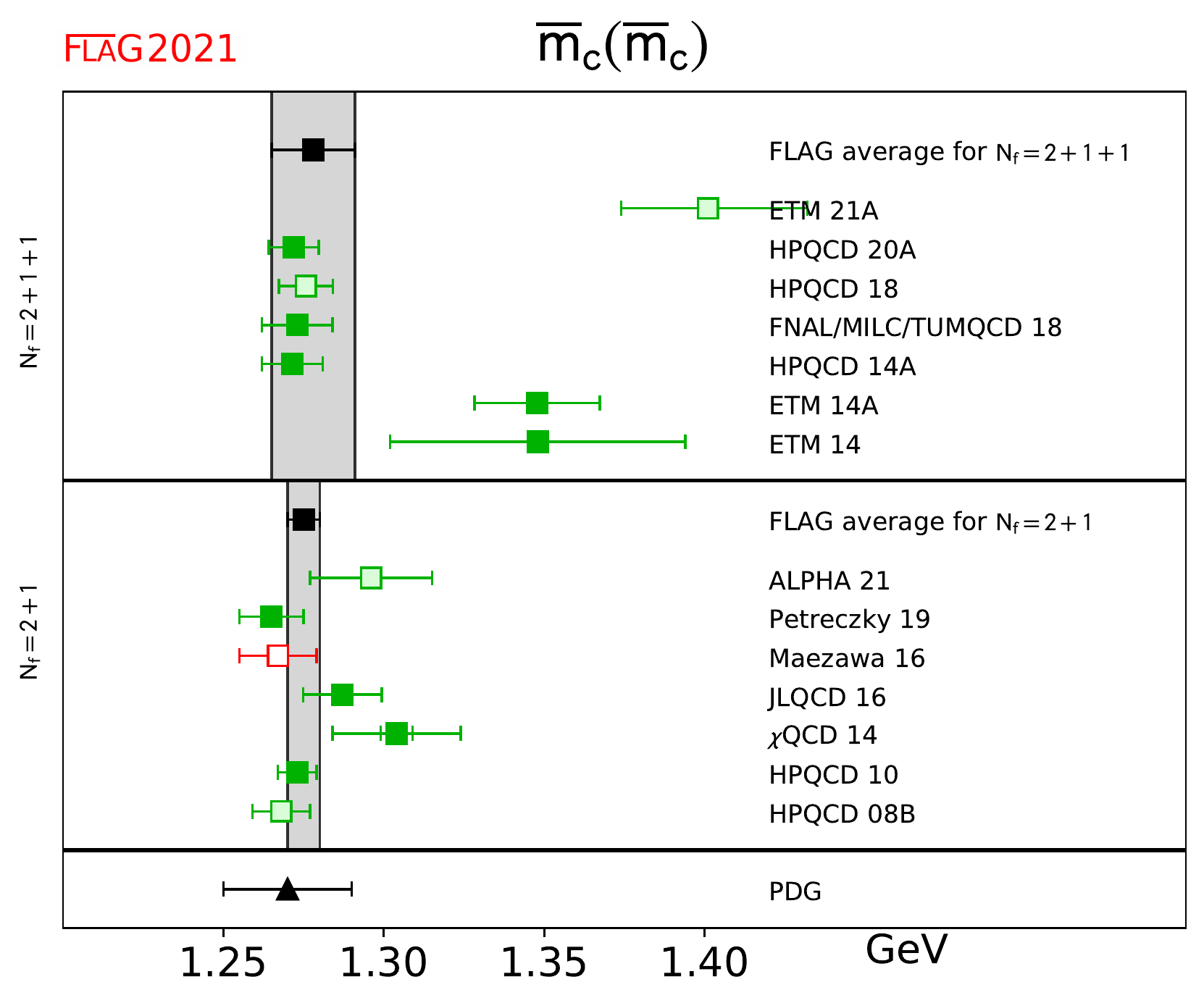}
\end{center}
\vspace{-1cm}
\caption{\label{fig:mc} The charm-quark mass for $2+1$ and $2+1+1$ flavours. For the latter a large stretching factor is used for the FLAG average due to poor $\chi^2$ from our fit.}
\end{figure}

\subsubsection{Lattice determinations of the ratio $m_c/m_s$}
\label{sec:mcoverms}
Because some of the results for quark masses given in this review are obtained via the quark-mass ratio $m_c/m_s$, we review these lattice calculations, which are listed in Tab.~\ref{tab:mcoverms}, as well.

\begin{table}[!htb]
\vspace{3cm}
{\footnotesize{
\begin{tabular*}{\textwidth}[l]{l@{\extracolsep{\fill}}rllllll}
Collaboration & Ref. & $\Nf$ & \hspace{0.15cm}\begin{rotate}{60}{publication status}\end{rotate}\hspace{-0.15cm}  &
 \hspace{0.15cm}\begin{rotate}{60}{chiral extrapolation}\end{rotate}\hspace{-0.15cm} &
 \hspace{0.15cm}\begin{rotate}{60}{continuum  extrapolation}\end{rotate}\hspace{-0.15cm}  &
 \hspace{0.15cm}\begin{rotate}{60}{finite volume}\end{rotate}\hspace{-0.15cm}  & \rule{0.1cm}{0cm} 
$m_c/m_s$ \\
&&&&&& \\[-0.1cm]
\hline
\hline
&&&&&& \\[-0.1cm]
ETM 21A & \cite{Alexandrou:2021gqw} & 2+1+1 & \oP & \good &  \good & \good & 11.48(12)($^{+25}_{-19}$)\\ 
FNAL/MILC/TUMQCD 18  & \cite{Bazavov:2018omf} & 2+1+1 & \gA & \good &  \good & \good & 11.784(11)(17)(00)(08) \\ 
HPQCD 14A  & \cite{Chakraborty:2014aca} & 2+1+1  & \gA & \good & \good & \good  & 11.652(35)(55) \\
FNAL/MILC 14A & \cite{Bazavov:2014wgs}  & 2+1+1  & \gA & \good & \good & \good  & 11.747(19)($^{+59}_{-43}$) \\
ETM 14 & \cite{Carrasco:2014cwa}  & 2+1+1  & \gA & \soso & \good & \soso & 11.62(16) \\
&&&&&& \\[-0.1cm]  
\hline 
&&&&&& \\[-0.1cm]
Maezawa 16  & \cite{Maezawa:2016vgv} & 2+1 & \gA & \bad & \good & \good  & 11.877(91)  \\
$\chi$QCD 14 & \cite{Yang:2014sea} & 2+1  & \gA & \soso & \soso & \soso & 11.1(8) \\
HPQCD 09A & \cite{Davies:2009ih}  & 2+1  & \gA & \soso & \good & \good & 11.85(16) \\
&&&&&& \\[-0.1cm]  
\hline
\hline
\end{tabular*}
}}
\caption{Lattice results for the quark-mass ratio $m_c/m_s$, together with the colour coding of the calculations used to obtain them.}
\label{tab:mcoverms}
\end{table}

The $N_f = 2+1$ results from $\chi$QCD 14 and HPQCD 09A~\cite{Davies:2009ih} are from the same calculations that were described for the charm-quark mass in the previous review. Maezawa 16 does not pass our chiral-limit test (see the previous review), though we note that it is quite consistent with the other values. Combining $\chi$QCD 14 and HPQCD 09A,  we obtain the same result reported in FLAG 19, 
 \be
      \label{eq:mcmsnf3} 
      \mbox{$N_f = 2+1$:} \qquad\FLAGAVBEGIN m_c / m_s = 11.82(16)\FLAGAVEND\qquad\Refs~\mbox{\cite{Yang:2014sea,Davies:2009ih}},
 \ee
with a $\chi^2/\mbox{dof} \simeq 0.85$.

Turning to $N_f = 2+1+1$, there is a new result from ETM 21A. The errors have actually increased compared to ETM 14, due to larger uncertainties in the baryon sector which enter their average with the meson sector. Again, ETM 21A does not yet enter the average since it was not published by the deadline for the review. See the earlier reviews for a discussion of previous results.

We note that some tension exists between the HPQCD 14A and FNAL/MILC/TUMQCD results. 
Combining these with ETM 14  yields

 \be
      \label{eq:mcmsnf4} 
      \mbox{$N_f = 2+1+1$:} \qquad \FLAGAVBEGIN m_c / m_s = 11.768(34)\FLAGAVEND\qquad\Refs~\mbox{\cite{Chakraborty:2014aca,Carrasco:2014cwa,Bazavov:2018omf}},
 \ee
where the error includes the stretching factor $\sqrt{\chi^2/\mbox{dof}} \simeq 1.5$. We have assumed a 100\% correlation of statistical errors for FNAL/MILC/TUMQCD 18 and HPQCD 14A.

Results for $m_c/m_s$ are shown in Fig.~\ref{fig:mc over ms} together with the FLAG averages for $N_f = 2+1$ and $2+1+1$ flavours. 

\begin{figure}[!htb]
\begin{center}
\includegraphics[width=11cm]{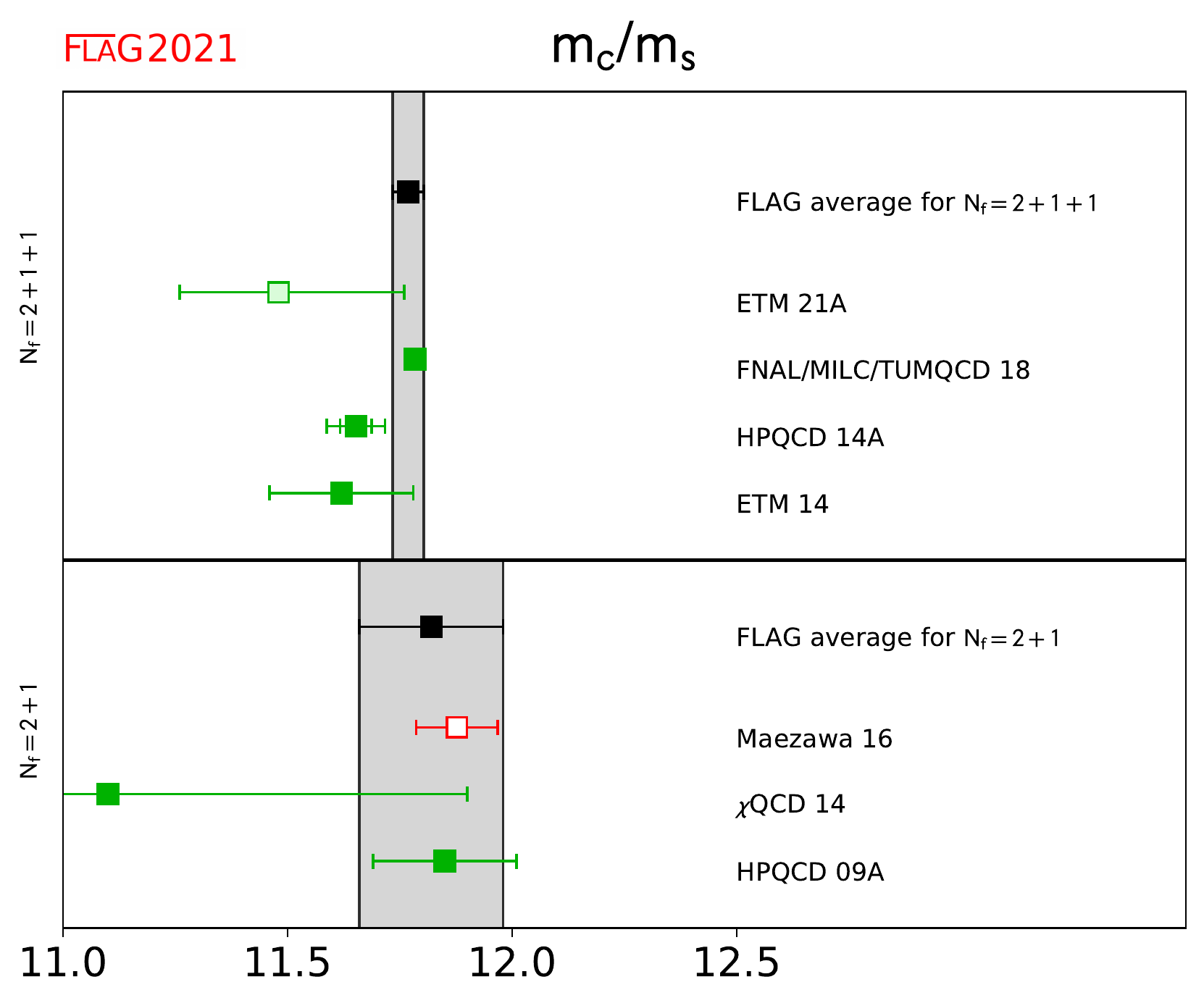}
\end{center}
\begin{center}
\caption{\label{fig:mc over ms}Lattice results for the ratio $m_c / m_s$ listed in Tab.~\ref{tab:mcoverms} and the FLAG averages corresponding to $2+1$ and $2+1+1$ quark flavours. The latter average includes a large stretching factor on the error due a poor $\chi^2$ from our fit.}
\end{center}
\end{figure}

\newpage

\subsection{Bottom-quark mass}
\label{s:bmass}

Now we review the lattice results for the $\overline{\rm MS}$ bottom-quark mass $\overline{m}_b$.  Related heavy-quark actions and observables have been discussed 
in previous FLAG reviews \cite{Aoki:2013ldr,Aoki:2016frl,Aoki:2019cca}, and descriptions
can be found in Sec.~A.1.3 of FLAG 19 \cite{Aoki:2019cca}.  In Tab.~\ref{tab:mb} we collect results for
$\overline{m}_b(\overline{m}_b)$ obtained with $N_f =2+1$ and
$2+1+1$ sea-quark flavours.  Available results for the quark-mass ratio $m_b / m_c$ are also reported.
After discussing the new results we evaluate the corresponding FLAG averages.

\begin{table}[!htb]
\vspace{3cm}
{\footnotesize{
\begin{tabular*}{\textwidth}[l]{l@{\extracolsep{\fill}}rl@{\hspace{0mm}}l@{\hspace{0mm}}l@{\hspace{0mm}}l@{\hspace{0mm}}l@{\hspace{0mm}}l@{\hspace{0mm}}lll}
Collaboration & Ref. & $N_f$ & \hspace{0.15cm}\begin{rotate}{60}{publication status}\end{rotate}\hspace{-0.15cm} &
 \hspace{0.15cm}\begin{rotate}{60}{chiral extrapolation}\end{rotate}\hspace{-0.15cm} &
 \hspace{0.15cm}\begin{rotate}{60}{continuum extrapolation}\end{rotate}\hspace{-0.15cm} &
 \hspace{0.15cm}\begin{rotate}{60}{finite volume}\end{rotate}\hspace{-0.15cm} &  
 \hspace{0.15cm}\begin{rotate}{60}{renormalization}\end{rotate}\hspace{-0.15cm} &  
 \hspace{0.15cm}\begin{rotate}{60}{heavy-quark treatment}\end{rotate}\hspace{-0.15cm} & 
 \rule{0.2cm}{0cm}$\overline{m}_b(\overline{m}_b)$ & 
 \rule{0.2cm}{0cm}$m_b / m_c$ \\
&&&&&&&&&& \\[-0.1cm]
\hline
\hline
&&&&&&&&&& \\[-0.1cm]
HPQCD 21  & \cite{Hatton:2021syc} & 2+1+1 & \gA & \good &\good & \good & $-$ & \okay & 4.209(21)$^{++}$ & 4.586(12)$^{**}$ \\ 
FNAL/MILC/TUM 18  & \cite{Bazavov:2018omf} & 2+1+1 & A & \good &  \soso & \good & $-$ & \okay & 4.201(12)(1)(8)(1)  & 4.578(5)(6)(0)(1)  \\ 
Gambino 17 & \cite{Gambino:2017vkx} & 2+1+1 & A & \soso & \good  & \soso &\good  &\okay & 4.26(18) & \\ 
ETM 16B & \cite{Bussone:2016iua} & 2+1+1 & A &\soso &\good & \soso &\good & \okay & 4.26(3)(10)$^+$ & 4.42(3)(8) \\ 
HPQCD 14B  & \cite{Colquhoun:2014ica} & 2+1+1 & \gA &\good & \good & \good & \good & \okay & 4.196(0)(23)$^\dagger$ & \\
ETM 14B & \cite{Bussone:2014cha} & 2+1+1 & \rC &\soso &\good & \soso &\good & \okay & 4.26(7)(14) & 4.40(6)(5) \\ 
HPQCD 14A  & \cite{Chakraborty:2014aca} & 2+1+1 & \gA & \good &\good & \good & $-$ & \okay & 4.162(48) & 4.528(14)(52) \\ 
&&&&&&&&&& \\[-0.1cm]
\hline
&&&&&&&&&& \\[-0.1cm]
Petreczky19  & \cite{Petreczky:2019ozv} & 2+1 & \gA & \good & \good & \good  & $\good$ &\okay & 4.188(37)  & 4.586(43)  \\
Maezawa 16  & \cite{Maezawa:2016vgv} & 2+1 & \gA & \bad & \good & \good  & $\good$ &\okay & 4.184(89)  & 4.528(57)  \\
HPQCD 13B  & \cite{Lee:2013mla} & 2+1 & \gA &\bad &\soso & $-$ & $-$ & \okay & 4.166(43) & \\ 
HPQCD 10 & \cite{McNeile:2010ji} & 2+1 & \gA &\good & \good & \good& $-$ & \okay & 4.164(23)$^\star$ & 4.51(4) \\ 
&&&&&&&&&& \\[-0.1cm]
\hline
&&&&&&&&&& \\[-0.1cm]
ETM 13B & \cite{Carrasco:2013zta} & 2 & \gA & \soso & \good & \soso & \good &\okay & 4.31(9)(8) & \\
ALPHA 13C & \cite{Bernardoni:2013xba} & 2 & \gA & \good & \good & \good & \good & \okay & 4.21(11) & \\
ETM 11A & \cite{Dimopoulos:2011gx} & 2 & \gA & \soso & \good & \soso & \good & \okay & 4.29(14) & \\[1.0ex]
\hline \hline
&&&&&&&&&& \\[-0.1cm]
PDG & \cite{Zyla:2020zbs} & & & & & & & & 4.18$^{+0.02}_{-0.03}$ & \\[1.0ex]
\hline \hline
&&&&&&&&&& \\
\end{tabular*}\\[-0.2cm]
}}
\begin{minipage}{\linewidth}
{\footnotesize 
\begin{itemize}
\item[$^{++}$] We quote the four-flavour result. For $N_f=5$, value is 4.202(21). \\[-5mm]
\item[$^{**}$] The ratio is quoted in the $\overline{\rm MS}$ scheme for $\mu=3$ GeV because of the different charges of the bottom and charm quarks. \\[-5mm]
\item[$^+$] The lattice spacing used in ETM 14B has been updated here. \\[-5mm]
\item[$^\dagger$] Only two pion points are used for chiral extrapolation. \\[-5mm]
\item[$^{\star}$] The number that is given is $m_b(10~\gev, N_f = 5) = 3.617(25)~\gev$.
\end{itemize}
}
\end{minipage}
\caption{\label{tab:mb} Lattice results for the $\msbar$ bottom-quark mass $\overline{m}_b(\overline{m}_b)$ in GeV,
  together with the systematic error ratings for each. Available results for the quark-mass ratio $m_b / m_c$ are also reported.}
\end{table}

\subsubsection{$N_f=2+1$}

There is one new three-flavour result since the last review, Petreczky 19, which was described already in the charm-quark section. The new result rates green stars, so our new average with HPQCD 10 is (both works quote values in the $N_f
= 5$ theory, so we simply use those values),
\begin{align}
&N_f= 2+1 :  &\FLAGAVBEGIN\overline{m}_b(\overline{m}_b)& = 4.171 (20)  \FLAGAVEND ~ \gev&&\Ref ~\mbox{\cite{McNeile:2010ji,Petreczky:2019ozv}}\,.
\end{align}
The corresponding four-flavour RGI average is
\begin{align}
&N_f= 2+1 :  & M_b^{\rm RGI} & = 6.881(33)_m(54)_\Lambda  ~ \gev &&\Ref ~\mbox{\cite{McNeile:2010ji,Petreczky:2019ozv}}\,.
\end{align}

\subsubsection{$N_f=2+1+1$}

HPQCD 21 \cite{Hatton:2021syc} is an update of HPQCD 14A (and replaces it in our average), including EM corrections for the first time for the $b$-quark mass. Four flavours of HISQ quarks are used on MILC ensembles with lattice spacings from about 0.09 to 0.03 fm. Ensembles with physical and unphysical mass sea-quarks are used. Quenched QED is used to obtain the dominant $\cO(\alpha)$ effect. The ratio of bottom- to charm-quark masses is computed in a completely nonperturbative formulation, and the $b$-quark mass is extracted using the value of $\overline{m}_c(3~\rm GeV)$ from HPQCD 20A. Since EM effects are included, the QED renormalization scale enters the ratio which is quoted for 3 GeV and $N_f=4$. The total error on the new result is more than two times smaller than for HPQCD 14A, but is only slightly smaller compared to the NRQCD result reported in HPQCD 14B. The inclusion of QED shifts the ratio $m_b/m_c$ up slightly from the pure QCD value by about one standard deviation, and the value of $\overline{m}_b(m_b)$ is consistent, within errors, to the other pure QCD results entering our average. Therefore we quote a single average.

HPQCD 14B employs the NRQCD
action~\cite{Colquhoun:2014ica} to treat the $b$ quark.  The
$b$-quark mass is computed with the moments method, that is, from
Euclidean-time moments of two-point, heavy-heavy-meson correlation
functions (see also Sec.~\ref{s:curr} for a description of the method).

In HPQCD 14B the $b$-quark mass is computed from ratios of the moments
$R_n$ of heavy current-current correlation functions, namely, 
\be
\left[\frac{R_n r_{n-2}}{R_{n-2}r_n}\right]^{1/2} \frac{\bar{M}_{\rm
    kin}}{2 m_b} = \frac{\bar{M}_{\Upsilon,\eta_b}}{2 \bar m_b(\mu)} ~
,
      \label{eq:moments}
\ee 
where $r_n$ are the perturbative moments calculated at N$^3$LO,
$\bar{M}_{\rm kin}$ is the spin-averaged kinetic mass of the
heavy-heavy vector and pseudoscalar mesons and
$\bar{M}_{\Upsilon,\eta_b}$ is the expe\-rimental spin average of the
$\Upsilon$ and $\eta_b$ masses.  The average kinetic mass $\bar{M}_{\rm kin}$
is chosen since in the lattice calculation the splitting of the
$\Upsilon$ and $\eta_b$ states is inverted.  In Eq.~(\ref{eq:moments}),
the bare mass $m_b$ appearing on the left-hand side is tuned so that
the spin-averaged mass agrees with experiment, while the mass
$\overline{m}_b$ at the fixed scale $\mu = 4.18$ GeV is extra\-polated
to the continuum limit using three HISQ (MILC) ensembles with $a
\approx$ 0.15, 0.12 and 0.09 fm and two pion masses, one of which is
the physical one. Their final result is
$\overline{m}_b(\mu = 4.18\, \gev) = 4.207(26)$ GeV, where the error is
from adding systematic uncertainties in quadrature only (statistical
errors are smaller than $0.1 \%$ and ignored). The errors arise from
renormalization, perturbation theory, lattice spacing, and NRQCD
systema\-tics. The finite-volume uncertainty is not estimated, but at
the lowest pion mass they have $ m_\pi L \simeq 4$, which leads to the
tag \good\ .

The next four-flavour result~\cite{Bussone:2016iua} is from the ETM collaboration and updates their preliminary result
appearing in a conference proceedings~\cite{Bussone:2014cha}. The calculation is performed on a set of configurations
generated with twisted-Wilson fermions with three lattice spacings in
the range 0.06 to 0.09 fm and with pion masses in the range 210 to 440
MeV. The $b$-quark mass is determined from a ratio of heavy-light
pseudoscalar meson masses designed to yield the quark pole mass in the
static limit. The pole mass is related to the $\overline{\rm MS}$ mass
through perturbation theory at N$^3$LO. The key idea is that by taking
ratios of ratios, the $b$-quark mass is accessible through fits to
heavy-light(strange)-meson correlation functions computed on the
lattice in the range $\sim 1$--$2\times m_c$ and the static limit, the
latter being exactly 1. By simulating below $\overline{m}_b$, taking
the continuum limit is easier. They find
$\overline{m}_b(\overline{m}_b) = 4.26(3)(10)$ GeV, where the first
error is statistical and the second systematic. The dominant errors
come from setting the lattice scale and fit systematics.

Gambino {\it et al.}~\cite{Gambino:2017vkx} use twisted-mass-fermion ensembles from the ETM collaboration and the ETM ratio method as in ETM 16. Three values of the lattice spacing are used, ranging from 0.062 to 0.089 fm. Several volumes are also used. The light-quark masses produce pions with masses from 210 to 450 MeV. The main difference with ETM 16 is that the authors use the kinetic mass defined in the heavy-quark expansion (HQE)  to extract the $b$-quark mass instead of the pole mass.

The final $b$-quark mass result is FNAL/MILC/TUM 18~\cite{Bazavov:2018omf}. The mass is extracted from the same fit and analysis done for the charm quark mass. 
{\color{black}
Note that relativistic HISQ valence masses reach the physical $b$ mass on
the two finest lattice spacings ($a = 0.042$ fm, 0.03 fm) at physical and
0.2 $m_s$ light-quark mass, respectively. In lattice units the heavy valence
masses correspond to $aM^{\rm RGI} > 0.90$, making the continuum extrapolation
challenging, but the authors investigated the effect of
leaving out the heaviest points from the fit, and the result did not
noticeably change. Their results are also consistent with an analysis
dropping the finest lattice from the fit. Since the $b$-quark mass region
is only reached with two lattice spacings, we rate this
work with a green circle for the continuum extrapolation. Note however
that for other values of the quark masses they use up to five values of the
lattice spacing (cf.~their charm-quark mass determination).}

All of the above results enter our average. We note that here the ETM 16 result is consistent with the average and a stretching factor on the error is not used. The average and error is dominated by the very precise FNAL/MILC/TUM 18 value,
\begin{align}
&N_f= 2+1+1 :&\FLAGAVBEGIN\overline{m}_b(\overline{m}_b)& = 4.203 (11)  \FLAGAVEND ~ \gev&&\Refs~\mbox{\cite{Hatton:2021syc,Chakraborty:2014aca,Colquhoun:2014ica,Bussone:2016iua,Gambino:2017vkx,Bazavov:2018omf}}\,.
\end{align}
We have included a 100\%
correlation on the statistical errors of ETM 16 and Gambino 17, since
the same ensembles are used in both. While FNAL/MILC/TUM 18 and HPQCD 21 also use the same MILC HISQ ensembles, the statistical error in the HPQCD 21 analysis is negligible, so we do not include a correlation between them. The average has $\chi^2/{\rm dof}=0.02$.

The above translates to the RGI average
\begin{align}
&N_f= 2+1+1 :& M_b^{\rm RGI} & = 6.934(18)_m(55)_\Lambda ~\gev&&\Refs~\mbox{\cite{Hatton:2021syc,Chakraborty:2014aca,Colquhoun:2014ica,Bussone:2016iua,Gambino:2017vkx,Bazavov:2018omf}}\,.
\end{align}

All the results for $\overline{m}_b(\overline{m}_b)$ discussed above
are shown in Fig.~\ref{fig:mb} together with the FLAG averages
corresponding to $N_f=2+1$ and $2+1+1$ quark flavours.
\begin{figure}[!htb]
\begin{center}
\includegraphics[width=11cm]{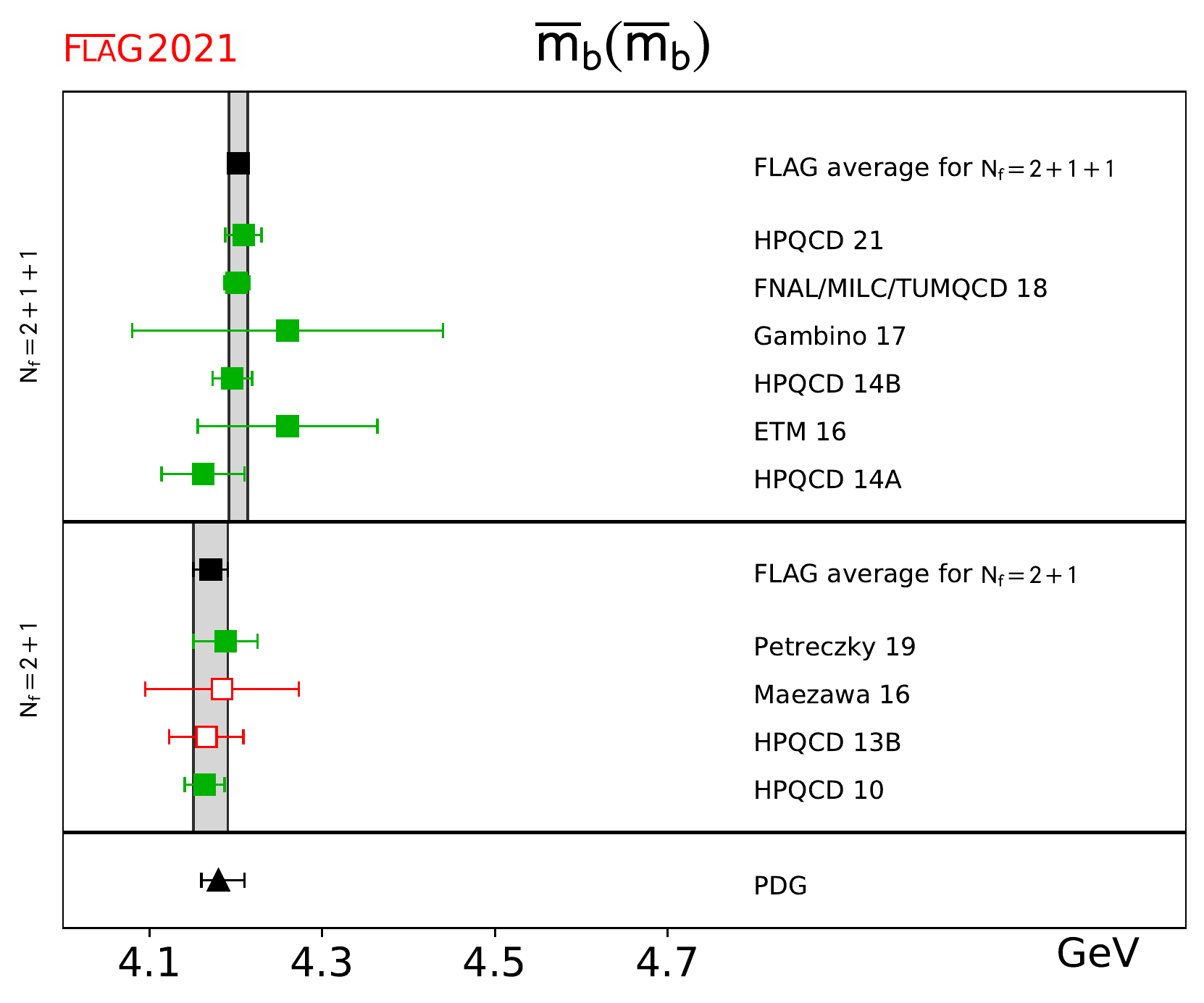}
\end{center}
\vspace{-1cm}
\caption{ \label{fig:mb} The $b$-quark mass for $N_f =2+1$ and $2+1+1$ flavours. The updated PDG value from
  Ref.~\cite{Zyla:2020zbs} is reported for comparison.}
\end{figure}

\clearpage
\setcounter{section}{3}
\section{Leptonic and semileptonic kaon and pion decay and $|V_{ud}|$ and $|V_{us}|$}
\label{sec:vusvud}
Authors: T.~Kaneko, J.~N.~Simone, S.~Simula, N.~Tantalo\\

This section summarizes state-of-the-art lattice calculations of the leptonic kaon and pion decay constants and the kaon semileptonic-decay form factor and provides an analysis in view of the Standard Model.
With respect to the previous edition of the FLAG review \cite{Aoki:2019cca} the data in this section has been updated.
As in Ref.~\cite{Aoki:2019cca}, when combining lattice data with experimental results, we take into account the strong $SU(2)$ isospin correction, either obtained in lattice calculations or estimated by using chiral perturbation theory ({\Ch}PT), both for the kaon leptonic decay constant $f_{K^\pm}$ and for the ratio $f_{K^\pm} / f_{\pi^\pm}$.

\subsection{Experimental information concerning $|V_{ud}|$, $|V_{us}|$, $f_+(0)$ and $\fKfpichargedr$}\label{sec:Exp} 

The following review relies on the fact that precision 
experimental data on kaon decays
very accurately determine the product $|V_{us}|f_+(0)$ \cite{Moulson:2017ive} and the ratio
$|V_{us}/V_{ud}|f_{K^\pm}/f_{\pi^\pm}$ \cite{Moulson:2017ive,Zyla:2020zbs}: 
\be\label{eq:products}
|V_{us}| f_+(0) = 0.2165(4)\co \hspace{1cm} \;
\left|\frac{V_{us}}{V_{ud}}\right|\frac{ f_{K^\pm}}{ f_{\pi^\pm}} \;
=0.2760(4)\fs\ee 
Here and in the following, $f_{K^\pm}$ and $f_{\pi^\pm}$ are the isospin-broken 
decay constants, respectively, in QCD. We will refer to the decay 
constants in the $SU(2)$ isospin-symmetric limit as $f_K$ and $f_\pi$ 
(the latter at leading order in the mass difference ($m_u - m_d$) coincides with $f_{\pi^\pm}$).
The parameters $|V_{ud}|$ and $|V_{us}|$ are
elements of the Cabibbo-Kobayashi-Maskawa matrix and $f_+(q^2)$ represents
one of the form factors relevant for the semileptonic decay
$K^0\rightarrow\pi^-\ell\,\nu$, which depends on the momentum transfer $q$
between the two mesons.  What matters here is the value at $q^2 = 0$:
$f_+(0) \equiv f_+^{K^0\pi^-}(0) = f_0^{K^0\pi^-}(0) = q^\mu \langle \pi^-(p^\prime) | \bar{s} \gamma_\mu u | K^0(p) \rangle / (M_K^2 - M_\pi^2)
\,\rule[-0.15cm]{0.02cm}{0.5cm}_{\;q^2\rightarrow 0}$. 
  The pion and kaon decay constants are defined by\footnote{The pion
  decay constant represents a QCD matrix element---in the full Standard
  Model, the one-pion state is not a meaningful notion: the correlation
  function of the charged axial current does not have a pole at
  $p^2=M_{\pi^+}^2$, but a branch cut extending from $M_{\pi^+}^2$ to
  $\infty$. The analytic properties of the correlation function and the
  problems encountered in the determination of $f_\pi$ are thoroughly
  discussed in Ref.~\cite{Gasser:2010wz}. The ``experimental'' value of $f_\pi$
  depends on the convention used when splitting the sum ${\cal
    L}_{\mbox{\tiny QCD}}+{\cal L}_{\mbox{\tiny QED}}$ into two parts. The lattice
  determinations of $f_\pi$ do not yet reach the accuracy where this is of
  significance, but at the precision claimed by the Particle Data Group
  \cite{Agashe:2014kda,Patrignani:2016xqp}, the numerical value does depend on the convention 
  used~\cite{Gasser:2003hk,Rusetsky:2009ic,Gasser:2007de,Gasser:2010wz}. 
  }  \bdm
\lvac \dbar\gamma_\mu\gamma_5 \hspace{0.05cm}u|\pi^+(p)\rangle=i
\hspace{0.05cm}p_\mu f_{\pi^+}\co\hspace{1cm} \lvac \sbar\gamma_\mu\gamma_5
\hspace{0.05cm} u|K^+(p)\rangle=i \hspace{0.05cm}p_\mu f_{K^+}\fs\edm In this
normalization, $f_{\pi^\pm} \simeq 130$~MeV, $f_{K^\pm}\simeq 155$~MeV.
 
 In Eq.~(\ref{eq:products}), the
electromagnetic effects have already been subtracted in the experimental
analysis using {\Ch}PT.
Recently, a new method~\cite{Carrasco:2015xwa} has been proposed
for calculating the leptonic decay rates of hadrons including both QCD and QED on the lattice, 
and successfully applied to the case of the ratio of the leptonic decay rates of kaons and 
pions~\cite{Giusti:2017dwk,DiCarlo:2019thl}.
The correction to the tree-level $K_{\mu2} / \pi_{\mu 2}$ decay rate, including both electromagnetic and strong 
isospin-breaking effects, is found to be equal to $-1.26 (14) \%$~\footnote{This has been updated in Ref.~\cite{DiCarlo:2019thl}
after the previous edition of this review.
See also the extended discussion concerning the isospin correction in Sec.~\ref{sec:scalesetting} on the scale setting.}
to be compared to the estimate $-1.12 (21) \%$ based on {\Ch}PT \cite{Rosner:2015wva,Cirigliano:2011tm}. 
Using the experimental values of the $K_{\mu2} $ and $\pi_{\mu 2}$ decay rates the result of 
Ref.~\cite{DiCarlo:2019thl} implies
 \be\label{eq:VusVud_new}
\left|\frac{V_{us}}{V_{ud}}\right|\frac{f_K}{f_\pi} = 0.27683 \, (29)_{\rm exp} \, (20)_{\rm th} \, [35] ~ , \ee 
where the last error in brackets is the sum in quadrature of the experimental and theoretical uncertainties, 
and the ratio of the decay constants is the one corresponding to isosymmetric QCD.
A large part of the theoretical uncertainty comes from the statistics and continuum and chiral extrapolation of lattice data, which can be systematically reduced by a more realistic simulation with high statistics.
We also note that an independent study of the electromagnetic effects
is in progress~\cite{Boyle:2019rdx}.
Therefore, it is feasible to more precisely determine $|V_{us} / V_{ud}|$ using only lattice-QCD+QED for $f_{K^\pm} / f_{\pi^\pm}$
and the ratio of the experimental values of the $K_{\mu2} $ and $\pi_{\mu 2}$ decay rates.

At present,
the superallowed nuclear $\beta$ transitions provide the most precise determination of $|V_{ud}|$.
Its accuracy has been limited by hadronic uncertainties in 
the universal electroweak radiative correction $\Delta_R^V$.
A recent analysis in terms of a dispersion relation~\cite{Seng:2018qru,Seng:2018yzq}
found $\Delta_R^V$ larger than the previous estimate~\cite{Marciano:2005ec}.
A more straightforward update of Ref.~\cite{Marciano:2005ec} also reported 
larger $\Delta_R^V$~\cite{Czarnecki:2019mwq}.
In the PDG review,
the fourteen precisely measured transitions~\cite{Hardy:2016vhg}
with the dispersive estimate of $\Delta_R^V$ yield~\cite{Zyla:2020zbs}
\be\label{eq:Vud beta}
|V_{ud}| =  0.97370(14),\ee
which differs by $\approx 3 \sigma$ from the previous estimate~\cite{Hardy:2016vhg}.
However, 
it is not a trivial matter to properly take account of the nuclear corrections
at this precision~\cite{Towner:2007np,Miller:2008my,Auerbach:2008ut,Liang:2009pf,Miller:2009cg,Towner:2010bx,Hardy:2014qxa,Hardy:2016vhg,Seng:2018qru,Gorchtein:2018fxl}.
For example, the dispersive approach has been applied in a recent update
of the so-called inner radiative correction
due to quenching of the axial-vector and isoscalar spin-magnetic-moment
couplings in nuclei~\cite{Seng:2018qru},
and in a recent estimate of a novel correction due to the distortion of
the emitted electron energy spectrum
by nuclear polarizabilities~\cite{Gorchtein:2018fxl}.
A recent reanalysis of twenty-three $\beta$ decays~\cite{Hardy:2020qwl}
obtained
\be\label{eq:Vud beta:new NC}
|V_{ud}| = 0.97373(31),
\ee
where the two nuclear corrections tend to cancel with each other
and, hence, leave the central value basically unchanged.
Their uncertainties, however, doubles that of $|V_{ud}|$.
In Secs.~\ref{sec:testing} and \ref{sec:SM},
we mainly use the PDG value~(\ref{eq:Vud beta})
but also test Eq.~(\ref{eq:Vud beta:new NC}) as an alternative input.

The matrix element $|V_{us}|$ can be determined from semi-inclusive 
$\tau$ decays
\cite{Gamiz:2002nu,Gamiz:2004ar,Maltman:2008na,Pich_Kass}. By separating the
inclusive decay $\tau\rightarrow \mbox{hadrons}+\nu$ into nonstrange and
strange final states, e.g.,~HFLAV 18~\cite{Amhis:2019ckw} obtains
\be\label{eq:Vus tau:hflav18}
|V_{us}| = 0.2195(19),
\ee and both Maltman {\em et
al.}~\cite{Maltman:2008ib,Maltman:2008na,Maltman:2009bh} and Gamiz {\em et al.}~\cite{Gamiz:2007qs,Gamiz:2013wn}
arrived at very similar values.
Inclusive hadronic $\tau$ decay offers an interesting way to measure
$|V_{us}|$, but the above value of $|V_{us}|$ differs from
the result one obtains from the kaon decays by about three standard deviations
(see Sec.~\ref{sec:SM}).
This apparent tension has been recently solved in Ref.~\cite{Hudspith:2017vew} 
thanks to the use of a different experimental input and to a new treatment of higher orders in the operator product expansion and 
of violations of quark-hadron duality.
A larger value of $|V_{us}|$ is obtained, 
namely, $|V_{us}| = 0.2231 (27)$,
which is in much better agreement with the results from the kaon decays.
This result is also stable against the choice of the upper limit and weight function of the experimental spectral integrals.~\footnote{A recent update can be found in Ref.~\cite{10.21468/SciPostPhysProc.1.006}}

Recently, Ref.~\cite{Boyle:2018ilm} proposed a new method
to determine $|V_{us}|$ from inclusive strange $\tau$ decays.
Through generalized dispersion relations, 
this method evaluates the spectral integral from lattice-QCD data
of the hadronic vacuum polarization function
at Euclidean momentum squared in the few-to-several 0.1~GeV$^2$ region.
This method, therefore, does not rely on the operator product expansion,
and obtained $|V_{us}|$ consistent with that from the kaon decays.
A later analysis yields~\cite{10.21468/SciPostPhysProc.1.006}
\be\label{eq:Vus tau}
|V_{us}| = 0.2240(18),
\ee
by taking account of updates on experimental strange $\tau$
branching fractions in 2018.
We quote Eqs.~(\ref{eq:Vus tau:hflav18}) and (\ref{eq:Vus tau})
as $|V_{us}|$ from the inclusive hadronic $\tau$ decays
in Sec.~\ref{sec:SM}.

The experimental results in Eq.~(\ref{eq:products}) are for the 
semileptonic decay of a neutral kaon into a negatively charged pion and the
charged pion and kaon leptonic decays, respectively, in QCD. In the case of
the semileptonic decays the corrections for strong
and electromagnetic isospin breaking in {\Ch}PT
at NLO have allowed for averaging the different experimentally
measured isospin channels~\cite{Antonelli:2010yf}. 
This is quite a convenient procedure as long as lattice-QCD simulations do not include
strong or QED isospin-breaking effects. 
Several lattice results for $f_K/f_\pi$ are quoted for QCD with (squared)
pion and kaon masses of $M_\pi^2=M_{\pi^0}^2$ and $M_K^2=\frac 12
	\left(M_{K^\pm}^2+M_{K^0}^2-M_{\pi^\pm}^2+M_{\pi^0}^2\right)$
for which the leading strong and electromagnetic isospin violations cancel.
For these results,
contact with experimental results is made
by correcting leading $SU(2)$ isospin breaking 
guided either by {\Ch}PT or by lattice calculations. 
We note, however, that
the modern trend for the leptonic decays is
to include strong and electromagnetic isospin breaking in the lattice simulations
(e.g.,~Refs.~\cite{Aoki:2008sm,deDivitiis:2011eh,Ishikawa:2012ix,TakuLat12,deDivitiis:2013xla,Tantalo:2013maa,Portelli:2015gda,Carrasco:2015xwa,Giusti:2017dwk}).
After the previous edition,
this trend has been extended to the semileptonic decays.
Reference~\cite{Sachrajda:2019uhh} discusses an extension of the method
in Refs.~\cite{Giusti:2017dwk,DiCarlo:2019thl},
which led to Eq.~(\ref{eq:VusVud_new}),
for the semileptonic decays.
References~\cite{Seng:2020jtz,Ma:2021azh,Seng:2021wcf} pursue  
an effective field theory setup supplemented by nonperturbative lattice-QCD
inputs to estimate the radiative corrections.

\subsection{Lattice results for $f_+(0)$ and $f_{K^\pm}/f_{\pi^\pm}$}

The traditional way of determining $|V_{us}|$ relies on using estimates for
the value of $f_+(0)$, invoking the Ademollo-Gatto theorem
\cite{Ademollo_Gatto}.  Since this theorem only holds to leading order of
the expansion in powers of $m_u$, $m_d$, and $m_s$, theoretical models are
used to estimate the corrections. Lattice methods have now reached the
stage where quantities like $f_+(0)$ or $f_K/f_\pi$ can be determined to
good accuracy. As a consequence, the uncertainties inherent in the
theoretical estimates for the higher order effects in the value of $f_+(0)$
do not represent a limiting factor any more and we shall therefore not
invoke those estimates. Also, we will use the experimental results based on
nuclear $\beta$ decay and inclusive hadronic $\tau$ decay exclusively
for comparison---the
main aim of the present review is to assess the information gathered with
lattice methods and to use it for testing the consistency of the SM and its
potential to provide constraints for its extensions.

The database underlying the present review of the semileptonic form factor 
and the ratio of decay constants is
listed in Tabs.~\ref{tab:f+(0)} and \ref{tab:FKFpi}. The properties of the
lattice data play a crucial role for the conclusions to be drawn from these
results: range of $M_\pi$, size of $L M_\pi$, continuum extrapolation,
extrapolation in the quark masses, finite-size effects, etc. The key
features of the various data sets are characterized by means of the 
colour code specified in Sec.~\ref{sec:color-code}.  
More detailed information
on individual computations are compiled in Appendix~\ref{app:VusVud}, 
which in this edition is limited to new results and to those entering the FLAG 
averages. For other calculations the reader should refer to the Appendix B.2 
of Ref.~\cite{Aoki:2016frl}.

The quantity $f_+(0)$ represents a matrix element of a strangeness-changing
null-plane charge, $f_+(0)=\langle K|Q^{\bar{u}s}|\pi \rangle$ (see Ref.~\cite{Gasser:1984ux}). The vector charges obey the
commutation relations of the Lie algebra of $SU(3)$, in particular
$[Q^{\bar{u}s},Q^{\bar{s}u}]=Q^{\bar{u}u-\bar{s}s}$. This relation implies the sum rule $\sum_n
|\langle K|Q^{\bar{u}s}|n \rangle|^2-\sum_n |\langle K|Q^{\bar{s}u}|n \rangle|^2=1$. Since the contribution from
the one-pion intermediate state to the first sum is given by $f_+(0)^2$,
the relation amounts to an exact representation for this quantity
\cite{Furlan}: \be \label{eq:Ademollo-Gatto} f_+(0)^2=1-\sum_{n\neq \pi}
|\langle K|Q^{\bar{u}s}|n \rangle|^2+\sum_n |\langle K |Q^{\bar{s}u}|n \rangle|^2\fs\ee While the first sum on the
right extends over nonstrange intermediate states, the second runs over
exotic states with strangeness $\pm 2$ and is expected to be small compared
to the first.

The expansion of $f_+(0)$ in $SU(3)$ {\Ch}PT
in powers of $m_u$, $m_d$, and $m_s$ starts with
$f_+(0)=1+f_2+f_4+\ldots\,$ \cite{Gasser:1984gg}.  Since all of the low-energy constants occurring in $f_2$ can be expressed in terms of $M_\pi$,
$M_K$, $M_\eta$ and $f_\pi$ \cite{Gasser:1984ux}, the NLO correction is
known. In the language of the sum rule (\ref{eq:Ademollo-Gatto}), $f_2$
stems from nonstrange intermediate states with three mesons. Like all
other nonexotic intermediate states, it lowers the value of $f_+(0)$:
$f_2=-0.023$ when using the experimental value of $f_\pi$ as input.  
The corresponding expressions have also been derived in
quenched or partially quenched (staggered) {\Ch}PT
\cite{Bernard:2013eya,Bazavov:2012cd}.  At the same order in the $SU(2)$ expansion
\cite{Flynn:2008tg}, $f_+(0)$ is parameterized in terms of $M_\pi$ and two
\textit{a priori} unknown parameters. The latter can be determined from the
dependence of the lattice results on the masses of the quarks.  Note that
any calculation that relies on the {\Ch}PT formula for $f_2$ is subject to
the uncertainties inherent in NLO results: instead of using the physical
value of the pion decay constant $f_\pi$, one may, for instance, work with
the constant $f_0$ that occurs in the effective Lagrangian and represents
the value of $f_\pi$ in the chiral limit. Although trading $f_\pi$ for
$f_0$ in the expression for the NLO term affects the result only at NNLO,
it may make a significant numerical difference in calculations where the
latter are not explicitly accounted for. (Lattice results concerning the
value of the ratio $f_\pi/f_0$ are reviewed in Sec.~\ref{sec:SU3results}.)

The lattice results shown in Fig.~\ref{fig:lattice data semileptonic}
indicate that the higher order contributions $\Delta f\equiv
f_+(0)-1-f_2$ are negative and thus amplify the effect generated by $f_2$.
This confirms the expectation that the exotic contributions are small. The
entries in the lower part of the left panel represent various model
estimates for $f_4$. In Ref.~\cite{Leutwyler:1984je}, the symmetry-breaking
effects are estimated in the framework of the quark model. The more recent
calculations are more sophisticated, as they make use of the known explicit
expression for the $K_{\ell3}$ form factors to NNLO in {\Ch}PT
\cite{Post:2001si,Bijnens:2003uy}. The corresponding formula for $f_4$
accounts for the chiral logarithms occurring at NNLO and is not subject to
the ambiguity mentioned above.\footnote{Fortran programs for the
  numerical evaluation of the form factor representation in
  Ref.~\cite{Bijnens:2003uy} are available on request from Johan Bijnens.} 
 The numerical result, however, depends on
the model used to estimate the low-energy constants occurring in $f_4$
\cite{Bijnens:2003uy,Jamin:2004re,Cirigliano:2005xn,Kastner:2008ch}. The
figure indicates that the most recent numbers obtained in this way
correspond to a positive or an almost vanishing rather than a negative value for $\Delta f$.
We note that FNAL/MILC 12I~\cite{Bazavov:2012cd},
JLQCD 17~\cite{Aoki:2017spo},
FNAL/MILC 18~\cite{Bazavov:2018kjg},
and Ref.~\cite{Bernard:2007tk} have made an attempt 
at determining a combination of some of the low-energy constants appearing 
in $f_4$ from lattice data.

\subsection{Direct determination of $f_+(0)$ and $f_{K^\pm}/f_{\pi^\pm}$}\label{sec:Direct} 

Many lattice results for the form factor $f_+(0)$ and for the ratio of decay constants, which we summarize here in Tabs.~\ref{tab:f+(0)} and~\ref{tab:FKFpi}, respectively, have been computed in isospin-symmetric QCD. 
The reason for this unphysical parameter choice is that there are only  a few simulations of isospin-breaking effects in lattice QCD, which is ultimately the cleanest way for predicting these effects
~\cite{Duncan:1996xy,Basak:2008na,Blum:2010ym,Portelli:2010yn,deDivitiis:2011eh,deDivitiis:2013xla,Tantalo:2013maa,Portelli:2015gda,Carrasco:2015xwa,Giusti:2017dwk}. 
In the meantime, one relies either on {\Ch}PT~\cite{Gasser:1984gg,Aubin:2004fs} to estimate the correction to the isospin limit or one calculates the breaking at leading order in $(m_u-m_d)$ in the valence quark sector by extrapolating the lattice data for the charged kaons to the physical value of the $up$($down$)-quark mass (the result for the pion decay constant is always extrapolated to the value of the average light-quark mass $\hat m$).
This defines the prediction for $f_{K^\pm}/f_{\pi^\pm}$.

Since the majority of results that qualify for inclusion into the FLAG average include the strong $SU(2)$ isospin-breaking correction, we confirm the choice made in the previous edition of the FLAG review \cite{Aoki:2019cca} and we provide in Fig.~\ref{fig:lattice data leptonic} the overview of the world data of $f_{K^\pm}/f_{\pi^\pm}$.
For all the results of Tab.~\ref{tab:FKFpi} provided only in the isospin-symmetric limit we apply individually an isospin correction that will be described later on (see Eqs.~(\ref{eq:convert})\,--\,(\ref{eq:iso})).

The plots in Figs.~\ref{fig:lattice data semileptonic} and \ref{fig:lattice data leptonic} illustrate our compilation of data for $f_+(0)$ and $f_{K^\pm}/f_{\pi^\pm}$.
The lattice data for the latter quantity is largely consistent even when comparing simulations with different $N_f$, while in the case of $f_+(0)$ a slight tendency to get higher values when increasing $N_f$ seems to be visible, even if it does not exceed one standard deviation.
We now proceed to form the corresponding averages, separately for the data with $\Nf=2+1+1$, $\Nf=2+1$, and $\Nf=2$ dynamical flavours, and in the following we will refer to these averages as the ``direct'' determinations.

\begin{table}[t]
\centering 
\vspace{2.8cm}
{\footnotesize\noindent
\begin{tabular*}{\textwidth}[l]{@{\extracolsep{\fill}}llllllll}
Collaboration & Ref. & $\Nf$ & 
\hspace{0.15cm}\begin{rotate}{60}{publication status}\end{rotate}\hspace{-0.15cm}&
\hspace{0.15cm}\begin{rotate}{60}{chiral extrapolation}\end{rotate}\hspace{-0.15cm}&
\hspace{0.15cm}\begin{rotate}{60}{continuum extrapolation}\end{rotate}\hspace{-0.15cm}&
\hspace{0.15cm}\begin{rotate}{60}{finite-volume errors}\end{rotate}\hspace{-0.15cm}&\rule{0.3cm}{0cm}
$f_+(0)$ \\
&&&&&&& \\[-0.1cm]
\hline
\hline&&&&&&& \\[-0.1cm]
FNAL/MILC 18               &\cite{Bazavov:2018kjg} &2+1+1  &\gA&\good&\good&\good& {0.9696(15)(12)}\\
ETM 16                     &\cite{Carrasco:2016kpy} &2+1+1  &\gA&\soso&\good&\soso& 0.9709(45)(9)\\
FNAL/MILC 13E               &\cite{Bazavov:2013maa} &2+1+1  &\gA&\good&\good&\good& {0.9704(24)(22)}\\
&&&&&&& \\[-0.1cm]
\hline
&&&&&&& \\[-0.1cm]
PACS 19             & \cite{Kakazu:2019ltq} &2+1  &\gA&\soso&\tbr&\good& 0.9603(16)($^{+50}_{-48}$)\\
JLQCD 17               & \cite{Aoki:2017spo} &2+1  &\gA&\soso&\tbr&\soso& 0.9636(36)($^{+57}_{-35}$)\\
RBC/UKQCD 15A              &\cite{Boyle:2015hfa}  &2+1  &\gA&\good&\soso&\soso& {0.9685(34)(14)}\\
RBC/UKQCD 13              & \cite{Boyle:2013gsa}  &2+1  &\gA&\good&\soso&\soso& 0.9670(20)($^{+18}_{-46}$)\\
FNAL/MILC 12I                 & \cite{Bazavov:2012cd} &2+1  &\gA&\soso&\soso&\tbg& {0.9667(23)(33)}\\
JLQCD 12                        & \cite{Kaneko:2012cta} &2+1  &\rC&\soso&\tbr&\tbg& 0.959(6)(5)\\
JLQCD 11                        & \cite{Kaneko:2011rp}  &2+1  &\rC&\soso&\tbr&\tbg& 0.964(6)\\
RBC/UKQCD 10              & \cite{Boyle:2010bh}   &2+1  &\gA&\soso&\tbr&\tbg& 0.9599(34)($^{+31}_{-47}$)(14)\rule{0cm}{0.4cm}\\ 
RBC/UKQCD 07              & \cite{Boyle:2007qe}   &2+1  &\gA&\soso&\tbr&\tbg& 0.9644(33)(34)(14)\\
&&&&&&& \\[-0.1cm]
\hline
&&&&&&& \\[-0.1cm]
ETM 10D                   & \cite{Lubicz:2010bv}  &2 &\rC&\soso&\tbg&\soso& 0.9544(68)$_{stat}$\\
ETM 09A 	                 & \cite{Lubicz:2009ht}  &2 &\gA&\soso&\soso&\soso& {0.9560(57)(62)}\\	
 &&&&&&& \\[-0.1cm]
\hline
\hline
\end{tabular*}}
\caption{Colour code for the data on $f_+(0)$. In this and previous editions~\cite{Aoki:2019cca}, old results with two red tags have been dropped.\hfill}\label{tab:f+(0)}
\end{table}

\begin{figure}[ht]
\centering
\psfrag{y}{\tiny $\star$}
\includegraphics[height=9.00cm]{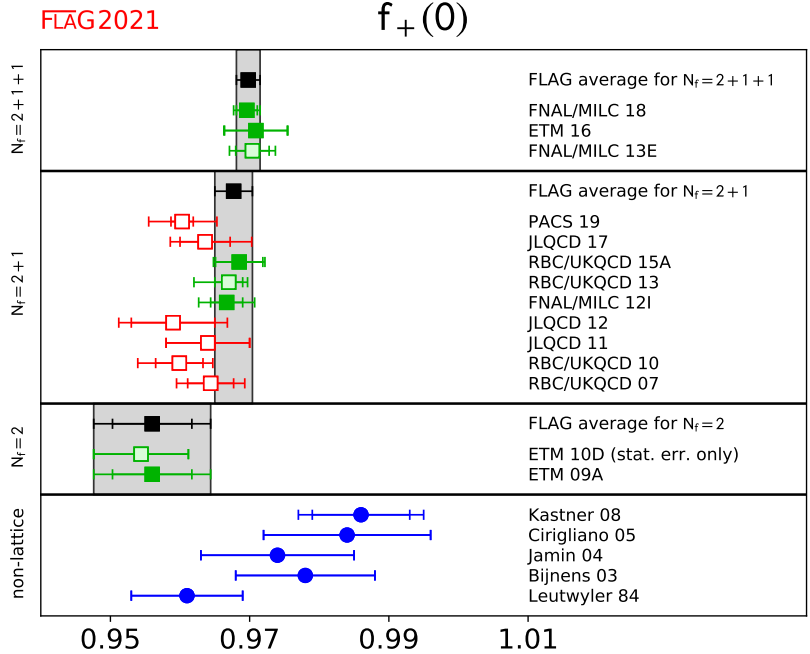}

\caption{\label{fig:lattice data semileptonic} 
Comparison of lattice results (squares) for $f_+(0)$ with various model estimates based on {\Ch}PT~\cite{Kastner:2008ch,Cirigliano:2005xn,Jamin:2004re,Bijnens:2003uy,Leutwyler:1984je} (blue circles). The black squares and grey bands indicate our averages (\ref{eq:fplus_direct_2p1p1})\,--\,(\ref{eq:fplus_direct_2}). The significance of the colours is explained in Sec.~\ref{sec:qualcrit}.}

\end{figure}

\subsubsection{Results for $f_+(0)$}
 
For $f_+(0)$ there are currently two computational strategies: 
FNAL/MILC uses the Ward identity to relate the $K\to\pi$ form factor at zero momentum transfer to the matrix element $\langle \pi|S|K\rangle$ of the flavour-changing scalar current $S = \bar{s} u$. 
Peculiarities of the staggered fermion discretization used by FNAL/MILC (see Ref.~\cite{Bazavov:2012cd}) makes this the favoured choice. 
The other collaborations are instead computing the vector current matrix element $\langle \pi | \bar{s} \gamma_\mu u |K\rangle$. 
Apart from FNAL/MILC 13E, RBC/UKQCD 15A, and FNAL/MILC 18, all simulations in Tab.~\ref{tab:f+(0)} involve unphysically heavy quarks and, therefore, the lattice data needs to be extrapolated to the physical pion and kaon masses corresponding to the $K^0\to\pi^-$ channel. 
We note also that the recent computations of $f_+(0)$ obtained by the FNAL/MILC and RBC/UKQCD collaborations make use of the partially-twisted boundary conditions to determine the form-factor results directly at the relevant kinematical point $q^2=0$ \cite{Guadagnoli:2005be,Boyle:2007wg}, avoiding in this way any uncertainty due to the momentum dependence of the vector and/or scalar form factors. 
The ETM collaboration uses partially-twisted boundary conditions to compare the momentum dependence of the scalar and vector form factors with the one of the experimental data \cite{Lubicz:2010bv,Carrasco:2016kpy}, while keeping at the same time the advantage of the high-precision determination of the scalar form factor at the kinematical end-point $q_{max}^2 = (M_K - M_\pi)^2$ \cite{Becirevic:2004ya,Lubicz:2009ht} for the interpolation at $q^2 = 0$.

According to the colour codes reported in Tab.~\ref{tab:f+(0)} and to the FLAG rules of Sec.~\ref{sec:averages}, only the result ETM 09A with $\Nf =2$, the results FNAL/MILC 12I and  RBC/UKQCD 15A with $\Nf=2+1$, and the results ETM 16 and FNAL/MILC 18 with $\Nf=2+1+1$ dynamical flavours of fermions, respectively, can enter the FLAG averages.
We note that the new entry in this edition is 
FNAL/MILC 18 for $N_f=2+1+1$, which did not enter the previous
FLAG average due to its publication status~\cite{Aoki:2019cca}.

At $\Nf=2+1+1$ the result from the FNAL/MILC collaboration, $f_+(0) = 0.9704 (24) (22)$ (FNAL/MILC 13E), is based on the use of the Highly Improved Staggered Quark (HISQ) action (for both valence and sea quarks), which has been tailored to reduce staggered taste-breaking effects, and includes simulations with three lattice spacings and physical light-quark masses.
These features allow to keep the uncertainties due to the chiral extrapolation and to the discretization artifacts well below the statistical error.
The remaining largest systematic uncertainty comes from finite-size effects, which have been investigated in Ref.~\cite{Bernard:2017scg} using one-loop {\Ch}PT (with and without taste-violating effects).
In Ref.~\cite{Bazavov:2018kjg}, the FNAL/MILC collaboration presented a more precise determination of $f_+(0)$, $f_+(0) = 0.9696 (15) (11)$ (FNAL/MILC 18).
In this update, 
their analysis is extended to two smaller lattice spacings $a = 0.06$ and 0.042~fm.
The physical light-quark mass is simulated at four lattice spacings.
They also added a simulation at a small volume to study the finite-size effects.
The improvement of the precision with respect to FNAL/MILC 13E is obtained mainly
by an estimate of finite-size effects,
which is claimed to be controlled at the level of $\sim 0.05$\,\% by comparing
two analyses with and without the one-loop correction.
The total uncertainty is largely reduced to $\sim 0.2$\,\%.
An independent calculation of such high precision would be highly welcome
to solidify the lattice prediction of $f_+(0)$,
which currently suggests a tension with CKM unitarity
with the updated value of $|V_{ud}|$ (see Sec.~\ref{sec:testing}).

The result from the ETM collaboration, $f_+(0) = 0.9709 (45) (9)$ (ETM 16), makes use of the twisted-mass discretization adopting three values of the lattice spacing in the range $0.06 - 0.09$ fm and pion masses simulated in the range $210 - 450$ MeV.  
The chiral and continuum extrapolations are performed in a combined fit together with the momentum dependence, using both a $SU(2)$-{\Ch}PT inspired ansatz (following Ref.~\cite{Lubicz:2010bv}) and a modified z-expansion fit.
The uncertainties coming from the chiral extrapolation, the continuum extrapolation and the finite-volume effects turn out to be well below the dominant statistical error, which includes also the error due to the fitting procedure.
A set of synthetic data points, representing both the vector and the scalar semileptonic form factors at the physical point for several selected values of $q^2$, is provided together with the corresponding correlation matrix.

The PACS collaboration obtained a new result for $N_f\!=\!2+1$,
$f_+(0) = 0.9603(16)\left(^{+50}_{-48}\right)$,
by creating an ensemble with the physical light-quark mass on a large lattice volume of $(10.9\,\mbox{fm})^4$~\cite{Kakazu:2019ltq}.
Such a large lattice enables them to interpolate $f_+(q^2)$ to zero momentum transfer and study the momentum-transfer dependence of the form factors 
without using partially-twisted boundary conditions.
Their result, however, does not enter the FLAG average, because they only use
a single lattice spacing, which is the source of the largest uncertainty in their calculation.

For $\Nf=2+1$, the two results eligible to enter the FLAG average are the one from RBC/UKQCD 15A, $f_+(0) = 0.9685 (34) (14)$~\cite{Boyle:2015hfa}, and the one from FNAL/MILC 12I, $f_+(0)=0.9667(23)(33)$~\cite{Bazavov:2012cd}. 
These results, based on different  fermion discretizations (staggered fermions in the case of FNAL/MILC and domain wall fermions in the case of RBC/UKQCD) are in nice agreement.
Moreover, in the case of FNAL/MILC the form factor has been determined from the scalar current matrix element, while in the case of RBC/UKQCD it has been determined including also the matrix element of the vector current. 
To a certain extent both simulations are expected to be affected by different systematic effects.

RBC/UKQCD 15A has analyzed results on ensembles with pion masses down to 140~MeV, mapping out the complete range from the $SU(3)$-symmetric limit to the physical point. 
No significant cut-off effects (results for two lattice spacings) were observed in the simulation results.
Ensembles with unphysical light-quark masses are weighted to work as a guide for small corrections toward the physical point, reducing in this way the model dependence in the fitting ansatz.
The systematic uncertainty turns out to be dominated by finite-volume effects, for which an estimate based on effective theory arguments is provided. 

The result FNAL/MILC 12I is from simulations reaching down to a lightest RMS pion mass of about 380~MeV (the lightest valence pion mass for one of their ensembles is about 260~MeV).
Their combined chiral and continuum extrapolation (results for two lattice spacings) is based on NLO staggered {\Ch}PT supplemented by the continuum NNLO expression~\cite{Bijnens:2003uy} and a phenomenological parameterization of the breaking of the Ademollo-Gatto theorem at finite lattice spacing inherent in their approach.
The $p^4$ low-energy constants entering the NNLO expression have been fixed in terms of external input~\cite{Amoros:2001cp}. 

The ETM collaboration uses the twisted-mass discretization and provides at $\Nf=2$ a comprehensive study of the systematics \cite{Lubicz:2009ht,Lubicz:2010bv}, by presenting results for four lattice spacings and by simulating at light pion masses (down to $M_\pi = 260$~MeV).  
This makes it possible to constrain the chiral extrapolation, using both $SU(3)$ \cite{Gasser:1984ux} and $SU(2)$ \cite{Flynn:2008tg} {\Ch}PT. 
Moreover, a rough estimate for the size of the effects due to quenching the strange quark is given, based on the comparison of the result for $\Nf=2$ dynamical quark flavours \cite{Blossier:2009bx} with the one in the quenched approximation, obtained earlier by the SPQcdR collaboration \cite{Becirevic:2004ya}. 

We now compute the $N_f = 2+1+1$ FLAG average for $f_+(0)$ using the FNAL/MILC 18 and ETM 16 (uncorrelated) results, the $N_f =2+1$ FLAG average based on FNAL/MILC 12I and RBC/UKQCD 15A, which we consider uncorrelated, while for $N_f = 2$ we consider directly the  ETM 09A result, respectively:
\begin{align}
&\label{eq:fplus_direct_2p1p1}
\mbox{direct},\,\Nf=2+1+1:&\FLAGAVBEGIN f_+(0) & = 0.9698(17)\FLAGAVEND  && \Refs~\mbox{\cite{Carrasco:2016kpy,Bazavov:2018kjg}},\\
&\label{eq:fplus_direct_2p1}                                                               
\mbox{direct},\,\Nf=2+1:  &\FLAGAVBEGIN f_+(0) &= 0.9677(27) \FLAGAVEND     &&\Refs~\mbox{\cite{Bazavov:2012cd,Boyle:2015hfa}},   \\
&\label{eq:fplus_direct_2}                                                                  
\mbox{direct},\,\Nf=2:    &\FLAGAVBEGIN f_+(0) &= 0.9560(57)(62)\FLAGAVEND  &&\Ref~\mbox{\cite{Lubicz:2009ht}},
\end{align}
where the parentheses in the third line indicate the statistical and systematic errors, respectively.
We stress that the results (\ref{eq:fplus_direct_2p1p1}) and (\ref{eq:fplus_direct_2p1}), corresponding to $N_f = 2+1+1$ and $N_f = 2+1$, respectively, include already simulations with physical light-quark masses.

\begin{table}[!htb]
\centering
\vspace{3.0cm}{\footnotesize\noindent
\begin{tabular*}{\textwidth}[l]{@{\extracolsep{\fill}}lrlllllll}
Collaboration & Ref. & $\Nf$ &
\hspace{0.15cm}\begin{rotate}{60}{publication status}\end{rotate}\hspace{-0.15cm}&
\hspace{0.15cm}\begin{rotate}{60}{chiral extrapolation}\end{rotate}\hspace{-0.15cm}&
\hspace{0.15cm}\begin{rotate}{60}{continuum extrapolation}\end{rotate}\hspace{-0.15cm}&
\hspace{0.15cm}\begin{rotate}{60}{finite-volume errors}\end{rotate}\hspace{-0.15cm}&
\rule{0.2cm}{0cm} $f_K/f_\pi$ &
\rule{0.2cm}{0cm} $f_{K^\pm}/f_{\pi^\pm}$ \\  
&&&&&&& \\[-0.1cm]
\hline
\hline
&&&&&&& \\[-0.1cm]
ETM 21 &\cite{Alexandrou:2021bfr}	     &2+1+1&\oP&\good &\good&\good    		&1.1995(44)(7)    & 1.1957(44)(7)\\
CalLat 20 &\cite{Miller:2020xhy}	     &2+1+1&\gA&\good &\good&\good    		&1.1964(32)(30)    & 1.1942(32)(31) \\
FNAL/MILC 17 &\cite{Bazavov:2017lyh}	     &2+1+1&\gA&\good &\good&\good    		&{1.1980(12)($_{-15}^{+5}$)}    &{1.1950(15)($_{-18}^{+6}$)} \\
ETM 14E       &\cite{Carrasco:2014poa}          &2+1+1&\gA&\soso &\good&\soso    		&	1.188(11)(11)   &{1.184(12)(11)} \\
FNAL/MILC 14A &\cite{Bazavov:2014wgs}	     &2+1+1&\gA&\good &\good&\good    		&					      &{1.1956(10)($_{-18}^{+26}$)} \\
ETM 13F       &\cite{Dimopoulos:2013qfa}      &2+1+1&\rC&\soso &\good&\soso    		&	 1.193(13)(10)    	      &1.183(14)(10)	\\
HPQCD 13A       &\cite{Dowdall:2013rya}	     &2+1+1&\gA&\good &\soso&\good    		&	 1.1948(15)(18)&{1.1916(15)(16)} \\
MILC 13A        &\cite{Bazavov:2013cp}	     &2+1+1&\gA&\good &\good&\good    		&					      &1.1947(26)(37) \\
MILC 11        &\cite{Bazavov:2011fh}	             &2+1+1&\rC&\soso &\soso&\soso    		&					      &1.1872(42)$^\dagger_{\rm stat.}$ \\
ETM 10E       &\cite{Farchioni:2010tb}            &2+1+1&\rC&\soso&\soso&\soso		        &       1.224(13)$_{\rm stat}$   &						\\
&&&&&&& \\[-0.1cm]                                                                                                              
\hline                                                                                                                          
&&&&&&& \\[-0.1cm]                                                                                                              
QCDSF/UKQCD 16  &\cite{Bornyakov:2016dzn}  &2+1&\gA&\soso&\good&\soso     & 1.192(10)(13)			        &    1.190(10)(13) \\
BMW 16          &\cite{Durr:2016ulb,Scholz:2016kcr}  &2+1&\gA&\good&\good&\good     & 1.182(10)(26)			        &    1.178(10)(26)                 \\
RBC/UKQCD 14B   &\cite{Blum:2014tka}    &2+1&\gA&\good    & \good	 &  \good  	&1.1945(45)					&					\\
RBC/UKQCD 12   &\cite{Arthur:2012opa}    &2+1&\gA&\good    & \soso	 &  \good  	&{1.199(12)(14)}				&					\\
Laiho 11       &\cite{Laiho:2011np}       &2+1&\rC&\soso    & \good   &  \soso  	&                                       	&$1.202(11)(9)(2)(5)$$^{\dagger\dagger}$	\\
MILC 10        &\cite{Bazavov:2010hj}&2+1&\rC&\soso&\good&\good			&                             			&{1.197(2)($^{+3}_{-7}$)}			\\
JLQCD/TWQCD 10 &\cite{Noaki:2010zz}&2+1&\rC&\soso&\tbr&\tbg			&1.230(19)					&                               		\\
RBC/UKQCD 10A  &\cite{Aoki:2010dy}   &2+1&\gA&\soso&\soso&\good			&1.204(7)(25)					&                               		\\
BMW 10         &\cite{Durr:2010hr}         &2+1&\gA&\good &\tbg&\tbg			&{1.192(7)(6)}					&                               		\\
MILC 09A       &\cite{Bazavov:2009fk}&2+1&\rC&\soso&\tbg&\tbg			&                                               &1.198(2)($^{\hspace{0.01cm}+6}_{-8}$)	\\
MILC 09        &\cite{Bazavov:2009bb}&2+1&\gA&\soso&\tbg&\tbg			&                                               &1.197(3)($^{\;+6}_{-13}$)		\\
Aubin 08       &\cite{Aubin:2008ie}  &2+1&\rC&\soso&\soso&\soso			&                                               &1.191(16)(17)					\\
RBC/UKQCD 08   &\cite{Allton:2008pn} &2+1&\gA&\soso&\tbr&\tbg			&1.205(18)(62)					&                                               \\
HPQCD/UKQCD 07 &\cite{Follana:2007uv}&2+1&\gA&\soso&\soso&\soso	&{1.189(2)(7)}		&                                               \\
MILC 04 &\cite{Aubin:2004fs}&2+1&\gA&\soso&\soso&\soso				&						&1.210(4)(13)				\\
&&&&&&& \\[-0.1cm]                                                                                                              
\hline                                                                                                                          
&&&&&&& \\[-0.1cm]                                                                                                              
ETM 14D         &\cite{Abdel-Rehim:2014nka} &2  &\rC&\good&\tbr&\soso			&1.203(5)$_{\rm stat}$		&                                       	\\
ALPHA 13A       &\cite{Lottini:2013rfa}&2  &\rC&\tbg    &\tbg   &\tbg    	&1.1874(57)(30)					&                                       	\\
ETM 10D        &\cite{Lubicz:2010bv} &2  &\rC&\soso&\tbg&\soso			&1.190(8)$_{\rm stat}$ 				&                                       	\\
ETM 09         &\cite{Blossier:2009bx}         &2  &\gA&\soso&\tbg&\soso			&{1.210(6)(15)(9)}				&                                       	\\
QCDSF/UKQCD 07 &\cite{QCDSFUKQCD}    &2  &\rC&\soso&\soso&\tbg			&1.21(3)					&                                       	\\
&&&&&&& \\[-0.1cm]
\hline
\hline
&&&&&&& \\[-0.1cm]
\end{tabular*}}\\[-2mm]
\begin{minipage}{\linewidth}
{\footnotesize 
\begin{itemize}
\item[$^\dagger$] Result with statistical error only from polynomial interpolation to the physical point.\\[-5mm]
\item[$^{\dagger\dagger}$] This work is the continuation of Aubin 08.
\end{itemize}
}
\end{minipage}
\vspace{-0.3cm}
\caption{Colour code for the data on the ratio of decay constants: $f_K/f_\pi$ is the pure QCD $SU(2)$-symmetric ratio, while $f_{K^\pm}/f_{\pi^\pm}$ is in pure QCD including
the $SU(2)$ isospin-breaking correction. In this and previous editions~\cite{Aoki:2019cca}, old results with two red tags have been dropped.\hfill}
\label{tab:FKFpi}
\end{table}

\subsubsection{Results for $f_{K^\pm}/f_{\pi^\pm}$}

In the case of the ratio of decay constants the data sets that meet the criteria formulated in the introduction are HPQCD 13A~\cite{Dowdall:2013rya}, ETM 14E~\cite{Carrasco:2014poa}, FNAL/MILC 17~\cite{Bazavov:2017lyh} (which updates FNAL/MILC 14A~\cite{Bazavov:2014wgs}) and CalLat 20 \cite{Miller:2020xhy} with $N_f=2+1+1$, HPQCD/UKQCD 07~\cite{Follana:2007uv}, MILC 10~\cite{Bazavov:2010hj}, BMW 10~\cite{Durr:2010hr}, RBC/UKQCD 14B~\cite{Blum:2014tka}, BMW 16~\cite{Durr:2016ulb,Scholz:2016kcr}, and QCDSF/UKQCD 16~\cite{Bornyakov:2016dzn} with $\Nf=2+1$ and ETM 09 \cite{Blossier:2009bx} with $\Nf=2$ dynamical flavours.
Note that only CalLat 20 for $N_f=2+1+1$ is the new entry for the FLAG average in this edition.

CalLat 20
employs a mixed action setup with the M\"obius domain-wall
valence quarks on gradient-flowed HISQ ensembles at four lattice spacings
$a = 0.06$\,--\,0.15~fm.
The valence pion mass reaches the physical point at three lattice spacings,
and the smallest valence-sea and sea pion masses are below 200~MeV.
Finite-volume corrections are studied on 
three lattice volumes at $a = 0.12$~fm and $M_\pi \sim 220$~MeV.
Their extrapolation to the continuum limit and the physical point is based on
NNLO {\Ch}PT~\cite{Ananthanarayan_2018}.
A comprehensive study of systematic uncertainties is performed
by exploring several options including
the use of the mixed-action effective theory expression,
and the inclusion of N$^3$LO counter terms.
They obtain
$f_{K^\pm}/f_{\pi^\pm}=1.1942(32)_{\rm stat}(12)_{\chi}(20)_{a^2}(1)_{FV}(12)_{M}(7)_{IB}$,
where the errors are statistical, due to the extrapolation in pion and kaon masses,
extrapolation in $a^2$, finite-size effects, choice of the fitting form
and isospin breaking corrections.

ETM 14E uses the twisted-mass discretization and provides a comprehensive study of the systematics by presenting results for three lattice spacings in the range $0.06 - 0.09$ fm and for pion masses in the range $210 - 450$ MeV.  
This makes it possible to constrain the chiral extrapolation, using both $SU(2)$ \cite{Flynn:2008tg} {\Ch}PT and polynomial fits.
The ETM collaboration includes the spread in the central values obtained from different ans\"atze into the systematic errors.
The final result of their analysis is $\fKfpichargedr = 1.184(12)_{\rm stat+fit}(3)_{\rm Chiral}(9)_{\rm a^2}(1)_{Z_P}(3)_{FV}(3)_{IB}$ where the errors are (statistical + the error due to the fitting procedure), due to the chiral extrapolation, the continuum extrapolation, the mass-renormalization constant, the finite-volume and (strong) isospin-breaking effects.

In ETM~21~\cite{Alexandrou:2021bfr},  
the ETM collaboration presented an independent estimate of $f_K/f_\pi$
in isosymmetric QCD with 2+1+1 dynamical flavours of
the twisted-mass quarks.
Their new set of gauge ensembles reaches the physical pion mass.
The quark action includes the Sheikoleslami-Wohlert term 
for a better control of discretization effects.
The finite-volume effects are examined by simulating three spatial volumes,
and are corrected by $SU(2)$ {\Ch}PT formulae~\cite{Colangelo:2005gd}.
Their new estimate
$f_K/f_\pi = 1.1995(44)_{\rm stat+fit}(7)_{\rm sys}$
is consistent with ETM 14E
with the total uncertainty reduced by a factor of $\sim$~3.5.
While ETM~21 satisfies all criteria on simulation parameters,
it does not enter the FLAG average in this edition
due to the publication status.

\begin{figure}[t]
\centering
\psfrag{y}{\tiny $\star$}
\includegraphics[height=9.0cm]{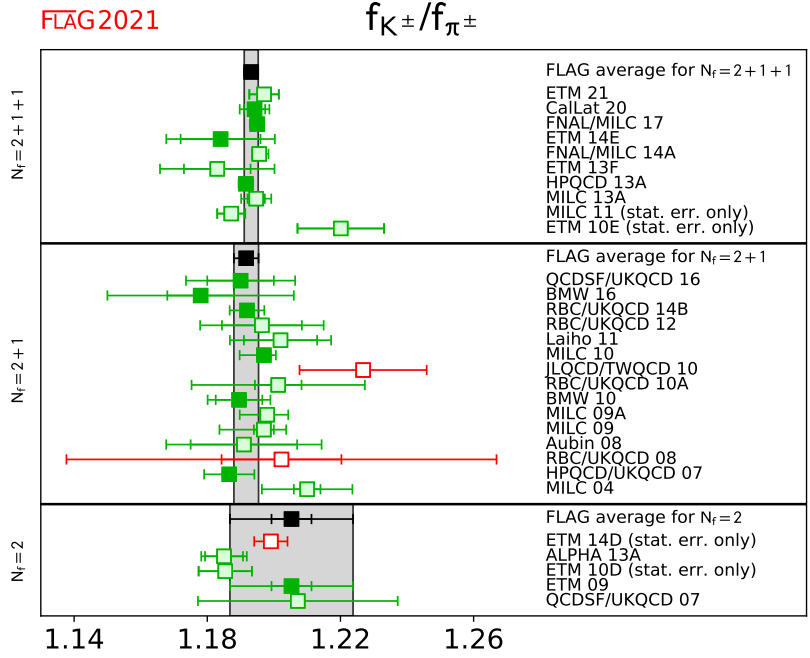}
  
\caption{\label{fig:lattice data leptonic}
Comparison of lattice results for $f_{K^\pm}/ f_{\pi^\pm}$.
This ratio is obtained in pure QCD including the $SU(2)$ isospin-breaking correction
(see Sec.~\ref{sec:Direct}).
The black squares and grey bands indicate our averages in Eqs.~(\ref{eq:fKfpi_direct_broken_2p1p1})\,--\,(\ref{eq:fKfpi_direct_broken_2}).
}

\end{figure}

FNAL/MILC 17 has determined the ratio of the decay constants from a comprehensive set of HISQ ensembles with $N_f = 2+1+1$ dynamical flavours. 
They have generated 24 ensembles for six values of the lattice spacing ($0.03 - 0.15$ fm, scale set with $f_{\pi^+}$) and with both physical and unphysical values of the light sea-quark masses, controlling in this way the systematic uncertainties due to chiral and continuum extrapolations.
With respect to FNAL/MILC 14A they have increased the statistics and added three ensembles at very fine lattice spacings, $a \simeq 0.03$ and $0.042$ fm, including for the latter case also a simulation at the physical value of the light-quark mass.
The final result of their analysis is $\fKfpichargedr=1.1950(14)_{\rm stat}($$_{-17}^{+0}$$)_{\rm a^2} (2)_{FV} (3)_{f_\pi, PDG} (3)_{EM} (2)_{Q^2}$, where the errors are statistical, due to the continuum extrapolation, finite-volume, pion decay constant from PDG, electromagnetic effects and sampling of the topological charge distribution.\footnote{To form the average in Eq.~(\ref{eq:fKfpi_direct_broken_2p1p1}), we have symmetrized the asymmetric systematic error and shifted the central value by half the difference as will be done throughout this section.}

HPQCD 13A has analyzed ensembles generated by MILC and therefore its study of $\fKfpichargedr$ is based on the same set of ensembles bar the ones at the finest lattice spacings (namely, only $a = 0.09 - 0.15$ fm, scale set with $f_{\pi^+}$ and relative scale set with the Wilson flow~\cite{Luscher:2010iy,Borsanyi:2012zs}) supplemented by some simulation points with heavier quark masses.
HPQCD employs a global fit based on continuum NLO $SU(3)$ {\Ch}PT for the decay constants supplemented by a model for higher-order terms including discretization and finite-volume effects (61 parameters for 39 data points supplemented by Bayesian priors). 
Their final result is $f_{K^\pm}/f_{\pi^\pm}=1.1916(15)_{\rm stat}(12)_{\rm a^2}(1)_{FV}(10)$, where the errors are statistical, due to the continuum extrapolation, due to finite-volume effects and the last error contains the combined uncertainties from the chiral extrapolation, the scale-setting uncertainty, the experimental input in terms of $f_{\pi^+}$ and from the uncertainty in $m_u/m_d$.

Because CalLat 20,
FNAL/MILC 17 and HPQCD 13A partly share their gauge ensembles,
we assume a 100\,\% correlation among their statistical errors.
A 100\,\% correlation on the total systematic uncertainty 
is also assumed between FNAL/MILC 17 and HPQCD 13A with the HISQ valence quarks.

For $N_f=2+1$ the results BMW 16 and QCDSF/UKQCD 16 are eligible to enter the FLAG average.
BMW 16 has analyzed the decay constants evaluated for 47 gauge ensembles generated using tree-level clover-improved fermions with two HEX-smearings and the tree-level Symanzik-improved gauge action. 
The ensembles correspond to five values of the lattice spacing ($0.05 - 0.12$ fm, scale set by $\Omega$ mass), to pion masses in the range $130 - 680$ MeV and to values of the lattice size from $1.7$ to $5.6$ fm, obtaining a good control over the interpolation to the physical mass point and the extrapolation to the continuum and infinite volume limits.

QCDSF/UKQCD 16 has used the nonperturbatively ${\cal{O}}(a)$-improved clover action for the fermions (mildly stout-smeared) and the tree-level Symanzik action for the gluons.
Four values of the lattice spacing ($0.06 - 0.08$ fm) have been simulated with pion masses down to $\sim 220$ MeV and values of the lattice size in the range $2.0 - 2.8$ fm.
The decay constants are evaluated using an expansion around the symmetric $SU(3)$ point $m_u = m_d = m_s = (m_u + m_d + m_s)^{phys} / 3$.

Note that for $N_f=2+1$ MILC 10 and HPQCD/UKQCD 07 are based on staggered fermions, BMW 10, BMW 16 and QCDSF/UKQCD 16 have used improved Wilson fermions and RBC/UKQCD 14B's result is based on the domain-wall formulation. 
In contrast to RBC/UKQCD 14B and BMW 16, the other simulations are for unphysical values of the light-quark masses (corresponding to smallest pion masses in the range $220 - 260$ MeV in the case of MILC 10, HPQCD/UKQCD 07, and QCDSF/UKQCD 16) and therefore slightly more sophisticated extrapolations needed to be controlled.
Various ans\"atze for the mass and cutoff dependence comprising $SU(2)$ and $SU(3)$ {\Ch}PT or simply polynomials were used and compared in order to estimate the model dependence.
While BMW 10, RBC/UKQCD 14B, and QCDSF/UKQCD 16 are entirely independent computations, subsets of the MILC gauge ensembles used by MILC 10 and HPQCD/UKQCD 07 are the same.
MILC 10 is certainly based on a larger and more advanced set of gauge configurations than HPQCD/UKQCD 07. 
This allows them for a more reliable estimation of systematic effects. 
In this situation we consider both statistical and systematic uncertainties to be correlated.

For $N_f=2$ no new result enters the corresponding FLAG average with respect to the previous edition of the FLAG review \cite{Aoki:2019cca}, which therefore remains the ETM 09 result, which has simulated twisted-mass fermions down to (charged) pion masses equal to 260 MeV.

We note that the overall uncertainties quoted by ETM 14E at $\Nf=2+1+1$ and by BMW 16 and QCDSF/UKQCD 16 at $\Nf=2+1$ are much larger than the overall uncertainties obtained with staggered (HPQCD 13A, FNAL/MILC 17 at $\Nf=2+1+1$, and MILC 10, HPQCD/UKQCD 07 at $\Nf=2+1$) and domain-wall fermions (RBC/UKQCD 14B at $\Nf=2+1$).

Before determining the average for $f_{K^\pm}/f_{\pi^\pm}$, which should be used for applications to Standard Model phenomenology, we apply the strong-isospin correction individually to all those results that have been published only in the isospin-symmetric limit, i.e.,~BMW 10, HPQCD/UKQCD 07 and RBC/UKQCD 14B at $N_f = 2+1$ and ETM 09 at $N_f = 2$. 
To this end, as in the previous editions of the FLAG reviews \cite{Aoki:2013ldr,Aoki:2016frl,Aoki:2019cca}, we make use of NLO $SU(3)$ {\Ch}PT~\cite{Gasser:1984gg,Cirigliano:2011tm}, which predicts
\begin{equation}\label{eq:convert}
	\fKfpicharged = \frac{f_K}{f_\pi} ~ \sqrt{1 + \delta_{SU(2)}} ~ ,
\end{equation}
where~\cite{Cirigliano:2011tm}
\begin{equation}\label{eq:iso}
 \begin{array}{rcl}
	 \delta_{SU(2)}& \approx&
	\sqrt{3}\,\epsilon_{SU(2)}
	\left[-\frac{4}{3} \left(f_K/f_\pi-1\right)+\frac 2{3 (4\pi)^2 f_0^2}
        \left(M_K^2-M_\pi^2-M_\pi^2\ln\frac{M_K^2}{M_\pi^2}\right)
        \right]\,.
  \end{array}
 \end{equation}
We use as input $\epsilon_{SU(2)} = \sqrt{3} / (4 R)$ with the FLAG result for $R$ of Eq.~(\ref{eq:RQres}), $F_0 = f_0 / \sqrt{2} = 80\,(20)$ MeV,
$M_\pi = 135$ MeV and $M_K = 495$ MeV (we decided to choose a conservative uncertainty on $f_0$ in order to reflect the magnitude of potential higher-order 
corrections).
The results are reported in Tab.~\ref{tab:correctedfKfPi}, where in the last column the last error is due to the isospin correction (the remaining errors are quoted in the same order as in the original data).

\begin{table}[!htb]
\begin{center}
\begin{tabular}{llll}
\hline\hline\\[-4mm]
		&$f_K/f_\pi$	&$\delta_{SU(2)}$&$f_{K^\pm}/f_{\pi^\pm}$\\
\hline\\[-4mm]
HPQCD/UKQCD 07	&1.189(2)(7)	&-0.0040(7)&1.187(2)(7)(2)\\
BMW 10		        &1.192(7)(6)	&-0.0041(7)&1.190(7)(6)(2)\\
RBC/UKQCD 14B	&1.1945(45)	&-0.0043(9)&1.1919(45)(26)\\
\hline\hline
\end{tabular}
\caption{Values of the $SU(2)$ isospin-breaking correction $\delta_{SU(2)}$ applied to the lattice data for $f_K/f_\pi$, entering the FLAG average at $N_f=2+1$, for obtaining the corrected charged ratio $f_{K^\pm}/f_{\pi^\pm}$.
The last error in the last column is due to a 100\,\% uncertainty assumed for
$\delta_{SU(2)}$ from $SU(3)$ {\Ch}PT.
}
\label{tab:correctedfKfPi}
\end{center}
\end{table}

For $N_f=2$ and $N_f=2+1+1$ dedicated studies of the strong-isospin correction in lattice QCD do exist. 
The updated $N_f=2$ result of the RM123 collaboration~\cite{deDivitiis:2013xla} amounts to $\delta_{SU(2)}=-0.0080(4)$ and we use this result for the isospin correction of the ETM 09 result.
Note that the above RM123 value for the strong-isospin correction is incompatible with the results based on $SU(3)$ {\Ch}PT, $\delta_{SU(2)}=-0.004(1)$ (see Tab.~\ref{tab:correctedfKfPi}).
Moreover, for $N_f=2+1+1$ HPQCD~\cite{Dowdall:2013rya}, FNAL/MILC~\cite{Bazavov:2017lyh} and ETM~\cite{Giusti:2017xrv} estimate a value for $\delta_{SU(2)}$ equal to $-0.0054(14)$, $-0.0052(9)$ and $-0.0073(6)$, respectively.
Note that the RM123 and ETM results are obtained using the insertion of the isovector scalar current according to the expansion method of Ref.~\cite{deDivitiis:2011eh}, while the HPQCD and FNAL/MILC results correspond to the difference between the values of the decay constant ratio extrapolated to the physical $u$-quark mass $m_u$ and to the average $(m_u + m_d) / 2$ light-quark mass. 

One would not expect the strange and heavier sea-quark contributions to be responsible for such a large effect. 
Whether higher-order effects in {\Ch}PT or other sources are responsible still needs to be understood. 
More lattice-QCD simulations of $SU(2)$ isospin-breaking effects are therefore required.
To remain on the conservative side we add a $100 \%$ error to the correction based on $SU(3)$ {\Ch}PT. 
For further analyses we add (in quadrature) such an uncertainty to the systematic error.

\begin{figure}[t]
\vspace{0.2cm}
\begin{center}
\hspace{0.5cm}\includegraphics[height=9.0cm]{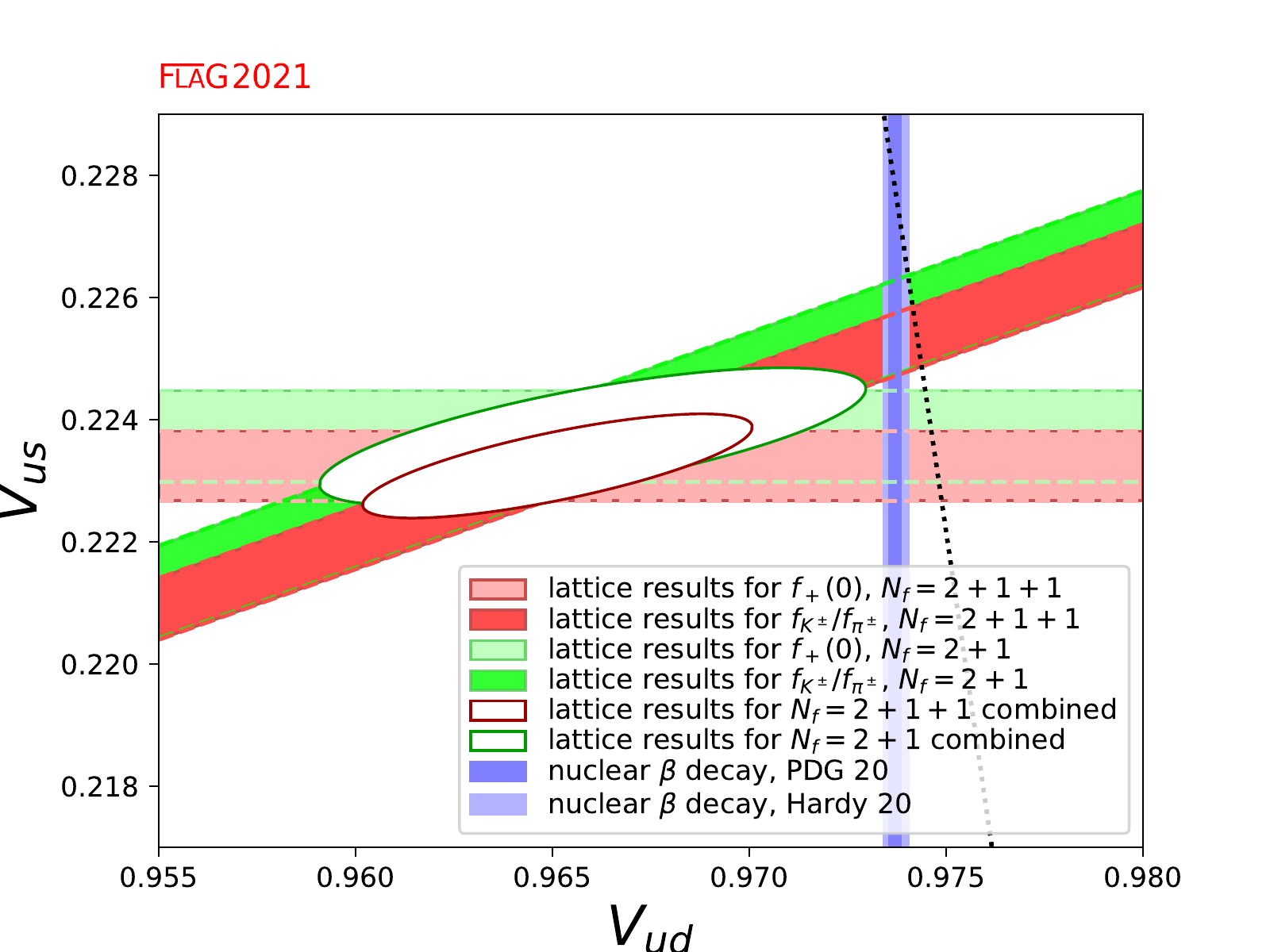}
\end{center}  
\vspace{-2.35cm}\hspace{10.7cm}\parbox{6cm}{\sffamily\tiny \cite{Zyla:2020zbs}\\

\vspace{-0.8em}\cite{Hardy:2020qwl}}
\vspace{0.75cm}
\caption{\label{fig:VusVersusVud} The plot compares the information for $|V_{ud}|$, $|V_{us}|$ obtained on the lattice for $N_f = 2+1$ and $N_f = 2+1+1$
with $|V_{ud}|$ extracted from nuclear $\beta$ transitions Eqs.~(\ref{eq:Vud beta}) and (\ref{eq:Vud beta:new NC}).
The dotted line indicates the correlation between $|V_{ud}|$ and $|V_{us}|$ that follows if the CKM-matrix is unitary. For the $N_f = 2$ results see the 2016 edition~\cite{Aoki:2016frl}.}
\end{figure}

\begin{figure}[h]
\vspace{0.2cm}
\begin{center}
\hspace{0.5cm}\includegraphics[height=9.0cm]{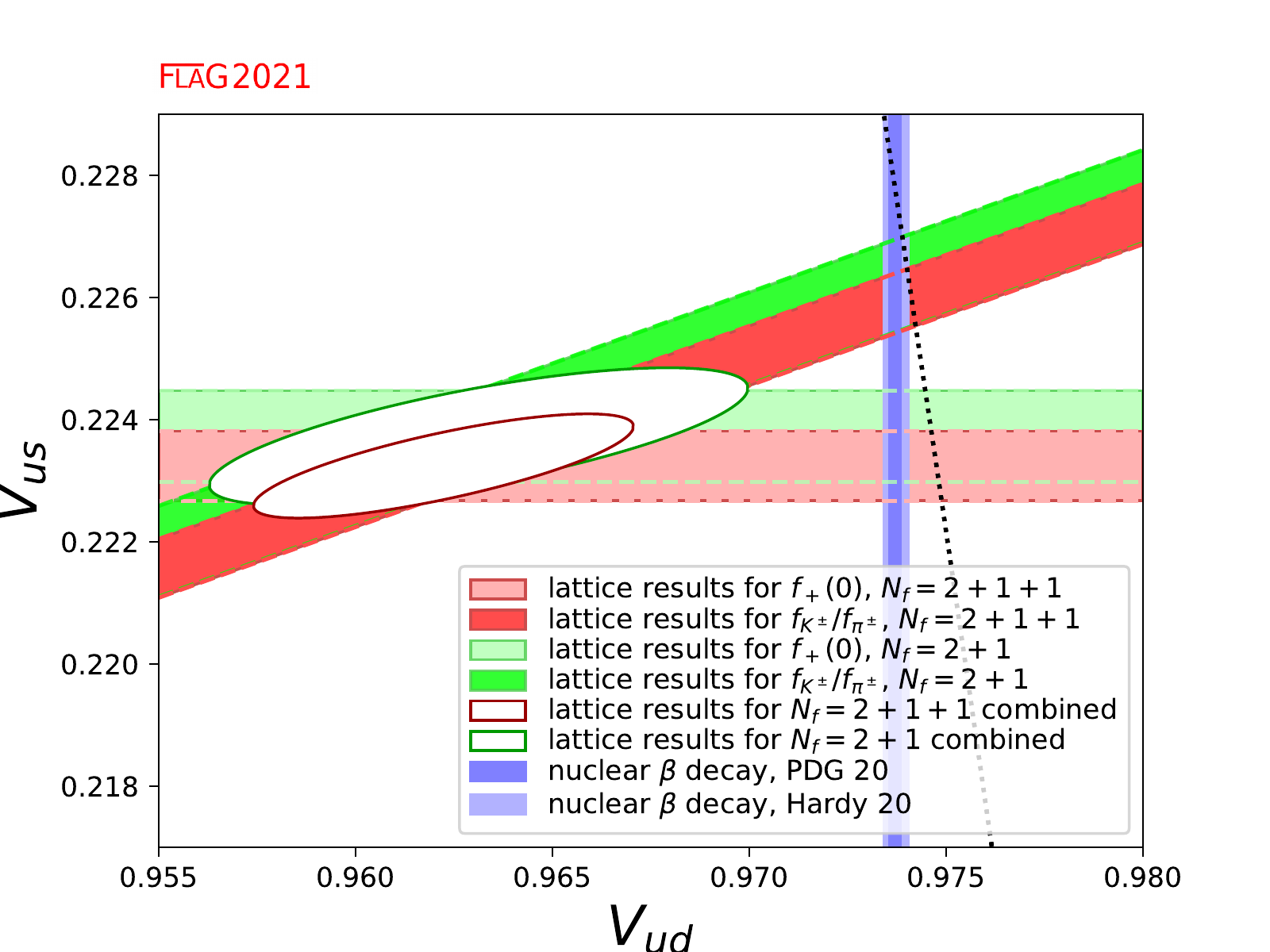}
\end{center}  
\vspace{-2.35cm}\hspace{10.7cm}\parbox{6cm}{\sffamily\tiny \cite{Zyla:2020zbs}\\

\vspace{-0.8em}\cite{Hardy:2020qwl}}
\vspace{0.75cm}
\caption{\label{fig:VusVersusVud_em-lat}
Same as Fig.~\protect\ref{fig:VusVersusVud}
but with $|V_{us}|/|V_{ud}|$ through Eq.~(\ref{eq:VusVud_new}).
}
\end{figure}

Using the results of Tab.~\ref{tab:correctedfKfPi} for $N_f = 2+1$ we obtain
\begin{align}
    \label{eq:fKfpi_direct_broken_2p1p1} 
&\mbox{direct},\,\Nf=2+1+1:&\FLAGAVBEGIN f_{K^\pm} / f_{\pi^\pm} & =  1.1932(21)\FLAGAVEND     && \Refs~\mbox{\cite{Dowdall:2013rya,Carrasco:2014poa,Bazavov:2017lyh,Miller:2020xhy}},        \\
    \label{eq:fKfpi_direct_broken_2p1}                                                                             
&\mbox{direct},\,\Nf=2+1:  &\FLAGAVBEGIN f_{K^\pm} / f_{\pi^\pm} & =  1.1917(37)\FLAGAVEND     &&\Refs~\mbox{\cite{Follana:2007uv,Bazavov:2010hj,Durr:2010hr,Blum:2014tka,Durr:2016ulb,Bornyakov:2016dzn}},\\
    \label{eq:fKfpi_direct_broken_2}                                                                               
&\mbox{direct},\,\Nf=2:    &\FLAGAVBEGIN f_{K^\pm} / f_{\pi^\pm} & =  1.205(18)\FLAGAVEND &&\Ref~\mbox{\cite{Blossier:2009bx}},
\end{align}
for QCD with broken isospin.

The averages obtained for $f_+(0)$ and $\fKfpichargedr$ at $\Nf=2+1$ and $\Nf=2+1+1$ [see Eqs.~(\ref{eq:fplus_direct_2p1p1}-\ref{eq:fplus_direct_2p1}) and (\ref{eq:fKfpi_direct_broken_2p1p1}-\ref{eq:fKfpi_direct_broken_2p1})] exhibit a precision better than $\sim 0.3 \%$.
At such a level of precision QED effects cannot be ignored and a consistent lattice treatment of both QED and QCD effects in leptonic and semileptonic decays becomes mandatory.

\subsubsection{Extraction of $|V_{ud}|$ and $|V_{us}|$}

It is instructive to convert the averages for $f_+(0)$ and $\fKfpichargedr$ into a corresponding range for the CKM matrix elements $|V_{ud}|$ and $|V_{us}|$, using the relations (\ref{eq:products}). 
Consider first the results for $\Nf=2+1+1$. 
The average for $f_+(0)$ in Eq.~(\ref{eq:fplus_direct_2p1p1}) is mapped into the interval $|V_{us}|=0.2232(6)$,
depicted as a horizontal red band in Fig.~\ref{fig:VusVersusVud}.
The one for $\fKfpichargedr$ in Eq.~(\ref{eq:fKfpi_direct_broken_2p1p1})
and $|V_{us}/V_{ud}|(\fKfpichargedr)$ in Eq.~(\ref{eq:products})
is converted into $|V_{us}|/|V_{ud}|= 0.2313(5)$,
shown as a tilted red band. 
The red ellipse is the intersection of these two bands and represents the 68\% likelihood contour,\footnote{Note that the ellipses shown in Fig.~5 of both Ref.~\cite{Colangelo:2010et} and Ref.~\cite{Aoki:2013ldr} correspond instead to the 39\% likelihood contours. Note also that in Ref.~\cite{Aoki:2013ldr} the likelihood was erroneously stated to be $68 \%$ rather than $39 \%$.} obtained by treating the above two results as independent measurements. 
Repeating the exercise for $\Nf=2+1$ leads to the green ellipse.
The vertical light and dark blue bands show $|V_{ud}|$ from nuclear $\beta$ decay, Eqs.~(\ref{eq:Vud beta}) and (\ref{eq:Vud beta:new NC}), respectively.
The PDG value (\ref{eq:Vud beta}) indicates a tension with 
both the $N_f=2+1+1$ and $N_f=2+1$ results from lattice QCD.

As we mentioned, QED radiative corrections are becoming relevant
for the extraction of the CKM elements at the current precision of
lattice QCD inputs.
We obtain a slightly larger value of $|V_{us}|/|V_{ud}|= 0.2320(5)$
by inputting $|V_{us}/V_{ud}|(\fKfpichargedr)$ in Eq.~(\ref{eq:VusVud_new})
with the QED corrections on the lattice.
Figure~\ref{fig:VusVersusVud_em-lat} suggests that
the kaon (semi)leptonic decays favour a slightly smaller value of $|V_{ud}|$
than the nuclear transitions.

\subsection{Tests of the Standard Model}\label{sec:testing}
  
In the Standard Model, the CKM matrix is unitary. In particular, the elements of the first row obey
\be
   \label{eq:CKM unitarity}
   |V_u|^2\equiv |V_{ud}|^2 + |V_{us}|^2 + |V_{ub}|^2 = 1\fs
\ee 
The tiny contribution from $|V_{ub}|$ is known much better than needed in the present context: $|V_{ub}|= 3.82 (24) \cdot 10^{-3}$ \cite{Zyla:2020zbs}. 
In the following, we test the first row unitarity Eq.~(\ref{eq:CKM unitarity}) by calculating $|V_u|^2$ and by analyzing the lattice data within the Standard Model.

In Fig.~\ref{fig:VusVersusVud}, the correlation between $|V_{ud}|$ and $|V_{us}|$ imposed by the unitarity of the CKM matrix is indicated by a dotted line (more precisely, in view of the uncertainty in $|V_{ub}|$, the correlation corresponds to a band of finite width, but the effect is too small to be seen here).
The plot shows that there is a tension with unitarity in the data for $N_f = 2 + 1 + 1$: Numerically, the outcome for the sum of the squares of the first row of the CKM matrix reads $|V_u|^2 = 0.9813(66)$, which deviates from unity at the level of $\simeq 2.8$ standard deviations. 
Still, it is fair to say that at this level the Standard Model passes a nontrivial test that exclusively involves lattice data and well-established kaon decay branching ratios.

The test sharpens considerably by
combining the lattice results for $f_+(0)$
with the $\beta$ decay value of $|V_{ud}|$:
$f_+(0)$ in Eq.~(\ref{eq:fplus_direct_2p1p1})
and the PDG estimate of $|V_{ud}|$ in Eq.~(\ref{eq:Vud beta})
lead to $|V_u|^2 = 0.99794(37)$, which highlights
a $\simeq 5.6~\sigma$ deviation with unitarity.
A lower tension at the three-$\sigma$ level is suggested
either from $\fKfpichargedr$ in Eq.~(\ref{eq:fKfpi_direct_broken_2p1p1})
($|V_u|^2 = 0.99883(37)$)
or $|V_{ud}|$ in Eq.~(\ref{eq:Vud beta:new NC}) with the updated nuclear corrections
($|V_u|^2 = 0.99800(65)$).
Unitarity is fulfilled with $\fKfpichargedr$
and $|V_{ud}|$~(\ref{eq:Vud beta:new NC}) ($|V_u|^2 = 0.99890(68)$).
Note that, when the PDG value of $|V_{ud}|$~(\ref{eq:Vud beta}) is employed,
the uncertainties on $|V_u|^2$ coming from the errors of $|V_{ud}|$ and
$|V_{us}|$ are of similar magnitude with each other.

The situation is similar for $\Nf=2+1$: with the lattice data alone one has $|V_u|^2 = 0.9832(89)$, which deviates from unity at the level of $\simeq 1.9$ standard deviations.
The lattice results for $f_+(0)$ in Eqs.~(\ref{eq:fplus_direct_2p1}) with the PDG value of $|V_{ud}|$~(\ref{eq:Vud beta}) lead to $|V_u|^2 = 0.99816(43)$, implying a $\simeq 4.3\,\sigma$ deviation from unitarity,
whereas the deviation is reduced to 2.3\,--\,2.6\,$\sigma$
with $\fKfpichargedr$ in Eq.~(\ref{eq:fKfpi_direct_broken_2p1})
($|V_u|^2 = 0.99896(45)$) and $|V_{ud}|$ in Eq.~(\ref{eq:Vud beta:new NC})
($|V_u|^2 = 0.99822(69)$).

For the analysis corresponding to $N_f = 2$ the reader should refer to the 2016 edition~\cite{Aoki:2016frl}.


\subsection{Analysis within the Standard Model} \label{sec:SM} 
 
The Standard Model implies that the CKM matrix is unitary. 
The precise experimental constraints quoted in Eq.~(\ref{eq:products}) and the unitarity condition Eq.~(\ref{eq:CKM unitarity}) then reduce the four quantities $|V_{ud}|,|V_{us}|,f_+(0),\fKfpichargedr$ to a single unknown: any one of these determines the other three within narrow uncertainties.
 
As Fig.~\ref{fig:Vus Vud} shows, the results obtained for $|V_{us}|$ and $|V_{ud}|$ from the data on $\fKfpichargedr$ (squares) are consistent with the determinations via $f_+(0)$ (triangles), while there is a tendency that 
$|V_{us}|$ ($|V_{ud}|$) from $f_+(0)$ is systematically smaller (larger)
than that from $\fKfpichargedr$.
In order to calculate the corresponding average values, we restrict ourselves to those determinations that enter the FLAG average in Sec.~\ref{sec:Direct}.
The corresponding results for $|V_{us}|$ are listed in Tab.~\ref{tab:Vus} (the error in the experimental numbers used to convert the values of $f_+(0)$ and $\fKfpichargedr$ into values for $|V_{us}|$ is included in the statistical error).
   
\begin{figure}[!htb]
\psfrag{y}{\tiny $\star$}
\begin{center}
\vspace{-0.5cm} 
\hspace*{-1.2cm} 
\includegraphics[height=13.0cm]{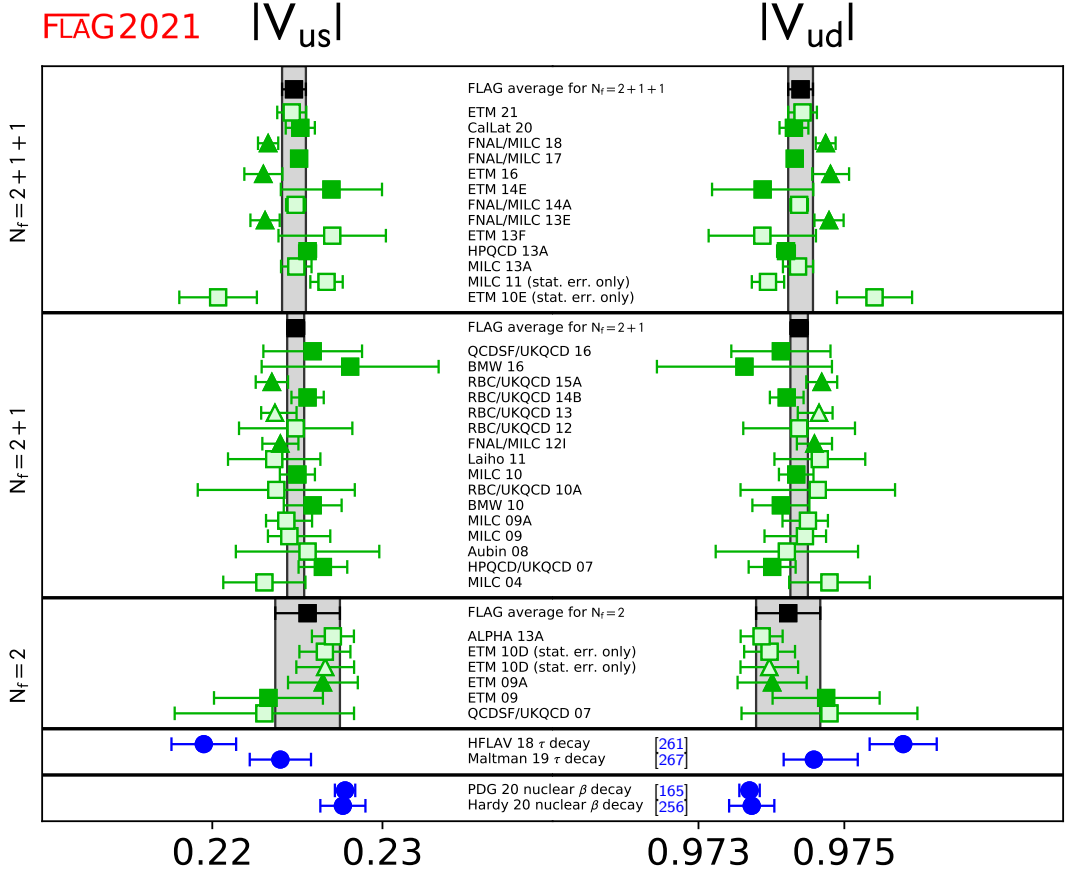}
\end{center}
\vspace{-3.0cm}\hspace{9.1cm}\parbox{6cm}{\sffamily\tiny  \cite{Amhis:2019ckw}\\

\vspace{-1.27em}\cite{10.21468/SciPostPhysProc.1.006}\\




\vspace{-0.38em}\hspace{0em}\cite{Zyla:2020zbs}

\vspace{-0.15em}\hspace{0em}\cite{Hardy:2020qwl}
}
\vspace{0.75cm}
\caption{\label{fig:Vus Vud} Results for $|V_{us}|$ and $|V_{ud}|$ that follow from the lattice data for $f_+(0)$ (triangles) and $\fKfpichargedr$ (squares), on the basis of the assumption that the CKM matrix is unitary. 
The black squares and the grey bands represent our averages, obtained by combining these two different ways of measuring $|V_{us}|$ and $|V_{ud}|$ on a lattice.
For comparison, the figure also indicates the results obtained if the data on nuclear $\beta$ decay and inclusive hadronic $\tau$ decay is analyzed within the Standard Model.}
\end{figure}  

\begin{table}[!htb]
\centering
\noindent
\begin{tabular*}{\textwidth}[l]{@{\extracolsep{\fill}}lclcll}
Collaboration & Ref. &\rule{0.5cm}{0cm}$\Nf$&from&\rule{0.6cm}{0cm}$|V_{us}|$&\rule{0.6cm}{0cm}$|V_{ud}|$\\
&&&&& \\[-2ex]
\hline \hline &&&&&\\[-2ex]
FNAL/MILC 18 & \cite{Bazavov:2018kjg}& $2+1+1$& $f_+(0)$  \rule{0cm}{0.45cm} & 0.2233(5)(3)& 0.97474(12)(6)\\
ETM 16 &\cite{Carrasco:2016kpy}&$2+1+1$&$f_+(0)$ \rule{0cm}{0.45cm} &0.2230(11)(2)&0.97481(25)(5)\\
 CalLat 20& \cite{Miller:2020xhy}& $2+1+1$& $\fKfpichargedr$ \rule{0cm}{0.45cm} & 0.2252(7)(6)& 0.97431(15)(13)\\
FNAL/MILC 17&\cite{Bazavov:2017lyh}&$2+1+1$&$\fKfpichargedr$ \rule{0cm}{0.45cm} &0.2251(4)(2)&0.97432(9)(5)\\
ETM 14E &\cite{Carrasco:2014poa}&$2+1+1$&$\fKfpichargedr$ \rule{0cm}{0.45cm} &0.2270(22)(20)&0.97388(51)(47)\\
HPQCD 13A &\cite{Dowdall:2013rya}&$2+1+1$&$\fKfpichargedr$ \rule{0cm}{0.45cm} &0.2256(4)(3)&0.97420(10)(7)\\
&&&&& \\[-2ex]
\hline
&&&&& \\[-2ex]
RBC/UKQCD 15A &\cite{Boyle:2015hfa}&$2+1$&$f_+(0)$          \rule{0cm}{0.45cm} &0.2235(9)(3)&0.97469(20)(7)\\
FNAL/MILC 12I    &\cite{Bazavov:2012cd}&$2+1$&$f_+(0)$       \rule{0cm}{0.45cm} &0.2240(7)(8)&0.97459(16)(18)\\
QCDSF/UKQCD 16  &\cite{Bornyakov:2016dzn}   &$2+1$&$\fKfpichargedr$ \rule{0cm}{0.45cm} &0.2259(18)(23)&0.97413(42)(54)\\ 
BMW 16 &\cite{Durr:2016ulb,Scholz:2016kcr}&$2+1$&$\fKfpichargedr$ \rule{0cm}{0.45cm} &0.2281(19)(48)&0.97363(44)(112)\\
RBC/UKQCD 14B  &\cite{Blum:2014tka}   &$2+1$&$\fKfpichargedr$ \rule{0cm}{0.45cm} &0.2256(3)(9)&0.97421(7)(22)\\ 
MILC 10 &\cite{Bazavov:2010hj}&$2+1$&$\fKfpichargedr$ \rule{0cm}{0.45cm} &0.2250(5)(9)&0.97434(11)(21)\\
BMW 10 &\cite{Durr:2010hr}  & $2+1$ \rule{0cm}{0.45cm}& $\fKfpichargedr$ & $0.2259(13)(11)$&0.97413(30)(25)\\
HPQCD/UKQCD 07 &\cite{Follana:2007uv}\rule{0cm}{0.4cm}& $2+1$ & $\fKfpichargedr$&  $  0.2265(6)(13)$&0.97401(14)(29)\\
&&&&& \\[-2ex]
\hline
&&&&& \\[-2ex]
ETM 09A & \cite{Lubicz:2009ht}\rule{0cm}{0.4cm}&2&$f_+(0)$&   $ 0.2265 (14) (15)$&0.97401(33)(34)\\
ETM 09  &\cite{Blossier:2009bx}\rule{0cm}{0.4cm}&2&$\fKfpichargedr$& $ 0.2233 (11) (30)$&0.97475(25)(69)\\
&&&&& \\[-2ex]
\hline \hline 
\end{tabular*}
\caption{\label{tab:Vus} Values of $|V_{us}|$ and $|V_{ud}|$ obtained from the lattice determinations of either $f_+(0)$ or $\fKfpichargedr$ assuming CKM unitarity. 
The first number in brackets represents the statistical error including the experimental uncertainty, whereas the second is the systematic one.
} 
\end{table} 

For $\Nf=2+1+1$ we consider the data both for $f_+(0)$ and $\fKfpichargedr$, treating ETM 16 and ETM 14E on the one hand and FNAL/MILC 18, CalLat 20, FNAL/MILC 17, and HPQCD 13A on the other hand, as statistically correlated according to the prescription of Sec.~\ref{sec:error_analysis}. 
As shown in Tab~\ref{tab:Final results}, we obtain $|V_{us}|=0.2248(7)$,
where the error is stretched
by a factor $\sqrt{\chi^2/{\rm dof}}\!\sim\!\sqrt{2.6}$.
This result is indicated on the left hand side of Fig.~\ref{fig:Vus Vud} by the narrow vertical band. 
In the case $N_f = 2+1$ we consider MILC 10, FNAL/MILC 12I and HPQCD/UKQCD 07 on the one hand and RBC/UKQCD 14B and RBC/UKQCD 15A on the other hand, as mutually statistically correlated, since the analysis in the two cases starts from partly the same set of gauge ensembles.
In this way we arrive at $|V_{us}| = 0.2249(5)$ with $\chi^2/{\rm dof} \simeq 0.8$. 
For $\Nf=2$ we consider ETM 09A and ETM 09 as statistically correlated, obtaining $|V_{us}|=0.2256(19)$ with $\chi^2/{\rm dof} \simeq 0.7$.
The figure shows that the results obtained for the data with $\Nf=2$, $\Nf=2+1$, and $\Nf=2+1+1$ are consistent with each other.
However, the larger error for $\Nf=2+1+1$ due to the stretch factor
$\sqrt{\chi^2/{\rm dof}}$ suggests
a slight tension between the estimates
from the semileptonic and leptonic decays.

Alternatively, we can solve the relations for $|V_{ud}|$ instead of $|V_{us}|$. 
Again, the result $|V_{ud}|=0.97440(17)$, which follows from the lattice data with $\Nf=2+1+1$, is perfectly consistent with the values $|V_{ud}|=0.97438(12)$ and $|V_{ud}|=0.97423(44)$ obtained from the data with $\Nf=2+1$ and $\Nf=2$, respectively.
We observe the difference of about 3 $\sigma$ from Eq.~(\ref{eq:Vud beta})
from the superallowed nuclear transitions.
It is, however, reduced to $\lesssim \! 2~\sigma$
with Eq.~(\ref{eq:Vud beta:new NC})
based on the updated nuclear corrections.

As mentioned in Sec.~\ref{sec:Exp},
the HFLAV value of $|V_{us}|$ from the inclusive hadronic $\tau$ decays
differs from those obtained from the kaon decays
by about three standard deviations.
Assuming the first row unitarity~(\ref{eq:CKM unitarity})
leads to a larger value of $|V_{ud}|$ than those
from the kaon and nuclear decays.
Such a tension does not appear with $|V_{us}|$ in Eq.~(\ref{eq:Vus tau})
from strange hadronic $\tau$ decay data
and lattice QCD data of the hadronic vacuum polarization function.

\begin{table}[!htb]
\centering
\begin{tabular*}{\textwidth}[l]{@{\extracolsep{\fill}}llllll}
  \rule[-0.2cm]{0cm}{0.5cm}& Ref. & \rule{0.3cm}{0cm} $|V_{us}|$&
  \rule{0.3cm}{0cm} $|V_{ud}|$ \\
\\[-2ex]
\hline \hline
\\[-2ex]
$\Nf= 2+1+1$& &\rule{0cm}{0.4cm} 0.2248(7)&  0.97440(17) \\
\\[-2ex]
\hline
$\Nf= 2+1$&   &\rule{0cm}{0.4cm}0.2249(5)& 0.97438(12)   \\
\\[-2ex]
\hline
\\[-2ex]
$\Nf=2$ & &\rule{0cm}{0.4cm}0.2256(19) &0.97423(44)      \\
\\[-2ex]
\hline\hline
\\[-2ex]
nuclear $\beta$ decay & \cite{Zyla:2020zbs} & 0.2278(6)  & 0.97370(14) \\
\\[-2ex]
nuclear $\beta$ decay & \cite{Hardy:2020qwl}& 0.2277(13) & 0.97373(31) \\
\\[-2ex]
inclusive $\tau$ decay &\cite{Amhis:2019ckw}&0.2195(19) & 0.97561(43)   \\
\\[-2ex]
\\[-2ex]
inclusive $\tau$ decay  &\cite{10.21468/SciPostPhysProc.1.006}&0.2240(18)&0.97458(40)\\
\\[-2ex]
\hline\hline
\end{tabular*}
\caption{\label{tab:Final results}The upper half of the table shows our final results for $|V_{us}|$, $|V_{ud}|$, $f_+(0)$ and $\fKfpichargedr$ that are obtained by analysing the lattice data within the Standard Model (see text). 
For comparison, the lower half lists the values that follow if the lattice results are replaced by the experimental results on nuclear $\beta$ decay and inclusive hadronic $\tau$ decay, respectively.}
\end{table}

\subsection{Direct determination of $f_{K^\pm}$ and $f_{\pi^\pm}$}\label{sec:fKfpi}

It is useful for flavour-physics studies to provide not only the lattice average of $f_{K^\pm} / f_{\pi^\pm}$, but also the average of the decay constant $f_{K^\pm}$. 
The case of the decay constant $f_{\pi^\pm}$ is different, since the 
the PDG value~\cite{Patrignani:2016xqp} of this quantity, based on the use of the value of $|V_{ud}|$ obtained from superallowed nuclear $\beta$ decays \cite{Hardy:2016vhg},
 is often used for setting the scale in lattice QCD (see Sec.~\ref{sec:scalesetting} on the scale setting).
However, the physical  scale can be set in different ways, namely, by using as input the mass of the $\Omega$ baryon($m_\Omega$) or the $\Upsilon$-meson spectrum ($\Delta M_\Upsilon$), which are less sensitive to the uncertainties of the chiral extrapolation in the light-quark mass with respect to $f_{\pi^\pm}$. 
In such cases the value of the decay constant $f_{\pi^\pm}$ becomes a direct prediction of the lattice-QCD simulations.
It is therefore interesting to provide also the average of the decay constant $f_{\pi^\pm}$, obtained when the physical scale is set through another hadron observable, in order to check the consistency of different scale-setting procedures.

Our compilation of the values of $f_{\pi^\pm}$ and $f_{K^\pm}$ with the corresponding colour code is presented in Tab.~\ref{tab:FK Fpi} and it is unchanged from the corresponding one in the previous FLAG review \cite{Aoki:2019cca}.

In comparison to the case of $f_{K^\pm} / f_{\pi^\pm}$ we have added two columns indicating which quantity is used to set the physical scale and the possible use of a renormalization constant for the axial current.
For several lattice formulations the use of the nonsinglet axial-vector Ward identity allows to avoid the use of any renormalization constant.

One can see that the determinations of $f_{\pi^\pm}$ and $f_{K^\pm}$ suffer from larger uncertainties with respect to the ones of the ratio $f_{K^\pm} / f_{\pi^\pm}$, which is less sensitive to various systematic effects (including the uncertainty of a possible renormalization constant) and, moreover, is not exposed to the uncertainties of the procedure used to set the physical scale.

According to the FLAG rules, for $N_f = 2 + 1 + 1$ three data sets can form the average of $f_{K^\pm}$ only: ETM 14E \cite{Carrasco:2014poa}, FNAL/MILC 14A \cite{Bazavov:2014wgs}, and HPQCD 13A \cite{Dowdall:2013rya}.
Following the same procedure already adopted in Sec.~\ref{sec:Direct} for the ratio of the decay constants,
we assume 100\,\% statistical and systematic correlation between
FNAL/MILC 14A and HPQCD 13A.
For $N_f = 2 + 1$ three data sets can form the average of $f_{\pi^\pm}$ and $f_{K^\pm}$ : RBC/UKQCD 14B \cite{Blum:2014tka} (update of RBC/UKQCD 12), HPQCD/UKQCD 07 \cite{Follana:2007uv}, and MILC 10 \cite{Bazavov:2010hj}, which is the latest update of the MILC program.
We consider HPQCD/UKQCD 07 and MILC 10 as statistically correlated and use the prescription of Sec.~\ref{sec:error_analysis} to form an average.
For $N_f = 2$ the average cannot be formed for $f_{\pi^\pm}$, and only one data set (ETM 09) satisfies the FLAG rules for $f_{K^\pm}$.

Thus, our averages read
\begin{align}
  \label{eq:fPi}
&N_f = 2 + 1:     &\FLAGAVBEGIN f_{\pi^\pm}&= 130.2 ~ (0.8)\FLAGAVEND  ~ \mbox{MeV} &&\Refs~\mbox{\cite{Follana:2007uv,Bazavov:2010hj,Blum:2014tka}},\\ \nonumber 
                \\                                                       
&N_f = 2 + 1 + 1: &\FLAGAVBEGIN f_{K^\pm} & = 155.7 ~ (0.3)\FLAGAVEND  ~ \mbox{MeV} &&\Refs~\mbox{\cite{Dowdall:2013rya,Bazavov:2014wgs,Carrasco:2014poa}}         ,\nonumber\\ 
&N_f = 2 + 1:     &\FLAGAVBEGIN f_{K^\pm} & = 155.7 ~ (0.7)\FLAGAVEND  ~ \mbox{MeV} &&\Refs~\mbox{\cite{Follana:2007uv,Bazavov:2010hj,Blum:2014tka}},\label{eq:fK}\\ 
\nonumber
&N_f = 2:         &\FLAGAVBEGIN f_{K^\pm} & = 157.5 ~ (2.4)\FLAGAVEND  ~ \mbox{MeV} &&\Ref~\mbox{\cite{Blossier:2009bx}}.\\\nonumber
 \end{align}
The lattice results of Tab.~\ref{tab:FK Fpi} and our averages (\ref{eq:fPi}-\ref{eq:fK}) are reported in Fig.~\ref{fig:latticedata_decayconstants}. 
Note that the FLAG averages of $f_{K^\pm}$ for $N_f = 2$ and $N_f = 2 + 1 + 1$ are based on calculations in which $f_{\pi^\pm}$ is used to set the lattice scale, while the $N_f = 2 + 1$ average does not rely on that. 

\begin{table}[!htb]
       {\centering
\vspace{2.0cm}{\footnotesize\noindent
\begin{tabular*}{\textwidth}[l]{@{\extracolsep{\fill}}l@{\hspace{1mm}}r@{\hspace{1mm}}l@{\hspace{1mm}}l@{\hspace{1mm}}l@{\hspace{1mm}}l@{\hspace{1mm}}l@{\hspace{3mm}}l@{\hspace{1mm}}l@{\hspace{1mm}}l@{\hspace{5mm}}l@{\hspace{1mm}}l}
Collaboration & Ref. & $\Nf$ &
\hspace{0.15cm}\begin{rotate}{40}{publication status}\end{rotate}\hspace{-0.15cm}&
\hspace{0.15cm}\begin{rotate}{40}{chiral extrapolation}\end{rotate}\hspace{-0.15cm}&
\hspace{0.15cm}\begin{rotate}{40}{continuum extrapolation}\end{rotate}\hspace{-0.15cm}&
\hspace{0.15cm}\begin{rotate}{40}{finite-volume errors}\end{rotate}\hspace{-0.15cm}& 
\hspace{0.15cm}\begin{rotate}{40}{renormalization}\end{rotate}\hspace{-0.15cm}&
\hspace{0.05cm}\begin{rotate}{40}{physical scale}\end{rotate}\hspace{-0.15cm}&\rule{0cm}{0cm}
\hspace{0.0cm}\begin{rotate}{40}{$SU(2)$ breaking}\end{rotate}\hspace{-0.15cm}&\rule{0.5cm}{0cm}
$f_{\pi^\pm}$&\rule{0.5cm}{0cm}$f_{K^\pm}$ \\
&&&&&&& \\[-0.1cm]
\hline
\hline
&&&&&&& \\[-0.1cm]
ETM 14E &\cite{Carrasco:2014poa}&2+1+1&\gA&\soso&\good&\soso&na&$f_\pi$&&--&{154.4(1.5)(1.3)}\\
FNAL/MILC 14A&\cite{Bazavov:2014wgs}&2+1+1&\gA&\good&\good&\good&na&$f_\pi$&&--&{155.92(13)($_{-23}^{+34}$)}\\
HPQCD 13A&\cite{Dowdall:2013rya}&2+1+1&\gA&\good&\soso&\good&na&$f_\pi$&&--&{155.37(20)(27)}\\
MILC 13A&\cite{Bazavov:2013cp}&2+1+1&\gA&\good&\soso&\good&na&$f_\pi$&&--&155.80(34)(54)\\
ETM 10E &\cite{Farchioni:2010tb}&2+1+1&\rC&\soso&\soso&\soso&na&$f_\pi$&\checkmark&--&159.6(2.0)\\
&&&&&&& \\[-0.1cm]
\hline
&&&&&&& \\[-0.1cm]
JLQCD 15C         &\cite{Fahy:2015xka}&2+1&\rC&\soso&\tbg&\tbg&NPR&$t_0$& &125.7(7.4)$_{\rm stat}$&\\
RBC/UKQCD 14B   &\cite{Blum:2014tka}&2+1&\gA&\good&\good&\good&NPR&$m_\Omega$ &\checkmark& 130.19(89) & 155.18(89) \\
RBC/UKQCD 12   &\cite{Arthur:2012opa}&2+1&\gA&\tbg&\soso&\good&NPR&$m_\Omega$ &\checkmark& 127.1(2.7)(2.7)& 152.1(3.0)(1.7) \\
Laiho 11       &\cite{Laiho:2011np}   &2+1&\rC&\soso&\good&\soso&na&${}^\dagger$ && $130.53(87)(2.10)$&$156.8(1.0)(1.7)$\\
MILC 10 &\cite{Bazavov:2010hj}&2+1&\rC&\soso&\good&\good&na&${}^\dagger$ & &{129.2(4)(1.4)}&--\\
MILC 10 &\cite{Bazavov:2010hj}&2+1&\rC&\soso&\good&\good&na&$f_\pi$ &&--          &{156.1(4)($_{-9}^{+6}$)}\\
JLQCD/TWQCD 10 &\cite{Noaki:2010zz}&2+1&\rC&\soso&\tbr&\tbg&na&$m_\Omega$&\checkmark&118.5(3.6)$_{\rm stat}$&145.7(2.7)$_{\rm stat}$\\
RBC/UKQCD 10A  &\cite{Aoki:2010dy} &2+1&\gA&\soso&\soso&\good&NPR&$m_\Omega$&\checkmark&124(2)(5)&148.8(2.0)(3.0)\\
MILC 09A &\cite{Bazavov:2009fk}&2+1&\rC&\soso&\tbg&\tbg &na&$\Delta M_\Upsilon$ &&128.0(0.3)(2.9)&          153.8(0.3)(3.9)\\
MILC 09A &\cite{Bazavov:2009fk}&2+1&\rC&\soso&\tbg&\tbg &na&$f_\pi$&&--&156.2(0.3)(1.1)\\
MILC 09 &\cite{Bazavov:2009bb}&2+1&\gA&\soso&\tbg&\tbg &na&$\Delta M_\Upsilon$&&128.3(0.5)($^{+2.4}_{-3.5}$)&154.3(0.4)($^{+2.1}_{-3.4}$) \\
MILC 09 &\cite{Bazavov:2009bb}&2+1&\gA&\soso&\tbg&\tbg &na&$f_\pi$&&&156.5(0.4)($^{+1.0}_{-2.7}$)\\
Aubin 08       &\cite{Aubin:2008ie} &2+1&\rC&\soso&\soso&\soso&na&$\Delta M_\Upsilon$     && 129.1(1.9)(4.0)   & 153.9(1.7)(4.4)  \\
RBC/UKQCD 08   &\cite{Allton:2008pn} &2+1&\gA&\soso&\tbr&\tbg&NPR&$m_\Omega$&\checkmark&124.1(3.6)(6.9) &        149.4(3.6)(6.3)\\
HPQCD/UKQCD 07 &\cite{Follana:2007uv}&2+1&\gA&\soso&\soso&\soso&na&$\Delta M_\Upsilon$&\checkmark& {132(2)}                & {156.7(0.7)(1.9)}\\
MILC 04 &\cite{Aubin:2004fs}&2+1&\gA&\soso&\soso&\soso&na&$\Delta M_\Upsilon$&&129.5(0.9)(3.5)     &     156.6(1.0)(3.6)\\[-1mm]
&&&&&&& \\[-0.1cm]
\hline
&&&&&&& \\[-0.1cm]
ETM 14D &\cite{Abdel-Rehim:2014nka}&2&\rC&\good&\tbr&\soso&na&$f_\pi$&\checkmark&--&153.3(7.5)$_{\rm stat}$\\
ETM 09         &\cite{Blossier:2009bx}         &2 &\gA&\soso&\tbg&\soso&na&$f_\pi$&\checkmark& --& {157.5(0.8)(2.0)(1.1)}$^{\dagger\dagger}$\\
&&&&&&& \\[-0.1cm]
\hline
\hline
&&&&&&& \\[-0.1cm]
\end{tabular*}}\\[-2mm]
}

\begin{minipage}{\linewidth}
\footnotesize The label 'na' indicates the lattice calculations that do not require the use of any renormalization constant for the axial current, while the label 'NPR' ('1lp') signals the use of a renormalization constant calculated nonperturbatively (at 1-loop order in perturbation theory).  
\begin{itemize}
{\footnotesize 
\item[$^{\dagger}$] The ratios of lattice spacings within the ensembles were determined using the quantity $r_1$. 
	The conversion to physical units was made on the basis of Ref.~\cite{Davies:2009tsa} and we note that such a determination depends on the PDG value~\cite{Patrignani:2016xqp} of the pion decay constant\\[-5mm]
\item[$^{\dagger\dagger}$] Errors are (stat+chiral)($a\neq 0$)(finite size).
\\[-5mm]
}
\end{itemize}
\end{minipage}
\caption{Colour code for the lattice data on $f_{\pi^\pm}$ and $f_{K^\pm}$ together with information on the way the lattice spacing was converted to physical units and on whether or not an isospin-breaking correction has been applied to the quoted result (see Sec.~\ref{sec:Direct}). The numerical values are listed in MeV units. In this and previous editions~\cite{Aoki:2019cca}, old results with two red tags have been dropped.\hfill}
\label{tab:FK Fpi}
\end{table}

\begin{figure}[!htb]
\begin{center}
\includegraphics[height=9.0cm]{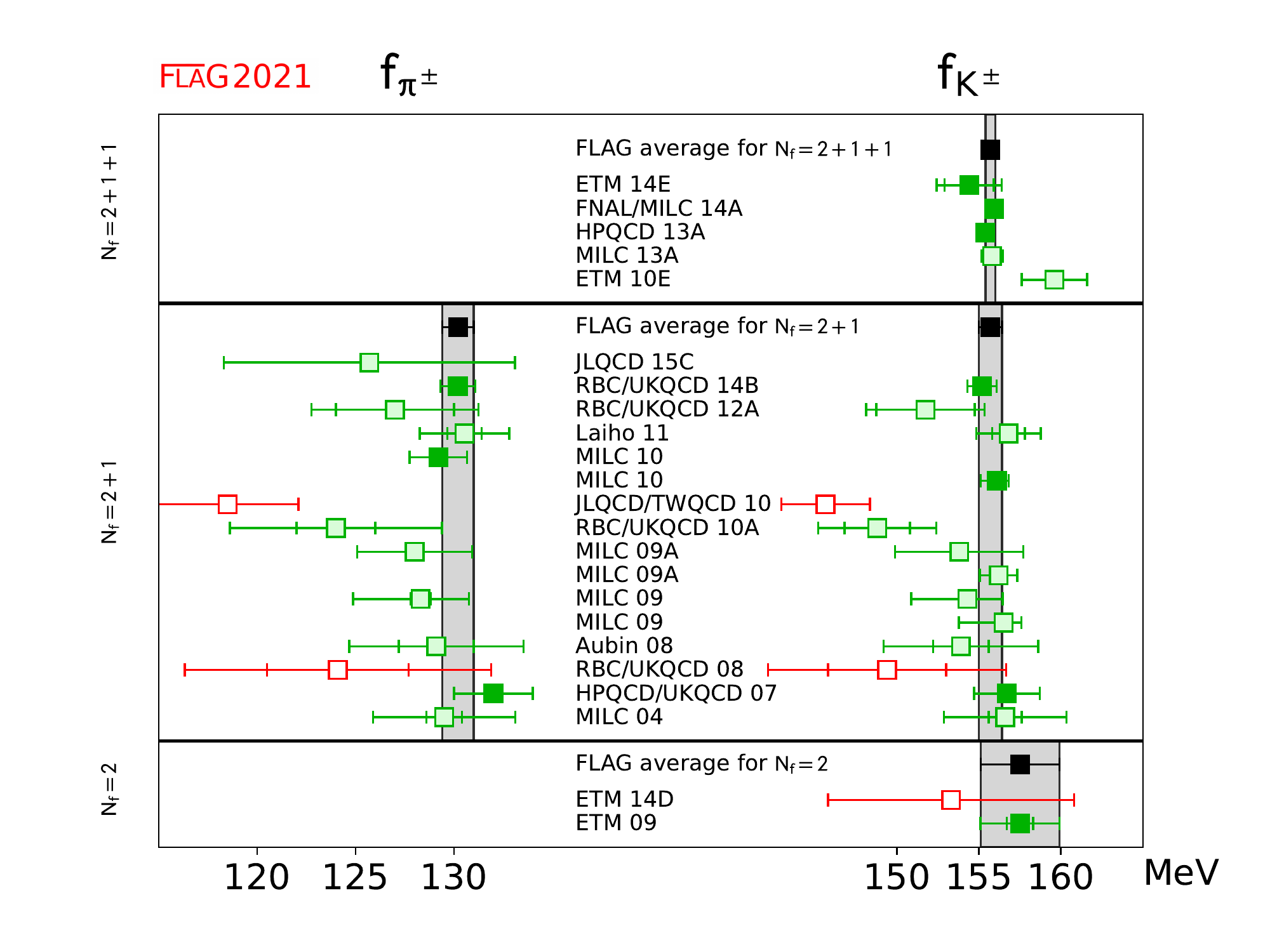}
\end{center}
\vspace{-0.75cm}
\caption{\label{fig:latticedata_decayconstants}
Values of $f_\pi$ and $f_K$.
The black squares and grey bands indicate our averages (\ref{eq:fPi}) and (\ref{eq:fK}).
}
\end{figure}

\clearpage

\clearpage
\setcounter{section}{4}

\newcommand{\pubA}{\gA} 
\newcommand{\pubP}{\oP} 
\newcommand{\pubC}{\rC} 


\section{Low-energy constants\label{sec:LECs}}
Authors: S.~D\"urr, H.~Fukaya, U.~M.~Heller\\


\subsection{Chiral perturbation theory and lattice QCD\label{sec:chPT}}


In the study of the quark-mass dependence of QCD observables calculated on the lattice, it is beneficial to use chiral perturbation theory ({\Ch}PT).
This framework predicts the nonanalytic quark-mass dependence of hadron masses and matrix elements, and it provides symmetry relations among such observables.
These predictions invoke a set of linearly independent and universal (i.e., process-independent) low-energy constants (LECs), defined as coefficients of the polynomial terms (in $m_q$ or $\Mpi^2$) of different observables.

{\Ch}PT is an effective field theory approach to the low-energy properties of QCD based on the spontaneous breaking of chiral symmetry,
$SU(\Nf)_L \times SU(\Nf)_R \to SU(\Nf)_V$, and its soft explicit breaking by quark-mass terms.
In its original implementation (i.e., in infinite volume) it is an expansion in powers of $m_q$ and $p^2$ with the counting rule $\Mpi^2 \sim m_q \sim p^2$.

If one expands around the $SU(2)$ chiral limit, two LECs appear at order $p^2$ in the chiral effective Lagrangian,
\be
F\equiv \Fpi\,\rule[-0.3cm]{0.01cm}{0.7cm}_{\;m_u,m_d\rightarrow 0}
\quad \mbox{and} \qquad
B\equiv \frac{\Sigma}{F^2} \; , \quad\mbox{where} \quad
\Sigma\equiv-\<\ubar u\>\,\Big|_{\;m_u,m_d\rightarrow 0} \; ,
\ee
and seven more at order $p^4$, called $\bar\ell_i$ with $i=1,\ldots,7$.
In the analysis of the $SU(3)$ chiral limit there are again%
\footnote{Here and in the following, we stick to the notation used in the papers where the {\Ch}PT formulae were established, i.e., we work with $\Fpi\equiv \fpi/\sqrt{2}=92.2(1)\MeV$ and $\Fka\equiv f_K/\sqrt{2}$.
The occurrence of different normalization conventions is not convenient, but avoiding it by reformulating the formulae in terms of $\fpi$, $f_K$ is not a good way out.
Since we are using different symbols, confusion cannot arise.\label{foot:fpi}}
two LECs at order $p^2$,
\be
F_0\equiv \Fpi\,\rule[-0.3cm]{0.01cm}{0.7cm}_{\;m_u,m_d,m_s\rightarrow 0}
\quad \mbox{and} \qquad
B_0\equiv \frac{\Sigma_0}{F_0^2} \; , \quad\mbox{where} \quad
\Sigma_0\equiv-\<\ubar u\>\,\Big|_{\;m_u,m_d,m_s\rightarrow 0} \; ,
\ee
but ten more at order $p^4$, indicated by the symbols $L_i(\mu)$ with $i=1,\ldots,10$.
These ``constants'' are independent of the quark masses%
\footnote{More precisely, they are independent of the 2 or 3 light-quark masses that are explicitly considered in the respective framework.
However, all low-energy constants depend on the masses of the remaining quarks $s$, $c$, $b$, $t$ or $c$, $b$, $t$ in the $SU(2)$ and $SU(3)$ framework,
respectively, although the dependence on the masses of the $c$, $b$, $t$ quarks is expected to be small \cite{Gasser:1983yg,Gasser:1984gg}.},
but they become scale dependent after renormalization (sometimes a superscript $r$ is used).
The $SU(2)$ constants $\bar\ell_i$ are $\mu$-independent, since they are defined at scale $\mu=\Mpi^\mr{phys}$ (as indicated by the bar).
The $SU(3)$ constants $L_i(\mu)$ are usually quoted at the renormalization scale $\mu=770\MeV$.
For the precise definition of these constants and their scale dependence we refer the reader to Refs.~\cite{Gasser:1983yg,Gasser:1984gg}.

In the previous four versions of the FLAG review, we summarized the {\Ch}PT formulae for the quark-mass dependence of the pion and kaon mass and decay constant, as well as the scalar and vector pion charge radius.
We briefly discussed the different regimes of {\Ch}PT, touched on partially quenched and mixed action formulations,
collected and colour-coded the available lattice results for the LECs considered, and formed FLAG estimates or averages, where possible.

Since the fourth edition in 2019 \cite{Aoki:2019cca} (referred to as FLAG 19 below) only a handful of papers
appeared with results on the set of LECs covered in our report, but none that qualifies to be included in an average.
We therefore decided to shorten the section on LECs considerably, referring the reader to the 2019 FLAG review for the {\Ch}PT formulae,
description of the results covered there, and the details and explanation of the FLAG estimates and averages.
In this edition, we will concentrate on the description of the new results and, for the convenience of our readers,
list the FLAG estimates and averages, asking the reader to consult FLAG 19 \cite{Aoki:2019cca} for the details.

In the 2019 edition, we introduced a section on $\pi\pi$ scattering in the context of $SU(2)$ {\Ch}PT and collected results, from finite-volume lattice calculations, of the isospin $I=0$ and $I=2$ scattering lengths.
In this edition, we will keep this section and describe the new results that appeared since the 2019 FLAG review.
We will, further, add a section on $\pi K$ and $KK$ scattering in the context of $SU(3)$ {\Ch}PT and collect the available results for the scattering lengths from finite-volume lattice calculations.


\subsubsection{$\pi\pi$ scattering}

The scattering of pseudoscalar octet mesons off each other (mostly $\pi\pi$ and $\pi K$ scattering) is a useful approach to determine {\Ch}PT low-energy constants \cite{Weinberg:1966kf,Gasser:1983kx,Bijnens:1995yn,Colangelo:2001df,Nebreda:2010wv}.
This statement holds true both in experiment and on the lattice.
We would like to point out the main difference between these two approaches is not so much the discretization of space-time, but rather the Minkowskian versus Euclidean setup.

In infinite-volume Minkowski space-time, 4-point Green's functions can be evaluated (e.g., in experiment) for a continuous range of (on-shell) momenta, as captured, for instance, by the Mandelstam variable $s$.
For a given isospin channel $I=0$ or $I=2$ the $\pi\pi$ scattering phase shift $\de^{I}(s)$ can be determined for a variety of $s$ values, and by matching to {\Ch}PT some low-energy constants can be determined (see below).
In infinite-volume Euclidean space-time, such 4-point Green's functions can only be evaluated at kinematic thresholds; this is the content of the so-called Maiani-Testa theorem \cite{Maiani:1990ca}.
However, in the Euclidean case, the finite volume comes to our rescue, as first pointed out by L\"uscher \cite{Luscher:1985dn,Luscher:1986pf,Luscher:1990ux,Luscher:1991cf}.
By comparing the energy of the (interacting) two-pion system in a box with finite spatial extent $L$ to twice the energy of a pion (with identical bare parameters) in infinite volume information on the scattering length can be obtained.
In particular, in the (somewhat idealized) situation where one can ``scan'' through a narrowly spaced set of box-sizes $L$ such information can be reconstructed in an efficient way.

We begin with a brief summary of the relevant formulae in $SU(2)$ {\Ch}PT terminology.
In the $x$-expansion the formulae for $a_\ell^I$ with $\ell=0$ and $I=0,2$ are found in Ref.~\cite{Gasser:1983yg}
\bea
a_0^0\Mpi&=&+\frac{7M^2}{32\pi F^2}
\bigg\{
1+\frac{5M^2}{84\pi^2 F^2}\Big[ \bar\ell_1+2\bar\ell_2-\frac{9}{10}\bar\ell_3 +\frac{21}{8}\Big]+\cO(x^2)
\bigg\}
\;,
\label{eq:pipi_ell0_I0_x}
\\
a_0^2\Mpi&=&-\frac{ M^2}{16\pi F^2}
\bigg\{
1-\frac{ M^2}{12\pi^2 F^2}\Big[ \bar\ell_1+2\bar\ell_2                        +\frac{ 3}{8}\Big]+\cO(x^2)
\bigg\}
\;,
\label{eq:pipi_ell0_I2_x}
\eea
where $x\equiv M^2/(4\pi F)^2$ with $M^2=(m_u+m_d) \Sigma /F^2$ is one possible expansion parameter of {\Ch}PT.
Throughout this report we deviate from the {\Ch}PT habit of absorbing a factor $-\Mpi$ into the scattering length (relative to the convention used in quantum mechanics);
we include just a minus sign but not the factor $\Mpi$.
Hence, our $a_\ell^I$ have the dimension of a length so that all quark- or pion-mass dependence is explicit (as is most convenient for the lattice community).
But the sign convention is the one of the chiral community (where $a_\ell^I\Mpi>0$ means attraction and $a_\ell^I\Mpi<0$ indicates repulsion).

An important difference between the two $S$-wave scattering lengths is evident already at tree-level.
The isospin-0 scattering length (\ref{eq:pipi_ell0_I0_x}) is large and positive at this order, while the isospin-2 counterpart (\ref{eq:pipi_ell0_I2_x}) is by a factor $\sim3.5$ smaller (in absolute magnitude) and negative.
Hence, in the channel with $I=0$ the interaction is \emph{attractive}, while in the channel with $I=2$ the interaction is \emph{repulsive} and significantly weaker.
In this convention, experimental results, evaluated with the unitarity constraint germane to any local quantum field theory, read $a_0^0\Mpi=0.2198(46)_\mr{stat}(16)_\mr{syst}(64)_\mr{theo}$
and $a_0^2\Mpi=-0.0445(11)_\mr{stat}(4)_\mr{syst}(8)_\mr{theo}$ \cite{Roy:1971tc,Ananthanarayan:2000ht,Colangelo:2001df,Caprini:2011ky}.
The ratio between the two (absolute) central values is about $4.9$, i.e., a bit larger than $3.5$.
This, in turn, suggests that NLO contributions to $a_0^0$ and $a_0^2$ are sizeable, but the expansion seems well behaved.

Equations~(\ref{eq:pipi_ell0_I0_x}, \ref{eq:pipi_ell0_I2_x}) may be recast in the $\xi$-expansion, with $\xi\equiv\Mpi^2/(4\pi\Fpi)^2$, as
\bea
a_0^0\Mpi&=&+\frac{7\Mpi^2}{32\pi\Fpi^2}
\bigg\{
1 +\xi\frac{1}{2}\bar\ell_3 +\xi2\bar\ell_4 +\xi\Big[ \frac{20}{21}\bar\ell_1+\frac{40}{21}\bar\ell_2-\frac{18}{21}\bar\ell_3 +\frac{ 5}{ 2}\Big] +\cO(\xi^2)
\bigg\}
\;,
\\
a_0^2\Mpi&=&-\frac{ \Mpi^2}{16\pi\Fpi^2}
\bigg\{
1 +\xi\frac{1}{2}\bar\ell_3 +\xi2\bar\ell_4 -\xi\Big[ \frac{ 4}{ 3}\bar\ell_1+\frac{ 8}{ 3}\bar\ell_2                         +\frac{ 1}{ 2}\Big] +\cO(\xi^2)
\bigg\}
\;,
\eea
where $M^2/(4\pi F)^2=\Mpi^2/(4\pi\Fpi)^2\{1+\frac{1}{2}\xi\bar\ell_3+2\xi\bar\ell_4+\cO(\xi^2)\}$ has been used.
Finally, this expression can be summarized as
\bea
a_0^0\Mpi&=&+\frac{7\Mpi^2}{32\pi\Fpi^2}
\bigg\{
1+\frac{9\Mpi^2}{32\pi^2\Fpi^2}\ln\frac{(\lambda_0^0)^2}{\Mpi^2}+\cO(\xi^2)
\bigg\}
\;,
\label{eq:pipi_ell0_I0_xi}
\\
a_0^2\Mpi&=&-\frac{ \Mpi^2}{16\pi\Fpi^2}
\bigg\{
1-\frac{3\Mpi^2}{32\pi^2\Fpi^2}\ln\frac{(\lambda_0^2)^2}{\Mpi^2}+\cO(\xi^2)
\bigg\}
\;,
\label{eq:pipi_ell0_I2_xi}
\eea
with the abbreviations
\bea
\frac{9}{2}\ln\frac{(\lambda_0^0)^2}{M_{\pi,\mr{phys}}^2}&=&
\frac{20}{21}\bar\ell_1 +\frac{40}{21}\bar\ell_2 -\frac{5}{14}\bar\ell_3 +2\bar\ell_4 +\frac{5}{2}
\;,
\label{eq:pipi_scale_00}
\\
\frac{3}{2}\ln\frac{(\lambda_0^2)^2}{M_{\pi,\mr{phys}}^2}&=&
\frac{ 4}{ 3}\bar\ell_1 +\frac{ 8}{ 3}\bar\ell_2 -\frac{1}{ 2}\bar\ell_3 -2\bar\ell_4 +\frac{1}{2}
\;,
\label{eq:pipi_scale_02}
\eea
where $\lambda_\ell^I$ with $\ell=0$ and $I=0,2$ are scales like the $\Lambda_i$ in $\bar\ell_i=\ln(\Lambda_i^2/M_{\pi,\mr{phys}}^2)$ for $i\in\{1,2,3,4\}$ (albeit they are not independent from the latter).
Here, we made use of the fact that $\Mpi^2/M_{\pi,\mr{phys}}^2=1+\cO(\xi)$ and thus $\xi\ln(\Mpi^2/M_{\pi,\mr{phys}}^2)=\cO(\xi^2)$.
In the absence of any knowledge on the $\bar\ell_i$, one would assume $\lambda_0^0\simeq\lambda_0^2$, and with this input Eqs.~(\ref{eq:pipi_ell0_I0_xi}, \ref{eq:pipi_ell0_I2_xi}) suggest that the NLO contribution to $|a_0^0|$ is by a factor $\sim10.5$ larger than the NLO contribution to $|a_0^2|$.
The experimental numbers quoted before clearly support this view.

Given that all of this sounds like a complete success story for the determination of the scattering lengths $a_0^0$ and $a_0^2$, one may wonder whether lattice QCD is helpful at all.
It is, because the ``experimental'' evaluation of these scattering lengths builds on a constraint between these two quantities that, in turn, is based on a (rather nontrivial) dispersive evaluation of scattering phase shifts \cite{Roy:1971tc,Ananthanarayan:2000ht,Colangelo:2001df,Caprini:2011ky}.
Hence, to overcome this possible loophole, an independent lattice determination of $a_0^0$ and/or $a_0^2$ is highly welcome.

On the lattice $a_0^2$ is much easier to determine than $a_0^0$, since the former quantity does not involve quark-line disconnected contributions.
The main upshot (to be reviewed below) is that the lattice determination of $a_0^2\Mpi$ at the physical mass point is in perfect agreement with the experimental numbers quoted before, thus supporting the view that the scalar condensate is---at least in the $SU(2)$ case---the dominant order parameter, and the original estimate $\bar\ell_3=2.9\pm2.4$ is correct (see below).
Still, from a lattice perspective it is natural to see a determination of $a_0^0\Mpi$ and/or $a_0^2\Mpi$ as a means to access the specific linear combinations of $\bar\ell_i$ with $i\in\{1,2,3,4\}$ defined in Eqs.~(\ref{eq:pipi_scale_00}, \ref{eq:pipi_scale_02}).

In passing, we note that an alternative version of Eqs.~(\ref{eq:pipi_ell0_I0_xi}, \ref{eq:pipi_ell0_I2_xi}) is used in the literature, too.
For instance, Refs.~\cite{Beane:2007xs,Feng:2009ij,Fu:2013ffa,Helmes:2015gla,Liu:2016cba} give their results in the form
\bea
a_0^0\Mpi&=&+\frac{7\Mpi^2}{32\pi\Fpi^2}
\bigg\{
1+\frac{\Mpi^2}{32\pi^2\Fpi^2}\Big[\ell^{I=0}_{\pi\pi}+5-9\ln\frac{\Mpi^2}{2\Fpi^2}\Big]+\cO(\xi^2)
\bigg\}
\;,
\label{alt_pipi0}
\\
a_0^2\Mpi&=&-\frac{\Mpi^2}{16\pi\Fpi^2}
\bigg\{
1-\frac{\Mpi^2}{32\pi^2\Fpi^2}\Big[\ell^{I=2}_{\pi\pi}+1-3\ln\frac{\Mpi^2}{2\Fpi^2}\Big]+\cO(\xi^2)
\bigg\}
\;,
\label{alt_pipi2}
\eea
where the quantities (used to quote the results of the lattice calculation)
\bea
\ell^{I=0}_{\pi\pi} &=&
\frac{40}{21}\bar{\ell_1}+\frac{80}{21}\bar{\ell_2}-\frac{5}{7}\bar{\ell_3}+4\bar{\ell_4}+9\ln\frac{M_{\pi,\mr{phys}}^2}{2F^2_{\pi,\mr{phys}}}
\;,
\label{def_lpipi0}
\\
\ell^{I=2}_{\pi\pi} &=&
\frac{ 8}{ 3}\bar{\ell_1}+\frac{16}{ 3}\bar{\ell_2}-           \bar{\ell_3}-4\bar{\ell_4}+3\ln\frac{M_{\pi,\mr{phys}}^2}{2F^2_{\pi,\mr{phys}}}
\;,
\label{def_lpipi2}
\eea
amount to linear combinations of the $\ell_i^\mr{ren}(\mu^\mr{ren})$ that, due to the explicit logarithms in Eqs.~(\ref{def_lpipi0}, \ref{def_lpipi2}),
are effectively renormalized at the scale $\mu_\mr{ren}=f_\pi^\mr{phys}=\sqrt{2}\Fpi^\mr{phys}=130.41(20)\MeV$ \cite{Agashe:2014kda}.
Note that in these equations the dependence on the \emph{physical} pion mass in the logarithms cancels the one that comes from the $\bar\ell_i$,
so that the right-hand-sides bear no knowledge of $\Mpi^\mr{phys}$.
This alternative form is slightly different from Eqs.~(\ref{eq:pipi_ell0_I0_xi}, \ref{eq:pipi_ell0_I2_xi}).
Exact equality would be reached upon substituting $\Fpi^2 \rightarrow F_{\pi,\mr{phys}}^2$ in the logarithms of Eqs.~(\ref{alt_pipi0}, \ref{alt_pipi2}).
Upon expanding $\Fpi^2/F_{\pi,\mr{phys}}^2$ and subsequently the logarithm, one realizes that this difference amounts to a term $\cO(\xi)$ within the square bracket.
It thus makes up for a difference at the NNLO, which is beyond the scope of these formulae.

We close by mentioning a few works that elaborate on specific issues in $\pi\pi$ scattering relevant to the lattice.
Reference~\cite{Chen:2005ab} does mixed action {\Ch}PT for 2 and 2+1 flavours of staggered sea quarks and Ginsparg-Wilson valence quarks,
Refs.~\cite{Buchoff:2008ve,Aoki:2008gy} work out scattering formulae in Wilson fermion {\Ch}PT, and
Ref.~\cite{Acharya:2017zje} lists connected and disconnected contractions in $\pi\pi$ scattering.


\subsubsection{$\pi K$ and $KK$ scattering}

The discussion of $\pi\pi$ scattering in the previous subsection carries over, without material changes, to the case of $\pi K$ and $KK$ scattering.
The one (tiny) difference is that results, if contact with {\Ch}PT is desired, must be matched against the $SU(3)$ version of this framework.
In other words, for $\pi\pi$ scattering there is a choice between $SU(2)$ and $SU(3)$, while for $\pi K$ and $KK$ scattering matching to the $SU(3)$ version of {\Ch}PT is mandatory%
\footnote{Note that this could be circumvented if one used a heavy-meson extended version of {\Ch}PT, in particular $SU(2)$ {\Ch}PT with an extra (heavy) strange quark
\cite{Burdman:1992gh,Wise:1992hn,Yan:1992gz}. However, we have the original Gasser-Leutwyler versions of $SU(2)$ and $SU(3)$ {\Ch}PT in mind.}.

For completeness we also include, below, the $SU(3)$ {\Ch}PT result for $I=2$ $\pi\pi$ scattering.
Since, as in the FLAG 19 review, we tabulate the $S$-wave scattering length with combined isospin $I$ in the dimensionless variable $a^I_0 M_\pi$,
where the physical pion mass is meant, the result can be converted into specific linear combinations of NLO {\Ch}PT coefficients in either the $SU(2)$ or $SU(3)$ {\Ch}PT framework.
In this conversion, an extra piece to the systematic error is to be included, to account for higher-order terms in the chiral expansion.

Below, we continue this tradition by summarizing results in the dimensionless variable $a_0^I\mu_{\pi K}$ for $\pi K$ scattering and $a_0^I\Mka$ for $KK$ scattering.
Throughout this report, $\mu_{\pi K}\equiv\Mpi\Mka/(\Mpi+\Mka)$ is the reduced mass of the kaon-pion system at the physical mass point.
Again, these results can be converted into linear combinations of the $L_i$, with proper adjustment of the systematic uncertainty, due to the chiral expansion.
In doing so, one should keep in mind that the $SU(3)$ framework does not converge as swiftly as the $SU(2)$ frameork, since $m_{ud}\ll m_s$.

We basically follow Ref.~\cite{Sasaki:2013vxa}, but we adopt, for masses and decay constants, the conventions of the LEC section in the FLAG 19 report.
We consider the {\Ch}PT formulae at ${\cal O}(p^4)$ in the chiral expansion, as given in Refs.~\cite{Gasser:1984gg,Bernard:1990kw,Bernard:1990kx,Kubis:2001bx,Chen:2006wf,Beane:2006gj}.
The scattering lengths of the $\pi\pi(I=2)$, $KK(I=1)$, $\pi K(I=\frac{3}{2})$ and $\pi K(I=\frac{1}{2})$ systems can be written as
\begin{eqnarray}
a_{0,\pi\pi}^2\Mpi
\!&\!=\!&\!
\frac{\Mpi^2}{16\pi\Fpi^2}
\bigg\{
-1+\frac{16}{\Fpi^2}
\Big[
\Mpi^2 L_\mr{scat}(\mu)
-\frac{\Mpi^2}{2}L_5(\mu)
+\chi_{\pi\pi}^2(\mu)
\Big]
\bigg\}
\;,
\label{eqn:SU3_chpt_a2_pipi}
\\
a_{0,KK}^1\Mka
\!&\!=\!&\!
\frac{\Mka^2}{16\pi\Fka^2}
\bigg\{
-1+\frac{16}{\Fka^2}
\Big[
\Mka^2 L_\mr{scat}(\mu)
-\frac{\Mka^2}{2}L_5(\mu)
+\chi_{KK}^1(\mu)
\Big]
\bigg\}
\;,
\label{eqn:SU3_chpt_a_KK}
\\
a_{0,\pi K}^{3/2}\mu_{\pi K}
\!&\!=\!&\!
\frac{\mu_{\pi K}^2}{8\pi\Fpi\Fka}
\bigg\{
-1+\frac{16}{\Fpi\Fka}
\Big[
\Mpi\Mka L_\mr{scat}(\mu)
-\frac{\Mpi^2+\Mka^2}{4}L_5(\mu)
+\chi_{\pi K}^{3/2}(\mu)
\Big]
\bigg\}
\;,
\label{eqn:SU3_chpt_a32_piK}
\\
a_{0,\pi K}^{1/2} \mu_{\pi K}
\!&\!=\!&\!
\frac{\mu_{\pi K}^2}{8\pi \Fpi\Fka}
\bigg\{
2+\frac{16}{\Fpi\Fka}
\Big[
\Mpi\Mka L_\mr{scat}(\mu)
+2\frac{\Mpi^2+\Mka^2}{4}L_5(\mu)
+\chi_{\pi K}^{1/2}(\mu)
\Big]
\bigg\}
\;.
\label{eqn:SU3_chpt_a12_piK}
\end{eqnarray}
These formulae are written in terms of ${\cal O}(p^4)$ values of the masses and decay constants ($\Mpi$, $\Mka$, $\Fpi$ and $\Fka$) of the Nambu-Goldstone bosons (which, in turn, depend on the quark masses).
We recall that the ``Bernese'' normalization for the pion decay constant at the physical point is adopted (cf.\ footnote\,\ref{foot:fpi}).
The constants $L_5(\mu)$ and
\begin{equation}
L_\mr{scat}(\mu)= 2L_1(\mu)+2L_2(\mu)+L_3(\mu)-2L_4(\mu)-\frac{1}{2}L_5(\mu)+2L_6(\mu)+L_8(\mu)
\label{def:L_scat}
\end{equation}
are the $SU(3)$ low-energy constants (LECs) at the renormalization scale $\mu$.
The objects $\chi_{PQ}^{(I)}(\mu)$ are known functions with chiral logarithmic terms and dependence on the scale $\mu$.
In terms of these objects the functions $\chi_{PQ}^I(\mu)$ in Eqs.~(\ref{eqn:SU3_chpt_a2_pipi})-(\ref{eqn:SU3_chpt_a12_piK}) read%
\footnote{\label{foot:buggy}
There is a typo in the original version of Ref.~\cite{Sasaki:2013vxa} which made us mistakenly give the last term in the square bracket of Eq.~(\ref{buggyformula}) as $\frac{10\Mka^2}{9}$ in the arXiv:2111.09849\_v1 version of this report.
The correct expression with the last term $\frac{7\Mka^2}{9}$ agrees with Eq.~(32) in \cite{Chen:2006wf} which, to the best of our knowledge, is the earliest reference for this quantity.
Moreover, in the $SU(3)$ limit $(16\pi)^2\chi_{\pi\pi}^2(\mu)\to-\frac{14}{9}\Mpi^2\log(\frac{\Mpi^2}{\mu^2})+\frac{4}{9}\Mpi^2$, while the Gell-Mann-Oakes-Renner relation and the substitution $\Mka^2=\Mpi^2+\ep$ yield
$(16\pi)^2\chi_{KK}^1(\mu)\to\frac{\Mpi^2(\Mpi^2+\ep)}{4\ep}\log(\frac{\Mpi^2}{\mu^2})-(\Mpi^2+\ep)\log(\frac{\Mpi^2+\ep}{\mu^2})+\frac{(\Mpi^2+\ep)(-20\Mpi^2-20\ep+11\Mpi^2)}{36\ep}\log(\frac{\Mpi^2+4\ep/3}{\mu^2})+\frac{7}{9}(\Mpi^2+\ep)$.
In this expression the terms $O(\ep^{-1})$ cancel, and with $\log(\frac{\Mpi^2+4\ep/3}{\mu^2})=\log(\frac{\Mpi^2}{\mu^2})+\frac{4\ep}{3\Mpi^2}$ one obtains $(16\pi)^2\chi_{KK}^1(\mu)\to-\frac{14}{9}\Mpi^2\log(\frac{\Mpi^2}{\mu^2})+\frac{4}{9}\Mpi^2$
in the limit $\ep\to0$.
Hence $\chi_{\pi\pi}^2(\mu)=\chi_{KK}^1(\mu)$ in the $SU(3)$ limit.
We are indebted to Andr\'e Walker-Loud and Kiyoshi Sasaki for pointing this out to us and for clarifying details, respectively.}
\begin{eqnarray}
\chi_{\pi\pi}^2(\mu)\hspace{3mm}
&=&
\frac{1}{(16\pi)^2}
\Big[
-\frac{3\Mpi^2}{ 2}\log(\frac{\Mpi^2}{\mu^2})
-\frac{ \Mpi^2}{18}\log(\frac{\Met^2}{\mu^2})
+\frac{4\Mpi^2}{ 9}
\Big]
\;,
\\
\chi_{KK}^1(\mu)\hspace{1.5mm}
&=&
\frac{1}{(16\pi)^2}
\Big[
\frac{ \Mpi^2 \Mka^2 }{ 4 ( \Mka^2 - \Mpi^2 )}
\log(\frac{\Mpi^2}{\mu^2})
-\ \Mka^2
\log(\frac{\Mka^2}{\mu^2})
\nonumber  \\
& &
+\ \frac{ - 20 \Mka^4 + 11 \Mpi^2 \Mka^2 }{ 36( \Mka^2 - \Mpi^2 )}
\log(\frac{\Met^2}{\mu^2})
+\ \frac{7 \Mka^2}{9} 
\Big]
\label{buggyformula}
\;,
\\
\chi_{\pi K}^{3/2}(\mu)
&=&
\frac{1}{(16\pi)^2}
\Big[
\frac{ 22 \Mpi^3 \Mka + 11 \Mpi^2 \Mka^2 - 5 \Mpi^4}
{ 8 ( \Mka^2 - \Mpi^2 )}
\log(\frac{\Mpi^2}{\mu^2})
\nonumber  \\
& &
+\ \frac{ 9 \Mka^4 - 134 \Mpi \Mka^3 + 16 \Mpi^3 \Mka - 55 \Mpi^2 \Mka^2 }{ 36( \Mka^2 - \Mpi^2 )}
\log(\frac{\Mka^2}{\mu^2})
\nonumber  \\
& &
+\ \frac{ 36 \Mka^4 + 48 \Mpi \Mka^3 - 10 \Mpi^3 \Mka + 11 \Mpi^2 \Mka^2 - 9 \Mpi^4 }{ 72( \Mka^2 - \Mpi^2 )}
\log(\frac{\Met^2}{\mu^2})
\nonumber  \\
& &
+\ \frac{43 \Mpi \Mka}{9}
-\ \frac{ 8 \Mpi \Mka}{9} t_1( \Mpi, \Mka )
\Big]
\;,
\\
\chi_{\pi K}^{1/2}(\mu)
&=&
\frac{1}{(16\pi)^2}
\Big[
\frac{ 11 \Mpi^3 \Mka - 11 \Mpi^2 \Mka^2 + 5 \Mpi^4}{ 4 ( \Mka^2 - \Mpi^2 )}
\log(\frac{\Mpi^2}{\mu^2})
\nonumber  \\
& &
+\ \frac{ - 9 \Mka^4 - 67 \Mpi \Mka^3 + 8 \Mpi^3 \Mka + 55 \Mpi^2 \Mka^2 }{ 18( \Mka^2 - \Mpi^2 )}
\log(\frac{\Mka^2}{\mu^2})
\nonumber  \\
& &
+\ \frac{ - 36 \Mka^4 + 24 \Mka^3 \Mpi - 5 \Mka \Mpi^3- 11 \Mka^2 \Mpi^2 + 9 \Mpi^4 }{ 36( \Mka^2 - \Mpi^2 )}
\log(\frac{\Met^2}{\mu^2})
\nonumber  \\
& &
+\ \frac{43 \Mpi \Mka}{9}
+\ \frac{ 4 \Mpi \Mka}{9}\; t_1( \Mpi, \Mka )
-\ \frac{12 \Mpi \Mka}{9}\; t_2( \Mpi, \Mka )
\Big]
\;,
\nonumber\\
\end{eqnarray}
where $t_1( \Mpi, \Mka )$, $t_2( \Mpi, \Mka )$ can be written as
\begin{eqnarray}
\!t_1( \Mpi, \Mka )
\!&\!=\!&\!
\frac{ \sqrt{( \Mka + \Mpi )( 2 \Mka - \Mpi )} }{ \Mka - \Mpi }
\arctan
\bigg(
\frac{ 2( \Mka - \Mpi )}{ \Mka + 2 \Mpi }
\sqrt{\frac{ \Mka + \Mpi }{ 2 \Mka - \Mpi }}
\bigg)
\;,
\nonumber\\ \\
\!t_2( \Mpi, \Mka )
\!&\!=\!&\!
\frac{ \sqrt{( \Mka - \Mpi )( 2 \Mka + \Mpi )} }{ \Mka + \Mpi }
\arctan
\bigg(
\frac{ 2( \Mka + \Mpi )}{ \Mka - 2 \Mpi }
\sqrt{\frac{ \Mka - \Mpi }{ 2 \Mka + \Mpi }}
\bigg)
\;.
\nonumber\\
\end{eqnarray}
In short, these formulae show that --~in the $SU(3)$ framework~-- the four scattering lengths $a_0^{1}\Mpi$, $a_0^{2}\Mka$, $a_0^{3/2}\mu_{\pi K}$, $a_0^{1/2}\mu_{\pi K}$
determine three linear combinations of $L_5(\mu)$ and $L_\mr{scat}(\mu)$.
Recall that Eq.~(\ref{def:L_scat}) shows that the latter object is itself a linear combination of the $L_i(\mu)$.
Interestingly, $\pi\pi$ and $KK$ scattering determine the same linear combination $L_\mr{scat}(\mu)-\frac{1}{2}L_5(\mu)$,
while $a_0^{3/2}\mu_{\pi K}$ and $a_0^{1/2}\mu_{\pi K}$ determine two more ($m_s/m_{ud}$-dependent) linear combinations.
In the last few lines, we established the habit of omitting the particle subscript in $a_{0,\pi K}^I$ and $a_{0,KK}^I$,
since the value of $I$ together with the factor $\Mpi$, $\mu_{\pi K}$ or $\Mka$ already tells the particles involved in the scattering process.
The remaining zero subscript is meant to indicate the $S$-wave component.


\subsection{Extraction of SU(2) low-energy constants \label{sec:SU2results}}


\subsubsection{New results for individual LO SU(2) LECs \label{sec:SU2_LO}}

We are aware of four new papers with results on individual $SU(2)$ LECs plus an additional one which we overlooked in FLAG 19 \cite{Aoki:2019cca}.
They all give results on the LO LECs, $B$ and/or $F$, where $B$ is frequently traded for the condensate $\Sigma\equiv BF^2$ (both $B$ and $\Sigma$ are renormalized at the scale $\mu=2\GeV$).
We start by briefly mentioning their details.

The paper ETM 20A~\cite{Fischer:2020fvl} 
presents an $\Nf=2$ calculation with twisted mass fermions, using three pion masses down to the physical value
at a single lattice spacing $a=0.0914(15)\,\fm$.
They report a value of $F$ as given in Tab.~\ref{tab:f} and a value of $\bar\ell_4$ discussed in Sec.~\ref{sec:SU2_NLO} below.
The publication status changed from ``preprint'' to ``accepted'' after our closing date (as did the quoted uncertainty).
In practical terms this change is insignificant, since the quoted number (due to a red tag) would not contribute to the $\Nf=2$ average.

The paper $\chi$QCD 21~\cite{Liang:2021pql} 
employs $\Nf=2+1$ QCD with domain wall fermions and RI/MOM renormalization.
They have two ensembles with physical pion mass ($139\MeV$) at lattice spacings $a=0.114\fm$ and $a=0.084\fm$, one ensemble with $\Mpi=234\MeV$ at $a=0.071\fm$,
and one with $\Mpi=371\MeV$ at $a=0.063\fm$ that is only used to test the lattice spacing dependence of the scalar renormalization factor.
They report the value of $\Sigma^{1/3}$ as listed in Tab.~\ref{tab:sigma}.

The paper ETM 21~\cite{Alexandrou:2021bfr} 
uses $\Nf=2+1+1$ flavours of twisted mass fermions,
ten ensembles, three lattice spacings ($a=0.092,0.080,0.068\fm$), up to four pion masses $\Mpi\in[135\MeV,346\MeV]$,
up to two volumes, and $L(M_{\pi,\mr{min}})=5.55\,\fm$.
The scale is set by $f_\pi^\mr{phys}=\sqrt{2}\Fpi^\mr{phys}=130.4(2)\MeV$ \cite{Agashe:2014kda}.
They analyze the quark mass dependence of both $\Fpi$ and the (chiral and finite-volume) log-free quantity $X_\pi=(\Fpi\Mpi^4)^{1/5}$ \cite{Durr:2014oba},
to determine $F$ and $\bar\ell_4$ in two different ways.
The two fitting procedures yield nearly identical results for $F$.
The two central values agree exactly, as do the two systematic uncertainties; only the combined statistical plus fitting uncertainty differs a bit among the two approaches.
Since the paper does not give preference to one of the fitting procedures, we take the liberty to condense them,
assuming 100\% correlation, into the single result $F=87.7(6)(5)\MeV$ as listed in Tab.~\ref{tab:f}.
They also report a value of $\bar\ell_4$ to be mentioned in Sec.~\ref{sec:SU2_NLO} below.

The paper ETM 21A~\cite{Alexandrou:2021gqw} 
is again based on $\Nf=2+1+1$ flavours of twisted mass fermions, ten ensembles, three lattice spacings,
$a=0.095,0.082,0.069\fm$, up to four pion masses $\Mpi\in[134\MeV,346\MeV]$,
up to two volumes, and $L(M_{\pi,\mr{min}})=5.52\,\fm$.
The scale is set by $f_\pi^\mr{phys}=\sqrt{2}\Fpi^\mr{phys}=130.4(2)\MeV$ \cite{Agashe:2014kda}, and cross-checked with the nucleon mass.
From the analysis of the pion sector they determine values of $F$ and $\Sigma^{1/3}$ as listed in Tab.~\ref{tab:f} and Tab.~\ref{tab:sigma}, respectively.

Finally, we should mention Ref.~\cite{Wang:2016lsv} which, regrettably, escaped our attention when preparing the last FLAG report \cite{Aoki:2019cca}.
The authors extract the quark condensate from an OPE analysis of the Landau-gauge quark propagator.
They use overlap valence quarks on three ensembles with (2+1)-flavor domain-wall fermions with $a^{-1}=1.75\GeV$ and sea pion masses of 331, 419 and 557\,MeV from the RBC/UKQCD collaboration.
Their eight valence pion masses range from 220 to 600\,MeV.
Their result for $\Sigma^{1/3}$ is listed in Tab.~\ref{tab:sigma}.
With only a single lattice spacing, their result does not contribute to the FLAG average.


\begin{table}[!tbp] 
\vspace*{3cm}
\centering
\footnotesize
\begin{tabular*}{\textwidth}[l]{l@{\extracolsep{\fill}}rlllllll}
Collaboration & Ref. & $\Nf$ &
\hspace{0.15cm}\begin{rotate}{60}{publication status}\end{rotate}\hspace{-0.15cm} &
\hspace{0.15cm}\begin{rotate}{60}{chiral extrapolation}\end{rotate}\hspace{-0.15cm}&
\hspace{0.15cm}\begin{rotate}{60}{cont.\ extrapolation}\end{rotate}\hspace{-0.15cm} &
\hspace{0.15cm}\begin{rotate}{60}{finite volume}\end{rotate}\hspace{-0.15cm} &
\hspace{0.15cm}\begin{rotate}{60}{renormalization}\end{rotate}\hspace{-0.15cm} & \rule{0.4cm}{0cm}$\Sigma^{1/3}$ \\[2mm]
\hline
\hline
\\[-2mm]
ETM 21A                 & \cite{Alexandrou:2021gqw}  &2+1+1& \pubP & \good & \soso & \good & \good & 267.6(1.8)(1.1)                      \\
ETM~17E                 & \cite{Alexandrou:2017bzk}  &2+1+1& \pubA & \soso & \good & \soso & \good & 318(21)(21)                          \\
ETM~13                  & \cite{Cichy:2013gja}       &2+1+1& \pubA & \soso & \good & \good & \good & 280(8)(15)                           \\[2mm]
\hline
\\[-2mm]
$\chi$QCD 21            & \cite{Liang:2021pql}       & 2+1 & \pubP & \good & \good & \good & \good & 260.3(0.7)(1.7)                      \\
JLQCD~17A               & \cite{Aoki:2017paw}        & 2+1 & \pubA & \soso & \good & \good & \good & 274(13)(29)                          \\
Wang~16                 & \cite{Wang:2016lsv}        & 2+1 & \pubA & \soso & \bad  & \bad  & \good & 305(15)(21)                          \\
JLQCD~16B               & \cite{Cossu:2016eqs}       & 2+1 & \pubA & \soso & \good & \good & \good & 270.0(1.3)(4.8)                      \\
RBC/UKQCD~15E           & \cite{Boyle:2015exm}       & 2+1 & \pubA & \good & \good & \good & \good & 274.2(2.8)(4.0)                      \\
RBC/UKQCD~14B           & \cite{Blum:2014tka}        & 2+1 & \pubA & \good & \good & \good & \good & 275.9(1.9)(1.0)                      \\
BMW~13                  & \cite{Durr:2013goa}        & 2+1 & \pubA & \good & \good & \good & \good & 271(4)(1)                            \\
Borsanyi~12             & \cite{Borsanyi:2012zv}     & 2+1 & \pubA & \soso & \soso & \good & \good & 272.3(1.2)(1.4)                      \\
JLQCD/TWQCD~10A         & \cite{Fukaya:2010na}       & 2+1 & \pubA & \good & \bad  & \bad  & \good & 234(4)(17)                           \\
MILC~10A                & \cite{Bazavov:2010yq}      & 2+1 & \pubC & \soso & \good & \good & \soso & 281.5(3.4)$\binom{+2.0}{-5.9}$(4.0)  \\
RBC/UKQCD~10A           & \cite{Aoki:2010dy}         & 2+1 & \pubA & \soso & \soso & \bad  & \good & 256(5)(2)(2)                         \\
JLQCD~09                & \cite{Fukaya:2009fh}       & 2+1 & \pubA & \good & \bad  & \bad  & \good & 242(4)$\binom{+19}{-18}$             \\
MILC~09A, $SU(3)$-fit   & \cite{Bazavov:2009fk}      & 2+1 & \pubC & \soso & \good & \good & \soso & 279(1)(2)(4)                         \\
MILC~09A, $SU(2)$-fit   & \cite{Bazavov:2009fk}      & 2+1 & \pubC & \soso & \good & \good & \soso & 280(2)$\binom{+4}{-8}$(4)            \\
MILC~09                 & \cite{Bazavov:2009bb}      & 2+1 & \pubA & \soso & \good & \good & \soso & 278(1)$\binom{+2}{-3}$(5)            \\
TWQCD~08                & \cite{Chiu:2008jq}         & 2+1 & \pubA & \bad  & \bad  & \bad  & \good & 259(6)(9)                            \\
PACS-CS~08, $SU(3)$-fit & \cite{Aoki:2008sm}         & 2+1 & \pubA & \good & \bad  & \bad  & \bad  & 312(10)                              \\
PACS-CS~08, $SU(2)$-fit & \cite{Aoki:2008sm}         & 2+1 & \pubA & \good & \bad  & \bad  & \bad  & 309(7)                               \\
RBC/UKQCD~08            & \cite{Allton:2008pn}       & 2+1 & \pubA & \soso & \bad  & \soso & \good & 255(8)(8)(13)                        \\[2mm]
\hline
\\[-2mm]
Engel~14                & \cite{Engel:2014eea}       &  2  & \pubA & \good & \good & \good & \good & 263(3)(4)                            \\
Brandt~13               & \cite{Brandt:2013dua}      &  2  & \pubA & \soso & \good & \soso & \good & 261(13)(1)                           \\
ETM~13                  & \cite{Cichy:2013gja}       &  2  & \pubA & \soso & \good & \soso & \good & 283(7)(17)                           \\
ETM~12                  & \cite{Burger:2012ti}       &  2  & \pubA & \soso & \good & \soso & \good & 299(26)(29)                          \\
Bernardoni~11           & \cite{Bernardoni:2011kd}   &  2  & \pubC & \soso & \bad  & \bad  & \good & 306(11)                              \\
TWQCD~11                & \cite{Chiu:2011bm}         &  2  & \pubA & \soso & \bad  & \bad  & \good & 230(4)(6)                            \\
TWQCD~11A               & \cite{Chiu:2011dz}         &  2  & \pubA & \soso & \bad  & \bad  & \good & 259(6)(7)                            \\
JLQCD/TWQCD~10A         & \cite{Fukaya:2010na}       &  2  & \pubA & \good & \bad  & \bad  & \good & 242(5)(20)                           \\
Bernardoni~10           & \cite{Bernardoni:2010nf}   &  2  & \pubA & \soso & \bad  & \bad  & \good & 262$\binom{+33}{-34}\binom{+4}{-5}$  \\
ETM~09C                 & \cite{Baron:2009wt}        &  2  & \pubA & \soso & \good & \soso & \good & 270(5)$\binom{+3}{-4}$               \\
ETM~08                  & \cite{Frezzotti:2008dr}    &  2  & \pubA & \soso & \soso & \soso & \good & 264(3)(5)                            \\
CERN~08                 & \cite{Giusti:2008vb}       &  2  & \pubA & \soso & \bad  & \soso & \good & 276(3)(4)(5)                         \\
Hasenfratz~08           & \cite{Hasenfratz:2008ce}   &  2  & \pubA & \soso & \bad  & \soso & \good & 248(6)                               \\
JLQCD/TWQCD~08A         & \cite{Noaki:2008iy}        &  2  & \pubA & \soso & \bad  & \bad  & \good & 235.7(5.0)(2.0)$\binom{+12.7}{-0.0}$ \\
JLQCD/TWQCD~07          & \cite{Fukaya:2007pn}       &  2  & \pubA & \soso & \bad  & \bad  & \good & 239.8(4.0)                           \\
JLQCD/TWQCD~07A         & \cite{Aoki:2007pw}         &  2  & \pubA & \good & \bad  & \bad  & \good & 252(5)(10)                           \\[2mm]
\hline
\hline
\end{tabular*}
\normalsize
\vspace*{-2mm}
\caption{\label{tab:sigma}
Cubic root of the $SU(2)$ quark condensate $\Sigma\equiv-\lim_{m_u,m_d\to0}\langle\ubar u\rangle$ in $\MeV$ units, in the $\overline{\rm MS}$-scheme, at the renormalization scale $\mu=2\GeV$.
All ETM values that were available only in $r_0$ units were converted on the basis of $r_0=0.48(2)\,\fm$ \cite{Aoki:2009sc,Bazavov:2014pvz,Abdel-Rehim:2015pwa}, with this error being added in quadrature to any existing systematic error.}
\end{table}

\begin{table}[!tbp] 
\vspace*{3cm}
\centering
\footnotesize
\begin{tabular*}{\textwidth}[l]{l@{\extracolsep{\fill}}rlllllll}
Collaboration & Ref. & $\Nf$ &
\hspace{0.15cm}\begin{rotate}{60}{publication status}\end{rotate}\hspace{-0.15cm}&
\hspace{0.15cm}\begin{rotate}{60}{chiral extrapolation}\end{rotate}\hspace{-0.15cm}&
\hspace{0.15cm}\begin{rotate}{60}{cont.\ extrapolation}\end{rotate}\hspace{-0.15cm} &
\hspace{0.15cm}\begin{rotate}{60}{finite volume}\end{rotate}\hspace{-0.15cm} &
\rule{0.2cm}{0cm} $F$ &\rule{0.2cm}{0cm} $\Fpi/F$ \\[2mm]
\hline
\hline
\\[-2mm]
ETM 21A                 & \cite{Alexandrou:2021gqw}  &2+1+1& \pubP & \good & \soso & \good & 86.85(23)(46)                       & {\sl 1.062(3)(6)}                     \\
ETM 21                  & \cite{Alexandrou:2021bfr}  &2+1+1& \pubP & \good & \soso & \good & 87.7(6)(5)                          & {\sl 1.051(7)(6)}                     \\
ETM~11                  & \cite{Baron:2011sf}        &2+1+1& \pubC & \soso & \good & \soso & 85.60(4){\sl(13)}                   & {\sl 1.077(2)}{\sl(2)}                \\
ETM~10                  & \cite{Baron:2010bv}        &2+1+1& \pubA & \soso & \bad  & \good & 85.66(6)(13)                        & 1.076(2)(2)                           \\[2mm]
\hline
\\[-2mm]
RBC/UKQCD~15E           & \cite{Boyle:2015exm}       & 2+1 & \pubA & \good & \good & \good & 85.8(1.1)(1.5)                      & 1.0641(21)(49)                        \\
RBC/UKQCD~14B           & \cite{Blum:2014tka}        & 2+1 & \pubA & \good & \good & \good & 86.63(12)(13)                       & 1.0645(15)(0)                         \\
BMW~13                  & \cite{Durr:2013goa}        & 2+1 & \pubA & \good & \good & \good & 88.0(1.3)(0.3)                      & 1.055(7)(2)                           \\
Borsanyi~12             & \cite{Borsanyi:2012zv}     & 2+1 & \pubA & \soso & \soso & \good & 86.78(05)(25)                       & 1.0627(06)(27)                        \\
NPLQCD~11               & \cite{Beane:2011zm}        & 2+1 & \pubA & \soso & \soso & \soso & {\sl 86.8(2.1)$\binom{+3.3}{-3.4}$} & 1.062(26)$\binom{+42}{-40}$           \\
MILC~10                 & \cite{Bazavov:2010hj}      & 2+1 & \pubC & \soso & \good & \good & 87.0(4)(5)                          & {\sl 1.060(5)(6)}                     \\
MILC~10A                & \cite{Bazavov:2010yq}      & 2+1 & \pubC & \soso & \good & \good & 87.5(1.0)$\binom{+0.7}{-2.6}$       & {\sl 1.054(12)$\binom{+31}{-09}$}     \\
MILC~09A, $SU(3)$-fit   & \cite{Bazavov:2009fk}      & 2+1 & \pubC & \soso & \good & \good & 86.8(2)(4)                          & 1.062(1)(3)                           \\
MILC~09A, $SU(2)$-fit   & \cite{Bazavov:2009fk}      & 2+1 & \pubC & \soso & \good & \good & 87.4(0.6)$\binom{+0.9}{-1.0}$       & {\sl 1.054(7)$\binom{+12}{-11}$}      \\
MILC~09                 & \cite{Bazavov:2009bb}      & 2+1 & \pubA & \soso & \good & \good & {\sl 87.66(17)$\binom{+28}{-52}$}  & 1.052(2)$\binom{+6}{-3}$               \\
PACS-CS~08, $SU(3)$-fit & \cite{Aoki:2008sm}         & 2+1 & \pubA & \good & \bad  & \bad  & 90.3(3.6)                           & 1.062(8)                              \\
PACS-CS~08, $SU(2)$-fit & \cite{Aoki:2008sm}         & 2+1 & \pubA & \good & \bad  & \bad  & 89.4(3.3)                           & 1.060(7)                              \\
RBC/UKQCD~08            & \cite{Allton:2008pn}       & 2+1 & \pubA & \soso & \bad  & \soso & 81.2(2.9)(5.7)                      & 1.080(8)                              \\[2mm]
\hline
\\[-2mm]
ETM 20A                 & \cite{Fischer:2020fvl}     &  2  & \pubA & \good & \bad  & \soso & 86.46(0.06)(2.40)                   & {\sl 1.067(1)(30)}                    \\ 
ETM~15A                 & \cite{Abdel-Rehim:2015pwa} &  2  & \pubA & \good & \bad  & \soso & 86.3(2.8)                           & {\sl 1.069(35)}                       \\
Engel~14                & \cite{Engel:2014eea}       &  2  & \pubA & \good & \good & \good & 85.8(0.7)(2.0)                      & {\sl 1.075(09)(25)}                   \\
Brandt~13               & \cite{Brandt:2013dua}      &  2  & \pubA & \soso & \good & \soso & 84(8)(2)                            & 1.080(16)(6)                          \\
QCDSF~13                & \cite{Horsley:2013ayv}     &  2  & \pubA & \good & \soso & \soso & 86(1)                               & {\sl 1.07(1)}                         \\
TWQCD~11                & \cite{Chiu:2011bm}         &  2  & \pubA & \soso & \bad  & \bad  & 83.39(35)(38)                       & {\sl 1.106(5)(5)}                     \\
ETM~09C                 & \cite{Baron:2009wt}        &  2  & \pubA & \soso & \good & \soso & 85.91(07)$\binom{+78}{-07}$         & 1.0755(6)$\binom{+08}{-94}$           \\
ETM~08                  & \cite{Frezzotti:2008dr}    &  2  & \pubA & \soso & \soso & \soso & 86.6(7)(7)                          & 1.067(9)(9)                           \\
Hasenfratz~08           & \cite{Hasenfratz:2008ce}   &  2  & \pubA & \soso & \bad  & \soso & 90(4)                               & {\sl 1.02(5)}                         \\
JLQCD/TWQCD~08A         & \cite{Noaki:2008iy}        &  2  & \pubA & \soso & \bad  & \bad  & 79.0(2.5)(0.7)$\binom{+4.2}{-0.0}$  & {\sl 1.167(37)(10)$\binom{+02}{-62}$} \\
JLQCD/TWQCD~07          & \cite{Fukaya:2007pn}       &  2  & \pubA & \soso & \bad  & \bad  & 87.3(5.6)                           & {\sl 1.06(7)}                         \\[2mm]
\hline
\\[-2mm]
Colangelo~03            & \cite{Colangelo:2003hf}    &     &       &       &       &       & 86.2(5)                             & 1.0719(52)                            \\[2mm]
\hline
\hline
\end{tabular*}
\normalsize
\vspace*{-2mm}
\caption{\label{tab:f}
Results for the $SU(2)$ low-energy constant $F$ (in MeV) and for the ratio $\Fpi/F$.
All ETM values that were available only in $r_0$ units were converted on the basis of $r_0=0.48(2)\,\fm$ \cite{Aoki:2009sc,Bazavov:2014pvz,Abdel-Rehim:2015pwa}, with this error being added in quadrature to any existing systematic error.
Numbers in slanted fonts have been calculated by us, based on $\sqrt{2}\Fpi^\mr{phys}=130.41(20)\MeV$ \cite{Agashe:2014kda}, with this error being added in quadrature to any existing systematic error (otherwise to the statistical error).
The systematic error in ETM~11 has been carried over from ETM~10.}
\end{table}

\begin{figure}[!tb]
\centering
\includegraphics[width=13.0cm]{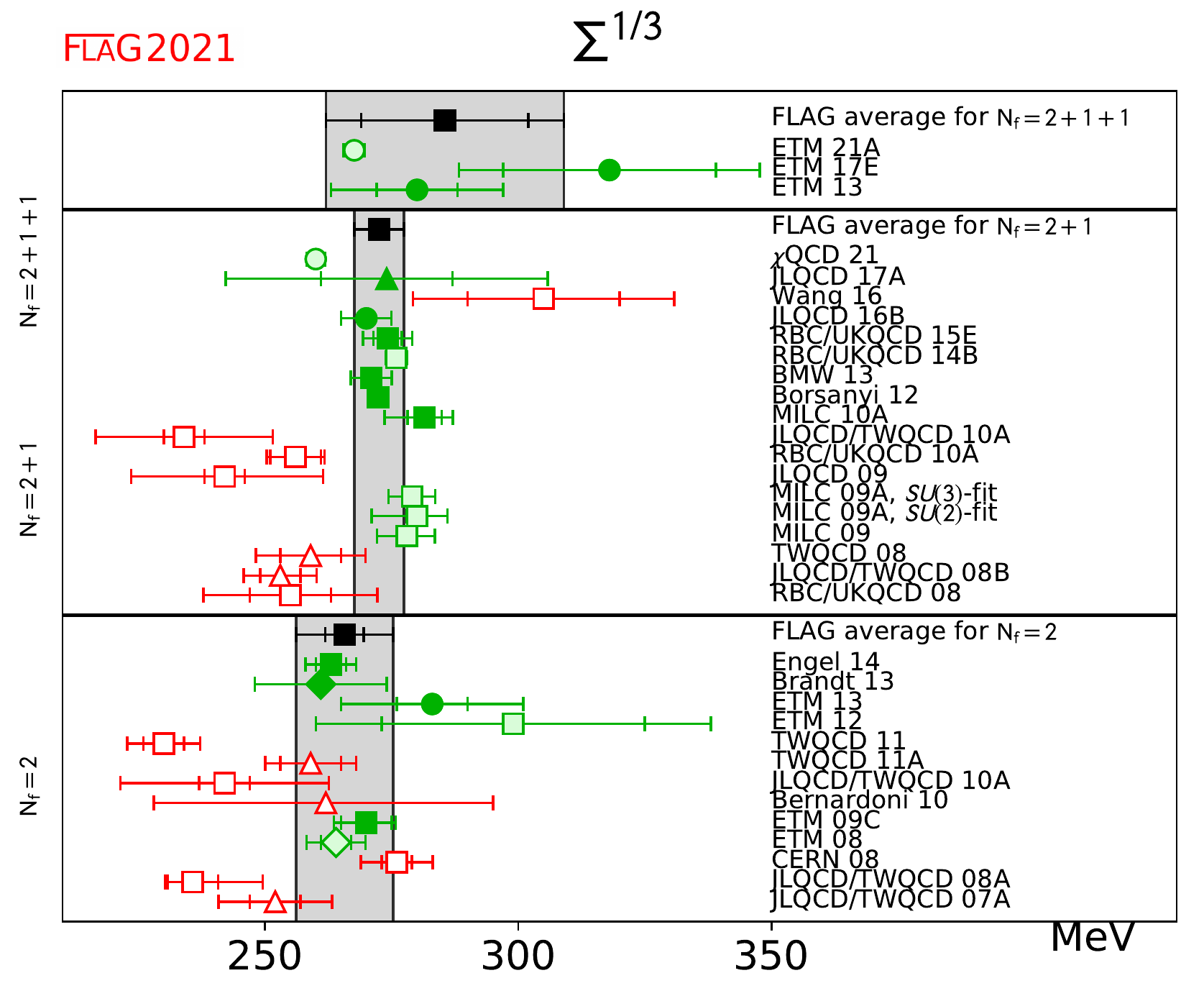}%
\vspace*{-2mm}
\caption{\label{fig:sigma}
Cubic root of the $SU(2)$ quark condensate $\Sigma\equiv-\lim_{m_u,m_d\to0}\langle\ubar u\rangle$ in the $\overline{\rm MS}$-scheme, at the renormalization scale $\mu=2\GeV$.
Square symbols indicate determinations from correlators in the $p$-regime, up triangles refer to extractions from the topological susceptibility,
diamonds to determinations from the pion form factor, and bullet points refer to the spectral density method.}
\end{figure}

\begin{figure}[!tb]
\centering
\includegraphics[width=13.0cm]{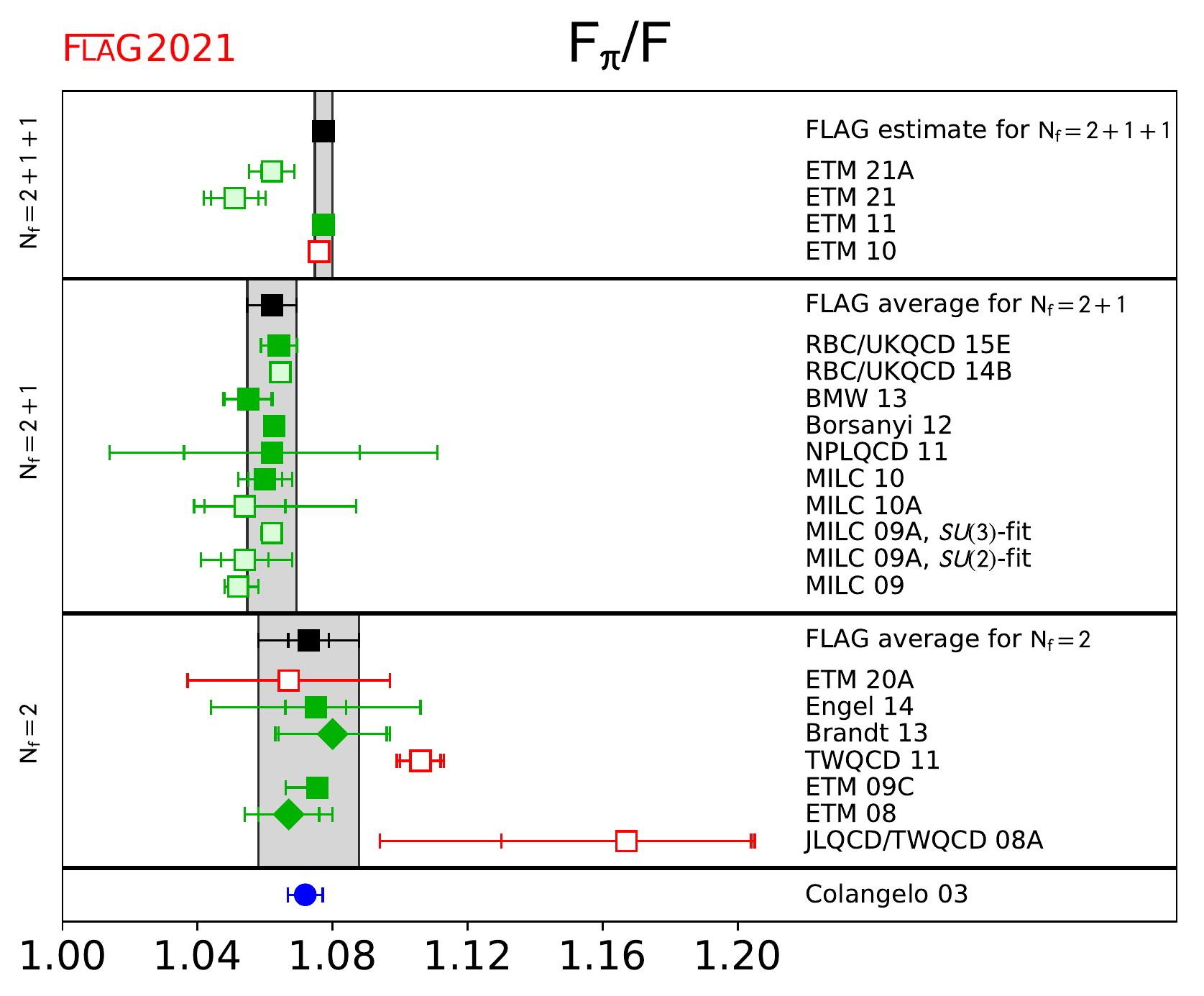}%
\vspace*{-2mm}
\caption{\label{fig:f}
Comparison of the results for the ratio of the physical pion decay constant $\Fpi$ and the leading-order $SU(2)$ low-energy constant $F$.
Square symbols indicate determinations from correlators in the $p$-regime, and diamonds from the pion form factor.}
\end{figure}

Perhaps it is worth comparing the results for $f\equiv\sqrt{2}F$ in Refs.~\cite{Alexandrou:2021bfr,Alexandrou:2021gqw}.
Carrying all errors along, one finds $\Delta\!f[\mr{MeV}]=124.0(0.9)(0.7)-122.82(32)(65)=1.18(1.35)$, which is less than one standard deviation.
Given that the two studies were carried out on largely the same ensemble basis, it is perhaps reasonable to assume the statistical error is $\sim100\%$ correlated.
In this case, the difference would be $\Delta\!f\,[\mr{MeV}]=124.0(0.7)-122.82(65)=1.18(0.96)$, which is $1.24\sigma$ and thus perfectly acceptable.
The chiral analysis in the two papers is treated somewhat differently, which would lead to differences in the neglected NNLO terms, and thus reflects a systematic effect.

The new results for $\Sigma^{1/3}$ and $F_{\pi}/F$, together with the previous ones, are shown in Fig.~\ref{fig:sigma} and Fig.~\ref{fig:f}, respectively.


\subsubsection{New results for individual NLO SU(2) LECs \label{sec:SU2_NLO}}

Two of the aforementioned papers contain new results on $\bar\ell_4$, i.e., a specific LEC at NLO of the $SU(2)$ framework.
ETM 20A~\cite{Fischer:2020fvl} quotes $\bar\ell_4=4.31(4)(2)(11)(5)$ for $\Nf=2$, while
ETM 21~\cite{Alexandrou:2021bfr} finds $\bar\ell_4=3.44(28)(36)$ for $\Nf=2+1+1$.
These results are listed in Tab.~\ref{tab:l3and4}.


\begin{table}[!tbp] 
\vspace*{3cm}
\centering
\footnotesize
\begin{tabular*}{\textwidth}[l]{l@{\extracolsep{\fill}}rlllllll}
Collaboration & Ref. & $\Nf$ &
\hspace{0.15cm}\begin{rotate}{60}{publication status}\end{rotate}\hspace{-0.15cm} &
\hspace{0.15cm}\begin{rotate}{60}{chiral extrapolation}\end{rotate}\hspace{-0.15cm}&
\hspace{0.15cm}\begin{rotate}{60}{cont.\ extrapolation}\end{rotate}\hspace{-0.15cm} &
\hspace{0.15cm}\begin{rotate}{60}{finite volume}\end{rotate}\hspace{-0.15cm} &\rule{0.3cm}{0cm} $\lbar_3$ & \rule{0.3cm}{0cm}$\lbar_4$ \\[2mm]
\hline
\hline
\\[-2mm]
ETM 21                  & \cite{Alexandrou:2021bfr}  &2+1+1& \pubP & \good & \soso & \good &                                & 3.44(28)(36)                   \\
ETM~11                  & \cite{Baron:2011sf}        &2+1+1& \pubC & \soso & \good & \soso & 3.53(5){\sl(26)}               & 4.73(2){\sl(10)}               \\
ETM~10                  & \cite{Baron:2010bv}        &2+1+1& \pubA & \soso & \bad  & \good & 3.70(7)(26)                    & 4.67(3)(10)                    \\[2mm]
\hline
\\[-2mm]
RBC/UKQCD~15E           & \cite{Boyle:2015exm}       & 2+1 & \pubA & \good & \good & \good & 2.81(19)(45)                   & 4.02(8)(24)                    \\
RBC/UKQCD~14B           & \cite{Blum:2014tka}        & 2+1 & \pubA & \good & \good & \good & 2.73(13)(0)                    & 4.113(59)(0)                   \\
BMW~13                  & \cite{Durr:2013goa}        & 2+1 & \pubA & \good & \good & \good & 2.5(5)(4)                      & 3.8(4)(2)                      \\
RBC/UKQCD~12            & \cite{Arthur:2012opa}      & 2+1 & \pubA & \good & \soso & \good & 2.91(23)(07)                   & 3.99(16)(09)                   \\
Borsanyi~12             & \cite{Borsanyi:2012zv}     & 2+1 & \pubA & \soso & \soso & \good & 3.16(10)(29)                   & 4.03(03)(16)                   \\
NPLQCD~11               & \cite{Beane:2011zm}        & 2+1 & \pubA & \soso & \soso & \soso & 4.04(40)$\binom{+73}{-55}$     & 4.30(51)$\binom{+84}{-60}$     \\
MILC~10                 & \cite{Bazavov:2010hj}      & 2+1 & \pubC & \soso & \good & \good & 3.18(50)(89)                   & 4.29(21)(82)                   \\
MILC~10A                & \cite{Bazavov:2010yq}      & 2+1 & \pubC & \soso & \good & \good & 2.85(81)$\binom{+37}{-92}$     & 3.98(32)$\binom{+51}{-28}$     \\
RBC/UKQCD~10A           & \cite{Aoki:2010dy}         & 2+1 & \pubA & \soso & \soso & \bad  & 2.57(18)                       & 3.83(9)                        \\
MILC~09A, $SU(3)$-fit   & \cite{Bazavov:2009fk}      & 2+1 & \pubC & \soso & \good & \good & 3.32(64)(45)                   & 4.03(16)(17)                   \\
MILC~09A, $SU(2)$-fit   & \cite{Bazavov:2009fk}      & 2+1 & \pubC & \soso & \good & \good & 3.0(6)$\binom{+9}{-6}$         & 3.9(2)(3)                      \\
PACS-CS~08, $SU(3)$-fit & \cite{Aoki:2008sm}         & 2+1 & \pubA & \good & \bad  & \bad  & 3.47(11)                       & 4.21(11)                       \\
PACS-CS~08, $SU(2)$-fit & \cite{Aoki:2008sm}         & 2+1 & \pubA & \good & \bad  & \bad  & 3.14(23)                       & 4.04(19)                       \\
RBC/UKQCD~08            & \cite{Allton:2008pn}       & 2+1 & \pubA & \soso & \bad  & \soso & 3.13(33)(24)                   & 4.43(14)(77)                   \\[2mm]
\hline
\\[-2mm]
ETM 20A                 & \cite{Fischer:2020fvl}     &  2  & \pubA & \good & \bad  & \soso &                                & 4.31(4)(2)(11)(5)              \\
ETM~15A                 & \cite{Abdel-Rehim:2015pwa} &  2  & \pubA & \good & \bad  & \soso &                                & 3.3(4)                         \\
G\"ulpers~15            & \cite{Gulpers:2015bba}     &  2  & \pubA & \good & \good & \good &                                & 4.54(30)(0)                    \\
G\"ulpers~13            & \cite{Gulpers:2013uca}     &  2  & \pubA & \soso & \bad  & \soso &                                & 4.76(13)                       \\
Brandt~13               & \cite{Brandt:2013dua}      &  2  & \pubA & \soso & \good & \soso & 3.0(7)(5)                      & 4.7(4)(1)                      \\
QCDSF~13                & \cite{Horsley:2013ayv}     &  2  & \pubA & \good & \soso & \soso &                                & 4.2(1)                         \\
Bernardoni~11           & \cite{Bernardoni:2011kd}   &  2  & \pubC & \soso & \bad  & \bad  & 4.46(30)(14)                   & 4.56(10)(4)                    \\
TWQCD~11                & \cite{Chiu:2011bm}         &  2  & \pubA & \soso & \bad  & \bad  & 4.149(35)(14)                  & 4.582(17)(20)                  \\
ETM~09C                 & \cite{Baron:2009wt}        &  2  & \pubA & \soso & \good & \soso & 3.50(9)$\binom{+09}{-30}$      & 4.66(4)$\binom{+04}{-33}$      \\
JLQCD/TWQCD~09          & \cite{JLQCD:2009qn}        &  2  & \pubA & \soso & \bad  & \bad  &                                & 4.09(50)(52)                   \\
ETM~08                  & \cite{Frezzotti:2008dr}    &  2  & \pubA & \soso & \soso & \soso & 3.2(8)(2)                      & 4.4(2)(1)                      \\
JLQCD/TWQCD~08A         & \cite{Noaki:2008iy}        &  2  & \pubA & \soso & \bad  & \bad  & 3.38(40)(24)$\binom{+31}{-00}$ & 4.12(35)(30)$\binom{+31}{-00}$ \\
CERN-TOV~06             & \cite{DelDebbio:2006cn}    &  2  & \pubA & \soso & \bad  & \bad  & 3.0(5)(1)                      &                                \\[2mm]
\hline
\\[-2mm]
Colangelo~01            & \cite{Colangelo:2001df}    &     &       &       &       &       &                                & 4.4(2)                         \\
Gasser~84               & \cite{Gasser:1983yg}       &     &       &       &       &       & 2.9(2.4)                       & 4.3(9)                         \\[2mm]
\hline
\hline
\end{tabular*}
\normalsize
\vspace*{-2mm}
\caption{\label{tab:l3and4}
Results for the $SU(2)$ NLO low-energy constants $\lbar_3$ and $\lbar_4$.
For comparison, the last two lines show results from phenomenological analyses.
The systematic error in ETM~11 has been carried over from ETM~10.}
\end{table}

\begin{figure}[!tb]
\centering
\includegraphics[width=13cm]{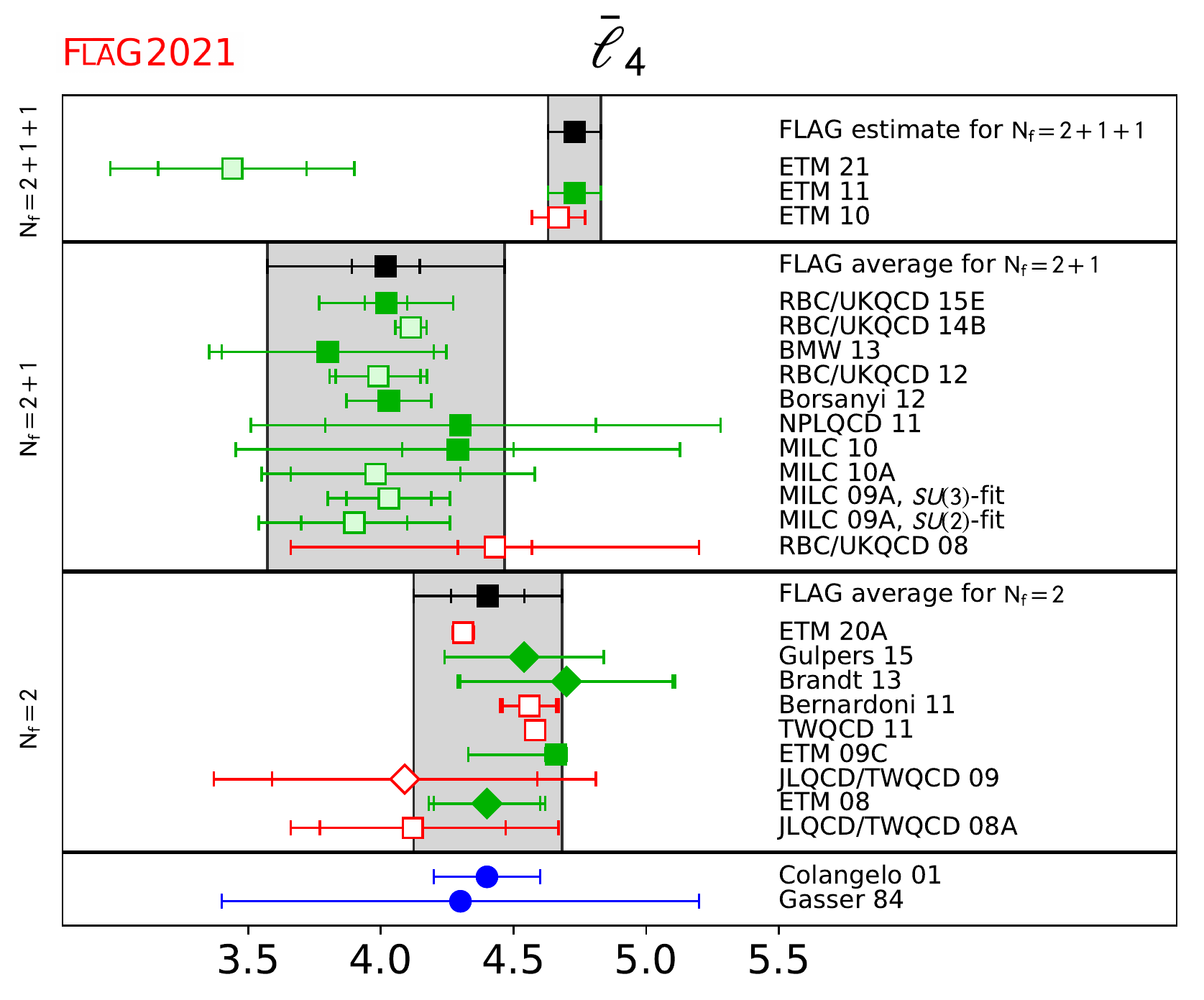}%
\vspace*{-2mm}
\caption{\label{fig:l3}
Effective coupling constant $\lbar_4$.
Squares indicate determinations from correlators in the $p$-regime, diamonds refer to determinations from the pion form factor.}
\end{figure}

If one were to ignore $\Nf$, the two new results would appear inconsistent.
While an implicit dependence on the strange- (and highly suppressed) charm-quark mass in the sea is a logical possibility,
it seems to us these results should be considered in conjunction with the FLAG 19 averages for the quantity $\bar\ell_4$.
The FLAG 19 average for $\Nf=2$, based on four papers, was $4.40(28)$,
the average for $\Nf=2+1$, based on five papers, was $4.02(45)$, and
the estimate for $\Nf=2+1+1$, based on a single paper, was $4.73(10)$.
In terms of standard deviations the difference ``old average minus new result'' is
$4.40(28)-4.31(13)=0.09(31)$ or $0.3\sigma$ for $\Nf=2$, while it is
$4.73(10)-3.44(46)=1.29(47)$ or $2.7\sigma$ for $\Nf=2+1+1$.
Hence, the new $\Nf=2$ result of ETM 20A~\cite{Fischer:2020fvl} is in perfect agreement with the corresponding FLAG 19 average.
On the other hand, the new $\Nf=2+1+1$ result of ETM 21~\cite{Alexandrou:2021bfr} is largely \emph{inconsistent} with the corresponding FLAG 19 estimate,
which was taken from Ref.~\cite{Baron:2011sf}.
Perhaps one should take a step back at this point, and consider the option that the implicit $\Nf$-dependence (through a dynamical strange and charm quark) is smaller than some unaccounted-for systematic effects in at least one of the works considered.
On the practial side neither one of the new results qualifies for a FLAG average
(ETM 20A~\cite{Fischer:2020fvl} has a red tag, ETM 21~\cite{Alexandrou:2021bfr} is still unpublished).
In summary, the time is not ripe to give an update on the $\bar\ell_4$ average given in FLAG 19.

The two new results on $\bar\ell_4$ in Tab.~\ref{tab:l3and4} are displayed in Fig.~\ref{fig:l3}, along with all previous determinations with systematic error bars.
Since there is no new entry in the first column of the table, there is no analogous figure for $\bar\ell_3$.


\begin{table}[!tb] 
\vspace*{3cm}
\centering
\footnotesize
\begin{tabular*}{\textwidth}[l]{l@{\extracolsep{\fill}}rlllllll}
Collaboration & Ref. & $\Nf$ &
\hspace{0.15cm}\begin{rotate}{60}{publication status}\end{rotate}\hspace{-0.15cm} &
\hspace{0.15cm}\begin{rotate}{60}{chiral extrapolation}\end{rotate}\hspace{-0.15cm}&
\hspace{0.15cm}\begin{rotate}{60}{continuum extrapolation}\end{rotate}\hspace{-0.15cm} &
\hspace{0.15cm}\begin{rotate}{60}{finite volume}\end{rotate}\hspace{-0.15cm} &
\rule{0.3cm}{0cm}$\<r^2\>_V^\pi\,[\mr{fm}^2]$ & \rule{0.3cm}{0cm}$\lbar_6$ \\[2mm]
\hline
\hline
\\[-2mm]
HPQCD~15B               & \cite{Koponen:2015tkr}     &2+1+1& \pubA & \good & \soso & \good & 0.403(18)(6)       &                \\[2mm]
\hline
\\[-2mm]
Gao 21                  & \cite{Gao:2021xsm}         & 2+1 & \pubP & \soso & \bad  & \good & $0.42(2)_\mr{tot}$ &                \\
$\chi$QCD 20            & \cite{Wang:2020nbf}        & 2+1 & \pubA & \good & \soso & \good & $0.430(5)(13)$     & 17.1(1.4)      \\
Feng 19                 & \cite{Feng:2019geu}        & 2+1 & \pubA & \good & \bad  & \good & $0.434(20)(13)$    &                \\
JLQCD~15A, $SU(2)$-fit  & \cite{Aoki:2015pba}        & 2+1 & \pubA & \soso & \bad  & \soso & 0.395(26)(32)      & 13.49(89)(82)  \\
JLQCD~14                & \cite{Fukaya:2014jka}      & 2+1 & \pubA & \good & \bad  & \bad  & 0.49(4)(4)         &  7.5(1.3)(1.5) \\
PACS-CS~11A             & \cite{Nguyen:2011ek}       & 2+1 & \pubA & \soso & \bad  & \soso & 0.441(46)          &                \\
RBC/UKQCD~08A           & \cite{Boyle:2008yd}        & 2+1 & \pubA & \bad  & \bad  & \soso & 0.418(31)          & 12.2(9)        \\
LHP~04                  & \cite{Bonnet:2004fr}       & 2+1 & \pubA & \bad  & \bad  & \bad  & 0.310(46)          &                \\[2mm]
\hline
\\[-2mm]
ETM~17F                 & \cite{Alexandrou:2017blh}  &  2  & \pubA & \good & \bad  & \good & 0.443(21)(20)      & 16.21(76)(70)  \\
Brandt~13               & \cite{Brandt:2013dua}      &  2  & \pubA & \soso & \good & \soso & 0.481(33)(13)      & 15.5(1.7)(1.3) \\
JLQCD/TWQCD~09          & \cite{JLQCD:2009qn}        &  2  & \pubA & \soso & \bad  & \bad  & 0.409(23)(37)      & 11.9(0.7)(1.0) \\
ETM~08                  & \cite{Frezzotti:2008dr}    &  2  & \pubA & \soso & \soso & \soso & 0.456(30)(24)      & 14.9(1.2)(0.7) \\
QCDSF/UKQCD~06A         & \cite{Brommel:2006ww}      &  2  & \pubA & \soso & \good & \bad  & 0.441(19)(63)      &                \\[2mm]
\hline
\\[-2mm]
Bijnens~98              & \cite{Bijnens:1998fm}      &     &       &       &       &       & 0.437(16)          & 16.0(0.5)(0.7) \\
NA7~86                  & \cite{Amendolia:1986wj}    &     &       &       &       &       & 0.439(8)           &                \\
Gasser~84               & \cite{Gasser:1983yg}       &     &       &       &       &       &                    & 16.5(1.1)      \\[2mm]
\hline
\hline
\end{tabular*}
\normalsize
\vspace*{-2mm}
\caption{\label{tab:radii}
Vector form factor of the pion: Lattice results for the charge radius $\<r^2\>_V^\pi$ and the chiral coupling
constant $\lbar_6$ are compared with the experimental value, as obtained by NA7, and some phenomenological estimates.
The publication status of $\chi$QCD 20~\cite{Wang:2020nbf} changed from ``preprint'' to ``accepted'' after our closing date.}
\end{table}

\begin{figure}[!tb]
\centering
\includegraphics[width=13cm]{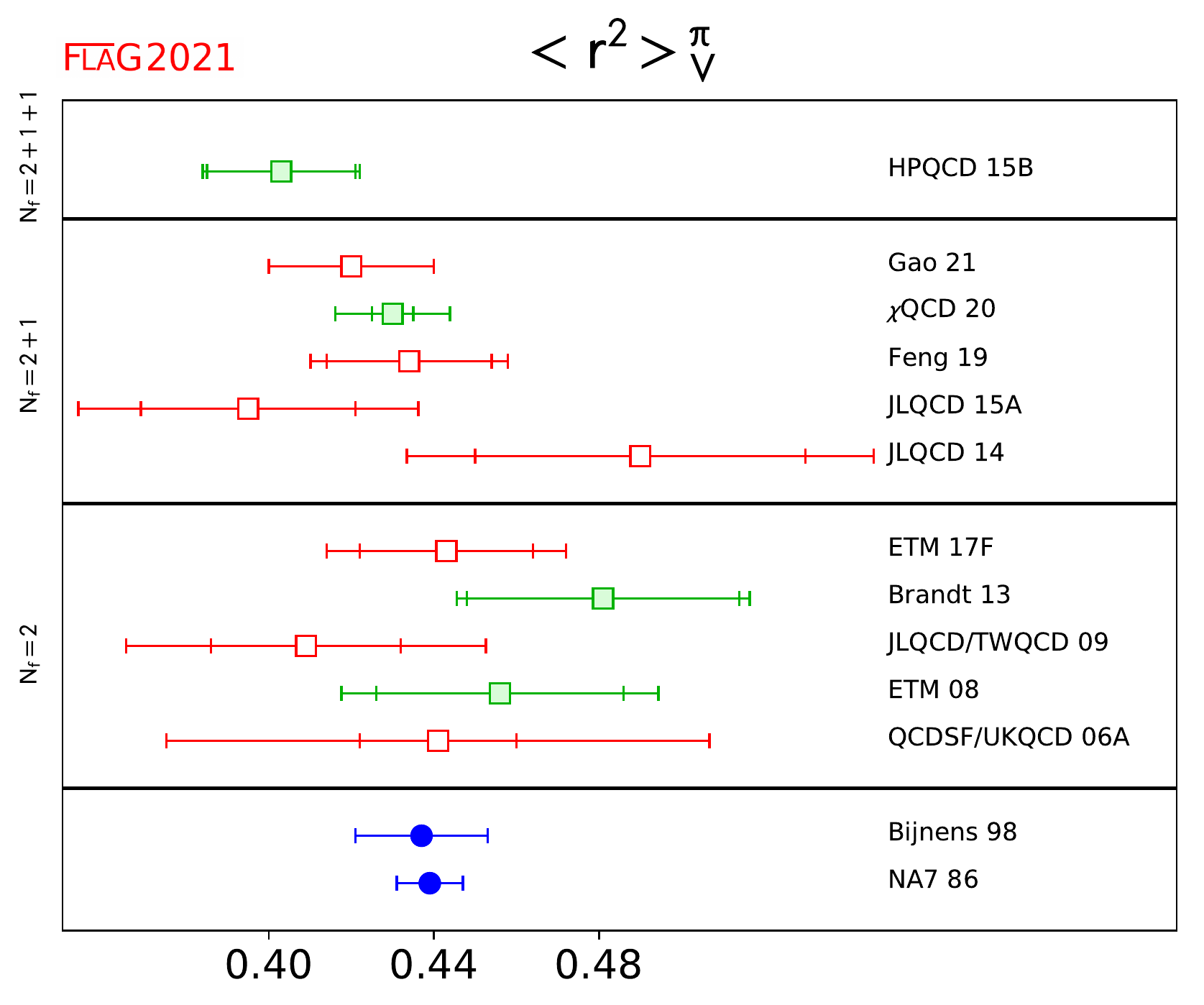}%
\vspace*{-2mm}
\caption{\label{fig:rsqu}
Summary of the pion form factor $\<r^2\>_V^\pi$.
The publication status of $\chi$QCD 20~\cite{Wang:2020nbf} changed from ``preprint'' to ``accepted'' after our closing date.}
\end{figure}

There is also new information on $\bar\ell_6$.
It appears in three new papers on the slope of the vector form factor at $q^2=0$ (``charge radius'') of the pion.
We follow our tradition of quoting and comparing results in terms of $\<r^2\>_V^\pi$ rather than $\bar\ell_6$.
As mentioned before, we start with a brief discussion of the particulars of these papers.

The paper Feng 19~\cite{Feng:2019geu} 
is based on $\Nf=2+1$ flavours of domain-wall valence quarks on domain-wall sea.
This collaboration uses four ensembles essentially at the physical mass point%
\footnote{This earns them a green box on ``chiral extrapolation'', but the criterion was crafted with the idea of a global fit which takes all available information into account.
In the setup of Feng 19~\cite{Feng:2019geu} it is barely possible to disentangle a small $\Mpi$ dependence in the vicinity of $\Mpi^\mr{phys}$ from cut-off effects.}
and another one at $\Mpi=341\MeV$.
At the physical mass point they have three lattice spacings in the range $a^{-1}=1.015-1.73\GeV$, i.e., none of them satisfies $a<0.1\fm$.
The respective box sizes are $L=[6.22,4.58,5.48]\fm$, hence $L(M_{\pi,\mr{min}})=6.22\fm$.

The paper $\chi$QCD 20~\cite{Wang:2020nbf} 
employs overlap valence quarks on $\Nf=2+1$ ensembles with domain-wall sea quarks.
They use a total of seven ensembles, with three of them being at the physical point.
They cover five lattice spacings $a=0.083-0.195\fm$, of which only one is below $0.1\fm$.
The relevant box size is $6.24\fm$ at the physical point, where they have $\Mpi L=4.45$.
Renormalization is done nonperturbatively.

The paper Gao 21~\cite{Gao:2021xsm} 
is based on $\Nf=2+1$ HISQ (staggered) ensembles on which they invert clover valence quarks.
They have $M_{\pi,\mr{sea}}=M_{\pi,\mr{val}}=140\MeV$ at $a=0.076\fm$ in a $64^3\times64$ volume.
In addition, they have $M_{\pi,\mr{sea}}=160\MeV,M_{\pi,\mr{val}}=300\MeV$ at $a=0.06\fm$ (in a $48^3\times64$ box),
and essentially the same sea-valence mass combination at $a=0.04\fm$ (in a $64^3\times64$ box).
The vector form factor is renormalized nonperturbatively.
Unfortunately, no continuum extrapolation is performed; they quote the result from the $a\simeq0.076\fm$ physical pion mass ensemble as listed in Tab.~\ref{tab:radii}.
The error quoted is a total error, comprising systematic uncertainties unrelated to cut-off effects.

The available information on $\<r^2\>_V^\pi$ is summarized in Fig.~\ref{fig:rsqu}.
It is obvious that the lattice computations for this quantity do not achieve the precision of the experimental result (NA\,7) yet.


\subsubsection{New results for an SU(2) linear combination linked to $\pi\pi$ scattering \label{sec:SU2_pipi}}

We are aware of four new papers on $\pi\pi$ scattering (in the isospin $I=2$ and/or $I=0$ state).
As before, we begin with a brief description of their specifics.

Reference~\cite{Horz:2019rrn} by B.~H\"orz and A.~Hanlon 
uses one CLS ensemble of $\Nf=2+1$ nonperturbatively improved Wilson (clover) fermions.
Since it is away from the physical mass point and no extrapolation to the latter is attempted,
we refrain from applying the FLAG criteria, and there will be no listing in tables and/or plots.
We add that this procedure is in strict analogy to our treatment of Ref.~\cite{Bulava:2016mks} in FLAG 19.
A sequel publication, based on the same data, is Ref.~\cite{Blanton:2019vdk}.
They find that the $\pi\pi$ ($I=2$) spectrum is fit well by an $S$-wave phase shift that incorporates the expected Adler zero.
Obviously, the same comment regarding the applicability of the FLAG criteria applies.

The paper Culver 19~\cite{Culver:2019qtx} 
uses $\Nf=2$ flavours of nHYP clover fermions at $a=0.12\fm$, $\Mpi=315\MeV$ on
$48\times24^2\times\{24,30,48\}$ and $\Mpi=226\MeV$ on $64\times24^2\times\{24,28,32\}$.
With a conventional analysis technique they find $a_0^2\Mpi=-0.0455(16)$, after extrapolation to physical pion mass.
From an inverse amplitude method, they obtain $a_0^2\Mpi=-0.0436\binom{+0.0013}{-0.0012}$, again at the physical pion mass.
Since the paper does not give preference to one of the analysis methods, we take the liberty to
condense the two numbers into the result $a_0^2\Mpi=-0.0445(14)(19)$, as shown in Tab.~\ref{tab:pipi}.
Here, the systematic error reflects the full difference between the two central values given in the paper.

The paper Mai 19~\cite{Mai:2019pqr} 
employs $\Nf=2$ nHYP clover fermions at a single lattice spacing ($a=0.12\fm$), with
$\Mpi=315\MeV$ on $48\times24^2\times\{24,30,48\}$ lattices and
$\Mpi=224\MeV$ on $64\times24^2\times\{24,28,32\}$ lattices.
They quote, extrapolated to the physical pion mass,
$a_0^0\Mpi=0.2132\binom{+0.0008}{-0.0009}$ and $a_0^2\Mpi=-0.0433\pm0.0002$ for $I=0$ and $I=2$, respectively.
With statistical error only, these results go into Tab.~\ref{tab:pipi}, but not into a plot.

The paper ETM 20B~\cite{Fischer:2020jzp} 
is based on $\Nf=2$ QCD with twisted mass fermions at $a=0.0914(15)\fm$, and with $c_\mr{SW}=1.57551$.
They have three pion masses ($\Mpi=340\MeV$ on $32^3\times64$ and $\Mpi=242\MeV$ and $\Mpi=134\MeV$ on $48^3\times96$).
They find, for $I=2$, at the pion masses considered, $a_0^2\Mpi=-0.2061(49),-0.156(15),-0.0481(86)$,
with the last being at physical pion mass, but finite $a$.
Accordingly, we take $a_0^2\Mpi=-0.0481(86)$ with unknown systematic error.
With statistical error only, this result goes into Tab.~\ref{tab:pipi}, but not into a plot.

\begin{table}[!tbp] 
\vspace*{3cm}
\centering
\footnotesize
\begin{tabular*}{\textwidth}[l]{l@{\extracolsep{\fill}}rlllllll}
Collaboration & Ref. & $\Nf$ &
\hspace{0.15cm}\begin{rotate}{60}{publication status}\end{rotate}\hspace{-0.15cm} &
\hspace{0.15cm}\begin{rotate}{60}{chiral extrapolation}\end{rotate}\hspace{-0.15cm}&
\hspace{0.15cm}\begin{rotate}{60}{cont.\ extrapolation}\end{rotate}\hspace{-0.15cm} &
\hspace{0.15cm}\begin{rotate}{60}{finite volume}\end{rotate}\hspace{-0.15cm} &
\rule{0.2cm}{0cm} $a_0^0\Mpi$ & $\ell_{\pi\pi}^0$ \\[2mm]
\hline
\hline
\\[-2mm]
Fu 17         & \cite{Fu:2017apw}           & 2+1 & \pubA & \bad  & \soso & \good & $0.217(9)(5)$           & $45.6(7.6)(3.8)$           \\
Fu 13         & \cite{Fu:2013ffa}           & 2+1 & \pubA & \bad  & \bad  & \good & $0.214(4)(7)$           & $43.2(3.5)(5.6)$           \\
Fu 11         & \cite{Fu:2011bz}            & 2+1 & \pubA & \bad  & \bad  & \good & $0.186(2)$              & $18.7(1.2)$                \\
\hline
\\[-2mm]
Mai 19        & \cite{Mai:2019pqr}          &  2  & \pubP & \bad  & \bad  & \soso & $0.2132(9)$             &                            \\
ETM 16C       & \cite{Liu:2016cba}          &  2  & \pubA & \good & \bad  & \good & $0.198(9)(6)$           & $30(8)(6)$                 \\
\hline
\\[-2mm]
Caprini 11    & \cite{Caprini:2011ky}       &     &       &       &       &       & $0.2198(46)(16)(64)$    &                            \\
Colangelo 01  & \cite{Colangelo:2001df}     &     &       &       &       &       & $0.220(5)_\mr{tot}$     &                            \\
\hline
\hline
\end{tabular*}
\\[3.5cm]
\begin{tabular*}{\textwidth}[l]{l@{\extracolsep{\fill}}rlllllll}
Collaboration & Ref. & $\Nf$ &
\hspace{0.15cm}\begin{rotate}{60}{publication status}\end{rotate}\hspace{-0.15cm} &
\hspace{0.15cm}\begin{rotate}{60}{chiral extrapolation}\end{rotate}\hspace{-0.15cm}&
\hspace{0.15cm}\begin{rotate}{60}{cont.\ extrapolation}\end{rotate}\hspace{-0.15cm} &
\hspace{0.15cm}\begin{rotate}{60}{finite volume}\end{rotate}\hspace{-0.15cm} &
\rule{0.2cm}{0cm} $a_0^2\Mpi$ & $\ell_{\pi\pi}^2$ \\[2mm]
\hline
\hline
\\[-2mm]
ETM 15E       & \cite{Helmes:2015gla}       &2+1+1& \pubA & \soso & \good & \good & $-0.0442(2)(^4_0)$      & $3.79(0.61)\binom{+1.34}{-0.11}$ \\[2mm]
\hline
\\[-2mm]
PACS-CS 13    & \cite{Sasaki:2013vxa}       & 2+1 & \pubA & \good & \bad  & \bad  & $-0.04243(22)(43)$      &                                  \\ 
Fu 13         & \cite{Fu:2013ffa}           & 2+1 & \pubA & \bad  & \bad  & \good & $-0.04430(25)(40)$      & $3.27(0.77)(1.12)$               \\
Fu 11         & \cite{Fu:2011bz}            & 2+1 & \pubA & \bad  & \bad  & \good & $-0.0416(2)$            & $11.6(9)$                        \\
NPLQCD 11A    & \cite{Beane:2011sc}         & 2+1 & \pubA & \bad  & \bad  & \good & $-0.0417(07)(02)(16)$   &                                  \\
NPLQCD 07     & \cite{Beane:2007xs}         & 2+1 & \pubA & \bad  & \bad  & \bad  & $-0.04330(42)_\mr{tot}$ &                                  \\
NPLQCD 05     & \cite{Beane:2005rj}         & 2+1 & \pubA & \bad  & \bad  & \bad  & $-0.0426(06)(03)(18)$   &                                  \\[2mm]
\hline
\\[-2mm]
ETM 20B       & \cite{Fischer:2020jzp}      &  2  & \pubA & \soso & \bad  & \soso & $-0.0481(86)$           &                                  \\
Mai 19        & \cite{Mai:2019pqr}          &  2  & \pubP & \bad  & \bad  & \soso & $-0.0433(2)$            &                                  \\
Culver 19     & \cite{Culver:2019qtx}       &  2  & \pubP & \bad  & \bad  & \soso & $-0.0445(14)(19)$       &                                  \\
Yagi 11       & \cite{Yagi:2011jn}          &  2  & \pubP & \soso & \bad  & \bad  & $-0.04410(69)(18)$      &                                  \\
ETM 09G       & \cite{Feng:2009ij}          &  2  & \pubA & \soso & \soso & \soso & $-0.04385(28)(38)$      & $4.65(0.85)(1.07)$               \\
CP-PACS 04    & \cite{Yamazaki:2004qb}      &  2  & \pubA & \bad  & \bad  & \good & $-0.0413(29)$           &                                  \\[2mm]
\hline
\\[-2mm]
Caprini 11    & \cite{Caprini:2011ky}       &     &       &       &       &       & $-0.0445(11)(4)(8)$     &                                  \\
Colangelo 01  & \cite{Colangelo:2001df}     &     &       &       &       &       & $-0.0444(10)_\mr{tot}$  &                                  \\
\hline
\hline
\end{tabular*}
\normalsize
\vspace*{-2mm}
\caption{\label{tab:pipi}
Summary of $\pi\pi$ scattering data in the $I=0$ (top) and $I=2$ (bottom) channels.
Some of the results have been adapted to our sign convention.
The results of Refs.~\cite{Colangelo:2001df,Caprini:2011ky} allow for a cross-check with phenomenology.}
\end{table}

\begin{figure}[!tbp]
\centering
\includegraphics[width=12cm]{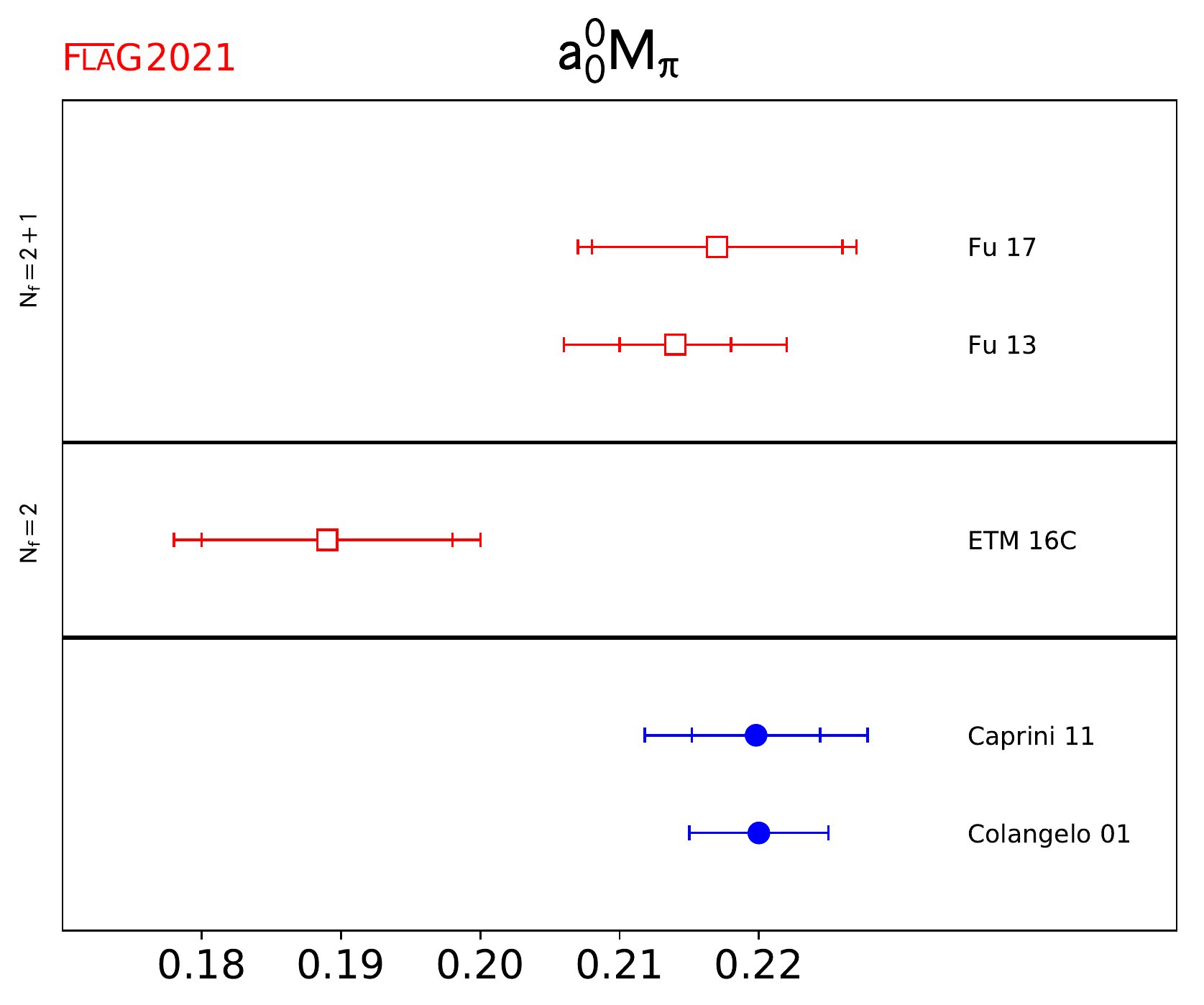}\\
\includegraphics[width=12cm]{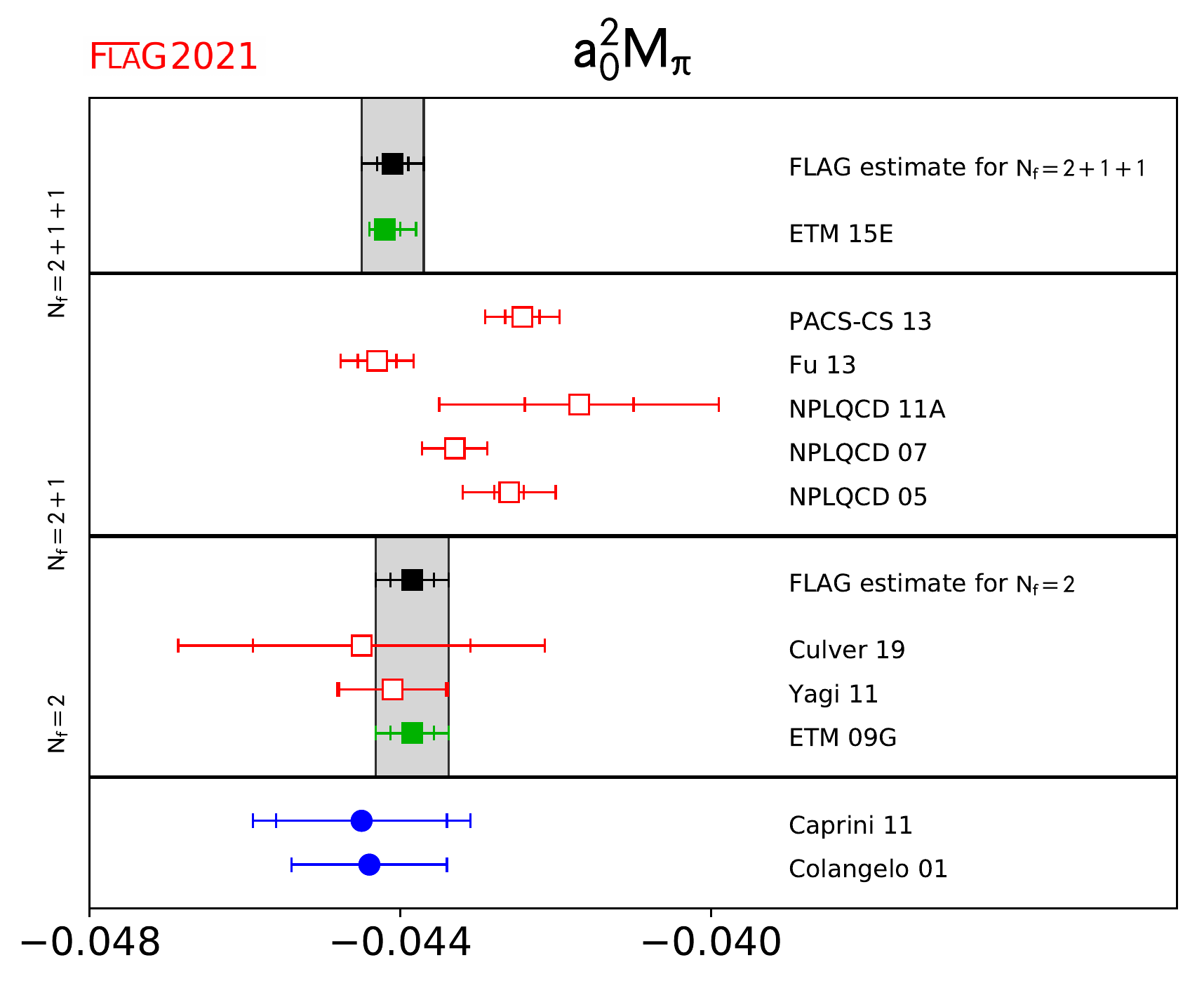}%
\vspace*{-2mm}
\caption{\label{fig:pionscattering}
Summary of the $\pi\pi$ scattering lengths $a_0^0\Mpi$ (top) and $a_0^2\Mpi$ (bottom).
Results in Tab.~\ref{tab:pipi} with statistical error only are not shown.}
\end{figure}

These four works, when combined with the information listed in FLAG 19, represent the information from the lattice on the $\pi\pi$ scattering lengths $a_0^0$ and $a_0^2$ in the isopin channels $I=0$ and $I=2$, respectively.
As can be seen from Eqs.~(\ref{eq:pipi_ell0_I0_xi},~\ref{eq:pipi_scale_00}), the $I=0$ scattering length carries information about $\frac{20}{21}\bar\ell_1+\frac{40}{21}\bar\ell_2-\frac{5}{14}\bar\ell_3+2\bar\ell_4$.
And from Eqs.~(\ref{eq:pipi_ell0_I2_xi},~\ref{eq:pipi_scale_02}) it follows that the $I=2$ counterpart carries information about the linear combination $\frac{4}{3}\bar\ell_1+\frac{8}{3}\bar\ell_2-\frac{1}{2}\bar\ell_3-2\bar\ell_4$.
Still, we prefer quoting the dimensionless products $a_0^{I}\Mpi$ (at the physical mass point) over the aforementioned linear combinations to ease comparison with phenomenology.

The updated Tab.~\ref{tab:pipi} summarizes the present lattice information on $a_0^{I=0}\Mpi$ and $a_0^{I=2}\Mpi$ at the physical mass point,
and the results are displayed in Fig.~\ref{fig:pionscattering}.
We remind the reader that a lattice computation of $a_0^{I=0}\Mpi$ involves quark-loop disconnected contributions, which tend to be very noisy and thus require large statistics.
Compared to the situation in FLAG 19 the number of computations has increased from three to five, but still none of them is free of red tags.
The situation is somewhat better for $a_0^{I=2}\Mpi$ which is computed from quark-line connected contributions only.
In this case there is one computation at $\Nf=2$ and one at $\Nf=2+1+1$ that qualifies for a FLAG average.
We quote these numbers in subsection \ref{sec:SU2_averages} below.

The available information on $a_0^{I=0}\Mpi$ and $a_0^{I=2}\Mpi$ is summarized in Fig.~\ref{fig:pionscattering}.
It is obvious that the former quantity (due to quark-loop disconnected contributions) is much harder to calculate on the lattice than the latter one.
Nonetheless, the good news is that in both cases the lattice determinations are in reasonable agreement with EFT results.


\subsubsection{LO and NLO SU(2) estimates and averages\label{sec:SU2_averages}}

As promised in an earlier section, here we list our FLAG 19 estimates and averages \cite{Aoki:2019cca} that all remain unchanged.
We refer the reader to that review for details and explanations.

For the $SU(2)$ LEC $\Sigma$, in the $\msbar$ scheme, at the renormalization scale $\mu=2\GeV$, we obtained the averages and/or estimate
\begin{align}
&N_f=2+1+1:&\FLAGAVBEGIN \Sigma^{1/3}&= 286(23) \FLAGAVEND \MeV &&\Refs~\mbox{\cite{Cichy:2013gja,Alexandrou:2017bzk}},\nonumber\\[-3mm]
\nonumber \\[-3mm]
&N_f=2+1  :&\FLAGAVBEGIN \Sigma^{1/3}&= 272( 5) \FLAGAVEND \MeV &&\Refs~\mbox{\cite{Bazavov:2010yq,Borsanyi:2012zv,Durr:2013goa,Boyle:2015exm,Cossu:2016eqs,Aoki:2017paw}},\\[-3mm]
\nonumber \\[-3mm]
&N_f=2    :&\FLAGAVBEGIN \Sigma^{1/3}&= 266(10) \FLAGAVEND \MeV &&\Refs~\mbox{\cite{Baron:2009wt,Cichy:2013gja,Brandt:2013dua,Engel:2014eea}} ,\nonumber
\label{eq:condensates}
\end{align}
where the errors include both statistical and systematic uncertainties.

For the ratio of the pion decay constant at the physical point, $\Fpi$, to its value in the $SU(2)$ chiral limit (zero up- and down-quark mass but physical strange-quark mass), $F$, we obtained the averages and/or estimate
\begin{align}
&N_f=2+1+1:&\FLAGAVBEGIN {\Fpi}/{F}&=1.077( 3) \FLAGAVEND &&\Refs~\mbox{\cite{Baron:2011sf}}                                                          ,\nonumber\\[-3mm]
\nonumber \\[-3mm]
&N_f=2+1  :&\FLAGAVBEGIN {\Fpi}/{F}&=1.062( 7) \FLAGAVEND &&\Refs~\mbox{\cite{Bazavov:2010hj,Beane:2011zm,Borsanyi:2012zv,Durr:2013goa,Boyle:2015exm}},\\[-3mm]
\nonumber \\[-3mm]
&N_f=2    :&\FLAGAVBEGIN {\Fpi}/{F}&=1.073(15) \FLAGAVEND &&\Refs~\mbox{\cite{Frezzotti:2008dr,Baron:2009wt,Brandt:2013dua,Engel:2014eea}}            .\nonumber
\label{eq:decayratios}
\end{align}

For $SU(2)$ NLO LECs we obtained the averages and/or estimates
\begin{align}
&N_f=2+1+1:&\FLAGAVBEGIN \lbar_3&=3.53(26) \FLAGAVEND &&\Refs~\mbox{\cite{Baron:2011sf}},                                                 \nonumber\\[-3mm]
\nonumber \\[-3mm]
&N_f=2+1  :&\FLAGAVBEGIN \lbar_3&=3.07(64) \FLAGAVEND &&\Refs~\mbox{\cite{Bazavov:2010hj,Beane:2011zm,Borsanyi:2012zv,Durr:2013goa,Boyle:2015exm}},\\[-3mm]
\nonumber \\[-3mm]
&N_f=2    :&\FLAGAVBEGIN \lbar_3&=3.41(82) \FLAGAVEND &&\Refs~\mbox{\cite{Frezzotti:2008dr,Baron:2009wt,Brandt:2013dua}},                 \nonumber
\label{results_l3}
\end{align}
\begin{align}
&N_f=2+1+1:&\FLAGAVBEGIN \lbar_4&=4.73(10) \FLAGAVEND &&\Refs~\mbox{\cite{Baron:2011sf}},                                                 \nonumber\\[-3mm]
\nonumber \\[-3mm]
&N_f=2+1  :&\FLAGAVBEGIN \lbar_4&=4.02(45) \FLAGAVEND &&\Refs~\mbox{\cite{Bazavov:2010hj,Beane:2011zm,Borsanyi:2012zv,Durr:2013goa,Boyle:2015exm}},\\[-3mm]
\nonumber \\[-3mm]
&N_f=2    :&\FLAGAVBEGIN \lbar_4&=4.40(28) \FLAGAVEND &&\Refs~\mbox{\cite{Frezzotti:2008dr,Baron:2009wt,Brandt:2013dua,Gulpers:2015bba}}, \nonumber
\label{results_l4}
\end{align}
as well as the estimate
\beq
N_f=2:\hspace{1cm}\FLAGAVBEGIN \lbar_6=15.1(1.2) \FLAGAVEND \qquad \Refs~\mbox{\cite{Frezzotti:2008dr,Brandt:2013dua}}.
\eeq

For the scattering length extracted from $\pi\pi$ scattering in the $I=2$ channel we quote
\begin{align}
&N_f=2+1+1:&\FLAGAVBEGIN a_0^2\Mpi&=-0.0441(4)   \FLAGAVEND &&\Refs~\mbox{\cite{Helmes:2015gla}},\nonumber\\[-3mm]
\nonumber \\[-3mm]
&N_f=2    :&\FLAGAVBEGIN a_0^2\Mpi&=-0.04385(47) \FLAGAVEND &&\Refs~\mbox{\cite{Feng:2009ij}},
\label{results_a02Mpi}
\end{align}
where the errors include both statistical and systematic uncertainties.
We remark that our preprocessing procedure%
\footnote{There are two naive procedures to symmetrize an asymmetric systematic error:
($i$) keep the central value untouched and enlarge the smaller error, ($ii$) shift the central value by half of the difference between the two original errors and enlarge/shrink both errors by the same amount.
Our procedure ($iii$) is to average the results of ($i$) and ($ii$).
In other words a result $c(s)\binom{+u}{-\ell}$ with $\ell>u$ is changed into $c+(u-\ell)/4$ with statistical error $s$ and a symmetric systematic error $(u+3\ell)/4$.
The case $\ell<u$ is handled accordingly.\label{foot:symmetrization}}
symmetrizes the asymmetric errors with a slight adjustment of the central value.

In all cases the references shown are the papers with the contributing results, and we ask the readers to cite those papers when quoting these averages.


\subsection{Extraction of SU(3) low-energy constants \label{sec:SU3results}}


\subsubsection{New results for individual LO SU(3) LECs\label{sec:SU3_LO}}

We are unaware of any new paper that determines a large number of LECs in the $SU(3)$ framework (as was done, in the past, by the MILC collaboration).
However, there is one paper, $\chi$QCD 21~\cite{Liang:2021pql}, with a new result on two $SU(3)$ LECs at LO.
They find $F_0=67.8(1.2)(3.2)$ and $\Sigma_0=232.6(0.9)(2.7)$ in the 3-flavour chiral limit%
\footnote{We use $\Sigma=\lim_{m_u,m_d\to0}\Sigma(m_u,m_d,m_s,m_c,...)$, $\Sigma_0=\lim_{m_u,m_d,m_s\to0}\Sigma(m_u,m_d,m_s,m_c,...)$,
and likewise for $B$, $B_0$, $F$ and $F_0$. The quantities $\Sigma,\Sigma_0,B,B_0$ are renormalized at the scale $\mu=2\GeV$.}.
They also quote $\Sigma/\Sigma_0=1.40(2)(2)$ which we consider iteresting for reasons detailed in Sec.~\ref{sec:SU3_Zweig}.


\begin{table}[!tbp] 
\vspace*{3cm}
\centering
\footnotesize
\begin{tabular*}{\textwidth}[l]{l@{\extracolsep{\fill}}rlllllllllll}
 Collaboration & Ref. & $\Nf$ &
\hspace{0.15cm}\begin{rotate}{60}{publication status}\end{rotate}\hspace{-0.15cm} &
\hspace{0.15cm}\begin{rotate}{60}{chiral extrapolation}\end{rotate}\hspace{-0.15cm} &
\hspace{0.15cm}\begin{rotate}{60}{cont. extrapolation}\end{rotate}\hspace{-0.15cm} &
\hspace{0.15cm}\begin{rotate}{60}{finite volume}\end{rotate}\hspace{-0.15cm} &
\rule{0.3cm}{0cm}$F_0\,[\mr{MeV}]$ & \rule{0.1cm}{0cm} $F/F_0$ & \rule{0.2cm}{0cm}$B/B_0$ & \hspace{2.5cm} \\[2mm]
\hline
\hline
\\[-2mm]
JLQCD/TWQCD~10A         & \cite{Fukaya:2010na}       &  3  & \pubA & \bad  & \bad  & \bad  & 71(3)(8)       &                           &                                  \\[2mm]
\hline
\\[-2mm]
$\chi$QCD 21            & \cite{Liang:2021pql}       & 2+1 & \pubP & \good & \good & \good & 67.8(1.2)(3.2) & \\
MILC~10                 & \cite{Bazavov:2010hj}      & 2+1 & \pubC & \soso & \good & \good & 80.3(2.5)(5.4) &                           &                                  \\
MILC~09A                & \cite{Bazavov:2009fk}      & 2+1 & \pubC & \soso & \good & \good & 78.3(1.4)(2.9) & {\sl 1.104(3)(41)}        & {\sl 1.21(4)$\binom{+5}{-6}$}    \\
MILC~09                 & \cite{Bazavov:2009bb}      & 2+1 & \pubA & \soso & \good & \good &                & 1.15(5)$\binom{+13}{-03}$ & {\sl 1.15(16)$\binom{+39}{-13}$} \\
PACS-CS~08              & \cite{Aoki:2008sm}         & 2+1 & \pubA & \good & \bad  & \bad  & 83.8(6.4)      & 1.078(44)                 & 1.089(15)                        \\
RBC/UKQCD~08            & \cite{Allton:2008pn}       & 2+1 & \pubA & \soso & \bad  & \soso & 66.1(5.2)      & 1.229(59)                 & 1.03(05)                         \\[2mm]
\hline
\hline
\end{tabular*}
\newline
\vspace*{3.5cm}
\begin{tabular*}{\textwidth}[l]{l@{\extracolsep{\fill}}rllllllll}
 Collaboration & Ref. & $\Nf$ &
\hspace{0.15cm}\begin{rotate}{60}{publication status}\end{rotate}\hspace{-0.15cm} &
\hspace{0.15cm}\begin{rotate}{60}{chiral extrapolation}\end{rotate}\hspace{-0.15cm} &
\hspace{0.15cm}\begin{rotate}{60}{cont. extrapolation}\end{rotate}\hspace{-0.15cm} &
\hspace{0.15cm}\begin{rotate}{60}{finite volume}\end{rotate}\hspace{-0.15cm} &
\hspace{0.15cm}\begin{rotate}{60}{renormalization}\end{rotate}\hspace{-0.15cm} &
\rule{0.3cm}{0cm}$\Sigma_0^{1/3}\,[\mr{MeV}]$ & \rule{0.1cm}{0cm} $\Sigma/\Sigma_0$ \\[2mm]
\hline
\hline
\\[-2mm]
JLQCD/TWQCD~10A         & \cite{Fukaya:2010na}       &  3  & \pubA & \bad  & \bad  & \bad  & \good & 214(6)(24)                  & {\sl 1.31(13)(52)}         \\[2mm]
\hline
\\[-2mm]
$\chi$QCD 21            & \cite{Liang:2021pql}       & 2+1 & \pubP & \good & \good & \good & \good & 232.6(0.9)(2.7)             & 1.40(2)(2)                 \\
MILC~09A                & \cite{Bazavov:2009fk}      & 2+1 & \pubC & \soso & \good & \good & \soso & 245(5)(4)(4)                & {\sl 1.48(9)(8)(10)}       \\
MILC~09                 & \cite{Bazavov:2009bb}      & 2+1 & \pubA & \soso & \good & \good & \soso & 242(9)$\binom{+05}{-17}$(4) & 1.52(17)$\binom{+38}{-15}$ \\
PACS-CS~08              & \cite{Aoki:2008sm}         & 2+1 & \pubA & \good & \bad  & \bad  & \bad  & 290(15)                     & 1.245(10)                  \\
RBC/UKQCD~08            & \cite{Allton:2008pn}       & 2+1 & \pubA & \soso & \bad  & \soso & \good &                             & 1.55(21)                   \\[2mm]
\hline
\hline
\end{tabular*}
\normalsize
\vspace*{-2mm}
\caption{\label{tab:SU3_overview}
Lattice results for the low-energy constants $F_0$, $B_0$ and $\Sigma_0\!\equiv\!F_0^2B_0$, which specify the effective $SU(3)$ Lagrangian at leading order.
The ratios $F/F_0$, $B/B_0$, $\Sigma/\Sigma_0$, which compare these with their $SU(2)$ counterparts, indicate the strength of the Zweig-rule violations in these quantities (in the large-$N_c$ limit, they tend to unity).
Numbers in slanted fonts are calculated by us, from the information given in the references.}
\end{table}

These values are listed, together with those of FLAG 19, in Tab.~\ref{tab:SU3_overview}.
The paper has been discussed and color coded in Sec.~\ref{sec:SU2results}.
As they are not published yet, there is no update to the FLAG averages/estimates here.


\subsubsection{New results for individual NLO SU(3) LECs\label{sec:SU3_NLO}}

There are a number of new results on $L_5$, for instance in Refs.~\cite{Beane:2006gj,Fu:2011wc,Fu:2012ng} to be discussed below in the context of $\pi K$ scattering.
This is not so surprising, since Eqns.~(\ref{eqn:SU3_chpt_a2_pipi}, \ref{eqn:SU3_chpt_a_KK}, \ref{eqn:SU3_chpt_a32_piK}, \ref{eqn:SU3_chpt_a12_piK}) indicate that the observables
$a_0^2\Mpi$, $a_0^1\Mka$, $a_0^{3/2}\mu_{\pi K}$, $a_0^{1/2}\mu_{\pi K}$ jointly determine the combination $L_\mr{scat}$ and $L_5$ (both of which are conventionally quoted at the scale $\mu=770\MeV$).
Determining any of these two LECs is afflicted with an extra uncertainty, compared to the four scattering lengths, due to the convergence of the $SU(3)$ chiral series%
\footnote{One of the issues is whether the convergence in the LECs pertinent to $a_0^1\Mka$, i.e., with two strange quarks involved, is visibly slower than for $a_0^{3/2}\mu_{\pi K}$ and $a_0^{1/2}\mu_{\pi K}$,
where only one strange quark appears.}.
Therefore we give preference to reviewing the scattering lengths and converting, once they exist, the pertinent FLAG averages into numerical values of $L_\mr{scat}$ and $L_5$,
over collecting values of $L_\mr{scat}$ and $L_5$ as converted by the individual collaborations.

On the other hand, there is no new result on those LECs at the NLO in the $SU(3)$ expansion which were covered in previous editions of FLAG ($L_4,L_6,L_9,L_{10}$).


\subsubsection{Results for SU(3) linear combinations linked to $\pi K$, $KK$ scattering\label{sec:SU3_KpiKK}}

Since $\pi K$, $KK$ scattering were not covered in previous editions of the FLAG report, we list here all works which include such results.
Following the example of the section on $\pi\pi$ scattering, where all results were given in the dimensionless variable $a_0^I\Mpi$,
we give the results on $\pi K$ scattering in the form $a_0^I\mu_{\pi K}$, where $\mu_{\pi K}$ is the pertinent reduced mass,
and the results on $KK$ scattering are given in the form $a_0^I\Mka$.
We start with a brief mentioning of all papers we are aware of.

The paper NPLQCD 06B~\cite{Beane:2006gj}
uses asqtad (staggered) sea quarks with $\Nf=2+1$ at a single lattice spacing ($a=0.125\fm$ with $L\simeq2.5\fm$) with $\Mpi=[290,350,490,600]\MeV$.
The domain-wall valence fermions come with quark masses such that the resulting pion masses match the aforementioned Nambu-Goldstone boson masses.
After chiral extrapolation they find
$a_0^{1/2} \mu_{\pi K} = 0.1346(13)\binom{+18}{-122}$ and 
$a_0^{3/2} \mu_{\pi K} = -0.0448(12)\binom{+19}{-45}$,    
with $L_5$ pinned down at a value extracted from the analysis of the quark mass dependence of $f_K/f_\pi$.
The color coding in Tab.~\ref{tab:KpiKK} is based on $M_{\pi,\mr{min}}(\mr{RMS})=488\MeV$.

The paper NPLQCD 07B~\cite{Beane:2007uh}
uses asqtad (staggered) sea quarks with $\Nf=2+1$ in conjunction with domain-wall valence quarks.
They have two lattice spacings ($a=0.125\,\fm,0.09\,\fm$) with somehat unequal span in quark masses.
At $a=0.125\,\fm$ they cover $\Mpi\simeq290,350,490,590\MeV$ with $L\simeq2.5\fm$.
At $a=0.09\,\fm$ they do not quote $\Mpi[\mr{MeV}]$, but from $a\Mpi=0.1453$ in Tab.II and $a\simeq0.09\,\fm$ one would conclude $\Mpi\simeq320\MeV$.
After chiral extrapolation, they find $a_0^1\Mka=-0.352(16)_\mr{tot}$.
The color coding in Tab.~\ref{tab:KpiKK} is based on $M_{\pi,\mr{min}}(\mr{RMS})=413\MeV$. 

The paper Fu 11A~\cite{Fu:2011wc}
employs one ensemble of $\Nf=2+1$ asqtad (staggered) quarks at $a\simeq0.15\,\fm$, $m_l/m_s=0.2$, $m_s\simeq m_s^\mr{phys}$ with $L=2.5\,\fm$.
It uses six valence pion masses $\Mpi=334-466\MeV$ to study $S$-wave scattering.
It quotes, after chiral extrapolation, 
$a_0^{1/2}\mu_{\pi K}=0.1425(29)$ and 
$a_0^{3/2}\mu_{\pi K}=-0.0394(15)$.   
The color coding in Tab.~\ref{tab:KpiKK} is based on $M_{\pi,\mr{min}}(\mr{RMS})=590\MeV$. 

We are also aware of Ref.~\cite{Lang:2012sv} which is based on a single ensemble of $\Nf=2$ clover quarks.
Since it is away from the physical mass point and no extrapolation to the latter is attempted,
we feel it would be unfair (or misleading) to quote its results in Tab.~\ref{tab:KpiKK}.

Reference PACS-CS 13~\cite{Sasaki:2013vxa}
uses five ensembles of $\Nf=2+1$ nonpertubative clover fermions with $a=0.09\fm$, $L=2.9\fm$, and $\Mpi=166,297,414,575,707\MeV$.
They quote, after extrapolation with {\Ch}PT:
$a_0^2\Mpi=-0.04243(22)(43)$ (see Tab.~\ref{tab:pipi}), 
$a_0^1\Mka=-0.312(17)(31)$, 
$a_0^{3/2}\mu_{\pi K}=-0.0477(27)(20)$ and 
$a_0^{1/2}\mu_{\pi K}=0.150(16)(37)$ (listed in Tab.~\ref{tab:KpiKK}).   
These figures reflect the final numbers quoted in the Erratum of Ref.~\cite{Sasaki:2013vxa}.
The reason for the change is the mishap reported in footnote~\ref{foot:buggy}; fortunately it turns out that it affected the final analysis only very mildly.
We thank the collaboration for keeping us up-to-date with all aspects of the revision.
Since there are no FLAG averages for scattering lengths for $\Nf=2+1$, these small changes have no impact on the quoted FLAG averages.

The paper HS 14A~\cite{Wilson:2014cna}
is based on $\Nf=2+1$ anisotropic clover fermions at $a_s\simeq0.12\,\fm$, $a_t\simeq0.035\,\fm$,
with $\Mpi=391\MeV$ in $\{16^3,20^3,24^3\}\times128$ boxes, i.e.\ with $L=1.9,2.4,2.9\,\fm$.
These parameters yield $\Mka=549\MeV$ thus $\mu_{\pi K}=228\MeV$.
They quote various resonance parameters and, in the $S$-wave $I=3/2$ channel,
$a_0^{3/2}\Mpi=-0.278(15)$ which we convert to $a_0^{3/2}\mu_{\pi K}=-0.161(9)$ at the given $\Mpi$.
Since this work does not extrapolate to $\Mpi^\mr{phys}$, we stay away from color coding.

The paper ETM 17G~\cite{Helmes:2017smr}
uses $\Nf=2+1+1$ twisted mass fermions at three lattice spacings, $a=0.089,0.082,0.062\,\fm$,
with up to five $\Mpi=230-450\MeV$, and $L(M_{\pi,\mr{min}})\simeq2.8\,\fm$.
In the $I=1$ channel they find
$a_0^1\Mka=-0.385(16)\binom{+0}{-12}\binom{+0}{-5}(4)$. 
We take the liberty to combine the various non-statistical errors in quadrature, using $a_0^1\Mka=-0.385(16)\binom{+4}{-14}$ as quoted in Tab.~\ref{tab:KpiKK}.

Reference~\cite{Brett:2018jqw} by R.~Brett \textit{et al.} 
uses one ensemble of $\Nf=2+1$ anisotropic clover fermions with $a_s=0.115\,\fm$,
$\Mpi=233\MeV$, in a $32^3\times256$ box, hence $L=3.7\fm$.
These parameters yield $\Mka=494\MeV$ and thus $\mu_{\pi K}=158\MeV$.
Their result for $I=1/2$ $S$-wave scattering reads $a_0^{1/2}\Mpi=-0.353(25)$, or $a_0^{1/2}\mu_{\pi K}=-0.240(17)$ in our notation.
Since this work does not extrapolate to $\Mpi^\mr{phys}$, we stay away from color coding.

The paper ETM 18B~\cite{Helmes:2018nug}
uses $\Nf=2+1+1$ twisted mass fermions at three lattice spacings, $a=0.089,0.082,0.062\,\fm$,
with up to five pion masses $\Mpi=230-450\MeV$ and up to two volumes.
From the tables, one finds $M_{\pi,\mr{min}}=276,302,311\MeV$ at the three lattice spacings.
They find, after chiral extrapolation,
$a_0^{1/2}\mu_{\pi K}=0.127(2)_\mr{tot}$ and 
$a_0^{3/2}\mu_{\pi K}=-0.0463(17)_\mr{tot}$  
as quoted in Tab.~\ref{tab:KpiKK}.

\begin{table}[!tb] 
\vspace*{3cm}
\centering
\footnotesize
\begin{tabular*}{\textwidth}[l]{l@{\extracolsep{\fill}}rllllllll}
Collaboration & Ref. & $\Nf$ &
\hspace{0.15cm}\begin{rotate}{60}{publication status}\end{rotate}\hspace{-0.15cm} &
\hspace{0.15cm}\begin{rotate}{60}{chiral extrapolation}\end{rotate}\hspace{-0.15cm}&
\hspace{0.15cm}\begin{rotate}{60}{cont.\ extrapolation}\end{rotate}\hspace{-0.15cm} &
\hspace{0.15cm}\begin{rotate}{60}{finite volume}\end{rotate}\hspace{-0.15cm} &
\rule{0.2cm}{0cm} $a_0^{1/2}\mu_{\pi K}$ & $a_0^{3/2}\mu_{\pi K}$ & $a_0^1\Mka$ \\[2mm]
\hline
\hline
\\[-2mm]
ETM 18B         & \cite{Helmes:2018nug}       & 2+1+1 & \pubA & \soso & \good & \soso & $0.127(2)_\mr{tot}$           & $-0.0463(17)_\mr{tot}$        &                             \\
ETM 17G         & \cite{Helmes:2017smr}       & 2+1+1 & \pubA & \soso & \good & \soso &                               &                               & $-0.385(16)\binom{+4}{-14}$ \\
\hline
\\[-2mm]
PACS-CS 13      & \cite{Sasaki:2013vxa}       & 2+1   & \pubA & \good & \bad  & \bad  & $0.150(16)(37)$               & $-0.0477(27)(20)$             & $-0.312(17)(31)$            \\
Fu 11A          & \cite{Fu:2011wc}            & 2+1   & \pubA & \bad  & \bad  & \good & $0.1425(29)$                  & $-0.0394(15)$                 &                             \\
NPLQCD 07B      & \cite{Beane:2007uh}         & 2+1   & \pubA & \bad  & \soso & \soso &                               &                               & $-0.352(16)_\mr{tot}$       \\
NPLQCD 06B      & \cite{Beane:2006gj}         & 2+1   & \pubA & \bad  & \bad  & \good & $0.1346(13)\binom{+18}{-122}$ & $-0.0448(12)\binom{+19}{-45}$ &                             \\
\hline
\hline
\end{tabular*}
\normalsize
\vspace*{-2mm}
\caption{\label{tab:KpiKK}
Summary of $\pi K$ scattering data in the $I=\frac{1}{2},\frac{3}{2}$ channels, and of $KK$ scattering with $I=1$.
Some of the results have been adapted to our sign convention.}
\end{table}

\begin{figure}[!tbp]
\centering
\includegraphics[width=8.2cm]{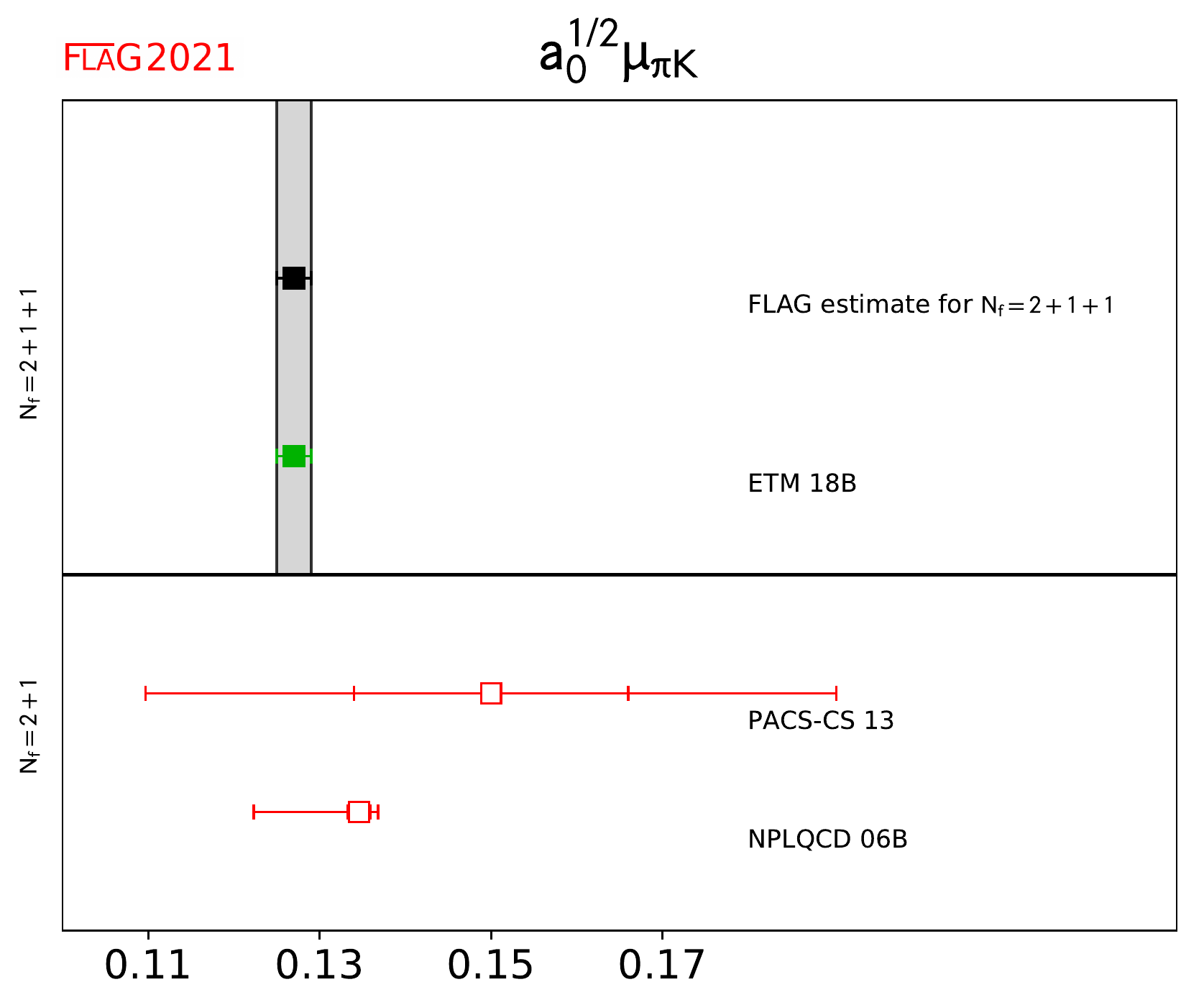}\\[-0mm]
\includegraphics[width=8.2cm]{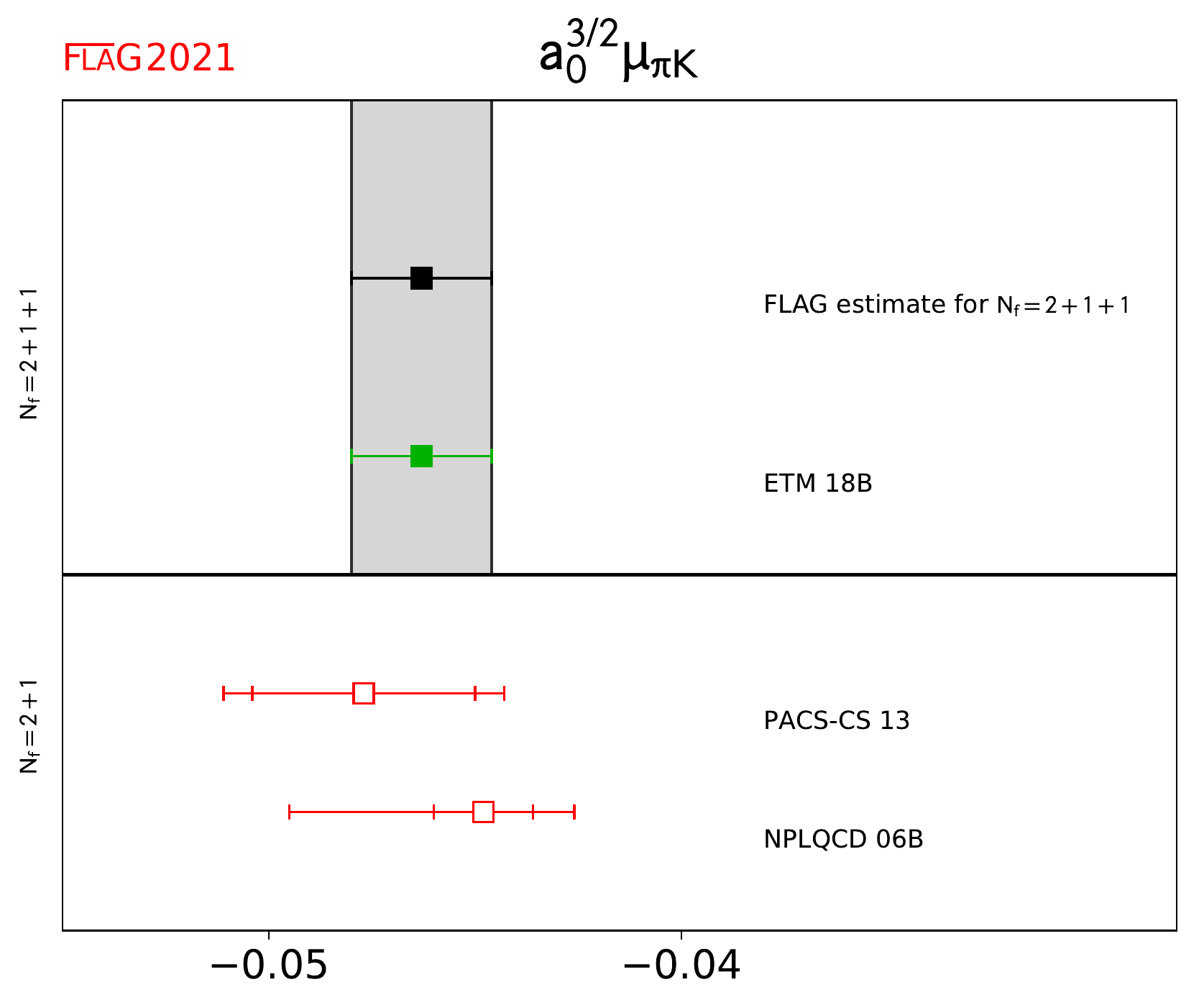}\\[-0mm]
\includegraphics[width=8.2cm]{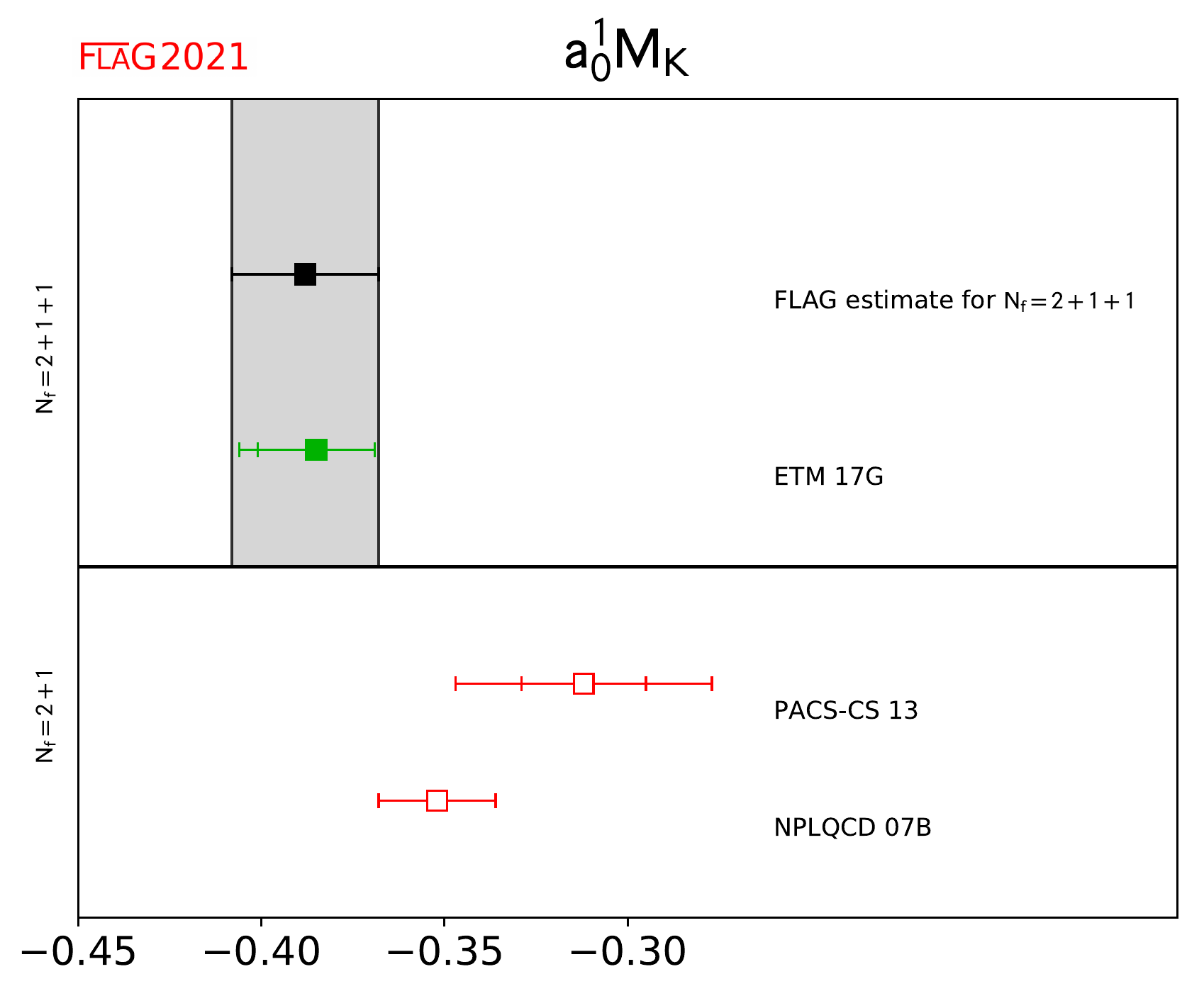}%
\vspace*{-2mm}
\caption{\label{fig:kaonscattering}
Summary of the $\pi K$ scattering lengths $a_0^{1/2}\mu_{\pi K}$ (top), $a_0^{3/2}\mu_{\pi K}$ (middle) and of the $KK$ scattering length $a_0^1\Mka$ (bottom).
Results in Tab.~\ref{tab:KpiKK} with statistical error only are not shown.}
\end{figure}

An overview of all scattering lengths with at least one kaon involved is shown in Fig.~\ref{fig:kaonscattering}.
As usual we refrain from displaying data with statistical error only.

In passing, we note that there is an additional paper by Z.~Fu,
  Ref.~\cite{Fu:2012ng},
which deals with $K\bar{K}$ scattering.
It employs one ensemble of $\Nf=2+1$ asqtad (staggered) quarks at $a\simeq0.15\,\fm$, $m_l/m_s=0.2$, $m_s\simeq m_s^\mr{phys}$ with $L=2.5\,\fm$
together with six valence pion masses $\Mpi=334-466\MeV$.
Extrapolating to the physical point, the result for $K\bar{K}$ scattering in the $I=1$ state is $a_0^1\Mka =0.211(33)$.
Hence the interaction for $K\bar{K}$ in the $S$-wave $I=1$ state is found to be attractive, in agreement with LO {\Ch}PT.

In summary, for the quantities $a_0^{1/2}\mu_{\pi K}$, $a_0^{3/2}\mu_{\pi K}$ and $a_0^{1}\Mka$ Refs.~\cite{Helmes:2017smr,Helmes:2018nug} are the only sources without red tags.
Since they appeared in refereed journals and no other works qualify, we take the results quoted in the top two lines of Tab.~\ref{tab:KpiKK} as the current FLAG averages.
For the reader's convenience we list them at the end of Sec.~\ref{sec:SU3_averages}.

Last but not least we like to remind the reader that $KK$ scattering might be outside the validity of $SU(3)$ {\Ch}PT, since it involves a scale around $2\Mka\simeq1\GeV$.
However, our review focuses on the scattering length $a_0^{1}\Mka$, where this issue does not feature prominently.
But it is a key topic in the subsequent conversion of such a scattering length to the low-energy constants $L_i$.
We hope that forthcoming high-quality data will allow a future edition of FLAG to address this topic.


\subsubsection{Implication on Zweig rule violations\label{sec:SU3_Zweig}}

Let us spend a minute to explain why we consider the result on $\Sigma/\Sigma_0$ of $\chi$QCD 21~\cite{Liang:2021pql} particularly interesting.
The reason is linked to the question of how close real-world QCD with $N_c=3$ is to the large-$N_c$ limit of 't\,Hooft (see also Ref.~\cite{Moussallam:1999aq}).
In the large-$N_c$ limit the Zweig rule becomes exact, and the NLO LECs $L_4$ and $L_6$ tend to zero.
As discussed in FLAG 19, the available lattice data are consistent with the view that these two couplings approximately satisfy the Zweig rule.
Also the ratios $F/F_0$, $B/B_0$ and $\Sigma/\Sigma_0$ (note that they are linearly dependent, since $\Sigma=BF^2$ and $\Sigma_0=B_0F_0^2$) test the validity of this rule.

The available data seem to confirm the paramagnetic inequalities of Ref.~\cite{DescotesGenon:1999uh}, which require $\Sigma/\Sigma_0>1$ and $F/F_0>1$.
There is much less information concerning $B/B_0$, and this is the point where the new result of $\chi$QCD 21~\cite{Liang:2021pql} comes in handy.
Let us assume, for the sake of an argument, $F/F_0=1.15(5)(5)$.
Together with $\Sigma/\Sigma_0=1.40(2)(2)$ \cite{Liang:2021pql}, this would imply $B/B_0=1.06(9)(9)$.
This numerical example illustrates how much precision is lost in forming the ratio $(\Sigma/\Sigma_0)/(F/F_0)^2$; with these numbers it would not be clear whether $B/B_0>1$.
Therefore we plead with all collaborations to calculate the numbers $F/F_0$, $B/B_0$ and $\Sigma/\Sigma_0$ in their analysis framework to take advantage of correlations.


\subsubsection{LO and NLO $SU(3)$ estimates\label{sec:SU3_averages}}

For each of the $SU(3)$ LO and NLO LECs discussed in the 2019 FLAG review \cite{Aoki:2019cca} exactly one paper contributed and hence constituted the FLAG average.
The present status is that this situation is unchanged.
For the convenience of the reader, we list the results here but refer to the 2019 FLAG review for the details and explanations.

The LO LECs in the $SU(3)$ chiral limit ($m_u,m_d,m_s\to0$) are denoted by a subscript $0$ to distinguish them from their $SU(2)$ chiral limit counterparts.
The parameters $\Sigma_0,B_0$ are in the $\msbar$ scheme at the renormalization scale $\mu=2\GeV$.
We quote
\begin{align}
&N_f=2+1  :& \FLAGAVBEGIN \Sigma_0^{1/3}      &= 245(8)    \FLAGAVEND \MeV && \Ref~\mbox{\cite{Bazavov:2009fk}} ,\\[-3mm]
\nonumber \\[-3mm] 
&N_f=2+1  :& \FLAGAVBEGIN \Sigma/\Sigma_0     &= 1.48(16)  \FLAGAVEND      && \Ref~\mbox{\cite{Bazavov:2009fk}} ,\\[-3mm]
\nonumber \\[-3mm] 
&N_f=2+1  :& \FLAGAVBEGIN F_0                 &= 80.3(6.0) \FLAGAVEND \MeV && \Ref~\mbox{\cite{Bazavov:2010hj}} ,\\[-3mm]
\nonumber \\[-3mm] 
&N_f=2+1  :& \FLAGAVBEGIN F/F_0               &= 1.104(41) \FLAGAVEND      && \Ref~\mbox{\cite{Bazavov:2009fk}} ,\\[-3mm]
\nonumber \\[-3mm] 
&N_f=2+1  :& \FLAGAVBEGIN B/B_0               &= 1.21(7)   \FLAGAVEND      && \Ref~\mbox{\cite{Bazavov:2009fk}} ,
\label{eq:SU3_LO_LECs}
\end{align}
where the errors include both statistical and systematic uncertainties.
The references shown are the papers from which the results are taken.

For $SU(3)$ NLO LECs we display the results for individual low-energy constants
\begin{align}
&N_f=2+1+1 :& \FLAGAVBEGIN L_4 &= +0.09(34) \times 10^{-3} \FLAGAVEND && \Ref~\mbox{\cite{Dowdall:2013rya}} ,\nonumber\\[-3mm]
\nonumber \\[-3mm]
&N_f=2+1   :& \FLAGAVBEGIN L_4 &= -0.02(56) \times 10^{-3} \FLAGAVEND && \Ref~\mbox{\cite{Bazavov:2010hj}}  ,\\[-3mm]
\nonumber \\[-3mm]
&N_f=2+1+1 :& \FLAGAVBEGIN L_5 &= +1.19(25) \times 10^{-3} \FLAGAVEND && \Ref~\mbox{\cite{Dowdall:2013rya}} ,\nonumber\\[-3mm]
\nonumber \\[-3mm]
&N_f=2+1   :& \FLAGAVBEGIN L_5 &= +0.95(41) \times 10^{-3} \FLAGAVEND && \Ref~\mbox{\cite{Bazavov:2010hj}}  ,\\[-3mm]
\nonumber \\[-3mm]
&N_f=2+1+1 :& \FLAGAVBEGIN L_6 &= +0.16(20) \times 10^{-3} \FLAGAVEND && \Ref~\mbox{\cite{Dowdall:2013rya}} ,\nonumber\\[-3mm]
\nonumber \\[-3mm]
&N_f=2+1   :& \FLAGAVBEGIN L_6 &= +0.01(34) \times 10^{-3} \FLAGAVEND && \Ref~\mbox{\cite{Bazavov:2010hj}}  ,\\[-3mm]
\nonumber \\[-3mm]
&N_f=2+1+1 :& \FLAGAVBEGIN L_8 &= +0.55(15) \times 10^{-3} \FLAGAVEND && \Ref~\mbox{\cite{Dowdall:2013rya}} ,\nonumber\\[-3mm]
\nonumber \\[-3mm]
&N_f=2+1   :& \FLAGAVBEGIN L_8 &= +0.43(28) \times 10^{-3} \FLAGAVEND && \Ref~\mbox{\cite{Bazavov:2010hj}}  ,
\label{eq:SU3_NLO_LECs}
\end{align}
at the chiral scale $\mu=770\MeV$, where again all errors quoted are total errors.
For details of the symmetrization of asymmetric error bars see footnote~\ref{foot:symmetrization}.

For the scattering lengths involving at least one kaon
\begin{align}
&N_f=2+1+1  :& \FLAGAVBEGIN a_0^{1/2}\mu_{\pi K}  &=  0.127(2)   \FLAGAVEND && \Ref~\mbox{\cite{Helmes:2018nug}} ,\\[-3mm]
\nonumber \\[-3mm] 
&N_f=2+1+1  :& \FLAGAVBEGIN a_0^{3/2}\mu_{\pi K}  &= -0.0463(17) \FLAGAVEND && \Ref~\mbox{\cite{Helmes:2018nug}} ,\\[-3mm]
\nonumber \\[-3mm] 
&N_f=2+1+1  :& \FLAGAVBEGIN a_0^1 M_K             &= -0.388(20)  \FLAGAVEND && \Ref~\mbox{\cite{Helmes:2017smr}} ,
\label{eq:scatteringlengths}
\end{align}
represent the FLAG estimates with all errors added in quadrature.
For details of the symmetrization of asymmetric error bars see footnote~\ref{foot:symmetrization}.
Throughout we ask the reader to cite the original references when using these values.

\clearpage
\setcounter{section}{5}
\section{Kaon mixing}
\label{sec:BK}
Authors: P.~Dimopoulos, X. Feng, G.~Herdo\'iza\\

The mixing of neutral pseudoscalar mesons plays an important role in the understanding of the physics of quark-flavour mixing and CP violation. In this section we discuss $K^0 - \bar K^0$ oscillations, which probe the physics of indirect CP violation. Extensive reviews on this subject can be found in  Refs.~\cite{Branco:1999fs,Sozzi:2008bookcp,Buras2020bookcp,Buchalla:1995vs,Buras:1998raa,Lellouch:2011qw}.
 With respect to the FLAG 19 report, in the new Sec.~\ref{sec:Kpipi_amplitudes} of the present
  edition the reader will find  an updated discussion regarding the
  lattice determination of the $K \to \pi\pi$ decay amplitudes and
  related quantities. Discussions concerning the kaon
  mixing within the Standard Model (SM) and Beyond the Standard Model
  (BSM) are presented in Secs.~\ref{sec:BK lattice} and \ref{sec:Bi},
  respectively. We note that FLAG averages for SM and BSM bag
  parameters have not changed with respect to the FLAG 19 report.

\subsection{Indirect CP violation and $\epsilon_{K}$ in the
  SM \label{sec:indCP}} 

Indirect CP violation arises in $K_L \rightarrow \pi \pi$ transitions
through the decay of the $\rm CP=+1$ component of $K_L$ into two pions
(which are also in a $\rm CP=+1$ state). Its measure is defined as
\be 
\epsilon_{K} \,\, = \,\, \dfrac{{\cal A} [ K_L \rightarrow
(\pi\pi)_{I=0}]}{{\cal A} [ K_S \rightarrow (\pi\pi)_{I=0}]} \,\, ,
\ee
with the final state having total isospin zero. The parameter
$\epsilon_{K}$ may also be expressed in terms of $K^0 - \bar K^0$
oscillations.  In the Standard Model, $\epsilon_{K}$ receives
  contributions from: (i) short-distance (SD) physics given by $\Delta
  S = 2$ ``box diagrams" involving $W^\pm$ bosons and $u,c$ and $t$
  quarks; (ii) the long-distance (LD) physics from light hadrons
  contributing to the imaginary part of the dispersive amplitude
  $M_{12}$ used in the two component description of $K^0-\bar{K}^0$
  mixing; (iii) the imaginary part of the absorptive amplitude
  $\Gamma_{12}$ from $K^0-\bar{K}^0$ mixing; and (iv)
  $\text{Im}(A_0)/\text{Re}(A_0)$, where $A_0$ is the $K \to
  (\pi\pi)_{I=0}$ decay amplitude. The various factors in this
  decomposition can vary with phase conventions. In terms of the
  $\Delta
  S = 2$ effective Hamiltonian, ${\cal H}_\text{eff}^{\Delta S = 2}$, it is
  common to represent contribution~(i) by
\be
 \text{Im}(M_{12}^\text{SD}) \equiv \frac{1}{2m_K}\text{Im} [ \langle
   \bar{K}^0 | {\cal H}_\text{eff}^{\Delta S = 2} | K^0 \rangle] ,
\ee
and contribution~(ii) by
$\text{Im}\,(M_{12}^\text{LD})$. Contribution~(iii) can be related to
$\text{Im}(A_0)/\text{Re}(A_0)$ since $(\pi\pi)_{I=0}$ states provide
the dominant contribution to absorptive part of the integral in
$\Gamma_{12}$. Collecting the various pieces yields the following
expression for the $\epsilon_{K}$
factor~\cite{Buras:1998raa,Anikeev:2001rk,Nierste:2009wg,Buras:2008nn,Buras:2010pz}
\be
\epsilon_K \,\,\, = \,\,\, \exp(i \phi_\epsilon) \, \sin(\phi_\epsilon)
  \left[
   \frac{\text{Im}(M_{12}^\text{SD})}{\Delta M_K}
 + \frac{\text{Im}(M_{12}^\text{LD})}{\Delta M_K}
 + \frac{\text{Im}(A_0)}{\text{Re}(A_0)}
  \right] ,
  \label{eq:epsK}
\ee
where the phase of $\epsilon_{K}$ is
given by
\be
\phi_\epsilon \,\,\, = \,\,\, \arctan \frac{\Delta M_{K}}{\Delta
  \Gamma_{K}/2} \,\,\, . 
\ee
The quantities $\Delta M_K$ and $\Delta \Gamma_K$ are the mass and
decay width differences between long- and short-lived neutral kaons.
The experimentally known values of the above quantities
read\,\cite{Zyla:2020zbs}:
\begin{eqnarray}
\vert \epsilon_{K} \vert \,\, &=& \,\, 2.228(11) \times 10^{-3} \,\, , \label{eq:epsilonK_exp}
 \\
\phi_\epsilon \,\, &=& \,\, 43.52(5)^\circ \,\, , \label{eq:phi_epsilonK}
 \\
\Delta M_{K} \,\, &\equiv& \,\, M_{K_{L}} - M_{K_{S}} \,\, = \,\,  3.484(6) \times 10^{-12}\, {\rm MeV} \,\, ,
 \\
\Delta \Gamma_{K}  \,\, &\equiv& \,\ \Gamma_{K_{S}} - \Gamma_{K_{L}} ~\, \,\, = \,\,  7.3382(33) \times 10^{-12} \,{\rm MeV} 
\,\,, 
\end{eqnarray}
where the latter three measurements have been obtained by imposing CPT
symmetry.

We will start by discussing the short-distance effects (i) since they
 provide the dominant contribution to $\epsilon_K$. To lowest
order in the electroweak theory, the contribution to $K^0 -
  \bar K^0$ oscillations arises from the so-called box diagrams, in which
two $W$ bosons and two ``up-type" quarks (i.e., up, charm, top) are
exchanged between the constituent down and strange quarks of the $K$
mesons. The loop integration of the box diagrams can be performed
exactly. In the limit of vanishing external momenta and external quark
masses, the result can be identified with an effective four-fermion
interaction, expressed in terms of the effective Hamiltonian
\be
  {\cal H}_{\rm eff}^{\Delta S = 2} \,\, = \,\,
  \frac{G_F^2 M_{\rm{W}}^2}{16\pi^2} {\cal F}^0 Q^{\Delta S=2} \,\, +
  \,\, {\rm h.c.} \,\,.
  \label{eq:HDeltaS2}
\ee
In this expression, $G_F$ is the Fermi coupling, $M_{\rm{W}}$ the
$W$-boson mass, and
\be
   Q^{\Delta S=2} =
   \left[\bar{s}\gamma_\mu(1-\gamma_5)d\right]
   \left[\bar{s}\gamma_\mu(1-\gamma_5)d\right]
   \equiv O_{\rm VV+AA}-O_{\rm VA+AV} \,\, ,
\label{eq:Q1def}
\ee
is a dimension-six, four-fermion operator. The subscripts V and A denote vector ($\bar{s}\gamma_{\mu} d$) and axial-vector ($\bar{s} \gamma_{\mu} \gamma_5 d$) bilinears, respectively. The function ${\cal F}^0$
is given by
\be
{\cal F}^0 \,\, = \,\, \lambda_c^2 S_0(x_c) \, + \, \lambda_t^2
S_0(x_t) \, + \, 2 \lambda_c  \lambda_t S_0(x_c,x_t)  \,\, , 
\label{eq:F0InamiLin}
\ee
where $\lambda_a = V^\ast_{as} V_{ad}$, and $a=c\,,t$ denotes a
flavour index. The quantities $S_0(x_c),\,S_0(x_t)$ and $S_0(x_c,x_t)$
with $x_c=m_c^2/M_{\rm{W}}^2$, $x_t=m_t^2/M_{\rm{W}}^2$ are the
Inami-Lim functions \cite{Inami:1980fz}, which express the basic
electroweak loop contributions without QCD corrections. The
contribution of the up quark, which is taken to be massless in this
approach, has been taken into account by imposing the unitarity
constraint $\lambda_u + \lambda_c + \lambda_t = 0$.

When strong interactions are included, $\Delta{S}=2$ transitions can
no longer be discussed at the quark level. Instead, the effective
Hamiltonian must be considered between mesonic initial and final
states. Since the strong coupling is large at typical hadronic scales,
the resulting weak matrix element cannot be calculated in perturbation
theory. The operator product expansion (OPE) does, however, factorize
long- and short- distance effects. For energy scales below the charm
threshold, the $K^0-\bar K^0$ transition amplitude of the effective
Hamiltonian can be expressed as
\begin{eqnarray}
\label{eq:Heff}
\langle \bar K^0 \vert {\cal H}_{\rm eff}^{\Delta S = 2} \vert K^0
\rangle  \,\, = \,\, \frac{G_F^2 M_{\rm{W}}^2}{16 \pi^2}  
\Big [ \lambda_c^2 S_0(x_c) \eta_1  \, + \, \lambda_t^2 S_0(x_t)
  \eta_2 \, + \, 2 \lambda_c  \lambda_t S_0(x_c,x_t) \eta_3
  \Big ]  \nn \\ 
\times 
  \left(\frac{\gbar(\mu)^2}{4\pi}\right)^{-\gamma_0/(2\beta_0)}
  \exp\bigg\{ \int_0^{\gbar(\mu)} \, dg \, \bigg(
  \frac{\gamma(g)}{\beta(g)} \, + \, \frac{\gamma_0}{\beta_0g} \bigg)
  \bigg\} 
   \langle \bar K^0 \vert  Q^{\Delta S=2}_{\rm R} (\mu) \vert K^0
   \rangle \,\, + \,\, {\rm h.c.} \,\, ,
\end{eqnarray}
where $\gbar(\mu)$ and $Q^{\Delta S=2}_{\rm R}(\mu)$ are the
renormalized gauge coupling and four-fermion operator in some
renormalization scheme. The factors $\eta_1, \eta_2$ and $\eta_3$
depend on the renormalized coupling $\gbar$, evaluated at the various
flavour thresholds $m_t, m_b, m_c$ and $ M_{\rm{W}}$, as required by
the OPE and Renormalization-Group (RG) running procedure that separate high- and low-energy
contributions. Explicit expressions can be found
in Refs.~\cite{Buchalla:1995vs} and references therein, except that $\eta_1$
and $\eta_3$ have been  calculated to NNLO in
Refs.~\cite{Brod:2011ty} and \cite{Brod:2010mj}, respectively.
We follow the same conventions for the RG equations as in
Ref.~\cite{Buchalla:1995vs}. Thus the Callan-Symanzik function and the
anomalous dimension $\gamma(\gbar)$ of $Q^{\Delta S=2}$ are defined by
\be
\dfrac{d \gbar}{d \ln \mu} = \beta(\gbar)\,,\qquad
\dfrac{d Q^{\Delta S=2}_{\rm R}}{d \ln \mu} =
-\gamma(\gbar)\,Q^{\Delta S=2}_{\rm R} \,\,,  
\label{eq:four_quark_operator_anomalous_dimensions}
\ee
with perturbative expansions
\begin{eqnarray}
\beta(g)  &=&  -\beta_0 \dfrac{g^3}{(4\pi)^2} \,\, - \,\, \beta_1
\dfrac{g^5}{(4\pi)^4} \,\, - \,\, \cdots , 
\label{eq:four_quark_operator_anomalous_dimensions_perturbative}
\\
\gamma(g)  &=&  \gamma_0 \dfrac{g^2}{(4\pi)^2} \,\, + \,\,
\gamma_1 \dfrac{g^4}{(4\pi)^4} \,\, + \,\, \cdots \,.\nn
\end{eqnarray}
We stress that $\beta_0, \beta_1$ and $\gamma_0$ are universal,
i.e., scheme independent. As for $K^0-\bar K^0$ mixing, this is usually considered
in the naive dimensional regularization (NDR) scheme of $\msbar$, and
below we specify the perturbative coefficient $\gamma_1$ in that
scheme:
\begin{eqnarray}
& &\beta_0 = 
         \left\{\frac{11}{3}N-\frac{2}{3}\Nf\right\}, \qquad
   \beta_1 = 
         \left\{\frac{34}{3}N^2-\Nf\left(\frac{13}{3}N-\frac{1}{N}
         \right)\right\}, \label{eq:RG-coefficients}\\[0.3ex]
& &\gamma_0 = \frac{6(N-1)}{N}, \qquad
         \gamma_1 = \frac{N-1}{2N} 
         \left\{-21 + \frac{57}{N} - \frac{19}{3}N + \frac{4}{3}\Nf
         \right\}\,.\nn
\end{eqnarray}
Note that for QCD the above expressions must be evaluated for $N=3$
colours, while $\Nf$ denotes the number of active quark flavours. As
already stated, Eq.~(\ref{eq:Heff}) is valid at scales below the charm
threshold, after all heavier flavours have been integrated out,
i.e., $\Nf = 3$.

In Eq.~(\ref{eq:Heff}), the terms proportional to $\eta_1,\,\eta_2$
and $\eta_3$, multiplied by the contributions containing
$\gbar(\mu)^2$, correspond to the Wilson coefficient of the OPE,
computed in perturbation theory. Its dependence on the renormalization
scheme and scale $\mu$ is canceled by that of the weak matrix element
$\langle \bar K^0 \vert Q^{\Delta S=2}_{\rm R} (\mu) \vert K^0
\rangle$. The latter corresponds to the long-distance effects of the
effective Hamiltonian and must be computed nonperturbatively. For
historical, as well as technical reasons, it is convenient to express
it in terms of the $B$-parameter $B_{K}$, defined as
\be
   B_{K}(\mu)= \frac{{\left\langle\bar{K}^0\left|
         Q^{\Delta S=2}_{\rm R}(\mu)\right|K^0\right\rangle} }{
     {\frac{8}{3}f_{K}^2m_{K}^2}} \,\, .
   \label{eq:defBK}
\ee
The four-quark operator $Q^{\Delta S=2}(\mu)$ is renormalized at scale $\mu$
in some regularization scheme, for instance, NDR-$\msbar$. Assuming that
$B_{K}(\mu)$ and the anomalous dimension $\gamma(g)$ are both known in
that scheme, the renormalization group independent (RGI) $B$-parameter
$\hat{B}_{K}$ is related to $B_{K}(\mu)$ by the exact formula
\be
  \hat{B}_{K} = 
  \left(\frac{\gbar(\mu)^2}{4\pi}\right)^{-\gamma_0/(2\beta_0)}
  \exp\bigg\{ \int_0^{\gbar(\mu)} \, dg \, \bigg(
  \frac{\gamma(g)}{\beta(g)} \, + \, \frac{\gamma_0}{\beta_0g} \bigg)
  \bigg\} 
\, B_{K}(\mu) \,\,\, .
\ee
At NLO in perturbation theory the above reduces to
\be
   \hat{B}_{K} =
   \left(\frac{\gbar(\mu)^2}{4\pi}\right)^{- \gamma_0/(2\beta_0)}
   \left\{ 1+\dfrac{\gbar(\mu)^2}{(4\pi)^2}\left[
   \frac{\beta_1\gamma_0-\beta_0\gamma_1}{2\beta_0^2} \right]\right\}\,
   B_{K}(\mu) \,\,\, .
\label{eq:BKRGI_NLO}
\ee
To this order, this is the scale-independent product of all
$\mu$-dependent quantities in Eq.~(\ref{eq:Heff}).

Lattice-QCD calculations provide results for $B_K(\mu)$.
However, these
results are usually obtained in intermediate schemes other
than the continuum $\msbar$ scheme used to calculate the Wilson
coefficients appearing in Eq.~(\ref{eq:Heff}). Examples of
intermediate schemes are the RI/MOM scheme \cite{Martinelli:1994ty}
(also dubbed the ``Rome-Southampton method'') and the Schr\"odinger
functional (SF) scheme \cite{Luscher:1992an}. These schemes are used
as they allow a nonperturbative renormalization of the four-fermion
operator, using an auxiliary lattice simulation.  This allows
$B_K(\mu)$ to be calculated with percent-level accuracy, as described
below.

In order to make contact with phenomenology, however, and in
particular to use the results presented above, one must convert from
the intermediate scheme to the $\msbar$ scheme or to the RGI quantity
$\hat{B}_{K}$. This conversion relies on 1- or
2-loop
perturbative matching calculations, the truncation errors in which
are, for many recent calculations, the dominant source of error in
$\hat{B}_{K}$ (see, for instance,
Refs.~\cite{Laiho:2011np,Arthur:2012opa,Bae:2014sja,Blum:2014tka,Jang:2015sla}).
While this scheme-conversion error is not, strictly speaking, an error
of the lattice calculation itself, it must be included in results for
the quantities of phenomenological interest, namely,
$B_K(\msbar,2\,{\rm GeV})$ and $\hat{B}_{K}$. Incidentally,
  we remark that this truncation error is estimated in different ways
  and that its relative contribution to the total error can
  considerably differ among the various lattice calculations.  We
note that this error can be minimized by matching between the
intermediate scheme and $\msbar$ at as large a scale $\mu$ as possible
(so that the coupling which determines the rate of convergence is
minimized). Recent calculations have pushed the matching $\mu$ up to
the range $3-3.5\,$GeV. This is possible because of the use of
nonperturbative RG running determined on the
lattice~\cite{Durr:2011ap,Arthur:2012opa,Blum:2014tka}. The
Schr\"odinger functional offers the possibility to run
nonperturbatively to scales $\mu\sim M_{\rm{W}}$ where the truncation
error can be safely neglected. However, so far this has been applied
only for two flavours for $B_K$ in Ref.~\cite{Dimopoulos:2007ht} and for
the case of the BSM bag parameters in Ref.~\cite{Dimopoulos:2018zef}, see
more details in Sec.~\ref{sec:Bi}.

Perturbative truncation errors in Eq.~(\ref{eq:Heff}) also affect the
Wilson coefficients $\eta_1$, $\eta_2$ and~$\eta_3$. It turns out that
the largest uncertainty arises from the charm quark contribution
$\eta_1=1.87(76)$~\cite{Brod:2011ty}. Although it is now calculated at
NNLO, the series shows poor convergence. 
 The net effect from the uncertainty on $\eta_1$ on the amplitude in Eq.~(\ref{eq:Heff}) is larger than that of present 
lattice  calculations of $B_K$.  Exploiting an idea presented
  in Ref.~\cite{Christ:2012se}, it has been recently
  shown in Ref.~\cite{Brod:2019rzc} that, by using the $u-t$ instead of the usual  
$c-t$ unitarity in the $\epsilon_{K}$ computation, the perturbative uncertainties associated   with residual  short-distance quark contributions can be reduced.

 We will now proceed to discuss the remaining contributions to
  $\epsilon_K$ in Eq.~(\ref{eq:epsK}). An analytical estimate of the
  leading contribution from $\Im(M_{12}^\text{LD})$ based on $\chi$PT, shows
  that it is approximately proportional to $\xi \equiv \Im(A_0)/\Re(A_0)$
  so that Eq.~(\ref{eq:epsK}) can be written as follows~\cite{Buras:2008nn,Buras:2010pz}
\be
    \epsilon_{K} \,\,\, = \,\,\, \exp(i \phi_\epsilon) \,\,
    \sin(\phi_\epsilon) \,\, \Big [ \frac{\text{Im}(M_{12}^{\rm SD})} {\Delta M_K }
                \,\,\, + \,\,\, \rho \,\xi \,\, \Big ] \, ,
                \label{eq:epsK-phenom}
\ee
 where the deviation of $\rho$ from one parameterizes the
  long-distance effects in $\Im(M_{12})$.

    In order to facilitate the subsequent
  discussions about the status of the lattice studies of $K \to
  \pi\pi$ and of the current estimates of $\xi$, we proceed by providing a brief
  account  of the parameter $\epsilon^{\prime}$ that describes  
  direct CP-violation in the kaon sector. The definition of $\epsilon^{\prime}$ is given by: 
\be 
\epsilon^{\prime} \equiv \dfrac{1}{\sqrt{2}}\dfrac{{\cal A} [ K_S
    \rightarrow (\pi\pi)_{I=2}] }{{\cal A} [ K_S \rightarrow (\pi\pi)_{I=0}]} \left( \dfrac{{\cal A}[ K_L \rightarrow
        (\pi\pi)_{I=2}]}{{\cal A} [ K_S \rightarrow (\pi\pi)_{I=2}]} - \dfrac{{\cal A}[ K_L \rightarrow
            (\pi\pi)_{I=0}]}{{\cal A} [ K_S \rightarrow (\pi\pi)_{I=0}]} \right)\,.
\ee 
By selecting appropriate phase conventions for the mixing
parameters between $K^0$ and $\bar{K}^0$ CP-eigenstates (see
e.g. Ref.~\cite{Sozzi:2008bookcp} for further details), the expression
of $\epsilon^{\prime}$ can be expressed in terms of the real and
imaginary parts of the isospin amplitudes, as follows
\be
     \epsilon^{\prime} \,\,\, = \,\,\, \dfrac{i \omega \, e^{i (\delta_2 - \delta_0)}}{\sqrt{2}} \,
    \Big [ \dfrac{\text{Im}(A_2)}{\text{Re}(A_2)} - \xi  \, \Big ] \, ,
                    \label{eq:epsprime-phenom}
\ee
where $\omega = \text{Re}(A_2) / \text{Re}(A_0)$, $A_2$
denotes the $\Delta{I}=3/2$ $K\to\pi\pi$ decay amplitude, and $\delta_I$
denotes the strong scattering phase shifts in the corresponding,
$I=0,2$, $K\to(\pi\pi)_I$ decays. Given that the phase,
$\phi_\epsilon^{\prime}=\delta_2 - \delta_0 + \pi/2 =
42.3(1.5)^\circ$~\cite{Zyla:2020zbs} is nearly equal to
$\phi_\epsilon$ in
Eq.~(\ref{eq:phi_epsilonK}), 
the ratio of parameters 
characterizing the direct and indirect CP-violation in the kaon sector
can be approximated in the following way,
\be
\epsilon^{\prime} / \epsilon \,\,\, \approx \,\,\, \Re(\epsilon^{\prime} / \epsilon) \,\,\, = \,\,\, \dfrac{\omega}{\sqrt{2}\, |\epsilon_{K}|} \,
    \Big [ \dfrac{\text{Im}(A_2)}{\text{Re}(A_2)} - \xi  \, \Big ] \, ,
\label{eq:epsprimeOVeps-phenom}
\ee
where  on the left hand side we have set $\epsilon \equiv \epsilon_{K}$.
The experimentally measured value reads~\cite{Zyla:2020zbs},
\be
\Re(\epsilon^{\prime} / \epsilon) = 16.6(2.3) \times
    10^{-4}\,. 
\label{eq:epspovepsexp}
\ee
We remark that isospin breaking and electromagnetic effects (see
Refs.~\cite{Cirigliano:2003nn,Cirigliano:2019cpi}, and the discussion
in Ref.~\cite{Buras2020bookcp}) introduce additional correction terms
into Eq.~(\ref{eq:epsprimeOVeps-phenom}).

\subsection{Lattice-QCD studies of the $K\to(\pi\pi)_I$ decay amplitudes, $\xi$ and $\epsilon^{\prime}/\epsilon$}
\label{sec:Kpipi_amplitudes}

 As a preamble to this section, it should be noted that the study of
 $K \to \pi\pi$ decay amplitudes requires the development of
 computational strategies that are at the forefront of lattice QCD
 techniques. These studies represent a significant
   advance in the study of kaon physics. However, at present, they
   have not yet reached the same level of maturity of most of the
   quantities analyzed in the FLAG report, where, for instance,
   independent results by various lattice collaborations are being
   compared and averaged. In the present version of this section we
   will therefore review the current status of $K \to \pi\pi$ lattice
   computations, but we will provide a FLAG average only for the case
   of the decay amplitude $A_2$.

We start by reviewing the determination of the parameter
$\xi = \Im(A_0)/\Re(A_0)$.
An estimate of $\xi$ has been obtained  from a direct evaluation of the
  ratio of amplitudes $\Im(A_0)/\Re(A_0)$ where $\Im(A_0)$ is
  determined from a lattice-QCD computation  by 
    RBC/UKQCD 20~\cite{Abbott:2020hxn} employing $N_f=2+1$ M\"obius domain wall fermions at
  a single value of the lattice spacing  while $\Re(A_0) \simeq |A_0|$ and   the value $|A_0| = 3.320(2) \times 10^{-7}$ GeV are used based on
  the relevant experimental input~\cite{Zyla:2020zbs} from the decay
  to two pions. This leads to a result for $\xi$ with a rather large
  relative error, 
\begin{equation}
   \xi = -2.1(5)\cdot10^{-4}.
    \label{eq:xilat1}
\end{equation}
Following a similar procedure, an estimate of $\xi$ was
  obtained through the use of a previous lattice QCD determination of
  $\Im(A_0)$ by RBC/UKQCD 15G~\cite{Bai:2015nea}. We refer to
Tab.~\ref{tab_A0_nf21} for further details about these computations of $\Im(A_0)$. The
comparison of the estimates of $\xi$ based on lattice QCD input are
collected in Tab.~\ref{tab_xi_nf21}. 

Another estimate for $\xi$  can be  obtained through a
lattice-QCD computation of the ratio of amplitudes
$\Im(A_2)/\Re(A_2)$ by RBC/UKQCD 15F~\cite{Blum:2015ywa} where the continuum-limit
result is based on computations at two values of the lattice spacing
employing $N_f=2+1$ M\"obius domain wall fermions. Further details about the lattice computations of $A_2$ are collected in Tab.~\ref{tab_A2_nf21}. 
To obtain the value of $\xi$, the expression in Eq.~(\ref{eq:epsprimeOVeps-phenom}) together with the experimental values of
$\Re(\epsilon^{\prime}/\epsilon)$, $|\epsilon_K|$ and $\omega$
are used.   In this case we obtain   $\xi = -1.6  (2)\cdot10^{-4}$.
The use of the updated value of $\text{Im}(A_2)=
  -8.34(1.03) \times 10^{-13}$\,GeV from
  Ref.~\cite{Abbott:2020hxn},\footnote{The update in
      $\text{Im}(A_2)$ is due to a change in the value of the imaginary part of the ratio of CKM matrix elements, $\tau =
  	-V^\ast_{ts}V_{td}/V^\ast_{us}V_{ud}$, as given in
        Ref.~\cite{Tanabashi:2018oca}. The lattice QCD input is
    therefore the one reported in Ref.~\cite{Blum:2015ywa}.}
  in combination with the experimental value of $\text{Re}(A_2) =
  1.479(4) \times 10^{-8}$\,GeV, introduces a small change with
  respect to the above result. The value for $\xi$ reads\,\footnote{  
  The current estimates for  the
  corrections owing to isospin breaking and electromagnetic effects~\cite{Cirigliano:2019cpi}  imply a 
  relative change on the theoretical value for $\epsilon^{\prime} / \epsilon$ by about -20\% 
  with respect to the determination based on Eq.~(\ref{eq:epsprimeOVeps-phenom}). The size 
of these isospin breaking and electromagnetic corrections is related
to the enhancement of the decay amplitudes between the $I=0$ and the
$I=2$ channels. As a consequence, one obtains a similar reduction on $\xi$, leading to a value that is close
  to the result of Eq.~(\ref{eq:xilat1}).} 
\begin{equation}
   \xi = -1.7 (2)\cdot10^{-4}.
\label{eq:xilat2}   
\end{equation}

A phenomenological estimate can also
be obtained from the relationship of $\xi$ to
$\Re (\epsilon^\prime/\epsilon)$, using the experimental value of the latter and further 
assumptions concerning the estimate of hadronic contributions.
The corresponding value of $\xi$ reads~\cite{Buras:2008nn,Buras:2010pz}
\begin{equation}
   \xi = -6.0(1.5)\cdot10^{-2}\sqrt{2}\,|\epsilon_K| 
       = -1.9(5)\cdot10^{-4}. 
       \label{eq:xipheno}
\end{equation}
We note that the use of the experimental value for $\Re(\epsilon^\prime/\epsilon)$ is based on the assumption that it is
free from New Physics contributions. The value of $\xi$ can then be combined with a ${\chi}\rm PT$-based
estimate for the long-range contribution,
$\rho=0.6(3)$~\cite{Buras:2010pz}. Overall, the combination $\rho\xi$
appearing in Eq.~(\ref{eq:epsK-phenom}) leads to a suppression of the
SM prediction of $|\epsilon_K|$ by about $3(2)\%$ relative to the
experimental measurement of $|\epsilon_K|$ given in
Eq.~(\ref{eq:epsilonK_exp}), regardless of whether the
phenomenological estimate of $\xi$ [see Eq.~(\ref{eq:xipheno})] or the
most precise lattice result [see Eq.~(\ref{eq:xilat1})] are used. The
uncertainty in the suppression factor is dominated by the error on
$\rho$.
Although this is a small correction, we note that its
contribution to the error of $\epsilon_K$ is larger than that arising
from the value of $B_{K}$ reported below.

Efforts are under way to compute the long-distance contributions to
$\epsilon_{K}$\,\cite{Bai:2016gzv} and to the $K_L-K_S$ mass
difference in lattice
QCD\,\cite{Christ:2012se,Bai:2014cva,Christ:2015pwa,Wang:2020jpi}. However,
the results are not yet precise enough to improve the accuracy in the
determination of the parameter $\rho$.

The lattice-QCD study of $K \to \pi\pi$ decays provides crucial
input to the SM prediction of $\epsilon_{K}$. We now
  proceed to describe the current status of these computations. In recent years, the RBC/UKQCD collaboration has undertaken a series of lattice-QCD
calculations of $K \to \pi\pi$ decay amplitudes~\cite{Blum:2015ywa,Bai:2015nea,Abbott:2020hxn}.
In 2015, the first calculation of the $K\to(\pi\pi)_{I=0}$ decay
amplitude $A_0$ was performed using physical kinematics on a
$32^3\times64$ lattice with an inverse lattice spacing of $a^{-1}=1.3784(68)$
GeV~\cite{Bai:2015nea, Bai:2016ocm}. 
The main features of the RBC/UKQCD 15G calculation included, fixing the $I=0$
$\pi\pi$ energy very close to the kaon mass by imposing G-parity boundary conditions, a continuum-like operator mixing pattern through the use of a domain wall fermion action with accurate chiral symmetry, and the construction of the complete set of correlation functions by computing seventy-five distinct diagrams. Results for the real and the imaginary parts of the decay amplitude $A_0$ from the RBC/UKQCD 15G computation are collected in Tab.~\ref{tab_A0_nf21}, where the first error is statistical and the second one is systematic. 
\begin{table}[!t]
	\mbox{} \\[3.0cm]
	\footnotesize
	\begin{tabular*}{\textwidth}[l]{@{\extracolsep{\fill}}l@{\hspace{1mm}}r@{\hspace{1mm}}l@{\hspace{1mm}}l@{\hspace{1mm}}l@{\hspace{1mm}}l@{\hspace{1mm}}l@{\hspace{1mm}}l@{\hspace{1mm}}l@{\hspace{5mm}}l@{\hspace{1mm}}c@{\hspace{1mm}}c}
		Collaboration & Ref. & $\Nf$ & 
		\hspace{0.15cm}\begin{rotate}{60}{publication status}\end{rotate}\hspace{-0.15cm} &
		\hspace{0.15cm}\begin{rotate}{60}{continuum extrapolation}\end{rotate}\hspace{-0.15cm} &
		\hspace{0.15cm}\begin{rotate}{60}{chiral extrapolation}\end{rotate}\hspace{-0.15cm}&
		\hspace{0.15cm}\begin{rotate}{60}{finite volume}\end{rotate}\hspace{-0.15cm}&
		\hspace{0.15cm}\begin{rotate}{60}{renormalization}\end{rotate}\hspace{-0.15cm}  &
		\hspace{0.15cm}\begin{rotate}{60}{running/matching}\end{rotate}\hspace{-0.15cm} & 
		\rule{0.15cm}{0cm} $\text{Re}(A_0)$ & 
		$\text{Im}(A_0)$ \\
		&&&&&&&&&  {\scriptsize $[10^{-7}$~GeV$]$} & {\scriptsize $[10^{-11}$~GeV$]$} \\[0.5ex]
		&&&&&&&&&&\\[-0.1cm]
		\hline
		\hline
		&&&&&&&&&& \\[-0.1cm]
		RBC/UKQCD 20 & \cite{Abbott:2020hxn} & 2+1 & \gA & \tbr & \soso & \soso
		& \good&  $\,a$ &2.99(0.32)(0.59) &$-6.98(0.62)(1.44)$ \\[0.5ex]
		RBC/UKQCD 15G & \cite{Bai:2015nea} & 2+1 & \gA & \tbr & \soso & \soso
		& \good&  $\,b$ &4.66(1.00)(1.26) &$-1.90(1.23)(1.08)$ \\[0.5ex]

		&&&&&&&&&& \\[-0.1cm]
		\hline
		\hline\\[-0.1cm]
	\end{tabular*}
	\begin{minipage}{\linewidth}
		{\footnotesize 
			\begin{itemize}
				\item[$a$] Nonperturbative renormalization with the RI/SMOM scheme
				at a scale of 1.53\,GeV and running to 4.01\,GeV employing a nonperturbatively determined step-scaling function. Conversion to $\msbar$ at 1-loop order.     \\[-5mm]
		       	\item[$b$] Nonperturbative renormalization with the RI/SMOM scheme
		       at a scale of 1.53\,GeV. Conversion to $\msbar$ at 1-loop order.     \\[-5mm]	
		\end{itemize}
		}
	\end{minipage}
	\caption{Results for the real and imaginary parts of the
           $K \to \pi\pi$  decay
		amplitude $A_0$ from lattice-QCD computations with $\Nf=2+1$
		dynamical flavours. Information about the renormalization, running
		and matching to the $\msbar$ scheme is indicated in
		the column ``running/matching", with details given at the bottom of
		the table. We refer to the text for
                  further details about the main differences between
                  the lattice computations in Refs.~\cite{Abbott:2020hxn}~and~\cite{Bai:2015nea}.}
	\label{tab_A0_nf21}
\end{table}

\begin{table}[h]
	\mbox{} \\[3.0cm]
	\footnotesize
	\begin{tabular*}{\textwidth}[l]{@{\extracolsep{\fill}}l@{\hspace{1mm}}r@{\hspace{1mm}}l@{\hspace{1mm}}l@{\hspace{1mm}}l@{\hspace{1mm}}l@{\hspace{1mm}}l@{\hspace{1mm}}l@{\hspace{1mm}}l@{\hspace{5mm}}l@{\hspace{1mm}}c@{\hspace{1mm}}c}
		Collaboration & Ref. & $\Nf$ & 
		\hspace{0.15cm}\begin{rotate}{60}{publication status}\end{rotate}\hspace{-0.15cm} &
		\hspace{0.15cm}\begin{rotate}{60}{continuum extrapolation}\end{rotate}\hspace{-0.15cm} &
		\hspace{0.15cm}\begin{rotate}{60}{chiral extrapolation}\end{rotate}\hspace{-0.15cm}&
		\hspace{0.15cm}\begin{rotate}{60}{finite volume}\end{rotate}\hspace{-0.15cm}&
		\hspace{0.15cm}\begin{rotate}{60}{renormalization}\end{rotate}\hspace{-0.15cm}  &
		\hspace{0.15cm}\begin{rotate}{60}{running/matching}\end{rotate}\hspace{-0.15cm} & 
		\rule{0.15cm}{0cm} $\text{Re}(A_2)$ & 
		$\text{Im}(A_2)$ \\
		&&&&&&&&&  {\scriptsize $[10^{-8}$~GeV$]$} & {\scriptsize $[10^{-13}$~GeV$]$} \\[0.5ex]
		&&&&&&&&&&\\[-0.1cm]
		\hline
		\hline
		&&&&&&&&&& \\[-0.1cm]
		RBC/UKQCD 15F & \cite{Blum:2015ywa} & 2+1 & \gA & \soso & \soso & \good
		& \good&  $\,a$ & 1.50(0.04)(0.14)&  $-8.34(1.03) ^\diamond$  \\[0.5ex]
		&&&&&&&&&& \\[-0.1cm]
		\hline
		\hline\\[-0.1cm]
	\end{tabular*}
	\begin{minipage}{\linewidth}
		{\footnotesize 
			\begin{itemize}
				\item[$a$] Nonperturbative renormalization with the RI/SMOM scheme
				at a scale of 3 GeV. Conversion to $\msbar$ at 1-loop order.     \\[-5mm]
				\item[$^\diamond$] This value of $\text{Im}(A_2)$ is an update  
					reported in
                                        Ref.~\cite{Abbott:2020hxn}
                                        which is based on the lattice QCD computation in Ref.~\cite{Blum:2015ywa} but where a change in the
					value of the imaginary part of the ratio of CKM matrix elements $\tau =
					-V^\ast_{ts}V_{td}/V^\ast_{us}V_{ud}$ reported in Ref.~\cite{Tanabashi:2018oca} has been applied.
				\\[-5mm]
			\end{itemize}
		}
	\end{minipage}
	\caption{Results for the real and the imaginary parts of the
          $K \to \pi\pi$ decay
		amplitude $A_2$ from lattice-QCD computations with $\Nf=2+1$
		dynamical flavours. Information about the renormalization
		and matching to the $\msbar$ scheme is indicated in
		the column ``running/matching", with details given at the bottom of
		the table. }
	\label{tab_A2_nf21}
\end{table}

\begin{table*}[t!]
	\begin{center}
		\mbox{} \\[3.0cm]
		{\footnotesize{
				\vspace*{-2cm}\begin{tabular*}{\textwidth}[l]{l
                                    @{\extracolsep{\fill}} c c c c }
					Collaboration & Ref. &
                                         $\Nf$ &  \rule{0.3cm}{0cm}$\xi$  		 \\
					&& \\[-0.1cm]
					\hline
					\hline
					&& \\[-0.1cm]
					
					RBC/UKQCD 20$^\dagger$ &
                                        \cite{Abbott:2020hxn} & 2+1 & $-2.1(5)\cdot10^{-4}$  \\[0.5ex]
				    RBC/UKQCD 15G$^\diamond$  &
				    \cite{Bai:2015nea} & 2+1 &  $-0.6(5)\cdot10^{-4}$  \\[0.5ex]
					RBC/UKQCD 15F$^\ast$  &
                                        \cite{Blum:2015ywa} & 2+1 &  $-1.7(2)\cdot10^{-4}$  \\[0.5ex]
                  	&& \\[-0.1cm]
					\hline
					\hline\\[-0.1cm]
				\end{tabular*}
		}}
		\begin{minipage}{\linewidth}
			{\footnotesize 
				\begin{itemize}
					\item[ $^\dagger$] Estimate for $\xi$  obtained  from a direct  evaluation of the
					ratio of amplitudes $\Im(A_0)/\Re(A_0)$ where $\Im(A_0)$ is
					determined from the
					lattice-QCD computation of
					Ref.~\cite{Abbott:2020hxn}
					while for $\Re(A_0) \simeq
					|A_0|$ is taken from  the experimental value for  $|A_0|$.  
						\item[ $^\diamond$] Estimate for $\xi$  obtained  from a direct  evaluation of the
					ratio of amplitudes $\Im(A_0)/\Re(A_0)$ where $\Im(A_0)$ is
					determined from the
					lattice-QCD computation of
					Ref.~\cite{Bai:2015nea}
					while for $\Re(A_0) \simeq
					|A_0|$ is taken from  the experimental value for  $|A_0|$.
					\item[ $^\ast$]  Estimate for
						$\xi$ based on the use of
						Eq.~(\ref{eq:epsprimeOVeps-phenom}).
						The new value of
                                                $\text{Im}(A_2)$
                                                reported in
                                                Ref.~\cite{Abbott:2020hxn}---based on the
                                                lattice-QCD
                                                computation of
                                                Ref.~\cite{Blum:2015ywa}
                                                following an update of
                                                a nonlattice input---is used in combination with
						the experimental values for
						$\text{Re}(A_2)$,
						$\Re(\epsilon^{\prime}/\epsilon)$,
						$|\epsilon_K|$ and $\omega$.

				\end{itemize}
			}
		\end{minipage}
		\caption{Results for the parameter
			$\xi=\Im(A_0)/\Re(A_0)$ obtained through the
			combination of lattice-QCD determinations of $K \to
			\pi\pi$ decay amplitudes with $\Nf=2+1$ dynamical
			flavours and experimental
			inputs.  \label{tab_xi_nf21}}
	\end{center}
\end{table*}

The latest 2020 calculation RBC/UKQCD 20~\cite{Abbott:2020hxn} using the same lattice setup has improved the 2015 calculation RBC/UKQCD 15G~\cite{Bai:2015nea} in three
important aspects: (i) an increase by a factor of 3.4 in statistics; (ii) the inclusion of
a scalar two-quark operator and the addition of another pion-pion
operator to isolate the ground state, and (iii) the use of step 
scaling techniques
to raise the renormalization
scale from 1.53 GeV to 4.01 GeV.
The updated determinations of  the real and the imaginary
    parts of $A_0$ in Ref.~\cite{Abbott:2020hxn} are shown in Tab.~\ref{tab_A0_nf21}.

As previously discussed, the 
determination of $\operatorname{Im}(A_0)$ from Ref.~\cite{Abbott:2020hxn}  has been used to 
  obtain the value of the parameter $\xi$ in Eq.~(\ref{eq:xilat1}). 
A first-principles computation of $\operatorname{Re}(A_0)$ is
  essential to address the so-called $\Delta I=1/2$ puzzle associated
  to the enhancement of $\Delta I=1/2$ over $\Delta I=3/2$ transitions
  owing, crucially, to long distance effects. Indeed, short-distance enhancements in the Wilson coefficients
are not large enough to explain the $\Delta I=1/2$
rule~\cite{Gaillard:1974nj,Altarelli:1974exa}.
Lattice-QCD calculations do provide a method to study such a
long-distance enhancement. The combination of the result 
for $A_0$ in Tab.~\ref{tab_A0_nf21}  with the earlier lattice calculation of $A_2$ in Ref.~\cite{Blum:2015ywa} leads to the ratio,
$\operatorname{Re}(A_0)/\operatorname{Re}(A_2)=19.9(5.0)$, which
agrees with the experimentally measured value,
$\operatorname{Re}(A_0)/\operatorname{Re}(A_2)=22.45(6)$. 
In Ref.~\cite{Abbott:2020hxn}, the lattice
determination of relative size of direct CP violation was updated as follows,
\be
\operatorname{Re}(\epsilon'/\epsilon)=21.7(2.6)(6.2)(5.0)\times10^{-4},
 \label{eq:epspoveps}
\ee
where the first two errors are statistical and systematic, respectively. 
The third error arises from the omitted strong and electromagnetic
isospin breaking effects. 
The value of
$\operatorname{Re}(\epsilon'/\epsilon)$ in Eq.~(\ref{eq:epspoveps}) uses the experimental values
of $\Re(A_0)$ and $\Re(A_2)$.  The lattice determination of $\operatorname{Re}(\epsilon'/\epsilon)$ 
is in good agreement with the experimental result in Eq.~(\ref{eq:epspovepsexp}). However,  while the result  in Eq.~(\ref{eq:epspoveps}) represents a significant step forward, it is 
important to keep in mind that the calculation of $A_0$ is currently based on a single value of the lattice spacing. 
It is expected that future work with additional values of the lattice spacing will contribute to improve the precision. 
For a description of the computation of the $\pi\pi$ scattering phase shifts entering in 
the determination of $\operatorname{Re}(\epsilon'/\epsilon)$ in Eq.~(\ref{eq:epspoveps}), we refer to Ref.~\cite{Blum:2021fcp}.

The real and imaginary values of the amplitude $A_2$
  have been determined by RBC/UKQCD 15F~\cite{Blum:2015ywa}
  employing $N_f=2+1$ M\"obius domain wall fermions at two values of
  the lattice spacing, namely $a=0.114$\,fm and $0.083$\,fm, and
  performing simulations  at the physical pion mass with $M_{\pi}L
  \approx 3.8$.

A compilation of  lattice results  for the real and imaginary parts of the $K \to \pi\pi$ decay amplitudes, $A_0$ and $A_2$, with $\Nf=2+1$  flavours of
	dynamical quarks is shown in
        Tabs.~\ref{tab_A0_nf21}~and~\ref{tab_A2_nf21}. In Appendix~\ref{app-Kpipi} we collect the corresponding information about  the lattice QCD
        simulations, including  the values of some of the most relevant
        parameters. The results for the parameter $\xi$, determined
        through the combined use of $K \to \pi \pi$ amplitudes
        computed on the lattice and experimental inputs,  are
        presented  in Tab.~\ref{tab_xi_nf21}. As previously discussed,
        we remark that the total uncertainty on the reported values of
        $\xi$ depends on the specific way in which the
        lattice and experimental inputs are selected.
    
The determination of the real and imaginary parts
  of $A_2$ by RBC/UKQCD 15F shown in Tab.~\ref{tab_A2_nf21} is free
  of red tags. We therefore quote the following FLAG averages:
%
%
\begin{align}
&&\FLAGAVBEGIN \text{Re}(A_2) &= 1.50(0.04)(0.14) \times 10^{-8}\FLAGAVEND~\text{GeV},  &\nonumber\\[-3mm]
& \Nf=2+1: & &&\Ref~\mbox{\cite{Blum:2015ywa}}.\\[-3mm]
&&\FLAGAVBEGIN \text{Im}(A_2) &= -8.34(1.03) \times 10^{-13}\FLAGAVEND~\text{GeV},  & \nonumber
\end{align}
%
Besides the RBC/UKQCD collaboration programme~\cite{Blum:2015ywa,Bai:2015nea,Abbott:2020hxn} using
domain-wall fermions, an approach based on improved Wilson
fermions~\cite{Ishizuka:2015oja, Ishizuka:2018qbn} has presented a
determination of the $K \to \pi\pi$ decay amplitudes,  $A_0$ and
$A_2$, at unphysical quark masses.
For an analysis of the scaling with the number of colours of $K \to \pi\pi$
decay amplitudes using lattice-QCD computations, we refer to
Refs.~\cite{Donini:2016lwz,Donini:2020qfu}.

Recent proposals aiming at the inclusion of electromagnetism in
lattice-QCD calculations of $K \to \pi\pi$ decays are being
explored~\cite{Christ:2017pze,Cai:2018why} in order to reduce the
uncertainties associated   with isospin breaking effects.

Finally, we notice that $\epsilon_{K}$ receives a contribution from
$|V_{cb}|$ through the $\lambda_t$ parameter in
Eq.~(\ref{eq:F0InamiLin}). The present uncertainty on $|V_{cb}|$ has a
significant impact on the error of $\epsilon_{K}$ (see,
e.g.,~Refs.~\cite{Bailey:2015tba,Bailey:2018feb} and the recent update in Ref.~\cite{Kim:2019vic}).

\subsection{Lattice computation of $B_{K}$}
\label{sec:BK lattice}

Lattice calculations of $B_{K}$ are affected by the same type of
systematic effects discussed in previous sections of this review. However, the issue
of renormalization merits special attention. The reason is that the
multiplicative renormalizability of the relevant operator $Q^{\Delta
S=2}$ is lost once the regularized QCD action ceases to be invariant
under chiral transformations.  As a result, the
  renormalization pattern of $B_K$ depends on the specific choice
  of the fermionic discretization.

In the case of Wilson fermions, $Q^{\Delta S=2}$
mixes with four additional dimension-six operators, which belong to
different representations of the chiral group, with mixing
coefficients that are finite functions of the gauge coupling. This
complicated renormalization pattern was identified as the main source
of systematic error in earlier, mostly quenched calculations of
$B_{K}$ with Wilson quarks. It can be bypassed via the
implementation of specifically designed methods, which are either
based on Ward identities~\cite{Becirevic:2000cy} or on a modification
of the Wilson quark action, known as twisted-mass
QCD~\cite{Frezzotti:2000nk,Dimopoulos:2006dm,Dimopoulos:2007cn}.

An advantage of staggered fermions is the presence of a remnant $U(1)$
chiral symmetry. However, at nonvanishing lattice spacing, the
symmetry among the extra unphysical degrees of freedom (tastes) is
broken. As a result, mixing with other dimension-six operators cannot
be avoided in the staggered formulation, which complicates the
determination of the $B$-parameter.   In general, taste conserving
  mixings are implemented directly in the lattice computation of the
  matrix element.  The effects of the broken taste symmetry are
usually treated  through an effective field theory, staggered
  Chiral Perturbation Theory
  (S$\chi$PT)~\cite{VandeWater:2005uq,Bailey:2012wb},
  parameterizing the quark-mass and lattice-spacing dependences.

Fermionic lattice actions based on the Ginsparg-Wilson
relation~\cite{Ginsparg:1981bj} are invariant under the chiral group,
and hence four-quark operators such as $Q^{\Delta S=2}$ renormalize
multiplicatively. However, depending on the particular formulation of
Ginsparg-Wilson fermions, residual chiral symmetry breaking effects
may be present in actual calculations. 
For instance, in the case of domain-wall fermions, the finiteness of the extra 5th dimension implies that the decoupling of modes with different chirality is not exact, which produces a residual nonzero quark mass in the chiral limit. The mixing with dimension-six operators of different chirality is expected to be an $\cO(m_{\rm{res}}^2)$ suppressed effect~\cite{Aoki:2005ga, Christ:2005xh} that should be investigated on a case-by-case basis.

Before proceeding to the description and compilation of the results of $B_K$, we would like to reiterate a discussion presented in the previous FLAG report about an issue related to the
computation of the kaon bag parameters through lattice-QCD simulations with $\Nf=2+1+1$ dynamical quarks. In practice, this only concerns the calculations of the kaon $B$-parameters including dynamical charm-quark effects in Ref.~\cite{Carrasco:2015pra}, that
were examined in the FLAG 16 report. As described in Sec.~\ref{sec:indCP}, the effective Hamiltonian in Eq.~(\ref{eq:HDeltaS2}) depends solely on the operator $Q^{\Delta
S=2}$ in Eq.~(\ref{eq:Q1def}) ---which appears in the definition of $B_K$ in Eq.~(\ref{eq:defBK})--- at energy scales below the charm threshold where charm-quark contributions are absent. As a result, a computation of $B_K$ based on $\Nf=2+1+1$ dynamical simulations will include an extra sea-quark contribution from charm-quark loop effects for which there is at present no direct evaluation in the literature.

When the matrix element of $Q^{{\Delta}S=2}$ is evaluated in a theory
that contains a dynamical
charm quark, the resulting estimate for $B_K$ must then be matched to
the three-flavour theory that underlies the effective four-quark
interaction.\footnote{We thank Martin L\"uscher for an interesting
  discussion on this issue.} In general, the matching of $2+1$-flavour
QCD with the theory containing $2+1+1$ flavours of sea quarks 
is performed around the charm threshold.  It is usually accomplished by requiring that the coupling and quark  masses are equal in the two theories at a renormalization scale $\mu$ around $m_c$. In addition, $B_K$ should be renormalized and run, in the four-flavour theory, to the value of $\mu$ at which the two theories are matched, as described in Sec.~\ref{sec:indCP}. The corrections associated with this matching are of
order $(E/m_c)^2$, where $E$ is a typical energy in the process under study, since the subleading operators have dimension eight
\cite{Cirigliano:2000ev}.

When the kaon-mixing amplitude is considered, the matching also
involves the relation between the relevant box diagrams and the
effective four-quark operator. In this case, corrections of order
$(E/m_c)^2$ arise not only from the charm quarks in the sea, but also
from the valence sector, since the charm quark propagates in the box
diagrams. We note that the original derivation of the effective
four-quark interaction is valid up to corrections of order
$(E/m_c)^2$. The kaon-mixing amplitudes evaluated in the
  $\Nf=2+1$ and $2+1+1$ theories are thus subject to corrections of
  the same order in $E/m_c$ as the derivation of the conventional
  four-quark interaction.

Regarding perturbative QCD corrections at the scale of the charm-quark mass on the amplitude in Eq.~(\ref{eq:Heff}), 
the uncertainty on  $\eta_1$ and
$\eta_3$ factors is of $\cO(\alpha_s(m_c)^3)$~\cite{Brod:2011ty,Brod:2010mj}, while that on $\eta_2$ is 
of
$\cO(\alpha_s(m_c)^2)$~\cite{Buras:1990fn}.\,\footnote{The recent results~\cite{Brod:2019rzc} based on the use of $u-t$ unitarity for the two corresponding perturbative factors, also have an uncertainty of $\cO(\alpha_s(m_c)^2)$ and $\cO(\alpha_s(m_c)^3)$. The estimates for the     missing higher-order contributions are, however, expected to be reduced with respect to the more traditional case where $c-t$ unitarity is used.}
On the other hand, the corrections of order $(E/m_c)^2$ 
due to dynamical charm-quark effects in the matching of the amplitudes are further suppressed by powers of $\alpha_s(m_c)$ and by a factor of
$1/N_c$, given that they arise from quark-loop diagrams.
In order to make progress in resolving this so far uncontrolled
systematic uncertainty, it is essential that any future calculation of
$B_K$ with $\Nf=2+1+1$ flavours properly addresses the size of these
residual dynamical charm effects in a quantitative way.

Another issue in this context is how the lattice scale and the
physical values of the quark masses are determined in the $2+1$ and
$2+1+1$ flavour theories. Here it is important to consider in which
way the quantities used to fix the bare parameters are affected by a
dynamical charm quark.

A recent study~\cite{Hollwieser:2020qri} using three degenerate light
quarks, together with a charm quark, indicates that the deviations
between the $\Nf=3+1$ and the $\Nf=3$ theories are considerably below the 1\%
level in dimensionless quantities constructed from ratios of gradient flow
observables, such as $t_0$ and $w_0$, used for scale
setting. This study extends the nonperturbative investigations with
two heavy mass-degenerate
quarks~\cite{Bruno:2014ufa,Athenodorou:2018wpk} which indicate that
dynamical charm-quark effects in low-energy hadronic observables are
considerably smaller than the expectation from a naive power counting
in terms of $\alpha_s(m_c)$. For an additional discussion on this
point, we refer to Ref.~\cite{Carrasco:2015pra}.
Given the hierarchy of scales between the charm-quark
  mass and that of $B_K$, we expect these errors to be modest, but a more
quantitative understanding is needed as statistical errors on $B_K$
are reduced. Within this review we will not discuss this issue
further. However, we wish to point out that the present discussion also
applies to $\Nf=2+1+1$ computations of the kaon BSM $B$-parameters discussed in Sec.~\ref{sec:Bi}.

A compilation of  results  for $B_{K}$ with $\Nf=2, 2+1$ and $2+1+1$ flavours of
dynamical quarks is shown in Tabs.~\ref{tab_BKsumm}
and~\ref{tab_BKsumm_nf2}, as well as Fig.~\ref{fig_BKsumm}. An
overview of the quality of systematic error studies is represented by
the colour coded entries in Tabs.~\ref{tab_BKsumm}
and~\ref{tab_BKsumm_nf2}. 
The values of the most relevant lattice parameters, and comparative tables on the various
estimates of systematic errors have been  collected in the
corresponding Appendices of the previous FLAG editions~\cite{Aoki:2019cca,
  Aoki:2016frl, Aoki:2013ldr}.

Since the last edition of the FLAG report no new results for $B_{K}$ have appeared in the bibliography.  
We mention here an ongoing 
work related to the $B_{K}$ computation where the relevant operators are defined in the gradient flow 
framework. In a first publication~\cite{Suzuki:2020zue} the small flow time expansion method is applied in order to compute, 
to 1-loop approximation, the finite matching coefficients between
the gradient flow and the
$\overline{\rm MS}$ schemes for the operators entering the 
$B_K$ computation.
 
For a detailed description of previous $B_K$ calculations---and in particular those considered in the computation of 
the average values---we refer
the reader to the FLAG 19~\cite{Aoki:2019cca}, FLAG 16~\cite{Aoki:2016frl} and FLAG 13~\cite{Aoki:2013ldr} reports.

We now give the global averages for $B_K$ for $N_f=2+1+1, 2+1 $ and 2 dynamical flavours. 
The details about the calculation of these averages can be found in  FLAG 19~\cite{Aoki:2019cca}. 

\begin{figure}[ht]
\centering
\includegraphics[width=13cm]{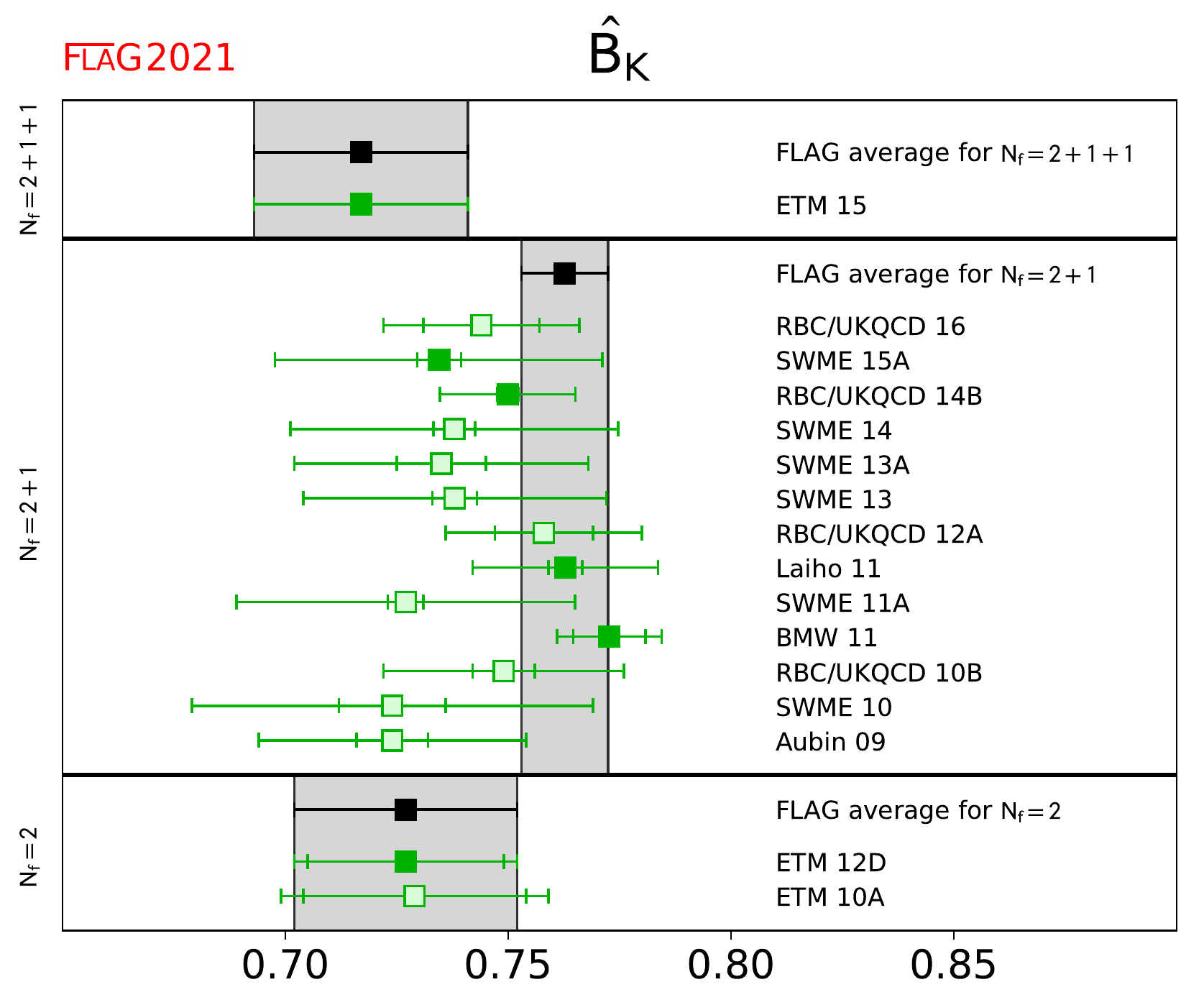}
\caption{Recent unquenched lattice results for the RGI $B$-parameter
  $\hat{B}_{K}$. The grey bands indicate our global averages
  described in the text. For $\Nf=2+1+1$ and $\Nf=2$ the global averages coincide
  with the results by ETM 15 and ETM 12D, respectively. \label{fig_BKsumm}}
\end{figure}

We begin with the $N_f=2+1$ global average since it is estimated by employing four different $B_K$ results, namely BMW 11~\cite{Durr:2011ap}, Laiho 11~\cite{Laiho:2011np}, RBC/UKQCD 14B~\cite{Blum:2014tka} and
SWME 15A~\cite{Jang:2015sla}. Note also that the expression of $\epsilon_K$ in terms of $B_K$ is obtained in the three-flavour  theory (see Sec.~\ref{sec:indCP}). After constructing the global covariance matrix according to Schmelling~\cite{Schmelling:1994pz}, we arrive at:
%
\begin{equation}
  \Nf=2+1:\hspace{2.5cm}\FLAGAVBEGIN \hat{B}_{K} = 0.7625(97)\FLAGAVEND\qquad\Refs~\mbox{\cite{Durr:2011ap,Laiho:2011np,Blum:2014tka,Jang:2015sla}},
\end{equation}
%
with $\chi^2/{\rm dof}=0.675$. After applying the NLO conversion
factor $\hat{B}_{K}/B_{K}^\msbar (2\,{\rm GeV})=1.369$,\footnote{We refer to the FLAG 19 report~\cite{Aoki:2019cca} for a discussion about the 
estimates of  these conversion factors.} this
translates into
\begin{equation}
  \Nf=2+1:\hspace{1cm} B_{K}^\msbar(2\,{\rm GeV})=0.5570(71)\qquad\Refs~\mbox{\cite{Durr:2011ap,Laiho:2011np,Blum:2014tka,Jang:2015sla}}.
\end{equation}
Note that the statistical
errors of each calculation entering the global average are small
enough to make their results statistically incompatible. It is
only because of the relatively large systematic errors that the
weighted average produces a value of $\cO(1)$ for the reduced $\chi^2$.

There is only a single  result for $N_f=2+1+1$, computed by the ETM
collaboration\,\cite{Carrasco:2015pra}. Since it is free of red tags,
it qualifies as the currently best global average, i.e., 
%
\begin{equation}
\Nf=2+1+1:\hspace{.2cm}\FLAGAVBEGIN\hat{B}_{K} = 0.717(18)(16)\FLAGAVEND\, ,
\hspace{.2cm}B_{K}^\msbar (2\,{\rm GeV}) = 0.524(13)(12)\hspace{.3cm}\Ref~\mbox{\cite{Carrasco:2015pra}}.
\end{equation}
%
For $\Nf=2$ flavours the  best global average is given by a single result, that 
of ETM 12D\,\cite{Bertone:2012cu}: 
%
\begin{equation}
\Nf=2:\hspace{.3cm}\FLAGAVBEGIN\hat{B}_{K} = 0.727(22)(12)\FLAGAVEND ,
\hspace{.3cm}B_{K}^\msbar (2\,{\rm GeV}) = 0.531(16)(19)\hspace{.4cm}\Ref~\mbox{\cite{Bertone:2012cu}}.
\end{equation}
%
The result in the $\msbar$ scheme has been obtained by applying the
same conversion factor of 1.369 as in the three-flavour theory.

\begin{table*}[ht]
\begin{center}
\mbox{} \\[3.0cm]
{\footnotesize{
\vspace*{-2cm}\begin{tabular*}{\textwidth}[l]{l @{\extracolsep{\fill}} r@{\hspace{1mm}}l@{\hspace{1mm}}l@{\hspace{1mm}}l@{\hspace{1mm}}l@{\hspace{1mm}}l@{\hspace{1mm}}l@{\hspace{1mm}}l@{\hspace{1mm}}l@{\hspace{1mm}}l}
Collaboration & Ref. & $\Nf$ & 
\hspace{0.15cm}\begin{rotate}{60}{publication status}\end{rotate}\hspace{-0.15cm} &
\hspace{0.15cm}\begin{rotate}{60}{continuum extrapolation}\end{rotate}\hspace{-0.15cm} &
\hspace{0.15cm}\begin{rotate}{60}{chiral extrapolation}\end{rotate}\hspace{-0.15cm}&
\hspace{0.15cm}\begin{rotate}{60}{finite volume}\end{rotate}\hspace{-0.15cm}&
\hspace{0.15cm}\begin{rotate}{60}{renormalization}\end{rotate}\hspace{-0.15cm}  &
\hspace{0.15cm}\begin{rotate}{60}{running}\end{rotate}\hspace{-0.15cm} & 
\rule{0.3cm}{0cm}$B_{{K}}(\overline{\rm MS},2\,{\rm GeV})$ 
& \rule{0.3cm}{0cm}$\hat{B}_{{K}}$ \\
&&&&&&&&&& \\[-0.1cm]
\hline
\hline
&&&&&&&&&& \\[-0.1cm]

ETM 15 & \cite{Carrasco:2015pra} & 2+1+1 & \gA & \good & \soso & \soso
& \good&  $\,a$ &   0.524(13)(12)  & 0.717(18)(16)$^1$ \\[0.5ex]
&&&&&&&&&& \\[-0.1cm]
\hline
&&&&&&&&&& \\[-0.1cm]
RBC/UKQCD 16 & \cite{Garron:2016mva} & 2+1 & \gA & \soso & \soso &
\soso & \good & $\,b$ & 0.543(9)(13)$^2$ & 0.744(13)(18)$^3$ \\[0.5ex]

SWME 15A & \cite{Jang:2015sla} & 2+1 & \gA & \good & \soso &
\good & \soso$^\ddagger$  & $-$ & 0.537(4)(26) & 0.735(5)(36)$^4$ \\[0.5ex]

RBC/UKQCD 14B
& \cite{Blum:2014tka} & 2+1 & \gA & \good & \good &
     \good  & \good & $\,b$  & 0.5478(18)(110)$^2$ & 0.7499(24)(150) \\[0.5ex]  

SWME 14 & \cite{Bae:2014sja} & 2+1 & \gA & \good & \soso &
\good & \soso$^\ddagger$  & $-$ & 0.5388(34)(266) & 0.7379(47)(365) \\[0.5ex]

SWME 13A & \cite{Bae:2013tca} & 2+1 & \gA & \good & \soso  &
\good & \soso$^\ddagger$  & $-$ & 0.537(7)(24) & 0.735(10)(33) \\[0.5ex]

SWME 13 & \cite{Bae:2013lja} & 2+1 & \rC & \good & \soso &
\good & \soso$^\ddagger$ & $-$ & 0.539(3)(25) & 0.738(5)(34) \\[0.5ex]

RBC/UKQCD 12A
& \cite{Arthur:2012opa} & 2+1 & \gA & \soso & \good &
     \soso & \good & $\,b$ & 0.554(8)(14)$^2$ & 0.758(11)(19) \\[0.5ex]  

Laiho 11 & \cite{Laiho:2011np} & 2+1 & \rC & \good & \soso &
     \soso & \good & $-$ & 0.5572(28)(150)& 0.7628(38)(205)$^4$ \\[0.5ex]  

SWME 11A & \cite{Bae:2011ff} & 2+1 & \gA & \good & \soso &
\soso & \soso$^\ddagger$ & $-$ & 0.531(3)(27) & 0.727(4)(38) \\[0.5ex]

BMW 11 & \cite{Durr:2011ap} & 2+1 & \gA & \good & \good & \good & \good
& $\,c$ & 0.5644(59)(58) & 0.7727(81)(84) \\[0.5ex]

RBC/UKQCD 10B & \cite{Aoki:2010pe} & 2+1 & \gA & \soso & \soso & \good &
\good & $\,d$ & 0.549(5)(26) & 0.749(7)(26) \\[0.5ex] 

SWME 10 & \cite{Bae:2010ki} & 2+1 & \gA & \good & \soso & \soso & \soso
& $-$ & 0.529(9)(32) &  0.724(12)(43) \\[0.5ex] 

Aubin 09 & \cite{Aubin:2009jh} & 2+1 & \gA & \soso & \soso &
     \soso & \tbg & $-$ & 0.527(6)(21)& 0.724(8)(29) \\[0.5ex]  

&&&&&&&&&& \\[-0.1cm]
\hline
\hline\\[-0.1cm]
\end{tabular*}
}}
\begin{minipage}{\linewidth}
{\footnotesize 
\begin{itemize}
\item[$^\ddagger$] The renormalization is performed using perturbation
        theory at 1-loop, with a conservative estimate of the uncertainty. \\[-5mm]
\item[$a$]  $B_K$ is renormalized nonperturbatively at scales $1/a \sim 2.2-3.3\,\gev$ in the $\Nf = 4$ RI/MOM scheme 
     using two different lattice momentum scale intervals, the first around $1/a$ while the second around  3.5 GeV. 
        The impact of the two ways to the final 
        result is taken into account  in the error budget. Conversion to $\msbar$ is at 1-loop at 3 GeV.  \\[-5mm]
\item[$b$] $B_K$ is renormalized nonperturbatively at a scale of 1.4 GeV
        in two RI/SMOM schemes for $\Nf = 3$, and 
	then run to 3 GeV using a nonperturbatively determined step-scaling
        function. 
	Conversion to $\msbar$ is at 1-loop order at 3 GeV.\\[-5mm]
\item[$c$] $B_K$ is renormalized and run nonperturbatively to a scale of
        $3.5\,\gev$ in the RI/MOM scheme. At the same scale conversion at 1-loop to $\msbar$ is applied. 
	Nonperturbative and NLO
        perturbative running agrees down to scales of $1.8\,\gev$ within
        statistical
	uncertainties of about 2\%.\\[-5mm]
\item[$d$] $B_K$ is renormalized nonperturbatively at a scale of 2\,GeV
        in two RI/SMOM schemes for $\Nf = 3$, and then 
	run to 3 GeV using a nonperturbatively determined step-scaling
        function. Conversion to $\msbar$ is at 
	1-loop order at 3 GeV.\\[-5mm]
\item[$^1$] $B_{K}(\msbar, 2\,\gev)$ and $\hat{B}_{{K}}$ are related
        using the conversion factor  1.369, i.e., the one obtained
        with $N_f=2+1$.  \\[-5mm]
\item[$^2$] $B_{K}(\msbar, 2\,\gev)$ is obtained from the estimate for
        $\hat{B}_{{K}}$ using the conversion factor 1.369.   \\[-5mm]
\item[$^3$] $\hat{B}_{{K}}$ is obtained from $B_{K}(\msbar, 3\,\gev)$ using the conversion factor 
            employed in  Ref.~\cite{Blum:2014tka}.           \\[-5mm]             
\item[$^4$] $\hat{B}_{{K}}$ is obtained from the estimate for
        $B_{K}(\msbar, 2\,\gev)$ using the conversion factor 1.369. 
\end{itemize}
}
\end{minipage}
\caption{Results for the kaon $B$-parameter in QCD with $\Nf=2+1+1$
  and $\Nf=2+1$ dynamical flavours, together with a summary of
  systematic errors.
      Information about nonperturbative 
  running is indicated in the column ``running", with details given at
  the bottom of the table.\label{tab_BKsumm}}
\end{center}
\end{table*}

\clearpage

\begin{table*}[h]
\begin{center}
\mbox{} \\[3.0cm]
{\footnotesize{
\vspace*{-2cm}\begin{tabular*}{\textwidth}[l]{l @{\extracolsep{\fill}} r l l l l l l l l l}
Collaboration & Ref. & $\Nf$ & 
\hspace{0.15cm}\begin{rotate}{60}{publication status}\end{rotate}\hspace{-0.15cm} &
\hspace{0.15cm}\begin{rotate}{60}{continuum extrapolation}\end{rotate}\hspace{-0.15cm} &
\hspace{0.15cm}\begin{rotate}{60}{chiral extrapolation}\end{rotate}\hspace{-0.15cm}&
\hspace{0.15cm}\begin{rotate}{60}{finite volume}\end{rotate}\hspace{-0.15cm}&
\hspace{0.15cm}\begin{rotate}{60}{renormalization}\end{rotate}\hspace{-0.15cm}  &
\hspace{0.15cm}\begin{rotate}{60}{running}\end{rotate}\hspace{-0.15cm} & 
\rule{0.3cm}{0cm}$B_{{K}}(\overline{\rm MS},2\,{\rm GeV})$ 
& \rule{0.3cm}{0cm}$\hat{B}_{{K}}$ \\
&&&&&&&&&& \\[-0.1cm]
\hline
\hline
&&&&&&&&&& \\[-0.1cm]

ETM 12D & \cite{Bertone:2012cu} & 2 & \gA & \good & \soso & \soso
& \good&  $\,e$ &   0.531(16)(9)  & 0.727(22)(12)$^1$ \\[0.5ex]
ETM 10A & \cite{Constantinou:2010qv} & 2 & \gA & \good & \soso & \soso
& \good&  $\,f$ &   0.533(18)(12)$^1$  & 0.729(25)(17) \\[0.5ex]
&&&&&&&&&& \\[-0.1cm]
\hline
\hline\\[-0.1cm]
\end{tabular*}
}}
\begin{minipage}{\linewidth}
{\footnotesize 
\begin{itemize}
\item[$e$] $B_K$ is renormalized nonperturbatively at scales $1/a \sim 2
        - 3.7\,\gev$ in the $\Nf = 2$ RI/MOM scheme. In this
        scheme, nonperturbative and NLO
        perturbative running are shown to agree from 4 GeV down to 2 GeV to
        better than 3\%
        \cite{Constantinou:2010gr,Constantinou:2010qv}.  \\[-5mm]
\item[$f$] $B_K$ is renormalized nonperturbatively at scales $1/a \sim 2
        - 3\,\gev$ in the $\Nf = 2$ RI/MOM scheme. In this
        scheme, nonperturbative and NLO
        perturbative running are shown to agree from 4 GeV down to 2 GeV to
        better than 3\%
        \cite{Constantinou:2010gr,Constantinou:2010qv}.  \\[-5mm]
        
\item[$^1$] $B_{K}(\msbar, 2\,\gev)$ and $\hat{B}_{{K}}$ are related using the conversion factor  1.369, i.e., the one obtained with $N_f=2+1$. 
\end{itemize}
}
\end{minipage}
\caption{Results for the kaon $B$-parameter in QCD with $\Nf=2$
  dynamical flavours, together with a summary of systematic
  errors.  Information about nonperturbative 
  running is
  indicated in the column ``running", with details given at the bottom
  of the table.\label{tab_BKsumm_nf2}}
\end{center}
\end{table*}

\subsection{Kaon BSM $B$-parameters}
\label{sec:Bi}

We now report on lattice results concerning the matrix elements of
operators that encode the effects of physics beyond the Standard Model
(BSM) to the mixing of neutral kaons. In this theoretical framework
both the SM and BSM contributions add up to reproduce the
experimentally observed value of $\epsilon_K$. Since BSM contributions
involve heavy but unobserved particles they are short-distance
dominated. The effective Hamiltonian for generic ${\Delta}S=2$
processes including BSM contributions reads
\begin{equation}
  {\cal H}_{\rm eff,BSM}^{\Delta S=2} = \sum_{i=1}^5
  C_i(\mu)Q_i(\mu),
\end{equation}
where $Q_1$ is the four-quark operator of Eq.~(\ref{eq:Q1def}) that
gives rise to the SM contribution to $\epsilon_K$. In the so-called
SUSY basis introduced by Gabbiani et al.~\cite{Gabbiani:1996hi} the
 operators $Q_2,\ldots,Q_5$ read\,\footnote{Thanks to QCD
  parity invariance lattice computations for three more dimension-six operators,
  whose parity conserving parts coincide with the corresponding parity
  conserving contributions of the operators $Q_1, Q_2$ and $Q_3$, can be ignored.}
\begin{eqnarray}
 & & Q_2 = \big(\bar{s}^a(1-\gamma_5)d^a\big)
           \big(\bar{s}^b(1-\gamma_5)d^b\big), \nonumber\\
 & & Q_3 = \big(\bar{s}^a(1-\gamma_5)d^b\big)
           \big(\bar{s}^b(1-\gamma_5)d^a\big), \nonumber\\
 & & Q_4 = \big(\bar{s}^a(1-\gamma_5)d^a\big)
           \big(\bar{s}^b(1+\gamma_5)d^b\big), \nonumber\\
 & & Q_5 = \big(\bar{s}^a(1-\gamma_5)d^b\big)
           \big(\bar{s}^b(1+\gamma_5)d^a\big),
\end{eqnarray}
where $a$ and $b$ denote colour indices.  In analogy to the case of
$B_{K}$ one then defines the $B$-parameters of $Q_2,\ldots,Q_5$
according to
\be
   B_i(\mu) = \frac{\left\langle \bar{K}^0\left| Q_i(\mu)\right|K^0
     \right\rangle}{N_i\left\langle\bar{K}^0\left|\bar{s}\gamma_5
     d\right|0\right\rangle \left\langle0\left|\bar{s}\gamma_5
     d\right|K^0\right\rangle}, \quad i=2,\ldots,5.
\ee
The factors $\{N_2,\ldots,N_5\}$ are given by $\{-5/3, 1/3, 2, 2/3\}$,
and it is understood that $B_i(\mu)$ is specified in some
renormalization scheme, such as $\msbar$ or a variant of the
regularization-independent momentum subtraction (RI-MOM) scheme.

The SUSY basis has been adopted in
Refs.~\cite{Boyle:2012qb,Bertone:2012cu,Carrasco:2015pra,Garron:2016mva}. Alternatively,
one can employ the chiral basis of Buras, Misiak and
Urban\,\cite{Buras:2000if}. The SWME collaboration prefers the latter
since the anomalous dimension that enters the RG running has been
calculated to 2-loops in perturbation
theory\,\cite{Buras:2000if}. Results obtained in the chiral basis can
be easily converted to the SUSY basis via
\be
   B_3^{\rm SUSY}={\textstyle\frac{1}{2}}\left( 5B_2^{\rm chiral} -
   3B_3^{\rm chiral} \right).
\ee
The remaining $B$-parameters are the same in both bases. In the
following we adopt the SUSY basis and drop the superscript.

Older quenched results for the BSM $B$-parameters can be found in
Refs.~\cite{Allton:1998sm, Donini:1999nn, Babich:2006bh}. For a nonlattice approach 
to get estimates for the BSM $B$-parameters see Ref.~\cite{Buras:2018lgu}.  

Estimates for $B_2,\ldots,B_5$ have been reported for QCD with $\Nf=2$
(ETM 12D~\cite{Bertone:2012cu}), $\Nf=2+1$
(RBC/UKQCD 12E\,\cite{Boyle:2012qb}, SWME 13A\,\cite{Bae:2013tca},
SWME 14C\,\cite{Jang:2014aea}, SWME 15A\,\cite{Jang:2015sla}, \\
RBC/UKQCD 16 \cite{Garron:2016mva,Boyle:2017skn})  and $\Nf=2+1+1$
(ETM 15\,\cite{Carrasco:2015pra}) flavours of dynamical quarks. 
Since the publication of the FLAG 19 report~\cite{Aoki:2019cca} 
no new results for the BSM $B$-parameters have appeared 
in the bibliography. The available results 
are listed and compared in Tab.~\ref{tab_Bi} and
Fig.~\ref{fig_Bisumm}. In general one finds that the BSM $B$-parameters computed by different collaborations do not show the same
level of consistency as the SM kaon-mixing parameter $B_K$ discussed
previously. 
Control over the systematic uncertainties from chiral and continuum extrapolations as well as finite-volume effects in $B_2,\ldots,B_5$
is expected to be at  a commensurate level as for $B_{K}$, as far as the
results by ETM 12D, ETM 15, SWME 15A and RBC/UKQCD 16 are concerned,
since the set of gauge ensembles employed in both kinds of computations is the same. 
The calculation by RBC/UKQCD 12E has been performed at a single value of
the lattice spacing and a minimum pion mass of 290\,MeV. 

Let us notice that as reported in  RBC/UKQCD 16~\cite{Garron:2016mva} the comparison of 
results obtained in the conventional RI-MOM and two RI-SMOM
schemes  shows significant discrepancies for $B_4$ and
$B_5$ in the $\overline{\rm{MS}}$ scheme at the scale of 3\,GeV, which
amount up to 2.8$\sigma$ in the case of $B_5$. By contrast, the
agreement for $B_2$ and $B_3$ determined for different intermediate
scheme is much better. 
The RBC/UKQCD collaboration has presented an ongoing
study~\cite{Boyle:2018eor} in which simulations with two values of
the lattice spacing at the physical point and with a third finer
lattice spacing at $M_\pi = 234$ MeV are employed in order to obtain
the BSM matrix elements in the continuum limit.  Results are still
preliminary.

The findings by RBC/UKQCD 16 \cite{Garron:2016mva,Boyle:2017skn}  provide evidence
that the nonperturbative determination of the matching factors depends
strongly on the details of the implementation of the Rome-Southampton
method. The use of nonexceptional momentum configurations in the
calculation of the vertex functions produces a significant
modification of the renormalization factors, which affects the matching
between $\overline{\rm{MS}}$ and the intermediate momentum subtraction
scheme. This effect is most pronounced in $B_4$ and $B_5$.  Furthermore,  
it can be noticed that the estimates for $B_4$ and $B_5$ from RBC/UKQCD 16 are much
closer to those of SWME 15A. At the same time, the results for $B_2$
and $B_3$ obtained by ETM 15, SWME 15A and RBC/UKQCD 16 are in good
agreement within errors.

\begin{table}[!h]
\begin{center}
\mbox{} \\[3.0cm]
{\footnotesize{
\begin{tabular*}{\textwidth}[l]{l @{\extracolsep{\fill}}r@{\hspace{1mm}}l@{\hspace{1mm}}l@{\hspace{1mm}}l@{\hspace{1mm}}l@{\hspace{1mm}}l@{\hspace{1mm}}l@{\hspace{1mm}}l@{\hspace{1mm}}l@{\hspace{1mm}}l@{\hspace{1mm}}l@{\hspace{1mm}}l}
Collaboration & Ref. & $\Nf$ & 
\hspace{0.15cm}\begin{rotate}{60}{publication status}\end{rotate}\hspace{-0.15cm} &
\hspace{0.15cm}\begin{rotate}{60}{continuum extrapolation}\end{rotate}\hspace{-0.15cm} &
\hspace{0.15cm}\begin{rotate}{60}{chiral extrapolation}\end{rotate}\hspace{-0.15cm}&
\hspace{0.15cm}\begin{rotate}{60}{finite volume}\end{rotate}\hspace{-0.15cm}&
\hspace{0.15cm}\begin{rotate}{60}{renormalization}\end{rotate}\hspace{-0.15cm}  &
\hspace{0.15cm}\begin{rotate}{60}{running}\end{rotate}\hspace{-0.15cm} & 
$B_2$ & $B_3$ & $B_4$ & $B_5$ \\
&&&&&&&&& \\[-0.1cm]
\hline
\hline
&&&&&&&&& \\[-0.1cm]

ETM 15 & \cite{Carrasco:2015pra} & 2+1+1 & \gA & \good & \soso & \soso
& \good&  $\,a$ & 0.46(1)(3) & 0.79(2)(5) & 0.78(2)(4) & 0.49(3)(3)  \\[0.5ex]
&&&&&&&&& \\[-0.1cm]

\hline

&&&&&&&&&& \\[-0.1cm]
RBC/UKQCD 16 & \cite{Garron:2016mva} & 2+1 & \gA & \soso & \soso &
\soso & \good & $\,b$ & 0.488(7)(17) & 0.743(14)(65) & 0.920(12)(16) &
0.707(8)(44) \\[0.5ex]

&&&&&&&&& \\[-0.1cm]
SWME 15A & \cite{Jang:2015sla} & 2+1 & \gA & \good & \soso &
\good & \soso$^\dagger$ & $-$ & 0.525(1)(23) & 0.773(6)(35) & 0.981(3)(62) & 0.751(7)(68)  \\[0.5ex]
&&&&&&&&& \\[-0.1cm]

SWME 14C & \cite{Jang:2014aea} & 2+1 & C & \good & \soso &
\good & \soso$^\dagger$ & $-$ & 0.525(1)(23) & 0.774(6)(64) & 0.981(3)(61) & 0.748(9)(79)  \\[0.5ex]
&&&&&&&&& \\[-0.1cm]

SWME 13A$^\ddagger$ & \cite{Bae:2013tca} & 2+1 & \gA & \good & \soso  &
\good & \soso$^\dagger$ & $-$ & 0.549(3)(28)  & 0.790(30) & 1.033(6)(46) & 0.855(6)(43)   \\[0.5ex]
&&&&&&&&& \\[-0.1cm]

RBC/
& \cite{Boyle:2012qb} & 2+1 & \gA & \tbr & \soso & \good &
\good & $\,b$ & 0.43(1)(5)  & 0.75(2)(9)  & 0.69(1)(7)  & 0.47(1)(6)
\\
UKQCD 12E & & & & & & & & & & & & \\[0.5ex]  
&&&&&&&&& \\[-0.1cm]

\hline

&&&&&&&&& \\[-0.1cm]
ETM 12D & \cite{Bertone:2012cu} & 2 & \gA & \good & \soso & \soso
& \good&  $\,c$ & 0.47(2)(1)  & 0.78(4)(2)  & 0.76(2)(2)  & 0.58(2)(2)  \\[0.5ex]

&&&&&&&&& \\[-0.1cm]
\hline
\hline\\[-0.1cm]
\end{tabular*}
}}
\begin{minipage}{\linewidth}
{\footnotesize 
\begin{itemize}
\item[$^\dagger$] The renormalization is performed using perturbation
        theory at 1-loop, with a conservative estimate of
         the uncertainty. \\[-5mm]
\item[$a$] $B_i$ are renormalized nonperturbatively at scales $1/a
        \sim 2.2-3.3\,\gev$ in the $\Nf = 4$ RI/MOM scheme 
        using two different lattice momentum scale intervals, with
        values around $1/a$ for the first and around
        3.5~GeV for the second one. The impact of
        these two ways to the final result is taken into account
         in the error budget. Conversion to $\msbar$ is at 1-loop at 3~GeV.\\[-5mm]
\item[$b$] The $B$-parameters are renormalized nonperturbatively at a scale of 3~GeV. \\[-5mm]
\item[$c$] $B_i$ are renormalized nonperturbatively at scales $1/a
        \sim 2-3.7\,\gev$ in the $\Nf = 2$ RI/MOM scheme using
        two different lattice momentum scale intervals,  
        with values around $1/a$ for the first and around 3~GeV
        for the second one.\\[-5mm]
\item[$^\ddagger$] The computation of $B_4$ and $B_5$ has been
        revised in Refs.~\cite{Jang:2015sla} and \cite{Jang:2014aea}. 
\end{itemize}
}
\end{minipage}
\caption{Results for the BSM $B$-parameters $B_2,\ldots,B_5$ in the
  $\msbar$ scheme at a reference scale of 3\,GeV.    Information about nonperturbative 
  running is indicated in the column
  ``running", with details given at the bottom of the
  table.~\label{tab_Bi}}
\end{center}
\end{table}
\clearpage

 A nonperturbative computation of the running of the four-fermion
operators contributing to the $B_2$, \dots , $B_5$ parameters has been
carried out with two dynamical flavours using the Schr\"odinger
functional renormalization
scheme~\cite{Dimopoulos:2018zef}. Renormalization matrices of the
operator basis are used to build step-scaling functions governing the
continuum-limit running between hadronic and electroweak scales. A
comparison to perturbative results using NLO (2-loops) for the
four-fermion operator anomalous dimensions indicates that, at scales
of about 3\,GeV, nonperturbative effects can induce a sizeable
contribution to the running.

A detailed look at the 
calculations reported in the works of ETM 15
\cite{Carrasco:2015pra}, SWME 15A \cite{Jang:2015sla} and
RBC/UKQCD 16 \cite{Garron:2016mva} reveals that cutoff effects appear
to be larger for the BSM $B$-parameters compared to $B_K$. Depending
on the details of the renormalization procedure and/or the fit ansatz 
for the combined chiral and continuum extrapolation, the
results obtained at the coarsest lattice spacing differ by
15--30\%. At the same time the available range of lattice spacings is
typically much reduced compared to the corresponding calculations of
$B_K$, as can be seen by comparing the quality criteria in Tabs.~\ref{tab_BKsumm} and \ref{tab_Bi}. Hence, the impact of the
renormalization procedure and the continuum limit on the BSM $B$-parameters certainly requires further investigation.

Finally we present our estimates for the BSM $B$-parameters, quoted in
the $\msbar$-scheme at scale 3\,GeV. For
$N_f=2+1$ our estimate is given by the average between the results
from SWME 15A and RBC/UKQCD 16, i.e.,
%
%
\begin{align}
  & \Nf=2+1: \\ 
  &\FLAGAVBEGIN B_2=0.502(14)\FLAGAVEND,\quad 
     \FLAGAVBEGIN B_3=0.766(32)\FLAGAVEND,\quad 
     \FLAGAVBEGIN B_4=0.926(19)\FLAGAVEND,\quad
      \FLAGAVBEGIN B_5=0.720(38)\FLAGAVEND,
  \quad\Refs~\mbox{\cite{Jang:2015sla,Garron:2016mva}}.  \nonumber
\end{align}
%
For $N_f=2+1+1$ and $N_f=2$, our estimates coincide with the ones by
ETM\ 15 and ETM 12D, respectively, since there is only one computation
for each case. Thus we quote
\begin{align}
  & \Nf=2+1+1: \\ 
  & \FLAGAVBEGIN B_2=0.46(1)(3)\FLAGAVEND,\quad
      \FLAGAVBEGIN B_3=0.79(2)(5)\FLAGAVEND,\quad 
      \FLAGAVBEGIN B_4=0.78(2)(4)\FLAGAVEND,\quad 
      \FLAGAVBEGIN B_5=0.49(3)(3)\FLAGAVEND, \quad\Ref~\mbox{\cite{Carrasco:2015pra}}, \nonumber\\  \nonumber\\ 
%
%
  & \Nf=2:  \\ 
  &\FLAGAVBEGIN B_2=0.47(2)(1)\FLAGAVEND,\quad
     \FLAGAVBEGIN B_3=0.78(4)(2)\FLAGAVEND,\quad
     \FLAGAVBEGIN B_4=0.76(2)(2)\FLAGAVEND,\quad 
     \FLAGAVBEGIN B_5=0.58(2)(2)\FLAGAVEND, 
  \quad\Ref~\mbox{\cite{Bertone:2012cu}}. \nonumber
\end{align}
%
Based on the above discussion on the effects of employing different
intermediate momentum subtraction schemes in the nonperturbative
renormalization of the operators, the discrepancy for $B_4$ and $B_5$
results between $N_f=2, 2+1+1$ and $N_f=2+1$ computations should not
be considered an effect associated with the number of dynamical
flavours. To clarify the present situation, it would be
  important to perform a direct comparison of results by the ETM
  collaboration obtained both with RI-MOM and RI-SMOM
  methods. Furthermore, extending the computation of the BSM-$B$
  parameters to include physical point simulations with improved
  continuum-limit extrapolations would also provide valuable
  information. As a closing remark, we encourage authors to provide
the correlation matrix of the $B_i$ parameters since this information
is required in phenomenological studies of New Physics scenarios.

\begin{figure}[ht]
\centering
\leavevmode
\includegraphics[width=\textwidth]{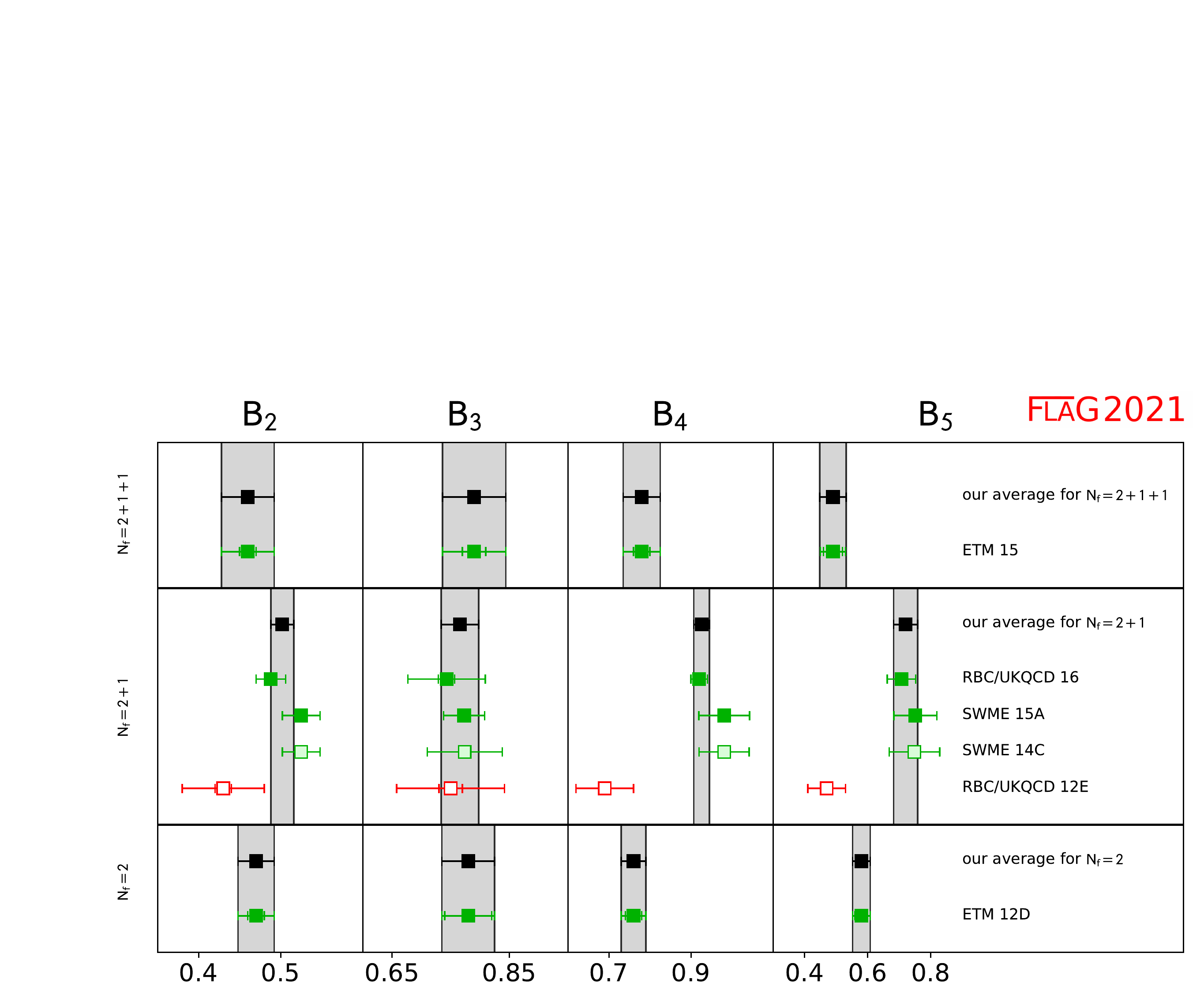}
\caption{Lattice results for the BSM $B$-parameters defined in the
  $\msbar$ scheme at a reference scale of 3\,GeV, see Tab.~\ref{tab_Bi}.
\label{fig_Bisumm}}
\end{figure}

\clearpage
\input{HQ/macros_static.sty}
\setcounter{section}{6}
\clearpage

\section{Charm hadron decay constants and form factors}
\label{sec:DDecays}
Authors: Y.~Aoki, M.~Della~Morte, E.~Lunghi, S.~Meinel, C.~Monahan, C.~Pena\\

Leptonic and semileptonic decays of charmed $D$ and $D_s$ mesons or $\Lambda_c$ and other charm baryons occur
via charged $W$-boson exchange, and are sensitive probes of $c \to d$
and $c \to s$ quark flavour-changing transitions.  Given experimental
measurements of the branching fractions combined with sufficiently
precise theoretical calculations of the hadronic matrix elements, they
enable the determination of the CKM matrix elements $|V_{cd}|$ and
$|V_{cs}|$ (within the Standard Model) and a precise test of the
unitarity of the second row of the CKM matrix.  Here, we summarize the
status of lattice-QCD calculations of the charmed leptonic decay
constants.  Significant progress has
been made in charm physics on the lattice in recent years,
largely due to the availability of gauge configurations produced using
highly-improved lattice-fermion actions that enable treating the
$c$ quark with the same action as for the $u$, $d$, and $s$ quarks.

This section updates the corresponding one in the last FLAG review~\cite{Aoki:2019cca} for results that
appeared before April 30, 2021.
As already done in Ref.~\cite{Aoki:2019cca}, we limit our
review to results based on modern simulations with reasonably light
pion masses (below approximately 500~MeV). 

Following our review of lattice-QCD calculations of $D_{(s)}$-meson
leptonic decay constants and charm-hadron semileptonic form factors, we then
interpret our results within the context of the Standard Model.  We
combine our best-determined values of the hadronic matrix elements
with the most recent experimentally-measured branching fractions to
obtain $|V_{cd(s)}|$ and test the unitarity of the second row of the
CKM matrix.

\subsection{Leptonic decay constants $f_D$ and $f_{D_s}$}
\label{sec:fD}

In the Standard Model, and up to electromagnetic corrections,
the decay constant $f_{D_{(s)}}$ of a
pseudoscalar $D$ or $D_s$ meson is related to the branching ratio for
leptonic decays mediated by a $W$ boson through the formula
\be
{\mathcal{B}}(D_{(s)} \to \ell\nu_\ell)= {{G_F^2|V_{cq}|^2 \tau_{D_{(s)}}}\over{8 \pi}} f_{D_{(s)}}^2 m_\ell^2 
m_{D_{(s)}} \left(1-{{m_\ell^2}\over{m_{D_{(s)}}^2}}\right)^2\;,
 \label{eq:Dtoellnu}
\ee
where $q$ is $d$ or $s$ and $V_{cd}$ ($V_{cs}$) is the appropriate CKM matrix element for a
$D$ ($D_s$) meson.  The branching fractions have been experimentally
measured by CLEO, Belle, Babar and BES with a precision around 4--5$\%$ for
both the $D$ and the $D_s$-meson
decay modes~\cite{Rosner:2015wva}.  When
combined with lattice results for the decay constants, they allow for
determinations of $|V_{cs}|$ and $|V_{cd}|$.

In lattice-QCD calculations, the decay constants $f_{D_{(s)}}$ are extracted from 
Euclidean  matrix elements of the axial current
\be
\langle 0| A^{\mu}_{cq} | D_q(p) \rangle = if_{D_q}\;p_{D_q}^\mu  \;,
\label{eq:dkconst}
\ee
with $q=d,s$ and $ A^{\mu}_{cq} =\bar{c}\gamma_\mu \gamma_5
q$. Results for $N_f=2,\; 2+1$ and $2+1+1$ dynamical flavours are
summarized in Tab.~\ref{tab_FDsummary} and Fig.~\ref{fig:fD}.
Since the publication of the last FLAG review, a handful of results
for  $f_D$ and $f_{D_s}$ have appeared, as described below.
We consider isospin-averaged quantities, although, in a few cases, results for $f_{D^+}$ are quoted
(see, for example, the FNAL/MILC~11,14A and 17 computations, where the difference between $f_D$ and $f_{D^+}$
has been estimated to be around 0.5 MeV).

Only one new computation appeared for $N_f=2$. Reference~\cite{Balasubramanian:2019net}, Balasubramanian~19, updates the result for $f_{D_s}$  in Blossier~18~\cite{Blossier:2018jol} (discussed in the previous review)
by including in the analysis two additional ensembles at a coarser lattice spacing ($a=0.075$ fm, compared to $0.065$ fm and $0.048$ fm used
in Ref.~\cite{Blossier:2018jol}). Pion masses at this coarser resolution reach 282 MeV and $M_\pi L$ is always kept larger than 4.
\begin{table}[h!]
\begin{center}
\mbox{} \\[3.0cm]
\footnotesize
\begin{tabular*}{\textwidth}[l]{@{\extracolsep{\fill}}l@{\hspace{1mm}}r@{\hspace{1mm}}l@{\hspace{1mm}}l@{\hspace{1mm}}l@{\hspace{1mm}}l@{\hspace{1mm}}l@{\hspace{1mm}}l@{\hspace{1mm}}l@{\hspace{1mm}}l@{\hspace{1mm}}l@{\hspace{1mm}}l}
Collaboration & Ref. & $\Nf$ & 
\hspace{0.15cm}\begin{rotate}{60}{publication status}\end{rotate}\hspace{-0.15cm} &
\hspace{0.15cm}\begin{rotate}{60}{continuum extrapolation}\end{rotate}\hspace{-0.15cm} &
\hspace{0.15cm}\begin{rotate}{60}{chiral extrapolation}\end{rotate}\hspace{-0.15cm}&
\hspace{0.15cm}\begin{rotate}{60}{finite volume}\end{rotate}\hspace{-0.15cm}&
\hspace{0.15cm}\begin{rotate}{60}{renormalization/matching}\end{rotate}\hspace{-0.15cm}  &
\hspace{0.15cm}\begin{rotate}{60}{heavy-quark treatment}\end{rotate}\hspace{-0.15cm} & 
\rule{0.4cm}{0cm}$f_D$ & \rule{0.4cm}{0cm}$f_{D_s}$  & 
 \rule{0.3cm}{0cm}$f_{D_s}/f_D$ \\[0.2cm]
\hline
\hline
&&&&&&&&&&& \\[-0.1cm]
FNAL/MILC 17 $^{\nabla\nabla}$ & \cite{Bazavov:2017lyh} & 2+1+1 & \gA & \good & \good &\good & \good & \okay & 
212.1(0.6)   & 249.9(0.5) &  1.1782(16) \\[0.5ex]

FNAL/MILC 14A$^{**}$ & \cite{Bazavov:2014wgs} & 2+1+1 & \gA & \good & \good &\good & \good & \okay & 
212.6(0.4) $+1.0 \choose -1.2$   & 249.0(0.3)$+1.1 \choose -1.5$ &  1.1745(10)$+29 \choose -32$ \\[0.5ex]

ETM 14E$^{\dagger}$ & \cite{Carrasco:2014poa} & 2+1+1 & \gA & \good & \soso  &  \soso & \good  &  \okay &
207.4(3.8)   & 247.2(4.1) &  1.192(22) \\[0.5ex]

ETM 13F & \cite{Dimopoulos:2013qfa} & 2+1+1 & \rC & \soso & \soso  &  \soso & \good  &  \okay &
202(8)   & 242(8) &  1.199(25) \\[0.5ex]

FNAL/MILC 13$^\nabla$ & \cite{Bazavov:2013nfa} & 2+1+1 & \rC & \good    & \good    & \good     
&\good & \okay  & 212.3(0.3)(1.0)   & 248.7(0.2)(1.0) & 1.1714(10)(25)\\[0.5ex]

FNAL/MILC 12B & \cite{Bazavov:2012dg} & 2+1+1 & \rC & \good    & \good    & \good     
&\good & \okay  & 209.2(3.0)(3.6)   & 246.4(0.5)(3.6) & 1.175(16)(11)\\[0.5ex]

&&&&&&&&&&& \\[-0.1cm]
\hline
&&&&&&&&&&& \\[-0.1cm]
$\chi$QCD 20A$^{\dagger\dagger}$ &\cite{Chen:2020qma} & 2+1 & \gA & \tbr & \good & \good & \good & \okay & 213(5) & 249(7) &1.16(3) \\[0.5ex]
RBC/UKQCD 18A$^{\square\nabla}$ &\cite{Boyle:2018knm} & 2+1 & \oP & \good & \good & \good & \good & \okay &  &  &1.1740(51)$+68 \choose -68$ \\[0.5ex]
RBC/UKQCD 17 &\cite{Boyle:2017jwu} & 2+1 & \gA & \good & \good &\soso & \good & \okay
& 208.7(2.8)$+2.1 \choose -1.8$ & 246.4(1.3)$+1.3 \choose -1.9$    & 1.1667(77)$+57 \choose -43$\\[0.5ex] 
$\chi$QCD~14$^{\dagger\square}$ &\cite{Yang:2014sea} & 2+1 &  \gA &\soso &\soso & \soso & \good & \okay
& & 254(2)(4) & \\[0.5ex]
HPQCD 12A &\cite{Na:2012iu} & 2+1 & \gA &\soso  &\soso &\soso &\good &\okay 
& 208.3(1.0)(3.3) & 246.0(0.7)(3.5) & 1.187(4)(12)\\[0.5ex]

FNAL/MILC 11& \cite{Bazavov:2011aa} & 2+1 & \gA & \soso &\soso &\soso  & 
 \soso & \okay & 218.9(11.3) & 260.1(10.8)&   1.188(25)   \\[0.5ex]  

PACS-CS 11 & \cite{Namekawa:2011wt} & 2+1 & \gA & \tbr & \good & \tbr  & 
\soso & \okay & 226(6)(1)(5) & 257(2)(1)(5)&  1.14(3)   \\[0.5ex] 

HPQCD 10A & \cite{Davies:2010ip} & 2+1 & \gA & \good  & \soso  & 
\good & \good & \okay & 213(4)$^{*}$ & 248.0(2.5)  \\[0.5ex]

HPQCD/UKQCD 07 & \cite{Follana:2007uv} & 2+1 &  \gA & \soso & \soso & 
\soso & \good  & \okay & 207(4) & 241 (3)& 1.164(11)  \\[0.5ex] 

FNAL/MILC 05 & \cite{Aubin:2005ar} & 2+1 & \gA &\soso &   \soso    &
\tbr      & \soso    &  \okay       & 201(3)(17) & 249(3)(16)  & 1.24(1)(7) \\[0.5ex]

&&&&&&&&&&& \\[-0.1cm]
\hline
&&&&&&&&&&& \\[-0.1cm]
 Balasubramanian 19 & \cite{Balasubramanian:2019net} & 2 &  \gA & \good & \good & \good & \good & \okay &
 &     244(4)(2)             &            \\[0.5ex]
 
Blossier 18 & \cite{Blossier:2018jol} & 2 &  \gA & \soso & \good & \soso & \good & \okay &
                 &     238(5)(2)             &            \\[0.5ex]

TWQCD 14$^{\square\square}$ & \cite{Chen:2014hva} & 2 & \gA & \tbr & \soso  &  \tbr & \good  &  \okay &
202.3(2.2)(2.6)   & 258.7(1.1)(2.9) &  1.2788(264) \\[0.5ex]

ALPHA 13B & \cite{Heitger:2013oaa} & 2 & \rC & \soso & \good & \soso & \good & \okay &
216(7)(5)   & 247(5)(5) &  1.14(2)(3) \\[0.5ex]

ETM 13B$^\square$ & \cite{Carrasco:2013zta} & 2 & \gA & \good & \soso  &  \soso & \good  &  \okay &
208(7)   & 250(7) &  1.20(2) \\[0.5ex]

ETM 11A & \cite{Dimopoulos:2011gx} & 2 & \gA & \good & \soso  &  \soso & \good  &  \okay &
212(8)   & 248(6) &  1.17(5) \\[0.5ex]

ETM 09 & \cite{Blossier:2009bx} & 2 & \gA & \soso & \soso  &  \soso & \good  &  \okay & 
197(9)   & 244(8) &  1.24(3) \\[0.5ex]

&&&&&&&&&&& \\[-0.1cm]
\hline
\hline
\end{tabular*}
\begin{tabular*}{\textwidth}[l]{l@{\extracolsep{\fill}}lllllllll}
  \multicolumn{10}{l}{\vbox{\begin{flushleft} 
$^{\dagger}$ Update of ETM 13F.\\
$^{\nabla}$ Update of FNAL/MILC 12B.\\
$^{*}$ This result is obtained by using the central value for $f_{D_s}/f_D$ from HPQCD/UKQCD~07 
and increasing the error to account for the effects from the change in the physical value of $r_1$. \\
$^{\square}$ Update of ETM 11A and ETM 09. \\
$^{\square\square}$ One lattice spacing $\simeq 0.1$ fm only. $m_{\pi,{\rm min}}L=1.93$.\\
$^{**}$ 
At $\beta = 5.8$, $m_{\pi, \rm min}L=3.2$ but this lattice spacing is not used in the final cont./chiral extrapolations.\\
$^{\nabla\nabla}$ Update of FNAL/MILC~14A. The ratio quoted is $f_{D_s}/f_{D^+}=1.1749(16)$. In order to compare with
results from other collaborations, we rescale the number by the ratio of central values for $f_{D+}$ and $f_D$.
We use the same rescaling in FNAL/MILC~14A. At the finest lattice spacing the
finite-volume criterium would produce an empty green circle, however, as checked by the authors, results would not significantly change by excluding this ensemble, which instead sharpens the continuum limit extrapolation.\\
$\square\nabla$ Update of RBC/UKQCD 17. \\
$\dagger\square$ Two values of sea pion masses.\\
$\dagger\dagger$ Four valence pion masses between 208 MeV and 114 MeV have been used at one value of the sea pion mass of 139 MeV.
\end{flushleft}}}
\end{tabular*}

\vspace{-0.5cm}
\caption{Decay constants of the $D$ and $D_{s}$ mesons (in MeV) and their ratio.
}
\label{tab_FDsummary}
\end{center}
\end{table}
%
%

\begin{figure}[tb]
\hspace{-0.8cm}\includegraphics[width=0.58\linewidth]{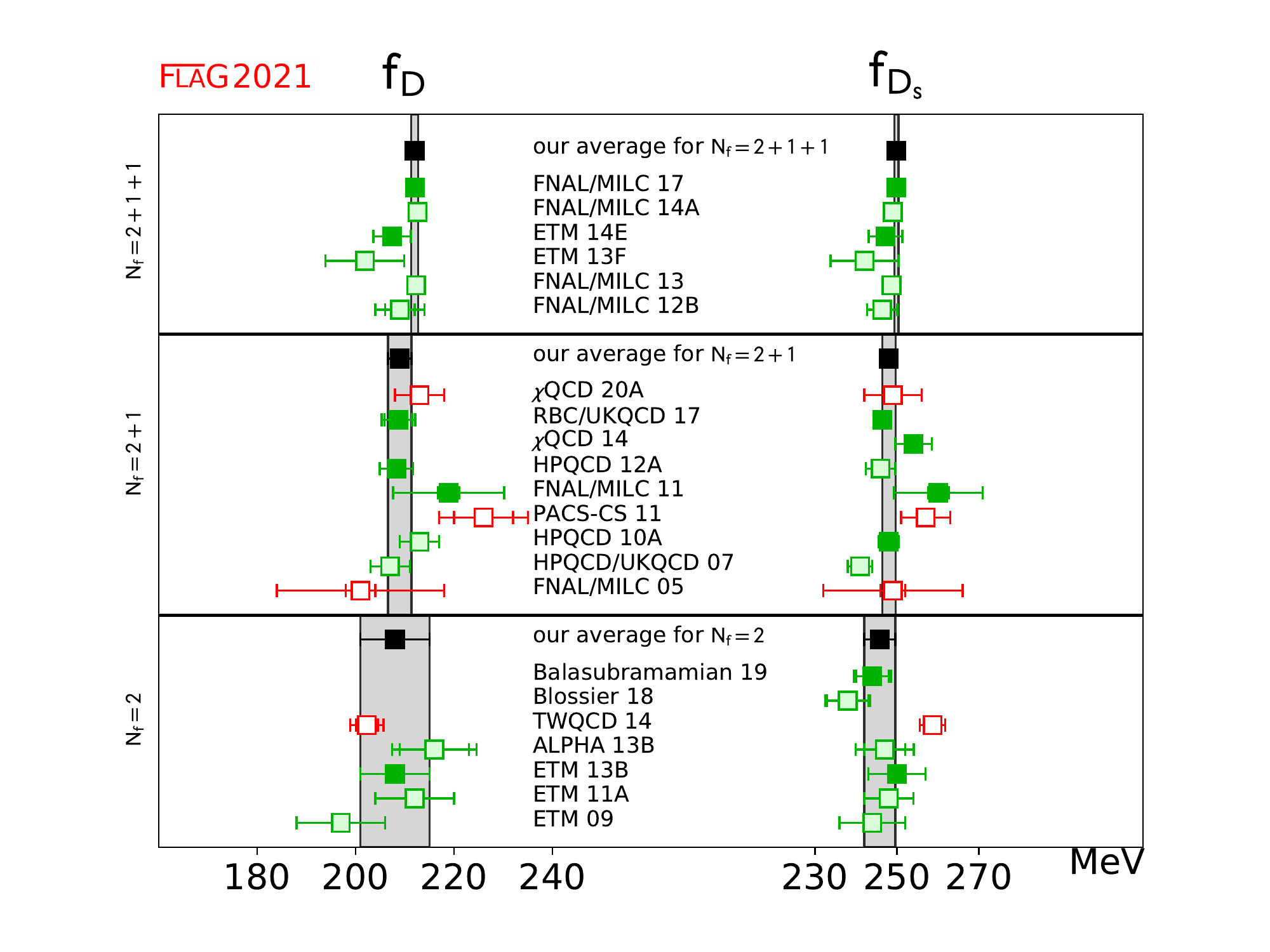} \hspace{-1cm}
\includegraphics[width=0.58\linewidth]{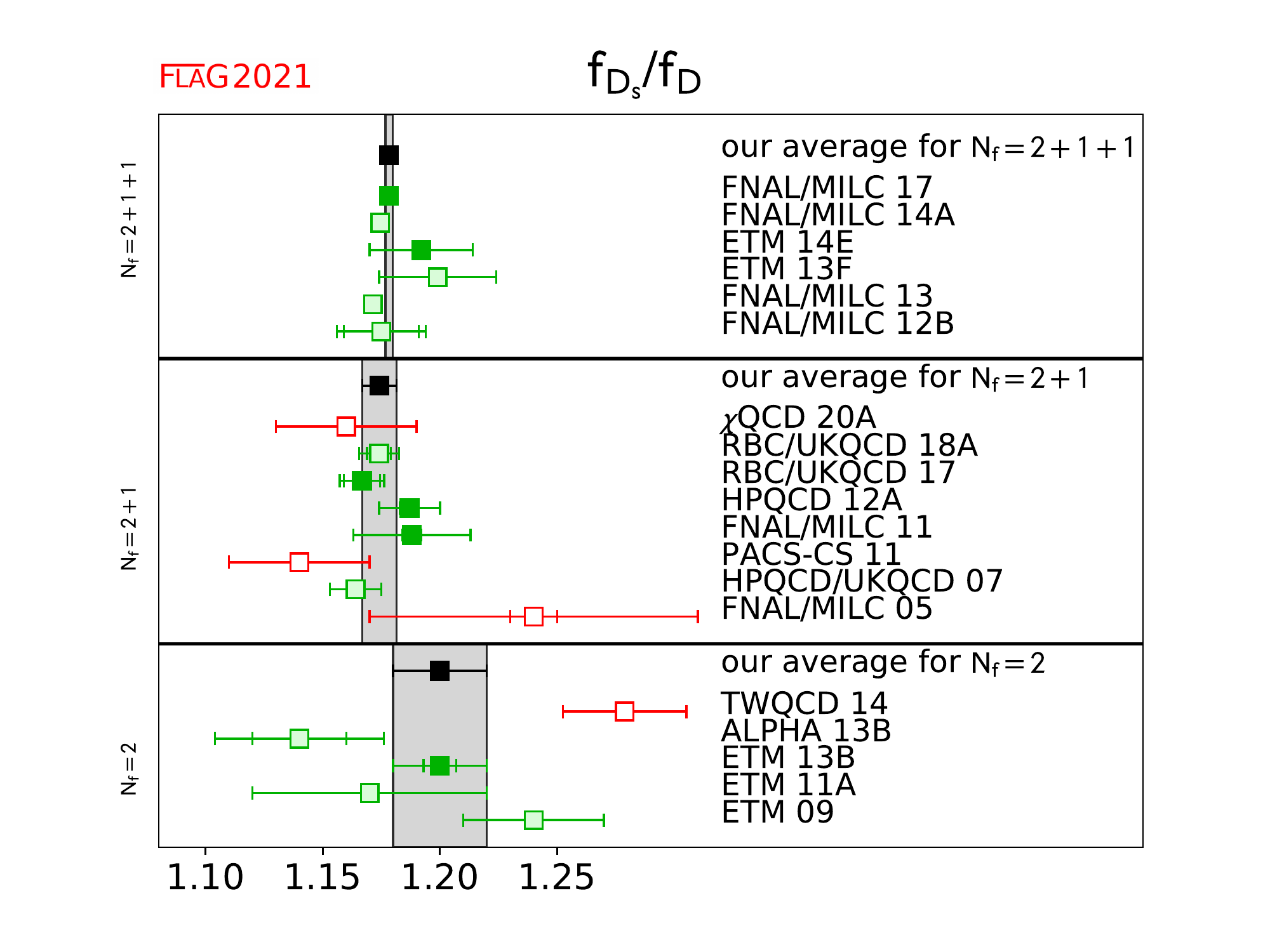}

\vspace{-2mm}
\caption{Decay constants of the $D$ and $D_s$ mesons [values in Tab.~\ref{tab_FDsummary} 
and  Eqs.~(\ref{eq:fD2}-\ref{eq:fDratio2+1+1})].
As usual, full green squares are used in the averaging procedure,
    pale green squares have been superseded by later
    determinations, while pale red squares do not
    satisfy the criteria.
The black squares and grey bands
  indicate our averages. }
\label{fig:fD}
\end{figure}
The $N_f=2$ averages for $f_D$ and ${{f_{D_s}}/{f_D}}$ coincide with those in the previous FLAG review and are given by the values in ETM~13B~\cite{Carrasco:2013zta},
while the estimate for $f_{D_s}$ is the result of the weighted average of the numbers in ETM~13B~\cite{Carrasco:2013zta} and Balasubramanian~19~\cite{Balasubramanian:2019net}. They read
\begin{align}
&\label{eq:fD2}
\Nf=2:&\FLAGAVBEGIN f_D &= 208(7) \FLAGAVEND\;{\rm MeV}
&&\Ref~\mbox{\cite{Carrasco:2013zta}},\\
&\label{eq:fDs2}
\Nf=2: &\FLAGAVBEGIN f_{D_s} &= 246(4)  \FLAGAVEND\; {\rm MeV}
&&\Refs~\mbox{\cite{Carrasco:2013zta,Balasubramanian:2019net}}, \\
&\label{eq:fDratio2}
\Nf=2: &\FLAGAVBEGIN f_{D_s}\over{f_D} &= 1.20(0.02)\FLAGAVEND
&&\Ref~\mbox{\cite{Carrasco:2013zta}}.
\end{align}

Turning to $N_f=2+1$ results, the $\chi$QCD collaboration presented in $\chi$QCD 20A~\cite{Chen:2020qma} a calculation of the  $D_s^{(*)}$, $D^{(*)}$
and $\phi$ meson decay constants. The couplings of the vector mesons to the tensor current are also computed. The computation is performed at a single
lattice spacing with $a^{-1} \approx 1.7$ GeV on a $2+1$ domain wall fermion ensemble generated by the RBC/UKQCD Collaboration. The sea pion mass is at its physical
value and the spatial extension is 5.5 fm. Overlap valence fermions are used with different values of the light, strange and (quenched) charm
quark masses. For the light quarks the corresponding pion masses range between 114 and 208 MeV. The setup follows very closely the one in $\chi$QCD~14~\cite{Yang:2014sea}
(presented in the 2016 FLAG review).
The decay constants $f_D$ and $f_{D_s}$ are obtained from an exactly conserved PCAC Ward identity so they do not depend on renormalization factors.
The results, however, do not enter the FLAG average as the simulations do not meet the quality criteria concerning the number of lattice spacings
used in the continuum extrapolation.

A new result (RBC/UKQCD~18A) for the SU(3)-breaking ratio $f_{D_s}/f_{D}$ has been reported in Ref.~\cite{Boyle:2018knm}.
The setup includes $2+1$ dynamical flavors of Domain Wall fermions.
This new result essentially supersedes RBC/UKQCD~17~\cite{Boyle:2017jwu} (discussed in the previous FLAG review) by implementing a number of improvements.
One level of stout smearing for the gauge fields has been introduced before performing the charm-quark inversions, which has allowed them to
simulate directly at the physical charm mass. At the same time, the valence the strange-quark mass has been tuned to
its physical value in order to eliminate a small correction needed previously. Finally, the number of source positions has been doubled
on a few ensembles.
As of April~30, 2021 the article has not been published in a journal. Therefore, the result does not contribute to the FLAG estimates.

The $N_f=2+1$ FLAG estimates remain unchanged and read
\begin{align}
&\label{eq:fD2+1}
\Nf=2+1:&\FLAGAVBEGIN f_D &= 209.0(2.4) \FLAGAVEND\;{\rm MeV}
&&\Refs~\mbox{\cite{Na:2012iu,Bazavov:2011aa,Boyle:2017jwu}},\\
&\label{eq:fDs2+1}
\Nf=2+1: &\FLAGAVBEGIN f_{D_s} &= 248.0(1.6) \FLAGAVEND\; {\rm MeV} 
&&\Refs~\mbox{\cite{Davies:2010ip,Bazavov:2011aa,Boyle:2017jwu,Yang:2014sea}}, \\
&\label{eq:fDratio2+1}
\Nf=2+1: &\FLAGAVBEGIN f_{D_s}\over{f_D} &= 1.174(0.007)\FLAGAVEND
&&\Refs~\mbox{\cite{Na:2012iu,Bazavov:2011aa,Boyle:2017jwu}},
\end{align}
where the error on the $\Nf=2+1$ average of $f_{D_s}$ has been rescaled by the factor $\sqrt{\chi^2/\mbox{dof}}=1.1$.
Those come from the results in HPQCD~12A~\cite{Na:2012iu}, FNAL/MILC~11~\cite{Bazavov:2011aa} as well as RBC/UKQCD~17 \cite{Boyle:2017jwu}
concerning $f_D$ while for $f_{D_s}$ 
also the $\chi$QCD~14~\cite{Yang:2014sea} result contributes, and instead of the value in HPQCD~12A~\cite{Na:2012iu}
the one in HPQCD~10A~\cite{Davies:2010ip} is used.
In addition, the statistical errors between the results of FNAL/MILC and HPQCD have been everywhere treated as 100\% correlated since
the two collaborations use overlapping sets of configurations. The same procedure had been used in the past reviews.

No new result appeared for $N_f=2+1+1$ since the last FLAG review. Our estimates, therefore, coincide with those in Ref.~\cite{Aoki:2019cca}, namely
\begin{align}
&\label{eq:fD2+1+1}
\Nf=2+1+1:&\FLAGAVBEGIN f_D &= 212.0(0.7) \FLAGAVEND\;{\rm MeV}
&&\Refs~\mbox{\cite{Bazavov:2017lyh,Carrasco:2014poa}},\\
&\label{eq:fDs2+1+1}
\Nf=2+1+1: &\FLAGAVBEGIN f_{D_s} &= 249.9(0.5) \FLAGAVEND\; {\rm MeV} 
&&\Refs~\mbox{\cite{Bazavov:2017lyh,Carrasco:2014poa}}, \\
&\label{eq:fDratio2+1+1}
\Nf=2+1+1: &\FLAGAVBEGIN f_{D_s}\over{f_D} &= 1.1783(0.0016)\FLAGAVEND
&&\Refs~\mbox{\cite{Bazavov:2017lyh,Carrasco:2014poa}},
\end{align}
where the error on the average of $f_{D}$ has been rescaled by the factor $\sqrt{\chi^2/\mbox{dof}}=1.22$.

On a general note, an important recent theoretical development is represented by the nonperturbative
calculation of the form factors $F_A$ and $F_V$ contributing to the radiative
leptonic decays of a charged pseudoscalar meson $P$. As discussed in Ref.~\cite{Carrasco:2015xwa}, those
 appear in the decomposition of the hadronic matrix element
\begin{equation}
H_W^{\alpha r}(k, {\bold{p}})=\epsilon_\mu^r(k)\int d^4y\, e^{iky}\, {\rm T}\langle0|j_W^\alpha(0) j_{em}^\mu(y)| P({\bold{p}})\rangle\;,
\end{equation}
with $\epsilon_\mu^r(k)$ the polarisation vector of the outgoing photon (with momentum $k$) and 
$j_W^\alpha$ and $j_{em}^\mu$ the weak and electromagnetic currents, respectively. With general kinematics four 
form factors together with the pseudoscalar decay constant $f_P$ are needed; however, for $k^2=0$, by choosing in 
addition a physical basis for the polarisation such that $\epsilon_r({\bold{k}})\cdot k=0$, the deacy rate can be calculated once $F_A$, $F_V$, and $f_P$ are known.
A preliminary study  has
been presented in Ref.~\cite{Kane:2019jtj} in the theory with $2+1$ dynamical
flavors. While a more complete calculation at three different lattice spacings
(in the range $0.09$--$0.06$ fm) and for $N_f=2+1+1$ appeared in Ref.~\cite{Desiderio:2020oej}.
The form factors, once used in combination with the nonperturbative calculation
of the corrections to $P\to \ell \bar{\nu}_\ell$ due to the exchange of a virtual photon, allow for a complete determination of the QED corrections to
semileptonic decays of mesons. In Ref.~\cite{Desiderio:2020oej} the form factors
are defined after removing the point-like, infrared divergent contribution,
in order to highlight the interesting structure dependent part.
Restricting attention to on-shell photons, the behaviour of discretisation effects is studied in Ref.~\cite{Desiderio:2020oej} as the photon momentum is changed
and heavy quarks are considered. A prescription is also given to nonperturbatively subtract infrared divergent cutoff effects.
Still, for charmed mesons discretization effects turned
out to be rather large, relative to the size of the form factors, suggesting
that very fine lattice spacings will be needed in the case of $B$ mesons.

\FloatBarrier

\subsection{Form factors for $D\to \pi \ell\nu$ and $D\to K \ell \nu$ semileptonic decays}
 \label{sec:DtoPiK}


The SM prediction for the differential decay rate of the semileptonic processes $D\to \pi \ell\nu$ and
$D\to K \ell \nu$ can be written as
\begin{align}
  \frac{d\Gamma(D\to P\ell\nu)}{dq^2} =
  &
    \frac{G_{\rm\scriptscriptstyle F}^2 |V_{cx}|^2}{24 \pi^3}
    \,\frac{(q^2-m_\ell^2)^2\sqrt{E_P^2-m_P^2}}{q^4m_{D}^2}
    \nonumber\\
  & \times
    \left[ \left(1+\frac{m_\ell^2}{2q^2}\right)
    m_{D}^2(E_P^2-m_P^2)|f_+(q^2)|^2 
    + \frac{3m_\ell^2}{8q^2}(m_{D}^2-m_P^2)^2|f_0(q^2)|^2\right]
    \label{eq:DtoPiKFull}
\end{align}
where $x = d, s$ is the daughter light quark, $P= \pi, K$ is the
daughter light-pseudoscalar meson, $E_P$ is the light-pseudoscalar meson energy 
in the rest frame of the decaying $D$, and $q = (p_D - p_P)$ is the
momentum of the outgoing lepton pair; in this section, the charged lepton $\ell$
will either be an electron (resp. positron) or (anti)muon.  The vector and scalar form
factors $f_+(q^2)$ and $f_0(q^2)$ parameterize the hadronic matrix
element of the heavy-to-light quark flavour-changing vector current
$V_\mu = \overline{x} \gamma_\mu c$,
\begin{equation}
\langle P| V_\mu | D \rangle  = f_+(q^2) \left( {p_D}_\mu+ {p_P}_\mu - \frac{m_D^2 - m_P^2}{q^2}\,q_\mu \right) + f_0(q^2) \frac{m_D^2 - m_P^2}{q^2}\,q_\mu \,,
\end{equation}
and satisfy the kinematic constraint $f_+(0) = f_0(0)$.  Because the contribution to the decay width from
the scalar form factor is proportional to $m_\ell^2$, within current precision standards it can be
neglected for $\ell = e, \mu$, and Eq.~(\ref{eq:DtoPiKFull})
simplifies to
\begin{equation}
\frac{d\Gamma \!\left(D \to P \ell \nu\right)}{d q^2} = \frac{G_{\rm\scriptscriptstyle F}^2}{24 \pi^3} |\vec{p}_{P}|^3 {|V_{cx}|^2 |f_+ (q^2)|^2} \,. \label{eq:DtoPiK}
\end{equation}
In models of new physics, decay rates may also receive contributions from matrix elements of other
parity-even currents. In the case of the scalar density, partial vector current conservation allows one
to write matrix elements of the latter in terms of $f_+$ and $f_0$, while for tensor currents $T_{\mu\nu}=\bar x\sigma_{\mu\nu}c$
a new form factor has to be introduced, viz.,
\begin{equation}
\langle P| T_{\mu\nu} | D \rangle  = \frac{2}{m_D+m_P}\left[p_{P\mu}p_{D\nu}-p_{P\nu}p_{D\mu}\right]f_T(q^2)\,.
\end{equation}
Recall that, unlike the Noether current $V_\mu$, the operator $T_{\mu\nu}$ requires
a scale-dependent renormalization.


Lattice-QCD computations of $f_{+,0}$ allow for comparisons to experiment
to ascertain whether the SM provides the correct prediction for the $q^2$-dependence of
$d\Gamma(D\to P\ell\nu)/dq^2$;
and, subsequently, to determine the CKM matrix elements $|V_{cd}|$ and $|V_{cs}|$
from Eq.~(\ref{eq:DtoPiKFull}). The inclusion of $f_T$ allows for analyses to
constrain new physics. Currently, state-of-the-art experimental results by
CLEO-c~\cite{Besson:2009uv} and BESIII~\cite{Ablikim:2017oaf,Ablikim:2018frk}
provide data for the differential rates in the whole $q^2$ range available,
with a precision of order 2--3\% for the total branching fractions in both
the electron and muon final channels.


Calculations of the $D\to \pi \ell\nu$ and $D\to
K \ell \nu$ form factors typically use the same light-quark and
charm-quark actions as those of the leptonic decay constants $f_D$ and
$f_{D_s}$. Therefore, many of the same issues arise; in particular,
considerations about cutoff effects coming from the large charm-quark mass,
or the normalization of weak currents, apply.
Additional complications arise,
however, due to the necessity of covering a sizeable range of values in $q^2$:
\begin{itemize}

\item Lattice kinematics imposes restrictions on the values
of the hadron momenta.
Because lattice calculations are performed
in a finite spatial volume, the pion or kaon three-momentum can only
take discrete values in units of $2\pi/L$ when periodic boundary
conditions are used.  For typical box sizes in recent lattice $D$- and
$B$-meson form-factor calculations, $L \sim 2.5$--3~fm; thus, the
smallest nonzero momentum in most of these analyses lies in the range
$|\vec{p}_P| \sim 400$--$500$~MeV.  The largest momentum in lattice
heavy-light form-factor calculations is typically restricted to
$ |\vec{p}_P| \leq 4\pi/L$. 
For $D \to \pi \ell \nu$ and $D \to
K \ell \nu$, $q^2=0$ corresponds to $|\vec{p}_\pi| \sim 940$~MeV and $|\vec{p}_K| \sim
1$~GeV, respectively, and the full recoil-momentum region is within
the range of accessible lattice momenta.
This has implications for both the accuracy of the study of the $q^2$-dependence,
and the precision of the computation, since statistical errors and cutoff effects
tend to increase at larger meson momenta.
As a consequence, many recent studies have incorporated the use of
nonperiodic (``twisted'') boundary conditions (tbc)~\cite{Bedaque:2004kc,Sachrajda:2004mi}
in the valence fields used for the computation of observables,
as a means to alleviate some of these difficulties. In particular, while they will
not necessarily lead to a decrease of numerical noise or cutoff effects, the use
of tbc allows not only for a better momentum resolution, but also to better
control the $q^2=0$ endpoint~\cite{DiVita:2011py,Koponen:2011ev,Koponen:2012di,Koponen:2013tua,Lubicz:2017syv,Lubicz:2018rfs}.
\item Final-state pions and kaons can have energies
$\gtrsim 1~{\rm GeV}$, given the available kinematical range $0 \lesssim q^2 \leq q_{\rm\scriptscriptstyle max}^2=(m_D-m_P)^2$.
This makes the use of (heavy-meson) chiral perturbation theory to extrapolate to physical
light-quark masses potentially problematic.

\item Accurate comparisons to experiment, including the determination of CKM parameters,
requires good control of systematic uncertainties in the parameterization of the $q^2$-dependence of form factors. While this issue is far more important for semileptonic
$B$ decays, where existing lattice computations cover just a fraction of the kinematic range,
the increase in experimental precision requires accurate work in the charm sector
as well. The parameterization of semileptonic form factors is discussed in detail
in Appendix \ref{sec:zparam}.

\end{itemize}


The most advanced $N_f = 2$ lattice-QCD calculation of the
$D \to \pi \ell \nu$ and $D \to K \ell \nu$ form factors is by the ETM
collaboration~\cite{DiVita:2011py}. This work, which did not proceed
beyond the preliminary stage, uses
the twisted-mass Wilson action for both the light and charm quarks,
with three lattice spacings down to $a \approx 0.068$~fm and (charged)
pion masses down to $m_\pi \approx 270$~MeV.  The calculation employs
the method of Ref.~\cite{Becirevic:2007cr} to avoid the need to
renormalize the vector current, by introducing double-ratios of lattice three-point correlation functions
in which the vector current renormalization cancels. Discretization
errors in the double ratio are of ${\mathcal O}((am_c)^2)$,
due to the automatic ${\mathcal O}(a)$ improvement at maximal twist.
The vector and scalar form factors $f_+(q^2)$ and
$f_0(q^2)$ are obtained by taking suitable linear combinations of
these double ratios.
Extrapolation to physical light-quark masses is performed
using $SU(2)$ heavy-light meson $\chi$PT.  The ETM collaboration simulates with twisted boundary
conditions for the valence quarks to access arbitrary momentum values
over the full physical $q^2$ range, and interpolate to $q^2=0$ using
the Be{\v{c}}irevi{\'c}-Kaidalov ansatz~\cite{Becirevic:1999kt}.  The
statistical errors in $f_+^{D\pi}(0)$ and $f_+^{DK}(0)$ are 9\% and
7\%, respectively, and lead to rather large systematic uncertainties
in the fits to the light-quark mass and energy dependence (7\% and
5\%, respectively).  Another significant source of uncertainty is from
discretization errors (5\% and 3\%, respectively).  On the finest
lattice spacing used in this analysis $am_c \sim 0.17$, so $\cO((am_c)^2)$ cutoff errors are expected to be about 5\%.  This can be
reduced by including the existing $N_f = 2$ twisted-mass ensembles
with $a \approx 0.051$~fm discussed in Ref.~\cite{Baron:2009wt}.


The first published $N_f = 2+1$ lattice-QCD calculation of the $D \to
\pi \ell \nu$ and $D \to K \ell \nu$ form factors came from the
Fermilab Lattice, MILC, and HPQCD
collaborations~\cite{Aubin:2004ej}.\footnote{Because only two of the
  authors of this work are members of HPQCD, and to distinguish it
  from other more recent works on the same topic by HPQCD, we
  hereafter refer to this work as ``FNAL/MILC.''}  This work uses
asqtad-improved staggered sea quarks and light ($u,d,s$) valence
quarks and the Fermilab action for the charm quarks, with a single
lattice spacing of $a \approx 0.12$~fm, and a minimum RMS-pion
mass of $\approx 510$~MeV, dictated by the presence of fairly large
staggered taste splittings. The vector current is normalized using a
mostly nonperturbative approach, such that the perturbative truncation
error is expected to be negligible compared to other
systematics. Results for the form factors are provided over the full
kinematic range, rather than focusing just at $q^2=0$ as was customary
in previous work, and fitted to a Be{\v{c}}irevi{\'c}-Kaidalov ansatz.
In fact, the publication of this result predated the precise
measurements of the $D\to K \ell\nu$ decay width by the
FOCUS~\cite{Link:2004dh} and Belle experiments~\cite{Abe:2005sh}, and
showed good agreement with the experimental determination of the shape
of $f_+^{DK}(q^2)$.  Progress on extending this work was reported
in~\cite{Bailey:2012sa}; efforts are aimed at reducing both the
statistical and systematic errors in $f_+^{D\pi}(q^2)$ and
$f_+^{DK}(q^2)$ by increasing the number of configurations analyzed,
simulating with lighter pions, and adding lattice spacings as fine as
$a \approx 0.045$~fm.

The most precise published calculations of the
$D \to \pi \ell \nu$~\cite{Na:2011mc} and $D \to
K \ell \nu$~\cite{Na:2010uf} form factors in $N_f=2+1$ QCD
are by the HPQCD collaboration. They are also based on $N_f = 2+1$
asqtad-improved staggered MILC configurations, but use two lattice spacings
$a \approx 0.09$ and 0.12~fm, and a HISQ action for the valence
$u,d,s$, and $c$ quarks. In these mixed-action calculations, the HISQ
valence light-quark masses are tuned so that the ratio $m_l/m_s$ is
approximately the same as for the sea quarks; the minimum RMS sea-pion
mass $\approx 390$~MeV. Form factors are determined only at $q^2=0$,
by using a Ward identity to relate matrix elements of vector
currents to matrix elements of the absolutely normalized quantity
$(m_{c} - m_{x} ) \langle P | \bar{x}c | D \rangle$ (where $x=u,d,s$),
and exploiting the kinematic identity $f_+(0) = f_0(0)$
to yield
$f_+(q^2=0) = (m_{c} - m_{x} ) \langle P | \bar{x}c | D \rangle / (m^2_D - m^2_P)$.
A modified $z$-expansion
(cf.~Appendix \ref{sec:zparam})
is employed to simultaneously extrapolate to the physical
light-quark masses and the continuum and to interpolate to $q^2 = 0$, and
allow the coefficients of the series expansion to vary with the light-
and charm-quark masses.  The form of the light-quark dependence is
inspired by $\chi$PT, and includes logarithms of the form $m_\pi^2
{\rm log} (m_\pi^2)$ as well as polynomials in the valence-, sea-, and
charm-quark masses.  Polynomials in $E_{\pi(K)}$ are also included to
parameterize momentum-dependent discretization errors.
The number of terms is increased until the result for $f_+(0)$
stabilizes, such that the quoted fit error for $f_+(0)$ not only contains
statistical uncertainties, but also reflects relevant systematics.  The
largest quoted uncertainties in these calculations are from statistics and
charm-quark discretization errors. Progress towards extending the computation
to the full $q^2$ range have been reported in Ref.~\cite{Koponen:2011ev,Koponen:2012di};
however, the information contained in these conference proceedings
is not enough to establish an updated value of $f_+(0)$ with respect
to the previous journal publications.

The most recent $N_f=2+1$ computation of $D$ semileptonic form factors has
been carried out by the JLQCD collaboration, and so far only published in conference
proceedings; most recently in Ref.~\cite{Kaneko:2017xgg}.
They use their own M\"obius domain-wall configurations at three values
of the lattice spacing $a=0.080, 0.055, 0.044~{\rm fm}$, with several
pion masses ranging from 226 to 501~MeV (though there is so far only one
ensemble, with $m_\pi=284~{\rm MeV}$, at the finest lattice spacing).
The vector and scalar form factors are computed at four values of the momentum transfer for each ensemble.
The computed form factors are observed to depend mildly on both the
lattice spacing and the pion mass.
The momentum dependence of the form factors is fitted to a BCL
$z$-parameterization (see Appendix~\ref{sec:zparam}) with a Blaschke factor that contains the measured
value of the $D_{(s)}^*$ mass in the vector channel,
and a trivial Blaschke factor in the scalar channel. The systematics
of this latter fit is assessed by a BCL fit with the experimental value
of the scalar resonance mass in the Blaschke factor.
Continuum and chiral extrapolations are carried out through a
linear fit in the squared lattice spacing and the squared pion and $\eta_c$ masses.
A global fit that uses hard-pion HM$\chi$PT to model the mass dependence
is furthermore used for a comparison of the form factor shapes with experimental data.\footnote{It is important
to stress the finding in Ref.~\cite{Colangelo:2012ew} that
the factorization of chiral logs in hard-pion $\chi$PT breaks down,
implying that it does not fulfill the expected requisites for a proper
effective field theory. Its use to model the mass dependence of form
factors can thus be questioned. \label{footnote:hardpion}}
Since the computation is only published in proceedings so far, it will not
enter our $N_f=2+1$ average.\footnote{The ensemble parameters
quoted in Ref.~\cite{Kaneko:2017xgg} appear to show that the volumes
employed at the lightest pion masses are insufficient to meet our criteria
for finite-volume effects. There is, however, a typo in the table which results
in a wrong assignment of lattice sizes, whereupon the criteria are indeed met.
We thank T.~Kaneko for correspondence on this issue.}

The first full computation of both the vector and scalar form factors
in $N_f=2+1+1$ QCD was achieved by the
ETM collaboration~\cite{Lubicz:2017syv}.  Furthermore, they have provided a separate
determination of the tensor form factor, relevant for new physics analyses~\cite{Lubicz:2018rfs}.
Both works use the available $N_f = 2+1+1$ twisted-mass Wilson lattices~\cite{Baron:2010bv},
totaling three lattice spacings down to $a\approx 0.06$~fm,
and a minimal pion mass of 220~MeV.
Matrix elements are extracted from suitable double ratios of correlation functions
that avoid the need of nontrivial current normalizations.
The use of twisted boundary conditions allows both for imposing
several kinematical conditions, and considering arbitrary frames
that include moving initial mesons. 
After interpolation to the physical strange- and charm-quark masses,
the results for form factors are fitted to a modified $z$-expansion
that takes into account both the light-quark mass dependence through
hard-pion $SU(2)$ $\chi$PT~\cite{Bijnens:2010ws}, and the
lattice-spacing dependence. In the latter  case,
a detailed study of Lorentz-breaking effects due to the breaking of
rotational invariance down to the hypercubic subgroup
is performed, leading to a nontrivial momentum-dependent parameterization
of cutoff effects.
The $z$-parameterization (see Appendix~\ref{sec:zparam}) itself includes a single-pole Blaschke factor
(save for the scalar channel in $D\to K$, where the Blaschke factor is trivial),
with pole masses treated as free parameters.
The final quoted uncertainty on the form factors
is about 5--6\% for $D\to\pi$, and 4\% for $D\to K$.
The dominant source of uncertainty is quoted as statistical+fitting procedure+input parameters ---
the latter referring to the values of quark masses, the lattice spacing (i.e., scale setting),
and the LO $SU(2)$ LECs.

Another $N_f=2+1+1$ computation of $f_+$ and $f_0$ in the full kinematical range for the $D\to Kl\nu$ mode,
performed by HPQCD, has recently been published --- HPQCD 21A (Ref.~\cite{Chakraborty:2021qav}).
This work uses MILC's HISQ ensembles at five values of the lattice spacing, 
and pion masses reaching to the physical point for the three coarsest values
of~$a$. Vector currents are normalized nonpertubatively by imposing that form factors
satisfy Ward identities exactly at zero recoil.
Results for the form factors are fitted to a modified $z$-expansion ansatz,
with all sub-threshold poles removed by using the experimental value of the
mass shifted by a factor that matches the corresponding result at finite lattice spacing.
The accuracy of the description of the $q^2$ dependence is crosschecked by comparing
to a fit based on cubic splines.
Finite-volume effects are expected to be small, and chiral-perturbation-theory-based
estimates for them are included in the chiral fit. However, the impact of frozen
topology at the finest lattice spacing is neglected.
The final uncertainty from the form factors in the determination of $|V_{cs}|$ quoted
in HPQCD 21A is at the 0.5\% level, and comparable to the rest of the uncertainty
(due to the experimental error, as well as weak and electromagnetic corrections);
in particular, the precision of the form factors is around seven times higher than 
that of the other existing $N_f=2+1+1$ determination by ETMC.
The work also provides an accurate prediction for the lepton flavour universality ratio
between the muon and electron modes, where the uncertainty is overwhelmingly dominated
by the electromagnetic corrections.

The FNAL/MILC collaboration has also reported ongoing work on extending their computation
to $N_f=2+1+1$, using MILC HISQ ensembles at four values
of the lattice spacing down to $a=0.042~{\rm fm}$ and pion masses down to the
physical point. The latest updates on this computation, focusing on the form factors
at $q^2=0$, but without explicit values of the latter yet, can be found in Refs.~\cite{Primer:2015qpz,Primer:2017xzs}.


\begin{table}[h]
\begin{center}
\mbox{} \\[3.0cm]
\footnotesize
\begin{tabular*}{\textwidth}[l]{l @{\extracolsep{\fill}} r ll  l @{\hspace{1mm}}l @{\hspace{1mm}}l@{\hspace{1mm}} l@{\hspace{1mm}} l c c}
Collaboration & Ref. & $\Nf$ & 
\hspace{0.15cm}\begin{rotate}{60}{publication status}\end{rotate}\hspace{-0.15cm} &
\hspace{0.15cm}\begin{rotate}{60}{continuum extrapolation}\end{rotate}\hspace{-0.15cm} &
\hspace{0.15cm}\begin{rotate}{60}{chiral extrapolation}\end{rotate}\hspace{-0.15cm}&
\hspace{0.15cm}\begin{rotate}{60}{finite volume}\end{rotate}\hspace{-0.15cm}&
\hspace{0.15cm}\begin{rotate}{60}{renormalization}\end{rotate}\hspace{-0.15cm}  &
\hspace{0.15cm}\begin{rotate}{60}{heavy-quark treatment}\end{rotate}\hspace{-0.15cm}  &
$f_+^{D\pi}(0)$ & $f_+^{DK}(0)$\\
&&&&&&&&& \\[-0.1cm]
\hline
\hline
&&&&&&&&& \\[-0.1cm]
HPQCD 21A & \cite{Chakraborty:2021qav} & 2+1+1 & \oP	 & \good & \good & \soso$^\dagger$ & \good & \okay & n/a & 0.7380(44) \\[0.5ex]
HPQCD 20 & \cite{Cooper:2020wnj} & 2+1+1 & \gA	 & \good & \soso & \good & \good & \okay & n/a & n/a \\[0.5ex]
ETM 17D, 18 & \cite{Lubicz:2017syv,Lubicz:2018rfs} & 2+1+1 & \gA	 & \good & \soso & \soso & \good & \okay & 0.612(35) & 0.765(31) \\[0.5ex]
&&&&&&&&& \\[-0.1cm]
\hline\\[0.5ex]
JLQCD 17B & \cite{Kaneko:2017xgg} & 2+1 & \rC & \good & \good & \soso & \good & \okay & 0.615(31)($^{+17}_{-16}$)($^{+28}_{-7}$)$^*$ & 0.698(29)(18)($^{+32}_{-12}$)$^*$ \\[0.5ex]
HPQCD 11 & \cite{Na:2011mc} & 2+1 & \gA  & \soso & \soso & \soso & \good &  \okay & 0.666(29) &\\[0.5ex]
HPQCD 10B & \cite{Na:2010uf} & 2+1 & \gA  & \soso & \soso & \soso & \good &  \okay & & 0.747(19)  \\[0.5ex]
FNAL/MILC 04 & \cite{Aubin:2004ej} & 2+1 & \gA  & \tbr & \tbr & \soso & \soso & \okay & 0.64(3)(6)& 0.73(3)(7)
\\[0.5ex]
&&&&&&&&& \\[-0.1cm]
\hline\\[0.5ex]
ETM 11B & \cite{DiVita:2011py} & 2 & \rC  & \soso & \soso & \good & \good &  \okay & 0.65(6)(6) & 0.76(5)(5)\\[0.5ex]
&&&&&&&&& \\[-0.1cm]
\hline
\hline
\end{tabular*}\\
\begin{minipage}{\linewidth}
{\footnotesize 
\begin{itemize}
   \item[$^*$] The first error is statistical, the second from the $q^2\to 0$ extrapolation, the third from the chiral-continuum extrapolation.
   \item[$^\dagger$] The volumes used in the computation satisfy the nominal criterion for finite-volume effects. However, the impact of the topologically frozen ensemble at $a \simeq 0.044~{\rm fm}$
is neglected. We therefore assign a \soso rating here, as a mark of caution.
\end{itemize}
}
\end{minipage}
\caption{Summary of computations of charmed-meson semileptonic form factors. Note that HPQCD 20 (discussed in Sec.~\protect\ref{sec:charmheavyspectator}) addresses the $B_c \to B_s$ and $B_c \to B_d$ transitions---hence the absence of quoted values for $f_+^{D\pi}(0)$ and $f_+^{DK}(0)$---while ETM~18 provides a computation of tensor form factors.}
\label{tab_DtoPiKsumm2}
\end{center}
\end{table}

Table \ref{tab_DtoPiKsumm2} contains our summary of the existing
calculations of the $D \to \pi \ell \nu$ and $D \to K \ell \nu$
semileptonic form factors.  Additional tables in
Appendix~\ref{app:DtoPi_Notes} provide further details on the
simulation parameters and comparisons of the error estimates. Recall
that only calculations without red tags that are published in a
refereed journal are included in the FLAG average. We will quote
no FLAG estimate for $N_f=2$, since the results by ETM have only appeared
in conference proceedings. For $N_f=2+1$, only HPQCD~10B,11 qualify,
which provides our estimate for $f_+(q^2=0)=f_0(q^2=0)$.
For $N_f=2+1+1$, we quote as the FLAG estimate for $f_+^{D\pi}(0)$ the
only published result by ETM 17D, while for $f_+^{DK}(0)$ we quote
the weighted average of the values published by ETM 17D and HPQCD 21A: 

%
\begin{align}
	&&\FLAGAVBEGIN f_+^{D\pi}(0)&=  0.666(29)\FLAGAVEND&&\Ref~\mbox{\cite{Na:2011mc}},\nonumber\\[-3mm]
&N_f=2+1:&\label{eq:Nf=2p1Dsemi}\\[-3mm]
        &&\FLAGAVBEGIN f_+^{DK}(0)  &= 0.747(19)\FLAGAVEND &&\Ref~\mbox{\cite{Na:2010uf}},\nonumber
\end{align}
%

%
\begin{align}
	&&\FLAGAVBEGIN f_+^{D\pi}(0)&=  0.612(35)\FLAGAVEND&&\Ref~\mbox{\cite{Lubicz:2017syv}},\nonumber\\[-3mm]
&N_f=2+1+1:&\label{eq:Nf=2p1p1Dsemi}\\[-3mm]
        &&\FLAGAVBEGIN f_+^{DK}(0)  &= 0.7385(44)\FLAGAVEND &&\Refs~\mbox{\cite{Lubicz:2017syv,Chakraborty:2021qav}}.\nonumber
\end{align}
%

It is worth noting that, at the current level of precision, no significant effect of
the dynamical charm quark is observed.  However, given the paucity of results, it is
premature to infer strong conclusions on this point.

In Fig.~\ref{fig:DtoPiK}, we display the existing $N_f =2$, $N_f = 2+1$, and $N_f=2+1+1$
results for $f_+^{D\pi}(0)$ and $f_+^{DK}(0)$; the grey bands show our
estimates of these quantities.  Section~\ref{sec:Vcd} discusses the
implications of these results for determinations of the CKM matrix
elements $|V_{cd}|$ and $|V_{cs}|$ and tests of unitarity of the
second row of the CKM matrix.

In the case of $N_f=2+1+1$, we can also provide a complete result for the $q^2$
dependence of $f_+$ and $f_0$. In the case of the $D \to \pi\ell\nu$ channel,
the latter is provided by the fit given in ETM 17D (Ref.~\cite{Lubicz:2017syv}),
to which we refer the reader.
For $D \to K\ell\nu$, we can average the results in ETM 17D (Ref.~\cite{Lubicz:2017syv}),
and HPQCD 21A (Ref.~\cite{Chakraborty:2021qav}). To that purpose, we use the parameterizations
provided in the papers to produce synthetic data for both $f_+(q^2)$ and $f_0(q^2)$ at
a number of values of $q^2$. The large correlations involved make covariance matrices
ill-behaved as the number of values of $q^2$ considered increases; we have settled
for two $q^2$ values for ETM 17D and three $q^2$ values for HPQCD 21A, in both cases
including the kinematical endpoints $q^2=0$ and $q^2=(m_D-m_K)^2$ of the semileptonic interval.
This choice allows us to obtain well-behaved covariance matrices.
We fit the resulting dataset to a BCL ansatz (cf. Eqs.~(\ref{eq:bcl_c},\ref{eq:bcl_f0}))
for a number of combinations of the highest orders $N_+$ and $N_0$ considered for either form
factor; the constraint $f_+(0)=f_0(0)$ is used to rewrite the highest-order coefficient $a^0_{N_0-1}$
in $f_0$ in terms of the other $N_++N_0-1$ coefficients. In both form factors, we include
non-trivial Blaschke factors, with pole masses set to the experimental values of the $D_s^*$
(for the vector channel) and $D_{s0}$ (scalar channel) masses found in the PDG~\cite{Zyla:2020zbs}.
We take flavour averages of charged and neutral states for the $D$ and $K$ masses.
Our external input is thus $m_D=1.87265~{\rm GeV}$, $m_K=495.644~{\rm MeV}$,
$m_{D_s^*}=2.1122~{\rm GeV}$, and $m_{D_{s0}}=2.317~{\rm GeV}$. With this setup,
we observe stable fits beyond the linear approximation in $z$ for the form factors,
although precision is rapidly lost for coefficients of terms of $\mathcal{O}(z^3)$ and higher.
We quote as our preferred fit, and, therefore, FLAG average, the $N_+=N_0=3$ result,
quoted in full in Tab.~\ref{tab:FFDK}, and illustrated in Fig.~\ref{fig:LQCDzfitDK}.
As clearly shown in the figure, there is some tension between the two datasets, that grows
with $q^2$ to reach the $\sim 2\sigma$ level. This results in a relatively poor $\chi^2/{\rm d.o.f.}=9.17/3$,
which has resulted in our rescaling the errors of our average fit accordingly.

\begin{table}[t]
\begin{center}
\begin{tabular}{|c|r|rrrrr|}
\multicolumn{7}{l}{$D\to K\ell\nu \; (N_f=2+1+1)$} \\[0.2em]\hline
        & \multicolumn{1}{c}{values} & \multicolumn{5}{|c|}{correlation matrix} \\[0.2em]\hline
$a_0^+$  & 0.7877(87)  & 1.000000 & $-$0.498440 & 0.073805 & 0.687417 & 0.363513\\[0.2em]
$a_1^+$  & $-$0.97(18) & $-$0.498440 & 1.000000 & $-$0.609159 & $-$0.063023 & 0.309377\\[0.2em]
$a_2^+$  & $-$0.3(2.0) & 0.073805 & $-$0.609159 & 1.000000 & 0.020575 & 0.007175\\[0.2em]
$a_0^0$  & 0.6959(47)  & 0.687417 & $-$0.063023 & 0.020575 & 1.000000 & 0.273019\\[0.2em]
$a_1^0$  & 0.775(69)   & 0.363513 & 0.309377 & 0.007175 & 0.273019 & 1.000000\\[0.2em]
\hline
\end{tabular}
\end{center}
\caption{Coefficients for the $N^+ =3, N^0=3$ $z$-expansion of the $N_f=2+1+1$ FLAG average for the $D\to K$ form factors $f_+$ and $f_0$, and their correlation matrix.\label{tab:FFDK}}
\end{table}

\begin{figure}[h]
\begin{center}
\includegraphics[width=0.7\linewidth]{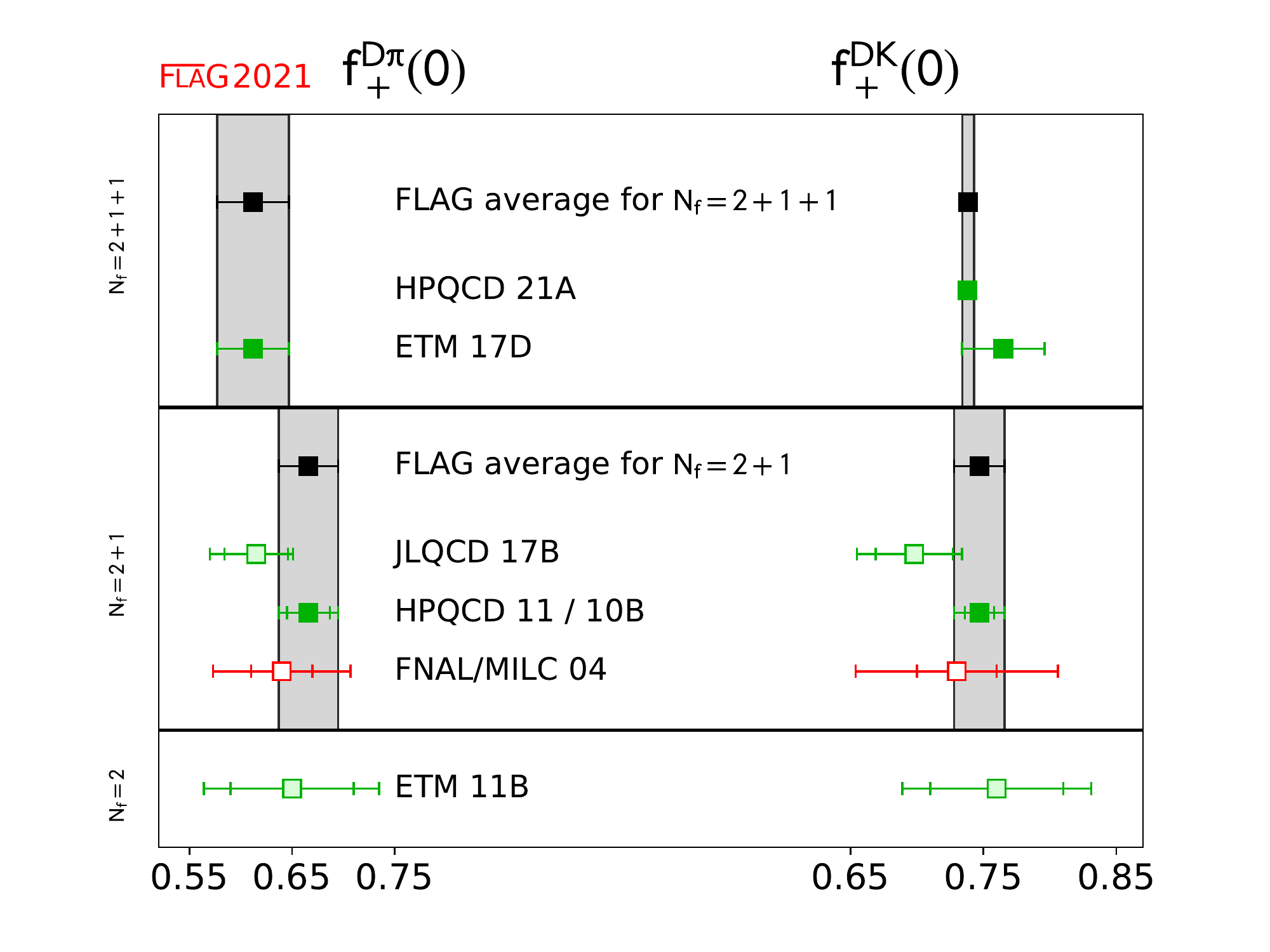}

\vspace{-2mm}
\caption{$D\to\pi \ell\nu$ and $D\to K\ell\nu$ semileptonic form
  factors at $q^2=0$. The $N_f=2+1$ HPQCD result for
  $f_+^{D\pi}(0)$ is from HPQCD 11, the one for $f_+^{DK}(0)$
  represents HPQCD 10B (see Tab.~\ref{tab_DtoPiKsumm2}). \label{fig:DtoPiK}}
 \end{center}
\end{figure}

\begin{figure}[tbp]
\begin{center}
\includegraphics[width=0.49\textwidth]{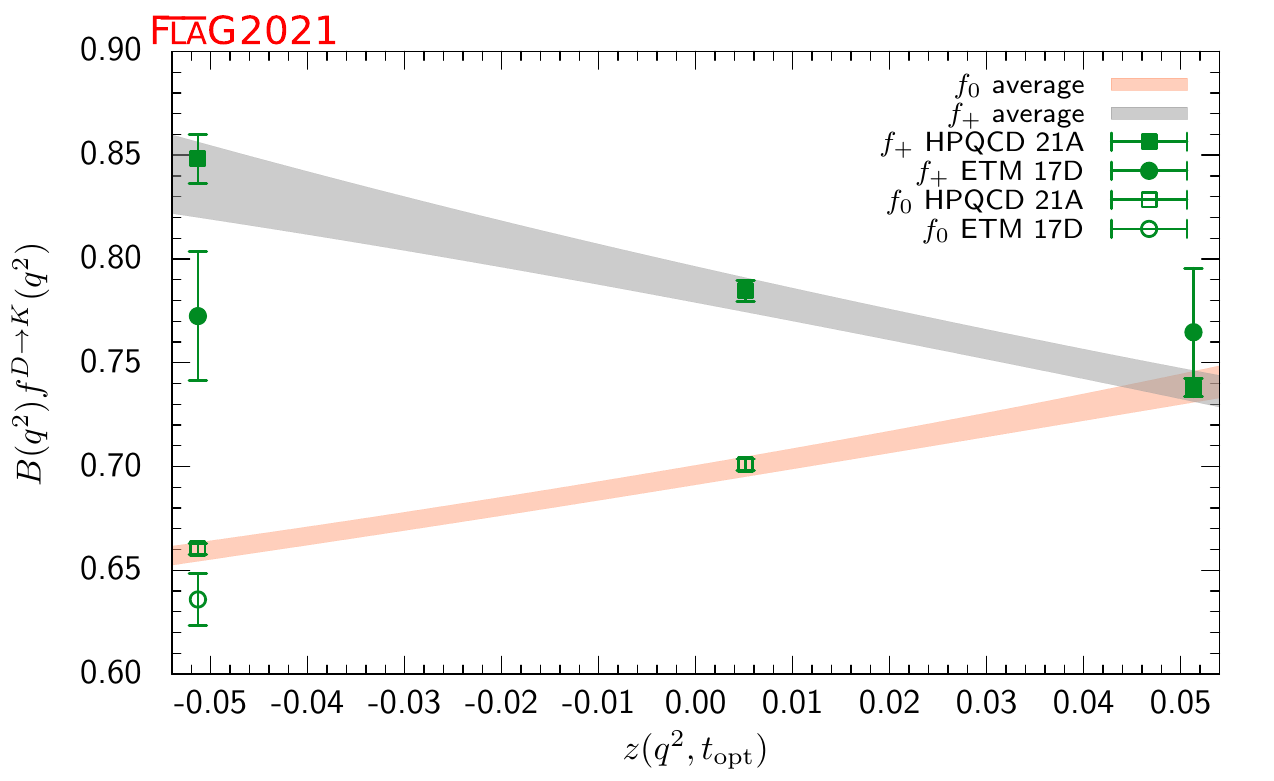}
\includegraphics[width=0.49\textwidth]{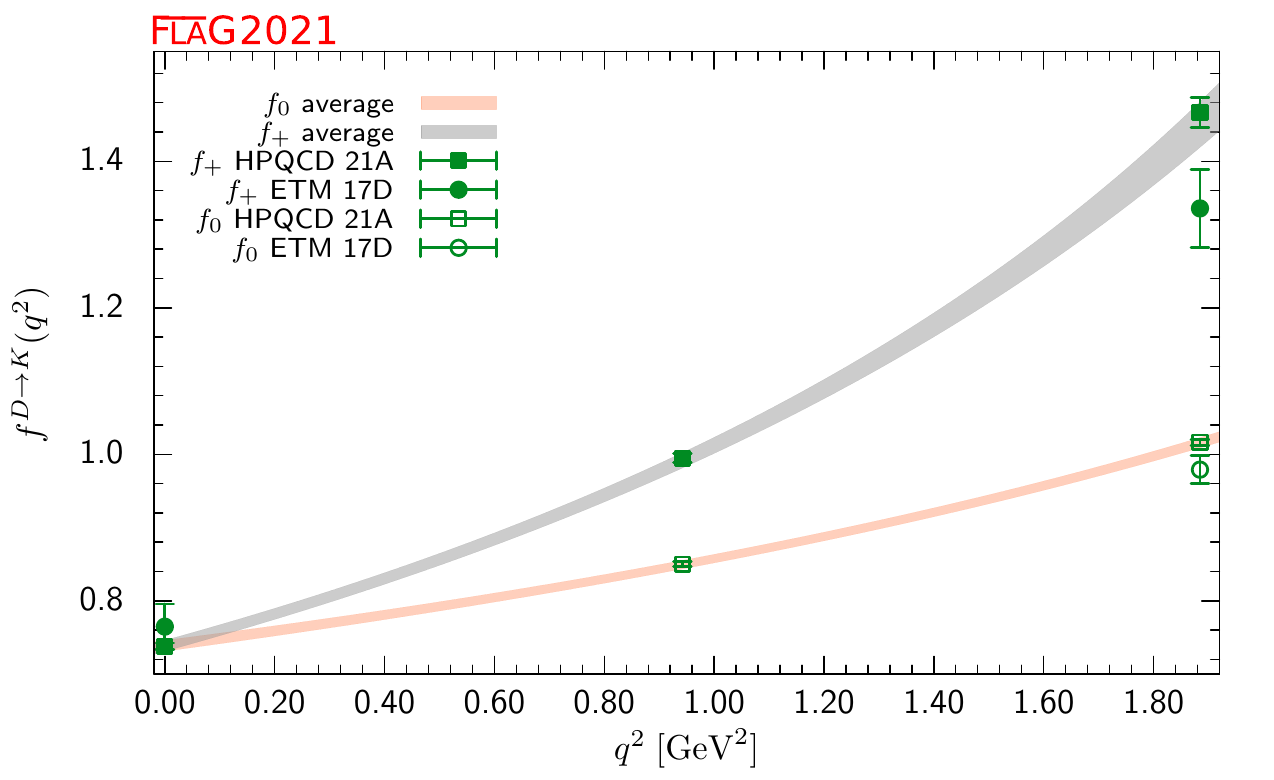}
\caption{The form factors $f_+(q^2)$ and $f_0 (q^2)$ for $D \to K\ell\nu$ plotted versus $z$ (left panel) and $q^2$ (right panel).
In the left plot, we removed the Blaschke factors.
See text for a discussion of the data set. The grey and salmon bands display our preferred $N^+=N^0=3$ BCL fit (five parameters).}\label{fig:LQCDzfitDK}
\end{center}
\end{figure}

\FloatBarrier

\subsection{Form factors for $\Lambda_c$ and $\Xi_c$ semileptonic decays}

The motivation for studying charm-baryon semileptonic decays is two-fold. First, these decays
allow for independent determinations of $|V_{cs}|$.
Second, given that possible new-physics contributions to the $c\to s\ell\nu$ weak effective Hamiltonian
are already constrained to be much smaller compared to $b\to u\ell\bar{\nu}$ and $b\to s \ell\ell$,
charm-baryon semileptonic decays allow testing the lattice techniques for baryons that are also
employed for bottom-baryon semileptonic decays (see Sec.~\ref{sec:Lambdab}) in a better-controlled
environment.


The amplitudes of the decays $\Lambda_c\to \Lambda\ell\nu$
receive contributions from both the vector and the axial components of the current
in the matrix element
$\langle \Lambda|\bar s\gamma^\mu(\mathbf{1}-\gamma_5)c|\Lambda_c\rangle$,
and can be parameterized in terms of six different form factors  $f_+$, $f_0$, $f_\perp$, $g_+$, $g_0$, $g_\perp$ --- see, e.g., Ref.~\cite{Feldmann:2011xf} for a complete description.

The computation in Meinel 16~\cite{Meinel:2016dqj} uses RBC/UKQCD $N_f=2+1$ DWF ensembles,
and treats the $c$ quarks within the Columbia RHQ approach.
Two values of the lattice spacing ($a\approx 0.11,~0.085~{\rm fm}$) are considered,
with the absolute scale set from the $\Upsilon(2S)$--$\Upsilon(1S)$ splitting.
In one ensemble, the pion mass $m_\pi\approx 139~{\rm MeV}$ is at the physical point,
while for other ensembles it ranges from 295 to 352 MeV.
Results for the form factors are obtained from suitable three-point functions,
and fitted to a modified $z$-expansion ansatz that combines the $q^2$-dependence
with the chiral and continuum extrapolations. The paper 
predicts for the total rates in the $e$ and $\mu$ channels
\begin{gather}
\begin{split}
\frac{\Gamma(\Lambda_c\to \Lambda e^+\nu_e)}{|V_{cs}|^2} &= 0.2007(71)(74)~{\rm ps}^{-1}\,,\\
\frac{\Gamma(\Lambda_c\to \Lambda\mu^+\nu_\mu)}{|V_{cs}|^2} &= 0.1945(69)(72)~{\rm ps}^{-1}\,,
\end{split}
\end{gather}
where the uncertainties are statistical and systematic, respectively. In combination with the recent experimental determination of the total branching fractions
by BESIII~\cite{Ablikim:2015prg,Ablikim:2016vqd}, it is possible to extract $|V_{cs}|$
as discussed in Sec.~\ref{sec:Vcd} below.

Lattice results are also available for the $\Lambda_c \to N$ form factors, where $N$ is a neutron or proton \cite{Meinel:2017ggx}.
This calculation uses the same lattice actions but a different set of ensembles with parameters matching those used in the
2015 calculation of the $\Lambda_b \to p$ form factors in Ref.~\cite{Detmold:2015aaa} (cf. Sec.~\ref{sec:Lambdab}). Predictions
are given for the rates of the $c\to d$ semileptonic decays $\Lambda_c\to n \ell^+\nu_\ell$; these modes have not yet
been observed. Reference~\cite{Meinel:2017ggx} also studies the phenomenology of the flavour-changing neutral-current decay
$\Lambda_c \to p \mu^+\mu^-$. As is typical for rare charm decays to charged leptons, this mode is dominated by long-distance effects
that have not yet been calculated on the lattice and whose description is model-dependent.

Recently, the authors of Zhang 21 \cite{Zhang:2021oja} also performed a first lattice calculation of the $\Xi_c \to \Xi$ form factors 
and extracted $|V_{cs}|$, with still large uncertainties, from the recent Belle measurement of the $\Xi_c \to \Xi \ell^+ \nu_\ell$ branching fractions \cite{Li:2021uhk}. This calculation uses only two ensembles with $2+1$ flavours of clover fermions, with lattice spacings of $0.108$ and $0.080$ fm and nearly identical pion masses of 290 and 300 MeV. The results are extrapolated to the continuum limit but are not extrapolated to the physical pion
mass. No systematic uncertainty is estimated for the effect of the missing chiral extrapolation.

A summary of the lattice calculations of charm-baryon semileptonic decay form factors is given in Tab.~\ref{tab_CharmBaryonSLsumm2}.

\begin{table}[h]
\begin{center}
\mbox{} \\[3.0cm]
\footnotesize
\begin{tabular}{l l @{\extracolsep{\fill}} r l l l l l l l}
Process & Collaboration & Ref. & $\Nf$ & 
\hspace{0.15cm}\begin{rotate}{60}{publication status}\end{rotate}\hspace{-0.15cm} &
\hspace{0.15cm}\begin{rotate}{60}{continuum extrapolation}\end{rotate}\hspace{-0.15cm} &
\hspace{0.15cm}\begin{rotate}{60}{chiral extrapolation}\end{rotate}\hspace{-0.15cm}&
\hspace{0.15cm}\begin{rotate}{60}{finite volume}\end{rotate}\hspace{-0.15cm}&
\hspace{0.15cm}\begin{rotate}{60}{renormalization}\end{rotate}\hspace{-0.15cm}  &
\hspace{0.15cm}\begin{rotate}{60}{heavy-quark treatment}\end{rotate}\hspace{-0.15cm} \\
&&&&&&&& \\[-0.1cm]
\hline
\hline
&&&&&&&& \\[-0.1cm]
$\Xi_c\to\Xi \ell\nu$           & Zhang  21  & \cite{Zhang:2021oja}     & 2+1 & \oP & \soso & \tbr  & \soso & \good & \tbr \\[0.5ex] 
$\Lambda_c\to  n \ell\nu$       & Meinel 17 & \cite{Meinel:2017ggx} & 2+1 & \gA & \soso & \soso & \tbr  & \soso & \okay \\[0.5ex]
$\Lambda_c\to  \Lambda \ell\nu$ & Meinel 16 & \cite{Meinel:2016dqj} & 2+1 & \gA & \soso & \good & \good & \soso & \okay \\[0.5ex]
&&&&&&&& \\[-0.1cm]
\hline
\hline
\end{tabular}
\caption{Summary of computations of charmed-baryon semileptonic form factors.}
\label{tab_CharmBaryonSLsumm2}
\end{center}
\end{table}

\subsection{Form factors for charm semileptonic decays with heavy spectator quarks}
\label{sec:charmheavyspectator}

Two other decays mediated by the $c\to s\ell\nu$ and $c\to d\ell\nu$ transitions are $B_c \to B_s \ell\nu$ and $B_c \to B^0 \ell\nu$, respectively. At present, there are no experimental results for these processes, but it may be possible to produce them at LHCb in the future. The HPQCD Collaboration has recently computed the form factors for both of these $B_c$ decay modes with $N_f=2+1+1$~\cite{Cooper:2020wnj}. The calculation uses six different MILC ensembles with HISQ light, strange, and charm quarks, and employs the PCAC 
Ward identity to nonperturbatively renormalize the $c\to s$ and $c\to d$ currents. Data were generated for two different choices of lattice action for the spectator $b$ quark: lattice NRQCD on five of the six ensembles, and HISQ on three of the six ensembles (cf.~Sec.~\ref{sec:BDecays} for a discussion of different lattice approaches used for the $b$ quark). For the NRQCD calculation, two of the ensembles have a physical light-quark mass, and the lattice spacings are 0.15 fm, 0.12 fm, and 0.09 fm. The heavy-HISQ calculation is performed only at $m_l/m_s=0.2$, and at lattice spacings of 0.12 fm, 0.09 fm, and 0.06 fm. The largest value of the heavy-HISQ mass used is 0.8 in lattice units on all three ensembles, which does not reach the physical $b$-quark mass even at the finest lattice spacing.

Form-factor fits are performed using $z$-expansions (see Appendix~\ref{sec:zparam}) modified to include dependence on the lattice spacing and quark masses, including an expansion in the inverse heavy quark mass in the case of the heavy-HISQ approach. The parameters $t_+$ are set to $(m_{B_c}+m_{B_{(s)}})^2$ even though the branch cuts start at $(m_D+m_K)^2$ or $(m_D+m_\pi)^2$, as also noted by the authors. The variable $z$ is rescaled by a constant. The lowest charmed-meson poles are removed before the $z$-expansion, but this still leaves the branch cuts and higher poles below $t_+$. As a consequence of this structure, the good convergence properties
of the $z$-expansion are not necessarily expected to apply. Fits are performed (i) using the NRQCD data only, (ii) using the HISQ data only, and (iii) using the NRQCD data, but with priors on the continuum-limit form-factor parameters equal to the results of the HISQ fit. The results from fits (i) and (ii) are mostly consistent, with the NRQCD fit having smaller uncertainties than the HISQ fit. Case (iii) then results in the smallest uncertainties and gives the predictions (for massless leptons)
\begin{gather}
\begin{split}
\frac{\Gamma(B_c \to B_s \ell^+\nu_\ell)}{|V_{cs}|^2} &= 1.738(55)\times10^{-11}~{\rm MeV}\,,\\
\frac{\Gamma(B_c \to B^0 \ell^+\nu_\ell)}{|V_{cd}|^2} &= 2.29(12)\times10^{-11}~{\rm MeV}\,.
\end{split}
\end{gather}
We note that there is a discrepancy between the NRQCD and HISQ results in the case of $f_0(B_c\to B^0)$, and the uncertainty quoted for method (iii) does not cover this discrepancy. However, this form factor does not enter in the decay rate for massless leptons.

\subsection{Determinations of $|V_{cd}|$ and $|V_{cs}|$ and test of  second-row CKM unitarity}
\label{sec:Vcd}

We now interpret the lattice-QCD results for the $D_{(s)}$ meson decays
as determinations of the CKM
matrix elements $|V_{cd}|$ and $|V_{cs}|$ in the Standard Model.

For the leptonic decays, we use the latest experimental averages from
the Particle Data Group~\cite{Zyla:2020zbs} (see Sec.~71.3.1)
\begin{equation}
f_D |V_{cd}| = 46.2(1.2)~{\rm MeV} \,, \qquad f_{D_s} |V_{cs}| = 245.7(4.6)~{\rm MeV}, \,
\end{equation}
where the errors include those from nonlattice theory, e.g., estimates of radiative corrections to lifetimes~\cite{Dobrescu:2008er}.
By combining these with the average values of $f_D$ and $f_{D_s}$ from
the individual $N_f = 2$, $N_f = 2+1$ and $N_f=2+1+1$ lattice-QCD 
calculations that
satisfy the FLAG criteria, we obtain the results for the CKM
matrix elements $|V_{cd}|$ and $|V_{cs}|$ in
Tab.~\ref{tab:VcdVcsIndividual}.  
For our preferred values we use the
averaged $N_f=2$, $2+1$,  and $2+1+1$ results for $f_D$ and $f_{D_s}$ in
Eqs.~(\ref{eq:fD2}-\ref{eq:fDratio2+1+1}).
We obtain
\begin{align}
  &{\rm leptonic~decays}, N_f=2+1+1:\hspace{-4mm} & |V_{cd}| &= 0.2179(7)(57)\,,  &|V_{cs}| &= 0.983(2)(18) \,, \\
  &\textrm{Refs.~\cite{Bazavov:2017lyh,Carrasco:2014poa}}\;, &&&& \nonumber \\
  &{\rm leptonic~decays}, N_f=2+1:                & |V_{cd}| &= 0.2211(25)(57)\,, &|V_{cs}| &= 0.991  (7)(19) \,, \\
  &\textrm{Refs.~\cite{Bazavov:2011aa,Davies:2010ip,Na:2012iu,Yang:2014sea,Boyle:2017jwu}}\;, &&&& \nonumber \\
  &{\rm leptonic~decays}, N_f=2:                & |V_{cd}| &= 0.2221(74)(57)\,, &|V_{cs}| &= 0.998 (16)(19) \,, \\
  &\textrm{Refs.~\cite{Carrasco:2013zta,Balasubramanian:2019net}}\;, &&&& \nonumber
\end{align}
where the errors shown are from the lattice calculation and experiment
(plus nonlattice theory), respectively.  For the $N_f = 2+1$ and the $N_f=2+1+1$
determinations, the uncertainties from the lattice-QCD calculations of
the decay constants are significantly smaller than the
experimental uncertainties in the branching fractions.

The leptonic determinations of these CKM matrix elements have uncertainties that are reaching the few-percent level.
However, higher-order electroweak and hadronic-structure dependent corrections to the rate have not been computed for
the case of $D_{(s)}$ mesons,
whereas they have been estimated to be around 1--2\% for pion and kaon decays~\cite{Cirigliano:2007ga}. 
Therefore, it is important that such theoretical calculations are tackled soon, perhaps directly on the lattice, as proposed
in Ref.~\cite{Carrasco:2015xwa}. 

\begin{table}[tb]
\begin{center}
\noindent
\begin{tabular*}{\textwidth}[l]{@{\extracolsep{\fill}}lrlcr}
Collaboration & Ref. &$\Nf$&from&\rule{0.8cm}{0cm}$|V_{cd}|$ or $|V_{cs}|$\\
&&&& \\[-2ex]
\hline \hline &&&&\\[-2ex]
FNAL/MILC~17 & \cite{Bazavov:2017lyh} & 2+1+1 & $ f_D$ & 0.2179(6)(57) \\
ETM~17D/Riggio 17 & \cite{Lubicz:2017syv,Riggio:2017zwh} & 2+1+1 & $D\to\pi\ell\nu$ & 0.2341(74) \\
ETM~14E & \cite{Carrasco:2014poa} &  2+1+1 & $ f_D$ & 0.2228(41)(57) \\
RBC/UKQCD~17 & \cite{Boyle:2017jwu} & 2+1 & $f_D$ & 0.2214(36)(57) \\
HPQCD 12A & \cite{Na:2012iu} & 2+1 & $f_{D}$  & 0.2218(36)(57) \\
HPQCD 11 & \cite{Na:2011mc} & 2+1 & $D \to \pi \ell \nu$  & 0.2140(93)(29) \\
FNAL/MILC 11  & \cite{Bazavov:2011aa} & 2+1 & $f_{D}$  &  0.2110(108)(55)  \\
ETM 13B  & \cite{Carrasco:2013zta} & 2 & $f_{D}$  &  0.2221(74)(57)  \\
&&&& \\[-2ex]
 \hline
&&&& \\[-2ex]
HPQCD 21A & \cite{Chakraborty:2021qav} & 2+1+1 & $D\to K\ell\nu$ & 0.9750(54)(45)$^\dagger$ \\
FNAL/MILC~17 & \cite{Bazavov:2017lyh} & 2+1+1 & $ f_{D_s}$ & 0.983(2)(18) \\
ETM~17D/Riggio 17 & \cite{Lubicz:2017syv,Riggio:2017zwh} & 2+1+1 & $D\to K\ell\nu$ & 0.970(33) \\
ETM~17D ($q^2=0$) & \cite{Lubicz:2017syv} & 2+1+1 & $D\to K\ell\nu$ & 0.939(38) \\
ETM~14E & \cite{Carrasco:2014poa} &  2+1+1 & $ f_{D_s}$ & 0.994(17)(19) \\
RBC/UKQCD~17 & \cite{Boyle:2017jwu} & 2+1 & $f_{D_s}$ & 0.997(9)(19) \\
Meinel~16 & \cite{Meinel:2016dqj} & 2+1 & $\Lambda_c\to\Lambda\ell\nu$ & 0.949(24)(51) \\
$\chi$QCD~14 & \cite{Yang:2014sea} & 2+1 & $f_{D_s}$  &  0.968(17)(19) \\
FNAL/MILC 11 & \cite{Bazavov:2011aa} & 2+1 & $f_{D_s}$  &  0.945(40)(19) \\
HPQCD 10A & \cite{Davies:2010ip} & 2+1 & $f_{D_s}$  & 0.991(10)(19)  \\
HPQCD 10B & \cite{Na:2010uf} & 2+1 & $D \to K \ell \nu$  & 0.975(25)(7) \\
Balasubramanian~19 & \cite{Balasubramanian:2019net} & 2 &  $f_{D_s}$  &  1.007(18)(19) \\
ETM 13B & \cite{Carrasco:2013zta} & 2 & $f_{D_s}$  &  0.983(28)(19) \\
&&&& \\[-2ex]
 \hline \hline
\end{tabular*}
\begin{minipage}{\linewidth}
{\footnotesize 
 $^\dagger$ The value quoted in HPQCD 21A is actually $|V_{cs}|= 0.9663(53)_{\rm latt}(39)_{\rm exp}(19)_{\eta_{EW}} (40)_{\rm EM}$,
 and takes into account an electroweak correction $\eta_{EW}=1.009(2)$ that we have eliminated to allow for a 
 straight comparison with the other results. The three remaining errors have been combined in quadrature. Note also
 that the other computations in the table do not incorporate estimates of electroweak and soft electromagnetic corrections.
 HPQCD 21A also quotes a value for $|V_{cs}|$ obtained from the total branching fraction that results in a very small
 decrease in the total error due to a reduction in the estimate of electromagnetic corrections.
 }
 \end{minipage}
\caption{Determinations of $|V_{cd}|$ (upper panel) and $|V_{cs}|$
  (lower panel) obtained from lattice calculations of $D$-meson
  leptonic decay constants
 and semileptonic form factors.
The errors
  shown are from the lattice calculation and experiment (plus
  nonlattice theory), respectively, save for ETM~17D/Riggio 17,
  where the joint fit to lattice and experimental data does
  not provide a separation of the two sources of error (although the latter is largely theory dominated,
  like other results using $D\to\pi$ and $D\to K$ decays). \label{tab:VcdVcsIndividual}}
\end{center}
\end{table}

\vskip 5mm

For $D$ meson semileptonic decays, there are still no $N_f=2$ results, and
for $N_f=2+1$ the only works entering the FLAG averages are still
HPQCD~10B/11~\cite{Na:2010uf,Na:2011mc}.
For $N_f=2+1+1$, on the other hand, there is a new work that enters FLAG averages,
HPQCD~21A (Ref.~\cite{Chakraborty:2021qav}).
There is also a new experimental result by BESIII~\cite{Ablikim:2018evp},
in which the muon mode $D^0 \to K^-\mu^+\nu_\mu$ has been measured for the first time.
This has two consequences. First, HFLAV has updated their averages for the combinations $f_+(0)|V_{cx}|$~\cite{Amhis:2019ckw}.
They now find
\begin{equation}
\label{eq:fpDtoPiandKexp}
	f_+^{D\pi}(0) |V_{cd}| =  0.1426(18) \,, \qquad f_+^{DK}(0) |V_{cs}| =  0.7180(33)  \,
\end{equation}
The previous HFLAV average $f_+^{DK}(0) |V_{cs}| =  0.7226(34)$ differed
from the new one by 1.4 standard deviations.
Second, we now determine $|V_{cs}|$ using the full $q^2$ dependence of the form factors
provided by both HPQCD~21A and ETM~17D (Ref.~\cite{Lubicz:2017syv}).
Using both the new lattice and new experimental input, we perform
a joint lattice+experimental fit to determine the
CKM matrix elements.  This reduces the error on the CKM matrix elements 
significantly compared with just using the form factor at $q^2=0$,
especially for $|V_{cd}|$ (cf. Fig.~\ref{fig:VcdVcs}).
This was, indeed, the strategy to extract $|V_{cd}|$ and $|V_{cs}|$ pursued in a companion paper to ETM 17D,
Ref.~\cite{Riggio:2017zwh}, as as well as in HPQCD 21A (for $|V_{cs}|$ only).\footnote{Notice that the estimate for $|V_{cs}|$ in Ref.~\cite{Riggio:2017zwh} does not
include the later experimental result in Ref.~\cite{Ablikim:2018evp}. The value obtained in
Ref.~\cite{Riggio:2017zwh} is however completely dominated by the uncertainty of the lattice
form factors, and changes very little once the full experimental information is incorporated
into the determination.}

The result for $|V_{cd}|$ in Ref.~\cite{Riggio:2017zwh} is still state-of-the-art, and we will
quote it as the FLAG estimate. In the case of $|V_{cs}|$, we have performed joint lattice+experiment
fits using the same ansatz as described for the lattice average of form factors
in Sec.~\ref{sec:DtoPiK}, including $|V_{cs}|^2$ as an additional coefficient that provides the
normalization of the experimental data.
The experimental datasets we include are three different measurements of the $D^0 \to K^- e^+\nu_e$
mode by BaBar (BaBar 07, Ref.~\cite{Aubert:2007wg}),
CLEO-c (CLEO 09/$0$, Ref.~\cite{Besson:2009uv}), and BESIII (BESIII 15, Ref.~\cite{Ablikim:2015ixa});
CLEO-c (CLEO 09/$+$, Ref.~\cite{Besson:2009uv}) and BESIII measurements of the $D^+ \to \bar K^0 e^+\nu_e$ mode (BESIII 17, Ref.~\cite{Ablikim:2017lks});
and the recent first measurement of the $D^0 \to K^-\mu^+\nu_\mu$ mode by BESIII, Ref.~\cite{Ablikim:2018evp}.
There is also a Belle dataset available in Ref.~\cite{Widhalm:2006wz}, but it provides
results for parameterized form factors rather than partial widths, which implies that
reverse modeling of the $q^2$ dependence of the form factor would be needed
to add them to the fit, which involves an extra source of systematic uncertainty;
it is, furthermore, the measurement with the largest error.  Thus, we will 
drop it. The CLEO collaboration provides correlation matrices for the systematic uncertainties
across the channels in their two measurements; the latter are, however, not available for BESIII,
and, therefore, we will conservatively treat their systematics with a 100\% correlation, following
the same prescription as in the HFLAV review~\cite{Amhis:2019ckw}. Since all lattice results have been obtained in the
isospin limit, we will average over the $D^0$ and $D^+$ electronic modes.

We observe that the error of the
final result for $|V_{cs}|$ is independent of the specific ansatz, while the central values
differ by at most one standard deviation. From the lattice point of view, HPQCD 21A dominates
the result completely, because of its much smaller uncertainties than in ETM 17D.
The precision of the data does not allow us
to consistently resolve the higher-order coefficients of the $z$-expansion beyond $N_+=N_0=3$,
at which point the result for $|V_{cs}|$ becomes insensitive to increasing the order.
Thus, we quote the result from the latter fit, provided in full detail in Table~\ref{tab:FFVCSPI}
and illustrated in Fig.~\ref{fig:DtoKdGammadqsqr}, as the $N_f=2+1+1$ FLAG average.
The $\chi^2/{\rm d.o.f.}$ of our preferred fit is $1.46$, and we have rescaled the full covariance
matrix with that value to obtain conservative error estimates.

\begin{figure}
\begin{center}
\includegraphics[width=0.8\linewidth]{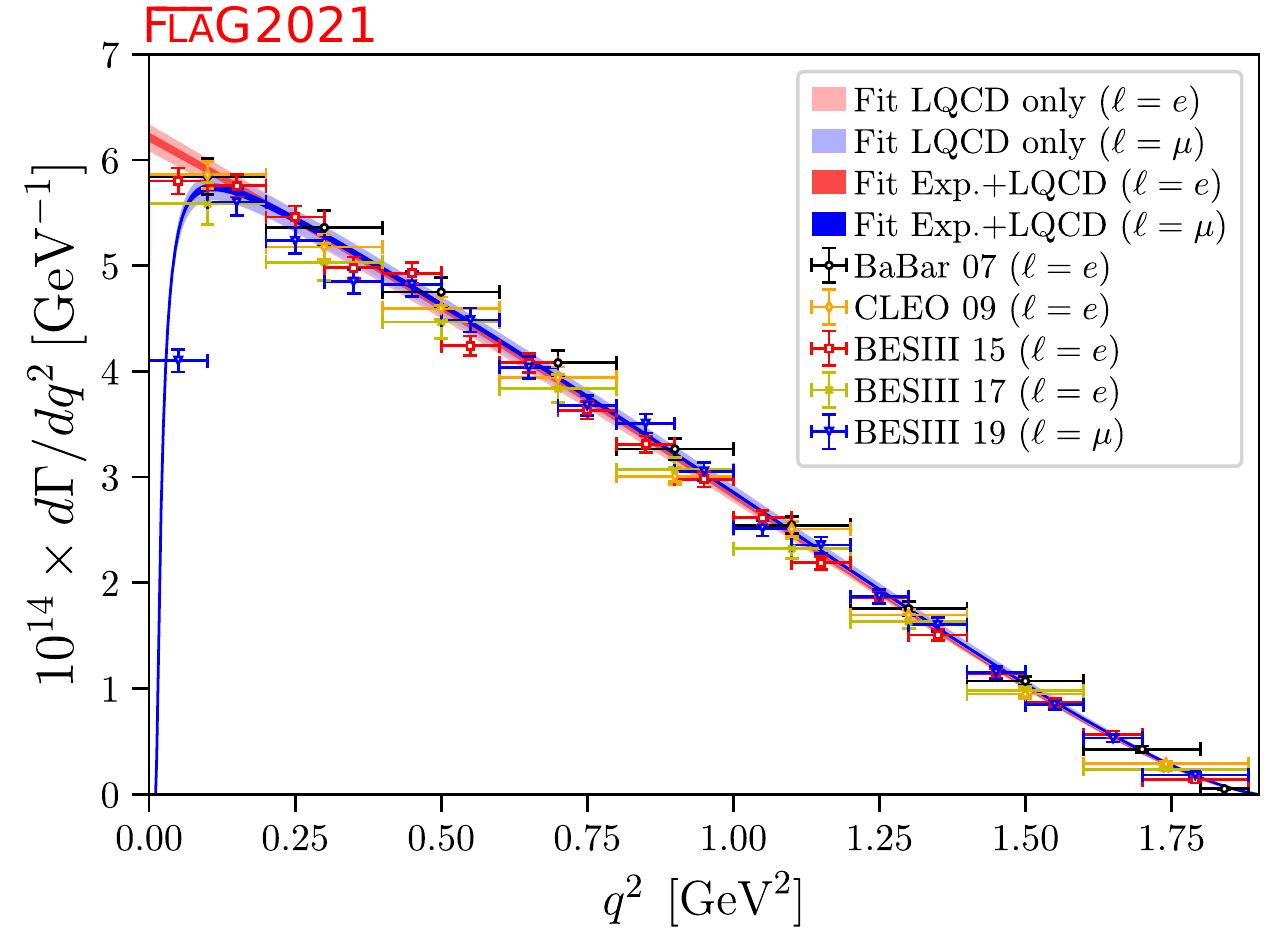}
\end{center}
\caption{The $D\to K \ell \nu$ differential decay rates.\label{fig:DtoKdGammadqsqr}}
\end{figure}

Notice that, notwithstanding the fact that HPQCD 21A dominates the fit, our final value $|V_{cs}|=0.9714(69)$
is slightly higher than their quoted value $|V_{cs}|=0.9663(66)$ (where for the error we have combined
in quadrature their lattice and experiment error, in order to allow for a direct comparison, and dropped
the estimated systematic uncertainties due to electroweak and electromagnetic corrections also provided in HPQCD 21A).
This is due to the fact that HPQCD 21A has applied the structure-independent electroweak correction factor $\eta_{EW}=1.009(2)$ in their
analysis, which we are not doing for consistency with other determinations in this review;
if we had applied the same procedure, our final result would be $|V_{cs}|=0.9628(68)$.

\begin{table}[t]
\begin{center}
\begin{tabular}{|c|r|rrrrrr|}
\multicolumn{8}{l}{$D\to K\ell\nu \; (N_f=2+1+1)$} \\[0.2em]\hline
        & \multicolumn{1}{c}{values} & \multicolumn{6}{|c|}{correlation matrix} \\[0.2em]\hline
$a_0^+$  & 0.7864(54)   & 1 & $-$0.282248 & $-$0.052775 & 0.760032 &  0.631483 & $-$0.899274 \\[0.2em]
$a_1^+$  & $-$0.849(68) & $-$0.282248 & 1 & $-$0.640953 & $-$0.088377 &  0.041977 & 0.128087 \\[0.2em]
$a_2^+$  & $-$1.5(1.1)  & $-$0.052775 & $-$0.640953 & 1 & 0.018139 &  0.115382 & 0.020790 \\[0.2em]
$a_0^0$  & 0.6958(32)   & 0.760032 & $-$0.088377 & 0.018139 & 1 & 0.300343 & $-$0.734376 \\[0.2em]
$a_1^0$  & 0.781(45)    & 0.631483 & 0.041977 & 0.115382 &  0.300343 & 1 & $-$0.664113 \\[0.2em]
$|V_{cs}|$ & 0.9714(69) & $-$0.899274 & 0.128087 & 0.020790 & $-$0.734376 & $-$0.664113 & 1 \\[0.2em]
\hline
\end{tabular}
\end{center}
\caption{Coefficients for the $N^+ = N^0=3$ $z$-expansion of the $D\to K$ form factors $f_+$ and $f_0$, $|V_{cs}|$, and their correlation matrix.\label{tab:FFVCSPI}}
\end{table}

Meinel 16 has also determined the form factors for $\Lambda_c\to\Lambda\ell\nu$
decays for $N_f=2+1$, which results in a determination of  $|V_{cs}|$ in combination with the
experimental measurement of the branching fractions for the $e^+$ and $\mu^+$ channels
in Refs.~\cite{Ablikim:2015prg,Ablikim:2016vqd}.
In Ref.~\cite{Meinel:2016dqj} the value $|V_{cs}|=0.949(24)(14)(49)$ is quoted, where
the first error comes from the lattice computation, the second from the $\Lambda_c$ lifetime,
and the third from the branching fraction of the decay.
While the lattice uncertainty is competitive with meson channels (for $N_f=2+1$),
the experimental uncertainty is far larger.

Our estimates for $|V_{cd}|$ and $|V_{cs}|$ from semileptonic decay are
\begin{align}
&& |V_{cd}| &=  0.2141(93)(29) &&\Ref~\mbox{\cite{Na:2011mc}},\nonumber\\[-0mm]
&\mbox{SL~averages~for}~N_f=2+1:&\label{eq:Nf=2p1VcdVcsSL}\\[-6mm]
&& |V_{cs}| &=  0.967(25)(5) &&\Ref~\mbox{\cite{Na:2010uf}},\nonumber\\[-0mm]
&& |V_{cs}|(\Lambda_c) &=  0.949(24)(51) &&\Ref~\mbox{\cite{Meinel:2016dqj}},\nonumber\\[3mm]
&& |V_{cd}| &=  0.2341(74) &&\Refs~\mbox{\cite{Lubicz:2017syv,Riggio:2017zwh}},\nonumber\\[-3mm]
&\mbox{SL~averages~for}~N_f=2+1+1:&\label{eq:Nf=2p1p1VcdVcsSL}\\[-3mm]
&& |V_{cs}| &=  0.9714(69) &&\Refs~\mbox{\cite{Lubicz:2017syv,Chakraborty:2021qav}},\hspace{4ex}\nonumber
\end{align}
where the errors for $N_f=2+1$ are lattice and experimental (plus nonlattice theory), respectively.
It has to be stressed that for meson decay errors are largely theory-dominated,
save for the $D \to K$ mode for $N_f=2+1+1$ where the lattice contribution to the error is only slightly
larger than the experimental one;
while in the baryon mode for $|V_{cs}|$ the dominant error is experimental.
The above values are compared
with individual leptonic determinations in Tab.~\ref{tab:VcdVcsIndividual}.

\vskip 5mm

In Tab.~\ref{tab:VcdVcsSummary}, we summarize the results for $|V_{cd}|$
and $|V_{cs}|$ from leptonic 
and semileptonic
 decays, and compare
them to determinations from neutrino scattering (for $|V_{cd}|$ only)
and global fits assuming CKM unitarity.  These results are also plotted in
Fig.~\ref{fig:VcdVcs}.  
For both $|V_{cd}|$ and $|V_{cs}|$, the errors in the direct determinations from
leptonic 
and semileptonic 
decays are approximately one order of magnitude larger
than the indirect determination from CKM unitarity.
The direct and indirect determinations are still always
compatible within at most $1.2\sigma$, save for the leptonic
determinations of $|V_{cs}|$---that show a $\sim 2\sigma$ deviation
for all values of $N_f$---and $|V_{cd}|$ using the $N_f=2+1+1$ lattice result,
where the difference is $1.8\sigma$.

In order to provide final estimates, we average all the available results
separately for each value of $N_f$. Whenever two results share ensembles, we have
conservatively fully correlated their statistical uncertainties.
This is a particularly sensitive issue in the average for $|V_{cs}|$, that is dominated by the FNAL/MILC~17
and HPQCD~21A results, and for which precision has been greatly improved by the latter;
however, the uncertainty of the leptonic determination is completely dominated by the
experimental uncertainty, and therefore the impact of the statistical correlation is all but negligible.
We have also 100\% correlated the errors from the heavy-quark discretization and scale setting
in HPQCD's $N_f=2+1$ results.
Finally, we include a 100\% correlation in the fraction of the error
of $|V_{cd(s)}|$ leptonic determinations that comes from the experimental input,
to avoid an artificial reduction of the experimental uncertainty in the averages.
Our results thus are
%
%
\begin{align}
  &{\rm our~average}, N_f=2+1+1 \hspace{-12mm}&\!\!\!\FLAGAVBEGIN |V_{cd}| &= 0.2236(37) \FLAGAVEND\,,& |V_{cs}| &= 0.9741(65)\,, \\ &\textrm{Refs.~\cite{Bazavov:2017lyh,Carrasco:2014poa,Lubicz:2017syv,Riggio:2017zwh,Chakraborty:2021qav}}\;, &&&& \nonumber \\
  &{\rm our~average}, N_f=2+1:  &\!\!\!\FLAGAVBEGIN |V_{cd}| &= 0.2192(54) \FLAGAVEND\,,&|V_{cs}| &= 0.982(16) \,, \\ &\textrm{Refs.~\cite{Bazavov:2011aa,Davies:2010ip,Na:2012iu,Yang:2014sea,Boyle:2017jwu,Na:2011mc,Na:2010uf,Meinel:2016dqj}}\;, &&&& \nonumber \\
  &{\rm our~average}, N_f=2:    &\!\!\!\FLAGAVBEGIN |V_{cd}| &= 0.2221(93) \FLAGAVEND\,,& |V_{cs}| &= 0.998(24) \,, \\ &\textrm{Refs.~\cite{Carrasco:2013zta,Balasubramanian:2019net}}\;, &&&& \nonumber
\label{eq:Vcdsfinal}
\end{align}
%
where the errors include both theoretical and experimental
uncertainties. These averages also appear in Fig.~\ref{fig:VcdVcs}.
The mutual consistency between the various lattice results is
good except for the case of $|V_{cd}|$ with $N_f=2+1+1$, where a $\sim 2\sigma$
tension between the leptonic and semileptonic determinations is observed. 
Currently, the leptonic and semileptonic determinations of $V_{cd}$
are controlled by experimental and lattice uncertainties, respectively. The leptonic error will 
be reduced by Belle~II and BES~III. It would be valuable to have other lattice calculations of the 
semileptonic form factors.

Using the lattice determinations of $|V_{cd}|$ and $|V_{cs}|$ in
Tab.~\ref{tab:VcdVcsSummary}, we can test the unitarity of the second row
of the CKM matrix.  We obtain
\begin{align}
&N_f=2+1+1:   &|V_{cd}|^2 + |V_{cs}|^2 + |V_{cb}|^2 - 1 &= -0.001(8) \,,\\  
&N_f=2+1:     &|V_{cd}|^2 + |V_{cs}|^2 + |V_{cb}|^2 - 1 &= 0.01(3) \,,  \\
&N_f=2:       &|V_{cd}|^2 + |V_{cs}|^2 + |V_{cb}|^2 - 1 &= 0.05(6) \,.  
\end{align}
The much-improved precision in $|V_{cs}|$ ---cf.~the value 0.025(22) quoted in the latest PDG review,
Ref.~\cite{Zyla:2020zbs}--- has thus not resulted in any tension with CKM unitarity.
Note that, given the current level of precision, this result does not depend on 
 $|V_{cb}|$, which is of $\cO(10^{-2})$. Notice, on the other hand, that the final
 quoted precision of 0.7\% makes the incorporation of electromagnetic corrections
 from first principles a necessary step for the near future, similarly to the ongoing
 developments in the light-meson sector.

\begin{table}[tb]
\begin{center}
\noindent
\begin{tabular*}{\textwidth}[l]{@{\extracolsep{\fill}}lcrcc}
& from & Ref. &\rule{0.8cm}{0cm}$|V_{cd}|$ & \rule{0.8cm}{0cm}$|V_{cs}|$\\
&& \\[-2ex]
\hline \hline &&\\[-2ex]
$N_f = 2+1+1$ &  $f_D$ \& $f_{D_s}$ &\cite{Bazavov:2017lyh,Carrasco:2014poa}& 0.2179(57) & 0.983(18) \\
$N_f = 2+1$   &  $f_D$ \& $f_{D_s}$ &\cite{Na:2012iu,Bazavov:2011aa,Boyle:2017jwu,Davies:2010ip,Yang:2014sea}& 0.2211(62) & 0.991(20) \\
$N_f = 2$     &  $f_D$ \& $f_{D_s}$ &\cite{Carrasco:2013zta,Balasubramanian:2019net}& 0.2220(93) & 0.999(25) \\
&& \\[-2ex]
 \hline
&& \\[-2ex]
$N_f = 2+1+1$ & $D \to \pi \ell\nu$ and $D\to K \ell\nu$ &\cite{Lubicz:2017syv,Riggio:2017zwh,Chakraborty:2021qav} & 0.2341(74) & 0.9714(69) \\
$N_f = 2+1$   & $D \to \pi \ell\nu$ and $D\to K \ell\nu$ &\cite{Na:2011mc,Na:2010uf}& 0.2141(97) & 0.967(25) \\
$N_f = 2+1$   & $\Lambda_c \to \Lambda\ell\nu$           &\cite{Meinel:2016dqj}&    n/a     & 0.949(56) \\
&& \\[-2ex]
 \hline
&& \\[-2ex]
PDG & neutrino scattering & \cite{Zyla:2020zbs} & 0.230(11)&  \\
PDG & CKM unitarity & \cite{Zyla:2020zbs} & 0.2265(5) & 0.9732(1) \\
&& \\[-2ex]
 \hline \hline 
\end{tabular*}
\caption{Comparison of determinations of $|V_{cd}|$ and $|V_{cs}|$
  obtained from lattice methods with nonlattice determinations and
  the Standard Model prediction from global fits assuming CKM
  unitarity. Experimental and lattice errors have been combined in quadrature.
  The PDG figures quoted are taken from the ``CKM Quark-Mixing Matrix'' review.
\label{tab:VcdVcsSummary}}
\end{center}
\end{table}

\begin{figure}[h]

\begin{center}
\includegraphics[width=0.7\linewidth]{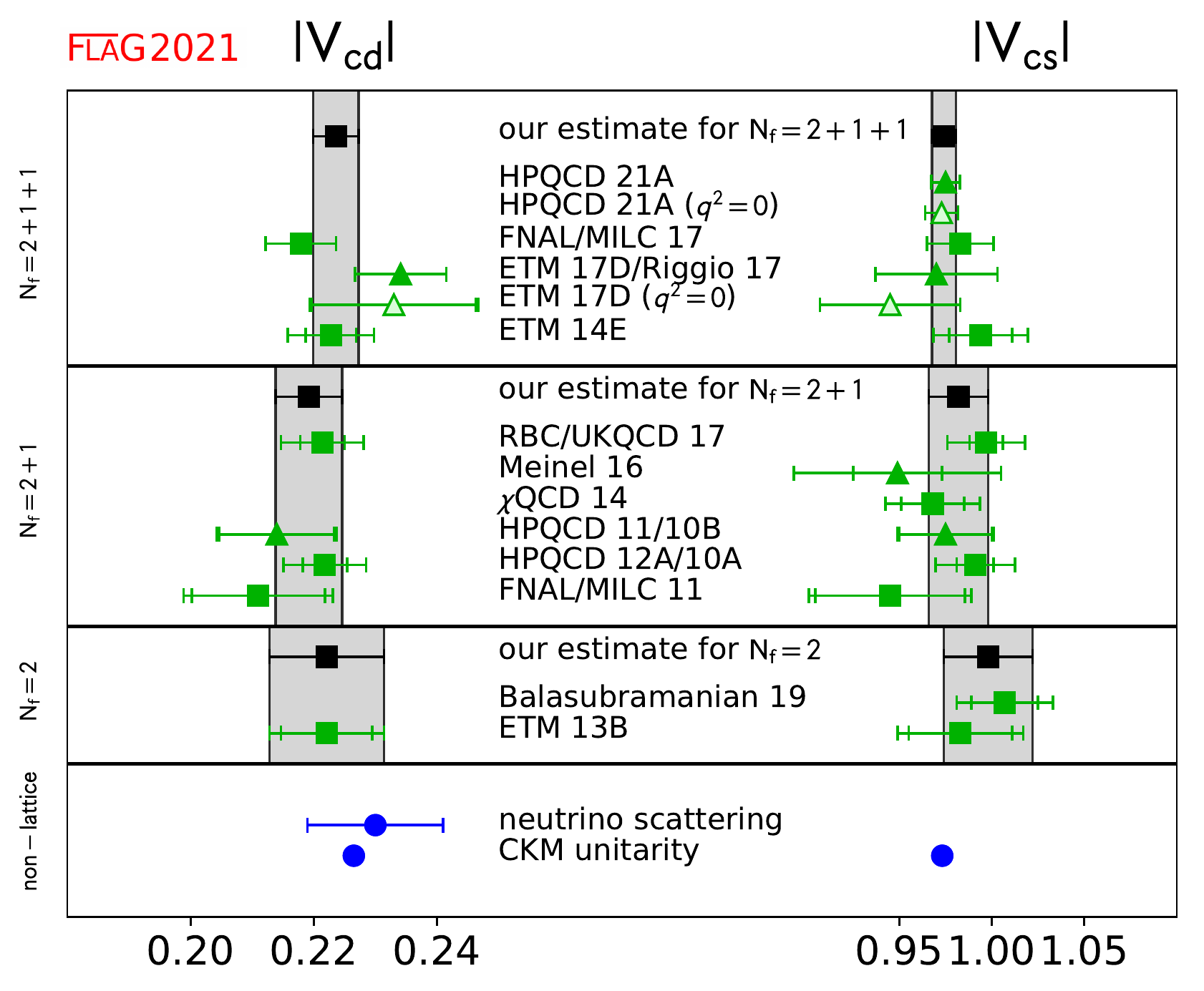}

\vspace{-2mm}
\caption{Comparison of determinations of $|V_{cd}|$ and $|V_{cs}|$
  obtained from lattice methods with nonlattice determinations and
  the Standard Model prediction based on CKM unitarity.  When two
  references are listed on a single row, the first corresponds to the
  lattice input for $|V_{cd}|$ and the second to that for $|V_{cs}|$.
  The results denoted by squares are from leptonic decays, while those
  denoted by triangles are from semileptonic
  decays. The points indicated as ($q^2=0$) do not contribute
  to the average, and are shown to stress the decrease in the final uncertainty
  obtained by considering the full $q^2$ dependence. Notice that the HPQCD 21A
  point includes estimates of the electroweak and soft electromagnetic uncertainties
  that we have not incorporated into our average.
\label{fig:VcdVcs}}
\end{center}
\end{figure}
\clearpage

\clearpage
\setcounter{section}{7}
\section{Bottom hadron decays and mixings}
\label{sec:BDecays}
Authors: Y.~Aoki, M.~Della~Morte, E.~Lunghi, S.~Meinel, C.~Monahan, C.~Pena\\


The (semi)leptonic decay and mixing processes of $B_{(s)}$ mesons have been playing
a crucial role in flavour physics.   In particular, they contain
important information for the investigation of the $b{-}d$ unitarity
triangle in the Cabibbo-Kobayashi-Maskawa (CKM) matrix, and can be
ideal probes of physics beyond the Standard Model.
The charged-current decay channels $B^{+} \rightarrow l^{+}
\nu_{l}$ and $B^{0} \rightarrow \pi^{-} l^{+} \nu_{l}$, where $l^{+}$
is a charged lepton with $\nu_{l}$ being the corresponding neutrino, are
essential in extracting the CKM matrix element $|V_{ub}|$.  Similarly,
the $B$ to $D^{(\ast)}$ semileptonic transitions can be used to
determine $|V_{cb}|$.   The flavour-changing neutral current (FCNC)
processes, such as $B\to K^{(*)} \ell^+
\ell^-$ and $B_{d(s)} \to \ell^+ \ell^-$,  occur only beyond the tree level in weak interactions and are suppressed in the Standard
Model. Therefore, these processes can be sensitive to new
physics, since heavy particles can contribute to the loop diagrams.
They are also suitable channels for the extraction of the CKM matrix
elements involving the top quark that can appear in the loop.
The decays $B\to D^{(*)}\ell\nu$ and $B\to K^{(*)} \ell\ell$ can also be used 
to test lepton flavour universality by comparing results for $\ell = e$, $\mu$ and $\tau$. 
In particular, anomalies have been seen in the ratios $R(D^{(*)}) = {\cal B} (B\to D^{(*)}\tau\nu) /{\cal B} (B\to D^{(*)}\ell\nu)_{\ell=e,\mu}$ and ${R}(K^{(*)}) = {\cal B} (B\to K^{(*)}\mu\mu) /{\cal B} (B\to K^{(*)}ee)$.
In addition, the neutral $B_{d(s)}$-meson mixings are FCNC processes and
are dominated by the 1-loop ``box'' diagrams containing the top quark
and the $W$ bosons.  Thus, using the experimentally measured neutral $B^0_{d(s)}$-meson oscillation
frequencies, $\Delta M_{d(s)}$, and the theoretical calculations for
the relevant hadronic mixing matrix elements, one can obtain
$|V_{td}|$ and $|V_{ts}|$ in the Standard Model.

At the Large Hadron Collider, decays of $b$ quarks can also be probed with $\Lambda_b$ and other bottom baryons, which
can provide complementary constraints on physics beyond the Standard Model. The most important processes are the charged-current
decays $\Lambda_b \to p \ell\bar{\nu}$ and $\Lambda_b \to \Lambda_c \ell\bar{\nu}$, and the neutral-current
decay $\Lambda_b \to \Lambda \ell^+\ell^-$.

Accommodating the light quarks and the $b$ quark simultaneously in
lattice-QCD computations is a challenging endeavour. To incorporate
the pion and the $b$ hadrons with their physical masses, the simulations have to be performed using the lattice
size $\hat{L} = L/a \sim \cO(10^{2})$, where $a$ is the lattice spacing and $L$
is the physical (dimensionful) box size.   
The most ambitious calculations are now using such volumes; 
however, many ensembles are smaller.
Therefore, in addition to employing Chiral Perturbation Theory for the extrapolations in the
light-quark mass, current lattice calculations for quantities involving
$b$ hadrons often make use of effective theories that allow one to
expand in inverse powers of $m_{b}$. In this regard, two general
approaches are widely adopted.  On the one hand, effective field theories
such as Heavy-Quark Effective Theory (HQET) and Nonrelativistic
QCD (NRQCD) can be directly implemented in numerical computations. On
the other hand, a relativistic quark action can be improved {\it \`{a} la}
Symanzik to suppress cutoff errors, and then re-interpreted in a manner
that is suitable for heavy-quark physics calculations.   
This latter strategy is often referred to as the method of the Relativistic
Heavy-Quark Action (RHQA).
The utilization of such effective theories inevitably introduces systematic
uncertainties that are not present in light-quark calculations.  These
uncertainties 
can arise from the truncation of the expansion in constructing the
effective theories (as in HQET and NRQCD),
or from more intricate
cutoff effects (as in NRQCD and RHQA).  They can also be introduced
through more complicated renormalization
procedures which often lead to significant systematic effects in
matching the lattice operators to their continuum counterparts.  For
instance, due to the use of different actions for the heavy and the
light quarks, it is more difficult to construct absolutely 
normalized bottom-light currents.  

Complementary to the above ``effective theory approaches'', 
another popular method is to simulate the heavy and the light quarks
using the same (normally improved) lattice action at several values of
the heavy-quark mass $m_{h}$ with $a m_{h} < 1$ and $m_{h} < m_{b}$.   
This enables one to employ HQET-inspired relations to extrapolate the
computed quantities to the physical $b$ mass.  When combined with
results obtained in the static heavy-quark limit, this approach can be
rendered into an interpolation, instead of extrapolation, in
$m_{h}$. The discretization errors are the main source of the
systematic effects in this method, and very small lattice spacings are
needed to keep such errors under control.

In recent years, it has also been
possible to perform lattice simulations at very fine lattice
spacings and treat heavy quarks as
fully relativistic fermions without resorting to effective field
theories.  
Such simulations are, of course, very demanding in computing
resources.  

Because of the challenge described above, the efforts that have been
made to obtain reliable, accurate lattice-QCD results for physics of the $b$ quark
have been enormous.   These efforts include significant theoretical progress in
formulating QCD with heavy quarks on the lattice. This aspect is
briefly reviewed in Appendix A.1.3 of FLAG 19 \cite{Aoki:2019cca}.

In this section, we summarize the results of the $B$-meson leptonic
decay constants, the neutral $B$-mixing parameters, and the
semileptonic form factors of $B$ mesons and $\Lambda_b$ baryons, 
from lattice QCD.  To focus on the
calculations that have strong phenomenological impact, we limit the
review to results based on modern simulations containing dynamical
fermions with reasonably light pion masses (below
approximately 500~MeV).

Following our review of $B_{(s)}$-meson
leptonic decay constants, the neutral $B$-meson mixing parameters, and
semileptonic form factors, we then interpret our results within the
context of the Standard Model.  We combine our best-determined values
of the hadronic matrix elements with the most recent
experimentally-measured branching fractions to obtain $|V_{ub}|$
and  $|V_{cb}|$,
and compare these results to those obtained from inclusive
semileptonic $B$ decays.


\subsection{Leptonic decay constants $f_B$ and $f_{B_s}$}
\label{sec:fB}
The $B$- and $B_s$-meson decay constants are crucial inputs for
extracting information from leptonic $B$ decays.  Charged $B$ mesons
can decay to a lepton-neutrino final state  through the
charged-current weak interaction.  On the other hand, neutral
$B_{d(s)}$ mesons can decay to a charged-lepton pair via a
flavour-changing neutral current (FCNC) process.

In the Standard Model, the decay rate for $B^+ \to \ell^+ \nu_{\ell}$
is described by a formula identical to Eq.~(\ref{eq:Dtoellnu}), with $D_{(s)}$ replaced by $B$, and the 
relevant CKM matrix element $V_{cq}$ replaced by $V_{ub}$,
\be
\Gamma ( B \to \ell \nu_{\ell} ) =  \frac{ m_B}{8 \pi} G_F^2  f_B^2 |V_{ub}|^2 m_{\ell}^2 
           \left(1-\frac{ m_{\ell}^2}{m_B^2} \right)^2 \;. \label{eq:B_leptonic_rate}
\ee
The only two-body charged-current $B$-meson decay that has been observed so far is 
$B^{+} \to \tau^{+} \nu_{\tau}$, which has been measured by the Belle
and Babar collaborations~\cite{Lees:2012ju,Kronenbitter:2015kls}.
Both collaborations have reported results with errors around $20\%$. These measurements can be used to 
determine $|V_{ub}|$ when combined with lattice-QCD predictions of the corresponding
decay constant.

Neutral $B_{d(s)}$-meson decays to a charged-lepton pair $B_{d(s)}
\rightarrow l^{+} l^{-}$ is a FCNC process, and can only occur at
one loop in the Standard Model.  Hence these processes are expected to
be rare, and are sensitive to physics beyond the Standard Model.
The corresponding expression for the branching fraction has the form 
\be
B ( B_q \to \ell^+ \ell^-) = \tau_{B_q} \frac{G_F^2}{\pi} \, Y \,
\left(  \frac{\alpha}{4 \pi \sin^2 \Theta_W} \right)^2
m_{B_q} f_{B_q}^2 |V_{tb}^*V_{tq}|^2 m_{\ell}^2 
           \sqrt{1- 4 \frac{ m_{\ell}^2}{m_B^2} }\;, 
\ee
where the light quark $q=s$ or $d$, and the function $Y$ includes NLO QCD and electro-weak
corrections \cite{Inami:1980fz,Buchalla:1993bv}. Evidence for the $B_s \to \mu^+ \mu^-$ decay was first observed
by the CMS and the LHCb collaborations, and a combined analysis was
presented in 2014 in Ref.~\cite{CMS:2014xfa}.  In 2020, the ATLAS, CMS and LHCb collaborations 
reported their measurements from a preliminary combined analysis as~\cite{ATLAS:2020acx}
\begin{eqnarray} 
   B(B_d \to \mu^+ \mu^-) &<& (1.9) \times \,10^{-10} \;\mathrm{at}\; 95\% \;\mathrm{CL}, \nonumber\\
   B(B_s \to \mu^+ \mu^-) &=& (2.69^{+0.37}_{-0.35}) \times \,10^{-9} ,
\label{eq:B_to_mumu_ATLAS_2020}
\end{eqnarray}
which are compatible with the Standard Model predictions within approximately 2 standard deviations~\cite{Beneke:2019slt}.
We note that the errors of these results are currently too large to enable a precise determination of $|V_{td}|$ and $|V_{ts}|$.

The decay constants $f_{B_q}$ (with $q=u,d,s$) parameterize the matrix
elements of the corresponding axial-vector currents $A^{\mu}_{bq}
= \bar{b}\gamma^{\mu}\gamma^5q$ analogously to the definition of
$f_{D_q}$ in Sec.~\ref{sec:fD}:
\be
\langle 0| A^{\mu} | B_q(p) \rangle = i p_B^{\mu} f_{B_q} \;.
\label{eq:fB_from_ME}
\ee
For heavy-light mesons, it is convenient to define and analyse the quantity 
\be
 \Phi_{B_q} \equiv f_{B_q} \sqrt{m_{B_q}} \;,
\ee
which approaches a constant (up to logarithmic corrections) in the
$m_B \to \infty$ limit, because of heavy-quark symmetry.
In the following discussion, we denote lattice data for $\Phi$, and the corresponding decay constant $f$,
obtained at a heavy-quark mass $m_h$ and light valence-quark mass
$m_{\ell}$ as $\Phi_{h\ell}$ and $f_{hl}$, to differentiate them from
the corresponding quantities at the physical $b$- and light-quark
masses.

The $SU(3)$-breaking ratio $f_{B_s}/f_B$ is of phenomenological
interest, because many systematic effects can be partially reduced in lattice-QCD calculations of this ratio.  The discretization errors, heavy-quark mass
tuning effects, and renormalization/matching errors may all be partially reduced. 
This $SU(3)$-breaking ratio is, however, still sensitive to the chiral
extrapolation. Provided the chiral extrapolation is under control,
one can then adopt $f_{B_s}/f_B$ as an input in extracting
phenomenologically-interesting quantities.  In addition, it often
happens to be easier to obtain lattice results for $f_{B_{s}}$ with
smaller errors than direct calculations of $f_{B}$.  Therefore, one can combine the $B_{s}$-meson
decay constant with the $SU(3)$-breaking ratio to calculate $f_{B}$.  Such
a strategy can lead to better precision in the computation of the
$B$-meson decay constant, and has been adopted by the
ETM~\cite{Carrasco:2013zta, Bussone:2016iua} and the
HPQCD collaborations~\cite{Na:2012sp}. An alternative strategy, used in Ref.~\cite{Balasubramanian:2019net}, is to obtain the $B_s$-meson decay constant by combining the $D_{s}$-meson decay constant with the ratio $f_{B_s}/f_{D_s}$.

It is clear that the decay constants for charged and neutral $B$
mesons play different roles in flavour-physics phenomenology.  Knowledge of the $B^{+}$-meson decay constant
$f_{B^{+}}$ is essential for extracting $|V_{ub}|$ from
leptonic $B^{+}$ decays.   The neutral $B$-meson decay constants
$f_{B^{0}}$ and $f_{B_{s}}$ are inputs to searches for new physics in rare leptonic $B^{0}$
decays.  In
view of this, it is desirable to include isospin-breaking effects in
lattice computations for these quantities, and have results for
$f_{B^{+}}$ and $f_{B^{0}}$.   
With the increasing precision of recent lattice calculations, isospin splittings for $B$-meson decay constants can be significant, 
and will play an important role in the foreseeable
future.    A few collaborations have reported $f_{B^{+}}$ and $f_{B^{0}}$
separately by taking into account strong isospin effects in the
valence sector, and estimated the corrections from
electromagnetism. The $N_{f}=2+1+1$ strong isospin-breaking effect was
computed in HPQCD 13~\cite{Dowdall:2013tga} (see 
Tab.~\ref{tab:FBssumm} in this subsection).  However, since only
unitary points (with equal sea- and valence-quark masses) were considered in
HPQCD 13~\cite{Dowdall:2013tga}, this procedure only correctly accounts for the effect from the
valence-quark masses, while introducing a spurious sea-quark
contribution. The decay constants $f_{B^{+}}$ and $f_{B^{0}}$ are also
separately reported in FNAL/MILC 17~\cite{Bazavov:2017lyh} by
taking into account the strong-isospin effect.  The new FNAL/MILC
results were obtained by keeping the averaged light sea-quark
mass fixed when varying the quark masses in their analysis procedure.
Their finding indicates that the strong isospin-breaking effects, $f_{B^+}-f_B\sim 0.5$ MeV, could be smaller than those suggested by previous computations. One would have to take into
account QED effects in the $B$-meson leptonic decay rates to properly use these results for extracting phenomenologically relevant information.\footnote{See Ref.~\cite{Carrasco:2015xwa} for a strategy that
has been proposed to account for QED effects.}  Currently, errors on the
experimental measurements on these decay rates are still very large.   In this review, we
will therefore concentrate on the isospin-averaged result $f_{B}$ and the
$B_{s}$-meson decay constant, as well as the $SU(3)$-breaking ratio
$f_{B_{s}}/f_{B}$.

The status of lattice-QCD computations for $B$-meson decay constants
and the $SU(3)$-breaking ratio, using gauge-field ensembles
with light dynamical fermions, is summarized in Tabs.~\ref{tab:FBssumm}
and~\ref{tab:FBratsumm}, while Figs.~\ref{fig:fB} and~\ref{fig:fBratio} contain the graphical
presentation of the collected results and our averages. Most results in these tables and plots have been reviewed in detail in
FLAG 19~\cite{Aoki:2019cca}. Here, we only describe the new results published after January 2019.  
\begin{table}[!htb]
\mbox{} \\[3.0cm]
\footnotesize
\begin{tabular*}{\textwidth}[l]{@{\extracolsep{\fill}}l@{\hspace{1mm}}r@{\hspace{1mm}}l@{\hspace{1mm}}l@{\hspace{1mm}}l@{\hspace{1mm}}l@{\hspace{1mm}}l@{\hspace{1mm}}l@{\hspace{1mm}}l@{\hspace{5mm}}l@{\hspace{1mm}}l@{\hspace{1mm}}l@{\hspace{1mm}}l@{\hspace{1mm}}l}
Collaboration & Ref. & $\Nf$ & 
\hspace{0.15cm}\begin{rotate}{60}{publication status}\end{rotate}\hspace{-0.15cm} &
\hspace{0.15cm}\begin{rotate}{60}{continuum extrapolation}\end{rotate}\hspace{-0.15cm} &
\hspace{0.15cm}\begin{rotate}{60}{chiral extrapolation}\end{rotate}\hspace{-0.15cm}&
\hspace{0.15cm}\begin{rotate}{60}{finite volume}\end{rotate}\hspace{-0.15cm}&
\hspace{0.15cm}\begin{rotate}{60}{renormalization/matching}\end{rotate}\hspace{-0.15cm}  &
\hspace{0.15cm}\begin{rotate}{60}{heavy-quark treatment}\end{rotate}\hspace{-0.15cm} & 
 $f_{B^+}$ & $f_{B^0}$   & $f_{B}$ & $f_{B_s}$  \\
&&&&&&&&&&&&\\[-0.1cm]
\hline
\hline
&&&&&&&&&&&& \\[-0.1cm]

FNAL/MILC 17  & \cite{Bazavov:2017lyh} & 2+1+1 & \gA & \good & \good & \good 
& \good &  \okay &  189.4(1.4) & 190.5(1.3) & 189.9(1.4) & 230.7(1.2) \\[0.5ex]

HPQCD 17A & \cite{Hughes:2017spc} & 2+1+1 & \gA & \soso & \good & \good 
& \soso &  \okay &  $-$ & $-$ & 196(6) & 236(7) \\[0.5ex]

ETM 16B & \cite{Bussone:2016iua} & 2+1+1 & \gA & \good & \soso & \soso 
& \soso &  \okay &  $-$ & $-$ & 193(6) & 229(5) \\[0.5ex]

ETM 13E & \cite{Carrasco:2013naa} & 2+1+1 & \rC & \good & \soso & \soso 
& \soso &  \okay &  $-$ & $-$ & 196(9) & 235(9) \\[0.5ex]

HPQCD 13 & \cite{Dowdall:2013tga} & 2+1+1 & \gA & \soso & \good & \good & \soso
& \okay &  184(4) & 188(4) &186(4) & 224(5)  \\[0.5ex]

&&&&&&&&&& \\[-0.1cm]
\hline
&&&&&&&&&& \\[-0.1cm]

RBC/UKQCD 14 & \cite{Christ:2014uea} & 2+1 & \gA & \soso & \soso & \soso 
  & \soso & \okay & 195.6(14.9) & 199.5(12.6) & $-$ & 235.4(12.2) \\[0.5ex]

RBC/UKQCD 14A & \cite{Aoki:2014nga} & 2+1 & \gA & \soso & \soso & \soso 
  & \soso & \okay & $-$ & $-$ & 219(31) & 264(37) \\[0.5ex]

RBC/UKQCD 13A & \cite{Witzel:2013sla} & 2+1 & \rC & \soso & \soso & \soso 
  & \soso & \okay & $-$ & $-$ &  191(6)$_{\rm stat}^\diamond$ & 233(5)$_{\rm stat}^\diamond$ \\[0.5ex]

HPQCD 12 & \cite{Na:2012sp} & 2+1 & \gA & \soso & \soso & \soso & \soso
& \okay & $-$ & $-$ & 191(9) & 228(10)  \\[0.5ex]

HPQCD 12 & \cite{Na:2012sp} & 2+1 & \gA & \soso & \soso & \soso & \soso
& \okay & $-$ & $-$ & 189(4)$^\triangle$ &  $-$  \\[0.5ex]

HPQCD 11A & \cite{McNeile:2011ng} & 2+1 & \gA & \good & \soso &
 \good & \good & \okay & $-$ & $-$ & $-$ & 225(4)$^\nabla$ \\[0.5ex] 

FNAL/MILC 11 & \cite{Bazavov:2011aa} & 2+1 & \gA & \soso & \soso &
     \good & \soso & \okay & 197(9) & $-$ & $-$ & 242(10) &  \\[0.5ex]  

HPQCD 09 & \cite{Gamiz:2009ku} & 2+1 & \gA & \soso & \soso & \soso &
\soso & \okay & $-$ & $-$ & 190(13)$^\bullet$ & 231(15)$^\bullet$  \\[0.5ex] 

&&&&&&&&&& \\[-0.1cm]
\hline
&&&&&&&&&& \\[-0.1cm]

Balasubramamian 19$^\dagger$ & \cite{Balasubramanian:2019net} & 2 & \gA & \good & \good & \good & \soso & \okay & $-$ & $-$ & $-$ & 215(10)(2)({\raisebox{0.5ex}{\tiny$\substack{+2 \\ -5}$}})\\[0.5ex]

ALPHA 14 & \cite{Bernardoni:2014fva} & 2 & \gA & \good & \good &\good 
& \good & \okay &  $-$ & $-$ & 186(13) & 224(14) \\[0.5ex]

ALPHA 13 & \cite{Bernardoni:2013oda} & 2 & \rC  & \good   & \good   &
\good    &\good  & \okay   & $-$ & $-$ & 187(12)(2) &  224(13) &  \\[0.5ex] 

ETM 13B, 13C$^\ddagger$ & \cite{Carrasco:2013zta,Carrasco:2013iba} & 2 & \gA & \good & \soso & \good
& \soso &  \okay &  $-$ & $-$ & 189(8) & 228(8) \\[0.5ex]

ALPHA 12A& \cite{Bernardoni:2012ti} & 2 & \rC  & \good      & \good      &
\good          &\good  & \okay   & $-$ & $-$ & 193(9)(4) &  219(12) &  \\[0.5ex] 

ETM 12B & \cite{Carrasco:2012de} & 2 & \rC & \good & \soso & \good
& \soso &  \okay &  $-$ & $-$ & 197(10) & 234(6) \\[0.5ex]

ALPHA 11& \cite{Blossier:2011dk} & 2 & \rC  & \good      & \soso      &
\good          &\good  & \okay  & $-$ & $-$ & 174(11)(2) &  $-$ &  \\[0.5ex]  

ETM 11A & \cite{Dimopoulos:2011gx} & 2 & \gA & \good & \soso & \good
& \soso &  \okay & $-$ & $-$ & 195(12) & 232(10) \\[0.5ex]

ETM 09D & \cite{Blossier:2009hg} & 2 & \gA & \good & \soso & \soso
& \soso &  \okay & $-$ & $-$ & 194(16) & 235(12) \\[0.5ex]
&&&&&&&&&& \\[-0.1cm]
\hline
\hline
\end{tabular*}
\begin{tabular*}{\textwidth}[l]{l@{\extracolsep{\fill}}lllllllll}
  \multicolumn{10}{l}{\vbox{\begin{flushleft} 
	$^\diamond$Statistical errors only. \\
        $^\triangle$Obtained by combining $f_{B_s}$ from HPQCD 11A with $f_{B_s}/f_B$ calculated in this work.\\
        $^\nabla$This result uses one ensemble per lattice spacing with light to strange sea-quark mass 
        ratio $m_{\ell}/m_s \approx 0.2$. \\
        $^\bullet$This result uses an old determination of  $r_1=0.321(5)$ fm from Ref.~\cite{Gray:2005ur} that 
        has since been superseded. \\
        $^\ddagger$Obtained by combining $f_{D_s}$, updated in this work, with $f_{B_s}/f_{D_s}$,  calculated in this work.\\
        $^\ddagger$Update of ETM 11A and 12B. 
\end{flushleft}}}
\end{tabular*}
\vspace{-0.5cm}
\caption{Decay constants of the $B$, $B^+$, $B^0$ and $B_{s}$ mesons
  (in MeV). Here $f_B$ stands for the mean value of $f_{B^+}$ and
  $f_{B^0}$, extrapolated (or interpolated) in the mass of the light
  valence-quark to the physical value of $m_{ud}$.}
\label{tab:FBssumm}
\end{table}

\begin{table}[!htb]
\begin{center}
\mbox{} \\[3.0cm]
\footnotesize
\begin{tabular*}{\textwidth}[l]{@{\extracolsep{\fill}}l@{\hspace{1mm}}r@{\hspace{1mm}}l@{\hspace{1mm}}l@{\hspace{1mm}}l@{\hspace{1mm}}l@{\hspace{1mm}}l@{\hspace{1mm}}l@{\hspace{1mm}}l@{\hspace{5mm}}l@{\hspace{1mm}}l@{\hspace{1mm}}l@{\hspace{1mm}}l}
Collaboration & Ref. & $\Nf$ & 
\hspace{0.15cm}\begin{rotate}{60}{publication status}\end{rotate}\hspace{-0.15cm} &
\hspace{0.15cm}\begin{rotate}{60}{continuum extrapolation}\end{rotate}\hspace{-0.15cm} &
\hspace{0.15cm}\begin{rotate}{60}{chiral extrapolation}\end{rotate}\hspace{-0.15cm}&
\hspace{0.15cm}\begin{rotate}{60}{finite volume}\end{rotate}\hspace{-0.15cm}&
\hspace{0.15cm}\begin{rotate}{60}{renormalization/matching}\end{rotate}\hspace{-0.15cm}  &
\hspace{0.15cm}\begin{rotate}{60}{heavy-quark treatment}\end{rotate}\hspace{-0.15cm} & 
 $f_{B_s}/f_{B^+}$  & $f_{B_s}/f_{B^0}$  & $f_{B_s}/f_{B}$  \\
&&&&&&&&&& \\[-0.1cm]
\hline
\hline
&&&&&&&&&& \\[-0.1cm]

FNAL/MILC 17  & \cite{Bazavov:2017lyh} & 2+1+1 & \gA & \good & \good
                                                                                   & \good 
& \good &  \okay &  1.2180(49) & 1.2109(41) & $-$ \\[0.5ex]

HPQCD 17A & \cite{Hughes:2017spc} & 2+1+1 & \gA & \soso & \good & \good 
& \soso &  \okay &  $-$ & $-$ & 1.207(7) \\[0.5ex]

ETM 16B & \cite{Bussone:2016iua} & 2+1+1 & \gA & \good & \soso & \soso 
& \soso &  \okay &  $-$ & $-$& 1.184(25) \\[0.5ex]

ETM 13E & \cite{Carrasco:2013naa} & 2+1+1 & \rC & \good & \soso & \soso
& \soso &  \okay &  $-$ & $-$ & 1.201(25) \\[0.5ex]

HPQCD 13 & \cite{Dowdall:2013tga} & 2+1+1 & \gA & \soso & \good & \good & \soso
& \okay & 1.217(8) & 1.194(7) & 1.205(7)  \\[0.5ex]

&&&&&&&&&& \\[-0.1cm]
\hline
&&&&&&&&&& \\[-0.1cm]
RBC/UKQCD 18A & \cite{Boyle:2018knm} & 2+1 & \oP & \good & \good & \good & \good & \okay & $-$ & $-$ & 1.1949(60)({\raisebox{0.5ex}{\tiny$\substack{+95 \\ -175}$}})
\\[0.5ex]

RBC/UKQCD 14 & \cite{Christ:2014uea} & 2+1 & \gA & \soso & \soso & \soso 
  & \soso & \okay & 1.223(71) & 1.197(50) & $-$ \\[0.5ex]

RBC/UKQCD 14A & \cite{Aoki:2014nga} & 2+1 & \gA & \soso & \soso & \soso 
  & \soso & \okay & $-$ & $-$ & 1.193(48) \\[0.5ex]

RBC/UKQCD 13A & \cite{Witzel:2013sla} & 2+1 & \rC & \soso & \soso & \soso 
  & \soso & \okay & $-$ & $-$ &  1.20(2)$_{\rm stat}^\diamond$ \\[0.5ex]

HPQCD 12 & \cite{Na:2012sp} & 2+1 & \gA & \soso & \soso & \soso & \soso
& \okay & $-$ & $-$ & 1.188(18) \\[0.5ex]

FNAL/MILC 11 & \cite{Bazavov:2011aa} & 2+1 & \gA & \soso & \soso &
     \good& \soso & \okay & 1.229(26) & $-$ & $-$ \\[0.5ex]  
     
RBC/UKQCD 10C & \cite{Albertus:2010nm} & 2+1 & \gA & \tbr & \tbr & \tbr 
  & \soso & \okay & $-$ & $-$ & 1.15(12) \\[0.5ex]

HPQCD 09 & \cite{Gamiz:2009ku} & 2+1 & \gA & \soso & \soso & \soso &
\soso & \okay & $-$ & $-$ & 1.226(26)  \\[0.5ex] 

&&&&&&&&&& \\[-0.1cm]
\hline
&&&&&&&&&& \\[-0.1cm]
ALPHA 14 \al \cite{Bernardoni:2014fva} & 2 & \gA & \good & \good & \good 
& \good &  \okay &  $-$ \al $-$ & 1.203(65)\\[0.5ex]

ALPHA 13 & \cite{Bernardoni:2013oda} & 2 & \rC  & \good  & \good  &
\good   &\good  & \okay   & $-$ & $-$ & 1.195(61)(20)  &  \\[0.5ex] 

ETM 13B, 13C$^\dagger$ & \cite{Carrasco:2013zta,Carrasco:2013iba} & 2 & \gA & \good & \soso & \good
& \soso &  \okay &  $-$ & $-$ & 1.206(24)  \\[0.5ex]

ALPHA 12A & \cite{Bernardoni:2012ti} & 2 & \rC & \good & \good & \good
& \good &  \okay & $-$ & $-$ & 1.13(6)  \\ [0.5ex]

ETM 12B & \cite{Carrasco:2012de} & 2 & \rC & \good & \soso & \good
& \soso &  \okay & $-$ & $-$ & 1.19(5) \\ [0.5ex]

ETM 11A & \cite{Dimopoulos:2011gx} & 2 & \gA & \soso & \soso & \good
& \soso &  \okay & $-$ & $-$ & 1.19(5) \\ [0.5ex]
&&&&&&&&&& \\[-0.1cm]
\hline
\hline
\end{tabular*}
\begin{tabular*}{\textwidth}[l]{l@{\extracolsep{\fill}}lllllllll}
  \multicolumn{10}{l}{\vbox{\begin{flushleft}
 	 $^\diamond$Statistical errors only. \\
          $^\dagger$Update of ETM 11A and 12B. 
\end{flushleft}}}
\end{tabular*}
\vspace{-0.5cm}
\caption{Ratios of decay constants of the $B$ and $B_s$ mesons (for details see Tab.~\ref{tab:FBssumm}).}
\label{tab:FBratsumm}
\end{center}
\end{table}
\begin{figure}[!htb]
\centering	
\includegraphics[width=0.48\linewidth]{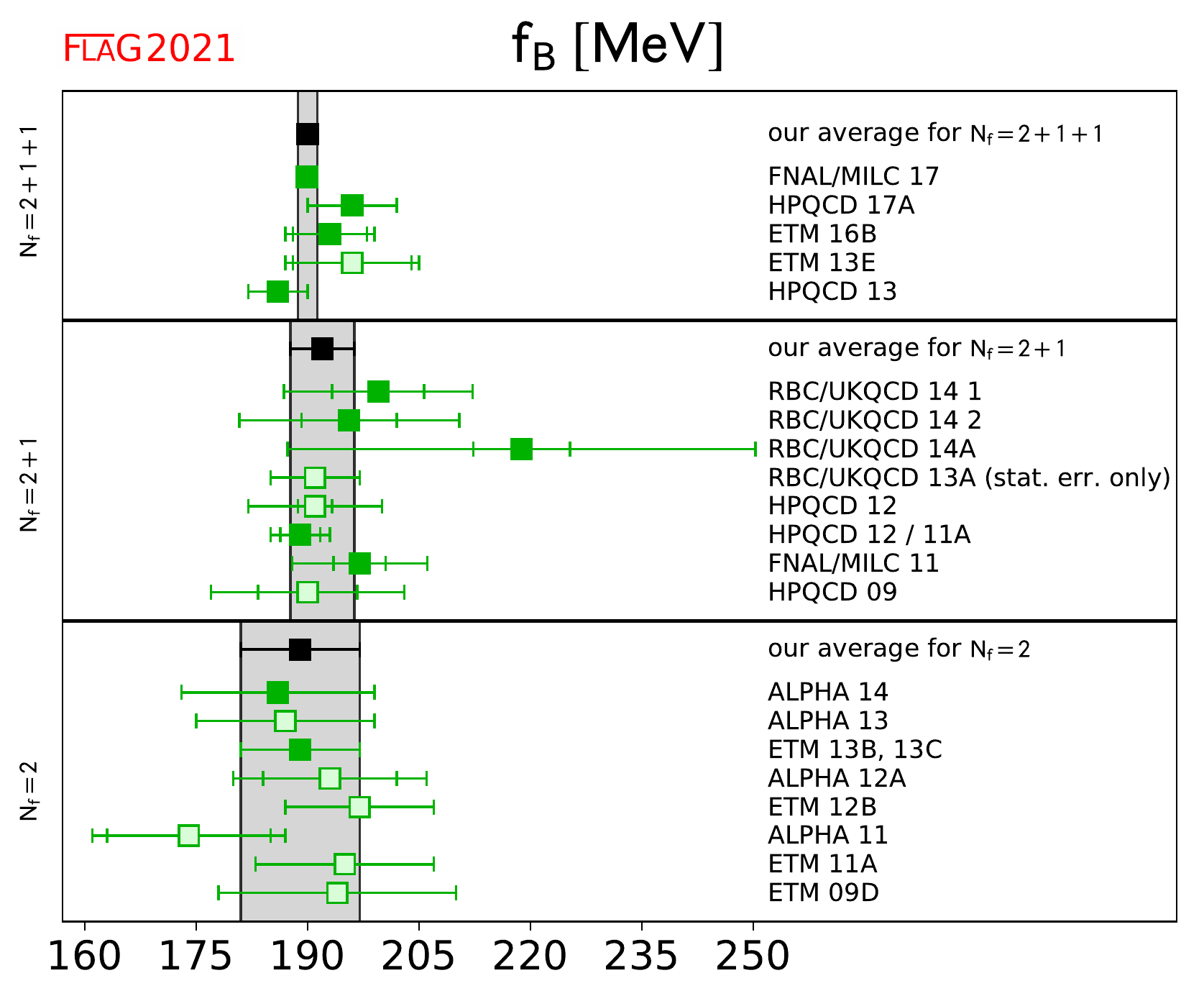}
\includegraphics[width=0.48\linewidth]{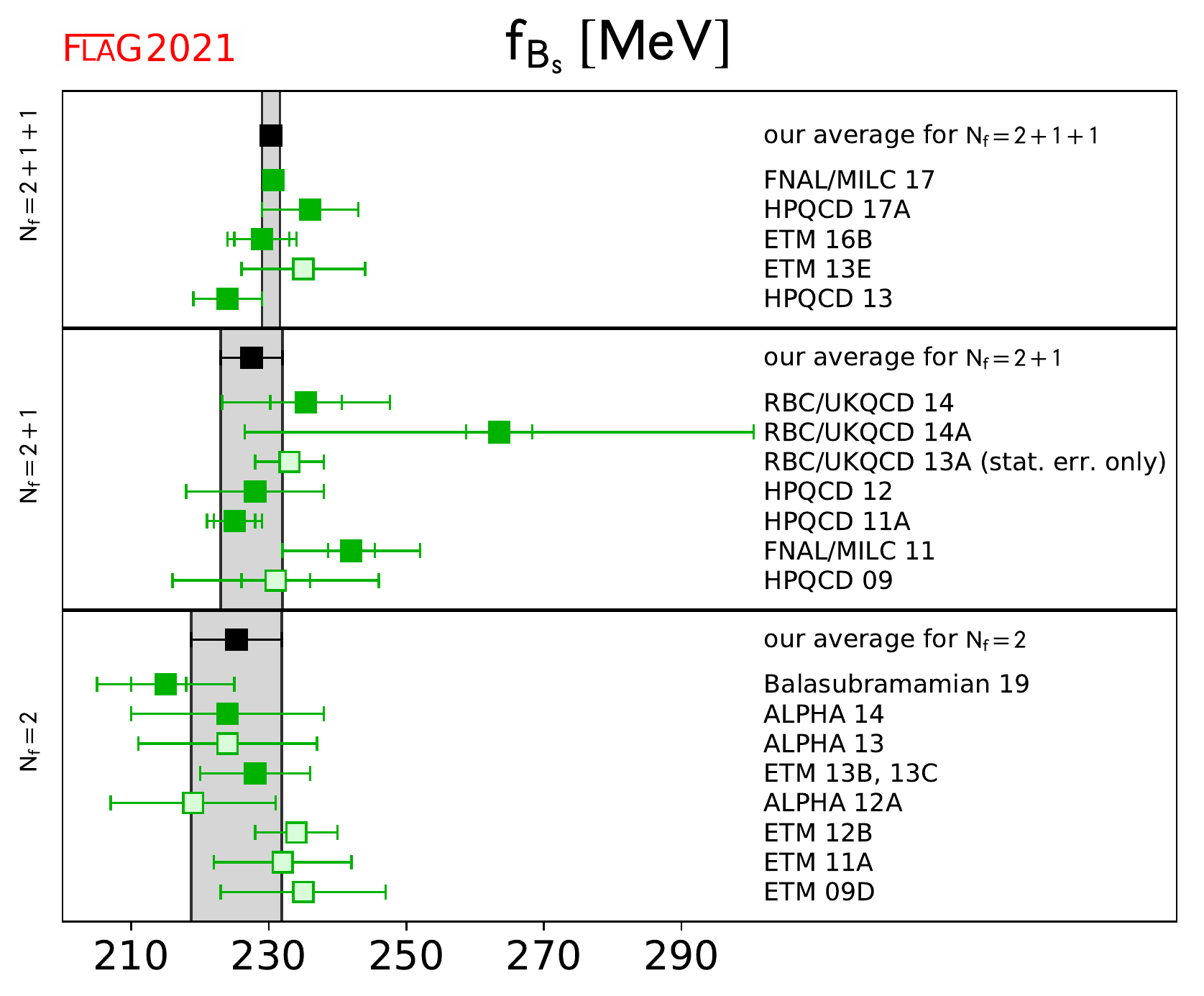}
 \vspace{-2mm}
\caption{Decay constants of the $B$ and $B_s$ mesons. The values are taken from Tab.~\ref{tab:FBssumm} 
(the $f_B$ entry for FNAL/MILC 11 represents $f_{B^+}$). The
significance of the colours is explained in Sec.~\ref{sec:qualcrit}.
The black squares and grey bands indicate
our averages in Eqs.~(\ref{eq:fB2}), (\ref{eq:fB21}),
(\ref{eq:fB211}), (\ref{eq:fBs2}), (\ref{eq:fBs21}) and
(\ref{eq:fBs211}).}
\label{fig:fB}
\end{figure}
\begin{figure}[!htb]
\begin{center}
\includegraphics[width=0.7\linewidth]{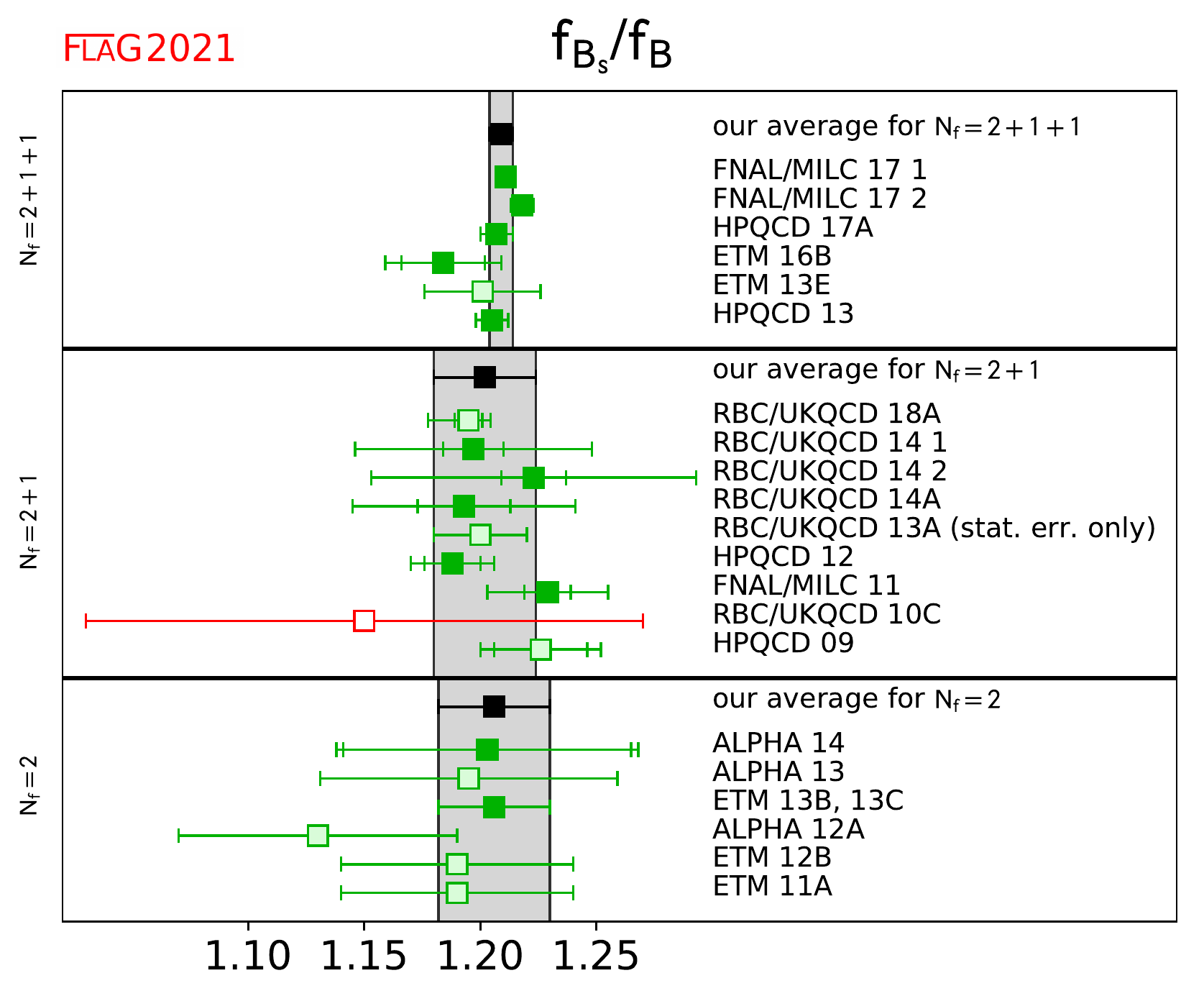}
\vspace{-2mm}
\caption{Ratio of the decay constants of the $B$ and $B_s$ mesons. The
  values are taken from Tab.~\ref{tab:FBratsumm}.  Results labelled
  as FNAL/MILC 17 1 and FNAL/MILC 17 2 correspond to
  those for $f_{B_{s}}/f_{B^{0}}$ and $f_{B_{s}}/f_{B^{+}}$ reported in FNAL/MILC
  17.  The
significance of the colours is explained in
Sec.~\ref{sec:qualcrit}.
The black squares and grey bands indicate
our averages in Eqs.~(\ref{eq:fBratio2}), (\ref{eq:fBratio21}), and
(\ref{eq:fBratio211}).}
\label{fig:fBratio}
\end{center}
\end{figure}
%

One new $N_{f}=2$ calculation of $f_{B_{s}}$ has appeared after the publication of the previous FLAG review~\cite{Aoki:2019cca}. In Tab.~\ref{tab:FBssumm}, this result is labelled Balasubramamian 19 \cite{Balasubramanian:2019net}.

In Balasubramamian 19~\cite{Balasubramanian:2019net}, simulations at three values of the lattice spacing, $a=0.0751$, 0.0653 and 0.0483 fm were performed with nonperturbatively ${\cal O}(a)$-improved Wilson-clover fermions and the Wilson plaquette gauge action. The pion masses in this work range from 194 to 439 MeV, and the
lattice sizes are between 2.09 and 4.18 fm.  A key feature of this calculation is the use of a variant of the ratio method~\cite{Blossier:2009hg}, applied for the first time to Wilson-clover fermions. This variant is required because, in contrast to twisted-mass Wilson fermions, there is no simple relationship between the heavy quark pole mass and the bare quark mass. In
the application of this approach to the $B_s$-decay constant,  one
first computes the quantity ${\mathcal{F}}_{hq} \equiv f_{hq}/M_{hq}$,
where $f_{hq}$ and $M_{hq}$ are the decay constant and mass of the
pseudoscalar meson composed of valence (relativistic) heavy quark $h$
and light (or strange) quark $q$.   The matching between the lattice
and the continuum heavy-light currents for extracting the above
$f_{hq}$ is straightforward because 
the valence heavy quark is also described by Wilson-clover fermions.  In the second step,
the ratio $z_{q} (M_{hq}, \lambda) \equiv
[{\mathcal{F}}_{hq}C_{A}^{{\mathrm{stat}}}(M_{h^\prime q})M_{hq}^{3/2}]/[{\mathcal{F}}_{h^{\prime}q}C_{A}^{{\mathrm{stat}}}(M_{hq})M_{h^{\prime}q}^{3/2}]$
is calculated, where $C_{A}^{{\mathrm{stat}}}(M_{hq})$ is the matching coefficient for the $(hq)$-meson decay constant in QCD and its counterpart in HQET, and $M_{hq}
= \lambda M_{h^{\prime}q}$.  The authors of Balasubramamian 19 \cite{Balasubramanian:2019net} use the NNLO perturbative result of
$C_{A}^{{\mathrm{stat}}}(M_{hq})$ \cite{Broadhurst:1994se,Chetyrkin:2003vi,Bekavac:2009zc} and $\lambda=1.18$.  By
starting from a ``triggering'' point with the heavy-meson mass around that
of the $D_s$ meson, one can proceed with the calculations in steps, such that
$M_{hq}$ is increased by a factor of $\lambda$ at each step.
The authors simulate up to heavy-quark mass
around 4.5 GeV, but observed significant $(aM_{H_s})^2$ cutoff effects on ensembles with lattice spacings $a = 0.0751$ and 0.0653 fm and so simulate up to 3.2 GeV on these lattices. 
In this formulation of the ratio method, the ratio obeys
$z_{q} (M_{hq}, \lambda) \rightarrow 1/\sqrt{\lambda}$ in the limit
$M_{hq} \rightarrow \infty$.   Designing the computations in
such a way that in the last step $M_{hq}$ is equal to the physical
$B_s$ mass, one obtains
$f_{B_{(s)}}/f_{D_{(s)}}$.  Combining this ratio with results for
$f_{D_{(s)}}$, updated with a third lattice spacing, the decay constant of the $B_s$ meson can be extracted. The authors estimated the systematic uncertainty associated with their generic fit form, which combines chiral-continuum extrapolation with heavy quark discretization effects, and quote a single systematic uncertainty. The systematic uncertainty associated with scale-setting is estimated from $f_{D_s}$.

There have been no new $N_{f}=2$ calculations of $f_B$ or $f_{B_{s}}/f_B$. Therefore, our averages for these two cases stay the same as those in Ref.~\cite{Aoki:2019cca}. We update our average of $f_{B_{s}}$ to include the new calculation of Balasubramamian 19 \cite{Balasubramanian:2019net}:
%
\begin{align}
&\label{eq:fB2}
\Nf=2:&\FLAGAVBEGIN f_{B} &= 188(7) \FLAGAVEND\;{\rm MeV}
&&\Refs~\mbox{\cite{Carrasco:2013zta,Bernardoni:2014fva}},\\
&\label{eq:fBs2}
\Nf=2: &\FLAGAVBEGIN f_{B_{s}} &= 225.3(6.6)\FLAGAVEND \; {\rm MeV} 
&&\Refs~\mbox{\cite{Carrasco:2013zta,Bernardoni:2014fva,Balasubramanian:2019net}}, \\
&\label{eq:fBratio2}
\Nf=2: &\FLAGAVBEGIN f_{B_{s}}\over{f_B} &= 1.206(0.023)\FLAGAVEND
&&\Refs~\mbox{\cite{Carrasco:2013zta,Bernardoni:2014fva}}.
\end{align}

One new $N_{f}=2+1$ calculation of $f_{B_{s}}/f_B$ was completed after the publication of the previous FLAG review~\cite{Aoki:2019cca}.  In Tab.~\ref{tab:FBratsumm}, this result is labelled RBC/UKQCD 18A~\cite{Boyle:2018knm}.

The RBC/UKQCD collaboration presented in RBC/UKQCD 18A~\cite{Boyle:2018knm} the ratio of
decay constants, $f_{B_{s}}/f_B$, using $N_f=2+1$ dynamical ensembles generated using Domain Wall Fermions (DWF). Three lattice spacings, of $a=0.114$, 0.0835 and 0.0727 fm, were used, with pion masses ranging from 139 to 431 MeV, and lattice sizes between 2.65 and 5.47 fm. Two different Domain Wall discretizations (M\"obius and Shamir) have been used for both valence and sea 
quarks.  These discretizations correspond to two different choices for the DWF kernel. The M\"obius DWF are loosely equivalent to Shamir DWF at twice the 
extension in the fifth dimension~\cite{Blum:2014tka}. The bare parameters for these discretizations were chosen to lie on the same scaling trajectory, to enable a combined continuum extrapolation. Heavy quark masses between the charm and approximately half the bottom quark mass were used, with a linear extrapolation in $1/m_H$ applied to reach the physical $B_s$ mass, where $m_H$ is the mass of the heavy meson used to set the heavy quark mass. For the central fit, the authors set the heavy quark mass through the pseudoscalar heavy-strange meson $H_s$, and estimate systematic uncertainties by comparing these results to those obtained with $H$ a heavy-light meson or a heavy-heavy meson.
For the quenched heavy quark M\"obius DWF are always used, with a domain-wall height slightly different from the one adopted
for light valence quarks. The choice helps to keep cutoff effects under control, according to the study in Ref.~\cite{Boyle:2016imm}. The chiral-continuum extrapolations are
performed with a Taylor expansion in $a^2$ and $m_\pi^2-(m_\pi^{\mathrm{phys}})^2$ and the associated systematic error is estimated by varying the fit function to apply cuts in the pion mass. The corresponding systematic error is estimated as approximately $0.5\%$, which is roughly equal to the statistical uncertainty and to the systematic uncertainties associated with extrapolation to the physical $m_{B_s}$ mass and with higher-order corrections to the static limit. These latter corrections take the form ${\cal O}(\Lambda^2/m_{B_s}^2)$. The error estimate comes from assuming the coefficient of such terms is up to five times larger than the fitted ${\cal O}(\Lambda/m_{B_s})$ coefficient. Isospin corrections and heavy-quark discretization effects are estimated to be less than $0.1\%$.

At time of writing, RBC/UKQCD 18A~\cite{Boyle:2018knm} has not been published and therefore is not included in our average. Thus, our averages for these quantities remain the same as in Ref.~\cite{Aoki:2019cca},
%
\begin{align}
&\label{eq:fB21}
\Nf=2+1:&\FLAGAVBEGIN f_{B} &= 192.0(4.3) \FLAGAVEND\;{\rm MeV}
&&\Refs~\mbox{\cite{Bazavov:2011aa,McNeile:2011ng,Na:2012sp,Aoki:2014nga,Christ:2014uea}},\\
&\label{eq:fBs21}
\Nf=2+1: &\FLAGAVBEGIN f_{B_{s}} &= 228.4(3.7)\FLAGAVEND \; {\rm MeV} 
&&\Refs~\mbox{\cite{Bazavov:2011aa,McNeile:2011ng,Na:2012sp,Aoki:2014nga,Christ:2014uea}}, \\
&\label{eq:fBratio21}
\Nf=2+1: &\FLAGAVBEGIN f_{B_{s}}\over{f_B} &= 1.201(0.016)\FLAGAVEND
&&\Refs~\mbox{\cite{Bazavov:2011aa,Na:2012sp,Aoki:2014nga,Christ:2014uea,Boyle:2018knm}}.
\end{align}

No new $N_{f}=2+1+1$ calculations of $f_{B}$, $f_{B_{s}}/f_{B}$ or $f_{B_{(s)}}$ have appeared since the last FLAG review. Therefore, our averages for these quantities remain the same as in Ref.~\cite{Aoki:2019cca},
%
\begin{align}
&\label{eq:fB211}
\Nf=2+1+1:&\FLAGAVBEGIN f_{B} &= 190.0(1.3) \FLAGAVEND\;{\rm MeV}
&&\Refs~\mbox{\cite{Dowdall:2013tga,Bussone:2016iua,Hughes:2017spc,Bazavov:2017lyh}},\\
&\label{eq:fBs211}
\Nf=2+1+1: &\FLAGAVBEGIN f_{B_{s}} &= 230.3(1.3)  \FLAGAVEND\; {\rm MeV}
&&\Refs~\mbox{\cite{Dowdall:2013tga,Bussone:2016iua,Hughes:2017spc,Bazavov:2017lyh}}, \\
&\label{eq:fBratio211}
\Nf=2+1+1: &\FLAGAVBEGIN f_{B_{s}}\over{f_B} &= 1.209(0.005)\FLAGAVEND
&&\Refs~\mbox{\cite{Dowdall:2013tga,Bussone:2016iua,Hughes:2017spc,Bazavov:2017lyh}}.
\end{align}

The PDG presented averages for the $N_{f}=2+1$ and $N_{f}=2+1+1$ lattice-QCD determinations of the isospin-averaged $f_{B}$, $f_{B_{s}}$ and $f_{B_{s}}/f_{B}$ in 2020~\cite{Zyla:2020zbs}.  The $N_{f}=2+1$ and $N_{f}=2+1+1$ lattice-computation results used in Ref.~\cite{Zyla:2020zbs} are identical to those included in our current work, and the averages quoted in Ref.~\cite{Zyla:2020zbs} are those determined in \cite{Aoki:2019cca}.

\subsection{Neutral $B$-meson mixing matrix elements}
\label{sec:BMix}

Neutral $B$-meson mixing is induced in the Standard Model through
1-loop box diagrams to lowest order in the electroweak theory,
similar to those for short-distance effects in neutral kaon mixing. The effective Hamiltonian
is given by
\begin{equation}
  {\cal H}_{\rm eff}^{\Delta B = 2, {\rm SM}} \,\, = \,\,
  \frac{G_F^2 M_{\rm{W}}^2}{16\pi^2} ({\cal F}^0_d {\cal Q}^d_1 + {\cal F}^0_s {\cal Q}^s_1)\,\, +
   \,\, {\rm h.c.} \,\,,
   \label{eq:HeffB}
\end{equation}
with
\begin{equation}
 {\cal Q}^q_1 =
   \left[\bar{b}\gamma_\mu(1-\gamma_5)q\right]
   \left[\bar{b}\gamma_\mu(1-\gamma_5)q\right],
   \label{eq:Q1}
\end{equation}
where $q=d$ or $s$. The short-distance function ${\cal F}^0_q$ in
Eq.~(\ref{eq:HeffB}) is much simpler compared to the kaon mixing case
due to the hierarchy in the CKM matrix elements. Here, only one term
is relevant,
\begin{equation}
 {\cal F}^0_q = \lambda_{tq}^2 S_0(x_t)
\end{equation}
where
\begin{equation}
 \lambda_{tq} = V^*_{tq}V_{tb},
\end{equation}
and where $S_0(x_t)$ is an Inami-Lim function with $x_t=m_t^2/M_W^2$,
which describes the basic electroweak loop contributions without QCD
\cite{Inami:1980fz}. The transition amplitude for $B_q^0$ with $q=d$
or $s$ can be written as
\begin{align}
  \langle \bar B^0_q\vert{\cal H}_{\rm eff}^{\Delta B = 2}\vert B^0_q\rangle
  =&
     \frac{G_F^2 M_W^2}{16\pi^2}  
     \left[\lambda_{tq}^2 S_0(x_t) \eta_{2B} \right]
     \nn \\ 
  & \times 
    \left(\frac{\gbar(\mu)^2}{4\pi}\right)^{-\gamma_0/(2\beta_0)}
    \exp\left\{ \int_0^{\gbar(\mu)}dg \left(
    \frac{\gamma(g)}{\beta(g)}+\frac{\gamma_0}{\beta_0g} \right) \right\}
    \nn \\
   & \times
     \langle \bar B^0_q \vert  Q^q_{\rm R} (\mu) \vert B^0_q\rangle \,\, + \,\, {\rm h.c.} \,\, ,
   \label{eq:BBME}
\end{align}
where $Q^q_{\rm R} (\mu)$ is the renormalized four-fermion operator
(usually in the NDR scheme of $\msbar$). The running coupling
$\gbar$, the $\beta$-function $\beta(g)$, and the anomalous
dimension of the four-quark operator $\gamma(g)$ are defined in
Eqs.~(\ref{eq:four_quark_operator_anomalous_dimensions})~and~(\ref{eq:four_quark_operator_anomalous_dimensions_perturbative}).
The product of $\mu$-dependent terms on the second line of
Eq.~(\ref{eq:BBME}) is, of course, $\mu$-independent (up to truncation
errors arising from the use of perturbation theory). The explicit expression for
the short-distance QCD correction factor $\eta_{2B}$ (calculated to
NLO) can be found in Ref.~\cite{Buchalla:1995vs}.

For historical reasons the $B$-meson mixing matrix elements are often
parameterized in terms of bag parameters defined as
\begin{equation}
 B_{B_q}(\mu)= \frac{{\left\langle\bar{B}^0_q\left|
   Q^q_{\rm R}(\mu)\right|B^0_q\right\rangle} }{
         {\frac{8}{3}f_{B_q}^2\mB^2}} \,\, .
         \label{eq:bagdef}
\end{equation}
The renormalization group independent (RGI) $B$ parameter $\hat{B}$ is defined as in the case of the kaon,
and expressed to 2-loop order as
\begin{equation}
 \hat{B}_{B_q} = 
   \left(\frac{\gbar(\mu)^2}{4\pi}\right)^{- \gamma_0/(2\beta_0)}
   \left\{ 1+\dfrac{\gbar(\mu)^2}{(4\pi)^2}\left[
   \frac{\beta_1\gamma_0-\beta_0\gamma_1}{2\beta_0^2} \right]\right\}\,
   B_{B_q}(\mu) \,\,\, ,
\label{eq:BBRGI_NLO}
\end{equation}
with $\beta_0$, $\beta_1$, $\gamma_0$, and $\gamma_1$ defined in
Eq.~(\ref{eq:RG-coefficients}). Note, as Eq.~(\ref{eq:BBME}) is
evaluated above the bottom threshold ($m_b<\mu<m_t$), the active number
of flavours here is $N_f=5$.

Nonzero transition amplitudes result in a mass difference between the
CP eigenstates of the neutral $B$-meson system. Writing the mass
difference for a $B_q^0$ meson as $\Delta m_q$, its Standard Model
prediction is
\begin{equation}
 \Delta m_q = \frac{G^2_Fm^2_W m_{B_q}}{6\pi^2} \,
  |\lambda_{tq}|^2 S_0(x_t) \eta_{2B} f_{B_q}^2 \hat{B}_{B_q}.
\end{equation}
Experimentally, the mass difference is determined from the oscillation
frequency of the CP eigenstates. The frequencies are measured
precisely with an error of less than a percent. Many different
experiments have measured $\Delta m_d$, but the current average
\cite{Zyla:2020zbs} is dominated by the 
LHC$b$ experiment. For $\Delta m_s$ the experimental average is again dominated by results
from LHC$b$
\cite{Zyla:2020zbs} and the precision reached is about one per mille.
With these experimental results and
lattice-QCD calculations of $f_{B_q}^2\hat{B}_{B_q}$,
$\lambda_{tq}$ can be determined.  In lattice-QCD calculations the
flavour $SU(3)$-breaking ratio
\begin{equation}
 \xi^2 = \frac{f_{B_s}^2B_{B_s}}{f_{B_d}^2B_{B_d}}
 \label{eq:xidef}
\end{equation} 
can be obtained more precisely than the individual $B_q$-mixing matrix
elements because statistical and systematic errors cancel in part.
From $\xi^2$, the ratio $|V_{td}/V_{ts}|$ can be determined and used
to constrain the apex of the CKM triangle.

Neutral $B$-meson mixing, being loop-induced in the Standard Model, is
also a sensitive probe of new physics. The most general $\Delta B=2$
effective Hamiltonian that describes contributions to $B$-meson mixing
in the Standard Model and beyond is given in terms of five local
four-fermion operators:
\be
  {\cal H}_{\rm eff, BSM}^{\Delta B = 2} = \sum_{q=d,s}\sum_{i=1}^5 {\cal C}_i {\cal Q}^q_i \;,
\ee
where ${\cal Q}_1$ is defined in Eq.~(\ref{eq:Q1}) and where
\begin{align}
{\cal Q}^q_2 & = \left[\bar{b}(1-\gamma_5)q\right]
   \left[\bar{b}(1-\gamma_5)q\right], \qquad
{\cal Q}^q_3  = \left[\bar{b}^{\alpha}(1-\gamma_5)q^{\beta}\right]
   \left[\bar{b}^{\beta}(1-\gamma_5)q^{\alpha}\right],\nonumber \\
{\cal Q}^q_4 & = \left[\bar{b}(1-\gamma_5)q\right]
   \left[\bar{b}(1+\gamma_5)q\right], \qquad
{\cal Q}^q_5 = \left[\bar{b}^{\alpha}(1-\gamma_5)q^{\beta}\right]
   \left[\bar{b}^{\beta}(1+\gamma_5)q^{\alpha}\right], 
   \label{eq:Q25}
\end{align}
with the superscripts $\alpha,\beta$ denoting colour indices, which
are shown only when they are contracted across the two bilinears.
There are three other basis operators in the $\Delta
B=2$ effective Hamiltonian. When evaluated in QCD, however, 
they give identical matrix elements to the ones already listed due to
parity invariance in QCD.
The short-distance Wilson coefficients ${\cal C}_i$ depend on the
underlying theory and can be calculated perturbatively.  In the
Standard Model only matrix elements of ${\cal Q}^q_1$ contribute to
$\Delta m_q$, while all operators do, for example, for general SUSY
extensions of the Standard Model~\cite{Gabbiani:1996hi}.
The matrix elements or bag parameters for the non-SM operators are also 
useful to estimate the width difference $\Delta \Gamma_q$ 
between the CP eigenstates of the neutral $B$ meson in the Standard Model,
where combinations of matrix elements of ${\cal Q}^q_1$,
${\cal Q}^q_2$, and ${\cal Q}^q_3$ contribute to $\Delta \Gamma_q$ 
at $\cO(1/m_b)$~\cite{Lenz:2006hd,Beneke:1996gn}.  

In this section, we report on results from lattice-QCD calculations for
the neutral $B$-meson mixing parameters $\hat{B}_{B_d}$,
$\hat{B}_{B_s}$, $f_{B_d}\sqrt{\hat{B}_{B_d}}$,
$f_{B_s}\sqrt{\hat{B}_{B_s}}$ and the $SU(3)$-breaking ratios
$B_{B_s}/B_{B_d}$ and $\xi$ defined in Eqs.~(\ref{eq:bagdef}),
(\ref{eq:BBRGI_NLO}), and (\ref{eq:xidef}).  The results are
summarized in Tabs.~\ref{tab_BBssumm} and \ref{tab_BBratsumm} and in
Figs.~\ref{fig:fBsqrtBB2} and \ref{fig:xi}. Additional details about
the underlying simulations and systematic error estimates are given in
Appendix~\ref{app:BMix_Notes}.  Some collaborations do not provide the
RGI quantities $\hat{B}_{B_q}$, but quote instead
$B_B(\mu)^{\overline{MS},NDR}$. In such cases, we convert the results
using Eq.~(\ref{eq:BBRGI_NLO})
to the RGI quantities quoted in Tab.~\ref{tab_BBssumm}
with a brief description for each case.
More detailed descriptions for these cases are provided in FLAG13 \cite{Aoki:2013ldr}.
We do not provide the $B$-meson matrix elements of the other operators
${\cal Q}_{2-5}$ in this report. They have been calculated in
Ref.~\cite{Carrasco:2013zta} for the $N_f=2$ case and 
in Refs.~\cite{Bouchard:2011xj,Bazavov:2016nty} for $N_f=2+1$.

\begin{table}[!htb]
\begin{center}
\mbox{} \\[3.0cm]
\footnotesize
\begin{tabular*}{\textwidth}[l]{l@{\extracolsep{\fill}}@{\hspace{1mm}}r@{\hspace{1mm}}c@{\hspace{1mm}}l@{\hspace{1mm}}l@{\hspace{1mm}}l@{\hspace{1mm}}l@{\hspace{1mm}}l@{\hspace{1mm}}l@{\hspace{5mm}}l@{\hspace{1mm}}l@{\hspace{1mm}}l@{\hspace{1mm}}l@{\hspace{1mm}}l}
Collaboration \al Ref. \al $\Nf$ \al
\hspace{0.15cm}\begin{rotate}{60}{publication status}\end{rotate}\hspace{-0.15cm} \al
\hspace{0.15cm}\begin{rotate}{60}{continuum extrapolation}\end{rotate}\hspace{-0.15cm} \al
\hspace{0.15cm}\begin{rotate}{60}{chiral extrapolation}\end{rotate}\hspace{-0.15cm}\al
\hspace{0.15cm}\begin{rotate}{60}{finite volume}\end{rotate}\hspace{-0.15cm}\al
\hspace{0.15cm}\begin{rotate}{60}{renormalization/matching}\end{rotate}\hspace{-0.15cm}  \al
\hspace{0.15cm}\begin{rotate}{60}{heavy-quark treatment}\end{rotate}\hspace{-0.15cm} \al 
\rule{0.12cm}{0cm}
\parbox[b]{1.2cm}{$f_{\rm B_d}\sqrt{\hat{B}_{\rm B_d}}$} \al
\rule{0.12cm}{0cm}
\parbox[b]{1.2cm}{$f_{\rm B_s}\sqrt{\hat{B}_{\rm B_s}}$} \al
\rule{0.12cm}{0cm}
$\hat{B}_{\rm B_d}$ \al 
\rule{0.12cm}{0cm}
$\hat{B}_{\rm B_{\rm s}}$ \\
&&&&&&&&&& \\[-0.1cm]
\hline
\hline
&&&&&&&&&& \\[-0.1cm]

HPQCD 19A \al \cite{Dowdall:2019bea} \al 2+1+1 \al \gA \al \soso \al \soso \al \good \al \soso
       \al \okay & 210.6(5.5) \al 256.1(5.7) \al 1.222(61) \al 1.232(53)\\[0.5ex]
&&&&&&&&&& \\[-0.1cm]
\hline
&&&&&&&&&& \\[-0.1cm]
FNAL/MILC 16 \al \cite{Bazavov:2016nty} \al 2+1 \al \gA \al \good \al \soso \al
     \good \al \soso
	\al \okay & 227.7(9.5) \al 274.6(8.4) \al 1.38(12)(6)$^\odot$ \al 1.443(88)(48)$^\odot$\\[0.5ex]

        RBC/UKQCD 14A \al \cite{Aoki:2014nga} \al 2+1 \al \gA \al \soso \al \soso \al
     \soso \al \soso
	\al \okay & 240(15)(33) \al 290(09)(40) \al 1.17(11)(24) \al 1.22(06)(19)\\[0.5ex]

FNAL/MILC 11A \al \cite{Bouchard:2011xj} \al 2+1 \al \rC \al \good \al \soso \al
     \good \al \soso
	\al \okay & 250(23)$^\dagger$ \al 291(18)$^\dagger$ \al $-$ \al $-$\\[0.5ex]

HPQCD 09 \al \cite{Gamiz:2009ku} \al 2+1 \al \gA \al \soso \al \soso$^\nabla$ \al \soso \al
\soso 
\al \okay & 216(15)$^\ast$ \al 266(18)$^\ast$ \al 1.27(10)$^\ast$ \al 1.33(6)$^\ast$ \\[0.5ex] 

HPQCD 06A \al \cite{Dalgic:2006gp} \al 2+1 \al \gA \al \tbr \al \tbr \al \good \al 
\soso 
	\al \okay & $-$ \al  281(21) \al $-$ \al 1.17(17) \\
&&&&&&&&&& \\[-0.1cm]
\hline
&&&&&&&&&& \\[-0.1cm]
ETM 13B \al \cite{Carrasco:2013zta} \al 2 \al \gA \al \good \al \soso \al \soso \al
    \good \al \okay & 216(6)(8) \al 262(6)(8) \al  1.30(5)(3) \al 1.32(5)(2) \\[0.5ex]

ETM 12A, 12B \al \cite{Carrasco:2012dd,Carrasco:2012de} \al 2 \al \rC \al \good \al \soso \al \soso \al
    \good \al \okay & $-$ \al $-$ \al  1.32(8)$^\diamond$ \al 1.36(8)$^\diamond$ \\[0.5ex]
&&&&&&&&&& \\[-0.1cm]
\hline
\hline\\
\end{tabular*}\\[-0.2cm]
\begin{minipage}{\linewidth}
{\footnotesize 
\begin{itemize}
 \item[$^\odot$] PDG averages of decay constant $f_{B^0}$ and $f_{B_s}$ \cite{Rosner:2015wva} are used to obtain these values.\\[-5mm]
   \item[$^\dagger$] Reported $f_B^2B$ at $\mu=m_b$ is converted to RGI by
	multiplying the 2-loop factor
	1.517.\\[-5mm]
   \item[$^\nabla$] While wrong-spin contributions are not included in
		the HMrS$\chi$PT fits, the effect is expected to be
		small for these quantities (see description in FLAG 13
		\cite{Aoki:2013ldr}). \\[-5mm] 
        \item[$^\ast$] This result uses an old determination of
		     $r_1=0.321(5)$~fm from Ref.~\cite{Gray:2005ur} that
		     has since been superseded, which however has
		     only a small effect in the total error budget (see
		     description in FLAG 13 \cite{Aoki:2013ldr}) .\\[-5mm]
        \item[$^\diamond$] Reported $B$ at $\mu=m_b=4.35$ GeV is converted to
     RGI by multiplying the 2-loop factor 1.521.
\end{itemize}
}
\end{minipage}
\caption{Neutral $B$- and $B_{\rm s}$-meson mixing matrix
 elements (in MeV) and bag parameters.}
\label{tab_BBssumm}
\end{center}
\end{table}

\begin{figure}[!htb]
\hspace{-0.8cm}\includegraphics[width=0.57\linewidth]{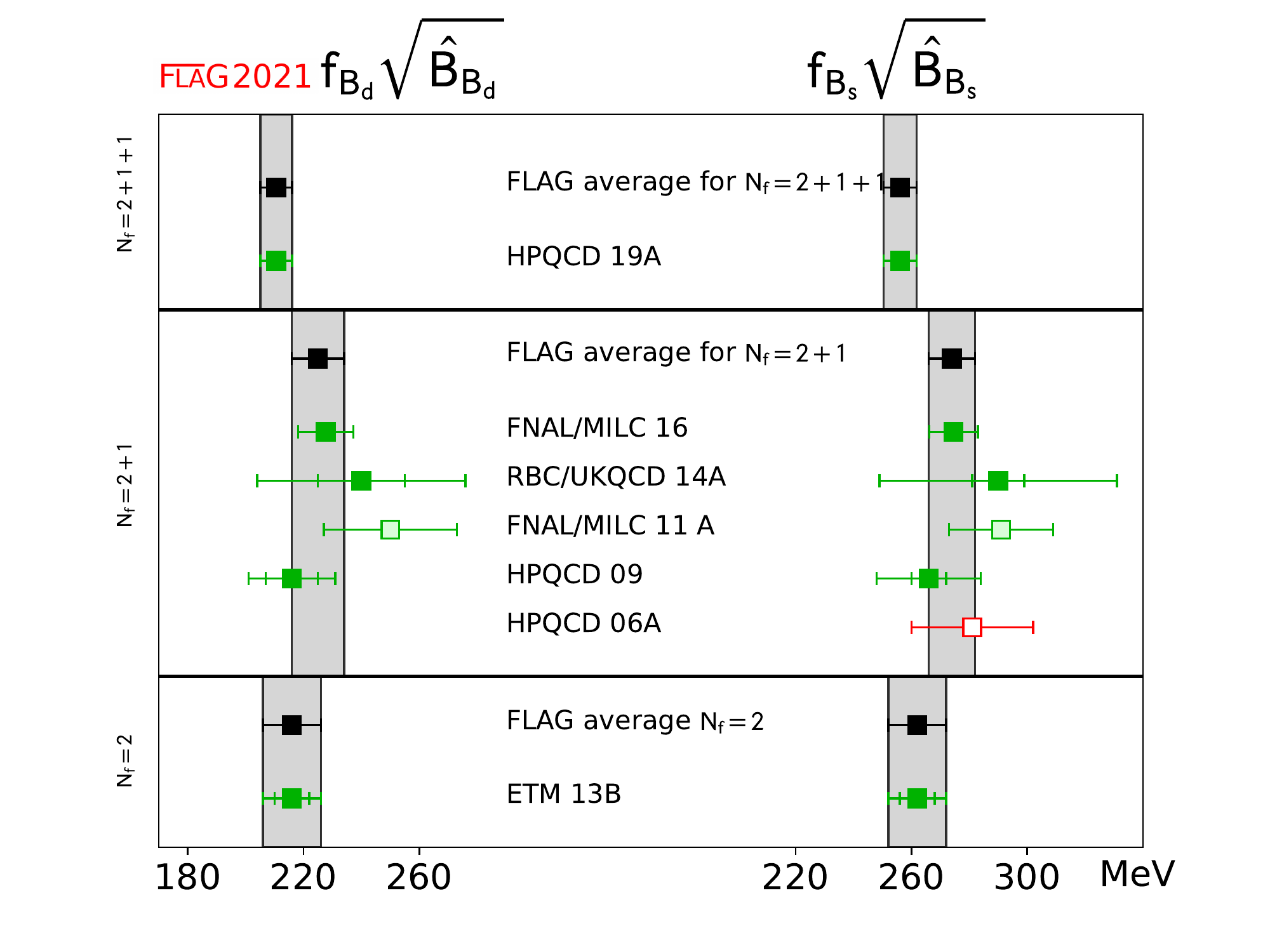}\hspace{-0.8cm}
\includegraphics[width=0.57\linewidth]{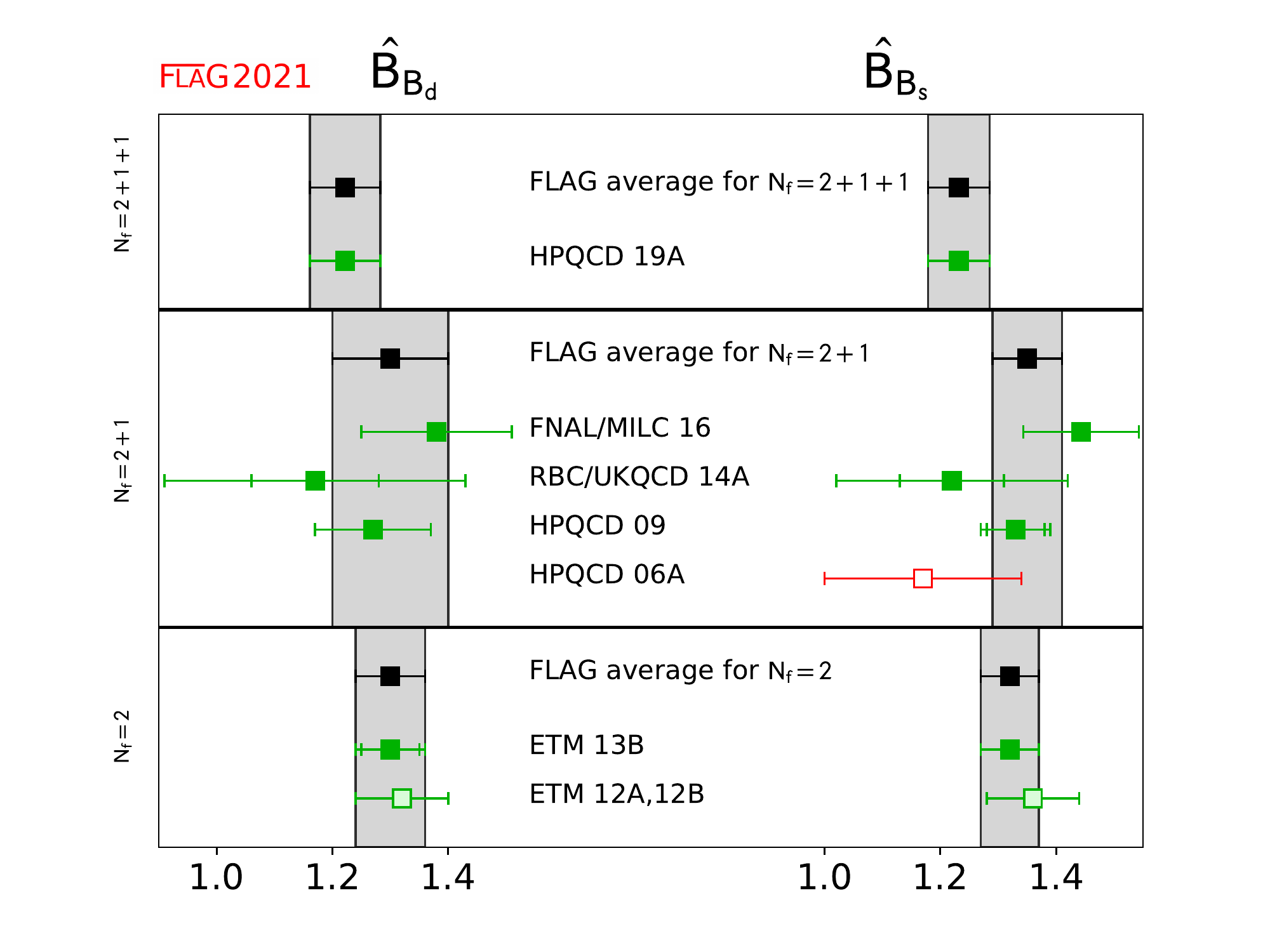}

\vspace{-5mm}
\caption{Neutral $B$- and $B_{\rm s}$-meson mixing matrix
 elements and bag parameters [values in Tab.~\ref{tab_BBssumm} and
 Eqs.~(\ref{eq:avfBB2}), (\ref{eq:avfBB}), (\ref{eq:avfBB4}), (\ref{eq:avBB2}), (\ref{eq:avBB}), (\ref{eq:avBB4})].
 \label{fig:fBsqrtBB2}}
\end{figure}

\begin{table}[!htb]
\begin{center}
\mbox{} \\[3.0cm]
\footnotesize
\begin{tabular*}{\textwidth}[l]{l @{\extracolsep{\fill}} r l l l @{\hspace{-0.1mm}}l @{\hspace{-0.1mm}}l @{\hspace{-0.1mm}}l @{\hspace{-0.1mm}}l l l}
Collaboration & Ref. & $\Nf$ & 
\hspace{0.15cm}\begin{rotate}{60}{publication status}\end{rotate}\hspace{-0.15cm} &
\hspace{0.15cm}\begin{rotate}{60}{continuum extrapolation}\end{rotate}\hspace{-0.15cm} &
\hspace{0.15cm}\begin{rotate}{60}{chiral extrapolation}\end{rotate}\hspace{-0.15cm}&
\hspace{0.15cm}\begin{rotate}{60}{finite volume}\end{rotate}\hspace{-0.15cm}&
\hspace{0.15cm}\begin{rotate}{60}{renormalization/matching}\end{rotate}\hspace{-0.15cm}  &
\hspace{0.15cm}\begin{rotate}{60}{heavy-quark treatment}\end{rotate}\hspace{-0.15cm} & 
\rule{0.12cm}{0cm}$\xi$ &
 \rule{0.12cm}{0cm}$B_{\rm B_{\rm s}}/B_{\rm B_d}$ \\
&&&&&&&&&& \\[-0.1cm]
\hline
\hline
&&&&&&&&&& \\[-0.1cm]

HPQCD 19A \al \cite{Dowdall:2019bea} & 2+1+1 & \gA & \soso & \soso & \good & \soso
       & \okay & 1.216(16) & 1.008(25) \\[0.5ex]

&&&&&&&&&& \\[-0.1cm]

\hline

&&&&&&&&&& \\[-0.1cm]

RBC/UKQCD 18A \al \cite{Boyle:2018knm} & 2+1 & \oP & \good & \good & 
     \good & \good & \okay & 1.1939(67)($^{+95}_{-177}$) & 0.9984(45)($^{+80}_{-63}$) \\[0.5ex]

FNAL/MILC 16 & \cite{Bazavov:2016nty} & 2+1 & \gA & \good & \soso &
     \good & \soso & \okay & 1.206(18) & 1.033(31)(26)$^\odot$ \\[0.5ex]

RBC/UKQCD 14A & \cite{Aoki:2014nga} & 2+1 & \gA & \soso & \soso &
     \soso & \soso & \okay & 1.208(41)(52) & 1.028(60)(49) \\[0.5ex]

FNAL/MILC 12 & \cite{Bazavov:2012zs} & 2+1 & \gA & \soso & \soso &
     \good & \soso & \okay & 1.268(63) & 1.06(11) \\[0.5ex]

RBC/UKQCD 10C
 & \cite{Albertus:2010nm} & 2+1 & \gA & \tbr & \tbr & \tbr
  & \soso & \okay & 1.13(12) & $-$ \\[0.5ex]

HPQCD 09 & \cite{Gamiz:2009ku} & 2+1 & \gA & \soso & \soso$^\nabla$ & \soso &
\soso & \okay & 1.258(33) & 1.05(7) \\[0.5ex] 

&&&&&&&&&& \\[-0.1cm]

\hline

&&&&&&&&&& \\[-0.1cm]

ETM 13B & \cite{Carrasco:2013zta} & 2 & \gA & \good & \soso & \soso & \good
			     & \okay & 1.225(16)(14)(22) & 1.007(15)(14) \\

ETM 12A, 12B & \cite{Carrasco:2012dd,Carrasco:2012de} & 2 & \rC & \good & \soso & \soso & \good
			     & \okay & 1.21(6) & 1.03(2) \\
&&&&&&&&&& \\[-0.1cm]
\hline
\hline\\
\end{tabular*}\\[-0.2cm]
\begin{minipage}{\linewidth}
{\footnotesize 
\begin{itemize}
 \item[$^\odot$] PDG average of the ratio of decay constants
	      $f_{B_s}/f_{B^0}$ \cite{Rosner:2015wva} is used to obtain
	      the value.\\[-5mm] 
   \item[$^\nabla$] Wrong-spin contributions are not included in the
		HMrS$\chi$PT fits. As the effect may not be negligible,
		these results are excluded from the average (see
		description in FLAG 13 \cite{Aoki:2013ldr}).
\end{itemize}
}
\end{minipage}
\caption{Results for $SU(3)$-breaking ratios of neutral $B_{d}$- and 
 $B_{s}$-meson mixing matrix elements and bag parameters.}
\label{tab_BBratsumm}
\end{center}
\end{table}

\begin{figure}[!htb]
\begin{center}
\includegraphics[width=11.5cm]{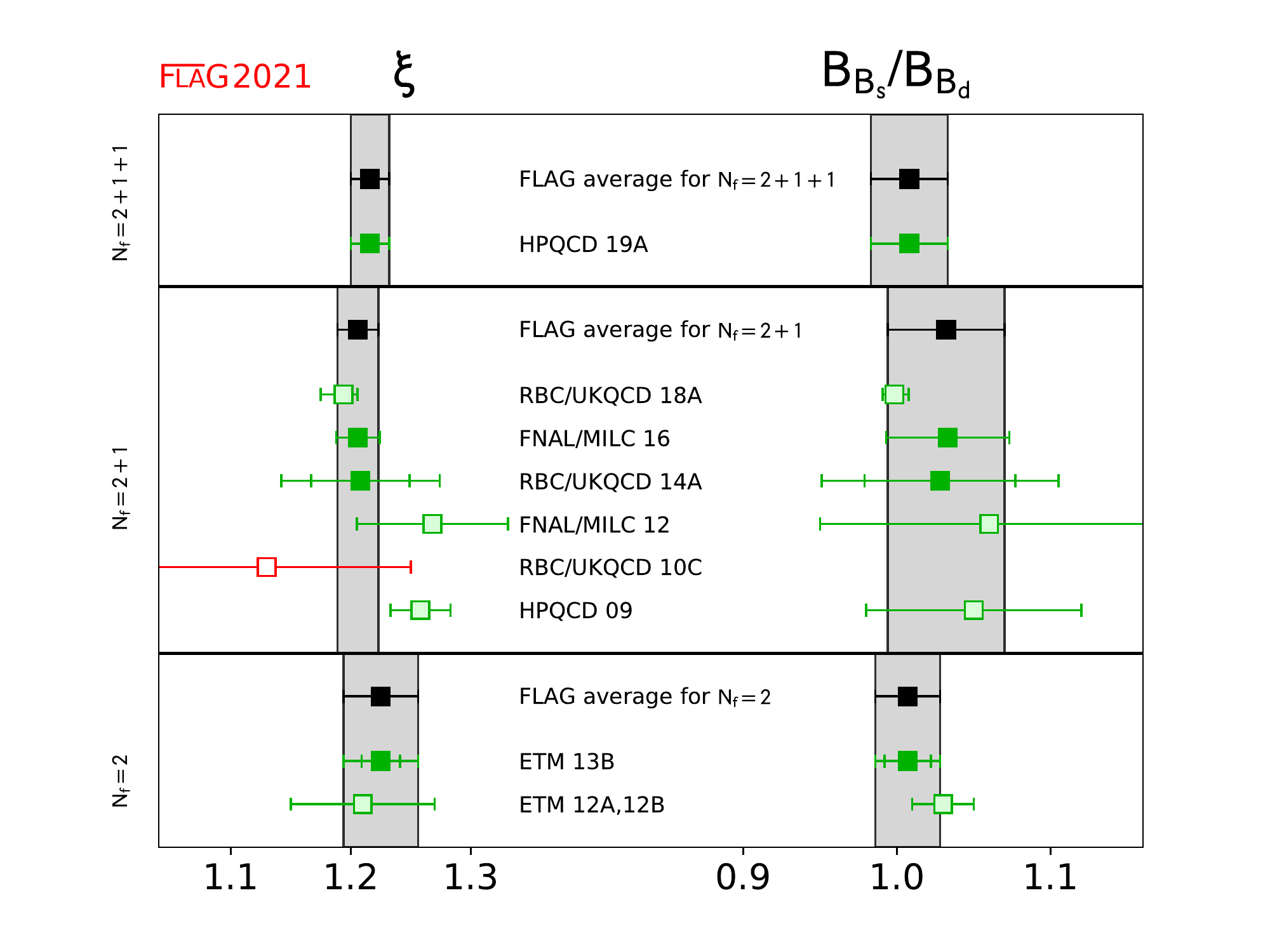}

\vspace{-2mm}
\caption{The $SU(3)$-breaking quantities $\xi$ and $B_{B_s}/B_{B_d}$
 [values in Tab.~\ref{tab_BBratsumm} and Eqs.~(\ref{eq:avxiBB2}), (\ref{eq:avxiBB}), (\ref{eq:avxiBB4})].}\label{fig:xi} 
\end{center}
\end{figure}

There are no new results for $N_f=2$ reported after the previous FLAG
review. 
In this category, one work (ETM~13B)~\cite{Carrasco:2013zta} passes
the quality criteria. 
A description of this work can be found in the FLAG 13 review
\cite{Aoki:2013ldr} where it did not enter the average as it had not
appeared in a journal. 
Because this is the only result 
available for $N_f=2$, we quote their values as our estimates  
\begin{align}
      &&  \FLAGAVBEGIN f_{B_d}\sqrt{\hat{B}_{b_d}}&= 216(10)\FLAGAVEND{\rm MeV}
         &\FLAGAVBEGIN f_{B_s}\sqrt{\hat{B}_{B_s}}&= 262(10)\FLAGAVEND{\rm MeV}
         &\Ref~\mbox{\cite{Carrasco:2013zta}},  \label{eq:avfBB2}\\
N_f=2:&&\FLAGAVBEGIN \hat{B}_{B_d}&= 1.30(6)\FLAGAVEND 
         &\FLAGAVBEGIN \hat{B}_{B_s}&= 1.32(5)\FLAGAVEND 
	 &\Ref~\mbox{\cite{Carrasco:2013zta}},  \label{eq:avBB2}\\
      &&  \FLAGAVBEGIN \xi &=  1.225(31)\FLAGAVEND  
  	& \FLAGAVBEGIN B_{B_s}/B_{B_d} & =  1.007(21)\FLAGAVEND
 	&\Ref~\mbox{\cite{Carrasco:2013zta}}. \label{eq:avxiBB2}
\end{align}

For the $N_f=2+1$ case 
the RBC/UKQCD collaboration reported their new results on the
flavour SU(3) breaking ratio of neutral $B$-meson mixing parameters in 2018.
Their paper \cite{Boyle:2018knm} has not been published yet, thus the results will not be
included in our averages presented here.
Their computation uses ensembles generated by the $2+1$ flavour domain-wall fermion (DWF) formulation.
The use of the DWFs also for the heavy quarks makes the renormalization structure simple.
Because of the chiral symmetry, the mixing is the same as in the continuum theory.
The operators for standard model mixing matrix elements are multiplicatively renormalized.
Since they only report the SU(3) breaking ratio, the renormalization of the operators is not needed.
The lattice spacings employed are not as fine as some of the recent results reported here.
But, by applying successive stout link smearings in the heavy DWF, the reach to heavy mass is improved,
which allows them to simulate up to half of the physical bottom mass.
Two ensembles are of physical $ud$ quark mass at $a=0.11$ and $0.09$ fm, and there is yet another ensemble off the physical point but with finer lattice spacing, $a=0.07$ fm. This is the first computation using physical light-quark mass for these quantities, which yields a drastic reduction of the chiral extrapolation error.

The results that enter our averages for $N_f=2+1$ are 
FNAL/MILC~16~\cite{Bazavov:2016nty}, which had been included in the averages at FLAG 19
\cite{Aoki:2019cca},
RBC/UKQCD~14A~\cite{Aoki:2014nga},
included in the averages at FLAG 16 \cite{Aoki:2016frl},
and HPQCD~09~\cite{Gamiz:2009ku} for which a description is available in
FLAG 13 \cite{Aoki:2013ldr}.
Thus, the averages for $N_f=2+1$ are unchanged:\\

$N_f=2+1:$
\begin{align}
        && \FLAGAVBEGIN f_{B_d}\sqrt{\hat{B}_{B_d}} &=  225(9)\FLAGAVEND \, {\rm MeV}  
         & \FLAGAVBEGIN f_{B_s}\sqrt{\hat{B}_{B_s}} &=  274(8)\FLAGAVEND \, {\rm MeV}
	  &\Refs~\mbox{\cite{Gamiz:2009ku,Aoki:2014nga,Bazavov:2016nty}},  \label{eq:avfBB}\\ 
&& \FLAGAVBEGIN \hat{B}_{B_d}  &= 1.30(10)\FLAGAVEND 
         & \FLAGAVBEGIN \hat{B}_{B_s} &=  1.35(6)\FLAGAVEND 
          &\Refs~\mbox{\cite{Gamiz:2009ku,Aoki:2014nga,Bazavov:2016nty}}, \label{eq:avBB}\\ 
        && \FLAGAVBEGIN \xi  &=  1.206(17)\FLAGAVEND 
        & \FLAGAVBEGIN B_{B_s}/B_{B_d}  &=  1.032(38)\FLAGAVEND 
          &\Refs~\mbox{\cite{Aoki:2014nga,Bazavov:2016nty}}. \label{eq:avxiBB} 
\end{align}
Here all the above equations have not been changed from the FLAG 19. 
The averages were obtained using the nested averaging scheme described in Sec.~\ref{sec:nested_average},
due to a nested correlation structure among the results. Details are discussed in
the FLAG 19 report \cite{Aoki:2019cca}.


We have the first $N_f=2+1+1$ calculation for these quantities by the HPQCD collaboration
HPQCD 19A \cite{Dowdall:2019bea}, using the MILC collaboration's HISQ ensembles.
The lattice spacings used are $0.15$, $0.12$ and $0.09$ fm, among which the mass of the Nambu-Goldstone
pion (lightest in the staggered taste multiplets) is as small as $130$ MeV for two coarser lattices.
However, the smallest root-mean-squared pion mass through all taste multiplets is
241 MeV, which is a similar size as the FNAL/MILC 16 result \cite{Bazavov:2016nty}
with $N_f=2+1$ and makes the rating on the chiral extrapolation a green circle.
The heavy quark formulation used is non-relativistic QCD (NRQCD).
The NRQCD action employed is improved from that used in older calculations, 
especially by including one-loop radiative corrections to most of the coefficients
of the $\cO(v_b^4)$ terms \cite{Dowdall:2011wh}.
The $b$-quark mass is pre-tuned with the spin-averaged kinetic mass of the $\Upsilon$ 
and $\eta_b$ states. Therefore, there is no need for extrapolation or interpolation
on the b-quark mass.
The HISQ-NRQCD four-quark operators are matched through $\cO(1/M)$ and renormalized to 
one-loop, which includes the effects of $\cO(\alpha_s)$, $\cO(\Lambda_{\rm QCD}/M)$, $\cO(\alpha_s/a M)$, 
and $\cO(\alpha_s \, \Lambda_{\rm QCD}/M)$.
The remaining error is dominated by $\cO(\alpha_s \Lambda_{\rm QCD}/M)~2.9$\% 
and $\cO(\alpha_s^2)~2.1$\% for individual bag parameters.
The bag parameters are the primary quantities calculated in this work.
The mixing matrix elements are obtained by combining the so-obtained bag parameters
with the $B$-meson decay constants
calculated by Fermilab-MILC collaboration (FNAL/MILC 17 \cite{Bazavov:2017lyh}).

Because this is the only result 
available for $N_f=2+1+1$, we quote their values as the FLAG estimates \\

$N_f=2+1+1$:
\begin{align}
  &&  \FLAGAVBEGIN f_{B_d}\sqrt{\hat{B}_{b_d}}&= 210.6(5.5)\FLAGAVEND \;{\rm MeV}\;\;
         &\FLAGAVBEGIN f_{B_s}\sqrt{\hat{B}_{B_s}}&= 256.1(5.7)\FLAGAVEND\; {\rm MeV}
         &\Ref~\mbox{\cite{Dowdall:2019bea}},  \label{eq:avfBB4}\\
      &&\FLAGAVBEGIN \hat{B}_{B_d}&= 1.222(61)\FLAGAVEND 
         &\FLAGAVBEGIN \hat{B}_{B_s}&= 1.232(53)\FLAGAVEND 
	 &\Ref~\mbox{\cite{Dowdall:2019bea}},  \label{eq:avBB4}\\
      &&  \FLAGAVBEGIN \xi &=  1.216(16)\FLAGAVEND  
  	& \FLAGAVBEGIN B_{B_s}/B_{B_d} & =  1.008(25)\FLAGAVEND
 	&\Ref~\mbox{\cite{Dowdall:2019bea}}. \label{eq:avxiBB4}
\end{align}
We note that the above results 
within same $N_f$ 
(e.g., those in Eqs.~(\ref{eq:avfBB4}-\ref{eq:avxiBB4}))
are all correlated with each other,
due to the use of the same 
gauge field ensembles for different quantities.
The results are also correlated with the averages obtained in 
Sec.~\ref{sec:fB} and shown in
Eqs.~(\ref{eq:fB2})--(\ref{eq:fBratio2}) for $N_f=2$,
Eqs.~(\ref{eq:fB21})--(\ref{eq:fBratio21}) for $N_f=2+1$ and
Eqs.~(\ref{eq:fB211})--(\ref{eq:fBratio211}) for $N_f=2+1+1$.
This is because the calculations of $B$-meson decay constants and  
mixing quantities 
are performed on the same (or on similar) sets of ensembles, and results obtained by a 
given collaboration 
use the same actions and setups. These correlations must be considered when 
using our averages as inputs to unitarity triangle (UT) fits. 
For this reason, if one were for example to estimate $f_{B_s}\sqrt{\hat{B}_s}$ from the separate averages of $f_{B_s}$ (Eq.~(\ref{eq:fBs21})) and $\hat{B}_s$ (Eq.~(\ref{eq:avBB})) for $N_f=2+1$, one would obtain a value about one standard deviation below the one quoted above in Eq.~(\ref{eq:avfBB}).  While these two estimates lead to compatible results, giving us confidence that all uncertainties have been properly addressed, we do not recommend combining averages this way, as many correlations would have to be taken into account to properly assess the errors. We recommend instead using the numbers quoted above.
In the future, as more independent 
calculations enter the averages, correlations between the lattice-QCD inputs to UT fits will become less significant.

\FloatBarrier
\subsection{Semileptonic form factors for $B$ decays to light flavours}
\label{sec:BtoPiK}

The Standard Model differential rate for the decay $B_{(s)}\to
P\ell\nu$ involving a quark-level $b\to u$ transition is given, at
leading order in the weak interaction, by a formula analogous to the
one for $D$ decays in Eq.~(\ref{eq:DtoPiKFull}), but with $D \to
B_{(s)}$ and the relevant CKM matrix element $|V_{cq}| \to |V_{ub}|$:
\begin{align}
  \frac{d\Gamma(B_{(s)}\to P\ell\nu)}{dq^2} =
  & \frac{G_F^2 |V_{ub}|^2}{24 \pi^3}
    \frac{(q^2-m_\ell^2)^2\sqrt{E_P^2-m_P^2}}{q^4m_{B_{(s)}}^2}
    \nn\\
  & \times
    \left[ \left(1+\frac{m_\ell^2}{2q^2}\right)
    m_{B_{(s)}}^2(E_P^2-m_P^2)|f_+(q^2)|^2
    \right.
    \nn\\
  & \left. \;\;\;\;\;
   + \frac{3m_\ell^2}{8q^2}(m_{B_{(s)}}^2-m_P^2)^2|f_0(q^2)|^2
    \right].
    \label{eq:B_semileptonic_rate}
\end{align}
Again, for $\ell=e,\mu$ the contribution from the scalar form factor
$f_0$ can be neglected, and one has a similar expression to
Eq.~(\ref{eq:DtoPiK}), which, in principle, allows for a direct
extraction of $|V_{ub}|$ by matching theoretical predictions to
experimental data.  However, while for $D$ (or $K$) decays the entire
physical range $0 \leq q^2 \leq q^2_{\rm max}$ can be covered with
moderate momenta accessible to lattice simulations, in
$B \to \pi \ell\nu$ decays one has $q^2_{\rm max} \sim 26~{\rm GeV}^2$
and only part of the full kinematic range is reachable.
As a consequence, obtaining $|V_{ub}|$ from $B\to\pi\ell\nu$ is more
complicated than obtaining $|V_{cd(s)}|$ from semileptonic $D$-meson
decays.

In practice, lattice computations are restricted
to large values of the momentum transfer $q^2$ (see Sec.~\ref{sec:DtoPiK})
where statistical and momentum-dependent discretization errors can be
controlled,\footnote{The variance of hadron correlation functions at
nonzero three-momentum is dominated at large Euclidean times by
zero-momentum multiparticle states~\cite{DellaMorte:2012xc}; therefore
the noise-to-signal grows more rapidly than for the vanishing three-momentum
case.} which in existing calculations roughly cover the upper third of
the kinematically allowed $q^2$ range.
Since, on the other hand, the decay rate is
suppressed by phase space at large $q^2$, most of the semileptonic $B\to
\pi$ events are observed in experiment at lower values of $q^2$, leading
to more accurate experimental results for the binned differential rate
in that region.\footnote{Upcoming data from Belle~II are expected to
significantly improve the precision of experimental results,
in particular, for larger values of $q^2$.}
It is, therefore, a challenge to find a window of
intermediate values of $q^2$ at which both the experimental and
lattice results can be reliably evaluated.

State-of-the-art determinations of CKM matrix elements, say, e.g., $|V_{ub}|$, are obtained
from joint fits to lattice and experimental results, keeping the relative normalization
$|V_{ub}|^2$ as a free parameter.
This requires, in particular, that both experimental and lattice data for the
$q^2$-dependence be parameterized by fitting data to specific ans\"atze,
with the ultimate aim of minimizing the systematic uncertainties involved.
This plays a key role in assessing the systematic uncertainties of CKM determinations,
and will be discussed extensively in this section.
A detailed discussion of the parameterization of form factors as a function of $q^2$ can be found
in Appendix \ref{sec:zparam}.

%

\subsubsection{Form factors for $B\to\pi\ell\nu$}
\label{sec:BtoPi}

The semileptonic decay process $B\to\pi\ell\nu$ enables determination of the CKM matrix element $|V_{ub}|$
within the Standard Model via Eq.~(\ref{eq:B_semileptonic_rate}).
Early results for
$B\to\pi\ell\nu$ form factors came from the HPQCD~\cite{Dalgic:2006dt}
and FNAL/MILC~\cite{Bailey:2008wp} collaborations.
Only HPQCD provided results for the scalar form factor $f_0$.
Our 2016 review featured a significantly extended
calculation of $B\to\pi\ell\nu$ from FNAL/MILC~\cite{Lattice:2015tia}
and a new computation from  RBC/UKQCD~\cite{Flynn:2015mha}.
All the above computations employ $N_f=2+1$ dynamical configurations,
and provide values for both form factors $f_+$ and $f_0$.
In addition, HPQCD using MILC ensembles had published the first
$N_f=2+1+1$ results for the $B\to\pi\ell\nu$ scalar
form factor, working at zero recoil ($q^2=q^2_{\rm max}$) and pion masses down to the physical value~\cite{Colquhoun:2015mfa};
this adds to previous reports on ongoing work to upgrade their 2006
computation~\cite{Bouchard:2012tb,Bouchard:2013zda}. Since this latter
result has no immediate impact on current $|V_{ub}|$ determinations,
which come from the vector-form-factor-dominated decay channels into light leptons,
we will from now on concentrate on the $N_f=2+1$ determinations of the
$q^2$-dependence of $B\to\pi$ form factors.

Several groups are working on new calculations of the $B\to\pi$ form factors
and have reported on their progress at the annual Lattice conferences and the 2020
Asia-Pacific Symposium for Lattice Field Theory. The results are preliminary or blinded, so
not yet ready for inclusion in this review.
The JLQCD collaboration is using M\"obius Domain Wall
fermions (including for the heavy quark) with $a\approx 0.08$, 0.055, and 0.044 fm and pion masses
down to 225 MeV to study this process~\cite{Colquhoun:2017gfi,Colquhoun:2018kwj,Colquhoun:2019tyq}.
FNAL/MILC is using $N_f=2+1+1$ HISQ ensembles with $a\approx 0.15$,
0.12, 0.088 fm, 0.057 fm, with Goldstone pion mass down to its physical
value~\cite{Gelzer:2017edb,Gelzer:2019zwx}.
The RBC/UKQCD Collaborations have added a new M\"obius-domain-wall-fermion ensemble with 
$a\approx 0.07$ fm and $m_\pi \approx 230$ MeV to their analysis \cite{Flynn:2019jbg}.

Returning to the calculations that contribute to our averages (with no new results since FLAG 19),
both the HPQCD and the FNAL/MILC computations of $B\to\pi\ell\nu$
amplitudes use ensembles of gauge configurations with $N_f=2+1$
flavours of rooted staggered quarks produced by the MILC collaboration;
however, the latest FNAL/MILC work makes a much more extensive
use of the currently available ensembles, both in terms of
lattice spacings and light-quark masses.
HPQCD have results at two values of the lattice spacing
($a\approx0.12,~0.09~{\rm fm}$), while FNAL/MILC employs four values
($a\approx0.12,~0.09,~0.06,~0.045~{\rm fm}$).
Lattice-discretization
effects are estimated within heavy-meson rooted staggered chiral perturbation theory (HMrS$\chi$PT) in the FNAL/MILC
computation, while HPQCD quotes the results at $a\approx 0.12~{\rm fm}$
as central values and uses the $a\approx 0.09~{\rm fm}$ results to quote
an uncertainty.
The relative scale is fixed in both cases through the quark-antiquark potential-derived ratio $r_1/a$.
HPQCD set the absolute scale through the $\Upsilon$ $2S$--$1S$ splitting,
while FNAL/MILC uses a combination of $f_\pi$ and the same $\Upsilon$
splitting, as described in Ref.~\cite{Bazavov:2011aa}.
The spatial extent of the lattices employed by HPQCD is $L\simeq 2.4~{\rm fm}$,
save for the lightest mass point (at $a\approx 0.09~{\rm fm}$) for which $L\simeq 2.9~{\rm fm}$.
FNAL/MILC, on the other hand, uses extents up to $L \simeq 5.8~{\rm fm}$, in order
to allow for light-pion masses while keeping finite-volume effects under
control. Indeed, while in the 2006 HPQCD work the lightest RMS pion mass is $400~{\rm MeV}$,
the latest FNAL/MILC work includes pions as light as $165~{\rm MeV}$---in both cases
the bound $m_\pi L \gtrsim 3.8$ is kept.
Other than the qualitatively different range of MILC ensembles used
in the two computations, the main difference between HPQCD and FNAL/MILC lies in the treatment of
heavy quarks. HPQCD uses the NRQCD formalism, with a 1-loop matching
of the relevant currents to the ones in the relativistic
theory. FNAL/MILC employs the clover action with the Fermilab
interpretation, with a mostly nonperturbative renormalization of the
relevant currents, within which the overall renormalization factor of the heavy-light current
is written as a product of the square roots of the renormalization factors of the
light-light and heavy-heavy temporal vector currents
(which are determined nonperturbatively) and a residual factor that is computed
using 1-loop perturbation theory. (See Tab.~\ref{tab_BtoPisumm2};
full details about the computations are provided in tables in
Appendix~\ref{app:BtoPi_Notes}.)

The RBC/UKQCD computation is based on $N_f=2+1$ DWF ensembles at two
values of the lattice spacing ($a\approx0.12,~0.09~{\rm fm}$), and pion masses
in a narrow interval ranging from slightly above $400~{\rm MeV}$ to slightly below $300~{\rm MeV}$,
keeping $m_\pi L \gtrsim 4$.
The scale is set using the $\Omega^-$ baryon mass. Discretization effects
coming from the light sector
are estimated in the $1\%$ ballpark using HM$\chi$PT supplemented with effective higher-order
interactions to describe cutoff effects.
The $b$ quark is treated using the Columbia RHQ action, with
a mostly nonperturbative renormalization of the relevant currents. Discretization
effects coming from the heavy sector are estimated with power-counting
arguments to be below $2\%$.

Given the large kinematical range available in the $B\to\pi$ transition,
chiral extrapolations are an important source of systematic uncertainty:
apart from the eventual need to reach physical pion masses in the extrapolation,
the applicability of $\chi$PT is not guaranteed for large values of the pion energy $E_\pi$.
Indeed, in all computations $E_\pi$ reaches values in the $1~{\rm GeV}$ ballpark,
and chiral extrapolation systematics is the dominant source of errors.
FNAL/MILC uses $SU(2)$ NLO HMrS$\chi$PT for the continuum-chiral extrapolation,
supplemented by NNLO analytic terms
and hard-pion $\chi$PT terms~\cite{Bijnens:2010ws};\footnote{It is important
to stress the finding in Ref.~\cite{Colangelo:2012ew} that
the factorization of chiral logs in hard-pion $\chi$PT breaks down,
implying that it does not fulfill the expected requisites for a proper
effective field theory. Its use to model the mass dependence of form
factors can thus be questioned.} systematic uncertainties
are estimated through an extensive study of the effects of varying the
specific fit ansatz and/or data range. RBC/UKQCD uses
$SU(2)$ hard-pion HM$\chi$PT to perform its combined continuum-chiral
extrapolation, and obtains estimates for systematic uncertainties
by varying the ans\"{a}tze and ranges used in fits. HPQCD performs chiral
extrapolations using HMrS$\chi$PT formulae, and estimates systematic
uncertainties by comparing the result with the ones from fits to a
linear behaviour in the light-quark mass, continuum HM$\chi$PT, and
partially quenched HMrS$\chi$PT formulae (including also data with
different sea and valence light-quark masses).

FNAL/MILC and RBC/UKQCD describe the $q^2$-dependence of
$f_+$ and $f_0$ by applying a BCL parameterization to
the form factors extrapolated to the continuum
limit, within the range of values of $q^2$ covered by data. (A discussion of the
various parameterizations can be found in Appendix~\ref{sec:zparam}.)
RBC/UKQCD generate synthetic data for the form factors at some values
of $q^2$ (evenly spaced in $z$) from the continuous function of $q^2$ obtained
from the joint chiral-continuum extrapolation,
which are then used as input for the fits. After having checked that the
kinematical constraint $f_+(0)=f_0(0)$ is satisfied within errors by the extrapolation
to $q^2=0$ of the results of separate fits, this constraint is imposed
to improve fit quality. In the case of FNAL/MILC, rather than producing
synthetic data a functional method is used to extract the $z$-parameterization
directly from the fit functions employed in the continuum-chiral extrapolation.
In the case of HPQCD, the parameterization of the $q^2$-dependence of form factors is
somewhat intertwined with chiral extrapolations: a set of fiducial
values $\{E_\pi^{(n)}\}$ is fixed for each value of the light-quark
mass, and $f_{+,0}$ are interpolated to each of the $E_\pi^{(n)}$;
chiral extrapolations are then performed at fixed $E_\pi$
(i.e., $m_\pi$ and $q^2$ are varied subject to $E_\pi$=constant). The
interpolation is performed using a Ball-Zwicky (BZ) ansatz~\cite{Ball:2005tb}.  The $q^2$-dependence of
the resulting form factors in the chiral limit is then described by
means of a BZ ansatz, which is cross-checked
against Becirevic-Kaidalov (BK) \cite{Becirevic:1999kt}, Richard Hill (RH) \cite{Hill:2005ju},
and Boyd-Grinstein-Lebed (BGL) \cite{Boyd:1994tt} parameterizations
(see Appendix \ref{sec:zparam}),
finding agreement
within the quoted uncertainties. Unfortunately, the correlation matrix for the values
of the form factors at different $q^2$ is not provided, which severely
limits the possibilities of combining them with other computations into
a global $z$-parameterization.

\begin{table}[t]
\begin{center}
\mbox{} \\[3.0cm]
\footnotesize
\begin{tabular}{l @{\extracolsep{\fill}} c @{\hspace{2mm}} c l l l l l l l }
Collaboration & Ref. & $\Nf$ &
\hspace{0.15cm}\begin{rotate}{60}{publication status}\end{rotate}\hspace{-0.15cm} &
\hspace{0.15cm}\begin{rotate}{60}{continuum extrapolation}\end{rotate}\hspace{-0.15cm} &
\hspace{0.15cm}\begin{rotate}{60}{chiral extrapolation}\end{rotate}\hspace{-0.15cm}&
\hspace{0.15cm}\begin{rotate}{60}{finite volume}\end{rotate}\hspace{-0.15cm}&
\hspace{0.15cm}\begin{rotate}{60}{renormalization}\end{rotate}\hspace{-0.15cm}  &
\hspace{0.15cm}\begin{rotate}{60}{heavy-quark treatment}\end{rotate}\hspace{-0.15cm}  &
\hspace{0.15cm}\begin{rotate}{60}{$z$-parameterization}\end{rotate}\hspace{-0.15cm} \\%
&&&&&&&&& \\[-0.0cm]
\hline
\hline
&&&&&&&&& \\[-0.0cm]
\SLfnalmilcBpi & \cite{Lattice:2015tia} & 2+1 & \gA  & \good & \soso & \good & \soso & \okay &
 BCL\\[-0.0cm]
\SLrbcukqcdBpi & \cite{Flynn:2015mha} & 2+1 & \gA  & \soso & \soso & \soso & \soso & \okay &
 BCL \\[-0.0cm]
HPQCD 06 & \cite{Dalgic:2006dt} & 2+1 & \gA  & \soso & \soso & \soso
& \soso & \okay &
 n/a \\[-0.0cm]
&&&&&&&&& \\[-0.0cm]
\hline
\hline
\end{tabular}
\caption{Results for the $B \to \pi\ell\nu$ semileptonic form factor.
\label{tab_BtoPisumm2}}
\end{center}
\end{table}

The different ways in which the current results are presented do not
allow a straightforward averaging procedure.
RBC/UKQCD only provides synthetic
values of $f_+$ and $f_0$ at a few values of $q^2$ as an illustration
of their results, and FNAL/MILC does not quote synthetic values at all.
In both cases, full results for BCL $z$-parameterizations defined by
Eq.~(\ref{eq:bcl_c}) are quoted.
In the case of HPQCD~06, unfortunately,
a fit to a BCL $z$-parameterization is not possible, as discussed above.

In order to combine these form factor calculations, we start from sets
of synthetic data for several $q^2$ values. HPQCD and RBC/UKQCD
directly provide this information; FNAL/MILC present only fits to a
BCL $z$-parameterization from which we can easily generate an
equivalent set of form factor values. It is important to note that in
both the RBC/UKQCD synthetic data and the FNAL/MILC
$z$-parameterization fits the kinematic constraint at $q^2=0$ is
automatically included (in the FNAL/MILC case the constraint is
manifest in an exact degeneracy of the $(a_n^+ ,a_n^0)$ covariance
matrix). Due to these considerations, in our opinion, the most accurate
procedure is to perform a simultaneous fit to all synthetic data for
the vector and scalar form factors. Unfortunately, the absence of
information on the correlation in the HPQCD result between the vector
and scalar form factors even at a single $q^2$ point makes it
impossible to include consistently this calculation in the overall
fit. In fact, the HPQCD and FNAL/MILC statistical uncertainties are
highly correlated (because they are based on overlapping subsets of
MILC $N_f=2+1$ ensembles) and, without knowledge of the $f_+ - f_0$
correlation we are unable to construct the HPQCD-FNAL/MILC
off-diagonal entries of the overall covariance matrix.

In conclusion, we will present as our best result a combined vector
and scalar form factor fit to the FNAL/MILC and RBC/UKQCD results that
we treat as completely uncorrelated. For sake of completeness, we will
also show the results of a vector form factor fit alone in which we
include one HPQCD datum at $q^2=17.34~\GeV^2$ assuming conservatively
a 100\% correlation between the statistical error of this point and of
all FNAL/MILC synthetic data. In spite of contributing just one point,
the HPQCD datum has a significant weight in the fit due to its small
overall uncertainty. We stress again that this procedure is slightly
inconsistent because FNAL/MILC and RBC/UKQCD include information on
the kinematic constraint at $q^2=0$ in their $f_+$ results.

The resulting data set is then fitted to the BCL parameterization in
Eqs.~(\ref{eq:bcl_c}) and (\ref{eq:bcl_f0}). We assess the systematic
uncertainty due to truncating the series expansion by considering fits
to different orders in $z$.  In Fig.~\ref{fig:LQCDzfit}, we show the
FNAL/MILC, RBC/UKQCD, and HPQCD data points for $(1-q^2/m_{B^*}^2)
f_+(q^2)$ and $f_0 (q^2)$ versus $z$.  The data is highly linear and
we get $\chi^2/{\rm dof} = 0.82$ with $N^+ = N^0 = 3$. Note that
this implies three independent parameters for $f_+$ corresponding to a
polynomial through $\cO(z^3)$ and two independent parameters for $f_0$
corresponding to a polynomial through $\cO(z^2)$ (the coefficient
$a_2^0$ is fixed using the $q^2=0$ kinematic constraint). We cannot
constrain the coefficients of the $z$-expansion beyond this order; for
instance, including a fourth parameter in $f_+$ results in 100\%
uncertainties on $a_2^+$ and $a_3^+$. The outcome of the
five-parameter $N^+ =N^0=3$ BCL fit to the FNAL/MILC and RBC/UKQCD
calculations is shown in Tab.~\ref{tab:FFPI}. The uncertainties on
$a_0^{+,0}$, $a_1^{+,0}$ and $a_2^+$ encompass the central values
obtained from $N^+=2,4$ and $N^0=2,4,5$ fits and thus adequately
reflect the systematic uncertainty on those series coefficients. The fit shown in Tab.~\ref{tab:FFPI}
can therefore be used as the averaged FLAG result for the lattice-computed form
factor $f_+(q^2)$. The coefficient $a_3^+$ can be obtained from the
values for $a_0^+$--$a_2^+$ using Eq.~(\ref{eq:red_coeff}). The
coefficient $a_2^0$ can be obtained from all other coefficients
imposing the $f_+(q^2=0) = f_0(q^2=0)$ constraint. The fit is
illustrated in Fig.~\ref{fig:LQCDzfit}.
\begin{table}[t]
\begin{center}
\begin{tabular}{|c|c|ccccc|}
\multicolumn{7}{l}{$B\to \pi \; (N_f=2+1)$} \\[0.2em]\hline
        & Central Values & \multicolumn{5}{|c|}{Correlation Matrix} \\[0.2em]\hline
$a_0^+$ & 0.404 (13)  &   1 & 0.404 & 0.118 & 0.327 & 0.344 \\[0.2em]
$a_1^+$ & $-$0.68 (13)  &   0.404 & 1 & 0.741 & 0.310 & 0.900  \\[0.2em]
$a_2^+$ & $-$0.86 (61)  &   0.118 & 0.741 & 1 & 0.363 & 0.886 \\[0.2em]
$a_0^0$ & 0.490 (21)  &     0.327 & 0.310 & 0.363 & 1 & 0.233 \\[0.2em]
$a_1^0$ & $-$1.61 (16)  &    0.344 & 0.900 & 0.886 & 0.233 & 1 \\[0.2em]
\hline
\end{tabular}
\end{center}
\caption{Coefficients and correlation matrix for the $N^+ =N^0=3$ $z$-expansion fit of the $B\to \pi$ form factors $f_+$ and $f_0$. The coefficient $a_2^0$ is fixed by the $f_+(q^2=0) = f_0(q^2=0)$ constraint. The chi-square per degree of freedom is $\chi^2/{\rm dof} = 0.82$. The lattice calculations that enter this fit are taken from \SLfnalmilcBpi~\cite{Lattice:2015tia} and \SLrbcukqcdBpi~\cite{Flynn:2015mha}. The parameterizations are defined in Eqs.~(\ref{eq:bcl_c}) and (\ref{eq:bcl_f0}).
\label{tab:FFPI} }
\end{table}
%
We emphasize that future
lattice-QCD calculations of semileptonic form factors should publish
their full statistical and systematic correlation matrices to enable
others to use the data. It is also preferable to  present a
set of synthetic form factors data equivalent to the $z$-fit results,
 since this allows for an independent
analysis that avoids further assumptions about the compatibility of
the procedures to arrive at a given $z$-parameterization.\footnote{
Note that generating synthetic data is a trivial task, but less so is choosing
the
number of required points and the $q^2$ values that lead to an optimal description of the form factors. }
 It is also preferable to present covariance/correlation matrices with enough significant digits to calculate correctly all their eigenvalues.

For the sake of completeness, we present also a standalone $z$-fit to the vector form factor. In this fit, we are able to include the single $f_+$ point at $q^2 = 17.34\; {\rm GeV}^2$ that we mentioned above. This fit uses the FNAL/MILC and RBC/UKQCD results that do make use of the kinematic constraint at $q^2=0$, but is otherwise unbiased. The results of the three-parameter BCL fit to the HPQCD, FNAL/MILC and RBC/UKQCD calculations of the vector form factor are:
 \begin{gather}
N_f=2+1:  \qquad
a_0^+ = 0.421(13)\,,~~~~
a_1^+ = -0.35(10)\,,~~~~
a_2^+ = -0.41(64)\,;
\\[1.0ex]
\nonumber \qquad\qquad
{\rm corr}(a_i,a_j)=\left(\begin{array}{rrr}
 1.000 &  0.306 &  0.084 \\
 0.306 &  1.000 &  0.856 \\
 0.084 &  0.856 &  1.000
\end{array}\right)\,.
\end{gather}
Note that the $a_0^+$ coefficient, that is the most relevant for input to the extraction of $V_{ub}$ from semileptonic $B\to \pi \ell \nu_\ell (\ell=e,\mu)$ decays, shifts by  about  a standard deviation.

\begin{figure}[tbp]
\begin{center}
\includegraphics[width=0.49\textwidth]{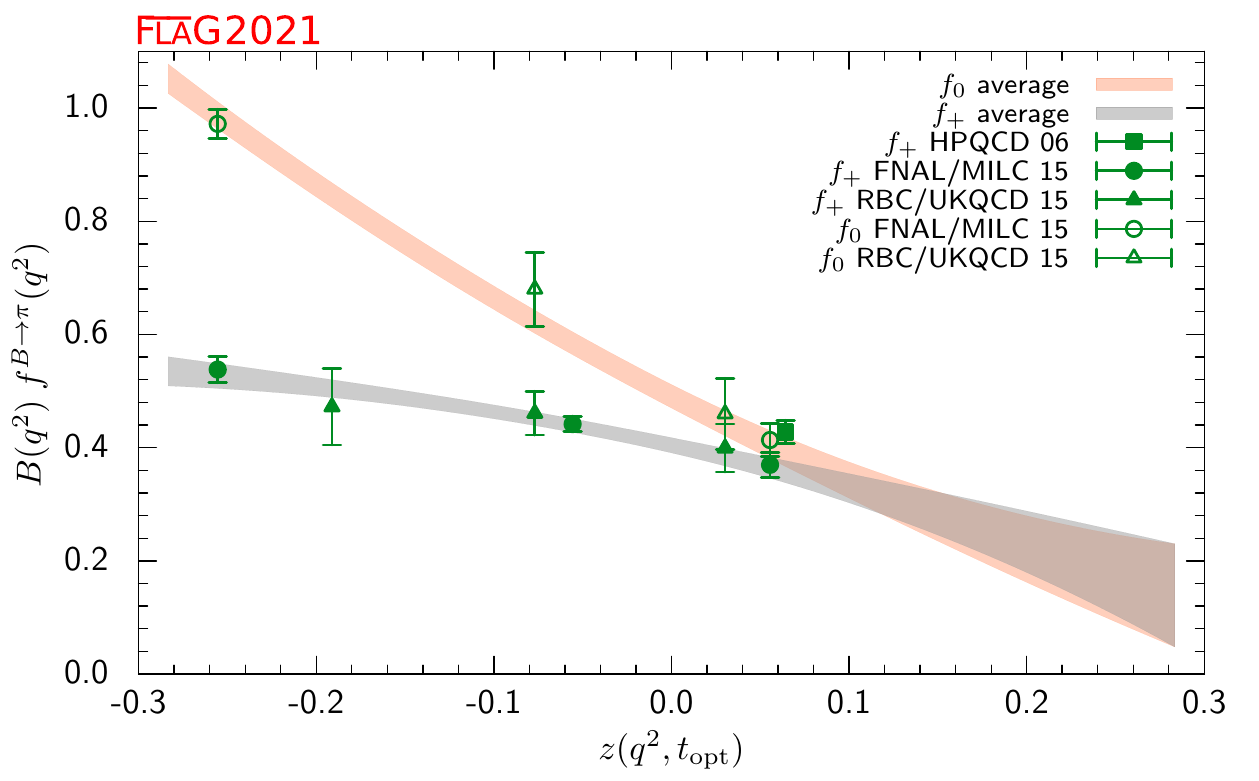}
\includegraphics[width=0.49\textwidth]{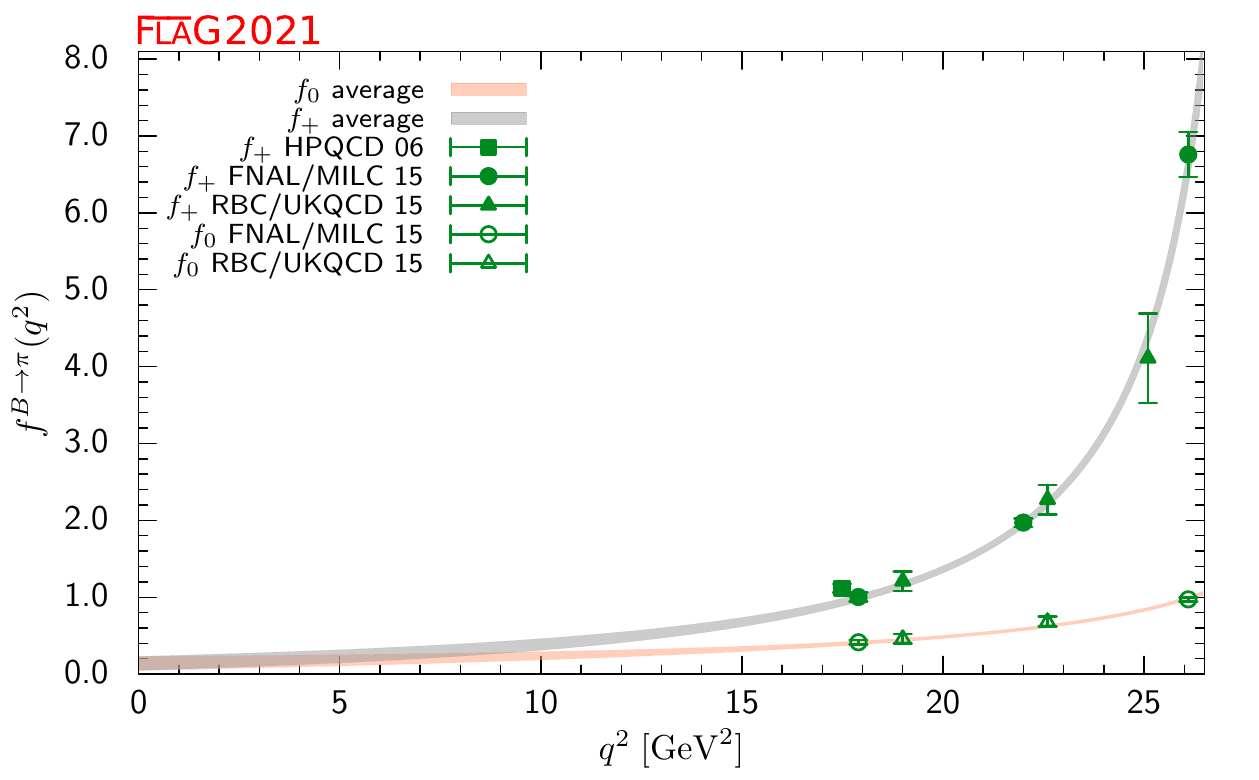}
\caption{The form factors $f_+(q^2)$ and $f_0 (q^2)$ for $B \to \pi\ell\nu$ plotted versus $z$ (left panel) and $q^2$ (right panel). In the left plot, we removed the Blaschke factors. See text for a discussion of the data set. The grey and salmon bands display our preferred $N^+=N^0=3$ BCL fit (five parameters).}\label{fig:LQCDzfit}
\end{center}
\end{figure}

\subsubsection{Form factors for $B_s\to K\ell\nu$}
\label{sec:BstoK}

Similar to $B\to\pi\ell\nu$, measurements of $B_s\to K\ell\nu$ decay rates enable determinations
of the CKM matrix element $|V_{ub}|$
within the Standard Model via Eq.~(\ref{eq:B_semileptonic_rate}).
From the lattice point of view, the two channels are very similar.  As
a matter of fact, $B_s\to K\ell\nu$ is actually somewhat simpler,
in that the kaon mass region is easily accessed by all simulations
making the systematic uncertainties related to chiral extrapolation
smaller. 

At the time of our FLAG 19 review \cite{Aoki:2019cca},
results for $B_s\to K\ell\nu$ form factors were
provided by HPQCD~\cite{Bouchard:2014ypa} and RBC/UKQCD~\cite{Flynn:2015mha}
for both form factors $f_+$ and $f_0$, in both cases using $N_f=2+1$ dynamical configurations.
HPQCD has recently emphasized the value of using ratios of form factors for the processes
$B_s\to K\ell\nu$ and $B_s\to D_s\ell\nu$ for the determination of
$|V_{ub}/V_{cb}|$~\cite{Monahan:2018lzv}.
In the FLAG Review 19 \cite{Aoki:2019cca}, FNAL/MILC preliminary
results had been reported for both $N_f=2+1$~\cite{Lattice:2017vqf}
and $N_f=2+1+1$~\cite{Gelzer:2017edb}, but were not included in the average
due to their non-final status. The $N_f=2+1$ results have since been published~\cite{Bazavov:2019aom};
we will, therefore, include them in the average here.

The RBC/UKQCD computation has been published together with the $B\to\pi\ell\nu$
computation discussed in Sec.~\ref{sec:BtoPi}, all technical details being
practically identical. The main difference is that errors are significantly smaller,
mostly due to the reduction of systematic uncertainties due to the chiral extrapolation.
Detailed information is provided in Appendix~\ref{app:BtoPi_Notes}. The RBC/UKQCD collaboration is also working on an improved determination of $B_s\to K$ form factors that includes a finer lattice spacing,
with preliminary results shown in Ref.~\cite{Flynn:2019jbg}, but these results cannot yet be included in the average.
The HPQCD computation uses ensembles of gauge configurations with $N_f=2+1$
flavours of asqtad rooted staggered quarks produced by the MILC collaboration
at two values of the lattice spacing ($a\approx0.12,~0.09~{\rm fm}$), for three
and two different sea-pion masses, respectively, down to a value of $260~{\rm MeV}$.
The $b$ quark is treated within the NRQCD formalism, with a 1-loop matching
of the relevant currents to the ones in the relativistic theory, omitting terms
of $\cO(\alpha_s\Lambda_{\rm QCD}/m_b)$. The HISQ action
is used for the valence $s$ quark. The continuum-chiral extrapolation
is combined with the description of the $q^2$-dependence of the form factors
into a modified $z$-expansion
(cf.~Appendix \ref{sec:zparam})
that formally coincides
in the continuum with the BCL ansatz. The dependence of
form factors on the pion energy and quark masses is fitted to a 1-loop ansatz
inspired by hard-pion $\chi$PT~\cite{Bijnens:2010ws},
that factorizes out the chiral logarithms describing soft physics.
The FNAL/MILC computation coincides with HPQCD's in using ensembles of gauge configurations with $N_f=2+1$
flavours of asqtad rooted staggered quarks produced by the MILC collaboration, but
only one ensemble is shared, and a different valence regularization is employed;
we will thus treat the two results as fully independent from the statistics point of view.
FNAL/MILC uses three values of the lattice spacing ($a\approx0.12,~0.09,~0.06~{\rm fm}$);
only one value of the sea pion mass and the volume is available at the extreme values
of the lattice spacing, while four different masses and volumes are considered at $a=0.09~{\rm fm}$.
Heavy quarks are treated within the Fermilab approach.
HMrS$\chi$PT expansion is used at next-to-leading
order in $SU(2)$ and leading order in $1/M_B$, including next-to-next-to-leading-order
(NNLO) analytic and generic discretization terms, to perform continuum-chiral extrapolations.
Hard kaons are assumed to decouple, i.e., their effect is reabsorbed in the $SU(2)$ LECs.
Continuum- and chiral-extrapolated values of the form factors are fitted to a $z$-parametrization
imposing the kinematical constraint $f_+(0)=f_0(0)$. See Tab.~\ref{tab_BstoKsumm} and the tables in Appendix~\ref{app:BtoPi_Notes} for full details.

\begin{table}[t]
\begin{center}
\mbox{} \\[3.0cm]
\footnotesize
\begin{tabular}{l  c @{\hspace{2mm}} c l l l l l l l l }
Collaboration & Ref. & $\Nf$ &
\hspace{0.15cm}\begin{rotate}{60}{publication status}\end{rotate}\hspace{-0.15cm} &
\hspace{0.15cm}\begin{rotate}{60}{continuum extrapolation}\end{rotate}\hspace{-0.15cm} &
\hspace{0.15cm}\begin{rotate}{60}{chiral extrapolation}\end{rotate}\hspace{-0.15cm}&
\hspace{0.15cm}\begin{rotate}{60}{finite volume}\end{rotate}\hspace{-0.15cm}&
\hspace{0.15cm}\begin{rotate}{60}{renormalization}\end{rotate}\hspace{-0.15cm}  &
\hspace{0.15cm}\begin{rotate}{60}{heavy-quark treatment}\end{rotate}\hspace{-0.15cm}  &
\hspace{0.15cm}\begin{rotate}{60}{$z$-parameterization}\end{rotate}\hspace{-0.15cm}\\%
&&&&&&&&& \\[-0.0cm]
\hline
\hline
&&&&&&&&& \\[-0.0cm]
\SLfnalmilcBsK & \cite{Bazavov:2019aom} & 2+1 & \gA  & \good & \soso & \good & \soso & \okay &
BCL \\[-0.0cm]
\SLrbcukqcdBpi & \cite{Flynn:2015mha} & 2+1 & \gA  & \soso & \soso & \soso & \soso & \okay &
BCL \\[-0.0cm]
\SLhpqcdBsK & \cite{Bouchard:2014ypa} & 2+1 & \gA  & \soso & \soso & \soso & \soso & \okay &
BCL$^\dagger$   \\[-0.0cm]
&&&&&&&&& \\[-0.0cm]
\hline
\hline\\
\multicolumn{4}{l}{$^\dagger$ Results from modified $z$-expansion.}\\
\end{tabular}\\[-0.2cm]
\begin{minipage}{\linewidth}
\end{minipage}
\caption{Summary of lattice calculations of the $B_s \to K\ell\nu$ semileptonic form factors.
\label{tab_BstoKsumm}}
\end{center}
\end{table}

In order to combine the results for the $q^2$ dependence of the form factors from the
three collaborations, we will follow a similar approach to the one
adopted above for $B\to\pi\ell\nu$, and produce synthetic data
from the preferred fits quoted in the papers, to obtain a dataset
to which a joint fit can be performed. Note that the kinematic
constraint at $q^2=0$ is included in all three cases; we will 
impose it in our fit as well, since the synthetic data will implicitly
depend on that fitting choice.  However, it is worth mentioning that
the systematic uncertainty of the resulting extrapolated value
$f_+(0)=f_0(0)$ can be fairly large, the main reason being the required long extrapolation
from the high-$q^2$ region covered by lattice data. While we stress that the
average far away from the high-$q^2$ region has to be used carefully, it is possible that 
increasing the number of $z$ coefficients beyond what is sufficient for a good description of
the lattice data and using unitarity constraints to control the size of additional terms, might 
yield fits with a more stable extrapolation at very low $q^2$. We plan to include said unitarity analysis
into the next edition of the FLAG review. It is, however, important to emphasize
that joint fits with experimental data, where the latter accurately
map the $q^2$ region, are expected to be safe.

Our fits employ a BCL ansatz with $t_+=(M_{B}+M_{\pi})^2$ and
$t_0=t_+-\sqrt{t_+(t_+-t-)}$, with $t_-=(M_{B_s}-M_K)^2$.
Our pole factors will contain a single pole in both the vector and scalar
channels, for which we take the mass values $M_{B^*}=5.32465~\GeV$
and $M_{B^*(0^+)}=5.68~\GeV$.\footnote{These are the values used in the
FNAL/MILC determination, while HPQCD and RBC/UKQCD use $M_{B^*(0+)}=5.6794(10)~\GeV$
and $M_{B^*(0+)}=5.63~\GeV$, respectively. They also employ different values of $t_+$
and $t_0$ than employed here, which again coincide with FNAL/MILC's choice.}
The constraint $f_+(0)=f_0(0)$ is imposed by expressing the coefficient $b^0_{N^0-1}$ in
terms of all others.
The outcome of the seven-parameter $N^+ = N^0 = 4$ BCL fit, which we
quote as our preferred result, is shown in Tab.~\ref{tab:FFBSK}. The fit has a chi-square per degree of freedom $\chi^2/{\rm dof} = 1.54$. Following the PDG recommendation, we rescale the whole covariance matrix by $\chi^2/{\rm dof}$: the errors on the $z$-parameters are increased by $\sqrt{\chi^2/{\rm dof}} = 1.24$ and the correlation matrix is unaffected. The parameters shown in Tab.~\ref{tab:FFBSK} provide the averaged FLAG results
for the lattice-computed form factors $f_+(q^2)$ and $f_0(q^2)$. The
coefficient $a_4^+$ can be obtained from the values for
$a_0^+$--$a_3^+$ using Eq.~(\ref{eq:red_coeff}). The fit is
illustrated in Fig.~\ref{fig:LQCDzfitBsK}.\footnote{Note that in FLAG 19 \cite{Aoki:2019cca} we had adopted the threshold $t_+=(M_{B_s}+M_{K})^2$ rather than $t_+=(M_{B}+M_{\pi})^2$. This change impacted the $z$-range which the physical $q^2$ interval maps onto. We also point out that, in the FLAG 19 version of Fig.~\ref{fig:LQCDzfitBsK}, the three synthetic $f_0$ data points from HPQCD were plotted incorrectly, but this did not affect the fit.}

We will conclude by pointing out progress in the application of the npHQET method to the
extraction of semileptonic form factors, reported for $B_s \to K$ transitions in Ref.~\cite{Bahr:2019eom},
which extends the work of Ref.~\cite{Bahr:2014iqa}.
This is a methodological study based on CLS $N_f=2$ ensembles at two different values of the
lattice spacing and pion masses, and full $1/m_b$ corrections are incorporated within the npHQET
framework. Emphasis is on the role of excited states in the extraction of the bare form factors,
which are shown to pose an impediment to reaching precisions better than a few percent.
\begin{table}[t]
\begin{center}
\begin{tabular}{|c|c|ccccccc|}
\multicolumn{9}{l}{$B_s\to K \; (N_f=2+1)$} \\[0.2em]\hline
        & Central Values & \multicolumn{7}{|c|}{Correlation Matrix} \\[0.2em]\hline
$a_0^+$ & 0.374  (12) & 1 & 0.2471 & -0.1715 & -0.2396 & 0.6445 & 0.3791 & 0.2857 \\[0.2em]
$a_1^+$ & -0.672  (64) & 0.2471 & 1 & 0.4198 & 0.1724 & 0.4626 & 0.8183 & 0.7948 \\[0.2em]
$a_2^+$ & 0.07    (31) &  -0.1715 & 0.4198 & 1 & 0.8136 & 0.3804 & 0.7293 & 0.7481 \\[0.2em]
$a_3^+$ & 1.34    (52) &  -0.2396 & 0.1724 & 0.8136 & 1 & 0.2823 & 0.5120 & 0.5529 \\[0.2em]
$a_0^0$ & 0.2203   (68) & 0.6445 & 0.4626 & 0.3804 & 0.2823 & 1 & 0.6570 & 0.4837 \\[0.2em]
$a_1^0$ & 0.089   (57) & 0.3791 & 0.8183 & 0.7293 & 0.5120 & 0.6570 & 1 & 0.9220 \\[0.2em]
$a_2^0$ & 0.24    (23) &  0.2857 & 0.7948 & 0.7481 & 0.5529 & 0.4837 & 0.9220 & 1 \\[0.2em]
\hline
\end{tabular}
\end{center}
\caption{Coefficients and correlation matrix for the $N^+ =N^0=4$ $z$-expansion of the $B_s\to K$ form factors $f_+$ and $f_0$. The coefficient $a_3^0$ is fixed by the $f_+(q^2=0) = f_0(q^2=0)$ constrain. The chi-square per degree of freedom is $\chi^2/{\rm dof} = 1.54$ and the errors on the $z$-parameters have been rescaled by  $\sqrt{\chi^2/{\rm dof}} = 1.24$.
\label{tab:FFBSK}}
\end{table}

\begin{figure}[tbp]
\begin{center}
\includegraphics[width=0.49\textwidth]{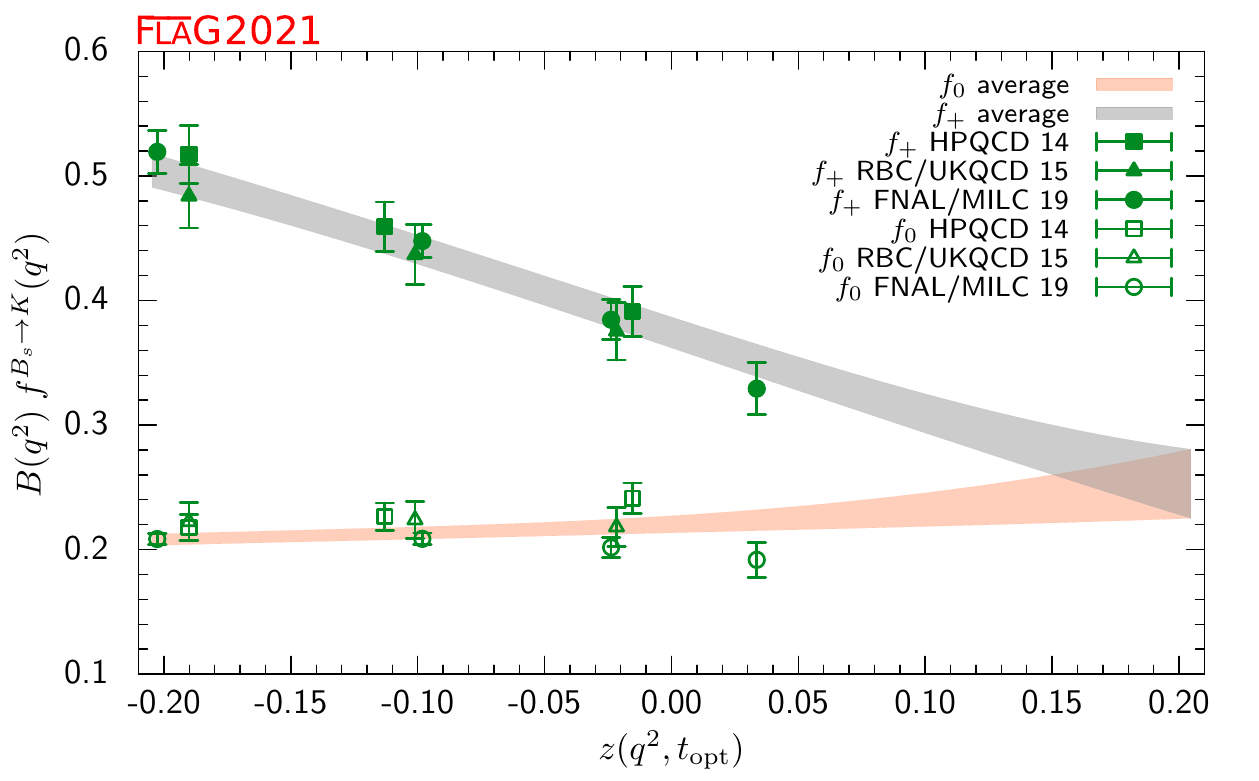}
\includegraphics[width=0.49\textwidth]{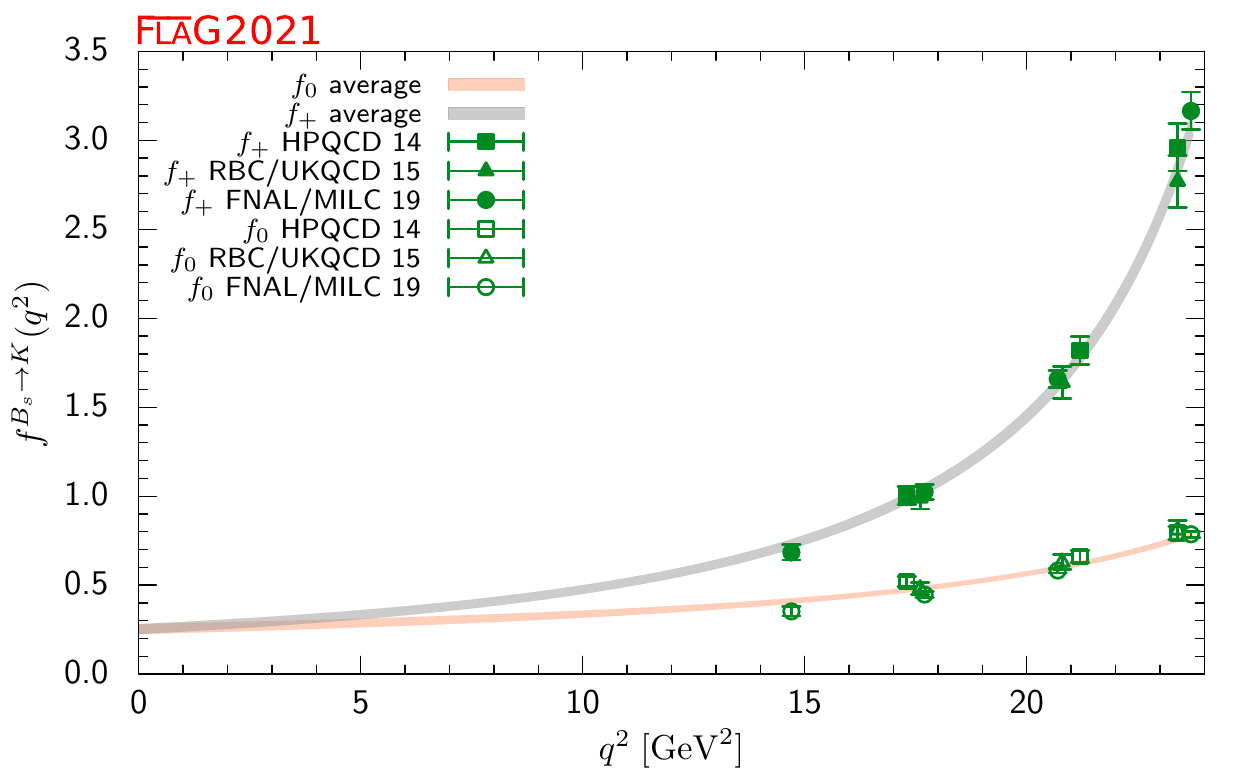}
\caption{The form factors $f_+(q^2)$ and $f_0(q^2)$ for $B_s \to K\ell\nu$ plotted versus $z$ (left panel) and $q^2$ (right panel). In the left plot, we remove the Blaschke factors. See text for a discussion of the data sets. The grey and salmon bands display our preferred $N^+=N^0=4$ BCL fit (seven parameters).}
\label{fig:LQCDzfitBsK}
\end{center}
\end{figure}

\subsubsection{Form factors for rare and radiative $B$-semileptonic decays to light flavours}

Lattice-QCD input is also available for some exclusive semileptonic
decay channels involving neutral-current $b\to q$ transitions at the
quark level, where $q=d,s$. Being forbidden at tree level in the SM,
these processes allow for stringent tests of potential new physics;
simple examples are $B\to K^*\gamma$, $B\to K^{(*)}\ell^+\ell^-$, or
$B\to\pi\ell^+\ell^-$ where the $B$ meson (and therefore the light
meson in the final state) can be either neutral or charged.

The corresponding SM effective weak Hamiltonian is considerably more
complicated than the one for the tree-level processes discussed above:
after integrating out the top quark and the W boson, as many as ten
dimension-six operators formed by the product of two hadronic currents
or one hadronic and one leptonic current appear.\footnote{See, e.g.,
  Ref.~\cite{Antonelli:2009ws} and references therein.}  Three of the
latter, coming from penguin and box diagrams, dominate at short
distances and have matrix elements that, up to small QED corrections,
are given entirely in terms of $B\to (\pi,K,K^*)$ form factors. The
matrix elements of the remaining seven operators can be expressed, up
to power corrections whose size is still unclear, in terms of form
factors, decay constants and light-cone distribution amplitudes (for
the $\pi$, $K$, $K^*$ and $B$ mesons) by employing OPE arguments (at
large di-lepton invariant mass) and results from Soft Collinear
Effective Theory (at small di-lepton invariant mass). In conclusion,
the most important contributions to all of these decays are expected
to come from matrix elements of current operators (vector, tensor, and
axial-vector) between one-hadron states, which in turn can be
parameterized in terms of a number of form factors (see
Ref.~\cite{Liu:2009dj} for a complete description).

\begin{table}[t]
\begin{center}
\mbox{} \\[3.0cm]
\footnotesize
\begin{tabular}{l  c @{\hspace{2mm}} c l l l l l l l l }
Collaboration & Ref. & $\Nf$ &
\hspace{0.15cm}\begin{rotate}{60}{publication status}\end{rotate}\hspace{-0.15cm} &
\hspace{0.15cm}\begin{rotate}{60}{continuum extrapolation}\end{rotate}\hspace{-0.15cm} &
\hspace{0.15cm}\begin{rotate}{60}{chiral extrapolation}\end{rotate}\hspace{-0.15cm}&
\hspace{0.15cm}\begin{rotate}{60}{finite volume}\end{rotate}\hspace{-0.15cm}&
\hspace{0.15cm}\begin{rotate}{60}{renormalization}\end{rotate}\hspace{-0.15cm}  &
\hspace{0.15cm}\begin{rotate}{60}{heavy-quark treatment}\end{rotate}\hspace{-0.15cm}  &
\hspace{0.15cm}\begin{rotate}{60}{$z$-parameterization}\end{rotate}\hspace{-0.15cm}\\%
&&&&&&&&& \\[-0.0cm]
\hline
\hline
&&&&&&&&& \\[-0.0cm]
\SLfnalmilcBK & \cite{Bailey:2015dka} & 2+1 & \gA  & \good & \soso & \good & \soso & \okay &
 BCL \\[-0.0cm]
\SLhpqcdBK & \cite{Bouchard:2013pna} & 2+1 & \gA  & \soso & \soso & \soso & \soso & \okay &
BCL   \\[-0.0cm]
&&&&&&&&& \\[-0.0cm]
\hline
\hline
\end{tabular}
\caption{Summary of lattice calculations of the $B \to K$ semileptonic form factors.
\label{tab_BtoKsumm}}
\end{center}
\end{table}

In channels with pseudoscalar mesons in the final state, the level of
sophistication of lattice calculations is similar to the $B\to \pi$
case and there are results for the vector, scalar, and tensor form
factors for $B\to K\ell^+\ell^-$ decays by
HPQCD~\cite{Bouchard:2013pna}, and more recent results for both
$B\to\pi\ell^+\ell^-$~\cite{Bailey:2015nbd} and $B\to
K\ell^+\ell^-$~\cite{Bailey:2015dka} from FNAL/MILC.  Full details
about these two calculations are provided in Tab.~\ref{tab_BtoKsumm}
and in Appendix~\ref{app:BtoK_Notes}. 
Both computations
employ MILC $N_f=2+1$ asqtad ensembles.  HPQCD~\cite{Bouchard:2013mia}
and FNAL/MILC~\cite{Du:2015tda} have also companion papers in which
they calculate the Standard Model predictions for the differential
branching fractions and other observables and compare to experiment.
The HPQCD computation employs NRQCD $b$ quarks and HISQ valence light
quarks, and parameterizes the form factors over the full kinematic
range using a model-independent $z$-expansion as in
Appendix~\ref{sec:zparam},
including the covariance matrix of the fit
coefficients.  In the case of the (separate) FNAL/MILC computations,
both of them use Fermilab $b$ quarks and asqtad light quarks, and a
BCL $z$-parameterization of the form factors.

Reference~\cite{Bailey:2015nbd} includes results for the tensor form factor
for $B\to\pi\ell^+\ell^-$ not included in previous publications on the
vector and scalar form factors~\cite{Lattice:2015tia}.
Nineteen ensembles from four lattice
spacings are used to control continuum and chiral extrapolations.
The results for $N_z=4$ $z$-expansion of the tensor form factor and its
correlations with the expansions for the vector and scalar form factors, which we consider
the FLAG estimate, are shown in Tab.~\ref{tab:FFPIT}.
Partial decay widths for decay into light leptons or
$\tau^+\tau^-$ are presented as a function of $q^2$.  The former is
compared with results from LHCb~\cite{Aaij:2015nea}, while the
latter is a prediction.
\begin{table}[t]
\begin{center}
\begin{tabular}{|c|c|cccc|}
\multicolumn{6}{l}{$B\to \pi \; (N_f=2+1)$} \\[0.2em]\hline
        & Central Values & \multicolumn{4}{|c|}{Correlation Matrix} \\[0.2em]\hline
$a_0^T$ & 0.393(17)   & 1.000  & 0.400 & 0.204 & 0.166 \\[0.2em]
$a_1^T$ & $-$0.65(23) & 0.400  & 1.000 & 0.862 & 0.806 \\[0.2em]
$a_2^T$ & $-$0.6(1.5) & 0.204  & 0.862 & 1.000 & 0.989 \\[0.2em]
$a_3^T$ & 0.1(2.8)    & 0.166  & 0.806 & 0.989 & 1.000 \\[0.2em]
\hline
\end{tabular}
\end{center}
\caption{Coefficients and correlation matrix for the $N^+ =N^0=3$ $z$-expansion of the $B\to \pi$ form factor $f_T$. \label{tab:FFPIT}}
\end{table}

The averaging of the HPQCD and FNAL/MILC results for the $B\to K$ form factors is similar to our
treatment of the $B\to \pi$ and $B_s\to K$ form factors. In this case,
even though the statistical uncertainties are partially correlated
because of some overlap between the adopted sets of MILC ensembles, we
choose to treat the two calculations as independent. The reason is
that, in $B\to K$, statistical uncertainties are subdominant and
cannot be easily extracted from the results presented by HPQCD and
FNAL/MILC. Both collaborations provide only the outcome of a
simultaneous $z$-fit to the vector, scalar and tensor form factors,
that we use to generate appropriate synthetic data. We then impose the
kinematic constraint $f_+(q^2=0) = f_0(q^2=0)$ and fit to $(N^+ = N^0
= N^T = 3)$ BCL parameterization. The functional forms of the form
factors that we use are identical to those adopted in
Ref.~\cite{Du:2015tda}.\footnote{Note in particular that not much is
  known about the sub-threshold poles for the scalar form
  factor. FNAL/MILC includes one pole at the $B_{s0}^*$ mass as taken
  from the calculation in Ref.~\cite{Lang:2015hza}.} The results of the fit are
  presented in Tab.~\ref{tab:FFK}. The fit is illustrated in Fig.~\ref{fig:LQCDzfitBK}. Note that the
average for the $f_T$ form factor appears to prefer the FNAL/MILC
synthetic data. This happens because we perform a correlated fit of
the three form factors simultaneously (both FNAL/MILC and HPQCD
present covariance matrices that include correlations between all form
factors). We checked that the average for the $f_T$ form factor,
obtained neglecting correlations with $f_0$ and $f_+$, is a little
lower and lies in between the two data sets.
There is still a noticeable tension between the FNAL/MILC and HPQCD data
for the tensor form factor; indeed, a standalone fit to these data results
in $\chi^2_{\rm\scriptscriptstyle red}=7.2/3=2.4$, while a similar standalone
joint fit to $f_+$ and $f_0$ has $\chi^2_{\rm\scriptscriptstyle red}=9.2/7=1.3$.
Finally, the global fit that is shown in the figure has $\chi^2_{\rm\scriptscriptstyle red}=18.6/10=1.86$.
\begin{table}[t]
\begin{center}
\begin{tabular}{|c|c|cccccccc|}
\multicolumn{10}{l}{$B\to K \; (N_f=2+1)$} \\[0.2em]\hline
        & Central Values & \multicolumn{8}{|c|}{Correlation Matrix} \\[0.2em]\hline
$a_0^+$ & 0.471 (14)   & 1 & 0.513 & 0.128 & 0.773 & 0.594 & 0.613 & 0.267 & 0.118   \\[0.2em]
$a_1^+$ & $-$0.74 (16) & 0.513 & 1 & 0.668 & 0.795 & 0.966 & 0.212 & 0.396 & 0.263   \\[0.2em]
$a_2^+$ & 0.32 (71)    & 0.128 & 0.668 & 1 & 0.632 & 0.768 & -0.104 & 0.0440 & 0.187 \\[0.2em]
$a_0^0$ & 0.301 (10)   & 0.773 & 0.795 & 0.632 & 1 & 0.864 & 0.393 & 0.244 & 0.200   \\[0.2em]
$a_1^0$ & 0.40 (15)    & 0.594 & 0.966 & 0.768 & 0.864 & 1 & 0.235 & 0.333 & 0.253   \\[0.2em]
$a_0^T$ & 0.455 (21)   & 0.613 & 0.212 & -0.104 & 0.393 & 0.235 & 1 & 0.711 & 0.608  \\[0.2em]
$a_1^T$ & $-$1.00 (31) & 0.267 & 0.396 & 0.0440 & 0.244 & 0.333 & 0.711 & 1 & 0.903  \\[0.2em]
$a_2^T$ & $-$0.9 (1.3) & 0.118 & 0.263 & 0.187 & 0.200 & 0.253 & 0.608 & 0.903 & 1   \\[0.2em]
\hline
\end{tabular}
\end{center}
\caption{Coefficients and correlation matrix for the $N^+ =N^0=N^T=3$ $z$-expansion of the $B\to K$ form factors $f_+$, $f_0$ and $f_T$. The coefficient $a_2^0$ is fixed by the $f_+(q^2=0) = f_0(q^2=0)$ constraint. The chi-square per degree of freedom is $\chi^2/{\rm dof} = 1.86$ and the errors on the $z$-parameters have been rescaled by  $\sqrt{\chi^2/{\rm dof}} = 1.36$.\label{tab:FFK}}
\end{table}
\begin{figure}[tbp]
\begin{center}
\includegraphics[width=0.49\textwidth]{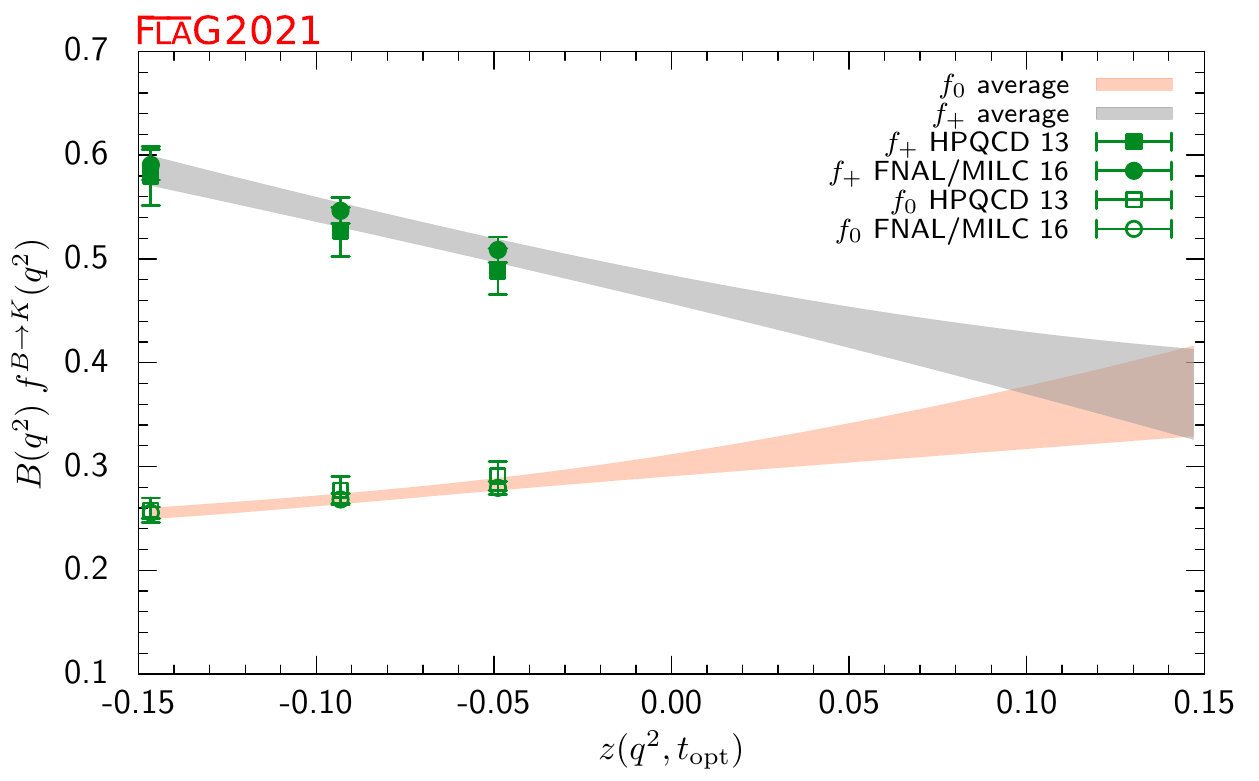}
\includegraphics[width=0.49\textwidth]{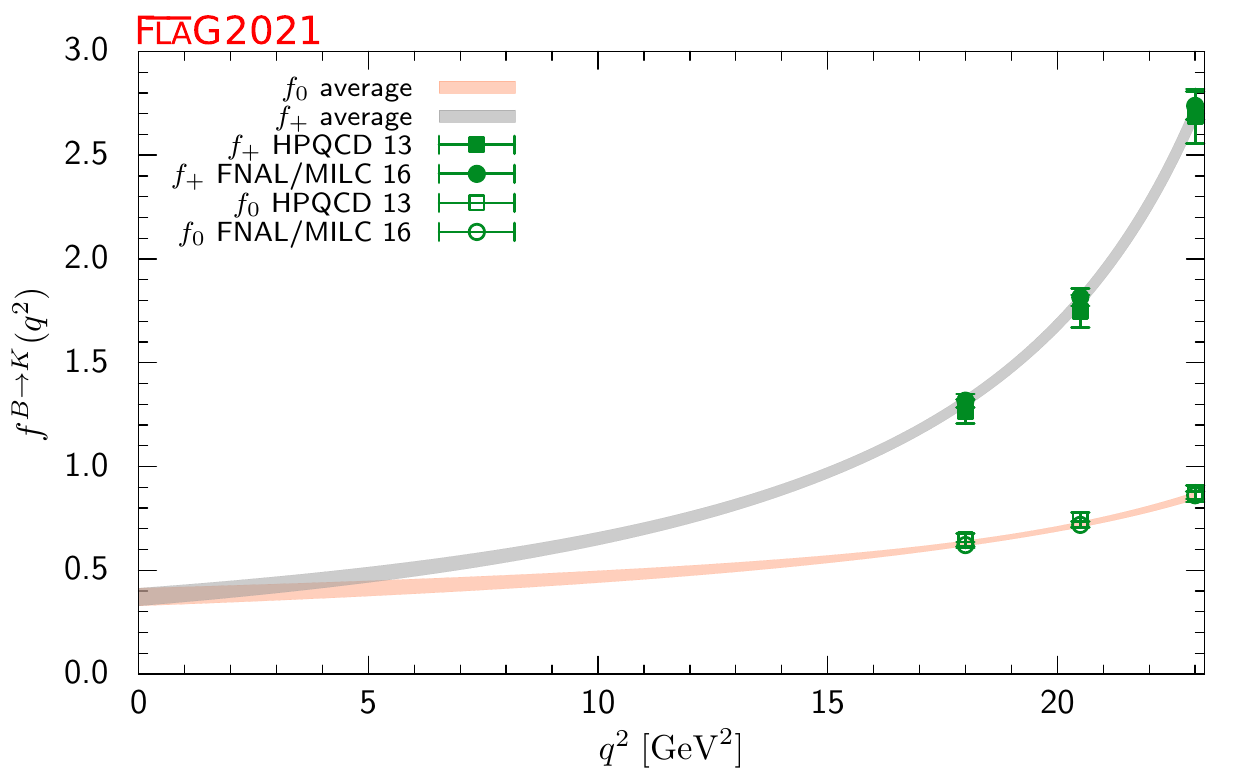}
\includegraphics[width=0.49\textwidth]{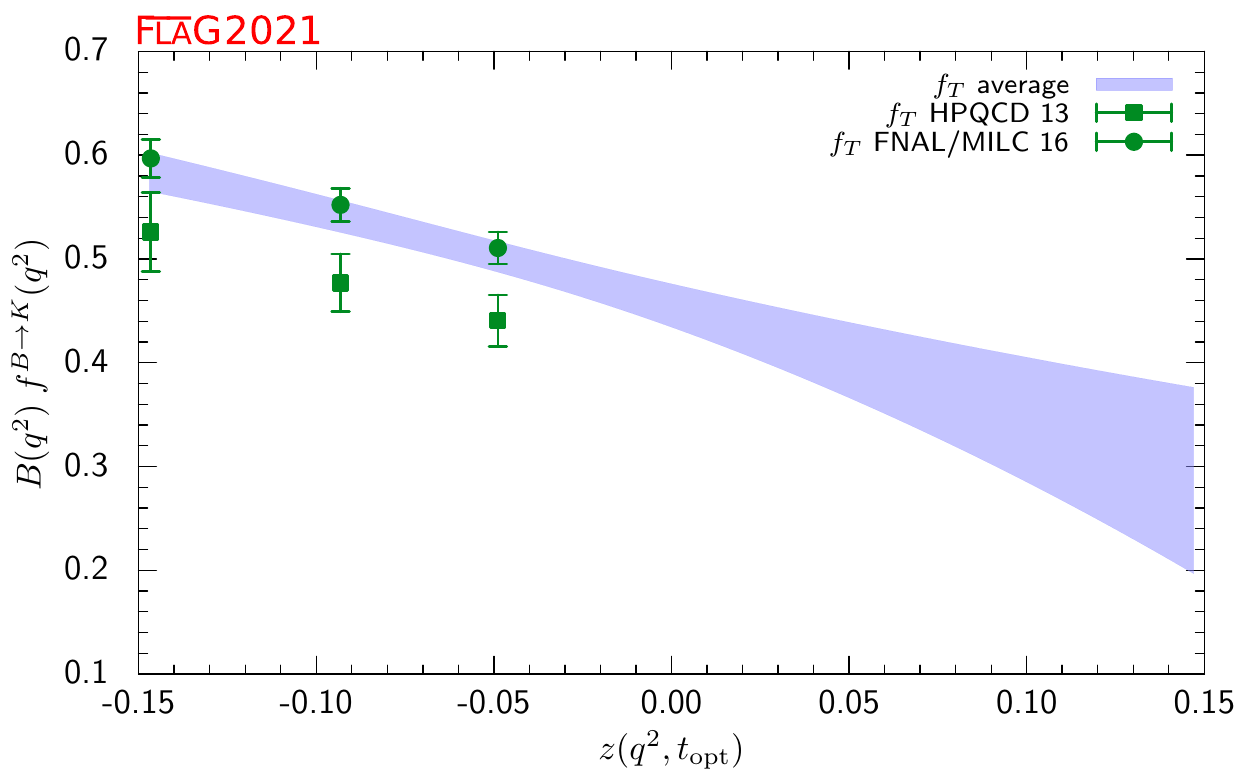}
\includegraphics[width=0.49\textwidth]{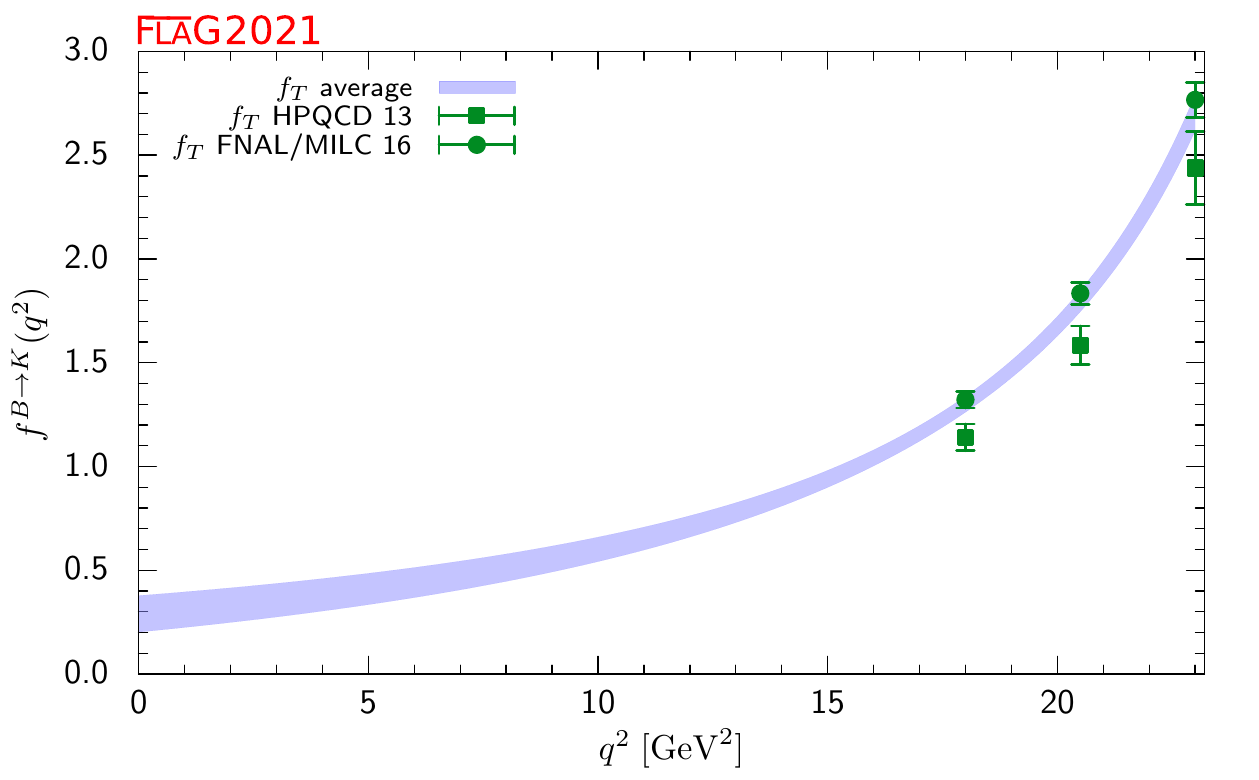}
\caption{The $B\to K$ form factors $f_+(q^2)$, $f_0(q^2)$ and $f_T(q^2)$
  plotted versus $z$ (left panels) and $q^2$ (right panels). In the plots as a function of $z$, we remove the Blaschke factors. See text for a discussion of the data sets. The grey, salmon and blue bands display our preferred $N^+=N^0=N^T=3$ BCL fit (eight parameters).}
\label{fig:LQCDzfitBK}
\end{center}
\end{figure}

Lattice computations of form factors in channels with a vector meson
in the final state face extra challenges with respect to the case of a
pseudoscalar meson: the state is unstable, and the extraction of the
relevant matrix element from correlation functions is significantly
more complicated; $\chi$PT cannot be used as a guide to extrapolate
results at unphysically heavy pion masses to the chiral limit. While
field-theory procedures to take resonance effects into account are
available~\cite{Luscher:1986pf,Luscher:1990ux,Luscher:1991cf,Lage:2009zv,Bernard:2010fp,Doring:2011vk,Hansen:2012tf,Briceno:2012yi,Dudek:2014qha,Briceno:2014uqa,Briceno:2015csa},
they have not yet been implemented in the existing preliminary
computations, which therefore suffer from uncontrolled systematic
errors in calculations of weak decay form factors into unstable vector
meson final states, such as the $K^*$ or $\rho$ mesons.\footnote{In
  cases such as $B\to D^*$ transitions, that will be discussed below,
  this is much less of a practical problem due to the very narrow
  nature of the resonance.}

As a consequence of the complexity of the problem, the level of maturity
of these computations is significantly below the one present for pseudoscalar form factors.
Therefore, we only provide a short guide to the existing results.
Horgan {\it et al.} have obtained the seven form factors governing $B \to K^* \ell^+
\ell^-$ (as well as those for $B_s \to \phi\, \ell^+ \ell^-$ and for the charged-current decay $B_s \to K^* \ell\nu$) in
Ref.~\cite{Horgan:2013hoa} using NRQCD $b$ quarks and asqtad staggered
light quarks.  In this work, they use a modified $z$-expansion to
simultaneously extrapolate to the physical light-quark masses and
fit the $q^2$-dependence.  As discussed above, the unstable nature of the vector mesons was not taken
into account. Horgan {\it et al.} use
their form-factor results to calculate the differential branching
fractions and angular distributions and discuss the implications for
phenomenology in a companion paper~\cite{Horgan:2013pva}. An update
of the form factor fits that enforces endpoint relations and also provides
the full correlation matrices can be found in Ref.~\cite{Horgan:2015vla}.
Finally, preliminary results on $B\to K^*\ell^+\ell^-$ and $B_s\to \phi\ell^+\ell^-$
by RBC/UKQCD have been reported in Refs.~\cite{Flynn:2015ynk,Flynn:2016vej,Lizarazo:2016myv}.

\subsection{Semileptonic form factors for $B_{(s)} \to D_{(s)} \ell \nu$ and $B_{(s)} \to D^*_{(s)}  \ell \nu$}
\label{sec:BtoD}

The semileptonic processes $ B_{(s)} \rightarrow D_{(s)} \ell \nu$ and
$B_{(s)} \rightarrow D^*_{(s)} \ell \nu$ have been studied
extensively by experimentalists and theorists over the years.  They
allow for the determination of the CKM matrix element $|V_{cb}|$, an
extremely important parameter of the Standard Model. The matrix
element $V_{cb}$
appears in many quantities that serve as inputs to CKM unitarity
triangle analyses and reducing its uncertainties is of paramount
importance.  For example, when $\epsilon_K$, the measure of indirect
CP violation in the neutral kaon system, is written in terms of the
parameters $\rho$ and $\eta$ that specify the apex of the unitarity
triangle, a factor of $|V_{cb}|^4$ multiplies the dominant term.  As a
result, the errors coming from $|V_{cb}|$ (and not those from $B_K$)
are now the dominant uncertainty in the Standard Model (SM) prediction
for this quantity.

The decay rate for $B \rightarrow D\ell\nu$ can be parameterized in terms of
vector and scalar form factors in the same way as, e.g., $B\to\pi\ell\nu$ (see Sec.~\ref{sec:BtoPiK}).  The decay rate for $B \rightarrow D^*\ell\nu$
is different because the final-state hadron is spin-1.
There are four form factors used to describe the vector and axial-vector
current matrix elements that are needed to calculate this decay.
We define the 4-velocity of the meson $P$ as $v_P= p_P/m_P$
and the polarization
vector of the $D^*$ as $\epsilon$.  When the light lepton $\ell=e$, or $\mu$,
it is traditional to use
$ w=v_B\cdot v_{D^{(*)}}$ rather than $q^2$ as the variable upon which
the form factors depend.
Then, the form factors $h_V$ and $h_{A_i}$, with $i=1$, 2 or 3 are defined by
\begin{align}
  \langle D^* | V_\mu | B \rangle  &= \sqrt{m_B m_{D^*}} h_V(w)
   \varepsilon_{\mu\nu\alpha\beta} \epsilon^{*\nu} v_{D^*}^\alpha v_B^\beta \,, \\
  \langle D^* | A_\mu | B \rangle  &= i \sqrt{m_B m_{D^*}} \left[
    h_{A_1}(w) (1+w) \epsilon^{*\mu} -
    h_{A_2}(w) \epsilon^*\cdot v_B {v_B}_\mu -
    h_{A_3}(w) \epsilon^*\cdot v_B {v_{D^*}}_\mu \right ].
\label{eq:BtoDstarAxialFormFactor}
\end{align}
The differential decay rates can then be written as\footnote{These are the only meson decay channels dealt with in this
review where we apply the Sirlin correction factor $\eta_{EW}$, that incorporates leading-order, structure-independent
corrections. This is in keeping with common practice.  While
including $\eta_{EW}$ in the analysis of $b \to c$ transitions is nearly universal in the literature, this is not
so in other flavour-changing decays. It is worth stressing that this is just part of the expected corrections ---cf.~the 
discussion of QED corrections in the sections of this review dealing with light meson decay--- and therefore
its inclusion is largely arbitrary, insofar as a precise control of the full corrections, including the structure-dependent
ones, is unavailable for a given channel. It is also necessary to remark, on the other hand, that different practices
contribute to a small ambiguity in the comparison of CKM matrix elements determined from different decays, precisely
of the order of the typically neglected electromagnetic corrections.}
\begin{align}
    \frac{d\Gamma_{B^-\to D^{0} \ell^-\bar{\nu}}}{dw} = &
        \frac{G^2_{\rm F} m^3_{D}}{48\pi^3}(m_B+m_{D})^2(w^2-1)^{3/2}  |\eta_\mathrm{EW}|^2|V_{cb}|^2 |\mathcal{G}(w)|^2,
    \label{eq:vxb:BtoD} \\
    \frac{d\Gamma_{B^-\to D^{0*}\ell^-\bar{\nu}}}{dw} = &
        \frac{G^2_{\rm F} m^3_{D^*}}{4\pi^3}(m_B-m_{D^*})^2(w^2-1)^{1/2}  |\eta_\mathrm{EW}|^2|V_{cb}|^2\chi(w)|\mathcal{F}(w)|^2 ,
    \label{eq:vxb:BtoDstar}
\end{align}
where $w = v_B \cdot v_{D^{(*)}}$ (depending on whether
the final-state meson is $D$ or $D^*$)
and $\eta_\mathrm{EW}=1.0066$
 is the 1-loop electroweak correction~\cite{Sirlin:1981ie}. The
 function $\chi(w)$ in Eq.~(\ref{eq:vxb:BtoDstar}) depends on the
 recoil $w$ and the meson masses, and reduces to unity at zero
 recoil~\cite{Antonelli:2009ws}.\footnote{The reason to keep the factor
 $\chi(w)$ outside the combination of form factors that defines $\mathcal{F}(w)$
 is conventional, and inspired by the heavy-quark limit. One particular consequence
 of this notation is that at zero recoil $\mathcal{F}(1)=h_{A_1}(1)$.}
 These formulas do not include terms
 that are proportional to the lepton mass squared, which can be
 neglected for $\ell = e, \mu$.  Further details of the definitions of
${\cal F}$ and ${\cal G}$ (which can be expressed in terms of the form factors $h_V$ and $h_{A_i}$) may be found, e.g., in Ref.~\cite{Antonelli:2009ws}.
Until recently, most unquenched lattice calculations for $B \rightarrow D^* \ell \nu$ and
$B \rightarrow D \ell \nu$ decays focused on the form
factors at zero recoil ${\cal F}^{B \rightarrow D^*}(1)$ and ${\cal G}^{B \rightarrow D}(1)$;
these can then be combined with experimental input to extract $|V_{cb}|$.
The main reasons for concentrating on the zero recoil point are that
(i) the decay rate then depends on a single form factor, and (ii) for
$B \rightarrow D^*\ell\nu$, there are no $\cO(\Lambda_{QCD}/m_Q)$
contributions due to Luke's theorem~\cite{Luke:1990eg}. Further, the zero recoil form
factor can be computed via a double ratio in which most of the current
renormalization cancels and heavy-quark discretization errors are
suppressed by an additional power of $\Lambda_{QCD}/m_Q$.
Recent work on $B \rightarrow D^{(*)}\ell\nu$ transitions
has started to explore the dependence of the relevant form factors on the
momentum transfer, using a similar methodology to the one employed
in $B\to\pi\ell\nu$ transitions; see Sec.~\ref{sec:BtoPiK}
for a detailed discussion.

Early computations of the form factors for $B \rightarrow D\ell\nu$ decays include $N_f=2+1$ results by FNAL/MILC~\cite{Okamoto:2004xg,Qiu:2013ofa}
for ${\cal G}^{B \rightarrow D}(1)$ and the
$N_f=2$ study
by Atoui {\it et al.}~\cite{Atoui:2013zza}, that in addition to providing
${\cal G}^{B \rightarrow D}(1)$ explored the $w>1$ region.
This latter work also
provided the first results for $B_s \rightarrow D_s\ell\nu$
amplitudes, again including information about the momentum-transfer dependence.
The first published unquenched results for ${\cal F}^{B \rightarrow D^*}(1)$,
obtained by FNAL/MILC, date from 2008~\cite{Bernard:2008dn}.
In 2014 and 2015, significant progress was achieved in $N_f=2+1$ computations:
the FNAL/MILC value for ${\cal F}^{B \rightarrow D^*}(1)$ was updated
in Ref.~\cite{Bailey:2014tva}, and full results for $B \rightarrow D\ell\nu$
at $w \geq 1$ were published by FNAL/MILC~\cite{Lattice:2015rga} and HPQCD~\cite{Na:2015kha}.
These works also provided full results for the scalar form factor, allowing
analysis of the decay with a final-state $\tau$.
In the FLAG 19 review \cite{Aoki:2019cca}, we included new results for
$B_s \rightarrow D_s\ell\nu$ form factors over the full kinematic
range for $N_f=2+1$ from HPQCD~\cite{Monahan:2016qxu,Monahan:2017uby}, and for
${{B}_{(s)}\to D_{(s)}^{*}\ell{\nu}}$ form factors at zero recoil with
$N_f=2+1+1$ also from HPQCD~\cite{Harrison:2016gup,Harrison:2017fmw}.
Most recently, HPQCD published further new calculations of the $B_s \to D_s^*$ form factor
at zero recoil \cite{McLean:2019sds} and of the $B_s \to D_s$ form factors in the full kinematic range \cite{McLean:2019qcx},
now using MILC's HISQ $N_f=2+1+1$ ensembles and using the HISQ action also for the $b$ quark. Both of these
calculations have recently been used by LHCb to determine $|V_{cb}|$ \cite{LHCb:2020cyw,LHCb:2021qbv}, as discussed further in Sec.~\ref{sec:Vcb}.
Improved calculations of the $B \to D$ and  $B_s \to D_s$ form factors are also underway by RBC/UKQCD \cite{Flynn:2019jbg},
and the Fermilab/MILC computation of the $B\to D^*$ form factors at nonzero recoil is nearing completion \cite{Vaquero:2019ary}.
The JLQCD collaboration also presented preliminary results for the $B\to D$ and  $B\to D^*$ form factors, both
at nonzero recoil \cite{Kaneko:2019vkx}.

In the discussion below, we mainly concentrate on the
latest generation
of results, which supersedes previous $N_f=2+1$ determinations and allows
for an extraction of $|V_{cb}|$ that incorporates information about the $q^2$-dependence
of the decay rate (cf.~Sec.~\ref{sec:Vcb}).

\subsubsection{ $B_{(s)} \rightarrow D_{(s)}$ decays}
\label{sec:BstoDsFFs}

We will first discuss the $N_f=2+1$ computations of $B \rightarrow D \ell \nu$
by FNAL/MILC and HPQCD mentioned above, both based on MILC asqtad ensembles.
Full details about all the computations are provided in Tab.~\ref{tab_BtoDStarsumm2}
and in the tables in Appendix~\ref{app:BtoD_Notes}.

The FNAL/MILC study~\cite{Lattice:2015rga} employs ensembles at four values of the lattice
spacing ranging between approximately $0.045~{\rm fm}$
and $0.12~{\rm fm}$, and several values of the light-quark mass corresponding to pions
with RMS masses ranging between $260~{\rm MeV}$ and $670~{\rm MeV}$ (with just
one ensemble with $M_\pi^{\rm RMS} \simeq 330~{\rm MeV}$ at the finest lattice spacing).
The $b$ and $c$ quarks are treated using the Fermilab approach.
The quantities directly studied are the form factors $h_\pm$
defined by
\begin{equation}
\frac{\langle D(p_D)| i\bar c \gamma_\mu b| B(p_B)\rangle}{\sqrt{m_D m_B}} =
h_+(w)(v_B+v_D)_\mu\,+\,h_-(w)(v_B-v_D)_\mu\,,
\end{equation}
which are related to the standard vector and scalar form factors by
\begin{align}
  f_+(q^2) &= \frac{1}{2\sqrt{r}}
             \left[(1+r)h_+(w)-(1-r)h_-(w)\right],
  \\
  f_0(q^2) &= \sqrt{r}
             \left[\frac{1+w}{1+r} h_+(w) + \frac{1-w}{1-r}h_-(w) \right],
\end{align}
with $r=m_D/m_B$. (Recall that
$q^2=(p_B-p_D)^2=m_B^2+m_D^2-2 w m_B m_D$.)  The hadronic form factor
relevant for experiment, $\mathcal{G}(w)$, is then obtained from the
relation $\mathcal{G}(w)=\sqrt{4r}f_+(q^2)/(1+r)$. The form factors are
obtained from double ratios of three-point functions in which the
flavour-conserving current renormalization factors cancel. The
remaining matching factor to the flavour-changing normalized current is estimated with
1-loop lattice perturbation theory.
In order to obtain $h_\pm(w)$, a joint continuum-chiral fit is performed
to an ansatz that
contains the light-quark mass and lattice-spacing dependence predicted
by next-to-leading order HMrS$\chi$PT,
and the leading dependence on $m_c$
predicted by the heavy-quark expansion ($1/m_c^2$ for $h_+$ and
$1/m_c$ for $h_-$). The $w$-dependence, which allows for an
interpolation in $w$, is given by analytic terms up to $(1-w)^2$, as
well as a contribution from the logarithm proportional to $g^2_{D^*D\pi}$.
The total resulting systematic error, determined as a function of $w$ and
quoted at the representative point $w=1.16$ as $1.2\%$ for $f_+$ and $1.1\%$ for $f_0$,
dominates the final error budget for the form factors.
After $f_+$ and $f_0$ have been determined as functions of $w$ within the interval
of values of $q^2$ covered by the computation, synthetic data points are
generated to be subsequently fitted to a $z$-expansion of the BGL form, cf.~Sec.~\ref{sec:BtoPiK},
with pole factors set to unity.
This in turn enables one to determine $|V_{cb}|$ from a joint fit of this $z$-expansion
and experimental data. The value of the zero-recoil form factor resulting
from the $z$-expansion is
\begin{equation}
{\cal G}^{B \rightarrow D}(1)= 1.054(4)_{\rm stat}(8)_{\rm sys}\,.
\end{equation}

The HPQCD computations~\cite{Na:2015kha,Monahan:2017uby} use ensembles at two values of the lattice
spacing, $a=0.09,~0.12~{\rm fm}$, and two and three values of light-quark masses, respectively.
The $b$ quark is treated using NRQCD, while for the $c$ quark the HISQ action is used.
The form factors studied, extracted from suitable three-point functions, are
\begin{equation}
\langle D_{(s)}(p_{D_{(s)}})| V^0 | B_{(s)}\rangle = \sqrt{2M_{B_{(s)}}}f^{(s)}_\parallel\,,~~~~~~~~
\langle D_{(s)}(p_{D_{(s)}})| V^k | B_{(s)}\rangle = \sqrt{2M_{B_{(s)}}}p^k_{D_{(s)}} f^{(s)}_\perp\,,
\end{equation}
where $V_\mu$ is the relevant vector current and the $B_{(s)}$ rest frame is chosen.
The standard vector and scalar form factors are retrieved as
\begin{align}
  f^{(s)}_+ =& \frac{1}{\sqrt{2M_{B_{(s)}}}}
               \left[ f^{(s)}_\parallel +
               (M_{B_{(s)}}-E_{D_{(s)}})f^{(s)}_\perp \right],
  \\
  f^{(s)}_0 =& \frac{\sqrt{2M_{B_{(s)}}}}{M_{B_{(s)}}^2-M_{D_{(s)}}^2}
               \left[(M_{B_{(s)}}-E_{D_{(s)}})f^{(s)}_\parallel
               +(M_{B_{(s)}}^2-E_{D_{(s)}}^2)f^{(s)}_\perp\right].
\end{align}
The currents in the effective theory are matched at 1-loop to their continuum
counterparts. Results for the form factors are then fitted to a modified BCL $z$-expansion
ansatz, that takes into account simultaneously the lattice spacing, light-quark masses,
and $q^2$-dependence. For the mass dependence, NLO chiral logarithms are included, in the
form obtained in hard-pion $\chi$PT (see footnote~\ref{footnote:hardpion}). As in the case of the FNAL/MILC computation,
once $f_+$ and $f_0$ have been determined as functions of $q^2$, $|V_{cb}|$ can
be determined from a joint fit of this $z$-expansion and experimental data.
The papers quote for the zero-recoil vector form factor the result
\begin{equation}
{\cal G}^{B \rightarrow D}(1)=1.035(40)\,~~~~{\cal G}^{B_s \rightarrow D_s}(1)=1.068(40)\,.
\end{equation}
The HPQCD and FNAL/MILC results for $B\to D$ differ by less than half a standard
deviation (assuming they are uncorrelated, which they are not as some of
the ensembles are common) primarily because of lower precision of the former
result.
The HPQCD central value is smaller by 1.8 of the FNAL/MILC standard deviations than the FNAL/MILC value.
The dominant source of errors in the $|V_{cb}|$ determination by HPQCD are discretization
effects and the systematic uncertainty associated with the perturbative matching.

In order to combine the form factor determinations of HPQCD and FNAL/MILC
into a lattice average, we proceed in a similar way as with $B\to\pi\ell\nu$
and $B_s\to K\ell\nu$ above. FNAL/MILC quotes synthetic values for each
form factor at three values of $w$ (or, alternatively, $q^2$) with a full
correlation matrix, which we take directly as input. In the case of HPQCD,
we use their preferred modified $z$-expansion parameterization to produce
synthetic values of the form factors at five different values of $q^2$ (three for $f_+$ and two for $f_0$).
This leaves us with a total of six (five) data points in the kinematical
range $w\in[1.00,1.11]$ for the form factor $f_+$ ($f_0$). As in the case of $B\to\pi\ell\nu$, we conservatively
assume a 100\% correlation of statistical uncertainties between HPQCD
and FNAL/MILC. We then fit this data set to a BCL ansatz, using
$t_+=(M_{B^0}+M_{D^\pm})^2 \simeq 51.12~\GeV^2$ and
$t_0=(M_{B^0}+M_{D^\pm})(\sqrt{M_{B^0}}-\sqrt{M_{D^\pm}})^2 \simeq 6.19~\GeV^2$.
In our fits, pole factors have been set to unity, i.e., we do not
take into account the effect of sub-threshold poles, which is then
implicitly absorbed into the series coefficients. The reason for this
is our imperfect knowledge of the relevant resonance spectrum in this channel,
which does not allow us to decide the precise number of poles needed.\footnote{As noted
above, this is the same approach adopted by FNAL/MILC in their fits to a BGL
ansatz. HPQCD, meanwhile, uses one single pole in the pole factors that
enter their modified $z$-expansion, using their spectral studies to fix
the value of the relevant resonance masses.}
This, in turn, implies that unitarity bounds do not rigorously apply,
which has to be taken into account when interpreting the results
(cf.~Appendix \ref{sec:zparam}).

With a procedure similar to what we adopted for the $B\to \pi$ and
$B_s\to K$ cases, we impose the kinematic constraint at $q^2=0$ by
expressing the $a^0_{N^0-1}$ coefficient in the $z$-expansion of $f_0$
in terms of all the other coefficients. As mentioned above, FNAL/MILC
provides synthetic data for $f_+$ and $f_0$ including correlations;
HPQCD presents the result of simultaneous $z$-fits to the two form
factors including all correlations, thus enabling us to generate a
complete set of synthetic data for $f_+$ and $f_0$. Since both
calculations are based on MILC ensembles, we then reconstruct the
off-diagonal HPQCD-FNAL/MILC entries of the covariance matrix by
conservatively assuming that statistical uncertainties are 100\%
correlated. The Fermilab/MILC (HPQCD) statistical error is 58\% (31\%)
of the total error for every $f_+$ value, and 64\% (49\%) for every
$f_0$ one. Using this information we can easily build the off-diagonal
block of the overall covariance matrix (e.g., the covariance between
$[f_+(q_1^2)]_{\rm FNAL}$ and $[f_0(q_2^2)]_{\rm HPQCD}$ is $(\delta
[f_+(q_1^2)]_{\rm FNAL} \times 0.58)\; (\delta [f_0(q_2^2)]_{\rm
  HPQCD} \times 0.49)$, where $\delta f$ is the total error). 

For our central value, we choose an $N^+ =N^0=3$ BCL fit, shown in Tab.~\ref{tab:FFD}. The coefficient $a_3^+$ can be obtained from the values for $a_0^+$--$a_2^+$ using Eq.~(\ref{eq:red_coeff}).  We find $\chi^2/{\rm dof} = 4.6/6 = 0.77$. The fit, which is dominated by the FNAL/MILC calculation, is illustrated in Fig.~\ref{fig:LQCDzfitBD}.

\begin{table}[t]
\begin{center}
\begin{tabular}{|c|c|ccccc|}
\multicolumn{7}{l}{$B\to D \; (N_f=2+1)$} \\[0.2em]\hline
$a_n^i$ & Central Values & \multicolumn{5}{|c|}{Correlation Matrix} \\[0.2em]\hline
$a_0^+$ & 0.896 (10) & 1 & 0.423 & -0.231 & 0.958 & 0.596 \\[0.2em]
$a_1^+$ & -7.94 (20) & 0.423 & 1 & 0.325 & 0.498 & 0.919 \\[0.2em]
$a_2^+$ & 51.4 (3.2) & -0.231 & 0.325 & 1 & -0.146 & 0.317 \\[0.2em]
$a_0^0$ & 0.7821 (81) & 0.958 & 0.498 & -0.146 & 1 & 0.593 \\[0.2em]
$a_1^0$ & -3.28 (20) & 0.596 & 0.919 & 0.317 & 0.593 & 1 \\[0.2em]
\hline
\end{tabular}
\end{center}
\caption{Coefficients and correlation matrix for the $N^+ =N^0=3$ $z$-expansion of the $B\to D$ form factors $f_+$ and $f_0$. The chi-square per degree of freedom is $\chi^2/{\rm dof} = 4.6/6=0.77$. The lattice calculations that enter this fit are taken from FNAL/MILC~\cite{Lattice:2015rga} and HPQCD~\cite{Na:2015kha}. \label{tab:FFD}}
\end{table}

\begin{figure}[tbp]
\begin{center}
\includegraphics[width=0.49\textwidth]{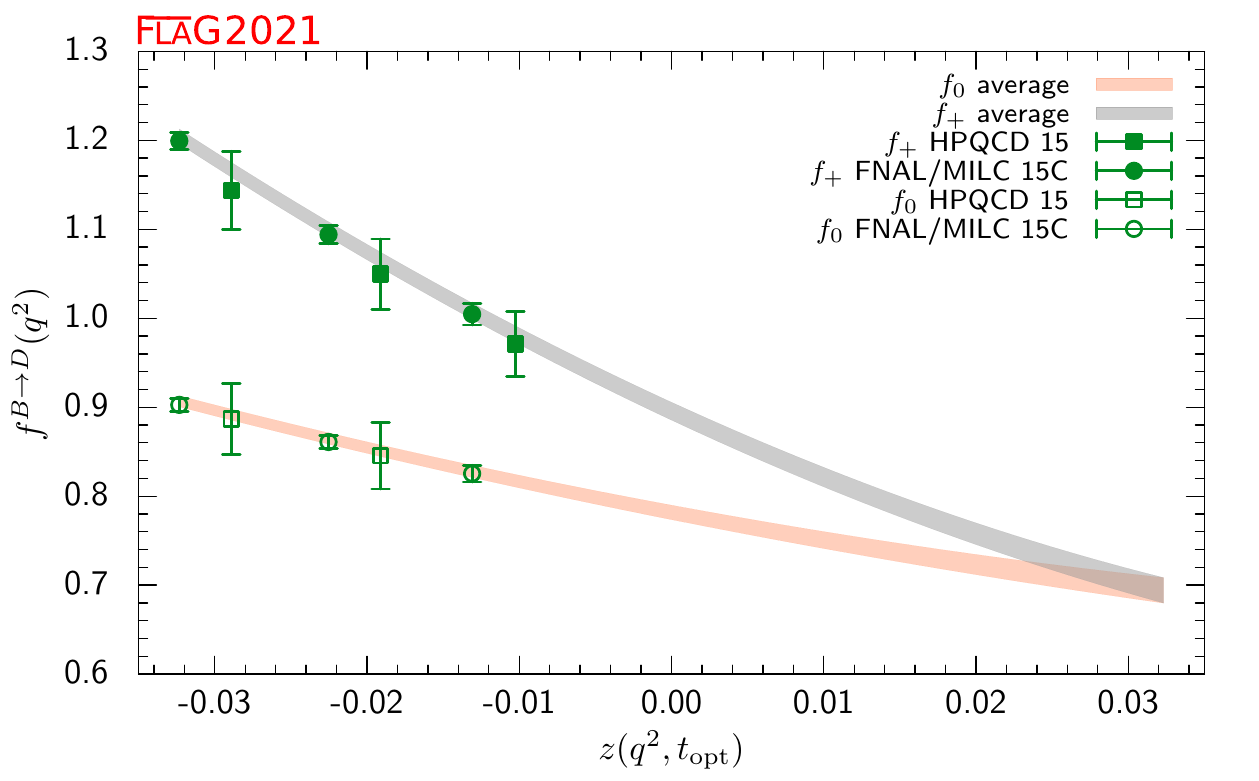}
\includegraphics[width=0.49\textwidth]{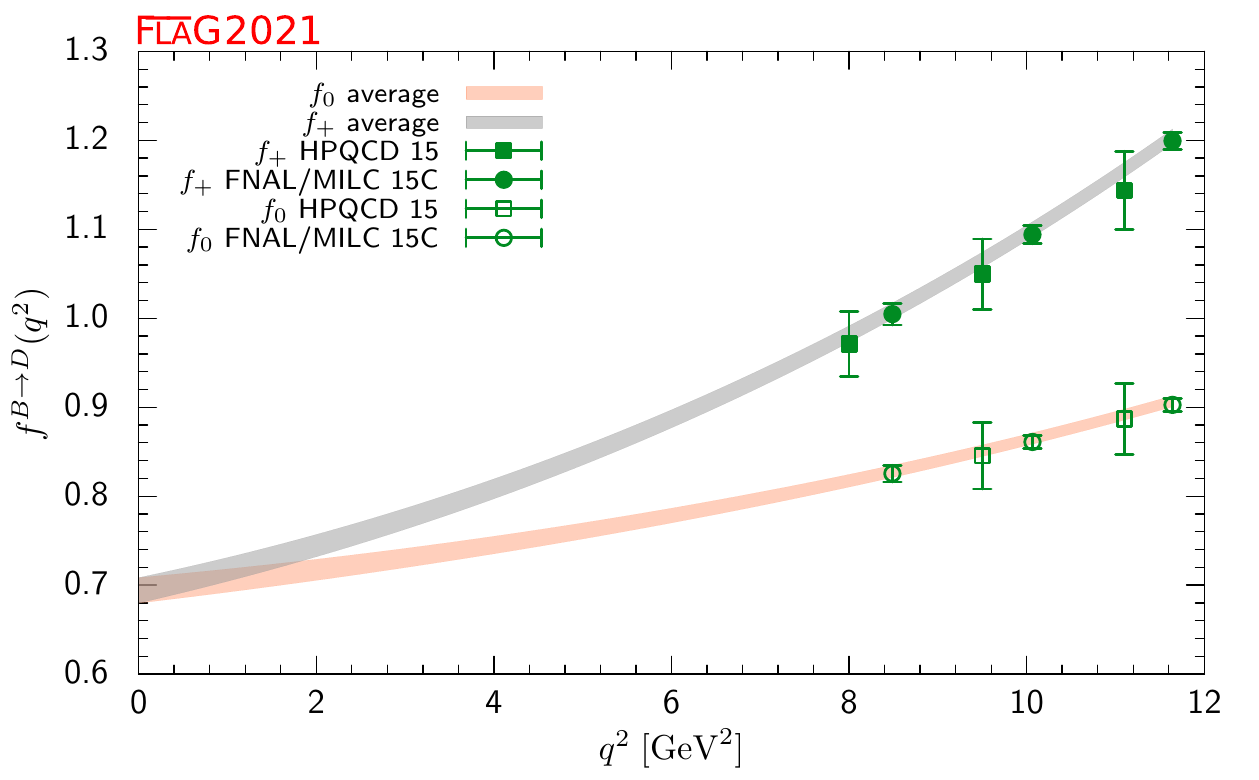}
\caption{The form factors $f_+(q^2)$ and $f_0(q^2)$ for $B \to D\ell\nu$ plotted versus $z$ (left panel) and $q^2$ (right panel). See text for a discussion of the data sets. The grey and salmon bands display our preferred $N^+=N^0=3$ BCL fit (five parameters).}
\label{fig:LQCDzfitBD}
\end{center}
\end{figure}

Reference~\cite{Atoui:2013zza} is the only existing $N_f=2$ work on $B \rightarrow D\ell\nu$
transitions, that furthermore provided the first
available results for $B_s \rightarrow D_s\ell\nu$.
This computation uses the publicly available ETM configurations
obtained with the twisted-mass QCD action at maximal twist.  Four
values of the lattice spacing, ranging between $0.054~{\rm fm}$ and
$0.098~{\rm fm}$, are considered, with physical box lengths ranging
between $1.7~{\rm fm}$ and $2.7~{\rm fm}$.  At two values of the
lattice spacing two different physical volumes are available.
Charged-pion masses range between $\approx 270~{\rm MeV}$ and $\approx
490~{\rm MeV}$, with two or three masses available per lattice spacing
and volume, save for the $a \approx 0.054~{\rm fm}$ point at which
only one light mass is available for each of the two volumes. The
strange- and heavy-valence quarks are also treated with maximally
twisted-mass QCD.

The quantities of interest are again the form factors $h_\pm$ defined above.
In order to control discretization effects from the heavy quarks, a strategy
similar to the one employed by the ETM collaboration in their studies of
$B$-meson decay constants (cf.~Sec.~\ref{sec:fB}) is employed: the value of
${\cal G}(w)$ is computed at a fixed value of $m_c$ and several values of
a heavier quark mass $m_h^{(k)}=\lambda^k m_c$, where $\lambda$ is a fixed
scaling parameter, and step-scaling functions are built as
\begin{equation}
\Sigma_k(w) = \frac{{\cal G}(w,\lambda^{k+1} m_c,m_c,a^2)}{{\cal G}(w,\lambda^k m_c,m_c,a^2)}\,.
\end{equation}
Each ratio is extrapolated to the continuum limit,
$\sigma_k(w)=\lim_{a \to 0}\Sigma_k(w)$.  One then exploits the fact
that the $m_h \to \infty$ limit of the step-scaling is fixed.  In
particular, it is easy to find from the heavy-quark expansion that
$\lim_{m_h\to\infty}\sigma(1)=1$. In this way, the physical result at
the $b$-quark mass can be reached by interpolating $\sigma(w)$ between
the charm region (where the computation can be carried out with
controlled systematics) and the known static limit value.

In practice, the values of $m_c$ and $m_s$ are fixed at each value of
the lattice spacing such that the experimental kaon and $D_s$ masses
are reached at the physical point, as determined
in Ref.~\cite{Blossier:2010cr}.  For the scaling parameter, $\lambda=1.176$
is chosen, and eight scaling steps are performed, reaching
$m_h/m_c=1.176^9\simeq 4.30$, approximately corresponding to the ratio
of the physical $b$- and $c$-masses in the $\overline{\rm MS}$ scheme
at $2~{\rm GeV}$.  All observables are obtained from ratios that do
not require (re)normalization.  The ansatz for the continuum and
chiral extrapolation of $\Sigma_k$ contains a constant and linear
terms in $m_{\rm sea}$ and $a^2$.  Twisted boundary conditions in
space are used for valence-quark fields for better momentum
resolution.  Applying this strategy, the form factors are finally
obtained at four reference values of $w$ between $1.004$ and $1.062$,
and, after a slight extrapolation to $w=1$, the result is
\begin{equation}
{\cal G}^{B_s \rightarrow D_s}(1) = 1.052(46)\,.
\end{equation}

The authors also provide values for the form factor relevant for the
meson states with light-valence quarks, obtained from a similar
analysis to the one described above for the $B_s\rightarrow D_s$ case.
Values are quoted from fits with and without a linear $m_{\rm
  sea}/m_s$ term in the chiral extrapolation. The result in the former
case, which safely covers systematic uncertainties, is
\begin{equation}
{\cal G}^{B \rightarrow D}(1)=1.033(95)\,.
\label{eq:avBDnf2}
\end{equation}
Given the identical strategy, and the small sensitivity of the ratios
used in their method to the light valence- and sea-quark masses, we
assign this result the same ratings in Tab.~\ref{tab_BtoDStarsumm2}
as those for their calculation of ${\cal G}^{B_s \rightarrow D_s}(1)$.
Currently, the precision of this calculation is not competitive with
that of $N_f=2+1$ works, but this is due largely to the small number of
configurations analyzed by Atoui {\it et al.}  The viability of their method
has been clearly demonstrated, however, which leaves significant room
for improvement on the errors of both the $B \to D$ and $B_s \to D_s$
form factors with this approach by including either additional
two-flavour data or analysing more recent ensembles with $N_f>2$.

Atoui {\it et al.} also study the scalar and tensor form factors, as well as the
momentum-transfer dependence of $f_{+,0}$. The value of the ratio $f_0(q^2)/f_+(q^2)$
is provided at a reference value of $q^2$ as a proxy for the slope of ${\cal G}(w)$
around the zero-recoil limit.

Let us finally discuss the most recent results for $B_s \to D_s$ form factors,
obtained by the HPQCD collaboration using MILC's $N_f=2+1+1$ ensembles in
Ref.~\cite{McLean:2019qcx}. Three values of the lattice spacing are
used, including a very fine ensemble at $a \simeq 0.044~{\rm fm}$;
the pion mass is kept fixed at around 300~{\rm MeV}, and in addition at the
coarser $a \simeq 0.09~{\rm fm}$ lattice an ensemble with the physical pion
mass is included. The scalar current needs no renormalization because of the PCVC relation,
while the vector current is nonperturbatively normalized by imposing a condition
based on the PCVC relation at zero recoil.
Heavy quarks are treated in a fully relativistic fashion
through the use of the HISQ regularization, employing bare values of the quark
mass up to $am_h=0.8$ for the extrapolation to the physical $b$ point.

Results for the form factors are fitted to a modified $z$-expansion ansatz,
based on a BCL ansatz with a Blaschke factor containing one sub-threshold pole, tuned
to reproduce the lattice-spacing and heavy-quark-mass-dependent mass of the corresponding resonance.
The final error budget is equally dominated by statistics and the combined effect
of the continuum and heavy quark mass extrapolations, which correspond to 1.1\% and 1.2\%
uncertainties, respectively, for the scalar form factor at zero recoil. The total
uncertainty of the latter is thus below 2\%, which remains true in the whole $q^2$
range. The uncertainty of $f_+$ is somewhat larger, starting at around 2\% at $q^2=0$
and increasing up to around 3.5\% at zero recoil.

One important matter of concern with this computation is the use of the $a \simeq 0.044~{\rm fm}$
ensemble with periodic boundary conditions, which suffers from severe topology freezing.
Other than possible implications for statistical uncertainties, the lack of topology
fluctuations are expected to significantly enhance finite-volume effects, which are no
longer exponential in $m_\pi L$, but become power-like in the spatial volume.
The authors neglect the impact of finite-volume effects in the computation, with a
twofold argument: for the two coarser lattice spacings, the impact of pion-mass-related
corrections on the heavy-meson states involved is presumably negligible; and, for the finest
ensemble, the estimate of finite-volume effects on the $D_s$ decay constant obtained
in Ref.~\cite{Bernard:2017npd} turns out to be very small, a result which is presumed to extend
to form factors. It is however unclear whether the latter argument would really hold,
since the computation in Ref.~\cite{Bernard:2017npd} does show that the expected effect
is heavily observable-dependent, reaching, e.g., more than 1\% for $f_D$.
We have, therefore, concluded that our standard criteria for finite-volume effects
cannot be applied at the finest lattice spacing, and opted to assign $\soso$ rating
to them.

We thus proceed to quote the final result of HPQCD 19 as the FLAG estimate for the $N_f=2+1+1$
$B_s \to D_s$ form factors. The preferred fit is a constrained BCL form with the imposition
of the kinematical constraint $f_+(0)=f_0(0)$, carried through $z^2$ for $f_0$ and $z^3$ for $f_+$.
Both form factors contain just one sub-threshold pole, to which the masses
$M_{B_c^*}= 6.329~\GeV$ and $M_{B_{c0}}= 6.704~\GeV$, respectively, have been assigned.
The fit parameters and covariance matrix, quoted in Table~VIII of Ref.~\cite{McLean:2019qcx},
are reproduced in Table~\ref{tab:BsDs}.

\begin{table}[t]
\begin{center}
\begin{tabular}{|c|c|cccccc|}
\multicolumn{7}{l}{$B_s\to D_s \; (N_f=2+1+1)$} \\[0.2em]\hline
$a_n^i$ & Central Values & \multicolumn{6}{|c|}{Correlation Matrix} \\[0.2em]\hline
$a_0^0$ &  0.666(12) & 1 & 0.62004 & 0.03149 & 1 & 0.03973 & 0.00122 \\[0.2em]
$a_1^0$ & -0.26(25)  & 0.62004 & 1 & 0.36842 & 0.62004 & 0.12945 & 0.00002 \\[0.2em]
$a_2^0$ & -0.1(1.8)  & 0.03149 & 0.36842 & 1 & 0.03149 & 0.22854 & -0.00168 \\[0.2em]
$a_0^+$ & -0.075(12) & 1 & 0.62004 & 0.03149 & 1 & 0.03973 & 0.00122 \\[0.2em]
$a_1^+$ & -3.24(45)  & 0.03973 & 0.12945 & 0.22854 & 0.03973 & 1 & 0.11086 \\[0.2em]
$a_2^+$ &  0.7(2.0)  & 0.00122 & 0.00002 & -0.00168 & 0.00122 & 0.11086 & 1 \\[0.2em]
\hline
\end{tabular}
\end{center}
\caption{Coefficients and correlation matrix for the $z$-expansion of the $B_s\to D_s$ form factors $f_+$ and $f_0$. \label{tab:BsDs}}
\end{table}

\subsubsection{Lepton-flavour-universality ratios $R(D)$ and $R(D_s)$}

The availability of results for the scalar form factor $f_0$
for $B\to D\ell\nu$ amplitudes allows us to study
interesting observables that involve the decay in the $\tau$ channel.
One such quantity is the ratio
\begin{equation}
  R(D) = \frac{{\cal B}(B \rightarrow D \tau \nu)}{{\cal B}(B \rightarrow D \ell \nu)}
  \;\;\;\;\mbox{with}\;\;\;\; \ell=e,\mu\,,
\end{equation}
which, in the Standard Model, depends only on the form factors and hadron and lepton masses.
Indeed, the recent availability of experimental results for $R(D)$ has made
this quantity particularly relevant in the search for possible physics beyond
the Standard Model. The most recent HFLAV average reads~\cite{Amhis:2016xyh}:
\begin{align}
R(D)_{\rm exp} &= 0.340 (27)(13) \; .
\end{align}

HPQCD provides a Standard-Model prediction for $R(D)$ using the form factors from their 2015 lattice computation \cite{Na:2015kha},
\begin{equation}
R(D)_{\rm lat} = 0.300(8)\;\;\Ref~\mbox{~\cite{Na:2015kha}}\,.
\end{equation}
The FNAL/MILC collaboration computed $R(D)$ using form factors from a combined fit of their lattice data and $B\to D\ell\nu$ experimental data:
\begin{equation}
R(D)_{\rm lat+exp} = 0.299(11)\;\;\Ref~\mbox{~\cite{Lattice:2015rga}}\,.
\end{equation}
Note that other authors have obtained even smaller uncertainties from fits including $B\to D\ell\nu$ experimental data, cf.~the value $R(D)_{\rm lat+exp}=0.299(3)$ quoted in Ref.~\cite{Bigi:2016mdz}. One important reason for this is that,
while Refs.~\cite{Na:2015kha,Lattice:2015rga} only use the 2009~BaBar data in Ref.~\cite{BaBar:2007cke}, Ref.~\cite{Bigi:2016mdz}
also incorporates the more precise Belle 2015 data in Ref.~\cite{Belle:2015pkj}.

If instead we take the lattice-only form factors from FNAL/MILC, we obtain
\begin{equation}
R(D)_{\rm lat} = 0.285(15)\;\;\Ref~\mbox{~\cite{Lattice:2015rga}}\,.
\end{equation}
(We average over electrons and muons in the denominator, which only affects the last digit.) 

Finally, using the FLAG average of the $B\to D$ form factors discussed above, we find $R(D)_{\rm lat}^{\rm FLAG} = 0.2934(38)$. The ratio $R(D)$ requires the integral of the branching ratios for $\ell=e,\mu,\tau$ over the whole phase space. Since lattice simulations are sensitive mostly to relatively large $q^2$ values, lattice-only calculations of $R(D)$ rely on the extrapolation of the form factors to low $q^2$ and are especially sensitive to the choice of parameterization. In order to estimate this source of systematics, we repeated the fit using the parameterization adopted by HPQCD in Ref.~\cite{Na:2015kha}. The main difference with respect to our default paremeterization is the inclusion of Blaschke factors for the form factors $f_+$ and $f_0$ located at $M_+ = M_{B_c^*} = 6.330(9)$ GeV and $M_0 = 6.420(9)$ GeV; additionally, the parameter $t_0$ is set to $(m_B-m_D)^2$. Using five coefficients ($a_{1,2,3}^+$ and $a_{1,2}^0$ with $a_3^0$ fixed by the $f_+(q^2=0) = f_0(q^2=0)$ condition) we find $R(D)_{\rm lat}^{\rm HPQCD} = 0.3009(38)$ which deviates from $R(D)_{\rm lat}^{\rm FLAG}$ by 1.4 $\sigma$. To take this potential source of systematic uncertainty into account we rescale accordingly the uncertainty of our default fit and obtain:
\begin{equation}
R(D)_{\rm lat} = 0.2934(53)\,,~~~~\mbox{our average.}
\label{HQeq:RDlat}
\end{equation}
This result is about 1.5$\sigma$ lower than the current experimental average \cite{Amhis:2019ckw}
for this quantity. It has to be stressed that achieving this level of precision
critically depends on the reliability with which the low-$q^2$ region is
controlled by the parameterizations of the form factors.

After including the $B\to D\ell\nu \; (\ell= e,\mu)$ data in the fit, as discussed at the end of Sec.~\ref{sec:Vcb}, we obtain the following combined lattice plus experiment result:
\begin{equation}
R(D)_{\rm lat+exp} = 0.2951(31)\,,~~~~\mbox{our average.}
\end{equation}

HPQCD also computes values for $R(D_s)$, the analog of
$R(D)$ with both heavy-light mesons containing a strange quark. The earlier calculation using NRQCD $b$ quarks gives
\begin{gather}
R(D_s)_{\rm lat} = 0.301(6)\,,~~~~~N_f=2+1\:\:\mbox{\cite{Monahan:2017uby}}.
\end{gather}
The newer calculation with HISQ $b$ quarks yields the somewhat more precise value
\begin{gather}
R(D_s)_{\rm lat} = 0.2987(46)\,,~~~~~N_f=2+1+1\:\:\mbox{\cite{McLean:2019qcx}}.
\end{gather}

A similar ratio $R(D^*)$ can be considered for $B \rightarrow D^*$ transitions. As a matter of fact, the experimental value of $R(D^*)$ is significantly more accurate than the one of $R(D)$. A preliminary lattice-QCD-only prediction of $R(D^*)$ was shown by A.~Vaquero~\cite{Vaquero:MIT}.

\subsubsection{Fragmentation fraction ratio $f_s / f_d$}

Another area of immediate interest in searches for physics
beyond the Standard Model is the measurement of $B_s \rightarrow \mu^+ \mu^-$ decays,
recently studied at the LHC.
%
%
One of the inputs required by the LHCb analysis is the ratio
of $B_q$ meson ($q = d,s$) fragmentation fractions $f_s / f_d$,
where $f_q$ is the probability that a $q$ quark hadronizes into a $B_q$.
This ratio can be measured by writing it as a product of ratios that involve
experimentally measurable quantities, cf. Refs.~\cite{Fleischer:2010ay,LHCb:2011ldp}.
One of the factors is the ratio $f_0^{(s)}(M_\pi^2) / f_0^{(d)}(M_K^2)$
of scalar form factors for the corresponding semileptonic meson decay,
which is where lattice input becomes useful.

A dedicated $N_f=2+1$ study by FNAL/MILC\footnote{This work also provided
a value for $R(D)$, now superseded by Ref.~\cite{Lattice:2015rga}.} \cite{Bailey:2012rr} addresses the
ratios of scalar form factors $f_0^{(q)}(q^2)$, and quotes:
\begin{equation}
f_0^{(s)}(M_\pi^2) / f_0^{(d)}(M_K^2) = 1.046(44)(15),
\qquad
f_0^{(s)}(M_\pi^2) / f_0^{(d)}(M_\pi^2) = 1.054(47)(17),
\end{equation}
where the first error is statistical and the second systematic.
The more recent results from HPQCD~\cite{Monahan:2017uby} are:
\begin{equation}
f_0^{(s)}(M_\pi^2) / f_0^{(d)}(M_K^2) = 1.000(62),
\qquad
f_0^{(s)}(M_\pi^2) / f_0^{(d)}(M_\pi^2) = 1.006(62).
\end{equation}
Results from both groups lead to fragmentation fraction
ratios $f_s/f_d$ that are
consistent with LHCb's measurements via other methods~\cite{LHCb:2011ldp}.

\subsubsection{$B_{(s)} \rightarrow D^*_{(s)}$ decays}

The most precise computation of the zero-recoil form
factors needed for the determination of $|V_{cb}|$ from exclusive $B$
semileptonic decays comes from the $B \rightarrow D^* \ell \nu$ form
factor at zero recoil ${\cal F}^{B \rightarrow D^*}(1)$, calculated
by the FNAL/MILC collaboration. The original computation, published
in Ref.~\cite{Bernard:2008dn}, has been updated~\cite{Bailey:2014tva}
by employing a much more extensive set of gauge ensembles and
increasing the statistics of the ensembles originally considered, while
preserving the analysis strategy. There are currently no final unquenched
results for the relevant form factors at nonzero recoil, but work
is in progress by FNAL/MILC \cite{Vaquero:2019ary} and JLQCD \cite{Kaneko:2019vkx}.

Reference \cite{Bailey:2014tva} uses
the MILC $N_f = 2 + 1$ ensembles.  The bottom and charm quarks are
simulated using the clover action with the Fermilab interpretation and
light quarks are treated via the asqtad staggered fermion action.  Recalling
the definition of the form factors in Eq.~(\ref{eq:BtoDstarAxialFormFactor}),
at zero recoil ${\cal F}^{B \rightarrow D^*}(1)$ reduces to a single form
factor $h_{A_1}(1)$ coming from the axial-vector current
\begin{equation}
\langle D^*(v,\epsilon^\prime)| {\cal A}_\mu | \overline{B}(v) \rangle = i \sqrt{2m_B 2 m_{D^*}} \; {\epsilon^\prime_\mu}^\ast h_{A_1}(1),
\end{equation}
where $\epsilon^\prime$ is the polarization of the $D^*$.
The form factor is accessed through a ratio of three-point
correlators, viz.,
\begin{equation}
{\cal R}_{A_1} = \frac{\langle D^*|\bar{c} \gamma_j \gamma_5 b | \overline{B}
\rangle \; \langle \overline{B}| \bar{b} \gamma_j \gamma_5 c | D^* \rangle}
{\langle D^*|\bar{c} \gamma_4 c | D^*
\rangle \; \langle \overline{B}| \bar{b} \gamma_4 b | \overline{B} \rangle}
= |h_{A_1}(1)|^2.
\end{equation}
Simulation data is obtained on
MILC ensembles with five lattice spacings, ranging from $a \approx 0.15~{\rm fm}$
to $a \approx 0.045~{\rm fm}$, and as many as five values of the light-quark masses
per ensemble (though just one at the finest lattice spacing).
Results are then extrapolated to the physical, continuum/chiral, limit
employing staggered $\chi$PT.

The $D^*$ meson is not a stable particle in QCD and decays
 predominantly into a $D$ plus a pion.  Nevertheless, heavy-light
 meson $\chi$PT can be applied to extrapolate lattice simulation
 results for the $B\to D^*\ell\nu$ form factor to the physical
 light-quark mass.  The $D^*$ width is quite narrow, 0.096 MeV for the
 $D^{*\pm}(2010)$ and less than 2.1 MeV for the $D^{*0}(2007)$, making
 this system much more stable and long lived than the $\rho$ or the
 $K^*$ systems. The fact that the $D^* - D$ mass difference is close
 to the pion mass leads to the well-known ``cusp'' in ${\cal
 R}_{A_1}$ just above the physical pion
 mass~\cite{Randall:1993qg,Savage:2001jw,Hashimoto:2001nb}. This cusp
 makes the chiral extrapolation sensitive to values used in the
 $\chi$PT formulas for the $D^*D \pi$ coupling $g_{D^*D\pi}$.  The
 error budget in Ref.~\cite{Bailey:2014tva} includes a separate
 error of 0.3\% coming from the uncertainty in $g_{D^*D \pi}$ in
 addition to general chiral extrapolation errors in order to take this
 sensitivity into account.

The final value presented in Ref.~\cite{Bailey:2014tva} is
\begin{equation}
\Nf=2+1: \;\;  {\cal F}^{B \rightarrow D^*}(1) =  0.906(4)(12)\,,
\label{eq:BDstarFNAL}
\end{equation}
where the first error is statistical, and the second the sum of systematic errors
added in quadrature, making up a total error of $1.4$\% (down from the original
$2.6$\% of Ref.~\cite{Bernard:2008dn}). The largest systematic
uncertainty comes from discretization errors followed by effects of
higher-order corrections in the chiral perturbation theory ansatz.

In 2017, the HPQCD collaboration has
published the first study of  ${{B}_{(s)}\to D_{(s)}^{*}\ell{\nu}}$
form factors at zero recoil
for $N_f=2+1+1$ using eight MILC ensembles with
lattice spacing $a\approx 0.15$ fm, 0.12 fm, and 0.09 fm~\cite{Harrison:2017fmw}.  There are three ensembles
with varying light-quark masses for the two coarser lattice spacings and two
choices of light-quark mass for the finest lattice spacing.  In each case,
there is one ensemble for which the light-quark mass is very close to the
physical value.  The $b$ quark is treated using NRQCD and the light quarks
are treated using the HISQ action.  The resulting zero-recoil form factors
are:
\begin{equation}
\Nf=2+1+1: \;\;
{\mathcal F}^{B \rightarrow D^*}(1) = 0.895(10)(24)\,,~~~~
{\mathcal F}^{B_s\to D_s^*}(1) = 0.883(12)(28)\,.
\label{eq:BDstarHPQCD}
\end{equation}
In 2019, the HPQCD collaboration published a new $N_f=2+1+1$ calculation of the $B_s \to D_s^*$ form factor
at zero recoil, now using the HISQ action also for the $b$ quark \cite{McLean:2019sds}. The lattice methodology
and ensembles used are the same as in their 2019 calculation of the $B_s \to D_s$ form factors \cite{McLean:2019qcx},
which was discussed in detail in Sec.~\ref{sec:BstoDsFFs}. The resulting form factor is:
\begin{equation}
\Nf=2+1+1: \;\;
{\mathcal F}^{B_s\to D_s^*}(1) = 0.9020(96)(90)\,.
\label{eq:BsDsstarHPQCD}
\end{equation}
The calculations in Refs.~\cite{Harrison:2017fmw, McLean:2019sds} use different $b$-quark actions and share only 
two ensambles at $a$ = 0.09 fm and can be considered essentially independent, yielding the average:
\begin{equation}
\Nf=2+1+1: \;\;
{\mathcal F}^{B_s\to D_s^*}(1) = 0.899(12)\,,~~~~\mbox{our average.}
\label{eq:BsDsstarFLAG}
\end{equation}

In recent years, the FNAL/MILC, HPQCD, and JLQCD collaborations have
periodically reported about their efforts to determine the full momentum
dependence of $B_{(s)}\to D_{(s)}^* \ell\nu$ form factors at Lattice conferences.
JLQCD efforts are based on $N_f=2+1$ M\"obius domain-wall ensembles and a relativistic heavy quark action.
The latest status update published in conference proceedings
can be found in Ref.~\cite{Kaneko:2019vkx}.
At the time of finalizing this review, FNAL/MILC and HPQCD have produced preprints
with their final results for $B \to D^*$ and $B_{s}\to D_{s}^*$ form factors, respectively.
The FNAL/MILC computation, Ref.~\cite{FermilabLattice:2021cdg} is based on $N_f=2+1$ asqtad sea-quark ensembles
with lattice spacing between approximately 0.15 and 0.045 fm, and uses a relativistic heavy-quark action with
the Fermilab interpretation.
The HPQCD computation, Ref.~\cite{Harrison:2021tol}, is based on $N_f=2+1+1$ HISQ ensembles, and uses
the same regularization for heavy quarks.
Upon publication of both works, we intend to include
full details for them in an upcoming intermediate update of this section.


\begin{table}[h]
\begin{center}
\mbox{} \\[3.0cm]
\footnotesize\hspace{-0.2cm}
\begin{tabular*}{\textwidth}[l]{l @{\extracolsep{\fill}} r l l l l l l l c l}
Collaboration & Ref. & $\Nf$ &
\hspace{0.15cm}\begin{rotate}{60}{publication status}\end{rotate}\hspace{-0.15cm} &
\hspace{0.15cm}\begin{rotate}{60}{continuum extrapolation}\end{rotate}\hspace{-0.15cm} &
\hspace{0.15cm}\begin{rotate}{60}{chiral extrapolation}\end{rotate}\hspace{-0.15cm}&
\hspace{0.15cm}\begin{rotate}{60}{finite volume}\end{rotate}\hspace{-0.15cm}&
\hspace{0.15cm}\begin{rotate}{60}{renormalization}\end{rotate}\hspace{-0.15cm}  &
\hspace{0.15cm}\begin{rotate}{60}{heavy-quark treatment}\end{rotate}\hspace{-0.15cm}  &
\multicolumn{2}{l}{\hspace{0mm} $w=1$ form factor / ratio}\\
&&&&&&&&&& \\[-0.1cm]
\hline
\hline
&&&&&&&&& \\[-0.1cm]
\SLhpqcdBD, HPQCD 17 & \cite{Na:2015kha,Monahan:2017uby} & 2+1 & \gA & \soso &  \soso &  \soso & \soso & \okay & ${\mathcal G}^{B\to D}(1)$ & 1.035(40) \\[0.5ex]
\SLfnalmilcBD & \cite{Lattice:2015rga} & 2+1 & \gA & \good &  \soso &  \good & \soso & \okay & ${\mathcal G}^{B\to D}(1)$  & $1.054(4)(8)$  \\[0.5ex]
Atoui 13 & \cite{Atoui:2013zza} & 2 & \gA & \good & \soso & \good & --- & \okay & ${\mathcal G}^{B\to D}(1)$  & 1.033(95) \\[0.5ex]
&&&&&&&&& \\[-0.1cm]
\hline
&&&&&&&&& \\[-0.1cm]
HPQCD 19 & \cite{McLean:2019qcx} & 2+1+1 & \gA & \good & \soso & \soso$^*$ & \okay & \okay & ${\mathcal G}^{B_s\to D_s}(1)$ & 1.071(37) \\[0.5ex]
\SLhpqcdBD, HPQCD 17 & \cite{Na:2015kha,Monahan:2017uby} & 2+1 & \gA & \soso &  \soso &  \soso & \soso & \okay & ${\mathcal G}^{B_s\to D_s}(1)$ & 1.068(40) \\[0.5ex]
Atoui 13 & \cite{Atoui:2013zza} & 2 & \gA & \good & \soso & \good & --- & \okay & ${\mathcal G}^{B_s\to D_s}(1)$  & 1.052(46) \\[0.5ex]
&&&&&&&&& \\[-0.1cm]
\hline
&&&&&&&&& \\[-0.1cm]
HPQCD 17B & \cite{Harrison:2017fmw} & 2+1+1 & \gA & \soso &  \good &  \good & \soso & \okay & ${\mathcal F}^{B\to D^*}(1)$   & 0.895(10)(24) \\[0.5ex]
\SLfnalmilcBDstar & \cite{Bailey:2014tva} & 2+1 & \gA & \good &  \soso &  \good & \soso & \okay&${\mathcal F}^{B\to D^*}(1)$   & 0.906(4)(12) \\[0.5ex]
&&&&&&&&& \\[-0.1cm]
\hline
&&&&&&&&& \\[-0.1cm]
HPQCD 17B & \cite{Harrison:2017fmw} & 2+1+1 & \gA & \soso &  \good &  \good & \soso & \okay & ${\mathcal F}^{B_s\to D_s^*}(1)$ & 0.883(12)(28) \\[0.5ex]
HPQCD 19B & \cite{McLean:2019sds} & 2+1+1 & \gA & \good & \soso & \soso$^*$ & \okay & \okay & ${\mathcal F}^{B_s\to D_s^*}(1)$ & 0.9020(96)(90) \\[0.5ex]
&&&&&&&&& \\[-0.1cm]
\hline
&&&&&&&&& \\[-0.1cm]
\SLhpqcdBD, HPQCD 17 & \cite{Na:2015kha,Monahan:2017uby} & 2+1 & \gA & \soso &  \soso &  \soso & \soso & \okay & ${\mathcal G}^{B_s\to D_s}(1)$ & 1.068(40) \\[0.5ex]
&&&&&&&&& \\[-0.1cm]
\hline
&&&&&&&&& \\[-0.1cm]
HPQCD 20B & \cite{Harrison:2020gvo} & 2+1+1 & \gA & \good & \soso & \soso$^*$ & \okay & \okay & n/a & n/a    \\[0.5ex]
&&&&&&&&& \\[-0.1cm]
\hline
&&&&&&&&& \\[-0.1cm]
\SLhpqcdBD, HPQCD 17 & \cite{Na:2015kha,Monahan:2017uby} & 2+1 & \gA & \soso &  \soso &  \soso & \soso & \okay &   $R(D)$ & 0.300(8) \\[0.5ex]
\SLfnalmilcBD & \cite{Lattice:2015rga} & 2+1 & \gA & \good &  \soso &  \good & \soso & \okay &  $R(D)$  & 0.299(11) \\[0.5ex]
&&&&&&&&& \\[-0.1cm]
\hline
\hline
\end{tabular*}\\
\begin{minipage}{\linewidth}
{\footnotesize 
\begin{itemize}
   \item[$^*$] The rationale for assigning a \soso rating is discussed in the text.
\end{itemize}
}
\end{minipage}
\caption{Lattice results for mesonic processes involving $b \to c$ transitions. \label{tab_BtoDStarsumm2}}
\end{center}
\end{table}

\subsection{Semileptonic form factors for $B_c\to (\eta_c, J/\psi)\ell\nu$ decays}
\label{sec:Bcdecays}

In a recent publication, HPQCD 20B \cite{Harrison:2020gvo} 
provided the first full determination of $B_c\to J/\psi$ form factors,
extending earlier preliminary work that also covered $B_c \to \eta_c$,
Refs.~\cite{Lytle:2016ixw, Colquhoun:2016osw}. While the latter employed
both NRQCD and HISQ actions for the valence $b$ quark,
and the HISQ action for the $c$ quark, in HPQCD 20B the HISQ action is used
throughout for all flavors. The setup is the same as for the $B_s\to D_s$ computation
discussed above, HPQCD 19; we refer to the entries for the latter paper in summary
tables for details. The flavor singlet nature of the final state
means that there are contributions to the relevant three-point functions from disconnected
Wick contractions, which are not discussed in the paper.

There are however some relevant differences with $B_s \to D_s$ decays.
In the $J/\psi$ case, since the hadron in the final state has vector quantum numbers, the description of the
hadronic amplitude requires four independent form factors $V$, $A_0$, $A_1$, $A_2$.  Specifically,
\begin{gather}
\label{eq:BcJpsiFF}
\begin{split}
 \langle  J/\psi(p',\lambda)|\bar{c}\gamma^\mu  b|B_c^-(p)\rangle =&
 \frac{2i V(q^2)}{M_{B_c} + M_{J/\psi}} \varepsilon^{\mu\nu\rho\sigma}\epsilon^*_\nu(p',\lambda) p'_\rho p_\sigma\,, \\[2.0ex]
\langle  J/\psi(p',\lambda)|\bar{c}\gamma^\mu \gamma^5 b|B_c^-(p)\rangle =&
 2M_{J/\psi}A_0(q^2)\frac{\epsilon^*(p',\lambda)\cdot q}{q^2} q^\mu\\
&\quad +(M_{B_c}+M_{J/\psi})A_1(q^2)\Big[ \epsilon^{*\mu}(p',\lambda) - \frac{\epsilon^*(p',\lambda)\cdot q}{q^2} q^\mu \Big] \\
&\quad - A_2(q^2)\frac{\epsilon^*(p',\lambda)\cdot q}{M_{B_c}+M_{J/\psi}}\Big[ p^\mu + p'^\mu - \frac{M_{B_c}^2-M_{J/\psi}^2}{q^2}q^\mu \Big],
\end{split}
\end{gather}
where $\epsilon_\mu$ is the polarization vector of the $J/\psi$ state.
The computed form factors are fitted
to a $z$-parameterization-inspired ansatz, where coefficients are modified to model
the lattice-spacing and the heavy- and light-mass dependences, for a total of 280~fit parameters.
In the continuum and at physical kinematics only 16 parameters survive, as each form factor
is parameterized by an expression of the form
\begin{gather}
\label{eq:FFpsi}
F(q^2) = \frac{1}{P(q^2)} \sum_{n=0}^3 a_n z^n\,,
\end{gather}
where the pole factor is given by
\begin{gather}
P(q^2)=\prod_k z(q^2,M_k^2)
\end{gather}
with $\{M_k\}$ a different set of pole energies below the $BD^*$ threshold for each set of $J^P$ quantum numbers,
taken from a mixture of experimental results, lattice determinations, and model estimates.
The values used (in GeV) are
\begin{gather}
\begin{split}
&0^-:~6.275,~6.872,~7.25;\\
&1^-:~6.335,~6.926,~7.02,~7.28;\\
&1^+:~6.745,~6.75,~7.15,~7.15.
\end{split}
\end{gather}
The outcome of the fit, that we quote as a FLAG estimate, is
\begin{center}
\begin{tabular}{ | c | c c c c | }
\hline
& $a_0$ & $a_1$ & $a_2$ & $a_3$ \\\hline
${V}$&  0.1057(55)&     -0.746(92)&     0.10(98)&       0.006(1.000)\\
${A0}$& 0.1006(37)&     -0.731(72)&     0.30(90)&       -0.02(1.00)\\
${A1}$& 0.0553(19)&     -0.266(40)&     0.31(70)&       0.11(99)\\
${A2}$& 0.0511(91)&     -0.22(19)&      -0.36(82)&      -0.05(1.00)\\
\hline
\end{tabular}
\end{center}
The correlation matrix for the coefficients is provided in Tables~XIX-XXVII of
Ref.~\cite{Harrison:2020gvo}

\subsection{Semileptonic form factors for $\Lambda_b\to (p,\Lambda_c^{(*)})\ell\bar{\nu}$ decays}
\label{sec:Lambdab}

The $b\to c\ell\bar{\nu}$ and $b\to u\ell\bar{\nu}$ transitions can also be probed
in decays of $\Lambda_b$ baryons. With the LHCb experiment, the final state of $\Lambda_b\to p\mu\bar{\nu}$
is easier to identify than that of $B\to\pi\mu\bar{\nu}$ \cite{LHCbRICHGroup:2012mgd},
and the first determination of $|V_{ub}|/|V_{cb}|$ at the Large Hadron Collider
was performed using a ratio of $\Lambda_b\to p\mu\bar{\nu}$ and $\Lambda_b\to \Lambda_c\mu\bar{\nu}$
decay rates~\cite{Aaij:2015bfa} (cf.~Sec.~\ref{sec:VubVcb}).

The amplitudes of the decays $\Lambda_b\to p\ell\bar{\nu}$ and $\Lambda_b\to \Lambda_c\ell\bar{\nu}$
receive contributions from both the vector and the axial-vector components of the current
in the matrix elements $\langle p|\bar u\gamma^\mu(\mathbf{1}-\gamma_5)b|\Lambda_b\rangle$
and $\langle \Lambda_c|\bar c\gamma^\mu(\mathbf{1}-\gamma_5)b|\Lambda_b\rangle$.
The matrix elements split into three form factors $f_+$, $f_0$, $f_\perp$
mediated by the vector component of the current, and another three form factors $g_+$, $g_0$, $g_\perp$
mediated by the axial-vector component---see, e.g., Ref.~\cite{Feldmann:2011xf}
for a complete description. Given the sensitivity to all Dirac structures,
measurements of the baryonic decay rates also provides useful complementary constraints on right-handed
couplings beyond the Standard Model \cite{Aaij:2015bfa}.

To date, only one unquenched lattice-QCD computation of the  $\Lambda_b\to p$
and $\Lambda_b\to \Lambda_c$ form factors with physical heavy-quark masses has been published: Detmold 15 \cite{Detmold:2015aaa}.
This computation uses RBC/UKQCD $N_f=2+1$ DWF ensembles,
and treats the $b$ and $c$ quarks within the Columbia RHQ approach. The renormalization of
the currents is carried out using a mostly nonperturbative method, with residual matching factors computed
at one loop.
Two values of the lattice spacing ($a\approx0.11,~0.085~{\rm fm}$) are considered,
with the absolute scale set from the $\Upsilon(2S)$--$\Upsilon(1S)$ splitting.
Sea pion masses lie in a narrow interval ranging from slightly above
$400~{\rm MeV}$ to slightly below $300~{\rm MeV}$, keeping $m_\pi L \gtrsim 4$;
however, lighter pion masses are considered in the valence DWF action
for the $u,d$ quarks. The lowest valence-valence pion mass is 227(3) MeV,
which leads to a \tbr~ rating of finite-volume effects.
Results for the form factors are obtained from suitable three-point functions,
and fitted to a modified $z$-expansion ansatz that combines the $q^2$-dependence
with the chiral and continuum extrapolations. The main results of the paper are
the predictions (errors are statistical and systematic, respectively)
\begin{align}
\zeta_{p\mu\bar\nu}(15{\rm GeV}^2) &\equiv \frac{1}{|V_{ub}|^2}\int_{15~{\rm GeV}^2}^{q^2_{\rm max}}\frac{{\rm d}\Gamma(\Lambda_b\to p\mu^-\bar\nu_\mu)}{{\rm d}q^2}\,{\rm d}q^2 &= 12.31(76)(77)~{\rm ps}^{-1}\,,\\
\zeta_{\Lambda_c \mu\bar\nu}(7{\rm GeV}^2) &\equiv\frac{1}{|V_{cb}|^2}\int_{7~{\rm GeV}^2}^{q^2_{\rm max}}\frac{{\rm d}\Gamma(\Lambda_b\to \Lambda_c\mu^-\bar\nu_\mu)}{{\rm d}q^2}\,{\rm d}q^2 &= 8.37(16)(34)~{\rm ps}^{-1}\,,\\
\displaystyle \frac{\zeta_{p\mu\bar\nu}(15{\rm GeV}^2)}{\zeta_{\Lambda_c \mu\bar\nu}(7{\rm GeV}^2)} &= 1.471(95)(109)\,,
\end{align}
which are the input for the LHCb analysis. Predictions for the total rates in all possible
lepton channels, as well as for ratios similar to $R(D)$ (cf.~Sec.~\ref{sec:BtoD}) between the $\tau$
and light-lepton channels are also available, in particular,
\begin{equation}
 R(\Lambda_c)=\frac{\Gamma (\Lambda_b \to \Lambda_c\: \tau^- \bar{\nu}_\tau)}{\Gamma (\Lambda_b \to \Lambda_c\: \mu^- \bar{\nu}_\mu)} = 0.3328(74)(70).
\end{equation}
Datta 2017 \cite{Datta:2017aue} additionally includes results for the ${\Lambda}_b\to {\Lambda}_c$ tensor
form factors $h_+$, $h_\perp$, $\widetilde{h}_+$, $\widetilde{h}_\perp$, based on the same lattice computation
as Detmold 15 \cite{Detmold:2015aaa}. The main focus of Datta 2017 is the phenomenology of
the $ {\Lambda}_b\to {\Lambda}_c\tau {\overline{\nu}}_{\tau } $ decay
and how it can be used to constrain contributions from beyond the Standard Model
physics. Unlike in the case of the vector and axial-vector currents, the residual matching factors
of the tensor currents are set to their tree-level value. While the matching systematic uncertainty is augmented to
take this fact into account, the procedure implies that the tensor current
retains an uncanceled logarithmic divergence at $\mathcal{O}(\alpha_s)$.

Recently, first lattice calculations have also been completed for $\Lambda_b$ semileptonic decays to negative-parity baryons in the final state. Such calculations are substantially more challenging and have not yet reached the same level of precision. Meinel 21 \cite{Meinel:2021rbm} considers the decays $\Lambda_b \to \Lambda_c^*(2595)\ell\bar{\nu}$ and $\Lambda_b \to \Lambda_c^*(2625)\ell\bar{\nu}$, where the $\Lambda_c^*(2595)$ and $\Lambda_c^*(2625)$ are the lightest charm baryons with isospin 0 and $J^P=\frac12^-$ and $J^P=\frac32^-$, respectively.
 These decay modes may eventually provide new opportunities to test lepton-flavor universality at the LHC, but are also very interesting from a theoretical point of view. The lattice results for the form factors may help tighten dispersive constraints in global analyses of $b\to c$ semileptonic decays \cite{Cohen:2019zev}, and may provide new insights into the internal structure of the negative-parity heavy baryons and their description in heavy-quark-effective-theory. The $\Lambda_c^*(2595)$ and $\Lambda_c^*(2625)$ are very narrow resonances decaying through the strong interaction into $\Lambda_c \pi\pi$. The strong decays are neglected in Meinel 21 \cite{Meinel:2021rbm}. The calculation was performed using the same lattice actions as previously for $\Lambda_b \to \Lambda_c$, albeit with newly tuned RHQ parameters. Only three ensembles are used, with $a\approx0.11,~0.08~{\rm fm}$ and pion masses in the range from approximately 300 to 430 MeV, with valence-quark masses equal to the sea-quark masses. Chiral-continuum extrapolations linear in $m_\pi^2$ and $a^2$ are performed, with systematic uncertainties estimated using higher-order fits. Finite-volume effects and effects associated with the strong decays of the $\Lambda_c^*$'s are not quantified. The calculation is done in the $\Lambda_c^*$ rest frame, where the cubic symmetry is sufficient to avoid mixing with unwanted lower-mass states. As a consequence, the calculation is limited to a small kinematic region near the zero-recoil point $w=1$. On each ensemble, lattice data were produced for two values of $w-1$ of approximately 0.01 and 0.03. The final results for the form factors are parameterized as linear functions of $w-1$ and can be found in Meinel 21 \cite{Meinel:2021rbm} and associated supplemental files. 

\subsection{Semileptonic form factors for $\Lambda_b\to \Lambda^{(*)}\ell\ell$}
\label{sec:LambdabLambda}

The decays $\Lambda_b\to \Lambda\ell^+\ell^-$ are mediated by the same underlying $b\to s\ell^+\ell^-$ FCNC transition as, for
example, $B\to K\ell^+\ell^-$ and $B\to K^*\ell^+\ell^-$, and can therefore provide additional information on the hints for
physics beyond the Standard Model seen in the meson decays. The $\Lambda$ baryon in the final state decays through the weak
interaction into $p \pi^-$ (or $n \pi^0$), leading to a wealth of angular observables even for unpolarized $\Lambda_b$. When
including the effects of a nonzero $\Lambda_b$ polarization, $\Lambda_b\to \Lambda(\to p \pi^-)\ell^+\ell^-$ decays are
characterized by five angles leading to 34 angular observables \cite{Blake:2017une}, which have been measured by LHCb in
the bin $q^2\in[15,20]\:{\rm GeV}^2$ \cite{Aaij:2018gwm}. Given that the $\Lambda$ is stable under the strong
interactions, the $\Lambda_b \to \Lambda$ form factors parametrizing the matrix elements of local $\bar{s}\Gamma b$
currents can be calculated on the lattice with high precision using standard methods. Of course, the process
$\Lambda_b\to \Lambda\ell^+\ell^-$ also receives contributions from nonlocal matrix elements of four-quark and
quark-gluon operators in the weak effective Hamiltonian combined with the electromagnetic current. As with the
mesonic $b\to s\ell^+\ell^-$ decays, these contributions cannot easily be calculated  on the lattice and one
relies on other theoretical tools for them, including the local OPE at high $q^2$ and a light-cone OPE / QCD
factorization at low $q^2$.

Following an early calculation with static $b$ quarks \cite{Detmold:2012vy}, Detmold 16 \cite{Detmold:2016pkz} provides results for
all ten relativistic $\Lambda_b\to \Lambda$ form factors parametrizing the matrix elements of the local vector, axial-vector and
tensor $b\to s$ currents.  The lattice setup is identical to that used in the 2015 calculation of the $\Lambda_b \to p$ form
factors in Detmold 15 \cite{Detmold:2015aaa}, and similar considerations as in the previous section thus apply. The lattice data cover the
upper 60\% of the $q^2$ range, and the form factors are extrapolated to the full $q^2$ range using BCL $z$-expansion fits.
This extrapolation is done simultaneously with the chiral and continuum extrapolations. The caveat regarding the
renormalization of the tensor currents also applies here.

Reference \cite{Blake:2019guk} uses the lattice results for the $\Lambda_b\to \Lambda$ form factors together with the
experimental results for $\Lambda_b\to \Lambda(\to p \pi^-)\mu^+\mu^-$ from LHCb \cite{Aaij:2015xza,Aaij:2018gwm} to perform
fits of the $b\to s\mu^+\mu^-$ Wilson coefficients and of the $\Lambda_b$ polarization parameter. Given the uncertainties
(which are still dominated by experiment), the results for the Wilson coefficients are presently consistent both with the
Standard-Model values and with the deviations seen in global fits that include all mesonic decays
\cite{Alguero:2019ptt,Altmannshofer:2021qrr}.		 

As with the $b\to c$ semileptonic form factors, a first lattice calculation, Meinel 2020 \cite{Meinel:2020owd}, 
was also recently completed for a $b\to s$ transition to a negative-parity baryon in the final state, in this case the $\Lambda^*(1520)$ with $J^P=\frac32^-$ 
(no calculation has yet been published for the strange $J^P=\frac12^-$ final states, which would
be the broader and even more challenging $\Lambda^*(1405)/\Lambda^*(1380)$ \cite{Zyla:2020zbs}). The $\Lambda^*(1520)$
decays primarily to $ pK^-/n \bar{K}^0$, $\Sigma \pi$, and $\Lambda\pi\pi$ with a total width of $15.6\pm 1.0$ MeV
\cite{Zyla:2020zbs} . The analysis of the lattice data again neglects the strong decays and does not quantify
finite-volume effects, and is again limited to a small kinematic region near $q^2_{\rm max}$.

\begin{table}[h]
\begin{center}
\mbox{} \\[3.0cm]
\footnotesize
\begin{tabular}{l l @{\extracolsep{\fill}} r l l l l l l l}
Process & Collaboration & Ref. & $\Nf$ & 
\hspace{0.15cm}\begin{rotate}{60}{publication status}\end{rotate}\hspace{-0.15cm} &
\hspace{0.15cm}\begin{rotate}{60}{continuum extrapolation}\end{rotate}\hspace{-0.15cm} &
\hspace{0.15cm}\begin{rotate}{60}{chiral extrapolation}\end{rotate}\hspace{-0.15cm}&
\hspace{0.15cm}\begin{rotate}{60}{finite volume}\end{rotate}\hspace{-0.15cm}&
\hspace{0.15cm}\begin{rotate}{60}{renormalization}\end{rotate}\hspace{-0.15cm}  &
\hspace{0.15cm}\begin{rotate}{60}{heavy-quark treatment}\end{rotate}\hspace{-0.15cm} \\
&&&&&&&& \\[-0.1cm]
\hline
\hline
&&&&&&&& \\[-0.1cm]
$\Lambda_b\to  \Lambda_c^*(2625) \,\ell^- \bar{\nu}_\ell$  & Meinel 21                         & \cite{Meinel:2021rbm}                & 2+1 & \gA & \soso & \soso & \tbr  & \soso & \okay \\[0.5ex]
$\Lambda_b\to  \Lambda_c^*(2595) \,\ell^- \bar{\nu}_\ell$  & Meinel 21                         & \cite{Meinel:2021rbm}                & 2+1 & \gA & \soso & \soso & \tbr  & \soso & \okay \\[0.5ex]
$\Lambda_b\to  \Lambda^*(1520) \,\ell^+\ell^-$    & Meinel 20                         & \cite{Meinel:2020owd}                & 2+1 & \gA & \soso & \soso & \tbr  & \soso & \okay \\[0.5ex]
$\Lambda_b\to  \Lambda \,\ell^+\ell^-$            & Detmold 16                        & \cite{Detmold:2016pkz}               & 2+1 & \gA & \soso & \soso & \tbr  & \soso & \okay \\[0.5ex]
$\Lambda_b\to  p \,\ell^- \bar{\nu}_\ell$         & Detmold 15 \hspace{1ex}           & \cite{Detmold:2015aaa}               & 2+1 & \gA & \soso & \soso & \tbr  & \soso & \okay \\[0.5ex]
$\Lambda_b\to  \Lambda_c \,\ell^- \bar{\nu}_\ell$ & Detmold 15, Datta 17 \hspace{1ex} & \cite{Detmold:2015aaa,Datta:2017aue} & 2+1 & \gA & \soso & \soso & \tbr  & \soso & \okay \\[0.5ex]
&&&&&&&& \\[-0.1cm]
\hline
\hline
\end{tabular}
\caption{
Summary of computations of bottom baryon semileptonic form factors (see also 
Refs.~\cite{Detmold:2012vy,Detmold:2013nia} 
for calculations with static $b$ quarks). The rationale for the \tbr~ rating of finite-volume effects in Meinel 20 and 21 
(despite meeting the {\color{green}\Large$\circ$} criterion based on the minimum pion mass) 
is that the unstable nature of the final-state baryons was neglected in the analysis.
}
\label{tab_BottomBaryonSLsumm2}
\end{center}
\end{table}

\FloatBarrier

\subsection{Determination of $|V_{ub}|$}
\label{sec:Vub}

We now use the lattice-determined Standard Model transition amplitudes
for leptonic (Sec.~\ref{sec:fB}) and semileptonic
(Sec.~\ref{sec:BtoPiK}) $B$-meson decays to obtain exclusive
determinations of the CKM matrix element $|V_{ub}|$.
In this section, we describe the aspect of our work
that involves experimental input for the relevant charged-current
exclusive decay processes.
The relevant
formulae are Eqs.~(\ref{eq:B_leptonic_rate})
and~(\ref{eq:B_semileptonic_rate}). Among leptonic channels the only
input comes from $B\to\tau\nu_\tau$, since the rates for decays to $e$
and $\mu$ have not yet been measured.  In the semileptonic case, we
only consider $B\to\pi\ell\nu$ transitions (experimentally
measured for $\ell=e,\mu$).

We first investigate the determination of $|V_{ub}|$ through the
$B\to\tau\nu_\tau$ transition.  This is the only experimentally
measured leptonic decay channel of the charged $B$ meson.
The experimental measurements of the branching fraction of
this channel, $B(B^{-} \to \tau^{-} \bar{\nu})$, have not been
updated since the publication of the FLAG Review in 2016~\cite{Aoki:2016frl}. The status of the experimental results for this branching fraction, summarized in Tab.~\ref{tab:leptonic_B_decay_exp}, is unchanged from FLAG Review 16~\cite{Aoki:2016frl}. Our corresponding values of $|V_{ub}|$ are unchanged from FLAG Review 19~\cite{Aoki:2019cca}.
\begin{table}[h]
\begin{center}
\noindent
\begin{tabular*}{\textwidth}[l]{@{\extracolsep{\fill}}lll}
Collaboration & Tagging method  & $B(B^{-}\to \tau^{-}\bar{\nu})
                                  \times 10^{4}$\\
&& \\[-2ex]
\hline \hline &&\\[-2ex]
Belle~\cite{Adachi:2012mm} &  Hadronic  & $0.72^{+0.27}_{-0.25}\pm 0.11$ \\
Belle~\cite{Kronenbitter:2015kls} &  Semileptonic & $1.25 \pm 0.28 \pm
                                                    0.27$ \\
&& \\[-2ex]
 \hline
&& \\[-2ex]
BaBar~\cite{Lees:2012ju} & Hadronic & $1.83^{+0.53}_{-0.49}\pm 0.24$\\
BaBar~\cite{Aubert:2009wt} & Semileptonic  & $1.7 \pm 0.8 \pm 0.2$\\
&& \\[-2ex]
\hline \hline && \\[-2ex]
\end{tabular*}
\caption{Experimental measurements for $B(B^{-}\to \tau^{-}\bar{\nu})$.
  The first error on each result is statistical, while the second
  error is systematic.}
\label{tab:leptonic_B_decay_exp}
\end{center}
\end{table}

It is obvious that all the measurements listed in Tab.~\ref{tab:leptonic_B_decay_exp} have significance smaller than
$5\sigma$, and the large uncertainties are dominated by statistical errors. These measurements lead to the averages
of experimental measurements for $B(B^{-}\to \tau \bar{\nu})$~\cite{Kronenbitter:2015kls,Lees:2012ju},
\begin{eqnarray}
 B(B^{-}\to \tau \bar{\nu} )\times 10^4 &=& 0.91 \pm 0.22 \mbox{ }{\rm from}\mbox{ } {\rm Belle,} \label{eq:leptonic_B_decay_exp_belle}\\
                             &=& 1.79 \pm 0.48 \mbox{ }{\rm from }\mbox{ } {\rm BaBar,} \label{eq:leptonic_B_decay_exp_babar}\\
                             &=& 1.06 \pm 0.33 \mbox{ }{\rm average,}
\label{eq:leptonic_B_decay_exp_ave}
\end{eqnarray}
where, following our standard procedure we perform a weighted average and rescale the uncertainty by the square root of the reduced chi-squared. Note that the Particle Data Group~\cite{Rosner:2015wva} did not inflate the uncertainty in the calculation of the averaged branching ratio.

Combining the results in Eqs.~(\ref{eq:leptonic_B_decay_exp_belle}--\ref{eq:leptonic_B_decay_exp_ave}) with
the experimental measurements of the mass of the $\tau$-lepton and the
$B$-meson lifetime and mass we get
\begin{eqnarray}
 |V_{ub}| f_{B} &=& 0.72 \pm 0.09 \mbox{ }{\rm MeV}\mbox{ }{\rm from}\mbox{ } {\rm Belle,}\label{eq:Vub_fB_belle}\\
                &=& 1.01 \pm 0.14 \mbox{ }{\rm MeV}\mbox{ }{\rm from }\mbox{ } {\rm BaBar,}\label{eq:Vub_fB_babar} \\
                &=& 0.77 \pm 0.12 \mbox{ }{\rm MeV}\mbox{ } {\rm average,}\label{eq:Vub_fB}
\end{eqnarray}
which can be used to extract $|V_{ub}|$, viz.,
\begin{align}
&N_f=2    &\mbox{Belle}~B\to\tau\nu_\tau:   && |V_{ub}| &= 3.83(14)(48) \times 10^{-3}  \,,  \\
&N_f=2+1  &\mbox{Belle}~B\to\tau\nu_\tau:   && |V_{ub}| &= 3.75(8)(47) \times 10^{-3}   \,,  \\
&N_f=2+1+1&\mbox{Belle}~B\to\tau\nu_\tau:   && |V_{ub}| &= 3.79(3)(47) \times 10^{-3}   \,;  \\
&         & \nonumber \\
&N_f=2    &\mbox{Babar}~B\to\tau\nu_\tau:   && |V_{ub}| &=  5.37(20)(74) \times 10^{-3} \,,  \\
&N_f=2+1  &\mbox{Babar}~B\to\tau\nu_\tau:   && |V_{ub}| &=  5.26(12)(73) \times 10^{-3} \,,  \\
&N_f=2+1+1&\mbox{Babar}~B\to\tau\nu_\tau:   && |V_{ub}| &= 5.32(4)(74) \times 10^{-3}  \,,  \\
&         & \nonumber \\
&N_f=2    &\mbox{average}~B\to\tau\nu_\tau:   && |V_{ub}| &=  4.10(15)(64) \times 10^{-3} \,,  \\
&N_f=2+1  &\mbox{average}~B\to\tau\nu_\tau:   && |V_{ub}| &=  4.01(9)(63) \times 10^{-3} \,,  \\
&N_f=2+1+1&\mbox{average}~B\to\tau\nu_\tau:   && |V_{ub}| &= 4.05(3)(64) \times 10^{-3}  \,,
\end{align}
where the first error comes from the uncertainty in $f_B$ and the second comes from experiment.

Let us now turn our attention to semileptonic decays. The experimental
value of $|V_{ub}|f_+(q^2)$ can be extracted from the measured
branching fractions for $B^0\to\pi^\pm\ell\nu$ and/or
$B^\pm\to\pi^0\ell\nu$ applying
Eq.~(\ref{eq:B_semileptonic_rate});\footnote{Since $\ell=e,\mu$ the
  contribution from the scalar form factor in
  Eq.~(\ref{eq:B_semileptonic_rate}) is negligible.}  $|V_{ub}|$ can
then be determined by performing fits to the constrained BCL $z$-parameterization of the form factor $f_+(q^2)$ given in
Eq.~(\ref{eq:bcl_c}). This can be done in two ways: one option is to
perform separate fits to lattice and experimental results, and extract
the value of $|V_{ub}|$ from the ratio of the respective $a_0$
coefficients; a second option is to perform a simultaneous fit to
lattice and experimental data, leaving their relative normalization
$|V_{ub}|$ as a free parameter. We adopt the second strategy, because
it combines the lattice and experimental input in a more efficient
way, leading to a smaller uncertainty on $|V_{ub}|$.

The available state-of-the-art experimental input consists of five
data sets: three untagged measurements by BaBar
(6-bin~\cite{delAmoSanchez:2010af} and 12-bin~\cite{Lees:2012vv}) and
Belle~\cite{Ha:2010rf}, all of which assume isospin symmetry and
provide combined $B^0\to\pi^-$ and $B^+\to\pi^0$ data; and the two
tagged Belle measurements of $\bar B^0\to\pi^+$ (13-bin) and
$B^-\to\pi^0$ (7-bin)~\cite{Sibidanov:2013rkk}.  Including all of
them, along with the available information about cross-correlations,
will allow us to obtain a meaningful final error
estimate.\footnote{See, e.g., Sec.~V.D of Ref.~\cite{Lattice:2015tia} for
  a detailed discussion.} The lattice input data set will be the same
discussed in Sec.~\ref{sec:BtoPiK}.

We perform a constrained BCL fit of the vector and scalar form factors (this is necessary in order to take into account the $f_+(q^2=0) = f_0 (q^2=0)$ constraint) together with the combined experimental data sets. We find that the error on $|V_{ub}|$ stabilizes for $N^+ = N^0 = 3$. The result of the combined fit is presented in
Tab.~\ref{tab:FFVUBPI}. The fit has a chi-square per degree of freedom $\chi^2/{\rm dof} = 78.7/56=1.41$. Following the PDG recommendation we rescale the whole covariance matrix by $\chi^2/{\rm dof}$: the errors on the $z$-parameters are increased by $\sqrt{\chi^2/{\rm dof}} = 1.19$ and the correlation matrix is unaffected.
\begin{table}[t]
\begin{center}
\begin{tabular}{|c|c|cccccc|}
\multicolumn{8}{l}{$B\to \pi\ell\nu \; (N_f=2+1)$} \\[0.2em]\hline
        & Central Values & \multicolumn{6}{|c|}{Correlation Matrix} \\[0.2em]\hline
$V_{ub}^{} \times 10^3$ & 3.74 (17)   &  1 & $-$0.851 & $-$0.349 & 0.375 & $-$0.211 & $-$0.246 \\[0.2em]
$a_0^+$                 & 0.415 (14)  &   $-$0.851 & 1 & 0.155 & $-$0.454 & 0.260 & 0.144 \\[0.2em]
$a_1^+$                 & $-$0.488 (53) &  $-$0.349 & 0.155 & 1 & $-$0.802 & $-$0.0962 & 0.220\\[0.2em]
$a_2^+$                 & $-$0.31 (18)  & 0.375 & $-$0.454 & $-$0.802 & 1 & 0.0131 & $-$0.100 \\[0.2em]
$a_0^0$                 & 0.500 (23)  &  $-$0.211 & 0.260 & $-$0.0962 & 0.0131 & 1 & $-$0.453  \\[0.2em]
$a_1^0$                 & $-$1.424 (54) &  $-$0.246 & 0.144 & 0.220 & $-$0.100 & $-$0.453 & 1 \\[0.2em]
\hline
\end{tabular}
\end{center}
\caption{$|V_{ub}|$, coefficients for the $N^+ =N^0=N^T=3$ $z$-expansion of the $B\to \pi$ form factors $f_+$ and $f_0$, and their correlation matrix. The chi-square per degree of freedom is $\chi^2/{\rm dof} = 78.7/56=1.41$ and the errors on the fit parameters have been rescaled by $\sqrt{\chi^2/{\rm dof}} = 1.19$. The lattice calculations that enter this fit are taken from FNAL/MILC~\cite{Lattice:2015tia} and RBC/UKQCD~\cite{Flynn:2015mha}. The experimental inputs are taken from BaBar~\cite{delAmoSanchez:2010af,Lees:2012vv} and Belle~\cite{Ha:2010rf,Sibidanov:2013rkk}.
\label{tab:FFVUBPI}}
\end{table}
In Fig.~\ref{fig:Vub_SL_fit}, we show both the lattice and experimental data for
$(1-q^2/m_{B^*}^2)f_+(q^2)$ as a function of $z(q^2)$, together with our preferred fit;
experimental data has been rescaled by the resulting value for $|V_{ub}|^2$.
It is worth noting the good consistency between the form factor shapes from
lattice and experimental data. This can be quantified, e.g., by computing the ratio of the
two leading coefficients in the constrained BCL parameterization: the fit to lattice form
factors yields  $a_1^+/a_0^+=-1.67(35)$   (cf.~the results presented in Sec.~\ref{sec:BtoPi}),
while the above lattice+experiment fit yields  $a_1^+/a_0^+=-1.19(13)$.

\begin{figure}[tbp]
\begin{center}
\includegraphics[width=0.49\textwidth]{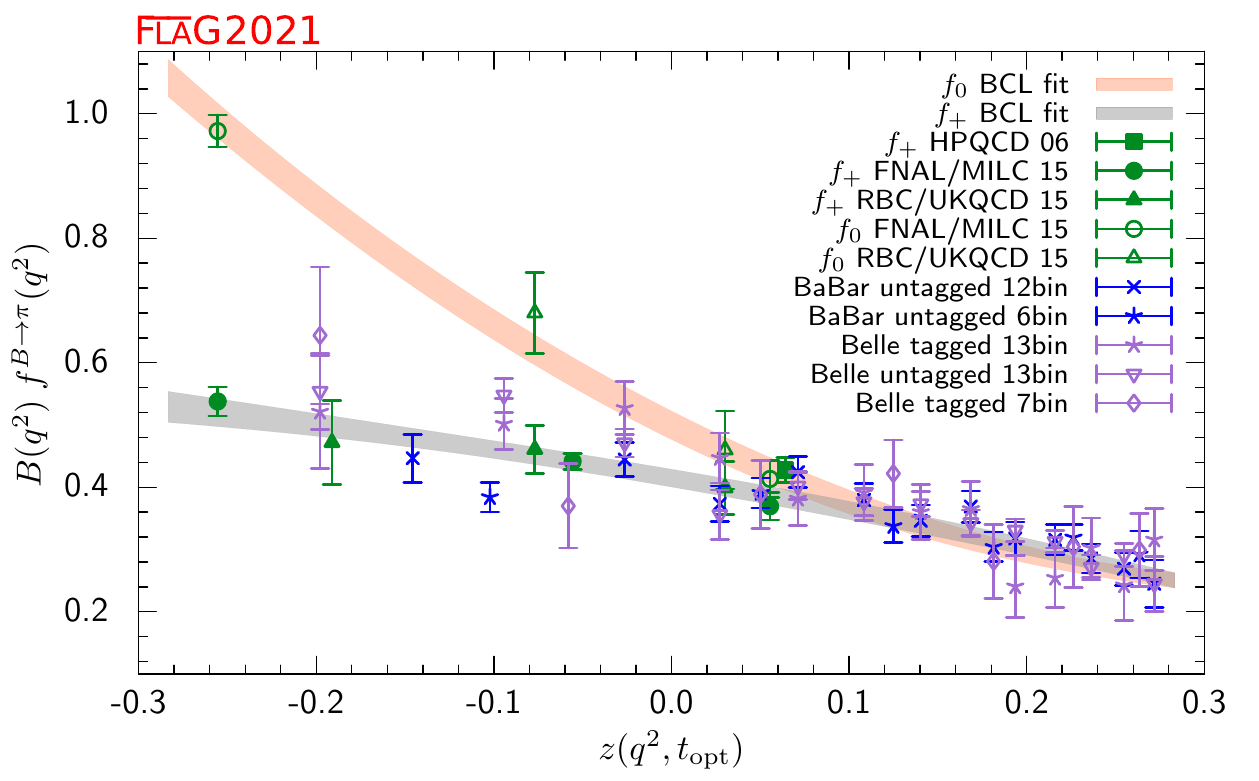}
\includegraphics[width=0.49\textwidth]{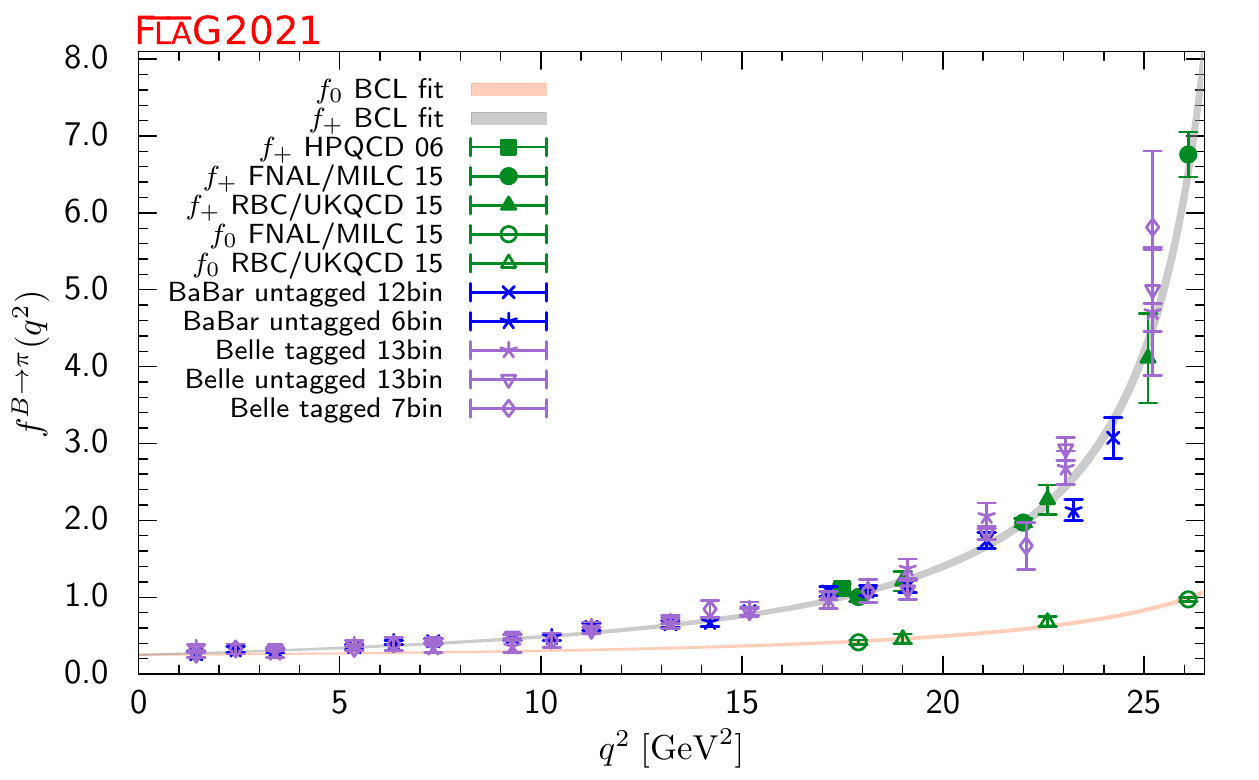}
\caption{
Lattice and experimental data for $f_+^{B\to\pi}(q^2)$ and $f_0^{B\to \pi} (q^2)$ versus $z$ (left panel) and $q^2$ (right panel). Experimental data has been rescaled by the value for $|V_{ub}|$ found from the joint fit. Green symbols denote lattice-QCD points included in the fit, while blue and indigo points show experimental data divided by the value of $|V_{ub}|$ obtained from the fit. The grey and orange bands display the preferred $N^+ = N^0 = 3$ BCL fit (five $z$-parameters and $|V_{ub}|$).}
\label{fig:Vub_SL_fit}
\end{center}
\end{figure}

We plot the values of $|V_{ub}|$ we have obtained in
Fig.~\ref{fig:Vxbsummary}, where the GGOU~\cite{Gambino:2007rp} determination through inclusive decays, $|V_{ub}^{}|_{\rm incl} = (4.32 \pm 0.12_{\rm exp} \pm 0.13_{\rm theo} \pm 0.23_{\Delta BF}) \times 10^{-3}$~\cite{Amhis:2019ckw, Zyla:2020zbs} (the $\Delta BF$ error has been added in Ref.~\cite{Zyla:2020zbs} to account for the spread in results obtained using different theoretical models), is also shown for comparison.\footnote{Note that a recent Belle measurement of partial $B\to X_u\ell^+\nu_\ell$ branching fractions which superseeds their previous result and which yields the somewhat lower value $|V_{ub}| =  4.10(9)(22)(15) \times 10^{-3}$, has not been included in the HFLAV average yet. } 
In this plot the tension between the BaBar and the Belle measurements of $B(B^{-} \to
\tau^{-} \bar{\nu})$ is manifest. As discussed above, it is for this
reason that we do not extract $|V_{ub}|$ through the average of
results for this branching fraction from these two collaborations. In
fact this means that a reliable determination of $|V_{ub}|$ using
information from leptonic $B$-meson decays is still absent;
the situation will only clearly improve with the more precise experimental
data expected from Belle~II~\cite{Urquijo:2015qsa,Kou:2018nap}.
The value for $|V_{ub}|$ obtained from semileptonic $B$ decays for $N_f=2+1$, on the other hand,
is significantly more precise than both the leptonic and the inclusive determinations,
and exhibits a $\sim 1.7\sigma$ tension with the latter.

\subsection{Determination of $|V_{cb}|$}
\label{sec:Vcb}

We will now use the lattice-QCD results for the $B \to D^{(*)}\ell\nu$ form factors
in order to obtain determinations of the CKM matrix element $|V_{cb}|$ in the Standard Model.
The relevant formulae are given in~Eq.~(\ref{eq:vxb:BtoDstar}).

Let us summarize the lattice input that satisfies FLAG requirements for the control
of systematic uncertainties, discussed in Sec.~\ref{sec:BtoD}.
In the (experimentally more precise) $B\to D^*\ell\nu$ channel, there is only one
$N_f=2+1$ lattice computation of the relevant form factor $\mathcal{F}^{B\to D^*}$ at zero recoil.
Concerning the $B \to D\ell\nu$ channel, for $N_f=2$ there is one determination
of the relevant form factor $\mathcal{G}^{B\to D}$ at zero recoil, while
for $N_f=2+1$ there are two determinations of the $B \to D$ form factor as a function
of the recoil parameter in roughly the lowest third of the kinematically allowed region.
In this latter case, it is possible to replicate the analysis carried out for $|V_{ub}|$
in~Sec.~\ref{sec:Vub}, and perform a joint fit to lattice and experimental data;
in the former, the value of $|V_{cb}|$ has to be extracted by matching to the
experimental value for $\mathcal{F}^{B\to D^*}(1)\eta_{\rm EW}|V_{cb}|$ and
$\mathcal{G}^{B\to D}(1)\eta_{\rm EW}|V_{cb}|$.

The latest experimental average by HFLAV~\cite{Amhis:2016xyh} for the $B\to D^*$ form factor at zero recoil makes use of the CLN~\cite{Caprini:1997mu} parameterization of the $B\to D^*$ form factor and is
\begin{gather}
[\mathcal{F}^{B\to D^*}(1)\eta_{\rm EW}|V_{cb}|]_{\rm CLN,HFLAV} = 35.61(43)\times 10^{-3}\,.
\label{eq:BDsHFLAFVCLN}
\end{gather}
Recently the Belle collaboration presented an updated measurement of the $B\to D^* \ell\nu$ branching ratio~\cite{Belle:2018ezy} in which, as suggested in Refs.~\cite{Bigi:2017njr, Bernlochner:2017xyx, Grinstein:2017nlq},  the impact of the form factor parameterization has been studied by comparing the CLN~\cite{Caprini:1997mu} and BGL~\cite{Boyd:1994tt, Boyd:1997kz} ans\"atze. The fit results using the two parameterizations are now consistent. In light of the fact that the BGL parameterization has a much stronger theoretical standing than the CLN one and that it imposes  less stringent constraints on the shape of the form factors, we do not consider the CLN determination any further and focus on the BGL fit:
\begin{align}
[\mathcal{F}^{B\to D^*}(1)\eta_{\rm EW}|V_{cb}|]_{\rm BGL, \; Belle}\ &= 35.44(23)(60) \times 10^{-3}\,,
\label{eq:BDsBelleBCL}
\end{align}
where the first error is statistical and the second is systematic.\footnote{Note that the BGL fit employed by Belle uses very few $z$ parameters and that this could lead to an underestimation of the error on $[\mathcal{F}^{B\to D^*}(1)\eta_{\rm EW}|V_{cb}|$. See Ref.~\cite{Gambino:2019sif} for a through review of this point.} Given the fact that the two determinations in Eqs.~(\ref{eq:BDsHFLAFVCLN}) and (\ref{eq:BDsBelleBCL}) are quite compatible and that the BGL parameterization is on firmer theoretical ground, in the following we present the determination of $|V_{cb}|$ obtained from Eq.~(\ref{eq:BDsBelleBCL}). We refer to the discussion presented at the end of Sec.~8.8 of the previous edition of this review~\cite{Aoki:2019cca} for further comments on the CLN and BGL parameterizations. 

By using $\eta_{\rm EW}=1.00662$~\footnote{Note that this determination does not include the electromagnetic Coulomb correction roughly estimated in Ref.~\cite{Bailey:2014tva}. Currently the numerical impact of this correction is negligible.} and the $N_f = 2 +1$ lattice value for $\mathcal{F}^{B\to D^*}(1)$ in~Eq.~(\ref{eq:BDstarFNAL})~\footnote{In light of our policy not to average $N_f=2+1$ and $N_f = 2+1+1$ calculations and of the controversy over the use of the CLN vs.\ BGL parameterizations, we prefer to simply use only the more precise $N_f=2+1$ determination of $\mathcal{F}^{B\to D^*}(1)$ in Eq.~(\ref{eq:BDstarFNAL}) for the extraction of $V_{cb}$.}, we thus extract the average
\begin{align}
& N_f=2+1 & [B\to D^*\ell\nu]_{\rm BGL, Belle}: && |V_{cb}| &= 38.86(54)(70) \times 10^{-3} \,,
\label{eq:vcbdsav}
\end{align}
where the first uncertainty comes from the lattice computation and the second from the
experimental input. 

For the zero-recoil $B \to D$ form factor, HFLAV~\cite{Amhis:2016xyh} quotes
\begin{gather}
\mbox{HFLAV:} \qquad \mathcal{G}^{B\to D}(1)\eta_{\rm EW}|V_{cb}| =
41.57(45)(89) \times 10^{-3}\,,
\label{eq:BDHFLAV}
\end{gather}
yielding the following average for $N_f=2$:
\begin{align}
&N_f=2&B\to D\ell\nu: && |V_{cb}| &= 40.0(3.7)(1.0) \times 10^{-3} \,,
\end{align}
where the first uncertainty comes from the lattice computation and the second from the experimental input.

Finally, for $N_f=2+1$ we perform, as discussed above, a joint fit to the available
lattice data, discussed in Sec.~\ref{sec:BtoD}, and state-of-the-art experimental determinations.
In this case, we will combine the aforementioned Belle measurement~\cite{Glattauer:2015teq},
which provides partial integrated decay rates in 10 bins in the recoil parameter $w$,
with the 2010 BaBar data set in Ref.~\cite{Aubert:2009ac}, which quotes the value of
$\mathcal{G}^{B\to D}(w)\eta_{\rm EW}|V_{cb}|$ for  ten values of $w$.\footnote{We thank Marcello Rotondo for providing the ten bins result of the BaBar analysis.}
The fit is dominated by the more precise Belle data; given this, and the fact that only partial
correlations among systematic uncertainties are to be expected, we will treat both data sets
as uncorrelated.\footnote{ We have checked that results using just one experimental data set
are compatible within $1\sigma$. In the case of BaBar, we have
taken into account the introduction of some EW corrections in the data.}

A constrained $(N^+ = N^0 = 3)$ BCL fit using the same ansatz as for lattice-only data
in Sec.~\ref{sec:BtoD}, yields our average, which we present in Tab.~\ref{tab:FFVCBD}. The chi-square per degree of freedom is $\chi^2/{\rm dof} = 20.0/25=0.80$. The fit is illustrated in Fig.~\ref{fig:Vcb_SL_fit}. In passing, we
note that, if correlations between the FNAL/MILC and HPQCD
calculations are neglected, the $|V_{cb}|$ central value rises to $40.3
\times 10^{-3}$ in nice agreement with the results presented in
Ref.~\cite{Bigi:2016mdz}.
\begin{table}[t]
\begin{center}
\begin{tabular}{|c|c|cccccc|}
\multicolumn{8}{l}{$B\to D\ell\nu \; (N_f=2+1)$} \\[0.2em]\hline
        & Central Values & \multicolumn{6}{|c|}{Correlation Matrix} \\[0.2em]\hline
$|V_{cb}^{}| \times 10^3$ & 40.0 (1.0) &   1.00 & -0.525 & -0.339 & 0.0487 & -0.521 & -0.433 \\[0.2em]
$a_0^+$                   & 0.8946 (94) &   -0.525 & 1.00 & 0.303 & -0.351 & 0.953 & 0.529 \\[0.2em]
$a_1^+$                   & -8.03 (16)  &  -0.339 & 0.303 & 1.00 & 0.203 & 0.375 & 0.876 \\[0.2em]
$a_2^+$                   & 50.1 (3.1)  &  0.0487 & -0.351 & 0.203 & 1.00 & -0.276 & 0.196 \\[0.2em]
$a_0^0$                   & 0.7804 (75) &   -0.521 & 0.953 & 0.375 & -0.276 & 1.0 & 0.502 \\[0.2em]
$a_1^0$                   & -3.38 (16)  &   -0.433 & 0.529 & 0.876 & 0.196 & 0.502 & 1.0 \\[0.2em]
\hline
\end{tabular}
\end{center}
\caption{$|V_{cb}|$, coefficients for the $N^+ =N^0$ $z$-expansion of the $B\to D$ form factors $f_+$ and $f_0$, and their correlation matrix. The coefficient $a_2^0$ is fixed by the $f_+(q^2=0) = f_0(q^2=0)$ constrain. The chi-square per degree of freedom is $\chi^2/{\rm dof} = 20.0/25=0.80$. The lattice calculations that enter this fit are taken from FNAL/MILC~\cite{Lattice:2015rga} and HPQCD~\cite{Na:2015kha}. The experimental inputs are taken from BaBar~\cite{Aubert:2009ac} and Belle~\cite{Glattauer:2015teq}.
\label{tab:FFVCBD}}
\end{table}

Before discussing the combination of the above $|V_{cb}|$ results, we note that the LHCb Collaboration recently reported
the first determination of $|V_{cb}|$ at the Large Hadron Collider
using $B_s\to D_s^- \mu^+\nu_\mu$ and $B_s\to D^{*-}_s \mu^+\nu_\mu$ decays \cite{LHCb:2020cyw,LHCb:2021qbv}. The differential decay
rates, in combination with the $N_f=2+1+1$ HPQCD 19 \cite{McLean:2019qcx} and HPQCD 19B \cite{McLean:2019sds} lattice results for $f_+^{B_s\to D_s}$
and ${\mathcal F}^{B_s\to D_s^*}(1)$, were analyzed using either the CLN or BGL form-factor parameterizations. The result for $|V_{cb}|$
from the BGL fit is \cite{LHCb:2021qbv}
\begin{align}
|V_{cb}^{}|\times 10^3 &= (41.7 \pm 0.8 \pm 0.9 \pm 1.1)\;\quad B_s\to D^{(*)-}_s \mu^+\nu_\mu,\: {\rm BGL,LHCb}\;. \label{eq:VcbLHCb}
\end{align}
The LHCb analysis used ratios to the reference decay modes $B^0\to D^- \mu^+\nu_\mu$ and $B^0\to D^{*-} \mu^+\nu_\mu$, whose branching fractions
are used as input in the form of the Particle Data Group averages of measurements by other experiments \cite{Tanabashi:2018oca}. The result (\ref{eq:VcbLHCb})
is therefore correlated with the determinations of $|V_{cb}|$ from $B\to D$ and $B\to D^*$ semileptonic decays.
Given the challenges
involved in performing our own fit to the LHCb data, we do not, at present, include the LHCb results for $B_s\to D_s^- \mu^+\nu_\mu$ and $B_s\to D^{*-}_s \mu^+\nu_\mu$ in our combination of $|V_{cb}|$.

We now proceed to combine the determinations of $|V_{cb}|$
from exclusive $B\to D$ and $B\to D^*$ semileptonic decays. To this end, we need to estimate
the correlation between the lattice uncertainties in the two modes.
We assume conservatively that the statistical component of the lattice
error in both determinations are 100\% correlated because they are based
on the same MILC configurations (albeit on different subsets).
We obtain:
\begin{align}
|V_{cb}^{}|\times 10^3 &=   39.36 (68) \;\quad {\rm BGL,Belle} \; .
\end{align}

Our results are summarized in Tab.~\ref{tab:Vcbsummary}, which also
shows the HFLAV inclusive determination of $|V_{cb}|=42.00(64) \times 10^{-3}$~\cite{Gambino:2016jkc} for comparison, and illustrated in Fig.~\ref{fig:Vxbsummary}. Finally, using the fit results in Tab.~\ref{tab:Vcbsummary}, we extract a value for $R(D)$ which includes both lattice and experimental information:
\begin{align}
R(D)_{\rm lat+exp} & = 0.2951 (31) \,,~~~~\mbox{our average.}
\end{align} 
Note that we do not need to rescale the uncertainty on $R(D)_{\rm lat+exp}$ because, after the inclusion of experimental $B\to D \ell\nu \; (\ell =e,\mu)$ results, the shift in central value caused by using a different parameterization is negligible (see the discussion above Eq.~(\ref{HQeq:RDlat})). 

\begin{table}[!t]
\begin{center}
\noindent
\begin{tabular*}{\textwidth}[l]{@{\extracolsep{\fill}}lcc}
 & from  & $|V_{cb}| \times 10^3$\\
&& \\[-2ex]
\hline \hline &&\\[-2ex]
our average for $N_f=2+1$ (BGL) & $B \to D^*\ell\nu$ & $38.86(54)(70)$ \\
our average for $N_f=2+1$ & $B \to D\ell\nu$ &  $40.0(1.0)$  \\
our average for $N_f=2+1$ (BGL) & $B \to (D,D^*)\ell\nu$  & $39.36(68)$ \\
&& \\[-2ex]
 \hline
our average for $N_f=2$ & $B \to D\ell\nu$ & $40.0(3.7)(1.0)$ \\
&& \\[-2ex]
 \hline
LHCb result for $N_f=2+1+1$ (BGL) & $B_s \to D_s^{(*)}\ell\nu$ & $41.7(0.8)(0.9)(1.1)$ \\
&& \\[-2ex]
 \hline
Gambino et al. & $B \to X_c\ell\nu$ & $42.00(64)$ \\
&& \\[-2ex]
\hline \hline && \\[-2ex]
\end{tabular*}
\caption{Results for $|V_{cb}|$. When two errors are quoted in
our averages, the first one comes from the lattice form factor, and the
second from the experimental measurement. The LHCb result using $B_s \to D_s^{(*)}\ell\nu$ decays \cite{LHCb:2020cyw,LHCb:2021qbv,McLean:2019qcx,McLean:2019sds},
as well as the inclusive average obtained in the kinetic scheme from Ref.~\cite{Gambino:2016jkc} are shown for comparison.}
\label{tab:Vcbsummary}
\end{center}
\end{table}
\begin{figure}[!t]
\begin{center}
\includegraphics[width=0.49\textwidth]{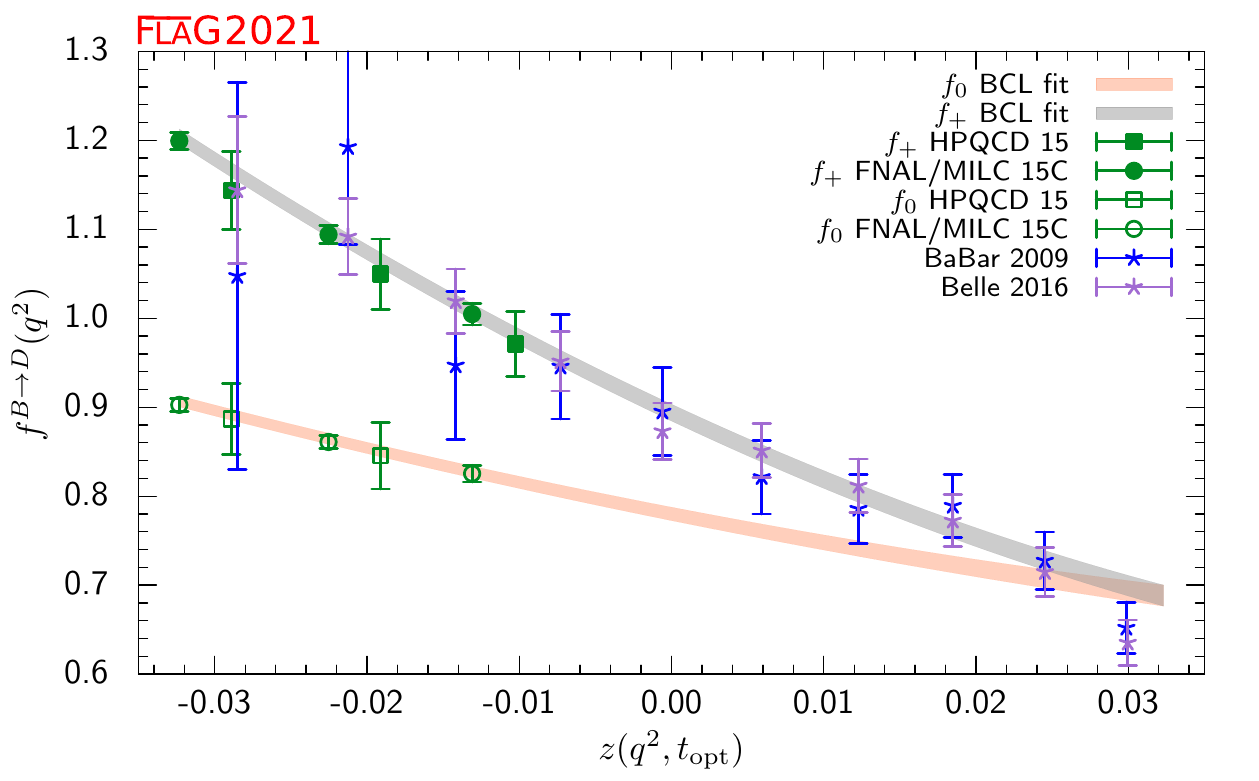}
\includegraphics[width=0.49\textwidth]{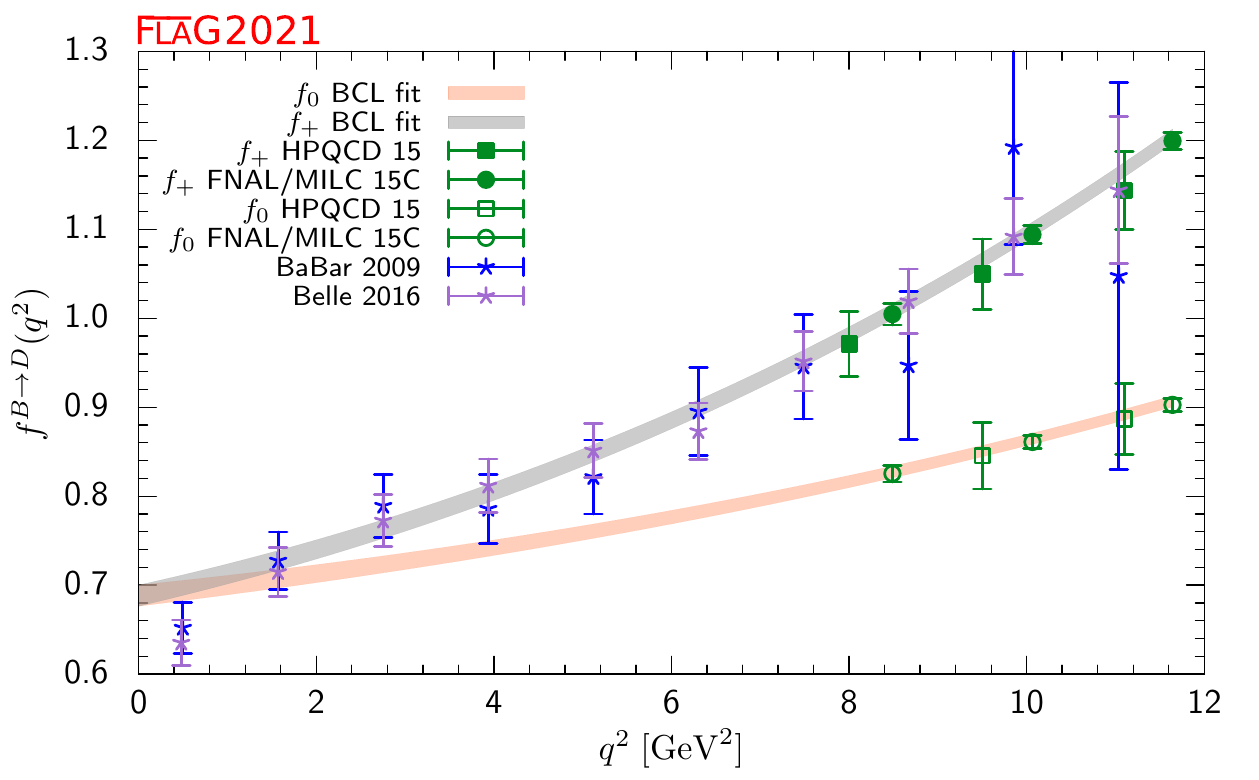}
\caption{
Lattice and experimental data for $f_+^{B\to D}(q^2)$ and $f_0^{B\to D}(q^2)$ versus $z$ (left panel) and $q^2$ (right panel). Green symbols denote lattice-QCD points included in the fit, while blue and indigo
points show experimental data divided by the value of $|V_{cb}|$ obtained from the fit. The grey and orange bands display the preferred $N^+=N^0=3$ BCL fit (five $z$-parameters and $|V_{cb}|$).
}
\label{fig:Vcb_SL_fit}
\end{center}
\end{figure}

\subsection{Determination of $|V_{ub}/V_{cb}|$ from $\Lambda_b$ decays}
\label{sec:VubVcb}

In 2015, the LHCb Collaboration reported a measurement of the ratio \cite{Aaij:2015bfa}
\begin{equation}
R_{\rm BF}(\Lambda_b)=\frac{\displaystyle\int_{15~{\rm GeV}^2}^{q^2_{\rm max}}\frac{{\rm d}\mathcal{B}(\Lambda_b\to p\mu^-\bar\nu_\mu)}{{\rm d}q^2}\,{\rm d}q^2}{\displaystyle\int_{7~{\rm GeV}^2}^{q^2_{\rm max}}\frac{{\rm d}\mathcal{B}(\Lambda_b\to \Lambda_c\mu^-\bar\nu_\mu)}{{\rm d}q^2}\,{\rm d}q^2}, 
\end{equation}
which, combined with the lattice QCD prediction \cite{Detmold:2015aaa} discussed in Sec.~\ref{sec:Lambdab} yields a determination of $|V_{ub}/V_{cb}|$. The LHCb analysis uses the decay $\Lambda_c \to p K \pi$ to reconstruct the $\Lambda_c$ and requires the branching fraction $\mathcal{B}(\Lambda_c \to p K \pi)$ of this decay as an external input. Using the latest world average of $\mathcal{B}(\Lambda_c \to p K \pi)=(6.28\pm0.32)\%$ \cite{Zyla:2020zbs} to update the LHCb measurement gives \cite{Amhis:2019ckw}
\begin{equation}
R_{\rm BF}(\Lambda_b) = (0.92 \pm 0.04 \pm 0.07) \times 10^{-2},
\end{equation}
and, combined with the lattice QCD prediction for $\frac{\zeta_{p\mu\bar\nu}(15{\rm GeV}^2)}{\zeta_{\Lambda_c \mu\bar\nu}(7{\rm GeV}^2)}$ discussed in Sec.~\ref{sec:Lambdab},
\begin{equation}
 |V_{ub}/V_{cb}| = 0.079 \pm 0.004_{\rm\, lat.} \pm 0.004_{\rm\, exp.}.
\end{equation}

\subsection{Determination of $|V_{ub}/V_{cb}|$ from $B_s$ decays}

More recently, LHCb reported the measurements \cite{Aaij:2020nvo}
\begin{eqnarray}
\nonumber R_{\rm BF}(B_s, \text{low}) &=& \frac{\displaystyle\int_{q^2_{\rm min}=m_\mu^2}^{7\:{\rm GeV}^2}\frac{{\rm d}\mathcal{B}(B_s\to K^- \mu^+\nu_\mu)}{{\rm d}q^2}\,{\rm d}q^2}{\mathcal{B}(B_s\to D_s^-\mu^+\nu_\mu)} \\
 &=& (1.66\pm0.12)\times10^{-3}, \\
\nonumber  && \\
\nonumber R_{\rm BF}(B_s, \text{high}) &=& \frac{\displaystyle\int_{7\:{\rm GeV}^2}^{q^2_{\rm max}=(m_{B_s}-m_{K})^2}\frac{{\rm d}\mathcal{B}(B_s\to K^- \mu^+\nu_\mu)}{{\rm d}q^2}\,{\rm d}q^2}{\mathcal{B}(B_s\to D_s^-\mu^+\nu_\mu)} \\
&=& (3.25\pm0.28)\times10^{-3}, \\
 \nonumber && \\
\nonumber R_{\rm BF}(B_s, \text{all}) &=& \frac{\mathcal{B}(B_s\to K^- \mu^+\nu_\mu)}{\mathcal{B}(B_s\to D_s^-\mu^+\nu_\mu)}\\
&=& (4.89\pm0.33)\times10^{-3}.
\end{eqnarray}
Using our average of the $B_s \to K$ form factors from lattice QCD as discussed in Sec.~\ref{sec:BstoK}, we obtain the Standard-Model predictions
\begin{eqnarray}
 \frac{1}{|V_{ub}|^2}\int_{q^2_{\rm min}=m_\mu^2}^{7\:{\rm GeV}^2}\frac{{\rm d}\Gamma(B_s\to K^- \mu^+\nu_\mu)}{{\rm d}q^2} &=& (2.26 \pm 0.38) \:{\rm ps}^{-1}, \\
 \frac{1}{|V_{ub}|^2}\int_{7\:{\rm GeV}^2}^{q^2_{\rm max}=(m_{B_s}-m_{K})^2}\frac{{\rm d}\Gamma(B_s\to K^- \mu^+\nu_\mu)}{{\rm d}q^2} &=& (4.02 \pm 0.31) \:{\rm ps}^{-1}, \\
 \frac{1}{|V_{ub}|^2}\Gamma(B_s\to K^- \mu^+\nu_\mu) &=& (6.28 \pm 0.67) \:{\rm ps}^{-1}.
\end{eqnarray}
For the denominator, we use the $B_s\to D_s$ form factors from Ref.~\cite{McLean:2019qcx}, which yields
\begin{eqnarray}
  \frac{1}{|V_{cb}|^2}\Gamma(B_s\to D_s^- \mu^+\nu_\mu) &=& (9.15 \pm 0.37) \:{\rm ps}^{-1}.
\end{eqnarray}
Combined with the LHCb measurements we obtain
\begin{eqnarray}
 \frac{|V_{ub}|}{|V_{cb}|}(\text{low}) &=& 0.0819  \pm 0.0072_{\rm\, lat.} \pm 0.0029_{\rm\, exp.}, \\
 \frac{|V_{ub}|}{|V_{cb}|}(\text{high}) &=& 0.0860 \pm 0.0037_{\rm\, lat.} \pm 0.0038_{\rm\, exp.} , \\
 \frac{|V_{ub}|}{|V_{cb}|}(\text{all}) &=& 0.0844 \pm 0.0048_{\rm\, lat.} \pm 0.0028_{\rm\, exp.} .
\end{eqnarray}
We note the excellent compatibility of the results in the high and low $q^2$ regions. Nevertheless, we will use the result from the high-$q^2$ region in our combination in Sec.~\ref{sec:VubVcbsummary}, as this is the region in which the form factor shape is most reliably constrained by the lattice data.

\subsection{Summary: $|V_{ub}|$ and $|V_{cb}|$}
\label{sec:VubVcbsummary}

\begin{figure}[!h]
\begin{center}
\includegraphics[width=0.49\linewidth]{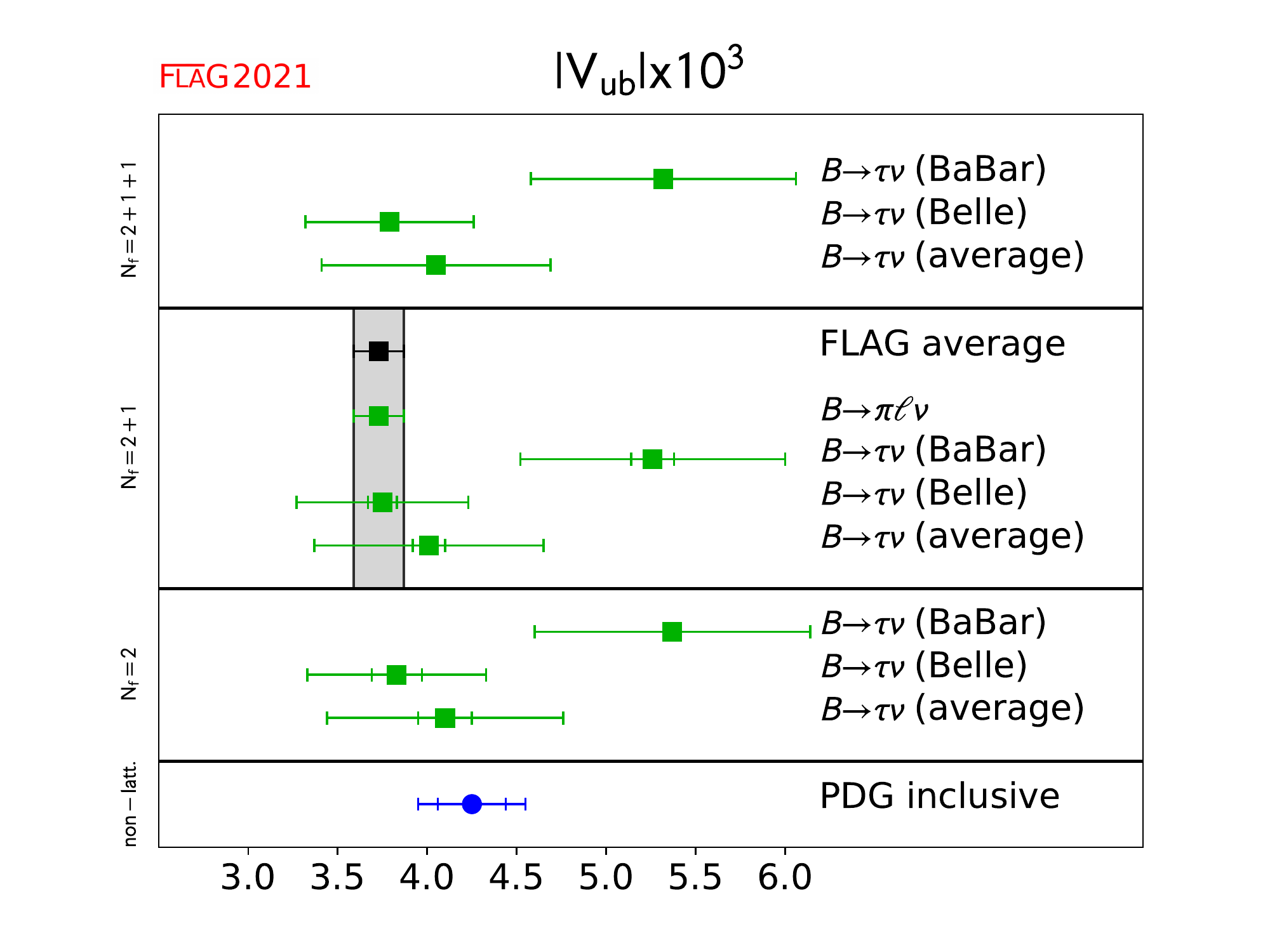}
\includegraphics[width=0.49\linewidth]{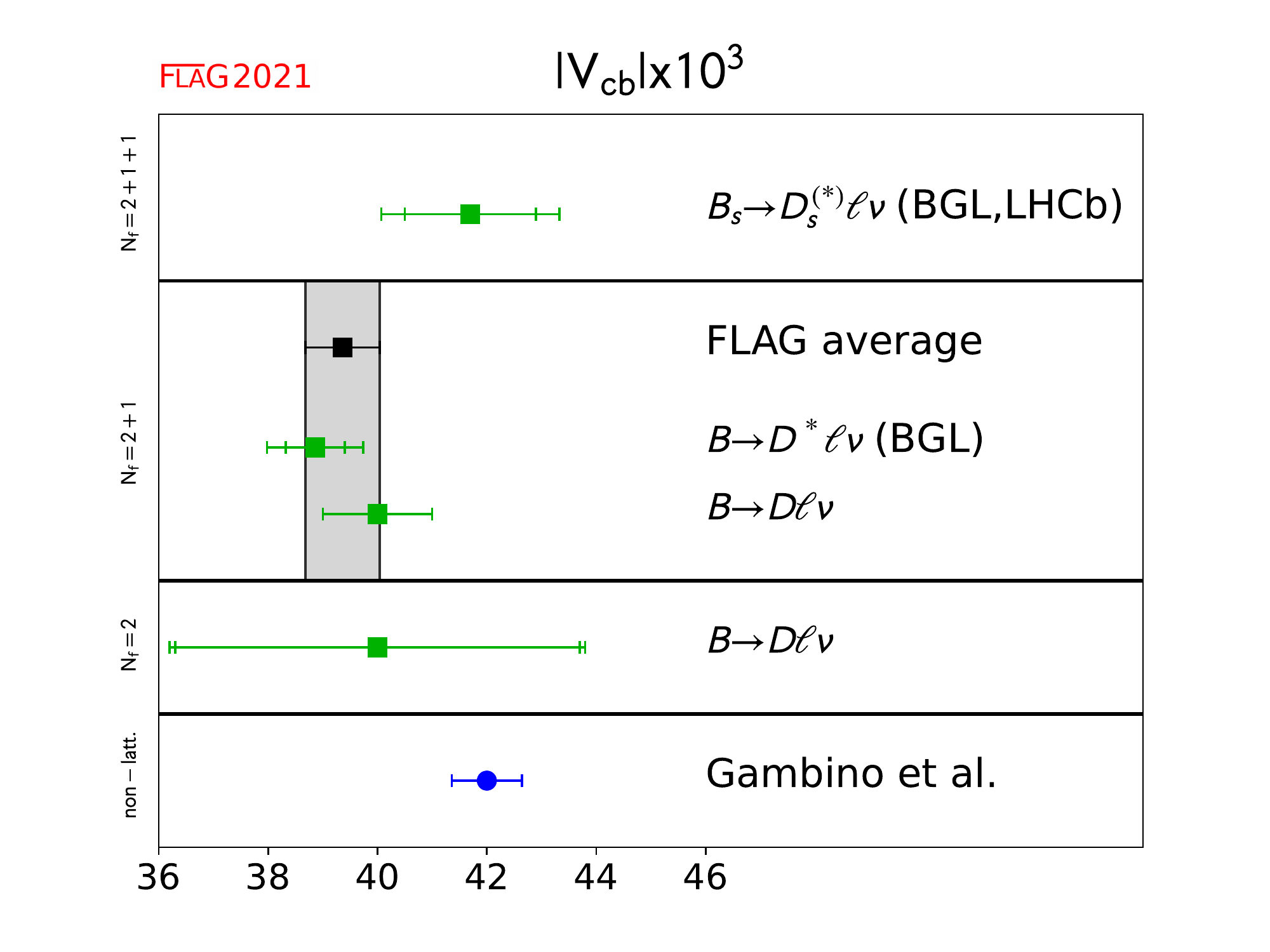}
\caption{Left: Summary of $|V_{ub}|$ determined using: i) the $B$-meson leptonic
decay branching fraction, $B(B^{-} \to \tau^{-} \bar{\nu})$, measured
at the Belle and BaBar experiments, and our averages for $f_{B}$ from
lattice QCD; and ii) the various measurements of the $B\to\pi\ell\nu$
decay rates by Belle and BaBar, and our averages for lattice determinations
of the relevant vector form factor $f_+(q^2)$.
Right: Same for determinations of $|V_{cb}|$ using semileptonic decays.
The inclusive results are taken from Refs.~\cite{Amhis:2016xyh, Gambino:2016jkc}.}
\label{fig:Vxbsummary}
\end{center}
\end{figure}

In Fig.~\ref{fig:VubVcb}, we present a summary of determinations of $|V_{ub}|$ and $|V_{cb}|$ from $B\to (\pi,D^{(*)})\ell\nu$, $B_s\to (K,D_s) \ell\nu$ (high $q^2$ only), $B\to \tau \nu$ and $\Lambda_b\to (p,\Lambda_c)\ell\nu$, as well as the results from inclusive $B\to X_{u,c} \ell\nu$ decays. Note that constraints on $\left| V_{ub}^{}/V_{cb}^{} \right|$ from baryon modes are displayed but, in view of the rating in Tab.~\ref{tab_BottomBaryonSLsumm2}, are not included in the global fit. As discussed in Sec.~\ref{sec:Vcb}, experimental inputs used in the extraction of $|V_{cb}|$ from $B_s \to D_s^{(*)} \ell\nu$ decays~\cite{LHCb:2020cyw,LHCb:2021qbv} given in Eq.~(\ref{eq:VcbLHCb}) are highly correlated with those entering the global $(|V_{ub}|,|V_{cb}|)$ fit described in this section. Given these correlations and the challenges in reproducing the LHCb analysis, for the time being we do not include the result Eq.~(\ref{eq:VcbLHCb}) into the global fit.

Currently, the determinations of $V_{cb}$ from $B\to D^*$ and $B\to D$ decays are quite compatible; however, a sizeable tension involving the extraction of $V_{cb}$ from inclusive decays remains. In the determination of the $1\sigma$ and $2\sigma$ contours for our average, we have included an estimate of the correlation between $|V_{ub}|$ and $|V_{cb}|$ from semileptonic $B$ decays: the lattice inputs to these quantities are dominated by results from the Fermilab/MILC and HPQCD collaborations that are both based on MILC $N_f=2+1$ ensembles, leading to our conservatively introducing a 100\% correlation between the lattice statistical uncertainties of the three computations involved. The results of the fit are
\begin{align}
|V_{cb}^{}| & =  39.48 (68) \times 10^{-3}\;, \\
|V_{ub}^{}| & = 3.63 (14) \times 10^{-3}  \;, \\
p{\rm -value} & = 0.39 \; .
\end{align}
For reference, the inclusive determinations read $|V_{cb}^{}|_{\rm incl} = (42.00 \pm 0.64 ) \times 10^{-3}$~\cite{Gambino:2016jkc} and $|V_{ub}^{}|_{\rm incl} = (4.32 \pm 0.12_{\rm exp} \pm 0.13_{\rm theo} \pm 0.23_{\Delta BF}) \times 10^{-3}$~\cite{Amhis:2019ckw, Zyla:2020zbs} (the $\Delta BF$ error has been added in Ref.~\cite{Zyla:2020zbs} to account for the spread in results obtained using different theoretical models). Note that a recent Belle analysis~\cite{Belle:2021eni} of partial $B\to X_u \ell^+ \nu_\ell$ branching fractions finds a slighly lower central value $|V_{ub}^{}|_{\rm incl, Belle} = (4.10 \pm 0.09_{\rm stat} \pm 0.22_{\rm syst} \pm 0.15_{\rm theo}) \times 10^{-3}$.
\begin{figure}[!h]
\begin{center}
\includegraphics[width=0.6\linewidth]{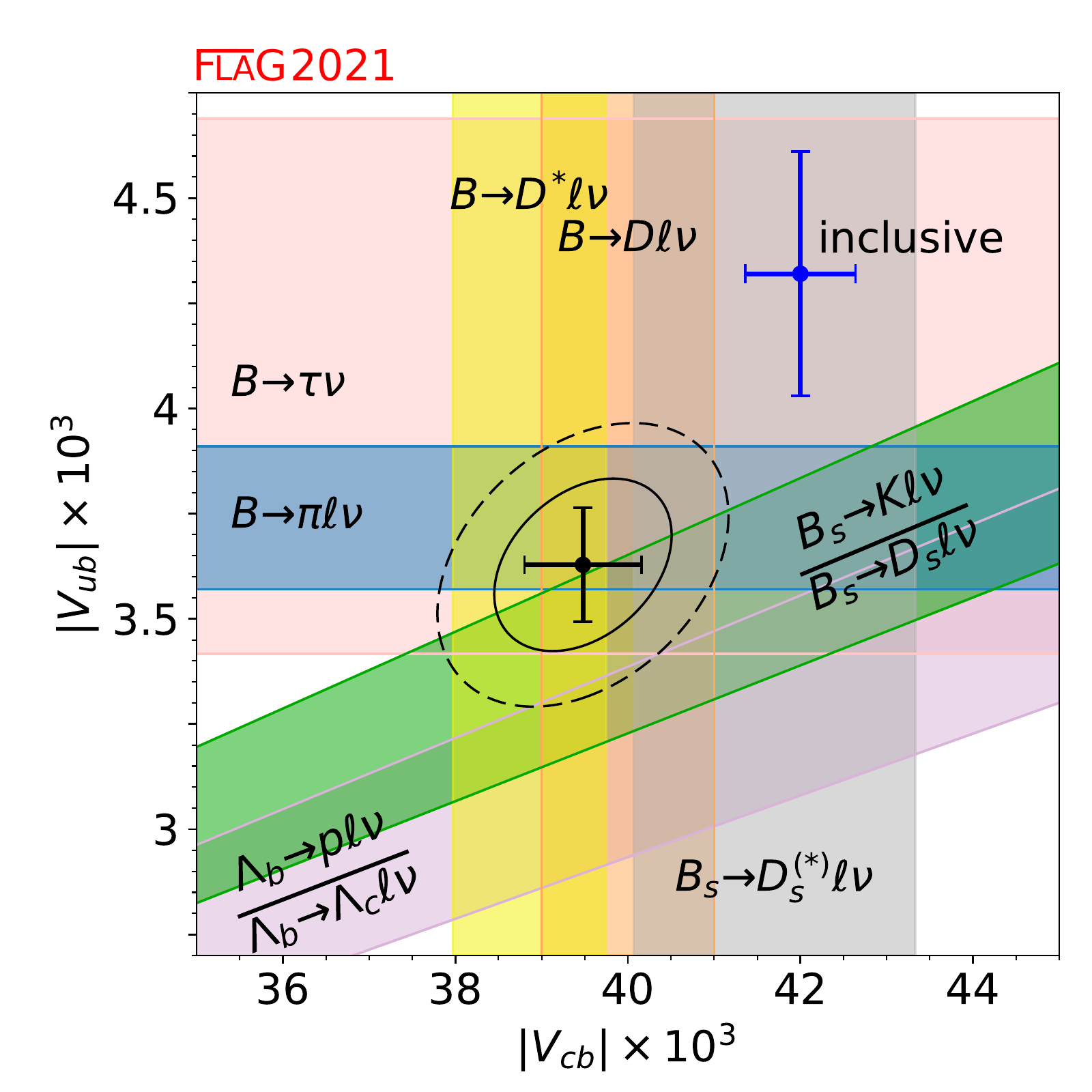}
\caption{Summary of $|V_{ub}|$ and $|V_{cb}|$ determinations. The black solid and dashed lines correspond to 68\% and 95\% C.L. contours, respectively. The result of the global fit (which does not include $\left| V_{ub}^{}/V_{cb}^{} \right|$ from baryon modes or $|V_{cb}|$ from $B_s \to D_s^{(*)}\ell\nu$) is $(|V_{cb}^{}|,|V_{ub}^{}|) = (39.48\pm 0.68, 3.63 \pm 0.14) \times 10^{-3}$ with a $p$-value of 0.39. 
The lattice and experimental results that contribute to the various contours are the following. 
$B\to\pi\ell\nu$: lattice (FNAL/MILC~\cite{Lattice:2015tia} and RBC/UKQCD~\cite{Flynn:2015mha}) and experiment (BaBar~\cite{delAmoSanchez:2010af,Lees:2012vv} and Belle~\cite{Ha:2010rf,Sibidanov:2013rkk}). 
$B\to D\ell\nu$: lattice (FNAL/MILC~\cite{Lattice:2015rga} and HPQCD~\cite{Na:2015kha}) and experiment (BaBar~\cite{Aubert:2009ac} and Belle~\cite{Glattauer:2015teq}).
$B\to D^*\ell\nu$: lattice (FNAL/MILC~\cite{Bailey:2014tva}) and experiment (Belle~\cite{Belle:2018ezy}). 
$B\to \tau\nu$: lattice ($f_B$ determinations in Fig~\ref{fig:fB}) and experiment (BaBar~\cite{Kronenbitter:2015kls} and Belle~\cite{Lees:2012ju}).
$B_s\to K\ell\nu/B_s\to D_s\ell\nu$: lattice (HPQCD~\cite{Bouchard:2014ypa}, RBC/UKQCD~\cite{Lattice:2015tia}, FNAL/MILC~\cite{Bazavov:2019aom}, HPQCD~\cite{McLean:2019qcx}) and experiment (LHCb~\cite{Aaij:2020nvo}).
$\Lambda_b\to p\ell\nu/\Lambda_b\to \Lambda_c \ell\nu$: lattice (Detmold 15~\cite{Detmold:2015aaa}) and experiment (LHCb~\cite{Aaij:2015bfa}).
$B_s \to D_s^* \ell\nu / B_s \to D_s \ell\nu$: lattice (HPQCD 19 \cite{McLean:2019qcx} and HPQCD 19B \cite{McLean:2019sds}) and experiment (LHCb~\cite{LHCb:2020cyw,LHCb:2021qbv}).
The inclusive determinations are taken from Refs.~\cite{Gambino:2016jkc, Amhis:2019ckw, Zyla:2020zbs} and read  $(|V_{cb}^{}|,|V_{ub}^{}|)_{\rm incl}  = (42.00\pm 0.64, 4.32 \pm 0.29) \times 10^{-3}$. \label{fig:VubVcb}}
\end{center}
\end{figure}

\clearpage
\setcounter{section}{8}
\section{The strong coupling $\alpha_{\rm s}$}
\label{sec:alpha_s}

Authors\footnote{There is a strong overlap with the  FLAG 19 report's section on $\alpha_s$, authored by
R. Horsley, T. Onogi and R. Sommer~\cite{Aoki:2019cca}. In particular the introduction,
and the description of methods without new data have been taken over almost unchanged.} : R.~Horsley, P.~Petreczky, S.~Sint\\

\newcommand{\todoalpha}[1]{{\color{magenta} \bf [ todo: #1]}}

\subsection{Introduction}

\label{introduction}


The strong coupling $\gbar_s(\mu)$ defined at scale $\mu$, plays a key
role in the understanding of QCD and in its application to collider
physics. For example, the parametric uncertainty from $\alpha_s$ is one of
the dominant sources of uncertainty in the Standard-Model prediction for
the $H \to b\bar{b}$ partial width, and the largest source of uncertainty
for $H \to gg$.  
Thus higher precision determinations of $\alpha_s$ are
needed to maximize the potential of experimental measurements at the LHC,
and for high-precision Higgs studies at future
colliders
and the study of the stability of the 
vacuum~\cite{Dittmaier:2012vm,Heinemeyer:2013tqa,Adams:2013qkq,Dawson:2013bba,Accardi:2016ndt,Lepage:2014fla,Buttazzo:2013uya,Espinosa:2013lma}. 
The value of $\alpha_s$ also yields one of the essential boundary conditions
for completions of the Standard Model at high energies. 

In order to determine the running coupling at scale $\mu$
\begin{eqnarray}
   \alpha_s(\mu) = { \gbar^2_{s}(\mu) \over 4\pi} \,,
\end{eqnarray}
we should first ``measure'' a short-distance quantity ${\oO}$ at scale
$\mu$ either experimentally or by lattice calculations, and then 
match it to a perturbative expansion in terms of a running coupling,
conventionally taken as $\alpha_{\overline{\rm MS}}(\mu)$,
\begin{eqnarray}
   {\oO}(\mu) = c_1 \alpha_{\overline{\rm MS}}(\mu)
              +  c_2 \alpha_{\overline{\rm MS}}(\mu)^2 + \cdots \,.
\label{eq:alpha_MSbar}
\end{eqnarray}
The essential difference between continuum determinations of
$\alpha_s$ and lattice determinations is the origin of the values of
$\oO$ in \eq{eq:alpha_MSbar}.

The basis of continuum determinations are 
experimentally measurable cross sections or decay widths from which $\oO$ is
defined. These cross sections have to be sufficiently inclusive 
and at sufficiently high scales such that perturbation theory 
can be applied. Often hadronization corrections have to be used
to connect the observed hadronic cross sections to the perturbative
ones. Experimental data at high $\mu$, where perturbation theory
is progressively more precise, usually have increasing experimental errors, 
and it is  not easy to find processes that allow one
to follow the $\mu$-dependence of a single $\oO(\mu)$ over
a range where $\alpha_s(\mu)$ changes significantly and precision is 
maintained.

In contrast, in lattice gauge theory, one can design $\oO(\mu)$ as
Euclidean short-distance quantities that are not directly related to
experimental observables. This allows us to follow the $\mu$-dependence until the perturbative regime is reached and
nonperturbative ``corrections'' are negligible.  The only
experimental input for lattice computations of $\alpha_s$ is the
hadron spectrum which fixes the overall energy scale of the theory and
the quark masses. Therefore experimental errors are completely
negligible and issues such as hadronization do not occur.  We can
construct many short-distance quantities that are easy to calculate
nonperturbatively in lattice simulations with small statistical
uncertainties.  We can also simulate at parameter values that do not
exist in nature (for example, with unphysical quark masses between
bottom and charm) to help control systematic uncertainties.  These
features mean that precise results for $\alpha_s$ can be achieved
with lattice-gauge-theory computations.  Further, as in the continuum,
the different methods available to determine $\alpha_s$ in
lattice calculations with different associated systematic
uncertainties enable valuable cross-checks.  Practical limitations are
discussed in the next section, but a simple one is worth mentioning
here. Experimental results (and therefore the continuum
determinations) of course have all quarks present, while in lattice
gauge theories in practice only the lighter ones are included and one
is then forced to use the matching at thresholds, as discussed in the following
subsection.

It is important to keep in mind that the dominant source of uncertainty
in most present day lattice-QCD calculations of $\alpha_s$ are from
the truncation of continuum/lattice perturbation theory and from
discretization errors. Perturbative truncation errors are of particular concern because they often cannot easily be estimated
from studying the data itself. Further, the size
of higher-order coefficients in the perturbative series can sometimes
turn out to be larger than naive expectations based on power counting
from the behaviour of lower-order terms.  We note
that perturbative truncation errors are also the dominant
source of uncertainty in several of the phenomenological
determinations of $\alpha_s$.  

The various phenomenological approaches to determining the running
coupling constant, $\alpha^{(5)}_{\overline{\rm MS}}(M_Z)$ are summarized by the
Particle Data Group \cite{Zyla:2020zbs}.
The PDG review lists five categories of phenomenological results 
used to obtain the running coupling: using hadronic
$\tau$ decays, hadronic final states of $e^+e^-$ annihilation,
deep inelastic lepton--nucleon scattering, electroweak precision data, and high energy hadron collider data.
Excluding lattice results, the PDG quotes the weighted average as
\begin{eqnarray}
   \alpha^{(5)}_{\overline{\rm MS}}(M_Z) &=& 0.1176(11) \,, \quad 
   \mbox{PDG 20 \cite{Zyla:2020zbs}}
\label{PDG_nolat}
\end{eqnarray}
compared to 
$
   \alpha^{(5)}_{\overline{\rm MS}}(M_Z) = 0.1174(16) 
$
of the older PDG 2018 \cite{Tanabashi:2018oca}.
For a general overview of the various phenomenological
and lattice approaches 
see, e.g., Ref.~
 \cite{Salam:2017qdl}. 
The extraction of $\alpha_s$ from $\tau$ data, which is one of the most precise and thus
has a large impact on the nonlattice average in Eq.~(\ref{PDG_nolat}), is
especially sensitive to the treatment of higher-order perturbative
terms as well as the treatment of nonperturbative effects.  
This is important to keep in mind when comparing our chosen
range for $\alpha^{(5)}_{\overline{\rm MS}}(M_Z)$ from lattice
determinations in Eq.~(\ref{eq:alpmz}) with the nonlattice average
from the PDG.

\subsubsection{Scheme and scale dependence of $\alpha_s$ and $\Lambda_{\rm QCD}$}

Despite the fact that the notion of the QCD coupling is 
initially a perturbative concept, the associated $\Lambda$ parameter
is nonperturbatively defined
\begin{eqnarray}
   \Lambda 
      &\equiv& \mu\,\varphi_s(\gbar_s(\mu)),\nonumber\\
      \varphi_s(\gbar_s) &=&
      (b_0\gbar_s^2)^{-b_1/(2b_0^2)} 
              e^{-1/(2b_0\gbar_s^2)}
             \exp\left[ -\int_0^{\gbar_s}\,dx 
                        \left( {1\over \beta(x)} + {1 \over b_0x^3} 
                                                 - {b_1 \over b_0^2x}
                        \right) \right] \,,
\nonumber\\ \label{eq:Lambda}
\end{eqnarray}
where $\beta(\gbar_s) = \mu\frac{\partial\gbar_s(\mu)}{\partial\mu}$ is the full renormalization group function in the scheme
which defines $\gbar_s$, and $b_0$ and $b_1$ are the first two
scheme-independent coefficients of the perturbative expansion
\begin{eqnarray}
   \beta(x) \sim -b_0 x^3 -b_1 x^5 + \ldots \,,
\label{eq:beta_pert}
\end{eqnarray}
with
\begin{eqnarray}
   b_0 = {1\over (4\pi)^2}
           \left( 11 - {2\over 3}N_f \right) \,, \qquad
   b_1 = {1\over (4\pi)^4}
           \left( 102 - {38 \over 3} N_f \right) \,.
\label{b0+b1}
\end{eqnarray}
Thus the $\Lambda$ parameter is renormalization-scheme-dependent but in an
exactly computable way, and lattice gauge theory is an ideal method to
relate it to the low-energy properties of QCD.
In the $\overline{\rm MS}$ scheme presently $b_{n_l}$
up to $n_l = 4$ are known~\cite{vanRitbergen:1997va,Czakon:2004bu,Luthe:2016ima,Herzog:2017ohr,Baikov:2016tgj}.

The change in the coupling from one scheme $S$ to another (taken here
to be the $\overline{\rm MS}$ scheme) is perturbative,
\begin{eqnarray}
   g_{\overline{\rm MS}}^2(\mu) 
      = g_{\rm S}^2(\mu) (1 + c^{(1)}_g g_{\rm S}^2(\mu) + \ldots ) \,,
\label{eq:g_conversion}
\end{eqnarray}
where $c^{(i)}_g, \, i\geq 1$ are finite renormalization coefficients.  The
scale $\mu$ must be taken high enough for the error in keeping only
the first few terms in the expansion to be small.
On the other hand, the conversion to the $\Lambda$ parameter
in the $\overline{\rm MS}$ scheme is given exactly by
\begin{eqnarray}
   \Lambda_{\overline{\rm MS}} 
      = \Lambda_{\rm S} \exp\left[ c_g^{(1)}/(2b_0)\right] \,.
      \label{eq:Lambdaconversion}
\end{eqnarray}

The fact that $\Lambda_\msbar$ can be obtained exactly from
$\Lambda_S$ in any scheme $S$ where $c^{(1)}_g$ is known 
together with the high-order knowledge (5-loop by now) of
$\beta_\msbar$ means that the errors in $\alpha_\msbar(m_\mathrm{Z})$
are dominantly due to the  errors of $\Lambda_S$. We will therefore
mostly discuss them in that way. 
Starting from \eq{eq:Lambda}, we have to consider (i) the
error of $\gbar_S^2(\mu)$ (denoted as $\left(\frac{\Delta \Lambda}{\Lambda}\right)_{\Delta \alpha_S}$ ) and (ii) the truncation error in $\beta_S$ (denoted as $\left( \frac{\Delta \Lambda}{\Lambda}\right)_{\rm trunc}$).
Concerning (ii), note that knowledge of $c_g^{(n_l)}$ for the scheme $S$ means that $\beta_S$ is known to $n_l+1$ loop order; $b_{n_l}$ is known. We thus see that in the region where 
perturbation theory can be applied, the following errors of $\Lambda_S$ (or consequently $\Lambda_{\overline{\rm MS}}$) have to be considered
\begin{eqnarray}
  \left(\frac{\Delta \Lambda}{\Lambda}\right)_{\Delta \alpha_S} &=& \frac{\Delta \alpha_{S}(\mu)}{ 8\pi b_0 \alpha_{S}^2(\mu)} \times \left[1 + \cO(\alpha_S(\mu))\right]\,,
  \label{eq:i}\\
 \left( \frac{\Delta \Lambda}{\Lambda}\right)_{\rm trunc} &=& k \alpha_{S}^{n_\mathrm{l}}(\mu) + \cO(\alpha_S^{n_\mathrm{l}+1}(\mu))\,,
  \label{eq:ii}  
\end{eqnarray}
where $k$ depends on $b_{n_\mathrm{l}+1}$ and in typical 
good schemes such as $\msbar$ it is numerically of order one. 
Statistical and systematic errors such as discretization
effects contribute to  $\Delta \alpha_{S}(\mu)$.  In the above we dropped a 
scheme subscript for the $\Lambda$-parameters because of
      \eq{eq:Lambdaconversion}.

By convention $\alpha_\msbar$ is usually quoted at a scale $\mu=M_Z$
where the appropriate effective coupling is the one in the
5-flavour theory: $\alpha^{(5)}_{\overline{\rm MS}}(M_Z)$.  In
order to obtain it from a result with fewer flavours, one connects effective
theories with different number of flavours as discussed by Bernreuther
and Wetzel~\cite{Bernreuther:1981sg}.  For example, one considers the
$\msbar$ scheme, matches the 3-flavour theory to the 4-flavour
theory at a scale given by the charm-quark mass~\cite{Chetyrkin:2005ia,Schroder:2005hy,Kniehl:2006bg}, runs with the
5-loop $\beta$-function~\cite{vanRitbergen:1997va,Czakon:2004bu,Luthe:2016ima,Herzog:2017ohr,Baikov:2016tgj} of the 4-flavour theory to a scale given by
the $b$-quark mass, and there matches to the 5-flavour theory, after
which one runs up to $\mu=M_Z$ with the 5-loop $\beta$ function.
For the matching relation at a given
quark threshold we use the mass $m_\star$ which satisfies $m_\star=
\overline{m}_\msbar(m_\star)$, where $\overline{m}$ is the running
mass (analogous to the running coupling). Then
\begin{eqnarray}
\label{e:convnfm1}
 \gbar^2_{N_f-1}(m_\star) =  \gbar^2_{N_f}(m_\star)\times 
      [1+ 0\times\gbar^{2}_{N_f}(m_\star) + \sum_{n\geq 2}t_n\,\gbar^{2n}_{N_f}(m_\star)]
\label{e:grelation}
\end{eqnarray}
{with  \cite{Grozin:2011nk,Kniehl:2006bg,Chetyrkin:2005ia} }
\def\nli{(N_f-1)}
\begin{eqnarray}
  t_2 &=&  {1 \over (4\pi^2)^2} {11\over72}\,,\\
  t_3 &=&  {1 \over (4\pi^2)^3} \left[- {82043\over27648}\zeta_3 + 
                     {564731\over124416}-{2633\over31104}(N_f-1)\right]\,, \\
  t_4 &=& {1 \over (4\pi^2)^4} \big[5.170347 - 1.009932 \nli - 0.021978 \,\nli^2\big]\,,
\end{eqnarray}
(where $\zeta_3$ is the Riemann zeta-function) provides the matching
at the thresholds in the $\msbar$ scheme.  Often the package {\tt RunDec}
is used for quark-threshold matching and running in the $\msbar$-scheme \cite{Chetyrkin:2000yt,Herren:2017osy}.

While $t_2,\,t_3,\,t_4$ are
numerically small coefficients, the charm-threshold scale is also
relatively low and so there are nonperturbative
uncertainties in the matching procedure, which are difficult to
estimate but which we assume here to be negligible.
Obviously there is no perturbative matching formula across
the strange ``threshold''; here matching is entirely nonperturbative.
Model dependent extrapolations of $\gbar^2_{N_f}$ from $N_f=0,2$ to
$N_f=3$ were done in the early days of lattice gauge theory. We will
include these in our listings of results but not in our estimates,
since such extrapolations are based on untestable assumptions.

\subsubsection{Overview of the review of $\alpha_s$}

We begin by explaining lattice-specific difficulties in \sect{s:crit}
and the FLAG criteria designed to assess whether the
associated systematic uncertainties can be controlled and estimated in
a reasonable manner. These criteria are taken over unchanged from 
the FLAG 19 report, as there has not yet been sufficiently broad progress
to make these criteria more stringent. We would also like to point to
a recent review~\cite{DelDebbio:2021ryq} of lattice methodology and systematic uncertainties 
for $\alpha_s$. There, a systematic scale variation is advocated to assess systematic
errors due to the truncation of the perturbative series and such a procedure may indeed 
be incorporated into future FLAG criteria, as it can be applied without change
to most lattice approaches. 

We then discuss, in \sect{s:SF} -- \sect{s:glu},
the various lattice approaches and results 
from calculations with $N_f = 0$, 2, 2+1, and 2+1+1 flavours.

Besides new results and upgrades of previous works,
a new strategy of nonperturbative renormalization by decoupling 
has been proposed by the ALPHA collaboration~\cite{DallaBrida:2019mqg},
which shifts the perspective on results with unphysical flavour
numbers, in particular for $\Nf=0$. As these can be nonperturbatively
related to $\Nf>0$ results by a nonperturbative matching calculation,
it becomes very  important to obtain precise and controlled $\Nf=0$ results,
with obvious implications for this and future FLAG reports.
A short account of the decoupling strategy is given in \sect{s:dec}.

In Sec.~\ref{s:alpsumm}, we present averages together with our best
estimates for $\alpha_{\overline{\rm MS}}^{(5)}$. These are currently determined
from 3- and 4-flavour QCD simulations only, however, in the near future
the decoupling strategy is expected to link e.g. 3-flavour simulations with the pure
gauge theory simulations.
Therefore, for the $\Lambda$ parameter, we also give results for other numbers of flavours,
including $\Nf=0$ and $\Nf=2$.



\subsubsection{Additions with respect to the FLAG 19 report}
\label{sec:npapers}
The additional papers since the FLAG 19 report are:

\begin{itemize}
\item[]
    Dalla Brida 19 \cite{DallaBrida:2019wur}
    and Nada 20 \cite{Nada:2020jay}
    from step-scaling methods (\sect{s:SF}).
\item[]
    ALPHA 19A \cite{DallaBrida:2019mqg}
    from the decoupling method (\sect{s:dec}).
\item[]
    TUMQCD 19  \cite{Bazavov:2019qoo} and
    Ayala 20    \cite{Ayala:2020odx} and
    Husung 20 \cite{Husung:2020pxg}
    from the static quark potential (\sect{s:qq}).
\item[]
    Cali 20 \cite{Cali:2020hrj}
    from (light-quark) vacuum polarization in position space (\sect{s:vac}).
\item[] 
    Petreczky 20 \cite{Petreczky:2020tky},
    Petreczky 19 \cite{Petreczky:2019ozv},
    and Boito 20 \cite{Boito:2019pqp,Boito:2020lyp}
    from heavy-quark current two-point functions (\sect{s:curr}).
\item[]
    Zafeiropoulos 19 \cite{Zafeiropoulos:2019flq}
    from QCD vertices (\sect{s:glu}).
\end{itemize}
    
\subsection{General issues}

\subsubsection{Discussion of criteria for computations entering the averages}


\label{s:crit}


As in the PDG review, we only use calculations of $\alpha_s$ published
in peer-reviewed journals, and that use NNLO or higher-order
perturbative expansions, to obtain our final range in
Sec.~\ref{s:alpsumm}.  We also, however, introduce further
criteria designed to assess the ability to control important
systematics, which we describe here.  Some of these criteria, 
e.g., that for the continuum extrapolation, are associated with
lattice-specific systematics and have no continuum analogue.  Other
criteria, e.g., that for the renormalization scale, could in
principle be applied to nonlattice determinations.
Expecting that lattice calculations
will continue to improve significantly in the near future, our goal in
reviewing the state-of-the-art here is to be conservative and avoid
prematurely choosing an overly small range.

In lattice calculations, we generally take ${\oO}$ to be some
combination of physical amplitudes or Euclidean correlation functions
which are free from UV and IR divergences and have a well-defined
continuum limit.  Examples include the force between static quarks and
two-point functions of quark-bilinear currents.

In comparison to values of observables ${\oO}$ determined
experimentally, those from lattice calculations require two more
steps.  The first step concerns setting the scale $\mu$ in \mbox{GeV},
where one needs to use some experimentally measurable low-energy scale
as input. Ideally one employs a hadron mass. Alternatively convenient
intermediate scales such as $\sqrt{t_0}$, $w_0$, $r_0$, $r_1$,
\cite{Luscher:2010iy,Borsanyi:2012zs,Sommer:1993ce,Bernard:2000gd} can
be used if their relation to an experimental dimensionful observable
is established. The low-energy scale needs to be computed at the same
bare parameters where ${\oO}$ is determined, at least as long as
one does not use the step-scaling method (see below).  This induces a
practical difficulty given present computing resources.  In the
determination of the low-energy reference scale the volume needs to be
large enough to avoid finite-size effects. On the other hand, in order
for the perturbative expansion of Eq.~(\ref{eq:alpha_MSbar}) to be
reliable, one has to reach sufficiently high values of $\mu$,
i.e., short enough distances. To avoid uncontrollable discretization
effects the lattice spacing $a$ has to be accordingly small.  This
means
\begin{eqnarray}
   L \gg \mbox{hadron size}\sim \Lambda_{\rm QCD}^{-1}\quad 
   \mbox{and} \quad  1/a \gg \mu \,,
   \label{eq:scaleproblem}
\end{eqnarray}
(where $L$ is the box size) and therefore
\begin{eqnarray} 
   L/a \ggg \mu/\Lambda_{\rm QCD} \,.
   \label{eq:scaleproblem2}
\end{eqnarray}
The currently available computer power, however, limits $L/a$, 
typically to
$L/a = 32-96$. 
Unless one accepts compromises in controlling  discretization errors
or finite-size effects, this means one needs to set 
the scale $\mu$ according to
\begin{eqnarray}
   \mu \lll L/a \times \Lambda_{\rm QCD} & \sim 10-30\, \mbox{GeV} \,.
\end{eqnarray}
(Here $\lll$ or $\ggg$ means at least one order of magnitude smaller or larger.) 
Therefore, $\mu$ can be $1-3\, \mbox{GeV}$ at most.
This raises the concern whether the asymptotic perturbative expansion
truncated at $1$-loop, $2$-loop, or $3$-loop in Eq.~(\ref{eq:alpha_MSbar})
is sufficiently accurate. There is a finite-size scaling method,
usually called step-scaling method, which solves this problem by identifying 
$\mu=1/L$ in the definition of ${\oO}(\mu)$, see \sect{s:SF}. 

For the second step after setting the scale $\mu$ in physical units
($\mbox{GeV}$), one should compute ${\oO}$ on the lattice,
${\oO}_{\rm lat}(a,\mu)$ for several lattice spacings and take the continuum
limit to obtain the left hand side of Eq.~(\ref{eq:alpha_MSbar}) as
\begin{eqnarray}
   {\oO}(\mu) \equiv \lim_{a\rightarrow 0} {\oO}_{\rm lat}(a,\mu) 
              \mbox{  with $\mu$ fixed}\,.
\end{eqnarray}
This is necessary to remove the discretization error.

Here it is assumed that the quantity ${\oO}$ has a continuum limit,
which is regularization-independent. 
The method discussed in \sect{s:WL}, which is based on the perturbative
expansion of a lattice-regulated, divergent short-distance quantity
$W_{\rm lat}(a)$ differs in this respect and must be
treated separately.

In summary, a controlled determination of $\alpha_s$ 
needs to satisfy the following:
\begin{enumerate}

   \item The determination of $\alpha_s$ is based on a
         comparison of a
         short-distance quantity ${\oO}$ at scale $\mu$ with a well-defined
         continuum limit without UV and IR divergences to a perturbative
         expansion formula in Eq.~(\ref{eq:alpha_MSbar}).

   \item The scale $\mu$ is large enough so that the perturbative expansion
         in Eq.~(\ref{eq:alpha_MSbar}) is precise 
         to the order at which it is truncated,
         i.e., it has good {\em asymptotic} convergence.
         \label{pt_converg}

   \item If ${\oO}$ is defined by physical quantities in infinite volume,  
         one needs to satisfy \eq{eq:scaleproblem2}.
         \label{constraints}

   \item[] Nonuniversal quantities need a separate discussion, see
        \sect{s:WL}.

\end{enumerate}

Conditions \ref{pt_converg}. and \ref{constraints}. give approximate lower and
upper bounds for $\mu$ respectively. It is important to see whether there is a
window to satisfy \ref{pt_converg}. and \ref{constraints}. at the same time.
If it exists, it remains to examine whether a particular lattice
calculation is done inside the window or not. 

Obviously, an important issue for the reliability of a calculation is
whether the scale $\mu$ that can be reached lies in a regime where
perturbation theory can be applied with confidence. However, the value
of $\mu$ does not provide an unambiguous criterion. For instance, the
Schr\"odinger Functional, or SF-coupling (Sec.~\ref{s:SF}) is
conventionally taken at the scale $\mu=1/L$, but one could also choose
$\mu=2/L$. Instead of $\mu$ we therefore define an effective
$\alpha_{\rm eff}$.  For schemes such as SF (see Sec.~\ref{s:SF}) or
$qq$ (see Sec.~\ref{s:qq}) this is directly the coupling  of
the scheme. For other schemes such as the vacuum polarization we use
the perturbative expansion \eq{eq:alpha_MSbar} for the observable
${\oO}$ to define
\begin{eqnarray}
   \alpha_{\rm eff} =  {\oO}/c_1 \,.
   \label{eq:alpeff}
\end{eqnarray}
If there is an $\alpha_s$-independent term it should first be subtracted.
Note that this is nothing but defining an effective,
regularization-independent coupling,
a physical renormalization scheme.

Let us now comment further on the use of the perturbative series.
Since it is only an asymptotic expansion, the remainder $R_n({\oO})={\oO}-\sum_{i\leq n}c_i \alpha_s^i$ of a truncated
perturbative expression ${\oO}\sim\sum_{i\leq n}c_i \alpha_s^i$
cannot just be estimated as a perturbative error $k\,\alpha_s^{n+1}$.
The error is nonperturbative. Often one speaks of ``nonperturbative
contributions'', but nonperturbative and perturbative cannot be
strictly separated due to the asymptotic nature of the series (see,
e.g., Ref.~\cite{Martinelli:1996pk}).

Still, we do have some general ideas concerning the 
size of nonperturbative effects. The known ones such as instantons
or renormalons decay for large $\mu$ like inverse powers of $\mu$
and are thus roughly of the form 
\begin{eqnarray}
   \exp(-\gamma/\alpha_s) \,,
\end{eqnarray}
with some positive constant $\gamma$. Thus we have,
loosely speaking,
\begin{eqnarray}
   {\oO} = c_1 \alpha_s + c_2 \alpha_s^2 + \ldots + c_n\alpha_s^n
                  + \cO(\alpha_s^{n+1}) 
                  + \cO(\exp(-\gamma/\alpha_s)) \,.
   \label{eq:Owitherr}
\end{eqnarray}
For small $\alpha_s$, the $\exp(-\gamma/\alpha_s)$ is negligible.
Similarly the perturbative estimate for the magnitude of
relative errors in \eq{eq:Owitherr} is small; as an
illustration for $n=3$ and $\alpha_s = 0.2$ the relative error
is $\sim 0.8\%$ (assuming coefficients $|c_{n+1} /c_1 | \sim 1$).

For larger values of $\alpha_s$ nonperturbative effects can become
significant in Eq.~(\ref{eq:Owitherr}). An instructive example comes
from the values obtained from $\tau$
decays, for which $\alpha_s\approx 0.3$. Here, different applications
of perturbation theory (fixed order and contour improved)
each look reasonably asymptotically convergent~\footnote{See, however, the recent discussion in~\cite{Hoang:2021nlz}.} 
but the difference does not seem to decrease much with the order (see, e.g., the contribution
of Pich in Ref.~\cite{Bethke:2011tr}). In addition nonperturbative terms
in the spectral function may be nonnegligible even after the
integration up to $m_\tau$ (see, e.g., Refs.~\cite{Boito:2014sta}, \cite{Boito:2016oam}). 
All of this is because $\alpha_s$ is not really small.

Since the size of the nonperturbative effects is very hard to
estimate one should try to avoid such regions of the coupling.  In a
fully controlled computation one would like to verify the perturbative
behaviour by changing $\alpha_s$ over a significant range instead of
estimating the errors as $\sim \alpha_s^{n+1}$ .  Some computations
try to take nonperturbative power `corrections' to the perturbative
series into account by including such terms in a fit to the $\mu$-dependence. We note that this is a delicate procedure, both because
the separation of nonperturbative and perturbative is theoretically
not well defined and because in practice a term like, e.g.,
$\alpha_s(\mu)^3$ is hard to distinguish from a $1/\mu^2$ term when
the $\mu$-range is restricted and statistical and systematic errors
are present. We consider it safer to restrict the fit range to the
region where the power corrections are negligible compared to the
estimated perturbative error.

The above considerations lead us to the following special
criteria for the determination of $\alpha_s$: 

\begin{itemize}
   \item Renormalization scale         
         \begin{itemize}
            \item[\good] all points relevant in the analysis have
             $\alpha_\mathrm{eff} < 0.2$
            \item[\soso] all points have $\alpha_\mathrm{eff} < 0.4$
                         and at least one 
                         $\alpha_\mathrm{eff} \le 0.25$
            \item[\bad]  otherwise                                   
         \end{itemize}

   \item Perturbative behaviour 
        \begin{itemize}
           \item[\good] verified over a range of a factor $4$ change
                        in $\alpha_\mathrm{eff}^{n_\mathrm{l}}$ without power
                        corrections  or alternatively 
                        $\alpha_\mathrm{eff}^{n_\mathrm{l}} \le \frac12 \Delta \alpha_\mathrm{eff} / (8\pi b_0 \alpha_\mathrm{eff}^2) $ is reached
           \item[\soso] agreement with perturbation theory 
                        over a range of a factor
                        $(3/2)^2$ in $\alpha_\mathrm{eff}^{n_\mathrm{l}}$ 
                        possibly fitting with power corrections or
                        alternatively 
                        $\alpha_\mathrm{eff}^{n_\mathrm{l}} \le \Delta \alpha_\mathrm{eff} / (8\pi b_0 \alpha_\mathrm{eff}^2)$
                        is reached
           \item[\bad]  otherwise
       \end{itemize}
        Here {$\Delta \alpha_\mathrm{eff}$ is the accuracy cited for the determination of 
        $\alpha_\mathrm{eff}$}
        and $n_\mathrm{l}$ is the loop order to which the 
        connection of $\alpha_\mathrm{eff}$ to the $\msbar$ scheme is known.
        Recall the discussion around Eqs.~(\ref{eq:i},\ref{eq:ii});
        the $\beta$-function of $\alpha_\mathrm{eff}$ is then known to 
        $n_\mathrm{l}+1$ loop order.%
        \footnote{Once one is in the perturbative region with 
        $\alpha_{\rm eff}$, the error in 
        extracting the $\Lambda$ parameter due to the truncation of 
        perturbation theory scales like  $\alpha_{\rm eff}^{n_\mathrm{l}}$,
        as discussed around Eq.~(\ref{eq:ii}). In order to 
        detect/control such corrections properly, one needs to change
        the correction term significantly; 
        we require a factor of four for a $\good$ and a factor $(3/2)^2$
        for a $\soso$. 
        An exception to the above is the situation 
        where the correction terms are small anyway, i.e.,  
        $\alpha_{\rm eff}^{n_\mathrm{l}} \approx (\Delta \Lambda/\Lambda)_\text{trunc} < (\Delta \Lambda/\Lambda)_{\Delta \alpha} \approx \Delta \alpha_{\rm eff} / (8\pi b_0 \alpha_{\rm eff}^2)$ is reached.}
         
   \item Continuum extrapolation 
        
        At a reference point of $\alpha_{\rm eff} = 0.3$ (or less) we require
         \begin{itemize}
            \item[\good] three lattice spacings with
                         $\mu a < 1/2$ and full $\cO(a)$
                         improvement, \\
                         or three lattice spacings with
                         $\mu a \leq 1/4$ and $2$-loop $\cO(a)$
                         improvement, \\
                         or $\mu a \leq 1/8$ and $1$-loop $\cO(a)$
                         improvement 
            \item[\soso] three lattice spacings with $\mu a < 3/2$
                         reaching down to $\mu a =1$ and full
                         $\cO(a)$ improvement, \\
                         or three lattice spacings with
                         $\mu a \leq 1/4$ and 1-loop $\cO(a)$
                         improvement        
            \item[\bad]  otherwise 
         \end{itemize}
\end{itemize}  

We also need to specify what is meant by $\mu$. Here are our choices:
\begin{eqnarray}
   \text{step-scaling} &:& \mu=1/L\,,
   \nonumber  \\
   \text{heavy quark-antiquark potential} &:& \mu=2/r\,,
   \nonumber  \\
   \text{observables in position space}  &:& \mu=1/|x|\,,
   \nonumber \\
   \text{observables in momentum space} &:& \mu =q \,,
   \nonumber   \\ 
    \text{moments of heavy-quark currents} 
                                        &:& \mu=2\bar{m}_\mathrm{c} \,,
   \nonumber   \\ 
    \text{eigenvalues of the Dirac operator}
                                        &:& \mu= \lambda_\msbar
\label{mu_def}
\end{eqnarray}
where $|x|$ is the Euclidean norm of the 4-vector $x$, 
$q$ is the magnitude of the momentum, $\bar{m}_\mathrm{c}$ is
the heavy-quark mass (in the $\msbar$ scheme) 
and usually taken around the charm-quark mass and
$\lambda_\msbar$ is the eigenvalue of the Dirac operator, see 
\sect{s:eigenvalue}.  We note again that the above criteria cannot
be applied when regularization dependent quantities
$W_\mathrm{lat}(a)$ are used instead of ${\oO}(\mu)$. These cases
are specifically discussed in \sect{s:WL}.

In principle one should also
account for electro-weak radiative corrections. However, both in the
determination of $\alpha_{s}$ at intermediate scales $\mu$ and
in the running to high scales, we expect electro-weak effects to be
much smaller than the presently reached precision. Such effects are
therefore not further discussed.

The attentive reader will have noticed that bounds such as $\mu a <
3/2$ or at least one value of $\alpha_\mathrm{eff}\leq 0.25$
which we require for a $\soso$ are
not very stringent. There is a considerable difference between
$\soso$ and $\good$. We have chosen the above bounds, unchanged as compared 
to FLAG 16 and FLAG 19, since not too many
computations would satisfy more stringent ones at present.
Nevertheless, we believe that the \soso\ criteria already give
reasonable bases for estimates of systematic errors. An exception 
may be Cali 20, which is discussed in detail in Sec.~\ref{s:vac}.
In the future, we
expect that we will be able to tighten our criteria for inclusion in
the average, and that many more computations will reach the present
\good\ rating in one or more categories. 

In addition to our explicit criteria, the following effects may influence
the precision of results: 

{\em Topology sampling:}
    In principle a good way to improve the quality 
    of determinations of $\alpha_s$ is to push to very small lattice 
    spacings thus enabling large $\mu$. It is known  
    that the sampling of field space becomes very difficult for the 
    HMC algorithm when the lattice spacing is small and one has the
    standard periodic boundary conditions. In practice, for all known
    discretizations the topological charge slows down dramatically for
    $a\approx 0.05\,\fm$ and smaller 
    \cite{DelDebbio:2002xa,Bernard:2003gq,Schaefer:2010hu,Chowdhury:2013mea,Brower:2014bqa,Bazavov:2014pvz,Fukaya:2015ara}. Open boundary conditions solve the problem 
    \cite{Luscher:2011kk} but are not frequently used. Since the effect of
    the freezing on short distance observables is not known, we also do need to pay
    attention to this issue. Remarks are added in the text when appropriate.
      
{\em Quark-mass effects:} We assume that effects
of the finite masses of the light quarks (including strange) 
are negligible in the effective
coupling itself where large, perturbative, $\mu$ is considered.

{\em Scale setting:}
    The scale does not need
    to be very precise, since using the lowest-order $\beta$-function
    shows that a 3\% error in the scale determination corresponds to a
    $\sim 0.5\%$ error in $\alpha_s(M_Z)$.  As long as systematic
    errors from chiral extrapolation and finite-volume effects are  well below
    3\% we do not need to be concerned about those at the present level of 
    precision in $\alpha_s(M_Z)$. This may change in the future.


\subsubsection{Physical scale}


Since FLAG 19, a new FLAG working group on scale setting has
been established. We refer to Sec.~\ref{sec:scalesetting}
for definitions and the current status. Note that the error from scale setting is
sub-dominant for current $\alpha_s$ determinations.

A popular scale choice has been the intermediate $r_0$ scale, 
and its variant $r_1$, which both derive from the force between static quarks, see Eq.(\ref{eq:r01}). 
One should bear in mind that their determination from physical
observables also has to be taken into account.  The phenomenological
value of $r_0$ was originally determined as $r_0 \approx
0.49\,\mbox{fm}$ through potential models describing quarkonia
\cite{Sommer:1993ce}. Of course the quantity is precisely defined,
independently of such model considerations.
But a lattice computation with the correct sea-quark content is 
needed to determine a completely sharp value. When the quark 
content is not quite realistic, the value of $r_0$ may depend to
some extent on which experimental input is used to determine 
(actually define) it. 
 
The latest determinations from two-flavour QCD are
$r_0$ = 0.420(14)--0.450(14)~fm by the ETM collaboration
\cite{Baron:2009wt,Blossier:2009bx}, using as input $f_\pi$ and $f_K$
and carrying out various continuum extrapolations. On the other hand,
the ALPHA collaboration \cite{Fritzsch:2012wq} determined $r_0$ =
0.503(10)~fm with input from $f_K$, and the QCDSF
collaboration \cite{Bali:2012qs} cites 0.501(10)(11)~fm from
the mass of the nucleon (no continuum limit).  Recent determinations
from three-flavour QCD are consistent with $r_1$ = 0.313(3)~fm
and $r_0$ = 0.472(5)~fm
\cite{Davies:2009tsa,Bazavov:2010hj,Bazavov:2011nk}. Due to the
uncertainty in these estimates, and as many results are based directly
on $r_0$ to set the scale, we shall often give both the dimensionless
number $r_0 \Lambda_{\overline{\rm MS}}$, as well as $\Lambda_{\overline{\rm MS}}$.
In the cases where no physical $r_0$ scale is given in
the original papers or we convert to
the $r_0$ scale, we use the value $r_0$ = 0.472~fm. In case
$r_1 \Lambda_{\overline{\rm MS}}$ is given in the publications,
we use $r_0 /r_1 = 1.508$ \cite{Bazavov:2011nk}, to convert,
which remains well consistent with the update \cite{Bazavov:2014pvz} 
neglecting the error on this ratio. In some, mostly early,
computations the string tension, $\sqrt{\sigma}$ was used.
We convert to $r_0$ using $r_0^2\sigma = 1.65-\pi/12$,
which has been shown to be an excellent approximation 
in the relevant pure gauge theory \cite{Necco:2001xg,Luscher:2002qv}.

The new scales $t_0,w_0$ based on the gradient flow are very attractive
alternatives to $r_0$ but their
discretization errors are still under discussion 
\cite{Ramos:2014kka,Fodor:2014cpa,Bazavov:2015yea,Bornyakov:2015eaa}
and their values at the physical point are not yet determined 
with great precision.
We remain with $r_0$ as our main reference scale for now.
A general discussion of the various scales is given in 
\cite{Sommer:2014mea} and in the scale-setting section of this FLAG report, cf.~Sec.~\ref{sec:scalesetting}.


{
\subsubsection{Studies of truncation errors of perturbation theory}
\label{s:trunc}

As discussed previously, we have to determine $\alpha_s$ in a region
where the perturbative expansion for the $\beta$-function, 
Eq.~(\ref{eq:beta_pert}) in the integral Eq.~(\ref{eq:Lambda}),
is reliable. In principle this must be checked, however, this is
difficult to achieve as we need to reach up to a sufficiently high scale.
A frequently used recipe to estimate the size of truncation errors
of the perturbative series is to vary the renormalization-scale
dependence around the chosen `optimal' scale $\mu_*$, of an observable
evaluated at a fixed order in the coupling from $\mu=\mu_*/2$ to $2\mu_*$. 
For examples, see Ref.~\cite{DelDebbio:2021ryq}.

Alternatively, or in addition, the renormalization scheme chosen
can be varied, which investigates the perturbative
conversion of the chosen scheme to the perturbatively defined
$\overline{\rm MS}$ scheme and in particular `fastest apparent
convergence' when the `optimal' scale is chosen so that the
$\cO(\alpha_s^2)$ coefficient vanishes.

The ALPHA collaboration in Ref.~\cite{Brida:2016flw} and 
ALPHA 17~\cite{DallaBrida:2018rfy}, within the SF approach defined
a set of $\nu$-schemes for which the 3-loop (scheme-dependent)
coefficient of the $\beta$-function for $N_f = 2+1$ flavours was
computed to be $b_2^\nu = -(0.064(27)+1.259(1)\nu)/(4\pi)^3$. 
The standard SF scheme has $\nu = 0$. For comparison, $b_2^\msbar = 0.324/(4\pi)^3$.
A range of scales from about 
$4\,\mbox{GeV}$ to $128\,\mbox{GeV}$ was investigated.
It was found that while the procedure of varying the
scale by a factor 2 up and down gave a correct estimate
of the residual perturbative error for $\nu \approx 0 \ldots  0.3$,  
for negative values, e.g.,  $\nu = -0.5$, the estimated perturbative
error is much too small to account for the mismatch in the 
$\Lambda$-parameter of $\approx 8\%$ at $\alpha_s=0.15$.
This mismatch, however, did, as expected, still scale with $\alpha_s^{n_l}$ with $n_l=2$. In the schemes
with negative $\nu$, the coupling $\alpha_s$ has to be quite small for
scale-variations of a factor 2 to correctly signal the perturbative errors. 

For a systematic study of renormalization scale variations as a measure of
perturbative truncation errors in various lattice determinations
of $\alpha_s$  we refer to the recent 
review by Del Debbio and Ramos~\cite{DelDebbio:2021ryq}.


\subsection{$\alpha_s$ from Step-Scaling Methods}
\label{s:SF}

\subsubsection{General considerations}


The method of step-scaling functions avoids the scale problem,
\eq{eq:scaleproblem}. It is in principle independent of the particular
boundary conditions used and was first developed with periodic
boundary conditions in a two-dimensional model~\cite{Luscher:1991wu}.

The essential idea of the step-scaling strategy
is to split the determination of the running coupling at large
$\mu$ and of a hadronic scale into two lattice calculations and
connect them by `step-scaling'. In the former part, we determine the
running coupling constant in a finite-volume scheme
in which the renormalization scale is set by the inverse lattice size
$\mu = 1/L$. In this calculation, one takes a high renormalization scale
while keeping the lattice spacing sufficiently small as
\begin{eqnarray}
   \mu \equiv 1/L \sim 10\,\ldots\, 100\,\mbox{GeV}\,, \qquad a/L \ll 1 \,.
\end{eqnarray}
In the latter part, one chooses a certain 
$\gbar^2_\mathrm{max}=\gbar^2(1/L_\mathrm{max})$, 
typically such that $L_\mathrm{max}$ is around $0.5$--$1$~fm. With a 
common discretization, one then determines $L_\mathrm{max}/a$ and
(in a large volume $L \ge$ 2--3~fm) a hadronic scale
such as a hadron mass, $\sqrt{t_0}/a$ or $r_0/a$ at the same bare
parameters. In this way one gets numbers for, e.g., $L_\mathrm{max}/r_0$
and by changing the lattice spacing $a$ carries out a continuum
limit extrapolation of that ratio. 
 
In order to connect $\gbar^2(1/L_\mathrm{max})$ to $\gbar^2(\mu)$ at
high $\mu$, one determines the change of the coupling in the continuum
limit when the scale changes from $L$ to $L/s$, starting from
$L=L_{\rm max}$ and arriving at $\mu = s^k /L_{\rm max}$. This part of
the strategy is called step-scaling. Combining these results yields
$\gbar^2(\mu)$ at $\mu = s^k \,(r_0 / L_\mathrm{max})\, r_0^{-1}$,
where $r_0$ stands for the particular chosen hadronic scale.
Most applications use a scale factor $s=2$.

At present most applications in QCD use Schr\"odinger 
functional boundary conditions~\cite{Luscher:1992an,Sint:1993un}
and we discuss this below in a little more detail.
(However, other boundary conditions are also possible, such as
twisted boundary conditions 
and the discussion also applies to them.)
An important reason is that these boundary conditions avoid zero modes
for the quark fields and quartic modes \cite{Coste:1985mn} in the
perturbative expansion in the gauge fields. Furthermore the corresponding
renormalization scheme is well studied in perturbation
theory~\cite{Luscher:1993gh,Sint:1995ch,Bode:1999sm} with the
3-loop $\beta$-function and 2-loop cutoff effects (for the
standard Wilson regularization) known.

In order to have a perturbatively well-defined scheme,
the SF scheme uses Dirichlet boundary conditions at time 
$t = 0$ and $t = T$. These break translation invariance and permit
${\cO}(a)$ counter terms at the boundary through quantum corrections. 
Therefore, the leading discretization error is ${\cO}(a)$.
Improving the lattice action is achieved by adding
counter terms at the boundaries whose coefficients are denoted
as $c_t,\tilde c_t$. In practice, these coefficients are computed
with $1$-loop or $2$-loop perturbative accuracy.
A better precision in this step yields a better 
control over discretization errors, which is important, as can be
seen, e.g., in Refs.~\cite{Takeda:2004xha,Necco:2001xg}.

Also computations with Dirichlet boundary conditions do in principle
suffer from the insufficient change of topology in the HMC algorithm
at small lattice spacing. However, in a small volume the weight of
nonzero charge sectors in the path integral is exponentially
suppressed~\cite{Luscher:1981zf}~\footnote{We simplify here and assume
  that the classical solution associated with the used boundary
  conditions has charge zero.  In practice this is the case.} and in a Monte Carlo run of
  typical length very few configurations
  with nontrivial topology should appear. Considering
the issue quantitatively Ref.~\cite{Fritzsch:2013yxa} finds a
strong suppression below $L\approx 0.8\,\fm$. Therefore the lack of
topology change of the HMC is not a serious issue for the high energy regime
in step-scaling studies. However, the matching to hadronic observables
requires volumes where the problem cannot be ignored. Therefore,
Ref.~\cite{DallaBrida:2016kgh} includes a projection to zero topology 
into the {\em definition} of the coupling.
We note also that a mix of Dirichlet and open boundary conditions is
expected to 
remove the topology issue entirely \cite{Luscher:2014kea}
and may be
considered in the future.

Apart from the boundary conditions, the very definition
of the coupling needs to be chosen. 
We briefly discuss in turn, the two schemes used at present, 
namely, the `Schr\"odinger
Functional' (SF) and `Gradient Flow' (GF) schemes.

The SF scheme is the first one, which was used in step-scaling studies
in gauge theories \cite{Luscher:1992an}. Inhomogeneous
Dirichlet boundary conditions are imposed in time,
\begin{eqnarray}
    A_k(x)|_{x_0=0} = C_k\,,
    \quad
    A_k(x)|_{x_0=L} = C_k'\,,    
\end{eqnarray}
for $k=1,2,3$.
Periodic boundary conditions (up to a phase for the fermion fields)  with period $L$ are imposed in space.
The matrices 
\begin{align}
LC_k &= i \,{\rm diag}\big( \eta- \pi/3, -\eta/2 , -\eta/2  + \pi/3 \big) \,,
\nonumber \\
LC^\prime_k &= i \,{\rm diag}\big( -(\eta+\pi), \eta/2 + \pi/3,\eta/2 + 2\pi/3 \big)\,,
\nonumber
\end{align}
just depend on the dimensionless parameter $\eta$.
The coupling $\bar{g}_\mathrm{SF}$ is obtained from
the $\eta$-derivative of the effective action,
\begin{eqnarray}
  \langle \partial_\eta S|_{\eta=0} \rangle = \frac{12\pi}{\gbar^2_\mathrm{SF}}\,.
\end{eqnarray}
For this scheme, the finite $c^{(i)}_g$, \eq{eq:g_conversion}, are 
known for $i=1,2$ 
\cite{Sint:1995ch,Bode:1999sm}.

More recently, gradient-flow couplings have been used frequently
because of their small statistical errors at large couplings (in contrast to 
$\gbar_\mathrm{SF}$, which has small statistical errors at small couplings). 
The gradient flow is introduced as follows \cite{Narayanan:2006rf,Luscher:2010iy}.
Consider the flow gauge field $B_\mu(t,x)$ with the flow time $t$, 
which is a one parameter deformation of the bare gauge field 
$A_\mu(x)$, where $B_\mu(t,x)$ is the solution to the 
gradient-flow equation
\begin{eqnarray}
   \partial_t B_\mu(t,x) 
            &=& D_\nu G_{\nu\mu}(t,x)\,,
                                                      \nonumber \\
   G_{\mu\nu} &=& \partial_\mu B_\nu - \partial_\nu B_\mu + [B_\mu,B_\nu] \,,
\end{eqnarray}
with initial condition $B_\mu(0,x) = A_\mu(x)$.
The renormalized coupling is defined by \cite{Luscher:2010iy}
\begin{eqnarray}
   \bar{g}^2_{\rm GF}(\mu) 
      = \left. {\cal N} t^2 \langle E(t,x)\rangle
                                        \right|_{\mu=1/\sqrt{8t}} \,,
\end{eqnarray}
with ${\cal N} = 16\pi^2/3 + \cO((a/L)^2)$
and where $E(t,x)$ is the action density given by
\begin{eqnarray}
   E(t,x) = \frac{1}{4} G^a_{\mu\nu}(t,x) G^a_{\mu\nu}(t,x). 
                                        \label{eq:Et}
\end{eqnarray}
In a finite volume, one needs to specify additional conditions.
In order not to introduce two independent scales one sets 
\begin{eqnarray}
   \sqrt{8t} = cL \,,
\end{eqnarray}
for some fixed number $c$ \cite{Fodor:2012td}. 
Schr\"odinger functional boundary conditions~\cite{Fritzsch:2013je}
or twisted boundary conditions \cite{Ramos:2014kla,Ishikawa:2017xam}  
have been employed.
Matching of the GF coupling to the $\overline{\rm MS}$-scheme coupling
is known to 1-loop for twisted boundary conditions with zero
quark flavours and $SU(3)$ group \cite{Ishikawa:2017xam} and to 2-loop with SF boundary conditions with zero
quark flavours \cite{DallaBrida:2017tru}.
The former is based on a MC evaluation at small couplings\footnote{For a variant of the 
twisted periodic finite volume scheme the 1-loop matching has been computed analytically~\cite{Bribian:2019ybc}.} and the 
latter on numerical stochastic perturbation theory.



\subsubsection{Discussion of computations}


In Tab.~\ref{tab_SF3} we give results from various determinations
\begin{table}[!htb]
   \vspace{3.0cm}
   \footnotesize
   \begin{tabular*}{\textwidth}[l]{l@{\extracolsep{\fill}}rlllllllll}
      Collaboration & Ref. & $\Nf$ &
      \hspace{0.15cm}\begin{rotate}{60}{publication status}\end{rotate}
                                                       \hspace{-0.15cm} &
      \hspace{0.15cm}\begin{rotate}{60}{renormalization scale}\end{rotate}
                                                       \hspace{-0.15cm} &
      \hspace{0.15cm}\begin{rotate}{60}{perturbative behaviour}\end{rotate}
                                                       \hspace{-0.15cm} &
      \hspace{0.15cm}\begin{rotate}{60}{continuum extrapolation}\end{rotate}
                               \hspace{-0.25cm} & 
                         scale & $\Lambda_\msbar[\MeV]$ & $r_0\Lambda_\msbar$ \\
      & & & & & & & & \\[-0.1cm]
      \hline
      \hline
      & & & & & & & & \\
      ALPHA 10A & \cite{Tekin:2010mm} & 4 
                    & \gA &\good & \good & \good 
                    & \multicolumn{3}{l}{only running of $\alpha_s$ in Fig.~4}
                    \\  
      Perez 10 & \cite{PerezRubio:2010ke} & 4 
                    & \rC &\good & \good & \soso  
                    & \multicolumn{3}{l}{only step-scaling function in Fig.~4}
                    \\           
      & & & & & & & & & \\[-0.1cm]
      \hline
      & & & & & & & & & \\[-0.1cm]
      ALPHA 17   &  \cite{Bruno:2017gxd} &2+1 
                    & \gA & \good & \good & \good 
                    & $\sqrt{8t_0}= 0.415\,\mbox{fm}$ & 341(12) & 0.816(29)
                    \\  
      PACS-CS 09A& \cite{Aoki:2009tf} & 2+1 
                    & \gA &\good &\good &\soso
                    & $m_\rho$ & $371(13)(8)(^{+0}_{-27})$$^{\#}$
                    & $0.888(30)(18)(^{+0}_{-65})$$^\dagger$
                    \\ 
                    &&&\gA &\good &\good &\soso 
                    & $m_\rho$  & $345(59)$$^{\#\#}$
                    & $0.824(141)$$^\dagger$
                    \\ 
      & & & & & & & & \\[-0.1cm]
      \hline  \\[-1.0ex]
      & & & & & & & & \\[-0.1cm]
      ALPHA 12$^*$  & \cite{Fritzsch:2012wq} & 2 
                    & \gA &\good &\good &\good
                    &  $f_{\rm K}$ & $310(20)$ &  $0.789(52)$
                    \\
      ALPHA 04 & \cite{DellaMorte:2004bc} & 2 
                    & \gA &\bad &\good &\good
                    & $r_0 = 0.5\,\mbox{fm}$$^\S$  & $245(16)(16)^\S$ 
                                                   & $0.62(2)(2)^\S$
                    \\
      ALPHA 01A & \cite{Bode:2001jv} & 2 
                    &\gA & \good & \good & \good 
                    &\multicolumn{3}{l}{only running of $\alpha_s$  in Fig.~5}
                    \\
      & & & & & & & & \\[-0.1cm]
      \hline  \\[-1.0ex]
      & & & & & & & & \\[-0.1cm]
      Nada 20    & \cite{Nada:2020jay} & 0 
                    & \gA & \good & \good & \good
                    &\multicolumn{3}{l}{consistency checks for \cite{DallaBrida:2019wur}, same gauge configurations}
                    \\
 Dalla Brida 19 & \cite{DallaBrida:2019wur} & 0 
                    & \gA & \good & \good & \good
                    & $r_0=0.5\fm$ & 260.5(4.4)  & 0.660(11) 
                    \\
      Ishikawa 17   & \cite{Ishikawa:2017xam} & 0 
                    & \gA & \good & \good & \good
                    & $r_0$, $[\sqrt{\sigma}]$ & $253(4)(^{+13}_{-2})$$^\dagger$
                                              & $0.606(9)(^{+31}_{-5})^+$
                    \\
      CP-PACS 04$^\&$  & \cite{Takeda:2004xha} & 0 
                    & \gA & \good & \good & \soso  
                    & \multicolumn{3}{l}{only tables of $g^2_{\rm SF}$}
                    \\
      ALPHA 98$^{\dagger\dagger}$ & \cite{Capitani:1998mq} & 0 
                    & \gA & \good & \good & \soso 
                    &  $r_0=0.5\fm$ & $238(19)$ & 0.602(48) 
                    \\
      L\"uscher 93  & \cite{Luscher:1993gh} & 0 
                    & \gA & \good & \soso & \soso
                    & $r_0=0.5\fm$ & 233(23)  & 0.590(60)$^{\S\S}$ 
                    \\
      &&&&&&& \\[-0.1cm]
      \hline
      \hline\\
\end{tabular*}\\[-0.2cm]
\begin{minipage}{\linewidth}
{\footnotesize 
\begin{itemize}
\item[$^{\#}$] Result with a constant (in $a$) continuum extrapolation
              of the combination $L_\mathrm{max}m_\rho$.             \\[-5mm]
\item[$^\dagger$] In conversion from $\Lambda_\msbar$ to
                 $r_0\Lambda_{\overline{\rm MS}}$ and vice versa, $r_0$ is
                 taken to be $0.472\,\mbox{fm}$.                   \\[-5mm]
\item[$^{\#\#}$] Result with a linear continuum extrapolation
             in $a$ of the combination $L_\mathrm{max}m_\rho$.        \\[-5mm]
\item[$^*$]  Supersedes ALPHA 04.                                   \\[-5mm]
\item[$^\S$] The $N_f=2$ results were based on values for $r_0/a$
             which have later been found to be too small by
             \cite{Fritzsch:2012wq}. The effect will be of the order of
             10--15\%, presumably an increase in $\Lambda r_0$.
             We have taken this into account by a $\bad$ in the 
             renormalization scale.                                  \\[-5mm]
\item[$^\&$] This investigation was a precursor for PACS-CS 09A
          and confirmed two step-scaling functions as well as the
          scale setting of ALPHA~98.                              \\[-5mm]
\item[$^{\dagger\dagger}$] Uses data of L\"uscher~93 and therefore supersedes it.
                                                                  \\[-5mm]
\item[$^{\S\S}$] Converted from $\alpha_\msbar(37r_0^{-1})=0.1108(25)$.
\item[$^+$] Also $\Lambda_\msbar/\sqrt{\sigma} = 0.532(8)(^{+27}_{-5})$ is quoted.

\end{itemize}
}
\end{minipage}
\caption{Results for the $\Lambda$ parameter from computations using 
         step-scaling of the SF-coupling. Entries without values for $\Lambda$
         computed the running and established perturbative behaviour
         at large $\mu$. 
         }
\label{tab_SF3}
\end{table}
of the $\Lambda$ parameter. For a clear assessment of the $N_f$-dependence, the last column also shows results that refer to a common
hadronic scale, $r_0$. As discussed above, the renormalization scale
can be chosen large enough such that $\alpha_s < 0.2$ and the
perturbative behaviour can be verified.  Consequently only $\good$ is
present for these criteria except for early work
where the $n_l=2$ loop correction to $\msbar$ was not yet known and we assigned a $\bad$ concerning the renormalization scale.
With dynamical fermions, results for the
step-scaling functions are always available for at least $a/L = \mu a
=1/4,1/6, 1/8$.  All calculations have a nonperturbatively
$\cO(a)$ improved action in the bulk. For the discussed
boundary $\cO(a)$ terms this is not so. In most recent
calculations 2-loop $\cO(a)$ improvement is employed together
with at least three lattice spacings.\footnote{With 2-loop
  $\cO(a)$ improvement we here mean $c_t$ including
  the $g_0^4$ term and $\tilde c_\mathrm{t}$ with the $g_0^2$
  term. For gluonic observables such as the running coupling this is
  sufficient for cutoff effects being suppressed to $\cO(g^6
  a)$.} This means a \good\ for the continuum extrapolation.  In 
other computations only 1-loop $c_t$ was available and we arrive at \soso. We
note that the discretization errors in the step-scaling functions 
of the SF coupling are
usually found to be very small, at the percent level or
below. However, the overall desired precision is very high as well,
and the results in CP-PACS 04~\cite{Takeda:2004xha} show that
discretization errors at the below percent level cannot be taken for
granted.  In particular with staggered fermions (unimproved except for
boundary terms) few percent effects are seen in
Perez~10~\cite{PerezRubio:2010ke}.

In the work by PACS-CS 09A~\cite{Aoki:2009tf}, the continuum
extrapolation in the scale setting is performed using a constant
function in $a$ and with a linear function.
Potentially the former leaves a considerable residual discretization 
error. We here use, as discussed with the collaboration, 
the continuum extrapolation linear in $a$,
as given in the second line of PACS-CS 09A \cite{Aoki:2009tf}
results in Tab.~\ref{tab_SF3}.
After perturbative conversion from a three-flavour result to five flavours
(see \sect{s:crit}), they obtain
\begin{eqnarray}
 \alpha_\msbar^{(5)}(M_Z)=0.118(3)\,. 
\end{eqnarray}

In Ref.~\cite{Bruno:2017gxd}, the ALPHA collaboration determined 
$\Lambda^{(3)}_{\msbar}$ combining step-scaling in $\gbar^2_{\rm GF}$
in the lower scale region $\mu_{\rm had} \leq \mu \leq \mu_0$, and 
step-scaling in $\gbar^2_{\rm SF}$ for higher scales  
$\mu_0 \leq \mu \leq \mu_{\rm PT}$. 
Both schemes are defined with SF boundary conditions. For $\gbar^2_{\rm GF}$ a projection to the sector of zero 
topological charge is included, \eq{eq:Et} is restricted to the 
magnetic components, and $c=0.3$.
The scales $\mu_{\rm had}$, $\mu_0$, and 
$\mu_{\rm PT}$ are defined by $\gbar^2_{\rm GF} (\mu_{\rm had})= 11.3$,
$\gbar^2_{\rm SF}(\mu_0) = 2.012$, and $\mu_{\rm PT} = 16 \mu_0$ which
are roughly estimated as
\begin{eqnarray}
   1/L_\mathrm{max}\equiv \mu_{\rm had} \approx 0.2 \mbox{ GeV}, & \mu_0 \approx 4 \mbox{ GeV} \,, 
      & \mu_{\rm PT}\approx 70 \mbox{ GeV} \,.
\end{eqnarray}
Step-scaling is carried out with an $\cO(a)$-improved Wilson quark action
\cite{Bulava:2013cta}
and L\"uscher-Weisz gauge action \cite{Luscher:1984xn} in the low-scale region
and an $\cO(a)$-improved Wilson quark action
\cite{Yamada:2004ja}
and Wilson gauge action in the high-energy part. 
For the step-scaling using steps of
$L/a \,\to\,2L/a$, three lattice sizes $L/a=8,12,16$ were simulated for
$\gbar^2_{\rm GF}$ and four lattice sizes $L/a=(4,)\, 6, 8, 12$ for 
$\gbar^2_{\rm SF}$. The final results do not use the small lattices given
in parenthesis. The parameter $\Lambda^{(3)}_{\msbar}$ is then obtained via 
\begin{eqnarray}
   \Lambda^{(3)}_{\msbar} 
      = \underbrace{\frac{\Lambda^{(3)}_{\msbar}}{\mu_{\rm PT}}}_{\rm perturbation  ~ theory}
           \times \underbrace{\frac{\mu_{\rm PT}}{\mu_{\rm had}}}_{\rm step-scaling}
           \times \underbrace{\frac{\mu_{\rm had}}{f_{\pi K}}}_{\rm large ~ volume~ simulation}
           \times \underbrace{f_{\pi K}}_{\rm experimental ~data} \,, 
\label{eq:Lambda3}
\end{eqnarray}
where the hadronic scale $f_{\pi K}$ is 
$f_{\pi K}= \frac{1}{3}(2 f_K + f_\pi) = 147.6 (5)\mbox{ MeV}$.
The first factor on the right hand side of Eq.~(\ref{eq:Lambda3}) is 
obtained from $\alpha_{\rm SF}(\mu_{\rm PT})$ which is the output from
SF step-scaling using Eq.~(\ref{eq:Lambda}) with 
$\alpha_{\rm SF}(\mu_{\rm PT})\approx 0.1$ and
the 3-loop $\beta$-function 
and the exact conversion to the $\msbar$-scheme.
The second factor is essentially obtained
from step-scaling in the GF scheme and the measurement of 
$\gbar^2_{\rm SF}(\mu_0)$ (except for the trivial scaling factor of 16 
in the SF running). The third factor is obtained from a measurement
of the hadronic quantity at large volume.

A large-volume simulation is done for three lattice spacings with 
sufficiently large volume and reasonable control over the chiral
extrapolation so that the scale determination is precise enough.  
The step-scaling results in both schemes
satisfy renormalization criteria, perturbation theory criteria,
and continuum limit criteria just as previous studies using step-scaling.
So we assign green stars for these criteria.

The dependence of $\Lambda $, Eq.~(\ref{eq:Lambda}) with 3-loop $\beta$-function, on $\alpha_s$ and on the chosen scheme is discussed
in \cite{Brida:2016flw}. This investigation provides a warning on estimating the 
truncation error of perturbative series. Details are explained in \sect{s:trunc}.

The result for the $\Lambda$ parameter is  
$\Lambda^{(3)}_{\overline{\rm MS}} = 341(12)~\mbox{MeV}$, 
where the dominant error comes from the error of 
$\alpha_{\rm SF}(\mu_{\rm PT})$ after step-scaling in the SF scheme.
Using 4-loop matching at the charm and bottom thresholds 
and 5-loop running one finally obtains
\begin{eqnarray}
   \alpha^{(5)}_{\overline{\rm MS}}(M_Z) = 0.11852(84)\,.
\end{eqnarray}
Several other results do not have a sufficient number of
quark flavours  or do not yet contain the conversion
of the scale to physical units (ALPHA~10A \cite{Tekin:2010mm}, 
Perez~10 \cite{PerezRubio:2010ke}). Thus no value for $\alpha_\msbar^{(5)}(M_Z)$
is quoted.

The computation of Ishikawa et al. \cite{Ishikawa:2017xam} 
is based on the gradient flow coupling with twisted boundary conditions
\cite{Ramos:2014kla} (TGF coupling)
in the pure gauge theory. Again they use 
$c=0.3$. Step-scaling with a scale factor $s=3/2$ is employed,
covering a large range of couplings from $\alpha_s\approx 0.5$ to
$\alpha_s\approx 0.1$ and taking the continuum limit through global
fits to the step-scaling function on $L/a=12,16,18$ lattices with between 6 and 
8 parameters. Systematic errors due to variations of the fit functions
are estimated. Two physical scales are considered:
$r_0/a$ is taken from \cite{Necco:2001xg} and $\sigma a^2$ from 
\cite{Allton:2008pn} and \cite{GonzalezArroyo:2012fx}.  
As the ratio $\Lambda_\mathrm{TGF}/\Lambda_\mathrm{\msbar}$   
has not yet been computed analytically, Ref.~\cite{Ishikawa:2017xam}
determines the 1-loop relation between $\gbar_\mathrm{SF}$ and 
$\gbar_\mathrm{TGF}$ from  MC simulations performed
in the weak coupling region and then uses the known
$\Lambda_\mathrm{SF}/\Lambda_\mathrm{\msbar}$. Systematic errors 
due to variations of the fit functions dominate the overall uncertainty.
\par
Since FLAG 19 two new and quite extensive $\Nf=0$ step-scaling studies have been carried out in Dalla Brida 19~\cite{DallaBrida:2019wur}
and by Nada and Ramos~\cite{Nada:2020jay}. They use different strategies for the running from mid to high energies, 
but use the same gauge configurations and share the running at low energies and matching to the hadronic scales.
These results are therefore correlated. However, given the comparatively high value for $r_0\lms$, 
it is re-assuring that these conceptually different approaches yield perfectly compatible results within errors of
similar size of around 1.5\% for $\sqrt{8t_0}\lms=0.6227(98)$, or, alternatively $r_0\lms = 0.660(11)$.

In Dalla Brida 19 \cite{DallaBrida:2019wur} two GF-coupling definitions with SF-boundary conditions are considered, corresponding to (colour-) magnetic and
electric components of the action density respectively. The coupling definitions include the projection to 
$Q=0$, as was also done in~\cite{Bruno:2017gxd}. The flow time parameter is set to $c=0.3$, 
and both Zeuthen and Wilson flow are measured. Lattice sizes
range from $L/a=8$ to $L/a=48$, covering up to a factor of 3 in lattice spacings 
for the step-scaling function, where both $L/a$ and $2L/a$ are needed. Lattice effects in the 
step-scaling function are visible but can be extrapolated using global fits with $a^2$ errors.
Some remnant $\cO(a)$ effects from the boundaries are expected, as their perturbative
cancellation is incomplete. These $\cO(a)$ contaminations are treated as a systematic error on the data, 
following \cite{Bruno:2017gxd} and are found to be subdominant. An intermediate reference scale $\mu_\mathrm{ref}$ is defined where $\alpha=0.2$,
and the scales above and below are analyzed separately. Again this is similar to \cite{Bruno:2017gxd}, except
that here GF coupling data is available also at high energy scales.
The GF $\beta$-functions are then obtained by fitting to the continuum extrapolated data 
for the step-scaling functions. In addition, a nonperturbative
matching to the standard SF coupling is performed above $\mu_\mathrm{ref}$ 
for a range of couplings covering a factor 2. The nonperturbative $\beta$-function for the SF scheme 
can thus be inferred from the GF $\beta$-function. It turns out that GF schemes are very slow 
to reach the perturbative regime.  Particularly the $\Lambda$-parameter for the magnetic GF coupling
shows a large slope in $\alpha^2$, which is the parametric uncertainty with known 3-loop $\beta$-function.
Also, convincing contact with the 3-loop $\beta$-function is barely seen down to $\alpha = 0.08$.
This is likely to be related to the rather large 3-loop $\beta$-function coefficients, 
especially for the magnetic GF scheme~\cite{DallaBrida:2017tru}.
In contrast, once the GF couplings are matched nonperturbatively to the SF scheme the contact to 
perturbative running can be safely made. It is also re-assuring that in all
cases the extrapolations (linear in $\alpha^2$) to $\alpha=0$ for the $\Lambda$-parameters agree very well,
and the authors argue in favour of such extrapolations. Their data confirms that this procedure
yields consistent results with the SF scheme for $\nu=0$, where such an extrapolation is not required.

The low energy regime between $\mu_\mathrm{ref}$ and a hadronic scale $\mu_\mathrm{had}$ is
covered again using the nonperturbative step-scaling function and the derived $\beta$-function.
Finally, contact between $\mu_\mathrm{had}$ and hadronic scales $t_0$ and $r_0$ is established using 5 lattice spacings
covering a factor up to 2.7. The multitude of cross checks of both continuum limit
and perturbative truncation errors make this a study which passes all current FLAG criteria by some margin.
The comparatively high value for $r_0\lms$ found in this study must therefore be taken very seriously.

In Nada 20 \cite{Nada:2020jay}, Nada and Ramos provide further consistency checks of \cite{DallaBrida:2019wur}
for scales larger than $\mu_\mathrm{ref}$. The step scaling function for $c=0.2$ is constructed in 2 steps, by determining
first the relation between couplings for $c=0.2$ and $c=0.4$ at the same $L$ and
then increasing $L$ to $2L$ keeping the flow time fixed (in units of the lattice spacing), 
so that one arrives again at $c=0.2$ on the $2L$ volume.
The authors demonstrate  that the direct construction of the step-scaling function for $c=0.2$ would require
much larger lattices in order to control the continuum limit at the same level of precision. 
The consistency with \cite{DallaBrida:2019wur} for the $\Lambda$-parameter is therefore a highly non-trivial check on the systematic effects of
the continuum extrapolations. The study obtains results for the
$\Lambda$-parameter (again extrapolating to $\alpha=0$) with a similar error as in \cite{DallaBrida:2019wur}.
using the low-energy running and matching to the hadronic scale from that reference.
For this reason and since gauge configurations are shared between both papers, 
these results are not independent of \cite{DallaBrida:2019wur},
so Dalla Brida 19 will be taken as representative for both works.

%


\subsection{The decoupling method}

\label{s:dec}

\newcommand{\mudec}{\mu_\text{dec}}

The ALPHA collaboration  has proposed a
new strategy to compute the $\Lambda$ parameter in QCD with $\Nf \ge 3$ flavours 
based on simultaneous decoupling of $\Nf\ge 3$ heavy quarks with RGI mass $M$~\cite{DallaBrida:2019mqg}.
We refer to \cite{DallaBrida:2020pag} for a pedagogical introduction and to \cite{Brida:2021xwa} for recent results.
Generically, a running coupling in a mass-dependent renormalization scheme
\begin{equation}
\label{eq:dec-start}
   \gbar^2(\mu, M)^{(\Nf)} = \gbar^2(\mu)^{(\Nf=0)} + O\left(M^{-k}\right)
\end{equation}
can be represented by the corresponding $\Nf=0$ coupling, up to power corrections in
$1/M$. The leading power is usually $k=2$, however renormalization schemes 
in finite volume may have $k=1$, depending on the set-up. For example, 
this is the case with standard SF or open boundary conditions in combination 
with a standard mass term. In practice one may try to render such boundary
contributions numerically small by a careful choice of the scheme's parameters.
In principle, power corrections can be either $(\mu/M)^k$ or $(\Lambda/M)^k$.
Fixing $\mu=\mudec$, e.g.~by prescribing a value for the mass-independent coupling,
such that $\mudec/\Lambda = \cO(1)$ thus helps to reduce the need for very large $M$. 
Defining  $\gbar^2(\mudec,M) = u_M$ at fixed $\gbar^2(\mudec,M=0)$,
Eq.~(\ref{eq:dec-start}) translates to a relation 
between $\Lambda$-parameters, which can be cast in the form,
\begin{eqnarray}
 \frac{\lms^{(\Nf)}}{\mudec}\;P\left(\frac{M}\mudec \frac{\mudec}{\lms^{(\Nf)}}\right) = 
 \frac{\lms^{(0)}}{\Lambda_s^{(0)}}\, \varphi_s^{(\Nf=0)}\left( \sqrt{u_\mathrm{M}}\right) + \cO(M^{-k})\,, \nonumber \\[-2ex]
\label{eq:basic}
\end{eqnarray}
with the function $\varphi_s$  as defined in Eq.~(\ref{eq:Lambda}), for scheme $s$ and $\Nf=0$.
A crucial observation is that the function $P$, which gives the ratios of $\Lambda$-parameters $\lms^{(0)}/\lms^{(\Nf)}$,
can be evaluated perturbatively to a very good approximation~\cite{Bruno:2014ufa,Athenodorou:2018wpk}.
Eq.~(\ref{eq:dec-start}) also implies a relation between the couplings in mass-independent schemes, 
in the theories with $\Nf$ and zero flavours, respectively. In the $\msbar$ scheme
this relation is analogous to Eq.~(\ref{e:grelation}),
\begin{eqnarray}
 \gbar^2_\msbar(m_\star)^{(\Nf=0)} =  \gbar^2_\msbar(m_\star)^{(\Nf)}\times C\left(\gbar^{}_\msbar(m_\star)^{(\Nf)}\right)
\end{eqnarray}
and the function $C(g)$ is also known up to to 4-loop order~\cite{Grozin:2011nk,Chetyrkin:2005ia,Schroder:2005hy,Kniehl:2006bg,Gerlach:2018hen}. 
The function $P(y)$, with $y\equiv M/\lms^{(\Nf)}$ can therefore be evaluated perturbatively in the $\msbar$ scheme, as the ratio
\begin{equation}
 P(y) = \dfrac{\varphi_\msbar^{(\Nf=0)}\left(g^\star(y) \sqrt{C(g^\star(y))}\right) }
               {\varphi_\msbar^{(\Nf)}(g^\star(y))}, \qquad g^\star(y) = \gbar_\msbar^{(\Nf)}(m_\star)\,.
\end{equation}
Hence, perturbation theory is only required at the scale set by the heavy-quark mass, which works the better 
the larger $M$ can be chosen. 
Once $P$ is known, the LHS of (\ref{eq:basic}) can be inferred from a $\Nf=0$ computation of the RHS in 
the scheme $s$, assuming the ratio $\lms/\Lambda_s$ is known from a 1-loop calculation.

To put the decoupling strategy into practice, the ALPHA collaboration uses $\Nf=3$, 
so that information from \cite{Bruno:2017gxd} can be used. Using the massless GF coupling in finite
volume from this project, $\mudec$ is defined through $\gbar^2_\text{GF}(\mudec) = 3.95$, and thus known in physical
units, $\mudec = 789(15)\MeV$. Varying $L/a$ between 12 and 32 (five lattice spacings) 
defines a range of values for the bare coupling along a line of constant $\mudec$  and for vanishing quark mass.
Next, a mass-dependent GF coupling is defined at constant $\mudec$, using the available information on nonperturbative 
mass renormalization~\cite{Campos:2018ahf} and $\cO(a)$ improvement. In order to obtain a larger suppression of
the leading $1/M$ boundary correction term, the time extent $T$ is here set to $2L$, so as to maximize the
distance to the time boundaries. 
Choosing 4 values of $z= M/\mudec$ within the range from 2 to 8, with up to 5 lattice spacings\footnote{At the largest mass, $z=8$,
only the 2-3 finest lattice spacings are useful in a linear extrapolation in $a^2$.}
and using precision results for $\Nf=0$ from \cite{DallaBrida:2019wur} then leads to the result for $\lms^{(\Nf=3)}$,
up to power corrections in $1/z$, expected to be predominantly of order $1/z^2$.
\begin{figure}[!htb]\hspace{-2cm}\begin{center}
      \includegraphics[width=13.5cm]{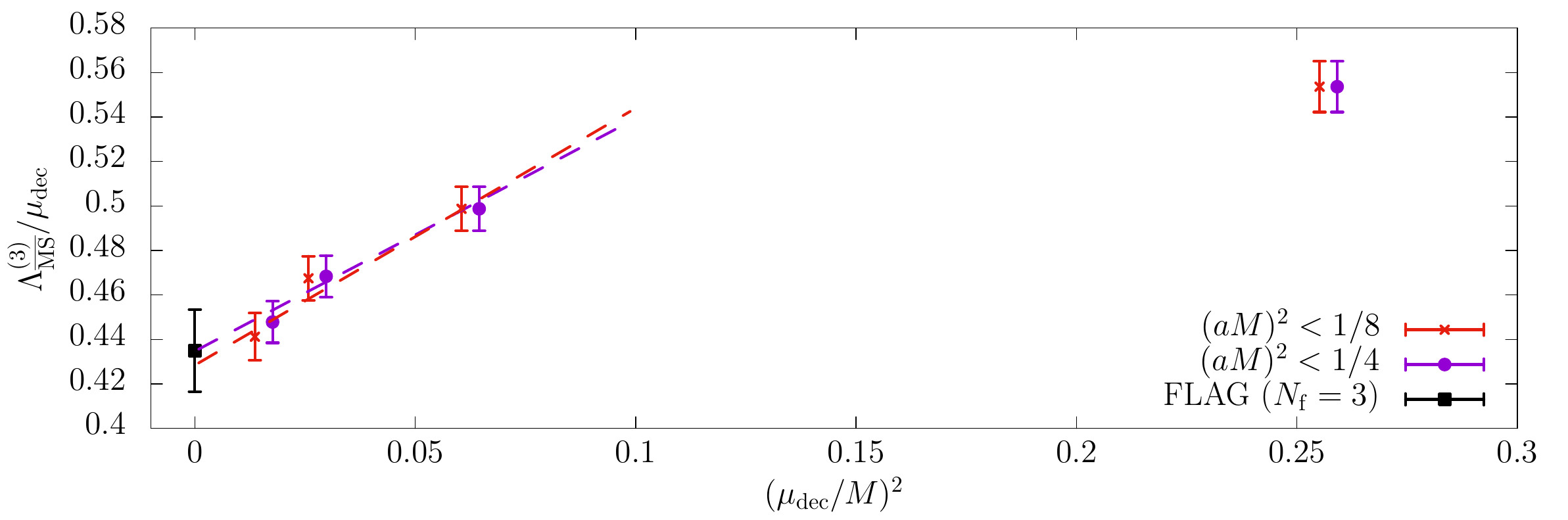}
      \end{center}
\vspace{-0.5cm}
\caption{Illustration of the decoupling method, taken from ref.~\cite{DallaBrida:2019mqg}.}
\label{fig:decoup}
\end{figure}
Figure~\ref{fig:decoup}, taken 
from \cite{DallaBrida:2019mqg} shows the continuum extrapolated results 
obtained for $\lms^{(3)}/\mudec$ at different values of $z$, together with the FLAG 19  average for three-flavour QCD. 
While the authors of \cite{DallaBrida:2019mqg} stopped short of quoting an extrapolated
value for the three-flavour $\Lambda$-parameter, the result 
$      \lms^{(3)} = 332(10)(2)\,\MeV $
is now given in the 2021 lattice conference proceedings~\cite{Brida:2021xwa}, compatible
with ALPHA 17 albeit with a somewhat smaller error. 
Despite some common elements with ALPHA 17, the authors emphasize that
the decoupling method is largely independent, with the 
overlap in squared error amounting to ca.~40 percent. 
This is due to the fact that the error in ALPHA 17 is dominated by the $\Nf=3$  step scaling procedure at {\em high} energy, 
and this part is completely replaced by the $\Nf=0$ result by Dalla Brida 19 \cite{DallaBrida:2019wur}.
The decoupling method thus seems to offer scope for a further error reduction, 
the major challenges being the continuum extrapolation for
the GF coupling at fixed and large RGI masses, followed by the large $M$ limit.

It is important to note that this new method relies 
on new precision results for $\Nf=0$ which have appeared in the last 
two years~\cite{DallaBrida:2019wur, Nada:2020jay}.
Therefore, the pure gauge theory acquires new relevance for $\alpha_s$ results,
beyond its traditional r\^ole as a test bed for the study of systematic errors.
FLAG will take account of this development by continuing to carefully monitor
$\Nf=0$ results. It is hoped that this will encourage more groups to undertake precision 
studies with $\Nf=0$. 

\subsection{$\alpha_s$ from the potential at short distances}
\label{s:qq}


\subsubsection{General considerations}


The basic method was introduced in Ref.~\cite{Michael:1992nj} and developed in
Ref.~\cite{Booth:1992bm}. The force or potential between an infinitely
massive quark and antiquark pair defines an effective coupling
constant via
\begin{eqnarray}
   F(r) = {d V(r) \over dr} 
        = C_F {\alpha_\mathrm{qq}(r) \over r^2} \,.
\label{force_alpha}
\end{eqnarray}
The coupling can be evaluated nonperturbatively from the potential
through a numerical differentiation, see below. In perturbation theory
one also defines couplings in different schemes $\alpha_{\bar{V}}$,
$\alpha_V$ via 
\begin{eqnarray}
   V(r) = - C_F {\alpha_{\bar{V}}(r) \over r} \,, 
   \qquad \mbox{or} \quad
   \tilde{V}(Q) = - C_F {\alpha_V(Q) \over Q^2} \,,
\label{pot_alpha}
\end{eqnarray}
where one fixes the unphysical constant in the potential
by $\lim_{r\to\infty}V(r)=0$ and $\tilde{V}(Q)$ is the
Fourier transform of $V(r)$. Nonperturbatively, the subtraction
of a constant in the potential introduces an additional 
renormalization constant, the value of $V(r_\mathrm{ref})$ at some 
distance $r_\mathrm{ref}$.  Perturbatively, it is believed to entail a 
renormalon ambiguity. In perturbation theory, the different definitions
are all simply related to each other, and their perturbative
expansions are known including the $\alpha_s^4,\,\alpha_s^4 \log\alpha_s$ 
and $\alpha_s^5 \log\alpha_s ,\,\alpha_s^5 (\log\alpha_s)^2$  terms
\cite{Fischler:1977yf,Billoire:1979ih,Peter:1997me,Schroder:1998vy,Brambilla:1999qa,Smirnov:2009fh,Anzai:2009tm,Brambilla:2009bi}.
 
The potential $V(r)$ is determined from ratios of Wilson loops,
$W(r,t)$, which behave as
\begin{eqnarray}
   \langle W(r, t) \rangle 
      = |c_0|^2 e^{-V(r)t} + \sum_{n\not= 0} |c_n|^2 e^{-V_n(r)t} \,,
      \label{e:vfromw}
\end{eqnarray}
where $t$ is taken as the temporal extension of the loop, $r$ is the
spatial one and $V_n$ are excited-state potentials.  To improve the
overlap with the ground state, and to suppress the effects of excited
states, $t$ is taken large. Also various additional techniques are
used, such as a variational basis of operators (spatial paths) to help
in projecting out the ground state.  Furthermore some
lattice-discretization effects can be reduced by averaging over Wilson
loops related by rotational symmetry in the continuum.

In order to reduce discretization errors it is of advantage 
to define the numerical derivative giving the force as
\begin{eqnarray}
   F(r_\mathrm{I}) = { V(r) - V(r-a) \over a } \,,
\end{eqnarray}
where $r_\mathrm{I}$ is chosen so that at tree level the force is the
continuum force. $F(r_\mathrm{I})$ is then a `tree-level improved' quantity
and similarly the tree-level improved potential can be defined
\cite{Necco:2001gh}.

Lattice potential results are in position space,
while perturbation theory is naturally computed in momentum space at
large momentum.
Usually, the Fourier transform 
of the perturbative expansion is then matched to  lattice data.

Finally, as was noted in Sec.~\ref{s:crit}, a determination
of the force can also be used to determine the scales $r_0,\,r_1$,
by defining them from the static force by
\begin{eqnarray}
   r_0^2 F(r_0) = {1.65} \,, \quad r_1^2 F(r_1) = 1\,.
   \label{eq:r01}
\end{eqnarray}


\subsubsection{Discussion of computations}
\label{short_dist_discuss}


\begin{table}[!htb]
   \vspace{3.0cm}
   \footnotesize
   \begin{tabular*}{\textwidth}[l]{l@{\extracolsep{\fill}}rllllll@{\hspace{1mm}}l@{\hspace{1mm}}ll}
      Collaboration & Ref. & $N_f$ &
      \hspace{0.15cm}\begin{rotate}{60}{publication status}\end{rotate}
                                                       \hspace{-0.15cm} &
      \hspace{0.15cm}\begin{rotate}{60}{renormalization scale}\end{rotate}
                                                       \hspace{-0.15cm} &
      \hspace{0.15cm}\begin{rotate}{60}{perturbative behaviour}\end{rotate}
                                                       \hspace{-0.15cm} &
      \hspace{0.15cm}\begin{rotate}{60}{continuum extrapolation}\end{rotate}
                               \hspace{-0.25cm} & 
                         scale & $\Lambda_\msbar[\MeV]$ & $r_0\Lambda_\msbar$ \\
      & & & & & & & & & \\[-0.1cm]
      \hline
      \hline
      & & & & & & & & & \\[-0.1cm]
    Ayala 20
                   & \cite{Ayala:2020odx} & 2+1  & \gA 
                   &  \soso     &  \good     & \soso 
                   & $r_1 = 0.3106(17)\,\mbox{fm}^c$ 
                   & $338(13)$
                   & $0.802(31)$                                    \\
    TUMQCD 19
                   & \cite{Bazavov:2019qoo} & 2+1  & \gA 
                   &  \soso     &  \good     & \soso 
                   & $r_1 = 0.3106(17)\,\mbox{fm}^c$ 
                   & $314^{+16}_{-8}$
                   & $0.745(^{+38}_{-19})$                                \\
  
     Takaura 18
                   & \cite{Takaura:2018lpw,Takaura:2018vcy} & 2+1  & \gA 
                   &  \bad     &  \soso     & \soso 
                   & $\sqrt{t_0}=0.1465(25)\fm^a$
                   & $334(10)(^{+20}_{-18})^b$
                   & $0.799(51)$$^+$                                 \\

      {Bazavov 14}
                   & \cite{Bazavov:2014soa}  & 2+1       & \gA & \soso
                   & \good  & \soso
                   & $r_1 = 0.3106(17)\,\mbox{fm}^c$
                   & $315(^{+18}_{-12})^d$
                   & $0.746(^{+42}_{-27})$                              \\

      {Bazavov 12}
                   & \cite{Bazavov:2012ka}   & 2+1       & \gA & \soso$^\dagger$
                   & \soso   & \soso$^\#$
                   & $r_0 = 0.468\,\mbox{fm}$ 
                   & $295(30)$\,$^\star$ 
                   & $0.70(7)$$^{\star\star}$                                   \\
      & & & & & & & & & \\[-0.1cm]
      \hline
      & & & & & & & & & \\[-0.1cm]

     Karbstein 18 
                   & \cite{Karbstein:2018mzo} & 2        & \gA
                   & \soso        &  \soso    & \soso
                   & $r_0 = 0.420(14)\,\mbox{fm}$$^e$
                   & $302(16)$
                   & $0.643(34)$                                       \\

     Karbstein 14 
                   & \cite{Karbstein:2014bsa} & 2        & \gA & \soso
                   & \soso & \soso
                   & $r_0 = 0.42\,\mbox{fm}$
                   & $331(21)$
                   & 0.692(31)                                         \\

      ETM 11C      & \cite{Jansen:2011vv}    & 2         & \gA & \soso  
                   & \soso  & \soso
                   & $r_0 = 0.42\,\mbox{fm}$
                   & $315(30)^\S$ 
                   & $0.658(55)$                                        \\
      & & & & & & & & & \\[-0.1cm]
      \hline
      & & & & & & & & & \\[-0.1cm]

      Husung 20    & \cite{Husung:2020pxg}  & 0  & C
                   & \soso & \good   & \good
                   &  \multicolumn{3}{c}{\text{no quoted value for} $\lms$}  \\
      
      Husung 17    & \cite{Husung:2017qjz}   & 0         & C
                   & \soso & \good   & \good
                   &  $r_0 = 0.50\,\mbox{fm}$ 
                   & 232(6) & $0.590(16)$  \\

      Brambilla 10 & \cite{Brambilla:2010pp} & 0         & \gA & \soso 
                   & \good\ & \soso$^{\dagger\dagger}$ &  & $266(13)$$^{+}$&
                   $0.637(^{+32}_{-30})$$^{\dagger\dagger}$                  \\
      UKQCD 92     & \cite{Booth:1992bm}    & 0         & \gA & \good 
                                  & \soso$^{++}$   & \bad   
                                  & $\sqrt{\sigma}=0.44\,\GeV$ 
                                  & $256(20)$
                                  & 0.686(54)                             \\
      Bali 92     & \cite{Bali:1992ru}    & 0         & \gA & \good 
                                  & \soso$^{++}$   & \bad 
                                  & $\sqrt{\sigma}=0.44\,\GeV$
                                  & $247(10)$                             
                                  & 0.661(27)                             \\
      & & & & & & & & & \\[-0.1cm]
      \hline
      \hline\\
\end{tabular*}\\[-0.2cm]
\begin{minipage}{\linewidth}
{\footnotesize 
\begin{itemize}
   \item[$^a$] Scale determined from $t_0$ in
               Ref.~\cite{Borsanyi:2012zs}.
   \item[$^b$]             
               $\alpha^{(5)}_{\overline{\rm MS}}(M_Z) = 0.1179(7)(^{+13}_{-12})$.  
   \item[$^c$]
   Determination on lattices with $m_\pi L=2.2 - 2.6$. 
   Scale from $r_1$ \cite{Bazavov:2014pvz}
   as determined from  $f_\pi$ in Ref.~\cite{Bazavov:2010hj}.      \\[-5mm]
   \item[$^d$]
         $\alpha^{(3)}_{\overline{\rm MS}}(1.5\,\mbox{GeV}) = 0.336(^{+12}_{-8})$, 
         $\alpha^{(5)}_{\overline{\rm MS}}(M_Z) = 0.1166(^{+12}_{-8})$.
         \\[-5mm]
   \item[$^e$] 
         Scale determined from $f_\pi$, see \cite{Baron:2009wt}.   \\[-5mm]
   \item[$^\dagger$]
   Since values of $\alpha_\mathrm{eff}$ within our designated range are used,
   we assign a \soso\ despite
   values of $\alpha_\mathrm{eff}$ up to $\alpha_\mathrm{eff}=0.5$ being used.  
   \\[-5mm]
   \item[$^\#$]Since values of $2a/r$ within our designated range are used,
   we assign a \soso\ although
   only values of $2a/r\geq1.14$ are used at $\alpha_\mathrm{eff}=0.3$.
   \\[-5mm]
   \item[$^\star$] Using results from Ref.~\cite{Bazavov:2011nk}.  \\[-5mm]
   \item[$^{\star\star}$]
         $\alpha^{(3)}_{\overline{\rm MS}}(1.5\,\mbox{GeV}) = 0.326(19)$, 
         $\alpha^{(5)}_{\overline{\rm MS}}(M_Z) = 0.1156(^{+21}_{-22})$.  \\[-5mm]
   \item[$^\S$] Both potential and $r_0/a$ are determined on a small 
   ($L=3.2r_0$) lattice.   \\[-5mm]
   \item[$^{\dagger\dagger}$] Uses lattice results of Ref.~\cite{Necco:2001xg}, 
   some of which have very small lattice spacings where 
   according to more recent investigations a bias due to the freezing of
   topology may be present.  \\[-5mm] 
   \item[$^+$] Our conversion using $r_0 = 0.472\,\mbox{fm}$.   \\[-5mm]
   \item[$^{++}$] We give a $\soso$ because only a NLO formula is used and
       the error bars are very large; our criterion does not apply 
       well to these very early calculations.           
\end{itemize}
}
\end{minipage}
\normalsize
\caption{Short-distance potential results.}
\label{tab_short_dist}
\end{table}

In Tab.~\ref{tab_short_dist}, we list results of determinations
of $r_0\Lambda_{\msbar}$ (together with $\Lambda_{\msbar}$
using the scale determination of the authors).

Since the last review, FLAG 19, there have been three new publications,
namely, TUMQCD 19 \cite{Bazavov:2019qoo}, Ayala 20 \cite{Ayala:2020odx}
and Husung 20 \cite{Husung:2020pxg}.

The first determinations in the three-colour Yang Mills theory are by
UKQCD 92 \cite{Booth:1992bm} and Bali 92 \cite{Bali:1992ru} who used
$\alpha_\mathrm{qq}$ as explained above, but not in the tree-level
improved form. Rather a phenomenologically determined lattice-artifact
correction was subtracted from the lattice potentials.  The comparison
with perturbation theory was on a more qualitative level on the basis
of a 2-loop $\beta$-function ($n_l=1$) and a continuum extrapolation
could not be performed as yet. A much more precise computation of
$\alpha_\mathrm{qq}$ with continuum extrapolation was performed in
Refs.~\cite{Necco:2001xg,Necco:2001gh}. Satisfactory agreement with
perturbation theory was found \cite{Necco:2001gh} but the stability of
the perturbative prediction was not considered sufficient to be able
to extract a $\Lambda$ parameter.

In Brambilla 10 \cite{Brambilla:2010pp} the same quenched lattice
results of Ref.~\cite{Necco:2001gh} were used and a fit was performed to
the continuum potential, instead of the force. Perturbation theory to
$n_l=3$ loop
was used including a resummation of terms $\alpha_s^3 (\alpha_s \ln\alpha_s)^n $ 
and $\alpha_s^4 (\alpha_s \ln\alpha_s)^n $. Close
agreement with perturbation theory was found when a renormalon
subtraction was performed. Note that the renormalon subtraction
introduces a second scale into the perturbative formula which is
absent when the force is considered.

Bazavov 14 \cite{Bazavov:2014soa} updates 
Bazavov 12 \cite{Bazavov:2012ka} and modifies this procedure
somewhat. They consider the 
perturbative expansion
for the force. 
They set $\mu = 1/r$
to eliminate logarithms and then integrate the force to obtain an
expression for the potential. 
The resulting integration constant is fixed by requiring
the perturbative potential to be equal to the nonperturbative 
one exactly at a reference distance $r_{\rm ref}$ and the two are then
compared at other values of $r$. As a further check,
the force is also used directly.

For the quenched calculation of Brambilla 10 \cite{Brambilla:2010pp}
very small lattice spacings,
$a \sim 0.025\,\mbox{fm}$, were available from Ref.~\cite{Necco:2001gh}.
For ETM 11C \cite{Jansen:2011vv}, Bazavov 12 \cite{Bazavov:2012ka},
Karbstein 14 \cite{Karbstein:2014bsa}
and Bazavov 14 \cite{Bazavov:2014soa} using dynamical
fermions such small lattice spacings are not yet realized 
(Bazavov 14 reaches down to $a \sim 0.041\,\mbox{fm}$). They
all use the tree-level improved potential as described above. 
We note that the value of $\Lambda_\msbar$ in physical units by
ETM 11C \cite{Jansen:2011vv} is based on a value of $r_0=0.42$~fm. 
This is at least 10\% smaller than the large majority of
other values of $r_0$. Also the values of $r_0/a$ 
on the finest lattices in ETM 11C \cite{Jansen:2011vv}
and $r_1/a$ for Bazavov 14 \cite{Bazavov:2014soa} come from
rather small lattices with $m_\pi L \approx 2.4$, $2.2$ respectively.

Instead of the procedure discussed previously, Karbstein 14 
\cite{Karbstein:2014bsa} reanalyzes the data of ETM 11C 
\cite{Jansen:2011vv} by first estimating
the Fourier transform $\tilde V(p)$ of $V(r)$ and then fitting  
the perturbative expansion of $\tilde V(p)$ in terms of 
$\alpha_\msbar(p)$. Of course, the Fourier transform requires
some modelling of the $r$-dependence of $V(r)$
at short and at large distances. The authors fit a linearly rising
potential at large distances together with string-like
corrections of order $r^{-n}$ and define the potential at large 
distances by this fit.\footnote{Note that at large distances,
where string breaking is known to occur, this is not 
any more the ground state potential defined by \eq{e:vfromw}.}
Recall that for observables in momentum space
we take the renormalization scale entering our criteria as $\mu=q$,
Eq.~(\ref{mu_def}). The analysis (as in ETM 11C \cite{Jansen:2011vv})
is dominated by the data at the smallest lattice spacing, where
a controlled determination of the overall scale  is difficult due to 
possible finite-size effects.
Karbstein 18  \cite{Karbstein:2018mzo} is a
reanalysis of Karbstein 14 and supersedes it. Some data with a different
discretization of the static quark is added (on the same configurations)
and the discrete lattice results for the static potential in position
space are first parameterized by a continuous function, which then
allows for an analytical Fourier transformation to momentum space.

Similarly also for Takaura 18~\cite{Takaura:2018lpw,Takaura:2018vcy}
the momentum space potential $\tilde{V}(Q)$ is the central object. 
Namely, they assume that renormalon/power-law effects are absent in
$\tilde{V}(Q)$ and only come in through the Fourier transformation.
They provide evidence that renormalon effects (both $u=1/2$ and $u=3/2$) can be 
subtracted and arrive at a nonperturbative term $k\,\Lambda_\msbar^3 r^2$. 
Two different analyses are carried out with the final result 
taken from ``Analysis II''. Our numbers including the evaluation of
the criteria refer to it. Together with the 
perturbative 3-loop (including the $ \alpha_s^4\log \alpha_s$ term) 
expression, this term is fitted to the 
nonperturbative results for the potential in the region
$0.04\,\fm \, \leq \, r \,\leq 0.35\,\fm$, where $0.04\,\fm$ is 
$r=a$ on the finest lattice.
The nonperturbative potential data originates from JLQCD ensembles (Symanzik-improved
gauge action and M\"obius domain-wall quarks) at three lattice spacings with 
a pion mass around $300\,\MeV$. Since at the maximal distance in the analysis
we find $\alpha_\msbar(2/r) = 0.43$, the renormalization scale 
criterion yields a \bad. 
The perturbative
behaviour is \soso\ because of the high orders in perturbation theory known. The 
continuum-limit criterion yields a $\soso$.

One of the main issues for all these computations is whether the
perturbative running of the coupling constant
has been reached.
While for $N_f=0$ fermions Brambilla~10
\cite{Brambilla:2010pp} reports agreement with perturbative behaviour
at the smallest distances, Husung~17 (which goes to
shorter distances) finds relatively large corrections beyond the 3-loop
$\alpha_\mathrm{qq}$.
For dynamical fermions,  
 Bazavov 12 \cite{Bazavov:2012ka}
and Bazavov 14 \cite{Bazavov:2014soa} report good agreement with perturbation
theory after the renormalon is subtracted or eliminated. 

A second issue is the coverage of configuration space in some of the
simulations, which use very small lattice spacings with periodic
boundary conditions. Affected are the smallest two lattice spacings
of Bazavov 14 \cite{Bazavov:2014soa} where very few tunnelings of
the topological charge occur \cite{Bazavov:2014pvz}.
With present knowledge, it also seems  possible that the older data
by Refs.~\cite{Necco:2001xg,Necco:2001gh} used by Brambilla 10 
\cite{Brambilla:2010pp} are partially obtained with (close to) frozen topology.

The computation in Husung 17~\cite{Husung:2017qjz},
for $N_f = 0$ flavours, first determines the coupling 
$\gbar_{\rm qq}^2(r,a)$ from the force and then performs a continuum extrapolation
on lattices down to $a \approx 0.015\,\mbox{fm}$, using a step-scaling method at short distances, $r/r_0 \lsim 0.5$. 
Using the $4$-loop $\beta^{\rm qq}$ function this allows $r_0\Lambda_{\rm qq}$
to be estimated, which is then converted to the $\overline{\rm MS}$ scheme.
$\alpha_{\rm eff} = \alpha_{\rm qq}$ ranges from $\sim 0.17$ to large 
values; we 
give $\soso$ for renormalization scale and \good\ for perturbative behaviour. The range
$a\mu = 2a/r \approx$ 0.37--0.14 leads to a $\good$
in the continuum extrapolation.
Recently these calculations have been extended in Husung 20~\cite{Husung:2020pxg}.
A finer lattice spacing of $a=0.01\,\fm$ (scale from $r_0=0.5~\fm$) is reached and lattice volumes up to $L/a=192$ are
simulated (in Ref.~\cite{Husung:2017qjz} the smallest lattice spacing is $0.015\,\fm$).
The Wilson action is used despite its significantly larger cutoff effects compared to Symanzik-improved actions;
this avoids unitarity violations, thus allowing for a clean ground state extraction via a generalized eigenvalue problem.
Open boundary conditions are used to avoid the topology-freezing problem.
Furthermore, new results for the continuum approach are employed, which determine the
cutoff dependence at $\cO(a^2)$ including the exact coupling-dependent terms, in the asymptotic region where
the Symanzik effective theory is applicable~\cite{Husung:2019ytz}. An ansatz for the remaining
higher order cutoff effects at $\cO(a^4)$ is propagated as a systematic error to the data, which
effectively discards data for $r/a < 3.5$. The large volume step-scaling function with step factor $3/4$ is
computed and compared to perturbation theory. For $\alpha_{qq}>0.2$ there is a noticeable difference between the 2-loop and 3-loop results. 
Furthermore, the ultra-soft contributions at 4-loop level give a significant
contribution to the static $Q\bar Q$ force. While this study is for $\Nf=0$ flavours 
it does raise the question whether the weak coupling expansion for the range of $r$-values 
used in present analyses of $\alpha_s$ is sufficiently reliable.
Around $\alpha_{\rm qq} \approx 0.21$ the differences get smaller but the error increases significantly, 
mainly due to the propagated lattice artifacts. 
The dependence of $\Lambda_{\overline{\rm MS}}^{n_f=0} \sqrt{8 t_0}$ on $\alpha_{\rm qq}^3$ is very similar to
the one observed in the previous study but no value for its $\alpha_{qq}\rightarrow 0$ limit
is quoted. Husung 20 \cite{Husung:2020pxg} is more pessimistic about the error on the $\Lambda$ parameter stating
the relative error has to be $5\%$ or larger, while Husung 17 quotes a relative error of $3 \%$.

In 2+1-flavor QCD two new papers appeared on the determination of the strong coupling constant
from the static quark anti-quark potential after the FLAG 19 report
\cite{Bazavov:2019qoo,Ayala:2020odx}. 
In TUMQCD 19~\cite{Bazavov:2019qoo}\footnote{The majority of authors are the same as in ~\cite{Bazavov:2014soa}.} 
the 2014 analysis of Bazavov~14~\cite{Bazavov:2014soa} 
has been extended by including three finer lattices with lattice
spacing $a=$ 0.035, 0.030 and 0.025~fm as well as lattice results
on the free energy of static quark anti-quark pair at non-zero temperature.
On the new fine lattices the effect of freezing topology has been observed, however,
it was verified that this does not affect the potential within the estimated errors
\cite{Bazavov:2017dsy,Weber:2018bam}. The comparison of the lattice result on
the static potential has been performed in the interval $r=[r_{\mathrm{min}},r_{\mathrm{max}}]$, with 
$r_{\mathrm{max}}$ = 0.131, 0.121, 0.098, 0.073 and 0.055~fm. The main result quoted in
the paper is based on the analysis with $r_{\mathrm{max}}=$ 0.073~fm \cite{Bazavov:2019qoo}. 
Since the new study employs a much wider range in $r$ than the previous one \cite{Bazavov:2014soa}
we give it a $\good$  for the perturbative behaviour. 
Since $\alpha_\text{eff}=\alpha_{qq}$ varies in the range 0.2--0.4 for the $r$ values used in
the main analysis we give $\soso$ for the renormalization scale.
Several values of $r_{\mathrm{min}}$ have been used
in the analysis, the largest being $r_{\mathrm{min}}/a=\sqrt{8}\simeq$ 2.82, which corresponds
to $a \mu \simeq$ 0.71. Therefore, we give a $\soso$ for continuum extrapolation
in this case. An important difference compared to the previous study \cite{Bazavov:2014soa} is
the variation of the renormalization scale. In Ref. \cite{Bazavov:2014soa} the renormalization
scale was varied by a factor of $\sqrt{2}$ around the nominal value of $\mu=1/r$, in order to
exclude very low scales, for which the running of the strong coupling constant is no longer
perturbative. In the new analysis the renormalization scale was varied by a factor of two.
As the result, despite the extended data set and shorter distances used in the new study
the perturbative error did not decrease \cite{Bazavov:2019qoo}. 
We also note that the scale dependence turned out to be non-monotonic in the range $\mu=1/(2r)$--$2/r$ \cite{Bazavov:2019qoo}.
The final result
reads (``us" stands for ``ultra-soft"),
\begin{eqnarray}
\lms^{\Nf=3}&=&314.0 \pm 5.8 (\text{stat}) \pm 3.0(\text{lat}) 
    \pm 1.7 (\text{scale})^{+13.4}_{-1.8} (\text{pert}) \pm 4.0 (\text{pert. us}) ~{\MeV} \nonumber\\
    &=& 314^{+16}_{-08}\,{\MeV}\,,
\label{eq:Lam-tumqcd}
\end{eqnarray}
where all errors were combined in quadrature. This is
in very good agreement with the previous determination \cite{Bazavov:2014soa}.

The analysis was also applied to the singlet static quark anti-quark free energy
at short distances. At short distances the free energy is expected to be the same
as the static potential. This is verified numerically in the lattice calculations
TUMQCD 19~\cite{Bazavov:2019qoo} for $rT < 1/4$ with $T$ being the temperature. Furthermore,
this is confirmed by the perturbative calculations at $T>0$ at NLO {\cite{Berwein:2017thy}.
The advantage of using the free energy is that it gives access to much shorter distances.
On the other hand, one has fewer data points because the condition $rT < 1/4$
has to be satisfied. The analysis based on the free energy gives
\begin{eqnarray}
\lms^{\Nf=3}&=&310.9 \pm 11.3 (\text{stat}) \pm 3.0(\text{lat}) 
\pm 1.7 (\text{scale})^{+5.6}_{-0.8}(\text{pert}) \pm 2.1 (\text{pert. us}) ~{\MeV} \nonumber \\
 &=& 311(13) \,\MeV,
\end{eqnarray}
in good agreement with the above result and thus, providing additional confirmation of it.

The analysis of Ayala 20~\cite{Ayala:2020odx} uses a subset of data presented in 
TUMQCD~19~\cite{Bazavov:2019qoo} with the same correction of the lattice effects. 
For this reason the continuum extrapolation gets $\soso$, too.
They match to perturbation theory for $1/r >2$ GeV, which corresponds to 
$\alpha_\text{eff}=\alpha_{qq}=$ 0.2--0.4. 
Therefore, we give $\soso$ for the renormalization scale.
They verify the perturbative behaviour in the region $1~{\rm GeV}<1/r<2.9~{\rm GeV}$, which
corresponds to variation of $\alpha_\text{eff}^3$ by a factor of 3.34. However, the relative
error on the final result has $\delta \Lambda/\Lambda \simeq 0.035$ which is larger
than $\alpha_\text{eff}^3=0.011$. Therefore, we give a $\good$ for the perturbative behaviour
in this case. The final result for the $\Lambda$-parameter reads:
\begin{equation}
\lms^{\Nf=3}=338 \pm 2 (\text{stat}) \pm 8 (\text{matching}) \pm 10 (\text{pert})~{\MeV} = 338(13)\MeV\,.
\label{eq:Lam-ayala}
\end{equation}
This is quite different from the above result. This difference is mostly due to the organization
of the perturbative series. The authors use ultra-soft (log) resummation, i.e.~they resum the terms
$\alpha_s^{3+n} \ln^n\alpha_s$ to all orders instead of using fixed-order perturbation theory.
They also include what is called the terminant of the perturbative series associated to the leading renormalon 
of the force \cite{Ayala:2020odx}.
When they use fixed order perturbation theory they obtain very similar 
results to Refs.~\cite{Bazavov:2014soa,Bazavov:2019qoo}. 
It has been argued that log resummation
cannot be justified since for the distance range available in the lattice studies $\alpha_s$
is not small enough and the logarithmic and non-logarithmic higher-order terms are of
a similar size \cite{Bazavov:2014soa}. On the other hand, the resummation of ultra-soft logs does not lead
to any anomalous behaviour of the perturbative expansion like large scale dependence or bad convergence \cite{Ayala:2020odx}.

To obtain the value of $\lms^{\Nf=3}$ from the static potential
we combine the results in Eqs.~(\ref{eq:Lam-tumqcd}) and (\ref{eq:Lam-ayala}) using the weighted
average with the weight given by the perturbative error and using the difference in
the central value as the error estimate. This leads to
\begin{equation}
\lms^{\Nf=3}=330(24) ~{\MeV}\,,
\end{equation}
from the static potential determination. In the case of TUMQCD 19, where the perturbative error
is very asymmetric we used the larger upper error for the calculation of the corresponding weight.


\subsection{$\alpha_s$ from the light-quark vacuum polarization in momentum/po\-sition space}


\label{s:vac}


\subsubsection{General considerations}


Except for the new calculation Cali 20 \cite{Cali:2020hrj}, where position space is used (see below),
the light-flavour-current 2-point function is usually evaluated in momentum 
space, in terms of the vacuum-polarization function.  For the flavour-nonsinglet 
currents $J^a_\mu$ ($a=1,2,3$) in the momentum representation this is
parametrized as 
\begin{eqnarray}
   \langle J^a_\mu J^b_\nu \rangle 
      =\delta^{ab} [(\delta_{\mu\nu}Q^2 - Q_\mu Q_\nu) \Pi_J^{(1)}(Q) 
                                     - Q_\mu Q_\nu\Pi_J^{(0)}(Q)] \,,
\end{eqnarray}
where $Q_\mu$ is a space-like momentum and $J_\mu\equiv V_\mu$
for a vector current and $J_\mu\equiv A_\mu$ for an axial-vector current. 
Defining $\Pi_J(Q)\equiv \Pi_J^{(0)}(Q)+\Pi_J^{(1)}(Q)$,
the operator product expansion (OPE) of  $\Pi_{V/A}(Q)$ is given by
\begin{eqnarray}
   \lefteqn{\Pi_{V/A}|_{\rm OPE}(Q^2,\alpha_s)}
      & &                                             \nonumber  \\
      &=& c + C_1^{V/A}(Q^2) + C_m^{V/A}(Q^2)
                       \frac{\bar{m}^2(Q)}{Q^2}
            + \sum_{q=u,d,s}C_{\bar{q}q}^{V/A}(Q^2)
                        \frac{\langle m_q\bar{q}q \rangle}{Q^4}
                                                      \nonumber  \\
      & &   + C_{GG}^{V/A}(Q^2) 
                \frac{\langle \alpha_s GG\rangle}{Q^4}+{\cO}(Q^{-6}) \,,
\label{eq:vacpol}
\end{eqnarray}
for large
$Q^2$. The perturbative coefficient functions $C_X^{V/A}(Q^2)$ for the
operators $X$ ($X=1$, $\bar{q}q$, $GG$) are given as $C_X^{V/A}(Q^2)=\sum_{i\geq0}\left( C_X^{V/A}\right)^{(i)}\alpha_s^i(Q^2)$  and $\bar m$ is the running 
mass of the mass-degenerate up and down quarks.
$C_1^{V/A}$ is known including $\alpha_s^4$
in a continuum renormalization scheme such as the
$\overline{\rm MS}$ scheme
\cite{Chetyrkin:1979bj,Surguladze:1990tg,Gorishnii:1990vf,Baikov:2008jh}.
Nonperturbatively, there are terms in $C_X^{V/A}$ that do not have a 
series expansion in $\alpha_s$. For an example for the unit
operator see Ref.~\cite{Balitsky:1993ki}.
The term $c$ is $Q$-independent and divergent in the limit of infinite
ultraviolet cutoff. However the Adler function defined as 
\begin{eqnarray}
   D(Q^2) \equiv - Q^2 { d\Pi(Q^2) \over dQ^2} \,,
\label{eq:adler}
\end{eqnarray}
is a scheme-independent finite quantity. Therefore one can determine
the running-coupling constant in the $\overline{\rm MS}$ scheme
from the vacuum-polarization function computed by a lattice-QCD
simulation. Of course, there is the choice whether to use the vector or the axial vector channel,
or both, the canonical choice being $\Pi_{V+A} = \Pi_V+\Pi_A$. 
While perturbation theory does not distinguish between
these channels, the nonperturbative contributions are different, and the 
quality of lattice data may differ, too.
For a given choice, the lattice data of the vacuum polarization is fitted with the 
perturbative formula Eq.~(\ref{eq:vacpol}) with fit parameter 
$\Lambda_{\overline{\rm MS}}$ parameterizing the running coupling 
$\alpha_{\overline{\rm MS}}(Q^2)$.  

While there is no problem in discussing the OPE at the
nonperturbative level, the `condensates' such as ${\langle \alpha_s
  GG\rangle}$ are ambiguous, since they mix with lower-dimensional
operators including the unity operator.  Therefore one should work in
the high-$Q^2$ regime where power corrections are negligible within
the given accuracy. Thus setting the renormalization scale as
$\mu\equiv \sqrt{Q^2}$, one should seek, as always, the window
$\Lambda_{\rm QCD} \ll \mu \ll a^{-1}$.

\subsubsection{Definitions in position space}

The 2-point current correlation functions in position space
contain the same physical information as in momentum space, but the technical 
details are sufficiently different to warrant a separate discussion.
The (Euclidean) current-current correlation function for $J^\mu_{ff'}$ (with flavour indices $f,f'$)
is taken to be either the flavour non-diagonal vector or axial vector current,
with the Lorentz indices contracted,
\begin{equation}
    C_\text{A,V}(x) = -\sum_{\mu}\left\langle J^\mu_{ff'\text{A,V}}(x)J^\mu_{f'f\text{A,V}}
(0)\right\rangle = \dfrac{6}{\pi^4(x^2)^3}\left( 1 + \frac{\alpha_s}{\pi} + \cO(\alpha^2)\right)\,.
\end{equation}
In the chiral limit, the perturbative expansion is known to $\alpha_s^4$~\cite{Chetyrkin:2010dx}, and is identical
for vector and axial vector correlators. The only scale is set by the Euclidean distance $\mu=1/|x|$ and 
the effective coupling can thus be defined as
\begin{equation}
  \alpha_\text{eff}(\mu=1/|x|) = \pi \left[(x^2)^3(\pi^4/6) C_\text{A,V}(x) - 1\right]\,.
\end{equation}
As communicated to us by the authors of \cite{Cali:2020hrj}, 
there is a typo in Eq.~(35) of~\cite{Chetyrkin:2010dx}.
For future reference, the numerical coefficients for the 3-loop conversion
\begin{equation}
   \alpha_\text{eff}(\mu) = \alpha_\msbar(\mu) +  c_1 \alpha^2_\msbar(\mu) + c_2\alpha^3_\msbar(\mu) + c_3\alpha^4_\msbar(\mu),
  \label{eq:CVAMSbar}
   \end{equation}
should read
\begin{equation}
   c_1 = -1.4346,\qquad c_2=0.16979,\qquad   c_3= 3.21120\,. 
\end{equation}


\subsubsection{Discussion of computations}


Results using this method in momentum space are, to date, only available using
overlap fermions or domain-wall fermions. Since the last review, FLAG 19,
there has been one new computation, Cali 20~\cite{Cali:2020hrj}, which uses
the vacuum polarization in position space, using $\cO(a)$ improved Wilson fermions.
The results are collected in Tab.~\ref{tab_vac} for
$N_f=2$, JLQCD/TWQCD 08C \cite{Shintani:2008ga} and for $N_f = 2+1$, JLQCD 10
\cite{Shintani:2010ph}, Hudspith 18~ \cite{Hudspith:2018bpz} and Cali 20 \cite{Cali:2020hrj}.

\begin{table}[!htb]
   \vspace{3.0cm}
   \footnotesize
   \begin{tabular*}{\textwidth}[l]{l@{\extracolsep{\fill}}rllllllll}
   Collaboration & Ref. & $\Nf$ &
   \hspace{0.15cm}\begin{rotate}{60}{publication status}\end{rotate}
                                                    \hspace{-0.15cm} &
   \hspace{0.15cm}\begin{rotate}{60}{renormalization scale}\end{rotate}
                                                    \hspace{-0.15cm} &
   \hspace{0.15cm}\begin{rotate}{60}{perturbative behaviour}\end{rotate}
                                                    \hspace{-0.15cm} &
   \hspace{0.15cm}\begin{rotate}{60}{continuum extrapolation}\end{rotate}
      \hspace{-0.25cm} & 
                         scale & $\Lambda_\msbar[\MeV]$ & $r_0\Lambda_\msbar$ \\
   & & & & & & & & & \\[-0.1cm]
   \hline
   \hline
   & & & & & & & & & \\[-0.1cm]
   Cali 20 & \cite{Cali:2020hrj} & 2+1 & \gA
            & \soso  & \good   & \good  
            &  $m_\Upsilon$$^\S$
            & $342(17)$
            & $0.818(41)$$^a$               \\
   & & & & & & & & & \\[-0.1cm]
   \hline
   & & & & & & & & & \\[-0.1cm]

   Hudspith 18 & \cite{Hudspith:2018bpz} & 2+1 & \oP
            & \soso  & \soso   & \bad  
            &  $m_\Omega$$^\star$
            & $337(40)$
            & $0.806(96)$$^b$               \\
   & & & & & & & & & \\[-0.1cm]
   \hline
   & & & & & & & & & \\[-0.1cm]

   Hudspith 15 & \cite{Hudspith:2015xoa} & 2+1 &\rC 
            & \soso  & \soso   & \bad  
            & $m_\Omega$$^\star$
            & $300(24)^+$
            & $0.717(58)$              \\
   & & & & & & & & & \\[-0.1cm]
   \hline
   & & & & & & & & & \\[-0.1cm]
   JLQCD 10 & \cite{Shintani:2010ph} & 2+1 &\gA & \bad 
            & \soso & \bad
            & $r_0 = 0.472\,\mbox{fm}$
            & $247(5)$$^\dagger$
            & $0.591(12)$              \\
   & & & & & & & & & \\[-0.1cm]
   \hline
   & & & & & & & & & \\[-0.1cm]
   JLQCD/TWQCD 08C & \cite{Shintani:2008ga} & 2 & \gA & \soso 
            & \soso & \bad
            & $r_0 = 0.49\,\mbox{fm}$
            & $234(9)(^{+16}_{-0})$
            & $0.581(22)(^{+40}_{-0})$    \\
            
   & & & & & & & & & \\[-0.1cm]
   \hline
   \hline
\end{tabular*}
\begin{tabular*}{\textwidth}[l]{l@{\extracolsep{\fill}}llllllll}
\multicolumn{8}{l}{\vbox{\begin{flushleft}
   $^\S$ via $t_0/a^2$, still unpublished. We use $r_0=0.472\,\fm$\\
   $^\star$ Determined in \cite{Blum:2014tka}.  \\
   $^a$ Evaluates to $\alpha_\msbar^{(5)}(M_Z)= 0.11864(114)$\\
   In conversion to $r_0\Lambda$ we used
        $r_0 = 0.472\,\mbox{fm}$.  \\
   $^b$ 
        $\alpha_\msbar^{(5)}(M_Z)=0.1181(27)(^{+8}_{-22})$. $\Lambda_\msbar$
        determined by us from $\alpha_\msbar^{(3)}(2\,\mbox{GeV})=0.2961(185)$.
        In conversion to $r_0\Lambda$ we used
        $r_0 = 0.472\,\mbox{fm}$.  \\
        $^+$ Determined by us from $\alpha_\msbar^{(3)}(2\,\GeV)=0.279(11)$. 
       Evaluates to $\alpha_\msbar^{(5)}(M_Z)=0.1155(18)$. \\
   $^\dagger$  $\alpha_\msbar^{(5)}(M_Z)=0.1118(3)(^{+16}_{-17})$. \\
   
\end{flushleft}}}
\end{tabular*}
\vspace{-0.3cm}
\normalsize
\caption{Results from the  vaccum polarization in both momentum and position space}
\label{tab_vac}
\end{table}

We first discuss the results of JLQCD/TWQCD 08C 
\cite{Shintani:2008ga} and JLQCD 10 \cite{Shintani:2010ph}.
The fit to \eq{eq:vacpol} is done with the 4-loop relation between
the running coupling and $\lms$. It is found that without introducing
condensate contributions, the momentum scale where the perturbative
formula gives good agreement with the lattice results is very narrow,
$aQ \simeq$ 0.8--1.0. When a condensate contribution is included the
perturbative formula gives good agreement with the lattice results for
the extended range $aQ \simeq$ 0.6--1.0. Since there is only a single
lattice spacing $a \approx$ 0.11~fm there is a 
\bad\ for the continuum limit. The renormalization scale $\mu$ is in
the range of $Q=$ 1.6--2~GeV. Approximating 
$\alpha_{\rm eff}\approx \alpha_{\overline{\rm MS}}(Q)$, we estimate that
$\alpha_{\rm eff}=$ 0.25--0.30 for $N_f=2$ and $\alpha_{\rm  eff}=$ 0.29--0.33
for $N_f=2+1$. Thus we give a \soso\ and \bad\ for $\Nf=2$ and 
$\Nf=2+1$, respectively, for the renormalization scale and a \bad\ for
the perturbative behaviour.

A further investigation of this method was initiated in
Hudspith 15 \cite{Hudspith:2015xoa} and completed by Hudspith 18 
\cite{Hudspith:2018bpz} (see also \cite{Hudspith:2018zlq}) based
on domain-wall fermion configurations at three lattice spacings, 
$a^{-1} =$ 1.78, 2.38, 3.15~GeV, with three different 
light-quark masses on the two coarser lattices and one on the fine lattice.
An extensive discussion of condensates, using continuum
finite-energy sum rules was employed to estimate where their contributions
might be negligible. It was found that even up to terms
of $O((1/Q^2)^8)$ (a higher order than depicted in Eq.~(\ref{eq:vacpol})
but with constant coefficients) 
no single condensate dominates
and apparent convergence was poor for low $Q^2$
due to cancellations between contributions of similar size
with alternating signs. (See, e.g., \ the list given by Hudspith 15
\cite{Hudspith:2015xoa}.) Choosing $Q^2$ to be at least
$\sim 3.8\,\mbox{GeV}^2$ mitigated the problem, but then the coarsest
lattice had to be discarded, due to large lattice artefacts.
So this gives a $\bad$ for continuum extrapolation.
With the higher $Q^2$ the quark-mass dependence of the
results was negligible, so ensembles with different quark masses were
averaged over.
A range of $Q^2$ from 3.8--16~GeV$^2$ gives 
$\alpha_{\rm eff}$ = 0.31--0.22,  so there is a $\soso$ for the
renormalization scale.
The value of $\alpha_{\rm eff}^3$ reaches
$\Delta \alpha_{\rm eff}/(8\pi b_0 \alpha_{\rm eff})$ and thus
gives a $\soso$ for perturbative behaviour.
In Hudspith 15 \cite{Hudspith:2015xoa} (superseded by Hudspith 18 
\cite{Hudspith:2018bpz}) about a 20\% difference in 
$\Pi_V(Q^2)$ was seen between the two lattice spacings and a
result is quoted only for the smaller $a$.

\subsubsection{Vacuum polarization in position space}

Cali 20~\cite{Cali:2020hrj} evaluate the light-current 2-point function in position space.
The 2-point functions for the nonperturbatively renormalized (non-singlet) flavour currents is computed for
distances $|x|$ between $0.1$ and $0.25$~fm and extrapolated to the chiral limit. 
The available CLS configurations are used for this work, with lattice spacings between $0.039$ and $0.086\,\fm$. 
Despite fully nonperturbative renormalization and $\cO(a)$ improvement,
the remaining $\cO(a^2)$ effects, as measured by $O(4)$ symmetry violations, are very large, 
even after subtraction of tree-level lattice effects. 
Therefore the authors performed a numerical stochastic perturbation theory (NSPT) simulation in order to determine the lattice artifacts at
$\cO(g^2)$. Only after subtraction of these effects the constrained continuum extrapolations from 3 different lattice directions
to the same continuum limit are characterized by reasonable $\chi^2$-values, so the feasibility of the study crucially depends
on this step. Interestingly, there is no subtraction performed of nonperturbative effects. For instance, 
chiral-symmetry breaking would manifest itself in a difference between the vector and the axial vector 2-point functions, 
and is invisible to perturbation theory, where these 2-point functions are known to $\alpha_s^4$~\cite{Chetyrkin:2010dx}. 
According to the authors, phenomenological estimates suggest that a difference of 1.5\% between 
the continuum correlators would occur around 0.3~fm and this difference would not be resolvable by their lattice data. 
Equality within their errors is confirmed for shorter distances. We note, however, that chiral symmetry breaking effects
are but one class of nonperturbative effects, and their smallness does not allow for the conclusion that
such effects are generally small. In fact, the need for explicit subtractions in momentum space analyses may
lead one to suspect that such effects are not negligible at the available distance scales.
For the determination of $\lms^{\Nf=3}$ the authors limit the range of distances to 0.13--0.19~fm,
where $\alpha_\text{eff} \in [0.2354,0.3075]$ (private communication by the authors).
These effective couplings are converted to $\msbar$ couplings at the same scales $\mu=1/|x|$ 
by solving Eq.~(\ref{eq:CVAMSbar}) numerically. Central values for the $\Lambda$-parameter thus obtained 
are in the range 325--370~MeV (using the $\beta$-function at 5-loop order) and a weighted average
yields the quoted result 342(17)~MeV, where the average emphasizes the data  around $|x|=$ 0.16~fm, 
or $\mu=$ 1.3~GeV. 

Applying the FLAG criteria the range of lattice spacings yields $\good$ for the continuum extrapolation. However, the FLAG criterion
implicitly assumes that the remaining cutoff effects after non-perturbative $\cO(a)$ improvement are small, which is 
not the case here. Some hypercubic lattice artefacts are still rather large even after 1-loop subtraction, but these are
not used for the analysis.
As for the renormalization scale, the lowest effective coupling entering the analysis is $0.235<0.25$, so we give $\soso$.
As for perturbative behaviour, for the range of couplings in the above interval $\alpha_\text{eff}^3$ 
changes by $(0.308/0.235)^3\approx 2.2$, marginally reaching $(3/2)^2=2.25$.
The errors $\Delta \alpha_\text{eff}$ after continuum and chiral extrapolations are 4--6\% (private
communication by the authors) and the induced uncertainty in $\Lambda$ is comfortably 
above $2\alpha_\text{eff}^3$, which gives a $\good$ according to FLAG criteria.

Although the current FLAG criteria are formally passed by this result, the quoted error of 5\% for $\Lambda$ 
seems very optimistic. We have performed a simple test, converting to the $\msbar$ scheme 
by inverting Eq.~(\ref{eq:CVAMSbar}) perturbatively (instead of solving the fixed-order equation numerically).
The differences between the couplings are of order $\alpha_s^5$ and thus indicative 
of the sensitivity to perturbative truncation
errors. The resulting $\Lambda$-parameter estimates are now in the range 409--468~MeV, i.e.~ca.~15--30\%
larger than before. While the difference between both estimates decreases proportionally to the expected 
$\alpha_\text{eff}^3$, an extraction of the $\Lambda$-parameter in this energy range is a priori affected by
systematic uncertainties corresponding to such differences. 
The FLAG criterion might fail to capture this e.g.~if the assumption of an $\cO(1)$ coefficient for
the asymptotic $\alpha_\text{eff}^3$ behaviour is not correct. Some indication
for a problematic behaviour is indeed seen when perturbatively inverting Eq.~(\ref{eq:CVAMSbar}) 
to order $\alpha_s^3$. The resulting $\msbar$ couplings are then closer to the values used in Cali~20,
although the difference is formally  $\cO(\alpha_s^4)$  rather than $\cO(\alpha_s^5)$. 

\subsection{$\alpha_s$ from observables at the lattice spacing scale}
\label{s:WL}


\subsubsection{General considerations}


The general method is to evaluate a short-distance quantity ${\oO}$
at the scale of the lattice spacing $\sim 1/a$ and then determine
its relationship to $\alpha_{\overline{\rm MS}}$ via a 
perturbative 
expansion.

This is epitomized by the strategy of the HPQCD collaboration
\cite{Mason:2005zx,Davies:2008sw}, discussed here for illustration,
which computes and then fits to a variety of short-distance quantities
\begin{eqnarray}
   Y = \sum_{n=1}^{n_{\rm max}} c_n \alphah^n(q^*) \,.
\label{Ydef}
\end{eqnarray}
The quantity $Y$ is taken as the logarithm of small Wilson loops (including some
nonplanar ones), Creutz ratios, `tadpole-improved' Wilson loops and
the tadpole-improved or `boosted' bare coupling ($\cO(20)$ quantities in
total). The perturbative coefficients $c_n$  (each depending on the
choice of $Y$) are known to $n = 3$ with additional coefficients up to
$n_{\rm max}$ being fitted numerically.   The running
coupling $\alphah$ is related to $\alphav$ from the static-quark potential
(see Sec.~\ref{s:qq}).\footnote{ $\alphah$ is defined by
  $\Lambda_\mathrm{V'}=\Lambda_\mathrm{V}$ and
  $b_i^\mathrm{V'}=b_i^\mathrm{V}$ for $i=0,1,2$ but $b_i^\mathrm{V'}=0$ for
  $i\geq3$. }

 The coupling
constant is fixed at a scale $q^* = d/a$.
The latter  is chosen as the mean value of $\ln q$ with the one-gluon loop
as measure
\cite{Lepage:1992xa,Hornbostel:2002af}. (Thus a different result
for $d$ is found for every short-distance quantity.)
A rough estimate yields $d \approx \pi$, and in general the
renormalization scale is always found to lie in this region.

For example, for the Wilson loop $W_{mn} \equiv \langle W(ma,na) \rangle$
we have
\begin{eqnarray}
   \ln\left( \frac{W_{mn}}{u_0^{2(m+n)}}\right)
      = c_1 \alphah(q^*) +  c_2 \alphah^2(q^*)  + c_3 \alphah^3(q^*)
        + \cdots \,,
\label{short-cut}
\end{eqnarray}
for the tadpole-improved version, where $c_1$, $c_2\,, \ldots$
are the appropriate perturbative coefficients and $u_0 = W_{11}^{1/4}$.
Substituting the nonperturbative simulation value in the left hand side,
we can determine $\alphah(q^*)$, at the scale $q^*$.
Note that one finds empirically that perturbation theory for these
tadpole-improved quantities have smaller $c_n$ coefficients and so
the series has a faster apparent convergence compared to the case
without tadpole improvement.

Using the $\beta$-function in the $\rm V'$ scheme,
results can be run to a reference value, chosen as
$\alpha_0 \equiv \alphah(q_0)$, $q_0 = 7.5\,\mbox{GeV}$.
This is then converted perturbatively to the continuum
$\msbar$ scheme
\begin{eqnarray}
   \alpha_{\overline{\rm MS}}(q_0)
      = \alpha_0 + d_1 \alpha_0^2 + d_2 \alpha_0^3 + \cdots \,,
\end{eqnarray}
where $d_1, d_2$ are known 1-and 2-loop coefficients.

Other collaborations have focused more on the bare `boosted'
coupling constant and directly determined its relationship to
$\alpha_{\overline{\rm MS}}$. Specifically, the boosted coupling is
defined by 
\begin{eqnarray}
   \alphap(1/a) = {1\over 4\pi} {g_0^2 \over u_0^4} \,,
\end{eqnarray}
again determined at a scale $\sim 1/a$. As discussed previously, 
since the plaquette expectation value in the boosted coupling
contains the tadpole-diagram contributions to all orders, which
are dominant contributions in perturbation theory,
there is an expectation that the perturbation theory using
the boosted coupling has 
smaller perturbative coefficients \cite{Lepage:1992xa}, and hence smaller 
perturbative errors.
 

\subsubsection{Continuum limit}


Lattice results always come along with discretization errors,
which one needs to remove by a continuum extrapolation.
As mentioned previously, in this respect the present
method differs in principle from those in which $\alpha_s$ is determined
from physical observables. In the general case, the numerical
results of the lattice simulations at a value of $\mu$ fixed in physical 
units can be extrapolated to the continuum limit, and the result can be 
analyzed as to whether it shows perturbative running as a function of 
$\mu$ in the continuum. For observables at the cutoff-scale ($q^*=d/a$),  
discretization effects cannot easily be separated out
from perturbation theory, as the scale for the coupling
comes from the lattice spacing. 
Therefore the restriction  $a\mu  \ll 1$ (the `continuum-extrapolation'
criterion) is not applicable here. Discretization errors of 
order $a^2$ are, however, present. Since 
$a\sim \exp(-1/(2b_0 g_0^2)) \sim \exp(-1/(8\pi b_0 \alpha(q^*))$, 
these errors now appear as power corrections to the perturbative 
running, and have to be taken into account in the study of the 
perturbative behaviour, which is to be verified by changing $a$. 
One thus usually fits with power corrections in this method.

In order to keep a symmetry with the `continuum-extrapolation' 
criterion for physical observables and to remember that discretization 
errors are, of course, relevant, 
we replace it here by one for the lattice spacings used:
\begin{itemize}
   \item Lattice spacings
         \begin{itemize}
            \item[\good] 
               3 or more lattice spacings, at least 2 points below
               $a = 0.1\,\mbox{fm}$
            \item[\soso]
               2 lattice spacings, at least 1 point below
               $a = 0.1\,\mbox{fm}$
            \item[\bad]
               otherwise 
         \end{itemize}
\end{itemize}


\subsubsection{Discussion of computations}

\begin{table}[!p]
   \vspace{3.0cm}
   \footnotesize
   \begin{tabular*}{\textwidth}[l]{l@{\extracolsep{\fill}}rllllllll}
   Collaboration & Ref. & $N_f$ &
   \hspace{0.15cm}\begin{rotate}{60}{publication status}\end{rotate}
                                                    \hspace{-0.15cm} &
   \hspace{0.15cm}\begin{rotate}{60}{renormalization scale}\end{rotate}
                                                    \hspace{-0.15cm} &
   \hspace{0.15cm}\begin{rotate}{60}{perturbative behaviour}\end{rotate}
                                                    \hspace{-0.15cm} &
   \hspace{0.15cm}\begin{rotate}{60}{lattice spacings}\end{rotate}
      \hspace{-0.25cm} & 
                         scale & $\Lambda_\msbar[\MeV]$ & $r_0\Lambda_\msbar$ \\
   & & & & & & & & \\[-0.1cm]
   \hline
   \hline
   & & & & & & & & \\[-0.1cm]
   HPQCD 10$^a$$^\S$& \cite{McNeile:2010ji}& 2+1 & \gA & \soso
            & \good & \good
            & $r_1 = 0.3133(23)\, \mbox{fm}$
            & 340(9) 
            & 0.812(22)                                   \\ 
   HPQCD 08A$^a$& \cite{Davies:2008sw} & 2+1 & \gA & \soso
            & \good & \good
            & $r_1 = 0.321(5)\,\mbox{fm}$$^{\dagger\dagger}$
            & 338(12)$^\star$
            & 0.809(29)                                   \\
   Maltman 08$^a$& \cite{Maltman:2008bx}& 2+1 & \gA & \soso
            & \soso & \good
            & $r_1 = 0.318\, \mbox{fm}$
            & 352(17)$^\dagger$
            & 0.841(40)                                   \\ 
   HPQCD 05A$^a$ & \cite{Mason:2005zx} & 2+1 & \gA & \soso
            & \soso & \soso
            & $r_1$$^{\dagger\dagger}$
            & 319(17)$^{\star\star}$
            & 0.763(42)                                   \\
   & & & & & & & & &  \\[-0.1cm]
   \hline
   & & & & & & & & &  \\[-0.1cm]
   QCDSF/UKQCD 05 & \cite{Gockeler:2005rv}  & 2 & \gA & \good
            & \bad  & \good
            & $r_0 = 0.467(33)\,\mbox{fm}$
            & 261(17)(26)
            & 0.617(40)(21)$^b$                           \\
   SESAM 99$^c$ & \cite{Spitz:1999tu} & 2 & \gA & \soso
            & \bad  & \bad
            & $c\bar{c}$(1S-1P)
            & 
            &                                             \\
   Wingate 95$^d$ & \cite{Wingate:1995fd} & 2 & \gA & \good
            & \bad  & \bad
            & $c\bar{c}$(1S-1P)
            & 
            &                                             \\
   Davies 94$^e$ & \cite{Davies:1994ei} & 2 & \gA & \good
            & \bad & \bad
            & $\Upsilon$
            & 
            &                                             \\
   Aoki 94$^f$ & \cite{Aoki:1994pc} & 2 & \gA & \good
            & \bad & \bad
            & $c\bar{c}$(1S-1P)
            & 
            &                                             \\
   & & & & & & & & &  \\[-0.1cm]
   \hline
   & & & & & & & & &  \\[-0.1cm]

   {Kitazawa 16}
            & \cite{Kitazawa:2016dsl}        & 0 & \gA
            & \good  & \good   & \good
            & $w_0$
            & $260(5)$$^j$
            & $0.621(11)$$^j$                            \\

   FlowQCD 15
            & \cite{Asakawa:2015vta}        & 0 & \oP 
            & \good  & \good   & \good
            & $w_{0.4}$$^i$
            & $258(6)$$^i$
            & 0.618(11)$^i$                             \\

   QCDSF/UKQCD 05 & \cite{Gockeler:2005rv}  & 0 & \gA & \good
            & \soso & \good
            & $r_0 = 0.467(33)\,\mbox{fm}$
            & 259(1)(20)
            & 0.614(2)(5)$^b$                              \\
   SESAM 99$^c$ & \cite{Spitz:1999tu} & 0 & \gA & \good
            & \bad  & \bad
            & $c\bar{c}$(1S-1P)
            & 
            &                                             \\
   Wingate 95$^d$ & \cite{Wingate:1995fd} & 0 & \gA & \good
            & \bad  & \bad
            & $c\bar{c}$(1S-1P)
            & 
            &                                             \\
   Davies 94$^e$ & \cite{Davies:1994ei}  & 0 & \gA & \good
            & \bad & \bad
            & $\Upsilon$
            & 
            &                                             \\
   El-Khadra 92$^g$ & \cite{ElKhadra:1992vn} & 0 & \gA & \good
            & \bad    & \soso
            & $c\bar{c}$(1S-1P)
            & 234(10)
            & 0.560(24)$^h$                               \\
   & & & & & & & & &  \\[-0.1cm]
   \hline
   \hline\\
\end{tabular*}\\[-0.2cm]
\begin{minipage}{\linewidth}
{\footnotesize 
\begin{itemize}
   \item[$^a$]The numbers for $\Lambda$ have been converted from the values for 
              $\alpha_s^{(5)}(M_Z)$. \\[-5mm]
   \item[$^{\S}$]     $\alpha_{\overline{\rm MS}}^{(3)}(5\ \mbox{GeV})=0.2034(21)$,
              $\alpha^{(5)}_{\overline{\rm MS}}(M_Z)=0.1184(6)$,
              only update of intermediate scale and $c$-, $b$-quark masses,
              supersedes HPQCD 08A.\\[-5mm]
   \item[$^\dagger$] $\alpha^{(5)}_{\overline{\rm MS}}(M_Z)=0.1192(11)$. \\[-4mm]
   \item[$^\star$]    $\alpha_V^{(3)}(7.5\,\mbox{GeV})=0.2120(28)$, 
              $\alpha^{(5)}_{\overline{\rm MS}}(M_Z)=0.1183(8)$,
              supersedes HPQCD 05. \\[-5mm]
   \item[$^{\dagger\dagger}$] Scale is originally determined from $\Upsilon$
              mass splitting. $r_1$ is used as an intermediate scale.
              In conversion to $r_0\Lambda_{\overline{\rm MS}}$, $r_0$ is
              taken to be $0.472\,\mbox{fm}$. \\[-5mm]
   \item[$^{\star\star}$] $\alpha_V^{(3)}(7.5\,\mbox{GeV})=0.2082(40)$,
              $\alpha^{(5)}_{\overline{\rm MS}}(M_Z)=0.1170(12)$. \\[-5mm]
   \item[$^b$]       This supersedes 
              Refs.~\cite{Gockeler:2004ad,Booth:2001uy,Booth:2001qp}.
              $\alpha^{(5)}_{\overline{\rm MS}}(M_Z)=0.112(1)(2)$.
              The $N_f=2$ results were based on values for $r_0 /a$
              which have later been found to be too 
              small~\cite{Fritzsch:2012wq}. The effect will  
              be of the order of 10--15\%, presumably an increase in 
              $\Lambda r_0$. \\[-5mm]
   \item[$^c$]       $\alpha^{(5)}_{\overline{\rm MS}}(M_Z)=0.1118(17)$. \\[-4mm]
   \item[$^d$]    
   $\alpha_V^{(3)}(6.48\,\mbox{GeV})=0.194(7)$ extrapolated from $\Nf=0,2$.
              $\alpha^{(5)}_{\overline{\rm MS}}(M_Z)=0.107(5)$.   \\[-4mm]
   \item[$^e$]  
              $\alpha_P^{(3)}(8.2\,\mbox{GeV})=0.1959(34)$ extrapolated
              from $N_f=0,2$. $\alpha^{(5)}_{\overline{\rm MS}}(M_Z)=0.115(2)$.
              \\[-5mm]
   \item[ $^f$]Estimated $\alpha^{(5)}_{\overline{\rm MS}}(M_Z)=0.108(5)(4)$. \\[-5mm]
   \item[$^g$]       This early computation violates our requirement that
              scheme conversions are done at the 2-loop level.
              $\Lambda_{\overline{\rm MS}}^{(4)}=160(^{+47}_{-37})\mbox{MeV}$, 
              $\alpha^{(4)}_{\overline{\rm MS}}(5\mbox{GeV})=0.174(12)$.
              We converted this number to give
              $\alpha^{(5)}_{\overline{\rm MS}}(M_Z)=0.106(4)$.  \\[-5mm]
   \item[$^h$]We used $r_0=0.472\,\mbox{fm}$ to convert to $r_0 \lms$. \\[-5mm]
   \item[$^i$]       Reference scale $w_{0.4}$ where $w_x$ is defined 
              by $\left. t\partial_t[t^2 \langle E(t)\rangle]\right|_{t=w_x^2}=x$
              in terms of the action density $E(t)$ at positive flow time $t$ 
              \cite{Asakawa:2015vta}. Our conversion to $r_0$ scale
              using \cite{Asakawa:2015vta} $r_0/w_{0.4}=2.587(45)$ and
              $r_0=0.472\,\mbox{fm}$. 
   \item[$^j$]{Our conversion from $w_0\Lambda_\msbar=0.2154(12)$ to $r_0$ scale
              using  $r_0/w_0=(r_0/w_{0.4}) \cdot   (w_{0.4}/w_0) = 2.885(50)$ 
              with the factors cited by the collaboration \cite{Asakawa:2015vta} 
              and with $r_0=0.472\,\mbox{fm}$. }
\end{itemize}
}
\end{minipage}
\normalsize
\caption{Wilson loop results. Some early results for $\Nf=0, 2$ did not determine  $\Lambda_\msbar$.
}
\label{tab_wloops}
\end{table}

Note that due to $\mu \sim 1/a$ being relatively large the
results easily have a $\good$ or $\soso$ in the rating on 
renormalization scale.

The work of El-Khadra 92 \cite{ElKhadra:1992vn} employs a 1-loop
formula to relate $\alpha^{(0)}_{\overline{\rm MS}}(\pi/a)$
to the boosted coupling for three lattice spacings
$a^{-1} = 1.15$, $1.78$, $2.43\,\mbox{GeV}$. (The lattice spacing
is determined from the charmonium 1S-1P splitting.) They obtain
$\Lambda^{(0)}_{\overline{\rm MS}}=234\,\mbox{MeV}$, corresponding
to $\alpha_{\rm eff} = \alpha^{(0)}_{\overline{\rm MS}}(\pi/a)
\approx$ 0.15--0.2. The work of Aoki 94 \cite{Aoki:1994pc}
calculates $\alpha^{(2)}_V$ and $\alpha^{(2)}_{\overline{\rm MS}}$
for a single lattice spacing $a^{-1}\sim 2\,\mbox{GeV}$,  again
determined from charmonium 1S-1P splitting in two-flavour QCD.
Using 1-loop perturbation theory with boosted coupling,
they obtain $\alpha^{(2)}_V=0.169$ and $\alpha^{(2)}_{\overline{\rm MS}}=0.142$.
Davies 94 \cite{Davies:1994ei} gives a determination of $\alphav$
from the expansion 
\begin{equation}
   -\ln W_{11} \equiv \frac{4\pi}{3}\alphav^{(N_f)}(3.41/a)
        \times [1 - (1.185+0.070N_f)\alphav^{(N_f)} ]\,,
\end{equation}
neglecting higher-order terms.  They compute the $\Upsilon$ spectrum
in $N_f=0$, $2$ QCD for single lattice spacings at $a^{-1} = 2.57$,
$2.47\,\mbox{GeV}$ and obtain $\alphav(3.41/a)\simeq$ 0.15, 0.18, respectively.  Extrapolating the inverse coupling linearly in $N_f$, a
value of $\alphav^{(3)}(8.3\,\mbox{GeV}) = 0.196(3)$ is obtained.
SESAM 99 \cite{Spitz:1999tu} follows a similar strategy, again for a
single lattice spacing. They linearly extrapolated results for
$1/\alphav^{(0)}$, $1/\alphav^{(2)}$ at a fixed scale of
$9\,\mbox{GeV}$ to give $\alphav^{(3)}$, which is then perturbatively
converted to $\alpha_{\overline{\rm MS}}^{(3)}$. This finally gave
$\alpha_{\overline{\rm MS}}^{(5)}(M_Z) = 0.1118(17)$.  Wingate 95
\cite{Wingate:1995fd} also follows this method.  With the scale
determined from the charmonium 1S-1P splitting for single lattice
spacings in $N_f = 0$, $2$ giving $a^{-1}\simeq 1.80\,\mbox{GeV}$ for
$N_f=0$ and $a^{-1}\simeq 1.66\,\mbox{GeV}$ for $N_f=2$, they obtain
$\alphav^{(0)}(3.41/a)\simeq 0.15$ and $\alphav^{(2)}\simeq 0.18$, 
respectively. Extrapolating the inverse coupling linearly in $N_f$, they
obtain $\alphav^{(3)}(6.48\,\mbox{GeV})=0.194(17)$.

The QCDSF/UKQCD collaboration, QCDSF/UKQCD 05
\cite{Gockeler:2005rv}, \cite{Gockeler:2004ad,Booth:2001uy,Booth:2001qp},
use the 2-loop relation (re-written here in terms of $\alpha$)
\begin{eqnarray}
   {1 \over \alpha_{\overline{\rm MS}}(\mu)} 
      = {1 \over \alphap(1/a)} 
        + 4\pi(2b_0\ln a\mu - t_1^{\rm P}) 
        + (4\pi)^2(2b_1\ln a\mu - t_2^{\rm P})\alphap(1/a) \,,
\label{gPtoMSbar}
\end{eqnarray}
where $t_1^{\rm P}$ and $t_2^{\rm P}$ are known. (A 2-loop relation corresponds
to a 3-loop lattice $\beta$-function.)  This was used to
directly compute $\alpha_{\rm \overline{\rm MS}}$, and the scale was
chosen so that the $\cO(\alphap^0)$ term vanishes, i.e., \
\begin{eqnarray}
   \mu^* = {1 \over a} \exp{[t_1^{\rm P}/(2b_0)] } 
        \approx \left\{ \begin{array}{cc}
                           2.63/a  & N_f = 0 \\
                           1.4/a   & N_f = 2 \\
                        \end{array}
                 \right. \,.
\label{amustar}
\end{eqnarray}
The method is to first compute $\alphap(1/a)$ and from this,  using
Eq.~(\ref{gPtoMSbar}) to find $\alpha_{\overline{\rm MS}}(\mu^*)$.
The RG equation, Eq.~(\ref{eq:Lambda}), then determines
$\mu^*/\Lambda_{\overline{\rm MS}}$ and hence using
Eq.~(\ref{amustar}) leads to the result for $r_0\Lambda_{\overline{\rm MS}}$.
This avoids giving the scale in $\mbox{MeV}$ until the end.
In the $\Nf=0$ case seven lattice spacings were used
\cite{Necco:2001xg}, giving a range $\mu^*/\Lambda_{\overline{\rm MS}}
\approx$ 24--72 (or $a^{-1} \approx$ 2--7~GeV) and
$\alpha_{\rm eff} = \alpha_{\overline{\rm MS}}(\mu^*) \approx$ 0.15--0.10. Neglecting higher-order perturbative terms (see discussion
after Eq.~(\ref{qcdsf:ouruncert}) below) in Eq.~(\ref{gPtoMSbar}) this
is sufficient to allow a continuum extrapolation of
$r_0\Lambda_{\overline{\rm MS}}$.
A similar computation for $N_f = 2$ by QCDSF/UKQCD~05 \cite{Gockeler:2005rv}
gave $\mu^*/\Lambda_{\overline{\rm MS}} \approx$ 12--17
(or roughly $a^{-1} \approx$ 2--3~GeV) 
and $\alpha_{\rm eff} = \alpha_{\overline{\rm MS}}(\mu^*)
\approx$ 0.20--0.18.
The $N_f=2$ results of QCDSF/UKQCD~05 \cite{Gockeler:2005rv} are affected by an 
uncertainty which was not known at the time of publication: 
It has been realized that the values of $r_0/a$ of Ref.~\cite{Gockeler:2005rv}
were significantly too low~\cite{Fritzsch:2012wq}. 
As this effect is expected to depend on $a$, it
influences the perturbative behaviour leading us to assign 
a \bad\ for that criterion. 

Since FLAG 13, there has been one new result for $N_f = 0$ 
by FlowQCD 15 \cite{Asakawa:2015vta}, later 
updated and published in Kitazawa 16 \cite{Kitazawa:2016dsl}.
They also use the techniques
as described in Eqs.~(\ref{gPtoMSbar}), (\ref{amustar}), but together
with the gradient flow scale $w_0$ (rather than the $r_0$ scale)
leading to a determination of $w_0\Lambda_{\overline{\rm MS}}$.
The continuum limit is estimated by extrapolating the data at $6$
lattice spacings linearly in $a^2$. The data range used is
$\mu^*/\Lambda_{\overline{\rm MS}} \approx$ 50--120 (or 
$a^{-1} \approx$ 5--11~GeV) and
$\alpha_{\overline{\rm MS}}(\mu^*) \approx$ 0.12--0.095.
Since a very small value of $\alpha_\msbar$ is reached, there is a $\good$ 
in the perturbative behaviour. Note that our conversion to the common
$r_0$ scale unfortunately leads to a significant increase of the error of the
$\Lambda$ parameter compared to using $w_0$ directly \cite{Sommer:2014mea}. 
Again we note that the results of QCDSF/UKQCD 05
\cite{Gockeler:2005rv} ($N_f = 0$) and Kitazawa 16 \cite{Kitazawa:2016dsl}
may be affected by frozen topology as they have
lattice spacings significantly below $a = 0.05\,\mbox{fm}$.
Kitazawa 16 \cite{Kitazawa:2016dsl} investigate this by evaluating
$w_0/a$ in a fixed topology and estimate any effect at about $\sim 1\%$.

The work of HPQCD 05A \cite{Mason:2005zx} (which supersedes
the original work \cite{Davies:2003ik}) uses three lattice spacings
$a^{-1} \approx 1.2$, $1.6$, $2.3\,\mbox{GeV}$ for $2+1$
flavour QCD. Typically the renormalization scale
$q \approx \pi/a \approx$ 3.50--7.10~GeV, corresponding to
$\alpha_\mathrm{V'} \approx$ 0.22--0.28. 

In the later update HPQCD 08A \cite{Davies:2008sw} twelve data sets
(with six lattice spacings) are now used reaching up to $a^{-1}
\approx 4.4\,\mbox{GeV}$,  corresponding to $\alpha_\mathrm{V'}\approx
0.18$. The values used for the scale $r_1$ were further updated in
HPQCD 10 \cite{McNeile:2010ji}. Maltman 08 \cite{Maltman:2008bx}
uses most of the same lattice ensembles as HPQCD
08A~\cite{Davies:2008sw}, but not the one at the smallest lattice spacing, $a\approx0.045$~fm. Maltman 08 \cite{Maltman:2008bx} also
considers a much smaller set of
quantities (three versus 22) that are less sensitive to condensates.
They also use different strategies for evaluating the condensates and
for the perturbative expansion, and a slightly different value for the
scale $r_1$. The central values of the final results from 
Maltman 08 \cite{Maltman:2008bx} and HPQCD 08A \cite{Davies:2008sw}
differ by 0.0009 (which would be decreased to 0.0007
taking into account a reduction of 0.0002 in the value of the $r_1$
scale used by Maltman 08 \cite{Maltman:2008bx}).
 
As mentioned before, the perturbative coefficients are computed
through $3$-loop order~\cite{Mason:2004zt}, while the higher-order
perturbative coefficients $c_n$ with $ n_{\rm max} \ge n > 3$ (with
$n_{\rm max} = 10$) are numerically fitted using the
lattice-simulation data for the lattice spacings with the help of
Bayesian methods.  It turns out that corrections in \eq{short-cut} are
of order $|c_i/c_1|\alpha^i=$ 5--15\% and 3--10\% for $i$ = 2, 3,
respectively.  The inclusion of a fourth-order term is necessary to
obtain a good fit to the data, and leads to a shift of the result by
$1$ -- $2$ sigma. For all but one of the 22 quantities, central values
of $|c_4/c_1|\approx$ 2--4 were found, with errors from the fits of
$\approx 2$. 
It should be pointed out that the description of lattice results for the short distance quantities 
does not require Bayesian priors, once the term proportional to $c_4$ is included \cite{Maltman:2008bx}. 
We also stress that different short distance quantities have quite different nonperturbative contributions \cite{Maltman:2010zza}.
Hence the fact that different observables lead to consistent $\alpha_s$ values is a nontrivial check of the
approach.

An important source of uncertainty is the truncation 
of perturbation theory. In HPQCD 08A \cite{Davies:2008sw}, 10
\cite{McNeile:2010ji} it
is estimated to be about $0.4$\% of $\alpha_\msbar(M_Z)$.  In \flagold\
we included a rather detailed discussion of the issue with the result
that we prefer for the time being a more conservative error
based on the above estimate $|c_4/c_1| = 2$. 
From Eq.~(\ref{Ydef}) this gives an estimate of the uncertainty
in $\alpha_{\rm eff}$ of
\begin{eqnarray}
  \Delta \alpha_{\rm eff}(\mu_1) = 
          \left|{c_4 \over c_1}\right|\alpha_{\rm eff}^4(\mu_1) \,,
\label{qcdsf:ouruncert}
\end{eqnarray}
at the scale $\mu_1$ where $\alpha_{\rm eff}$ is computed from
the Wilson loops. This can be used with a variation
in $\Lambda$ at lowest order of perturbation theory and also
applied to $\alpha_s$ evolved to a different scale $\mu_2$,%
\footnote{From Eq.~(\ref{e:grelation}) we see that at low order in PT
the coupling $\alpha_s$ is continuous and differentiable across
the mass thresholds (at the same scale). Therefore 
to leading order $\alpha_s$ and $\Delta \alpha_s$
are independent of $N_f$.}
\begin{eqnarray}
   {\Delta\Lambda \over \Lambda} 
      = {1\over 8\pi b_0 \alpha_s} 
                  {\Delta \alpha_s \over \alpha_s}
                                                         \,, \qquad
   {\Delta \alpha_s(\mu_2) \over \Delta \alpha_s(\mu_1)}
      = {\alpha_s^2(\mu_2) \over \alpha_s^2(\mu_1)} \,.
   \label{e:dLL}   
\end{eqnarray}
With $\mu_2 = M_Z$
and $\alpha_s(\mu_1)=0.2$ (a typical value extracted 
from Wilson loops in HPQCD 10 \cite{McNeile:2010ji}, HPQCD 08A
\cite{Davies:2008sw} at $\mu = 5\,\mbox{GeV}$) we have 
\begin{eqnarray}
  \Delta \alpha_\msbar(m_Z) = 0.0012 \,,
\label{hpqcd:ouruncert}
\end{eqnarray}
which we shall later use as the typical perturbative uncertainty 
of the method with $2+1$ fermions.
}

Tab.~\ref{tab_wloops} summarizes the results. Within the errors of 3--5\% $N_f=3$ determinations of $r_0 \Lambda$ nicely agree.


\subsection{$\alpha_s$ from heavy-quark current two-point functions}


\label{s:curr}


\subsubsection{General considerations}


The method has been introduced in HPQCD 08, Ref.~\cite{Allison:2008xk},
and updated in HPQCD 10, Ref.~\cite{McNeile:2010ji}, see also
Ref.~\cite{Bochkarev:1995ai}.  In addition
there is a 2+1+1-flavour result, HPQCD 14A \cite{Chakraborty:2014aca}.

The basic observable is constructed from a current,
\begin{eqnarray}
  J(x) = i am_c\overline\psi_c(x)\gamma_5\psi_{c'}(x)\,,
  \label{e:Jx}
\end{eqnarray}
of two mass-degenerate heavy-valence quarks, $c$, $c^\prime$,
usually taken to be at or around the charm-quark mass.
The pre-factor $m_c$ denotes the bare mass of the quark.
When the lattice discretization respects chiral symmetry, 
$J(x)$ is a renormalization group
invariant local field, i.e., it requires no renormalization.
Staggered fermions and twisted-mass fermions have such a residual
chiral symmetry. The (Euclidean) time-slice correlation function
\begin{eqnarray}
   G(x_0) = a^6 \sum_{\vec{x}} \langle J^\dagger(x) J(0) \rangle \,,
\end{eqnarray}
($J^\dagger(x) = i am_c\overline\psi_{c'}(x)\gamma_5\psi_c(x)$)
has a $\sim x_0^{-3}$  singularity at short distances and moments
\begin{eqnarray}
   G_n = a \sum_{x_0=-(T/2-a)}^{T/2-a} x_0^n \,G(x_0) \,
\label{Gn_smu}
\end{eqnarray}
are nonvanishing for even $n$ and furthermore finite for $n \ge 4$. 
Here $T$ is the time extent of the lattice.
The moments are dominated by contributions at $x_0$ of order $1/m_c$.
For large mass $m_c$ these are short distances and the moments
become increasingly perturbative for decreasing $n$.
Denoting the lowest-order perturbation theory moments by $G_n^{(0)}$,
one defines the
normalized moments 
\begin{eqnarray}
    R_n = \left\{ \begin{array}{cc}
          G_4/G_4^{(0)}          & \mbox{for $n=4$} \,, \\[0.5em]
          {am_{\eta_c}\over 2am_c} 
                \left( { G_n \over G_n^{(0)}} \right)^{1/(n-4)}
                               & \mbox{for $n \ge 6$} \,, \\
                 \end{array}
         \right.
\label{Rn}
\end{eqnarray}
of even order $n$. Note that \eq{e:Jx} contains the variable
(bare) heavy-quark mass $m_c$. 
The normalization $G_n^{(0)}$ is introduced to help in
reducing lattice artifacts.
In addition, one can also define moments with different normalizations,
\begin{eqnarray}
   \tilde R_n = 2 R_n / m_{\eta_c} \qquad \mbox{for $n \ge 6$}\,.
\end{eqnarray}
While $\tilde R_n$ also remains renormalization-group invariant,
it now also has a scale which might introduce an
additional ambiguity \cite{Nakayama:2016atf}.

The normalized moments can then be parameterized in terms of functions
\begin{eqnarray}
   R_n \equiv \left\{ \begin{array}{cc}
                         r_4(\alpha_s(\mu))
                                        & \mbox{for $n=4$} \,,     \\[0.5em]
                         \frac{m_{\eta_c}}{2 \bar{m}_c(\mu_m)} r_n(\alpha_s(\mu))
                                        & \mbox{for $n \ge 6$} \,, \\
                      \end{array}
              \right.
              \label{e:Rn}
\end{eqnarray}
with $\bar{m}_c(\mu_m)$ being the renormalized heavy-quark  mass. 
The scale $\mu_m$ at which the heavy-quark mass is defined could be different from the scale $\mu$ at which $\alpha_s$ is defined \cite{Dehnadi:2015fra}.
The HPQCD collaboration, however, used the choice $\mu=\mu_m=3 m_c(\mu)$. This ensures that the renormalization scale is never too small.
The reduced moments $r_n$ have a perturbative expansion
\begin{eqnarray}
   r_n = 1 + r_{n,1}\alpha_s + r_{n,2}\alpha_s^2 + r_{n,3}\alpha_s^3 + \ldots\,,
\label{rn_expan}
\end{eqnarray}
where the written terms $r_{n,i}(\mu/\bar{m}_c(\mu))$, $i \le 3$ are known
for low $n$ from Refs.~\cite{Chetyrkin:2006xg,Boughezal:2006px,Maier:2008he,
Maier:2009fz,Kiyo:2009gb}. In practice, the expansion is performed in
the $\overline{\rm MS}$ scheme. Matching nonperturbative lattice results
for the moments to the perturbative expansion, one determines an
approximation to $\alpha_{\overline{\rm MS}}(\mu)$ as well as $\bar m_c(\mu)$.
With the lattice spacing (scale) determined from some extra physical input,
this calibrates $\mu$. As usual suitable pseudoscalar masses
determine the bare-quark masses, here in particular the charm mass, 
and then through \eq{e:Rn} the renormalized charm-quark mass.

A difficulty with this approach is that large masses are needed to enter
the perturbative domain. Lattice artifacts can then be sizeable and
have a complicated form. The ratios in Eq.~(\ref{Rn}) use the
tree-level lattice results in the usual way for normalization.
This results in unity as the leading term in Eq.~(\ref{rn_expan}),
suppressing some of the kinematical lattice artifacts.
We note that in contrast to, e.g., the definition of $\alpha_\mathrm{qq}$,
here the cutoff effects are of order $a^k\alpha_s$, while there the
tree-level term defines $\alpha_s$ and therefore the cutoff effects
after tree-level improvement are of order $a^k\alpha_s^2$.
To obtain the continuum results for the moments it is important to perform
fits with high powers of $a$. This implies many fit parameters.
To deal with this problem the HPQCD collaboration
used Bayesian fits of their lattice results. More recent analyses of
the moments, however,  did not rely on Bayesian fits \cite{Nakayama:2016atf,Maezawa:2016vgv,Petreczky:2019ozv,Petreczky:2020tky}.

Finite-size effects (FSE) due to the omission of
$|x_0| > T /2$ in Eq.~(\ref{Gn_smu}) grow with $n$ as 
$(m_{\eta_c}T/2)^n\, \exp{(-m_{\eta_c} T/2)}$. 
In practice, however, since the (lower) moments
are short-distance dominated, the FSE are expected to be small
at the present level of precision. Possible exception could be the
ratio $R_8/R_{10}$, where the finite-volume effects could be significant
as discussed below.

Moments of correlation functions of the quark's electromagnetic
current can also be obtained from experimental data for $e^+e^-$
annihilation~\cite{Kuhn:2007vp,Chetyrkin:2009fv}.  This enables a
nonlattice determination of $\alpha_s$ using a similar analysis
method.  In particular, the same continuum perturbation-theory
computation enters both the lattice and the phenomenological determinations.


\subsubsection{Discussion of computations}

\begin{table}[!htb]
   \vspace{3.0cm}
   \footnotesize
   \begin{tabular*}{\textwidth}[l]{l@{\extracolsep{\fill}}rl@{\hspace{1mm}}l@{\hspace{1mm}}l@{\hspace{1mm}}l@{\hspace{1mm}}lll@{\hspace{1mm}}l}
      Collaboration & Ref. & $N_f$ &
      \hspace{0.15cm}\begin{rotate}{60}{publication status}\end{rotate}
                                                       \hspace{-0.15cm} &
      \hspace{0.15cm}\begin{rotate}{60}{renormalization scale}\end{rotate}
                                                       \hspace{-0.15cm} &
      \hspace{0.15cm}\begin{rotate}{60}{perturbative behaviour}\end{rotate}
                                                       \hspace{-0.15cm} &
      \hspace{0.15cm}\begin{rotate}{60}{continuum extrapolation}\end{rotate}
      \hspace{-0.25cm} & 
                         scale & $\Lambda_\msbar[\MeV]$ 
                       & $r_0\Lambda_\msbar$ \\
      &&&&&&&&& \\[-0.1cm]
      \hline
      \hline
      &&&&&&&&& \\[-0.1cm]
      HPQCD 14A   &  \cite{Chakraborty:2014aca}
                                              & 2+1+1   & \gA
                   & \soso & \good   & \soso
                   & $w_0=0.1715(9)\,\mbox{fm}^a$
                   & 294(11)$^{bc}$
                   & 0.703(26)             \\
      &&&&&&&&& \\[-0.1cm]
      \hline
      &&&&&&&&& \\[-0.1cm]

      {Petreczky 20}
                   & \cite{Petreczky:2020tky}  & 2+1   & \oP
                   & \soso & \soso  & \good   
                   & $r_1 = 0.3106(18)$ fm                   
                   & 332(17)$^h$             & 0.792(41)$^g$  \\

        {Boito 20}
                   & \cite{Boito:2020lyp}  & 2+1   & \gA 
                   & \bad & \bad  & \soso         
                   & $m_c(m_c)=1.28(2)$ GeV  
                   & 328(30)$^h$             & 0.785(72)  \\ 
    {Petrezcky 19, $\scriptstyle m_h=m_c$} 
                   & \textcolor{blue}{\cite{Petreczky:2019ozv}}  & 2+1   & \gA 
                   & \bad & \bad  & \good          
                   & {$r_1 = 0.3106(18)\,\mbox{fm}^{g}$}  
                   &  314(10)            &    0.751(24)$^g$  \\
     {Petrezcky 19, $\scriptstyle \frac{m_h}{m_c}=1.5$} 
                   & \textcolor{blue}{\cite{Petreczky:2019ozv}}  & 2+1   & \gA
                   & \bad & \bad  & \soso  
                   & {$r_1 = 0.3106(18)\,\mbox{fm}^{g}$}
                   &  310(10)            &    0.742(24)$^g$  \\
      {Maezawa 16}
                   & \textcolor{blue}{\cite{Maezawa:2016vgv}}  & 2+1   & \gA & \bad
                   & \bad  & \soso          
                   & {$r_1 = 0.3106(18)\,\mbox{fm}$$^{d}$}  
                   & 309(10)$^{e}$             & 0.739(24)$^{e}$  \\
       {JLQCD 16}   & \cite{Nakayama:2016atf}  
                   & 2+1     & \gA 
                   & \bad & \soso  & \soso           
                   & {$\sqrt{t_0} = 0.1465(25)\,\mbox{fm}$}
                   & {331(38)$^{f}$}  &  0.792(89)$^{f}$ \\
      HPQCD 10     & \cite{McNeile:2010ji}  & 2+1       & \gA 
                   & \soso & \good   & \soso           
                   & $r_1 = 0.3133(23)\, \mbox{fm}$$^\dagger$
                   & 338(10)$^\star$           &  0.809(25)           \\
      HPQCD 08B    & \cite{Allison:2008xk}  & 2+1       & \gA 
                   & \bad  & \bad  & \bad           
                   & $r_1 = 0.321(5)\,\mbox{fm}$$^\dagger$  
                   & 325(18)$^+$             &  0.777(42)            \\
      &&&&&&&&& \\[-0.1cm]
      \hline
      \hline\\
\end{tabular*}\\[-0.2cm]
\begin{minipage}{\linewidth}
{\footnotesize 
\begin{itemize}
   \item[$^a$]  Scale determined in \cite{Dowdall:2013rya} using $f_\pi$. \\[-5mm]
   \item[$^b$]  $\alpha^{(4)}_\msbar(5\,\mbox{GeV}) = 0.2128(25)$, 
         $\alpha^{(5)}_{\overline{\rm MS}}(M_Z) = 0.11822(74)$.         \\[-5mm]
   \item[$^c$] We evaluated $\Lambda_{\overline{\rm MS}}^{(4)}$ from $\alpha^{(4)}_\msbar$. 
         We also used $r_0 = 0.472\,\mbox{fm}$.\\[-5mm]
   \item[$^{d}$] 
   Scale is determined from $f_\pi$ . 
    \\[-5mm]
   \item[$^{e}$]       $\alpha^{(3)}_\msbar(m_c=1.267\,\mbox{GeV}) = 0.3697(85)$,
               $\alpha^{(5)}_\msbar(M_Z) = 0.11622(84)$. Our conversion with $r_0 = 0.472\,\mbox{fm}$.         
               \\[-5mm]
    \item[$^{f}$]  We evaluated $\Lambda_{\overline{\rm MS}}^{(3)}$ from the given $\alpha^{(4)}_\msbar(3\,\mbox{GeV}) = 0.2528(127)$.
          $\alpha^{(5)}_{\overline{\rm MS}}(M_Z) = 0.1177(26)$.
          We also used $r_0 = 0.472\,\mbox{fm}$ to convert. \\[-5mm]
    \item[$^{g}$] We used $r_0 = 0.472\,\mbox{fm}$ to convert. \\[-5mm]
     \item[$^{h}$] We back-engineered from $\alpha^{(5)}_\msbar(M_Z) = 0.1177(20)$. We used $r_0 = 0.472\,\mbox{fm}$ to convert. 
  \\[-5mm]
   \item[$^\star$]  $\alpha^{(3)}_\msbar(5\,\mbox{GeV}) = 0.2034(21)$,
            $\alpha^{(5)}_\msbar(M_Z) = 0.1183(7)$.         \\[-4mm]
   \item[$^\dagger$] Scale is determined from $\Upsilon$ mass splitting.    \\[-5mm]
   \item[$^+$]  We evaluated $\Lambda_{\overline{\rm MS}}^{(3)}$ from the given $\alpha^{(4)}_\msbar(3\,\mbox{GeV}) = 0.251(6)$. $\alpha^{(5)}_\msbar(M_Z) = 0.1174(12)$.       

\end{itemize}
}
\end{minipage}
\normalsize
\caption{Heavy-quark current two-point function results. Note that all analysis using $2+1$
  flavour simulations perturbatively add a dynamical charm quark.
  Partially they then quote results in $\Nf=4$-flavour 
  QCD, which we converted back to $\Nf=3$, corresponding to the
  nonperturbative sea quark content.}
\label{tab_current_2pt}
\end{table}

The determination of the strong-coupling constant from the moments of quarkonium correlators
by HPQCD collaboration have been discussed in detail in the FLAG 2016 and 2019 reports.
Therefore, we only give the summary of these determinations in Table \ref{tab_current_2pt}.

Two additional computations have appeared between the  FLAG 16 and the FLAG 19 reports.
We re-discuss them here (see also the summary section), as the assessment in FLAG 19 
was partially based on an inconsistent use of the FLAG criteria and has now been changed.
Maezawa and Petreczky, \cite{Maezawa:2016vgv} computed the two-point
functions of the $c\bar{c}$ pseudoscalar operator and obtained 
$R_4$, $R_6/R_8$ and $R_8/R_{10}$ based on the HotQCD collaboration
HISQ staggered ensembles, \cite{Bazavov:2014pvz}. The scale is set by measuring
$r_1=0.3106(18)$ fm. Continuum limits are taken fitting the 
lattice-spacing dependence with $a^2+a^4$ form as the best fit. For $R_4$,
they also employ other forms for fit functions such as $a^2$,
$\alpha_s^{\rm boosted} a^2+a^4$, etc., the results agreeing within errors.
Matching $R_4$ with the 3-loop formula Eq. (\ref{rn_expan}) 
through order $\alpha_\msbar^3$ \cite{Chetyrkin:2006xg}, 
where $\mu$ is fixed to $m_c$, they obtain
$\alpha^{(3)}_{\overline{\rm MS}}(\mu=m_c) = 0.3697(54)(64)(15)$. The 
first error is statistical, the second is the uncertainty 
in the continuum extrapolation, and the third is the truncation error
in the perturbative approximation of $r_4$. This last error is estimated
by the ``typical size'' of the missing 4-loop contribution,  
which they assume to be $\alpha^4_{\overline{\rm MS}}(\mu)$ multiplied by
2 times the 3-loop coefficient 
$2 \times r_{4,3} \times \alpha^4_{\msbar}(\mu) 
= 0.2364 \times \alpha^4_{\msbar}(\mu)$.
The result is converted to
\begin{eqnarray}
   \alpha^{(5)}_{\msbar}(M_Z) = 0.11622(84) \,.
\end{eqnarray}
Since $\alpha_{\rm eff}=0.38$ we assign
$\bad$ for the criterion of the renormalization scale. 
As $\Delta \Lambda / \Lambda < \alpha_{\rm eff}^2$, we assign
$\bad$ for the criterion of perturbative behaviour.
The lattice cutoff ranges as  $a^{-1} $ = 1.42--4.89~GeV with 
$\mu=2m_c\sim 2.6$ GeV so that we assign $\soso$ for continuum extrapolation.

JLQCD 16 \cite{Nakayama:2016atf} also computed the two-point functions
of the $c\bar{c}$ pseudoscalar operator and obtained $R_6$, $R_8$, $R_{10}$
and their ratios  based on 2+1-flavour QCD with M\"obius domain-wall
quark for three lattice cutoff $a^{-1}$ = 2.5, 3.6, 4.5~GeV.
The scale is set by $\sqrt{t_0}=0.1465(21)(13)\,\mbox{fm}$.
The continuum limit is taken assuming linear dependence on $a^2$.
They find a sizeable lattice-spacing dependence of $R_4$, which is 
therefore not used in their analysis, but for $R_6,R_8, R_{10}$
the dependence is mild giving reasonable control over the continuum limit.
They use the perturbative formulae for the vacuum polarization in the
pseudoscalar channel $\Pi_{PS}$ through order $\alpha_\msbar^3$ in the 
$\overline{\rm MS}$ scheme \cite{Maier:2008he, Maier:2009fz} to obtain 
$\alpha^{(4)}_{\overline{\rm MS}}$. Combining the matching of lattice
results with continuum perturbation theory for $R_6$, $R_6/R_8$ and $R_{10}$,
they obtain $\alpha^{(4)}_{\overline{\rm MS}}(\mu=3\,\GeV)=0.2528(127)$,
where the error is dominated by the perturbative truncation error.
To estimate the truncation error they study the dependence of the final result 
on the choice of the renormalization  scales $\mu, \;\mu_m$ which are used as renormalization scales for
$\alpha_s$ and the quark mass. Independently~\cite{Dehnadi:2015fra} the two scales
are varied in the range of 2~GeV to 4~GeV. 
The above result is converted to $\alpha^{(5)}_\msbar(M_Z)$ as
\begin{eqnarray}
   \alpha^{(5)}_{\overline{\rm MS}}(M_Z) = 0.1177(26) \,.
\end{eqnarray}%
Since $\alpha_{\rm eff} \simeq 0.37$, they have $\bad$ 
for the renormalization scale criterion. Since 
$\Delta \Lambda / \Lambda \simeq  \alpha_{\rm eff}^2$, we also assign
$\soso$ for the criterion of perturbative behaviour. The lattice cutoff
ranges over  $a^{-1}$ = 2.5--4.5~GeV with $\mu=3$ GeV so we also give
them a $\soso$ for continuum extrapolation. We note, however, that the $\chi^2/\mathrm{dof}$ of the
$a^2$ extrapolation was quite bad, namely between 2.1 and 5.1 \cite{Nakayama:2016atf}.
Please note that the 2019 FLAG review mistakenly took $\alpha_\msbar(2m_c)$ for $\alpha_{\rm eff}$. 
This resulted in a $\soso$ rating for the renormalization scale for both Maezawa 16 and JLQCD 16. 
With the consistent definition of $\alpha_{\rm eff}$ both 
determinations now have $\bad$ for the renormalization scale.

Three new determinations of $\alpha_s$ from the moments of quarkonium correlators
appeared since the 2019 FLAG review \cite{Petreczky:2019ozv,Boito:2020lyp,Petreczky:2020tky}.
Petreczky 19 \cite{Petreczky:2019ozv} extended the calculation of \cite{Maezawa:2016vgv} by considering
heavy-quark masses 
larger than the charm-quark mass, namely, $m_h=1.5m_c,~2 m_c$ and $3m_c$. Also three additional lattice
spacings, $a=$ 0.025, 0.03 and 0.035~fm have been added to the analysis. 
Another improvement compared to Maezawa 16 was the use of random-colour wall sources which greatly reduced the statistical errors. 
In fact, the statistical errors on the moments were completely negligible compared to other sources of errors.
The lattices corresponding to the three smallest lattice spacings
have been generated for the calculations of the QCD equation of state at high temperature \cite{Bazavov:2017dsy} at
light sea-quark masses corresponding to the pion mass of $300$ MeV in
the continuum limit, instead of the pion mass of $160$ MeV as in the previous calculations. However, it has been
checked that the effect of the larger light sea-quark masses is very small, about the size of the statistical errors \cite{Petreczky:2019ozv}.
Therefore, the calculations at the two light sea-quark masses have been combined into a single analysis \cite{Petreczky:2019ozv}.
For each value of the heavy-quark mass 
the continuum extrapolations have been performed using various fit ans\"atze, some of which included high powers of $a$. Due
to availability of many lattice spacing it was possible to perform such fits without using Bayesian priors. 
The variation of the continuum-extrapolated values with the variation of the fit range in $a^2$ and the fit
forms has been investigated and included as the systematic error of the continuum results.
The renormalization scale $\mu$ was fixed to the heavy-quark mass, and $\alpha_s(\mu= m_h)$ 
and the corresponding $\Lambda_{\overline{MS}}^{\Nf=3}$ has been
determined for each value of $m_h$ using continuum results for $R_4$, $R_6/R_8$ and $R_8/R_{10}$. 
The perturbative error was estimated as in Maezawa 16 but with the coefficient of the 4-loop term
being 1.6 times the coefficient of the 3-loop term.
The values  of  $\Lambda_{\overline{MS}}^{\Nf=3}$ obtained for $m_h=m_c$
and $m_h=1.5m_c$ were consistent with each other, $\Lambda_{\overline{MS}}^{\Nf=3}=314(10)$ MeV for $m_h=m_c$
and $\Lambda_{\overline{MS}}^{\Nf=3}=310(10)$ MeV for $m_h=1.5m_c$.
However, the $\Lambda_{\overline{MS}}^{\Nf=3}$ values
turned out to be significantly lower for $m_h=2 m_c$ and $3 m_c$. In Petreczky 20 \cite{Petreczky:2020tky}, 
it has been argued that reliable continuum extrapolations of $R_4$, $R_6/R_8$ and $R_8/R_{10}$ are not possible for $m_h \ge 2 m_c$.
Therefore, we only review the results obtained for $m_h=m_c$ and $m_h=1.5m_c$.
There are many lattice spacings available for analysis, including three lattice spacings $a \le 0.035$~fm, implying
that $a \mu< 0.5$. Therefore, we assign $\good$ for the continuum extrapolatiom. The value of $\alpha_\text{eff}$ is 0.38
and 0.31 for $m_h=m_c$ and $m_h=1.5 m_c$, respectively. So we assign $\bad$ for the renormalization scale. 
Since $(\Delta \Lambda/\Lambda)_{\Delta\alpha} < \alpha_\text{eff}^2$ we assign $\bad$ for the perturbative behaviour.

Petreczky 20 \cite{Petreczky:2020tky} used the same raw lattice data as Petreczky 19 but a different strategy
for continuum extrapolation and $\alpha_s$ extraction. The lattice spacing dependence of the results of $R_4$
at different quark masses was fitted simultaneously in a similar manner as in the HPQCD 10 and HPQCD 14 analyses, but
without using Bayesian priors. In extracting $\alpha_s$ several choices of the renormalization scale $\mu$ in the range 
$2/3 m_h$--$3 m_h$ have been considered. The perturbative error was estimated as in Petreczky 19 but the variation
of the results due to the scale variation was larger than the estimated perturbative error. The final error of the
result $\Lambda_{\overline{MS}}^{\Nf=3}=331(17)$ MeV comes mostly from the scale variation \cite{Petreczky:2020tky}.
Since there are three lattice spacing available with $a \mu<0.5$ we give $\good$ for continuum extrapolation.
Because $\alpha_\text{eff}=0.22-0.38$ we give $\soso$ for the renormalization scale. Finally, since 
$(\Delta \Lambda/\Lambda)_{\Delta \alpha}>\alpha_\text{eff}^2$ for the smallest $\alpha_\text{eff}$ value 
we give $\soso$ for the perturbative behaviour.
In addition to $R_4$ Petreczky 20 also considered using $R_6/R_8$ and $R_8/R_{10}$ for the $\alpha_s$ determination.
It was pointed out that the lattice spacing dependence of $R_6/R_8$ is quite subtle and therefore reliable continuum
extrapolations for this ratio are not possible for $m_h \ge 2 m_c$ \cite{Petreczky:2020tky}. 
For $m_h=m_c$ and $1.5m_c$ the ratio $R_6/R_8$ leads to $\alpha_s$ values that are consistent with the ones from $R_4$.
Furthermore, it was argued
that finite-volume effects in the case of $R_8/R_{10}$ are large for $m_h=m_c$ and therefore the corresponding
data are not suitable for extracting $\alpha_s$. 
This observation may explain why the central values of $\alpha_s$ extracted from $R_8/R_{10}$ in some previous studies
were systematically lower \cite{Allison:2008xk,Maezawa:2016vgv,Petreczky:2019ozv}.
On the other hand for $m_h \ge 1.5 m_c$ the finite-volume effects are sufficiently
small in the continuum extrapolated results if some small-volume lattice data are excluded from the analysis \cite{Petreczky:2020tky}.
The $\alpha_s$ obtained from $R_8/R_{10}$ with $m_h\ge 1.5 m_c$ were consistent with the ones obtained from $R_4$.

Boito 20 \cite{Boito:2020lyp} use published continuum extrapolated lattice results on $R_4$, $R_6/R_8$ and $R_8/R_{10}$ from various
groups combined with experimental results on $e^{+}e^{-}$ annihilation. They quote a separate result for
each lattice determinations of $R_4$, $R_6/R_8$ and $R_8/R_{10}$ for $m_h=m_c$ from different lattice groups.
They vary the scale $\mu$ and $\mu_m$ independently in the region between $m_c$ and 4~GeV. 
As the typical value they quote $\alpha_s(M_Z)=0.1177(20)$. 
The error is dominated by the perturbative uncertainty.
Since the effective coupling is around $0.38$ we give $\bad$ for the renormalization scale.
Because $(\Delta \Lambda/\Lambda)_{\Delta \alpha}<\alpha_\text{eff}^2$ we give this determination $\bad$ for perturbative behaviour.
The continuum results used in the analysis were rated as $\soso$ with the exception
of HPQCD 08B, which however, does not affect the quoted $\alpha_s$ value. Therefore we give
them $\soso$ for the continuum extrapolation. 
An interesting point of the Boito 20 analysis is that the $\alpha_s$ values extracted from $R_8/R_{10}$ are systematically
lower than the ones extracted from $R_4$. This confirms the above assertion that finite volume effects are significant
for $R_8/R_{10}$ at $m_h=m_c$.

Aside from the final results for $\alpha_s(m_Z)$ obtained by matching with perturbation theory, it is 
interesting to make a comparison of the short distance quantities 
in the continuum limit $R_n$ which are available from 
HPQCD 08~\cite{Allison:2008xk}, 
JLQCD 16 \cite{Nakayama:2016atf}, Maezawa 16 \cite{Maezawa:2016vgv}, Petreczky 19 \cite{Petreczky:2019ozv} 
and Petreczky 20 \cite{Petreczky:2020tky} (all using $2+1$ flavours). 
This comparison is shown in 
Tab.~\ref{Rn_moments}.
\begin{table}[h]
\begin{center}
\begin{tabular}{c|cccccc}
\hline
             & HPQCD 08    &  HPQCD 10  & Maezawa 16 & JLQCD 16  & Petreczky 19 & Petreczky 20      \\
\hline
$R_4$        & 1.272(5)    & 1.282(4)   & 1.265(7)  & -          & 1.279(4)     & 1.278(2)          \\
$R_6$        & 1.528(11)   & 1.527(4)   & 1.520(4)  & 1.509(7)   & 1.521(3)     & 1.522(2)          \\
$R_8$        & 1.370(10)   & 1.373(3)   & 1.367(8)  & 1.359(4)   & 1.369(3)     & 1.368(3)          \\
$R_{10}$     & 1.304(9)    & 1.304(2)   & 1.302(8)  & 1.297(4)   & 1.311(7)     & 1.301(3)          \\
$R_6/R_8$    & 1.113(2)    & -          & 1.114(2)  & 1.111(2)   & 1.1092(6)    & 1.10895(32)       \\
$R_8/R_{10}$ & 1.049(2)    & -          & 1.0495(7) & 1.0481(9)  & 1.0485(8)    & -                 \\
\hline
\end{tabular}
\end{center}
\caption{Moments and the ratios of the moments from $N_f=3$ simulations at the charm mass.}
\label{Rn_moments}
\end{table} 
The results are in quite good agreement with each other.
For future studies it is of course interesting to check
agreement of these numbers before turning to the more
involved determination of $\alpha_s$.


\subsection{$\alpha_s$ from QCD vertices}

\label{s:glu}


\subsubsection{General considerations}


The most intuitive and in principle direct way to determine the
coupling constant in QCD is to compute the appropriate
three- or four-point gluon vertices
or alternatively the
quark-quark-gluon vertex or ghost-ghost-gluon vertex (i.e., \ $
q\overline{q}A$ or $c\overline{c}A$ vertex, respectively).  A suitable
combination of renormalization constants then leads to the relation
between the bare (lattice) and renormalized coupling constant. This
procedure requires the implementation of a nonperturbative
renormalization condition and the fixing of the gauge. For the study
of nonperturbative gauge fixing and the associated Gribov ambiguity,
we refer to Refs.~\cite{Cucchieri:1997dx,Giusti:2001xf,Maas:2009ph} and
references therein.
In practice the Landau gauge is used and the
renormalization constants are defined 
by requiring that the vertex is equal to the tree-level
value at a certain momentum configuration.
The resulting renormalization
schemes are called `MOM' scheme (symmetric momentum configuration)
or `$\rm \widetilde{MOM}$' (one momentum vanishes), 
which are then converted perturbatively
to the $\overline{\rm MS}$ scheme.

A pioneering work to determine the three-gluon vertex in the $N_f = 0$
theory is Alles~96~\cite{Alles:1996ka} (which was followed by
Ref.~\cite{Boucaud:2001qz} for two flavour QCD); a more recent $N_f = 0$
computation was Ref.~\cite{Boucaud:2005gg} in which the three-gluon vertex
as well as the ghost-ghost-gluon vertex was considered.  (This
requires a computation of the propagator of the
Faddeev--Popov ghost on the lattice.) The latter paper concluded that
the resulting $\Lambda_{\overline{\rm MS}}$ depended strongly on the
scheme used, the order of perturbation theory used in the matching and
also on nonperturbative corrections \cite{Boucaud:2005xn}.

Subsequently in Refs.~\cite{Sternbeck:2007br,Boucaud:2008gn} a specific
$\widetilde{\rm MOM}$ scheme with zero ghost momentum for the
ghost-ghost-gluon vertex was used. In this scheme, dubbed
the `MM' (Minimal MOM) or `Taylor' (T) scheme, the vertex
is not renormalized, and so the renormalized coupling reduces to
\begin{eqnarray}
   \alpha_{\rm T}(\mu) 
      = D^{\rm gluon}_{\rm lat}(\mu, a) D^{\rm ghost}_{\rm lat}(\mu, a)^2 \,
                      {g_0^2 \over 4\pi} \,,
\end{eqnarray}
where $D^{\rm ghost}_{\rm lat}$ and $D^{\rm gluon}_{\rm lat}$ are the
(bare lattice) dressed ghost and gluon `form factors' of these
propagator functions in the Landau gauge,
\begin{eqnarray}
   D^{ab}(p) = - \delta^{ab}\, {D^{\rm ghost}(p) \over p^2}\,, \qquad
   D_{\mu\nu}^{ab}(p) 
      = \delta^{ab} \left( \delta_{\mu\nu} - {p_\mu p_\nu \over p^2} \right) \,
        {D^{\rm gluon}(p) \over p^2 } \,,
\end{eqnarray}
and we have written the formula in the continuum with 
$D^{\rm ghost/gluon}(p)=D^{\rm ghost/gluon}_{\rm lat}(p, 0)$.
Thus there is now no need to compute the ghost-ghost-gluon vertex,
just the ghost and gluon propagators.


\subsubsection{Discussion of computations}
\label{s:glu_discuss}

\begin{table}[!h]
   \vspace{3.0cm}
   \footnotesize
   \begin{tabular*}{\textwidth}[l]{l@{\extracolsep{\fill}}rl@{\hspace{1mm}}l@{\hspace{1mm}}l@{\hspace{1mm}}l@{\hspace{1mm}}llll}
   Collaboration & Ref. & $\Nf$ &
   \hspace{0.15cm}\begin{rotate}{60}{publication status}\end{rotate}
                                                    \hspace{-0.15cm} &
   \hspace{0.15cm}\begin{rotate}{60}{renormalization scale}\end{rotate}
                                                    \hspace{-0.15cm} &
   \hspace{0.15cm}\begin{rotate}{60}{perturbative behaviour}\end{rotate}
                                                    \hspace{-0.15cm} &
   \hspace{0.15cm}\begin{rotate}{60}{continuum extrapolation}\end{rotate}
      \hspace{-0.25cm} & 
                         scale & $\Lambda_\msbar[\MeV]$ & $r_0\Lambda_\msbar$ \\
      & & & & & & & & \\[-0.1cm]
      \hline
      \hline
      & & & & & & & & \\[-0.1cm]
       ETM 13D      & \cite{Blossier:2013ioa}   & {2+1+1} & {\gA}
                    & \soso & \soso  & \bad  
                    & $f_\pi$
                    & $314(7)(14)(10)$$^a$
                    & $0.752(18)(34)(81)$$^\dagger$                        \\
       ETM 12C        & \cite{Blossier:2012ef}   & 2+1+1 & \gA 
                    & \soso & \soso  & \bad  
                    & $f_\pi$
                    & $324(17)$$^\S$
                    & $0.775(41)$$^\dagger$                                \\
      ETM 11D       & \cite{Blossier:2011tf}   & 2+1+1 & \gA 
                    & \soso & \soso  & \bad  
                    & $f_\pi$
                    & $316(13)(8)(^{+0}_{-9})$$^\star$
                    & $0.756(31)(19)(^{+0}_{-22})$$^\dagger$                 \\
      & & & & & & & & \\[-0.1cm]
      \hline
      & & & & & & & & \\[-0.1cm]
      Zafeiropoulos 19
                    & \cite{Zafeiropoulos:2019flq}
                                               & 2+1  & \gA
                    & \bad & \bad & \bad 
                    & $m_\Omega$
                    & $320(4)(12)$$^b$
                    & $0.766(10)(29)$$^\dagger$                            \\
      Sternbeck 12  & \cite{Sternbeck:2012qs}  & 2+1  & \rC
                    &     &        & 
                    & \multicolumn{3}{l}{only running of 
                                         $\alpha_s$ in Fig.~4}            \\
      & & & & & & & & \\[-0.1cm]
      \hline
      & & & & & & & & \\[-0.1cm]
      Sternbeck 12  & \cite{Sternbeck:2012qs}  & 2  & \rC
                    &  &  & 
                    & \multicolumn{3}{l}{Agreement with $r_0\Lambda_\msbar$ 
                                         value of \cite{Fritzsch:2012wq} } \\
      Sternbeck 10  & \cite{Sternbeck:2010xu}  & 2  & \rC 
                    & \soso  & \good & \bad
                    &
                    & $251(15)$$^\#$
                    & $0.60(3)(2)$                                       \\
      ETM 10F       & \cite{Blossier:2010ky}   & 2  & \gA 
                    & \soso  & \soso  & \soso 
                    & $f_\pi$
                    & $330(23)(22)(^{+0}_{-33})$\hspace{-2mm}
                    & $0.72(5)$$^+$                                       \\
      Boucaud 01B    & \cite{Boucaud:2001qz}    & 2 & \gA 
                    & \soso & \soso  & \bad
                    & $K^{\ast}-K$
                    & $264(27)$$^{\star\star}$
                    & 0.669(69)                              \\
      & & & & & & & & \\[-0.1cm]
      \hline
      & & & & & & & & \\[-0.1cm]
      Sternbeck 12  & \cite{Sternbeck:2012qs}   & 0 & \rC 
                    &  &  &
                    &  \multicolumn{3}{l}{Agreement with $r_0\Lambda_\msbar$
                                          value of \cite{Brambilla:2010pp}} \\
      Sternbeck 10  & \cite{Sternbeck:2010xu}   & 0 & \rC
                    & \good & \good & \bad
                    &
                    & $259(4)$$^\#$
                    & $0.62(1)$                                            \\
      Ilgenfritz 10 & \cite{Ilgenfritz:2010gu}  & 0 & \gA
                    &    \good    &  \good      & \bad 
                    & \multicolumn{2}{l}{only running of
                                         $\alpha_s$ in Fig.~13}           \\
{Boucaud 08}    & \cite{Boucaud:2008gn}       & 0         &\gA  
                    & \soso & \good   & \bad 
                    & $\sqrt{\sigma} = 445\,\mbox{MeV}$
                    & $224(3)(^{+8}_{-5})$
                    & $0.59(1)(^{+2}_{-1})$         
 \\
{Boucaud 05}    & \cite{Boucaud:2005gg}       & 0       &\gA  
                    & \bad & \good   & \bad 
                    & $\sqrt{\sigma} = 445\,\mbox{MeV}$
                    & 320(32)
                    & 0.85(9)           
 \\
   Soto 01        & \cite{DeSoto:2001qx}        & 0         & \gA  
                    & \soso & \soso  & \soso
                    & $\sqrt{\sigma} = 445\,\mbox{MeV}$
                    & 260(18)
                    & 0.69(5)          
 \\
{Boucaud 01A}    & \cite{Boucaud:2001st}      & 0         &\gA  
                    & \soso & \soso  & \soso
                    & $\sqrt{\sigma} = 445\,\mbox{MeV}$
                    & 233(28)~MeV
                    & 0.62(7)       
 \\
{Boucaud 00B}   & \cite{Boucaud:2000nd}      & 0         &\gA  
                    & \soso & \soso  & \soso
                    & 
                    & \multicolumn{2}{l}{only running of
                                         $\alpha_s$}
 \\
{Boucaud 00A}     &\cite{Boucaud:2000ey}     &  0    &\gA  
                    & \soso & \soso  & \soso
                    & $\sqrt{\sigma} = 445\,\mbox{MeV}$
                    & $237(3)(^{+~0}_{-10})$
                    & $0.63(1)(^{+0}_{-3})$            
 \\
{Becirevic 99B}  & \cite{Becirevic:1999hj} & 0 &\gA  
                    & \soso & \soso  & \bad 
                    & $\sqrt{\sigma} = 445\,\mbox{MeV}$
                    & $319(14)(^{+10}_{-20})$
                    & $0.84(4)(^{+3}_{-5})$   
 \\
{Becirevic 99A}  & \cite{Becirevic:1999uc} & 0 &\gA  
                    & \soso & \soso  & \bad 
                    & $\sqrt{\sigma} = 445\,\mbox{MeV}$
                    & $\lesssim 353(2)(^{+25}_{-15})$
                    & $\lesssim 0.93 (^{+7}_{-4})$         
 \\
{Boucaud 98B}  & \cite{Boucaud:1998xi} & 0 &\gA  
                    & \bad  & \soso  & \bad 
                    & $\sqrt{\sigma} = 445\,\mbox{MeV}$
                    & 295(5)(15)
                    & 0.78(4)           
 \\
{Boucaud 98A}    & \cite{Boucaud:1998bq} & 0 &\gA  
                    & \bad  & \soso  & \bad 
                    & $\sqrt{\sigma} = 445\,\mbox{MeV}$
                    & 300(5)
                    & 0.79(1)         
\\
{Alles 96}    & \cite{Alles:1996ka} & 0 &\gA  
                    & \bad  & \bad   & \bad 
                    & $\sqrt{\sigma} = 440\,\mbox{MeV}$\hspace{0.3mm}$^{++}$\hspace{-0.3cm}        
                    & 340(50)
                    & 0.91(13)   
\\
      & & & & & & & & \\[-0.1cm]
      \hline
      \hline\\
\end{tabular*}\\[-0.2cm]
\begin{minipage}{\linewidth}
{\footnotesize 
\begin{itemize}
   \item[$^a$] $\alpha_{\overline{\rm MS}}^{(5)}(M_Z)=0.1196(4)(8)(6)$.                   \\[-5mm]
   \item[$^\dagger$] We use the 2+1 value $r_0=0.472$~fm.                        \\[-5mm]
   \item[$^\S$] $\alpha_{\overline{\rm MS}}^{(5)}(M_Z)=0.1200(14)$.                   \\[-5mm]
   \item[$^\star$] First error is statistical; second is due to the lattice
           spacing and third is due to the chiral extrapolation.
           $\alpha_{\overline{\rm MS}}^{(5)}(M_Z)=0.1198(9)(5)(^{+0}_{-5})$.    \\[-5mm]
   \item[$^b$] $\alpha_{\overline{\rm MS}}^{(5)}(M_Z)=0.1172(3)(9)(5)$. The first error
           is the uncertainty in the determination of $\alpha_T$, 
           the second due to the condensate while the third is due to higher order 
           nonperturbative corrections.                                        \\[-5mm]
   \item[$^\#$] In the paper only $r_0\Lambda_{\overline{\rm MS}}$ is given,
         we converted to $\MeV$ with $r_0=0.472$~fm.                    \\[-5mm]
   \item[$^+$] The determination of $r_0$
        from the $f_\pi$ scale is found in Ref.~\cite{Baron:2009wt}.          \\[-5mm]
   \item[$^{\star\star}$]  $\alpha_{\overline{\rm MS}}^{(5)}(M_Z)=0.113(3)(4)$.         \\[-5mm]
   \item[$^{++}$]  The scale is taken from the string tension computation
           of Ref.~\cite{Bali:1992ru}.
\end{itemize}
}
\end{minipage}
\normalsize
\caption{Results for the gluon--ghost vertex.}
\label{tab_vertex}
\end{table}

For the calculations considered here, to match to perturbative
scaling, it was first necessary to reduce lattice artifacts by an
$H(4)$ extrapolation procedure (addressing $O(4)$ rotational
invariance), e.g., ETM 10F \cite{Blossier:2010ky} or by lattice
perturbation theory, e.g., Sternbeck 12 \cite{Sternbeck:2012qs}.  To
match to perturbation theory, collaborations vary in their approach.
In ETM 10F \cite{Blossier:2010ky},  it was necessary to include the
operator $A^2$ in the OPE of the ghost and gluon propagators, while in
{Sternbeck 12 \cite{Sternbeck:2012qs}} very large momenta are used and
$a^2p^2$ and $a^4p^4$ terms are included in their fit to the momentum
dependence. A further later refinement was the introduction of
higher nonperturbative OPE power corrections in ETM 11D
\cite{Blossier:2011tf} and ETM 12C \cite{Blossier:2012ef}.
Although
the expected leading power correction, $1/p^4$, was tried, ETM finds
good agreement with their data only when they fit with the
next-to-leading-order term, $1/p^6$.  The update ETM 13D
\cite{Blossier:2013ioa} investigates this point in more detail, using
better data with reduced statistical errors.  They find that after
again including the $1/p^6$ term they can describe their data over a
large momentum range from about 1.75~GeV to 7~GeV.

In all calculations except for Sternbeck 10 \cite{Sternbeck:2010xu},
Sternbeck 12 \cite{Sternbeck:2012qs},
the matching with the perturbative formula is performed including
power corrections in the form of 
condensates, in particular $\langle A^2 \rangle$. 
Three lattice spacings are present in almost all 
calculations with $N_f=0$, $2$, but the scales $ap$ are rather large.
This mostly results in a $\bad$ on the continuum extrapolation
(Sternbeck 10 \cite{Sternbeck:2010xu},
  Boucaud 01B \cite{Boucaud:2001qz} for $N_f=2$.
 Ilgenfritz 10 \cite{Ilgenfritz:2010gu},   
 Boucaud 08 \cite{Boucaud:2008gn},
 Boucaud 05 \cite{Boucaud:2005gg}, 
 Becirevic 99B \cite{Becirevic:1999hj},
  Becirevic 99A \cite{Becirevic:1999uc},
 Boucaud 98B \cite{Boucaud:1998xi},
 Boucaud 98A \cite{Boucaud:1998bq},
 Alles 96 \cite{Alles:1996ka} for $N_f=0$).
A \soso\ is reached in the $\Nf=0$ computations 
Boucaud 00A \cite{Boucaud:2000ey}, 00B \cite{Boucaud:2000nd},
01A \cite{Boucaud:2001st}, Soto 01 \cite{DeSoto:2001qx} due to
a rather small lattice spacing,  but this is done on a lattice
of a small physical size. 
The $N_f=2+1+1$ calculation, fitting with condensates, 
is carried out for two lattice spacings
and with $ap>1.5$, giving $\bad$
for the continuum extrapolation as well. 
In ETM 10F \cite{Blossier:2010ky} we have
$0.25 < \alpha_{\rm eff} < 0.4$, while in ETM 11D \cite{Blossier:2011tf}, 
ETM 12C \cite{Blossier:2012ef} (and ETM 13 \cite{Cichy:2013gja})
we find $0.24 < \alpha_{\rm eff} < 0.38$,  which gives a $\soso$ 
in these cases for the renormalization scale.
In ETM 10F \cite{Blossier:2010ky} the values of $ap$ violate our criterion
for a continuum limit only slightly, and 
we give a \soso.

In {Sternbeck 10 \cite{Sternbeck:2010xu}}, the coupling ranges over
$0.07 \leq \alpha_{\rm eff} \leq 0.32$ for $N_f=0$ and $0.19 \leq
\alpha_{\rm eff} \leq 0.38$ for $N_f=2$ giving $\good$ and $\soso$ for
the renormalization scale,  respectively.  The fit with the perturbative
formula is carried out without condensates, giving a satisfactory
description of the data.  In {Boucaud 01A \cite{Boucaud:2001st}},
depending on $a$, a large range of $\alpha_{\rm eff}$ is used which
goes down to $0.2$ giving a $\soso$ for the renormalization scale and
perturbative behaviour, and several lattice spacings are used leading
to $\soso$ in the continuum extrapolation.  The $\Nf=2$ computation
Boucaud 01B \cite{Boucaud:2001st}, fails the continuum limit criterion
because both $a\mu$ is too large and an unimproved Wilson fermion
action is used.  Finally in the conference proceedings
Sternbeck 12 \cite{Sternbeck:2012qs}, the $N_f$ = 0, 2, 3 coupling
$\alpha_\mathrm{T}$ is studied.  Subtracting 1-loop lattice artifacts
and subsequently fitting with $a^2p^2$ and $a^4p^4$ additional lattice
artifacts, agreement with the perturbative running is found for large
momenta ($r_0^2p^2 > 600$) without the need for power corrections.  In
these comparisons, the values of $r_0\Lambda_\msbar$ from other
collaborations are used. As no numbers are given, we have not
introduced ratings for this study.

Since the previous FLAG review, there has been one new result, 
Zafeiropoulos 19 \cite{Zafeiropoulos:2019flq}, again based on the method
described in ETM 10F, \cite{Blossier:2010ky} but now for $N_f = 3$ flavours
rather than two. 
Again an $\langle A^2 \rangle$ condensate is included, but cannot be determined;
an estimate is used from ETM 10F ($N_f=2$) and ETM12C ($N_f=4$). The scale $\Lambda$
is determined from the largest momenta available (when a plateau appears),
and the error is estimated from the larger range $p\sim$ 3.0--3.7~GeV.
This is used to determine $\alpha_{\overline{MS}}$. 
In this work there is also some emphasis on being close 
to the physical-quark masses, using three domain-wall fermion data sets
and careful consideration of discretization effects following \cite{Boucaud:2018xup}.
The disadvantage is that a lower upper bound on the momenta is now reached.

The  range of effective couplings is $0.35 \lsim \alpha_{\rm eff} \lsim 0.42$,
and over this range we have \linebreak
$(\alpha_{\mathrm{eff}}(3.0\,\GeV)/\alpha_{\mathrm{eff}}(3.7\,\GeV))^{3} \sim 1.7$,
which leads to a $\bad$ for perturbative behaviour. With no $\alpha_\text{eff}$ 
at or below $0.3$ and only two lattice spacings, we also obtain a $\bad$ for both the renormalization
scale and the continuum extrapolation.

In Tab.~\ref{tab_vertex} we summarize the results. Presently there
are no $N_f \geq 3$ calculations of $\alpha_s$ from QCD vertices that
satisfy the FLAG criteria to be included in the range.



\subsection{$\alpha_s$ from the eigenvalue spectrum of the Dirac operator}


\label{s:eigenvalue}


\subsubsection{General considerations}

Consider the spectral density of the continuum
Dirac operator 
\begin{equation}
    \label{eq:def_rho}
    \rho(\lambda) = \frac{1}{V} \left\langle
    \sum_k (\delta(\lambda-i\lambda_k) + \delta(\lambda+i\lambda_k))  
    \right\rangle\,,
\end{equation}
where $V$ is the volume and $\lambda_k$ are the 
eigenvalues of the Dirac operator 
in a gauge background. 

Its perturbative expansion 
\begin{equation}
    \label{eq:exp_rho}
    \rho(\lambda) =
    \frac{3}{4\pi^2}
    \,\lambda^3 (1-\rho_1\gbar^2 
     -\rho_2\gbar^4 -\rho_3\gbar^6 + \cO(\gbar^8) )\,,
\end{equation}
is known including $\rho_3$  in the $\overline{\mathrm{MS}}$ scheme  
\cite{Chetyrkin:1994ex,Kneur:2015dda}.
In renormalization group improved form one sets the renormalization
scale $\mu$ to $\mu=s\lambda$ with $s=\cO(1)$ and
the $\rho_i$ are pure numbers. 
Nakayama 18~\cite{Nakayama:2018ubk} initiated a study 
of $\rho(\lambda)$ in the perturbative regime. They prefer to consider 
$\mu$ independent from $\lambda$. Then $\rho_i$ are 
polynomials in $\log(\lambda/\mu)$ of degree $i$.
One may consider 
\begin{equation}
    \label{eq:expF}
    F(\lambda) \equiv {\partial \log(\rho(\lambda)) 
                       \over \partial \log(\lambda)}
    = 3 -F_1\gbar^2 
     -F_2\gbar^4 -F_3\gbar^6 - F_4\gbar^8 +  \cO(\gbar^{10}) \,,
\end{equation}
where the coefficients, $F_i$, which are known for 
$i = 1, \ldots, 4$, are again polynomials of degree $i$ in 
$\log(\lambda/\mu)$. Choosing the alternate renormalization-group-improved form
with $\mu=s\lambda$ in \eq{eq:exp_rho}, \eq{eq:expF} would instead lead to
\begin{equation}
    \label{eq:expFRGimpr}
    F(\lambda) 
    = 3 -\bar F_2\gbar^4(\lambda) -\bar F_3\gbar^6(\lambda) - 
    \bar F_4\gbar^8(\lambda) +  \cO(\gbar^{10}) \,,
\end{equation}
with pure numbers $\bar F_i$ and $\bar F_1=0$. 
Determinations of $\alpha_s$ can be carried out by a computation
and continuum extrapolation of $\rho(\lambda)$ and/or 
$F(\lambda)$ at large $\lambda$. 
Such computations are made possible by the techniques of 
\cite{Giusti:2008vb,Cossu:2016eqs,Nakayama:2018ubk}.

We note that according to our general
discussions in terms of an effective coupling, we have $n_\mathrm{l}=2$;
the 3-loop $\beta$ function of a coupling defined from
\eq{eq:exp_rho} or \eq{eq:expFRGimpr} is known.~\footnote{In the present situation, 
Nakayama 18 \cite{Nakayama:2018ubk}, the effective coupling is defined by 
$\gbar^2_\lambda(\mu) = \bar F_2^{-1/2}\,(3 - F(\lambda))$ with 
$\mu=\lambda$. The alternative definition, \eq{eq:expFRGimpr},
would give $\gbar^2_\lambda(\mu) = \bar F_2^{-1/2}\,(3 - F(\lambda))^{1/2}$.}



\subsubsection{Discussion of computations}


There is one pioneering result to date using this method by
Nakayama 18 \cite{Nakayama:2018ubk}. They computed the eigenmode 
distributions of the Hermitian operator 
$a^2 D^\dagger_{\rm ov}D_{\rm ov}$
where $D_{\rm ov}=D_{\rm ov}(m_f=0,am_{\rm PV})$ is the overlap operator and $m_{\rm PV}$ is the
Pauli--Villars regulator on ensembles with 2+1 flavours using
M\"obius domain-wall quarks for three lattice cutoffs
$a^{-1}= 2.5, 3.6, 4.5 $ GeV, where $am_{\rm PV} = 3$ or $\infty$.
The bare eigenvalues are converted
to the $\msbar$ scheme at $\mu= 2\,\mbox{GeV}$ by multiplying with the 
renormalization constant $Z_m(2\,\mbox{GeV})$, which is then transformed
to those renormalized at $\mu=6\,\mbox{GeV}$ using the 
renormalization-group equation. The scale is set by $\sqrt{t_0}=0.1465(21)(13)\,\mbox{fm}$.
The continuum limit is taken assuming a linear dependence in $a^2$, while
the volume size is kept about constant: 2.6--2.8~fm.
Choosing the renormalization scale 
$\mu= 6\,\mbox{GeV}$, Nakayama~18~\cite{Nakayama:2018ubk} extracted
the strong coupling constant
$\alpha_{\overline{\rm MS}}^{(3)}(6\,\mbox{GeV})$ = 0.204(10).
The result is converted to 
\begin{eqnarray}
   \alpha^{(5)}_{\overline{\rm MS}}(M_Z) = 0.1226(36) \,.
\end{eqnarray}
Three lattice spacings in the range $a^{-1} =$ 2.5--4.5~GeV
with $\mu=\lambda=$ 0.8--1.25~GeV yield quite
small values $a\mu$. However, our continuum-limit criterion does not
apply as it requires us to consider $\alpha_s=0.3$. We thus deviate
from the general rule and give a \soso\,  which would result at the
smallest value  $\alpha_\msbar(\mu)=0.4$ considered by Nakayama~18~
\cite{Nakayama:2018ubk}. The values of $\alpha_\msbar$ lead to a $\bad$
for the renormalization scale, while perturbative behaviour is rated \soso.

In Tab.~\ref{tab_eigenvalue} we list this result.
\begin{table}[!htb]
   \vspace{3.0cm}
   \footnotesize
   \begin{tabular*}{\textwidth}[l]{l@{\extracolsep{\fill}}rllllllll}
   Collaboration & Ref. & $\Nf$ &
   \hspace{0.15cm}\begin{rotate}{60}{publication status}\end{rotate}
                                                    \hspace{-0.15cm} &
   \hspace{0.15cm}\begin{rotate}{60}{renormalization scale}\end{rotate}
                                                    \hspace{-0.15cm} &
   \hspace{0.15cm}\begin{rotate}{60}{perturbative behaviour}\end{rotate}
                                                    \hspace{-0.15cm} &
   \hspace{0.15cm}\begin{rotate}{60}{continuum extrapolation}\end{rotate}
      \hspace{-0.25cm} & 
                         scale & $\Lambda_\msbar[\MeV]$ & $r_0\Lambda_\msbar$ \\
   & & & & & & & & & \\[-0.1cm]
   \hline
   \hline
   & & & & & & & & & \\[-0.1cm] 
   Nakayama 18 & \cite{Nakayama:2018ubk} & 2+1 & \gA
            &     \bad   &  \soso      & \soso  
            & $\sqrt{t_0}$ 
            & $409(60)$\,$^*$   
            & $0.978(144)$                   \\ 
   & & & & & & & & & \\[-0.1cm]
   \hline
   \hline
\end{tabular*}
\begin{tabular*}{\textwidth}[l]{l@{\extracolsep{\fill}}llllllll}
\multicolumn{8}{l}{\vbox{\begin{flushleft}
   $^*$ $\alpha_\msbar^{(5)}(M_Z)=0.1226(36)$. $\Lambda_\msbar$ determined 
        by us using $\alpha_\msbar^{(3)}(6\,\mbox{GeV})=0.204(10)$.
        Uses $r_0 = 0.472\,\mbox{fm}$   \\
\end{flushleft}}}
\end{tabular*}
\vspace{-0.3cm}
\normalsize
\caption{Dirac eigenvalue result.}
\label{tab_eigenvalue}
\end{table}


\subsection{Summary}
\label{s:alpsumm}

\newcommand{\pp}{\phantom{0}}


After reviewing the individual computations, we are now in a position
to discuss the overall result. We first present the current status and
for that briefly consider $r_0\Lambda$ with its flavour dependence from
$N_f = 0$ to $4$ flavours. Then we discuss
the central $\alpha_{\overline{\rm MS}}(M_Z)$ results, which just 
use $N_f \geq 3$, give ranges for each sub-group discussed previously, and give final FLAG average
as well as an overall average together with the current PDG nonlattice numbers. 
Finally we return to $r_0\Lambda$, presenting our estimates for the various $N_f$.

{ 
\subsubsection{The present situation}


We first summarize the status of lattice-QCD calculations of the QCD
scale $\Lambda_\msbar$.  Fig.~\ref{r0LamMSbar15} shows all the results for
$r_0\Lambda_{\overline{\rm MS}}$ discussed in the previous sections.
\begin{figure}[!htb]\hspace{-2cm}\begin{center}
      \includegraphics[width=11.5cm]{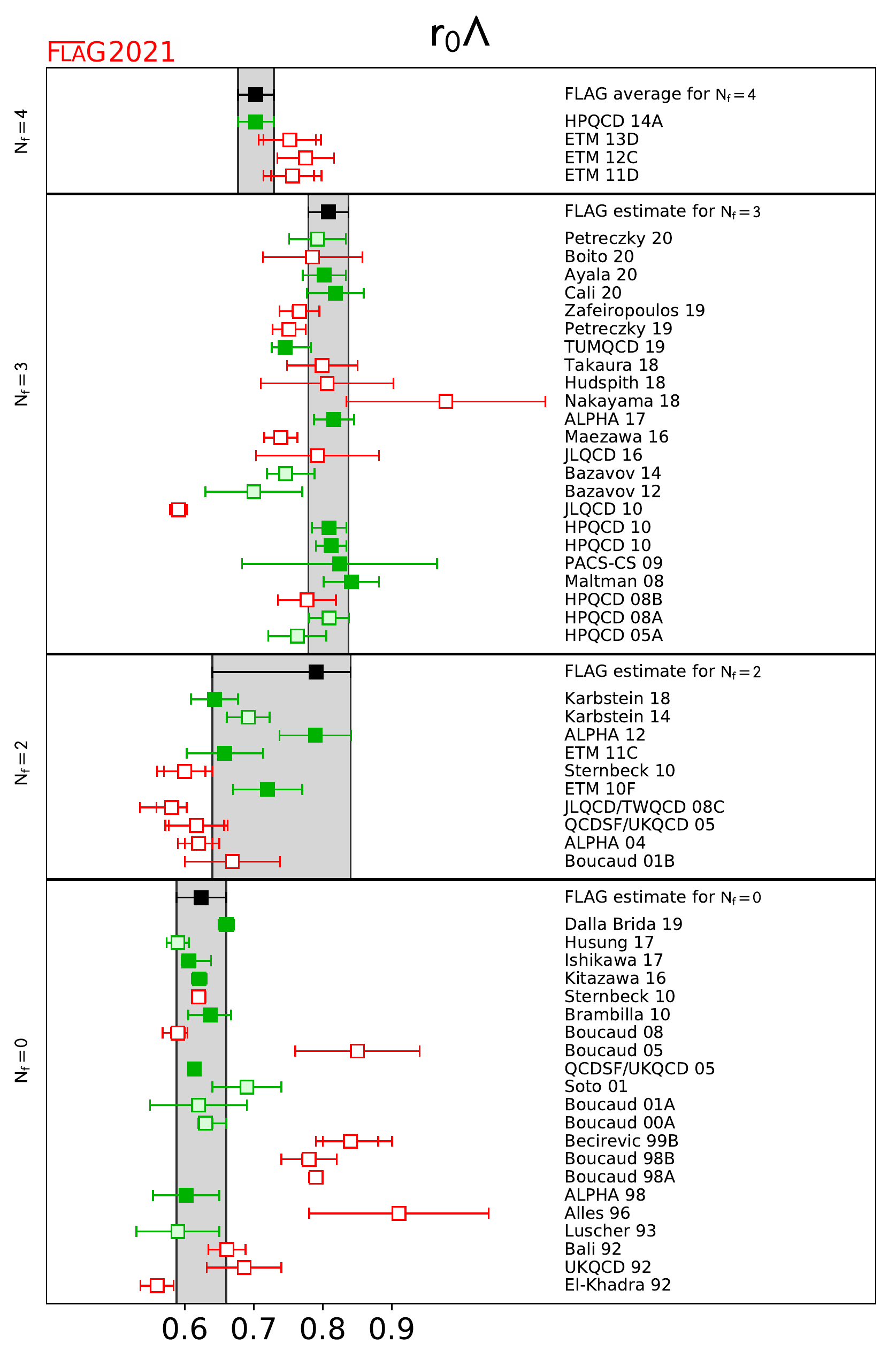}
      \end{center}
\vspace{-1cm}
\caption{$r_0\Lambda_{\overline{\rm MS}}$ estimates for
         $N_f = 0$, $2$, $3$, $4$ flavours.
         Full green squares are used in our final
         ranges, pale green squares also indicate that there are no
         red squares in the colour coding but the computations were
         superseded by later more complete ones or not
         published, while red open squares mean that there is at
         least one red square in the colour coding.}
\label{r0LamMSbar15}
\end{figure}

Many of the numbers are the ones given directly in the papers. 
However, when only $\Lambda_{\overline{\rm MS}}$ in physical units
($\mbox{MeV}$) is available, we have converted them by multiplying
with the value of $r_0$ in physical units. The notation used
is full green squares for results used in our final average,
while a lightly shaded green square indicates that there are no
red squares in the previous colour coding but the computation does
not enter the ranges because either it has been superseded by an update
or it is not published. Red open squares mean that there is at least
one red square in the colour coding.

For $N_f=0$ there is now some tension: 
the value of the new result, Dalla Brida 19~\cite{DallaBrida:2019wur} is
rather high compared to the previous FLAG average 
and yet it passes the FLAG 19 criteria by some margin.

When two flavours of quarks are included, the numbers extracted 
by the various groups show a considerable spread, as in particular
older computations did not yet control the systematics sufficiently. 
This illustrates the difficulty of the problem and emphasizes the 
need for strict criteria.  
The agreement among the more modern calculations with three or more flavours, 
however, is quite good.

We now turn to the status of the essential result for phenomenology,
$\alpha_{\overline{\rm MS}}^{(5)}(M_Z)$.  In Tab.~\ref{tab_alphmsbar18}
\begin{table}[!htb]
   \vspace{3.0cm}
   \tiny
   \begin{tabular*}{\textwidth}[l]{l@{\extracolsep{\fill}}rl@{\hspace{-1mm}}l@{\hspace{-1mm}}l@{\hspace{-1mm}}l@{\hspace{-1mm}}l@{\hspace{-1mm}}l@{\hspace{-1mm}}l@{\hspace{-3mm}}r}
   Collaboration & Ref. & $N_f$ &
   \hspace{0.15cm}\begin{rotate}{60}{publication status}\end{rotate}
                                                    \hspace{-0.15cm} &
   \hspace{0.15cm}\begin{rotate}{60}{renormalization scale}\end{rotate}
                                                    \hspace{-0.15cm} &
   \hspace{0.15cm}\begin{rotate}{60}{perturbative behaviour}\end{rotate}
                                                    \hspace{-0.15cm} &
   \hspace{0.15cm}\begin{rotate}{60}{continuum extrapolation}\end{rotate}
      \hspace{-0.25cm} & 
       $\alpha_\msbar(M_\mathrm{Z})$ & Remark  & Tab. \\
   & & & & & & & & & \\[-0.1cm]
   \hline
   \hline
   & & & & & & & & & \\[-0.1cm]
   {ALPHA 17}
            & \cite{Bruno:2017gxd}    & 2+1       & \gA 
            & \good   & \good    & \good 
            & $0.11852(\pp84)$
            & step-scaling
            & \ref{tab_SF3}                                   \\
  PACS-CS 09A& \cite{Aoki:2009tf} & 2+1 
            & \gA &\good &\good &\soso
            & $0.11800(300)$
            & step-scaling \hspace{-0.5cm}
            & \ref{tab_SF3}                                        \\[1ex]
  \multicolumn{3}{l}{pre-range (average)}  & & & & & 0.11848(\pp81)             & &     
  \\[1ex] \hline & & & & & & & & & \\[-0.1cm]
   {Ayala 20}
            & \cite{Ayala:2020odx}    & 2+1       & \gA & \soso
           & \good   & \soso
            & $0.11836(88)$
            & $Q$-$\bar{Q}$ potential
            & \ref{tab_short_dist}                            \\[1ex]
 
 {TUMQCD 19}
            & \cite{Bazavov:2019qoo}    & 2+1       & \gA 
            & \soso & \good   & \soso
            & $0.11671(^{+110}_{-57})$
            & $Q$-$\bar{Q}$ potential (and free energy)
            & \ref{tab_short_dist}                            \\[1ex]
   
  {Takaura 18}
            & \cite{Takaura:2018lpw,Takaura:2018vcy} & 2+1  & \gA 
            & \bad  & \soso  & \soso
            & $0.11790(70)(^{+130}_{-120})$
            & $Q$-$\bar{Q}$ potential
            & \ref{tab_short_dist}                            \\[1ex]
   {Bazavov 14}
            & \cite{Bazavov:2014soa}    & 2+1       & \gA 
            & \soso & \good   & \soso
            & $0.11660(100)$
            & $Q$-$\bar{Q}$ potential
            & \ref{tab_short_dist}                            \\[1ex]
   {Bazavov 12}
            & \cite{Bazavov:2012ka}   & 2+1       & \gA & \soso
            & \soso  & \soso
            & $0.11560(^{+210}_{-220})$ 
            & $Q$-$\bar{Q}$ potential
            & \ref{tab_short_dist}                            \\[1ex]
   \multicolumn{5}{l}{pre-range with estimated pert. error}    & & & 0.11782(165)  &      
 &      
   \\[1ex] \hline & & & & & & & & & \\[-0.1cm]
    {Cali 20} 
            & \cite{Cali:2020hrj}    & 2+1       & \gA
            & \soso  & \good  & \good
            & $0.11863(114)$
            & vacuum pol. (position space)
            & \ref{tab_vac}      \\
  {Hudspith 18} 
            & \cite{Hudspith:2018bpz}    & 2+1       & P
            & \soso  & \good     & \bad
            & $0.11810(270)(^{\pp+80}_{-220})$
            & vacuum polarization
            & \ref{tab_vac}      \\      
   JLQCD 10 & \cite{Shintani:2010ph} & 2+1 &\gA & \bad 
            & \soso & \bad
            & $0.11180(30)(^{+160}_{-170})$    
            & vacuum polarization  
            & \ref{tab_vac} \\[1ex]
  \multicolumn{5}{l}{pre-range with estimated pert. error}    & & & 0.11863(360)  &      
&      
  \\[1ex] \hline & & & & & & & & & \\[-0.1cm]

   HPQCD 10& \cite{McNeile:2010ji}& 2+1 & \gA & \soso
            & \good & \good
            & {0.11840(\pp60)}    
            & Wilson loops
            & \ref{tab_wloops}  
            \\
   Maltman 08& \cite{Maltman:2008bx}& 2+1 & \gA & \soso
            & \soso & \good
            & {$0.11920(110)$}
            & Wilson loops
            & \ref{tab_wloops}                               \\[1ex]
  \multicolumn{5}{l}{pre-range with estimated pert. error}    & & & 0.11871(128)  &      
&      
  \\[1ex] \hline & & & & & & & & & \\[-0.1cm]
{Petreczky 20} & \cite{Petreczky:2020tky}       & 2+1 & P &
           \soso & \soso & \good     
            & $0.11773(119)$. 
            & heavy current two points
             & \ref{tab_current_2pt}        \\
{Boito 20}
            & \cite{Boito:2019pqp,Boito:2020lyp}  & 2+1       & \gA 
            & \bad  &  \bad  & \soso
            & $0.1177(20)$
            & use published lattice data
            & \ref{tab_current_2pt}        \\
{Petreczky 19}
            & \cite{Petreczky:2019ozv}    & 2+1       & \gA 
            & \bad  &  \bad  & \good
            & $0.1159(12).$
            & heavy current two points
            & \ref{tab_current_2pt}      \\
{JLQCD 16}
            & \cite{Nakayama:2016atf}    & 2+1       & \gA 
            & \bad  &  \soso  & \soso
            & $0.11770(260)$
            & heavy current two points
            & \ref{tab_current_2pt}                                \\
  {Maezawa 16}
            & \cite{Maezawa:2016vgv}    & 2+1       & \gA 
            & \bad  &  \bad   & \soso
            & $0.11622(\pp84)$
            & heavy current two points
            & \ref{tab_current_2pt}                            \\
 HPQCD 14A 
                    &  \cite{Chakraborty:2014aca} & 2+1+1 & \gA 
                    & \soso & \good   & \soso
                    & 0.11822(\pp74)
                    & heavy current two points
                    & \ref{tab_current_2pt}                    \\

   HPQCD 10   & \cite{McNeile:2010ji}  & 2+1       & \gA & \soso
             & \good  & \soso          
             & 0.11830(\pp70)          
             & heavy current two points
             & \ref{tab_current_2pt} \\
   HPQCD 08B  & \cite{Allison:2008xk}  & 2+1       & \gA & \bad
             & \bad  & \bad
             & 0.11740(120) 
             & heavy current two points
             & \ref{tab_current_2pt}                                \\[1ex]
  \multicolumn{5}{l}{pre-range with estimated pert. error}    & & & 0.11826(200)  &      
&      
  \\[1ex] \hline & & & & & & & & & \\[-0.1cm]
   Zafeiropoulos 19 &  \cite{Zafeiropoulos:2019flq}   & 2+1& \gA
                    & \bad & \bad  & \bad 
                    & 0.1172(11)
                    & gluon-ghost vertex
                    & \ref{tab_vertex}                      \\
   ETM 13D    &  \cite{Blossier:2013ioa}   & 2+1+1& \gA
                    & \soso & \soso  & \bad 
                    & 0.11960(40)(80)(60)
                    & gluon-ghost vertex
                    & \ref{tab_vertex}                         \\
   ETM 12C    & \cite{Blossier:2012ef}   & 2+1+1 & \gA 
                    & \soso & \soso  & \bad  
                    & 0.12000(140)
		 & gluon-ghost vertex
                    & \ref{tab_vertex}                         \\
   ETM 11D   & \cite{Blossier:2011tf}   & 2+1+1 & \gA 
             & \soso & \soso & \bad  
                    & $0.11980(90)(50)(^{\pp+0}_{-50})$
                    & gluon-ghost vertex
                    & \ref{tab_vertex}                         
\\[1ex] \hline & & & & & & & & & \\[-0.1cm]
  
  {Nakayama 18}
            & \cite{Nakayama:2018ubk}    & 2+1       & \gA
            &     \good  &  \soso      & \bad
            & $0.12260(360)$
            & Dirac eigenvalues
            & \ref{tab_eigenvalue}                             \\[1ex]
   & & & & & & & & & \\[-0.1cm]
   \hline
   \hline
\end{tabular*}
\begin{tabular*}{\textwidth}[l]{l@{\extracolsep{\fill}}lllllll}
\multicolumn{8}{l}{\vbox{\begin{flushleft} 
\end{flushleft}}}
\end{tabular*}
\vspace{-0.8cm}
\caption{Results for $\alpha_\msbar(M_\mathrm{Z})$.
Different methods are listed separately and they are combined to a pre-range 
when computations are available without any \protect\bad.
A weighted average of the pre-ranges gives $0.11843(60)$, using the smallest pre-range 
uncertainty gives $0.11843(81)$ while the average uncertainty of the ranges 
used as an error gives $0.11843(187)$. Note that TUMQCD 19 supersedes Bazavov 14/12.  
}
\label{tab_alphmsbar18}
\end{table}  
and 
the upper plot in 
Fig.~\ref{alphasMSbarZ} we show all the results for
\begin{figure}[!htb]
   \begin{center}
      \includegraphics[width=11.5cm]{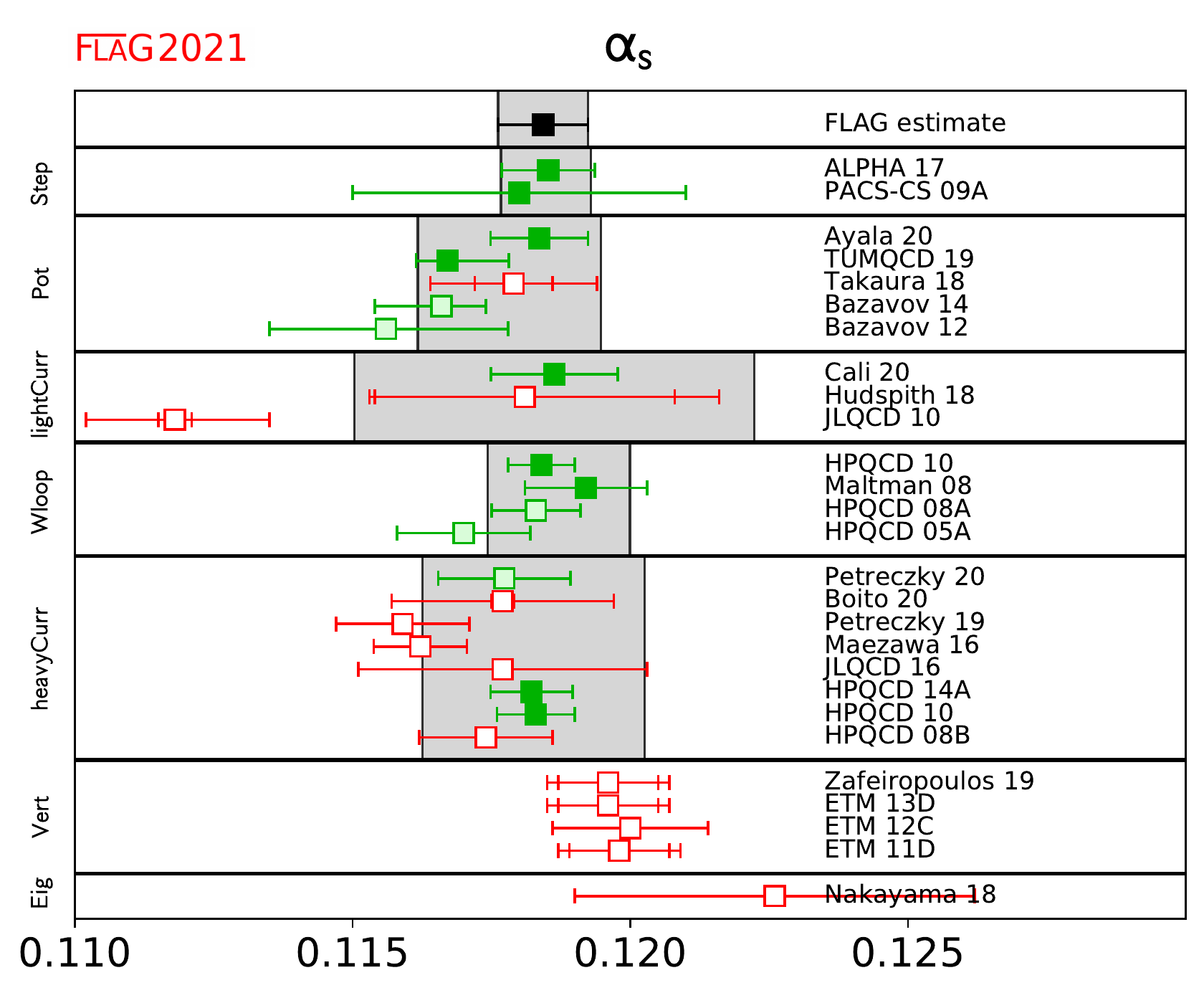}\\
      \includegraphics[width=11.5cm]{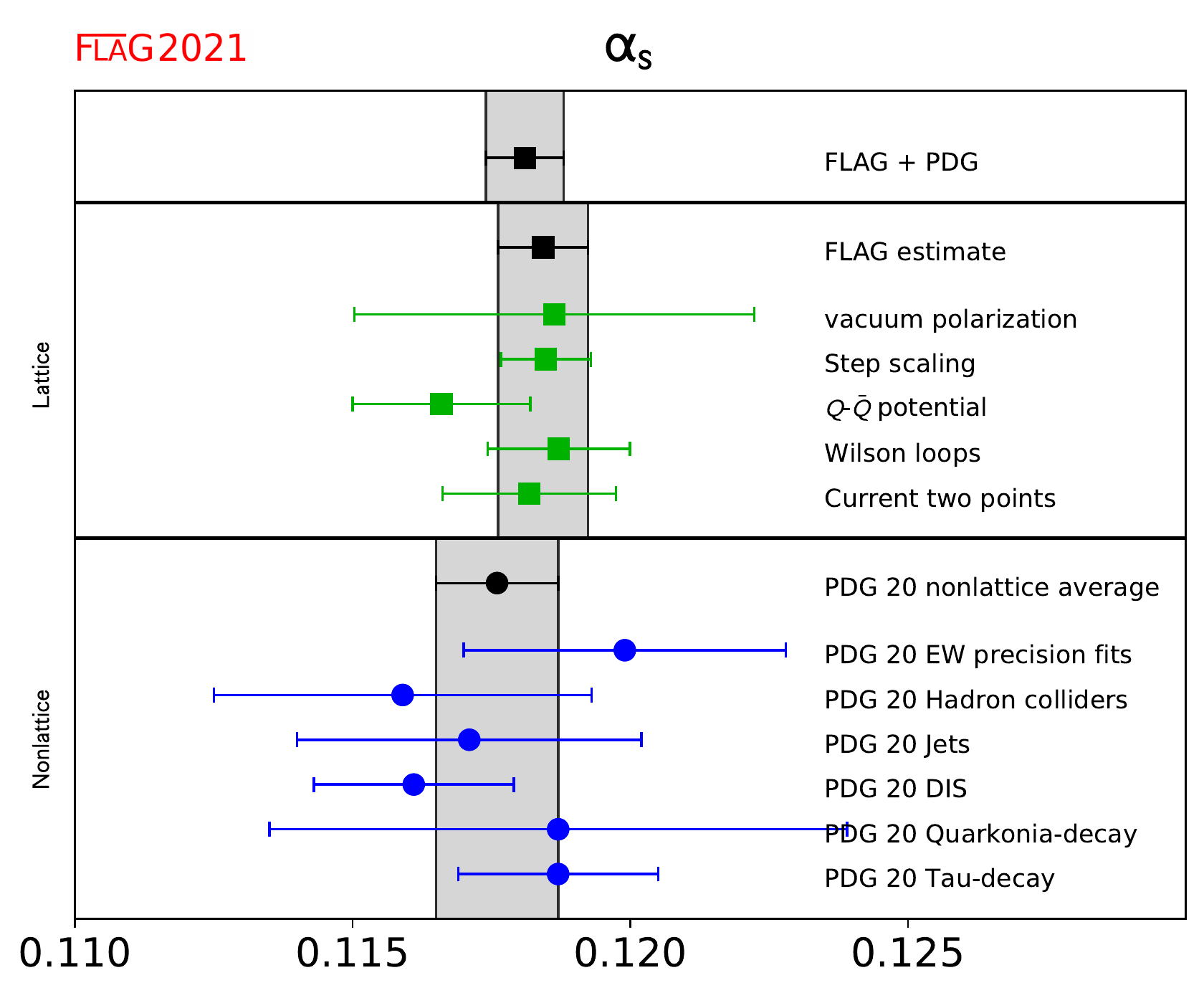}
   \end{center}
\caption{$\alpha_{\overline{\rm MS}}^{(5)}(M_Z)$, the coupling
  constant in the $\overline{\rm MS}$ scheme at the $Z$ mass. 
Top: lattice  results, pre-ranges from different calculation methods, and final average. 
Bottom: Comparison of the lattice pre-ranges and average with the nonlattice ranges and average.
The first PDG 20 entry gives the outcome of their analysis excluding lattice results
  (see Sec.~\ref{subsubsec:alpha_s_Conclusions}).}
\label{alphasMSbarZ}
\end{figure}
$\alpha_{\overline{\rm MS}}^{(5)}(M_Z)$ (i.e., \ $\alpha_{\overline{\rm MS}}$ at
the $Z$ mass) obtained from $N_f=2+1$ and $N_f = 2+1+1$
simulations. 
The conversion from $\Nf = 3$ or $\Nf = 4$
to $\Nf = 5$ is made by matching the coupling constant at the charm and
bottom quark thresholds and using the scale as determined or used by the
authors. 

As can be seen from the tables and figures, at present there are
several computations satisfying the criteria to be included in
the FLAG average. Since FLAG 19 four new computations of
$\alpha_{\overline{\rm MS}}^{(5)}(M_Z)$
pass all our criteria with at least a \soso . 
The results agree quite well within the stated
uncertainties, which vary significantly. 
}


\subsubsection{Our range for $\alpha_{\overline{\rm MS}}^{(5)}$}
\label{subsubsect:Our range}


We now explain the determination of our range.  We only include those
results without a red tag and that are published in a refereed journal.
We also do not include any numbers that were obtained by
extrapolating from theories with less than three flavours.  They are 
not controlled and can be looked up in the previous FLAG reviews.

A general issue with most determinations of $\alpha_\msbar$,
both lattice and nonlattice, is that they are dominated by
perturbative truncation errors, which are difficult to estimate.
Further, all results discussed here except for
those of Secs.~\ref{s:SF},~\ref{s:WL} are based on extractions of
$\alpha_\msbar$ that are largely influenced by data with
$\alpha_\mathrm{eff}\geq 0.3$.  At smaller $\alpha_s$ the momentum scale
$\mu$ quickly gets at or above $a^{-1}$. We have included computations
using $a\mu$ up to $1.5$ and $\alpha_\mathrm{eff}$ up to 0.4, but one
would ideally like to be significantly below that. Accordingly we
choose to not simply perform weighted averages with the individual
errors estimated by each group. 
Rather, we use our own more conservative estimates of the perturbative truncation errors in the weighted average.  

In the following we repeat aspects of the methods and 
calculations that  inform our estimates of the perturbative truncation errors.  
We also provide separate estimates for $\alpha_s$ obtained from step-scaling, 
the heavy-quark potential, Wilson loops, heavy-quark current 
two-point functions and vacuum polarization 
to enable a comparison of the different lattice approaches; 
these are summarized in \tab{tab_alphmsbar18}.

\begin{itemize}
\item
{\em Step-scaling\\} 
The step-scaling computations of PACS-CS~09A~\cite{Aoki:2009tf}
and ALPHA~17~\cite{Bruno:2017gxd} reach energies around the
$Z$-mass where perturbative uncertainties in the three-flavour theory 
are negligible. Perturbative errors do enter in the conversion of
the $\Lambda$-parameters from three to five flavours, but successive
order contributions decrease rapidly and can be neglected. 
We form a weighted average of the two results and obtain 
$\alpha_{\msbar}=0.11848(81)$.
\item
{\em Static-quark potential computations\\} 
Brambilla 10 \cite{Brambilla:2010pp},
ETM 11C \cite{Jansen:2011vv} and Bazavov 12 \cite{Bazavov:2012ka}  give
evidence that they have reached distances where perturbation theory
can be used. However, in addition to $\Lambda$, a scale
is introduced into the perturbative prediction by the process of
subtracting the renormalon contribution. 
This subtraction is avoided in
Bazavov 14 \cite{Bazavov:2014soa} by using the force and again 
agreement with perturbative running is reported.
Husung 17 \cite{Husung:2017qjz} (unpublished) studies
the reliability of perturbation theory
 in the pure gauge theory with lattice spacings down to 
$0.015\,\fm$ and finds that at weak coupling there is a downwards
trend in the $\Lambda$-parameter with a slope 
$\Delta \Lambda / \Lambda \approx 9 \alpha_s^3$. 
The downward trend is broadly confirmed in Husung 20~\cite{Husung:2020pxg} albeit with larger errors.

Bazavov 14 \cite{Bazavov:2014soa} satisfies all of the criteria to enter the FLAG average for $\alpha_s$
but has been superseded by TUMQCD 19~\cite{Bazavov:2019qoo}.
Moreover, there is another study, Ayala 20~\cite{Ayala:2020odx} who use the very
same data as TUMQCD 19, but treat perturbation theory differently, resulting
in a rather different central value. This shows that perturbative truncation 
errors are the main source of errors. We combine the results
for $\lms^{\Nf=3}$ from both groups as a weighted average (with the 
larger upward error of TUMQCD 19) and take
the difference of the central values as the uncertainty of the average.
We obtain $\lms^{\Nf=3} =$ 330(24)~MeV, which translates to $\alpha_s(m_Z) = 0.11782(165)$.


\item 
{\em Small Wilson loops\\} 
Here the situation is unchanged as compared to
FLAG~16. 
In the determination of $\alpha_s$ from
observables at the lattice spacing scale, there is an interplay
of higher-order perturbative terms and lattice artifacts.
In HPQCD 05A \cite{Mason:2005zx}, HPQCD 08A \cite{Davies:2008sw}
and Maltman 08 \cite{Maltman:2008bx} both lattice artifacts (which are
power corrections in this approach) and higher-order perturbative
terms are fitted.  We note that Maltman 08~\cite{Maltman:2008bx} and
HPQCD 08A~\cite{Davies:2008sw} analyze largely the same data set but
use different versions of the perturbative expansion and treatments of
nonperturbative terms.  After adjusting for the slightly different
lattice scales used, the values of $\alpha_\msbar(M_Z)$ differ by
$0.0004$ to $0.0008$ for the three quantities considered.  In fact the
largest of these differences ($0.0008$) comes from a tadpole-improved
loop, which is expected to be best behaved perturbatively.
 We therefore replace the perturbative-truncation errors from \cite{Maltman:2008bx} 
and \cite{McNeile:2010ji} with our estimate of the perturbative uncertainty 
\eq{hpqcd:ouruncert}. Taking the perturbative errors to be 100\% correlated between the results, we obtain for the weighted average 
$\alpha_{\msbar}=0.11871(128)$.

\item 
{\em Heavy quark current two-point functions\\}
Other computations with small errors are
HPQCD~10 \cite{McNeile:2010ji} and HPQCD~14A~\cite{Chakraborty:2014aca},
where correlation functions of heavy valence quarks are
used to construct short-distance quantities. Due to the large quark
masses needed to reach the region of small coupling, considerable
discretization errors are present, see Fig.~30 of FLAG~16. These
are treated by fits to the perturbative running (a 5-loop running
$\alpha_{\overline{\rm MS}}$ with a fitted 5-loop coefficient in
the $\beta$-function is used) with high-order terms in a double expansion
in $a^2\Lambda^2$ and $a^2 m_\mathrm{c}^2$ supplemented by priors
which limit the size of the coefficients.  The priors play an
especially important role in these fits given the much larger number
of fit parameters than data points.  We note, however, that the size
of the coefficients does not prevent high-order terms from
contributing significantly, since the data includes values of
$am_{\textrm c}$ that are rather close to 1.

We note that the result of JLQCD 16 was classified in FLAG 19 as having passed all FLAG criteria, 
although the scale is set by the charm-quark mass, implying $\alpha_\text{eff} \simeq 0.38$.
We now assign a red flag for renormalization scale, as we do for Petreczky 19 and Boito 20 (see below).
Since FLAG 19, there have been three new studies, Petreczky 19 \cite{Petreczky:2019ozv}, Petreczky 20 \cite{Petreczky:2020tky} 
and Boito 20~\cite{Boito:2020lyp} (Petreczky 19/Petreczky 20 supersede Maezawa 16~\cite{Maezawa:2016vgv}). 
While Petreczky 19/Petreczky 20 share the same lattice data for heavy quark masses in the range $m_h=m_c$--$4m_c$ 
they use a different strategy for continuum extrapolations and a different treatment of perturbative uncertainties. 
Petreczky 19 \cite{Petreczky:2019ozv} perform continuum extrapolation separately for each value of the valence-quark
mass, while Petreczky 20 rely on joint continuum extrapolations of the lattice data at different heavy-quark masses,
similar to the analysis of HPQCD, but without Bayesian priors. It is concluded that reliable continuum extrapolations 
for $m_h \ge 2 m_c$ require a joint fit to the data. This limits the eligible $\alpha_s$ 
determinations in Petreczky 19 \cite{Petreczky:2019ozv} to $m_h=m_c$ and $1.5m_c$, for which, however, 
the FLAG criteria are not satisfied.  There is also a difference in the choice of renormalization scale between both analyses: 
Petreczky 19 \cite{Petreczky:2019ozv} uses $\mu= m_h$, while Petreczky 20 \cite{Petreczky:2020tky} considers 
several choices of $\mu$ in the range $\mu=2/3 m_h$--$3 m_h$, which leads to larger perturbative uncertainties 
in the determination of $\alpha_s$ \cite{Petreczky:2020tky}. 
Boito 20 \cite{Boito:2020lyp} use published continuum extrapolated lattice results
for $m_h=m_c$ and performs its own extraction of $\alpha_s$. Limiting the choice of $m_h$ to the charm-quark mass means
that the FLAG criteria are not met ($\alpha_\text{eff} \simeq 0.38$). However, their analysis gives valuable insight
into the perturbative error. In addition to the renormalization scale $\mu$, Boito 20 also vary
the renormalization scale $\mu_m$ at which the charm quark mass is defined. The corresponding result
$\alpha_s(M_Z)=0.1177(20)$ agrees well with previous lattice determination but has a larger error, which is dominated
by the perturbative uncertainty due to the variation of both scales. 
This increased uncertainty suggests that the perturbative error estimated by HPQCD using a fixed scale $\mu=3 m_h$ may be too small. 
Therefore, we take the average of the HPQCD 10 and HPQCD 14A determinations and assign an error of $0.0020$, 
based on the analysis of Boito 20 \cite{Boito:2020lyp}. This results in the range $\alpha_s(M_Z)=0.11826(200)$.

\item
{\em Light quark vacuum polarization \\}
Since FLAG 19 a new study, Cali 20~\cite{Cali:2020hrj} appeared, which uses the light current two-point functions in position space,
evaluated on a subset of CLS configurations for lattice spacings in the range 0.038--0.076~fm, and for Euclidean 
distances 0.13--0.19~fm, corresponding to renormalization scales $\mu=$ 1--1.5~GeV.
Both flavour nonsinglet vector and axial vector currents are considered and their difference is shown to vanish within errors.
After continuum and chiral limits are taken, the effective coupling from the axial vector two-point function is converted
at 3-loop order to $\alpha_\msbar(\mu)$. The authors do this by numerical solution for $\alpha_\msbar$ and then
perform a weighted average of the $\Lambda$-parameter estimates for the available energy range, which yields $\lms^{\Nf=3}= 342(17)$~MeV. 
Note that this is the first calculation in the vacuum polarization category that passes the current FLAG criteria.
Yet the renormalization scales are rather low and one might suspect that other nonperturbative (i.e., non chiral-symmetry breaking)  effects may still be sizeable.  Our main issue is a rather optimistic estimate 
of perturbative truncation errors, based only on the variation of the $\Lambda$-parameter from the range of effective couplings considered. 
If the solution for the $\msbar$ coupling is done by series expansion in $\alpha_\text{eff}$, the differences in $\alpha_\msbar$, formally of order $\alpha_\text{eff}^5$, are still 
large at the scales considered. Hence, as a measure of the systematic uncertainty we take the difference $409-355$~MeV 
between $\lms^{\Nf=3}$ estimates at $\mu=1.5$~GeV as a proxy for the total error, i.e. $\lms^{\Nf=3} =$ 342(54)~MeV, 
which translates to our pre-range, $\alpha_s(m_Z)=0.11863(360)$, from vacuuum polarization.

\item 
{\em Other methods}
\\
Computations using other methods do not qualify for an average yet,
predominantly due to a lacking \soso\ in the continuum extrapolation. 
\end{itemize}

We obtain the central value for our range of $\alpha_s$ from the weighted average of the five pre-ranges listed in Tab.~\ref{tab_alphmsbar18}.  
The error of this weighted average is 0.0006, which is quite a bit smaller than the most precise entry.  Because, however, the errors on almost all of the $\alpha_s$ calculations that enter the average are dominated by perturbative truncation errors, which are especially difficult to estimate, we choose instead to take a larger range for $\alpha_s$ of $0.0008$.  This is the error on the pre-range for $\alpha_s$ from step-scaling, because perturbative-truncation errors are sub-dominant in this method.
Our final range is then given by 
\begin{eqnarray}
  \alpha_{\overline{\rm MS}}^{(5)}(M_Z) = 0.1184(8) \,.
 \label{eq:alpmz}
\end{eqnarray}
moving up by 2 in the last given digit compared to FLAG 19 and with the same uncertainty.
Of the eleven calculations that are included most are within 1$\sigma$ of this range, an exception being TUMQCD~19 (which supersedes Bazavov 14 and Bazavov 12).
Further, the range for $\alpha_{\overline{\rm MS}}^{(5)}(M_Z)$ presented here is based on results with rather different systematics (apart from the matching across the charm threshold).  We therefore believe that the true value is very likely to lie within this range.
 
All computations which enter this range, with the exception of
HPQCD~14A \cite{Chakraborty:2014aca}, rely on a perturbative 
inclusion of the charm and bottom quarks. Perturbation theory
for the matching of $\gbar^2_{N_f}$ and $\gbar^2_{N_f-1}$ looks very well
behaved even at the mass of the charm. Worries that still there may be
purely nonperturbative effects at this rather low scale 
have been removed by nonperturbative studies of the accuracy of 
perturbation theory. While the original study in Ref.~\cite{Bruno:2014ufa}
was not precise enough, the extended one in Ref.~\cite{Athenodorou:2018wpk}
estimates effects in the $\Lambda$-parameter to be significantly below
1\% and thus negligible for the present and near future accuracy.


\subsubsection{Ranges for $[r_0 \Lambda]^{(\Nf)}$ and $\lms$}

In the present situation, we give ranges for $[r_0 \Lambda]^{(\Nf)}$
and $\lms$, discussing their determination case by case.  We include
results with $\Nf<3$ because it is interesting to see the
$\Nf$-dependence of the connection of low- and high-energy QCD.  This
aids our understanding of the field theory and helps in finding
possible ways to tackle it beyond the lattice approach. It is also of
interest in providing an impression on the size of the 
vacuum-polarization effects of quarks, in particular with an eye on the still
difficult-to-treat heavier charm and bottom quarks. 
Most importantly, however, the decoupling strategy described in subsection~\ref{s:dec}
means that $\Lambda$-parameters at different $\Nf$ can be connected
by a nonperturbative matching computation. Thus, even results at unphysical flavour
numbers, in particular $\Nf=0$, may enter results for the physically
interesting case. Rather than phasing out results for ``unphysical flavour numbers",
continued scrutiny by FLAG will be necessary.
Having said this, we emphasize that results for $[r_0 \Lambda]^{(0)}$
and $[r_0 \Lambda]^{(2)}$ are {\em not}\/ meant to be used directly for
phenomenology. 

For the ranges we obtain: 
\begin{eqnarray}
   [r_0 \lms]^{(4)}   &=& 0.70(3)\,,               \label{eq:lms4} \\[0pt]
   [r_0 \lms]^{(3)}   &=& 0.808(29) \,,            \label{eq:lms3} \\[0pt]
   [r_0 \lms]^{(2)}   &=& 0.79(^{+~5}_{-{15}})\,,  \label{eq:lms2} \\[0pt]
   [r_0 \lms]^{(0)}   &=& 0.624(36) \,.            \label{eq:lms0}  
\end{eqnarray}  
No change has occurred since FLAG 19 for $\Nf=2,4$, so we take over the respective discussion from FLAG 19.

 For $\Nf=2+1+1$, we presently do not quote a range
as there is a single result: HPQCD 14A 
\cite{Chakraborty:2014aca} found $[r_0 \Lambda]^{(4)} = 0.70(3)$.

 For $\Nf=2+1$, we take as a central value the weighted average of 
Cali 20 \cite{Cali:2020hrj}, 
Ayala 20 \cite{Ayala:2020odx},  
TUMQCD 19\cite{Bazavov:2019qoo}, 
ALPHA 17 \cite{Bruno:2017gxd}
HPQCD~10 \cite{McNeile:2010ji} (Wilson loops and current two-point correlators),
PACS-CS~09A \cite{Aoki:2009tf} (with linear continuum extrapolation) and 
Maltman~08 \cite{Maltman:2008bx}.
Since the uncertainty in $r_0$ is small compared to that of $\Lambda$,
we can directly propagate the error from the analog of 
\eq{eq:alpmz} with the 2+1+1 number removed and arrive at 
\begin{eqnarray}
   [r_0 \lms]^{(3)} = 0.808(29) \,.
 \label{eq:lms3}
 \end{eqnarray}
 (The error of the straight weighted average is $0.012$.)
 It is in good agreement with all 2+1 results without red tags. 
 In physical units, using $r_0=$ 0.472~fm and neglecting
 its error, this means\footnote{In the FLAG 19 report \cite{Aoki:2019cca}, an inaccurate conversion of $[r_0 \lms]^{(3)}$ 
 in Eq.~(345) to physical units (using $r_0=0.472$~fm) led to $343\,\mbox{MeV}$ in Eqs.~(346,353). 
 However, using fm$\times$MeV$= 1/197.3$  gives $337\,\mbox{MeV}$ (Eqs.~(351) and (352) are however correct). Note: Equation references in this footnote are from FLAG 19~\cite{Aoki:2019cca}.}
 \begin{eqnarray}
   \lms^{(3)} = 338(12)\,\mbox{MeV}\,,
 \label{e:lms3}
 \end{eqnarray}
 where the error of the straight weighted average is less than $5\MeV$. 

 For $N_f=2$, at present there is one computation with a \good\ rating
 for all criteria, ALPHA 12 \cite{Fritzsch:2012wq}. We adopt it as our
 central value and enlarge the error to cover the central values of the
 other three results with filled green boxes. This results in an
 asymmetric error. Our range is unchanged as compared to \flagold,
 \begin{eqnarray}
    [r_0 \lms]^{(2)} = 0.79(^{+~5}_{-{15}}) \,, \quad
    \label{eq:lms2}
 \end{eqnarray}
 and in physical units, using $r_0=$ 0.472~fm, 
 \begin{eqnarray}
    \lms^{(2)} = 330(^{+21}_{-{63}}) \mbox{\,MeV}\,.  \quad 
 \end{eqnarray}
 A weighted average of the four eligible numbers would yield 
 $[r_0 \lms]^{(2)} = 0.689(23)$, not covering the best result and in
 particular leading to a smaller error than we feel is justified, given
 the issues discussed previously in
 Sec.~\ref{short_dist_discuss} (Karbstein 18 \cite{Karbstein:2018mzo},
 ETM 11C \cite{Jansen:2011vv}) and Sec.~\ref{s:glu_discuss}
 (ETM 10F \cite{Blossier:2010ky}). Thus we believe that our estimate
 is a conservative choice; the low values of ETM 11C \cite{Jansen:2011vv}
 and Karbstein 18 \cite{Karbstein:2018mzo} lead to a large downward error.
 We note that this can largely be explained by different values
 of $r_0$ between ETM 11C \cite{Jansen:2011vv} and
 ALPHA 12 \cite{Fritzsch:2012wq}.
 We still hope that future work will improve the situation.

For $\Nf=0$, the new result DallaBrida 19~\cite{DallaBrida:2019wur},
is quite large compared to the FLAG 19 average. We combine
it with those results which entered the FLAG 19 report, namely
ALPHA~98 \cite{Capitani:1998mq}, QCDSF/UKQCD~05 \cite{Gockeler:2005rv},
Brambilla~10 \cite{Brambilla:2010pp}, Kitazawa~16 \cite{Kitazawa:2016dsl}
and Ishikawa~17 \cite{Ishikawa:2017xam} for forming a range.%
\footnote{We have assigned a
  \soso\ for the continuum limit, in Boucaud 00A \cite{Boucaud:2000ey},
  00B \cite{Boucaud:2000nd}, 01A \cite{Boucaud:2001st},
  Soto~01 \cite{DeSoto:2001qx} but these results are from lattices of a
  very small physical size with finite-size effects that are not
  easily quantified.}  
Taking a weighted average of the six numbers, we obtain 
$[r_0 \lms]^{(0)} = 0.624(5)$, up from $0.615(5)$ for FLAG 19.

Clearly the errors are dominantly systematic, mostly due to perturbative truncation errors.
Since we do not change the FLAG 19 criteria for this edition, we give 
a range which encompasses all central values. Unfortunately, this requires
to double the error of the FLAG 19 result (which was given by $0.615(18)$), 
due to the large central value of $0.660$ by DallaBrida 19.
We arrive at our range for $\Nf=0$, 
\begin{equation}
 [r_0 \lms]^{(0)} =0.624(36) \,.
  \label{eq:lms0}
\end{equation}
This is clearly not very satisfactory, and, despite this large error, 
this still means that the high quality, and statistics dominated 
new step-scaling result Dalla Brida~19 is more than 3 sigma away 
from the central value of the new FLAG average.

Converting to physical units, again using $r_0=0.472\,\mbox{fm}$ yields
\begin{eqnarray}
   \lms^{(0)} =  261(15)\,\mbox{MeV}\,. \quad
\end{eqnarray}
While the conversion of the $\Lambda$ parameter to physical units is
quite unambiguous for $\Nf=2+1$, our choice of $r_0=0.472$~fm also for
smaller numbers of flavour amounts to a convention, in particular for
$\Nf=0$. Indeed, in the Tabs.~\ref{tab_SF3}--\ref{tab_vertex}
somewhat different numbers in MeV are found.


\subsubsection{Conclusions}
\label{subsubsec:alpha_s_Conclusions}


With the present results our range for the strong coupling is

(repeating Eq.~(\ref{eq:alpmz}))
\begin{eqnarray*}
 \FLAGAVBEGIN \alpha_{\overline{\rm MS}}^{(5)}(M_Z) = 0.1184(8)\FLAGAVEND\qquad\Refs~\mbox{\cite{Ayala:2020odx, Bazavov:2019qoo, Cali:2020hrj,Bruno:2017gxd,Chakraborty:2014aca,McNeile:2010ji,Aoki:2009tf,Maltman:2008bx}}, 
\end{eqnarray*}
and the associated $\Lambda$ parameters
\begin{eqnarray}
  \FLAGAVBEGIN \Lambda_{\overline{\rm MS}}^{(5)} = 214(10)\FLAGAVEND\,\MeV\hspace{5mm}\qquad\Refs~\mbox{\cite{Ayala:2020odx, Bazavov:2019qoo, Cali:2020hrj,Bruno:2017gxd,Chakraborty:2014aca,McNeile:2010ji,Aoki:2009tf,Maltman:2008bx}},
  \\
  \FLAGAVBEGIN \Lambda_{\overline{\rm MS}}^{(4)} = 297(12)\FLAGAVEND\,\MeV\hspace{5mm}\qquad\Refs~\mbox{\cite{Ayala:2020odx, Bazavov:2019qoo, Cali:2020hrj,Bruno:2017gxd,Chakraborty:2014aca,McNeile:2010ji,Aoki:2009tf,Maltman:2008bx}},
  \\
  \FLAGAVBEGIN \Lambda_{\overline{\rm MS}}^{(3)} = 339(12)\FLAGAVEND\,\MeV\hspace{5mm}\qquad\Refs~\mbox{\cite{Ayala:2020odx, Bazavov:2019qoo, Cali:2020hrj,Bruno:2017gxd,Chakraborty:2014aca,McNeile:2010ji,Aoki:2009tf,Maltman:2008bx}}\,.
\end{eqnarray} 
%
Compared with FLAG 19, the central values have moved slightly, with the errors remaining the same.

 It is interesting to compare 
 with the Particle Data Group average of nonlattice
 determinations of recent years,
 \begin{eqnarray}
  \alpha^{(5)}_{\overline{\rm MS}}(M_Z) &=& 0.1176(11) \,, \quad 
    \mbox{PDG 20, nonlattice \cite{Zyla:2020zbs},
    also appeared as Eq.~(\ref{PDG_nolat})}\nonumber\,,
 \\
  \alpha^{(5)}_{\overline{\rm MS}}(M_Z) &=& 0.1174(16) \,, \quad 
    \mbox{PDG 18, nonlattice \cite{Tanabashi:2018oca}},
 \\
    \alpha^{(5)}_{\overline{\rm MS}}(M_Z) &=& 0.1174(16) \,, \quad 
    \mbox{PDG 16, nonlattice \cite{Patrignani:2016xqp}} \,,
 \\
    \alpha^{(5)}_{\overline{\rm MS}}(M_Z) &=& 0.1175(17) \,, \quad 
    \mbox{PDG 14, nonlattice \cite{Agashe:2014kda}} \,,
 \\
    \alpha^{(5)}_{\overline{\rm MS}}(M_Z) &=& 0.1183(12) \,, \quad 
    \mbox{PDG 12, nonlattice \cite{Beringer:1900zz}}\,. \qquad\qquad
 \end{eqnarray}
 (there was no update in \cite{Tanabashi:2018oca}).
 There is good agreement with Eq.~(\ref{eq:alpmz}). Despite our very conservative error estimate,
 the FLAG lattice average has an error that is 30\% smaller than the PDG~20
 nonlattice-world average and a weighted average of the two [Eq.~(\ref{eq:alpmz})
 and Eq.~(\ref{PDG_nolat})] yields 
 \begin{eqnarray}
    \alpha^{(5)}_{\overline{\rm MS}}(M_Z) &=& 
  0.1181(7) \,, \quad 
    \mbox{FLAG 21 + PDG 20}.
 \label{PDG_FLAG_alpha}  
 \end{eqnarray}
 In  the lower plot in Fig.~\ref{alphasMSbarZ} we 
 show as blue circles 
 the various PDG pre-averages which 
 lead to the  PDG 20 nonlattice average. They are on a similar level
 as our pre-ranges 
 (green squares)
 : each one corresponds to an estimate
 (by the PDG) of $\alpha_s$ determined from one set of input quantities.
 Within each pre-average multiple groups did the analysis and published their 
 results as displayed in Ref.~\cite{Zyla:2020zbs}.
 
The fact that 
our range for the lattice determination of $\alpha_{\overline{\rm MS}}(M_Z)$
in Eq.~(\ref{eq:alpmz}) is in excellent agreement with
the PDG 20 nonlattice average  Eq.~(\ref{PDG_nolat}) is an excellent check for the 
subtle interplay of theory, phenomenology and experiments in the
nonlattice determinations. The work done on the lattice provides an
entirely independent determination, with negligible experimental 
uncertainty,  which reaches a better precision even with our quite conservative estimate of its uncertainty.

 We finish by commenting on perspectives for the future. 
 The step-scaling methods have been shown to yield a very precise result
 and to satisfy all criteria easily. A downside is that dedicated
 simulations have to be done and the method is thus hardly used.
 It would be desirable to have at least one more such 
 computation by an independent collaboration, as also requested
 in the review \cite{Salam:2017qdl}.
While this FLAG review does not report an error reduction compared to FLAG 19,
the understanding of some systematic errors has improved. 
With the exception of the step-scaling result, all determinations of $\alpha_s$,
appear to be limited by systematic uncertainties due to perturbative truncation errors. 
Similar conclusions have been drawn in the recent review article~\cite{DelDebbio:2021ryq}. 
In order to improve control of systematics it would be necessary 
to reach higher energy scales without incurring large cutoff effects.
This could be achieved by applying step-scaling methods in large (infinite) volume, provided that finite volume
effects are carefully controlled. Even a relatively modest increase by a scale factor 2--3 
could significantly enhance the scope  for some of the current approaches to determine $\alpha_s$.
Another hope for improvement are decoupling strategies, following the recent proposal by 
the ALPHA collaboration, cf.~Sec.~\ref{s:dec}. This in turn motivates further 
state-of-the-art studies in the pure gauge theory ($\Nf=0$), 
where it would be important to resolve the current tension between results in the literature.

\clearpage
\setcounter{section}{9}

\section{Nucleon matrix elements (NME)\label{sec:NME}}
Authors: S.~Collins, R.~Gupta, A.~Nicholson, H.~Wittig\\

A large number of experiments testing the Standard Model (SM) and searching
for physics Beyond the Standard Model (BSM) involve either free
nucleons (proton and neutron beams) or the scattering of electrons,
muons, neutrinos and dark matter off nuclear targets. Necessary
ingredients in the analysis of the experimental results are the matrix
elements of various probes (fundamental currents or operators in a low
energy effective theory) between nucleon or nuclear states. The goal
of lattice-QCD calculations in this context is to provide high
precision predictions of these matrix elements, the simplest of which
give the nucleon charges and form factors.  Determinations of the
charges are the most mature and in this review we summarize the
results for twelve quantities, the isovector and flavour diagonal axial
vector, scalar and tensor charges. Other quantities that are not being reviewed but for which
significant progress has been made in the last five years are the
nucleon axial vector and electromagnetic form
factors~\cite{Syritsyn:2014saa,Capitani:2015sba,Sufian:2016pex,Rajan:2017lxk,Green:2017keo,Chambers:2017tuf,Alexandrou:2017ypw,Alexandrou:2018zdf,Ishikawa:2018rew,Alexandrou:2018sjm,Shintani:2018ozy,Bali:2019yiy,Hasan:2019noy,Alexandrou:2020okk,Djukanovic:2021cgp}
and parton distribution functions~\cite{Lin:2017snn,Constantinou:2020pek,Constantinou:2020hdm,Cichy:2018mum,Monahan:2018euv}. The more
challenging calculations of nuclear matrix elements, that are needed, for example, to 
calculate the cross-sections of neutrinos or dark matter scattering off nuclear targets, are proceeding
along three paths. First is direct evaluation of matrix elements calculated
with initial and final states consisting of multiple nucleons~\cite{Savage:2011xk,Chang:2017eiq}. Second, convoluting
nucleon matrix elements with nuclear effects~\cite{Carlson:2014vla}, and third, determining two and
higher body terms in the nuclear potential via the direct or the HAL QCD methods~\cite{Wagman:2017tmp,Iritani:2017wvu}.  We
expect future FLAG reviews to include results on these quantities once a sufficient
level of control over all the systematics is reached.

\subsection{Isovector and flavour diagonal charges of the nucleon\label{sec:intro}}

The simplest nucleon matrix elements are composed of local quark
bilinear operators, $\overline{q_i} \Gamma_\alpha q_j$, where
$\Gamma_\alpha$ can be any of the sixteen Dirac matrices. In this
report, we consider two types of flavour structures: (a) when $i = u$
and $j = d$. These $\overline{u} \Gamma_\alpha d$ operators arise in
$W^\pm$ mediated weak interactions such as in neutron or pion decay.
We restrict the discussion to the matrix elements of the axial vector~($A$),
scalar~($S$) and tensor~($T$) currents, which give the isovector charges,
$g_{A,S,T}^{u-d}$.\footnote{In the isospin symmetric limit $\langle
  p|\bar{u}\Gamma d|n\rangle=\langle p|\bar{u}\Gamma u-\bar{d}\Gamma
  d|p\rangle=\langle n|\bar{d}\Gamma d-\bar{u}\Gamma u|n\rangle$ for
  nucleon and proton states $|p\rangle$ and $|n\rangle$,
  respectively. The latter two~(equivalent) isovector matrix elements are computed
  on the lattice.  } (b) When $i = j $ for $j \in \{u, d, s\}$,
there is no change of flavour, e.g., in processes mediated via the
electromagnetic or weak neutral interaction or dark matter.  These
$\gamma$ or $Z^0$ or possible dark matter mediated processes couple to all
flavours with their corresponding charges. Since these probes interact
with nucleons within nuclear targets, one has to include the effects
of QCD (to go from the couplings defined at the quark and gluon level
to those for nucleons) and nuclear forces in order to make contact with 
experiments. The isovector and flavour diagonal charges, given by the
matrix elements of the corresponding operators calculated between nucleon states,
are these nucleon level couplings. Here we review results for the
light and strange flavours, $g_{A,S,T}^{u}$, $g_{A,S,T}^{d}$, and
$g_{A,S,T}^{s}$ and the isovector charges $g_{A,S,T}^{u-d}$.

The isovector and flavour diagonal operators also arise in BSM
theories due to the exchange of novel force carriers or as effective
interactions due to loop effects.  The associated couplings are
defined at the energy scale $\Lambda_{\rm BSM}$,
while lattice-QCD calculations of matrix elements are carried out at a hadronic
scale, $\mu$, of a few GeV. The tool for connecting the couplings at
the two scales is the renormalization group. Since the operators of
interest are composed of quark fields~(and more generally also of gluon
fields), the predominant change in the corresponding couplings under a
scale transformation is due to QCD.  To define the operators and their
couplings at the hadronic scale $\mu$, one constructs renormalized
operators, whose matrix elements are finite in the continuum limit. This requires
calculating both multiplicative renormalization factors, including the
anomalous dimensions and finite terms, and the mixing with other
operators. We discuss the details of the renormalization factors
needed for each of the six operators reviewed in this report in
Sec.~\ref{sec:renorm}.

Once renormalized operators are defined, the matrix elements of
interest are extracted using expectation values of two-point and
three-point correlation functions illustrated in
Fig.~\ref{fig:feynman}, where the latter can have both quark-line connected 
and disconnected contributions. In order to isolate the
ground-state matrix element, these correlation functions are analyzed
using their spectral decomposition. The current practice is to fit the
$n$-point correlation functions (or ratios involving three- and
two-point functions) including contributions from one or two excited
states.  In some cases, such as axial and vector operators, Ward
identities provide relations between correlation functions, or ground
state matrix elements, or facilitate the calculation of renormalization
constants.  It is important to ensure that all such Ward identities
are satisfied in lattice calculations, especially as in the case of
axial form factors where they provide checks of whether excited state
contamination has been removed in obtaining matrix elements within
ground state nucleons~\cite{Liang:2018pis,Jang:2019vkm,Bali:2019yiy}.

The ideal situation occurs if the time separation $\tau$ between the
nucleon source and sink positions, and the distance of the operator
insertion time from the source and the sink, $t$ and $\tau - t$,
respectively, are large enough such that the contribution of all
excited states is negligible. In the limit of large $\tau$, the ratio
of noise to signal in the nucleon two- and three-point correlation
functions grows exponentially as $e^{(M_N -
  \frac{3}{2}M_\pi)\tau}$~\cite{Hamber:1983vu,Lepage:1989hd}, where
$M_N$ and $M_\pi$ are the masses of the nucleon and the pion,
respectively. Therefore, in particular at small pion masses,
maintaining reasonable errors for large $\tau$ is challenging, with
current calculations limited to $\tau \lesssim 1.5$~fm. In addition,
the mass gap between the ground and excited (including multi-particle)
states is smaller than in the meson sector and at these separations,
excited-state effects can be significant. The approach commonly taken
is to first obtain results with high statistics at multiple values of
$\tau$, using the methods described in Sec.~\ref{sec:technical}. Then,
as mentioned above, excited-state contamination is removed by fitting
the data using a fit form involving one or two excited states. The
different strategies that have been employed to minimize excited-state
contamination are discussed in Sec.~\ref{sec:ESC}.

\begin{figure}[tpb] 
\centerline{
\includegraphics[width=0.32\linewidth]{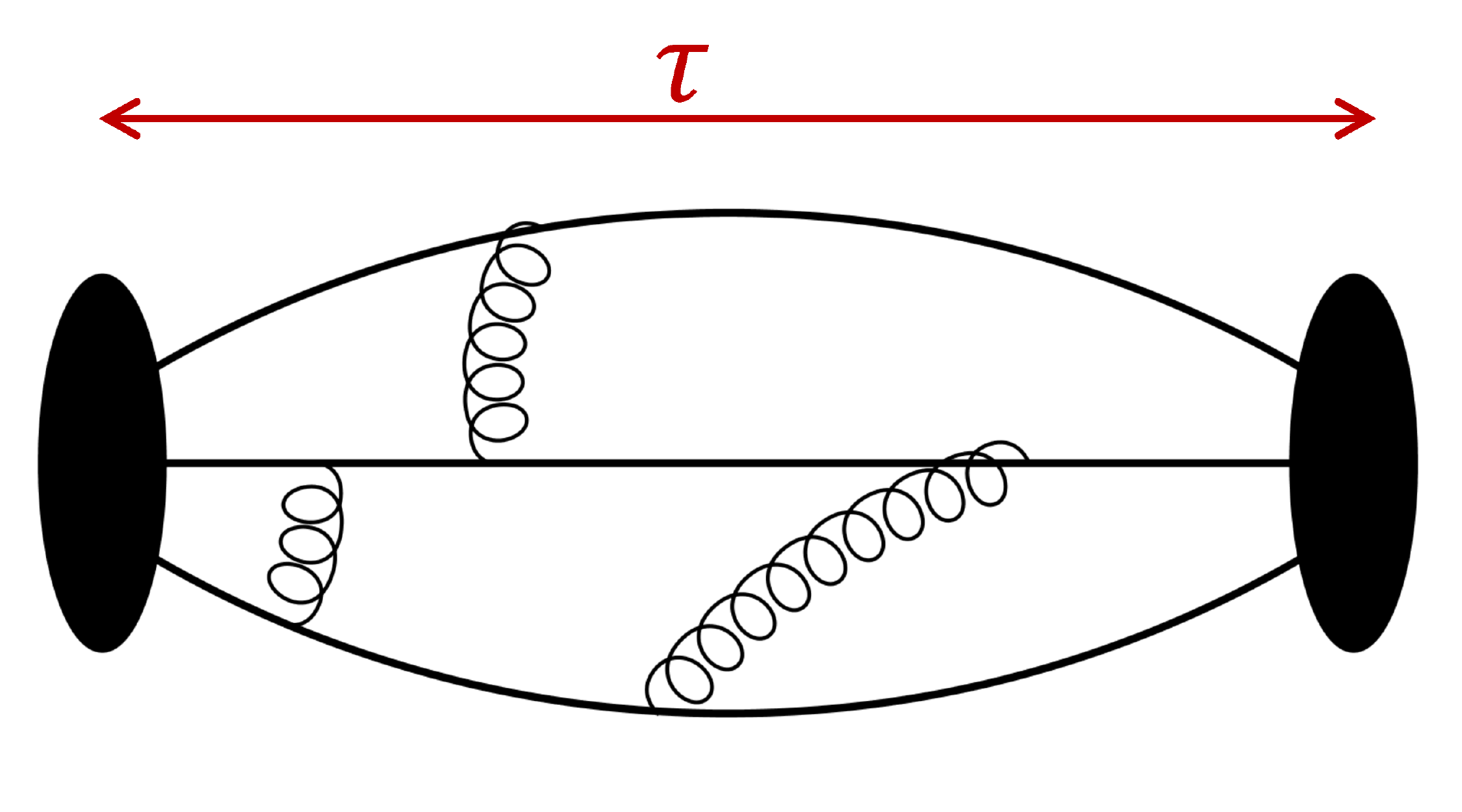} 
\includegraphics[width=0.32\linewidth]{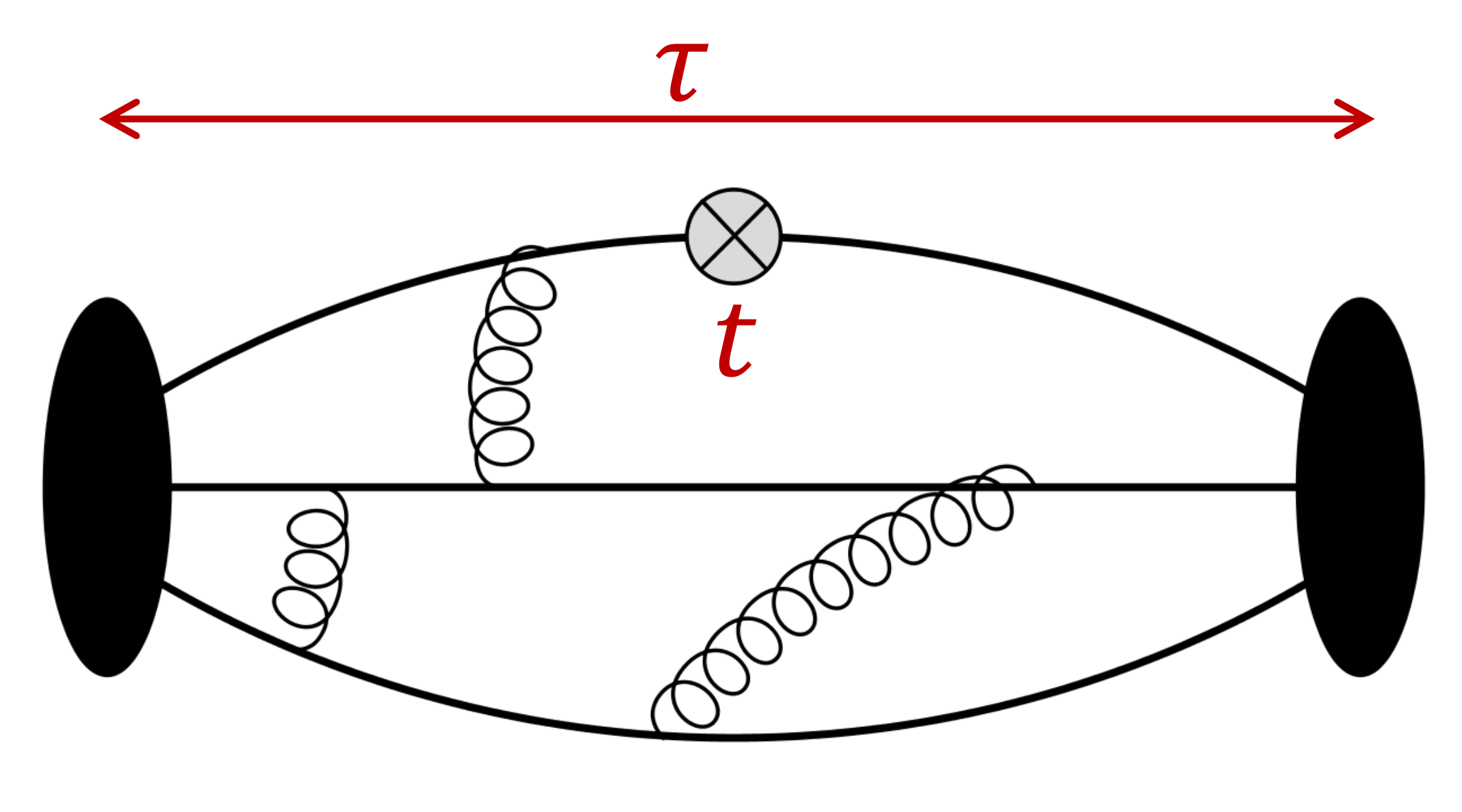}
\includegraphics[width=0.32\linewidth]{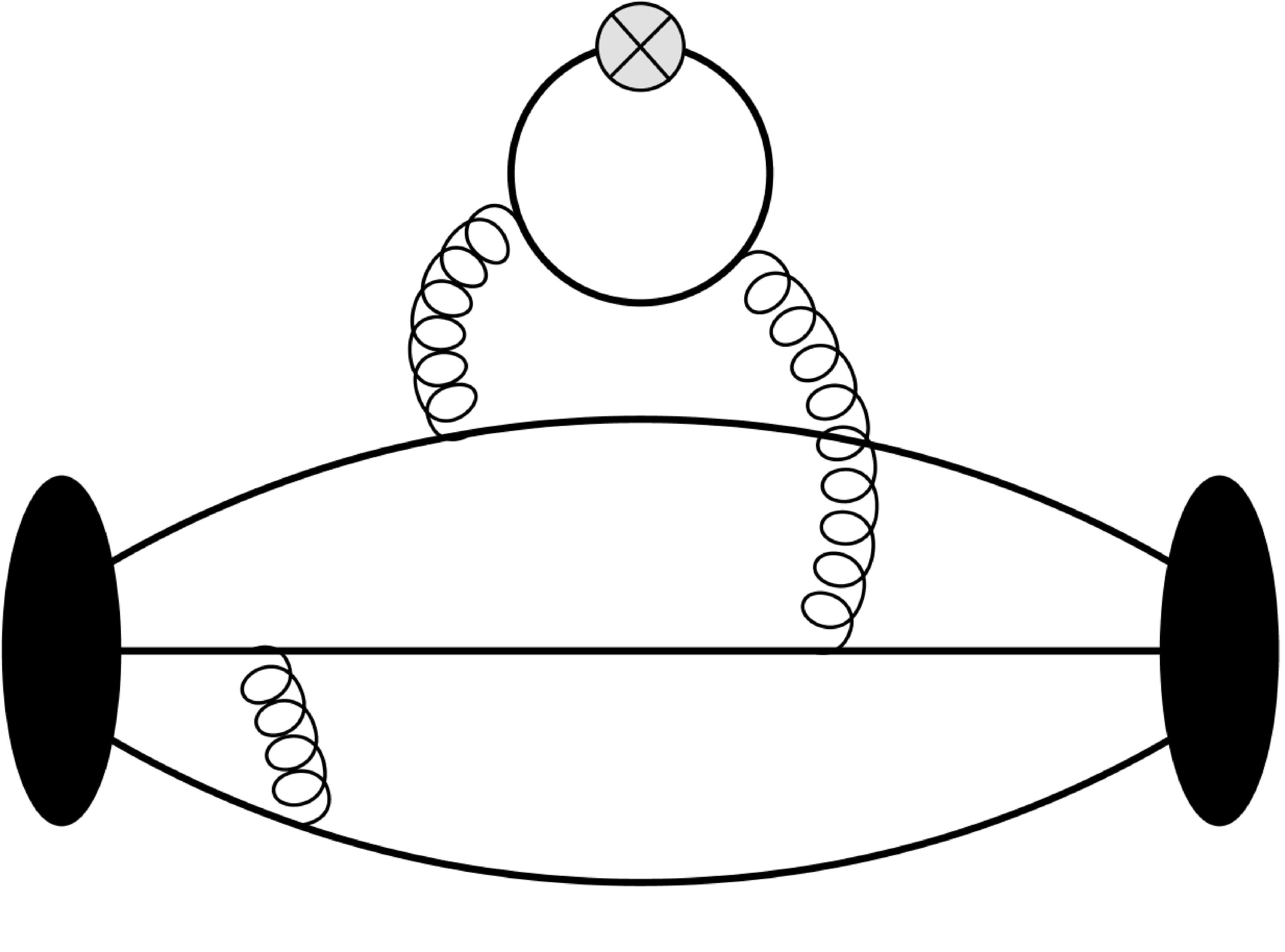}}
\caption{The two- and three-point correlation functions (illustrated by Feynman diagrams) that need to be
  calculated to extract the ground state nucleon matrix
  elements. (Left) the nucleon two-point function. (Middle) the
  connected three-point function with source-sink separation $\tau$
  and operator insertion time slice $t$. (Right) the disconnected
  three-point function with operator insertion at $t$. }
\label{fig:feynman}
\end{figure}

Usually, the quark-connected part of the three-point function
(corresponding to the plot in the centre of Fig.~\ref{fig:feynman}) is
computed via the so-called ``sequential propagator method'', which
uses the product of two quark propagators between the positions of the
initial and the final nucleons as a source term for another inversion
of the lattice Dirac operator. This implies that the position of the
sink timeslice is fixed at some chosen value. Varying the value of the
source-sink separation $\tau$ then requires the calculation of another
sequential propagator.

The evaluation of quark-disconnected contributions is computationally
more challenging as the disconnected loop~(which contains the operator
insertion, as illustrated in Fig.~\ref{fig:feynman} right) is needed
at all points on a particular timeslice or, in general, over the whole
lattice. The quark loop is computed stochastically and then correlated
with the nucleon two-point function before averaging this three-point
function over the ensemble of gauge configurations. The associated
statistical error, therefore, is a combination of that due to the
stochastic evaluation~(on each configuration) and that from the gauge
average. The number of stochastic sources employed on each
configuration is, typically, optimized to reduce the overall error for
a given computational cost. The statistical errors of the connected
contributions, in contrast, usually come only from the ensemble
average since they are often evaluated exactly on each configuration,
for a small number of source positions. If these positions are
well-separated in space and time, then each measurement is
statistically independent. The methodology applied for these
calculations and the variance reduction techniques are summarized in
Sec.~\ref{sec:technical}. By construction, arbitrary values of $\tau$
across the entire temporal extent of the lattice can be realized when
computing the quark-disconnected contribution, since the source-sink
separation is determined by the part of the diagram that corresponds
to the two-point nucleon correlator. However, in practice statistical
fluctuations of both the connected and disconnected contributions
increase sharply, so that the signal is lost in the statistical noise
for $\tau\gtrsim1.5$\,fm.

The lattice calculation is performed for a given number of quark
flavours and at a number of values of the lattice spacing $a$, the pion
mass $M_\pi$, and the lattice size, represented by $M_\pi L$. The
results need to be extrapolated to the physical point defined by
$a=0$, $M_\pi = 135$~MeV and $M_\pi L \to \infty$. This is done by
fitting the data simultaneously in these three variables using
a theoretically motivated ansatz. The ans\"atze used and the fitting
strategy are described in Sec.~\ref{sec:extrap}.

The procedure for rating the various calculations and the
criteria specific to this chapter are discussed in
Sec.~\ref{sec:rating}, which also includes a brief description of how
the final averages are constructed. The physics motivation for
computing the isovector charges, $g_{A,S,T}^{u-d}$, and the review of
the lattice results are presented in Sec.~\ref{sec:isovector}. This is
followed by a discussion of the relevance of the flavour diagonal
charges, $g_{A,S,T}^{u,d,s}$, and a presentation of the lattice results in Sec.~\ref{sec:FDcharges}.

\subsubsection{Technical aspects of the calculations of nucleon matrix elements\label{sec:technical}}

The calculation of $n$-point functions needed to extract nucleon matrix
elements requires making four essential choices. The first involves choosing between 
the suite of background gauge field ensembles one has access to. The range of lattice parameters 
should be large enough to facilitate the
extrapolation to the continuum and infinite-volume limits, and, ideally, the
evaluation at the physical pion mass taken to be~
$M_\pi=135$~MeV. Such ensembles have been generated with a variety of
discretization schemes for the gauge and fermion actions that have
different levels of improvement and preservation of continuum
symmetries. 
The actions employed at present
include (i) Wilson gauge with nonperturbatively improved
Sheikholeslami-Wohlert fermions~(nonperturbatively improved clover
fermions)~\cite{Khan:2006de,Bali:2012qs,Horsley:2013ayv,Capitani:2012gj,Bali:2014nma,Bali:2016lvx,Capitani:2017qpc}, (ii) Iwasaki gauge with nonperturbatively improved
clover fermions~\cite{Ishikawa:2009vc,Ishikawa:2018rew}, (iii) Iwasaki gauge with twisted-mass
fermions with a clover term~\cite{Abdel-Rehim:2015owa,Abdel-Rehim:2016won,Alexandrou:2017hac,Alexandrou:2017oeh,Alexandrou:2017qyt}, (iv) tadpole Symanzik improved
gauge with highly improved staggered quarks
(HISQ)~\cite{Bhattacharya:2013ehc,Bhattacharya:2015wna,Bhattacharya:2015esa,Bhattacharya:2016zcn,Berkowitz:2017gql,Gupta:2018lvp,Lin:2018obj,Gupta:2018qil,Chang:2018uxx}, (v) Iwasaki gauge with domain-wall fermions
(DW)~\cite{Yamazaki:2008py,Yamazaki:2009zq,Aoki:2010xg,Gong:2013vja,Yang:2015uis,Gong:2015iir,Liang:2018pis} and (vi) Iwasaki gauge with overlap fermions~\cite{Ohki:2008ff,Oksuzian:2012rzb,Yamanaka:2018uud}. For details of
the lattice actions, see the glossary in the Appendix A.1 of FLAG 19 \cite{Aoki:2019cca}.

The second choice is of the valence quark action. Here there are two
choices, to maintain a unitary formulation by choosing exactly the
same action as is used in the generation of gauge configurations or to
choose a different action and tune the quark masses to match the
pseudoscalar meson spectrum in the two theories.  Such mixed action
formulations are nonunitary but are expected to have the same
continuum limit as QCD. The reason for choosing a mixed action
approach is expediency. For example, the generation of 2+1+1 flavour
HISQ and 2+1 flavour DW ensembles with physical quark masses has been
possible even at the coarse lattice spacing of $a=0.15$~fm and
there are indications that cut-off effects are reasonably small. These
ensembles have been analyzed using clover-improved Wilson fermions, DW
and overlap fermions since the construction of baryon correlation
functions with definite spin and parity is much simpler compared to
staggered fermions.

The third choice is the combination of the algorithm for inverting the
Dirac matrix and variance reduction techniques. Efficient inversion
and variance reduction techniques are needed for the calculation of
nucleon correlation functions with high precision because the signal-to-noise ratio degrades exponentially as $e^{({\frac{3}{2}M_\pi-M_N}) \tau}$
with the source-sink separation $\tau$. Thus, the number of
measurements needed for high precision is much larger than in the
meson sector. Commonly used inversion algorithms include the
multigrid~\cite{Babich:2010qb} and the deflation-accelerated Krylov
solvers~\cite{Luscher:2007es}, which can handle linear systems with
large condition numbers very efficiently, thereby enabling
calculations of correlation functions at the physical pion mass.

The sampling of the path integral is limited by the number $N_{\rm
  conf}$ of gauge configurations generated. One requires sufficiently
large $N_{\rm conf}$ such that the phase space (for example, different
topological sectors) has been adequately sampled and all the
correlation functions satisfy the expected lattice symmetries such as
$C$, $P$, $T$, momentum and translation invariance. Thus, one needs
gauge field generation algorithms that give decorrelated large volume
configurations cost-effectively. On such large lattices, to reduce
errors one can exploit the fact that the volume is large enough to
allow multiple measurements of nucleon correlation functions that are
essentially statistically independent. Two other common variance
reduction techniques that reduce the cost of multiple measurements on
each configuration are: the truncated solver with bias correction
method~\cite{Bali:2009hu} and deflation of the Dirac matrix for
the low lying modes followed by sloppy solution with bias correction
for the residual matrix consisting predominately of the high frequency
modes~\cite{Bali:2009hu,Blum:2012uh}.

A number of other variance reduction methods are also being used and
developed. These include deflation with hierarchical probing for
disconnected diagrams~\cite{Stathopoulos:2013aci,Gambhir:2016jul}, the
coherent source sequential propagator
method~\cite{Bratt:2010jn,Yoon:2016dij}, low-mode
averaging~\cite{DeGrand:2004qw,Giusti:2004yp}, the hopping-parameter
expansion~\cite{Gupta:1989kx,Thron:1997iy} and
partitioning~\cite{Bernardson:1993he}~(also known as
dilution~\cite{Foley:2005ac}).

The final choice is of the interpolating operator used to create and
annihilate the nucleon state, and of the operator used to calculate
the matrix element. Along with the choice of the interpolating
operator (or operators if a variational method is used) one also
chooses a ``smearing'' of the source used to construct the quark
propagator. By tuning the width of the smearing, one can optimize the
spatial extent of the nucleon interpolating operator to reduce the
overlap with the excited states.  Two common smearing algorithms are
Gaussian (Wuppertal)~\cite{Gusken:1989ad} and
Jacobi~\cite{Alexandrou:1990dq} smearing.

Having made all the above choices, for which a reasonable recipe
exists, one calculates a statistical sample of correlation functions
from which the desired ground state nucleon matrix element is
extracted. Excited states, unfortunately, contribute significantly to
nucleon correlation functions in present studies.  To remove their
contributions, calculations are performed with multiple source-sink
separations $\tau$ and fits are made to the correlation functions using
their spectral decomposition as discussed in the next section.

\subsubsection{Controlling excited-state contamination\label{sec:ESC}}

Nucleon matrix elements are determined from a combination of two- and
three-point correlation functions. To be more specific, let
$B^\alpha(\vec{x},t)$ denote an interpolating operator for the
nucleon. Placing the initial state at time slice $t=0$, the two-point
correlation function of a nucleon with momentum $\vec{p}$ reads
\begin{equation}
\label{eq:nucl2pt}
   C_2(\vec{p};\tau) =
   \sum_{\vec{x},\vec{y}}\,e^{i\vec{p}\cdot(\vec{x}-\vec{y})}\,
   \mathbb{P}_{\beta\alpha}\,\left\langle 
   B^\alpha(\vec{x},\tau)\,\overline{B}^\beta(\vec{y},0) \right\rangle,
\end{equation}
where the projector $\mathbb{P}$ selects the polarization, and
$\alpha, \beta$ denote Dirac indices. The three-point function of two
nucleons and a quark bilinear operator $O_\Gamma$ is defined as
\begin{equation}
\label{eq:nucl3pt}
   C_3^\Gamma(\vec{q};t,\tau) = \sum_{\vec{x},\vec{y},\vec{z}}\,
   e^{ i\vec{p\,}^\prime\cdot(\vec{x}-\vec{z})}\,
   e^{-i\vec{p}\cdot(\vec{y}-\vec{z})}\,
   \mathbb{P}_{\beta\alpha}\,\left\langle 
   B^\alpha(\vec{x},\tau)\,O_\Gamma(\vec{z},t)\,
   \overline{B}^\beta(\vec{y},0) \right\rangle,
\end{equation}
where $\vec{p},\ \vec{p\,}^\prime$ denote the momenta of the nucleons at
the source and sink, respectively, and
$\vec{q}\equiv\vec{p\,}^\prime-\vec{p}$ is the momentum transfer. The
bilinear operator is inserted at time slice $t$, and $\tau$ denotes the
source-sink separation. Both $C_2$ and $C_3^\Gamma$ can be expressed
in terms of the nonperturbative quark propagators, $D^{-1}(y,x)$,
where $D$ denotes the lattice Dirac operator.

The framework for the analysis of excited-state contamination is based
on spectral decomposition. After inserting complete sets of
eigenstates of the transfer matrix, the expressions for the
correlators $C_2$ and $C_3^\Gamma$ read
\begin{eqnarray}
  \label{eq:specdec2pt}
  C_2(\vec{p};\tau) &=& \frac{1}{L^3}
  \sum_{n}\,\mathbb{P}_{\beta\alpha}\,\langle\Omega|B^\alpha|n\rangle
  \langle n|\overline{B}^\beta|\Omega\rangle\,
  e^{-E_n\tau}, \\ 
  \label{eq:specdec3pt}
  C_3^\Gamma(\vec{q};t,\tau) &=& \frac{1}{L^3}\sum_{n,m}\,
  \mathbb{P}_{\beta\alpha}\,
  \langle\Omega|B^\alpha|n\rangle\,
  \langle n|O_\Gamma|m\rangle\,
  \langle m|\overline{B}^\beta|\Omega\rangle\,
  e^{-E_n(\tau-t)}\,e^{-E_m t},
\end{eqnarray}
where $|\Omega\rangle$ denotes the vacuum state, and $E_n$ represents
the energy of the $n^{\rm th}$ eigenstate $|n\rangle$ in the nucleon
channel. Here we restrict the discussion to vanishing momentum
transfer, i.e., the forward limit $\vec{q}=0$, and label the ground state by $n=0$. The matrix
element of interest $g_\Gamma\equiv\langle0|O_\Gamma|0\rangle$ can,
for instance, be obtained from the asymptotic behaviour of the ratio
\begin{equation}
\label{eq:3ptratio}
  R_\Gamma(t,\tau) \equiv
  \frac{C_3^\Gamma(\vec{q}=0;t,\tau)}{C_2(\vec{p}=0;\tau)}
  \stackrel{t,(\tau-t)\to\infty}{\longrightarrow} g_{\Gamma} +
        {\rm O}(e^{-\Delta t},\,e^{-\Delta(\tau-t)},\,e^{-\Delta\tau}),
\end{equation}
where $\Delta\equiv E_1-E_0$ denotes the energy gap between the 
ground state and the first excitation. We also assume that the
bilinear operator $O_\Gamma$ is appropriately renormalized (see
Sec.~\ref{sec:renorm}).

Excited states with the same quantum numbers as the nucleon include
resonances such as a Roper-like state with a mass of about 1.5\,GeV,
or multi-particle states consisting of a nucleon and one or more pions
\cite{Tiburzi:2009zp,Bar:2017gqh}.  The latter can provide significant
contributions to the two- and three-point correlators in
Eqs.~(\ref{eq:nucl2pt}) and~(\ref{eq:nucl3pt}) or their
ratios~(\ref{eq:3ptratio}) as the pion mass approaches its physical
value. Ignoring the interactions between the individual hadrons, one
can easily identify the lowest-lying multi-particle states: they
include the $N\pi\pi$ state with all three particles at rest at
$\sim1.2$\,GeV, as well as $N\pi$ states with both hadrons having
nonzero and opposite momentum. Depending on the spatial box size $L$
in physical units (with the smallest nonzero momentum equal to $2\pi
/L$), there may be a dense spectrum of $N\pi$ states before the first
nucleon resonance is encountered. Corrections to nucleon correlation
functions due to the pion continuum have been studied using chiral
effective theory \cite{Tiburzi:2009zp, Bar:2017gqh, Bar:2016uoj,
  Bar:2016jof} and L\"uscher's finite-volume quantization condition
\cite{Hansen:2016qoz}.

The well-known noise problem of baryonic correlation functions implies
that the long-distance regime, $t, (\tau-t)\to\infty$, where the
correlators are dominated by the ground state, is difficult to
reach. Current lattice calculations of baryonic three-point functions
are typically confined to source-sink separations of
$\tau\lesssim1.5$\,fm, despite the availability of efficient noise
reduction methods. In view of the dense excitation spectrum
encountered in the nucleon channel, one has to demonstrate that the
contributions from excited states are sufficiently suppressed to
guarantee an unbiased determination of nucleon matrix elements. There
are several strategies to address this problem:
\begin{itemize}
\item Multi-state fits to correlator ratios or individual two- and
  three-point functions;
\item Three-point correlation functions summed over the operator
  insertion time $t$;
\item Increasing the projection of the interpolator $B^\alpha$ onto
  the ground state.
\end{itemize}
The first of the above methods includes excited state contributions
explicitly when fitting to the spectral decomposition of the
correlation functions, Eqs.~(\ref{eq:specdec2pt}, \ref{eq:specdec3pt})
or, alternatively, their ratio (see Eq.~(\ref{eq:3ptratio})). In its
simplest form, the resulting expression for $R_\Gamma$ includes the
contributions from the first excited state, i.e.,
\begin{equation}
\label{eq:multistate}
  R_\Gamma(t,\tau) = g_\Gamma +c_{01}\,e^{-{\Delta}t}
  +c_{10}\,e^{-{\Delta}(\tau-t)} +c_{11}\,e^{-{\Delta}\tau}+\ldots,
\end{equation}
where $c_{01}, c_{10}, c_{11}$ and $\Delta$ are treated as additional
parameters when fitting $R_\Gamma(t,\tau)$ simultaneously over
intervals in the source-sink separation $\tau$ and the operator
insertion timeslice~$t$. Multi-exponential fits become more difficult
to stabilize for a growing number of excited states, since an
increasing number of free parameters must be sufficiently constrained
by the data. Therefore, a high level of statistical precision at
several source-sink separations is required. One common way to address
this issue is to introduce Bayesian constraints, as described in
\cite{Yoon:2016jzj}. Alternatively, one may try to reduce the number
of free parameters, for instance, by determining the energy gap
$\Delta$ from nucleon two-point function and/or using a common gap for
several different nucleon matrix elements~\cite{Harris:2019bih}.

Ignoring the explicit contributions from excited states and fitting
$R_\Gamma(t,\tau)$ to a constant in $t$ for fixed $\tau$ amounts to
applying what is called the ``plateau method''. The name derives from
the ideal situation that sufficiently large source-sink separations $\tau$ 
can be realized, which would cause $R_\Gamma(t,\tau)$ to exhibit a
plateau in $t$ independent of $\tau$. The ability to control
excited-state contamination is rather limited in this approach, since
the only option is to check for consistency in the estimate of the
plateau as $\tau$ is varied. In view of the exponential degradation of
the statistical signal for increasing $\tau$, such stability checks
are difficult to perform reliably.

Summed operator insertions, originally proposed in
Ref.~\cite{Maiani:1987by}, have also emerged as a widely used method to
address the problem of excited-state contamination. One way to
implement this method \cite{Dong:1997xr,Capitani:2010sg} proceeds by
summing $R_\Gamma(t,\tau)$ over the insertion time $t$, resulting in the
correlator ratio $S_\Gamma(\tau)$,
\begin{equation}
  S_\Gamma(\tau) \equiv \sum_{t=a}^{\tau-a}\,R_\Gamma(t,\tau).
\end{equation}
The asymptotic behaviour of $S_\Gamma(\tau)$, including sub-leading
terms, for large source-sink separations $\tau$ can be easily derived
from the spectral decomposition of the correlators and is given by
\cite{Bulava:2011yz}
\begin{equation}
\label{eq:summation}
  S_\Gamma(\tau)\;\stackrel{\tau\gg1/\Delta}{\longrightarrow}\;
  K_\Gamma+(\tau-a)\,g_\Gamma+(\tau-a)\,e^{-\Delta\tau}d_\Gamma
  +e^{-\Delta\tau}f_\Gamma +\ldots,
\end{equation}
where $K_\Gamma$ is a constant, and the coefficients $d_\Gamma$ and
$f_\Gamma$ contain linear combinations of transition matrix elements
involving the ground and first excited states. Thus, the matrix
element of interest $g_\Gamma$ is obtained from the linear slope of
$S_\Gamma(\tau)$ with respect to the source-sink separation
$\tau$. While the leading corrections from excited states $e^{-\Delta\tau}$ are
smaller than those of the original ratio
$R_\Gamma(t,\tau)$ (see Eq.~(\ref{eq:3ptratio})), extracting the slope
from a linear fit to $S_\Gamma(\tau)$ typically results in relatively
large statistical errors. In principle, one could include the
contributions from excited states explicitly in the expression for
$S_\Gamma(\tau)$. However, in practice it is often difficult to
constrain an enlarged set of parameters reliably, in particular if one
cannot afford to determine $S_\Gamma(\tau)$ except for a handful of
source-sink separations.

The original summed operator insertion technique described in
Refs.~\cite{Maiani:1987by,Gusken:1988yi,Gusken:1989ad,Sommer:1989rf} avoids
the explicit summation over the operator insertion time $t$ at every
fixed value of $\tau$. Instead,
one replaces one of the quark propagators that appear in the
representation of the two-point correlation function $C_2(t)$ by a
``sequential'' propagator, according to
\begin{equation}
  D^{-1}(y,x) \to D_\Gamma^{-1}(y,x) = \sum_z D^{-1}(y,z)\Gamma
    D^{-1}(z,x).
\end{equation}
In this expression, the position $z\equiv(\vec{z},t)$ of the insertion
of the quark bilinear operator is implicitly summed over, by inverting
the lattice Dirac operator $D$ on the source field $\Gamma
D^{-1}(z,x)$. While this gives access to all source-sink separations
$0\leq\tau\leq T$, where $T$ is the temporal extent of the lattice,
the resulting correlator also contains contact terms, as well as
contributions from $\tau<t<T$ that must be controlled. This
method\footnote{In Ref.\,\cite{Bouchard:2016heu} it is shown that the
  method can be linked to the Feynman-Hellmann theorem. A direct
  implementation of the Feynman-Hellmann theorem by means of a
  modification of the lattice action is discussed and applied
  in Refs.~\cite{Chambers:2014qaa,Chambers:2015bka}.} has been adopted
recently by the CalLat collaboration in their calculation of the
isovector axial charge \cite{Berkowitz:2017gql,Chang:2018uxx}.

As in the case of explicitly summing over the operator insertion time,
the matrix element of interest is determined from the slope of the
summed correlator. For instance, in Ref.~\cite{Chang:2018uxx}, the
axial charge was determined from the summed three-point correlation
function, by fitting to its asymptotic
behaviour~\cite{Bouchard:2016heu} including sub-leading terms.

In practice, one often uses several methods simultaneously, e.g.,
  multi-state fits and the summation method based on
  Eq.~(\ref{eq:summation}), in order to check the robustness of the
result. All of the approaches for controlling excited-state
contributions proceed by fitting data obtained in a finite interval in
$\tau$ to a function that describes the approach to the asymptotic
behaviour derived from the spectral decomposition. Obviously, the
accessible values of $\tau$ must be large enough so that the model
function provides a good representation of the data that enter such a
fit. It is then reasonable to impose a lower threshold on $\tau$ above
which the fit model is deemed reliable. We will return to this issue
when explaining our quality criteria in Sec.~\ref{sec:rating}.

The third method for controlling excited-state contamination aims at
optimizing the projection onto the ground state in the two-point and
three-point correlation functions
\cite{Owen:2012ts,Bali:2014nma,Yoon:2016dij,Egerer:2018xgu}. The RQCD collaboration
has chosen to optimize the parameters in the Gaussian smearing
procedure, so that the overlap of the nucleon interpolating operator
onto the ground state is maximized \cite{Bali:2014nma}. In this way it
may be possible to use shorter source-sink separations without
incurring a bias due to excited states. 

The variational method, originally designed to provide detailed
information on energy levels of the ground and excited states in a
given channel \cite{Fox:1981xz,Michael:1985ne, Luscher:1990ck, Blossier:2009kd},
has also been adapted to the determination of hadron-to-hadron
transition elements \cite{Bulava:2011yz}. In the case of nucleon
matrix elements, the authors of Ref.\,\cite{Owen:2012ts} have employed
a basis of operators to construct interpolators that couple to
individual eigenstates in the nucleon channel. The method has produced
promising results when applied to calculations of the axial and other
forward matrix elements at a fixed value of the pion mass
\cite{Owen:2012ts,Dragos:2016rtx,Yoon:2016dij,Egerer:2018xgu}. However, a more
comprehensive study aimed at providing an estimate at the physical
point has, until now, not been performed.

The investigation of excited-state effects is an active subfield in
NME calculations, and many refinements and extensions have been
implemented since the previous edition of the FLAG report. For
instance, it has been shown that the previously observed failure of
the axial and pseudoscalar form factors to satisfy the PCAC relation linking them could be
avoided by including the enhanced contribution of $N\pi$
excitations, either by including additional information on the nucleon
excitation spectrum extracted from the three-point function of the
axial current \cite{Jang:2019vkm}, or with guidance from chiral effective field theory analyses
of nucleon three-point functions \cite{Bali:2019yiy}.

The variety of methods that are employed to address the problem of
excited-state contamination~(ESC) has greatly improved our
understanding of and control over excited-state effects in NME
calculations. However, there is still room for further improvement:
For instance, dedicated calculations of the excitation spectrum using
the variational method could replace the often rudimentary spectral
information gained from multi-state fits to the two- and three-point
functions used primarily for the determination of the matrix elements.
In general, the development of methods to explicitly include
multi-particle states, such as $N\pi$ and $N\pi\pi$ with appropriate
momentum configurations, coupled with the determination of the
associated (transition) matrix elements, is needed to significantly
enhance the precision of a variety of nucleon matrix elements.  Such
approaches would, to some extent, eliminate the need to extend the
source-sink separation $\tau$ into a regime that is currently
inaccessible due to the noise problem.

Since the ongoing efforts to study excited-state contamination are
producing deeper insights, we have decided to follow a more cautious
approach in the assessment of available NME calculations. This is
reflected in a modification of the quality criterion for excited-state
contamination that is described and discussed in
Sec.~\ref{sec:rating}.

\subsubsection{Renormalization and Symanzik improvement of local currents\label{sec:renorm}}

In this section we discuss the matching of the normalization of lattice operators to a
continuum reference scheme such as $\msbar$, and the application of
Symanzik improvement to remove $\cO(a)$ contributions. The relevant
operators for this review are the axial~($A_\mu$), tensor~($T_{\mu\nu}$) and
scalar~($S$) local operators of the form ${\cal
  O}_\Gamma=\overline{q}\Gamma q$, with $\Gamma=\gamma_\mu\gamma_5$,
$i\sigma_{\mu\nu}$ and $\mathbf{1}$, respectively, whose matrix
elements are evaluated in the forward limit. The general form for
renormalized operators in the isovector flavour combination, at a
scale $\mu$, reads
\begin{equation}
{\cal O}_\Gamma^{\msbar}(\mu) = Z_{\cal O}^{\msbar,{\rm Latt}}(\mu a,g^2)\left[{\cal O}_\Gamma(a) +ab_{\cal O}m{\cal O}_\Gamma(a)+ac_{\cal O}{\cal O}_\Gamma^{\rm imp}(a)\right] +\cO(a^2),\label{eq_op_improv}
\end{equation}
where $Z_{\cal O}^{\msbar,{\rm Latt}}(\mu a,g^2)$ denotes the
multiplicative renormalization factor determined in the chiral limit, $m \to 0$, 
and the second and third terms represent all possible quark-mass dependent
and independent Symanzik improvement terms,
respectively, at $\cO(a)$.\footnote{ Here $a(g^2)$ refers to the lattice spacing in
  the chiral limit, however, lattice simulations are usually carried
  out by fixing the value of $g^2$ while varying the quark
  masses. This means $a=a(\tilde{g}^2)$ where $\tilde{g}^2=g^2(1+b_g
  am_q)$~\cite{Jansen:1995ck,Luscher:1996sc} is the improved coupling
  that varies with the average sea-quark mass $m_q$. The difference
  between the $Z$ factors calculated with
  respect to $g^2$ and $\tilde{g}^2$ can effectively be absorbed into
  the $b_{\mathcal{O}}$
  coefficients~\cite{Bali:2016umi,Gerardin:2018kpy}.
}  The chiral properties of overlap,
domain-wall fermions~(with improvement up to $\cO(m_{\rm res}^n)$ where $m_{\rm res}$ is the residual mass) and twisted-mass
fermions~(at maximal twist~\cite{Frezzotti:2003ni,Frezzotti:2003xj})
mean that the $\cO(a)$ improvement terms are absent, while for
nonperturbatively improved Sheikholeslami-Wohlert-Wilson
(nonperturbatively-improved clover) fermions all terms appear in principle.  For
the operators of interest here there are several mass dependent terms
but at most one dimension-four ${\cal O}_\Gamma^{\rm imp}$; see,
e.g., Refs.~\cite{Capitani:2000xi,Bhattacharya:2005rb}.
However, the latter involve external derivatives whose 
corresponding matrix elements vanish in the forward limit.  Note that
no mention is made of staggered fermions as they are not, currently,
widely employed as valence quarks in nucleon matrix element calculations.

In order to illustrate the above remarks we consider the
renormalization and improvement of the isovector axial current. This
current has no anomalous dimension and hence the renormalization
factor, $Z_A=Z_A^{\msbar,{\rm Latt}}(g^2)$, is independent of the
scale.  The factor is usually computed nonperturbatively via the
axial Ward identity~\cite{Bochicchio:1985xa} or the Rome-Southampton
method~\cite{Martinelli:1994ty}~(see Sec.~A.3 of FLAG 19 \cite{Aoki:2019cca}
for details). In some studies, the ratio with the corresponding vector
renormalization factor, $Z_A/Z_V$, is determined for which some of the
systematics cancel. In this case, one constructs the combination $Z_A
g_A/(Z_V g_V)$, where $Z_V g_V=1$ and $g_A$ and $g_V$ are the lattice
forward matrix elements, to arrive at the renormalized axial
charge~\cite{Bhattacharya:2016zcn}. For domain-wall fermions the ratio is
employed in order to remove $\cO(am_{\rm res})$ terms and achieve
leading discretization effects starting at
$\cO(a^2)$~\cite{Blum:2014tka}. Thus, as mentioned above, $\cO(a)$
improvement terms are only present for nonperturbatively-improved
clover fermions.  For the axial current, Eq.~(\ref{eq_op_improv})
takes the explicit form,
\begin{equation}
A_\mu^{\msbar}(\mu) = Z_A^{\msbar,{\rm Latt}}(g^2)\left[
  \left(1+ ab_A m_{\rm val}+ 3a\tilde{b}_A m_{\rm sea}\right) A_\mu(a)+ac_A
  \partial_\mu P(a)\right] +\cO(a^2),
\end{equation}
where $m_{\rm val}$ and $m_{\rm sea}$ are the average valence- and sea-quark masses derived from the vector Ward identity~\cite{Bochicchio:1985xa,Luscher:1996sc,Bhattacharya:2005rb}, and $P$ is the
pseudoscalar operator $\overline{q}\gamma_5 q$. The matrix element of
the derivative term is equivalent to $q_\mu \langle
N(p^\prime)|P|N(p)\rangle$ and hence vanishes in the forward limit
when the momentum transfer $q_\mu=0$.  The improvement coefficients
$b_A$ and $\tilde{b}_A$ are known perturbatively for a variety of
gauge actions~\cite{Sint:1997jx,Taniguchi:1998pf,Capitani:2000xi} and
nonperturbatively for the tree-level Symanzik-improved gauge action for
$\Nf=2+1$~\cite{Korcyl:2016ugy}.

Turning to operators for individual quark flavours, these can mix 
under renormalization and the singlet and nonsinglet renormalization 
factors can differ.
For the axial current, such mixing occurs for all
fermion formulations just like in the continuum, where the singlet
combination acquires an anomalous dimension due to the $U_A(1)$
anomaly. The ratio of singlet to nonsinglet renormalization
factors, $r_{\cal O}=Z^{\rm s.}_{\cal O}/Z^{\rm n.s.}_{\cal O}$ for
${\cal O}=A$ differs from 1 at $\cO(\alpha_s^2)$ in perturbation
theory~(due to quark loops), suggesting that the mixing is a small
effect.  The nonperturbative determinations performed so far find $r_A\approx
1$~\cite{Alexandrou:2017hac,Green:2017keo}, supporting this.  For the
tensor current the disconnected diagram vanishes in the continuum due
to chirality and consequently on the lattice $r_T=1$ holds for overlap
and DW fermions~(assuming $m_{\rm res}=0$ for the latter). For
twisted-mass and clover fermions the mixing is expected to be small
with $r_T=1+\cO(\alpha_s^3)$~\cite{Constantinou:2016ieh} and this is
confirmed by the nonperturbative studies of
Refs.~\cite{Alexandrou:2017qyt,Bali:2017jyw}.

The scalar operators for the individual quark flavours,
$\overline{q}q$, are relevant not only for the corresponding scalar
charges, but also for the sigma terms $\sigma_q=m_q\langle
N|\overline{q}q|N\rangle$ when combined with the quark
masses~$m_q$. For overlap and DW fermions $r_S=1$, like in the
continuum and all $\overline{q}q$ renormalize multiplicatively with
the isovector $Z_S$. The latter is equal to the inverse of the mass
renormalization and hence $m_q\overline{q}q$ is renormalization
group~(RG) invariant. For twisted-mass fermions, through the use of
Osterwalder-Seiler valence fermions, the operators
$m_{ud}(\overline{u}u+\overline{d}d)$ and $m_s\overline{s}s$ are also
invariant~\cite{Dinter:2012tt}.\footnote{Note that for twisted-mass
  fermions the pseudoscalar renormalization factor is the relevant
  factor for the scalar operator. The isovector~(isosinglet) scalar
  current in the physical basis becomes the isosinglet~(isovector)
  pseudoscalar current in the twisted basis. Perturbatively
  $r_P=1+\cO(\alpha_s^3)$ and nonperturbative determinations have found
  $r_P\approx 1$~\cite{Alexandrou:2017qyt}.}  In contrast, the lack of
good chiral properties leads to significant mixing between quark
flavours for clover fermions. 
Nonperturbative determinations via the axial Ward
identity~\cite{Fritzsch:2012wq,Bali:2016lvx} have found the ratio
$r_S$ to be much larger than the perturbative expectation
$1+\cO(\alpha_s^2)$~\cite{Constantinou:2016ieh} may suggest.
While the sum
over the quark flavours which appear in the action $\sum^{\Nf}_q m_q
\overline{q}q$ is RG invariant, large cancellations between the
contributions from individual flavours can occur when evaluating,
e.g., the strange sigma term. Note that for twisted-mass and clover
fermions there is also an additive contribution $\propto
a^{-3}\mathbf{1}$~(or $\propto \mu a^{-2}\mathbf{1}$) to the scalar
operator.  This contribution is removed from the nucleon scalar matrix
elements by working with the subtracted current, $\overline{q}q -
\langle \overline{q}q\rangle$, where $\langle \overline{q}q\rangle$ is
the vacuum expectation value of the
current~\cite{Bhattacharya:2005rb}.

Symanzik improvement for the singlet currents follows the same pattern
as in the isovector case with $\cO(a)$ terms only appearing for
nonperturbatively-improved clover fermions. For the axial and tensor operators only
mass dependent terms are relevant in the forward limit while for the
scalar there is an additional gluonic operator ${\cal O}_S^{\rm
  imp}=\text{Tr}(F_{\mu\nu}F_{\mu\nu})$ with a coefficient of
$\cO(\alpha_s)$ in perturbation theory. When constructing the sigma terms
from the quark masses and the scalar operator, the improvement terms
remain and they must be included to remove all $\cO(a)$ effects for
nonperturbatively-improved clover fermions, see Ref.~\cite{Bhattacharya:2005rb} for a
discussion.

\subsubsection{Extrapolations in $a$, $M_\pi$ and $M_\pi L$\label{sec:extrap}}

To obtain physical results that can be used to compare to or make
predictions for experiment, all quantities must be extrapolated to the
continuum and infinite-volume limits. In general, either a chiral
extrapolation or interpolation must also be made to the physical pion
mass. These extrapolations need to be performed simultaneously since
discretization and finite-volume effects are themselves dependent upon
the pion mass. Furthermore, in practice it is not possible  to hold the
pion mass fixed while the lattice spacing is varied, as some variation
in $a$ occurs when tuning the quark masses at fixed gauge coupling. Thus, one 
performs a simultaneous extrapolation in all three 
variables using a theoretically motivated formula of the form,
\begin{eqnarray}
g(M_{\pi},a,L) = g_{\mathrm{phys}} + \delta_{M_{\pi}} + \delta_a + \delta_L \ ,
\end{eqnarray}
where $g_{\mathrm{phys}}$ is the desired extrapolated result, and
$\delta_{M_{\pi}}$, $\delta_a$, $\delta_L$ are the deviations due to the 
pion mass, the lattice spacing, and the volume, respectively. 
Below we outline the forms for each of these terms.

All observables discussed in this section are dimensionless, therefore
the extrapolation formulae may be parameterized by a set of
dimensionless variables:
\begin{eqnarray}
\epsilon_{\pi} = \frac{M_{\pi}}{\Lambda_{\chi}} \ , \qquad M_{\pi} L \ , \qquad \epsilon_a = \Lambda_a a \ .
\end{eqnarray}
Here, $\Lambda_{\chi} \sim 1$~GeV is a chiral symmetry breaking scale,
which, for example, can be set to $\Lambda_{\chi} = 4 \pi F_{\pi}$,
where $F_{\pi} = 92.2$~MeV is the pion decay constant, and $\Lambda_a$ is a
discretization scale, e.g., $\Lambda_a = \frac{1}{4\pi w_0}$,
where $w_0$ is a gradient-flow scale~\cite{Borsanyi:2012zs}.

Effective field theory methods may be used to determine the form of
each of these extrapolations. For the single nucleon charges, Heavy-Baryon $\chi$PT (HB$\chi$PT) is a common
choice~\cite{Jenkins:1990jv,Bernard:1995dp}, however, other variants, such as
unitarized~\cite{Truong:1988zp} or covariant~$\chi$PT~\cite{Becher:1999he,Fuchs:2003qc}, are also employed. Various 
formulations of HB$\chi$PT exist, including those
for two- and three-flavours, as well as with and without explicit
$\Delta$ baryon degrees of freedom. Two-flavour HB$\chi$PT is typically used
due to issues with convergence of the three-flavour 
theory~\cite{WalkerLoud:2008bp,Torok:2009dg,Ishikawa:2009vc,Jenkins:2009wv,WalkerLoud:2011ab}.
The convergence properties of all known formulations for baryon $\chi$PT,
even at the physical pion mass, have not been well-established.

To $\cO(\epsilon_{\pi}^2)$, the two-flavour chiral expansion for the nucleon charges is known to be of the form~\cite{Bernard:1992qa},
\begin{eqnarray}
\label{eq:chi}
g = g_0 + g_1 \epsilon_{\pi} + g_2 \epsilon_{\pi}^2 + \tilde{g}_2 \epsilon_{\pi}^2 \ln \left(\epsilon_{\pi}^2\right) \ ,
\end{eqnarray}
where $g_1=0$ for all charges $g$ except $g_S^{u,d}$. The
dimensionless coefficients $g_{0,1,2}, \tilde{g}_2$ are assumed to be
different for each of the different charges. The coefficients in front
of the logarithms, $\tilde{g}_2$, are known functions of the low-energy 
constants (LECs), and do not represent new, independent
LECs. Mixed action calculations will have further dependence upon the
mixed valence-sea pion mass, $m_{vs}$.

Given the potential difficulties with convergence of the chiral
expansion, known values of the $\tilde{g}_2$ in terms of LECs are not
typically used, but are left as free fit parameters. Furthermore, many
quantities have been found to display mild pion mass dependence, such
that Taylor expansions, i.e., neglecting logarithms in the above
expressions, are also often employed. The lack of a rigorously
established theoretical basis for the extrapolation in the pion mass
thus requires data close to the physical pion mass for obtaining high
precision extrapolated/interpolated results.

Discretization effects depend upon the lattice action used in a particular
calculation, and their form may be determined using the standard Symanzik
power counting. In general, for an unimproved action, the
corrections due to discretization effects $\delta_a$ include terms
of the form,
\begin{eqnarray}
\delta_a = c_1 \epsilon_a + c_2 \epsilon_a^2 + \cdots \ ,
\end{eqnarray}
where $c_{1,2}$ are dimensionless coefficients. Additional terms of
the form $\tilde{c}_n \left(\epsilon_{\pi} \epsilon_a\right)^n$, where
$n$ is an integer whose lowest value depends on the combined
discretization and chiral properties, will also appear. Improved
actions systematically remove correction terms, e.g., an
$\cO(a)$-improved action, combined with a similarly 
improved operator, will contain terms in the extrapolation ansatz beginning
at $\epsilon_a^2$ (see Sec.~\ref{sec:renorm}).

Finite volume corrections $\delta_L$ may be determined in the usual
way from effective field theory, by replacing loop integrals over
continuous momenta with discrete sums. Finite volume effects therefore
introduce no new undetermined parameters to the extrapolation. For
example, at next-to-leading order, and neglecting contributions from
intermediate $\Delta$ baryons, the finite-volume corrections for the
axial charge in two-flavour HB$\chi$PT take the
form~\cite{Beane:2004rf},
\begin{eqnarray}
\delta_L &\equiv& g_{A}(L) - g_{A}(\infty) = \frac{8}{3} \epsilon_{\pi}^2 \left[ g_0^3 F_1\left(M_{\pi}L\right) + g_0 F_3\left(M_{\pi} L\right)\right] \ ,
\label{eq:FVdeltaL}
\end{eqnarray}
where
\begin{eqnarray}
F_1\left(mL\right) &=& \sum_{\mathbf{n\neq 0}}\left[K_0\left(mL|\mathbf{n}|\right) - \frac{K_1\left(mL|\mathbf{n}|\right)}{mL|\mathbf{n}|}\right] \,, \cr
F_3\left(mL\right) &=& -\frac{3}{2} \sum_{\mathbf{n} \neq 0} \frac{K_1\left(mL|\mathbf{n}|\right)}{mL|\mathbf{n}|} \ ,
\end{eqnarray}
and $K_{\nu}(z)$ are the modified Bessel functions of the second
kind. Some extrapolations are performed using the form for
asymptotically large $M_{\pi} L$,
\begin{eqnarray}\label{eq:Vasymp}
K_0(z) \to \frac{e^{-z}}{\sqrt{z}} \ ,
\end{eqnarray}
and neglecting contributions due to $K_1$. Care must, however, be
taken to establish that these corrections are negligible for all
included values of $M_{\pi} L$. The numerical coefficients, for
example, $8/3$ in Eq.~\eqref{eq:FVdeltaL}, are often taken to be
additional free fit parameters, due to the question of convergence of
the theory discussed above.

Given the lack of knowledge about the convergence of the expansions and
the resulting plethora of possibilities for extrapolation models at
differing orders, it is important to include statistical tests of model selection 
for a given set of data. Bayesian model averaging
\cite{doi:10.1080/01621459.1995.10476572} or use of the Akaike
Information Criterion \cite{1100705} are common choices which penalize
over-parameterized models.

\subsection{Quality criteria for nucleon matrix elements and
  averaging procedure \label{sec:rating}}

There are two specific issues that call for a modification and
extension of the FLAG quality criteria listed in
Sec.~\ref{sec:qualcrit}. The first concerns the rating of the
chiral extrapolation: The FLAG criteria reflect the ability of
$\chi$PT to provide accurate descriptions of the pion mass dependence
of observables. Clearly, this ability is linked to the convergence
properties of $\chi$PT in a particular mass range. Quantities
extracted from nucleon matrix elements are extrapolated to the
physical pion mass using some variant of baryonic $\chi$PT, whose
convergence is not well established as compared to the mesonic
sector. Therefore, we have opted for stricter quality criteria, 
200 MeV $\le M_{\pi,\mathrm{min}} \le 300$ MeV, for a 
green circle in the chiral extrapolation of nucleon matrix elements, i.e.,\\

\noindent
\good \hspace{0.2cm} $M_{\pi,\mathrm{min}}< 200$ MeV with three or more pion masses used in the extrapolation\\ 
\noindent \rule{0.05em}{0em}\hspace{0.45cm} \underline{or}
two values of $M_\pi$ with one lying within 10 MeV of 135 MeV (the physical neutral\\
\noindent \rule{0.05em}{0em}\hspace{0.45cm} pion mass) and the other one below 200 MeV\\
\rule{0.05em}{0em}\soso \hspace{0.2cm} 200 MeV $\le M_{\pi,\mathrm{min}}
\le 300$ MeV with three or more pion masses used in the extrapolation;\\
\noindent \rule{0.05em}{0em}\hspace{0.45cm} \underline{or}
two values of $M_\pi$ with $M_{\pi,\mathrm{min}}< 200$ MeV;\\
\noindent \rule{0.05em}{0em}\hspace{0.45cm} \underline{or} a single
value of $M_\pi$ lying within 10 MeV of 135\,MeV  (the physical neutral pion mass)\\
\rule{0.05em}{0em}\bad \hspace{0.2cm} Otherwise \\

In Sec.~\ref{sec:ESC} we have discussed that insufficient control
over excited-state contributions, arising from the noise problem in 
baryonic correlation functions, may lead to a systematic bias in the
determination of nucleon matrix elements. We therefore introduce an
additional criterion that rates the efforts to suppress excited-state
contamination in the final result.
As described in Sec.~\ref{sec:ESC}, the applied methodology to control
excited-state contamination is quite diverse. Since a broad consensus
on the question which procedures should be followed has yet to emerge,
our criterion is expressed in terms of simulation parameters that can
be straightforwardly extracted on the basis of publications.
Furthermore, the criterion must also be readily applicable to a
variety of different local operators whose matrix elements are
discussed in this chapter. These requirements are satisfied by the
source-sink separation $\tau$, i.e., the
Euclidean distance between the initial and final nucleons.
The discussion at the end of Sec.~\ref{sec:ESC} shows that there is
room for improvement in the ability to control excited-state
contamination. Hence, we have reverted to a binary system, based on
the range of source-sink separations of a given calculations. While we
do not award the highest category---a green star---in this edition,
we stress that the adoption of the modified ESC criterion has not led
to a situation where calculations that were previously rated with a
green star are now excluded from FLAG averages. The rating scale
concerning control over excited-state contributions is thus \\

\noindent
\soso \hspace{0.2cm} Three or more source-sink separations $\tau$, at
least two of which must be above 1.0 fm. \\ 
\rule{0.05em}{0em}\bad \hspace{0.2cm} Otherwise \\

We will continue to monitor the situation concerning excited-state
contamination and, if necessary, adapt the criteria further in future
editions of the FLAG report.

As explained in Sec.~\ref{sec:qualcrit}, FLAG averages are
distinguished by the sea-quark content. Hence, for a given
configuration of the quark sea (i.e., for $N_f=2$, $2+1$, $2+1+1$, or $1+1+1+1$), we
first identify those calculations that pass the FLAG and the additional quality criteria
defined in this section, i.e. excluding any calculation that has a red tag
in one or more of the categories. We then add statistical and
systematic errors in quadrature and perform a weighted average. If the
fit is of bad quality (i.e., if $\chi^2_{\rm min}/{\rm dof}>1$), the
errors of the input quantities are scaled by $\sqrt{\chi^2/{\rm dof}}$. In the following step, correlations among different
calculations are taken into account in the error estimate by applying
Schmelling's procedure~\cite{Schmelling:1994pz}.

\subsection{Isovector charges\label{sec:isovector}}

The axial, scalar and tensor isovector charges are needed to interpret
the results of many experiments and phenomena mediated by weak
interactions, including probes of new physics.  The most natural process from
which isovector charges can be measured is neutron beta decay ($n \to
p^{+} e^{-} \overline{\nu}_e$).  At the quark level, this process
occurs when a down quark in a neutron transforms into an up quark due
to weak interactions, in particular due to the axial current
interaction. While scalar and tensor currents have not been observed
in nature, effective scalar and tensor interactions arise in the
SM due to loop effects. At the TeV and higher scales,
contributions to these three currents could arise due to new
interactions and/or loop effects in BSM theories. These super-weak
corrections to standard weak decays can be probed through high
precision measurements of the neutron decay distribution by examining
deviations from SM predictions as described in
Ref.~\cite{Bhattacharya:2011qm}. The lattice-QCD methodology for the
calculation of isovector charges is well-established, and the control
over statistical and systematic uncertainties is becoming robust.

The axial charge $g_A^{u-d}$ is an important parameter that
encapsulates the strength of weak interactions of nucleons. It enters
in many analyses of nucleon structure and of SM and BSM 
physics. For example, it enters in (i) the
extraction of $V_{ud}$ and tests of the unitarity of the
Cabibbo-Kobayashi-Maskawa (CKM) matrix; (ii) the analysis of
neutrinoless double-beta decay, (iii) neutrino-nucleus quasi-elastic 
scattering cross-section; (iv) the rate of proton-proton fusion,
the first step in the thermonuclear reaction chains that power
low-mass hydrogen-burning stars like the Sun; (v) solar and reactor
neutrino fluxes; (vi) muon capture rates, etc.. The current best
determination of the ratio of the axial to the vector charge,
$g_A/g_V$, comes from measurement of neutron beta decay using
polarized ultracold neutrons by the UCNA collaboration,
$1.2772(20)$~\cite{Mendenhall:2012tz,Brown:2017mhw}, and by PERKEO II,
$1.2761{}^{+14}_{-17}$~\cite{Mund:2012fq}. Note that, in the SM,
$g_V=1$ up to second-order corrections in isospin
breaking~\cite{Ademollo:1964sr,Donoghue:1990ti} as a result of the
conservation of the vector current.  Given the accuracy with which
$g_A^{u-d}$ has been measured in experiments, the goal of lattice-QCD
calculations is to calculate it directly with $\cO(1\%)$
accuracy.

Isovector scalar or tensor interactions contribute to the
helicity-flip parameters, called $b$ and $B$, in the neutron decay
distribution. By combining the calculation of the scalar and
tensor charges with the measurements of $b$ and $B$, one can put 
constraints on novel scalar and tensor interactions at the TeV scale
as described in Ref.~\cite{Bhattacharya:2011qm}.  To optimally
bound such scalar and tensor interactions using measurements of
$b$ and $B$ parameters in planned experiments targeting $10^{-3}$ 
precision~\cite{abBA,WilburnUCNB,Pocanic:2008pu}, 
we need to determine $g_S^{u-d}$ and $g_T^{u-d}$ at the $10\%$ level as 
explained in Refs.~\cite{Bhattacharya:2011qm,Bhattacharya:2016zcn}. 
Future higher-precision measurements of $b$ and $B$ would require
correspondingly higher-precision calculations of the matrix elements
to place even more stringent bounds on these couplings at the TeV-scale.  

One can estimate $g_S^{u-d}$ using the conserved vector current (CVC)
relation, $g_S/g_V = (M_{\mathrm{neutron}}-M_{\mathrm{proton}})^{\rm QCD}/ (m_d-m_u)^{\rm  QCD}$, as done by 
Gonzalez-Alonso {\it et al.}~\cite{Gonzalez-Alonso:2013ura}. In
their analysis, they took estimates of the two mass differences on the
right-hand side from the global lattice-QCD data~\cite{Aoki:2013ldr} and obtained 
$g_S^{u-d}=1.02(8)(7)$. 

The tensor charge $g_T^{u-d}$ can be extracted experimentally from
semi-inclusive deep-inelastic scattering (SIDIS)
data~\cite{Dudek:2012vr,Ye:2016prn,Lin:2017stx,Radici:2018iag}. A sample of these 
phenomenological estimates is shown in Fig.~\ref{fig:gt}, and the noteworthy feature is 
that the current uncertainty in these phenomenological estimates is large.

\subsubsection{Results for $g_A^{u-d}$\label{sec:gA-IV}}

\begin{table}[t!]
\begin{center}
\mbox{} \\[3.0cm]
\footnotesize
\begin{tabular*}{\textwidth}[l]{l @{\extracolsep{\fill}} r l l l l l l l l }
Collaboration & Ref. & $\Nf$ & 
\hspace{0.15cm}\begin{rotate}{60}{publication status}\end{rotate}\hspace{-0.15cm} &
\hspace{0.15cm}\begin{rotate}{60}{continuum extrapolation}\end{rotate}\hspace{-0.15cm} &
\hspace{0.15cm}\begin{rotate}{60}{chiral extrapolation}\end{rotate}\hspace{-0.15cm}&
\hspace{0.15cm}\begin{rotate}{60}{finite volume}\end{rotate}\hspace{-0.15cm}&
\hspace{0.15cm}\begin{rotate}{60}{renormalization}\end{rotate}\hspace{-0.15cm}  &
\hspace{0.15cm}\begin{rotate}{60}{excited states}\end{rotate}\hspace{-0.15cm}  &
$g^{u-d}_A$\\
&&&&&&&&& \\[-0.1cm]
\hline
\hline
&&&&&&&& \\[-0.1cm]

CalLat 19 & \cite{Walker-Loud:2019cif} & 2+1+1 & \rC & \soso & \good & \good & \good & \soso & 1.2642(93) \\[0.5ex]
ETM 19 & \cite{Alexandrou:2019brg} & 2+1+1 & \gA & \bad & \soso & \good & \good & \soso & 1.286(23) \\[0.5ex]
PNDME 18$^a$ & \cite{Gupta:2018qil} & 2+1+1 & \gA & \good$^\ddag$ & \good & \good & \good & \soso & 1.218(25)(30) \\[0.5ex]
CalLat 18 & \cite{Chang:2018uxx} & 2+1+1 & \gA & \soso & \good & \good & \good & \soso & 1.271(10)(7) \\[0.5ex]
CalLat 17 & \cite{Berkowitz:2017gql} & 2+1+1 & \oP & \soso & \good & \good & \good & \soso & 1.278(21)(26) \\[0.5ex]
PNDME 16$^a$ & \cite{Bhattacharya:2016zcn} & 2+1+1 & \gA & \soso$^\ddag$ & \good & \good & \good & \soso & 1.195(33)(20) \\[0.5ex]
\\[-0.1ex]\hline\\[0.2ex]
NME 21$^a$ & \cite{Park:2021ypf} & 2+1 & \oP & \soso$^\ddag$ & \good & \good & \good & \soso & 1.31(6)(5) \\[0.5ex]
LHPC 19 & \cite{Hasan:2019noy} & 2+1 & \gA & \bad$^\ddag$ & \good & \good & \good & \soso & 1.265(49) \\[0.5ex]
Mainz 19 & \cite{Harris:2019bih} & 2+1 & \gA & \good & \soso & \good & \good & \soso & 1.242(25)($^{+0}_{-0.030}$) \\[0.5ex]
PACS 18A & \cite{Shintani:2018ozy} & 2+1 & \gA & \bad & \good & \good & \good & \soso & 1.273(24)(5)(9) \\[0.5ex]
PACS 18 & \cite{Ishikawa:2018rew} & 2+1 & \gA & \bad & \bad & \good & \good & \bad & 1.163(75)(14) \\[0.5ex]
$\chi$QCD 18 & \cite{Liang:2018pis} & 2+1 & \gA & \soso & \good & \good & \good & \soso & 1.254(16)(30)$^\$$ \\[0.5ex]
JLQCD 18 & \cite{Yamanaka:2018uud} & 2+1 & \gA & \bad & \soso & \soso & \good & \soso & 1.123(28)(29)(90) \\[0.5ex]
LHPC 12A$^b$ & \cite{Green:2012ud} & 2+1 & \gA & \bad$^\ddag$ & \good & \good & \good & \soso & 0.97(8) \\[0.5ex]
LHPC 10 & \cite{Bratt:2010jn} & 2+1 & \gA & \bad & \soso & \bad & \good & \bad & 1.21(17) \\[0.5ex]
RBC/UKQCD 09B & \cite{Yamazaki:2009zq} & 2+1 & \gA & \bad & \bad & \soso & \good & \bad & 1.19(6)(4) \\[0.5ex]
RBC/UKQCD 08B & \cite{Yamazaki:2008py} & 2+1 & \gA & \bad & \bad & \soso & \good & \bad & 1.20(6)(4) \\[0.5ex]
LHPC 05 & \cite{Edwards:2005ym} & 2+1 & \gA & \bad & \bad & \good & \good & \bad & 1.226(84) \\[0.5ex]
\\[-0.1ex]\hline\\[0.2ex]
Mainz 17 & \cite{Capitani:2017qpc} & 2 & \gA & \good & \good & \good & \good & \bad & 1.278(68)($^{+0}_{-0.087}$) \\[0.5ex]
ETM 17B & \cite{Alexandrou:2017hac} & 2 & \gA & \bad & \soso & \soso & \good & \soso & 1.212(33)(22) \\[0.5ex]
ETM 15D & \cite{Abdel-Rehim:2015owa} & 2 & \gA & \bad & \soso & \soso & \good & \soso & 1.242(57) \\[0.5ex]
RQCD 14 & \cite{Bali:2014nma} & 2 & \gA & \soso & \good & \good & \good & \bad & 1.280(44)(46) \\[0.5ex]
QCDSF 13 & \cite{Horsley:2013ayv} & 2 & \gA & \soso & \good & \bad & \good & \bad & 1.29(5)(3) \\[0.5ex]
Mainz 12 & \cite{Capitani:2012gj} & 2 & \gA & \good & \soso & \soso & \good & \bad & 1.233(63)($^{+0.035}_{-0.060}$) \\[0.5ex]
RBC 08 & \cite{Lin:2008uz} & 2 & \gA & \bad & \bad & \bad & \good & \bad & 1.23(12) \\[0.5ex]
QCDSF 06 & \cite{Khan:2006de} & 2 & \gA & \soso & \bad & \bad & \good & \bad & 1.31(9)(7) \\[0.5ex]
&&&&&&&& \\[-0.1cm]
\hline
\hline
\end{tabular*}
\begin{minipage}{\linewidth}
{\footnotesize 
\begin{itemize}
\item[$^a$] The improvement coefficient in the valence quark action is
  set to its tadpole-improved tree-level value. \\[-5mm]
\item[$^b$] The quark action is tree-level improved. \\[-5mm]
\item[$^\ddag$]The rating takes into account that the action is not fully O($a$) improved by requiring an additional lattice spacing. \\[-5mm]\item[$^\$$] For this partially quenched analysis the criteria are applied to the unitary points.
\end{itemize}
}
\end{minipage}
\caption{Overview of results for $ g^{u-d}_A$. \label{tab:ga}}
\end{center}
\end{table}

Calculations of the isovector axial charge have a long history, as can
be seen from the compilation given in Tab.\,\ref{tab:ga} and plotted
in Fig.~\ref{fig:ga}.
The issue of excited-state contamination received little if any
attention before 2010. As a consequence, the range of source-sink
separations employed in many of the early calculations prior to that
year was rather limited, offering little control over this important
systematic effect. This concerns, in particular, the calculations by LHPC~05
\cite{Edwards:2005ym}, LHPC~10 \cite{Bratt:2010jn}, RBC~08
\cite{Lin:2008uz}, RBC/UKQCD~08B \cite{Yamazaki:2008py}, 
RBC/UKQCD~09B \cite{Yamazaki:2009zq} and
QCDSF~06 \cite{Khan:2006de}.
Since the last edition of the FLAG report, no new results in
two-flavour QCD have been published. An exception is the calculation
ETM 19~\cite{Alexandrou:2019brg}, which reanalyzed two ensembles
with $N_f=2$ around the physical pion mass to study finite-volume
effects, while the main result is quoted from a calculation with
$N_f=2+1+1$. These two-flavour calculations still do not 
qualify for inclusion in the FLAG average (see Table~\ref{tab:ga}). We thus
refrain from providing a detailed discussion of the results in Refs.
\cite{Khan:2006de,Lin:2008uz,Capitani:2012gj,Horsley:2013ayv,Bali:2014nma,Abdel-Rehim:2015owa,Alexandrou:2017hac,Capitani:2017qpc} and refer the reader to the corresponding chapter in
the previous edition of the FLAG report.

Estimates for the axial charge with $N_f=2+1$ have been published by
many collaborations, i.e., LHPC
\cite{Edwards:2005ym, Bratt:2010jn, Green:2012ud, Hasan:2019noy},
RBC/UKQCD \cite{Yamazaki:2008py, Yamazaki:2009zq}, JLQCD~18
\cite{Yamanaka:2018uud}, $\chi$QCD~18 \cite{Liang:2018pis}, PACS~18/PACS~18A
\cite{Ishikawa:2018rew,Shintani:2018ozy}, Mainz 19 \cite{Harris:2019bih} (superseding
the previously listed result in~\cite{Ottnad:2018fri}) and NME 21
\cite{Park:2021ypf}.

The calculations in LHPC~05 \cite{Edwards:2005ym} and LHPC~10
\cite{Bratt:2010jn} were based on a mixed-action setup,
combining domain-wall fermions in the valence sector with staggered
(asqtad) gauge ensembles generated by MILC. Although the dependence of
the results on the source-sink separation was studied to some extent
in LHPC~10, excited-state effects are not sufficiently controlled
according to our quality criteria described in Sec.\,\ref{sec:rating}.
A different discretization of the quark action was used in their later
studies (LHPC~12A \cite{Green:2012ud} and LHPC 19
\cite{Hasan:2019noy}), employing tree-level improved Wilson fermions
with smeared gauge links, both in the sea and valence sectors. While
this setup does not realize full $\cO(a)$ improvement, it was found that
smeared gauge links reduce the leading discretization effects of
$\cO(a)$ substantially.
The most recent publication (LHPC 19) is based on two ensembles
within 1.5\% of the physical pion, at two different values of the
lattice spacing. Results for $g_A^{u-d}$ were determined using the
summation and ratio methods, with and without including the first
excitation in the fit. LHPC quotes the result from the finer lattice
spacing, with an error that covers the spread of uncertainties on both
ensembles.

The RBC/UKQCD collaboration has employed $N_f=2+1$ flavours of domain-wall fermions. The results quoted in
RBC/UKQCD~08B \cite{Yamazaki:2008py} and RBC/UKQCD~09B
\cite{Yamazaki:2009zq} were obtained at relatively heavy pion masses
at a single value of the lattice spacing, with only limited control
over excited-state effects. While systematic investigation of different
source-sink separations has been recently performed on two
ensembles at the same lattice spacing and pion masses of 250 and
170\,MeV, respectively \cite{Abramczyk:2019fnf}, an estimate for
$g_A^{u-d}$ at the physical point has not been quoted.

The JLQCD collaboration (JLQCD~18 \cite{Yamanaka:2018uud}) has
performed a calculation using $N_f=2+1$ flavours of overlap fermions
and the Iwasaki gauge action. Owing to the large numerical cost of
overlap fermions, which preserve exact chiral symmetry at nonzero
lattice spacing, they have only simulated four light quark masses with
$290 < M_\pi < 540$~MeV and at a single lattice spacing so far. Their
simultaneous fit to the data for the correlator ratio $R_A(t,\tau)$
computed at six values of $\tau$ to a constant, gives a low value for
$g_A^{u-d}$ at the physical point. Overlap valence quarks were also
used by the $\chi$QCD collaboration in their study of various nucleon
matrix elements ($\chi$QCD~18 \cite{Liang:2018pis}), utilizing the
gauge ensembles generated by RBC/UKQCD with domain-wall fermions. The
quoted estimate for the axial charge was obtained from a combination
of two-state fits and the summation method, applied over a range of
source-sink separations.

Calculations with $N_f=2+1$ flavours of $\cO(a)$ improved Wilson
fermions have been performed by PACS, the Mainz group and NME.
The calculations by the PACS collaboration
(PACS~18 \cite{Ishikawa:2018rew} and PACS~18A
\cite{Shintani:2018ozy}) were performed on very
large volumes (8.2\,fm and 10.8\,fm, respectively) at or near the
physical pion mass. In PACS~18A, the ratio method without including
excited states was used to determine the isovector axial charge, which
was found to be in good agreement with the experimental value.
However, only a single lattice spacing was used in PACS~18 and
PACS~18A, so that these calculations lack control over
discretization effects.
The Mainz group (Mainz 19 \cite{Harris:2019bih}) has presented
results for the axial charge, obtained by performing two-state fits to
six different nucleon matrix elements (including the scalar and tensor
charges), assuming that the mass gap to the excited state can be more
reliably constrained in this way. Up to six source-sink separations
per ensemble have been studied. The final results are obtained from a
combined chiral, continuum and finite-volume extrapolation.
The NME collaboration (NME 21 \cite{Park:2021ypf}) has recently
published the results from a calculation of various nucleon form
factors and charges. Results were obtained from multi-state fits,
using up to four (three) states in the two-point (three-point)
correlation functions. In order to describe and control excited-state
effects, $N\pi$ and $N\pi\pi$ states with different relative momenta
were included in the analysis. The preferred result for $g_A^{u-d}$
was obtained from the axial form factor $G_A(Q^2)$ extrapolated to
$Q^2=0$.

Three groups, PNDME, CalLat and ETMC, have published results for $N_f=2+1+1$,
i.e. PNDME~16 \cite{Bhattacharya:2016zcn}, PNDME~18
\cite{Gupta:2018qil}, CalLat~17 \cite{Berkowitz:2017gql}
CalLat~18 \cite{Chang:2018uxx}, CalLat 19 \cite{Walker-Loud:2019cif}.
PNDME and CalLat share the staggered (HISQ) gauge ensembles generated
by the MILC collaboration, but employ different discretizations in the
valence quark sector: PNDME use $\cO(a)$ improved Wilson fermions with
the improvement coefficient $c_{\rm sw}$ set to its tadpole-improved
tree-level value. By contrast, CalLat use the M\"obius variant of
domain-wall fermions, which are fully $\cO(a)$ improved. The CalLat set
of ensembles includes three values of the lattice spacing, i.e.
$a$ = 0.09, 0.12, and 0.15~fm, while PNDME added another set of
ensembles at the finer lattice spacing of 0.06~fm to this collection.
Both groups have included physical pion mass ensembles in their
calculations. The operator matrix elements are renormalized
nonperturbatively, using the Rome-Southampton method.

In order to control excited-state contamination, PNDME perform
multi-state fits, including up to four (three) energy levels in the
two-point (three-point) correlation functions. By contrast, CalLat
have employed the Feynman-Hellmann-inspired implementation of summed
operator insertions described in Sec.\,\ref{sec:ESC}. Plotting the
summed correlator $S_A(\tau)$ as a function of the source-sink
separation, they find that excited-state effects cannot be detected for
$\tau\gtrsim1.0$\,fm at their level of statistics. After subtracting
the leading contributions from excited states determined from
two-state fits, they argue that the data for $S_A(\tau)$ can be
described consistently down to $\tau\simeq0.3$\,fm.

The recent calculation by ETMC (ETM 19 \cite{Alexandrou:2019brg})
with $N_f=2+1+1$ was performed using a single twisted-mass QCD
ensemble with $m_\pi\approx139$\,MeV. In order to control excited-state
effects, the summation method and multi-state fits were used.
No significant finite-volume effects were expected based on a similar 
analysis of two $N_f = 2$ ensembles with different spatial extents.  
The
quoted estimate is identified with the result obtained from a
two-state fit on the single $N_f=2+1+1$ ensemble, which agrees with
the value determined from the summation method.

We now proceed to discuss global averages for the axial charge, in
accordance with the procedures in Sec.~\ref{sec:rating}. For QCD with
$N_f=2+1+1$, the calculations of PNDME and CalLat pass all our quality
criteria, while the result of ETM 19 is excluded due to the fact that
it was performed at a single value of the lattice spacing. Hence the
results from PNDME~18 \cite{Gupta:2018qil} and CalLat 19
\cite{Walker-Loud:2019cif}, which is an update of CalLat~18
\cite{Chang:2018uxx}, qualify for being included in a global average.
Since both PNDME and CalLat use gauge ensembles produced by MILC, we
assume that the quoted
errors are 100\% correlated, even though the range of pion
masses and lattice spacings explored in Refs.~\cite{Gupta:2018qil} and
\cite{Chang:2018uxx, Walker-Loud:2019cif} is not exactly identical.
Performing a weighted average yields $g_A^{u-d} = 1.2617(126)$ with
$\chi^2/{\rm dof}=1.33$, where the error has been scaled 
by about 15\% because of the large $\chi^2/{\rm dof}$.
The result by CalLat dominates the weighted average due to its smaller
error.
Given that the calculations of PNDME~18 and CalLat 19 are correlated,
the large value of $\chi^2/{\rm dof}$ indicates a slight tension
between the two results. In this situation we adopt a more
conservative approach, by requiring that the uncertainty assigned to
the FLAG estimate encompasses the central value of PNDME~18.
As a result, we choose to represent the axial charge 
by the interval $1.218\leq g_A^{u-d}\leq 1.274$, where the lower bound
is identified with the result of PNDME~18, while the upper bound is
the weighted average plus the scaled 1$\sigma$ uncertainty. Hence, for
$N_f=2+1+1$ we quote $g_A^{u-d}=1.246(28)$ as the FLAG estimate, where
the central value marks the mid-point of the interval, and half the
width is taken to be the error.

\begin{figure}[!t]
\begin{center}
\includegraphics[width=11.5cm]{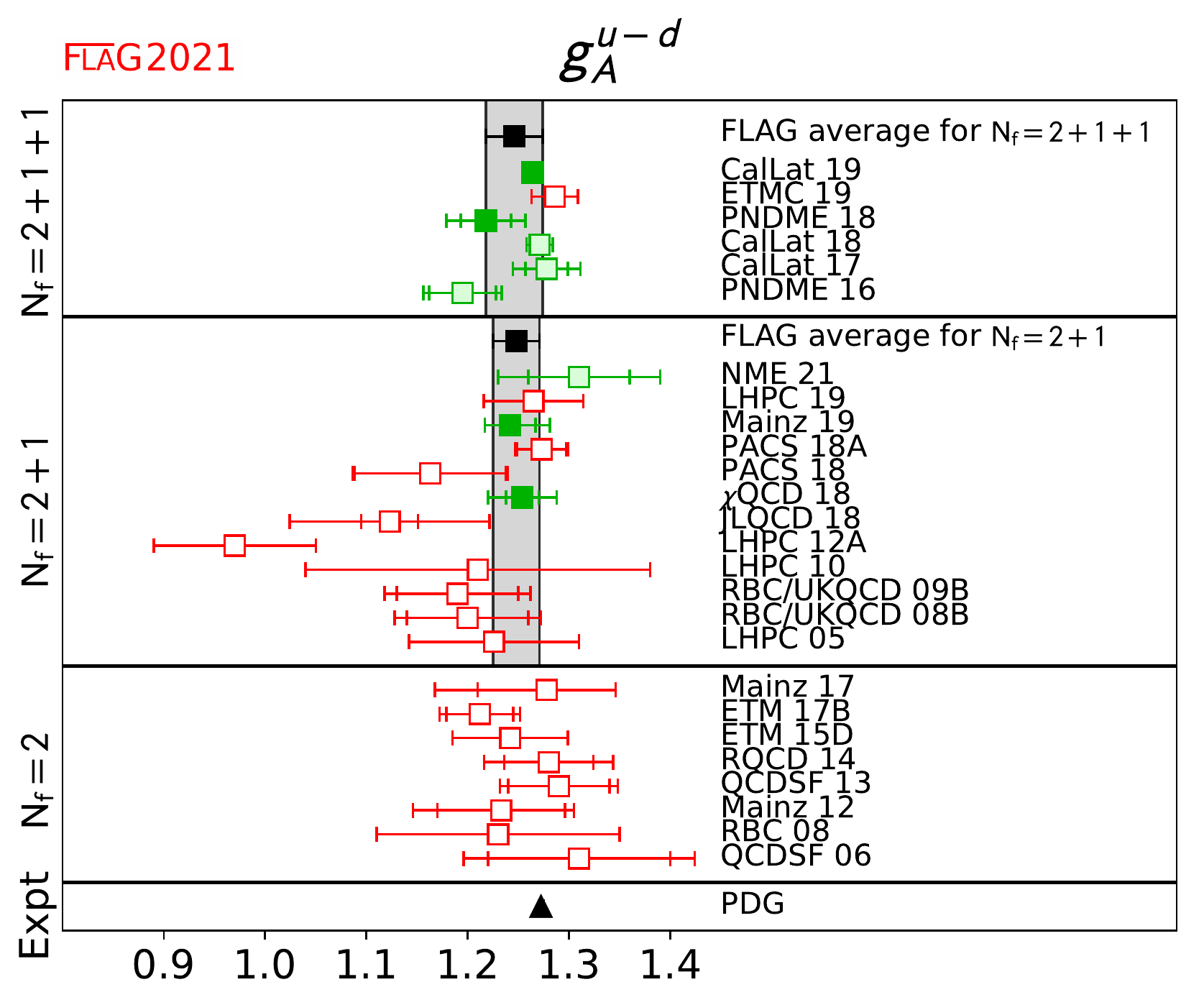}
\end{center}
\vspace{-1cm}
\caption{\label{fig:ga} Lattice results and FLAG averages for the
  isovector axial charge $g_A^{u-d}$ for $N_f=2$, $2+1$ and $2+1+1$
flavour calculations. Also shown is the experimental result as quoted
in the PDG \cite{Zyla:2020zbs}.}
\end{figure}

For QCD with $N_f=2+1$ dynamical quarks, the calculations of
$\chi$QCD~18 \cite{Liang:2018pis}, Mainz~19 \cite{Harris:2019bih} and
NME 21 \cite{Park:2021ypf} are free of red tags, while the
calculation by PACS~18A \cite{Shintani:2018ozy} and LHPC 19
\cite{Hasan:2019noy} do not offer enough control over lattice
artefacts according to the FLAG criteria. Since the result by NME 21
was published only as a preprint by the FLAG deadline, it does not
qualify for being included in a global average. Hence, for $N_f=2+1$
we compute a weighted average from $\chi$QCD~18 \cite{Liang:2018pis}
and Mainz 19 \cite{Harris:2019bih}, assuming no correlations between
the two calculations. This yields $g_A^{u-d}=1.248(23)$ with
$\chi^2/\rm dof=0.07$.

Due to the modified criteria for excited-state contamination, none of
the results obtained in two-flavour QCD qualify for a global average.
Nonetheless, we find it instructive to show the results for $N_f=2$
together with the calculations with $N_f=2+1$ and $2+1+1$ and the
respective FLAG estimates in Fig.~\ref{fig:ga}.


To summarize, the FLAG averages for the axial charge read
\begin{align}
&\label{eq:ga_2p1p1}
	\Nf=2+1+1:&\FLAGAVBEGIN g_A^{u-d} &= 1.246(28) \FLAGAVEND
  &&\Refs~\mbox{\cite{Gupta:2018qil,Chang:2018uxx,Walker-Loud:2019cif}},\\
&\label{eq:ga_2p1}
	\Nf=2+1:&\FLAGAVBEGIN g_A^{u-d} &= 1.248(23) \FLAGAVEND
  &&\Ref~\mbox{\cite{Liang:2018pis,Harris:2019bih}},
\end{align}
Within errors, these averages are compatible with the result of
$g_A^{u-d}=1.2724(23)$ quoted by the PDG. While the most recent
lattice calculations reproduce the axial charge at the level of a few
percent or even better, the experimental result is more precise by an
order of magnitude.

\subsubsection{Results for $g_S^{u-d}$\label{sec:gS-IV}}

\begin{table}[t!]
\begin{center}
\mbox{} \\[3.0cm]
\footnotesize
\begin{tabular*}{\textwidth}[l]{l @{\extracolsep{\fill}} r l l l l l l l l }
Collaboration & Ref. & $\Nf$ & 
\hspace{0.15cm}\begin{rotate}{60}{publication status}\end{rotate}\hspace{-0.15cm} &
\hspace{0.15cm}\begin{rotate}{60}{continuum extrapolation}\end{rotate}\hspace{-0.15cm} &
\hspace{0.15cm}\begin{rotate}{60}{chiral extrapolation}\end{rotate}\hspace{-0.15cm}&
\hspace{0.15cm}\begin{rotate}{60}{finite volume}\end{rotate}\hspace{-0.15cm}&
\hspace{0.15cm}\begin{rotate}{60}{renormalization}\end{rotate}\hspace{-0.15cm}  &
\hspace{0.15cm}\begin{rotate}{60}{excited states}\end{rotate}\hspace{-0.15cm}  &
$g^{u-d}_S$\\
&&&&&&&&& \\[-0.1cm]
\hline
\hline
&&&&&&&& \\[-0.1cm]

ETM 19 & \cite{Alexandrou:2019brg} & 2+1+1 & \gA & \bad & \soso & \good & \good & \soso & 1.35(17) \\[0.5ex]
PNDME 18 & \cite{Gupta:2018qil} & 2+1+1 & \gA & \good$^\ddag$ & \good & \good & \good & \soso & 1.022(80)(60) \\[0.5ex]
PNDME 16 & \cite{Bhattacharya:2016zcn} & 2+1+1 & \gA & \soso$^\ddag$ & \good & \good & \good & \soso & 0.97(12)(6) \\[0.5ex]
PNDME 13 & \cite{Bhattacharya:2013ehc} & 2+1+1 & \gA & \bad$^\ddag$ & \bad & \good & \good & \soso & 0.72(32) \\[0.5ex]
\\[-0.1ex]\hline\\[0.2ex]
NME 21 & \cite{Park:2021ypf} & 2+1 & P & \soso$^\ddag$ & \good & \good & \good & \soso & 1.06(10)(6) \\[0.5ex]
$\chi$QCD 21A & \cite{Liu:2021irg} & 2+1 & P & \good & \good & \good & \good & \soso & 0.94(10)(6) \\[0.5ex]
RBC/UKQCD 19 & \cite{Abramczyk:2019fnf} & 2+1 & \gA & \bad & \soso & \good & \good & \bad & 0.9(3) \\[0.5ex]
Mainz 19 & \cite{Harris:2019bih} & 2+1 & \gA & \good & \soso & \good & \good & \soso & 1.13(11)($^{7}_{6}$) \\[0.5ex]
LHPC 19 & \cite{Hasan:2019noy} & 2+1 & \gA & \bad$^\ddag$ & \good & \good & \good & \soso & 0.927(303) \\[0.5ex]
JLQCD 18 & \cite{Yamanaka:2018uud} & 2+1 & \gA & \bad & \soso & \soso & \good & \soso & 0.88(8)(3)(7) \\[0.5ex]
LHPC 12 & \cite{Green:2012ej} & 2+1 & \gA & \bad$^\ddag$ & \good & \good & \good & \soso & 1.08(28)(16) \\[0.5ex]
\\[-0.1ex]\hline\\[0.2ex]
ETM 17 & \cite{Alexandrou:2017qyt} & 2 & \gA & \bad & \soso & \soso & \good & \soso & 0.930(252)(48)(204) \\[0.5ex]
RQCD 14 & \cite{Bali:2014nma} & 2 & \gA & \soso & \good & \good & \good & \bad & 1.02(18)(30) \\[0.5ex]
&&&&&&&& \\[-0.1cm]
\hline
\hline
\end{tabular*}
\begin{minipage}{\linewidth}
{\footnotesize 
\begin{itemize}
\item[$^\ddag$]The rating takes into account that the action is not fully O(a) improved by requiring an additional lattice spacing.
\end{itemize}
}
\end{minipage}
\caption{Overview of results for $ g^{u-d}_S$. \label{tab:gs}}
\end{center}
\end{table}

Calculations of the isovector scalar charge have, in general, larger
errors than the isovector axial charge as can be seen from the
compilation given in Tab.\,\ref{tab:gs} and plotted in
Fig.~\ref{fig:gs}. The isovector scalar charge can also be determined indirectly via the
conserved vector current~(CVC) relation from results for the
neutron-proton mass difference~\cite{Walker-Loud:2012ift,Shanahan:2012wa,
Beane:2006fk,Horsley:2012fw,deDivitiis:2013xla,Budapest-Marseille-Wuppertal:2013rtp,
Borsanyi:2014jba,Horsley:2015eaa,Brantley:2016our} and the down and up quark mass
difference~(see Sec.~\ref{subsec:mumd}). For comparison, Fig.~\ref{fig:gs} also
shows an indirect determination obtained using lattice and
phenomenological input~\cite{Gonzalez-Alonso:2013ura}.

As in FLAG 19, for 2+1+1 flavours, only PNDME~18 \cite{Gupta:2018qil}, which supersedes 
PNDME~16 \cite{Bhattacharya:2016zcn} and 
PNDME~13 \cite{Bhattacharya:2013ehc}, meets all the 
criteria for inclusion in the average. The discussions for this and other past calculations are repeated from FLAG 19 for completion.

This mixed-action calculation was performed using the
MILC HISQ ensembles, with a clover valence action.  The 11 ensembles
used include three pion mass values, $M_{\pi} \sim$ 135, 225, 320~MeV,
and four lattice spacings, $a \sim$ 0.06, 0.09, 0.12, 0.15~fm. Note
that four lattice spacings are required to meet the green star
criteria, as this calculation is not fully $\cO(a)$ improved. Lattice
size ranges between $3.3 \lesssim M_{\pi} L \lesssim 5.5$. Physical
point extrapolations were performed simultaneously, keeping only the
leading-order terms in the various expansion parameters. For the finite-volume extrapolation, the asymptotic limit of
the $\chi$PT prediction, Eq.~(\ref{eq:Vasymp}), was used. Excited-state contamination is controlled using two-state fits to
between three and five source-sink time separations between $0.72 \lesssim \tau \lesssim 1.68$~fm. Renormalization was performed nonperturbatively using the
RI-SMOM scheme and converted to $\msbar$ at 2~GeV using 2-loop
perturbation theory.

The calculation performed in ETM 19 \cite{Alexandrou:2019brg} was generated using twisted-mass fermions with a clover term. The calculation utilized a single 2+1+1-flavour gauge configuration, with a pion mass near the physical point, $m_{\pi}\sim$ 139~MeV, lattice spacing of $a\sim$ 0.08~fm, and volume corresponding to $m_{\pi}L = 3.86$. Seven source-sink separations were used in the analysis, ranging from $t=$ 0.64--1.6~fm. Two further two-flavour ensembles were also explored, having the same pion mass, $m_{\pi}\sim$ 130~MeV  and lattice spacing $a\sim$ 0.09~fm, but with different volumes corresponding to $m_{\pi}L \sim 3$ and $m_{\pi}L \sim 4$. The final result is quoted from the single 2+1+1 flavour ensemble and does not include an assessment of discretization systematics, and therefore does not meet the continuum quality criterion for inclusion in the average.

Regarding 2+1-flavour calculations, a single calculation meets all criteria necessary for inclusion in the average.  
The Mainz 19~\cite{Harris:2019bih} calculation was
performed on the Wilson CLS ensembles, using four lattice spacings
($a\sim 0.05$~fm to $0.086$~fm),  several pion masses ranging from $\sim 200$~MeV to $\sim 350$~MeV, and volumes corresponding to $m_{\pi}L\sim 3$ to $\sim 5.4$. Physical point extrapolations were performed simultaneously in the lattice spacing, pion mass, and volume. Excited states were controlled using two-state simultaneous fits to multiple observables, and included several
source-sink separations typically in the range 1-1.5~fm. Renormalization was performed nonperturbatively using the
RI-SMOM scheme and converted to $\msbar$ at 2~GeV using 2-loop perturbation theory. 

The 2+1-flavour calculation of $\chi$QCD 21A \cite{Liu:2021irg} was performed using a mixed-action approach with domain-wall fermion gauge configurations generated by the RBC/UKQCD collaboration and overlap valence quarks.
They include five pion
masses ranging from $m_{\pi}\sim$ 140~MeV to 370~MeV, four lattice spacings ($a \sim$ 0.06, 0.08, 0.11, and 0.14~fm). Three to six different valence-quark masses are computed on each ensemble. The extrapolation to the physical pion mass, continuum and
infinite-volume limits is obtained by a global fit of all data to a partially quenched chiral perturbation theory ansatz. Excited-state contamination is assessed using three to five sink-source separations and multi-state fits. Renormalization is performed using RI/MOM and the final result quoted in $\msbar$ at 2 GeV. At the time of writing of this review, this calculation was unpublished and the results are therefore not included in the average.

The NME 21\cite{Park:2021ypf} 2+1-flavour calculation utilized seven ensembles of Wilson-clover fermions. Three lattice spacings, ranging from $a\sim 0.07$~fm to $0.13$~fm, several pion masses, $m_{\pi}\sim$ 165~MeV to 285~MeV, and volumes corresponding to $m_{\pi}L\sim$ 3.75 to 6.15 were used. Combined continuum, chiral, and infinite-volume extrapolations are performed to the physical point using leading-order fit functions. Several fitting strategies are explored using four to six source-sink separations ranging from 0.7--1.8~fm. Final results are quoted by averaging results from two of these fitting strategies, in which the excited-state energy for the three-point function is fixed using two different prior strategies. Renormalization is non-perturbative (RI-SMOM) using two strategies, and quoted in $\msbar$ at 2~GeV. This work was also unpublished at the time of writing of this review and is not included in the average.

The RBC/UKQCD 19~\cite{Abramczyk:2019fnf} calculation employed 2+1 flavours of domain-wall fermions using an Iwasaki and dislocation-suppressing-determinant-ratio gauge action. They utilized two values of the pion mass, $m_{\pi} \sim$ 250 and 170~MeV with volumes corresponding to $m_{\pi}L\sim$ 5.8 and 4.0, respectively. The results are quoted using only one lattice spacing of 0.14~fm, and a single source-sink separation of 1.3~fm and therefore do not meet the criteria for continuum or excited-state contamination. The LHPC 19 \cite{Hasan:2019noy} calculation used a 2+1 flavour 2-HEX-smeared Wilson-clover action with two ensembles near the physical pion mass, $m_{\pi}\sim$ 133 and 137~MeV. The lattice spacings corresponded to $a\sim$ 0.09 and 0.12~fm and volumes $m_{\pi}L\sim 4$. They used 3 and 8 different time separations for the two ensembles and compare ratio, summation, and multi-state methods to assess excited-state contamination. Because the calculation is not fully $\cO(a)$ improved, an additional lattice spacing would be necessary to meet the continuum criterion for inclusion in the average.

The JLQCD 18~\cite{Yamanaka:2018uud} calculation, performed using
overlap fermions on the Iwasaki gauge action, covered four pion masses
down to 290~MeV. The lattice size was adjusted to keep $M_\pi L \geq 4$ in all four cases. However, the single lattice spacing
of $a=0.11$~fm does not meet the criteria for continuum
extrapolation. The calculations presented in LHPC~12A used three
different lattice actions, Wilson-clover, domain-wall, and mixed
action. Pion masses ranged down to near the physical pion mass. Data at two
lattice spacings were produced with the domain-wall and Wilson
actions, however, the final result utilized only the single lattice
spacing of $a=0.116$~fm from the Wilson action. Because the action is
not fully $\cO(a)$ improved, two lattice spacings are not sufficient for
meeting the quality criteria for the continuum extrapolation.

The two-flavour calculations in Tab.~\ref{tab:gs} include
ETM~17, which employed twisted-mass fermions on the
Iwasaki gauge action\footnote{The earlier work, ETM~15D~\cite{Abdel-Rehim:2015owa}, did not give a final value for $g_S^{u-d}$ and is therefore not included in the tables.}. This work utilized a single physical pion mass ensemble with lattice spacing $a\sim 0.09$~fm, and therefore does not
meet the criteria for continuum extrapolation. The RQCD~14 calculation
included three lattice spacings down to 0.06~fm and several pion
masses down to near the physical point. While a study of excited-state
contamination was performed on some ensembles using multiple
source-sink separations, many ensembles included only a single time
separation, so it does not meet the criteria for excited states.\looseness-1

The final FLAG value for $g_S^{u-d}$ is
\begin{align}
&\label{eq:gs_2p1p1}
  \Nf=2+1+1:&\FLAGAVBEGIN g_S^{u-d} &=  1.02(10) \FLAGAVEND
  &&\Ref~\mbox{\cite{Gupta:2018qil}}, \\
&\label{eq:gs_2p1}    \Nf=2+1:&\FLAGAVBEGIN g_S^{u-d} &=  1.13(14)\FLAGAVEND
  &&\Ref~\mbox{\cite{Harris:2019bih}}.
\end{align}

\begin{figure}[t!]
\begin{center}
\includegraphics[width=11.5cm]{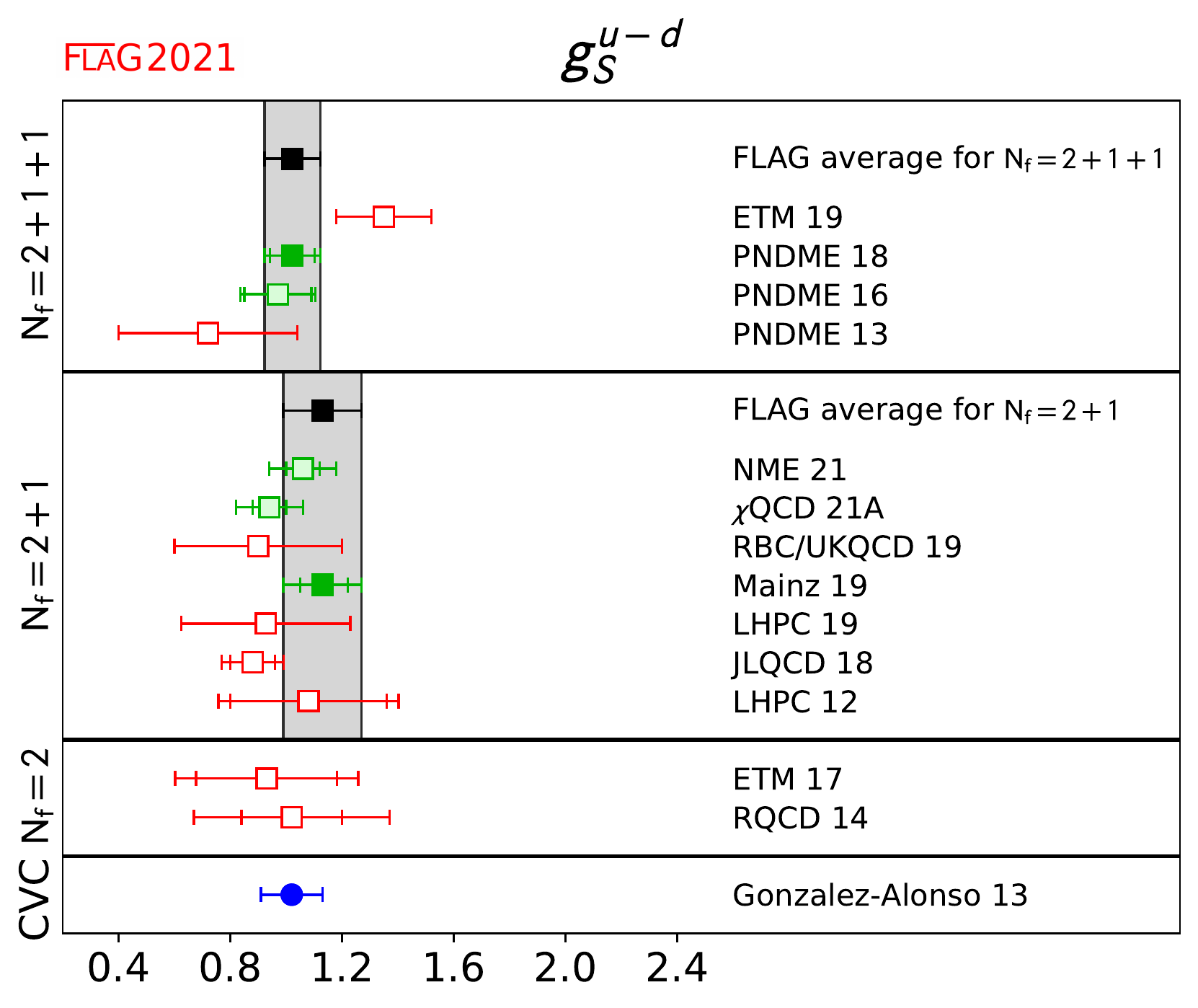}
\end{center}
\vspace{-1cm}
\caption{\label{fig:gs} Lattice results and FLAG averages for the isovector scalar charge $g^{u-d}_S$ 
for $\Nf = 2$, $2+1$, and $2+1+1$ flavour calculations. Also shown is a phenomenological result obtained using the conserved vector current~(CVC) relation~\cite{Gonzalez-Alonso:2013ura} (circle).}
\end{figure}

\subsubsection{Results for $g_T^{u-d}$\label{sec:gT-IV}}

\begin{table}[t!]
\begin{center}
\mbox{} \\[3.0cm]
\footnotesize
\begin{tabular*}{\textwidth}[l]{l @{\extracolsep{\fill}} r l l l l l l l l }
Collaboration & Ref. & $\Nf$ & 
\hspace{0.15cm}\begin{rotate}{60}{publication status}\end{rotate}\hspace{-0.15cm} &
\hspace{0.15cm}\begin{rotate}{60}{continuum extrapolation}\end{rotate}\hspace{-0.15cm} &
\hspace{0.15cm}\begin{rotate}{60}{chiral extrapolation}\end{rotate}\hspace{-0.15cm}&
\hspace{0.15cm}\begin{rotate}{60}{finite volume}\end{rotate}\hspace{-0.15cm}&
\hspace{0.15cm}\begin{rotate}{60}{renormalization}\end{rotate}\hspace{-0.15cm}  &
\hspace{0.15cm}\begin{rotate}{60}{excited states}\end{rotate}\hspace{-0.15cm}  &
$g^{u-d}_T$\\
&&&&&&&&& \\[-0.1cm]
\hline
\hline
&&&&&&&& \\[-0.1cm]

ETM 19 & \cite{Alexandrou:2019brg} & 2+1+1 & \gA & \bad & \soso & \good & \good & \soso & 0.936(25) \\[0.5ex]
PNDME 18 & \cite{Gupta:2018qil} & 2+1+1 & \gA & \good$^\ddag$ & \good & \good & \good & \soso & 0.989(32)(10) \\[0.5ex]
PNDME 16 & \cite{Bhattacharya:2016zcn} & 2+1+1 & \gA & \soso$^\ddag$ & \good & \good & \good & \soso & 0.987(51)(20) \\[0.5ex]
PNDME 15 & \cite{Bhattacharya:2015wna,Bhattacharya:2015esa} & 2+1+1 & \gA & \soso$^\ddag$ & \good & \good & \good & \soso & 1.020(76) \\[0.5ex]
PNDME 13 & \cite{Bhattacharya:2013ehc} & 2+1+1 & \gA & \bad$^\ddag$ & \bad & \good & \good & \soso & 1.047(61) \\[0.5ex]
\\[-0.1ex]\hline\\[0.2ex]
NME 21 & \cite{Park:2021ypf} & 2+1 & P & \soso$^\ddag$ & \good & \good & \good & \soso & 0.95(5)(2) \\[0.5ex]
RBC/UKQCD 19 & \cite{Abramczyk:2019fnf} & 2+1 & \gA & \bad & \soso & \good & \good & \bad & 1.04(5) \\[0.5ex]
Mainz 19 & \cite{Harris:2019bih} & 2+1 & \gA & \good & \soso & \good & \good & \soso & 0.965(38)($^{13}_{41}$) \\[0.5ex]
LHPC 19 & \cite{Hasan:2019noy} & 2+1 & \gA & \bad$^\ddag$ & \good & \good & \good & \soso & 0.972(41) \\[0.5ex]
JLQCD 18 & \cite{Yamanaka:2018uud} & 2+1 & \gA & \bad & \soso & \soso & \good & \soso & 1.08(3)(3)(9) \\[0.5ex]
LHPC 12 & \cite{Green:2012ej} & 2+1 & \gA & \bad$^\ddag$ & \good & \good & \good & \soso & 1.038(11)(12) \\[0.5ex]
RBC/UKQCD 10D & \cite{Aoki:2010xg} & 2+1 & \gA & \bad & \bad & \soso & \good & \bad & 0.9(2) \\[0.5ex]
\\[-0.1ex]\hline\\[0.2ex]
ETM 17 & \cite{Alexandrou:2017qyt} & 2 & \gA & \bad & \soso & \soso & \good & \soso & 1.004(21)(2)(19) \\[0.5ex]
ETM 15D & \cite{Abdel-Rehim:2015owa} & 2 & \gA & \bad & \soso & \soso & \good & \soso & 1.027(62) \\[0.5ex]
RQCD 14 & \cite{Bali:2014nma} & 2 & \gA & \soso & \good & \good & \good & \bad & 1.005(17)(29) \\[0.5ex]
RBC 08 & \cite{Lin:2008uz} & 2 & \gA & \bad & \bad & \bad & \good & \bad & 0.93(6) \\[0.5ex]
&&&&&&&& \\[-0.1cm]
\hline
\hline
\end{tabular*}
\begin{minipage}{\linewidth}
{\footnotesize 
\begin{itemize}
\item[$^\ddag$]The rating takes into account that the action is not fully O(a) improved by requiring an additional lattice spacing.
\end{itemize}
}
\end{minipage}
\caption{Overview of results for $ g^{u-d}_T$. \label{tab:gt}}
\end{center}
\end{table}

Estimates of the isovector tensor charge are currently the most precise of the isovector charges with values that are 
stable over time, as can be seen from the
compilation given in Tab.\,\ref{tab:gt} and plotted in
Fig.~\ref{fig:gt}. This is a consequence of the smaller statistical
fluctuations in the raw data and the very mild dependence on $a$, $M_\pi$, and the
lattice size $M_\pi L$. As a result, the uncertainty due to the 
various extrapolations is small. Also shown for comparison in Fig.~\ref{fig:gt} are phenomenological results using measures of transversity~\cite{Radici:2015mwa,Kang:2015msa,Kang:pc2015,Goldstein:2014aja,Pitschmann:2014jxa}.

As in FLAG 19, for 2+1+1 flavours, only PNDME~18 \cite{Gupta:2018qil}, which supersedes 
PNDME~16 \cite{Bhattacharya:2016zcn}, PNDME 15 \cite{Bhattacharya:2015wna} and 
PNDME~13 \cite{Bhattacharya:2013ehc}, meets all the 
criteria for inclusion in the average. The details for this calculation 
are the same as those for $g_S^{u-d}$ described in the previous section (Sec.~\ref{sec:gS-IV}), 
except that three-state fits were used to remove excited-state effects. The details of the 2+1+1 flavour calculation by ETM 19, which does not meet the criteria for averaging, are also the same as those described in the previous section for $g_S^{u-d}$.

For 2+1-flavour calculations, only Mainz 19~\cite{Harris:2019bih} meets all criteria for inclusion in the averages. Details of this calculation are the same as for $g_S^{u-d}$, described in the previous section.  

Details for the 2+1-flavour NME 21, RBC/UKQCD 19, LHPC 19, Mainz~18,
JLQCD~18, and LHPC~12A, calculations are identical to those
presented previously in Sec.~\ref{sec:gS-IV}. The earlier RBC/UKQCD~10
calculation was performed using domain-wall fermions on the Iwasaki
gauge action, with two volumes and several pion masses. The lowest
pion mass used was $M_{\pi}\sim 330$~MeV and does not meet the
criteria for chiral extrapolation. In addition, the single lattice
spacing and single source-sink separation do not meet the criteria for
continuum extrapolation and excited states.

Two-flavour calculations include RQCD~14, with details identical to those described in Sec.~\ref{sec:gS-IV}. 
There are two calculations, ETM~15D~\cite{Abdel-Rehim:2015owa} and ETM~17~\cite{Alexandrou:2017qyt}, which employed
twisted-mass fermions on the Iwasaki gauge action. The earlier work
utilized three ensembles, with three volumes and two pion masses down
to the physical point. The more recent work used only the physical
pion mass ensemble. Both works used only a single lattice spacing
$a\sim 0.09$~fm, and therefore do not meet the criteria for continuum
extrapolation. The early work by RBC~08 with domain-wall fermions used three heavy values for the pion mass, and a single
value for the lattice spacing, volume, and source-sink separation,
and therefore do not meet many of the criteria.

The final FLAG value for $g_T^{u-d}$ is
\begin{align}
&\label{eq:gt_2p1p1}
  \Nf=2+1+1:&\FLAGAVBEGIN g_T^{u-d} &=  0.989(34)\FLAGAVEND
  &&\Ref~\mbox{\cite{Gupta:2018qil}}, \\
 &\label{eq:gt_2p1}   \Nf=2+1:&\FLAGAVBEGIN g_T^{u-d} &=  0.965(61)\FLAGAVEND
  &&\Ref~\mbox{\cite{Harris:2019bih}}.
\end{align}

\begin{figure}[t!]
\begin{center}
\includegraphics[width=11.5cm]{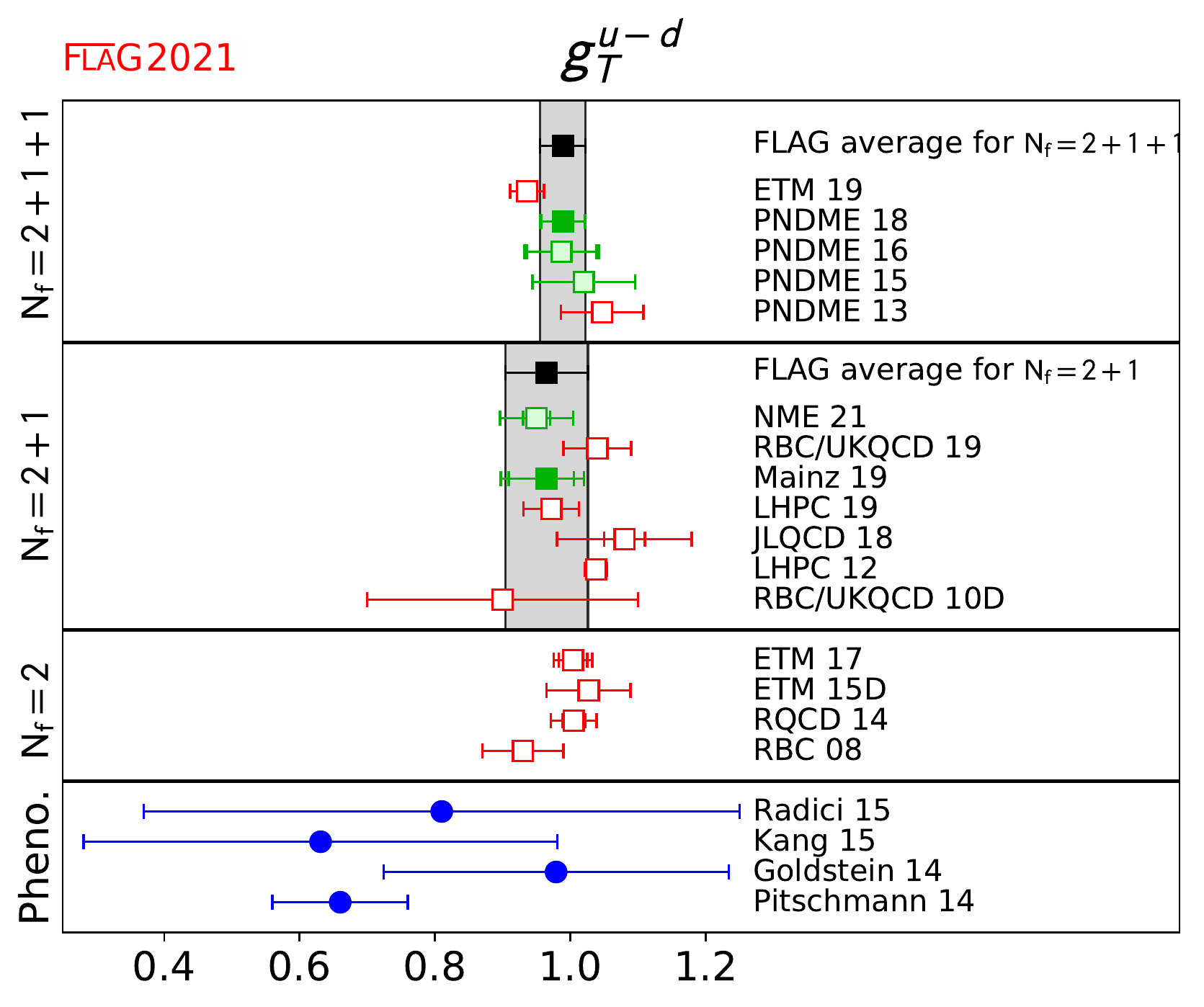}
\end{center}
\vspace{-1cm}
\caption{\label{fig:gt} Lattice results and FLAG averages for the isovector tensor charge $g^{u-d}_T$
for $\Nf = 2$, $2+1$, and $2+1+1$ flavour calculations. Also shown are phenomenological results using measures of transversity~\cite{Radici:2015mwa,Kang:2015msa,Kang:pc2015,Goldstein:2014aja,Pitschmann:2014jxa} (circles).}
\end{figure}

\subsection{Flavour Diagonal Charges\label{sec:FDcharges}}

Three examples of interactions for which matrix elements of
flavour-diagonal operators ($q \Gamma q$ where $\Gamma$ defines the
Lorentz structure of the bilinear quark operator) are needed are the
neutral current interactions of neutrinos, elastic scattering of
electrons off nuclei, and the scattering of dark matter off
nuclei. In addition, these
matrix elements also probe intrinsic properties of nucleons (the spin,
the nucleon sigma term and strangeness content, and the contribution of the 
electric dipole moment (EDM) of the quarks to the nucleon EDM) as
explained below. For brevity, all operators are assumed to be appropriately
renormalized as discussed in Sec.~\ref{sec:renorm}. 

The matrix elements of the scalar operator $\overline{q} q$ with
flavour $q$ give the rate of change in the nucleon mass due to
nonzero values of the corresponding quark mass. This relationship is
given by the Feynman-Hellmann theorem. The quantities of interest are
the nucleon $\sigma$-term, $\sigma_{\pi N}$, and the strange and charm
content of the nucleon, $\sigma_{s}$ and $\sigma_{c}$,
\begin{align}
\sigma_{\pi N} &= m_{ud} \langle N| \overline{u} u +  \overline{d} d | N \rangle  \,, \\
\sigma_{s}     &= m_s \langle N| \overline{s} s | N \rangle  \,, \\
\sigma_{c}     &= m_c \langle N| \overline{c} c | N \rangle  \,.
\label{eq:gSdef}
\end{align}
Here $m_{ud}$ is the average of the up and down quark masses and $m_s$
($m_c$) is the strange (charm) quark mass.  The $\sigma_{\pi N, s, c}$
give the shift in $M_N$ due to nonzero light-, strange- and charm-quark
masses.  The same matrix elements are also needed to quantify the spin
independent interaction of dark matter with nucleons. Note that, while
$\sigma_b$ and $\sigma_t$ are also phenomenologically interesting,
they are unlikely to be calculated on the lattice due to the expected tiny signal in the matrix elements. In principle, the
heavy sigma terms can be estimated using $\sigma_{u,d,s}$ by exploiting
the heavy-quark
limit~\cite{Shifman:1978zn,Chetyrkin:1997un,Hill:2014yxa}.

The matrix elements of the axial operator $\overline{q} \gamma_\mu
\gamma_5 q$ give the contribution $\Delta q$ of quarks of flavour
$q$ to the spin of the nucleon:
\begin{align}
\langle N| \overline{q} \gamma_\mu \gamma_5 q | N \rangle  &= g_A^q \overline{u}_N \gamma_\mu \gamma_5 u_N,  \nonumber \\ 
g_A^q \equiv \Delta q &= \int_0^1 dx (\Delta q(x) + \Delta \overline{q} (x) )  \,.
\label{eq:gAdefnme}
\end{align}
The charge $g_A^q$ is thus the contribution of the spin of a quark of
flavour $q$ to the spin of the nucleon.  It is also related to the
first Mellin moment of the polarized parton distribution function
(PDF) $\Delta q$ as shown in the second line in
Eq.~\eqref{eq:gAdefnme}.  Measurements by the European Muon
collaboration in 1987 of the spin asymmetry in polarized deep
inelastic scattering showed that the sum of the spins of the quarks
contributes less than half of the total spin of the
proton~\cite{Ashman:1987hv}.  To understand this unexpected result,
called the ``proton spin crisis'', it is common to start with Ji's sum
rule~\cite{Ji:1996ek}, which provides a gauge invariant decomposition of
the nucleon's total spin, as
\begin{equation}
\frac{1}{2} =  \sum_{q=u,d,s,c,\cdot} \left(\frac{1}{2} \Delta q + L_q\right) + J_g \,,
\label{eq:Ji}
\end{equation}
where $\Delta q /2 \equiv g_A^q /2 $ is the contribution of the
intrinsic spin of a quark with flavour $q$; $L_q$ is the orbital
angular momentum of that quark; and $J_g$ is the total angular
momentum of the gluons.  Thus, to obtain the spin of the proton
starting from QCD requires calculating the contributions of the three
terms: the spin and orbital angular momentum of the quarks, and the
angular momentum of the gluons. Lattice-QCD calculations of the
various matrix elements needed to extract the three contributions are
underway. An alternate decomposition of the spin of the proton has
been provided by Jaffe and Manohar~\cite{Jaffe:1989jz}. The two
formulations differ in the decomposition of the contributions of the
quark orbital angular momentum and of the gluons. The contribution of
the quark spin, which is the subject of this review and given in
Eq.~\eqref{eq:gAdefnme}, is the same in both formulations.

The tensor charges are defined as the matrix elements of the tensor
operator $\overline{q} \sigma^{\mu\nu} q$ with $\sigma^{\mu\nu} = 
\{\gamma_\mu,\gamma_\nu\}/2$:
\begin{align}
g_T^q \overline{u}_N \sigma_{\mu \nu} u_N &= \langle N| \overline{q} \sigma_{\mu \nu} q | N \rangle  \,.
\label{eq:gTdef}
\end{align}
These flavour-diagonal tensor charges $g_T^{u,d,s,c}$ quantify the
contributions of the $u$, $d$, $s$, $c$ quark EDM to the neutron electric dipole moment
(nEDM)~\cite{Bhattacharya:2015wna,Pospelov:2005pr}. Since 
particles can have an EDM only due to P and T (or CP assuming CPT is a good
symmetry) violating interactions, the nEDM is a very sensitive probe of
new sources of CP violation that arise in most extensions of the
SM designed to explain nature at the TeV scale. The
current experimental bound on the nEDM is $d_n < 2.9 \times
10^{-26}\ e$~cm~\cite{Baker:2006ts}, while the known CP violation in the SM
implies  $d_n < 10^{-31}\ e$~cm~\cite{Seng:2014lea}. A nonzero result over the
intervening five orders of magnitude would signal new physics.
Planned experiments aim to reduce the bound to around $
10^{-28}\ e$~cm. A discovery or reduction in the bound from these
experiments will put stringent constraints on many BSM theories,
provided the matrix elements of novel CP-violating interactions, of
which the quark EDM is one, are calculated with the required
precision.

One can also extract these tensor charges from the zeroth moment of the
transversity distributions that are measured in many experiments
including Drell-Yan and semi-inclusive deep inelastic scattering
(SIDIS). Of particular importance is the active program at Jefferson Lab (JLab) to measure
them~\cite{Dudek:2012vr,Ye:2016prn}. 
Transversity distributions describe the net transverse
polarization of quarks in a transversely polarized nucleon. Their 
extraction from the data taken over a limited range of $Q^2$ and
Bjorken $x$ is, however, not straightforward and requires additional
phenomenological modeling. At present, lattice-QCD estimates of
$g_T^{u,d,s}$ are the most accurate~\cite{Bhattacharya:2015wna,Radici:2018iag,Lin:2017stx} 
as can be deduced from Fig.~\ref{fig:gt}.  Future experiments will
significantly improve the extraction of the transversity
distributions.  Thus, accurate calculations of the tensor charges
using lattice QCD will continue to help elucidate the structure of the
nucleon in terms of quarks and gluons and provide a benchmark against
which phenomenological estimates utilizing measurements at JLab and
other experimental facilities worldwide can be compared.

The methodology for the calculation of flavour-diagonal charges is also 
well-established. The major challenges are the much larger statistical errors in the 
disconnected contributions for the same computational cost and the 
need for the additional calculations of the isosinglet renormalization factors. 

\subsubsection{Results for $g_A^{u,d,s}$\label{sec:gA-FD}}

\begin{table}[t!]
\begin{center}
\mbox{} \\[3.0cm]
\footnotesize
\begin{tabular*}{\textwidth}[l]{l @{\extracolsep{\fill}} r l l l l l l l l l}
Collaboration & Ref. & $\Nf$ & 
\hspace{0.15cm}\begin{rotate}{60}{publication status}\end{rotate}\hspace{-0.15cm} &
\hspace{0.15cm}\begin{rotate}{60}{continuum extrapolation}\end{rotate}\hspace{-0.15cm} &
\hspace{0.15cm}\begin{rotate}{60}{chiral extrapolation}\end{rotate}\hspace{-0.15cm}&
\hspace{0.15cm}\begin{rotate}{60}{finite volume}\end{rotate}\hspace{-0.15cm}&
\hspace{0.15cm}\begin{rotate}{60}{renormalization}\end{rotate}\hspace{-0.15cm}  &
\hspace{0.15cm}\begin{rotate}{60}{excited states}\end{rotate}\hspace{-0.15cm}  &
$\Delta u$ & $\Delta d$ \\
&&&&&&&&& & \\[-0.1cm]
\hline
\hline
&&&&&&&& &  \\[-0.1cm]

PNDME 20 & \cite{Park:2020axe} & 2+1+1 & \rC & \good$^\ddag$ & \good & \good & \good & \soso & 0.790(23)(30) & $-$0.425(15)(30)  \\[0.5ex]
ETM 19 & \cite{Alexandrou:2019brg} & 2+1+1 & \gA & \bad & \soso & \good & \good & \soso & 0.862(17) & $-$0.424(16) \\[0.5ex]
PNDME 18A & \cite{Lin:2018obj} & 2+1+1 & \gA & \good$^\ddag$ & \good & \good & \good & \soso & 0.777(25)(30)$^\#$ & $-$0.438(18)(30)$^\#$ \\[0.5ex]
\\[-0.1ex]\hline\\[0.2ex]
Mainz 19A & \cite{Djukanovic:2019gvi} & 2+1 & \rC & \good & \soso & \good & \good & \soso & 0.84(3)(4) & $-$0.40(3)(4)   \\[0.5ex]
$\chi$QCD 18 & \cite{Liang:2018pis} & 2+1 & \gA & \soso & \good & \good & \good & \soso & 0.847(18)(32)$^\$$ & $-$0.407(16)(18)$^\$$ \\[0.5ex]
\\[-0.1ex]\hline\\[0.2ex]
ETM 17C & \cite{Alexandrou:2017oeh} & 2 & \gA & \bad & \soso & \soso & \good & \soso & $0.830(26)(4)$ & $-0.386(16)(6)$ \\[0.5ex]
 & & & & & & & & & & \\[-0.1cm]
\hline
\hline
 & & & & & & & & & & \\[-0.1cm]
 & & & & & & & & & $\Delta s$& \\[-0.1cm]
 & & & & & & & & & &\\[-0.1cm]
\hline
\hline
     & & & & & & & & & &\\[-0.1cm]
PNDME 20 & \cite{Park:2020axe} & 2+1+1 & \rC & \good$^\ddag$ & \good & \good & \good & \soso &  $-$0.053(7) &  \\[0.5ex]
ETM 19 & \cite{Alexandrou:2019brg} & 2+1+1 & \gA & \bad & \soso & \good & \good & \soso & $-0.0458(73)$ & \\[0.5ex]
PNDME 18A & \cite{Lin:2018obj} & 2+1+1 & \gA & \good$^\ddag$ & \good & \good & \good & \soso &  $-$0.053(8)$^\#$ & \\[0.5ex]
\\[-0.1ex]\hline\\[0.2ex]
Mainz 19A & \cite{Djukanovic:2019gvi} & 2+1 & \rC & \good & \soso & \good & \good & \soso & $-$0.044(4)(5) &    \\[0.5ex]
$\chi$QCD 18 & \cite{Liang:2018pis} & 2+1 & \gA & \soso & \good & \good & \good & \soso &  $-$0.035(6)(7)$^\$$ & \\[0.5ex]
JLQCD 18 & \cite{Yamanaka:2018uud} & 2+1 & \gA & \bad & \soso & \soso & \good & \soso &  $-$0.046(26)(9)$^{\#}$ & \\[0.5ex]
$\chi$QCD 15 & \cite{Gong:2015iir} & 2+1 & \gA & \bad & \soso & \bad & \good & \soso &  $-$0.0403(44)(78)$^\#$ & \\[0.5ex]
Engelhardt 12 & \cite{Engelhardt:2012gd} & 2+1 & \gA & \bad & \soso & \bad & \good & \soso &  $-$0.031(17)$^\#$ & \\[0.5ex]
\\[-0.1ex]\hline\\[0.2ex]
ETM 17C & \cite{Alexandrou:2017oeh} & 2 & \gA & \bad & \soso & \soso & \good & \soso &  $-$0.042(10)(2) & \\[0.5ex]
&&&&&&&& \\[-0.1cm]
\hline
\hline
\end{tabular*}
\begin{minipage}{\linewidth}
{\footnotesize 
\begin{itemize}
\item[$^\#$] Assumed that $Z_A^{n.s.}=Z_A^{s}$. \\[-5mm]
\item[$^\ddag$] The rating takes into account that the action is not fully O(a) improved by requiring an additional lattice spacing. \\[-5mm]\item[$^\$$] For this partially quenched analysis the criteria are applied to the unitary points.
\end{itemize}
}
\end{minipage}
\caption{Overview of results for $g^q_A$.\label{tab:ga-singlet}}
\end{center}
\end{table}

\begin{figure}[!t]
\begin{center}
\includegraphics[width=7.5cm]{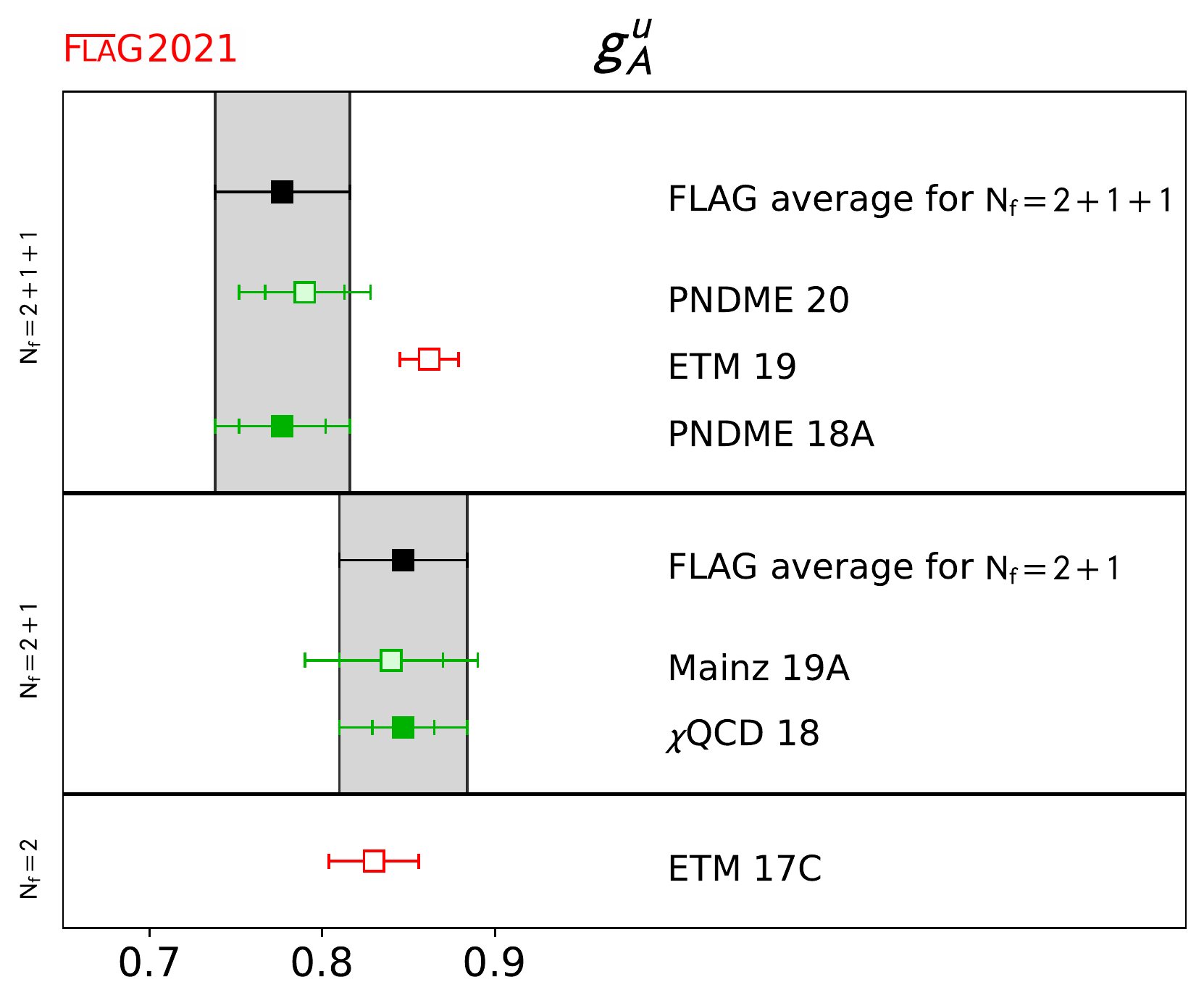}
\includegraphics[width=7.5cm]{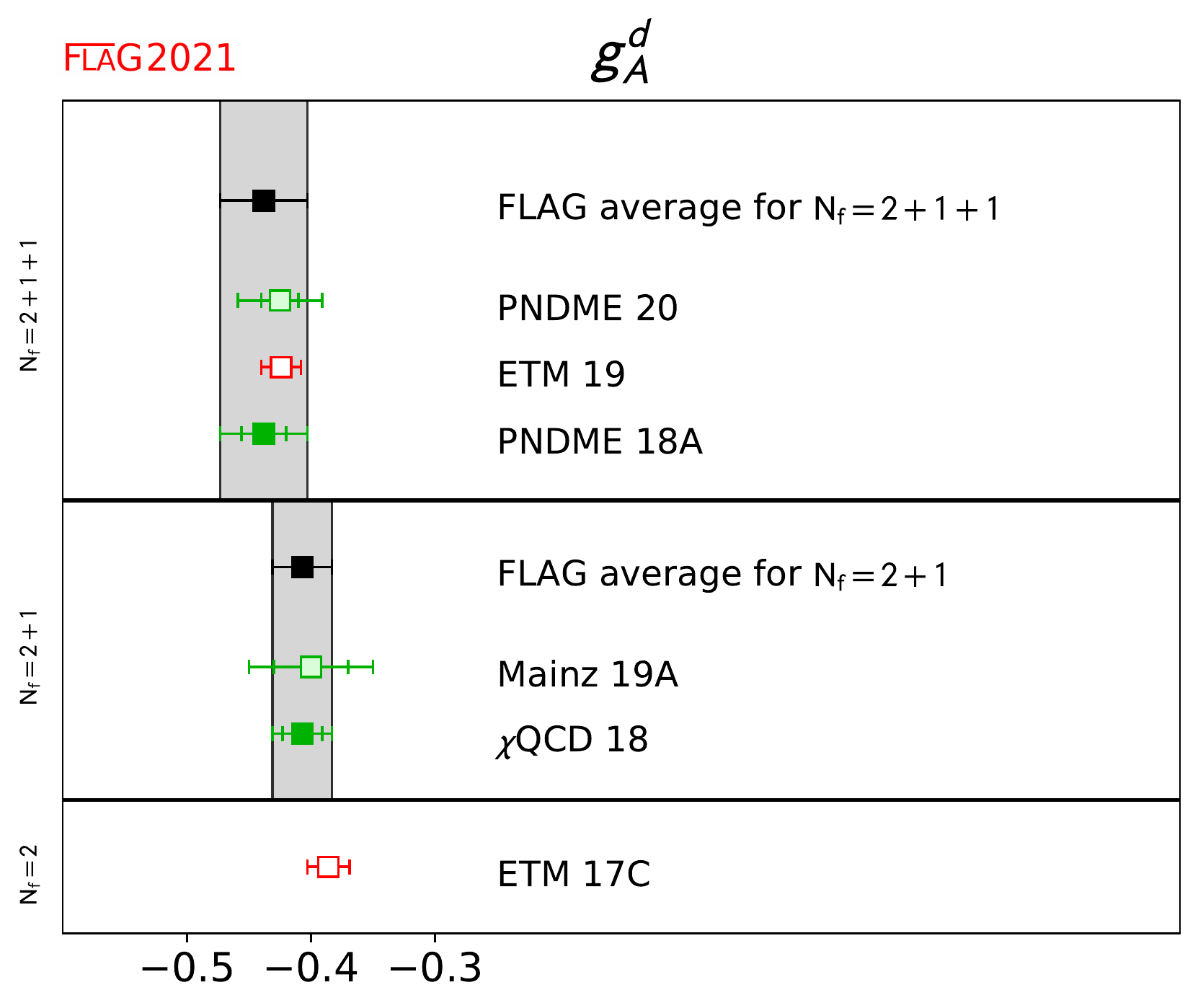}
\includegraphics[width=7.5cm]{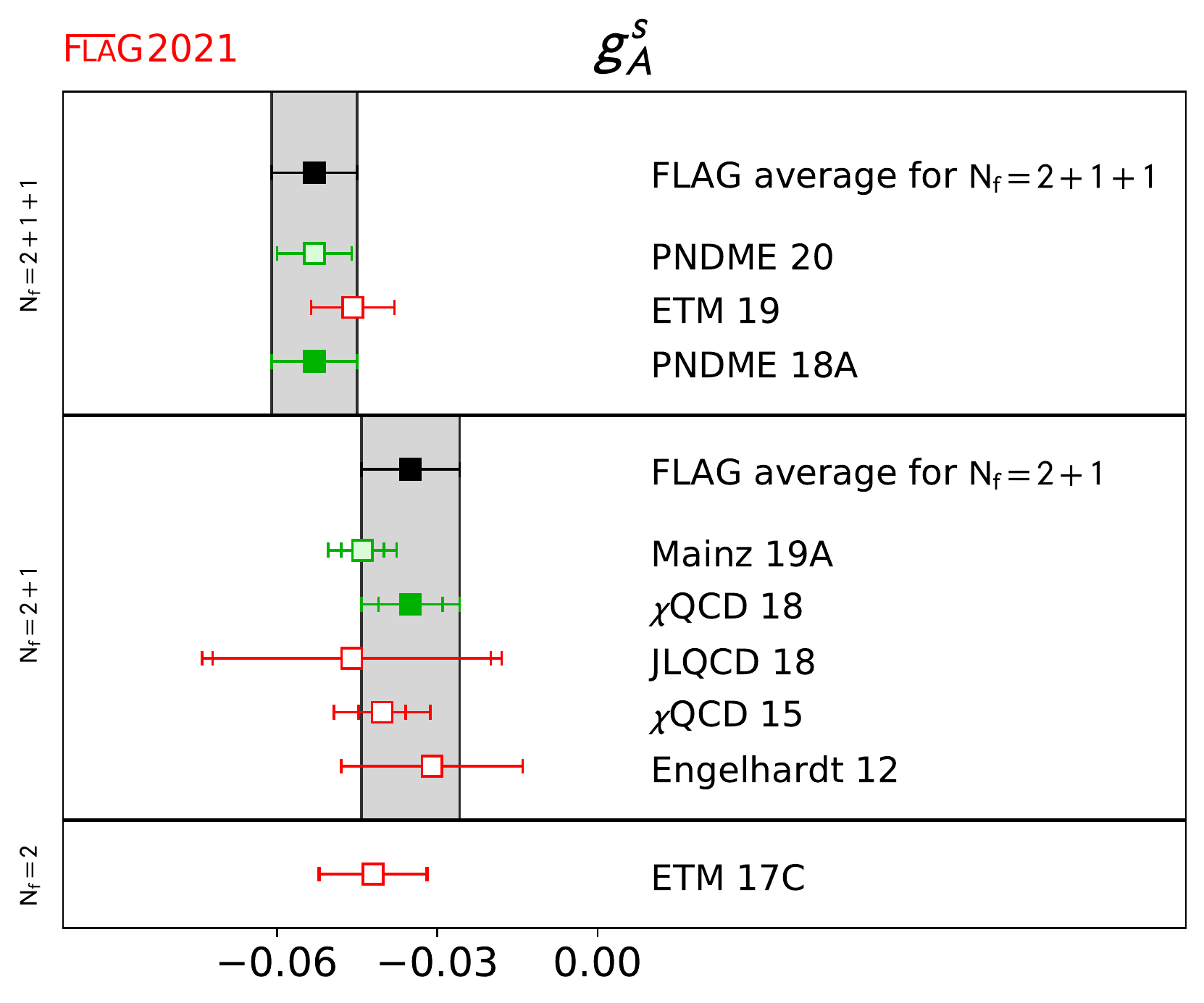}
\end{center}
\vspace{-1cm}
\caption{\label{fig:ga-singlet} Lattice results and FLAG averages for 
  $g_A^{u,d,s}$ for the $\Nf =
  2$, $2+1$, and $2+1+1$ flavour calculations.  }
\end{figure}

A compilation of results for the flavour-diagonal axial charges
for the proton is given in Tab.\,\ref{tab:ga-singlet} and plotted in
Fig.~\ref{fig:ga-singlet}.  Results for the neutron can be obtained by
interchanging the $u $ and $d$ flavour indices. Only two calculations 
already discussed in FLAG 19~\cite{Aoki:2019cca} 
qualify for global averages: the PNDME 18A \cite{Lin:2018obj} for
2+1+1 flavours and the $\chi$QCD 18 \cite{Liang:2018pis} for 2+1
flavours. 

The PNDME 18A~\cite{Lin:2018obj} results were obtained using the 2+1+1 flavour 
clover-on-HISQ formulation. The connected contributions were obtained
on 11 HISQ ensembles generated by the MILC collaboration with $a
\approx 0.057$, 0.87, 0.12 and 0.15~fm, $ M_\pi \approx 135$, 220 and
320~MeV, and $3.3 < M_\pi L < 5.5$. The light
disconnected contributions were obtained on six of these ensembles with the
lowest pion mass $M_\pi \approx 220$~MeV, while the strange disconnected
contributions were obtained on seven ensembles, i.e., including an additional one at $a \approx 0.087$~fm 
and $M_\pi \approx 135$~MeV. The excited state and the
chiral-continuum fits were done separately for the connected and
disconnected contributions, which introduces a systematic that is 
hypothesied  to be small as explained in Ref.~\cite{Lin:2018obj}. 
The analysis of the excited-state contamination, 
discussed in Sec.~\ref{sec:ESC}, was done using
three-state fits for the connected contribution and two-state fits for the
disconnected contributions. The chiral-continuum extrapolation was
done keeping the leading correction terms proportional to $M_\pi^2$ and $a$ in both cases, and 
the leading finite-volume correction in $M_\pi L$ was included in the analysis of the connected contributions.
Isovector renormalization constants,  
calculated on the lattice in the RI-SMOM scheme and converted to $\msbar$, 
are used for all three flavour diagonal operators. 

The PNDME 20~\cite{Park:2020axe} provided a status update to PNDME 18A~\cite{Lin:2018obj} 
and presented results showing that flavour mixing in the calculation of renormalization constants is small, and 
the isovector renormalization factor 
is a good approximation for renormalizing flavour diagonal axial charges as discussed in Sec.~\ref{sec:renorm}. 
It is not considered for the average as it is a conference proceeding. 

The ETM 19~\cite{Alexandrou:2019brg} presented new results for $g_A^{u,d,s,c}$ from 
a single ensemble with 2+1+1-flavour twisted-mass fermions with a clover term at $a=0.0801(4)$~fm 
and $M_\pi= 139.3(7)$~MeV. These are not
considered for the averages as they do not satisfy the criteria for
the continuum extrapolation.

The 2+1+1 flavour FLAG values for the axial charges $g_A^{u,d,s}$ of
the proton are, therefore, the same as the corresponding results given
in Tab.~\ref{tab:ga-singlet} and unchanged from FLAG 19~\cite{Aoki:2019cca}: 
\begin{align}
&  \mbox{}\Nf=2+1+1: &\FLAGAVBEGIN g_A^u  &= \phantom{-}0.777(25)(30)  \FLAGAVEND   &&\Ref~\mbox{\cite{Lin:2018obj}}, \\
&  \mbox{}\Nf=2+1+1: &\FLAGAVBEGIN g_A^d  &=           -0.438(18)(30)  \FLAGAVEND   &&\Ref~\mbox{\cite{Lin:2018obj}}, \\
&  \mbox{}\Nf=2+1+1: &\FLAGAVBEGIN g_A^s  &=           -0.053(8)   \FLAGAVEND   &&\Ref~\mbox{\cite{Lin:2018obj}}. 
\end{align}

There are also new results for $g_A^{u,d,s}$ from Mainz 19A \cite{Djukanovic:2019gvi} with 
2+1-flavour ensembles. While they satisfy all the criteria, they are not included in the 
averages as~\cite{Djukanovic:2019gvi} is a conference proceeding. 

The 2+1 flavour FLAG results from $\chi$QCD 18 \cite{Liang:2018pis} were obtained using
the overlap-on-domain-wall formalism. Three
domain-wall ensembles with lattice spacings 0.143, 0.11 and 0.083~fm
and sea-quark pion masses $M_\pi = 171$, 337 and 302~MeV,
respectively, were analyzed.  In addition to the three approximately
unitary points, the paper presents data for an additional 4--5 valence
quark masses on each ensemble, i.e., partially quenched data. Separate
excited-state fits were done for the connected and disconnected
contributions.  The continuum, chiral and volume extrapolation to the
combined unitary and nonunitary data is made including terms proportional 
to both $M_{\pi,{\rm valence}}^2$ and $M_{\pi,{\rm sea}}^2$, and two $\cO(a^2)$ discretization terms 
for the two different domain-wall actions. With just three unitary points, not all the coefficients  
are well constrained. The $M_{\pi,sea}$ dependence is omitted and considered as a systematic, 
and a prior is used for the coefficients of the $a^2$ terms to stabilize the fit.
These $\chi$QCD~18 2+1 flavour results for the proton,
which supersede the $\chi$QCD~15 \cite{Gong:2015iir} analysis, are
\begin{align}
&  \mbox{}\Nf=2+1: &\FLAGAVBEGIN g_A^u  &= \phantom{-}0.847(18)(32)  \FLAGAVEND   &&\Ref~\mbox{\cite{Liang:2018pis}}, 
\\
&  \mbox{}\Nf=2+1: &\FLAGAVBEGIN g_A^d  &=           -0.407(16)(18)  \FLAGAVEND   &&\Ref~\mbox{\cite{Liang:2018pis}}, 
\\
&  \mbox{}\Nf=2+1: &\FLAGAVBEGIN g_A^s  &=           -0.035(6)(7)    \FLAGAVEND   &&\Ref~\mbox{\cite{Liang:2018pis}}. 
\end{align}

The JLQCD~18 \cite{Yamanaka:2018uud},
ETM~17C \cite{Alexandrou:2017oeh} and
Engelhardt~12 \cite{Engelhardt:2012gd} calculations were not
considered for the averages as they did not satisfy the criteria for
the continuum extrapolation. All three calculations were done at a
single lattice spacing. The JLQCD~18 calculation used overlap
fermions and the Iwasaki gauge action. They perform a chiral fit using
data at four pion masses in the range 290--540~MeV. Finite volume
corrections are assumed to be negligible since each of the two pairs
of points on different lattice volumes satisfy $M_\pi L \geq 4$.  The
ETM~17C calculation is based on a single twisted-mass ensemble with
$M_\pi=130$~MeV, $a=0.094$ and a relatively small $M_\pi L = 2.98$.
Engelhardt~12 \cite{Engelhardt:2012gd} calculation was done on three asqtad ensembles with
$M_\pi = 293$, 356 and 495~MeV, but all at a single lattice spacing
$a=0.124$~fm.

Results for $g_A^s$ were also presented recently by LHPC in
Ref.~\cite{Green:2017keo}.  However, this calculation is not included in 
Tab.\,\ref{tab:ga-singlet} 
as it has been performed on a single ensemble with $a=$ 0.114~fm and a
heavy pion mass value of $M_\pi \approx 317$~MeV.

\subsubsection{Results for $g_S^{u,d,s}$ from direct and hybrid calculations of the matrix elements\label{sec:gS-FD}}

The sigma terms $\sigma_q=m_q\langle N|\bar{q}q|N\rangle=m_q g_S^q$ or
the quark mass fractions $f_{T_q}=\sigma_q/M_N$ are normally computed
rather than $g_S^q$.  These combinations have the advantage of being
renormalization group invariant in the continuum, and this holds on
the lattice for actions with good chiral properties, see
Sec.~\ref{sec:renorm} for a discussion. In order to aid comparison
with phenomenological estimates, e.g. from $\pi$-$N$
scattering~\cite{Alarcon:2011zs,Chen:2012nx,Hoferichter:2015dsa}, the
light quark sigma terms are usually added to give the $\pi N$ sigma
term, $\sigma_{\pi N}=\sigma_u+\sigma_d$. The direct evaluation of the
sigma terms involves the calculation of the corresponding three-point
correlation functions for different source-sink separations
$\tau$. For $\sigma_{\pi N}$ there are both connected and disconnected
contributions, while for most lattice fermion formulations only
disconnected contributions are needed for $\sigma_s$.  The techniques
typically employed lead to the availability of a wider range of $\tau$
for the disconnected contributions compared to the connected
ones~(both, however, suffer from signal-to-noise problems for large
$\tau$, as discussed in Sec.~\ref{sec:intro}) and we only comment on
the range of $\tau$ computed for the latter in the following.

Recent results for $\sigma_{\pi N}$ and for $\sigma_s$ from the direct
approach are compiled in Tab.~\ref{tab:gs-ud-s-direct}.  ETM
19~\cite{Alexandrou:2019brg} (discussed below) is the only new study
included in this table since the last FLAG
report. Ref.~\cite{Gupta:2021ahb} is also new, however, it was
submitted to the arXiv after the deadline and will be reviewed in the
next edition. For completeness, the descriptions of
other works are reproduced from FLAG 19.

For both $\sigma_{\pi N}$ and for
$\sigma_s$, only the results from $\chi$QCD~15A~\cite{Yang:2015uis}
qualify for global averaging.  In this mixed-action study, three
RBC/UKQCD $\Nf=2+1$ domain-wall ensembles are analyzed comprising two
lattice spacings, $a=0.08$~fm with $M_{\pi,\rm sea}=300$~MeV and
$a=0.11$~fm with $M_{\pi,\rm sea}=330$~MeV and $139$~MeV. Overlap
fermions are employed with a number of nonunitary valence quark
masses. The connected three-point functions are measured with three
values of $\tau$ in the range 0.9--1.4~fm. A combined chiral,
continuum and volume extrapolation is performed for all data with
$M_\pi<350$~MeV. The leading order expressions are taken for the
lattice-spacing and volume dependence while partially quenched $SU(2)$
HB$\chi$PT up to $M_\pi^3$ terms models the chiral behaviour for
$\sigma_{\pi N}$. The strange-quark sigma term has a milder dependence
on the pion mass and only the leading-order quadratic terms are
included in this case.

\begin{table}[t!]
\begin{center}
\mbox{} \\[3.0cm]
\footnotesize
\begin{tabular*}{\textwidth}[l]{l @{\extracolsep{\fill}} r l l l l l l l l l}
Collaboration & Ref. & $\Nf$ & 
\hspace{0.15cm}\begin{rotate}{60}{publication status}\end{rotate}\hspace{-0.15cm} &
\hspace{0.15cm}\begin{rotate}{60}{continuum extrapolation}\end{rotate}\hspace{-0.15cm} &
\hspace{0.15cm}\begin{rotate}{60}{chiral extrapolation}\end{rotate}\hspace{-0.15cm}&
\hspace{0.15cm}\begin{rotate}{60}{finite volume}\end{rotate}\hspace{-0.15cm}&
\hspace{0.15cm}\begin{rotate}{60}{renormalization}\end{rotate}\hspace{-0.15cm}  &
\hspace{0.15cm}\begin{rotate}{60}{excited states}\end{rotate}\hspace{-0.15cm}&
$\sigma_{\pi N}$~[MeV]  &
$\sigma_s$~[MeV]\\
&&&&&&&&& \\[-0.1cm]
\hline
\hline
&&&&&&&& \\[-0.1cm]
ETM 19 & \cite{Alexandrou:2019brg} & 2+1+1 & \gA & \bad & \soso & \good & na/na & \soso & 41.6(3.8) & 45.6(6.2) \\[0.5ex]
\\[-0.1ex]\hline\\[0.2ex]
JLQCD 18 & \cite{Yamanaka:2018uud} & 2+1 & \gA & \bad & \soso & \soso & na/na & \soso & 26(3)(5)(2) & 17(18)(9) \\[0.5ex]
$\chi$QCD 15A & \cite{Yang:2015uis} & 2+1 & \gA & \soso & \good & \good & na/na & \soso & 45.9(7.4)(2.8)$^\$$ & 40.2(11.7)(3.5)$^\$$ \\[0.5ex]
$\chi$QCD 13A & \cite{Gong:2013vja} & 2+1 & \gA & \bad & \bad & \soso & $-$/na & \soso & $-$ & 33.3(6.2)$^\$$ \\[0.5ex]
JLQCD 12A & \cite{Oksuzian:2012rzb} & 2+1 & \gA & \bad & \soso & \soso & $-$/na & \soso & $-$ & 0.009(15)(16)$\times m_N$$^\dagger$ \\[0.5ex]
Engelhardt 12 & \cite{Engelhardt:2012gd} & 2+1 & \gA & \bad & \soso & \bad & $-$/na & \soso & $-$ & 0.046(11)$\times m_N$$^\dagger$ \\[0.5ex]
\\[-0.1ex]\hline\\[0.2ex]
ETM 16A & \cite{Abdel-Rehim:2016won} & 2 & \gA & \bad & \soso & \soso & na/na & \soso & 37.2(2.6)($^{4.7}_{2.9}$) & 41.1(8.2)($^{7.8}_{5.8}$) \\[0.5ex]
RQCD 16 & \cite{Bali:2016lvx} & 2 & \gA & \soso & \good & \good & na/\good & \bad & 35(6) & 35(12) \\[0.5ex]
&&&&&&&& \\[-0.1cm]
\hline
\hline
&&&&&&&& \\[-0.1cm]
MILC 12C & \cite{Freeman:2012ry} & 2+1+1 & \gA & \good & \good & \good & $-$/\soso & \soso & $-$ & 0.44(8)(5)$\times m_s$$^{\P\S}$ \\[0.5ex]
\\[-0.1ex]\hline\\[0.2ex]
MILC 12C & \cite{Freeman:2012ry} & 2+1 & \gA & \good & \soso & \good & $-$/\soso & \soso & $-$ &0.637(55)(74)$\times m_s$$^{\P\S}$ \\[0.5ex]
MILC 09D & \cite{Toussaint:2009pz} & 2+1 & \gA & \good & \soso & \good & $-$/na & \soso & $-$ &59(6)(8)$^\S$ \\[0.5ex]
&&&&&&&& \\[-0.1cm]
\hline
\hline
\end{tabular*}
\begin{minipage}{\linewidth}
{\footnotesize The renormalization criteria is given for $\sigma_{\pi N}$~(first) and $\sigma_s$~(second). The
  label ’na’ indicates that no renormalization is required.
\begin{itemize}
\item[$^\$$] For this partially quenched analysis the criteria are applied to the unitary points. \\[-5mm] 
\item[$\dagger$] This study computes the strange quark fraction $f_{T_s}/m_N$. \\[-5mm] 
\item[$^\S$] This study employs a hybrid method, see Ref.~\cite{Toussaint:2009pz}. \\[-5mm] 
\item[$^\P$] The matrix element $\langle N|\bar{s}s|N\rangle$ at the scale $\mu=2$~GeV in the $\msbar$ scheme is computed.
\end{itemize}
}
\end{minipage}
\caption{Overview of results for $\sigma_{\pi N}$ and $\sigma_s$ from the direct approach~(above) and $\sigma_s$ from the hybrid approach~(below).  \label{tab:gs-ud-s-direct}}
\end{center}
\end{table}

The lack of other qualifying studies is an indication of the
difficulty and computational expense of performing these
calculations. Nonetheless, this situation is likely to improve in the
future. We note that although the recent analyses,
ETM~16A~\cite{Abdel-Rehim:2016won}, 
the new study ETM 19~\cite{Alexandrou:2019brg} and
JLQCD~18~\cite{Yamanaka:2018uud}, are at a single lattice
spacing~($a=0.09$~fm, 0.08~fm and 0.11~fm, respectively),
they satisfy the criteria for chiral extrapolation, finite volume and
excited states. ETM~16A is a single ensemble study with $\Nf=2$
twisted-mass fermions with a pion mass close to the physical point and
$M_\pi L=3.0$. Excited states are investigated utilizing $\tau=0.9$~fm
up to $\tau=1.7$~fm for the connected three-point functions. 
  In ETM 19 a high statistics analysis was carried out employing
  a $N_f=2+1+1$ physical point ensemble and seven source-sink
  separations in the range $\tau$ = 0.6--1.6~fm, improving the precision
  they obtain for both $\sigma_{\pi N}$ and $\sigma_s$ compared to their $N_f=2$ results. JLQCD in
JLQCD~18 utilize $\Nf=2+1$ overlap fermion ensembles with pion masses
reaching down to 293~MeV~($M_\pi L=4.0$) and apply techniques which
give a wide range of $\tau$ for the connected contribution, with the
final results extracted from $\tau\ge 1.2$~fm.

RQCD (RQCD~16~\cite{Bali:2016lvx}) investigates the continuum,
physical quark mass and infinite-volume limits, where the lattice
spacing spans the range 0.06--0.08~fm, the minimum $M_\pi$ is
$150$~MeV and $M_\pi L$ is varied between 3.4 to 6.7 at
$M_\pi=290$~MeV.  This $\Nf=2$ study has a red tag for the excited
state criterion as multiple source-sink separations for the connected
three-point functions are only computed on a subset of the
ensembles. Clover fermions are employed and the lack of good chiral
properties for this action means that there is mixing between quark
flavours under renormalization when determining $\sigma_s$ and a
gluonic term needs to be considered for full $\cO(a)$ improvement~(which
has not been included, see Sec.~\ref{sec:renorm} for a discussion).

Earlier work focuses only on $\sigma_s$. The analysis of
JLQCD~12A~\cite{Oksuzian:2012rzb}, is performed on the same set of
ensembles as the JLQCD~18 study discussed above and in addition
includes smaller volumes for the lightest two pion masses.\footnote{JLQCD  also determine $f_{T_s}$ in Ref.~\cite{Takeda:2010cw} in a single lattice
spacing study on small volumes with heavy pion masses.} No
significant finite-volume effects are
observed. Engelhardt~12~\cite{Engelhardt:2012gd} and
$\chi$QCD~13A~\cite{Gong:2013vja} have less control over the
systematics. The former is a single lattice spacing analysis restricted
to small spatial volumes while the latter is a partially quenched
study on a single ensemble with unitary $M_\pi>300$~MeV.

MILC has also computed $\sigma_s$ using a hybrid
method~\cite{Toussaint:2009pz} which makes use of the
Feynman-Hellmann~(FH) theorem and involves evaluating the nucleon matrix
element $\langle N|\int\! d^4\!x\, \bar{s}s|N\rangle$.\footnote{Note
  that in the direct method the matrix element $\langle N|\int\!
  d^3\!x\, \bar{s}s|N\rangle$, involving the spatial volume sum, is
  evaluated for a fixed timeslice.} This method is applied in
MILC~09D~\cite{Toussaint:2009pz} to the $\Nf=2+1$ asqtad ensembles
with lattice spacings $a=$ 0.06, 0.09, 0.12~fm and values of $M_\pi$
ranging down to 224~MeV. A continuum and chiral extrapolation is
performed including terms linear in the light-quark mass and quadratic
in $a$.  As the coefficient of the discretization term is poorly
determined, a Bayesian prior is used, with a width corresponding to a
10\% discretization effect between the continuum limit and the
coarsest lattice spacing.\footnote{This is consistent with
  discretization effects observed in other quantities at $a=0.12$~fm.}
A similar updated analysis is presented in
MILC~12C~\cite{Freeman:2012ry}, with an improved evaluation of
$\langle N|\int\! d^4\!x\, \bar{s}s|N\rangle$ on a subset of the $\Nf=2+1$
asqtad ensembles. The study is also extended to HISQ $\Nf=2+1+1$
ensembles comprising four lattice spacings with $a=$ 0.06--0.15~fm and a
minimum pion mass of 131~MeV.  Results are presented for
$g_S^s=\langle N|\bar{s}s|N\rangle$~(in the $\msbar$ scheme at 2~GeV)
rather than for $\sigma_s$. The scalar matrix element is renormalized
for both three and four flavours using the 2-loop factor for the
asqtad action~\cite{Mason:2005bj}. The error incurred by applying the
same factor to the HISQ results is expected to be small.\footnote{At
  least at 1-loop the $Z$ factors for HISQ and asqtad are very
  similar, cf.  Ref.~\cite{McNeile:2012xh}.}

Both MILC~09D and MILC~12C achieve green tags for all the criteria,
see Tab.~\ref{tab:gs-ud-s-direct}. As the same set of asqtad ensembles is
utilized in both studies we take MILC~12C as superseding MILC~09D
for the three-flavour case. The global averaging is discussed in
Sec.~\ref{sec:gS-sum}.

\subsubsection{Results for $g_S^{u,d,s}$ using the Feynman-Hellmann theorem\label{sec:gS-FD-FH}}

An alternative approach for accessing the sigma terms is to determine
the slope of the nucleon mass as a function of the quark masses, or
equivalently, the squared pseudoscalar meson masses. The Feynman-Hellman~(FH)
theorem gives
\begin{equation}
\sigma_{\pi N}=m_u\frac{\partial M_N}{\partial m_u}+ m_d\frac{\partial M_N}{\partial m_d}\approx M_\pi^2 \frac{\partial M_N}{\partial M_\pi^2},\hspace{0.7cm} \sigma_s = m_s \frac{\partial M_N}{\partial m_s}\approx M_{\bar{s}s}^2 \frac{\partial M_N}{\partial M_{\bar{s}s}^2},\label{eq:fheq1}
\end{equation}
where the fictitious $\bar{s}s$ meson has a mass squared
$M^2_{\bar{s}s}=2M_K^2-M_\pi^2$.  In principle this is a
straightforward method as the nucleon mass can be extracted from fits
to two-point correlation functions, and a further fit to $M_N$ as a
function of $M_\pi$~(and also $M_K$ for $\sigma_s$) provides the
slope. Nonetheless, this approach presents its own challenges: a
functional form for the chiral behaviour of the nucleon mass is
needed, and while baryonic $\chi$PT~(B$\chi$PT) is the natural choice,
the convergence properties of the different formulations are not well
established. Results are sensitive to the formulation chosen and the
order of the expansion employed. If there is an insufficient number of
data points when implementing higher order terms, the coefficients are
sometimes fixed using additional input, e.g. from analyses of
experimental data. This may influence the slope extracted. Simulations
with pion masses close to or bracketing the physical point can
alleviate these difficulties. In some studies the nucleon mass is used
to set the lattice spacing. This naturally forces the fit to reproduce
the physical nucleon mass at the physical point and may affect the
extracted slope. Note that, if the nucleon mass is fitted as
a  function of the pion and kaon masses, the dependence of the meson masses on
  the quark masses also, in principle, needs to be considered in order to extract the sigma terms.

\begin{table}[t!]
\begin{center}
\mbox{} \\[3.0cm]
\footnotesize
\begin{tabular*}{\textwidth}[l]{l @{\extracolsep{\fill}} r l l l l l l l }
Collaboration & Ref. & $\Nf$ & 
\hspace{0.15cm}\begin{rotate}{60}{publication status}\end{rotate}\hspace{-0.15cm} &
\hspace{0.15cm}\begin{rotate}{60}{continuum extrapolation}\end{rotate}\hspace{-0.15cm} &
\hspace{0.15cm}\begin{rotate}{60}{chiral extrapolation}\end{rotate}\hspace{-0.15cm}&
\hspace{0.15cm}\begin{rotate}{60}{finite volume}\end{rotate}\hspace{-0.15cm}&
$\sigma_{\pi N}$~[MeV] & $\sigma_s$~[MeV] \\
 & & & & & & & &\\[-0.1cm]
\hline
\hline
 & & & & & & & &\\[-0.1cm]
BMW 20A & \cite{Borsanyi:2020bpd} & 1+1+1+1 & \oP & \good$^\ddag$ & \good & \good & 0.0398(32)(44)$\times m_N$$^\dagger$ & 0.0577(46)(33)$\times m_N$$^\dagger$ \\[0.5ex]
ETM 14A & \cite{Alexandrou:2014sha} & 2+1+1 & \gA & \good & \soso & \soso & 64.9(1.5)(13.2)$^\triangle$ & $-$ \\[0.5ex]
\\[-0.1ex]\hline\\[0.2ex]
BMW 15 & \cite{Durr:2015dna} & 2+1 & \gA & \good$^\ddag$ & \good & \good & 38(3)(3) & 105(41)(37) \\[0.5ex]
Junnarkar 13 & \cite{Junnarkar:2013ac} & 2+1 & \gA & \soso & \soso & \soso & $-$ & 48(10)(15) \\[0.5ex]
Shanahan 12 & \cite{Shanahan:2012wh} & 2+1 & \gA & \bad & \soso & \soso & 45(6)/51(7)$^\star$ & 21(6)/59(6)$^\star$ \\[0.5ex]
JLQCD 12A & \cite{Oksuzian:2012rzb} & 2+1 & \gA & \bad & \soso & \soso & $-$ & 0.023(29)(28)$\times m_N$$^\dagger$ \\[0.5ex]
QCDSF 11 & \cite{Horsley:2011wr} & 2+1 & \gA & \bad & \bad & \soso & 31(3)(4) & 71(34)(59) \\[0.5ex]
BMW 11A & \cite{Durr:2011mp} & 2+1 & \gA & \soso$^\ddag$ & \good & \soso & 39(4)($^{18}_{7}$) & 67(27)($^{55}_{47}$) \\[0.5ex]
Martin~Camalich 10 & \cite{MartinCamalich:2010fp} & 2+1 & \gA & \bad & \good & \bad & 59(2)(17) & $-$4(23)(25) \\[0.5ex]
PACS-CS 09 & \cite{Ishikawa:2009vc} & 2+1 & \gA & \bad & \good & \bad & 75(15) & $-$ \\[0.5ex]
Walker-Loud 08 & \cite{Walker-Loud:2008rui} & 2+1 & \gA & \bad & \soso & \bad & 84(17)(20)/42(14)(9)$^\star$ & $-$ \\[0.5ex]
\\[-0.1ex]\hline\\[0.2ex]
QCDSF 12 & \cite{Bali:2012qs} & 2 & \gA & \soso & \good & \soso & 37(8)(6) & $-$ \\[0.5ex]
JLQCD 08B & \cite{Ohki:2008ff} & 2 & \gA & \bad & \soso & \bad & 53(2)($^{+21}_{-7}$) & $-$ \\[0.5ex]
&&&&&&&& \\[-0.1cm]
\hline
\hline
\end{tabular*}
\begin{minipage}{\linewidth}
{\footnotesize 
\begin{itemize}
 \item[$^\triangle$]  Two results for $\sigma_{\pi N}$ are quoted arising from different fit ans\"atze to the nucleon mass. The systematic error  is the same as in  Ref.~\cite{Alexandrou:2017xwd} for a combined $N_f=2$ and $N_f=2+1+1$ analysis~\cite{Kallidonis:pc2018}.
\\[-5mm] \item[$^\ddag$]The rating takes into account that the action is not fully O(a) improved by requiring an additional lattice spacing.
\\[-5mm] \item[$^\star$] Two results are quoted.  
\\[-5mm] \item[$^\dagger$] The quark fractions $f_{T_{ud}}=f_{T_{u}}+f_{T_{d}}=\sigma_{\pi N}/m_N$ and/or $f_{T_s}=\sigma_s/m_N$ are computed.
\end{itemize}
}
\end{minipage}
\caption{Overview of results for $\sigma_{\pi N}$ and $\sigma_s$ from the Feynman-Hellmann approach.  \label{tab:gs-singlet-fh}}
\end{center}
\end{table}

An overview of recent determinations of $\sigma_{\pi N}$ and
$\sigma_s$ is given in Tab.~\ref{tab:gs-singlet-fh}.  BMW
20A~\cite{Borsanyi:2020bpd} (discussed below) is the only new study
since the last FLAG report. For completeness, the descriptions of other works are
reproduced from FLAG 19.  Note that the renormalization criterion is
not included in Tab.~\ref{tab:gs-singlet-fh} as renormalization is not
normally required when computing the sigma terms in the
Feynman-Hellmann approach.\footnote{An exception to this is when
clover fermions are employed. In this case one must take care of the
mixing between quark flavours when renormalizing the quark masses that
appear in Eq.~\eqref{eq:fheq1}.}  At present, a rating indicating
control over excited state contamination is also not considered since
a wide range of source-sink separations are available for nucleon
two-point functions and ground state dominance is normally achieved.
This issue may be revisited in the future as statistical precision
improves and this systematic is further investigated.

There are several results for $\sigma_{\pi N}$ that can be included in
a global average. For $\Nf=2$, one study meets the selection
criteria.\footnote{The ETM collaboration also determine $\sigma_{\pi N}$ in
  Ref.~\cite{Alexandrou:2009qu} as part of an $\Nf=2$ analysis to
  determine the lattice spacing from the nucleon mass. However, no
  final result is given.} The analysis of
QCDSF~12~\cite{Bali:2012qs} employs nonperturbatively improved
clover fermions over three lattice spacings ($a=$ 0.06--0.08~fm) with pion
masses reaching down to around 160~MeV. Finite volume corrected
nucleon masses are extrapolated via $\cO(p^4)$ covariant B$\chi$PT with
three free parameters. The other coefficients are taken from
experiment, phenomenology or FLAG, with the corresponding
uncertainties accounted for in the fit for those coefficients that
are not well known. The nucleon mass is used to set the scale.  A
novel feature of this study is that a direct determination of
$\sigma_{\pi N}$ at around $M_\pi=290$~MeV was used as an additional
constraint on the slope.

Turning to $\Nf=2+1$, two studies performed by the BMW collaboration
and one by $\chi$QCD are relevant. In
BMW~11A~\cite{Durr:2011mp}, stout-smeared tree-level clover fermions
are employed on 15 ensembles with simulation parameters encompassing
$a$ = 0.06--0.12~fm, $M_\pi \sim$ 190--550~MeV and $M_\pi L \gsim 4$.
Taylor, Pad\'{e} and covariant $SU(3)$ B$\chi$PT fit forms are
considered. Due to the use of smeared gauge links, discretization
effects are found to be mild even though the fermion action is not
fully $\cO(a)$ improved. Fits are performed including an $\cO(a)$ or
$\cO(a^2)$ term and also without a lattice-spacing dependent term.
Finite volume effects were assessed to be small in an earlier
work~\cite{Durr:2008zz}. The final results are computed considering
all combinations of the fit ansatz weighted by the quality of the fit.
In BMW~15~\cite{Durr:2015dna}, a more extensive analysis on 47
ensembles is presented for HEX-smeared clover fermions involving five
lattice spacings and pion masses reaching down to 120~MeV. Bracketing
the physical point reduces the reliance on a chiral extrapolation.
Joint continuum, chiral and infinite-volume extrapolations are carried
out for a number of fit parameterisations with the final results
determined via the Akaike information criterion
procedure~\cite{1100705}. Although only $\sigma_{\pi N}$ is accessible
in the FH approach in the isospin limit, the individual quark
fractions $f_{T_q}=\sigma_q/M_N$ for $q=u,d$ for the proton and the
neutron are also quoted in BMW~15, using isospin
relations.\footnote{These isospin relations were also derived in
Ref.~\cite{Crivellin:2013ipa}.}

Regarding $\Nf=2+1+1$, there is only one recent study. In
ETM~14A~\cite{Alexandrou:2014sha}, fits are performed to the nucleon
mass utilizing $SU(2)$ $\chi$PT for data with $M_\pi \ge 213$~MeV as
part of an analysis to set the lattice spacing. The expansion is
considered to $\cO(p^3)$ and $\cO(p^4)$, with two and three of the coefficients
as free parameters, respectively.  The difference between the two fits
is taken as the systematic error. No discernable discretization or
finite-volume effects are observed where the lattice spacing is varied
over the range $a$ = 0.06--0.09~fm and the spatial volumes cover 
$M_\pi L=3.4$ up to $M_\pi L>5$. The results are unchanged when a near
physical point $\Nf=2$ ensemble is added to the analysis in
Ref.~\cite{Alexandrou:2017xwd}.

Since FLAG 19, BMW have performed a new $N_f=1+1+1+1$ study
 BMW 20A~\cite{Borsanyi:2020bpd}. A two step analysis is
followed: the dependence of the nucleon mass on the pion and kaon
masses is determined on HEX-smeared clover ensembles with
$a=$ 0.06--0.1~fm and pion masses in the range $M_\pi=$ 195--420~MeV.
The meson masses as a function of the quark masses are evaluated on
stout-staggered ensembles with a similar range in $a$ and quark masses
which bracket their physical values.  As \cite{Borsanyi:2020bpd} is a
preprint, the result is not considered for the average.

We note that
the $N_f=2+1$ study by $\chi$QCD~\cite{Yang:2018nqn} based on overlap valence
fermions on four domain-wall fermion ensembles with $a=$ 0.08--0.14~fm
and $M_\pi$ down to the physical point is also new. However, since
$\sigma_{\pi N}$ is determined from a single fit and the systematic
uncertainties are not estimated, we do not present the result in
the table.

Other determinations of $\sigma_{\pi N}$ in
Tab.~\ref{tab:gs-singlet-fh} receive one or more red tags.
Walker-Loud 08~\cite{Walker-Loud:2008rui}, JLQCD~08B~\cite{Ohki:2008ff}, PACS-CS~09~\cite{Ishikawa:2009vc} and
QCDSF~11~\cite{Horsley:2011wr} are single lattice spacing studies.
In addition, the volume for the minimum pion mass is rather small for
Walker-Loud 08, JLQCD~08B and PACS-CS~09, while QCDSF~11 is restricted to heavier
pion masses.

We also consider publications that are based on results for baryon
masses found in the literature. As different lattice setups~(in terms
of $\Nf$, lattice actions, etc.) will lead to different systematics,
we only include works in Tab.~\ref{tab:gs-singlet-fh}  which utilize a
single setup. These correspond to Shanahan~12~\cite{Shanahan:2012wh}
and Martin~Camalich~10~\cite{MartinCamalich:2010fp}, which fit
PACS-CS data~\cite{Aoki:2008sm}~(the PACS-CS~09 study is also based
on these results). Note that Shanahan~12 avoids a red tag for the
volume criterion as the lightest pion mass ensemble is omitted.
Recent studies which combine data from different setups/collaborations
are displayed for comparison in Figs.~\ref{fig:gs-sum-light}
and~\ref{fig:gs-sum-strange} in the next section.

Several of the above studies have also determined the strange quark
sigma term. This quantity is difficult to access via the
Feynman-Hellmann method since in most simulations the physical point
is approached by varying the light-quark mass, keeping $m_s$
approximately constant. While additional ensembles can be generated,
it is hard to resolve a small slope with respect to $m_s$. Such
problems are illustrated by the large uncertainties in the results
from BMW~11A and BMW~15. Alternative approaches have been pursued
in QCDSF~11, where the physical point is approached along a
trajectory keeping the average of the light- and strange-quark masses
fixed, and JLQCD~12A~\cite{Oksuzian:2012rzb}, where quark mass
reweighting is applied.  The latter is a single lattice spacing study.
One can also fit to the whole baryon octet and apply $SU(3)$ flavour
symmetry constraints as investigated in, e.g. Martin~Camalich~10,
Shanahan~12, QCDSF~11 and BMW~11A.

The determinations of $\sigma_s$ in BMW~11A and BMW~15 qualify for
averaging. The mixed action study of
Junnarkar~13~\cite{Junnarkar:2013ac} with domain-wall valence fermions
on MILC $\Nf=2+1$ asqtad ensembles also passes the FLAG criteria. The
derivative $\partial M_N/\partial m_s$ is determined from simulations
above and below the physical strange quark mass for $M_\pi$ around
240--675~MeV. The resulting values of $\sigma_s$ are extrapolated
quadratically in $M_\pi$. The quark fraction $f_{T_s}=\sigma_s/M_N$
exhibits a milder pion-mass dependence and extrapolations of this
quantity were also performed using ans\"atze linear and quadratic in
$M_\pi$. A weighted average of all three fits was used to form the
final result. Two lattice spacings were analyzed, with $a$ around
$0.09$~fm and $0.12$~fm, however, discretization effects could not be
resolved. We note that BMW in their $N_f=1+1+1+1$ study
\cite{Borsanyi:2020bpd} significantly improve the precision of their
estimate of $\sigma_s$. Even though all the criteria are satisfied, it is not considered for the
average as Ref.~\cite{Borsanyi:2020bpd} is a preprint.  The global averaging of all calculations that qualify is
discussed in the next section.

\subsubsection{Summary of Results for $g_S^{u,d,s}$\label{sec:gS-sum}}

We consider computing global averages of results determined via the
direct, hybrid and Feynman-Hellmann (FH) methods. These are unchanged
from FLAG 19.  Beginning with $\sigma_{\pi N}$,
Tabs.~\ref{tab:gs-ud-s-direct} and \ref{tab:gs-singlet-fh} show that
for $\Nf=2+1+1$ only ETM~14A~(FH) satisfies the selection criteria. We
take this value as our FLAG result for the four-flavour case.
\begin{align}
  &\label{eq:sigmaud_2p1p1} \mbox{}\Nf=2+1+1: &\FLAGAVBEGIN \sigma_{\pi N} &= 64.9(1.5)(13.2)\FLAGAVEND ~\mbox{MeV} &&\Ref~\mbox{\cite{Alexandrou:2014sha}}.
\end{align}
We remark that although the $N_f=1+1+1+1$ BMW 20A study also satisfies the
criteria, as Ref.~\cite{Borsanyi:2020bpd} is a preprint this work is not considered for averaging.
For $\Nf=2+1$ we form an average from the BMW~11A~(FH), BMW~15~(FH) and
$\chi$QCD~15A~(direct) results, yielding
\begin{align}
&  \mbox{}\Nf=2+1: &\FLAGAVBEGIN \sigma_{\pi N} &= 	39.7(3.6) \FLAGAVEND ~\mbox{MeV} 
  &&\Refs~\mbox{\cite{Durr:2011mp,Durr:2015dna,Yang:2015uis}}.
\end{align}
Note that both BMW results are included as they were obtained on
independent sets of ensembles~(employing different fermion
actions). The average is dominated by the BMW~15 calculation,
which has much smaller overall errors compared to the other two
studies.  

Turning to the results for $\Nf=2$, only QCDSF~12~(FH)
qualifies. This is taken as the FLAG result 
\begin{align}
  &\label{eq:sigmaud_2} \mbox{}\Nf=2: &\FLAGAVBEGIN \sigma_{\pi N} &= 37(8)(6)\FLAGAVEND ~\mbox{MeV} 
  &&\Ref~\mbox{\cite{Bali:2012qs}}.
\end{align}

Moving on to $\sigma_s$ and the calculations detailed in
Tab.~\ref{tab:gs-ud-s-direct}, for $N_f=2+1+1$ MILC~12C~(hybrid) 
and BMW 20A satisfy the quality criteria, however,
the latter is a preprint and is not considered for averaging. In order to convert
the result for $\langle N|\bar{s}s|N\rangle$ given in MILC~12C to a
value for $\sigma_s$, we multiply by the appropriate FLAG average for
$m_s$ given in Eq.~(35) 
of FLAG 19.  This gives our
result for four flavours.
\begin{align}
  &\label{eq:sigmas_2p1p1}
  \mbox{}\Nf=2+1+1:&\FLAGAVBEGIN \sigma_{s} &= 41.0(8.8)\FLAGAVEND ~\mbox{MeV} 
                       &&\Ref~\mbox{\cite{Freeman:2012ry}}.
\end{align}
For $\Nf=2+1$ we perform a weighted average of BMW~11A~(FH),
MILC~12C~(hybrid), Junnarkar~13~(FH), BMW~15~(FH) and
$\chi$QCD~15A (direct). MILC~09D~\cite{Toussaint:2009pz} also
passes the FLAG selection rules, however, this calculation is
superseded by MILC~12C. As for Eq.~\eqref{eq:sigmas_2p1p1}, the
strangeness scalar matrix element determined in the latter study is
multiplied by the three flavour FLAG average for $m_s$ given in
Eq.~(33) 
of FLAG 19.  There are correlations between the MILC~12C
and Junnarkar~13 results as there is some overlap between the sets
of asqtad ensembles used in both cases. To be conservative we take the
statistical errors for these two studies to be 100\% correlated. The
global average is
\begin{align}
  &\label{eq:sigmas_2p1}
  \mbox{}\Nf=2+1:  &\FLAGAVBEGIN \sigma_{s} &= 52.9(7.0) \FLAGAVEND ~\mbox{MeV} 
                       &&\Refs~\mbox{\cite{Durr:2011mp,Freeman:2012ry,Junnarkar:2013ac,Durr:2015dna,Yang:2015uis}}.
\end{align}
Given that all of the $\Nf=2$ studies have at least one red tag 
we are not able to give an average in this case.

\begin{figure}[!t]
\begin{center}
\includegraphics[width=11.5cm]{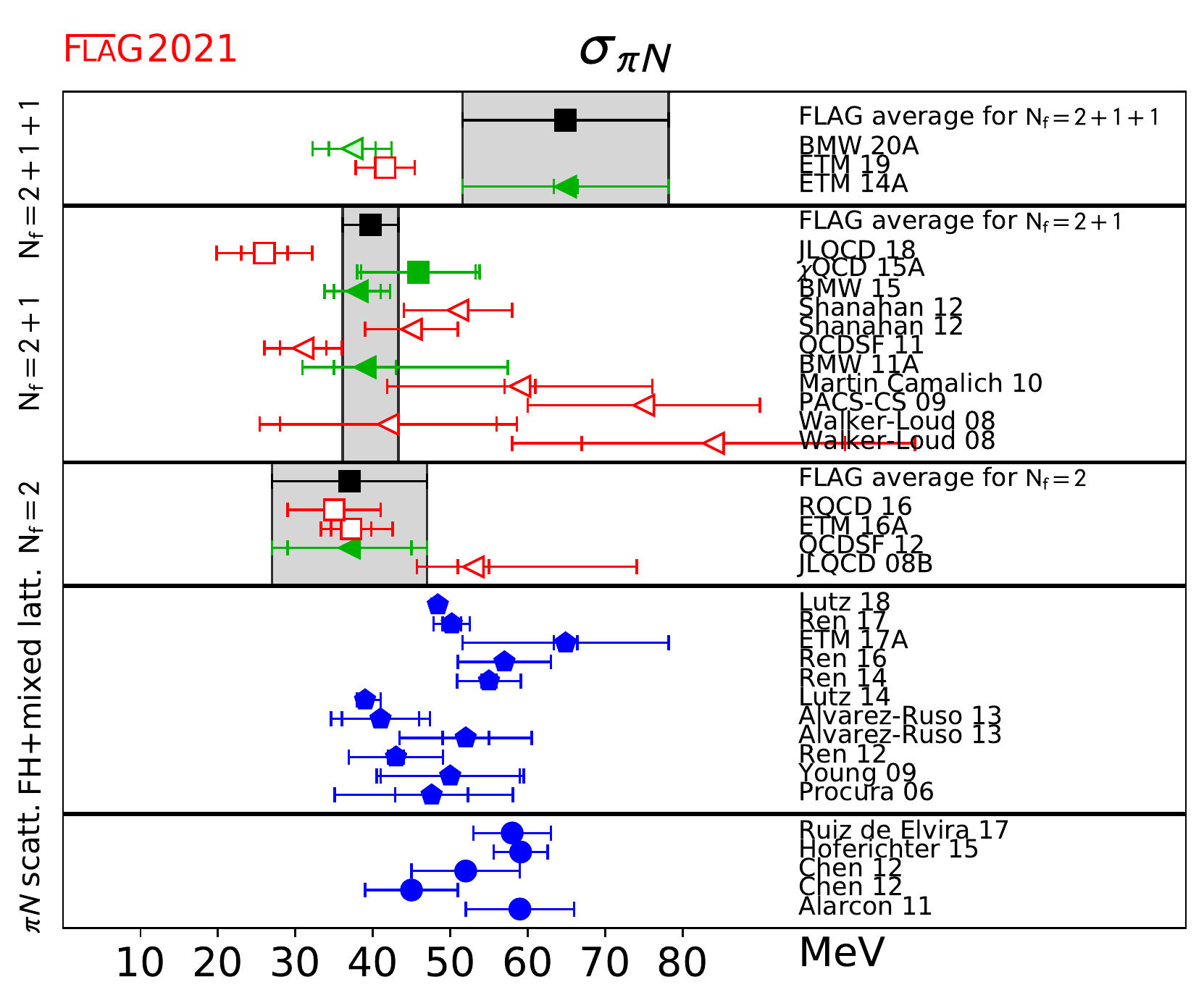}
\end{center}
\vspace{-1cm}
\caption{\label{fig:gs-sum-light} Lattice results and FLAG averages
  for the nucleon sigma term, $\sigma_{\pi N}$, for the $\Nf = 2$,
  $2+1$, and $2+1+1$ flavour calculations. Determinations via the
  direct approach are indicated by squares and the Feynman-Hellmann
  method by triangles. Results from calculations which analyze more
  than one lattice data set within the Feynman-Hellmann
  approach~\cite{Procura:2006bj,Young:2009zb,Ren:2012aj,Alvarez-Ruso:2013fza,Lutz:2014oxa,Ren:2014vea,Ren:2016aeo,Alexandrou:2017xwd,Ling:2017jyz,Lutz:2018cqo}
  are shown for comparison~(pentagons) along with those from recent
  analyses of $\pi$-$N$
  scattering~\cite{Alarcon:2011zs,Chen:2012nx,Hoferichter:2015dsa,RuizdeElvira:2017stg}~(circles).}
\end{figure}

\begin{figure}[!t]
\begin{center}
\includegraphics[width=11.5cm]{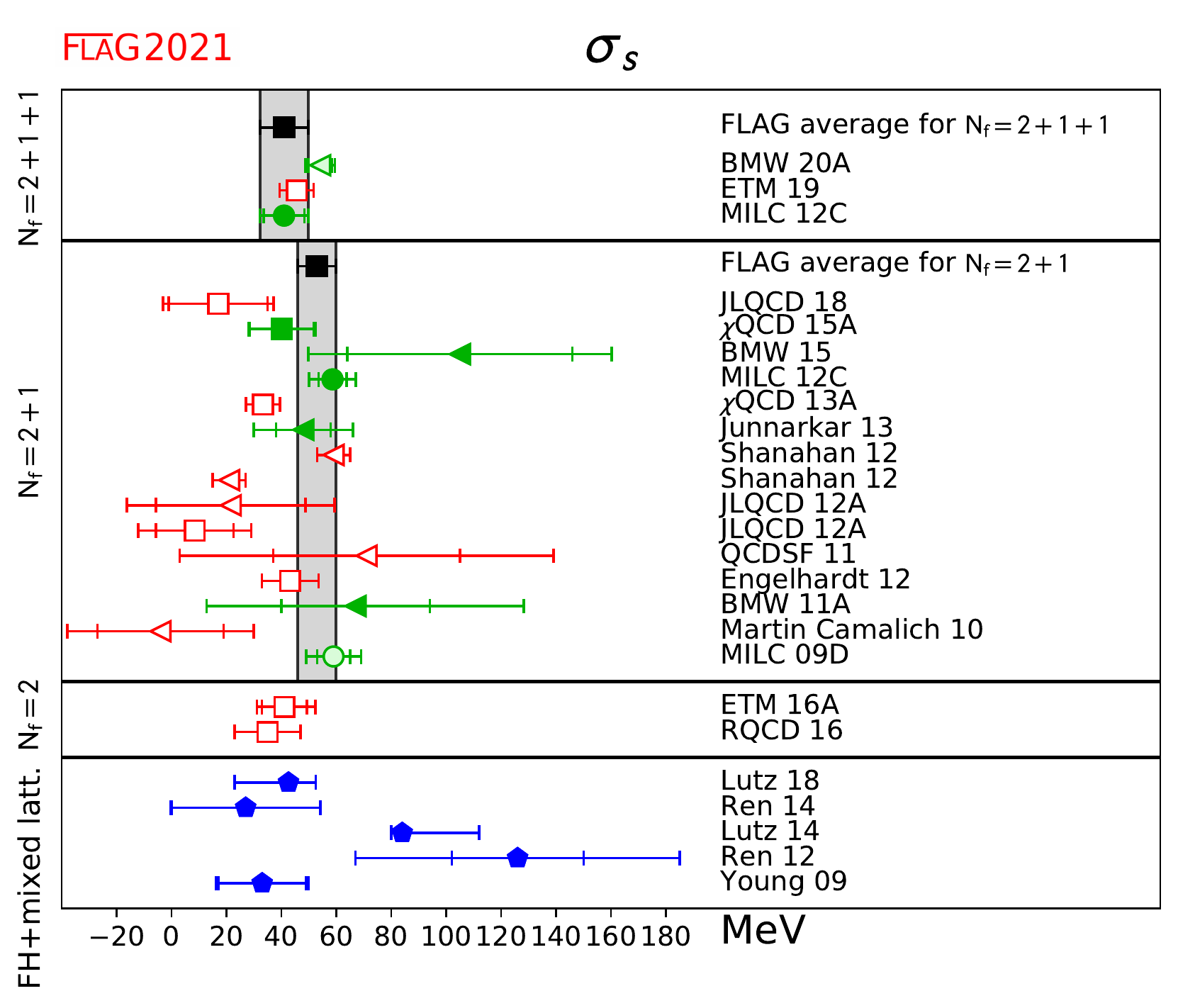}
\end{center}
\vspace{-1cm}
\caption{\label{fig:gs-sum-strange} Lattice results and FLAG averages
  for $\sigma_s$ for the $\Nf = 2$, $2+1$, and $2+1+1$ flavour
  calculations. Determinations via the direct approach are indicated
  by squares, the Feynman-Hellmann method by triangles and the hybrid
  approach by circles. Results from calculations which analyze more
  than one lattice data set within the Feynman-Hellmann
  approach~\cite{Young:2009zb,Ren:2012aj,Lutz:2014oxa,Ren:2014vea,Lutz:2018cqo}
  are shown for comparison~(pentagons). }
\end{figure}

All the results for $\sigma_{\pi N}$ and $\sigma_s$ are displayed in
Figs.~\ref{fig:gs-sum-light} and~\ref{fig:gs-sum-strange} along with
the averages given above. Note that where 
  $f_{T_{ud}}=f_{T_{u}}+f_{T_{d}}$ or $f_{T_s}$ is quoted in
Tabs.~\ref{tab:gs-ud-s-direct} and \ref{tab:gs-singlet-fh}, we
multiply by the experimental proton mass in order to include the
results in the figures. Those results which pass the FLAG criteria,
shown in green, are reasonably consistent.
However,
there is some fluctuation in the central values, in particular, when
taking the lattice results as a whole into account, and we caution the
reader that the averages may change as new results become available.

Also shown for comparison in the figures are determinations from the
FH method which utilize more than one lattice data
set~\cite{Procura:2006bj,Young:2009zb,Ren:2012aj,Alvarez-Ruso:2013fza,Lutz:2014oxa,Ren:2014vea,Ren:2016aeo,Alexandrou:2017xwd,Ling:2017jyz,Lutz:2018cqo}
as well as results for $\sigma_{\pi N}$ obtained from recent analyses
of $\pi$-$N$
scattering~\cite{Alarcon:2011zs,Chen:2012nx,Hoferichter:2015dsa,RuizdeElvira:2017stg}.
There is some tension, at the level of three to four standard
deviations, between the lattice average for $\Nf=2+1$ and Hoferichter et al.~\cite{Hoferichter:2015dsa}~(Hoferichter 15 in Fig.~\ref{fig:gs-sum-light}), who quote a precision similar to
that of the average.

Finally we remark that, by exploiting the heavy-quark limit, the
light- and strange-quark sigma terms can be used to estimate
$\sigma_q$ for the charm, bottom and top
quarks~\cite{Shifman:1978zn,Chetyrkin:1997un,Hill:2014yxa}. The
resulting estimate for the charm quark, see, e.g. the RQCD 16 $\Nf=2$
analysis of Ref.~\cite{Bali:2016lvx} that reports $f_{T_c}=0.075(4)$
or $\sigma_c=70(4)$~MeV is consistent with the direct determinations
of ETM 19~\cite{Alexandrou:2019brg} for $N_f=2+1+1$
  of $\sigma_c=107(22)$~MeV, ETM 16A~\cite{Abdel-Rehim:2016won} for
$\Nf=2$ of $\sigma_c=79(21)(^{12}_{8})$~MeV and $\chi$QCD
13A~\cite{Gong:2013vja} for $\Nf=2+1$ of $\sigma_c=94(31)$~MeV.  
  BMW in  BMW 20A~\cite{Borsanyi:2020bpd} employing the Feynman-Hellmann approach obtain
  $f_{T_c}=\sigma_c/m_N=0.0734(45)(55)$ for $N_f=1+1+1+1$. MILC in MILC
12C~\cite{Freeman:2012ry} find $\langle N|\bar{c}c|N\rangle=0.056(27)$
in the $\msbar$ scheme at a scale of 2~GeV for $\Nf=2+1+1$ via the
hybrid method. Considering the large uncertainty, this is consistent
with the other results once multiplied by the charm quark mass.

\subsubsection{Results for $g_T^{u,d,s}$\label{sec:gT-FD}}

A compilation of recent results for the flavour-diagonal tensor charges
$g_T^{u,d,s}$ for the proton in the $\msbar$ scheme at 2~GeV is given
in Tab.\,\ref{tab:gt-singlet} and plotted in
Fig.~\ref{fig:gt-singlet}.  Results for the neutron can be obtained by
interchanging the $u $ and $d$ flavour indices. Only the PNDME 2+1+1 flavour
calculations qualify for the global average. 

The FLAG values remain the same as in FLAG 19, i.e., the 
PNDME 18B \cite{Gupta:2018lvp} results, which supersede the
PNDME 16 \cite{Bhattacharya:2016zcn} and the
PNDME 15 \cite{Bhattacharya:2015wna} results: 

\begin{align}
&  \mbox{}\Nf=2+1+1: &\FLAGAVBEGIN g_T^u  &= \phantom{-}0.784(28)(10)  \FLAGAVEND   &&\Ref~\mbox{\cite{Gupta:2018lvp}},  \\
&  \mbox{}\Nf=2+1+1: &\FLAGAVBEGIN g_T^d  &=           -0.204(11)(10)  \FLAGAVEND   &&\Ref~\mbox{\cite{Gupta:2018lvp}}, \\
&  \mbox{}\Nf=2+1+1: &\FLAGAVBEGIN g_T^s  &=           -0.0027(16)  \FLAGAVEND   &&\Ref~\mbox{\cite{Gupta:2018lvp}}. 
\end{align}

The ensembles and the analysis strategy used in PNDME~18B is the
same as described in Sec.~\ref{sec:gA-FD} for $g_A^{u,d,s}$. The only
difference for the tensor charges was that a one-state (constant) fit
was used for the disconnected contributions as the data did not show
significant excited-state contamination. The isovector renormalization
constant, used for all three flavour-diagonal tensor operators, was
calculated on the lattice in the RI-SMOM scheme and converted to
$\msbar$ at 2~GeV using 2-loop perturbation theory. 

The PNDME 20 \cite{Park:2020axe} provided a status update
on $g_T^{u,d,s}$ to PNDME 18B \cite{Lin:2018obj} but is not considered 
for the average as it is a conference proceeding. It also presented results 
showing that flavour mixing in the calculation of tensor renormalization constants is small, and 
the isovector renormalization factor 
is a good approximation for renormalizing flavour-diagonal tensor charges as discussed in Sec.~\ref{sec:renorm}.

\begin{table}[t!]
\begin{center}
\mbox{} \\[3.0cm]
\footnotesize
\begin{tabular*}{\textwidth}[l]{l @{\extracolsep{\fill}} r l l l l l l l l@{\hspace{1mm}} l}
Collaboration & Ref. & $\Nf$ & 
\hspace{0.15cm}\begin{rotate}{60}{publication status}\end{rotate}\hspace{-0.15cm} &
\hspace{0.15cm}\begin{rotate}{60}{continuum extrapolation}\end{rotate}\hspace{-0.15cm} &
\hspace{0.15cm}\begin{rotate}{60}{chiral extrapolation}\end{rotate}\hspace{-0.15cm}&
\hspace{0.15cm}\begin{rotate}{60}{finite volume}\end{rotate}\hspace{-0.15cm}&
\hspace{0.15cm}\begin{rotate}{60}{renormalization}\end{rotate}\hspace{-0.15cm}  &
\hspace{0.15cm}\begin{rotate}{60}{excited states}\end{rotate}\hspace{-0.15cm}  &
$g_T^u$ & $g_T^d$ \\
&&&&&&&&& & \\[-0.1cm]
\hline
\hline
&&&&&&&& &  \\[-0.1cm]
PNDME 20 & \cite{Park:2020axe} & 2+1+1 & \rC & \good$^\ddag$ & \good & \good & \good & \soso & 0.783(27)(10) & $-$0.205(10)(10)   \\[0.5ex]
ETM 19 & \cite{Alexandrou:2019brg} & 2+1+1 & \gA & \bad & \soso & \good & \good & \soso & 0.729(22) & $-$0.2075(75) \\[0.5ex]
PNDME 18B & \cite{Gupta:2018lvp} & 2+1+1 & \gA & \good$^\ddag$ & \good & \good & \good & \soso & 0.784(28)(10)$^\#$ & $-$0.204(11)(10)$^\#$ \\[0.5ex]
PNDME 16 & \cite{Bhattacharya:2016zcn} & 2+1+1 & \gA & \soso$^\ddag$ & \good & \good & \good & \soso & 0.792(42)$^{\#\&}$ & $-$0.194(14)$^{\#\&}$ \\[0.5ex]
PNDME 15 & \cite{Bhattacharya:2015wna,Bhattacharya:2015esa} & 2+1+1 & \gA & \soso$^\ddag$ & \good & \good & \good & \soso & 0.774(66)$^\#$ & $-$0.233(28)$^\#$ \\[0.5ex]
\\[-0.1ex]\hline\\[0.2ex]
Mainz 19A & \cite{Djukanovic:2019gvi} & 2+1 & \rC & \good & \soso & \good & \good & \soso & 0.77(4)(6) & $-$0.19(4)(6)   \\[0.5ex]
JLQCD 18 & \cite{Yamanaka:2018uud} & 2+1 & \gA & \bad & \soso & \soso & \good & \soso & 0.85(3)(2)(7) & $-$0.24(2)(0)(2) \\[0.5ex]
\\[-0.1ex]\hline\\[0.2ex]
ETM 17 & \cite{Alexandrou:2017qyt} & 2 & \gA & \bad & \soso & \soso & \good & \soso & 0.782(16)(2)(13) & $-$0.219(10)(2)(13) \\[0.5ex]
 & & & & & & & & & & \\[-0.1cm]
\hline
\hline
 & & & & & & & & & & \\[-0.1cm]
 & & & & & & & & & $ g_T^s $& \\[-0.1cm]
 & & & & & & & & & &\\[-0.1cm]
\hline
\hline
     & & & & & & & & & &\\[-0.1cm]
PNDME 20 & \cite{Park:2020axe} & 2+1+1 & \rC & \good$^\ddag$ & \good & \good & \good & \soso &  $-$0.0022(12)  & \\[0.5ex]
ETM 19 & \cite{Alexandrou:2019brg} & 2+1+1 & \gA & \bad & \soso & \good & \good & \soso & $-0.00268(58)$ & \\[0.5ex]
PNDME 18B & \cite{Gupta:2018lvp} & 2+1+1 & \gA & \good$^\ddag$ & \good & \good & \good & \soso &  $-$0.0027(16)$^\#$ & \\[0.5ex]
PNDME 15 & \cite{Bhattacharya:2015wna,Bhattacharya:2015esa} & 2+1+1 & \gA & \soso$^\ddag$ & \good & \good & \good & \soso &  0.008(9)$^\#$ & \\[0.5ex]
\\[-0.1ex]\hline\\[0.2ex]
Mainz 19A & \cite{Djukanovic:2019gvi} & 2+1 & \rC & \good & \soso & \good & \good & \soso & $-$0.0026(73)(42) &   \\[0.5ex]
JLQCD 18 & \cite{Yamanaka:2018uud} & 2+1 & \gA & \bad & \soso & \soso & \good & \soso &  $-$0.012(16)(8) & \\[0.5ex]
\\[-0.1ex]\hline\\[0.2ex]
ETM 17 & \cite{Alexandrou:2017qyt} & 2 & \gA & \bad & \soso & \soso & \good & \soso &  $-$0.00319(69)(2)(22) & \\[0.5ex]
&&&&&&&& \\[-0.1cm]
\hline
\hline
\end{tabular*}
\begin{minipage}{\linewidth}
{\footnotesize 
\begin{itemize}
\item[$^\ddag$]The rating takes into account that the action is not fully O(a) improved by requiring an additional lattice spacing.\\[-5mm]\item[$^\#$] Assumed that $Z_T^{n.s.}=Z_T^{s}$. \\[-5mm] \item[$^\&$] Disconnected terms omitted.
\end{itemize}
}
\end{minipage}
\caption{Overview of results for $g^q_T$.\label{tab:gt-singlet} }
\end{center}
\end{table}

The ETM 19~\cite{Alexandrou:2019brg} presented new results for $g_T^{u,d,s,c}$ from 
a single ensemble with 2+1+1-flavour twisted-mass fermions with a clover term at $a=$ 0.0801(4)~fm 
and $M_\pi= 139.3(7)$~MeV. It was 
not considered for the final averages because it did not satisfy the criteria 
for the continuum extrapolation as already discussed in Sec.~\ref{sec:gA-FD}. 
The same applies to 
the JLQCD 18 \cite{Yamanaka:2018uud} and ETM 17 \cite{Alexandrou:2017qyt} calculations. 
The Mainz 19A \cite{Djukanovic:2019gvi} results with 
2+1-flavour ensembles of clover fermions are not included in the 
averages as Ref.~\cite{Djukanovic:2019gvi} is a conference proceeding.

\begin{figure}[!t]
\begin{center}
\includegraphics[width=7.5cm]{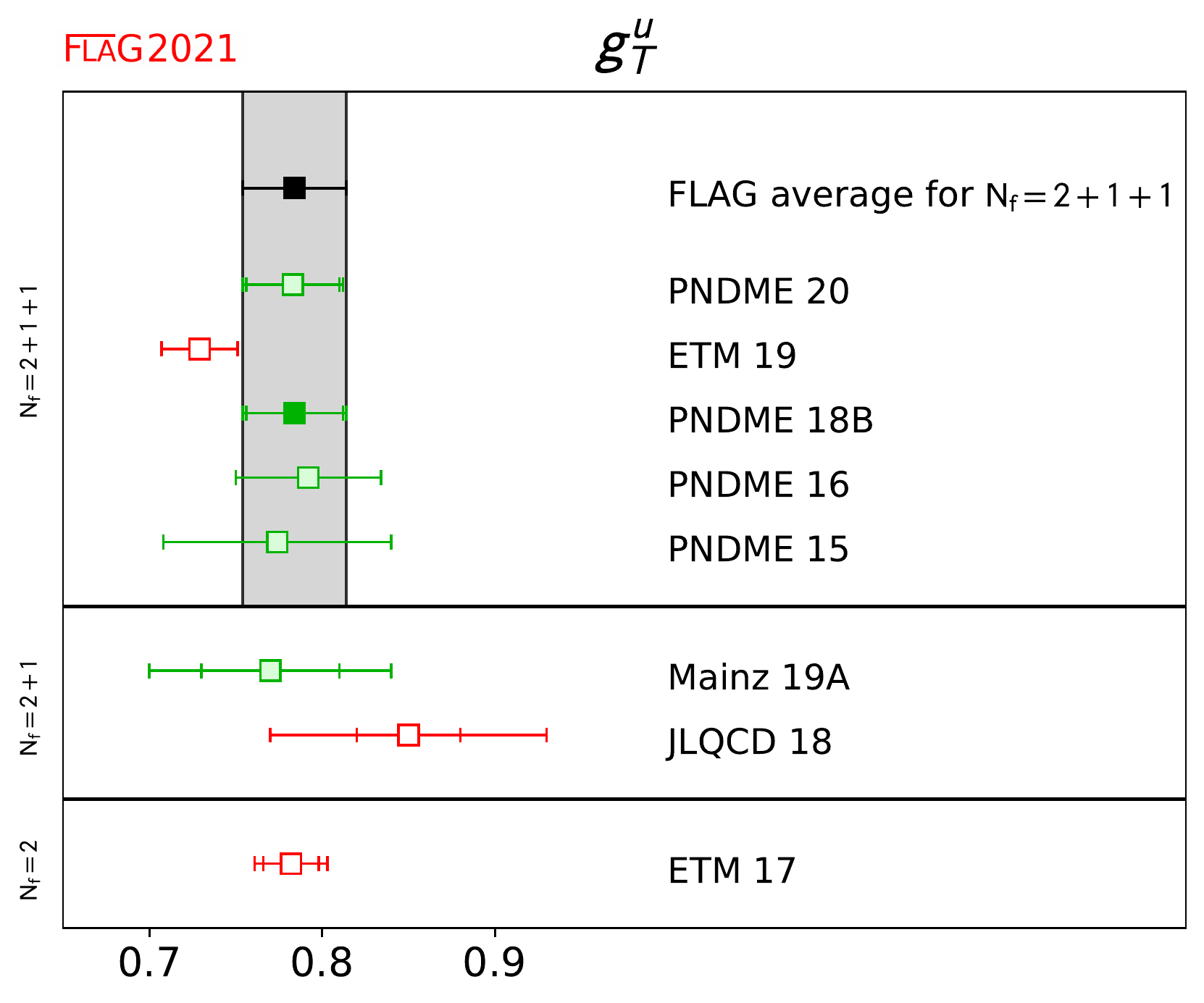}
\includegraphics[width=7.5cm]{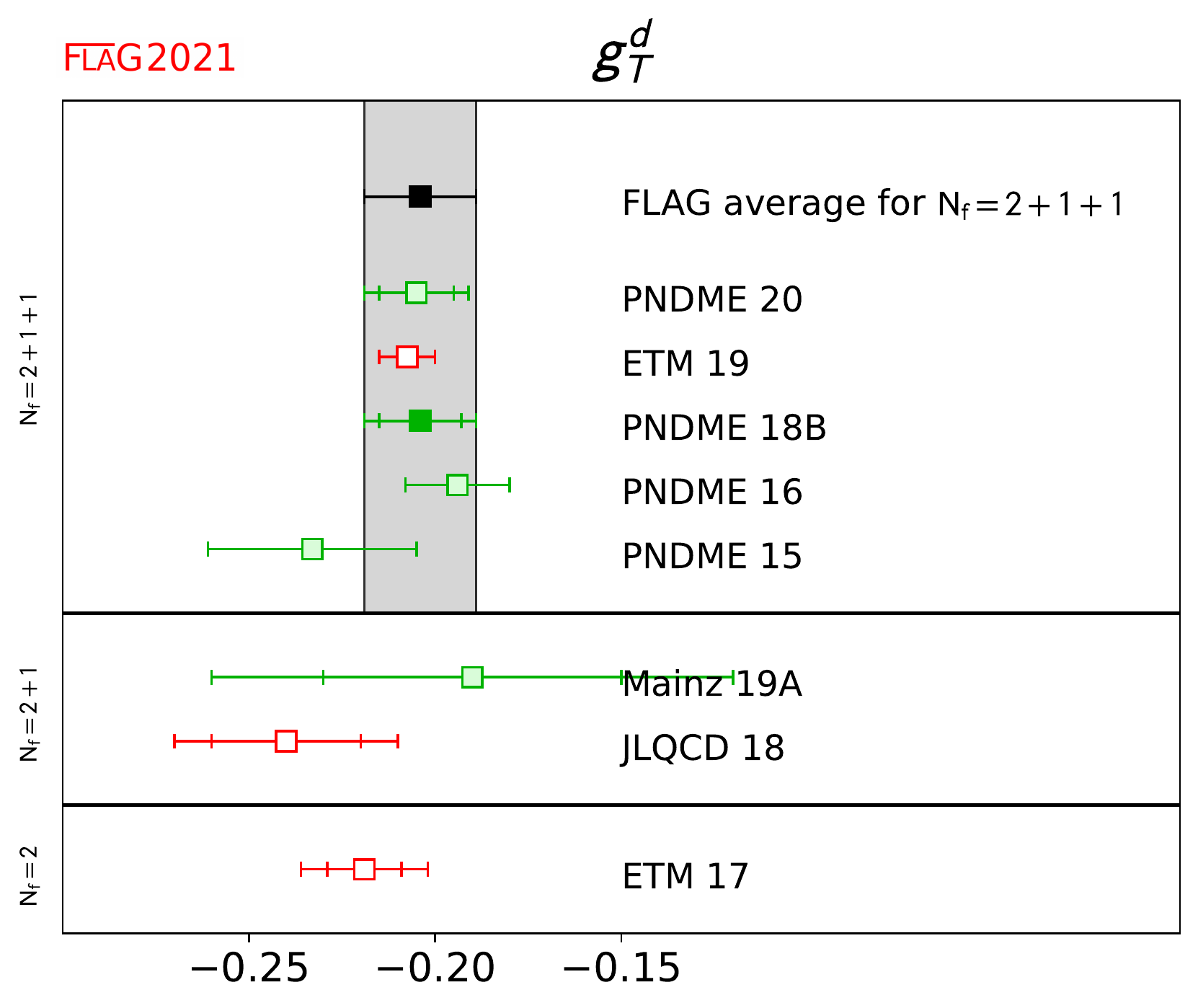}
\includegraphics[width=7.5cm]{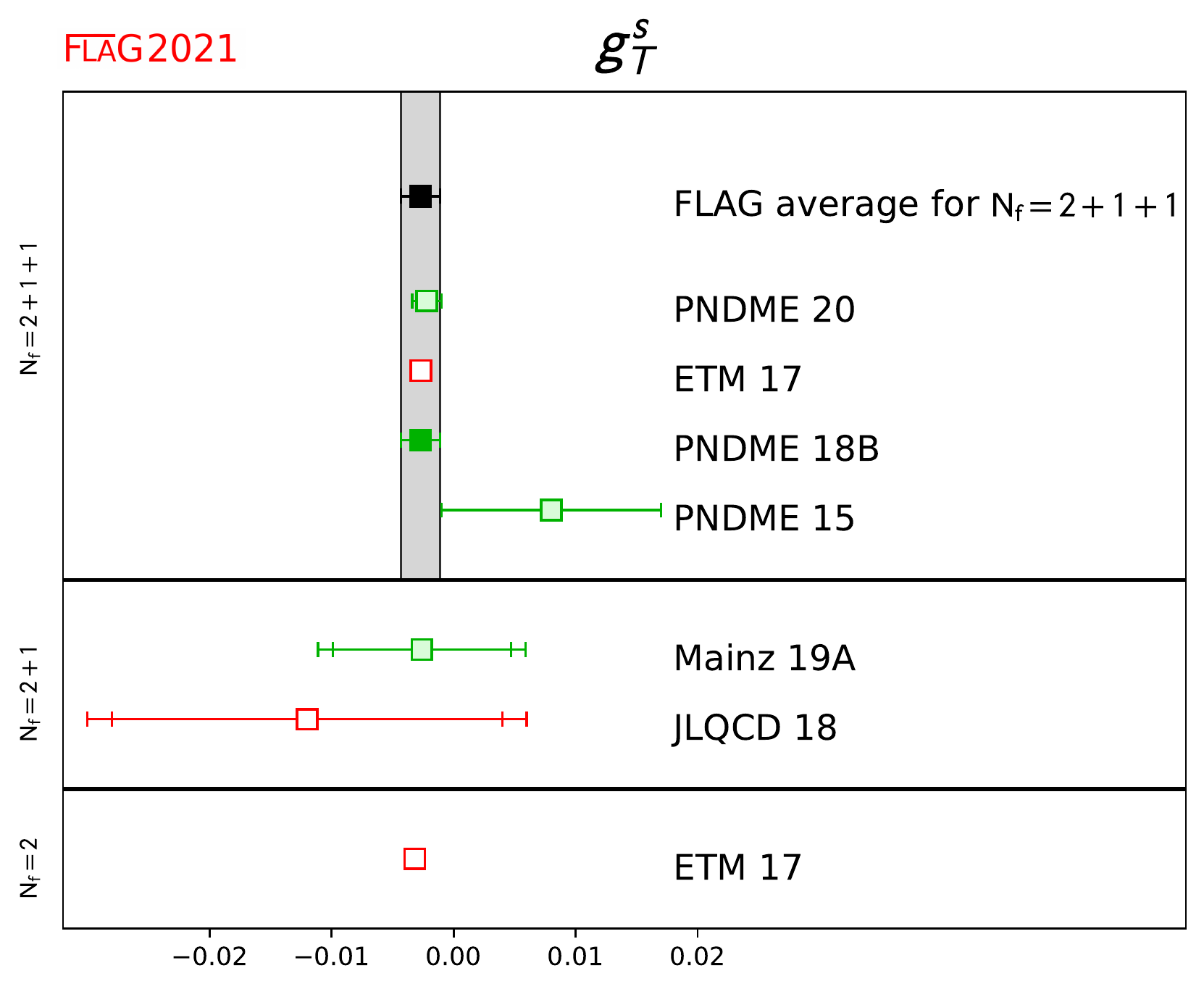}
\end{center}
\caption{\label{fig:gt-singlet} Lattice results and FLAG averages for 
  $g_T^{u,d,s}$ for the $\Nf =
  2$, $2+1$, and $2+1+1$ flavour calculations.  }
\end{figure}

\clearpage
\setcounter{section}{10}

\section{Scale setting\label{sec:scalesetting}}
Authors: R.~Sommer, N.~Tantalo, U.~Wenger\\

\def\qedl{$\rm QED_L$}
\def\U{\hfill {\cred Urs}}
\def\R{\hfill{\cred Rainer}}
\def\N{\hfill{\cred Nazario}}

\ifx\versionforCollabs\undefine

\renewcommand{\cred}{}

{\cred Matching QCD to Nature requires fixing the quark masses
and matching an overall scale to experiment. That overall
energy scale $\scal$ may be taken, for example, as the nucleon mass. 
This process is referred to as scale setting. }

\subsection{Impact}

The scale setting procedure, described in some detail below, is a rather technical step necessary to obtain predictions from QCD. What may easily be overlooked is that the exact predictions obtained may depend rather sensitively on the scale. 

As long as the theory is incomplete, e.g., because we have predictions from $\Nf=2+1$ QCD, results will depend on which physics scale is used. Whenever a theory scale (see Sec.~\ref{s:theoryscales}) is used, it matters which value one imposes. Thus, to know whether computations of a particular quantity agree or not, one should check which (value for a) scale was used. 

The sensitivity of predictions to the scale vary with the observable. For example, the $\Lambda$ parameter of the theory has a linear dependence,
\begin{equation}
 \frac{\delta\Lambda}{\Lambda}	\approx \frac{\delta\scal}{\scal}\,,
\end{equation}
because $\Lambda$ has mass dimension one and other hidden dependences on the scale are (usually) suppressed. Let us preview the results. The present precision  on the most popular theory scale, $w_0$ in \eq{eq:w0_2p1p1} is about 0.4\% and for $\sqrt{t_0}$ it is 0.6\%. On the $\Lambda$ parameter it is about 3\%. Thus, we would think that the scale uncertainty is irrelevant. However, in Sec.~\ref{s:Scalconcl} we will discuss that differences between $\Nf=2+1$ and 2+1+1 numbers for $\sqrt{t_0}$ are at around 2\% which {\em does matter}.

Also, light-quark masses  have an approximatively  linear dependence on the scale (roughly speaking one determines, e.g., $m_{ud} = \frac1{\scal}\times[m_\pi^2]_\mathrm{exp}\times[\frac{m_{ud}\, \scal}{m_\pi^2} ]_\mathrm{lat}$) and scale uncertainties may play an important r\^ole in the discussion of agreement vs.~disagreement of computations within their error budget.

The list of quantities where scale setting is very important may be continued;  we just want to mention an observable very much discussed at present, the hadronic vacuum polarisation contribution to the anomalous magnetic moment of the muon \cite{Aoyama:2020ynm}. It is easily seen that the dependence on the scale is about quadratic in that case \cite{DellaMorte:2017dyu},
\begin{equation}
 \frac{\delta a_\mu^\mathrm{HVP}}{a_\mu^\mathrm{HVP}}	\approx 2 \frac{\delta\scal}{\scal}\,.
\end{equation}
This fact means that scale setting has to be precise at the few per-mille precision 
to have an impact~\cite{Borsanyi:2020mff} on the discussion whether  or not
$a_\mu$ computed in the standard model shows a deviation from experiment.

\subsection{Scale setting as part of hadronic renormalization schemes}
\label{sec:QCDhadRen}
We consider QCD with $\Nf$ quarks and without a $\theta$-parameter. This theory is  completely defined by its coupling constant as well as $\Nf$ quark masses. After these parameters are specified all other properties of the theory are predictions. Coupling and  quark masses depend on a renormalization scale $\mu$ as well as on a renormalization scheme. The most popular scheme in the framework of perturbative computations is the $\msbar$ scheme, but one may also define nonperturbative renormalization schemes, see Secs.~\ref{sec:qmass} and \ref{sec:alpha_s}.

In principle, a lattice computation may, therefore, use these $\Nf+1$ parameters as input together with the renormalization scale $\mu$  to fix the bare quark masses and coupling of the discretized Lagrangian, perform continuum and infinite volume limit and obtain desired results, e.g., for decay rates.\footnote{At first sight this seems like too many inputs,  but note that it is the scale $\mu$, at which $\alpha(\mu)$ has a  particular value, which is the input. The coupling $\alpha$ by itself can have any (small) value as it runs.} However, there are various reasons why this strategy is inefficient. The most relevant one is that coupling and quark masses cannot be obtained from experiments without invoking perturbation theory and thus necessarily truncation errors. Moreover, these parameters are naturally short distance quantities, since this is where perturbation theory applies. Lattice QCD on the other hand is most effective at long distances, where the lattice spacing plays a minor role. Therefore, it is  more natural to proceed differently.

Namely, we may fix $\Nf+1$ nonperturbative, long-distance observables to have the values found in Nature. An obvious choice are $\Nf+1$ hadron masses that are stable in the absence of weak interactions. This hadronic renormalization scheme is defined by
\begin{equation}
	\frac{M_i(g_0,\{a m_\mathrm{0,j}\})}{
	M_1(g_0,\{a m_\mathrm{0,j}\})} 
	=\frac{M_i^\mathrm{exp}}{M_1^\mathrm{exp}}\,, \quad 
	i=2\ldots\Nf+1\,,
	\quad
	j=1\ldots\Nf\,.
	\label{e:hadscheme}
\end{equation}
Here, $M_i$ are the chosen hadron masses, $g_0$ is the bare coupling, and $am_{0,j}$ are the bare quark masses in lattice units. The ratio $M_i/M_1$ is, precisely speaking, defined through the hadron masses in lattice units, but in infinite volume. In QCD (without QED), all particles are massive. Therefore, the infinite volume limit of the properties of stable particles is approached with exponentially small corrections which are assumed to be estimated reliably. The power-like finite-volume corrections in QCD$+$QED are discussed in subsection~\ref{sec:isobreak}.  For fixed $g_0$,
Eq.~\eqref{e:hadscheme} needs to be solved for the bare quark masses,
\begin{equation}
	a m_\mathrm{0,j} = \mu_j(g_0)\,. \label{e:lcp}
\end{equation}
The functions $\mu_j$ define a line in the bare parameter space, called the line of constant physics. Its dependence on the set of masses $\{M_i\}$ is suppressed. The continuum limit is obtained as $g_0\to0$ with the lattice
spacing shrinking 
roughly as $aM_1 \sim \rme^{-1/(2b_0 g_0^2)}$. More precisely, consider 
observables $\cal O$ with mass dimension $d_{\cal O}$. One defines their
dimensionless ratio
\begin{equation}
 	\hat {\cal O}(aM_1)=\left. 
 	\frac{{\cal O}}{M_1^{d_{\cal O}}}\right|_{am_{0,j}=\mu_j(g_0)}\,,
\end{equation}
and obtains the continuum prediction as
\begin{equation}
  {\cal O}^\mathrm{cont} = \left(M_1^{\mathrm{exp}}\right)^{d_{\cal
      O}} \, \lim_{aM_1\to 0} \hat {\cal O}(aM_1)
\end{equation}
which explains why the determination and use of $aM_1$ is referred to as scale setting.

Equation (\ref{e:lcp}) has to be obtained from numerical results. Therefore, it 
is easiest and most transparent if the $i$-th mass ratio depends 
predominantly on the $i$-th quark mass.
Remaining for a while in the isospin-symmetric theory with $m_{0,1} = m_{0,2}$
(we enumerate the quark masses in the order up, down, strange, charm, bottom and ignore the top quark),
we have natural candidates for the numerators as the pseudoscalar masses in the
associated flavour sectors, i.e., $\pi,\,K,\,D,\,B$. The desired strong
dependence on light- (strange-)quark masses of $\pi$- ($K$-)meson masses derives
from their pseudo-Goldstone nature of the approximate $SU(3)_\mathrm{L}\times SU(3)_\mathrm{R}$ symmetry of the massless QCD Lagrangian  which
predicts that $M_\pi^2$ is roughly proportional to the light-quark mass and $M_{\mathrm K}^2$  to  the sum of light- and strange-quark masses. For $D$ and $B$ mesons approximate heavy-quark symmetry
predicts $M_{D}$ and $M_{B}$ to be proportional to charm- and bottom-quark masses.
Also other heavy-light bound states have this
property. There is another  important feature which singles out pseudoscalar masses. Because they are the lightest particles with the given flavour quantum numbers, their correlation functions have the least signal/noise problem in the
Monte Carlo evaluation of the path integral \cite{Lepage:1989hd,Luscher:2010ae}.

Still restricting ourselves to isospin-symmetric QCD (isoQCD), we thus take it for granted that the choice $M_i,\, i\geq 2$ is easy, and we do not need to discuss it in detail: the pseudoscalar meson masses are very good choices, and some variations for 
heavy quarks may provide further improvements.

The choice of $M_1$ is more difficult. 
From the point of view of physics, a natural choice is the nucleon mass, $M_1=M_\textrm{nucl}$. Unfortunately it has a rather bad signal/noise problem 
when quark masses are close to their physical values. 
The ratio of signal to noise of the correlation function
at time $x_0$ from $N$ measurements behaves as \cite{Lepage:1989hd}
\begin{eqnarray}
	 R_{S/N}^\mathrm{nucl} &\simas{x_0\; \mathrm{large}}& \sqrt{N}\, \exp(-(m_\mathrm{nucl}-\frac32\mpi)\,x_0) \approx \sqrt{N}\,\exp(-x_0 / 0.27\,\fm) \,,
\end{eqnarray}
where the numerical value of $0.27\,$fm uses the experimental
masses. The behaviour in practice, but at still favourably large quark
masses, is illustrated in Fig.~\ref{f:plateaux}.
\begin{figure}[ht!]
\centering
   \includegraphics[width=0.8\textwidth]{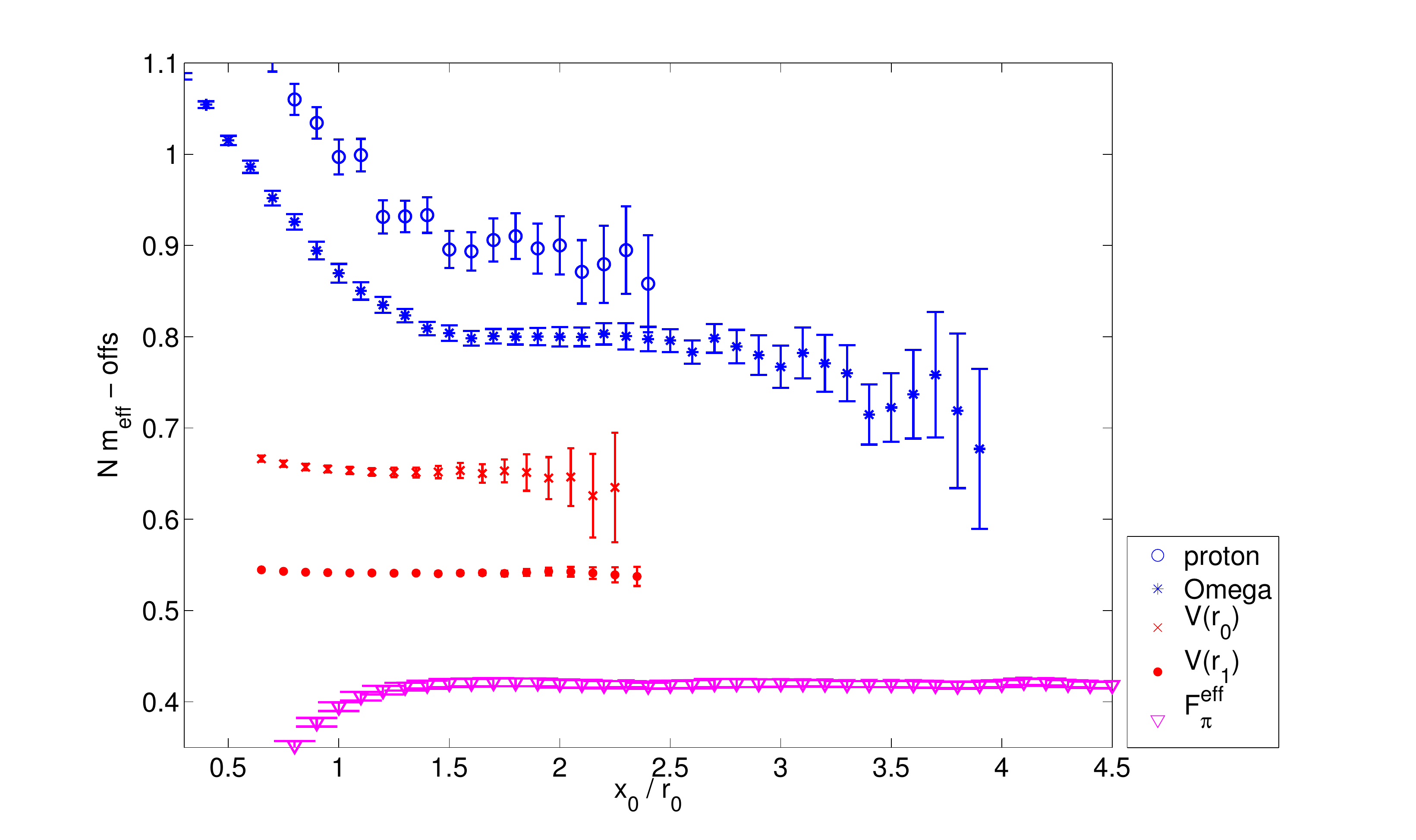}
   \vspace*{-2mm}
\caption{Effective masses for 
$M_\mathrm{proton}$ \cite{Jager:2013kha},
$M_\Omega$ \cite{Capitani:2011fg},
$V(\approx r_0)$, $V(\approx r_1)$~\cite{Fritzsch:2012wq}
and $\fpi$~\cite{Lottini:2013rfa} on $\Nf=2$ CLS ensemble N6 with 
$a=0.045\,\fm, M_\pi=340\,\MeV$ on a $48^3 \, 96$ lattice \cite{Lottini:2013rfa}. 
All effective ``masses'' have been scaled 
such that the errors in the graph reflect
directly the errors of the determined scales. They are shifted vertically
by arbitrary amounts. Figure from Ref.~\cite{Sommer:2014mea}. Note that this example is at still favourably large quark masses. The situation for
$M_\mathrm{proton}$ becomes worse closer to the physical point, but may be changed by algorithmic improvements.
\label{f:plateaux}
}
\end{figure}
Because
this property leads to large statistical errors and it is further difficult to control excited-state contaminations when statistical errors are large, it is useful to search for alternative 
physics scales. The community has gone this way, and we discuss some of them below. For illustration, here we just  give one example:
the decay constants of leptonic $\pi$ or $K$ decays have mass dimension one 
and can directly replace $M_1$ above. Figure \ref{f:plateaux} demonstrates their long and precise plateaux as a function of the Euclidean time. Advantages and disadvantages of this choice
and others are discussed more systematically in Sec.~\ref{s:physscal}. 

\subsubsection{Theory scales}
Since the signal/noise problem of physics scales is rather severe,
they were already replaced by theory scales  in the very first days
of lattice QCD.  These scales cannot be determined from experiment alone. Rather, their values have to be computed by lattice QCD using a physics scale as input.

Creutz already used the string tension in his seminal paper
on $SU(2)$ Yang Mills theory \cite{Creutz:1980zw}, because it is by far easier to determine
than glueball masses. A further step was made by the potential scale
 $r_0$, defined in terms of the static force $F(r)$ as \cite{Sommer:1993ce}
 \begin{equation}
 	r_0^2 F(r_0) =1.65\,.
 \end{equation}
Even though $r_0$ can vaguely be related to the phenomenology of
charmonium and bottomonium states, its precise definition is in terms
of $F(r)$ which can be obtained accurately 
from Monte Carlo lattice computations with (improvable) control
over the uncertainties, but not from experiment. In that sense, it is a prototype of a theory scale. \\[1ex]
Useful properties of a good theory scale are high statistical accuracy, easy to control systematics  (also large volume), quark mass dependence only due to the fermion determinant, and low numerical cost for its evaluation.
These properties are realized to varying degrees by the different theory scales 
covered in this section and, in this respect, they are much preferred
compared to physics scales. Consequently, the physics scale $M_1$ has often been replaced by a theory scale as, e.g., 
$\scal=r_0^{-1}$ in the form
\begin{equation}
  {\cal O}^\mathrm{cont} = \left(\scal^{\mathrm{phys}}\right)^{d_{\cal O}} \,
  \lim_{a\scal\to0} \hat {\cal O}_{\scal}(a\scal)\; \text{ with }\; \hat {\cal O}_{\scal}(a\scal)= \left[\scal^{-d_{\cal O}}\, \cal O\right]_{am_{0,j}=\mu_j(g_0)}\,,
\end{equation}
and
\begin{equation}
  {\scal}^\mathrm{phys} = \left(M_1^{\mathrm{exp}}\right) \, \lim_{aM_1\to0} \hat {\scal}_{M_1}(aM_1)\,. \label{e:scalesettbasic}
\end{equation}

In this section, we review the determination of numerical results for the values
of various theory scales in physical units,
\eq{e:scalesettbasic}. The main difficulty is that a physics scale $M_1$ has to
be determined first in order to connect to Nature and, in particular, that the continuum limit of the theory scale in units of the physics scale has to be taken.

\subsection{Isospin breaking, electromagnetism, and definition of hadronic schemes} 
\label{sec:isobreak}

\subsubsection{The approximate nature of QCD}

For simplicity and because it is a very good approximation, 
we have assumed above that all other interactions except for QCD can be ignored when hadron masses and many other properties of hadrons are considered. This is a
natural point of view because QCD is a renormalizable field theory and thus provides unique results. 

However, we must be aware that while it is true that the predictions
(e.g., for hadron masses $M_i, \, i>\Nf+1$) are unique once
\eq{e:hadscheme} is specified, they will change when we change the
inputs $M_i^\mathrm{exp}$. These ambiguities are due to the neglected
electroweak and gravitational interactions, namely because QCD is only an
approximate---even if  precise---theory of hadrons. At the sub-percent level, QED effects and isospin violations due to $m_{u} \ne m_{d}$ must be included. At that level one has 
a very precise description of Nature, where weak decays 
or weak effects, in general, can be included perturbatively and systematically
in an effective field theory description through the 
weak effective interaction Hamiltonian, while gravity may be ignored.

We now discuss how to handle the scale setting as part of
the renormalization of QCD$+$QED.  Note that a similar discussion with emphasis on quark masses can be found in Sec.~\ref{sec:qmass}. In the following discussion, we focus more on the issues related to the scale setting (see also Ref.~\cite{DiCarlo:2019thl}). In this connection, triviality of QED does not play a r\^ole at small enough $\alpha$: we may think of replacing the continuum limit $a\to0$ by a limit $a\to a_\mathrm{w}$ with $a_\mathrm{w}$ nonzero but very far below all QCD$+$QED scales.

\subsubsection{Hadronic renormalization of QCD$+$QED}
The definition and implementation of a hadronic renormalization scheme of QCD+QED defined on the lattice needs
some additions to subsection~\ref{sec:QCDhadRen} which we now discuss.

In addition to the $N_f+1$ parameters of the QCD action (without isospin symmetry), one  now also has the elementary electric charge e. This requires $N_f+2$ experimentally measurable observables to fix the bare parameters of the theory. A natural choice for the experimental inputs are again hadron masses. Indeed, hadron masses are infrared safe quantities also in QCD$+$QED, while in the cases of cross sections and decay rates, infrared divergences appear at intermediate stages of the calculations (see below). Therefore, we consider the generalization 
\begin{equation}
	\frac{M_i(g_0,e_0,\{a m_\mathrm{0,j}\})}{
	M_1(g_0,e_0,\{a m_\mathrm{0,j}\})} 
	=\frac{M_i^\mathrm{exp}}{M_1^\mathrm{exp}}\,, 
	\quad i=2\ldots\Nf+2\,,
	\quad
	j=1\ldots\Nf\,
	\label{e:hadschemeQED}
\end{equation}
of Eq.~\eqref{e:hadscheme}.
Here, $M_i$ are the chosen hadron masses, $g_0$ the bare strong coupling, $e_0$ the bare electric charge, and $am_{0,j}$ are the bare quark masses in lattice units.  
For fixed $g_0$, the system of equations \eqref{e:hadschemeQED} now needs to be solved for the bare quark masses and the bare
electric charge,
\begin{equation}
	a m_\mathrm{0,j} = \mu_j(g_0)\,,
	\qquad
	e_0=\mathrm{e}(g_0)\,,
	\label{e:lcpQED}
\end{equation}
to obtain the line of constant physics of the theory. Some observations are in order.

So far, we have assumed that QCD$+$QED is simulated nonperturbatively in the electromagnetic coupling constant $\alpha_\mathrm{em}$. In this case, the bare electric charge can be conveniently fixed by considering among the experimental inputs both the charged and neutral pion masses. Indeed, by neglecting terms of $\cO((m_u-m_d)^2)$ \cite{Dashen:1969eg} one has that $m_{\pi^+}^2-m_{\pi^0}^2\sim \alpha_\mathrm{em}$. If the theory is instead treated neglecting $\cO(\alpha_\mathrm{em}^2)$ contributions, the electric charge does not renormalize and it is consistent and convenient to fix it by the condition \cite{Zyla:2020zbs}
\begin{equation}
\frac{4\pi}{e_0^2}=\frac{1}{\alpha_\mathrm{em}^\mathrm{Thomson}} = 137.035999084(21) \, . 
\end{equation}

Another important difference from pure QCD concerns finite-volume effects. In contrast to the exponentially suppressed finite-volume effects of stable hadron masses at $\alpha_\mathrm{em}=0$, in QCD$+$QED with  $\alpha_\mathrm{em}>0$ finite-volume effects are
\begin{equation}
\frac{M_i(g_0,e_0,a m_\mathrm{0,j};L)}{M_i(g_0,e_0,a m_\mathrm{0,j};\infty)}
=1 
+
\frac{\alpha_\mathrm{em} q_i^2 \xi(1)}{L M_i(g_0,e_0,a m_\mathrm{0,j};\infty)}
+
\frac{\alpha_\mathrm{em} q_i^2 \xi(2)}{\left[L M_i(g_0,e_0,a m_\mathrm{0,j};\infty)\right]^2}
+O\left(L^{-n}, \alpha_\mathrm{em}^2\right)\,,
\label{eq:QEDFV}
\end{equation}
where $q_i$ is the electric charge of the hadron in units of the charge of the positron, the $\xi(i)$ are known numerical constants that depend on the spatial boundary conditions, and the remainder terms start with a power $n=3$ in the  QED$_\mathrm{L}$ formulation\cite{Borsanyi:2014jba,Davoudi:2014qua} and with $n=4$ in QED$_\mathrm{C}$~\cite{Lucini:2015hfa}. These two definitions of 
QED in a finite volume are discussed in Refs.~\cite{Hayakawa:2008an,Blum:2010ym,Fodor:2015pna} and Refs.~\cite{Wiese:1991ku,Polley:1993bn,Lucini:2015hfa}, respectively.
For other formulations \cite{Gockeler:1989wj,Duncan:1996xy,Endres:2015gda},
  we refer to Sec.~\ref{sec:qmass} and
for a discussion on open problems to Ref.~\cite{Patella:2017fgk}.  
Since the bare parameters need to be fixed through experimental observables, finite volume effects have to be removed from the $M_i$ and the behaviour Eq.~\eqref{eq:QEDFV} is crucial in this respect.
 
Another important observation concerns the use of observables associated with decay rates or cross sections in setting the scale. The issue is particularly subtle in QCD$+$QED because of the well known problem associated with the appearance of infrared divergences at intermediate stages of the calculations. The solution requires a proper definition
of infrared-safe observables, according to the
Bloch--Nordsieck mechanism~\cite{Bloch:1937pw}. These measurable
observables are obtained by including in the final state of a process
 any number of soft photons with total energy up to a given physical threshold. Once the infrared-safe measurable observable has been constructed, it can be used in the scale setting as any other measurable quantity. A particularly relevant example is the leptonic decay rate of the pion,
\begin{equation}
\Gamma^\mathrm{QCD+QED}[\pi^-\mapsto \mu\bar\nu_\mu(\gamma),E_\gamma] \,.
\label{eq:piellnudecayrate}
\end{equation}
Here, phase space is integrated over with the constraint that the total energy of the photons is below $E_\gamma$. The feasibility of using such an observable in place of a stable hadron mass has to be judged on the basis of the overall precision, statistical plus systematics, that is achievable in the lattice calculation (see Refs.~\cite{Giusti:2017dwk,DiCarlo:2019thl}).

\subsubsection{Hadronic definition of QCD and of QED corrections}
Under the assumption of negligible weak (and gravity) corrections QCD$+$QED is the complete theory and, therefore, the predictions obtained from lattice simulations for any observable ${\cal O}^\mathrm{QCD+QED}$, that has not already been used in the scale setting, are  unambiguous.  
On the contrary, what we call the QCD contribution ${\cal O}^\mathrm{QCD}$ and the associated radiative corrections,
\begin{equation}
\delta {\cal O}^\mathrm{QCD} = \frac{{\cal O}^\mathrm{QCD+QED}}{{\cal O}^\mathrm{QCD}} -1 \,,
\label{e:deltaQCD}
\end{equation}
\emph{do} depend upon the inputs used to define QCD.

Going back to Eq.~(\ref{e:hadscheme}), different hadronic definitions of QCD can be obtained by chosing different hadron masses and/or different values for the ``physical'' inputs. Once we have chosen which hadron masses to use, the different hadronic schemes can be identified by writing
\begin{equation}
	\frac{M_i(g_0,\{ a m_\mathrm{0,j}\})}{
	M_1(g_0,\{a m_\mathrm{0,j}\})} 
	=\frac{M_i^\mathrm{QCD}}{M_1^\mathrm{QCD}}\,, \quad 
	i=2\ldots\Nf+1\,,
	\quad
	j=1\ldots\Nf\,,
	\label{e:hadschemeQCD}
\end{equation}
and by specifying the values of the external inputs, for example parameterized by $\varepsilon_i^\mathrm{QCD}$ in\footnote{After having calibrated the full theory (QCD$+$QED) with physical hadronic inputs, one can compute the strong coupling constant and the quark masses in a given renormalization scheme. These can then be used to define QCD by matching the corresponding renormalized quantities. This is the so--called GRS approach originally introduced in Ref.~\cite{Gasser:2003hk}. We refer to Ref.~\cite{DiCarlo:2019thl} for a discussion concerning the connection of the $\varepsilon$--language used here and the GRS scheme, and to Sec.~\ref{sec:qmass} for a detailed discussion of the different schemes that have been used in the literature to define (iso)QCD including the original references on the subject.}
\begin{equation}
M_i^\mathrm{QCD}=M_i^\mathrm{exp}\left(1+ \varepsilon_i^\mathrm{QCD}\right)\,.
\label{eq:firsttimeeps}
\end{equation} 

A  ``natural'' choice is to set $\varepsilon_i^\mathrm{QCD}=0$, i.e., to define QCD by using exactly the experimental values for the stable hadron masses entering the calibration procedure. In this case, if the same hadron mass is used in the definition of the full theory, Eq.~(\ref{e:hadschemeQED}), and in the definition of QCD, then the radiative corrections on these quantities are zero by construction. 
Radiative corrections on any other predictable quantity are well defined and nonvanishing. 

In light of this observation, the introduction of the $\varepsilon_i^\mathrm{QCD}$ parameters might appear unnecessary. 
However, this is not the case for the following reasons. 
Isosymmetric QCD (isoQCD),  already introduced
in \sect{sec:QCDhadRen}, is another  good approximation of the real world. Due to $m_u=m_d=m_{ud}$, the theory only depends on $N_f$ parameters. In order to set the masses of the light and strange quarks in isoQCD the options of using the charged or the neutral pion and kaon masses are equally valid from the physical point of view. If one picks, e.g., the neutral meson masses, then 
one has nonzero $\varepsilon_i$ when the right hand side of Eq.~(\ref{eq:firsttimeeps}) is written in
terms of the charged ones. Furthermore, on the basis of symmetry
arguments and/or (chiral) effective theory calculations one may argue
that certain linear combinations of charged and neutral meson masses
are more ``natural'' than others (see the discussion in the quark-mass section, Sec.~\ref{sec:qmass}) because the resulting radiative corrections are smaller.

As a matter of fact, many of the existing lattice calculations have been performed in the isospin-symmetric limit, but not all the results considered in this review correspond to the very same definition of QCD. The commonly adopted values for the pion and $\Omega$ masses in isoQCD are
\begin{equation*}
M_\pi^\mathrm{isoQCD}=M_{\pi^0}^\mathrm{exp}\,,
\qquad
M_\Omega^\mathrm{isoQCD}=M_{\Omega^-}^\mathrm{exp}\quad \text{in Refs.~\cite{Miller:2020evg,Blum:2014tka,Aoki:2008sm}}\,.
\end{equation*}
For the kaon mass in isoQCD different collaborations made different choices, e.g., the values 
\begin{eqnarray*}
	M_K^\mathrm{isoQCD}&=&494.2(4)~\mev \quad \text{in Refs.~\cite{Miller:2020evg,Carrasco:2014cwa,Bazavov:2009fk} },
	\\
	M_K^\mathrm{isoQCD}&=&495.7~\mev \quad \text{in Ref.~\cite{Blum:2014tka} },
 \\	
	M_K^\mathrm{isoQCD}&=&497.6~\mev \quad \text{in Ref.~\cite{Aoki:2008sm} }.
\end{eqnarray*}
The different choices of experimental inputs are perfectly legitimate if QED radiative corrections are neglected, but
in principle predictions of isoQCD 
do depend on these choices, and it is not meaningful to average numbers obtained with different inputs. However,
at the present level of precision the sub-percent differences in the
inputs are most likely not relevant, and we will average and compare isoQCD results irrespective of these differences. The issue will become important when results become significantly more precise. Of course, it may not be ignored, when radiative corrections, Eq.~\eqref{e:deltaQCD}, 
are directly compared between collaborations. In this case, we strongly
suggest to compare results for the unambiguous full theory observable
or to stick to a standard.

Indeed, for the future, it is  highly desirable to define such a standard for the parameters used to define (iso)QCD. We suggest using\footnote{ We note that the $\pi^0$ is unstable in QCD$+$QED and, therefore, it is much more convenient to use $M_{\pi^+}^\mathrm{exp}$ to calibrate the full theory. Although it is perfectly consistent to use different observables in the calibration of the full theory and of isoQCD, in the specific case one can write $M_{\pi}^\mathrm{isoQCD}=M_{\pi^+}^\mathrm{exp}(1+\varepsilon_\pi^\mathrm{isoQCD})$ with $\varepsilon_\pi^\mathrm{isoQCD}=(M_{\pi^0}^\mathrm{exp}-M_{\pi^+}^\mathrm{exp})/M_{\pi^+}^\mathrm{exp}$. In that language, the same observable is used in both theories but with a nonvanishing $\varepsilon$.}
\begin{eqnarray}
M_\pi^\mathrm{isoQCD}&=&M_{\pi^0}^\mathrm{exp}\,,
\nonumber \\[-1ex]
\label{e:FLAGiso}
 \\[-1ex]
M_K^\mathrm{isoQCD}&=&M_{K^0}^\mathrm{exp}\,,
\nonumber
\end{eqnarray}
while it is difficult to define a standard scale $M_1$ right now.
Going by the majority of the large-scale computations, the two options
$m_\Omega$ and pion leptonic decay rate are equally popular at the moment (see Sec.~\ref{s:physscal}). 

Since leptonic decay rates of pion and kaon play a prominent r\^ole in scale setting, we discuss the (pure) QCD definition of these quantities and of the associated radiative corrections in some detail. There are no ambiguities in the definition of the physical observable in QCD$+$QED that, in this case, is the decay rate introduced in Eq.~(\ref{eq:piellnudecayrate}) above. We now assume that (iso)QCD has been \emph{already defined} by using as hadronic inputs hadron masses. It is then possible  to compute the leptonic decay rate in QCD,
\begin{eqnarray}
\Gamma^\mathrm{QCD}[\pi\mapsto \mu\bar\nu_\mu]
=
\frac{G_F^2}{8\pi}\vert V_{ud}\vert^2\, M_{\pi^-}^\mathrm{exp}\left(m_\mu^\mathrm{exp}\right)^2\left[
1-\frac{\left(m_\mu^\mathrm{exp}\right)^2}{\left(M_{\pi^-}^\mathrm{exp}\right)^2}
\right]\, \left(f_\pi^\mathrm{QCD}\right)^2
\label{eq:piellnuQCD}
\end{eqnarray} 
where the so--called decay constant of the pion is given by
\begin{eqnarray}
f_\pi^\mathrm{QCD}=\frac{\bra{0}\bar u \gamma^0 \gamma^5 d\ket{\pi}^\mathrm{QCD}}{M_\pi^\mathrm{QCD}}\;.
\end{eqnarray}
Radiative corrections to $f_\pi^\mathrm{QCD}$ are then defined by
\begin{eqnarray}
\delta f_\pi^\mathrm{QCD}(E_\gamma) = 
\sqrt{\frac{\Gamma^\mathrm{QCD+QED}[\pi^-\mapsto \mu\bar\nu_\mu(\gamma),E_\gamma]}
{\Gamma^\mathrm{QCD}[\pi\mapsto \mu\bar\nu_\mu]}}-1\;,
\end{eqnarray}
such that
\begin{eqnarray}
\Gamma^\mathrm{QCD+QED}[\pi^-\mapsto \mu\bar\nu_\mu(\gamma),E_\gamma]
=
\Gamma^\mathrm{QCD}[\pi\mapsto \mu\bar\nu_\mu]\left[
1+\delta f_\pi^\mathrm{QCD}(E_\gamma)
\right]^2\,.
\end{eqnarray}
We want to stress once again that the definition of $\delta f_\pi^\mathrm{QCD}(E_\gamma)$ is not unique. As in the case of any other observable, different values of $\delta f_\pi^\mathrm{QCD}(E_\gamma)$ are obtained if one changes the prescription used to define QCD. In this case, in addition, one has to specify the photon energy treshold $E_\gamma$ and, moreover, the exact expression used to define $\Gamma^\mathrm{QCD}$. Indeed, it would be perfectly legitimate to replace $M_{\pi^-}^\mathrm{exp}$ appearing in the kinematical factors of Eq.~(\ref{eq:piellnuQCD}) with $M_\pi^\mathrm{QCD}$. The effect of such a different definition of $\Gamma^\mathrm{QCD}$ would be compensated by a change in $\delta f_\pi^\mathrm{QCD}(E_\gamma)$ with no ambiguities in the full theory observable $\Gamma^\mathrm{QCD+QED}$.

We mentioned in Sec.~\ref{sec:QCDhadRen} that there are advantages from the numerical point of view in using the leptonic decay constants of the mesons in the QCD scale setting procedure. That observation can now be made more precise in light of the discussion of the previous paragraph. When  we say that we use $f_\pi$ to calibrate QCD, we mean that we choose a value for the $\delta f_\pi^\mathrm{QCD}(E_\gamma)$ and we \emph{define}
\begin{eqnarray}
f_\pi^\mathrm{QCD}=
\frac{1}{1+\delta f_\pi^\mathrm{QCD}(E_\gamma)}
\sqrt{
\frac{\Gamma^\mathrm{exp}[\pi^-\mapsto \mu\bar\nu_\mu(\gamma),E_\gamma]}
{
\frac{G_F^2}{8\pi}\vert V_{ud}\vert^2\, M_{\pi^-}^\mathrm{exp}\left(m_\mu^\mathrm{exp}\right)^2\left[
1-\frac{\left(m_\mu^\mathrm{exp}\right)^2}{\left(M_{\pi^-}^\mathrm{exp}\right)^2}
\right]
}
}\;.
\end{eqnarray}
In the notation of the $\varepsilon$ parameters introduced above, one has
\begin{eqnarray}
1+\varepsilon^\mathrm{QCD}_{f_\pi}= 
\frac{1}{1+\delta f_\pi^\mathrm{QCD}(E_\gamma)}\;.
\end{eqnarray}
Again, a possible choice would be to set $\varepsilon^\mathrm{QCD}_{f_\pi}$ to zero and to use directly the experimentally measured decay rate at a given value of $E_\gamma$.\footnote{This procedure unavoidably requires that one provides a value for the CKM matrix element $V_{ud}$ that has then to be considered an input of the lattice calculation and not a predictable quantity.} 
 Common practice among the different lattice collaborations is to set
\begin{eqnarray}
E_\gamma=E_\gamma^\mathrm{max}=\frac{M_{\pi^-}^\mathrm{exp}}{2}\left[
1-\frac{\left(m_\mu^\mathrm{exp}\right)^2}{\left(M_{\pi^-}^\mathrm{exp}\right)^2}
\right]\;,
\end{eqnarray}
the maximum energy allowed to a single photon in the case of negligible $\cO(\alpha_\mathrm{em}^2)$ corrections, and to use the value 
\begin{equation}
	\delta f_\pi^\mathrm{isoQCD}(E_\gamma^\mathrm{max}) = 0.0088(11) \label{e:deltafpi}
\end{equation}
obtained in Refs.~\cite{Cirigliano:2007ga,Ananthanarayan:2004qk,Cirigliano:2011tm} in chiral perturbation theory and using the standard definition \eq{e:FLAGiso}. The corresponding number for kaon decays is
\begin{equation}
	\delta f_K^\mathrm{isoQCD}(E_\gamma^\mathrm{max}) = 0.0053(11) \,.\label{e:deltafk}
\end{equation}
A recent lattice determination in the electro-quenched approximation~\cite{DiCarlo:2019thl} \begin{eqnarray}
	\delta f_\pi^\mathrm{isoQCD}(E_\gamma^\mathrm{max}) &=& 0.0076(9)\,,
\end{eqnarray}
agrees well with \eq{e:deltafpi}, 
while the number for Kaon decays,
\begin{eqnarray}
	\delta f_K^\mathrm{isoQCD}(E_\gamma^\mathrm{max}) &=& 0.0012(5)\,,		
\end{eqnarray}
differs by more than three (quadratically
combined) error bars from \eq{e:deltafk}. 
The scheme dependence can be neglected at the present 
level of accuracy.

\subsection{Physical scales}
\label{s:physscal}
The purpose of this short section is to summarize the most popular
scales and give a short discussion of their advantages and
disadvantages. We restrict ourselves to those used in more recent
computations and  thus have a rather short list.

\subsubsection{The mass of the $\Omega$ baryon}
As already discussed, masses of hadrons that are stable in QCD$+$QED
and have a small width, in general, are very good candidates for
physical scales since there are no QED infrared divergences to be discussed. Furthermore, 
remaining within this class, the radiative corrections $\delta M^\mathrm{QCD}_i$, \eq{e:deltaQCD}, are expected to be small. Furthermore, the $\Omega$ baryon has a significantly better noise/signal ratio than the nucleon (see \fig{f:plateaux}). It also has little dependence 
on up- and down-quark masses, since it is composed entirely of strange
valence quarks. 

Still, one has to be aware that
the mass is not extracted from the plateau
 region but from a modelling of the approach to a plateau in the form of fits \cite{Miller:2020evg,Borsanyi:2020mff,Blum:2014tka,Borsanyi:2012zs,Aoki:2010dy,Aoki:2008sm}. In this sense, the noise/signal ratio problem may persist.
The use of various interpolating fields for the 
$\Omega$ helps in constraining such analyses, but it
would be desirable to have a theoretical understanding of multi-hadron (or in QCD$+$QED multi-hadron $+$ photon) contributions as it exists for the nucleon~\cite{Bar:2017kxh} as discussed in \sect{sec:NME}. In the present review, we take the estimates of the collaborations at face value and do not try to apply a rating or an estimate of systematic error due to excited-state contributions.

\subsubsection{Pion and kaon leptonic decay rates}
These decay rates have been discussed above. Here, we just summarize
the main issues. In QCD$+$QED there is so far only one computation of
the decay rate in the electro-quenched approximation
\cite{DiCarlo:2019thl}. The derived estimate for the radiative
corrections agrees with the estimates from chiral perturbation theory (see
Eqs.~(\ref{e:deltafpi}) and (\ref{e:deltafk})). The quoted
uncertainties are at the level of 0.001. This directly sets a limit to
the achievable precision on the scale in isoQCD. At present, this limit
is not yet relevant. A second source of uncertainty is due to the
knowledge of $V_{ud}$ and $V_{us}$. For convenience, we summarize the
isoQCD values
\begin{align}
  f_\pi^\mathrm{isoQCD}\,|V_{ud}| &=127.13(2)_\mathrm{exp}(13)_\mathrm{QED}\,\mev  \,,\\
  f_\pi^\mathrm{isoQCD} &= 130.56(2)_\mathrm{exp}(13)_\mathrm{QED}(2)_{V_{ud}}\,\mev  \,,\\
& \nonumber \\
  f_K^\mathrm{isoQCD} \,|V_{us}| &= \phantom{0}35.09(4)_\mathrm{exp}(4)_\mathrm{QED} \,\mev \,, \\
  f_K^\mathrm{isoQCD} &=157.2(2)_\mathrm{exp}(2)_\mathrm{QED}(4)_{V_{us}}\,\mev \,,
\end{align}
where we have used the PDG values \cite{Zyla:2020zbs} for $f_x |V_y|$ (equivalent to Eqs.~(\ref{e:deltafpi}) and (\ref{e:deltafk})), 
and the values 
\begin{equation*}
	V_{ud}=0.97370(14)\,, \quad V_{us}=0.2232(6)\,.
\end{equation*}
Here, $V_{ud}$ is from the PDG  \cite{Zyla:2020zbs} (beta decays) and the latter from Sec.~\ref{sec:vusvud} ($f_+(0)$ for $N_f=2+1+1$).
Of course, the information on pion and kaon leptonic decays do not
enter the determinations of $V_{ud}$ and $V_{us}$ used here.
The uncertainties in the above values are in the following assumed to have been considered in the
estimates of the scale given by the collaborations.
This is analogous to the systematics due to excited-state contaminations in hadron masses, an issue which is irrelevant in the pseudoscalar channel (see \fig{f:plateaux}).

Depending on the lattice formulation, there is also a
nontrivial renormalization of the axial current. Since it is easily determined from a chiral Ward identity, it does not play an important r\^ole. 
When it is present, it is assumed to be accounted for in the statistical errors.

\subsubsection{Other physics scales}
Scales derived from bottomonium have been used in the past, in particular, the splitting $\Delta m_{\Upsilon} =  m_{\Upsilon(2s)}-m_{\Upsilon(1s)}$.
They have very little dependence on the light-quark masses, but need
an input for the $b$-quark mass. In all relevant cases, the $b$ quark is treated by NRQCD.

\subsection{Theory scales}
\label{s:theoryscales}
 
In the following, we consider in more detail the two classes of theory
scales that are most commonly used in typical lattice
computations. The first class consists of scales related to the static
quark-antiquark potential \cite{Sommer:1993ce}. The second class is related to the action density renormalized through the gradient flow~\cite{Luscher:2010iy}.

\subsubsection{Potential scales}
In this approach, lattice scales are derived from the properties of
the static quark-antiquark potential. In particular, a scale can be
defined by fixing the force $F(r)$ between a static quark and
antiquark separated by the distance $r$ in
physical units \cite{Sommer:1993ce}. Advantages of using the
potential include the ease and accuracy of its computation,
and its mild dependence on the valence-quark mass. In general, a
potential scale $r_c$ can be fixed through
the condition that the static force takes a predescribed value, i.e., 
\begin{equation}
\label{eq:r0 definition}
r_c^2 F(r_c) = X_c
\end{equation}
where $X_c$ is a suitably chosen number. Phenomenological and
computational considerations
suggest that the optimal choice for $X_c$ is in the region where the
static force turns over from Coulomb-like to linear behaviour and
before string breaking occurs. In the original work
\cite{Sommer:1993ce}, it was suggested to use $X_0=1.65$ leading to
the condition
\begin{equation}
\label{eq:r0 definition}
r_0^2 F(r_0) = 1.65 \, .
\end{equation}
In Ref.~\cite{Bernard:2000gd}, the value $X_1=1.0$ was proposed yielding the
scale $r_1$.

The static force is the derivative of the static quark-antiquark potential $V(r)$ which can be determined  through the calculation of Wilson loops. 
More specifically, the potential at distance $r$ is extracted from the asymptotic time dependence of the $r \times t$-sized Wilson loops $W(r,t)$,
\begin{equation}
\label{eq:V(r) Wilson loops}
V(r)=-\lim_{t\rightarrow \infty} \frac{d}{d t} \log \langle W(r,t) \rangle \,.
\end{equation}
The derivative of the potential needed for the force is then determined through the derivative of a suitable local parameterization of the potential as a function of $r$, e.g., 
\begin{equation}
V(r) =  C_-\frac{1}{r} + C_0 + C_+ r \,,
\label{eq:V parametrization}
\end{equation}
estimating uncertainties due to the parameterization.
In some calculations, the gauge field is fixed to Coulomb or temporal
gauge in order to ease the computation of the potential at arbitrary
distances.

In order to optimize the overlap of the Wilson loops with the ground
state of the potential, one can use different types and levels of
spatial gauge field smearing and extract the ground state energy from
the corresponding correlation matrix by solving
a generalized eigenvalue problem~\cite{Michael:1985ne,Luscher:1990ck,Niedermayer:2000yx}. 
Finally, one can also make use of the noise reduction proposed in Refs.~\cite{DellaMorte:2005nwx,Donnellan:2010mx}. It changes the definition of the discretized loops by a smearing of the temporal parallel transporter~\cite{Hasenfratz:2001hp} and thus yields a different discretization of the continuum force.

\subsubsection{Gradient flow scales}
\label{s:flowscales}
The gradient flow $B_\mu(t,x)$ of gauge fields is defined in the continuum by the flow equation
\begin{align}
\label{eq:GF equation continuum 1}
\dot B_\mu &= D_\nu G_{\nu\mu}, \quad \left. B_\mu \right|_{t=0} =
A_\mu\, ,\\
G_{\mu\nu} &= \partial_\mu B_\nu - \partial_\nu B_\mu + [B_\mu,B_\nu],
\quad D_\mu = \partial_\mu + [B_\mu, \cdot\,]\,,
\label{eq:GF equation continuum 2}
\end{align}
where $A_\mu$ is the fundamental gauge field, $G_{\mu\nu}$ the field strength tensor, and $D_\mu$ the covariant derivative~\cite{Luscher:2010iy}.
At finite lattice spacing, a possible form of Eqs.~(\ref{eq:GF equation continuum 1}) and (\ref{eq:GF equation continuum 2}) is 
\begin{equation}
  a^2\frac{d}{dt} V_t(x,\mu) = -g_0^2 \cdot {\partial_{x,\mu} S_G(V_t)} \cdot V_t(x,\mu)\,,
\label{eq:GF equation lattice}
\end{equation}
where $V_t(x,\mu)$ is the flow of the original gauge field $U(x,\mu)$
at flow time $t$, $S_G$ is an arbitrary lattice discretization of the
gauge action, and $\partial_{x,\mu}$ denotes the su$(3)$-valued
differential operator with respect to $V_t(x,\mu)$. An important point to
note is that the flow time $t$ has the dimension of a length squared,
i.e., $t \sim a^2$, and hence provides a means for setting the scale. 

One crucial property of the gradient flow is that any function of the
gauge fields evaluated at flow times $t>0$ is
renormalized~\cite{Luscher:2011bx} by just renormalizing the gauge
coupling. Therefore, one can define a  scale by keeping a suitable
gluonic observable defined at constant flow time $t$, e.g., the action
density $E=-\frac1{2} \Tr
\,G_{\mu\nu}G_{\mu\nu}$~\cite{Luscher:2010iy}, fixed in physical
units. This can, for example, be achieved through the condition
\begin{equation}
t_c^2 \langle E(t_c,x)\rangle = c \,, \quad E(t,x)=-\frac1{2} \Tr\, G_{\mu\nu}(t,x)G_{\mu\nu}(t,x)
\label{eq:GF t definition}
\end{equation}
where $ G_{\mu\nu}(t,x)$ is the field strength tensor evaluated on the
flown gauge field $V_t$.
Then, the lattice scale $a$ can be determined from the dimensionless
flow time in lattice units, $\hat t_c = a^2 t_c$. The original proposal in \cite{Luscher:2010iy} was to use $c = 0.3$
yielding the scale $t_0$,
\begin{equation}
t_0^2 \langle E(t_0)\rangle = 0.3 \, .
\label{eq:GF t0 definition}
\end{equation}
 For convenience one sometimes also defines
$s_0 = \sqrt{t_0}$. 

An alternative scale $w_0$ has been introduced in
Ref.~\cite{Borsanyi:2012zs}. It is defined by fixing a suitable
derivative of the action density,
\begin{equation}
\label{eq:GF w0 definition}
W(t_c) = t_c \cdot \partial_t \left(t^2 \langle
  E(t)\rangle\right)_{t=t_c}  = c\, .
\end{equation}
Setting $c=0.3$ yields the scale $w_0$ through
\begin{equation}
W(w_0^2) = 0.3 \, .
\end{equation}

In addition to the lattice scales from $t_0$ and $w_0$, one can also consider the scale from the dimensionful combination $t_0/w_0$. 
This combination has been found to have a very weak dependence on the quark mass~\cite{Deuzeman:2012jw,Abdel-Rehim:2015pwa,Alexandrou:2021bfr}.

A useful property of the gradient flow scales is the fact that their quark-mass dependence is known from $\chi$PT \cite{Bar:2013ora}.

Since the action density at $t \sim t_0 \sim w_0^2$ usually suffers from large autocorrelation~\cite{Deuzeman:2012jw,Schaefer:2012tq}, the calculation of the statistical error needs special care.

Lattice artefacts in the
gradient flow scales originate from different sources \cite{Ramos:2015baa}, which are systematically discussed by considering $t$ as a coordinate in a fifth dimension. First, there is the choice of the action $S_G$ for $t>0$.  Second, there is the discretization of $E(t,x)$. Third, there is the discretization of the 4-dimensional quantum action, which is always there, and fourth, there 
are also terms localized at the boundary $t=0_+$. 
The interplay
between the different sources of lattice artefacts  turns out to be rather
subtle~\cite{Ramos:2015baa}. 

Removing discretization errors due to the first two sources requires only classical ($g_0$-independent) improvement. 
Those due to the quantum
action are common to all $t=0$ observables, but the effects of the boundary terms are not easily removed in practice. At tree level, the Zeuthen flow~\cite{Ramos:2015baa}  does the complete job, but none of the computations reviewed here have used it. 
Discretization effects due to $S_G$
can be removed by using an improved action such as the tree-level
Symanzik-improved gauge action
\cite{Borsanyi:2012zs,Bazavov:2013gca}. More phenomenological attempts of improving the gradient flow scales consist of
applying a
$t$-shift \cite{Cheng:2014jba}, or tree-level improvement \cite{Fodor:2014cpa}.

\subsubsection{Other theory scales}
\label{subsubsec:other scales}
The MILC collaboration has been using another set of scales,  
the partially quenched pseudoscalar decay constant $f_{p4s}$ with
degenerate valence quarks with a mass $m_q=0.4 \cdot
m_\mathrm{strange}$,  and the corresponding partially quenched pseudoscalar mass $M_{p4s}$. So far it has been a quantity only used by the
MILC collaboration \cite{Bazavov:2014wgs,Bazavov:2017lyh,Bazavov:2012xda}. We  do not perform an in-depth discussion or an average but will list numbers in the results section.

Yet another scale that has been used is the leptonic decay constant of the $\eta_s$. This fictitious particle is 
a pseudoscalar made of a valence quark-antiquark pair with different
(fictitious) flavours which are mass-degenerate
with the strange quark \cite{Dowdall:2011wh,Davies:2009tsa,Gray:2005ur}. 

\else
\fi

\subsection{List of computations and results}

\subsubsection{Gradient flow scales}
We now turn to a review of the calculations of the gradient flow scales $\sqrt{t_0}$ and $w_0$. The results are compiled in Tab.~\ref{tab_GFscales} and shown in Fig.~\ref{fig_GFscales}. In the following, we briefly discuss the calculations in the order that they appear in the table and figure.
\begin{table}[!h]
  \vspace*{3cm}
\footnotesize{
\begin{tabular*}{1.0\textwidth}[l]{l@{\extracolsep{\fill}}rl@{\hspace{1mm}}l@{\hspace{1mm}}l@{\hspace{1mm}}l@{\hspace{1mm}}lll@{\hspace{1mm}}l}

Collaboration & Ref. & $\Nf$ & \begin{rotate}{60}{publication status}\end{rotate} &  \begin{rotate}{60}{chiral extrapolation}\end{rotate} & \begin{rotate}{60}{continuum extrapolation}\end{rotate} & \begin{rotate}{60}{finite volume}\end{rotate} &\begin{rotate}{60}{physical scale}\end{rotate}  & $\sqrt{t_0}$ [fm] & $w_0$ [fm] \\
\hline\hline
ETM 21 & \cite{Alexandrou:2021bfr} & 2+1+1 & \oP &\good&\good&\good& $f_\pi$  & 0.14436(61) & 0.17383(63) \\

CalLat 20A& \cite{Miller:2020evg} & 2+1+1 &  \gA &\good&\good&\good& $m_\Omega$ & 0.1422(14) & 0.1709(11) \\


BMW 20 & \cite{Borsanyi:2020mff} & 1+1+1+1 &\gA &\good&\good&\good& $m_\Omega$ &  &  0.17236(29)(63)[70] \\

ETM 20 & \cite{Dimopoulos:2020eqd} & 2+1+1 & \rC &\good&\good&\good& $f_\pi$  &  & 0.1706(18) \\
  
MILC 15 &\cite{Bazavov:2015yea} & 2+1+1 & \gA &\good&\good&\good& ${F_{p4s}(f_\pi)}^\#$ &  0.1416(+8/-5) &  0.1714(+15/-12) \\
  
HPQCD 13A & \cite{Dowdall:2013rya} &  2+1+1 & \gA & \good &\soso & \good & $f_\pi$ & 0.1420(8) & 0.1715(9) \\
\hline

CLS 16 & \cite{Bruno:2016plf} & 2+1 & \gA &\soso&\good&\good& $f_\pi,f_K$  & 0.1467(14)(7)  & \\

QCDSF/UKQCD 15B & \cite{Bornyakov:2015eaa} & 2+1 & \oP & \soso & \soso & \soso & $m_P^{SU(3)}$  & 0.1511(22)(6)(5)(3) & 0.1808(23)(5)(6)(4) \\

RBC/UKQCD 14B & \cite{Blum:2014tka} & 2+1 & \gA &\good&\good&\good& $m_\Omega$  & 0.14389(81) & 0.17250(91) \\

HotQCD 14 & \cite{Bazavov:2014pvz} &  2+1  & \gA &\good&\good&\good& ${r_1(f_\pi)}^\#$  & & 0.1749(14) \\

BMW 12A & \cite{Borsanyi:2012zs} & 2+1 &\gA & \good & \good & \good & $m_\Omega$ &  0.1465(21)(13)  &  0.1755(18)(4) \\

\hline\hline
\end{tabular*}\\[-0.2cm]

\caption{Results for gradient flow scales at the physical point, cf.~\eq{e:scalesettbasic}.
Note that BMW 20 \cite{Borsanyi:2020mff} take IB and QED corrections into account. 
   Some additional results for ratios of scales are:\newline
     ETM 21 \cite{Alexandrou:2021bfr}: $t_0/w_0 = 0.11969(62)$ fm.
\newline
$^\#$ These scales are not physical scales and have been determined from $f_\pi$. 
\label{tab_GFscales}
}
}\end{table}


ETM 21 \cite{Alexandrou:2021bfr}
  finalizes and supersedes ETM 20 discussed below. It determines the scales $\sqrt{t_0}, w_0$, also $t_0/w_0 = 0.11969(62)$ fm, and the ratio $\sqrt{t_0}/w_0 = 0.82930(65)$, cf.~also HPQCD 13A \cite{Dowdall:2013rya}.

  CalLat 20A \cite{Miller:2020evg} use M\"obius Domain-Wall valence fermions on HISQ ensembles generated by the MILC and CalLat collaborations. The gauge fields entering the M\"obius Domain-Wall operator are gradient-flow smeared with $t=a^2$. They compute the $\Omega$ mass and the scales  $w_0,\,t_0$ and perform global fits to determine $w_0 M_\Omega$ and $\sqrt{t_0} M_\Omega$ at the physical point. The flow is discretized with the Symanzik tree-level improved action and the clover discretization of $E(t)$ is used. A global fit with Bayesian priors is performed including terms derived from  {\Ch}PT for finite volume and quark-mass dependences, as well as $a^2$ and $a^2 \alpha_s(1.5/a)$ terms for discretization errors. Also, a tree-level improved definition of the GF scales is used where the leading-in-$g^2$ cutoff effects are removed up to and including
  ${\cal O}(a^8/t^4)$.

  BMW 20 \cite{Borsanyi:2020mff}
     presents a result for $w_0$ in the context of their staggered
     fermion calculation of the muon anomalous magnetic moment. It is
     the first computation that takes QED and isospin-breaking
     corrections into account.
  The simulations are performed by using staggered fermions with stout
  gauge field smearing with six lattice spacings
  and several pion masses around the physical
  point with $M_\pi$ between $110$ and $140$ MeV. Volumes are around
  $L=6$ fm. 
  At the largest lattice spacing, it is demonstrated how the effective masses of the $\Omega$ correlator almost reach the plateau value extracted from a four-state fit (two states per parity). Within the range where the data is fitted, the deviation 
  of data points from the estimated plateau is less than a percent. Isospin-breaking  corrections are computed by Taylor expansion around isoQCD with QED treated as \qedl. Finite volume effects in QED are taken from the $1/L, 1/L^2$ universal corrections and $\cO(1/L^3)$ effects are neglected. The results for $M_\Omega w_0$ are extrapolated to the continuum by a fit with $a^2$ and $a^4$ terms. 

   ETM 20 \cite{Dimopoulos:2020eqd} presents in their proceedings contribution a preliminary analysis of their $N_f=2+1+1$ Wilson twisted-mass fermion simulations at maximal twist (i.e., automatic ${\cal O}(a)$ improved), at three lattice spacings and pion masses at the physical point. Their determination of $w_0=0.1706(18)$ fm from  $f_\pi$  using an analysis in terms of $M_\pi$ is the value quoted above. They obtain the consistent value $w_0=0.1703(18)$ fm from an analysis in terms of the renormalized light quark mass.

   MILC 15 \cite{Bazavov:2015yea} sets the physical scale using the fictitious pseudoscalar decay constant $F_{p4s=}153.90(9)(+21/-28)$ MeV with degenerate valence quarks of mass $m_v = 0.4 m_s$ and physical sea-quark masses \cite{Bazavov:2012xda}. ($F_{p4s}$ has strong dependence on the valence-quark mass and is determined from $f_\pi$.) They use a definition of the flow scales where the tree-level lattice artefacts up to ${\cal O}(a^4/t^2)$ are divided out. Charm-quark mass mistunings are between 1\% and 11\%. They are taken into account at leading order in $1/m_c$  through $\Lambda^{(3)}_\text{QCD}$ applied directly to $F_{p4s}$ and $1/m_c$ corrections are included as terms in the fits. They use elaborate variations of fits in order to estimate extrapolation errors (both in GF scales and $F_{p4s}$). They include errors from   FV effects and experimental errors in $f_\pi$ in $F_{p4s}$.

  HPQCD~13A \cite{Dowdall:2013rya} uses
  eight MILC-HISQ ensembles with  lattice spacings $a$ = 0.088, 0.121, 0.151 fm.
Values of $L$ are between 2.5~fm and 5.8~fm with $M_\pi L =$ 3.3--4.6.
Pion masses range between $128$ and $306$~MeV.
QCD is defined by using the inputs 
$M_{\pi}=134.98(32)$ MeV,
$M_{K}=494.6(3)$ MeV,
$f_{\pi^+}=130.4(2)$ MeV
derived by model subtractions
of IB effects.
Additional scale ratios are given: $\sqrt{t_0}/w_0 = 0.835(8)$, $r_1/w_0=1.789(26)$.
%

CLS~13 \cite{Bruno:2016plf} uses CLS
   configurations of 2+1 nonperturbatively ${\cal O}(a)$-improved Wilson fermions. There are a few pion masses with the strange mass adjusted  along a line of $m_u+m_d+m_s=\mathrm{const}$. Three different lattice spacings are used. They determine $t_0$ at the physical point defined by  $\pi$ and $K$ masses and the linear combination $f_K+\frac12 f_\pi$. They use the Wilson flow with the clover definition of $E(t)$.

QCDSF 15B \cite{Bornyakov:2015eaa,Bornyakov:2015plz} 
  results, unpublished, are obtained by simulating $\Nf=2+1$ QCD with the tree-level Symanzik improved gauge action and clover Wilson fermions with single level stout smearing for the hopping terms together with unsmeared links for the clover term (SLiNC action). Simulations are performed at four different lattice spacings, in the range $[0.06,0.08]$~fm, with $M_{\pi,\text{min}}=228$~MeV and $M_{\pi,\text{min}}L=4.1$. The results for the gradient flow scales have been obtained by relying on the observation that flavour-symmetric quantities get corrections of $\cO((\Delta m_q)^2)$ where $\Delta m_q$ is the difference of the quark mass from the $SU(3)$-symmetric value. The $\cO(\Delta m_q^2)$ terms are not detected in the data and subsequently neglected.

    RBC/UKQCD 14B \cite{Blum:2014tka} presents results for $\sqrt{t_0}$ and $w_0$ obtained in QCD with $2+1$ dynamical flavours. The simulations are performed by using domain-wall fermions on six ensembles with lattice spacing $a^{-1}=1.38, 1.73, 1.78, 2.36, 2.38$, and 3.15 GeV, pion masses in the range $M_\pi^{unitary}\in[139,360]$~MeV. The simulated volumes are such that $M_\pi L> 3.9$. The effective masses of the $\Omega$ correlator are extracted with two-state fits and it is shown, by using two different nonlocal interpolating operators at the source, that the correlators almost reach a pleateau. In the calculation of $\sqrt{t_0}$ and $w_0$, the clover definition of $E(t)$ is used. The values given are $\sqrt{t_0}=0.7292(41) \GeV^{-1}$ and $w_0=0.8742(46) \GeV^{-1}$ which we converted to the values in Tab.~\ref{tab_GFscales}.  
%

    HotQCD 14 \cite{Bazavov:2014pvz}
  determines the equation of state with $N_f=2+1$ flavours using highly
  improved staggered quarks (HISQ/tree).
   As a byproduct, they update the results of HotQCD 11 \cite{Bazavov:2011nk} by adding simulations at four new values of $\beta$, for a total of 24 ensembles, with lattice spacings in the range $[0.04,0.25]$~fm and volumes in the range $[2.6,6.1]$~fm with $M_\pi= 160$ MeV. They obtain values for  the scale parameters $r_0$ and $w_0$, via the ratios $r_0/r_1, w_0/r_1$ and using $r_1 = 0.3106(14)(8)(4)$ fm  from MILC 10 \cite{Bazavov:2010hj}. They obtain for the ratios $(r_0/r_1)_{cont}=1.5092(39)$ and $(w_0/r_1)_{cont}  = 0.5619(21)$ in the continuum.  They crosscheck their   determination of the scale $r_1$ using the hadronic quantities $f_K$, $f_\eta$ from HPQCD 09B \cite{Davies:2009tsa} and the experimental value of $M_\varphi$, and find good agreement.

  BMW 12A  \cite{Borsanyi:2012zs} is the work in which $w_0$ was introduced.  Simulations with 2HEX smeared Wilson fermions and two-level stout-smeared rooted staggered fermions are done. The Wilson flow with clover $E(t)$ is used, and a test of the Symanzik flow is carried out. They take the results with Wilson fermions as their central value, because those ``do not rely on the `rooting' of the fermion determinant''. 
Staggered fermion results agree within uncertainties.


\begin{figure}
  \includegraphics[width=0.5\textwidth]{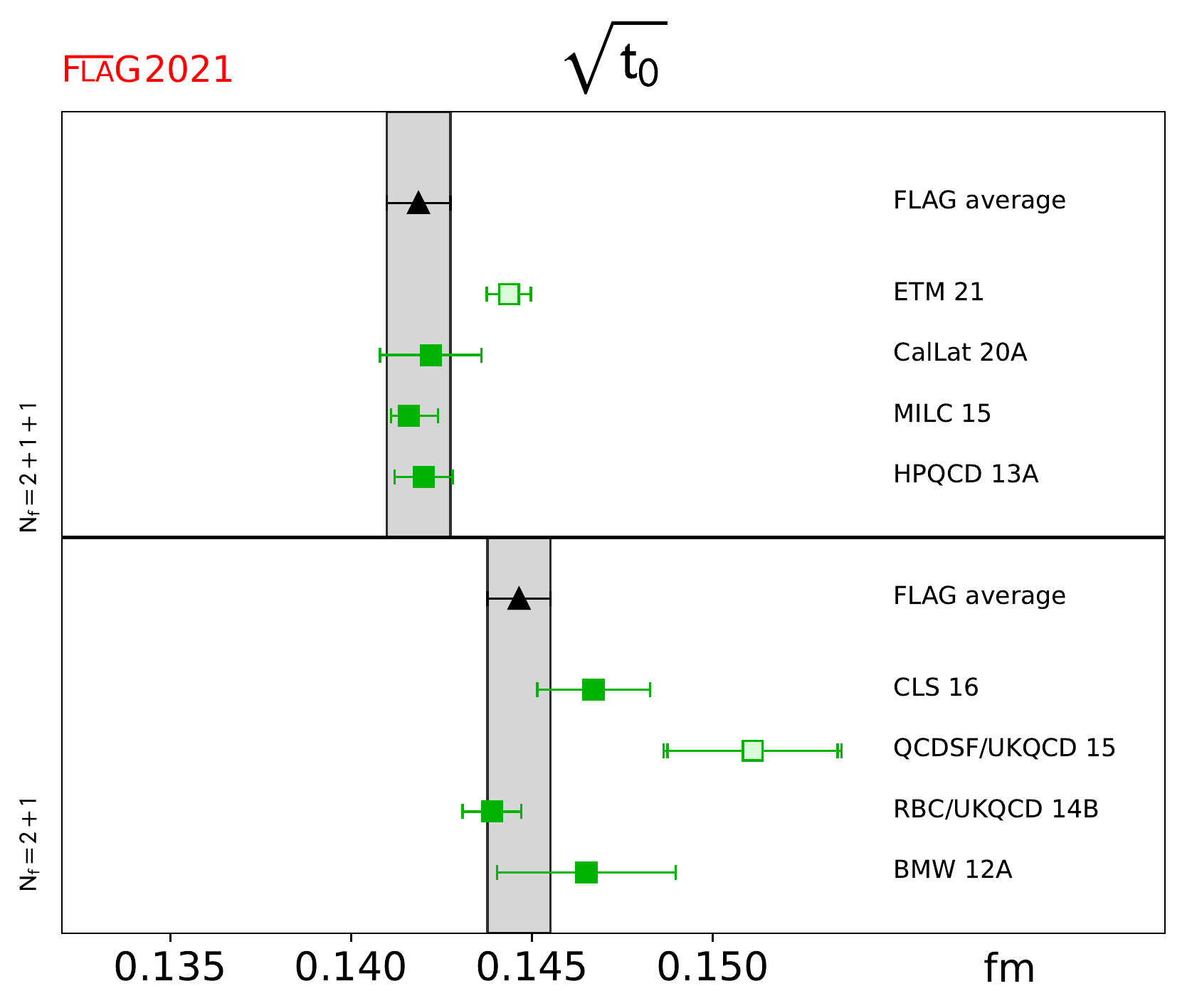}
  \includegraphics[width=0.5\textwidth]{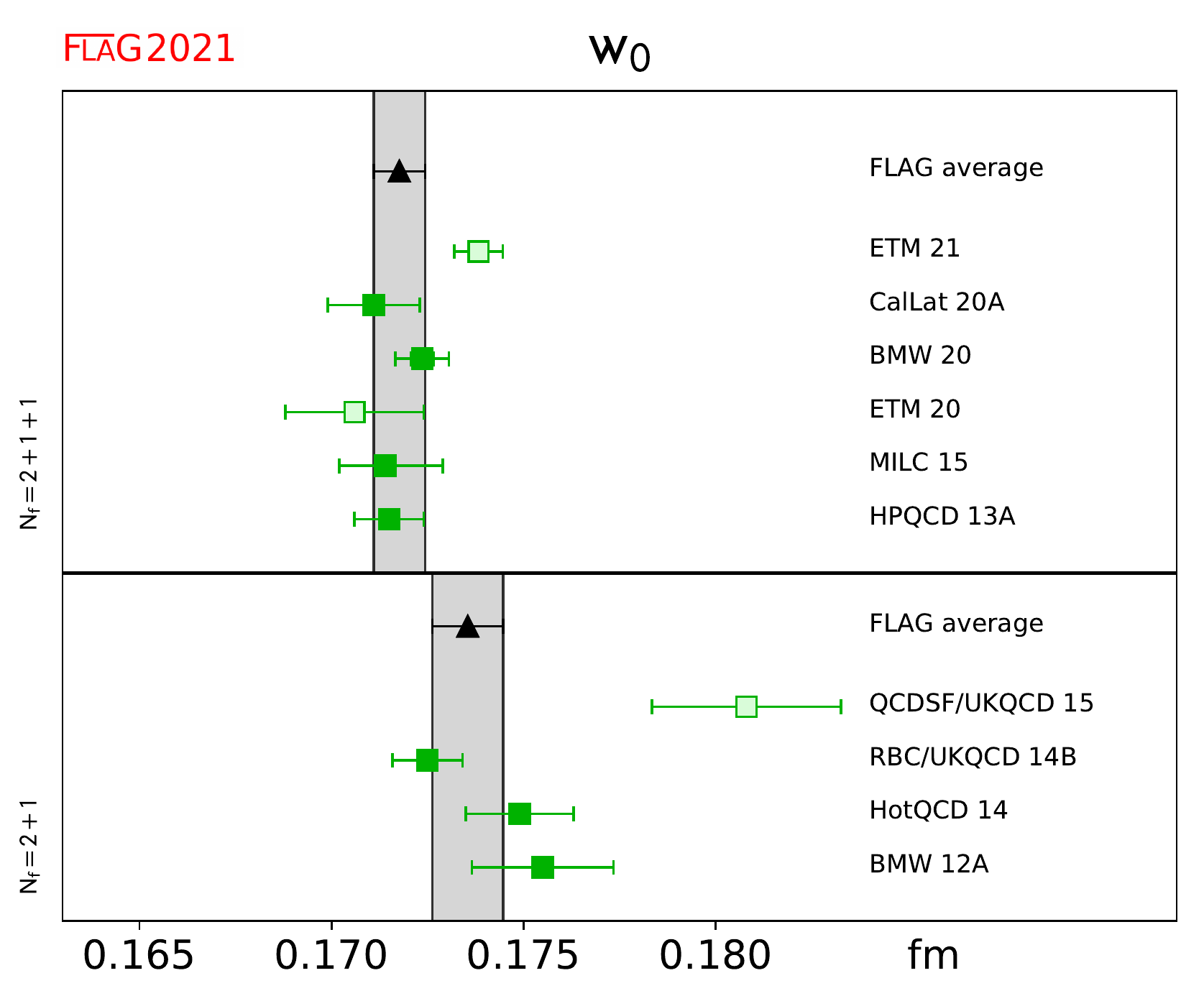}
\caption{\label{fig_GFscales} Results for gradient flow scales.}
\end{figure}

\subsubsection{Potential scales}

We now turn to a review of the calculations of the potential scales $r_0$ and $r_1$. The results are compiled in Tab.~\ref{tab_potentialscales} and shown in Fig.~\ref{fig_potentialscales}. The most recent calculations date back to 2014, and we discuss them in the order that they appear in the table and the figure.
\begin{table}[!h]
  \vspace*{3cm}
\footnotesize{
\begin{tabular*}{1.0\textwidth}[l]{l@{\extracolsep{\fill}}rlll@{\hspace{1mm}}l@{\hspace{1mm}}llll}

Collaboration & Ref. & $\Nf$ & \begin{rotate}{60}{publication status}\end{rotate} &  \begin{rotate}{60}{chiral extrapolation}\end{rotate} & \begin{rotate}{60}{continuum extrapolation}\end{rotate} & \begin{rotate}{60}{finite volume}\end{rotate} &\begin{rotate}{60}{physical scale}\end{rotate} & $r_0$ [fm] & $r_1$ [fm] \\
\hline\hline  

ETM 14 & \cite{Carrasco:2014cwa} & 2+1+1 & \gA &\soso&\good&\good& $f_\pi$ & 0.474(14) & \\
  
HPQCD 13A & \cite{Dowdall:2013rya} & 2+1+1 & \gA &\good&\soso&\good& $f_\pi$ & & 0.3112(30) \\

HPQCD 11B &  \cite{Dowdall:2011wh} & 2+1+1 & \gA &\soso&\soso&\soso& $\Delta M_{\Upsilon}$, $f_{\eta_s}$ & & 0.3209(26) \\
\hline

HotQCD 14 & \cite{Bazavov:2014pvz} & 2+1 & \gA &\good&\good&\good& $r_1$(\cite{Bazavov:2010hj})$^\#$ & 0.4688(41) & \\

$\chi$QCD 14 & \cite{Yang:2014sea} & 2+1 & \gA & \soso    & \soso & \soso    &  three inputs\footnote{$M_{D_s^*}, M_{D_s^*}-M_{D_s}, M_{J/\psi}$}     &  0.465(4)(9) & \\


HotQCD 11 & \cite{Bazavov:2011nk} & 2+1 & \gA   &\good&\good&\good& $f_\pi$ & 0.468(4) & \\
RBC/UKQCD 10A &  \cite{Aoki:2010dy} & 2+1 & \gA & \soso & \soso & \soso & $M_\Omega$  & 0.487(9) & 0.333(9) \\
MILC 10 & \cite{Bazavov:2010hj} & 2+1 & \rC &\soso &\good &\good &$f_\pi$ & &0.3106(8)(14)(4) \\

MILC 09 & \cite{Bazavov:2009bb}  & 2+1 & \gA &\soso&\good&\good& $f_\pi$ & & 0.3108(15)($^{+26}_{-79}$) \\

MILC 09A & \cite{Bazavov:2009fk} & 2+1 & \rC &\soso&\good&\good& $f_\pi$ &  & 0.3117(6)($^{+12}_{-31}$) \\

HPQCD 09B & \cite{Davies:2009tsa} & 2+1 &  \gA  &\soso&\good&\soso& three inputs &  & 0.3133(23)(3) \\

PACS-CS 08 & \cite{Aoki:2008sm} & 2+1 & \gA &\good&\bad&\bad& $M_\Omega$ & 0.4921(64)($^{+74}_{-2}$) &  \\

HPQCD 05B & \cite{Gray:2005ur} & 2+1 & \gA &\soso&\soso&\soso&$\Delta M_{\Upsilon}$ & 0.469(7) & 0.321(5) \\

Aubin 04 & \cite{Aubin:2004wf}  & 2+1  &\gA &\soso&\soso&\soso& $\Delta M_{\Upsilon}$ & 0.462(11)(4) & 0.317(7)(3) \\
\hline\hline
\end{tabular*}\\[-0.2cm]
}
\caption{Results for potential scales at the physical point, cf.~\eq{e:scalesettbasic}. $\Delta M_{\Upsilon} =  M_{\Upsilon(2s)}-M_{\Upsilon(1s)}$.
\newline $^\#$ This theory scale was determined in turn from 
$r_1$ \cite{Bazavov:2010hj}.
}    
\label{tab_potentialscales}
\end{table}


ETM 14 \cite{Carrasco:2014cwa} uses $N_f=2+1+1$ Wilson twisted-mass fermions at maximal twist (i.e., automatic ${\cal O}(a)$-improved), three lattice spacings and pion masses reaching down to $M_\pi = 211$ MeV. They determine the scale $r_0$ through $f_\pi=f_{\pi^+}=130.41$ MeV.
  A crosscheck of the so obtained lattice spacings with the ones obtained via the fictitious pseudoscalar meson $M_{s's'}$ made of two strange-like quarks gives consistent results. The crosscheck is done using the  dimensionless combinations $r_0 M_{s's'}$ (with $r_0$ in the chiral limit) and $f_\pi/M_{s's'}$ determined in the continuum, and then using $r_0/a$ and the value of $M_{s's'}$ obtained from the experimental value of $f_\pi$. We also note that in Ref.~\cite{Deuzeman:2012jw}  using the same ensembles the preliminary value $w_0=0.1782$ fm is determined, however, without error due to the missing or incomplete investigation of the systematic effects.  
  
  HPQCD~13A \cite{Dowdall:2013rya} was already discussed
  above in connection with the gradient flow scales. 

HPQCD~11B \cite{Dowdall:2011wh} uses five MILC-HISQ ensembles and determines $r_1$ from  $M_{\Upsilon(2s)}-M_{\Upsilon(1s)}$ and the decay constant
$f_{\eta_s}$ (see HPQCD~09B).
The valence $b$ quark is treated by NRQCD, while the light valence quarks have the HISQ discretization, identical to the sea quarks.
  %
  
HotQCD 14 \cite{Bazavov:2014pvz} was already discussed
in connection with the gradient flow scales. 

$\chi$QCD 14 \cite{Yang:2014sea} uses  overlap fermions as valence quarks on $N_f=2+1$ domain-wall fermion  gauge configurations generated by the RBC/UKQCD collaboration  \cite{Aoki:2010dy}. Using the physical masses of $D_s, D_s^*$ and  $J/\psi$ as inputs, the strange and charm quark masses and the decay  contant $f_{D_s}$ are determined as well as the scale $r_0$.

HotQCD 11 \cite{Bazavov:2011nk}  uses configurations with tree-level improved Symanzik gauge action and HISQ staggered quarks in addition to previously generated ensembles with p4 and asqtad staggered quarks. In this calculation, QCD is defined by generating lines of constant physics with $m_l/m_s=\{0.2,0.1,0.05, 0.025 \}$ and setting the strange quark mass by requiring that the mass of a fictious $\eta_{s\bar s}$ meson is $M_{\eta_{s\bar s}}=\sqrt{2M_K^2-M_\pi^2}$. The physical point is taken to be at $m_l/m_s=0.037$. The physical scale is set by using the value $r_1=0.3106(8)(18)(4)$~fm obtained in Ref.~\cite{Bazavov:2010hj} by using $f_\pi$ as physical input. In the paper, this result is shown to be consistent within the statistical and systematic errors with the choice of $f_K$ as physical input. {The result $r_0/r_1=1.508(5)$ is obtained by averaging over 12 ensembles at $m_l/m_s=0.05$ with lattice spacings in the range $[0.066,0.14]$~fm. This result is then used to get $r_0=0.468(4)$~fm. Finite volume effects have been monitored with 20 ensembles in the range $[3.2,6.1]$fm with $M_\pi L>2.6$.}

RBC/UKQCD 10A \cite{Aoki:2010dy} uses $N_f=2+1$ flavours of domain-wall quarks and the Iwasaki gauge action at two values of the lattice spacing with unitary pion masses in the approximate range $[290,420]$ MeV. They use the masses of $\pi$ and $K$ meson and of the $\Omega$ baryon to determine the physical quark masses and the lattice spacings, and so obtain estimates of the scales $r_0, r_1$ and the ratio $r_1/r_0$ from a combined chiral and continuum extrapolation.  

MILC 10  \cite{Bazavov:2010hj} presents a further update of $r_1$ with asqtad staggered quark ensembles with $a\in\{0.045,0.06,0.09\}$~fm. It supersedes MILC 09  \cite{Bazavov:2009bb,Bazavov:2009fk,Bazavov:2009tw}.

MILC 09  \cite{Bazavov:2009bb} presents an $N_f=2+1$ calculation of the potential scales on asqtad staggered quark ensembles with $a\in\{0.045,0.06,0.09,0.12,0.15,0.18\}$~fm. The continuum extrapolation is performed by using Goldstone boson pions as light as $M_\pi=224$~MeV (RMS pion mass of 258~MeV). The physical scale is set from $f_\pi$. The result for $r_1$ obtained in the published paper  \cite{Bazavov:2009bb} is then updated and, therefore, superseded by the conference proceedings  MILC 09A and 09B \cite{Bazavov:2009fk,Bazavov:2009tw}. 

HPQCD~09B \cite{Davies:2009tsa} is an extension of HPQCD~05B \cite{Gray:2005ur} and uses HISQ valence quarks instead of asqtad quarks. The scale $r_1$ is obtained from three different inputs. First $r_1=0.309(4)$~fm from the splitting of 2S and 1S $\Upsilon$ states as in Ref.~\cite{Gray:2005ur}, second $r_1=0.316(5)$~fm from $M_{D_s}- M_{\eta_s}/2$ and third $r_1=0.315(3)$~fm from the decay constant of the $\eta_s$. The ficitious $\eta_s$ state is operationally defined by setting quark masses to the s-quark mass and dropping disconnected diagrams. Its mass and decay constant are obtained from a partially quenched chiral perturbation theory analysis using the pion and kaon states from experiment together with various partially quenched lattice data. The three results are combined to $r_1=0.3133(23)(3)$~fm.

PACS-CS 08 \cite{Aoki:2008sm} presents a calculation of $r_0$ in $N_f=2+1$ QCD by using NP ${\cal O}(a)$-improved clover Wilson quarks and Iwasaki gauge action. The calculation is done at fixed lattice spacing  $a=0.09$~fm and is extrapolated to the physical point from (unitary) pion masses in the range $[156,702]$~MeV. The $N_f=2+1$ theory is defined by fixing $M_\pi$, $M_K$, and $M_{\Omega}$ to  $135.0,\; 497.6$, and $1672.25~\mev$, respectively. The effective masses of smeared-local $\Omega$ correlators   averaged over the four spin polarizations show quite good plateaux. 

HPQCD~05B \cite{Gray:2005ur} performed the first bottomonium spectrum calculation in full QCD with $N_f=2+1$ on MILC asqtad configurations and the $b$ quark treated by NRQCD. They find agreement of the low lying $\Upsilon$ states with experiment and also compare to quenched and $\Nf=2$ results. They determined $r_0$ and $r_1$ from the splitting of 2S and 1S states.
%

Aubin 04 \cite{Aubin:2004wf} presents an $N_f=2+1$ calculation of the potential scales by using asqtad staggered quark ensembles with $a=0.09$ and $0.12\}$~fm. The continuum extrapolation is performed by using Goldstone boson pions as light as $m_\pi=250$~MeV. The physical scale is set from the $\Upsilon$ 2S-1S and 1P-1S splittings computed with NRQCD by HPQCD \cite{Wingate:2003gm}.


\begin{figure}[htb]
  \includegraphics[width=0.5\textwidth]{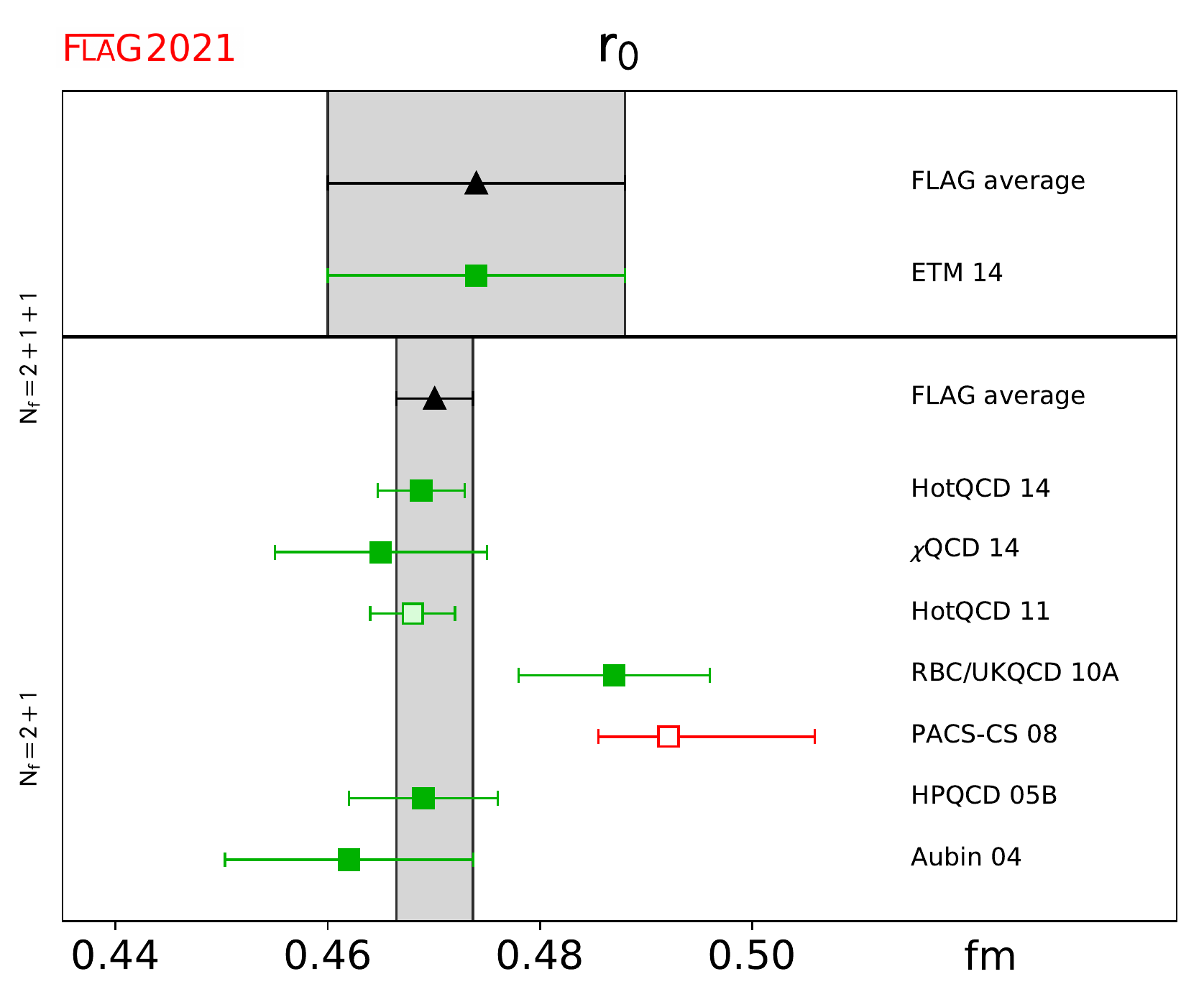}
  \includegraphics[width=0.5\textwidth]{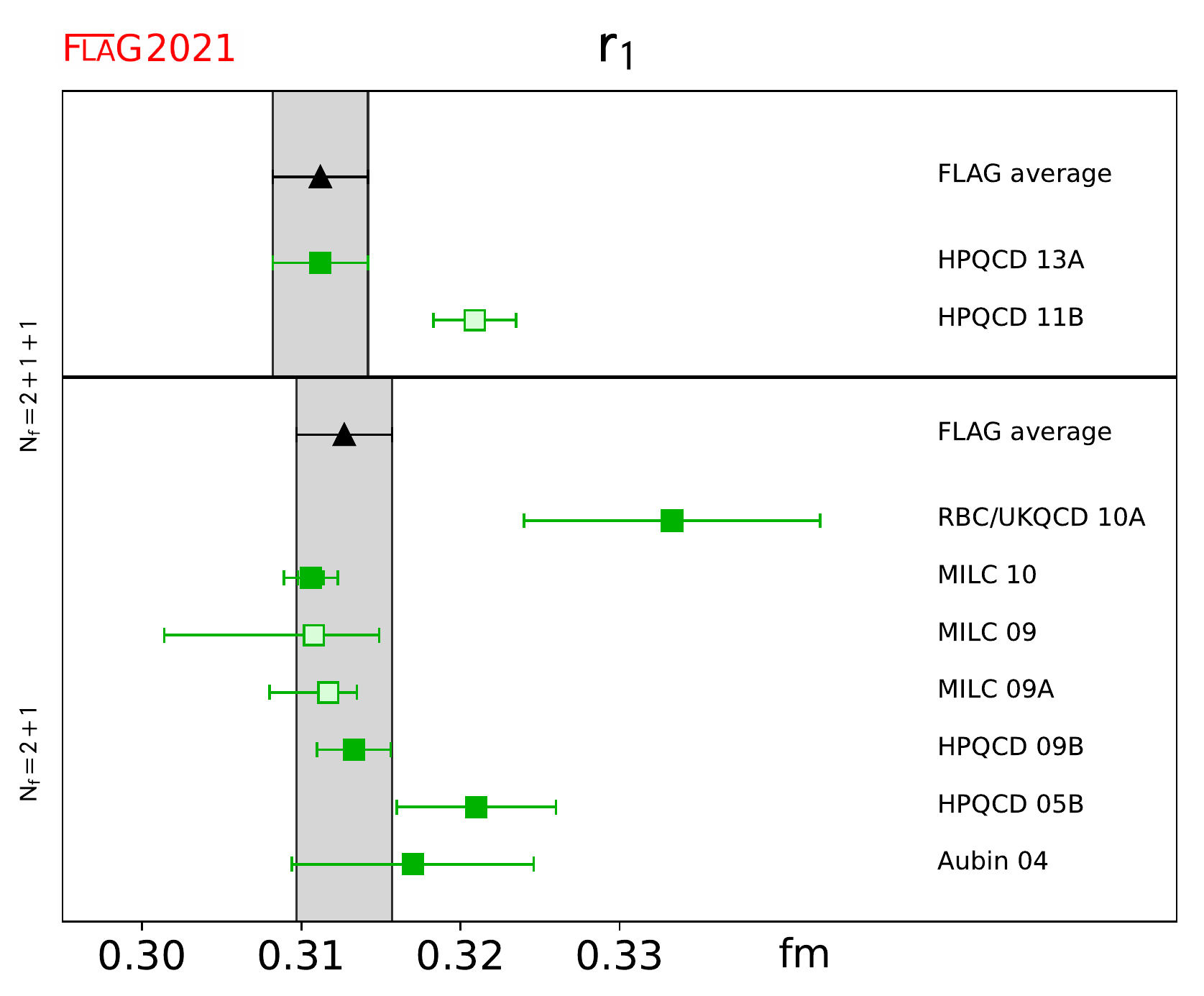}
\caption{\label{fig_potentialscales} Results for potential scales.}
\end{figure}

\off{
\begin{table}[!h]
\begin{tabular*}{\textwidth}[l]{l@{\extracolsep{\fill}}lllllllll}
Ref. & $\Nf$ & $M_\pi$  & $M_K$ & $M_\Omega$ & $f_\pi$ & $f_K$ & Description \\
\hline\hline \\ 

\cite{Miller:2020evg} & 2+1+1 & 134.8(3)$_*$ & 494.2(3)$_*$ & 1672.43(32)$_*$ &  & & \parbox[t]{2.5cm}{$D_s$ for charm} \\

\cite{Carrasco:2014cwa} & 2+1+1 & 134.98$_*$ & 494.2(4)$_*$$^\dagger$ & & 130.41$_*$ & & \parbox[t]{2.5cm}{$D$ and $D_s$ for charm} \\
  
\cite{Blum:2014tka} & 2+1 & 135.0$_*$ & 495.7$_*$ & 1672.25$_*$ & 130.2(9) & 155.5(8)\\

\cite{Bazavov:2009fk} & 1+1+1 & 135.0$_*$ & 494.4$_*$ &  & 130.4(2)$_*$ & 156.2(11) & \parbox[t]{2.5cm}{$m_{ud}$ defined using $M_K$ and $m_u$ fixed with $M_{K^+}$ with a partially quenched analysis}\\

\cite{Aoki:2008sm} & 2+1 & 135.0$_*$ & 497.6$_*$ & 1672.25$_*$ & 134.0(42) & 159.4(31) & \\

\end{tabular*}\\[-0.2cm]
\caption{Values in MeV for the hadronic quantities that can be used to identify the renormalization prescription adopted to define QCD in absence of QED corrections. In each line, the entries marked with the $_*$ subscript have been used by the collaboration as hadronic inputs in the renormalization procedure. Those without the $_*$ mark have been obtained as results of the simulations. \newline
${}^\dagger$Corrected for leading strong and electromagnetic isospin-breaking effects.
}
\label{tab_Hscheme}
\end{table}
}

\subsubsection{Ratios of scales}
It is convenient in many cases to also have ratios of scales at hand. In addition to translating from one scale to another, the ratios provide important crosschecks between different determinations. Results on ratios provided by the collaborations are compiled in Tab.~\ref{tab:scale_ratios} and Fig.~\ref{fig:scale_ratios}. The details of the computations were already discussed in the previous sections.
\begin{table}[!h]
  \vspace*{3cm}
\footnotesize{
\begin{tabular*}{1.0\textwidth}[l]{l@{\extracolsep{\fill}}rllllllll}

Collaboration & Ref. & $\Nf$ & \begin{rotate}{60}{publication status}\end{rotate} &  \begin{rotate}{60}{chiral extrapolation}\end{rotate} & \begin{rotate}{60}{continuum extrapolation}\end{rotate} & \begin{rotate}{60}{finite volume}\end{rotate} & $\sqrt{t_0}/w_0$ & $r_0/r_1$  & $r_1/w_0$\\
\hline\hline
ETM 21 & \cite{Alexandrou:2021bfr} & 2+1+1 & \oP & \good & \good & \good & 0.82930(65) &  &\\
HPQCD 13A & \cite{Dowdall:2013rya} &  2+1+1 & \gA & \good &\soso & \good & 0.835(8) &  & 1.789(26) \\
\hline
HotQCD 14 & \cite{Bazavov:2014pvz} & 2+1 & \gA & \good & \good & \good & & & 1.7797(67) \\
HotQCD 11 & \cite{Bazavov:2011nk} & 2+1 & \gA   & \good & \good & \good &  & 1.508(5) & \\
RBC/UKQCD 10A &  \cite{Aoki:2010dy} & 2+1 & \gA & \soso & \soso & \soso  & & 1.462(32)$^\#$ & \\
Aubin 04 & \cite{Aubin:2004wf}  & 2+1  &\gA & \soso & \soso & \soso &  & 1.474(7)(18) & \\
\hline\hline
\end{tabular*}
}
\caption{Results for dimensionless ratios of scales.
  \newline
 \;\;$^\#$This value is obtained from $r_1/r_0=0.684(15)(0)(0)$.
}
\label{tab:scale_ratios} 
\end{table}

\begin{figure}[htb]
  \centering
  \includegraphics[width=0.5\textwidth]{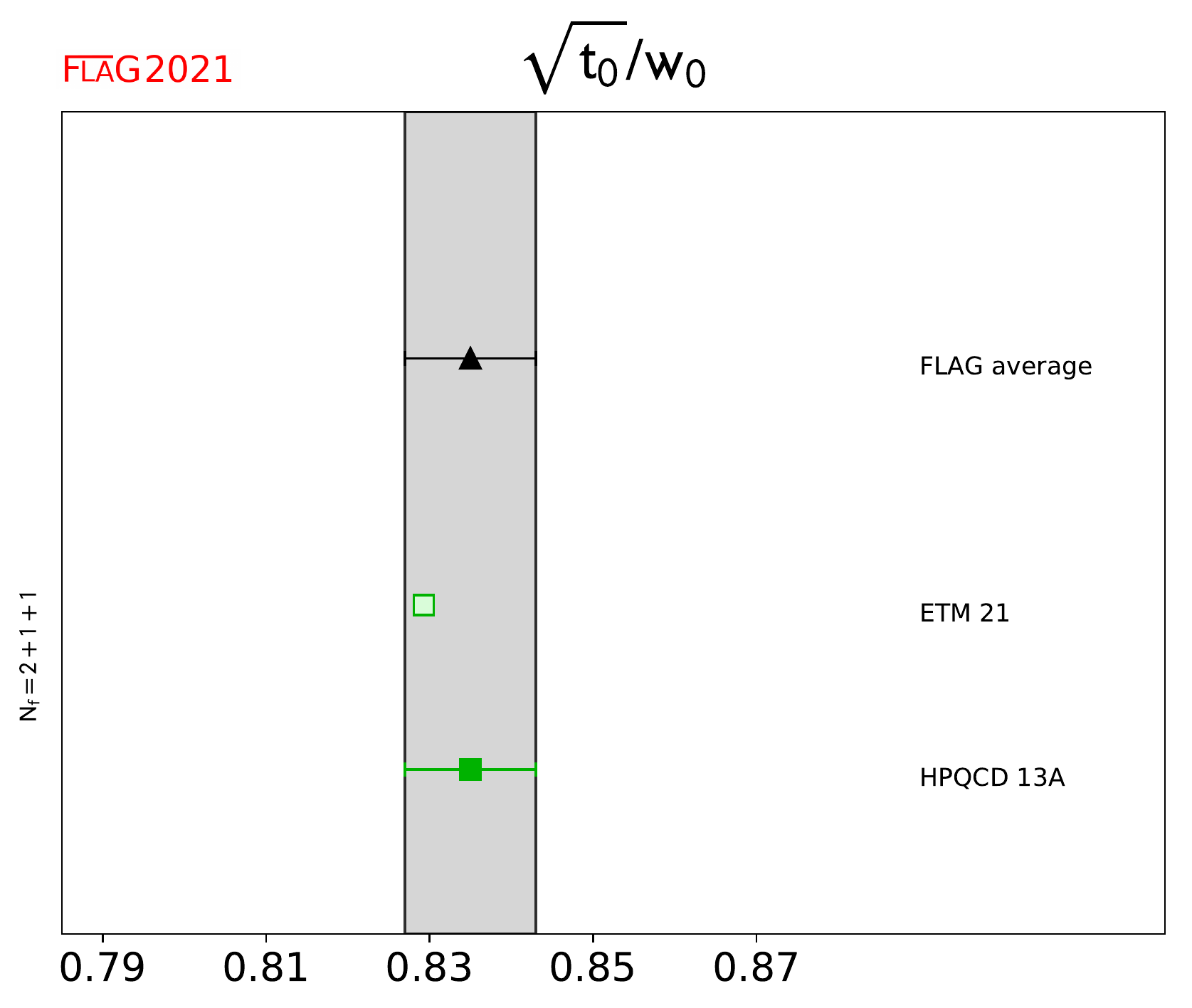}\\
  \includegraphics[width=0.49\textwidth]{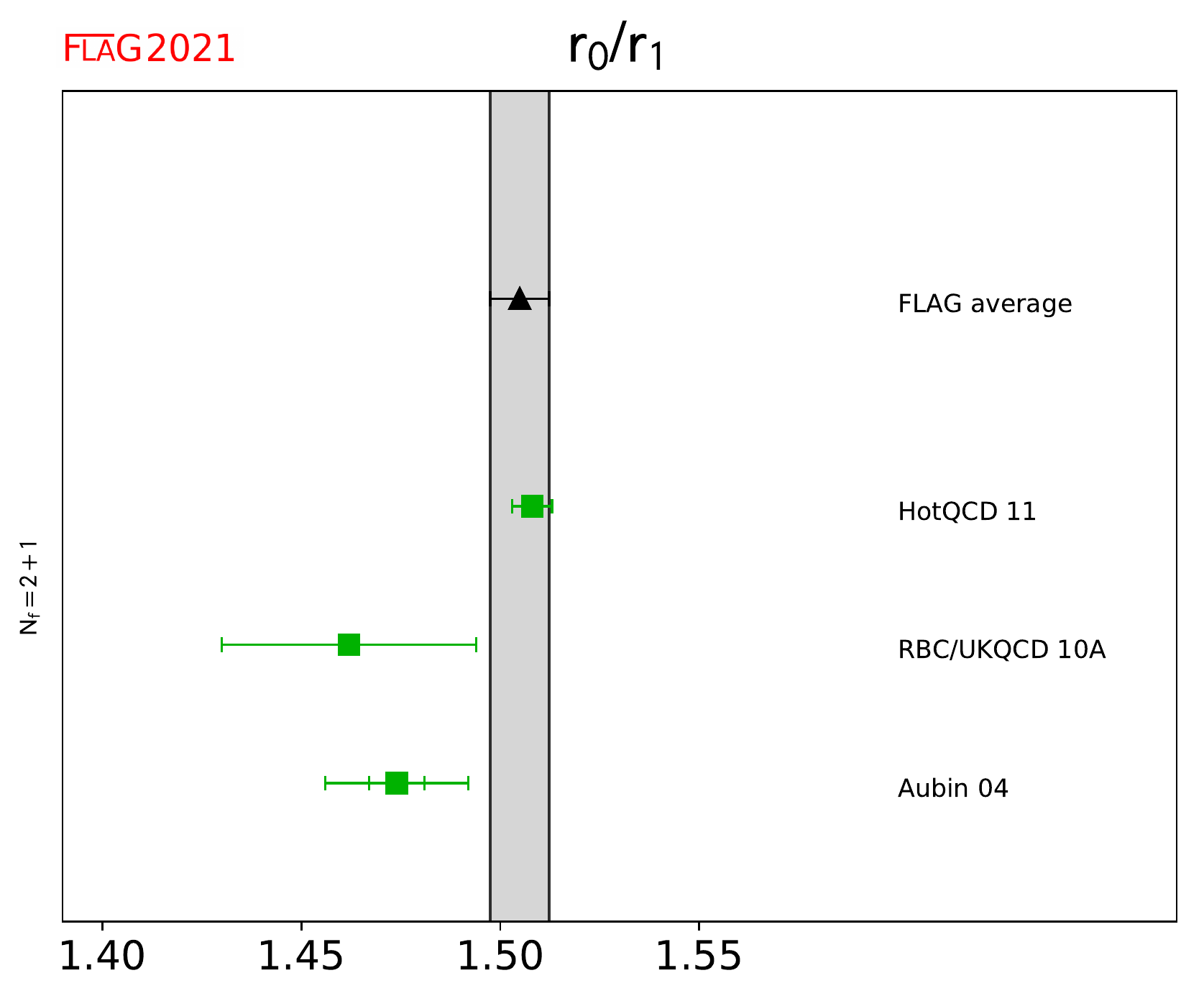}
  \includegraphics[width=0.49\textwidth]{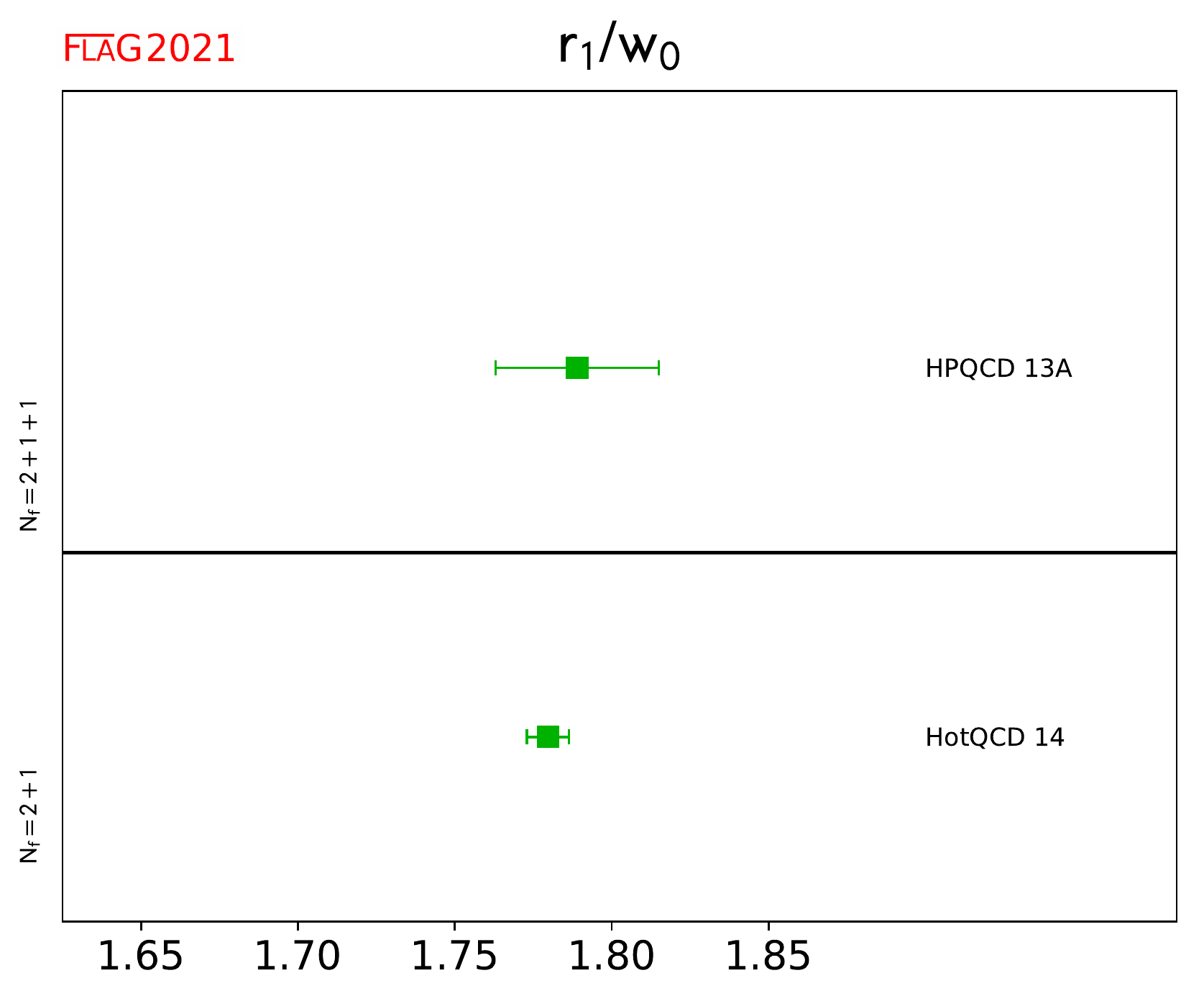}
\caption{\label{fig:scale_ratios} Results for dimensionless ratios of scales.}
\end{figure}
\clearpage

\subsection{Averages}

\noindent
{\it Gradient flow scale $\sqrt{t_0}$}

%

For $N_f=2+1+1$, we have two very recent calculations from ETM 21 \cite{Alexandrou:2021bfr},  CalLat 20A \cite{Miller:2020evg}, and two less recent ones from MILC 15 \cite{Bazavov:2015yea}, and HPQCD 13A \cite{Dowdall:2013rya} fulfilling the FLAG criteria to enter the average. The latter two and CalLat 20A are based on the same MILC-HISQ gauge field ensembles, hence we consider their statistical errors to be 100\% correlated. ETM 21 is not published, and, hence, does not enter the FLAG average.

For $N_f=2+1$, we have three calculations from CLS 16 \cite{Bruno:2016plf}, RBC/UKQCD 14B \cite{Blum:2014tka}, and BMW 12A \cite{Borsanyi:2012zs} which all enter the FLAG average. They are all independent computations, so there is no correlation to be taken into account. QCDSF/UKQCD 15B \cite{Bornyakov:2015eaa} does not contribute to the average, because it is not published.

Performing the weighted and correlated average we obtain
\begin{align}
&\label{eq:sqrtt0_2p1p1}
\Nf=2+1+1:&\FLAGAVBEGIN \sqrt{t_0} &= 0.14186(88) \FLAGAVEND\text{ fm}
&&\Refs~\mbox{\cite{Miller:2020evg,Bazavov:2015yea,Dowdall:2013rya}},\\ &\label{eq:sqrtt0_2p1}
\Nf=2+1: &\FLAGAVBEGIN \sqrt{t_0} &= 0.14464(87) \FLAGAVEND \text{ fm} &&\Refs~\mbox{\cite{Bruno:2016plf,Blum:2014tka,Borsanyi:2012zs}}.
\end{align}
We note that for the $N_f=2+1$ average the stretching factor based on the $\chi^2$-value from the weighted average is 1.25.
\\

\noindent
{\it Gradient flow scale $w_0$}

For $N_f=1+1+1+1$, including QED,  there is a single calculation, BMW 20 \cite{Borsanyi:2020mff} with the result
\begin{align}
&\label{eq:w0_1p1p1p1}
\Nf=1+1+1+1+\mathrm{QED}:&\FLAGAVBEGIN w_0 &= 0.17236(70)\FLAGAVEND \text{ fm}
&&\Ref~\mbox{\cite{Borsanyi:2020mff}}.
\end{align}

For $N_f=2+1+1$, we have four calculations ETM 21 \cite{Alexandrou:2021bfr},  CalLat 20A \cite{Miller:2020evg}, MILC 15 \cite{Bazavov:2015yea}, and HPQCD 13A \cite{Dowdall:2013rya} fulfilling the FLAG criteria to enter the average. However, ETM 20 is a proceedings contribution, while ETM 21 is not published, hence, they do not contribute to the FLAG average. As discussed above in connection with $\sqrt{t_0}$ we correlate the statistical errors of CalLat 20A, MILC 15, and HPQCD 13A. 

For $N_f=2+1$, we have three calculations RBC/UKQCD 14B \cite{Blum:2014tka}, HotQCD 14  \cite{Bazavov:2014pvz}, and BMW 12A \cite{Borsanyi:2012zs} that enter the FLAG average. These calculations are independent, and no correlation needs to be taken into account. QCDSF/UKQCD 15B \cite{Bornyakov:2015eaa} does not contribute to the average, because it is not published.

Performing the weighted and correlated average, we obtain
\begin{align}
&\label{eq:w0_2p1p1}
\Nf=2+1+1:&\FLAGAVBEGIN w_0 &= 0.17128(107) \FLAGAVEND\text{ fm}
&&\Refs~\mbox{\cite{Miller:2020evg,Bazavov:2015yea,Dowdall:2013rya}},\\ &\label{eq:w0_2p1}
\Nf=2+1: &\FLAGAVBEGIN w_0 &= 0.17355(92) \FLAGAVEND\text{ fm}  &&\Refs~\mbox{\cite{Blum:2014tka,Bazavov:2014pvz,Borsanyi:2012zs}}.
\end{align}
We note that for the $N_f=2+1$ average, the stretching factor based on the $\chi^2$-value from the weighted average is 1.23.

Isospin-breaking and electromagnetic corrections are expected to be small at the level of present uncertainties. This is also confirmed by the explicit computation by BMW~12A. Therefore, we also 
perform an average over all $\Nf>2+1$ computations and obtain
\begin{align}
&\label{eq:w0_2p1p1}
\Nf>2+1:&\FLAGAVBEGIN w_0 &= 0.17177(67) \FLAGAVEND\text{ fm}
&&\Ref~\mbox{\cite{Miller:2020evg,Borsanyi:2020mff,Bazavov:2015yea,Dowdall:2013rya}}.
\end{align}
\\


{\it Potential scale $r_0$}

For $N_f=2+1+1$, there is currently only one determination of $r_0$ from ETM 14 \cite{Carrasco:2014cwa}, namely $r_0=0.474(14)$~fm, which, therefore, represents the FLAG average.

For $N_f=2+1$, all but one calculation fulfill all the criteria to enter the FLAG average. HotQCD 14 \cite{Bazavov:2014pvz} is essentially an update of HotQCD 11 \cite{Bazavov:2011nk} by enlarging the set of ensembles used in the computation. Therefore, the result from HotQCD 14 supersedes the one from HotQCD 11 and, hence, we only use the former in the average. The computation of $\chi$QCD \cite{Yang:2014sea} is based on the configurations produced by RBC/UKQCD 10A \cite{Aoki:2010dy}, and we, therefore, assume a 100\% correlation between the statistical errors of the two calculations. HPQCD 05B \cite{Gray:2005ur} enhances the calculation of Aubin 04 \cite{Aubin:2004wf} by adding ensembles at a coarser lattice spacing and using the same discretization for the valence fermion. Therefore, we consider the full errors (statistical and systematic) on the results from Aubin 04 and HPQCD 05B to be 100\% correlated.

Performing the weighted and correlated average, we obtain
\begin{align}
&\label{eq:r0_2p1p1}
\Nf=2+1+1:&\FLAGAVBEGIN r_0 &= 0.474(14) \FLAGAVEND\text{ fm}
&&\Ref~\mbox{\cite{Carrasco:2014cwa}},\\ &\label{eq:r0_2p1}
\Nf=2+1: &\FLAGAVBEGIN r_0 &= 0.4701(36) \FLAGAVEND\text{ fm}  &&\Refs~\mbox{\cite{Bazavov:2014pvz,Yang:2014sea,Aoki:2010dy,Gray:2005ur,Aubin:2004wf}}.
\end{align}
We note that for the $N_f=2+1$ average, the stretching factor based on the $\chi^2$-value from the weighted average is 1.14.\\

\noindent
{\it Potential scale $r_1$}
    
For $\Nf=2+1+1$, there are two works that fulfill the criteria to enter the FLAG average, namely HPQCD 13A \cite{Dowdall:2013rya} and HPQCD 11B \cite{Dowdall:2011wh}. Both are based on MILC-HISQ ensembles, the former uses eight, the latter only five. The result from HPQCD 13A supersedes the result from HPQCD 11B (in line with a corresponding statement in HPQCD 13A) and forms the FLAG average.

For $\Nf=2+1$, all the results quoted in Tab.~\ref{tab_potentialscales} fulfill the FLAG criteria, but not all of them enter the average. The published result from MILC 09 \cite{Bazavov:2009bb} is superseded by the result in the proceedings MILC 10 \cite{Bazavov:2010hj}, while MILC 09A \cite{Bazavov:2009fk} is a proceedings contribution and does not enter the average. HPQCD 09B \cite{Davies:2009tsa} uses HISQ valence quarks instead of asqtad valence quarks as in HPQCD 05B \cite{Gray:2005ur}.
Therefore, we have RBC/UKQCD 10A \cite{Aoki:2010dy}, MILC 10, HPQCD 09B, HPQCD 05B, and Aubin 04 entering the average. However, since the latter four calculations are based on the aqtad MILC ensembles, we attribute 100\% correlation on the statistical error between them and 100\% correlation on the systematic error between HPQCD 05B and Aubin 04 as discussed above in connection with $r_0$.

Performing the weighted and correlated average, we obtain
\begin{align}
&\label{eq:r1_2p1p1}
\Nf=2+1+1:&\FLAGAVBEGIN r_1 &= 0.3112(30) \FLAGAVEND\text{ fm}
&&\Ref~\mbox{\cite{Dowdall:2013rya}},\\ &\label{eq:r1_2p1}
\Nf=2+1: &\FLAGAVBEGIN r_1 &= 0.3127(30) \FLAGAVEND\text{ fm}  &&\Refs~\mbox{\cite{Aoki:2010dy,Bazavov:2010hj,Davies:2009tsa,Gray:2005ur,Aubin:2004wf}}. 
\end{align}
We note that for the $N_f=2+1$ average the stretching factor based on the $\chi^2$-value from the weighted average is 1.57.\\

\noindent
{\it The scales $M_{p4s}$ and $f_{p4s}$}

    As mentioned in Sec.~\ref{subsubsec:other scales}, these scales have been used only by the MILC and FNAL/MILC collaborations \cite{Bazavov:2014wgs,Bazavov:2017lyh,Bazavov:2012xda}. The latest numbers from Ref.~\cite{Bazavov:2017lyh} are $f_{4ps} = 153.98(11)(^{+2}_{-12})(12)[4]$ MeV and $M_{p4s}=433.12(14)(^{+17}_{-6})(4)[40]$ MeV and, hence, we have 
\begin{align}
  &\Nf=2+1+1: & f_{4ps} &= 153.98(20) \text{ MeV} && \Ref~\mbox{\cite{Bazavov:2017lyh}}, \\
  &\Nf=2+1+1: & M_{4ps} &= 433.12(30) \text{ MeV} && \Ref~\mbox{\cite{Bazavov:2017lyh}}.
\end{align}

\noindent
    {\it Dimensionless ratios of scales}
    
We start with the ratio $\sqrt{t_0}/w_0$ for which two $N_f=2+1+1$ calculations from ETM 21 \cite{Alexandrou:2021bfr} and HPQCD 13A \cite{Dowdall:2013rya} are available. Since the former is not published, HPQCD 13A forms the FLAG value,
\begin{align}
  &\Nf=2+1+1: & \sqrt{t_0}/w_0 &= 0.835(8) && \Ref~\mbox{\cite{Dowdall:2013rya}}. 
\end{align}

For the ratio $r_0/r_1$ there are three calculations from HotQCD 11 \cite{Bazavov:2011nk}, RBC/UKQCD 10A \cite{Aoki:2010dy}, and Aubin 04 \cite{Aubin:2004wf} available. They all fulfill the FLAG criteria and enter the FLAG average of this ratio,
\begin{align}
  &\Nf=2+1: & r_0/r_1 &= 1.5049(74) && \Refs~\mbox{\cite{Bazavov:2011nk,Aoki:2010dy,Aubin:2004wf}}. 
\end{align}
We note that the stretching factor based on the $\chi^2$-value from the weighted average is 1.54.

Finally, for the ratio $r_1/w_0$ there is one computation from HotQCD 14  \cite{Bazavov:2014pvz} for $\Nf=2+1+1$, and one from HPQCD 13A \cite{Dowdall:2013rya} for $\Nf=2+1$ fulfilling the FLAG criteria, and, hence, forming the FLAG values
\begin{align}
  &\Nf=2+1+1: & r_1/w_0 &= 1.789(26) && \Ref~\mbox{\cite{Dowdall:2013rya}},\\
  &\Nf=2+1: & r_1/w_0 &= 1.7797(67) && \Ref~\mbox{\cite{Bazavov:2014pvz}}. 
\end{align}

\subsection{Observations and conclusions}
\label{s:Scalconcl}
The different computations for theory scales reported here are generally in good agreement within each set of 2+1+1 and 2+1  flavour content. Quantitatively, the computed stretching  factors are almost all below 1.3, a value which is insignificant considering the small number of computations that enter the averages.  The only exception are the potential scales, where the domain-wall fermion results  of RBC/UKQCD~10A are above the staggered fermion results, and also the ratio $r_0/r_1$ needs a stretching factor of 1.54. Of course, the limited number of large-scale QCD simulations that are available means that we have only a small number of truly independent determinations of the scales. For example, three out of the four computations entering our average for $w_0$ are based on the same HISQ rooted staggered fermion configurations and thus those differences are only due to the choice of the physical scale ($m_\Omega$ vs.~$f_\pi$), the valence quark action (M\"obius domain-wall valence fermions vs.~staggered fermions) employed to compute it and different analysis of continuum limit, etc.

Differences between $\Nf=2+1$ and 2+1+1 QCD are small but still larger than expectations \cite{Bruno:2014ufa,Knechtli:2017xgy}. A marginally significant difference is indicated by the figure and numbers on $w_0$ and a larger one by the ones for $\sqrt{t_0}$. The effect of the charm quark is $-1.0(7)$\% on $w_0$ and $-1.9(9)$\% on $\sqrt{t_0}$ as computed from the FLAG averages. In contrast, precision studies of the decoupling of charm quarks predict generic effects of a magnitude of only $\approx 0.2\%$  \cite{Bruno:2014ufa,Knechtli:2017xgy} for low energy quantities. We are looking forward to further and more precise results for clarifying whether these $-1.9(9)$\% hold up over time. In this respect, it is highly desirable for future computations to also publish ratios such as $\sqrt{t_0}/w_0$ where numbers are rare so far.

Such ratios of gradient flow scales are also of high interest in order to better understand the specific discretization errors of gradient flow observables. So far, systematic studies and information on the different contributions (see Sec.~\ref{s:flowscales} and Ref.~\cite{Ramos:2015baa}) are missing. A worrying result is, for example, the scale-setting study of Ref.~\cite{Hollwieser:2020qri} on ratios of scales. The authors find indications that the asymptotic $\sim a^2$ scaling does not set in before $a\approx0.05~\fm$ and the $a=0.04~\fm$ data has a relevant influence on their continuum extrapolations.

A final word concerns the physics scales that all results depend on. While the mass of the $\Omega$ baryon is more popular than the leptonic decay rate of the pion, both have systematics which are difficult to estimate. For the $\Omega$ baryon it is the contaminations by excited states and for the decay rates it is the QED effects $\delta f_\pi^\mathrm{isoQCD}$. The uncertainty in $V_{ud}$ is {\em not} relevant at this stage, but means that one is relying more on the standard model being an accurate low energy theory than in the case of the $\Omega$ mass. In principle, excited state effects are controlled by just going to large Euclidean time, but, in practice, this yields errors that are too large. One, therefore, performs fits with a very small number of excitations while theoretically  there is a multitude of multi-hadron states that are expected to contribute. For the leptonic decay rate of the pion, the situation is quite reversed, namely, the problematic QED contributions have a well-motivated theory: chiral perturbation theory. The needed combination of low-energy constants is not accessible from experiment but its large-$N$ estimate \cite{Cirigliano:2007ga} has been (indirectly) confirmed by the recent computation of $\delta f_\pi^\mathrm{isoQCD}$~\cite{DiCarlo:2019thl}. Unfortunately the same comparison is not so favourable for the leptonic Kaon decay.

\newpage

\clearpage
\section*{Acknowledgments}
\addcontentsline{toc}{section}{Acknowledgments}
We are very grateful to the external reviewers for providing detailed comments and suggestions on the draft of this review. These reviewers were
Gunnar Bali, Claude Bernard, Johann Bjinens,  J\'er\^ome Charles, Luigi Del Debbio, Paolo Gambino, Elvira G\'amiz, Christian Hoelbling, Marc Knecht, Bira van Kolck, Laurent Lellouch, Zoltan Ligeti, Kim Maltman, Emilie Passemar, Chris Sachrajda, Tilo Wettig, and Takeshi Yamazaki. 

The kick-off meeting for the present review was held in February 2020 at the University of Bern and was supported by 
the Albert Einstein Center for fundamental physics. 
The mid-review meeting was held virtually in December 2020.
We thank our hosts for their hospitality and financial support is gratefully acknowledged.

Members of FLAG were supported by funding agencies; in particular: 
\begin{itemize}
\item    Y.A. was supported by JSPS KAKENHI Grant No. 16K05320.

\item S.C.~acknowledges the support of the European Union’s Horizon 2020 research and innovation programme under the Marie Sk{\l}odowska-Curie grant agreement no. 813942 (ITN EuroPLEx).

\item M.D.M.~was supported by DFF Research project 1. Grant n. 8021-00122B.

\item P.D.~acknowledges support from the European Unions Horizon 2020 research and innovation programme under the Marie Sk\l{}odowska-Curie grant agreement No.~813942 (EuroPLEx) and from INFN under the research project INFN-QCDLAT.

\item X.F.~is supported in part by NSFC of China under Grant No. 11775002, 12070131001, 12125501 and National Key Research and Development Program of China under Contracts No.~2020YFA0406400.

\item S.G. acknowledges support from the U.S.~Department of Energy through
grant DE-SC0010120.

\item G.H.~and C.P.~acknowledge support from the EU H2020-MSCA-ITN-2018-813942 (EuroPLEx), Spanish MICINN grant PGC2018-094857-B-I00, Spanish Agencia Estatal de Investigaci\'on through the grant ''IFT Centro de Excelencia Severo Ochoa'' SEV-2016-0597 and CEX2020-001007-S.

\item S.H.~is supported in part by JSPS KAKENHI Grant Number 18H03710.
S.H. is supported in part by MEXT as "Program for Promoting Researches on the Supercomputer Fugaku"
(Simulation for basic science: from fundamental laws of particles to creation of nuclei,
JPMXP1020200105) through the Joint Institute for Computational Fundamental Science (JICFuS).

\item A.J.~received funding from the STFC consolidated grant ST/P000711/1 and ST/ T000775/1.

\item T.K.~is supported in part by JSPS KAKENHI Grant Number 21H01085.
T.K. is supported in part by MEXT as "Program for Promoting Researches on the Supercomputer Fugaku"
(Simulation for basic science: from fundamental laws of particles to creation of nuclei,
JPMXP1020200105) through the Joint Institute for Computational Fundamental Science (JICFuS).

\item S.M.~is supported by the U.S.~Department of Energy, Office of Science, Office of High Energy Physics under Award Number DE-SC0009913.

\item  C.J.M.~is supported in part by USDOE grant No. DE-AC05-06OR23177, under which Jefferson Science Associates, LLC, manages and operates Jefferson Lab.

\item A.P.~is supported in part by UK STFC grant ST/P000630/1 and also received funding from the European Research Council (ERC) under the European Union’s Horizon 2020 research and innovation programme under grant agreements No 757646 \& 813942.
  
\item A.R.~is supported by the Generalitat Valenciana through the plan GenT program (CIDEGENT/2019/040) and the Ministerio de Ciencia e Innovacion (PID2020-113644GB-I00).

\item S.R.S.~is supported in part by the U.S.~Department of Energy (grant DE-SC0011637).

\item S.S.~acknowledges support from the EU H2020-MSCA-ITN-2018-813942 (EuroPLEx).

\end{itemize}

\appendix

\begin{appendix}
\clearpage
\twocolumn
\section{List of acronyms}\label{app:acronyms}
\begin{supertabular}[ht]{p{.21\linewidth}@{\hspace{.02\linewidth}}p{.72\linewidth}}
B$\chi$PT&	  baryonic chiral perturbation theory\\
BCL&	          Bourrely-Caprini-Lellouch\\
BGL&	          Boyd-Grinstein-Lebed\\
BK&	          Becirevic-Kaidalov\\
BSM&	          beyond standard model\\
BZ&	          Ball-Zwicky\\
$\chi$PT&	  chiral perturbation theory\\
CKM&	          Cabibbo-Kobayashi-Maskawa\\
CLN&	          Caprini-Lellouch-Neubert\\
CP&	          charge-parity\\
CPT&	          charge-parity-time reversal\\
CVC&	          conserved vector current\\
DSDR&		  dislocation suppressing determinant ratio\\
DW&	          domain wall\\
DWF&	          domain wall fermion\\
EDM&	          electric dipole moment\\
EFT&	          effective field theory\\
EM&	 	  electromagnetic\\
ESC&	          excited state contributions\\
EW&		  electroweak\\
FCNC&	          flavor-changing neutral current\\
FH&	          Feynman-Hellman\\
FSE&	 	  finite-size effects\\
FV&	          finite volume\\
GF&	          gradient flow\\
GGOU&	          Gambino-Giordano-Ossola-Uraltsev\\
GRS&	          Gasser-Rusetsky-Scimemi\\
HEX&	          hypercubic stout\\
HISQ&	          highly-improved staggered quarks\\
HM$\chi$PT&	  heavy-meson chiral perturbation theory\\
HMC&	          hybrid Monte Carlo\\
HMrS$\chi$PT&	  heavy-meson rooted staggered chiral perturbation theory\\
HQET&	          heavy-quark effective theory\\
IR&	          infrared\\
isoQCD&	          isospin-symmetric QCD\\
LD&	          long distance\\
LEC&	          low-energy constant\\
LO&		  leading order\\
LW&	 	  L\"uscher-Weisz\\
MC&	          Monte Carlo\\
MM&	          minimal MOM\\
MOM&		  momentum subtraction\\
$\msbar$&	  modified  minimal substraction scheme \\
NDR&	          naive dimensional regularization\\
nEDM&	          nucleon electric dipole moment\\
NGB&	          Nambu-Goldstone bosons\\
NLO&	          next-to-leading order\\
NME&	          nucleon matrix elements\\
NNLO&	          next-to-next-to-leading order\\
NP&	          nonperturbative\\
npHQET&		  nonperturbative heavy-quark effective theory\\
NRQCD&	          nonrelativistic QCD\\
NSPT&	          numerical stochastic perturbation theory\\
OPE&	          operator product expansion\\
PCAC&	          partially-conserved axial current\\
PDF&	          parton distribution function\\
PDG&	          particle data group\\
QCD&	          quantum chromodynamics\\
QED&	          quantum electrodynamics\\
QED$_{\rm L}$&	  formulation of QED in finite volume (see~\cite{Hayakawa:2008an}) \\
QED$_{\rm TL}$&	  formulation of QED in finite volume (see~\cite{Duncan:1996be})\\
RG&	          renormalization group\\
RGI&	          renormalization group invariant\\
RH&	          R. Hill\\
RHQ&	          relativistic heavy-quark\\
RHQA&	          relativistic heavy-quark action\\
RI-MOM&		  regularization-independent momentum subtraction (also RI/MOM)\\
RI-SMOM&	  regularization-independent symmetric momentum (also RI/SMOM) \\
RMS&	          root mean square\\
S$\chi$PT&	  staggered chiral perturbation theory\\
SD&	          short distance\\
SF&	          Schr\"odinger functional\\
SIDIS&	          semi-inclusive deep-inelastic scattering\\
SM&	          standard model\\
SSF&		  step-scaling function\\
SUSY&	 	  supersymmetric\\
SW&		  Sheikholeslami-Wohlert\\
UT&	          unitarity triangle\\
UV&	          ultraviolet\\
\end{supertabular}
\onecolumn

\ifx\noglossary\undefined  
\clearpage
\section{Glossary}\label{comm}
\subsection{Lattice actions}\label{sec_lattice_actions}
In this appendix we give brief descriptions of the lattice actions
used in the simulations and summarize their main features.

\subsubsection{Gauge actions \label{sec_gauge_actions}}

The simplest and most widely used discretization of the Yang-Mills
part of the QCD action is the Wilson plaquette action\,\cite{Wilson:1974sk}:
\be
 S_{\rm G} = \beta\sum_{x} \sum_{\mu<\nu}\Big(
  1-\frac{1}{3}{\rm Re\,\Tr}\,W_{\mu\nu}^{1\times1}(x)\Big),
\label{eq_plaquette}
\ee
where $\beta \equiv 6/g_0^2$ (with $g_0$ the bare gauge coupling) and
the plaquette $W_{\mu\nu}^{1\times1}(x)$ is the product of
link variables around an elementary square of the lattice, i.e.,
\be
  W_{\mu\nu}^{1\times1}(x) \equiv U_\mu(x)U_\nu(x+a\hat{\mu})
   U_\mu(x+a\hat{\nu})^{-1} U_\nu(x)^{-1}.
\ee
This expression reproduces the Euclidean Yang-Mills action in the
continuum up to corrections of order~$a^2$.  There is a general
formalism, known as the ``Symanzik improvement programme''
\cite{Symanzik:1983dc,Symanzik:1983gh}, which is designed to cancel
the leading lattice artifacts, such that observables have an
accelerated rate of convergence to the continuum limit.  The
improvement programme is implemented by adding higher-dimensional
operators, whose coefficients must be tuned appropriately in order to
cancel the leading lattice artifacts. The effectiveness of this
procedure depends largely on the method with which the coefficients
are determined. The most widely applied methods (in ascending order of
effectiveness) include perturbation theory, tadpole-improved
(partially resummed) perturbation theory, renormalization group
methods, and the nonperturbative evaluation of improvement
conditions.

In the case of Yang-Mills theory, the simplest version of an improved
lattice action is obtained by adding rectangular $1\times2$ loops to
the plaquette action, i.e.,
\be
   S_{\rm G}^{\rm imp} = \beta\sum_{x}\left\{ c_0\sum_{\mu<\nu}\Big(
  1-\frac{1}{3}{\rm Re\,\Tr}\,W_{\mu\nu}^{1\times1}(x)\Big) +
   c_1\sum_{\mu,\nu} \Big(
  1-\frac{1}{3}{\rm Re\,\Tr}\,W_{\mu\nu}^{1\times2}(x)\Big) \right\},
\label{eq_Sym}
\ee
where the coefficients $c_0, c_1$ satisfy the normalization condition
$c_0+8c_1=1$. The {\sl Symanzik-improved} \cite{Luscher:1984xn},
{\sl Iwasaki} \cite{Iwasaki:1985we}, and {\sl DBW2}
\cite{Takaishi:1996xj,deForcrand:1999bi} actions are all defined
through \eq{eq_Sym} via particular choices for $c_0, c_1$. Details are
listed in Tab.~\ref{tab_gaugeactions} together with the
abbreviations used in the summary tables. Another widely used variant is the {\sl tadpole Symanzik-improved} \cite{Lepage:1992xa,Alford:1995hw} action which is obtained by adding additional 6-link parallelogram loops $W_{\mu\nu\sigma}^{1\times 1\times 1}(x)$ to the action in Eq.~(\ref{eq_Sym}), i.e.,
\be
S_{\rm G}^{\rm tadSym} = S_{\rm G}^{\rm imp} + \beta \sum_x c_2 \sum_{\mu<\nu<\sigma}\Big(1-\frac{1}{3} {\rm Re\,\Tr}\,W_{\mu\nu\sigma}^{1\times1\times1}(x)\Big),
\ee
where
\be
  W_{\mu\nu\sigma}^{1\times1\times1}(x) \equiv U_\mu(x)U_\nu(x+a\hat{\mu})U_\sigma(x+a\hat\mu+a\hat\nu)
   U_\mu(x+a\hat\sigma+a\hat{\nu})^{-1} U_\nu(x+a\hat\sigma)^{-1} U_\sigma(x)^{-1}
\ee
allows for 1-loop improvement \cite{Luscher:1984xn}.

\vspace{-0.07cm}
\begin{table}[!h]
\begin{center}
{\footnotesize
\begin{tabular*}{\textwidth}[l]{l @{\extracolsep{\fill}} c l}
\hline\hline \\[-1.0ex]
Abbrev. & $c_1$ & Description 
\\[1.0ex] \hline \hline \\[-1.0ex]
Wilson    & 0 & Wilson plaquette action \\[1.0ex] \hline \\[-1.0ex]
tlSym   & $-1/12$ & tree-level Symanzik-improved gauge action \\[1.0ex] \hline \\[-1.0ex]
tadSym  & variable & tadpole Symanzik-improved gauge action
 \\[1.0ex] \hline \\[-1.0ex]
Iwasaki & $-0.331$ & Renormalization group improved (``Iwasaki'')
action \\[1.0ex] \hline \\[-1.0ex]
DBW2 & $-1.4088$ & Renormalization group improved (``DBW2'') action 
\\ [1.0ex] 
\hline\hline
\end{tabular*}
}
\caption{Summary of lattice gauge actions. The leading lattice
 artifacts are $\cO(a^2)$ or better for all
  discretizations. \label{tab_gaugeactions}} 
\end{center}
\end{table}


\subsubsection{Light-quark actions \label{sec_quark_actions}}

If one attempts to discretize the quark action, one is faced with the
fermion doubling problem: the naive lattice transcription produces a
16-fold degeneracy of the fermion spectrum. \\

\noindent
{\it Wilson fermions}\\
\noindent

Wilson's solution to the fermion doubling problem is based on adding a
dimension-5 (irrelevant) operator to the lattice action. The
Wilson-Dirac operator for the massless case reads
\cite{Wilson:1974sk,Wilson:1975id}
\be
     D_{\rm w} = \half\gamma_\mu(\nabla_\mu+\nabla_\mu^*)
   +a\nabla_\mu^*\nabla_\mu,
\ee
where $\nabla_\mu,\,\nabla_\mu^*$ denote the covariant forward and
backward lattice derivatives, respectively.  The addition of the
Wilson term $a\nabla_\mu^*\nabla_\mu$, results in fermion doublers
acquiring a mass proportional to the inverse lattice spacing; close to
the continuum limit these extra degrees of freedom are removed from
the low-energy spectrum. However, the Wilson term also results in an
explicit breaking of chiral symmetry even at zero bare quark mass.
Consequently, it also generates divergences proportional to the UV
cutoff (inverse lattice spacing), besides the usual logarithmic
ones. Therefore the chiral limit of the regularized theory is not
defined simply by the vanishing of the bare quark mass but must be
appropriately tuned. As a consequence quark-mass renormalization
requires a power subtraction on top of the standard multiplicative
logarithmic renormalization.  The breaking of chiral symmetry also
implies that the nonrenormalization theorem has to be applied with
care~\cite{Karsten:1980wd,Bochicchio:1985xa}, resulting in a
normalization factor for the axial current which is a regular function
of the bare coupling.  On the other hand, vector symmetry is
unaffected by the Wilson term and thus a lattice (point split) vector
current is conserved and obeys the usual nonrenormalization theorem
with a trivial (unity) normalization factor. Thus, compared to lattice
fermion actions which preserve chiral symmetry, or a subgroup of it,
the Wilson regularization typically results in more complicated
renormalization patterns.

Furthermore, the leading-order lattice artifacts are of order~$a$.
With the help of the Symanzik improvement programme, the leading
artifacts can be cancelled in the action by adding the so-called
``Clover'' or Sheikholeslami-Wohlert (SW) term~\cite{Luscher:1996sc}.
The resulting expression in the massless case reads
\be
   D_{\rm sw} = D_{\rm w}
   +\frac{ia}{4}\,\csw\sigma_{\mu\nu}\widehat{F}_{\mu\nu},
\label{eq_DSW}
\ee
where $\sigma_{\mu\nu}=\frac{i}{2}[\gamma_\mu,\gamma_\nu]$, and
$\widehat{F}_{\mu\nu}$ is a lattice transcription of the gluon field
strength tensor $F_{\mu\nu}$. The coefficient $\csw$ can be determined
perturbatively at tree-level ($\csw = 1$; tree-level improvement or
tlSW for short), via a mean field approach \cite{Lepage:1992xa}
(mean-field improvement or mfSW) or via a nonperturbative approach
\cite{Luscher:1996ug} (nonperturbatively improved or npSW).
Hadron masses, computed using $D_{\rm sw}$, with the coefficient
$\csw$ determined nonperturbatively, will approach the continuum
limit with a rate proportional to~$a^2$; with tlSW for $\csw$ the rate
is proportional to~$g_0^2 a$.

Other observables require additional improvement
coefficients~\cite{Luscher:1996sc}.  A common example consists in the
computation of the matrix element $\langle \alpha \vert Q \vert \beta
\rangle$ of a composite field $Q$ of dimension-$d$ with external
states $\vert \alpha \rangle$ and $\vert \beta \rangle$. In the
simplest cases, the above bare matrix element diverges logarithmically
and a single renormalization parameter $Z_Q$ is adequate to render it
finite. It then approaches the continuum limit with a rate
proportional to the lattice spacing $a$, even when the lattice action
contains the Clover term. In order to reduce discretization errors to
${\cO}(a^2)$, the lattice definition of the composite operator $Q$
must be modified (or ``improved''), by the addition of all
dimension-$(d+1)$ operators with the same lattice symmetries as $Q$.
Each of these terms is accompanied by a coefficient which must be
tuned in a way analogous to that of $\csw$. Once these coefficients
are determined nonperturbatively, the renormalized matrix element of
the improved operator, computed with a npSW action, converges to the
continuum limit with a rate proportional to~$a^2$. A tlSW improvement
of these coefficients and $\csw$ will result in a rate proportional
to~$g_0^2 a$.

It is important to stress that the improvement procedure does not
affect the chiral properties of Wilson fermions; chiral symmetry
remains broken.

Finally, we mention ``twisted-mass QCD'' as a method which was
originally designed to address another problem of Wilson's
discretization: the Wilson-Dirac operator is not protected against the
occurrence of unphysical zero modes, which manifest themselves as
``exceptional'' configurations. They occur with a certain frequency in
numerical simulations with Wilson quarks and can lead to strong
statistical fluctuations. The problem can be cured by introducing a
so-called ``chirally twisted'' mass term. The most common formulation
applies to a flavour doublet $\bar \psi = ( u \quad d)$ of
mass-degenerate quarks, with the fermionic part of the QCD action in
the continuum assuming the form \cite{Frezzotti:2000nk}
\be
   S_{\rm F}^{\rm tm;cont} = \int d^4{x}\, \psibar(x)(\gamma_\mu
   D_\mu +
   m + i\mu_{\rm q}\gamma_5\tau^3)\psi(x).
\ee
Here, $\mu_{\rm q}$ is the twisted-mass parameter, and $\tau^3$ is a
Pauli matrix in flavour space. The standard action in the continuum
can be recovered via a global chiral field rotation. The physical
quark mass is obtained as a function of the two mass parameters $m$
and $\mu_{\rm q}$. The corresponding lattice regularization of twisted-mass QCD (tmWil) for $\Nf=2$ flavours is defined through the fermion
matrix
\be
   D_{\rm w}+m_0+i\mu_{\rm q}\gamma_5\tau^3 \,\, .
\label{eq_tmQCD}
\ee
Although this formulation breaks physical parity and flavour
symmetries, resulting in nondegenerate neutral and charged pions,
is has a number of advantages over standard Wilson
fermions. Firstly, the presence of the twisted-mass parameter
$\mu_{\rm q}$ protects the discretized theory against unphysical zero
modes. A second attractive feature of twisted-mass lattice QCD is the
fact that, once the bare mass parameter $m_0$ is tuned to its ``critical value''
(corresponding to massless pions in the standard Wilson formulation),
the leading lattice artifacts are of order $a^2$ without the
need to add the Sheikholeslami-Wohlert term in the action, or other
improving coefficients~\cite{Frezzotti:2003ni}. A third important advantage
is that, although the problem of explicit chiral
symmetry breaking remains, quantities computed with twisted fermions
with a suitable tuning of the mass parameter $\mu_{\rm q}$,
are subject to renormalization patterns which are simpler than the ones with
standard Wilson fermions. Well known examples are the pseudoscalar decay
constant  and $B_{\rm K}$.\\

\noindent
{\it Staggered Fermions}\\
\noindent

An alternative procedure to deal with the doubling problem is based on Kogut-Susskind fermions \cite{Kogut:1974ag,Banks:1975gq} and is now known under the name ``staggered''
 fermion formulation \cite{Susskind:1976jm, Kawamoto:1981hw,Sharatchandra:1981si}.
Here the degeneracy is only lifted partially, from 16 down to 4.  It has become customary
to refer to these residual doublers as ``tastes'' in order to distinguish them from physical
flavours.  Taste changing interactions 
can occur via the exchange of gluons with one or more components
  of momentum near the cutoff $\pi/a$.  This leads to the breaking of the $SU(4)$ vector symmetry among 
  tastes, thereby generating order $a^2$ lattice artifacts.

The residual doubling of staggered quarks (four tastes per
flavour) is removed by taking a fractional power of the fermion determinant \cite{Marinari:1981qf} --- the ``fourth-root 
procedure,'' or, sometimes, the ``fourth root trick.''  
This procedure would be unproblematic if
the action had full $SU(4)$ taste symmetry, which would give a
Dirac operator that was block-diagonal in taste space.  
However, the breaking of taste symmetry at nonzero lattice spacing leads to a
variety of problems. In fact, the fourth root of the determinant is not equivalent
to the determinant of any local lattice Dirac operator \cite{Bernard:2006ee}.
This in turn leads 
to violations of unitarity 
on the lattice \cite{Prelovsek:2005rf,Bernard:2006zw,Bernard:2007qf,Aubin:2008wk}.

According to standard renormalization group lore, the taste
violations, which are associated with lattice operators of dimension
greater than four, might be expected to go away in the continuum limit,
resulting in the restoration of locality and unitarity.  However,
there is a problem with applying the standard lore to this nonstandard
situation: the usual renormalization group reasoning assumes that the
lattice action is local.  Nevertheless, Shamir
\cite{Shamir:2004zc,Shamir:2006nj} shows that one may apply the
renormalization group to a ``nearby'' local theory, and thereby gives
a strong argument that that the desired local, unitary theory of QCD
is reproduced by the rooted staggered lattice theory in the continuum
limit.

A version of chiral perturbation that includes the lattice artifacts
due to taste violations and rooting (``rooted staggered chiral
perturbation theory'') can also be worked out
\cite{Lee:1999zxa,Aubin:2003mg,Sharpe:2004is} and shown to correctly
describe the unitarity-violating lattice artifacts in the pion sector
\cite{Bernard:2006zw,Bernard:2007ma}.  This provides additional
evidence that the desired continuum limit can be obtained. Further, it
gives a practical method for removing the lattice artifacts from
simulation results. Versions of rooted staggered chiral perturbation
theory exist for heavy-light mesons with staggered light quarks but
nonstaggered heavy quarks \cite{Aubin:2005aq}, heavy-light mesons with
staggered light and heavy quarks
\cite{Komijani:2012fq,Bernard:2013qwa}, staggered baryons
\cite{Bailey:2007iq}, and mixed actions with a staggered sea
\cite{Bar:2005tu,Bae:2010ki}, as well as the pion-only version
referenced above.

There is also considerable numerical evidence that the rooting
procedure works as desired.  This includes investigations in the
Schwinger model \cite{Durr:2003xs,Durr:2004ta,Durr:2006ze}, studies of
the eigenvalues of the Dirac operator in QCD
\cite{Follana:2004sz,Durr:2004as,Wong:2004nk,Donald:2011if}, and
evidence for taste restoration in the pion spectrum as $a\to0$
\cite{Aubin:2004fs,Bazavov:2009bb}.

Issues with the rooting procedure have led Creutz
\cite{Creutz:2006ys,Creutz:2006wv,Creutz:2007yg,Creutz:2007pr,Creutz:2007rk,Creutz:2008kb,Creutz:2008nk}
to argue that the continuum limit of the rooted staggered theory
cannot be QCD.  These objections have however been answered in
Refs.~\cite{Bernard:2006vv,Sharpe:2006re,Bernard:2007eh,Kronfeld:2007ek,Bernard:2008gr,Adams:2008db,Golterman:2008gt,Donald:2011if}. 
In particular, a claim that the continuum 't Hooft
vertex \cite{'tHooft:1976up,'tHooft:1976fv} could not be properly
reproduced by the rooted theory has been refuted
\cite{Bernard:2007eh,Donald:2011if}.

Overall, despite the lack of rigorous proof of the correctness of the
rooting procedure, we think the evidence is strong enough to consider staggered
QCD simulations on a par with simulations using other actions.
See the following reviews for further evidence and discussion:
Refs.~\cite{Durr:2005ax,Sharpe:2006re,Kronfeld:2007ek,Golterman:2008gt,Bazavov:2009bb}.
\\

\noindent
{\it Improved Staggered Fermions}\\
\noindent

An improvement program can be used to suppress taste-changing
interactions, leading to ``improved staggered fermions,'' with the
so-called ``Asqtad'' \cite{Orginos:1999cr}, ``HISQ''
\cite{Follana:2006rc}, ``Stout-smeared'' \cite{Aoki:2005vt}, and
``HYP'' \cite{Hasenfratz:2001hp} actions as the most common versions.
All these actions smear the gauge links in order to reduce the
coupling of high-momentum gluons to the quarks, with the main goal of
decreasing taste-violating interactions. In the Asqtad case, this is
accomplished by replacing the gluon links in the derivatives by
averages over 1-, 3-, 5-, and 7-link paths.  The other actions reduce
taste changing even further by smearing more.  In addition to the
smearing, the Asqtad and HISQ actions include a three-hop term in the
action (the ``Naik term'' \cite{Naik:1986bn}) to remove order $a^2$
errors in the dispersion relation, as well as a ``Lepage term''
\cite{Lepage:1998vj} to cancel other order $a^2$ artifacts introduced
by the smearing.  In both the Asqtad and HISQ actions, the leading
taste violations are of order $\alpha_S^2 a^2$, and ``generic''
lattices artifacts (those associated with discretization errors other
than taste violations) are of order $\alpha_S a^2$.  The overall
coefficients of these errors are, however, significantly smaller with
HISQ than with Asqtad.  With the Stout-smeared and HYP actions, the
errors are formally larger (order $\alpha_S a^2$ for taste violations
and order $a^2$ for generic lattices artifacts).  Nevertheless, the
smearing seems to be very efficient, and the actual size of errors at
accessible lattice spacings appears to be at least as small as with
HISQ.

Although logically distinct from the light-quark improvement program
for these actions, it is customary with the HISQ action to include an
additional correction designed to reduce discretization errors for
heavy quarks (in practice, usually charm quarks)
\cite{Follana:2006rc}. The Naik term is adjusted to remove leading
$(am_c)^4$ and $\alpha_S(am_c)^2$ errors, where $m_c$ is the
charm-quark mass and ``leading'' in this context means leading in
powers of the heavy-quark velocity $v$ ($v/c\sim 1/3$ for $D_s$).
With these improvements, the claim is that one can use the staggered
action for charm quarks, although it must be emphasized that it is not
obvious {\it a priori}\/ how large a value of $am_c$ may be tolerated
for a given desired accuracy, and this must be studied in the
simulations.  \\

\noindent
{\it Ginsparg-Wilson fermions}\\
\noindent

Fermionic lattice actions, which do not suffer from the doubling
problem whilst preserving chiral symmetry go under the name of
``Ginsparg-Wilson fermions''. In the continuum the massless Dirac
operator ($D$) anti-commutes with $\gamma_5$. At nonzero lattice spacing a 
chiral symmetry can be realized if this condition is relaxed
to \cite{Hasenfratz:1998jp,Hasenfratz:1998ri,Luscher:1998pqa}
\be
   \left\{D,\gamma_5\right\} = aD\gamma_5 D,
\label{eq_GWrelation}
\ee
which is now known as the Ginsparg-Wilson relation
\cite{Ginsparg:1981bj}. The Nielsen-Ninomiya
theorem~\cite{Nielsen:1981hk}, which states that any lattice
formulation for which $D$ anticommutes with $\gamma_5$ necessarily has
doubler fermions, is circumvented since $\{D,\gamma_5\}\neq 0$.

A lattice Dirac operator which satisfies \eq{eq_GWrelation} can be
constructed in several ways. The so-called ``overlap'' or
Neuberger-Dirac operator~\cite{Neuberger:1997fp} acts in four
space-time dimensions and is, in its simplest form, defined by
\be
   D_{\rm N} = \frac{1}{\abar} \left( 1-\epsilon(A)
   \right),\quad\mathrm{where}\quad\epsilon(A)\equiv A (A^\dagger A)^{-1/2}, \quad A=1+s-aD_{\rm w},\quad \abar=\frac{a}{1+s},
\label{eq_overlap}
\ee
$D_{\rm w}$ is the massless Wilson-Dirac operator and $|s|<1$
is a tunable parameter. The overlap operator $D_{\rm N}$ removes all
doublers from the spectrum, and can readily be shown to satisfy the
Ginsparg-Wilson relation. The occurrence of the sign function $\epsilon(A)$ in
$D_{\rm N}$ renders the application of $D_{\rm N}$ in a computer
program potentially very costly, since it must be implemented using,
for instance, a polynomial approximation.

The most widely used approach to satisfying the Ginsparg-Wilson
relation \eq{eq_GWrelation} in large-scale numerical simulations is
provided by \textit{Domain Wall Fermions}
(DWF)~\cite{Kaplan:1992bt,Shamir:1993zy,Furman:1994ky} and we
therefore describe this in some more detail. Following early
exploratory studies~\cite{Blum:1996jf}. this approach has been
developed into a practical formulation of lattice QCD with good chiral
and flavour symmetries leading to results which contribute
significantly to this review. In this formulation, the fermion fields
$\psi(x,s)$ depend on a discrete fifth coordinate $s=1,\ldots,N$ as well as
the physical 4-dimensional space-time coordinates $x_\mu,\,\mu=1\cdots
4$ (the gluon fields do not depend on $s$). The lattice on which the
simulations are performed, is therefore a five-dimensional one of size
$L^3\times T\times N$, where $L,\,T$ and $N$ represent the number of
points in the spatial, temporal and fifth dimensions respectively.
The remarkable feature of DWF is that for each flavour there exists a
physical light mode corresponding to the field $q(x)$:
\begin{eqnarray}
q(x)&=&\frac{1+\gamma^5}{2}\psi(x,1)+\frac{1-\gamma^5}{2}\psi(x,N)\\
\bar{q}(x)&=&\overline{\psi}(x,N)\frac{1+\gamma^5}{2} + \overline{\psi}(x,1)\frac{1-\gamma^5}{2}\,.
\end{eqnarray}
The left and right-handed modes of the physical field are located on
opposite boundaries in the 5th dimensional space which, for
$N\to\infty$, allows for independent transformations of the left and
right components of the quark fields, that is for chiral
transformations. Unlike Wilson fermions, where for each flavour the
quark-mass parameter in the action is fine-tuned requiring a
subtraction of contributions of $\cO(1/a)$ where $a$ is the lattice
spacing, with DWF no such subtraction is necessary for the physical
modes, whereas the unphysical modes have masses of $\cO(1/a)$ and
decouple.

In actual simulations $N$ is finite and there are small violations of
chiral symmetry which must be accounted for. The theoretical framework
for the study of the residual breaking of chiral symmetry has been a
subject of intensive investigation (for a review and references to the
original literature see, e.g., \cite{Sharpe:2007yd}). The breaking
requires one or more \emph{crossings} of the fifth dimension to couple
the left and right-handed modes; the more crossings that are required
the smaller the effect.  For many physical quantities the leading
effects of chiral symmetry breaking due to finite $N$ are parameterized
by a \emph{residual} mass, $m_{\mathrm{res}}$.  For example, the PCAC
relation (for degenerate quarks of mass $m$) $\partial_\mu A_\mu(x) =
2m P(x)$, where $A_\mu$ and $P$ represent the axial current and
pseudoscalar density respectively, is satisfied with
$m=m^\mathrm{DWF}+m_\mathrm{res}$, where $m^\mathrm{DWF}$ is the bare
mass in the DWF action. The mixing of operators which transform under
different representations of chiral symmetry is found to be negligibly
small in current simulations. The important thing to note is that the
chiral symmetry breaking effects are small and that there are
techniques to mitigate their consequences.

The main price which has to be paid for the good chiral symmetry is
that the simulations are performed in 5 dimensions, requiring
approximately a factor of $N$ in computing resources and resulting in
practice in ensembles at fewer values of the lattice spacing and quark
masses than is possible with other formulations. The current
generation of DWF simulations is being performed at physical quark
masses so that ensembles with good chiral and flavour symmetries are
being generated and analyzed~\cite{Arthur:2012opa}. For a discussion
of the equivalence of DWF and overlap fermions
see Refs.~\cite{Borici:1999zw,Borici:1999da}.

A third example of an operator which satisfies the Ginsparg-Wilson
relation is the so-called fixed-point action
\cite{Bietenholz:1995cy,Hasenfratz:2000xz,Hasenfratz:2002rp}. This
construction proceeds via a renormalization group approach. A related
formalism are the so-called ``chirally improved'' fermions
\cite{Gattringer:2000js}.\\

\begin{table}
\begin{center}
{\footnotesize
\begin{tabular*}{\textwidth}[l]{l @{\extracolsep{\fill}} l l l l}
\hline \hline  \\[-1.0ex]
\parbox[t]{1.5cm}{Abbrev.} & Discretization & \parbox[t]{2.2cm}{Leading lattice \\artifacts} & Chiral symmetry &  Remarks
\\[4.0ex] \hline \hline \\[-1.0ex]
Wilson     & Wilson & $\cO(a)$ & broken & 
\\[1.0ex] \hline \\[-1.0ex]
tmWil   & twisted-mass Wilson &  \parbox[t]{2.2cm}{$\cO(a^2)$
at\\ maximal twist} & broken & \parbox[t]{5cm}{flavour-symmetry breaking:\\ $(M_\text{PS}^{0})^2-(M_\text{PS}^\pm)^2\sim \cO(a^2)$}
\\[4.0ex] \hline \\[-1.0ex]
tlSW      & Sheikholeslami-Wohlert & $\cO(g^2 a)$ & broken & tree-level
impr., $\csw=1$
\\[1.0ex] \hline \\[-1.0ex]
\parbox[t]{1.0cm}{n-HYP tlSW}      & Sheikholeslami-Wohlert & $\cO(g^2 a)$ & broken & \parbox[t]{5cm}{tree-level
impr., $\csw=1$,\\
n-HYP smeared gauge links
}
\\[4.0ex] \hline \\[-1.0ex]
\parbox[t]{1.2cm}{stout tlSW}      & Sheikholeslami-Wohlert & $\cO(g^2 a)$ & broken & \parbox[t]{5cm}{tree-level
impr., $\csw=1$,\\
stout smeared gauge links
}
\\[4.0ex] \hline \\[-1.0ex]
\parbox[t]{1.2cm}{HEX tlSW}      & Sheikholeslami-Wohlert & $\cO(g^2 a)$ & broken & \parbox[t]{5cm}{tree-level
impr., $\csw=1$,\\
HEX smeared gauge links
}
\\[4.0ex] \hline \\[-1.0ex]
mfSW      & Sheikholeslami-Wohlert & $\cO(g^2 a)$ & broken & mean-field impr.
\\[1.0ex] \hline \\[-1.0ex]
npSW      & Sheikholeslami-Wohlert & $\cO(a^2)$ & broken & nonperturbatively impr.
\\[1.0ex] \hline \\[-1.0ex]
KS      & Staggered & $\cO(a^2)$ & \parbox[t]{3cm}{$\rm
  U(1)\times U(1)$ subgr.\\ unbroken} & rooting for $\Nf<4$
\\[4.0ex] \hline \\[-1.0ex]
Asqtad  & Staggered & $\cO(g^2a^2)$ & \parbox[t]{3cm}{$\rm
  U(1)\times U(1)$ subgr.\\ unbroken}  & \parbox[t]{5cm}{Asqtad
  smeared gauge links, \\rooting for $\Nf<4$}  
\\[4.0ex] \hline \\[-1.0ex]
HISQ  & Staggered & $\cO(g^2a^2)$ & \parbox[t]{3cm}{$\rm
  U(1)\times U(1)$ subgr.\\ unbroken}  & \parbox[t]{5cm}{HISQ
  smeared gauge links, \\rooting for $\Nf<4$}  
\\[4.0ex] \hline \\[-1.0ex]
DW      & Domain Wall & \parbox[t]{2.2cm}{asymptotically \\$\cO(a^2)$} & \parbox[t]{3cm}{remnant
  breaking \\exponentially suppr.} & \parbox[t]{5cm}{exact chiral symmetry and\\$\cO(a)$ impr. only in the limit \\
 $N\rightarrow \infty$}
\\[7.0ex] \hline \\[-1.0ex]
oDW      & optimal Domain Wall & \parbox[t]{2.2cm}{asymptotically \\$\cO(a^2)$} & \parbox[t]{3cm}{remnant
  breaking \\exponentially suppr.} & \parbox[t]{5cm}{exact chiral symmetry and\\$\cO(a)$ impr. only in the limit \\
 $N\rightarrow \infty$}
\\[7.0ex] \hline \\[-1.0ex]
M-DW      & Moebius Domain Wall & \parbox[t]{2.2cm}{asymptotically \\$\cO(a^2)$} & \parbox[t]{3cm}{remnant
  breaking \\exponentially suppr.} & \parbox[t]{5cm}{exact chiral symmetry and\\$\cO(a)$ impr. only in the limit \\
 $N\rightarrow \infty$}
\\[7.0ex] \hline \\[-1.0ex]
overlap    & Neuberger & $\cO(a^2)$ & exact
\\[1.0ex] 
\hline\hline
\end{tabular*}
}
\caption{The most widely used discretizations of the quark action
  and some of their properties. Note that in order to maintain the
  leading lattice artifacts of the action in nonspectral observables
  (like operator matrix elements)
  the corresponding nonspectral operators need to be improved as well. 
\label{tab_quarkactions}}
\end{center}
\end{table}

\noindent
{\it Smearing}\\
\noindent

A simple modification which can help improve the action as well as the
computational performance is the use of smeared gauge fields in the
covariant derivatives of the fermionic action. Any smearing procedure
is acceptable as long as it consists of only adding irrelevant (local)
operators. Moreover, it can be combined with any discretization of the
quark action.  The ``Asqtad'' staggered quark action mentioned above
\cite{Orginos:1999cr} is an example which makes use of so-called
``Asqtad'' smeared (or ``fat'') links. Another example is the use of
n-HYP smeared \cite{Hasenfratz:2001hp,Hasenfratz:2007rf}, stout smeared
\cite{Morningstar:2003gk,Durr:2008rw} or HEX (hypercubic stout) smeared \cite{Capitani:2006ni} gauge links in the tree-level clover improved
discretization of the quark action, denoted by ``n-HYP tlSW'',
``stout tlSW'' and ``HEX tlSW'' in the following.\\

\noindent
In Tab.~\ref{tab_quarkactions} we summarize the most widely used
discretizations of the quark action and their main properties together
with the abbreviations used in the summary tables. Note that in order
to maintain the leading lattice artifacts of the actions as given in
the table in nonspectral observables (like operator matrix elements)
the corresponding nonspectral operators need to be improved as well.

\subsubsection{Heavy-quark actions}
\label{app:HQactions}

Charm and bottom quarks are often simulated with different
lattice-quark actions than up, down, and strange quarks because their
masses are large relative to typical lattice spacings in current
simulations; for example, $a m_c \sim 0.4$ and $am_b \sim 1.3$ at
$a=0.06$~fm.  Therefore, for the actions described in the previous
section, using a sufficiently small lattice spacing to control generic
$(am_h)^n$ discretization errors at the physical $b$-quark mass is
computationally demanding and has so far not been possible, with the
first exception being the calculation of FNAL/MILC in
\cite{Bazavov:2017lyh} which uses the HISQ action and a lattice spacing of $a \approx 0.03$\,fm.

One alternative approach for lattice heavy quarks is direct application of
effective theory.  In this case the lattice heavy-quark action only
correctly describes phenomena in a specific kinematic regime, such as
Heavy-Quark Effective Theory
(HQET)~\cite{Isgur:1989vq,Eichten:1989zv,Isgur:1989ed} or
Nonrelativistic QCD (NRQCD)~\cite{Caswell:1985ui,Bodwin:1994jh}.  One
can discretize the effective Lagrangian to obtain, for example,
Lattice HQET~\cite{Heitger:2003nj} or Lattice
NRQCD~\cite{Thacker:1990bm,Lepage:1992tx}, and then simulate the
effective theory numerically.  The coefficients of the operators in
the lattice-HQET and lattice-NRQCD actions are free parameters that
must be determined by matching to the underlying theory (QCD) through
the chosen order in $1/m_h$ or $v_h^2$, where $m_h$ is the heavy-quark
mass and $v_h$ is the heavy-quark velocity in the the heavy-light
meson rest frame.

Another approach is to interpret a relativistic quark action such as
those described in the previous section in a manner suitable for heavy
quarks.  One can extend the standard Symanzik improvement program,
which allows one to systematically remove lattice cutoff effects by
adding higher-dimension operators to the action, by allowing the
coefficients of the dimension 4 and higher operators to depend
explicitly upon the heavy-quark mass.  Different prescriptions for
tuning the parameters correspond to different implementations: those
in common use are often called the Fermilab
action~\cite{ElKhadra:1996mp}, the relativistic heavy-quark action
(RHQ)~\cite{Christ:2006us}, and the Tsukuba
formulation~\cite{Aoki:2001ra}.  In the Fermilab approach, HQET is
used to match the lattice theory to continuum QCD at the desired order
in $1/m_h$.

More generally, effective theory can be used to estimate the size of
cutoff errors from the various lattice heavy-quark actions.  The power
counting for the sizes of operators with heavy quarks depends on the
typical momenta of the heavy quarks in the system.  Bound-state
dynamics differ considerably between heavy-heavy and heavy-light
systems.  In heavy-light systems, the heavy quark provides an
approximately static source for the attractive binding force, like the
proton in a hydrogen atom.  The typical heavy-quark momentum in the
bound-state rest frame is $|\vec{p}_h| \sim \Lambda_{\rm QCD}$, and
heavy-light operators scale as powers of $(\Lambda_{\rm QCD}/m_h)^n$.
This is often called ``HQET power-counting'', although it applies to
heavy-light operators in HQET, NRQCD, and even relativistic
heavy-quark actions described below.  Heavy-heavy systems are similar
to positronium or the deuteron, with the typical heavy-quark momentum
$|\vec{p}_h| \sim \alpha_S m_h$.  Therefore motion of the heavy quarks
in the bound state rest frame cannot be neglected.  Heavy-heavy
operators have complicated power counting rules in terms of
$v_h^2$~\cite{Lepage:1992tx}; this is often called ``NRQCD power
counting.''

Alternatively, one can simulate bottom or charm quarks with the same
action as up, down, and strange quarks provided that (1) the action is
sufficiently improved, and (2) the lattice spacing is sufficiently
fine.  These qualitative criteria do not specify precisely how large a
numerical value of $am_h$ can be allowed while obtaining a given
precision for physical quantities; this must be established
empirically in numerical simulations.  At present, both the HISQ and
twisted-mass Wilson actions discussed previously are being used to
simulate charm quarks.
Simulations with HISQ quarks have employed heavier-quark masses than
those with twisted-mass Wilson quarks because the action is more
highly improved, but neither action has been used to simulate at the
physical $am_b$ until the recent calculation of FNAL/MILC in
\cite{Bazavov:2017lyh}, where a lattice spacing of $a \approx 0.03$\,fm is available.
All other calculations
of heavy-light decay constants with these actions still rely on
effective theories: the ETM collaboration
interpolates between twisted-mass Wilson data generated near $am_c$
and the static point~\cite{Dimopoulos:2011gx}, while the HPQCD
collaboration, for the coarser lattice spacings,  
extrapolates HISQ data generated below $am_b$ up to the
physical point using an HQET-inspired series expansion in
$(1/m_h)^n$~\cite{McNeile:2011ng}.
\\


\noindent
{\it Heavy-quark effective theory}\\
\noindent

HQET was introduced by Eichten and Hill in
Ref.~\cite{Eichten:1989zv}. It provides the correct asymptotic
description of QCD correlation functions in the static limit
$m_{h}/|\vec{p}_h| \!\to\! \infty$. Subleading effects are described
by higher dimensional operators whose coupling constants are formally
of ${\cO}((1/m_{h})^n)$.  The HQET expansion works well for
heavy-light systems in which the heavy-quark momentum is small
compared to the mass.

The HQET Lagrangian density at the leading (static) order in the rest
frame of the heavy quark is given by
\be
{\mathcal L}^{\rm stat}(x) = \overline{\psi}_{h}(x) \,D_0\, \psi_{h}(x)\;,
\ee
with
\be
P_+ \psi_{h} = \psi_{h} \; , \quad\quad \overline{\psi}_{h} P_+=\overline{\psi}_{h} \;,  
\quad\quad P_+={{1+\gamma_0}\over{2}} \;.
\ee
A bare quark mass $m_{\rm bare}^{\rm stat}$ has to be added to the energy  
levels $E^{\rm stat}$ computed with this Lagrangian to obtain the physical ones.
 For example, the mass of the $B$ meson in the static approximation is given by
\be
m_{B} = E^{\rm stat} + m_{\rm bare}^{\rm stat} \;.
\ee
At tree-level $m_{\rm bare}^{\rm stat}$ is simply the (static approximation of
the) $b$-quark mass, but in the quantized lattice formulation it has
to further compensate a divergence linear in the inverse lattice spacing.
Weak composite fields  are also rewritten in terms of the static fields, e.g.,
\begin{equation}
A_0(x)^{\rm stat}=Z_{\rm A}^{\rm stat} \left( \overline{\psi}(x) \gamma_0\gamma_5\psi_h(x)\right)\;,
\end{equation}
where the renormalization factor of the axial current in the static
theory $Z_{\rm A}^{\rm stat}$ is scale-dependent.  Recent lattice-QCD
calculations using static $b$ quarks and dynamical light
quarks \cite{Albertus:2010nm,Dimopoulos:2011gx} perform the operator
matching at 1-loop in mean-field improved lattice perturbation
theory~\cite{Ishikawa:2011dd,Blossier:2011dg}.  Therefore the
heavy-quark discretization, truncation, and matching errors in these
results are of ${\cO}(a^2 \Lambda_{\rm QCD}^2)$, ${\cO}
(\Lambda_{\rm QCD}/m_h)$, and ${\cO}(\alpha_s^2, \alpha_s^2
a \Lambda_{\rm QCD})$.

In order to reduce heavy-quark truncation errors in $B$-meson masses
and matrix elements to the few-percent level, state-of-the-art
lattice-HQET computations now include corrections of ${\cO}(1/m_h)$.  Adding the $1/m_{h}$ terms, the HQET Lagrangian reads
\begin{eqnarray}
{\mathcal L}^{\rm HQET}(x) &=&  {\mathcal L}^{\rm stat}(x) - \omegakin{\mathcal{O}}_{\rm kin}(x)
        - \omegaspin{\mathcal{O}}_{\rm spin}(x)  \,, \\[2.0ex]
  \mathcal{O}_{\rm kin}(x) &=& \overline{\psi}_{h}(x){\bf D}^2\psi_{h}(x) \,,\quad
  \mathcal{O}_{\rm spin}(x) = \overline{\psi}_{h}(x){\boldsymbol\sigma}\!\cdot\!{\bf B}\psi_{h}(x)\,.
\end{eqnarray}
At this order, two other parameters appear in the Lagrangian,
$\omegakin$ and $\omegaspin$. The normalization is such that the
tree-level values of the coefficients are
$\omegakin=\omegaspin=1/(2m_{h})$.  Similarly the operators are
formally expanded in inverse powers of the heavy-quark mass.  The time
component of the axial current, relevant for the computation of
mesonic decay constants is given by
\begin{eqnarray}
A_0^{\rm HQET}(x) &=& Z_{\rm A}^{\rm HQET}\left(A_0^{\rm stat}(x) +\sum_{i=1}^2 c_{\rm A}^{(i)} A_0^{(i)}(x)\right)\;, \\
A_0^{(1)}(x)&=&\overline{\psi}\frac{1}{2}\gamma_5 \gamma_k  (\nabla_k-\overleftarrow{\nabla}_k)\psi_h(x), \qquad k=1,2,3\\
A_0^{(2)} &=& -\partial_kA_k^{\rm stat}(x)\;, \quad A_k^{\rm stat}=\overline{\psi}(x) \gamma_k\gamma_5\psi_h(x)\;,
\end{eqnarray}
and depends on two additional parameters $c_{\rm A}^{(1)}$ and $c_{\rm A}^{(2)}$.

A framework for nonperturbative HQET on the lattice has been
introduced in Refs.~\cite{Heitger:2003nj,Blossier:2010jk}.  As pointed out
in Refs.~\cite{Sommer:2006sj,DellaMorte:2007ny}, since $\alpha_s(m_h)$
decreases logarithmically with $m_h$, whereas corrections in the
effective theory are power-like in $\Lambda/m_h$, it is possible that
the leading errors in a calculation will be due to the perturbative
matching of the action and the currents at a given order
$(\Lambda/m_h)^l$ rather than to the missing ${\cO}((\Lambda/m_h)^{l+1})$ terms.  Thus, in order to keep matching
errors below the uncertainty due to truncating the HQET expansion, the
matching is performed nonperturbatively beyond leading order in
$1/m_{h}$. The asymptotic convergence of HQET in the limit
$m_h \to \infty$ indeed holds only in that case.

The higher dimensional interaction terms in the effective Lagrangian
are treated as space-time volume insertions into static correlation
functions.  For correlators of some multi-local fields ${\oO}$
and up to the $1/m_h$ corrections to the operator, this means
\begin{equation}
\langle {\oO} \rangle =\langle {\oO} \rangle_{\rm stat} +\omegakin a^4 \sum_x
\langle {\oO\mathcal{O}}_{\rm kin}(x) \rangle_{\rm stat} + \omegaspin a^4 \sum_x
\langle {\oO\mathcal{O}}_{\rm spin}(x) \rangle_{\rm stat} \;, 
\end{equation}
where $\langle {\oO} \rangle_{\rm stat}$ denotes the static
expectation value with ${\mathcal{L}}^{\rm stat}(x)
+{\mathcal{L}}^{\rm light}(x)$.  Nonperturbative renormalization of
these correlators guarantees the existence of a well-defined continuum
limit to any order in $1/m_h$.  The parameters of the effective action
and operators are then determined by matching a suitable number of
observables calculated in HQET (to a given order in $1/m_{h}$) and in
QCD in a small volume (typically with $L\simeq 0.5$ fm), where the
full relativistic dynamics of the $b$-quark can be simulated and the
parameters can be computed with good accuracy.
In Refs.~\cite{Blossier:2010jk,Blossier:2012qu} the Schr\"odinger Functional
(SF) setup has been adopted to define a set of quantities, given by
the small volume equivalent of decay constants, pseudoscalar-vector
splittings, effective masses and ratio of correlation functions for
different kinematics, that can be used to implement the matching
conditions.  The kinematical conditions are usually modified by
changing the periodicity in space of the fermions, i.e., by directly
exploiting a finite-volume effect.  The new scale $L$, which is
introduced in this way, is chosen such that higher orders in $1/m_hL$
and in $\Lambda_{\rm QCD}/m_h$ are of about the same size. At the end
of the matching step the parameters are known at lattice spacings
which are of the order of $0.01$ fm, significantly smaller than the
resolutions used for large volume, phenomenological, applications. For
this reason a set of SF-step scaling functions is introduced in the
effective theory to evolve the parameters to larger lattice spacings.
The whole procedure yields the nonperturbative parameters with an
accuracy which allows to compute phenomenological quantities with a
precision of a few percent
(see Refs.~\cite{Blossier:2010mk,Bernardoni:2012ti} for the case of the
$B_{(s)}$ decay constants).  Such an accuracy can not be achieved by
performing the nonperturbative matching in large volume against
experimental measurements, which in addition would reduce the
predictivity of the theory.  For the lattice-HQET action matched
nonperturbatively through ${\cO}(1/m_h)$, discretization and
truncation errors are of ${\cO}(a \Lambda^2_{\rm QCD}/m_h,
a^2 \Lambda^2_{\rm QCD})$ and ${\cO}((\Lambda_{\rm QCD}/m_h )^2)$.

The noise-to-signal ratio of static-light correlation functions grows
exponentially in Euclidean time, $\propto e^{\mu x_0}$ . The rate
$\mu$ is nonuniversal but diverges as $1/a$ as one approaches the
continuum limit. By changing the discretization of the covariant
derivative in the static action one may achieve an exponential
reduction of the noise to signal ratio. Such a strategy led to the
introduction of the $S^{\rm stat}_{\rm HYP1,2}$
actions~\cite{DellaMorte:2005yc}, where the thin links in $D_0$ are
replaced by HYP-smeared links~\cite{Hasenfratz:2001hp}.  These actions
are now used in all lattice applications of HQET.
\\


\noindent
{\it Nonrelativistic QCD}\\
\noindent

Nonrelativistic QCD (NRQCD) \cite{Thacker:1990bm,Lepage:1992tx} is an
 effective theory that can be matched to full QCD order by order in
 the heavy-quark velocity $v_h^2$ (for heavy-heavy systems) or in
 $\Lambda_{\rm QCD}/m_h$ (for heavy-light systems) and in powers of
 $\alpha_s$.  Relativistic corrections appear as higher-dimensional
 operators in the Hamiltonian.
 
 As an effective field theory, NRQCD is only useful with an
 ultraviolet cutoff of order $m_h$ or less. On the lattice this means
 that it can be used only for $am_h>1$, which means that $\cO(a^n)$
 errors cannot be removed by taking $a\to0$ at fixed $m_h$. Instead
 heavy-quark discretization errors are systematically removed by
 adding additional operators to the lattice Hamiltonian.  Thus, while
 strictly speaking no continuum limit exists at fixed $m_h$, continuum
 physics can be obtained at finite lattice spacing to arbitrarily high
 precision provided enough terms are included, and provided that the
 coefficients of these terms are calculated with sufficient accuracy.
 Residual discretization errors can be parameterized as corrections to
 the coefficients in the nonrelativistic expansion, as shown in
 Eq.~(\ref{deltaH}).  Typically they are of the form
 $(a|\vec{p}_h|)^n$ multiplied by a function of $am_h$ that is smooth
 over the limited range of heavy-quark masses (with $am_h > 1$) used
 in simulations, and can therefore can be represented by a low-order
 polynomial in $am_h$ by Taylor's theorem (see
 Ref.~\cite{Gregory:2010gm} for further discussion).  Power-counting
 estimates of these effects can be compared to the observed lattice-spacing dependence in simulations. Provided that these effects are
 small, such comparisons can be used to estimate and correct the
 residual discretization effects.

An important feature of the NRQCD approach is that the same action can
be applied to both heavy-heavy and heavy-light systems. This allows,
for instance, the bare $b$-quark mass to be fixed via experimental
input from $\Upsilon$ so that simulations carried out in the $B$ or
$B_s$ systems have no adjustable parameters left.  Precision
calculations of the $B_s$-meson mass (or of the mass splitting
$M_{B_s} - M_\Upsilon/2$) can then be used to test the reliability of
the method before turning to quantities one is trying to predict, such
as decay constants $f_B$ and $f_{B_s}$, semileptonic form factors or
neutral $B$ mixing parameters.

Given the same lattice-NRQCD heavy-quark action, simulation results
will not be as accurate for charm quarks as for bottom ($1/m_b <
1/m_c$, and $v_b < v_c$ in heavy-heavy systems).  For charm, however,
a more serious concern is the restriction that $am_h$ must be greater
than one.  This limits lattice-NRQCD simulations at the physical
$am_c$ to relatively coarse lattice spacings for which light-quark and
gluon discretization errors could be large.  Thus recent lattice-NRQCD
simulations have focused on bottom quarks because $am_b > 1$ in the
range of typical lattice spacings between $\approx$ 0.06 and 0.15~fm.

In most simulations with NRQCD $b$-quarks during the past decade one
has worked with an NRQCD action that includes tree-level relativistic
corrections through ${\cO}(v_h^4)$ and discretization corrections
through ${\cO}(a^2)$,
 \begin{eqnarray}
 \label{nrqcdact}
&&  S_{\rm NRQCD}  =
a^4 \sum_x \Bigg\{  {\Psi}^\dagger_t \Psi_t -
 {\Psi}^\dagger_t
\left(1 \!-\!\frac{a \delta H}{2}\right)_t
 \left(1\!-\!\frac{aH_0}{2n}\right)^{n}_t \nonumber \\
& \times &
 U^\dagger_t(t-a)
 \left(1\!-\!\frac{aH_0}{2n}\right)^{n}_{t-a}
\left(1\!-\!\frac{a\delta H}{2}\right)_{t-a} \Psi_{t-a} \Bigg\} \, ,
 \end{eqnarray}
where the subscripts $``t''$ and $``t-a''$ denote that the heavy-quark, gauge, $\bf{E}$,  and $\bf{B}$-fields are on time slices $t$ or $t-a$, respectively.
 $H_0$ is the nonrelativistic kinetic energy operator,
 \be
 H_0 = - {\delsq\over2m_h} \, ,
 \ee
and $\delta H$ includes relativistic and finite-lattice-spacing
corrections,
 \begin{eqnarray}
\delta H
&=& - c_1\,\frac{(\delsq)^2}{8m_h^3}
+ c_2\,\frac{i g}{8m_h^2}\left(\delv\cdot\Ev - \Ev\cdot\delv\right) \nl
& &
 - c_3\,\frac{g}{8m_h^2} \sigmav\cdot(\delvt\times\Ev - \Ev\times\delvt)\nl
& & - c_4\,\frac{g}{2m_h}\,\sigmav\cdot\Bv
  + c_5\,\frac{a^2\delfour}{24m_h}  - c_6\,\frac{a(\delsq)^2}
{16nm_h^2} \, .
\label{deltaH}
\end{eqnarray}
 $m_h$ is the bare heavy-quark mass, $\delsq$ the lattice Laplacian,
$\delv$ the symmetric lattice derivative and $\delfour$ the lattice
discretization of the continuum $\sum_i D^4_i$.  $\delvt$ is the
improved symmetric lattice derivative and the $\Ev$ and $\Bv$ fields
have been improved beyond the usual clover leaf construction. The
stability parameter $n$ is discussed in Ref.~\cite{Lepage:1992tx}.  In most
cases the $c_i$'s have been set equal to their tree-level values $c_i
= 1$.  With this implementation of the NRQCD action, errors in
heavy-light-meson masses and splittings are of ${\cO}(\alpha_S \Lambda_{\rm QCD}/m_h )$, ${\cO}(\alpha_S (\Lambda_{\rm
QCD}/m_h)^2 )$, ${\cO}((\Lambda_{\rm QCD}/m_h )^3)$, and ${\cO}(\alpha_s a^2 \Lambda_{\rm QCD}^2)$, with coefficients that are
functions of $am_h$.  1-loop corrections to many of the coefficients
in Eq.~(\ref{deltaH}) have now been calculated, and are starting to be
included in
simulations \cite{Morningstar:1994qe,Hammant:2011bt,Dowdall:2011wh}.

Most of the operator matchings involving heavy-light currents or
four-fermion operators with NRQCD $b$-quarks and Asqtad or HISQ light
quarks have been carried out at 1-loop order in lattice perturbation
theory.  In calculations published to date of electroweak matrix
elements, heavy-light currents with massless light quarks have been
matched through ${\cO}(\alpha_s, \Lambda_{\rm QCD}/m_h, \alpha_s/(a
m_h),
\alpha_s \Lambda_{\rm QCD}/m_h)$, and four-fermion operators through \\
 ${\cO}(\alpha_s, \Lambda_{\rm QCD}/m_h, 
\alpha_s/(a m_h))$.
NRQCD/HISQ currents with massive HISQ quarks are also of interest,
e.g.,  for the bottom-charm currents in $B \rightarrow D^{(*)} l \nu$
semileptonic decays and the relevant matching calculations have been
performed at 1-loop order in Ref.~\cite{Monahan:2012dq}.  Taking all
the above into account, the most significant systematic error in
electroweak matrix elements published to date with NRQCD $b$-quarks is
the ${\cO}(\alpha_s^2)$ perturbative matching uncertainty.  Work is
therefore underway to use current-current correlator methods combined
with very high order continuum perturbation theory to do current
matchings nonperturbatively~\cite{Koponen:2010jy}.
\\


\noindent
{\it Relativistic heavy quarks}\\
\noindent

An approach for relativistic heavy-quark lattice formulations was
first introduced by El-Khadra, Kronfeld, and Mackenzie in
Ref.~\cite{ElKhadra:1996mp}.  Here they showed that, for a general
lattice action with massive quarks and non-Abelian gauge fields,
discretization errors can be factorized into the form $f(m_h
a)(a|\vec{p}_h|)^n$, and that the function $f(m_h a)$ is bounded to be
of ${\cO}(1)$ or less for all values of the quark mass $m_h$.
Therefore cutoff effects are of ${\cO}(a \Lambda_{\rm QCD})^n$
and ${\cO}((a|\vec{p}_h|)^n)$, even for $am_h \gtapprox 1$, and
can be controlled using a Symanzik-like procedure.  As in the standard
Symanzik improvement program, cutoff effects are systematically
removed by introducing higher-dimension operators to the lattice
action and suitably tuning their coefficients.  In the relativistic
heavy-quark approach, however, the operator coefficients are allowed
to depend explicitly on the quark mass.  By including lattice
operators through dimension $n$ and adjusting their coefficients
$c_{n,i}(m_h a)$ correctly, one enforces that matrix elements in the
lattice theory are equal to the analogous matrix elements in continuum
QCD through $(a|\vec{p}_h|)^n$, such that residual heavy-quark
discretization errors are of ${\cO}(a|\vec{p}_h|)^{n+1}$.

The relativistic heavy-quark approach can be used to compute the
matrix elements of states containing heavy quarks for which the
heavy-quark spatial momentum $|\vec{p}_h|$ is small compared to the
lattice spacing.  Thus it is suitable to describe bottom and charm
quarks in both heavy-light and heavy-heavy systems.  Calculations of
bottomonium and charmonium spectra serve as nontrivial tests of the
method and its accuracy.

At fixed lattice spacing, relativistic heavy-quark formulations
recover the massless limit when $(am_h) \ll 1$, recover the static
limit when $(am_h) \gg 1$, and smoothy interpolate between the two;
thus they can be used for any value of the quark mass, and, in
particular, for both charm and bottom.  Discretization errors for
relativistic heavy-quark formulations are generically of the form
$\alpha_s^k f(am_h)(a |\vec{p}_h|)^n$, where $k$ reflects the order of
the perturbative matching for operators of ${\cO}((a |\vec{p}_h|)^n)$.
For each $n$, such errors are removed completely if the operator
matching is nonperturbative. When $(am_h) \sim 1$, this gives rise to
nontrivial lattice-spacing dependence in physical quantities, and it
is prudent to compare estimates based on power-counting with a direct
study of scaling behaviour using a range of lattice spacings.  At
fixed quark mass, relativistic heavy-quark actions possess a smooth
continuum limit without power-divergences.  Of course, as $m_h \to
\infty$ at fixed lattice spacing, the static limit is recovered and by
then taking the continuum limit the corresponding power divergences
are reproduced (see, e.g., Ref.~\cite{Harada:2001fi}).

The relativistic heavy-quark formulations in use all begin with the
asymmetric (or anisotropic) Sheikholeslami-Wohlert (``clover'')
action~\cite{Sheikholeslami:1985ij}:
\begin{equation}
S_\textrm{lat} = a^4 \sum_{x,x'} \bar{\psi}(x') \left( m_0 + \gamma_0 D_0 + \zeta \vec{\gamma} \cdot \vec{D} - \frac{a}{2} (D^0)^2 - \frac{a}{2} \zeta (\vec{D})^2+ \sum_{\mu,\nu} \frac{ia}{4} c_{\rm SW} \sigma_{\mu\nu} F_{\mu\nu} \right)_{x' x} \psi(x) \,,
\label{eq:HQAct}
\end{equation}
where $D_\mu$ is the lattice covariant derivative and $F_{\mu\nu}$ is
the lattice field-strength tensor.  Here we show the form of the
action given in Ref.~\cite{Christ:2006us}.  The introduction of a
space-time asymmetry, parameterized by $\zeta$ in
Eq.~(\ref{eq:HQAct}), is convenient for heavy-quark systems because
the characteristic heavy-quark four-momenta do not respect space-time
axis exchange ($\vec{p}_h < m_h$ in the bound-state rest frame).
Further, the Sheikoleslami-Wohlert action respects the continuum
heavy-quark spin and flavour symmetries, so HQET can be used to
interpret and estimate lattice discretization
effects~\cite{Kronfeld:2000ck,Harada:2001fi,Harada:2001fj}.  We
discuss three different prescriptions for tuning the parameters of the
action in common use below.  In particular, we focus on aspects of the
action and operator improvement and matching relevant for evaluating
the quality of the calculations discussed in the main text.

The meson energy-momentum dispersion relation plays an important role
in relativistic heavy-quark formulations:
\begin{equation}
	E(\vec{p}) = M_1 + \frac{\vec{p}^2}{2M_2} + {\cO}(\vec{p}^4) \,,
\end{equation}
where $M_1$ and $M_2$ are known as the rest and kinetic masses,
respectively.  Because the lattice breaks Lorentz invariance, there
are corrections proportional to powers of the momentum.  Further, the
lattice rest masses and kinetic masses are not equal ($M_1 \neq M_2$),
and only become equal in the continuum limit.

The Fermilab interpretation~\cite{ElKhadra:1996mp} is suitable for
calculations of mass splittings and matrix elements of systems with
heavy quarks.  The Fermilab action is based on the hopping-parameter
form of the Wilson action, in which $\kappa_h$ parameterizes the
heavy-quark mass.  In practice, $\kappa_h$ is tuned such that the the
kinetic meson mass equals the experimentally-measured heavy-strange
meson mass ($m_{B_s}$ for bottom and $m_{D_s}$ for charm).  In
principle, one could also tune the anisotropy parameter such that $M_1
= M_2$.  This is not necessary, however, to obtain mass splittings and
matrix elements, which are not affected by
$M_1$~\cite{Kronfeld:2000ck}.  Therefore in the Fermilab action the
anisotropy parameter is set equal to unity.  The clover coefficient in
the Fermilab action is fixed to the value $c_{\rm SW} = 1/u_0^3$ from
mean-field improved lattice perturbation theory~\cite{Lepage:1992xa}.
With this prescription, discretization effects are of ${\cO}(\alpha_sa|\vec{p}_h|, (a|\vec{p}_h|)^2)$.  Calculations of
electroweak matrix elements also require improving the lattice current
and four-fermion operators to the same order, and matching them to the
continuum.  Calculations with the Fermilab action remove tree-level
${\cO}(a)$ errors in electroweak operators by rotating the
heavy-quark field used in the matrix element and setting the rotation
coefficient to its tadpole-improved tree-level value (see, e.g.,
Eqs.~(7.8) and (7.10) of Ref.~\cite{ElKhadra:1996mp}).  Finally,
electroweak operators are typically renormalized using a mostly
nonperturbative approach in which the flavour-conserving light-light
and heavy-heavy current renormalization factors $Z_V^{ll}$ and
$Z_V^{hh}$ are computed nonperturbatively~\cite{ElKhadra:2001rv}.  The
flavour-conserving factors account for most of the heavy-light current
renormalization.  The remaining correction is expected to be close to
unity due to the cancellation of most of the radiative corrections
including tadpole graphs~\cite{Harada:2001fi}; therefore it can be
reliably computed at 1-loop in mean-field improved lattice
perturbation theory with truncation errors at the percent to
few-percent level.

The relativistic heavy-quark (RHQ) formulation developed by Li, Lin,
and Christ builds upon the Fermilab approach, but tunes all the
parameters of the action in Eq.~(\ref{eq:HQAct})
nonperturbatively~\cite{Christ:2006us}.  In practice, the three
parameters $\{m_0a, c_{\rm SW}, \zeta\}$ are fixed to reproduce the
experimentally-measured $B_s$ meson mass and hyperfine splitting
($m_{B_s^*}-m_{B_s}$), and to make the kinetic and rest masses of the
lattice $B_s$ meson equal~\cite{Aoki:2012xaa}.  This is done by
computing the heavy-strange meson mass, hyperfine splitting, and ratio
$M_1/M_2$ for several sets of bare parameters $\{m_0a, c_{\rm
SW}, \zeta\}$ and interpolating linearly to the physical $B_s$ point.
By fixing the $B_s$-meson hyperfine splitting, one loses a potential
experimental prediction with respect to the Fermilab formulation.
However, by requiring that $M_1 = M_2$, one gains the ability to use
the meson rest masses, which are generally more precise than the
kinetic masses, in the RHQ approach.  The nonperturbative
parameter-tuning procedure eliminates ${\cO}(a)$ errors from
the RHQ action, such that discretization errors are of ${\cO}((a|\vec{p}_h|)^2)$.  Calculations of $B$-meson decay constants and
semileptonic form factors with the RHQ action are in
progress~\cite{Witzel:2012pr,Kawanai:2012id}, as is the corresponding
1-loop mean-field improved lattice perturbation
theory~\cite{Lehner:2012bt}.  For these works, cutoff effects in the
electroweak vector and axial-vector currents will be removed through
${\cO}(\alpha_s a)$, such that the remaining discretization
errors are of ${\cO}(\alpha_s^2a|\vec{p}_h|,
(a|\vec{p}_h|)^2)$.  Matching the lattice operators to the continuum
will be done following the mostly nonperturbative approach described
above.

The Tsukuba heavy-quark action is also based on the
Sheikholeslami-Wohlert action in Eq.~(\ref{eq:HQAct}), but allows for
further anisotropies and hence has additional parameters: specifically
the clover coefficients in the spatial $(c_B)$ and temporal $(c_E)$
directions differ, as do the anisotropy coefficients of the $\vec{D}$
and $\vec{D}^2$ operators~\cite{Aoki:2001ra}.  In practice, the
contribution to the clover coefficient in the massless limit is
computed nonperturbatively~\cite{Aoki:2005et}, while the
mass-dependent contributions, which differ for $c_B$ and $c_E$, are
calculated at 1-loop in mean-field improved lattice perturbation
theory~\cite{Aoki:2003dg}.  The hopping parameter is fixed
nonperturbatively to reproduce the experimentally-measured
spin-averaged $1S$ charmonium mass~\cite{Namekawa:2011wt}.  One of the
anisotropy parameters ($r_t$ in Ref.~\cite{Namekawa:2011wt}) is also
set to its 1-loop perturbative value, while the other ($\nu$ in
Ref.~\cite{Namekawa:2011wt}) is fixed noperturbatively to obtain the
continuum dispersion relation for the spin-averaged charmonium $1S$
states (such that $M_1 = M_2$).  For the renormalization and
improvement coefficients of weak current operators, the contributions
in the chiral limit are obtained
nonperturbatively~\cite{Kaneko:2007wh,Aoki:2010wm}, while the
mass-dependent contributions are estimated using 1-loop lattice
perturbation theory~\cite{Aoki:2004th}.  With these choices, lattice
cutoff effects from the action and operators are of ${\cO}(\alpha_s^2 a|\vec{p}|, (a|\vec{p}_h|)^2)$.
\\


\noindent
{\it Light-quark actions combined with HQET}\\
\noindent

The heavy-quark formulations discussed in the previous sections use
effective field theory to avoid the occurence of discretization errors
of the form $(am_h)^n$.  In this section we describe methods that use
improved actions that were originally designed for light-quark systems
for $B$ physics calculations. Such actions unavoidably contain
discretization errors that grow as a power of the heavy-quark mass. In
order to use them for heavy-quark physics, they must be improved to at
least ${\cO}(am_h)^2$.  However, since $am_b > 1$ at the smallest
lattice spacings available in current simulations, these methods also
require input from HQET to guide the simulation results to the
physical $b$-quark mass.

The ETM collaboration has developed two methods, the ``ratio
method'' \cite{Blossier:2009hg} and the ``interpolation
method'' \cite{Guazzini:2006bn,Blossier:2009gd}. They use these
methods together with simulations with twisted-mass Wilson fermions,
which have discretization errors of $\cO(am_h)^2$.  In the interpolation
method $\Phi_{hs}$ and $\Phi_{h\ell}$ (or $\Phi_{hs}/\Phi_{h\ell}$)
are calculated for a range of heavy-quark masses in the charm region
and above, while roughly keeping $am_h \ltsim 0.5 $. The relativistic
results are combined with a separate calculation of the decay
constants in the static limit, and then interpolated to the physical
$b$ quark mass. In ETM's implementation of this method, the heavy
Wilson decay constants are matched to HQET using NLO in continuum
perturbation theory. The static limit result is renormalized using
1-loop mean-field improved lattice perturbation theory, while for
the relativistic data PCAC is used to calculate absolutely normalized
matrix elements. Both, the relativistic and static limit data are then
run to the common reference scale $\mu_b = 4.5 \GeV$ at NLO in
continuum perturbation theory.  In the ratio method, one constructs
physical quantities $P(m_h)$ from the relativistic data that have a
well-defined static limit ($P(m_h) \to$ const.~for $m_h \to \infty$)
and evaluates them at the heavy-quark masses used in the simulations.
Ratios of these quantities are then formed at a fixed ratio of heavy-quark masses, $z = P(m_h) / P(m_h/\lambda)$ (where $1 < \lambda \lsim
1.3$), which ensures that $z$ is equal to unity in the static limit.
Hence, a separate static limit calculation is not needed with this
method.  In ETM's implementation of the ratio method for the $B$-meson
decay constant, $P(m_h)$ is constructed from the decay constants and
the heavy-quark pole mass as $P(m_h) = f_{h\ell}(m_h) \cdot (m^{\rm
pole}_h)^{1/2}$. The corresponding $z$-ratio therefore also includes
ratios of perturbative matching factors for the pole mass to $\msbar$
conversion.  For the interpolation to the physical $b$-quark mass,
ratios of perturbative matching factors converting the data from QCD
to HQET are also included. The QCD-to-HQET matching factors improve
the approach to the static limit by removing the leading logarithmic
corrections. In ETM's implementation of this method (ETM 11 and 12)
both conversion factors are evaluated at NLO in continuum perturbation
theory. The ratios are then simply fit to a polynomial in $1/m_h$ and
interpolated to the physical $b$-quark mass.  The ratios constructed
from $f_{h\ell}$ ($f_{hs}$) are called $z$ ($z_s$).  In order to
obtain the $B$ meson decay constants, the ratios are combined with
relativistic decay constant data evaluated at the smallest reference
mass.

The HPQCD collaboration has introduced a method in
Ref.~\cite{McNeile:2011ng} which we shall refer to as the ``heavy
HISQ'' method.  The first key ingredient is the use of the HISQ action
for the heavy and light valence quarks, which has leading
discretization errors of ${\cO} \left(\alpha_s (v/c) (am_h)^2,
(v/c)^2 (am_h)^4\right)$.  With the same action for the heavy- and
light-valence quarks it is possible to use PCAC to avoid
renormalization uncertainties.  Another key ingredient at the time of formulation was the
availability of gauge ensembles over a large range of lattice
spacings, in this case the library of $N_f = 2+1$
asqtad ensembles made public by the MILC collaboration which include
lattice spacings as small as $a \approx 0.045$~fm.
Since the HISQ
action is so highly improved and with lattice spacings as small as
$0.045$~fm, HPQCD is able to use a large range of heavy-quark masses,
from below the charm region to almost up to the physical $b$-quark
mass with $am_h \ltsim 0.85$. They then fit their data in a combined
continuum and HQET fit (i.e., using a fit function that is motivated by
HQET) to a polynomial in $1/m_H$ (the heavy pseudoscalar-meson mass
of a meson containing a heavy ($h$) quark).

This approach has been extended in recent work by the HPQCD and
FNAL/MILC collaborations using the MILC-generated $N_f=2+1+1$ HISQ
ensembles with lattice spacings down to
$0.03$~fm~\cite{Bazavov:2017lyh}.  These are being used by the HPQCD
and the FNAL/MILC collaborations for their B-physics programmes and
the corresponding analyses include heavy-quark masses at the physical
$b$ quark mass.

\bigskip

In Tab.~\ref{tab_heavy_quarkactions} we list the discretizations of
the quark action most widely used for heavy $c$ and $b$ quarks
together with the abbreviations used in the summary tables.  We also
summarize the main properties of these actions and the leading lattice
discretization errors for calculations of heavy-light meson matrix
quantities with them.  Note that in order to maintain the leading
lattice artifacts of the actions as given in the table in nonspectral
observables (like operator matrix elements) the corresponding
nonspectral operators need to be improved as well.

\begin{table}
\begin{center}
{\footnotesize
\begin{tabular*}{\textwidth}[l]{l @{\extracolsep{\fill}} l l l}
\hline \hline  \\[-1.0ex]
\parbox[t]{1.5cm}{Abbrev.} & Discretization & 
\parbox[t]{4cm}{Leading lattice artifacts\\and truncation errors\\for heavy-light mesons} &  
Remarks
\\[7.0ex] \hline \hline \\[-1.0ex]
tmWil   & twisted-mass Wilson &  ${\cO}\big((am_h)^2\big)$ & \parbox[t]{4.cm}{PCAC relation for axial-vector current}  
\\[3.0ex] \hline \\[-1.0ex]
HISQ  & Staggered & \parbox[t]{4cm}{${\cO}\big (\alpha_S (am_h)^2 (v/c), \\(am_h)^4 (v/c)^2 \big)$}  & \parbox[t]{4.cm}{PCAC relation for axial-vector current; Ward identity for vector current}  
\\[6.0ex] \hline \\[-1.0ex]
static  & static effective action &  \parbox[t]{4cm}{${\cO}\big( a^2 \Lambda_{\rm QCD}^2, \Lambda_{\rm QCD}/m_h, \\ \alpha_s^2, \alpha_s^2 a \Lambda_{\rm QCD} \big)$}  & \parbox[t]{4.5cm}{implementations use APE, HYP1, and HYP2 smearing}  
\\[4.0ex] \hline \\[-1.0ex]
HQET  & Heavy-Quark Effective Theory &  \parbox[t]{4cm}{${\cO}\big( a \Lambda^2_{\rm QCD}/m_h,  a^2 \Lambda^2_{\rm QCD},\\
 (\Lambda_{\rm QCD}/m_h)^2 \big)$}  & \parbox[t]{4.5cm}{Nonperturbative matching through ${\cO}(1/m_h)$}  
\\[4.0ex] \hline \\[-1.0ex]
NRQCD  & Nonrelativistic QCD & \parbox[t]{4cm}{${\cO}\big(\alpha_S \Lambda_{\rm QCD}/m_h, \\ \alpha_S (\Lambda_{\rm QCD}/m_h)^2 , \\ (\Lambda_{\rm QCD}/m_h )^3,  \alpha_s a^2 \Lambda_{\rm QCD}^2 \big)$}  & \parbox[t]{4.5cm}{Tree-level relativistic corrections through 
${\cO}(v_h^4)$ and discretization corrections through ${\cO}(a^2)$}  
\\[9.5ex] \hline \\[-1.0ex] 
Fermilab  & Sheikholeslami-Wohlert & ${\cO}\big(\alpha_sa\Lambda_{\rm QCD}, (a\Lambda_{\rm QCD})^2\big)$  & \parbox[t]{4.5cm}{Hopping parameter tuned nonperturbatively; clover coefficient computed at tree-level in mean-field-improved lattice perturbation theory}  
\\[12.0ex] \hline \\[-1.0ex] 

RHQ       & Sheikholeslami-Wohlert & ${\cO}\big( \alpha_s^2 a\Lambda_{\rm QCD}, (a\Lambda_{\rm QCD})^2 \big)$  & \parbox[t]{4.5cm}{Hopping parameter, anisoptropy and clover coefficient tuned nonperturbatively by fixing the $B_s$-meson hyperfine splitting} \\[12.0ex] \hline \\[-1.0ex] 

Tsukuba  & Sheikholeslami-Wohlert & ${\cO}\big( \alpha_s^2 a\Lambda_{\rm QCD}, (a\Lambda_{\rm QCD})^2 \big)$  & \parbox[t]{4.5cm}{NP clover coefficient at $ma=0$ plus mass-dependent corrections calculated at 1-loop in lattice perturbation theory; $\nu$ calculated NP from dispersion relation; $r_s$ calculated at 1-loop in lattice perturbation theory}  
\\[20.0ex]
\hline\hline
\end{tabular*}
}
\caption{Discretizations of the quark action most widely used for heavy $c$ and $b$ quarks  and some of their properties.
\label{tab_heavy_quarkactions}}
\end{center}
\end{table}

\subsection{Matching and running \label{sec_match}}

The lattice formulation of QCD amounts to introducing a particular
regularization scheme. Thus, in order to be useful for phenomenology,
hadronic matrix elements computed in lattice simulations must be
related to some continuum reference scheme, such as the
$\msbar$-scheme of dimensional regularization. The matching to the
continuum scheme usually involves running to some reference scale
using the renormalization group. 

In principle, the matching factors which relate lattice matrix
elements to the $\msbar$-scheme, can be computed in perturbation
theory formulated in terms of the bare coupling. It has been known for
a long time, though, that the perturbative expansion is not under good 
control. Several techniques have been developed which allow for a
nonperturbative matching between lattice regularization and continuum
schemes, and are briefly introduced here.\\


\noindent
{\sl Regularization-independent Momentum Subtraction}\\
\noindent

In the {\sl Regularization-independent Momentum Subtraction}
(``RI/MOM'' or ``RI'') scheme \cite{Martinelli:1994ty} a
nonperturbative renormalization condition is formulated in terms of
Green functions involving quark states in a fixed gauge (usually
Landau gauge) at nonzero virtuality. In this way one relates operators
in lattice regularization nonperturbatively to the RI scheme. In a
second step one matches the operator in the RI scheme to its
counterpart in the $\msbar$-scheme. The advantage of this procedure is
that the latter relation involves perturbation theory formulated in
the continuum theory. The use of lattice perturbation theory can thus
be avoided, and the continuum perturbation theory, which is
technically more feasible for higher order calculations, could be
applied if more precision is required. A technical complication is
associated with the accessible momentum scales (i.e., virtualities),
which must be large enough (typically several $\gev$) in order for the
perturbative relation to $\msbar$ to be reliable. The momentum scales
in simulations must stay well below the cutoff scale (i.e., $2\pi$
over the lattice spacing), since otherwise large lattice artifacts are
incurred. Thus, the applicability of the RI scheme traditionally
relies on the existence of a ``window'' of momentum scales, which
satisfy
\be
   \Lambda_{\rm QCD} \;\lesssim\; p \;\lesssim\; 2\pi a^{-1}.
\ee
However, solutions for mitigating this limitation, which involve
continuum limit, nonperturbative running to higher scales in the
RI/MOM scheme, have recently been proposed and implemented
\cite{Arthur:2010ht,Durr:2010vn,Durr:2010aw,Aoki:2010pe}.

Within the RI/MOM framework one has some freedom in the choice of the
external momenta used in the Green functions. In the choice made in
the original work, the virtuality of each external leg is nonzero, but
that of the momentum transfer between different legs can
vanish~\cite{Martinelli:1994ty}.  This leads to enhanced
nonperturbative contributions that fall as powers of $p^2$.  An
alternative choice that reduces these issues is the symmetric MOM scheme, in which virtualities in all channels are
nonzero~\cite{Sturm:2009kb}.  This scheme is now widely used. To
distinguish it from the original choice of virtualities, it is
referred to as the RI/SMOM (or RI-SMOM) scheme, while the original choice is called
the RI/MON (or RI-MOM) scheme.\\

\noindent
{\it Schr\"odinger functional}\\
\noindent

Another example of a nonperturbative matching procedure is provided
by the Schr\"odinger functional (SF) scheme \cite{Luscher:1992an}. It
is based on the formulation of QCD in a finite volume. If all quark
masses are set to zero the box length remains the only scale in the
theory, such that observables like the coupling constant run with the
box size~$L$. The great advantage is that the RG running of
scale-dependent quantities can be computed nonperturbatively using
recursive finite-size scaling techniques. It is thus possible to run
nonperturbatively up to scales of, say, $100\,\gev$, where one is
sure that the perturbative relation between the SF and
$\msbar$-schemes is controlled.\\

\noindent
{\sl Perturbation theory}\\
\noindent

The third matching procedure is based on perturbation theory in which
higher order are effectively resummed \cite{Lepage:1992xa}. Although
this procedure is easier to implement, it is hard to estimate the
uncertainty associated with it.\\

\noindent
{\sl Mostly nonperturbative renormalization}\\
\noindent

Some calculations of heavy-light and heavy-heavy matrix elements adopt a mostly nonperturbative matching approach.  Let us consider a weak 
decay process mediated by a current with quark flavours $h$ and $q$, where $h$ is the initial heavy quark (either bottom or charm) and 
$q$ can be a light ($\ell = u,d$), strange, or charm quark. The matrix elements of lattice current  $J_{hq}$ are matched to the 
corresponding continuum matrix elements with continuum current ${\cal J}_{hq}$ by calculating the renormalization factor $Z_{J_{hq}}$. 
The mostly nonperturbative renormalization method takes advantage of rewriting the current renormalization factor as the following product:
\begin{align}
Z_{J_{hq}} = \rho_{J_{hq}} \sqrt{Z_{V^4_{hh}}Z_{V^4_{qq}}} \,
\label{eq:Zvbl}
\end{align}
The flavour-conserving renormalization factors $Z_{V^4_{hh}}$ and $Z_{V^4_{qq}}$ can be obtained nonperturbatively from standard heavy-light 
and light-light meson charge normalization conditions.  $Z_{V^4_{hh}}$ and $Z_{V^4_{qq}}$ account for  the bulk of the renormalization. The remaining 
correction $\rho_{J_{hq}}$ is expected to be close to unity because most of the radiative corrections, including self-energy corrections and 
contributions from tadpole graphs, cancel in the ratio~\cite{Harada:2001fj,Harada:2001fi}.  The 1-loop coefficients of $\rho_{J_{hq}}$  have been calculated for
heavy-light and heavy-heavy currents for Fermilab heavy and both (improved) Wilson light \cite{Harada:2001fj,Harada:2001fi} and 
asqtad light  \cite{ElKhadra:2007qe} quarks. In all cases the 1-loop coefficients are found to be very small, yielding sub-percent to few percent level corrections.

\bigskip
\noindent
In Tab.~\ref{tab_match} we list the abbreviations used in the
compilation of results together with a short description.

\begin{table}[ht]
{\footnotesize
\begin{tabular*}{\textwidth}[l]{l @{\extracolsep{\fill}} l}
\hline \hline \\[-1.0ex]
Abbrev. & Description
\\[1.0ex] \hline \hline \\[-1.0ex]
RI  &  regularization-independent momentum subtraction scheme 
\\[1.0ex] \hline \\[-1.0ex]
SF  &  Schr\"odinger functional scheme
\\[1.0ex] \hline \\[-1.0ex]
PT1$\ell$ & matching/running computed in perturbation theory at one loop
\\[1.0ex] \hline \\[-1.0ex]
PT2$\ell$ & matching/running computed in perturbation theory at two loops 
\\[1.0ex] \hline \\[-1.0ex]
mNPR & mostly nonperturbative renormalization 
%
\\[1.0ex]
\hline\hline
\end{tabular*}
}
\caption{The most widely used matching and running
  techniques. \label{tab_match} 
}
\end{table}

\subsection{Chiral extrapolation\label{sec_ChiPT}}
As mentioned in the introduction, Symanzik's framework can be combined 
with Chiral Perturbation Theory. The well-known terms occurring in the
chiral effective Lagrangian are then supplemented by contributions 
proportional to powers of the lattice spacing $a$. The additional terms are 
constrained by the symmetries of the lattice action and therefore 
depend on the specific choice of the discretization. 
The resulting effective theory can be used to analyze the $a$-dependence of 
the various quantities of interest -- provided the quark masses and the momenta
considered are in the range where the truncated chiral perturbation series yields 
an adequate approximation. Understanding the dependence on the lattice spacing 
is of central importance for a controlled extrapolation to the continuum limit.
 
For staggered fermions, this program has first been carried out for a
single staggered flavour (a single staggered field) \cite{Lee:1999zxa}
at $\cO(a^2)$. In the following, this effective theory is denoted by
S{\Ch}PT. It was later generalized to an arbitrary number of flavours
\cite{Aubin:2003mg,Aubin:2003uc}, and to next-to-leading order
\cite{Sharpe:2004is}. The corresponding theory is commonly called
Rooted Staggered chiral perturbation theory and is denoted by
RS{\Ch}PT.

For Wilson fermions, the effective theory has been developed in
\cite{Sharpe:1998xm,Rupak:2002sm,Aoki:2003yv}
and is called W{\Ch}PT, while the theory for Wilson twisted-mass
fermions \cite{Sharpe:2004ny,Aoki:2004ta,Bar:2010jk} is termed tmW{\Ch}PT.

Another important approach is to consider theories in which the
valence and sea quark masses are chosen to be different. These
theories are called {\it partially quenched}. The acronym for the
corresponding chiral effective theory is PQ{\Ch}PT
\cite{Bernard:1993sv,Golterman:1997st,Sharpe:1997by,Sharpe:2000bc}.

Finally, one can also consider theories where the fermion
discretizations used for the sea and the valence quarks are different. The
effective chiral theories for these ``mixed action'' theories are
referred to as MA{\Ch}PT  \cite{Bar:2002nr,Bar:2003mh,Bar:2005tu,Golterman:2005xa,Chen:2006wf,Chen:2007ug,Chen:2009su}. \\

\noindent
{\sl Finite-Volume Regimes of QCD}\\
\noindent

Once QCD with $\Nf$ nondegenerate flavours is regulated both in the UV and
in the IR, there are $3+\Nf$ scales in play: The scale
$\Lambda_\mathrm{QCD}$ that reflects ``dimensional transmutation''
(alternatively, one could use the pion decay constant or the nucleon mass,
in the chiral limit), the inverse lattice spacing $1/a$, the inverse box
size $1/L$, as well as $\Nf$ meson masses (or functions of meson masses)
that are sensitive to the $\Nf$ quark masses, e.g., $\Mpi^2$,
$2\Mka^2-\Mpi^2$ and the spin-averaged masses of ${}^1S$ states of
quarkonia.

Ultimately, we are interested in results with the two regulators
removed, i.e., physical quantities for which the limits $a \to 0$ and
$L \to \infty$ have been carried out. In both cases there is an
effective field theory (EFT) which guides the extrapolation.  For the
$a \to 0$ limit, this is a version of the Symanzik EFT which depends,
in its details, on the lattice action that is used, as outlined in
Sec.~\ref{sec_lattice_actions}.  The finite-volume effects are
dominated by the lightest particles, the pions.  Therefore, a chiral
EFT, also known as {\Ch}PT, is appropriate to parameterize the
finite-volume effects, i.e., the deviation of masses and other
observables, such as matrix elements, in a finite-volume from their
infinite volume, physical values.  Most simulations of
phenomenological interest are carried out in boxes of size $L \gg
1/M_\pi$, that is in boxes whose diameter is large compared to the
Compton wavelength that the pion would have, at the given quark mass,
in infinite volume. In this situation the finite-volume corrections
are small, and in many cases the ratio $M_\mathrm{had}(L) /
M_\mathrm{had}$ or $f(L) / f$, where $f$ denotes some generic matrix
element, can be calculated in {\Ch}PT, such that the leading
finite-volume effects can be taken out analytically. In the
terminology of {\Ch}PT this setting is referred to as the $p$-regime,
as the typical contributing momenta $p \sim M_\pi \gg 1/L$.  A
peculiar situation occurs if the condition $L \gg 1/\Mpi$ is violated
(while $L\Lambda_\mathrm{QCD}\gg1$ still holds), in other words if the
quark mass is taken so light that the Compton wavelength that the pion
would have (at the given $m_q$) in infinite volume, is as large or
even larger than the actual box size. Then the pion zero-momentum mode
dominates and needs to be treated separately. While this setup is
unlikely to be useful for standard phenomenological computations, the
low-energy constants of {\Ch}PT can still be calculated, by matching
to a re-ordered version of the chiral series, and following the
details of the reordering such an extreme regime is called the
$\epsilon$- or $\delta$-regime, respectively.  Accordingly, further
particulars of these regimes are discussed in Sec.~\ref{sec:chPT} of this report.

\subsection{Parameterizations of semileptonic form factors}\label{sec:zparam}

In this section, we discuss the description of the $q^2$-dependence of
form factors, using the vector form factor $f_+$ of $B\to\pi\ell\nu$ decays
as a benchmark case. Since in this channel the parameterization of the
$q^2$-dependence is crucial for the extraction of $|V_{ub}|$ from the existing
measurements (involving decays to light leptons), as explained
above, it has been studied in great detail in the literature. Some comments
about the generalization of the techniques involved will follow.

\paragraph{The vector form factor for $B\to\pi\ell\nu$}

All form factors are analytic functions of $q^2$ outside physical
poles and inelastic threshold branch points; in the case of
$B\to\pi\ell\nu$, the only pole expected below the $B\pi$ production
region, starting at $q^2 = t_+ = (m_B+m_\pi)^2$, is the $B^*$.  A
simple ansatz for the $q^2$-dependence of the $B\to\pi\ell\nu$
semileptonic form factors that incorporates vector-meson dominance is
the Be\v{c}irevi\'c-Kaidalov (BK)
parameterization~\cite{Becirevic:1999kt},
which for the vector form factor reads:
\begin{gather}
f_+(q^2) = \frac{f(0)}{(1-q^2/m_{B^*}^2)(1-\alpha q^2/m_{B^*}^2)}\,.
\label{eq:BKparam}
\end{gather}
Because the BK ansatz has few free parameters, it has been used
extensively to parameterize the shape of experimental
branching-fraction measurements and theoretical form-factor
calculations.  A variant of this parameterization proposed by Ball and
Zwicky (BZ) adds extra pole factors to the expressions in
Eq.~(\ref{eq:BKparam}) in order to mimic the effect of multiparticle
states~\cite{Ball:2004ye}. A similar idea, extending the use of effective
poles also to $D\to\pi\ell\nu$ decays, is explored in Ref.~\cite{Becirevic:2014kaa}.
Finally, yet another variant (RH) has been proposed by
Hill in Ref.~\cite{Hill:2005ju}. Although all of these parameterizations
capture some known properties of form factors, they do not manifestly
satisfy others.  For example,
perturbative QCD scaling constrains the
behaviour of $f_+$ in the deep Euclidean region~\cite{Lepage:1980fj,Akhoury:1993uw,Lellouch:1995yv}, and
angular momentum conservation constrains the asymptotic behaviour near
thresholds---e.g., ${\rm Im}\,f_+(q^2) \sim (q^2-t_+)^{3/2}$ (see, e.g., Ref.~\cite{Bourrely:2008za}).  Most importantly, these parameterizations do not allow for an easy
quantification of systematic uncertainties.

A more systematic approach that improves upon the use of simple models
for the $q^2$ behaviour exploits the positivity and analyticity
properties of two-point functions of vector currents to obtain optimal
parameterizations of form
factors~\cite{Bourrely:1980gp,Boyd:1994tt,Lellouch:1995yv,Boyd:1997kz,Boyd:1997qw,Arnesen:2005ez,Becher:2005bg}.
Any form factor $f$ can be shown to admit a series expansion of the
form
\begin{gather}
f(q^2) = \frac{1}{B(q^2)\phi(q^2,t_0)}\,\sum_{n=0}^\infty a_n(t_0)\,z(q^2,t_0)^n\,,
\end{gather}
where the squared momentum transfer is replaced by the variable
\begin{gather}
z(q^2,t_0) = \frac{\sqrt{t_+-q^2}-\sqrt{t_+-t_0}}{\sqrt{t_+-q^2}+\sqrt{t_+-t_0}}\,.
\end{gather}
This is a conformal transformation, depending on an arbitrary real
parameter $t_0<t_+$, that maps the $q^2$ plane cut for $q^2 \geq t_+$
onto the disk $|z(q^2,t_0)|<1$ in the $z$ complex plane. The function
$B(q^2)$ is called the {\it Blaschke factor}, and contains poles and
cuts below $t_+$ --- for instance, in the case of $B\to\pi$ decays,
\begin{gather}
B(q^2)=\frac{z(q^2,t_0)-z(m_{B^*}^2,t_0)}{1-z(q^2,t_0)z(m_{B^*}^2,t_0)}=z(q^2,m_{B^*}^2)\,.
\end{gather}
Finally, the quantity $\phi(q^2,t_0)$, called the {\em outer
function}, is some otherwise arbitrary function that does not introduce further
poles or branch cuts.  The crucial property of this series expansion
is that the sum of the squares of the coefficients
\begin{gather}
\sum_{n=0}^\infty a_n^{2} = \frac{1}{2\pi i}\oint \frac{dz}{z}\,|B(z)\phi(z)f(z)|^2\,,
\end{gather}
is a finite quantity. Therefore, by using this parameterization an
absolute bound to the uncertainty induced by truncating the series can
be obtained.  The aim in choosing $\phi$ is to obtain
a bound that is useful in practice, while
(ideally) preserving the correct behaviour of the form factor at high
$q^2$ and around thresholds.

The simplest form of the bound would correspond to $\sum_{n=0}^\infty
a_n^{2}=1$.  {\it Imposing} this bound yields the following ``standard''
choice for the outer function
\begin{gather}
\label{eq:comp_of}
\begin{split}
\phi(q^2,t_0)=&\sqrt{\frac{1}{32\pi\chi_{1^-}(0)}}\,
\left(\sqrt{t_+-q^2}+\sqrt{t_+-t_0}\right)\\
&\times\,\left(\sqrt{t_+-q^2}+\sqrt{t_+-t_-}\right)^{3/2}
\left(\sqrt{t_+-q^2}+\sqrt{t_+}\right)^{-5}
\,\frac{t_+-q^2}{(t_+-t_0)^{1/4}}\,,
\end{split}
\end{gather}
where $t_-=(m_B-m_\pi)^2$, and $\chi_{1^-}(0)$ is the derivative of the transverse component of
the polarization function (i.e., the Fourier transform of the vector
two-point function) $\Pi_{\mu\nu}(q)$ at Euclidean momentum
$Q^2=-q^2=0$. It is computed perturbatively, using operator product
expansion techniques, by relating the $B\to\pi\ell\nu$ decay amplitude
to $\ell\nu\to B\pi$ inelastic scattering via crossing symmetry and
reproducing the correct value of the inclusive $\ell\nu\to X_b$ amplitude.
We will refer to the series parameterization with the outer function
in Eq.~(\ref{eq:comp_of}) as Boyd, Grinstein, and Lebed (BGL).  The
perturbative and OPE truncations imply that the bound is not strict,
and one should take it as
\begin{gather}
\sum_{n=0}^N a_n^{2} \lesssim 1\,,
\end{gather}
where this holds for any choice of $N$.  Since the values of $|z|$ in
the kinematical region of interest are well below~1 for judicious
choices of $t_0$, this provides a very stringent bound on systematic
uncertainties related to truncation for $N\geq 2$. On the other hand,
the outer function in Eq.~(\ref{eq:comp_of}) is somewhat unwieldy and,
more relevantly, spoils the correct large $q^2$ behaviour and induces
an unphysical singularity at the $B\pi$ threshold.

A simpler choice of outer function has been proposed by Bourrely,
Caprini and Lellouch (BCL) in Ref.~\cite{Bourrely:2008za}, which leads to a
parameterization of the form
\begin{gather}
\label{eq:bcl}
f_+(q^2)=\frac{1}{1-q^2/m_{B^*}^2}\,\sum_{n=0}^N a_n^{+}(t_0) z(q^2,t_0)^n\,.
\end{gather}
This satisfies all the basic properties of the form factor, at the price
of changing the expression for the bound to
\begin{gather}
\sum_{j,k=0}^N B_{jk}(t_0)a_j^{+}(t_0)a_k^{+}(t_0) \leq 1\,.
\end{gather}
The constants $B_{jk}$ can be computed and shown to be
$|B_{jk}|\lesssim \cO(10^{-2})$ for judicious choices of
$t_0$; therefore, one again finds that truncating at $N\geq 2$
provides sufficiently stringent bounds for the current level of
experimental and theoretical precision.  It is actually possible to
optimize the properties of the expansion by taking
\begin{gather}
t_0 = t_{\rm opt} = (m_B+m_\pi)(\sqrt{m_B}-\sqrt{m_\pi})^2\,,
\end{gather}
which for physical values of the masses results in the semileptonic
domain being mapped onto the symmetric interval $|z| \ltapprox 0.279$
(where this range differs slightly for the $B^{\pm}$ and $B^0$ decay
channels), minimizing the maximum truncation error.  If one also
imposes that the asymptotic behaviour ${\rm Im}\,f_+(q^2) \sim
(q^2-t_+)^{3/2}$ near threshold is satisfied, then the highest-order
coefficient is further constrained as
\begin{gather}
\label{eq:red_coeff}
a_N^{+}=-\,\frac{(-1)^N}{N}\,\sum_{n=0}^{N-1}(-1)^n\,n\,a_n^{+}\,.
\end{gather}
Substituting the above constraint on $a_N^{+}$ into Eq.~(\ref{eq:bcl})
leads to the constrained BCL parameterization
\begin{gather}
\label{eq:bcl_c}
f_+(q^2)=\frac{1}{1-q^2/m_{B^*}^2}\,\sum_{n=0}^{N-1} a_n^{+}\left[z^n-(-1)^{n-N}\,\frac{n}{N}\,z^N\right]\,,
\end{gather}
which is the standard implementation of the BCL parameterization used
in the literature.

Parameterizations of the BGL and BCL kind, to which we will refer
collectively as ``$z$-parameterizations'', have already been adopted
by the BaBar and Belle collaborations to report their results, and
also by the Heavy Flavour Averaging Group (HFAG, later renamed HFLAV).
Some lattice
collaborations, such as FNAL/MILC and ALPHA, have already started to
report their results for form factors in this way.  The emerging trend
is to use the BCL parameterization as a standard way of presenting
results for the $q^2$-dependence of semileptonic form factors. Our
policy will be to quote results for $z$-parameterizations when the
latter are provided in the paper (including the covariance matrix of
the fits); when this is not the case, but the published form factors
include the full correlation matrix for values at different $q^2$, we
will perform our own fit to the constrained BCL ansatz
in Eq.~(\ref{eq:bcl_c}); otherwise no fit will be quoted.
We however stress the importance of providing, apart from parameterization
coefficients, values for the form factors themselves (in the continuum limit
and at physical quark masses) for a number of values of $q^2$, so that
the results can be independently parameterized by the readers if so wished.

\paragraph{The scalar form factor for $B\to\pi\ell\nu$}

The discussion of the scalar $B\to \pi$ form factor is very similar. The main differences are the absence of a constraint analogue to Eq.~(\ref{eq:red_coeff}) and the choice of the overall pole function. In our fits we adopt the simple expansion:
\begin{gather}
\label{eq:bcl_f0}
f_0 (q^2) = \sum_{n=0}^{N-1} a_n^0 \; z^n \, .
\end{gather}
We do impose the exact kinematical constraint $f_+ (0) = f_0 (0)$ by expressing the $a_{N-1}^0$ coefficient in terms of all remaining $a_n^+$ and $a_n^0$ coefficients. This constraint introduces important correlations between the $a_n^+$ and $a_n^0$ coefficients; thus only lattice calculations that present the correlations between the vector and scalar form factors can be used in an average that takes into account the constraint at $q^2 = 0$. 

Finally we point out that we do not need to use the same number of parameters for the vector and scalar form factors. For instance, with $(N^+ = 3, N^0 = 3)$ we have $a_{0,1,2}^+$ and $a_{0,1}^0$, while with $(N^+ = 3, N^0 = 4)$ we have $a_{0,1,2}^+$ and $a_{0,1,2}^0$ as independent fit parameters. In our average we will choose the combination that optimizes uncertainties.

\paragraph{Extension to other form factors}

The discussion above largely extends to form factors for other semileptonic transitions (e.g., $B_s\to K$ and $B_{(s)} \to D^{(*)}_{(s)}$,
and semileptonic $D$ and $K$ decays).
Details are discussed in the relevant sections.

A general discussion of semileptonic meson decay in this context can be found,
e.g., in Ref.~\cite{Hill:2006ub}. Extending what has been discussed above for
$B\to\pi$, the form factors for a generic $H \to L$
transition will display a cut starting at the production threshold $t_+$, and the optimal
value of $t_0$ required in $z$-parameterizations is $t_0=t_+(1-\sqrt{1-t_-/t_+})$
(where $t_\pm=(m_H\pm m_L)^2$).
For unitarity bounds to apply, the Blaschke factor has to include all sub-threshold
poles with the quantum numbers of the hadronic current --- i.e., vector (resp. scalar) resonances
in $B\pi$ scattering for the vector (resp. scalar) form factors of $B\to\pi$, $B_s\to K$,
or $\Lambda_b \to p$; and vector (resp. scalar) resonances
in $B_c\pi$ scattering for the vector (resp. scalar) form factors of $B\to D$
or $\Lambda_b \to \Lambda_c$.\footnote{A more complicated analytic structure
may arise in other cases, such as channels with vector mesons in the final state.
We will however not discuss form-factor parameterizations for any such process.}
Thus, as emphasized above, the control over systematic uncertainties brought in by using
$z$-parameterizations strongly depends on implementation details.
This has practical consequences, in particular, when the resonance spectrum
in a given channel is not sufficiently well-known. Caveats may also
apply for channels where resonances with a nonnegligible width appear.
A further issue is whether $t_+=(m_H+m_L)^2$ is the proper choice for the start of the cut in cases such as $B_s\to K\ell\nu$ and $B\to D\ell\nu$, where there are lighter two-particle states that project on the current ($B$,$\pi$ and $B_c$,$\pi$ for the two processes, respectively).\footnote{We are grateful
to G.~Herdo\'{\i}za, R.J.~Hill, A.~Kronfeld and A.~Szczepaniak for illuminating discussions
on this issue.}
In any such
situation, it is not clear a priori that a given $z$-parameterization will
satisfy strict bounds, as has been seen, e.g., in determinations of the proton charge radius
from electron-proton scattering~\cite{Hill:2010yb,Hill:2011wy,Epstein:2014zua}.

The HPQCD collaboration pioneered a variation on the $z$-parameterization
approach, which they refer to as a ``modified $z$-expansion,'' that
is used to simultaneously extrapolate their lattice simulation data
to the physical light-quark masses and the continuum limit, and to
interpolate/extrapolate their lattice data in $q^2$.  This entails
allowing the coefficients $a_n$ to depend on the light-quark masses,
squared lattice spacing, and, in some cases the charm-quark mass and
pion or kaon energy.  Because the modified $z$-expansion is not
derived from an underlying effective field theory, there are several
potential concerns with this approach that have yet to be studied.
The most significant is that there is no theoretical
derivation relating the coefficients of the modified $z$-expansion to
those of the physical coefficients measured in experiment; it
therefore introduces an unquantified model dependence in the
form-factor shape. As a result, the applicability of unitarity bounds has to be examined carefully.
Related to this, $z$-parameterization coefficients implicitly depend on quark masses,
and particular care should be taken in the event that some state can move
across the inelastic threshold as quark masses are changed (which would
in turn also affect the form of the Blaschke factor). Also, the lattice-spacing dependence of form factors provided by Symanzik effective theory
techniques may not extend trivially to $z$-parameterization coefficients.
The modified $z$-expansion is now being utilized by collaborations
other than HPQCD and for quantities other than $D \to \pi \ell \nu$
and $D \to K \ell \nu$, where it was originally employed.
We advise treating results that utilize the modified $z$-expansion to
obtain form-factor shapes and CKM matrix elements with caution,
however, since the systematics of this approach warrant further study.

\ifx\nosimulatedlatticeactiontables\undefined

\subsection{Summary of simulated lattice actions}
In the following Tabs.~\ref{tab:simulated Nf2 actions}--\ref{tab:simulated Nf4 bc actions}
we summarize the gauge and quark actions used
in the various calculations with $N_f=2, 2+1$ and $2+1+1$ quark
flavours. The calculations with $N_f=0$ quark flavours mentioned in
Sec.~\ref{sec:alpha_s} all used the Wilson gauge action and are not
listed. Abbreviations are explained in Secs.~\ref{sec_gauge_actions}, \ref{sec_quark_actions} and
\ref{app:HQactions}, and summarized in Tabs.~\ref{tab_gaugeactions},
\ref{tab_quarkactions} and \ref{tab_heavy_quarkactions}.

%
\hspace{-1.5cm}
\begin{table}[h]
{\footnotesize
\begin{tabular*}{\textwidth}[l]{l @{\extracolsep{\fill}} c c c c}
\hline \hline \\[-1.0ex]
Collab. & Ref. & $\Nf$ & \parbox{1cm}{gauge\\action} & \parbox{1cm}{quark\\action} 
\\[2.0ex] \hline \hline \\[-1.0ex]
ALPHA 01A, 04, 05, 12, 13A & \cite{Bode:2001jv,DellaMorte:2004bc,DellaMorte:2005kg,Fritzsch:2012wq,Lottini:2013rfa} & 2 & Wilson & npSW \\
[2.0ex] \hline \\[-1.0ex]
Aoki 94 & \cite{Aoki:1994pc} & 2 & Wilson &  KS \\
[2.0ex] \hline \\[-1.0ex]
Bernardoni 10 & \cite{Bernardoni:2010nf} & 2 & Wilson & npSW ${}^\dagger$ \\
[2.0ex] \hline \\[-1.0ex]
Bernardoni 11 & \cite{Bernardoni:2011kd} & 2 & Wilson & npSW \\
[2.0ex] \hline \\[-1.0ex]
Brandt 13 & \cite{Brandt:2013dua} & 2 & Wilson & npSW \\
[2.0ex] \hline \\[-1.0ex]
Boucaud 01B & \cite{Boucaud:2001qz} & 2 & Wilson &  Wilson \\
[2.0ex] \hline \\[-1.0ex]
CERN-TOV 06 & \cite{DelDebbio:2006cn} & 2 & Wilson & Wilson/npSW \\
[2.0ex] \hline \\[-1.0ex]
CERN 08 & \cite{Giusti:2008vb} & 2 & Wilson & npSW \\
[2.0ex] \hline \\[-1.0ex]
{CP-PACS 01, 04} & \cite{AliKhan:2001tx,Takeda:2004xha} & 2 & Iwasaki & mfSW  \\
[2.0ex] \hline \\[-1.0ex]
Davies 94  & \cite{Davies:1994ei} & 2 & Wilson  & KS \\
[2.0ex] \hline \\[-1.0ex]
D\"urr 11 & \cite{Durr:2011ed} & 2 & Wilson & npSW \\
[2.0ex] \hline \\[-1.0ex]
Engel 14 & \cite{Engel:2014eea} & 2 & Wilson & npSW \\
[2.0ex] 
\hline\hline \\
\end{tabular*}\\[-0.2cm]
\begin{minipage}{\linewidth}
{\footnotesize 
\begin{itemize}
   \item[${}^\dagger$] The calculation uses overlap fermions in the valence-quark sector.\\[-5mm]
\end{itemize}
}
\end{minipage}
}
\caption{Summary of simulated lattice actions with $\Nf=2$ quark
  flavours.
\label{tab:simulated Nf2 actions}}
\end{table}

\begin{table}[h]
\addtocounter{table}{-1}
{\footnotesize
\begin{tabular*}{\textwidth}[l]{l @{\extracolsep{\fill}} c c c c}
\hline \hline \\[-1.0ex]
Collab. & Ref. & $\Nf$ & \parbox{1cm}{gauge\\action} & \parbox{1cm}{quark\\action} 
\\[2.0ex] \hline \hline \\[-1.0ex]
\parbox[t]{4.0cm}{ETM 07, 07A, 08, 09, 09A-D, 09G 10B, 10D, 10F, 11C, 12, 13, 13A} & \parbox[t]{2.5cm}{
\cite{Blossier:2007vv,Boucaud:2007uk,Frezzotti:2008dr,Blossier:2009bx,Lubicz:2009ht,Jansen:2009tt,Baron:2009wt,Blossier:2009hg,Feng:2009ij,Blossier:2010cr,Lubicz:2010bv,Blossier:2010ky,Jansen:2011vv,Burger:2012ti,Cichy:2013gja,Herdoiza:2013sla}} & 2 &  tlSym & tmWil \\
[13.0ex] \hline \\[-1.0ex]
{ETM 10A, 12D} & \cite{Constantinou:2010qv,Bertone:2012cu} & 2 &  tlSym & tmWil ${}^*$ \\
[2.0ex] \hline \\[-1.0ex]
\parbox[t]{4.0cm}{ETM 14D, 15A, 16C} & \parbox[t]{2.5cm}{\cite{Abdel-Rehim:2014nka,Abdel-Rehim:2015pwa,Liu:2016cba}} & 2 &  Iwasaki & tmWil with npSW \\
[2.0ex] \hline \\[-1.0ex]
\parbox[t]{4.0cm}{ETM 15D, 16A, 17, 17B, 17C} & \parbox[t]{2.5cm}{\cite{Abdel-Rehim:2015owa,Abdel-Rehim:2016won,Alexandrou:2017hac,Alexandrou:2017oeh,Alexandrou:2017qyt}} & 2 &  Iwasaki & tmWil with npSW ${}^*$\\
[4.0ex] \hline \\[-1.0ex]
G\"ulpers 13, 15 & \cite{Gulpers:2013uca,Gulpers:2015bba} & 2 & Wilson & npSW \\
[2.0ex] \hline \\[-1.0ex]
Hasenfratz 08 & \cite{Hasenfratz:2008ce} & 2 & tadSym &
n-HYP tlSW\\
[2.0ex] \hline \\[-1.0ex]
JLQCD 08, 08B & \cite{Aoki:2008ss,Ohki:2008ff} & 2 & Iwasaki & overlap \\
[2.0ex] \hline \\[-1.0ex]
{JLQCD 02, 05} & \cite{Aoki:2002uc,Tsutsui:2005cj} & 2 & Wilson & npSW \\
[2.0ex] \hline \\[-1.0ex]
JLQCD/TWQCD 07, 08A, 08C, 10 & \cite{Fukaya:2007pn,Noaki:2008iy,Shintani:2008ga,Fukaya:2010na} & 2 & Iwasaki & overlap \\
[2.0ex] \hline \\[-1.0ex]
Mainz 12, 17 & \cite{Capitani:2012gj,Capitani:2017qpc} & 2 & Wilson & npSW \\
[2.0ex] \hline \\[-1.0ex]
QCDSF 06, 07, 12, 13 & \cite{Khan:2006de,Brommel:2007wn,Bali:2012qs,Horsley:2013ayv} & 2 & Wilson &  npSW \\
[2.0ex] \hline \\[-1.0ex]
QCDSF/UKQCD 04, 05, 06, 06A, 07 & \parbox[t]{2.5cm}{\cite{Gockeler:2004rp,Gockeler:2005rv,Gockeler:2006jt,Brommel:2006ww,QCDSFUKQCD}} & 2 & Wilson &  npSW \\
[5.0ex] \hline \\[-1.0ex]
{RBC 04, 06, 07, 08} & \cite{Aoki:2004ht,Dawson:2006qc,Blum:2007cy,Lin:2008uz} & 2 & DBW2 & DW \\
[2.0ex] \hline \\[-1.0ex]
RBC/UKQCD 07 & \cite{Boyle:2007qe} & 2 & Wilson & npSW \\ 
[2.0ex]\hline \\ [-1.0ex]
RM123 11, 13 & \cite{deDivitiis:2011eh,deDivitiis:2013xla}& 2 &  tlSym & tmWil  \\
[2.0ex] \hline \\[-1.0ex]
RQCD 14, 16 & \cite{Bali:2014nma,Bali:2016lvx} & 2 & Wilson & npSW \\
[2.0ex] \hline \\[-1.0ex]
SESAM 99 & \cite{Spitz:1999tu} & 2 & Wilson &  Wilson \\
[2.0ex] 
\hline\hline \\
\end{tabular*}\\[-0.2cm]
\begin{minipage}{\linewidth}
{\footnotesize 
\begin{itemize}
   \item[${}^*$] The calculation uses Osterwalder-Seiler fermions \cite{Osterwalder:1977pc} in the valence quark sector to treat strange and
 charm quarks.
\end{itemize}
}
\end{minipage}
}
\caption{(cntd.) Summary of simulated lattice actions with $\Nf=2$ quark
  flavours.}
\end{table}

\begin{table}[h]
\addtocounter{table}{-1}
{\footnotesize
\begin{tabular*}{\textwidth}[l]{l @{\extracolsep{\fill}} c c c c}
\hline \hline \\[-1.0ex]
Collab. & Ref. & $\Nf$ & \parbox{1cm}{gauge\\action} & \parbox{1cm}{quark\\action} 
\\[2.0ex] \hline \hline \\[-1.0ex]
Sternbeck 10, 12 & \cite{Sternbeck:2010xu,Sternbeck:2012qs}& 2 &  Wilson & npSW  \\
[2.0ex] \hline \\[-1.0ex]
{SPQcdR 05} & \cite{Becirevic:2005ta} & 2 & Wilson & Wilson \\
[2.0ex] \hline \\ [-1.0ex]
TWQCD 11, 11A  & \cite{Chiu:2011bm,Chiu:2011dz} & 2 & Wilson & optimal DW \\
[2.0ex] \hline \\ [-1.0ex]
UKQCD 04 & \cite{Flynn:2004au,Boyle:2007qe} & 2 & Wilson & npSW \\ 
[2.0ex]\hline \\ [-1.0ex]
Wingate 95 & \cite{Wingate:1995fd} & 2 & Wilson &  KS \\
[2.0ex] 
\hline\hline \\
\end{tabular*}\\[-0.2cm]
}
\caption{(cntd.) Summary of simulated lattice actions with $\Nf=2$ quark
  flavours.}
\end{table}
%
%
\begin{table}[h]
{\footnotesize
\begin{tabular*}{\textwidth}[l]{l @{\extracolsep{\fill}} c c c c}
\hline \hline \\[-1.0ex]
Collab. & Ref. & $\Nf$ & \parbox{1cm}{gauge\\action} & \parbox{1cm}{quark\\action} 
\\[2.0ex] \hline \hline \\[-1.0ex]
ALPHA 17 & \cite{Bruno:2017gxd} & $2+1$ & tlSym/Wilson & npSW\\
[2.0ex] \hline \\[-1.0ex]
Aubin 08, 09 & \cite{Aubin:2008ie,Aubin:2009jh} & $2+1$ & tadSym & Asqtad ${}^\dagger$\\
[2.0ex] \hline \\[-1.0ex]
Bazavov 12, 14 & \cite{Bazavov:2012ka,Bazavov:2014soa} & $2+1$ & tlSym & HISQ \\
[2.0ex] \hline \\[-1.0ex] 
Blum 10 & \cite{Blum:2010ym} & $2+1$ & Iwasaki & DW\\
[2.0ex] \hline \\[-1.0ex]
%
%
\parbox[t]{4.0cm}{BMW 10A-C, 11, 13, 15, 16, 16A} & \parbox[t]{2.5cm}{\cite{Durr:2010vn,Durr:2010aw,Portelli:2010yn,Durr:2011ap,Durr:2013goa,Durr:2015dna,Durr:2016ulb,Fodor:2016bgu}} & $2+1$ &  tlSym & 2-level HEX tlSW \\
[4.0ex] \hline \\[-1.0ex]
BMW 10, 11A & \cite{Durr:2010hr,Durr:2011mp} & $2+1$ & tlSym & 6-level stout tlSW\\
[2.0ex] \hline \\[-1.0ex]
Boyle 14 & \cite{Boyle:2014pja} & $2+1$ & \parbox[t]{1.5cm}{Iwasaki, Iwasaki+DSDR${}^*$} &  DW \\
[4.0ex] \hline \\[-1.0ex]
$\chi$QCD 13A, 15& \cite{Gong:2013vja,Gong:2015iir} & $2+1$ & Iwasaki & DW ${}^+$\\
[2.0ex] \hline \\[-1.0ex]
$\chi$QCD 15A& \cite{Yang:2015uis} & $2+1$ & Iwasaki & M-DW ${}^+$\\
[2.0ex] \hline \\[-1.0ex]
$\chi$QCD 18 & \cite{Liang:2018pis} & $2+1$ & Iwasaki & DW, M-DW ${}^+$ \\
[2.0ex] \hline \\[-1.0ex]
CP-PACS/JLQCD 07&  \cite{Ishikawa:2007nn} & $2+1$ & Iwasaki & npSW \\
[2.0ex] \hline \hline\\
\end{tabular*}\\[-0.2cm]
\begin{minipage}{\linewidth}
{\footnotesize 
\begin{itemize}
\item[${}^\dagger$] The calculation uses domain wall fermions in
  the valence-quark sector.\\[-5mm]
\item[${}^*$] An additional
weighting factor known as the dislocation suppressing determinant ratio (DSDR) is added to the gauge action \cite{Arthur:2012opa}.\\[-5mm]
\item[${}^+$] The calculation uses overlap fermions in the
  valence-quark sector.\\[-5mm]
\end{itemize}
}
\end{minipage}
}
\caption{Summary of simulated lattice actions with $\Nf=2+1$ or $\Nf=3$ quark flavours.
\label{tab:simulated Nf3 actions}}
\end{table}

\begin{table}[h]
\addtocounter{table}{-1}
{\footnotesize
\begin{tabular*}{\textwidth}[l]{l @{\extracolsep{\fill}} c c c c}
\hline \hline \\[-1.0ex]
Collab. & Ref. & $\Nf$ & \parbox{1cm}{gauge\\action} & \parbox{1cm}{quark\\action} 
\\[2.0ex] \hline \hline \\[-1.0ex]
Engelhardt 12 & \cite{Engelhardt:2012gd} & $2+1$ & tadSym & Asqtad $^\dagger$ \\
[2.0ex] \hline \\[-1.0ex]
FNAL/MILC 12, 12I  & \cite{Bazavov:2012zs,Bazavov:2012cd} & $2+1$ & tadSym & Asqtad \\
[2.0ex] \hline \\[-1.0ex]
HPQCD 05, 05A, 08A, 13A&  \cite{Mason:2005bj,Mason:2005zx,Davies:2008sw,Dowdall:2013rya}& $2+1$ & tadSym &  Asqtad \\
[2.0ex] \hline \\[-1.0ex]
HPQCD 10 & \cite{McNeile:2010ji} & $2+1$ & tadSym & Asqtad ${}^*$\\ 
[2.0ex] \hline \\[-1.0ex]
HPQCD/UKQCD 06 & \cite{Gamiz:2006sq} & $2+1$ & tadSym & Asqtad \\ 
[2.0ex] \hline \\[-1.0ex]
HPQCD/UKQCD 07 & \cite{Follana:2007uv} & $2+1$ & tadSym & Asqtad ${}^*$\\ 
[2.0ex] \hline \\[-1.0ex]
HPQCD/MILC/UKQCD 04&  \cite{Aubin:2004ck}& $2+1$ & tadSym &  Asqtad \\
[2.0ex] \hline \\[-1.0ex]
\parbox[t]{4.0cm}{Hudspith 15, 18}  &
\cite{Hudspith:2015xoa,Hudspith:2018bpz} & $2+1$ & \parbox[t]{2.0cm}{Iwasaki, Iwasaki+DSDR${}^+$ } &  DW, \mbox{M-DW} \\
[4.0ex] \hline \\[-1.0ex]
JLQCD 09, 10 & \cite{Fukaya:2009fh,Shintani:2010ph} & $2+1$ & Iwasaki & overlap \\
[2.0ex] \hline \\[-1.0ex]
\parbox[t]{4.0cm}{JLQCD 11, 12, 12A, 14, 15A, 17, 18} & \parbox[t]{2.5cm}{\cite{Kaneko:2011rp,Kaneko:2012cta,Oksuzian:2012rzb,Fukaya:2014jka,Aoki:2015pba,Aoki:2017spo,Yamanaka:2018uud}}          & $2+1$ & \parbox[t]{3.5cm}{Iwasaki (fixed topology)} & overlap\\
[6.0ex] \hline \\[-1.0ex]
\parbox[t]{4.0cm}{JLQCD 15B-C, 16, 16B, 17A} &
\parbox[t]{2.5cm}{\cite{Nakayama:2015hrn,Fahy:2015xka,Nakayama:2016atf,Cossu:2016eqs,Aoki:2017paw}}
& $2+1$ & tlSym & M-DW \\ 
[4.0ex] \hline \\[-1.0ex]
JLQCD/TWQCD 08B, 09A & \cite{Chiu:2008kt,JLQCD:2009sk} & $2+1$ & Iwasaki & overlap \\
[2.0ex] \hline \\[-1.0ex]
JLQCD/TWQCD 10 & \cite{Fukaya:2010na} & $2+1, 3$ & Iwasaki & overlap \\
[2.0ex] \hline \\[-1.0ex]
Junnarkar 13 & \cite{Junnarkar:2013ac} & $2+1$ & tadSym & Asqtad ${}^\dagger$\\
[2.0ex] \hline \\[-1.0ex]
Laiho 11 & \cite{Laiho:2011np} & $2+1$ & tadSym & Asqtad ${}^\dagger$\\
[2.0ex] \hline \\[-1.0ex]
LHP 04, LHPC 05, 10 & \parbox[t]{2.5cm}{\cite{Bonnet:2004fr,Edwards:2005ym,Bratt:2010jn}} & $2+1$ & tadSym & Asqtad $^\dagger$ \\
[4.0ex] \hline \\[-1.0ex]
LHPC 12, 12A & \cite{Green:2012ej,Green:2012ud} & $2+1$ & tlSym & 2-level HEX tlSW \\
[2.0ex] \hline \hline\\
\end{tabular*}\\[-0.2cm]
\begin{minipage}{\linewidth}
{\footnotesize 
\begin{itemize}
\item[${}^\dagger$] The calculation uses domain wall fermions in
  the valence-quark sector.\\[-5mm]
\item[${}^+$] An additional
weighting factor known as the dislocation suppressing determinant ratio (DSDR) is added to the gauge action \cite{Arthur:2012opa}.\\[-5mm]
\item[${}^*$] The calculation uses HISQ staggered fermions in the valence-quark sector.
\end{itemize}
}
\end{minipage}
}
\caption{(cntd.) Summary of simulated lattice actions with $\Nf=2+1$ or $\Nf=3$ quark flavours.
}
\end{table}

\begin{table}[h]
\addtocounter{table}{-1}
{\footnotesize
\begin{tabular*}{\textwidth}[l]{l @{\extracolsep{\fill}} c c c c}
\hline \hline \\[-1.0ex]
Collab. & Ref. & $\Nf$ & \parbox{1cm}{gauge\\action} & \parbox{1cm}{quark\\action} 
\\[2.0ex] \hline \hline \\[-1.0ex]
Mainz 18 & \cite{Ottnad:2018fri} & $2+1$ & tlSym & npSW \\
[2.0ex] \hline \\[-1.0ex]
Maltman 08 & \cite{Maltman:2008bx} & $2+1$ & tadSym &  Asqtad \\
[2.0ex] \hline \\[-1.0ex]
Martin Camalich 10 & \cite{MartinCamalich:2010fp} & $2+1$ & Iwasaki & npSW \\
[2.0ex] \hline \\[-1.0ex]
\parbox[t]{4.0cm}{MILC 04, 07, 09, 09A, 09D, 10, 10A, 12C, 16} & \parbox[t]{3.0cm}{\cite{Aubin:2004ck,Aubin:2004fs,Bernard:2007ps,Bazavov:2009bb,Toussaint:2009pz,Bazavov:2010hj,Bazavov:2010yq,Freeman:2012ry,Basak:2016jnn}}& $2+1$ &
tadSym & Asqtad \\
[4.0ex] \hline \\[-1.0ex]
\parbox[t]{4.0cm}{Nakayama 18} &
\cite{Nakayama:2018ubk}
& $2+1$ & tlSym & M-DW \\ 
[2.0ex] \hline \\[-1.0ex]
NPLQCD 06& \cite{Beane:2006kx}& $2+1$ & tadSym & Asqtad $^\dagger$ \\
[2.0ex] \hline \\[-1.0ex]
PACS 18 & \cite{Ishikawa:2018rew} & $2+1$ & Iwasaki & npSW \\
[2.0ex] \hline \\[-1.0ex]
\parbox[t]{4.0cm}{PACS-CS 08, 08A, 09, 09A, 10, 11A, 12, 13} & \parbox[t]{2.5cm}{\cite{Aoki:2008sm,Kuramashi:2008tb,Ishikawa:2009vc,Aoki:2009ix,Aoki:2009tf,Nguyen:2011ek,Aoki:2010wm,Sasaki:2013vxa}} & $2+1$ & Iwasaki & npSW \\
[4.0ex] \hline \\[-1.0ex]
QCDSF 11 & \cite{Bali:2011ks} & $2+1$ & tlSym & npSW \\
[2.0ex] \hline \\[-1.0ex]
QCDSF/UKQCD 15, 16 & \cite{Horsley:2015eaa,Bornyakov:2016dzn} & $2+1$ & tlSym & npSW \\
[2.0ex] \hline \\[-1.0ex]
\parbox[t]{4.0cm}{RBC/UKQCD 07, 08, 08A, 10, 10A-B, 11, 12, 13, 16}& \parbox[t]{2.5cm}{\cite{Antonio:2007pb,Allton:2008pn,Boyle:2008yd,Boyle:2010bh,Aoki:2010dy,Aoki:2010pe,Kelly:2012uy,Arthur:2012opa,Boyle:2013gsa,Garron:2016mva}} & $2+1$ & \parbox[t]{2.0cm}{Iwasaki, Iwasaki+DSDR${}^*$ } &  DW \\
[7.0ex] \hline \\[-1.0ex]
RBC/UKQCD 08B, 09B, 10D, 12E  & \cite{Yamazaki:2008py,Yamazaki:2009zq,Aoki:2010xg,Boyle:2012qb} & $2+1$ & Iwasaki &  DW \\
[2.0ex] \hline \\[-1.0ex]
\parbox[t]{4.0cm}{RBC/UKQCD 14B, 15A, 15E}  &
\parbox[t]{2.5cm}{\cite{Blum:2014tka,Boyle:2015hfa,Boyle:2015exm,Hudspith:2015xoa,Hudspith:2018bpz}} & $2+1$ & \parbox[t]{2.0cm}{Iwasaki, Iwasaki+DSDR${}^*$} &  DW, \mbox{M-DW} \\
[4.0ex] \hline \\[-1.0ex]
Shanahan 12 & \cite{Shanahan:2012wh} & $2+1$ & Iwasaki & npSW \\
[2.0ex] \hline \\[-1.0ex]
Sternbeck 12 & \cite{Sternbeck:2012qs} & $2+1$ & tlSym & npSW \\
[2.0ex] \hline \\[-1.0ex]
\parbox[t]{4.0cm}{SWME 10, 11, 11A, 13, 13A, 14A, 14C, 15A} & \parbox[t]{3.0cm}{\cite{Bae:2010ki,Kim:2011qg,Bae:2011ff,Bae:2013lja,Bae:2013tca,Bae:2014sja,Jang:2014aea,Jang:2015sla}} & $2+1$ & tadSym & Asqtad${}^+$ \\
[4.0ex] \hline \\[-1.0ex]
\parbox[t]{4.0cm}{Takaura 18} &
\cite{Takaura:2018lpw,Takaura:2018vcy}
& $2+1$ & tlSym & M-DW \\ 
[2.0ex] \hline \\[-1.0ex]
TWQCD 08 & \cite{Chiu:2008jq}  & $2+1$ & Iwasaki &  DW \\
[2.0ex] \hline\hline\\
\end{tabular*}\\[-0.2cm]
\begin{minipage}{\linewidth}
{\footnotesize 
\begin{itemize}
   \item[${}^\dagger$] The calculation uses domain wall fermions in the valence-quark sector.\\[-5mm]
\item[${}^*$] An additional
weighting factor known as the dislocation suppressing determinant ratio (DSDR) is added to the gauge action \cite{Arthur:2012opa}.\\[-5mm]
\item[${}^+$] The calculation uses HYP smeared improved staggered fermions in the valence-quark sector.
\end{itemize}
}
\end{minipage}
}
\caption{(cntd.) Summary of simulated lattice actions with $\Nf=2+1$ or $\Nf=3$ quark flavours.}
\end{table}

\begin{table}[h]
{\footnotesize
\begin{tabular*}{\textwidth}[l]{l @{\extracolsep{\fill}} c c c c}
\hline \hline \\[-1.0ex]
Collab. & Ref. & $\Nf$ & \parbox{1cm}{gauge\\action} & \parbox{1cm}{quark\\action} 
\\[2.0ex] \hline \hline \\[-1.0ex]
ALPHA 10A & \cite{Tekin:2010mm} & $4$ & Wilson &  npSW \\
[2.0ex] \hline \\[-1.0ex]
CalLat 17, 18 & \cite{Berkowitz:2017gql,Chang:2018uxx} & $2+1+1$ 
 & tadSym & HISQ ${}^*$\\
[2.0ex] \hline \\[-1.0ex]
\parbox[t]{4.0cm}{ETM 10, 10E, 11, 11D, 12C, 13, 13A, 13D, 15E, 16} &  \parbox[t]{3.0cm}{\cite{Baron:2010bv,Farchioni:2010tb,Baron:2011sf,Blossier:2011tf,Blossier:2012ef,Cichy:2013gja,Blossier:2013ioa,Herdoiza:2013sla,Helmes:2015gla,Carrasco:2016kpy}} & $2+1+1$ & Iwasaki & tmWil \\
[7.0ex] \hline \\[-1.0ex]
\parbox[t]{4.0cm}{ETM 14A, 14B, 14E, 15, 15C, 17E} &  \parbox[t]{3.0cm}{\cite{Alexandrou:2014sha,Bussone:2014cha,Carrasco:2014poa,Carrasco:2015pra,Carrasco:2016kpy,Lubicz:2017asp}} & $2+1+1$ & Iwasaki & tmWil ${}^+$ \\
[4.0ex] \hline \\[-1.0ex]
\parbox[t]{4.0cm}{FNAL/MILC 12B, 12C, 13, 13C, 13E, 14A, 17, 18} &\parbox[t]{3.0cm}{ \cite{Bazavov:2012dg,Bailey:2012rr,Bazavov:2013nfa,Gamiz:2013xxa,Bazavov:2013maa,Bazavov:2014wgs,Bazavov:2017lyh,Bazavov:2018kjg}} & $2+1+1$ & tadSym & HISQ \\
[4.0ex] \hline \\[-1.0ex] 
HPQCD 14A, 15B, 18  & \cite{Chakraborty:2014aca,Koponen:2015tkr,Lytle:2018evc} & $2+1+1$ & tadSym & HISQ \\
[2.0ex] \hline \\[-1.0ex] 
MILC 12C, 13A, 18 & \cite{Freeman:2012ry,Bazavov:2013cp,Basak:2018yzz} & $2+1+1$ & tadSym & HISQ \\
[2.0ex] \hline \\[-1.0ex] 
Perez 10 & \cite{PerezRubio:2010ke} & $4$ & Wilson &  npSW \\
[2.0ex] \hline \\[-1.0ex]
\parbox[t]{4.0cm}{PNDME 13, 15, 15A, 16, 18, 18A, 18B} & \cite{Bhattacharya:2013ehc,Bhattacharya:2015wna,Bhattacharya:2015esa,Bhattacharya:2016zcn,Gupta:2018qil,Lin:2018obj,Gupta:2018lvp} & $2+1+1$ 
 & tadSym & HISQ ${}^\dagger$\\
 [4.0ex] \hline \hline\\[-1.0ex]
\end{tabular*}\\[-0.2cm]
\begin{minipage}{\linewidth}
{\footnotesize 
\begin{itemize}
  \item[${}^*$] The calculation uses M\"obius domain-wall fermions
    (M-DW) in the valence sector.\\[-5mm]
   \item[${}^+$] The calculation uses Osterwalder-Seiler fermions
     \cite{Osterwalder:1977pc} in the valence-quark sector.\\[-5mm]
   \item[${}^\dagger$] The calculation uses mean-field improved clover
     fermions (mfSW) in the valence-quark sector.
\end{itemize}
}
\end{minipage}
}
\caption{Summary of simulated lattice actions with $\Nf=4$ or $\Nf=2+1+1$ quark flavours.\label{tab:simulated Nf4 actions}}
\end{table}

\begin{table}[!ht]
{\footnotesize
\begin{tabular*}{\textwidth}[l]{l @{\extracolsep{\fill}} c c c c c c}
\hline\hline \\[-1.0ex]
Collab. & Ref. & $\Nf$ & Gauge & \multicolumn{3}{c}{Quark actions}  \\
& & & action & sea & light valence & heavy \\[1.0ex] \hline \hline \\[-1.0ex]
\parbox[t]{3.5cm}{ALPHA 11, 12A, 13, 14, 14B} & \parbox[t]{2.5cm}{\cite{Blossier:2011dk,Bernardoni:2012ti,Bernardoni:2013oda,Bernardoni:2014fva,Bahr:2014iqa}} & 2 &  plaquette & npSW  & npSW & HQET
\\[4.0ex] \hline \\[-1.0ex]
ALPHA 13C & \cite{Heitger:2013oaa} & 2 &  plaquette & npSW  & npSW & npSW
\\[2.0ex] \hline \\[-1.0ex]
Blossier 18 & \cite{Blossier:2018jol} & 2 &  plaquette & npSW  & npSW & npSW
\\[2.0ex] \hline \\[-1.0ex]
\parbox[t]{2.5cm}{Atoui 13} & \cite{Atoui:2013zza} & 2 &  tlSym & tmWil & tmWil & tmWil
\\[2.0ex] \hline \\[-1.0ex]
\parbox[t]{3.5cm}{ETM 09, 09D, 11B, 12A, 12B, 13B, 13C} & \parbox[t]{2.5cm}{\cite{Blossier:2009bx,Blossier:2009hg,DiVita:2011py,Carrasco:2012dd,Carrasco:2012de,Carrasco:2013zta,Carrasco:2013iba}} & 2 &  tlSym & tmWil & tmWil & tmWil
\\[4.0ex] \hline \\[-1.0ex]
ETM  11A & \cite{Dimopoulos:2011gx} & 2 &  tlSym & tmWil & tmWil & tmWil, static
\\[4.0ex] \hline \\[-1.0ex]
TWQCD 14 & \cite{Chen:2014hva} & 2 & plaquette & oDW & oDW & oDW \\[2.0ex] \hline\hline
\end{tabular*}
\caption{Summary of lattice simulations $N_f=2$ sea-quark flavours and with $b$ and $c$ valence quarks.\label{tab:simulated Nf2 bc actions}}
}
\end{table}

\begin{table}[!ht]
{\footnotesize
\begin{tabular*}{\textwidth}[l]{l @{\extracolsep{\fill}} c c c c c c}
\hline\hline \\[-1.0ex]
Collab. & Ref. & $\Nf$ & Gauge & \multicolumn{3}{c}{Quark actions}  \\
& & & action & sea & light valence & heavy \\[1.0ex] \hline \hline \\[-2.0ex]
$\chi$QCD 14 & \cite{Yang:2014sea} & 2+1 & Iwasaki & DW & overlap & overlap \\[1.0ex] \hline \\[-2.0ex]
Datta 17 & \cite{Datta:2017aue} & 2+1 & \parbox[t]{1.0cm}{Iwasaki, Iwasaki +DSDR${}^+$} & DW     & DW     & RHQ 
\\[6.0ex] \hline \\[-2.0ex]
Detmold 16 & \cite{Detmold:2016pkz} & 2+1 & \parbox[t]{1.0cm}{Iwasaki, Iwasaki +DSDR${}^+$} & DW     & DW     & RHQ 
\\[6.0ex] \hline \\[-2.0ex]
\parbox[t]{3.5cm}{FNAL/MILC 04, 04A, 05, 08, 08A, 10, 11, 11A, 12, 13B} &  \parbox[t]{2.0cm}{\cite{Aubin:2004ej,Okamoto:2004xg,Aubin:2005ar,Bernard:2008dn,Bailey:2008wp,Bailey:2010gb,Bazavov:2011aa,Bouchard:2011xj,Bazavov:2012zs,Qiu:2013ofa}}  & 2+1 & tadSym & Asqtad & Asqtad & Fermilab
\\[9.0ex] \hline \\[-2.0ex]
FNAL/MILC 14, 15C, 16      &   \cite{Bailey:2014tva,Lattice:2015rga,Bazavov:2016nty}   & 2+1  & tadSym	&	Asqtad	& Asqtad${}^*$ &	Fermilab${}^*$\\[1.0ex] \hline \\[-2.0ex]
FNAL/MILC 15, 15D, 15E    &   \cite{Lattice:2015tia,Bailey:2015dka,Bailey:2015nbd}   & 2+1  & tadSym	&	Asqtad	& Asqtad &	Fermilab\\[1.0ex] \hline \\[-2.0ex]
\parbox[t]{3.5cm}{HPQCD 06, 06A, 08B, 09, 13B}& \parbox[t]{2.0cm}{\cite{Dalgic:2006dt,Dalgic:2006gp,Allison:2008xk,Gamiz:2009ku,Lee:2013mla}} & 2+1 &  tadSym & Asqtad & Asqtad & NRQCD
\\[4.0ex] \hline \\[-2.0ex]
HPQCD 12, 13E & \cite{Na:2012sp,Bouchard:2013pna} & 2+1 &  tadSym & Asqtad & HISQ & NRQCD
\\[1.0ex] \hline \\[-2.0ex]
HPQCD 15 & \cite{Na:2015kha} & 2+1 &  tadSym & Asqtad & HISQ${}^\dagger$ & NRQCD${}^\dagger$
\\[1.0ex] \hline \\[-2.0ex]
HPQCD 17 & \cite{Monahan:2017uby} & 2+1 &  tadSym & Asqtad & HISQ & \parbox[t]{1.2cm}{HISQ, NRQCD}
\\[4.0ex] \hline \\[-2.0ex]
\parbox[t]{3.5cm}{HPQCD/UKQCD 07, HPQCD 10A,  10B, 11, 11A, 12A, 13C}  & \parbox[t]{2.0cm}{\cite{Follana:2007uv,Davies:2010ip,Na:2010uf,Na:2011mc,McNeile:2011ng,Na:2012iu,Koponen:2013tua}} & 2+1 &  tadSym & Asqtad & HISQ & HISQ
\\[6.0ex] \hline \\[-2.0ex]
JLQCD 16 & \cite{Nakayama:2016atf} & 2+1 & tlSym & M-DW & M-DW & M-DW
\\[1.0ex] \hline \\[-2.0ex]
JLQCD 17B & \cite{Kaneko:2017xgg} & 2+1 & tlSym & DW & DW & DW
\\[1.0ex] \hline \\[-2.0ex]
Maezawa 16 & \cite{Maezawa:2016vgv} & 2+1 & tlSym & HISQ & HISQ & HISQ
\\[1.0ex] \hline \\[-2.0ex]
Meinel 16 & \cite{Meinel:2016dqj} & 2+1 & \parbox[t]{1.0cm}{Iwasaki, Iwasaki + DSDR${}^+$} & DW     & DW     & RHQ 
\\[6.0ex] \hline \\[-2.0ex]
PACS-CS 11 & \cite{Namekawa:2011wt} & 2+1 & Iwasaki & npSW & npSW & Tsukuba
\\[1.0ex] \hline \\[-2.0ex]  
RBC/UKQCD 10C, 14A & \cite{Albertus:2010nm,Aoki:2014nga} & 2+1 & Iwasaki & DW & DW & static
\\[1.0ex] \hline \\[-2.0ex]
RBC/UKQCD 13A, 14, 15 & \cite{Witzel:2013sla,Christ:2014uea,Flynn:2015mha} & 2+1 & Iwasaki & DW & DW & RHQ
\\[1.0ex] \hline \\[-2.0ex]
RBC/UKQCD 17 & \cite{Boyle:2017jwu} & 2+1 & Iwasaki & DW/M-DW & M-DW & M-DW
\\[1.0ex] \hline \\[-2.0ex]
\parbox[t]{3.5cm}{ETM 13E, 13F, 14E, 17D, 18}   & \parbox[t]{2.0cm}{\cite{Carrasco:2013naa,Dimopoulos:2013qfa,Carrasco:2014poa,Lubicz:2017syv,Lubicz:2018rfs}} & 2+1+1 &  Iwasaki & tmWil & tmWil & tmWil
\\[4.0ex] \hline \hline\\
\end{tabular*}\\[-0.2cm]
\begin{minipage}{\linewidth}
{\footnotesize 
\begin{itemize}
   \item[$^*$] Asqtad for $u$, $d$ and $s$ quark; Fermilab for $b$ and $c$ quark.\\[-5mm]
\item[${}^+$] An additional
weighting factor known as the dislocation suppressing determinant ratio (DSDR) is added to the gauge action \cite{Arthur:2012opa}.\\[-5mm]
   \item[$^\dagger$] HISQ for $u$, $d$, $s$  and $c$ quark; NRQCD for $b$ quark.
\end{itemize}
}
\end{minipage}
\caption{Summary of lattice simulations with $N_f=2+1$ sea-quark flavours and $b$ and $c$ valence quarks.  \label{tab:simulated Nf3 bc actions}
}
}
\end{table}

\begin{table}[!ht]
{\footnotesize
\begin{tabular*}{\textwidth}[l]{l @{\extracolsep{\fill}} c c c c c c}
\hline\hline \\[-1.0ex]
Collab. & Ref. & $\Nf$ & Gauge & \multicolumn{3}{c}{Quark actions}  \\
& & & action & sea & light valence & heavy \\[1.0ex] \hline \hline \\[-2.0ex]
ETM 16B   & \cite{Bussone:2016iua} & 2+1+1 &  Iwasaki & tmWil & tmWil & tmWil$^+$
\\[1.0ex] \hline \\[-2.0ex]
FNAL/MILC 12B, 13, 14A & \cite{Bazavov:2012dg,Bazavov:2013nfa,Bazavov:2014wgs} & 2+1+1 & tadSym & HISQ & HISQ & HISQ
\\[1.0ex] \hline \\[-2.0ex]
FNAL/MILC 17 & \cite{Bazavov:2017lyh} & 2+1+1 & tadSym & HISQ & HISQ & HISQ \\
[1.0ex] \hline \\[-2.0ex] 
FNAL/MILC/TUMQCD 18& \cite{Bazavov:2018omf} & 2+1+1 & tadSym & HISQ & HISQ & HISQ \\
[1.0ex] \hline \\[-2.0ex] 
Gambino 17& \cite{Gambino:2017vkx} & 2+1+1 & Iwasaki & tmWil&tmWil&tmWil$^+$ \\
[1.0ex] \hline \\[-2.0ex] 
\parbox[t]{2.5cm}{HPQCD 13, 17A} & \cite{Dowdall:2013tga,Hughes:2017spc} &  2+1+1 & tadSym & HISQ & HISQ & NRQCD
\\[1.0ex] \hline\\[-2.0ex]
\parbox[t]{2.5cm}{HPQCD 17B} & \cite{Harrison:2017fmw} &  2+1+1 & tadSym & HISQ & HISQ & HISQ, NRQCD
\\[1.0ex] \hline\\[-2.0ex]
 RM123 17 & \cite{Giusti:2017dmp} & 2+1+1 & Iwasaki &  tmWil&tmWil&tmWil$^+$\\
[1.0ex] \hline\hline\\
\end{tabular*}\\[-0.2cm]
\begin{minipage}{\linewidth}
{\footnotesize 
\begin{itemize}
   \item[${}^+$] The calculation uses Osterwalder-Seiler fermions \cite{Osterwalder:1977pc} in the valence-quark sector.\\[-5mm]
\end{itemize}
}
\end{minipage}
\caption{Summary of lattice simulations with  $N_f=2+1+1$ sea-quark flavours and $b$ and $c$ valence quarks.  \label{tab:simulated Nf4 bc actions}
}
}
\end{table}

\fi

\else
\section{Appendix}

\fi

\clearpage
\section{Notes}

In the following Appendices we provide more detailed information on the simulations used
to calculate the quantities discussed in Secs.~\ref{sec:qmass}--\ref{sec:NME}. 
\ifx\reducedapptables\undefined
\else
We present this information only for results
that are new w.r.t.~FLAG 19. For all other results the information is available
in the corresponding Appendices B.1--8 in FLAG 19 \cite{Aoki:2019cca} and B.1--7 of FLAG 16~\cite{Aoki:2016frl}. The complete information is available on the FLAG website \href{http://flag.unibe.ch}{{\tt
http://flag.unibe.ch}} \cite{FLAG:webpage}.
\fi

\subsection{Notes to Sec.~\ref{sec:qmass} on quark masses}

\begin{table}[!ht]
{\footnotesize

\caption{Continuum extrapolations/estimation of lattice artifacts in
  determinations of $m_{ud}$, $m_s$ and, in some cases $m_u$ and
  $m_d$, with $N_f=2+1+1$ quark flavours.}
}
\end{table}

\begin{table}[!ht]
{\footnotesize
\begin{tabular*}{\textwidth}[l]{l @{\extracolsep{\fill}} c c c l}
\hline\hline \\[-1.0ex]
Collab. & Ref. & $\Nf$ & $a$ [fm] & Description
\\[1.0ex] \hline \hline \\[-1.0ex]
ALPHA 19  & \cite{Bruno:2019vup} & 2+1 & 0.05, 0.064, 0.076, 0.086&\\[2.0ex]
%
\ifx\reducedapptables\undefined \\ [1.0ex] \hline
\\[-1.0ex]
{BMW 16}& \cite{Fodor:2016bgu} & 2+1 & 0.05, 0.06, 0.07, 0.09, 0.11 &  \parbox[t]{4.5cm}{Scale set through $M_{\Omega}$ and $M_{\Xi}$.}\\
[2.0ex] \hline \\[-1.0ex]
{MILC 16}& \cite{Basak:2016jnn} & 2+1 & 0.045, 0.06, 0.09, 0.12 &  \parbox[t]{4.5cm}{Scale set through $f_\pi$.}
 \hline \\[-1.0ex]
{QCDSF/UKQCD 15}& \cite{Horsley:2015eaa} & 2+1 & 0.07 &  \parbox[t]{4.5cm}{Scale set through the gradient flow parameter $w_0$.}\\
[4.0ex] \hline \\[-1.0ex]
{RBC/UKQCD 14B}& \cite{Blum:2014tka} & 2+1 & 0.063, 0.084,\, 0.114,\, 0.144  &  \parbox[t]{4.5cm}{Scale set through $M_\Omega$.}\\
[2.0ex] \hline \\[-1.0ex]
{RBC/UKQCD 12}& \cite{Arthur:2012opa} & 2+1 & 0.085,\, 0.113,\, 0.144  &  \parbox[t]{4.5cm}{Scale set through $M_\Omega$.}\\
[2.0ex] \hline \\[-1.0ex]
PACS-CS 12 & \cite{Aoki:2012st} & 1+1+1 & 0.09
& \parbox[t]{4.5cm}{Reweighting of PACS-CS 08 $N_f=2+1$ QCD
  configurations with e.m. and $m_u\ne m_d$.} \\
[7.0ex] \hline \\[-1.0ex]
Laiho 11 & \cite{Laiho:2011np} & 2+1 & 0.06, 0.09, 0.15
& \parbox[t]{4.5cm}{MILC staggered ensembles~\cite{Bazavov:2010yq}, scale set using
  $r_1$ determined by HPQCD with $\Upsilon$ splittings, pseudoscalar
  decay constants, through $r_1$~\cite{Davies:2009tsa}.} \\ [13.0ex]
\fi
\hline \hline
\\[-1.0ex]
\end{tabular*}
\caption{Continuum extrapolations/estimation of lattice artifacts in
  determinations of $m_{ud}$, $m_s$ and, in some cases $m_u$ and
  $m_d$, with $N_f=2+1$ quark flavours.}
}
\end{table}

\ifx\reducedapptables\undefined
\begin{table}[!ht]
\addtocounter{table}{-1}
{\footnotesize
 \begin{tabular*}{\textwidth}[l]{l @{\extracolsep{\fill}} c c c l}
 \hline\hline \\[-1.0ex]
 Collab. & Ref. & $\Nf$ & $a$ [fm] & Description
 \\[1.0ex] \hline \hline \\[-1.0ex]
PACS-CS 10 & \cite{Aoki:2010wm} &  2+1 &  0.09
& \parbox[t]{4.5cm}{cf.~PACS-CS 08} \\
[1.0ex] \hline \\[-1.0ex]
{MILC 10A}& \cite{Bazavov:2010yq} & 2+1 & & cf.~MILC 09, 09A \\
[1.0ex] \hline \\[-1.0ex]
{BMW 10A, 10B}& \cite{Durr:2010vn,Durr:2010aw}
& 2+1 & \parbox[t]{2.3cm}{0.054, 0.065, 0.077, 0.093, 0.116} &  Scale set via
$M_\pi, M_K, M_\Omega$.\\
\end{tabular*}
 \caption{(cntd.) Continuum extrapolations/estimation of lattice artifacts in
   determinations of $m_{ud}$, $m_s$ and, in some cases $m_u$ and
   $m_d$, with $N_f=2+1$ quark flavours.}
}
\end{table}

\newpage
\fi

\begin{table}[!ht]
{\footnotesize
%
\caption{Chiral extrapolation/minimum pion mass in determinations of
  $m_{ud}$, $m_s$ and, in some cases, $m_u$ and
  $m_d$,  with $N_f=2+1+1$ quark flavours.}
}
\end{table}

\begin{table}[!ht]
{\footnotesize
\begin{tabular*}{\textwidth}[l]{l @{\extracolsep{\fill}} c c c l}
\hline \hline  \\[-1.0ex]
Collab. & Ref. & $\Nf$ & $M_{\pi,\text{min}}$ [MeV] & Description
\\[1.0ex] \hline \hline \\[-1.0ex]
ALPHA 19  & \cite{Bruno:2019vup} & 2+1 & 198 & \\ [1.0ex]
\ifx\reducedapptables\undefined
\hline\\[-1.0ex]
\hline \\[-1.0ex]
{QCDSF/UKQCD 15}& \cite{Horsley:2015eaa} & 2+1 & 205 (val.) &  \parbox[t]{5.7cm}{Expansion around the symmetric point $m_u = m_d = m_s$.}\\
[4.0ex] \hline \\[-1.0ex]
{RBC/UKQCD 14B}& \cite{Blum:2014tka} & 2+1 & 139 &\parbox[t]{5.7cm}{NLO PQ $SU(2)$ $\chi$PT as well as analytic ans\"atze.}\\
[4.0ex] \hline \\[-1.0ex]
{RBC/UKQCD 12}& \cite{Arthur:2012opa} & 2+1 & 170 & \parbox[t]{5.7cm}{Combined fit to Iwasaki and Iwasaki+DSDR gauge action ensembles.} \\
[4.0ex] \hline \\[-1.0ex]
PACS-CS 12 & \cite{Aoki:2012st} &  1+1+1 &  & cf.~PACS-CS 08 \\
[1.0ex] \hline \\[-1.0ex]
Laiho 11 & \cite{Laiho:2011np} & 2+1 & \parbox[t]{3cm}{210 (val.)\\ 280 (sea-RMS)} &
\parbox[t]{5.7cm}{NLO $SU(3)$, mixed-action $\chi$PT \cite{Bar:2005tu}, with N$^2$LO-N$^4$LO
  analytic terms.} \\ [4.0ex] \hline \\[-1.0ex]
PACS-CS 10 & \cite{Aoki:2010wm} &  2+1 & & cf.~PACS-CS 08 \\
[1.0ex] \hline \\[-1.0ex]
{MILC 10A}& \cite{Bazavov:2010yq} & 2+1 & & \parbox[t]{5.7cm}{NLO $SU(2)$
S{\Ch}PT. cf.~also MILC 09A, 09.} \\
[4.0ex] \hline \\[-1.0ex]
{BMW 10A, 10B}& \cite{Durr:2010vn,Durr:2010aw}
& 2+1 & 135 & \parbox[t]{6.cm}{Interpolation to the physical point.} \\
[1.0ex] \hline \\[-1.0ex]
{RBC/UKQCD 10A}& \cite{Aoki:2010dy} & 2+1 & 290 & \parbox[t]{5.7cm}{NLO PQ $SU(2)$ $\chi$PT as well as analytic ans\"atze.}\\
[4.0ex] \hline \\[-1.0ex]
{Blum 10}& \cite{Blum:2010ym,Allton:2008pn}
& 2+1 & \parbox[t]{2cm}{242 (valence), 330 (sea)} &
\parbox[t]{5.7cm}{Extrapolation done on the basis of PQ$\chi$PT formulae with
  virtual photons.}
\\
[4.0ex] \fi
\hline \hline\\[-1.0ex]
\end{tabular*}
\caption{Chiral extrapolation/minimum pion mass in determinations of
  $m_{ud}$, $m_s$ and, in some cases $m_u$ and
  $m_d$,  with $N_f=2+1$ quark flavours.}
}
\end{table}

\ifx\reducedapptables\undefined
\begin{table}[!ht]
\addtocounter{table}{-1}
{\footnotesize
\begin{tabular*}{\textwidth}[l]{l @{\extracolsep{\fill}} c c c l}
\hline \hline  \\[-1.0ex]
Collab. & Ref. & $\Nf$ & $M_{\pi,\text{min}}$ [MeV] & Description
\\[1.0ex] \hline \hline \\[-1.0ex]

{PACS-CS 09}& \cite{Aoki:2009ix}& 2+1 & 135 & \parbox[t]{5.5cm}{Physical point reached by
reweighting technique, no chiral extrapolation needed.}\\
[7.0ex] \hline \\[-1.0ex]
{HPQCD 09A, 10}&  \cite{Davies:2009ih,McNeile:2010ji}& 2+1 &  &  cf.~MILC 09\\
[1.0ex] \hline \\[-1.0ex]
{MILC 09A, 09} & \cite{Bazavov:2009fk,Bazavov:2009bb} & 2+1 & 177, 224 & \parbox[t]{5.5cm}{
		NLO $SU(3)$ RS{\Ch}PT, continuum {\Ch}PT at NNLO and NNNLO and NNNNLO 
		analytic terms. The lightest Nambu-Goldstone mass
                is 177 MeV (09A) and 224 MeV (09)
		(at $a=$0.09fm) and the lightest RMS mass is 258MeV (at
                $a=$0.06fm).} \\ 
[18.0ex] \hline \\[-1.0ex]
PACS-CS 08 &\cite{Aoki:2008sm}& 2+1 &156& 
		\parbox[t]{5.5cm}{NLO $SU(2)$ {\Ch}PT and $SU(3)$
                  (Wilson){\Ch}PT.}\\
[4.0ex] \hline \\[-1.0ex]
RBC/UKQCD 08	&\cite{Allton:2008pn}	&2+1& \parbox[t]{2cm}{242 (valence), 330 (sea)} & \parbox[t]{5.5cm}{$SU(3)$ PQ{\Ch}PT 
and heavy kaon NLO $SU(2)$ PQ{\Ch}PT fits.} \\
[4.0ex] \hline \\[-1.0ex]
{CP-PACS/JLQCD 07}&  \cite{Ishikawa:2007nn} & 2+1 & 620 & \parbox[t]{5.5cm}{NLO Wilson {\Ch}PT
fits to meson masses.} \\
[1.0ex] \hline \\[-1.0ex]
{HPQCD 05}& \cite{Mason:2005bj}& 2+1 & 240 & \parbox[t]{5.5cm}{PQ RS{\Ch}PT fits.} \\
[1.0ex] \hline \\[-1.0ex]
\parbox[t]{4.5cm}{HPQCD/MILC/UKQCD 04,\\ MILC 04}
& \cite{Aubin:2004ck,Aubin:2004fs}& 2+1 & 240 & \parbox[t]{5.5cm}{PQ RS{\Ch}PT fits.} \\
[4.0ex] \hline \hline \\[-1.0ex]
\end{tabular*}
\caption{(cntd.) Chiral extrapolation/minimum pion mass in determinations of
  $m_{ud}$, $m_s$ and, in some cases $m_u$ and
  $m_d$,  with $N_f=2+1$ quark flavours.}
}
\end{table}

\newpage
\fi

\begin{table}
{\footnotesize
%
\caption{Finite-volume effects in determinations of $m_{ud}$, $m_s$
  and, in some cases $m_u$ and $m_d$, with $N_f=2+1+1$ quark flavours.}
}
\end{table}

\begin{table}
{\footnotesize
\begin{tabular*}{\textwidth}[l]{l @{\extracolsep{\fill}} c c c c l}
\hline\hline  \\[-1.0ex]
Collab. & Ref. & $\Nf$ & $L$ [fm] & $M_{\pi,\text{min}} L$ & Description
\\[1.0ex] \hline \hline \\[-1.0ex]
ALPHA 19  & \cite{Bruno:2019vup} & 2+1 & 2.4 -- 4.1 & $\gtrsim 4.0$ &\\ [1.0ex] 
\ifx\reducedapptables\undefined
\hline\\[-1.0ex]
{BMW 16}& \cite{Fodor:2016bgu} & 2+1 &  $\gtrsim 5.0$ & $\gtrsim 4.0$ &  \parbox[t]{4.5cm}{Universal $\mathrm{QED}_{\mathrm{TL}}$ FV effects subtracted and $\cO(1/L^3)$ fitted.}\\
[4.0ex] \hline \\[-1.0ex]
{MILC 16}& \cite{Basak:2016jnn} & 2+1 & 1.4 -- 5.5 & 2.7 & \parbox[t]{4.5cm}{Universal $\mathrm{QED}_{\mathrm{TL}}$ FV effects subtracted and higher order model using S$\chi$PT.}\\[7.0ex] 
\hline \\[-1.0ex]
{QCDSF/UKQCD 15}& \cite{Horsley:2015eaa} & 2+1 & 1.7,\,2.2,\,3.4& & \parbox[t]{4.5cm}{Effective field theory used to extrapolate to infinite volume.} \\
[4.0ex] \hline \\[-1.0ex]
{RBC/UKQCD 14B}& \cite{Blum:2014tka} & 2+1 & 2.0,\,2.7,\,4.6,\,5.4 &  $3.8$ &  
\parbox[t]{4.5cm}{Uses FV chiral perturbation theory to estimate the error, which is deemed negligible and omitted.}\\
[10.0ex] \hline \\[-1.0ex]
{RBC/UKQCD 12}& \cite{Arthur:2012opa} & 2+1 & 2.7, 4.6 &  $\gtrsim 4.0$
& \parbox[t]{4.5cm}{Uses FV chiral perturbation theory to estimate the error.} \\
[4.0ex] \hline \\[-1.0ex]
PACS-CS 12 & \cite{Aoki:2012st} &  1+1+1 & & & cf.~PACS-CS 08 \\
[1.0ex] \hline \\[-1.0ex]
Laiho 11 & \cite{Laiho:2011np} & 2+1 &
\parbox[t]{2cm}{2.5, 2.9, 3.0,\\ 3.6, 3.8, 4.8}&\parbox[t]{1.5cm}{4.1 (val.)\\
4.1 (sea)}&  \parbox[t]{4.5cm}{Data corrected using NLO $SU(3)$ $\chi$PT finite-V formulae.} \\
[4.0ex] 
\fi
\hline \hline\\[-1.0ex]
\end{tabular*}
\caption{Finite-volume effects in determinations of $m_{ud}$, $m_s$
  and, in some cases $m_u$ and $m_d$, with $N_f=2+1$ quark flavours.}
}
\end{table}
 
\ifx\reducedapptables\undefined

\begin{table}
\addtocounter{table}{-1}
{\footnotesize
\begin{tabular*}{\textwidth}[l]{l @{\extracolsep{\fill}} c c c c l}
\hline\hline  \\[-1.0ex]
Collab. & Ref. & $\Nf$ & $L$ [fm] & $M_{\pi,\text{min}} L$ & Description
\\[1.0ex] \hline \hline \\[-1.0ex]
PACS-CS 10 & \cite{Aoki:2010wm} &  2+1 & & & cf.~PACS-CS 08 \\
[1.0ex] \hline \\[-1.0ex]
{MILC 10A}& \cite{Bazavov:2010yq} & 2+1 & & & cf.~MILC 09A, 09 \\
[1.0ex] \hline \\[-1.0ex]
{BMW 10A, 10B}& \cite{Durr:2010vn,Durr:2010aw}
& 2+1 &  $\gtrsim 5.0$ & $\gtrsim 4.0$ & \parbox[t]{3.7cm}{FV corrections
  below 5 per mil on the largest lattices.}\\
\end{tabular*}
\caption{(cntd.) Finite-volume effects in determinations of $m_{ud}$, $m_s$
  and, in some cases $m_u$ and $m_d$, with $N_f=2+1$ quark flavours.}
}
\end{table}
 
\newpage
\fi

\begin{table}[!ht]
{\footnotesize
\begin{tabular*}{\textwidth}[l]{l @{\extracolsep{\fill}} c c l}
\hline\hline \\[-1.0ex]
Collab. & Ref. & $\Nf$  & Description
\\[1.0ex] \hline \hline \\[-1.0ex]
ETM 21A  & \cite{Alexandrou:2021gqw}  & 2+1+1 &  Nonperturbative (RI/MOM) \\[1.0ex]
\ifx\reducedapptables\undefined
[4.0ex]\hline\\[-1.0ex]
MILC 18  & \cite{Basak:2018yzz}  & 2+1+1 & \parbox[t]{4.7cm}{Renormalization not required for $m_u/m_{d}$.} \\
[4.0ex] \hline \\[-1.0ex]
FNAL/MILC/TUMQCD 18  & \cite{Bazavov:2018omf}  & 2+1+1 & \parbox[t]{4.7cm}{MRS scheme from~\citep{Brambilla:2017hcq}.} \\
[3.0ex] \hline \\[-1.0ex]
FNAL/MILC 17  & \cite{Bazavov:2017lyh}  & 2+1+1 & \parbox[t]{4.7cm}{Renormalization not required for quark mass ratios.}\\
[4.0ex] \hline \\[-1.0ex]
HPQCD 18  & \cite{Lytle:2018evc}  & 2+1+1 & \parbox[t]{4.7cm}{Nonperturbative renormalization (RI/SMOM).}\\
[4.0ex] \hline \\[-1.0ex]
RM123 17  & \cite{Giusti:2017dmp}  & 2+1+1 & \parbox[t]{4.7cm}{Nonperturbative renormalization (RI/MOM).}\\[4.0ex] 
\hline \\[-1.0ex]
HPQCD 14A & \cite{Chakraborty:2014aca} & 2+1+1 & \parbox[t]{4.7cm}{Renormalization not required through the use of the ratio $m_c/m_s$.} \\
[7.0ex] \hline \\[-1.0ex]
{FNAL/MILC 14A}& \cite{Bazavov:2014wgs} & 2+1+1 & \parbox[t]{4.7cm}{Renormalization not required for $m_s/m_{ud}$.} \\
[4.0ex] \hline \\[-1.0ex]
ETM 14& \cite{Carrasco:2014cwa} & 2+1+1 & \parbox[t]{4.7cm}{Nonperturbative renormalization (RI/MOM).} \\[4.0ex]
\fi
 \hline  \hline \\[-1.0ex]
\end{tabular*}
\caption{Renormalization in determinations of $m_{ud}$, $m_s$
  and, in some cases $m_u$ and $m_d$, with $N_f=2+1+1$ quark flavours.}
}
\end{table}

\begin{table}[!ht]
{\footnotesize
%
\caption{Renormalization in determinations of $m_{ud}$, $m_s$
  and, in some cases $m_u$ and $m_d$, with $N_f=2+1$ quark flavours.}
}
\end{table}
 

\clearpage
\begin{table}[!ht]
{\footnotesize
\begin{tabular*}{\textwidth}[l]{l @{\extracolsep{\fill}} c c c l}
\hline\hline \\[-1.0ex]
Collab. & Ref. & $\Nf$ & $a$ [fm] & Description
\\[1.0ex] \hline \hline \\[-1.0ex]
ETM 21A  & \cite{Alexandrou:2021gqw}  & 2+1+1 & \parbox[t]{3cm}{0.07, 0.08, 0.09}  & \parbox[t]{4.5cm}{Scale set from $w_0$. Twisted mass action for charm quarks.} \\
[4.0ex]\hline\\[-1.0ex]
HPQCD 20A  & \cite{Hatton:2020qhk}  & 2+1+1 & \parbox[t]{3cm}{0.03, 0.042,0.06, 0.09, 0.12}  & \parbox[t]{4.5cm}{Scale set from $w_0$ and $f_\pi$. HISQ action for charm quarks.} \\[4.0ex]
\ifx\reducedapptables\undefined
\hline\\[-1.0ex]
HPQCD 18  & \cite{Lytle:2018evc}  & 2+1+1 & \parbox[t]{3cm}{0.06, 0.09, 0.12}  & \parbox[t]{4.5cm}{Scale set from $w_0$ and $f_\pi$. HISQ action for charm quarks.} \\
[4.0ex]\hline\\[-1.0ex]
FNAL/MILC/TUMQCD 18  & \cite{Bazavov:2018omf}  & 2+1+1 & \parbox[t]{3cm}{0.03, 0.042, 0.06, 0.09, 0.12, 0.15}  & \parbox[t]{4.5cm}{Scale set from $f_\pi$. HISQ action for charm quarks.} \\
[4.0ex] 
\hline \\[-1.0ex]
HPQCD 14A & \cite{Chakraborty:2014aca} & 2+1+1 & 0.06, 0.09, 0.12, 0.15  & \parbox[t]{4.5cm}{Scale set through the Wilson flow parameter $w_0$.} \\
[4.0ex] \hline \\[-1.0ex]
ETM 14& \cite{Carrasco:2014cwa} & 2+1+1 & 0.062, 0.082, 0.089  & \parbox[t]{4.5cm}{Scale set through $F_\pi$.}\\
[1.0ex] \hline \\[-1.0ex]
ETM 14A & \cite{Alexandrou:2014sha} & 2+1+1 & 0.062, 0.082, 0.089 & \parbox[t]{4.5cm}{Scale set through the nucleon mass $M_N$.} \\
[4.0ex]
\fi
 \hline\hline \\[-1.0ex]
\end{tabular*}
\caption{Continuum extrapolations/estimation of lattice artifacts in the determinations of $m_c$ with $\Nf=2+1+1$ quark flavours.}
}
\end{table}

\begin{table}[!ht]
{\footnotesize
\begin{tabular*}{\textwidth}[l]{l @{\extracolsep{\fill}} c c c l}
\hline\hline \\[-1.0ex]
Collab. & Ref. & $\Nf$ & $a$ [fm] & Description
\\[1.0ex] \hline \hline \\[-1.0ex]
ALPHA 21  & \cite{Heitger:2021apz} & 2+1 & 0.039 to 0.087 fm & \parbox[t]{4.5cm}{Wilson-clover action for the charm quark.\\} \\ [1.0ex] \hline
\\[-1.0ex]
Petreczky 19  & \cite{Petreczky:2019ozv} & 2+1 & 0.025 to 0.11 fm & \parbox[t]{4.5cm}{HISQ action for the charm quark. Scale set from $r_1$ parameter of heavy quark potential and $f_\pi$.\\} \\ [1.0ex] 
\ifx\reducedapptables\undefined
\hline
\\[-1.0ex]
Maezawa 16  & \cite{Maezawa:2016vgv} & 2+1 & 0.04 to 0.14 fm & \parbox[t]{4.5cm}{HISQ action for the charm quark. Scale set from $r_1$ parameter of heavy quark potential.\\} \\ [1.0ex] \hline
\\[-1.0ex]
JLQCD 16 & \cite{Nakayama:2016atf} & 2+1 & 0.044, 0.055. 0.083 & \parbox[t]{4.5cm}{M\"obius domain wall fermions.} \\ [1.0ex] 
\hline
\\[-1.0ex]
$\chi$QCD 14 & \cite{Yang:2014sea} &  2+1 & 0.087, 0.11 & \parbox[t]{4.5cm}{Overlap valence fermions on domain-wall sea quarks from \cite{Aoki:2010dy}.
The lattice scale is set together with the strange and charm-quark masses using the experimental values of the $D_s$, $D_s^*$ and $J/\psi$ meson masses.} \\
[18.0ex] \hline \\[-1.0ex]
HPQCD 10 & \cite{McNeile:2010ji} & 2+1 & \parbox[t]{3cm}{0.044, 0.059, 0.085, 0.12, 0.15} & \parbox[t]{4.5cm}{Scale set through the static-quark potential parameter $r_1$.} \\
[4.0ex] \hline \\[-1.0ex]
HPQCD 08B & \cite{Allison:2008xk} & 2+1 & 0.06, 0.09, 0.12, 0.15 & \parbox[t]{4.5cm}{Scale set through the static-quark potential parameter $r_1$.} \\
[4.0ex]
\fi
 \hline\hline \\[-1.0ex]
\end{tabular*}
\caption{Continuum extrapolations/estimation of lattice artifacts in the determinations of $m_c$ with $\Nf=2+1$ quark flavours.}
}
\end{table}

\begin{table}[!ht]
{\footnotesize
\begin{tabular*}{\textwidth}[l]{l @{\extracolsep{\fill}} c c c l}
\hline \hline  \\[-1.0ex]
Collab. & Ref. & $\Nf$ & $M_{\pi,\text{min}}$ [MeV] & Description
\\[1.0ex] \hline \hline \\[-1.0ex]
ETM 21A  & \cite{Alexandrou:2021gqw}  & 2+1+1 &physical  & \parbox[t]{4.5cm}{} \\
[4.0ex]\hline\\[-1.0ex]
HPQCD 20A  & \cite{Hatton:2020qhk}  & 2+1+1 &  physical   & \parbox[t]{4.5cm}{} \\[2.0ex]
\ifx\reducedapptables\undefined
[4.0ex]\hline\\[-1.0ex]
HPQCD 18  & \cite{Lytle:2018evc}  & 2+1+1 & $130~\mathrm{MeV}$  & \parbox[t]{4.50cm}{Fits include sea quark mass tunings. $m_c$ set from $\eta_c$ meson.} \\[4.0ex]\hline\\[-1.0ex]
FNAL/MILC/TUMQCD 18  & \cite{Bazavov:2018omf} & 2+1+1 & $130$ ($173_{RMS}$)& \\
[1.0ex]
 \hline \\[-1.0ex]
HPQCD 14A & \cite{Chakraborty:2014aca} & 2+1+1 & $128$ ($173_{RMS}$)& \\
[1.0ex] \hline \\[-1.0ex]
ETM 14& \cite{Carrasco:2014cwa} & 2+1+1 & $180_{\pi^0}$ ($220_{\pi^\pm}$)  & \\
[1.0ex] \hline \\[-1.0ex]
ETM 14A & \cite{Alexandrou:2014sha} & 2+1+1 & 210 & cf.~ETM 14 \\
[1.0ex]
\fi
 \hline  \hline \\[-1.0ex]
\end{tabular*}
\caption{Chiral extrapolation/minimum pion mass in the determinations of $m_c$ with $\Nf=2+1+1$ quark flavours.}
}
\end{table}

\begin{table}[!ht]
{\footnotesize
\begin{tabular*}{\textwidth}[l]{l @{\extracolsep{\fill}} c c c l}
\hline \hline  \\[-1.0ex]
Collab. & Ref. & $\Nf$ & $M_{\pi,\text{min}}$ [MeV] & Description
\\[1.0ex] \hline \hline \\[-1.0ex]
ALPHA 21  & \cite{Heitger:2021apz} & 2+1 & 198 & \\
[2.0ex] \hline \\[-1.0ex]
\\[-2.0ex]
Petreczky 19  & \cite{Petreczky:2019ozv} & 2+1 &  161 &\\[2.0ex]
\ifx\reducedapptables\undefined
[3.0ex] \hline \\[-1.0ex]
\\[-1.0ex]
Maezawa 16  & \cite{Maezawa:2016vgv} & 2+1 & 160 &  \\
[1.0ex] \hline \\[-1.0ex]
JLQCD 16 & \cite{Nakayama:2016atf} & 2+1 & 230 &  \\
[1.0ex]
 \hline \\[-1.0ex]
$\chi$QCD 14 & \cite{Yang:2014sea} & 2+1 & 290 & \\
[3.0ex] \hline \\[-1.0ex]
HPQCD 10 & \cite{McNeile:2010ji} & 2+1 & 260 & \\
[1.0ex] \hline \\[-1.0ex]
HPQCD 08B & \cite{Allison:2008xk} & 2+1 &  & \\
[1.0ex]
\fi
 \hline  \hline \\[-1.0ex]
\end{tabular*}
\caption{Chiral extrapolation/minimum pion mass in the determinations of $m_c$ with $\Nf=2+1$ quark flavours.}
}
\end{table}

\newpage

\begin{table}
{\footnotesize
\begin{tabular*}{\textwidth}[l]{l @{\extracolsep{\fill}} c c c c l}
\hline\hline  \\[-1.0ex]
Collab. & Ref. & $\Nf$ & $L$ [fm] & $M_{\pi,\text{min}} L$ & Description
\\[1.0ex] \hline \hline \\[-1.0ex]
ETM 21A  & \cite{Alexandrou:2021gqw}  & 2+1+1 & 2.5 -- 5.6 & 3.6 & \parbox[t]{4.5cm}{} \\
[2.0ex]\hline\\[-1.0ex]
HPQCD 20A  & \cite{Hatton:2020qhk}  & 2+1+1 & 1.9 -- 5.76 & 3.8 & \parbox[t]{4.5cm}{} \\
[1.0ex]
\ifx\reducedapptables\undefined
[4.0ex]\hline\\[-1.0ex]
HPQCD 18  & \cite{Lytle:2018evc}  & 2+1+1 & 2.88 -- 5.76  & 3.8 & \parbox[t]{4.3cm}{Finite-volume effects negligible for charm-quark mass.} \\
[4.0ex] \hline \\[-1.0ex]
FNAL/MILC/TUMQCD 18  & \cite{Bazavov:2018omf} & 2+1+1 & 2.89 -- 6.12  & 3.7 & \\
 \hline \\[-1.0ex]
HPQCD 14A & \cite{Chakraborty:2014aca} & 2+1+1 & 2.5 -- 5.8  & 3.7 & \\
[3.0ex] \hline \\[-1.0ex]
ETM 14& \cite{Carrasco:2014cwa} & 2+1+1 & 2.0 -- 3.0  & $2.7_{\pi^0}$ ($3.3_{\pi^\pm}$) & \\
[3.0ex] \hline \\[-1.0ex]
ETM 14A & \cite{Alexandrou:2014sha} & 2+1+1 & 2.0 -- 3.0 & $2.7_{\pi^0}$ ($3.3_{\pi^\pm}$) & \\
[1.0ex]
\fi
 \hline  \hline \\[-1.0ex]
\end{tabular*}
\caption{Finite-volume effects in the determinations of $m_c$ with $\Nf=2+1+1$ quark flavours.}
}
\end{table}

\begin{table}
{\footnotesize
\begin{tabular*}{\textwidth}[l]{l @{\extracolsep{\fill}} c c c c l}
\hline\hline  \\[-1.0ex]
Collab. & Ref. & $\Nf$ & $L$ [fm] & $M_{\pi,\text{min}} L$ & Description
\\[1.0ex] \hline \hline \\[-1.0ex]
ALPHA 21  & \cite{Heitger:2021apz} & 2+1 & 2.5 -- 4.0 & 4.2  & \\
[2.0ex] \hline \\[-1.0ex]
\\[-2.0ex]
Petreczky 19  & \cite{Petreczky:2019ozv} & 2+1 & 1.6 -- 5.2 & 4.2  &\\
[1.0ex]
\ifx\reducedapptables\undefined
[3.0ex] \hline \\[-1.0ex]
\\[-1.0ex]
Maezawa 16  & \cite{Maezawa:2016vgv} & 2+1 & 2.6 -- 5.2 & 4.2 & \\ 
[1.0ex] \hline \\[-1.0ex]
JLQCD 16 & \cite{Nakayama:2016atf}& 2+1 & 2.6 -- 2.8 & 4.4 & \\ 
 \hline \\[-1.0ex]
$\chi$QCD 14 & \cite{Yang:2014sea} &  2+1 &  2.8 & 4.1 & \\
[1.0ex] \hline \\[-1.0ex]
HPQCD 10 & \cite{McNeile:2010ji} & 2+1 & 2.3 -- 3.4 & 3.8 & \\
%
[1.0ex]
\fi
 \hline  \hline \\[-1.0ex]
\end{tabular*}
\caption{Finite-volume effects in the determinations of $m_c$ with $\Nf=2+1$ quark flavours.}
}
\end{table}

\begin{table}[!ht]
{\footnotesize
\begin{tabular*}{\textwidth}[l]{l @{\extracolsep{\fill}} c c l}
\hline\hline \\[-1.0ex]
Collab. & Ref. & $\Nf$  & Description
\\[1.0ex] \hline \hline \\[-1.0ex]
ETM 21A  & \cite{Alexandrou:2021gqw}  & 2+1+1 &  Nonperturbative (RI/MOM) \\
[2.0ex]\hline\\[-1.0ex]
HPQCD 20A  & \cite{Hatton:2020qhk}  & 2+1+1 & Nonperturbative (RI/SMOM) \\
[2.0ex]
\ifx\reducedapptables\undefined
[4.0ex]\hline\\[-1.0ex]
HPQCD 18  & \cite{Lytle:2018evc}  & 2+1+1 & {Nonperturbative (RI/SMOM).}\\
[1.0ex]\hline\\[-1.0ex]
FNAL/MILC/TUMQCD 18 & \cite{Bazavov:2018omf} & 2+1+1 & not required. \\
 \hline \\[-1.0ex]
HPQCD 14A & \cite{Chakraborty:2014aca} & 2+1+1 & Renormalization not required. \\
[1.0ex] \hline \\[-1.0ex]
ETM 14& \cite{Carrasco:2014cwa} & 2+1+1 & Nonperturbative renormalization (RI/MOM). \\
[1.0ex] \hline \\[-1.0ex]
ETM 14A & \cite{Alexandrou:2014sha} & 2+1+1 & Nonperturbative renormalization (RI/MOM). \\
[1.0ex]
\fi
 \hline  \hline \\[-1.0ex]
\end{tabular*}
\caption{Renormalization in the determinations of $m_c$ with $\Nf=2+1+1$ quark flavours.}
}
\end{table}

\begin{table}[!ht]
{\footnotesize
\begin{tabular*}{\textwidth}[l]{l @{\extracolsep{\fill}} c c l}
\hline\hline \\[-1.0ex]
Collab. & Ref. & $\Nf$  & Description
\\[1.0ex] \hline \hline \\[-1.0ex]
ALPHA 21  & \cite{Heitger:2021apz} & 2+1 & Schr\"odinger functional\\
[2.0ex] \hline \\[-1.0ex]
\\[-2.0ex]
Petreczky 19  & \cite{Petreczky:2019ozv} & 2+1  & not required \\
[2.0ex]
\ifx\reducedapptables\undefined
[3.0ex] \hline \\[-1.0ex]
\\[-1.0ex]
Maezawa 16  & \cite{Maezawa:2016vgv} & 2+1 & not required \\ [1.0ex] \hline
\\[-1.0ex]
JLQCD 16 & \cite{Nakayama:2016atf}& 2+1 &  not required \\  \hline
\\[-1.0ex]
$\chi$QCD 14 & \cite{Yang:2014sea} &  2+1 & Nonperturbative renormalization (RI/MOM). \\
[1.0ex] \hline \\[-1.0ex]
HPQCD 10 & \cite{McNeile:2010ji} & 2+1 &  Renormalization not required. \\
[1.0ex] \hline \\[-1.0ex]
HPQCD 08B & \cite{Allison:2008xk} & 2+1 & Renormalization not required. \\
[1.0ex]
\fi
 \hline  \hline \\[-1.0ex]
\end{tabular*}
\caption{Renormalization in the determinations of $m_c$ with $\Nf=2+1$ quark flavours.}
}
\end{table}


\clearpage

\begin{table}[!ht]
{\footnotesize
\begin{tabular*}{\textwidth}[l]{l @{\extracolsep{\fill}} c c c l}
\hline\hline \\[-1.0ex]
Collab. & Ref. & $\Nf$ & $a$ [fm] & Description
\\[1.0ex] \hline \hline \\[-1.0ex]
%
HPQCD 21  & \cite{Hatton:2021syc}  & 2+1+1 & \parbox[t]{3cm}{0.03, 0.042, 0.06, 0.09}  & \parbox[t]{4.5cm}{Scale set from $w_0$ and $f_\pi$. HISQ action for charm quarks.} \\
[4.0ex]
\ifx\reducedapptables\undefined
\hline\\[-1.0ex]
FNAL/MILC/TUMQCD 18  & \cite{Bazavov:2018omf}  & 2+1+1 & \parbox[t]{3cm}{0.03, 0.042, 0.06, 0.09, 0.12, 0.15}  & \parbox[t]{4.5cm}{Scale set from $f_\pi$. HISQ action for charm quarks.} \\
[4.0ex] \hline \\[-1.0ex]
Gambino 17 & \cite{Gambino:2017vkx} & 2+1+1 & 0.062, 0.082, 0.089  & \parbox[t]{4.5cm}{Scale set from $f_{\pi^+}$} \\
[1.0ex] \hline \\[-1.0ex]
ETM 16B & \cite{Bussone:2016iua} & 2+1+1 &  0.0619, 0.0815, 0.0885  & \parbox[t]{4.5cm}{Scale set from $f_{\pi}$} \\
 \hline \\[-1.0ex]
HPQCD 14B  & \cite{Colquhoun:2014ica} & 2+1+1 & 0.09, 0.12, 0.15 & \parbox[t]{4.5cm}{Scale set through the $\Upsilon^\prime-\Upsilon$ mass splitting.} \\
[4.0ex] \hline \\[-1.0ex]
ETM 14B & \cite{Bussone:2014cha} & 2+1+1 & 0.062, 0.082, 0.089 & \parbox[t]{4.5cm}{Scale set through $F_\pi$.}\\
[1.0ex] \hline \\[-1.0ex]
HPQCD 14A  & \cite{Chakraborty:2014aca} & 2+1+1 & 0.06, 0.09, 0.12, 0.15  & \parbox[t]{4.5cm}{Scale set through the Wilson flow parameter $w_0$.} \\
[4.0ex]
\fi
 \hline \hline \\[-1.0ex]
\end{tabular*}
\caption{Continuum extrapolations/estimation of lattice artifacts in the determinations of $m_b$ with $\Nf=2+1+1$ quark flavours.}
}
\end{table}

\begin{table}[!ht]
{\footnotesize
\begin{tabular*}{\textwidth}[l]{l @{\extracolsep{\fill}} c c c l}
\hline\hline \\[-1.0ex]
Collab. & Ref. & $\Nf$ & $a$ [fm] & Description
\\[1.0ex] \hline \hline \\[-1.0ex]
%
Petreczky 19  & \cite{Petreczky:2019ozv} & 2+1 & 0.025 to 0.11 fm & \parbox[t]{4.5cm}{HISQ action for the charm quark. Scale set from $r_1$ parameter of heavy quark potential and $f_\pi$.\\}
\\[4.0ex]
\ifx\reducedapptables\undefined \\ [1.0ex] \hline
\\[-1.0ex]
Maezawa 16  & \cite{Maezawa:2016vgv} & 2+1 & 0.04 to 0.14 fm & \parbox[t]{4.5cm}{HISQ action for the charm quark. Scale set from $r_1$ parameter of heavy quark potential.\\} \\ 
 \hline
\\[-1.0ex]
HPQCD 13B  & \cite{Lee:2013mla} & 2+1 & 0.084, 0.12 & \parbox[t]{4.5cm}{Scale set through the static-quark potential parameter $r_1$.} \\
[4.0ex] \hline \\[-1.0ex]
HPQCD 10 & \cite{McNeile:2010ji} & 2+1 & \parbox[t]{3cm}{0.044, 0.059, 0.084, 0.12, 0.15} & \parbox[t]{4.5cm}{Scale set through the static-quark potential parameter $r_1$.} \\
[4.0ex]
\fi
 \hline \hline \\[-1.0ex]
\end{tabular*}
\caption{Continuum extrapolations/estimation of lattice artifacts in the determinations of $m_b$ with $\Nf=2+1$ quark flavours.}
}
\end{table}

\begin{table}[!ht]
{\footnotesize
\begin{tabular*}{\textwidth}[l]{l @{\extracolsep{\fill}} c c c}
\hline \hline  \\[-1.0ex]
Collab. & Ref. & $\Nf$ & $M_{\pi,\text{min}}$ [MeV] 
\\[1.0ex] \hline \hline \\[-1.0ex]
%
HPQCD 21  & \cite{Hatton:2021syc}  & 2+1+1 & physical \\
[1.0ex]
\ifx\reducedapptables\undefined
[4.0ex]\hline\\[-1.0ex]
FNAL/MILC/TUMQCD 18  & \cite{Bazavov:2018omf} & 2+1+1 & $130$ ($173_{RMS}$) \\
[1.0ex] \hline \\[-1.0ex]
Gambino 17 & \cite{Gambino:2017vkx} & 2+1+1 & $210 $ \\
[1.0ex] \hline \\[-1.0ex]
ETM 16B & \cite{Bussone:2016iua} &  2+1+1 & $210 $ \\
 \hline \\[-1.0ex]
HPQCD 14B  & \cite{Colquhoun:2014ica} & 2+1+1 & 306,  128 \\
[1.0ex] \hline \\[-1.0ex]
ETM 14B & \cite{Bussone:2014cha} & 2+1+1 &  210  \\
[1.0ex] \hline \\[-1.0ex]
HPQCD 14A  & \cite{Chakraborty:2014aca} & 2+1+1 & $128_{\pi, 5}$ ($173_{RMS}$) \\
[1.0ex]
\fi
 \hline \hline \\[-1.0ex]
\end{tabular*}
\caption{Chiral extrapolation/minimum pion mass in the determinations of $m_b$ with $\Nf=2+1+1$ quark flavours.}
}
\end{table}

\begin{table}[!ht]
{\footnotesize
\begin{tabular*}{\textwidth}[l]{l @{\extracolsep{\fill}} c c c}
\hline \hline  \\[-1.0ex]
Collab. & Ref. & $\Nf$ & $M_{\pi,\text{min}}$ [MeV] 
\\[1.0ex] \hline \hline \\[-1.0ex]
Petreczky 19  & \cite{Petreczky:2019ozv} & 2+1 & 161 \\ [1.0ex] 
\ifx\reducedapptables\undefined
\hline
\\[-1.0ex]
Maezawa 16  & \cite{Maezawa:2016vgv} & 2+1 & 160   \\
[1.0ex]
 \hline \\[-1.0ex]
HPQCD 13B  & \cite{Lee:2013mla} & 2+1 & 345  \\
[1.0ex] \hline \\[-1.0ex]
HPQCD 10 & \cite{McNeile:2010ji} & 2+1 & 260  \\
[1.0ex]
\fi
 \hline \hline \\[-1.0ex]
\end{tabular*}
\caption{Chiral extrapolation/minimum pion mass in the determinations of $m_b$ with $\Nf=2+1$ quark flavours.}
}
\end{table}

\newpage

\begin{table}[!ht]
{\footnotesize
\begin{tabular*}{\textwidth}[l]{l @{\extracolsep{\fill}} c c c c}
\hline\hline  \\[-1.0ex]
Collab. & Ref. & $\Nf$ & $L$ [fm] & $M_{\pi,\text{min}} L$ 
\\[1.0ex] \hline \hline \\[-1.0ex]
%
HPQCD 21  & \cite{Hatton:2021syc}  & 2+1+1 &  1.9 - 5.76 & 3.8\\
[1.0ex]
\ifx\reducedapptables\undefined
[4.0ex]\hline\\[-1.0ex]
FNAL/MILC/TUMQCD 18  & \cite{Bazavov:2018omf} & 2+1+1 & 2.89 - 6.12  & 3.7 \\
[1.0ex] \hline \\[-1.0ex]
Gambino 17 & \cite{Gambino:2017vkx} & 2+1+1 & $2.7_{\pi^0}$ ($3.3_{\pi^\pm}$) \\
[1.0ex] \hline \\[-1.0ex]
ETM 16B & \cite{Bussone:2016iua}  & 2+1+1 & $2.7_{\pi^0}$ ($3.3_{\pi^\pm}$) \\
 \hline \\[-1.0ex]
HPQCD 14B  & \cite{Colquhoun:2014ica} & 2+1+1 & 2.4-7.8 & 3.0-3.8 \\
[1.0ex] \hline \\[-1.0ex]
ETM 14B & \cite{Bussone:2014cha} & 2+1+1 &  1.9-2.8 & 3.0-5.8   \\
[1.0ex] \hline \\[-1.0ex]
HPQCD 14A  & \cite{Chakraborty:2014aca} & 2+1+1 & 2.5-5.8  & 3.7  \\
[1.0ex]
\fi
 \hline \hline \\[-1.0ex]
\end{tabular*}
\caption{Finite-volume effects in the determinations of $m_b$ with $\Nf=2+1+1$ quark flavours.}
}
\end{table}

\begin{table}[!ht]
{\footnotesize
\begin{tabular*}{\textwidth}[l]{l @{\extracolsep{\fill}} c c c c}
\hline\hline  \\[-1.0ex]
Collab. & Ref. & $\Nf$ & $L$ [fm] & $M_{\pi,\text{min}} L$ 
\\[1.0ex] \hline \hline \\[-1.0ex]
%
Petreczky 19  & \cite{Petreczky:2019ozv} & 2+1 &  1.6 -- 5.2 & 4.2 \\ [1.0ex]
\ifx\reducedapptables\undefined
 \hline
\\[-1.0ex]
Maezawa 16  & \cite{Maezawa:2016vgv} & 2+1 & 2.6 -- 5.2 & 4.2  \\ 
[1.0ex]
 \hline \\[-1.0ex]
HPQCD 13B  & \cite{Lee:2013mla} & 2+1 & 2.4, 3.4 & 4.1  \\
[1.0ex] \hline \\[-1.0ex]
HPQCD 10 & \cite{McNeile:2010ji} & 2+1 & 2.3 -- 3.4 & 3.8  \\
[1.0ex] 
 \fi
\hline \hline \\[-1.0ex]
\end{tabular*}
\caption{Finite-volume effects in the determinations of $m_b$ with $\Nf=2+1$ quark flavours.}
}
\end{table}

\ifx\reducedapptables\undefined

\begin{table}[!ht]
{\footnotesize
\begin{tabular*}{\textwidth}[l]{l @{\extracolsep{\fill}} c c c c}
\hline\hline  \\[-1.0ex]
Collab. & Ref. & $\Nf$ & $L$ [fm] & $M_{\pi,\text{min}} L$ 
\\[1.0ex] \hline \hline \\[-1.0ex]
ETM 13B & \cite{Carrasco:2013zta} & 2 & $\gtrsim 2.0$ & 3.5   \\
[1.0ex] \hline \\[-1.0ex]
ALPHA 13C& \cite{Bernardoni:2013xba} & 2 & 2.3--3.6 & 4.1 \\
[1.0ex] \hline \\[-1.0ex]
ETM 11A & \cite{Dimopoulos:2011gx} & 2 & $\gtrsim 2.0$ & 3.5  \\
[1.0ex]
 \hline \hline \\[-1.0ex]
\end{tabular*}
\caption{Finite-volume effects in the determinations of $m_b$ with $\Nf=2$ quark flavours.}
}
\end{table}
\clearpage
\fi

\begin{table}[!ht]
{\footnotesize
\begin{tabular*}{\textwidth}[l]{l @{\extracolsep{\fill}} c c l}
\hline\hline \\[-1.0ex]
Collab. & Ref. & $\Nf$  & Description
\\[1.0ex] \hline \hline \\[-1.0ex]
%
%
ETM 21A  & \cite{Alexandrou:2021gqw}  & 2+1+1 &  Nonperturbative (RI/MOM) \\
[2.0ex]
\ifx\reducedapptables\undefined
\hline\\[-1.0ex]
HPQCD 21 & \cite{Hatton:2021syc}  & 2+1+1 &  Scale dependence enters through QED\\
[4.0ex]\hline\\[-1.0ex]
FNAL/MILC/TUMQCD 18 & \cite{Bazavov:2018omf}& 2+1+1 & Renormalization not required. \\
[1.0ex] \hline \\[-1.0ex]
Gambino 17 & \cite{Gambino:2017vkx}& 2+1+1 & \parbox[t]{4.5cm}{Nonperturbative renormalization (RI/MOM). }\\
[4.0ex] \hline \\[-1.0ex]
ETM 16B & \cite{Bussone:2016iua} & 2+1+1 & \parbox[t]{4.5cm}{Nonperturbative renormalization (RI/MOM). }\\
[4.0ex]
 \hline \\[-1.0ex]
HPQCD 14B & \cite{Colquhoun:2014ica} & 2+1+1 & \parbox[t]{4.5cm}{Renormalization not required.}\\
[1.0ex] \hline \\[-1.0ex]
ETM 14B& \cite{Bussone:2014cha} & 2+1+1 & \parbox[t]{4.5cm}{Nonperturbative renormalization (RI/MOM). }\\
[4.0ex] \hline \\[-1.0ex]
HPQCD 14A & \cite{Chakraborty:2014aca} & 2+1+1 & \parbox[t]{4.5cm}{Renormalization not required. }\\
[1.0ex] 
\fi
\hline \hline \\[-1.0ex]
\end{tabular*}
\caption{Lattice renormalization in the determinations of $m_b$ with $\Nf=2+1+1$ flavours.}
}
\end{table}

\begin{table}[!ht]
{\footnotesize
\begin{tabular*}{\textwidth}[l]{l @{\extracolsep{\fill}} c c l}
\hline\hline \\[-1.0ex]
Collab. & Ref. & $\Nf$  & Description
\\[1.0ex] \hline \hline \\[-1.0ex]
%
Petreczky 19  & \cite{Petreczky:2019ozv} & 2+1  & not required \\
[2.0ex]
\ifx\reducedapptables\undefined
 \hline \\[-1.0ex]
\\[-1.0ex]
Maezawa 16  & \cite{Maezawa:2016vgv} & 2+1 & Renormalization not required. \\ [1.0ex] 
\hline
\\[-1.0ex]
HPQCD 13B  & \cite{Lee:2013mla} & 2+1 & \parbox[t]{4.5cm}{Renormalization not required.} \\
[1.0ex] \hline \\[-1.0ex]
HPQCD 10 & \cite{McNeile:2010ji} & 2+1 &  \parbox[t]{4.5cm}{Renormalization not required.} \\
[1.0ex]
\fi
 \hline \hline \\[-1.0ex]
\end{tabular*}
\caption{Lattice renormalization in the determinations of $m_b$ with $\Nf=2+1$ flavours.}
}
\end{table}


\setcounter{subsection}{1}

\clearpage


\subsection{Notes to Sec.~\ref{sec:vusvud} on $|V_{ud}|$ and  $|V_{us}|$}
\label{app:VusVud}

\begin{table}[!h]
{\footnotesize
\begin{tabular*}{\textwidth}[l]{l @{\extracolsep{\fill}} c c c l}
\hline\hline \\[-1.0ex]
Collab. & Ref. & $\Nf$ & $a$ [fm] & Description
\\[1.0ex] \hline \hline \\[-2.0ex]
FNAL/MILC 18 &\cite{Bazavov:2018kjg} & 2+1+1 & 0.042, 0.06, 0.09, 0.12, 0.15&
\parbox[t]{5.9cm}{HISQ quark action. Relative scale through $r_1$.}
\\
[3.0ex] \hline \\[-2.5ex]
PACS 19 &\cite{Kakazu:2019ltq} & 2+1 & 0.085 &
\parbox[t]{5.9cm}{Nonperturbative $\cO(a)$ clover quark action.
Scale set from $\Xi$-baryon mass.}
\\
[3.0ex] \hline \\[-2.5ex]
\ifx\reducedapptables\undefined
{ETM 16}& \cite{Carrasco:2016kpy} & 2+1+1 & 0.062, 0.082, 0.089  &
		    \parbox[t]{5.9cm}{ 
		    Scale set through $f_\pi$. Automatic $\cO(a)$ improvement.} \\
[3.0ex] \hline  \\[-2.5ex]
FNAL/MILC 13E &\cite{Bazavov:2013maa}			&2+1+1&0.06, 0.09, 0.12, 0.15&
	\parbox[t]{5.9cm}{HISQ action for both sea and valence quarks. Relative scale through $r_1$, physical scale from 
	         pseudoscalar decay constants calculated with Asqtad fermions. The ensemble with $a \simeq 0.06$~fm 
	         is used only for cross-checking discretization effects.}\\
[17.0ex] \hline \\[-2.5ex]
%
%
JLQCD 17 &\cite{Aoki:2017spo}			&2+1&0.112&
	\parbox[t]{5.9cm}{Scale set through $\Omega$ mass.}\\
[0.0ex] \hline \\[-2.5ex]
RBC/UKQCD 15A	&\cite{Boyle:2015hfa}			&2+1&0.08, 0.11&
	\parbox[t]{5.9cm}{Scale set through $\Omega$ mass.}\\
[0.0ex] \hline \\[-2.5ex]
FNAL/MILC 12I		&\cite{Bazavov:2012cd}			&2+1&0.09, 0.12&
	\parbox[t]{5.9cm}{Relative scale $r_1$, physical scale determined from a 
		mixture of $f_\pi$, $f_K$, radial excitation of $\Upsilon$
		and $m_{D_s}-\frac 12 m_{ {\eta}_c}$.}\\
[6.0ex] \hline \\[-2.5ex]
%
%
%
%
%
%
ETM 09A		&\cite{Lubicz:2009ht}	&2  & 0.07, 0.09, 0.10&
		\parbox[t]{5.9cm}{
		Scale set through $f_\pi$. Automatic $\cO(a)$ impr., flavour symmetry
		breaking:\\ $(M_{PS}^{0})^2 - (M_{PS}^\pm)^2\sim \cO(a^2)$. \\
  		Three lattice spacings only for pion mass 470\,MeV.}\\
[12.0ex] 
\hline \\[-2.5ex]
%
%
%
%
\fi
\hline
%
\end{tabular*}
\caption{Continuum extrapolations/estimation of lattice artifacts in the
determinations of $f_+(0)$.}
}
\end{table}

\noindent
\begin{table}[!h]
{\footnotesize
\begin{tabular*}{\textwidth}[l]{l @{\extracolsep{\fill}} c c c l}
\hline \hline  \\[-1.0ex]
Collab. & Ref. & $\Nf$ & $M_{\pi,\text{min}}$ [MeV] & Description
\\[1.0ex] \hline \hline \\[-2.0ex]
FNAL/MILC 18 &\cite{Bazavov:2018kjg}	&2+1+1&$144_{\rm RMS}(135_{\rm \pi,5})$&
\parbox[t]{6.5cm}{Chiral interpolation through NLO $SU(3)$ PQ staggered {\Ch}PT
                  with continuum {\Ch}PT at NNLO. Lightest Nambu-Goldstone 
	          mass is 135~MeV and lightest RMS mass is 144~MeV
		  at the same gauge ensemble with $a \simeq 0.06$~fm.}
\\
[12.0ex] \hline \\[-2.5ex]
PACS 19 & \cite{Kakazu:2019ltq}	&2+1&  135 &
\parbox[t]{6.3cm}{Physical point simulation at a single pion mass 135~MeV.}
\\
[3.0ex] \hline \\[-2.5ex]
\ifx\reducedapptables\undefined
{ETM 16}& \cite{Carrasco:2016kpy} & 2+1+1 & $180_{\rm \pi^0}(220_{\pi^\pm})$  & 
          \parbox[t]{6.3cm}{Chiral extrapolation performed through $SU(2)$ or $SU(3)$ {\Ch}PT.}\\
[3.0ex] \hline  \\[-2.5ex]
FNAL/MILC 13E &\cite{Bazavov:2013maa}			&2+1+1&$173_{\rm RMS}(128_{\rm \pi,5})$&
		\parbox[t]{6.3cm}{NLO $SU(3)$ PQ staggered {\Ch}PT with
		        continuum {\Ch}PT at NNLO. Lightest Nambu-Goldstone 
		        mass is 128~MeV and lightest RMS mass is 173~MeV
			for the same gauge ensemble with $a \simeq 0.09$~fm.
			}\\
[12.0ex] \hline \\[-2.5ex]
%
%
JLQCD 17 &\cite{Aoki:2017spo}			&2+1&290&
		\parbox[t]{6.3cm}{NNLO $SU(3)$ {\Ch}PT.}\\
[3.0ex] \hline \\[-2.5ex]
RBC/UKQCD 15A	&\cite{Boyle:2015hfa}			&2+1&140&
		\parbox[t]{6.3cm}{NLO $SU(3)$ {\Ch}PT with phenomenological ansatz 
			for higher orders or polynomial models.}\\
[6.0ex] \hline \\[-2.5ex]
FNAL/MILC 12I		&\cite{Bazavov:2012cd}			&2+1&$378_{\rm RMS}(263_{\rm \pi,5})$&
		\parbox[t]{6.3cm}{NLO $SU(3)$ PQ staggered {\Ch}PT with either 
			phenomenological NNLO ansatz or NNLO {\Ch}PT.
			Lightest Nambu-Goldstone mass is 263~MeV with 
			$a=0.12$~fm and lightest RMS mass is 378~MeV
			with $a=0.09$~fm.
			}\\
[12.0ex] \hline \\[-2.5ex]
%
%
%
%
%
%
ETM 09A		&\cite{Lubicz:2009ht}	&2  &$210_{\pi^0}(260_{\pi^\pm})$& 
		\parbox[t]{6.3cm}{NLO heavy kaon $SU(2)$ {\Ch}PT and
		NLO $SU(3)$ {\Ch}PT and phenomenological ansatz for higher orders.}\\
[6.0ex] 
\hline \\[-2.5ex]
%
%
%
%
\fi
\hline
%
\end{tabular*}
\caption{Chiral extrapolation/minimum pion mass in determinations of $f_+(0)$.
The subscripts RMS and $\pi,5$ in the case of staggered fermions indicate
the root-mean-square mass and the Nambu-Goldstone boson mass, respectively. 
}
}
\end{table}

\noindent
\begin{table}[!h]
{\footnotesize
\begin{tabular*}{\textwidth}[l]{l @{\extracolsep{\fill}} c c c c l}
\hline \hline  \\[-1.0ex]
Collab. & Ref. & $\Nf$ &$L$ [fm]& $M_{\pi,\text{min}}L$ & Description
\\[1.0ex] \hline \hline \\[-1.5ex]
FNAL/MILC 18 &\cite{Bazavov:2018kjg}   &2+1+1&2.6--5.8&$4.2_{\rm RMS}(3.9_{\pi,5})$& \parbox[t]{5cm}{The values correspond to $M_{\pi,\rm RMS}=144$~MeV and 
		$M_{\pi,5}=135$~MeV, respectively.}
		\parbox[t]{5.2cm}{}\\
[6.5ex] \hline \\[-2.5ex]
PACS 19 &\cite{Kakazu:2019ltq}   &2+1& 10.9 & 7.5 & \parbox[t]{5cm}{}
		\parbox[t]{5.2cm}{}\\
[1.0ex] \hline \\[-2.5ex]
\ifx\reducedapptables\undefined
{ETM 16}& \cite{Carrasco:2016kpy} & 2+1+1 & 2.0--3.0 & $2.7_{\pi^0}(3.3_{\pi^\pm})$ &
        \parbox[t]{5.2cm}{FSE observed only in the slopes of the vector and scalar form factors.} \\
[3.0ex]\hline  \\[-2.5ex]
FNAL/MILC 13E &\cite{Bazavov:2013maa} 		&2+1+1&2.9--5.8&$4.9_{\rm RMS}(3.6_{\pi,5})$& \parbox[t]{5cm}{The values correspond to $M_{\pi,\rm RMS}=173$~MeV and 
		$M_{\pi,5}=128$~MeV, respectively.}
		\parbox[t]{5.2cm}{}\\
[6.0ex] \hline \\[-2.5ex]
%
%
JLQCD 17 &\cite{Aoki:2017spo} 	&2+1&1.8, 2.7&4.1& 
		\parbox[t]{5.2cm}{}\\
[0.0ex] \hline \\[-2.5ex]
RBC/UKQCD 15A	&\cite{Boyle:2015hfa} 	&2+1&2.6, 5.2&3.9& 
		\parbox[t]{5.2cm}{}\\
[0.0ex] \hline \\[-2.5ex]
FNAL/MILC 12I		&\cite{Bazavov:2012cd} 		&2+1&2.4--3.4&$6.2_{\rm RMS}(3.8_{\pi,5})$& \parbox[t]{5cm}{The values correspond to $M_{\pi,\rm RMS}=378$~MeV and 
		$M_{\pi,5}=263$~MeV, respectively.}
		\parbox[t]{5.2cm}{}\\
[6.0ex] \hline \\[-2.5ex]
%
%
%
%
%
%
ETM 09A		&\cite{Lubicz:2009ht}	&2  &2.1, 2.8&$3.0_{\pi^0}(3.7_{\pi^\pm})$& 
		\parbox[t]{5.2cm}{Two volumes at $M_\pi=300$\,MeV and {\Ch}PT-motivated 
		estimate of the error due to FSE.}\\
[6.0ex] 
\hline \\[-2.5ex]
%
%
%
%
\fi
\hline
%
\end{tabular*}
\caption{Finite-volume effects in determinations of $f_+(0)$. The subscripts RMS and $\pi,5$ in the case of staggered fermions indicate
the root-mean-square mass and the Nambu-Goldstone boson mass, respectively. 
}
}
\end{table}

\begin{table}[!h]
{\footnotesize
\begin{tabular*}{\textwidth}[l]{l @{\extracolsep{\fill}} c c c l}
\hline \hline  \\[-1.0ex]
Collab. & Ref. & $\Nf$ & $a$ [fm] & Description
\\[1ex] \hline \hline \\[-2.0ex]
{ETM 21} & \cite{Alexandrou:2021bfr} & 2+1+1 & 0.07, 0.08, 0.09 &
\parbox[t]{5.2cm}{Wilson-clover twisted mass quark action. Relative scale through gradient flow scale $w_0$ and absolute scale though $f_\pi$.}\\
[9.0ex] \hline \\[-2.0ex]
{CalLat 20} & \cite{Miller:2020xhy} & 2+1+1 & 0.06, 0.09, 0.12, 0.15&
\parbox[t]{5.2cm}{M\"obius domain-wall valence quarks on gradient-flowed HISQ ensembles.  Relative scale through the gradient flow scale $w_0$.}\\
[9.0ex] \hline \\[-2.5ex]
\ifx\reducedapptables\undefined
{FNAL/MILC 17}& \cite{Bazavov:2017lyh} & 2+1+1 & 0.03, 0.042, 0.06,\,0.09,\,0.12,\,0.15  &  
	\parbox[t]{5.2cm}{ 
	HISQ action for both valence and sea quarks. Absolute scale though $f_\pi$.}\\
[3.0ex]  \hline  \\[-2.5ex]
{ETM 14E}& \cite{Carrasco:2014poa} & 2+1+1 & 0.062, 0.082, 0.089  &
		    \parbox[t]{5.2cm}{ 
		    Scale set through $f_\pi$. Automatic $\cO(a)$ improvement, flavour symmetry
		    breaking: $(M_{PS}^{0})^2-(M_{PS}^\pm)^2\sim \cO(a^2)$. Discretization and volume effects
		    due to the $\pi^0 - \pi^\pm$ mass splitting are taken into account through {\Ch}PT for
		    twisted-mass fermions.} \\
[17.0ex] \hline  \\[-2.5ex]
%
%
\hline  \\[-2.5ex]
%
%
%
\fi
\hline 
\end{tabular*}
\caption{Continuum extrapolations/estimation of lattice artifacts in
  determinations of $f_K/f_\pi$ for $N_f=2+1+1$ simulations.}
}
\end{table}

\ifx\reducedapptables\undefined
\begin{table}[!h]
{\footnotesize
\begin{tabular*}{\textwidth}[l]{l @{\extracolsep{\fill}} c c c l}
\hline \hline  \\[-1.0ex]
 
Collab. & Ref. & $\Nf$ & $a$ [fm] & Description
\\[1ex] \hline \hline \\[-2.5ex]
QCDSF/UKQCD 16  &\cite{Bornyakov:2016dzn}		&2+1&
		\parbox[t]{3.5cm}{\centering 0.059, 0.068, 0.074, 0.082}&
		\parbox[t]{5.7cm}{Scale set through $M_N$. Nonperturbative
		$\cO(a)$ clover improvement.}\\
[3.0ex] \hline \\[-2.5ex]
D\"urr 16  	&\cite{Durr:2016ulb,Scholz:2016kcr}		&2+1&
 		\parbox[t]{3.5cm}{\centering 0.054, 0.065, 0.077, \\ 0.092, 0.12}&
 		\parbox[t]{5.7cm}{Scale set through $M_\Omega$. Perturbative
 		$\cO(a)$-improvement.}\\
[3.0ex] \hline \\[-2.5ex]
 
{RBC/UKQCD 14B}& \cite{Blum:2014tka} & 2+1 & 0.063,\, 0.085,\, 0.114  &  
 	\parbox[t]{5.7cm}{ 
	Scale set through $m_\Omega$.}\\
\hline  \\[-2.5ex]

{RBC/UKQCD 12}& \cite{Arthur:2012opa} & 2+1 & 0.09,\, 0.11,\, 0.14  &  
	\parbox[t]{5.7cm}{ 
	Scale set through $m_\Omega$.}\\
\hline  \\[-2.5ex]

{Laiho 11}& \cite{Laiho:2011np} & 2+1 & 0.06, 0.09, 0.125 &  
	\parbox[t]{5.7cm}{Scale set through $r_1$ and $\Upsilon$ and 
	continuum extrapolation based on MA{\Ch}PT.}\\
[3.0ex] \hline  \\[-2.5ex]

{JLQCD/TWQCD 10}& \cite{Noaki:2010zz} & 2+1 & 0.112  & Scale set
through $M_\Omega$. \\
\hline  \\[-2.5ex]

{RBC/UKQCD 10A}& \cite{Aoki:2010dy} & 2+1 & 0.087, 0.114 & Scale set
through $M_\Omega$. \\
\hline  \\[-2.5ex]
 
MILC 10 & \cite{Bazavov:2010hj} & 2+1 & 0.045, 0.06, 0.09  & \parbox[t]{5.7cm}{3 lattice
spacings, continuum extrapolation by means of RS{\Ch}PT.}\\
[3.0ex] \hline  \\[-2.5ex]
BMW 10		&\cite{Durr:2010hr}		&2+1&
 		\parbox[t]{2.5cm}{\centering 0.07, 0.08,\\ 0.12}&
 		\parbox[t]{5.7cm}{Scale set through $M_{\Omega,\Xi}$. Perturbative
 		$\cO(a)$-improvement.}\\
[3.0ex] \hline \\[-2.5ex]
 
JLQCD/TWQCD 09A &\cite{Noaki:2008iy}	&2+1& 0.1184(3)(21) & 
		\parbox[t]{5.7cm}{Scale set through $F_\pi$. Automatic 
		$\cO(a)$-improvement due to chiral symmetry of action.}\\[6.0ex]
\hline \\[-2.5ex]

{PACS-CS 09}& \cite{Aoki:2009ix} & 2+1 & 0.0900(4)  & Scale set
through $M_\Omega$. \\
\hline  \\[-2.5ex]
 
MILC 09A	&\cite{Bazavov:2009fk}	&2+1& 0.045, 0.06, 0.09 & 
		\parbox[t]{5.7cm}{Scale set through $r_1$ and $\Upsilon$ and 
		continuum extrapolation based on RS{\Ch}PT.}\\
[3.0ex]  \hline \\[-2.5ex]
 
MILC 09		&\cite{Bazavov:2009bb}	&2+1& 	\parbox[t]{2.5cm}{\centering 0.045, 0.06, \\0.09, 0.12} & 
		\parbox[t]{5.7cm}{Scale set through $r_1$ and $\Upsilon$ and 
		continuum extrapolation based on RS{\Ch}PT.}\\
[3.0ex] \hline \\[-2.5ex]
 
Aubin 08	&\cite{Aubin:2008ie}	&2+1& 0.09, 0.12 & 	
		\parbox[t]{5.7cm}{Scale set through $r_1$ and $\Upsilon$ and 
		continuum extrapolation based on MA{\Ch}PT.}\\
[3.0ex]  \hline \\[-2.5ex]
 
PACS-CS 08, 08A &\cite{Aoki:2008sm,Kuramashi:2008tb}& 2+1 & 0.0907(13) & 
		\parbox[t]{5.7cm}{Scale set 
		through $M_\Omega$. Nonperturbatively $\cO(a)$-improved. }\\
[3.0ex]  \hline \\[-2.5ex]
 
HPQCD/UKQCD 07	&\cite{Follana:2007uv}	&2+1& 0.09, 0.12, 0.15 & 
 		\parbox[t]{5.7cm}{Scale set through $r_1$ and $\Upsilon$ and 
 		continuum extrapolation on continuum-{\Ch}PT motivated ansatz. Taste
 		breaking of sea quarks ignored.}\\
[9.0ex]  
\hline \\[-2.5ex]
 
RBC/UKQCD 08	&\cite{Allton:2008pn}	&2+1& 0.114(2) & 
		\parbox[t]{5.7cm}{Scale set through $M_\Omega$. Automatic 
		$\cO(a)$-improvement due to appoximate chiral symmetry. 
		$(\Lambda_{\rm QCD}a)^2\approx 4\%$ systematic error
		due to lattice artifacts added.}\\
[9.0ex]  \hline \\[-2.5ex]
 
NPLQCD 06	&\cite{Beane:2006kx}	&2+1& 0.125 & 
		\parbox[t]{5.7cm}{Scale set through $r_0$ and $F_\pi$. Taste breaking
		of sea quarks ignored.}\\
[3.0ex]  \hline \\[-2.5ex]
 
MILC 04	&\cite{Aubin:2004fs}	 &2+1& 0.09, 0.12 & \parbox[t]{5.7cm}{Scale
   set through $r_1$ and $\Upsilon$ and continuum extrapolation based on RS{\Ch}PT.} \\
[3.0ex] %
\hline
\hline
\end{tabular*}
\caption{Continuum extrapolations/estimation of lattice artifacts in
  determinations of $f_K/f_\pi$ for $N_f=2+1$ simulations.}
}
\end{table}
\fi

\ifx\reducedapptables\undefined
\begin{table}[!h]
{\footnotesize
\begin{tabular*}{\textwidth}[l]{l @{\extracolsep{\fill}} c c c l}
\hline \hline  \\[-1.0ex]
Collab. & Ref. & $\Nf$ & $a$ [fm] & Description
\\[1ex] \hline \hline \\[-1.5ex]
%
%
%
%
%
ETM 09		&\cite{Blossier:2009bx}		&2  & 0.07, 0.09, 0.10 & 
		\parbox[t]{5.7cm}{Scale set through $F_\pi$.
		Automatic $\cO(a)$ impr., flavour symmetry
		breaking: $(M_{PS}^{0})^2-(M_{PS}^\pm)^2\sim \cO(a^2)$. 
		  }\\
[6.0ex] 
\hline \\[-2.5ex]
%
%
\hline
%
\end{tabular*}
\caption{Continuum extrapolations/estimation of lattice artifacts in
  determinations of $f_K/f_\pi$ for $N_f=2$ simulations.}
}
\end{table}
\fi

\vfill
\noindent
\begin{table}[!h]
{\footnotesize
\begin{tabular*}{\textwidth}[l]{l @{\extracolsep{\fill}} c c c l}
\hline \hline  \\[-1.0ex]
Collab. & Ref. & $\Nf$ & $M_{\pi,\text{min}}$ [MeV] & Description
\\[1ex] \hline \hline \\[-2.0ex]
{ETM 21}& \cite{Alexandrou:2021bfr} & 2+1+1 & 134 & 
          \parbox[t]{6.2cm}{ 
            Chiral extrapolation based on NLO $SU(2)$ {\Ch}PT.
          }\\
[3.0ex]\hline  \\[-2.0ex]
{CalLat 20}& \cite{Miller:2020xhy} & 2+1+1 & 157 & 
          \parbox[t]{6.2cm}{ 
            Chiral extrapolation based on NNLO $SU(3)$ {\Ch}PT.
            We quote the root-mean-square (RMS) mass of the valence and
            valence-sea pions as $M_{\pi,\text{min}}$.
            The smallest mass is 176~MeV for the HISQ sea pions,
            which do not enter until NNLO in the {\Ch}PT expression.
          }\\
[17.0ex]\hline  \\[-2.5ex]
\ifx\reducedapptables\undefined
{FNAL/MILC 17}& \cite{Bazavov:2017lyh} & 2+1+1 & $143_{\rm RMS}(128_{\pi,5})$  & 
          \parbox[t]{6.2cm}{ 
	Linear interpolation to physical point. The lightest RMS mass is from the 
	$a=0.06$~fm ensemble and the lightest Nambu-Goldstone mass is from the
	$a=0.09$~fm ensemble.}\\
[9.0ex]\hline  \\[-2.5ex]
{ETM 14E}& \cite{Carrasco:2014poa} & 2+1+1 & $180_{\rm \pi^0}(220_{\pi^\pm})$  & 
          \parbox[t]{6.2cm}{Chiral extrapolation performed through $SU(2)$ {\Ch}PT or polynomial fit.}\\
[3.0ex]\hline  \\[-2.5ex]
%
%
{HPQCD 13A}& \cite{Dowdall:2013rya} & 2+1+1 &$173_{\rm RMS}(128_{\pi,5})$  &  
	\parbox[t]{6.2cm}{ 
	NLO {\Ch}PT supplemented by model for NNLO. Both the lightest RMS and 
	the lightest
	Nambu-Goldstone mass are from the $a=0.09$~fm ensemble.}\\
[9.0ex] 
\hline  \\[-2.5ex]
%
%
%
%
\fi
\hline
%
\end{tabular*}
\caption{Chiral extrapolation/minimum pion mass in determinations of $f_K / f_\pi$ for $N_f=2+1+1$ simulations. 
}
}
\end{table}

\ifx\reducedapptables\undefined
\noindent
\vspace{-3cm}
\begin{table}[!h]
{\footnotesize
\begin{tabular*}{\textwidth}[l]{l @{\extracolsep{\fill}} c c c l}
\hline \hline  \\[-1.0ex]
Collab. & Ref. & $\Nf$ & $M_{\pi,\text{min}}$ [MeV] & Description
\\[1ex] \hline \hline \\[-1.5ex]
QCDSF/UKQCD 16  &\cite{Bornyakov:2016dzn}		&2+1& 220 & 
		\parbox[t]{6.2cm}{Expansion around the $SU(3)$ symmetric point 
		$m_u = m_d = m_s = (m_u + m_d + m_s)^{phys} / 3$.}\\
[3.0ex] \hline \\[-2.5ex]
D\"urr 16   	&\cite{Durr:2016ulb,Scholz:2016kcr}			&2+1& 130 & 
		\parbox[t]{6.2cm}{Comparison between $SU(3)$ {\Ch}PT 
		and polynomial fit-ans\"atze. }\\
[3.0ex] \hline \\[-2.5ex]
{RBC/UKQCD 14B}& \cite{Blum:2014tka} & 2+1 & 139 &\parbox[t]{6.2cm}{
		NLO PQ $SU(2)$ {\Ch}PT as well as analytic ans\"atze.}\\
[3.0ex] \hline \\[-2.5ex]
%
%
%
%
%
MILC 10       & \cite{Bazavov:2010hj}  & 2+1 & $258_{\rm RMS}(177_{\pi,5})$ &  \parbox[t]{6.2cm}{Lightest
Nambu-Goldstone mass is 177\,MeV (at 0.09\,fm) and
 lightest RMS mass is 258\,MeV (at 0.06\,fm). NLO rS{\Ch}PT and NNLO {\Ch}PT.} \\
[9ex]\hline \\[-2.5ex]
BMW 10		&\cite{Durr:2010hr}		&2+1& 190 & 
		\parbox[t]{6.2cm}{Comparison of various fit-ans\"atze:
		$SU(3)$ {\Ch}PT, heavy kaon $SU(2)$ {\Ch}PT, polynomial. }\\
[3.0ex] \hline \\[-2.5ex]
%
%
%
%
HPQCD/UKQCD 07	&\cite{Follana:2007uv}	&2+1&$375_{\rm RMS}(263_{\pi,5})$& 
		\parbox[t]{6.2cm}{NLO $SU(3)$ chiral perturbation theory
		with NNLO and NNNLO analytic terms. 
		The lightest RMS mass is from the $a=0.09$~fm ensemble and 
		the lightest Nambu-Goldstone mass is from the 
		$a=0.12$~fm ensemble.}\\
[12.0ex] 
\hline \\[-2.5ex]
\hline\\
\end{tabular*}
\vskip -0.5cm
\caption{Chiral extrapolation/minimum pion mass in determinations of $f_K/f_\pi$ for $N_f=2+1$ simulations. The subscripts RMS and $\pi,5$ in the case of staggered fermions indicate
the root-mean-square mass and the Nambu-Goldstone boson mass. In the case
of twisted-mass fermions $\pi^0$ and $\pi^\pm$ indicate the neutral and
charged pion mass and where applicable, ``val'' and ``sea'' indicate valence
and sea pion masses.}
}
\end{table}
\fi

\ifx\reducedapptables\undefined
\noindent
\vspace{-3cm}
\begin{table}[!h]
\addtocounter{table}{-1}
{\footnotesize
\begin{tabular*}{\textwidth}[l]{l @{\extracolsep{\fill}} c c c l}
\hline \hline  \\[-1.0ex]
Collab. & Ref. & $\Nf$ & $M_{\pi,\text{min}}$ [MeV] & Description
\\[1ex] \hline \hline \\[-2.5ex]
Aubin 08	&\cite{Aubin:2008ie}	&2+1&$329_{\rm RMS}(246_{\pi,5})$& 
		\parbox[t]{6.2cm}{NLO MA{\Ch}PT. According to \cite{Bazavov:2009fk}
		the lightest sea Nambu-Goldstone mass is 246\,MeV (at $a=0.09$ fm)
		and the lightest RMS mass is 329\,MeV (at $a=0.09$ fm).}\\
[9.0ex] \hline \\[-2.5ex]
 
PACS-CS 08, 08A &\cite{Aoki:2008sm,Kuramashi:2008tb}& 2+1 &156& 
		\parbox[t]{6.2cm}{NLO $SU(2)$ {\Ch}PT and $SU(3)$ (Wilson){\Ch}PT.}\\
\hline \\[-2.5ex]
 
HPQCD/UKQCD 07	&\cite{Follana:2007uv}	&2+1&$375_{\rm RMS}(263_{\pi,5})$& 
		\parbox[t]{6.2cm}{NLO $SU(3)$ chiral perturbation theory
		with NNLO and NNNLO analytic terms. 
		The lightest RMS mass is from the $a=0.09$~fm ensemble and 
		the lightest Nambu-Goldstone mass is from the 
		$a=0.12$~fm ensemble.}\\
[12.0ex] \hline \\[-2.5ex]
RBC/UKQCD 08	&\cite{Allton:2008pn}	&2+1&$330_{\rm sea}$, $242_{\rm val}$& 
		\parbox[t]{6.2cm}{While $SU(3)$ PQ{\Ch}PT fits were studied,
		final results are based on 
		heavy kaon NLO $SU(2)$ PQ{\Ch}PT. }\\
[6.0ex] \hline \\[-2.5ex]
 
NPLQCD 06	&\cite{Beane:2006kx}	&2+1&300& 
		\parbox[t]{6.2cm}{NLO $SU(3)$ {\Ch}PT and some NNLO terms. 
		The sea RMS mass for the employed lattices is heavier.}\\
[6.0ex] \hline \\[-2.5ex]
 
MILC 04	&\cite{Aubin:2004fs}	 &2+1& $400_{\rm RMS}(260_{\pi,5})$ & \parbox[t]{6.2cm}{PQ RS{\Ch}PT fits. The lightest sea Nambu-Goldstone mass is 260\,MeV
		(at $a=0.12$ fm) and the lightest RMS mass is 400\,MeV (at $a=0.09$ fm).} \\
[9ex]\hline\hline\\

\end{tabular*}
\vskip -0.5cm
\caption{(cntd.) Chiral extrapolation/minimum pion mass in determinations of $f_K/f_\pi$ for $N_f=2+1$ simulations. The subscripts RMS and $\pi,5$ in the case of staggered fermions indicate
the root-mean-square mass and the Nambu-Goldstone boson mass. In the case
of twisted-mass fermions $\pi^0$ and $\pi^\pm$ indicate the neutral and
charged pion mass and where applicable, ``val'' and ``sea'' indicate valence
and sea pion masses.}
}
\end{table}
\fi

\ifx\reducedapptables\undefined
\begin{table}[!h]
{\footnotesize
\begin{tabular*}{\textwidth}[l]{l @{\extracolsep{\fill}} c c c l}
\hline \hline  \\[-1.0ex]
Collab. & Ref. & $\Nf$ & $M_{\pi,\text{min}}$ [MeV] & Description
\\[1ex] \hline \hline \\[-2.5ex]
%
\\
%
%
%
%
ETM 09		&\cite{Blossier:2009bx}		&2  &$210_{\pi^0}(260_{pi^\pm})$& 
		\parbox[t]{6.2cm}{NLO heavy meson $SU(2)$ {\Ch}PT and NLO
		$SU(3)$ {\Ch}PT.}\\
[4.0ex] \hline \\[-2.5ex]
%
%
\hline\\
\end{tabular*}
\vskip -0.5cm
\caption{Chiral extrapolation/minimum pion mass in determinations of $f_K/f_\pi$ for $N_f=2$ simulations. The subscripts RMS and $\pi,5$ in the case of staggered fermions indicate
the root-mean-square mass and the Nambu-Goldstone boson mass. In the case
of twisted-mass fermions $\pi^0$ and $\pi^\pm$ indicate the neutral and
charged pion mass and where applicable, ``val'' and ``sea'' indicate valence
and sea pion masses.}
}
\end{table}
\fi

\noindent
\begin{table}[!h]
{\footnotesize
\begin{tabular*}{\textwidth}[l]{l @{\extracolsep{\fill}} c c c c l}
\hline \hline  \\[-1.0ex]
Collab. & Ref. & $\Nf$ &$L$ [fm]& $M_{\pi,\text{min}}L$ & Description
\\[1ex] \hline \hline \\[-2.0ex]
{ETM 21}& \cite{Alexandrou:2021bfr} & 2+1+1 &2.0--5.6 & 3.8 &  
\parbox[t]{4.7cm}{Three different volumes at $M_\pi=253$~MeV
and $a=0.08$~fm.
}\\
[3.0ex] \hline  \\[-2.0ex]
{CalLat 20}& \cite{Miller:2020xhy} & 2+1+1 &2.4--7.2 & 3.8 &  
\parbox[t]{4.7cm}{Three different volumes at $M_\pi=220$~MeV
and $a=0.12$~fm.
}\\
[3.0ex] \hline  \\[-2.5ex]
\ifx\reducedapptables\undefined
{FNAL/MILC 17}& \cite{Bazavov:2017lyh} & 2+1+1 &2.4--6.1 & $3.9_{\rm RMS}(3.7_{\pi,5}) $ &  
	\parbox[t]{4.7cm}{ }\\
[0.0ex] \hline  \\[-2.5ex]
{ETM 14E}& \cite{Carrasco:2014poa} & 2+1+1 & 2.0 -- 3.0 & $2.7_{\pi^0}(3.3_{\pi^\pm})$ &
        \parbox[t]{4.7cm}{FSE for the pion is corrected through resummed NNLO {\Ch}PT for twisted-mass fermions,
        which takes into account the effects due to the $\pi^0 - \pi^\pm$ mass splitting.} \\
[12.0ex] \hline \\[-2.5ex]
%
	\parbox[t]{4.7cm}{ }\\
%
{HPQCD 13A}& \cite{Dowdall:2013rya} & 2+1+1 &2.5--5.8& $4.9_{\rm RMS}(3.7_{\pi,5})$  &  
	\parbox[t]{4.7cm}{ }\\
[1.0ex] \hline  \\[-2.5ex]
%
%
%
\fi
\hline\\
\end{tabular*}
\caption{Finite-volume effects in determinations of $f_K/f_\pi$ for $N_f=2+1+1$.
}
}
\end{table}

\ifx\reducedapptables\undefined
\begin{table}[!h]
{\footnotesize
\begin{tabular*}{\textwidth}[l]{l @{\extracolsep{\fill}} c c c c l}
\hline \hline  \\[-1.0ex]
Collab. & Ref. & $\Nf$ &$L$ [fm]& $M_{\pi,\text{min}}L$ & Description
\\[1ex] \hline \hline \\[-2.5ex]
QCDSF/UKQCD 16  &\cite{Bornyakov:2016dzn}	&2+1& 
{2.0--2.8} &3.0& 
		\parbox[t]{4.5cm}{...}\\
[3.0ex] \hline \\[-2.5ex]
D\"urr 16  	&\cite{Durr:2016ulb,Scholz:2016kcr}		&2+1& 
{1.5--5.5} &3.85& 
		\parbox[t]{4.5cm}{Various volumes for comparison and
		corrections for FSE from NLO {\Ch}PT with re-fitted coefficients.}\\
[9.0ex] \hline \\[-2.5ex]
\ifx\reducedapptables\undefined
{RBC/UKQCD 14B}& \cite{Blum:2014tka} & 2+1 & 2.0,\,2.7,\,4.6,\,5.4 &  $3.8$ & 
\\
[0.0ex] \hline \\[-2.5ex]
%
%
%
%
%
MILC 10       & \cite{Bazavov:2010hj}                 & 2+1 & 2.5-3.8 &
$7.0_{\rm RMS}(4.0_{\pi,5})$  & \parbox[t]{4.5cm}{$L\!\geq\!2.9\,\fm$ for the lighter masses.}\\
[0.0ex] \hline \\[-2.5ex]
BMW 10		&\cite{Durr:2010hr}		&2+1& 
{2.0--5.3} &4.0& 
		\parbox[t]{4.5cm}{Various volumes for comparison and
		correction for FSE from {\Ch}PT using \cite{Colangelo:2005gd}.}\\
[6.0ex] \hline \\[-2.5ex]
%
%
%
%
%
%
%
HPQCD/UKQCD 07	&\cite{Follana:2007uv}	&2+1&
{2.4--2.9}&$4.1_{\rm RMS}(3.8_{\pi,5})$& 
		\parbox[t]{4.5cm}{Correction for FSE from {\Ch}PT using 
		\cite{Colangelo:2005gd}.}\\
[3.0ex] \hline \\[-2.5ex]
%
%
%
%
%
%
%
%
%
ETM 09		&\cite{Blossier:2009bx}		&2  &2.0--2.7&$3.0_{\pi^0}(3.7_{\pi^\pm})$& 
		\parbox[t]{4.5cm}{Correction for FSE from {\Ch}PT 
		\cite{Gasser:1986vb,Gasser:1987ah,Colangelo:2005gd}.}\\
[3.0ex] 
\hline \\[-2.5ex]
%
%
\fi
\hline
\end{tabular*}
\caption{Finite-volume effects in determinations of $f_K/f_\pi$ for
  $N_f=2+1$ and $N_f=2$.
The subscripts RMS and $\pi,5$ in the case of staggered fermions indicate
the root-mean-square mass and the Nambu-Goldstone boson mass. In the case
of twisted-mass fermions $\pi^0$ and $\pi^\pm$ indicate the neutral and
charged pion mass and where applicable, ``val'' and ``sea'' indicate valence
and sea pion masses.}
}
\end{table}
\fi


\clearpage
\subsection{Notes to section \ref{sec:LECs} on low-energy constants}


\begin{table}[!htbp]
{\footnotesize
\begin{tabular*}{\textwidth}[l]{l l @{\extracolsep{\fill}} l c l}
\hline\hline
\\[-1.2ex]
Collab. & Ref. & $\Nf$ & $a\,[\rm{fm}]$ & Description
\\[1.0ex]
\hline \hline
\\[-1.2ex]
ETM 21A      & \cite{Alexandrou:2021gqw} & 2+1+1 & 0.095, 0.082, 0.069       & Scale set by $f_\pi$ = 130.4(2) MeV.
\\[2.0ex] \hline \\[-1.0ex]
ETM 21       & \cite{Alexandrou:2021bfr} & 2+1+1 & 0.092, 0.080, 0.068       & Scale set by $f_\pi$ = 130.4(2) MeV.
\\[2.0ex] \hline \\[-1.0ex]
$\chi$QCD 21 & \cite{Liang:2021pql}      & 2+1   & 0.063, 0.071, 0.084, 0.114 & Same configs.~as RBC/UKQCD 15E.
\\[2.0ex] \hline \\[-1.0ex]
Wang 16      & \cite{Wang:2016lsv}       & 2+1   & 0.113                   & Same configs.\ as RBC/UKQCD 08A.\\
\\[2.0ex] \hline \\[-1.0ex]
ETM 20A      & \cite{Fischer:2020fvl}    & 2     & 0.0914(15)              & Single lattice spacing.
\\[2.0ex]
\hline \hline
\end{tabular*}
}
\caption{Continuum extrapolations/estimation of lattice artifacts in
  determinations of the $SU(2)$ low-energy constants $\Sigma, F, \bar \ell_4, \bar \ell_6$, and $SU(3)$ low-energy constants $\Sigma_0, F_0$.
}
\end{table}

\begin{table}[!htbp]
{\footnotesize
\begin{tabular*}{\textwidth}[l]{l l @{\extracolsep{\fill}} l c l}
\hline\hline
\\[-1.2ex]
Collab. & Ref. & $\Nf$ & $a\,[\rm{fm}]$ or $a^{-1}\,[\rm{GeV}]$ & Description
\\[1.0ex]
\hline \hline
\\[-1.2ex]
Gao 21       & \cite{Gao:2021xsm}        & 2+1   & 0.04, 0.06, 0.076         & One lattice spacing at phys. pt.
\\[2.0ex] \hline \\[-1.0ex]
$\chi$QCD 20 & \cite{Wang:2020nbf}       & 2+1   & 0.083--0.195             & One lattice spacing below 0.1~fm.
\\[2.0ex] \hline \\[-1.0ex]
Feng 19      & \cite{Feng:2019geu}       & 2+1   & 1.015, 1.378, 1.730       & $a>0.1$~fm.
\\[2.0ex]
\hline \hline
\end{tabular*}
}
\caption{Continuum extrapolations/estimation of lattice artifacts in
  determinations of the low-energy constants related to the vector form factor of the pion.
}
\end{table}

\begin{table}[!htbp]
{\footnotesize
\begin{tabular*}{\textwidth}[l]{l l @{\extracolsep{\fill}} l c l}
\hline\hline
\\[-1.2ex]
Collab. & Ref. & $\Nf$ & $a\,[\rm{fm}]$ & Description
\\[1.0ex]
\hline \hline
\\[-1.2ex]
ETM 20B      & \cite{Fischer:2020jzp}    &  2    & 0.0914(15)              & Single lattice spacing.
\\[2.0ex] \hline \\[-1.0ex]
Mai 19       & \cite{Mai:2019pqr}        &  2    & 0.12                    & Single lattice spacing.
\\[2.0ex] \hline \\[-1.0ex]
Culver 19    & \cite{Culver:2019qtx}     &  2    & 0.12                    & Single lattice spacing.
\\[2.0ex]
\hline \hline
\end{tabular*}
}
\caption{Continuum extrapolations/estimation of lattice artifacts in
  determinations of the low-energy constants related to $\pi\pi$ scattering.
}
\end{table}

\begin{table}[!htbp]
{\footnotesize
\begin{tabular*}{\textwidth}[l]{l l @{\extracolsep{\fill}} l c l}
\hline\hline
\\[-1.2ex]
Collab. & Ref. & $\Nf$ & $a\,[\rm{fm}]$ & Description
\\[1.0ex]
\hline \hline
\\[-1.2ex]
ETM 18B      & \cite{Helmes:2018nug}     & 2+1+1 &  0.089, 0.082, 0.062      & Same configuration with ETM 17G.
\\[2.0ex] \hline \\[-1.0ex]
ETM 17G      & \cite{Helmes:2017smr}     & 2+1+1 &  0.089, 0.082, 0.062      & Scale set by the Sommer parameter $r_0$.
\\[2.0ex] \hline \\[-1.0ex]
PACS-CS 13   & \cite{Sasaki:2013vxa}     & 2+1   &  0.09                   & Single lattice spacing.
\\[2.0ex] \hline \\[-1.0ex]
Fu 11A       & \cite{Fu:2011wc}          & 2+1   &  0.15                   & Single lattice spacing.
\\[2.0ex] \hline \\[-1.0ex]
NPLQCD 07B   & \cite{Beane:2007uh}       & 2+1   &  0.09, 0.125             & Configurations generated by MILC. 
\\[2.0ex] \hline \\[-1.0ex]
NPLQCD 06B   & \cite{Beane:2006gj}       & 2+1   &  0.125                  & Single lattice spacing.
\\[2.0ex]
\hline
\hline
\end{tabular*}
}
\caption{Continuum extrapolations/estimation of lattice artifacts in
  determinations of the low-energy constants related to $\pi K$ scattering.
}
\end{table}


\begin{table}[!htbp]
{\footnotesize
\begin{tabular*}{\textwidth}[l]{l l @{\extracolsep{\fill}} l c l}
\hline\hline
\\[-1.2ex]
Collab. & Ref. & $\Nf$ & $M_{\pi,\mr{min}}\,[\rm{MeV}]$ & Description
\\[1.0ex]
\hline \hline
\\[-1.2ex]
ETM 21A      & \cite{Alexandrou:2021gqw} & 2+1+1 & 134       & 4 pion masses in $[134, 346]$ MeV.
\\[2.0ex] \hline \\[-1.0ex]
ETM 21       & \cite{Alexandrou:2021bfr} & 2+1+1 & 135       & 4 pion masses in $[134, 346]$ MeV.
\\[2.0ex] \hline \\[-1.0ex]
$\chi$QCD 21 & \cite{Liang:2021pql}      & 2+1   & 139       & 3 pion masses with different $a$.
\\[2.0ex] \hline \\[-1.0ex]
Wang 16      & \cite{Wang:2016lsv}       & 2+1   & 220       & 8 (3) pion masses in val (sea) sector.\\
\\[2.0ex] \hline \\[-1.0ex]
ETM 20A      & \cite{Fischer:2020fvl}    & 2     & 132       & 
\\[2.0ex]
\hline
\hline
\end{tabular*}
}
\caption{
  Chiral extrapolation/minimum pion mass in
  determinations of the $SU(2)$ low-energy constants $\Sigma, F, \bar \ell_4, \bar \ell_6$, and $SU(3)$ low-energy constants $\Sigma_0, F_0$.
}
\end{table}

\begin{table}[!htbp]
{\footnotesize
\begin{tabular*}{\textwidth}[l]{l l @{\extracolsep{\fill}} l c c}
\hline\hline
\\[-1.2ex]
Collab. & Ref. & $\Nf$ & $M_{\pi,\mr{min}}\,[\rm{MeV}]$ & Description
\\[1.0ex]
\hline \hline
\\[-1.2ex]
Gao 21       & \cite{Gao:2021xsm}        & 2+1   & 140       
\\[2.0ex] \hline \\[-1.0ex]
$\chi$QCD 20 & \cite{Wang:2020nbf}       & 2+1   & 139       
\\[2.0ex] \hline \\[-1.0ex]
Feng 19      & \cite{Feng:2019geu}       & 2+1   & $\sim135$ 
\\[2.0ex]
\hline
\hline
\end{tabular*}
}
\caption{
  Chiral extrapolation/minimum pion mass in
  determinations of the low-energy constants related to the vector form factor of the pion.
}
\end{table}

\begin{table}[!htbp]
{\footnotesize
\begin{tabular*}{\textwidth}[l]{l l @{\extracolsep{\fill}} l c l}
\hline\hline
\\[-1.2ex]
Collab. & Ref. & $\Nf$ & $M_{\pi,\mr{min}}\,[\rm{MeV}]$ & Description
\\[1.0ex]
\hline \hline
\\[-1.2ex]
ETM 20B      & \cite{Fischer:2020jzp}    & 2     & 134       & 2 pion masses.
\\[2.0ex] \hline \\[-1.0ex]
Mai 19       & \cite{Mai:2019pqr}        & 2     & 224       
\\[2.0ex] \hline \\[-1.0ex]
Culver 19    & \cite{Culver:2019qtx}     & 2     & 226       & 2 pion masses.
\\[2.0ex]
\hline
\hline
\end{tabular*}
}
\caption{
  Chiral extrapolation/minimum pion mass in
  determinations of the low-energy constants related to $\pi\pi$ scattering.
}
\end{table}

\begin{table}[!htbp]
{\footnotesize
\begin{tabular*}{\textwidth}[l]{l l @{\extracolsep{\fill}} l c l}
\hline\hline
\\[-1.2ex]
Collab. & Ref. & $\Nf$ & $M_{\pi,\mr{min}}\,[\rm{MeV}]$ & Description
\\[1.0ex]
\hline \hline
\\[-1.2ex]
ETM 18B      & \cite{Helmes:2018nug}     & 2+1+1 & 276       & 5 pion masses in $[230, 450]$ MeV.
\\[2.0ex] \hline \\[-1.0ex]
ETM 17G      & \cite{Helmes:2017smr}     & 2+1+1 & 276       & 5 pion masses in $[230, 450]$ MeV.
\\[2.0ex] \hline \\[-1.0ex]
PACS-CS 13   & \cite{Sasaki:2013vxa}     & 2+1   & 166       & 5 pion masses in $[166, 707]$ MeV.
\\[2.0ex] \hline \\[-1.0ex]
Fu 11A       & \cite{Fu:2011wc}          & 2+1   & 590 (RMS) & 6 valence pion masses.\\
\\[2.0ex] \hline \\[-1.0ex]
NPLQCD 07B   & \cite{Beane:2007uh}       & 2+1   & 413 (RMS) & 4 pion masses.
\\[2.0ex] \hline \\[-1.0ex]
NPLQCD 06B   & \cite{Beane:2006gj}       & 2+1   & 488 (RMS) & 4 pion masses.
\\[2.0ex]
\hline
\hline
\end{tabular*}
}
\caption{
  Chiral extrapolation/minimum pion mass in
  determinations of the low-energy constants related to $\pi K$ scattering.
}
\end{table}


\begin{table}[!htbp]
{\footnotesize
\begin{tabular*}{\textwidth}[l]{l l @{\extracolsep{\fill}} l c c}
\hline\hline
\\[-1.2ex]
Collab. & Ref. & $\Nf$ & $L\,[\rm{fm}]$ & \#V
\\[1.0ex]
\hline \hline
\\[-1.2ex]
ETM 21A      & \cite{Alexandrou:2021gqw} & 2+1+1 & 5.52  &2 \\
[2.0ex] \hline \\[-1.0ex]
ETM 21       & \cite{Alexandrou:2021bfr} & 2+1+1 & 5.55  &2 \\
[2.0ex] \hline \\[-1.0ex]
$\chi$QCD 21 & \cite{Liang:2021pql}      & 2+1   & 5.4   &2 at physical point. \\
[2.0ex] \hline \\[-1.0ex]
Wang 16      & \cite{Wang:2016lsv}       & 2+1   & 2.7   &1 \\
[2.0ex] \hline \\[-1.0ex]
ETM 20A      & \cite{Fischer:2020fvl}    & 2     & 4.39  &2  \\
[2.0ex]
\hline
\hline
\end{tabular*}
}
\caption{
  Finite-volume effects in determinations of the $SU(2)$ low-energy constants $\Sigma, F, \bar \ell_4, \bar \ell_6$, and $SU(3)$ low-energy constants $\Sigma_0, F_0$.
}
\end{table}

\begin{table}[!htbp]
{\footnotesize
\begin{tabular*}{\textwidth}[l]{l l @{\extracolsep{\fill}} l c c}
\hline\hline
\\[-1.2ex]
Collab. & Ref. & $\Nf$ & $L\,[\rm{fm}]$ & \#V
\\[1.0ex]
\hline \hline
\\[-1.2ex]
Gao 21       & \cite{Gao:2021xsm}        & 2+1   & 4.86  &1 \\
[2.0ex] \hline \\[-1.0ex]
$\chi$QCD 20 & \cite{Wang:2020nbf}       & 2+1   & 6.24  &3 \\
[2.0ex] \hline \\[-1.0ex]
Feng 19      & \cite{Feng:2019geu}       & 2+1   & 6.22  &3 \\
[2.0ex]
\hline
\hline
\end{tabular*}
}
\caption{
  Finite-volume effects in determinations of the low-energy constants related to the vector form factor of the pion.
}
\end{table}

\begin{table}[!htbp]
{\footnotesize
\begin{tabular*}{\textwidth}[l]{l l @{\extracolsep{\fill}} l c c}
\hline\hline
\\[-1.2ex]
Collab. & Ref. & $\Nf$ & $L\,[\rm{fm}]$ & \#V
\\[1.0ex]
\hline \hline
\\[-1.2ex]
ETM 20B      & \cite{Fischer:2020jzp}    & 2     & 2.92  &2 \\
[2.0ex] \hline \\[-1.0ex]
Mai 19       & \cite{Mai:2019pqr}        & 2     & 2.88  &1\\
[2.0ex] \hline \\[-1.0ex]
Culver 19    & \cite{Culver:2019qtx}     & 2     & 2.88  &3 \\
[2.0ex]
\hline
\hline
\end{tabular*}
}
\caption{
  Finite-volume effects in determinations of the low-energy constants related to $\pi\pi$ scattering.
}
\end{table}

\begin{table}[!htbp]
{\footnotesize
\begin{tabular*}{\textwidth}[l]{l l @{\extracolsep{\fill}} l c c}
\hline\hline
\\[-1.2ex]
Collab. & Ref. & $\Nf$ & $L\,[\rm{fm}]$ & \#V
\\[1.0ex]
\hline \hline
\\[-1.2ex]
ETM 18B      & \cite{Helmes:2018nug}     & 2+1+1 & 2.832 &2\\
[2.0ex] \hline \\[-1.0ex]
ETM 17G      & \cite{Helmes:2017smr}     & 2+1+1 & 2.832 &2\\
[2.0ex] \hline \\[-1.0ex]
PACS-CS 13   & \cite{Sasaki:2013vxa}     & 2+1   & 2.9   &1\\
[2.0ex] \hline \\[-1.0ex]
Fu 11A       & \cite{Fu:2011wc}          & 2+1   & 2.4   &1\\
[2.0ex] \hline \\[-1.0ex]
NPLQCD 07B   & \cite{Beane:2007uh}       & 2+1   & 2.52  &2\\
[2.0ex] \hline \\[-1.0ex]
NPLQCD 06B   & \cite{Beane:2006gj}       & 2+1   & 2.5   &2\\
[2.0ex]
\hline
\hline
\end{tabular*}
}
\caption{
  Finite-volume effects in determinations of the low-energy constants related to $\pi K$ scattering.
}
\end{table}

\clearpage

\setcounter{subsection}{3}

\clearpage
\subsection{Notes to Sec.~\ref{sec:BK} on kaon mixing}
\label{app-BK}


\subsubsection{$K \to \pi\pi$ decay amplitudes}
\label{app-Kpipi}


\begin{table}[!ht]

{\footnotesize
\begin{tabular*}{\textwidth}{l c c c l}
\hline\hline \\[-1.0ex]
Collab. & Ref. & $\Nf$ & $a$ [fm] & Description 
\\[1.0ex] \hline \hline \\[-1.0ex]

RBC/UKQCD 20 & \cite{Abbott:2020hxn} & 2+1 & 0.143  &
\parbox[t]{7.5cm}{Single lattice spacing.}
\\[4.0ex] \hline \\[-1.0ex]

RBC/UKQCD 15G & \cite{Bai:2015nea} & 2+1 & 0.143  &
\parbox[t]{7.5cm}{Single lattice spacing.}
\\[4.0ex] \hline \\[-1.0ex]

RBC/UKQCD 15F & \cite{Blum:2015ywa} & 2+1 & 0.114, 0.084  &
\parbox[t]{7.5cm}{Combined chiral-continuum fit based on two values of the lattice spacing. Systematic error associated with the extrapolation to the continuum limit is negligible
	with respect to the statistical errors.}
\\
\\[1.0ex] \hline \hline \\[-1.0ex]
\end{tabular*}
\caption{Continuum extrapolations/estimation of lattice artifacts in
  determinations of the $K \to \pi\pi$ decay amplitudes.}
}
\end{table}


\begin{table}[!ht]
{\footnotesize
\begin{tabular*}{\textwidth}{l @{\extracolsep{\fill}} c c c l}
\hline\hline \\[-1.0ex]
{Collab.} & {Ref.} & {$\Nf$} & {$M_{\pi,\rm min}\,[\mev]$} & {Description}  
\\[1.0ex] \hline \hline \\[-1.0ex]

RBC/UKQCD 20 & \cite{Abbott:2020hxn} & 2+1 & 
\parbox[t]{1.8cm}{142.3}
 & \parbox[t]{6cm}{Single pion mass value, close to the physical point.}
\\[6.0ex] \hline \\[-1.0ex]
RBC/UKQCD 15G & \cite{Bai:2015nea} & 2+1 & 
\parbox[t]{1.8cm}{143.1}
& \parbox[t]{6cm}{Single pion mass value, close to the physical point.}
\\[6.0ex] \hline \\[-1.0ex]
RBC/UKQCD 15F & \cite{Blum:2015ywa} & 2+1 & 
\parbox[t]{1.8cm}{139.1, 139.2}
& \parbox[t]{6cm}{Single pion mass value---close to the physical point---at each of the two values of the lattice spacing.}
\\
\\[1.0ex]\hline\hline
\end{tabular*}
\caption{Chiral extrapolation/minimum pion mass in
  determinations of the $K \to \pi\pi$ decay amplitudes.}
}
\end{table}


\begin{table}[!ht]
{\footnotesize
\begin{tabular*}{\textwidth}{l @{\extracolsep{\fill}} c c l c l}
\hline\hline \\[-1.0ex]
Collab. & Ref. & $\Nf$ & $L$ [fm] & ${M_{\pi,\rm min}}L$ & Description 
\\[1.0ex] \hline \hline \\[-1.0ex]

RBC/UKQCD 20 & \cite{Abbott:2020hxn} & 2+1
& 4.6 & $3.3$ 
& \parbox[t]{5cm}{Finite volume effects amount to a 7\% systematic error
	contribution to the final error budget of  $A_0$.  } 
\\[8.0ex] \hline \\[-1.0ex]
RBC/UKQCD 15G & \cite{Bai:2015nea} & 2+1
& 4.6 & $3.3$ 
& \parbox[t]{5cm}{Finite volume effects amount to a 7\% systematic error
	contribution to the final error budget of  $A_0$.  } 
\\[8.0ex] \hline \\[-1.0ex]
RBC/UKQCD 15F & \cite{Blum:2015ywa} & 2+1
& \parbox[t]{1.8cm}{5.5, \, 5.4} & $3.8$ 
& \parbox[t]{5cm}{Finite volume effects amount to a 3.5\% systematic error contribution to the  error budget of $\text{Im}(A_2)/\text{Re}(A_2)$. } 
\\
\\[1.0ex]\hline\hline \\[-1.0ex]
\end{tabular*}
\caption{Finite volume effects in determinations of  the $K \to \pi\pi$ decay amplitudes.}
}
\end{table}


\begin{table}[!ht]

  {\footnotesize
\begin{tabular*}{\textwidth}{l @{\extracolsep{\fill}} c c c c l}
\hline\hline \\[-1.0ex]
& & & & running & \\
\rb{Collab.} & \rb{Ref.} & \rb{$\Nf$} & \rb{Ren.} & match. & \rb{Description} 
\\[1.0ex] \hline \hline \\[-1.0ex]

RBC/UKQCD 20 & \cite{Abbott:2020hxn} & 2+1 & RI & PT1$\ell$ &
  \parbox[t]{5cm}{Two different RI-SMOM schemes are used.
     The relative systematic errors due to the
      renormalisation of the relevant operators amount to 4\%, while
      those  arising from the computation of the Wilson coefficients
         in the $\msbar$ scheme correspond to 12\%.}
    \\\\ \hline \\[-1.0ex]
    RBC/UKQCD 15G & \cite{Bai:2015nea} & 2+1 & RI & PT1$\ell$ &
    \parbox[t]{5cm}{Two different RI-SMOM schemes are used.
    	The relative systematic errors due to the
    	renormalisation of the relevant operators amount to 15\%, while
    	those arising from the computation of the Wilson coefficients
         in the $\msbar$ scheme correspond to 12\%.}
    \\\\ \hline \\[-1.0ex]

RBC/UKQCD 15F & \cite{Blum:2015ywa} & 2+1 & RI & PT1$\ell$ &
\parbox[t]{5cm}{Two different RI-SMOM schemes are
  used.  The relative systematic errors for the conversion to $\msbar$ are 2.9\% for $\text{Re}A_0$ and 7\% for $\text{Im}A_0$.}
\\

\\[1.0ex] \hline\hline \\[-1.0ex]
\end{tabular*}
\caption{Running and matching in determinations of  the $K \to \pi\pi$ decay amplitudes.}
}
\end{table}

\clearpage

\subsubsection{Kaon $B$-parameter $B_K$}
\label{app:BKSM}

No new calculations w.r.t. the previous FLAG report.
\ifx\reducedapptables\undefined
\begin{table}[!ht]

{\footnotesize
\begin{tabular*}{\textwidth}{l @{\extracolsep{\fill}}c c c l}
\hline\hline \\[-1.0ex]
Collab. & Ref. & $\Nf$ & $a$ [fm] & Description 
\\[1.0ex] \hline \hline \\[-1.0ex]

ETM 15      & \cite{Carrasco:2015pra} & 2+1+1 & 0.09, 0.08, 0.06 & 
 \parbox[t]{5.7cm}{Combined chiral and continuum extrapolation. Systematic error of 2.0\% is obtained from the distribution of results over analyses which differ by  
O$(a^2)$ effects.}
 \\[12.0ex] \hline\hline \\[-1.0ex] 
\end{tabular*}
\caption{Continuum extrapolations/estimation of lattice artifacts in
  determinations of $B_K$ with $\Nf=2+1+1$ quark flavours.}
}
\end{table}

\begin{table}[!ht]
{\footnotesize
\begin{tabular*}{\textwidth}{l @{\extracolsep{\fill}}c c c l}
\hline\hline \\[-1.0ex]
Collab. & Ref. & $\Nf$ & $a$ [fm] & Description 
\\[1.0ex] \hline \hline \\[-1.0ex]

RBC/UKQCD 16 & \cite{Garron:2016mva} & 2+1 & 0.111, 0.083  &
\parbox[t]{5.7cm}{Systematic uncertainty of 1.3\% obtained from half the 
difference between the results on the fine lattice spacing and the continuum limit.}
\\[10.0ex]
\hline \\[-1.0ex]
SWME 15A & \cite{Jang:2015sla} & 2+1 & 0.12, 0.09, 0.06, 0.045  &
\parbox[t]{5.7cm}{The three finest lattice spacings are used for the combined chiral and continnum extrapolation.  
Residual combined discretization, sea-quark extrapolation
and $\alpha_s$ matching error of 4.4\% from difference between linear
  fit in $a^2$, $m_{\rm sea}$ and a fit where $\alpha_s$ dependence is added.}
\\[18.0ex] \hline \\[-1.0ex]

RBC/UKQCD 14B & 
\cite{Blum:2014tka} & 2+1 & 
\parbox[t]{2.0cm}{0.111, 0.083, 0.063,  0.114, 0.084}  &
\parbox[t]{5.7cm}{The three first lattice spacings use different action from the 
last two ones. Combined continuum and chiral fits.}
\\[7.0ex] \hline \\[-1.0ex]

SWME 14 & \cite{Bae:2014sja} & 2+1 & 0.082, 0.059, 0.044  &
\parbox[t]{5.7cm}{Residual combined discretization and sea-quark extrapolationg
error of 0.9\% from difference between linear
  fit in $a^2$, $m_{\rm sea}$  and a constrained nine-parameter extrapolation.}
\\[12.0ex] \hline \\[-1.0ex]

SWME 13A & \cite{Bae:2013tca} & 2+1 & 0.09, 0.06, 0.045  &
\parbox[t]{5.7cm}{Residual combined discretization, sea-quark extrapolation
and $\alpha_s$ matching error of 4.4\% from difference between linear
  fit in $a^2$, $m_{\rm sea}$ and a fit where $\alpha_s$ dependence is added.}
\\[12.0ex]

\hline \hline \\[-1.0ex]
\end{tabular*}
\caption{Continuum extrapolations/estimation of lattice artifacts in
  determinations of $B_K$ with $\Nf=2+1$ quark flavours.}
}
\end{table}
\fi

\ifx\reducedapptables\undefined
\begin{table}[!t]
\addtocounter{table}{-1}
{\footnotesize
\begin{tabular*}{\textwidth}{l c c c l}
\hline\hline \\[-1.0ex]
Collab. & Ref. & $\Nf$ & $a$ [fm] & Description 
\\[1.0ex] \hline \hline \\[-1.0ex]

SWME 13 & \cite{Bae:2013lja} & 2+1 & 0.12, 0.09, 0.06, 0.045  &
\parbox[t]{5.7cm}{Continuum extrapolation with the coarsest lattice spacing
  omitted; residual combined discretization and sea-quark extrapolationg
error of 1.1\% from difference between linear
  fit in $a^2$, $m_{\rm sea}$ and a constrained nine-parameter extrapolation.}
\\[18.0ex] \hline \\[-1.0ex]
RBC/UKQCD 12A & 
\cite{Arthur:2012opa} & 2+1 & 0.146, 0.114, 0.087  &
\parbox[t]{5.7cm}{Coarsest lattice spacing uses different action.
Combined continuum and chiral fits.}
\\[7.0ex] \hline \\[-1.0ex]
%
%
Laiho 11 & \cite{Laiho:2011np} & 2+1 & 0.12, 0.09, 0.06  &
\parbox[t]{5.7cm}{Combined continuum and chiral extrapolation based on SU(3)
  mixed-action partially quenched {\Ch}PT.}
\\[7.0ex] \hline \\[-1.0ex]
SWME 11, 11A & \parbox[t]{1.0cm}{\cite{Kim:2011qg,Bae:2011ff}} & 2+1 & 0.12, 0.09, 0.06, 0.045  &
\parbox[t]{5.7cm}{Continuum extrapolation with the coarsest lattice spacing
  omitted; residual discretization error of 1.9\% from difference between
  fit to a constant and a constrained five-parameter extrapolation.}
\\[12.0ex] \hline \\[-1.0ex]
BMW 11 & \cite{Durr:2011ap} & 2+1 & \parbox[t]{2.5cm}{0.093, 0.077, \\0.065, 0.054}  &
\parbox[t]{5.7cm}{Combined continuum and chiral extrapolation;
  discretization error of 0.1\% from comparison of $O({\alpha_s}a)$
  and $O(a^2)$ extrapolations.} 
\\[9.0ex] \hline \\[-1.0ex]
RBC/UKQCD 10B & \cite{Aoki:2010pe} & 2+1 & 0.114, 0.087  &
\parbox[t]{5.7cm}{Two lattice spacings. Combined chiral and continuum
fits.} 
\\[4.0ex]  \hline \\[-1.0ex]

SWME 10   & \cite{Bae:2010ki} & 2+1 &  0.12, 0.09, 0.06 &
\parbox[t]{5.7cm}{Continuum extrapolation of results obtained at four
lattice spacings; residual discretization error of 0.21\% from
difference to result at smallest lattice spacing.}   
\\[9.0ex] \hline \\[-1.0ex]

Aubin 09 & \cite{Aubin:2009jh} & 2+1 & 0.12, 0.09 & 
\parbox[t]{5.7cm}{Two
   lattice spacings; quote 0.3\% discretization error, estimated from
   various $a^2$-terms in fit function}
\\[6.0ex] \hline \\[-1.0ex]

RBC/UKQCD 07A, 08 & \parbox[t]{1.0cm}{\cite{Antonio:2007pb,Allton:2008pn}} & 2+1 & 0.114(2) & 
\parbox[t]{5.7cm}{Single lattice spacing; quote 4\%
  discretization error, estimated from the difference between
  computed and experimental values of $f_\pi$.}
\\ [9.0ex] \hline \\[-1.0ex]
HPQCD/UKQCD 06 & \cite{Gamiz:2006sq} & {2+1} & 0.12 & \parbox[t]{5.5cm}{Single
  lattice spacing; 3\% discretization error quoted without providing
  details.}
 \\[7.0ex] \hline\hline \\[-1.0ex] 
\end{tabular*}
\caption{(cntd.) Continuum extrapolations/estimation of lattice artifacts in
  determinations of $B_K$ with $\Nf=2+1$ quark flavours.}
}
\end{table}
\fi

\ifx\reducedapptables\undefined
\begin{table}[!t]
{\footnotesize
\begin{tabular*}{\textwidth}{l @{\extracolsep{\fill}}c c c l}
\hline\hline \\[-1.0ex]
Collab. & Ref. & $\Nf$ & $a$ [fm] & Description 
\\[1.0ex] \hline \hline \\[-1.0ex]

ETM 12D      & \cite{Bertone:2012cu} & 2 & 0.1, 0.09, 0.07, 0.05 & 
 \parbox[t]{5.7cm}{Four lattice spacings; systematic quoted obtained from the 
difference between the finest lattice spacing and the continuum limit and comparing 
results using two evaluations of the RCs that differ by O$(a^2)$ effects.}
 \\[15.0ex] \hline \\[-1.0ex] 

ETM 10A      & \cite{Constantinou:2010qv} & 2 & 0.1, 0.09, 0.07 & 
 \parbox[t]{5.7cm}{Three lattice spacings; 1.2\% error quoted.}
 \\[4.0ex] \hline \\[-1.0ex] 
JLQCD 08 & \cite{Aoki:2008ss} & 2 & 0.118(1) & 
\parbox[t]{5.7cm}{Single
  lattice spacing; no error quoted.}
\\[1.0ex] \hline \\[-1.0ex]
RBC 04 & \cite{Aoki:2004ht} & 2 & 0.117(4) & 
\parbox[t]{5.7cm}{Single
   lattice spacing; no error quoted.}
\\[1.0ex] \hline \\[-1.0ex]
UKQCD 04 & \cite{Flynn:2004au} & 2 & 0.10 & \parbox[t]{5.7cm}{Single
   lattice spacing; no error quoted.}
\\[1.0ex]
\hline\hline
\end{tabular*}
\caption{Continuum extrapolations/estimation of lattice artifacts in
  determinations of $B_K$ with $\Nf=2$ quark flavours.}
}
\end{table}
\fi


\ifx\reducedapptables\undefined
\begin{table}[!ht]
{\footnotesize
\begin{tabular*}{\textwidth}{l @{\extracolsep{\fill}} c c c l}
\hline\hline \\[-1.0ex]
{Collab.} & {Ref.} & {$\Nf$} & {$M_{\pi,\rm min}\,[\mev]$} & {Description}  
\\[1.0ex] \hline \hline \\[-1.0ex]
%
ETM 15 & \cite{Carrasco:2015pra} & 2+1+1 & \parbox[t]{1.8cm}{245,\,239,\,211} & 
  \parbox[t]{6cm}{Each $M_{\pi,\rm min}$ entry corresponds to a
  different lattice spacing. Simultaneous chiral \& continuum
  extrapolations, based on polynomial and {\Ch}PT at NLO, are carried out leads to systematic
  error of 0.8\% .}
  \\ [12.0ex]
\hline\hline
\end{tabular*}
\caption{ Chiral extrapolation/minimum pion mass in
  determinations of $B_K$ with $\Nf=2+1+1$ quark flavours.}
}
\end{table}
\fi

\ifx\reducedapptables\undefined
\begin{table}[!ht]
{\footnotesize
\begin{tabular*}{\textwidth}{l @{\extracolsep{\fill}} c c c l}
\hline\hline \\[-1.0ex]
{Collab.} & {Ref.} & {$\Nf$} & {$M_{\pi,\rm min}\,[\mev]$} & {Description}  
\\[1.0ex] \hline \hline \\[-1.0ex]
RBC/UKQCD 16 & \cite{Garron:2016mva} & 2+1 & 
\parbox[t]{1.8cm}{337, 302}
 & \parbox[t]{6cm}{Chiral extrapolations based on polynomial and $SU(2)$-$\chi$PT fits at NLO.
A systematic uncertainty of 0.4\% is quoted, which is half the difference between the two results.\\}
\\

\hline \\[-1.0ex]
SWME 15A & \cite{Jang:2015sla} & 2+1 & 
\parbox[t]{1.8cm}{222/372,\, 206/174,\, 195/222, \, 206/316}
 & \parbox[t]{6cm}{Valence/sea RMS $M_{\pi,\rm min}$ entries correspond to the
four lattice spacings (the last three are used for the chiral-continuum extrapolation). Chiral extrapolations based on SU(2) staggered
{\Ch}PT at NNLO (with some coefficients fixed by Bayesian priors), and
also including one analytic NNNLO term. Residual error of 0.05\%  from
changing the Bayesian priors and fit method.}
\\[23.0ex] \hline \\[-1.0ex]

RBC/UKQCD 14B & \cite{Blum:2014tka} & 2+1 & 
\parbox[t]{1.8cm}{337,\,302,\, 371,\,139,\, 139}  &
\parbox[t]{6.0cm}{$M_{\pi,\rm min}$ entries correspond to the five lattice
spacings. Combined chiral \& continuum extrapolation, using
$M_\pi<260\;$MeV and $M_\pi<370\;$MeV.}
\\[9.0ex] \hline \\[-1.0ex]

SWME 14 & \cite{Bae:2014sja} & 2+1 & 
\parbox[t]{1.8cm}{206/174,\, 195/222,\, 207/316}
 & \parbox[t]{6cm}{Valence/sea RMS $M_{\pi,\rm min}$ entries correspond to the
three lattice spacings. Chiral extrapolations based on SU(2) staggered
{\Ch}PT at NNLO (with some coefficients fixed by Bayesian priors), and
also including one analytic NNNLO term. Residual error of 0.1\% error from
doubling the widths of Bayesian priors.}
\\[20.0ex] \hline \\[-1.0ex]

SWME 13A & \cite{Bae:2013tca} & 2+1 & 
\parbox[t]{1.8cm}{207/243,\, 196/262,\, 207/316}
 & \parbox[t]{6cm}{Valence/sea RMS $M_{\pi,\rm min}$ entries correspond to the
three lattice spacings. Chiral extrapolations based on SU(2) staggered
{\Ch}PT at NNLO (with some coefficients fixed by Bayesian priors), and
also including one analytic NNNLO term. Residual error of 0.1\%  from
doubling the widths of Bayesian priors.}
\\[20.0ex] \hline \\[-1.0ex]

SWME 13 & \cite{Bae:2013lja} & 2+1 & 
\parbox[t]{1.8cm}{442/445,\, 299/273,\, 237/256,\, 222/334}
 & \parbox[t]{6cm}{Valence/sea RMS $M_{\pi,\rm min}$ entries correspond to the
four lattice spacings. Chiral extrapolations based on SU(2) staggered
{\Ch}PT at NNLO (with some coefficients fixed by Bayesian priors), and
also including one analytic NNNLO term. Residual error of 0.33\% error from
doubling the widths of Bayesian priors.}
\\[20.0ex] \hline\\[-1.0ex]
RBC/UKQCD 12A & \cite{Arthur:2012opa} & 2+1 & 
\parbox[t]{1.8cm}{140/170,\, 240/330,\, 220/290}  &
\parbox[t]{6.0cm}{Valence/sea
$M_{\pi,\rm min}$ entries correspond to the three lattice
spacings. Combined chiral \& continuum extrapolation, using
$M_\pi<350\;$MeV.}
\\[10.0ex]
\hline\hline
\end{tabular*}
\caption{Chiral extrapolation/minimum pion mass in
  determinations of $B_K$ with $\Nf=2+1$ quark flavours.}
}
\end{table}
\fi

\ifx\reducedapptables\undefined
\begin{table}[!ht]
\addtocounter{table}{-1}
{\footnotesize
\begin{tabular*}{\textwidth}{l @{\extracolsep{\fill}} c c c l}
\hline\hline \\[-1.0ex]
{Collab.} & {Ref.} & {$\Nf$} & {$M_{\pi,\rm min}\,[\mev]$} & {Description}  
\\[1.0ex] \hline \hline \\[-1.0ex]

Laiho 11 & \cite{Laiho:2011np} & 2+1 & \parbox[t]{1.8cm}{210/280}
 & \parbox[t]{6cm}{$M_{\pi,\rm min}$ entries correspond to the smallest
  valence/sea quark masses. Chiral \& continuum fits based on NLO mixed action
  {\Ch}PT, including a subset of NNLO terms. Systematic error estimated
  from spread arising from variations in the fit function.}
\\[15.0ex] \hline \\[-1.0ex]
SWME 11, 11A & \cite{Kim:2011qg,Bae:2011ff} & 2+1 & 
\parbox[t]{1.8cm}{442/445,\, 299/325,\, 237/340,\, 222/334}
 & \parbox[t]{6cm}{Valence/sea RMS $M_{\pi,\rm min}$ entries correspond to the
four lattice spacings. Chiral extrapolations based on SU(2) staggered
{\Ch}PT at NNLO (with some coefficients fixed by Bayesian priors), and
also including one analytic NNNLO term. Residual error of 0.33\% error from
doubling the widths of Bayesian priors.}
\\[20.0ex] \hline \\[-1.0ex]

BMW 11 & \cite{Durr:2011ap} & 2+1 & 
\parbox[t]{1.8cm}{219,\, 182,\\ 120,\, 131}
 & \parbox[t]{6cm}{$M_{\pi,\rm min}$ entries correspond to the
four lattice spacings used in the final result. Combined fit to the chiral and continuum
behaviour. Systematics investigated by applying cuts to the maximum pion
mass used in fits. Uncertainty of 0.1\% assigned to chiral fit.} 
\\[18.0ex] \hline \\[-1.0ex]
RBC/UKQCD 10B & \cite{Aoki:2010pe} & 2+1 & 
\parbox[t]{1.8cm}{240/330,\, 220/290}  &
\parbox[t]{6cm}{Valence/sea $M_{\pi,\rm min}$ entries correspond to the two
lattice spacings. Combined chiral and continuum extrapolations.}
\\[7.0ex] \hline \\[-1.0ex]
SWME 10   & \cite{Bae:2010ki} & 2+1 &  
\parbox[t]{1.8cm}{442/445,\, 299/325,\, 237/340} &
\parbox[t]{6cm}{Valence/sea $M_{\pi,\rm min}$ entries correspond to the three
lattice spacings. Chiral extrapolations based on SU(2) staggered {\Ch}PT
at NLO, including some analytic NNLO terms. SU(3) staggered {\Ch}PT as
cross-check. Combined 1.1\% error from various different variations in
the fit procedure.}
\\[18.0ex] \hline \\[-1.0ex]
Aubin 09 & \cite{Aubin:2009jh} & 2+1 & \parbox[t]{1.8cm}{240/370} &
  \parbox[t]{6cm}{ $M_{\pi,\rm min}$ entries correspond to the
  smallest valence/sea quark masses. Chiral \& continuum fits based on
  NLO mixed action {\Ch}PT at NLO, including a subset of NNLO
  terms. Systematic error estimated from spread arising from
  variations in the fit function.}
\\[18.0ex] \hline\\[-1.0ex]
RBC/UKQCD 07A, 08 & {\cite{Antonio:2007pb,Allton:2008pn}} & {2+1} & 
 \parbox[t]{1.8cm}{330} & {\parbox[t]{6cm}{Fits based on SU(2) PQ{\Ch}PT
 at NLO. Effect of neglecting higher orders estimated at 6\% via
 difference between fits based on LO and NLO expressions.}} 
\\[10.0ex] \hline \\[-1.0ex]
HPQCD/UKQCD 06 & {\cite{Gamiz:2006sq}} & {2+1} & \parbox[t]{1.8cm}{360}
    & {\parbox[t]{6cm}{3\% uncertainty from chiral extrapolation 
    quoted, without giving further details.}}
\\[4.0ex] \hline\hline \\[-1.0ex]
\end{tabular*}
\caption{(cntd.) Chiral extrapolation/minimum pion mass in
  determinations of $B_K$ with $\Nf=2+1$ quark flavours.}
}
\end{table}
\fi

\ifx\reducedapptables\undefined
\begin{table}[!ht]
{\footnotesize
\begin{tabular*}{\textwidth}{l @{\extracolsep{\fill}} c c c l}
\hline\hline \\[-1.0ex]
{Collab.} & {Ref.} & {$\Nf$} & {$M_{\pi,\rm min}\,[\mev]$} & {Description}  
\\[1.0ex] \hline \hline \\[-1.0ex]
ETM 12D & \cite{Bertone:2012cu} & 2 & \parbox[t]{1.8cm}{400,
  280,\\ 300, 280} & 
  \parbox[t]{6cm}{Each $M_{\pi,\rm min}$ entry corresponds to a
  different lattice spacing. Simultaneous chiral \& continuum
  extrapolations, based on polynomial and {\Ch}PT at NLO, are carried out. Systematic
  error from several sources, including lattice calibration, quark
  mass calibration, chiral and continuum extrapolation etc., estimated
  at 3.0\%.} 
\\ [21.0ex] \hline \\[-1.0ex]

ETM 10A & \cite{Constantinou:2010qv} & 2 & \parbox[t]{1.8cm}{400,
  280,\\ 300} & 
  \parbox[t]{6cm}{Each $M_{\pi,\rm min}$ entry corresponds to a
  different lattice spacing. Simultaneous chiral \& continuum
  extrapolations, based on {\Ch}PT at NLO, are carried out. Systematic
  error from several sources, including lattice calibration, quark
  mass calibration, chiral and continuum extrapolation etc., estimated
  at 3.1\%.} 
\\ [21.0ex] \hline \\[-1.0ex]

JLQCD 08 & \cite{Aoki:2008ss} & 2 & 290 & \parbox[t]{6cm}{Fits based on
  NLO PQ{\Ch}PT. Range of validity investigated. Fit error included in
  statistical uncertainty.}  
\\[7.0ex] \hline \\[-1.0ex]
RBC 04 & \cite{Aoki:2004ht} & 2 & 490 & \parbox[t]{6cm}{Fits based on NLO
     PQ{\Ch}PT. Fit error included in statistical uncertainty.} 
\\[4.0ex] \hline \\[-1.0ex]
UKQCD 04 & \cite{Flynn:2004au} & 2 & 780 & \parbox[t]{6cm}{Fits to
  continuum chiral behaviour at fixed sea quark mass. Separate
  extrapolation in sea quark mass. Fit error included in overall
  uncertainty.} 
\\[10.0ex]
\hline\hline
\end{tabular*}
\caption{Chiral extrapolation/minimum pion mass in
  determinations of $B_K$ with $\Nf=2$ quark flavours.}
}
\end{table}
\fi


\ifx\reducedapptables\undefined
\begin{table}[!ht]
{\footnotesize
\begin{tabular*}{\textwidth}{l @{\extracolsep{\fill}} c c c c l}
\hline\hline \\[-1.0ex]
Collab. & Ref. & $\Nf$ & $L$ [fm] & ${M_{\pi,\rm min}}L$ & Description 
\\[1.0ex] \hline \hline \\[-1.0ex]

ETM 15 & \cite{Carrasco:2015pra} & 2+1+1 & \parbox[t]{1.8cm}{2.1--2.8, 
  2.6, 3.0}& \parbox[t]{1.8cm}{3.5,\,3.2,\,3.2} &  
  \parbox[t]{5cm}{Each $L$ entry corresponds to a different
  lattice spacing, with two volumes at the coarsest lattice
  spacing; results from these two volumes at $M_{\pi} \sim 280$ MeV are
  compatible.} 
\\ [12.0ex]
\hline\hline
\end{tabular*}
\caption{Finite-volume effects in determinations of $B_K$ with $\Nf=2+1+1$ quark flavours.
If partially-quenched fits are used,
the quoted $M_{\pi,\rm min}L$ is for lightest valence (RMS) pion.}
}
\end{table}
\fi

\ifx\reducedapptables\undefined

\begin{table}[!ht]
{\footnotesize
\begin{tabular*}{\textwidth}{l @{\extracolsep{\fill}} c c c c l}
\hline\hline \\[-1.0ex]
Collab. & Ref. & $\Nf$ & $L$ [fm] & ${M_{\pi,\rm min}}L$ & Description 
\\[1.0ex] \hline \hline \\[-1.0ex]
RBC/UKQCD 16 & \cite{Garron:2016mva} & 2+1
& \parbox[t]{1.8cm}{2.7, 2.7} & $4.5, \, 4.0$ 
& \parbox[t]{5cm}{Finite-volume effects are found to be negligible
  compared to the systematic errors and are thus omitted in the final
  error budget.\\} 
	\\
	\hline\\[-1ex]
SWME 15A & \cite{Jang:2015sla} & 2+1
& \parbox[t]{1.8cm}{2.4--3.4, 2.5--5.8, \,2.9--3.9,\,2.9} & 
$\gtrsim 3.8$ 
& \parbox[t]{5cm}{$L$ entries correspond to the four lattice spacings,
with several volumes in most cases. Finite-volume effects
estimated using NLO SU(2) S{\Ch}PT.}
\\[9.0ex] \hline \\[-1.0ex]

RBC/UKQCD 14B & \cite{Blum:2014tka} 
& 2+1 & \parbox[t]{1.9cm}{2.7,\,2.7, 2.0,\,5.5, 5.3} &
$\gtrsim 3.8$ %
& \parbox[t]{5cm}{
$L$ entries correspond to the five lattice
spacings. Finite-volume effects estimated using NLO {\Ch}PT; 
negligible with comparison to the statistical error.}
\\[12.0ex] \hline \\[-1.0ex]

SWME 14 & \cite{Bae:2014sja} & 2+1
& \parbox[t]{1.8cm}{2.8--5.4,\, 2.8--3.8, 2.8} & 
5.6,\,3.7,\,2.9
& \parbox[t]{5cm}{$L$ entries correspond to the three lattice spacings,
with several volumes in most cases. Finite-volume effects
estimated using NLO {\Ch}PT.}
\\[9.0ex] \hline \\[-1.0ex]

SWME 13A & \cite{Bae:2013tca} & 2+1
& \parbox[t]{1.8cm}{2.4--3.4, 2.8--3.8, 2.8} & 
3.5,\,3.3,\,2.9
& \parbox[t]{5cm}{$L$ entries correspond to the three lattice spacings,
with several volumes in most cases. Finite-volume effects
estimated using NLO {\Ch}PT.}
\\[9.0ex] \hline \\[-1.0ex]

SWME 13 & \cite{Bae:2013lja} & 2+1
& \parbox[t]{1.8cm}{2.4--3.3, 2.4--5.5, 2.8--3.8, 2.8} & 
$\gtrsim 3.2$ 
& \parbox[t]{5cm}{$L$ entries correspond to the four lattice spacings,
with several volumes in most cases. Finite-volume effects
estimated using NLO {\Ch}PT.}
\\[9.0ex] \hline \\[-1.0ex]
RBC/UKQCD 12A & \cite{Arthur:2012opa} 
& 2+1 & \parbox[t]{1.8cm}{4.6,\,2.7,\,2.8} &
$\gtrsim 3.2$ 
& \parbox[t]{5cm}{
$L$ entries correspond to the three lattice
spacings. Finite-volume effects estimated using NLO {\Ch}PT.}
\\[7.0ex] \hline \\[-1.0ex]
Laiho 11 & \cite{Laiho:2011np} & 2+1 & \parbox[t]{1.8cm}{2.4,\,3.4,\,3.8} &
$\gtrsim 3.5$
 & \parbox[t]{5cm}{$L$ entries correspond to the three lattice spacings.
Finite-volume effects estimated using NLO {\Ch}PT.}
\\[7.0ex] \hline \\[-1.0ex]
SWME 11, 11A & \parbox[t]{1.0cm}{\cite{Kim:2011qg,Bae:2011ff}} & 2+1
& \parbox[t]{1.8cm}{2.4/3.3, 2.4, 2.8, 2.8} & 
$\gtrsim 3.2$ 
& \parbox[t]{5cm}{$L$ entries correspond to the four lattice spacings,
with two volumes at the coarsest lattice. Finite-volume effects
estimated using NLO {\Ch}PT.}
\\[1.0ex] \\[-1.0ex]

\hline\hline
\end{tabular*}
\caption{Finite-volume effects in determinations of $B_K$.
If partially-quenched fits are used,
the quoted $M_{\pi,\rm min}L$ is for lightest valence (RMS) pion with $\Nf=2+1$ quark flavours.}
}
\end{table}
\fi

\ifx\reducedapptables\undefined
\begin{table}[!ht]
\addtocounter{table}{-1}
{\footnotesize
\begin{tabular*}{\textwidth}{l @{\extracolsep{\fill}} c c c c l}
\hline\hline \\[-1.0ex]
Collab. & Ref. & $\Nf$ & $L$ [fm] & ${M_{\pi,\rm min}}L$ & Description 
\\[1.0ex] \hline \hline \\[-1.0ex]
BMW 11 & \cite{Durr:2011ap} & 2+1 & \parbox[t]{1.8cm}{6.0, 4.9,
4.2, 3.5} & $\gtrsim 3.8, 3.0$ &
\parbox[t]{5cm}{$L$ entries correspond to the four lattice
spacings, and are the largest of several volumes at each $a$.
${M_{\pi,\rm min}}L\approx3.0$ for the ensemble
at $a\approx0.08\,\fm$. Finite-volume effects estimated in {\Ch}PT and
by combined fit to multiple volumes.}
\\[18.0ex] \hline  \\[-1.0ex]
RBC/UKQCD 10B & \cite{Aoki:2010pe} & 2+1 & \parbox[t]{1.8cm}{2.7,\,2.8} &  
$\gtrsim 3.1$ 
& \parbox[t]{5cm}{$L$ entries correspond to the three lattice spacings.
Finite-volume effects estimated using NLO {\Ch}PT.}
\\[7.0ex] \hline \\[-1.0ex]
SWME 10   & \cite{Bae:2010ki} & 2+1 & \parbox[t]{1.8cm}{2.4/3.3, 2.4,2.8} & 
$\gtrsim 3.4$ 
& \parbox[t]{5cm}{$L$ entries correspond to the three lattice spacings,
with two volumes for the coarsest spacing. 
Finite-volume error of 0.9\% estimated from difference
obtained these two volumes.}
\\[15.0ex]\hline  \\[-1.0ex]

Aubin 09 & \cite{Aubin:2009jh} & 2+1 & \parbox[t]{1.8cm}{2.4,\,3.4} & 3.5 &
   \parbox[t]{5cm}{$L$ entries correspond to the two lattice spacings.
Keep $m_{\pi}L\;\gtaeq\;3.5$; no
   comparison of results from different volumes; 0.6\% error
   estimated from mixed action {\Ch}PT correction.}
\\[12.0ex] \hline \\[-1.0ex]
RBC/UKQCD 07A, 08 & \parbox[t]{1.0cm}{\cite{Antonio:2007pb,Allton:2008pn}} & 2+1 &
  \parbox[t]{1.8cm}{1.83/2.74} & 4.60 & \parbox[t]{5cm}{Each $L$ entry
  corresponds to a different volume at the same lattice spacing; 1\%
  error from difference in results on two volumes.} 
\\[10.0ex] \hline \\[-1.0ex]
HPQCD/UKQCD 06 & \cite{Gamiz:2006sq} & 2+1 & \parbox[t]{1.8cm}{2.46} &
  \parbox[t]{1.8cm}{4.49} & {Single volume; no error quoted.}
\\  [1.0ex] 
\hline\hline
\end{tabular*}
	\caption{(cntd.) Finite-volume effects in determinations of $B_K$ with $\Nf=2+1$ quark flavours.
If partially-quenched fits are used,
the quoted $M_{\pi,\rm min}L$ is for lightest valence (RMS) pion.}
}
\end{table}
\fi

\ifx\reducedapptables\undefined
\begin{table}[!ht]
{\footnotesize
\begin{tabular*}{\textwidth}{l @{\extracolsep{\fill}} c c c c l}
\hline\hline \\[-1.0ex]
Collab. & Ref. & $\Nf$ & $L$ [fm] & ${M_{\pi,\rm min}}L$ & Description 
\\[1.0ex] \hline \hline \\[-1.0ex]

ETM 12D & \cite{Bertone:2012cu} & 2 & \parbox[t]{1.8cm}{2.1,\\
  2.2/2.9,\\ 2.2, 2.6}& \parbox[t]{1.8cm}{5,\\ 3.3/4.3,\\ 3.3, 3.5} &  
  \parbox[t]{5cm}{Each $L$ entry corresponds to a different
  lattice spacing, with two volumes at the second less coarse lattice
  spacing. Results from these two volumes at $M_{\pi} \sim 300$ MeV are
  compatible.} 
\\ [15.0ex] \hline \\[-1.0ex]

ETM 10A & \cite{Constantinou:2010qv} & 2 & \parbox[t]{1.8cm}{2.1,\\
  2.2/2.9,\\ 2.2}& \parbox[t]{1.8cm}{5,\\ 3.3/4.3,\\ 3.3} &  
  \parbox[t]{5cm}{Each $L$ entry corresponds to a different
  lattice spacing, with two volumes at the intermediate lattice
  spacing. Results from these two volumes at $M_{\pi} \sim 300$ MeV are
  compatible.} 
\\ [15.0ex] \hline \\[-1.0ex]
JLQCD 08 & \cite{Aoki:2008ss} & 2 & 1.89 & 2.75 & \parbox[t]{5cm}{Single
  volume; data points with $m_{\rm val} < m_{\rm sea}$ excluded; 5\%
  error quoted as upper bound of PQ{\Ch}PT estimate of the 
  effect.}
\\[10.0ex] \hline \\[-1.0ex]
RBC 04 & \cite{Aoki:2004ht} & 2 & 1.87 & 4.64 & Single volume; no
   error quoted.
\\[1.0ex] \hline \\[-1.0ex]
UKQCD 04 & \cite{Flynn:2004au} & 2 & 1.6 & 6.51 & Single volume; no
   error quoted.
\\[1.0ex] 
\hline\hline
\end{tabular*}
\caption{Finite-volume effects in determinations of $B_K$ with $\Nf=2$ quark flavours}
}
\end{table}
\fi


\ifx\reducedapptables\undefined
\begin{table}[!t]
{\footnotesize
\begin{tabular*}{\textwidth}{l @{\extracolsep{\fill}} c c c c l}
\hline\hline \\[-1.0ex]
& & & & running & \\
\rb{Collab.} & \rb{Ref.} & \rb{$\Nf$} & \rb{Ren.} & match. & \rb{Description} 
\\[1.0ex] \hline \hline \\[-1.0ex]

ETM 15 & \cite{Carrasco:2015pra} & 2+1+1 & RI & PT1$\ell$ &
  \parbox[t]{5cm}{Uncertainty from RI renormalization estimated at 2\%. Additional error of 0.6\% for
  the conversion to $\msbar$.}
\\ [9.0ex]
\hline\hline
\end{tabular*}
\caption{Running and matching in determinations of $B_K$ for $N_f=2+1+1$.}
}
\end{table}
\fi

\ifx\reducedapptables\undefined
\begin{table}[!ht]
{\footnotesize
\begin{tabular*}{\textwidth}{l @{\extracolsep{\fill}} c c c c l}
\hline\hline \\[-1.0ex]
& & & & running & \\
\rb{Collab.} & \rb{Ref.} & \rb{$\Nf$} & \rb{Ren.} & match. & \rb{Description} 
\\[1.0ex] \hline \hline \\[-1.0ex]

RBC/UKQCD 16 & \cite{Garron:2016mva} & 2+1 & RI & PT1$\ell$ &
  \parbox[t]{5cm}{Two different RI-SMOM schemes used to estimate
2\% systematic error in conversion to $\msbar$.}
	\\[8ex]
	\hline\\[-1ex]
SWME 15A & \cite{Jang:2015sla} & 2+1 & PT1$\ell$ & PT1$\ell$ &
\parbox[t]{5cm}{Uncertainty from neglecting higher orders estimated at
4.4\% by identifying the unknown 2-loop coefficient with result at the
  smallest lattice spacing.}
\\[12.0ex] \hline \\[-1.0ex]

RBC/UKQCD 14B & \cite{Blum:2014tka} & 2+1 & RI & PT1$\ell$ &
\parbox[t]{5cm}{Two different RI-SMOM schemes used to estimate
2\% systematic error in conversion to $\msbar$.}
\\[7.0ex] \hline \\[-1.0ex]

SWME 14 & \cite{Bae:2014sja} & 2+1 & PT1$\ell$ & PT1$\ell$ &
\parbox[t]{5cm}{Uncertainty from neglecting higher orders estimated at
4.4\% by identifying the unknown 2-loop coefficient with result at the
  smallest lattice spacing.}
\\[12.0ex] \hline \\[-1.0ex]

SWME 13A & \cite{Bae:2013tca} & 2+1 & PT1$\ell$ & PT1$\ell$ &
\parbox[t]{5cm}{Uncertainty from neglecting higher orders estimated at
4.4\% (in combination with systematic uncertainty from 
CL and chiral extrapolation fit) by identifying the unknown 2-loop coefficient with result at the
  smallest lattice spacing.}
\\[18.0ex] \hline \\[-1.0ex]

SWME 13 & \cite{Bae:2013lja} & 2+1 & PT1$\ell$ & PT1$\ell$ &
\parbox[t]{5cm}{Uncertainty from neglecting higher orders estimated at
4.4\% by identifying the unknown 2-loop coefficient with result at the
  smallest lattice spacing.}
\\[12.0ex] \hline \\[-1.0ex]
RBC/UKQCD 12A & \cite{Arthur:2012opa} & 2+1 & RI & PT1$\ell$ &
\parbox[t]{5cm}{Two different RI-SMOM schemes used to estimate
2\% systematic error in conversion to $\msbar$.}
\\[7.0ex] \hline \\[-1.0ex]
Laiho 11 & \cite{Laiho:2011np} & 2+1 & RI & PT1$\ell$ &
\parbox[t]{5cm}{Total uncertainty in matching \& running of 3\%.
  Perturbative truncation error in the conversion to $\msbar$, RGI schemes is
  dominant uncertainty.} 
\\[12.0ex]
\hline\hline \\[-1.0ex]
\end{tabular*}
\caption{Running and matching in determinations of $B_K$ with $\Nf=2+1$.}
}
\end{table}

\fi

\ifx\reducedapptables\undefined
\begin{table}[!ht]
\addtocounter{table}{-1}
{\footnotesize
\begin{tabular*}{\textwidth}{l @{\extracolsep{\fill}} c c c c l}
\hline\hline \\[-1.0ex]
& & & & running & \\
\rb{Collab.} & \rb{Ref.} & \rb{$\Nf$} & \rb{Ren.} & match. & \rb{Description} 
\\[1.0ex] \hline \hline \\[-1.0ex]
SWME 11, 11A & \cite{Kim:2011qg,Bae:2011ff} & 2+1 & PT1$\ell$ & PT1$\ell$ &
\parbox[t]{5cm}{Uncertainty from neglecting higher orders estimated at
4.4\% by identifying the unknown 2-loop coefficient with result at the
  smallest lattice spacing.}
\\[12.0ex] \hline \\[-1.0ex]
BMW 11 & \cite{Durr:2011ap} & 2+1 & RI & PT1$\ell$ &
\parbox[t]{5cm}{Uncertainty of 0.05\% in the determination of the
  renormalization factor included. 1\% error estimated due to truncation of perturbative
  matching to $\msbar$ and RGI schemes at NLO.}
\\[17.0ex]  \hline  \\[-1.0ex]
RBC/UKQCD 10B & \cite{Aoki:2010pe} & 2+1 & RI & PT1$\ell$ &
\parbox[t]{5cm}{Variety of different RI-MOM schemes including
non-exceptional momenta. Residual uncertainty of 2\% uncertainty in
running \& matching.}\\[12.0ex] \hline \\[-1.0ex] 
SWME 10  & \cite{Bae:2010ki} & 2+1 & PT1$\ell$ & PT1$\ell$ &
\parbox[t]{5cm}{Uncertainty from neglecting higher orders estimated at
5.5\% by identifying the unknown 2-loop coefficient with result at the
  smallest lattice spacing.}
\\[12.0ex] \hline \\[-1.0ex] 
Aubin 09 & \cite{Aubin:2009jh} & 2+1 & RI & PT1$\ell$ &
  \parbox[t]{5cm}{Total uncertainty in matching \& running of 3.3\%,
  estimated from a number of sources, including chiral extrapolation
  fit ansatz for n.p. determination, strange sea quark mass
  dependence, residual chiral symmetry breaking, perturbative matching
  \& running.}
\\[21.0ex] \hline \\[-1.0ex]
RBC/UKQCD 07A, 08 & {\cite{Antonio:2007pb,Allton:2008pn}} & 
 {2+1} & {RI} & {PT1$\ell$} & {\parbox[t]{5cm}{Uncertainty
 from n.p. determination of ren. factor included in statistical error; 
 2\% systematic error from perturbative matching to $\msbar$
 estimated via size of correction itself.}}
\\[15.0ex] \hline \\[-1.0ex]
HPQCD/UKQCD 06 & {\cite{Gamiz:2006sq}} & {2+1} & {PT1$\ell$}&
    {PT1$\ell$} & {\parbox[t]{5cm}{Uncertainty due to neglecting 2-loop
    order in perturbative matching and running estimated by
    multiplying result by $\alpha^2$.}}
\\[10.0ex] \hline \hline \\[-1.0ex]
\end{tabular*}
	\caption{(cntd.) Running and matching in determinations of $B_K$ with $\Nf=2+1$.}
}
\end{table}
\fi

\ifx\reducedapptables\undefined
\begin{table}[!ht]
{\footnotesize
\begin{tabular*}{\textwidth}{l @{\extracolsep{\fill}} c c c c l}
\hline\hline \\[-1.0ex]
& & & & running & \\
\rb{Collab.} & \rb{Ref.} & \rb{$\Nf$} & \rb{Ren.} & match. & \rb{Description} 
\\[1.0ex] \hline \hline \\[-1.0ex]
ETM 12D & \cite{Bertone:2012cu} & 2 & RI & PT1$\ell$ &
  \parbox[t]{5cm}{Uncertainty from RI renormalization estimated at 2.5\%.}
\\ [4.0ex] \hline \\[-1.0ex]

ETM 10A & \cite{Constantinou:2010qv} & 2 & RI & PT1$\ell$ &
  \parbox[t]{5cm}{Uncertainty from RI renormalization estimated at 2.5\%.}
\\ [4.0ex] \hline \\[-1.0ex]
JLQCD 08 & \cite{Aoki:2008ss} & 2 & RI & PT1$\ell$ &
 \parbox[t]{5cm}{Uncertainty from n.p. determination of ren. factor 
 included in statistical error; 2.3\% systematic error from
 perturbative matching to $\msbar$ estimated via size of correction
 itself.} 
\\[15.0ex] \hline \\[-1.0ex]
RBC 04 & \cite{Aoki:2004ht} & 2 & RI & PT1$\ell$ &
   \parbox[t]{5cm}{Uncertainty from n.p. determination of ren. factor
 included.}
\\[4.0ex] \hline \\[-1.0ex]
UKQCD 04 & \cite{Flynn:2004au} & 2 & PT1$\ell$ & PT1$\ell$ & No error quoted.
\\[1.0ex] 
\hline\hline
\end{tabular*}
\caption{Running and matching in determinations of $B_K$ for $N_f=2$.}
}
\end{table}

\fi
\subsubsection{Kaon BSM $B$-parameters}
\label{app-Bi}

No new calculations w.r.t. the previous FLAG report.

\ifx\reducedapptables\undefined
\begin{table}[!ht]

{\footnotesize
\begin{tabular*}{\textwidth}{l @{\extracolsep{\fill}} c c c l}
\hline\hline \\[-1.0ex]
Collab. & Ref. & $\Nf$ & $a$ [fm] & Description 
\\[1.0ex] \hline \hline \\[-1.0ex]
%
ETM 15      & \cite{Carrasco:2015pra} & 2+1+1 & 0.09, 0.08, 0.06 & 
 \parbox[t]{6.0cm}{Combined chiral and continuum extrapolation. Systematic errors to $B_{i}$ from about 4\% to 6\% are obtained 
 from the distribution of results over analyses which differ by  
O$(a^2)$ effects.}
 \\[12.0ex]
\hline\hline
\end{tabular*}
\caption{Continuum extrapolations/estimation of lattice artifacts in
  determinations of the BSM $B_i$ parameters with $\Nf=2+1+1$.}
}
\end{table}
\fi

\ifx\reducedapptables\undefined
\begin{table}[!ht]
{\footnotesize
\begin{tabular*}{\textwidth}{l @{\extracolsep{\fill}} c c c l}
\hline\hline \\[-1.0ex]
{Collab.} & {Ref.} & {$\Nf$} & $a$ [fm] & {Description}  
\\[1.0ex] \hline \hline \\[-1.0ex]
RBC/UKQCD 16 & \cite{Garron:2016mva} & 2+1 & 0.111, 0.083  &
\parbox[t]{6.0cm}{Systematic uncertainty of 1.3\% obtained from half the 
difference between the results on the fine lattice spacing and the
continuum limit.} 
\\[10.0ex] 
\hline \\[-1.0ex]
SWME 15A & \cite{Jang:2015sla} & 2+1 & 0.12, 0.09, 0.06, 0.045  &
\parbox[t]{6.0cm}{The three finest lattice spacings are used for the combined chiral and continnum extrapolation.  
Residual combined discretization, sea-quark extrapolation
and $\alpha_s$ matching error from about 4.4\% to 9.6\% is reported for $B_{i}$ and is obtained from the difference between linear
  fit in $a^2$, $m_{\rm sea}$ and a fit where $\alpha_s$ dependence is added.}
\\[21.0ex] \hline \\[-1.0ex]

SWME 14C & \cite{Jang:2014aea} & 2+1 & 0.082, 0.059, 0.044  &
\parbox[t]{6.0cm}{Residual combined discretization and sea-quark extrapolationg
error of 1--8\% from difference between linear
  fit in $a^2$, $m_{\rm sea}$  and a constrained nighteen-parameter extrapolation.}
\\[12.0ex] \hline \\[-1.0ex]

SWME 13A & \cite{Bae:2013tca} & 2+1 & 0.09, 0.06, 0.045  &
\parbox[t]{6.0cm}{Residual combined discretization, sea-quark extrapolation
and $\alpha_s$ matching error for $B_i$ varies from 4.5\% to -5.7\% , from difference between linear
  fit in $a^2$, $m_{\rm sea}$ and a fit where $\alpha_s$ dependence is added.}
\\[12.0ex] \hline \\[-1.0ex]

RBC/UKQCD 12E & 
\cite{Boyle:2012qb} & 2+1 & 0.087  &
\parbox[t]{6.0cm}{Computation at only one value of the lattice spacing. 
Estimate for the systematic discretisation error of about 1.5\% based on the 
corresponding estimate from the $B_K$ computation.}
\\[12.0ex]
\hline\hline
\end{tabular*}
\caption{Continuum extrapolations/estimation of lattice artifacts in
  determinations of the BSM $B_i$ parameters with $\Nf=2+1$.}
}
\end{table}
\fi

\ifx\reducedapptables\undefined
\begin{table}[!ht]
{\footnotesize
\begin{tabular*}{\textwidth}{l @{\extracolsep{\fill}} c c c l}
\hline\hline \\[-1.0ex]
{Collab.} & {Ref.} & {$\Nf$} & {$M_{\pi,\rm min}\,[\mev]$} & {Description}  
\\[1.0ex] \hline \hline \\[-1.0ex]

ETM 12D      & \cite{Bertone:2012cu} & 2 & 0.1, 0.09, 0.07, 0.05 & 
 \parbox[t]{6.0cm}{Four lattice spacings; Estimates of systematic uncertainties 
 obtained from the half difference of the distance between the finest lattice spacing and the 
 continuum limit.}
 \\[1.0ex]  \\[-1.0ex] 
\hline\hline
\end{tabular*}
\caption{Continuum extrapolations/estimation of lattice artifacts in
  determinations of the BSM $B_i$ parameters with $\Nf=2$.}
}
\end{table}
\fi


\ifx\reducedapptables\undefined
\begin{table}[!ht]
{\footnotesize
\begin{tabular*}{\textwidth}{l @{\extracolsep{\fill}} c c c l}
\hline\hline \\[-1.0ex]
{Collab.} & {Ref.} & {$\Nf$} & {$M_{\pi,\rm min}\,[\mev]$} & {Description}  
\\[1.0ex] \hline \hline \\[-1.0ex]
%
ETM 15 & \cite{Carrasco:2015pra} & 2+1+1 & \parbox[t]{1.8cm}{245,\,239,\,211} & 
  \parbox[t]{6cm}{Each $M_{\pi,\rm min}$ entry corresponds to a
  different lattice spacing. Simultaneous chiral \& continuum
  extrapolations, based on polynomial and {\Ch}PT at NLO, are carried out leads to systematic
  errors of $1.1-2.6$\% depending on the bag parameter.}
  \\[15.0ex] 
	\hline\hline
\end{tabular*}
\caption{Chiral extrapolation/minimum pion mass in
  determinations of the BSM $B_i$ parameters with $\Nf=2+1+1$.}
}
\end{table}
\fi

\ifx\reducedapptables\undefined
\begin{table}[!ht]
{\footnotesize
\begin{tabular*}{\textwidth}{l @{\extracolsep{\fill}} c c c l}
\hline\hline \\[-1.0ex]
{Collab.} & {Ref.} & {$\Nf$} & {$M_{\pi,\rm min}\,[\mev]$} & {Description}  
\\[1.0ex] \hline \hline \\[-1.0ex]
RBC/UKQCD 16 & \cite{Garron:2016mva} & 2+1 & 
\parbox[t]{1.8cm}{337, 302}
 & \parbox[t]{6cm}{Chiral extrapolations based on polynomial and $SU(2)$-$\chi$PT fits at NLO.
A systematic uncertainty of 0.4\% is quoted, which is half the difference between the two results.}
\\[10.0ex] 
\hline \\[-1.0ex]
SWME 15A & \cite{Jang:2015sla} & 2+1 & 
\parbox[t]{1.8cm}{222/372,\, 206/174,\, 195/222, \, 206/316}
 & \parbox[t]{6cm}{Valence/sea RMS $M_{\pi,\rm min}$ entries correspond to the
four lattice spacings (the last three are used for the chiral-continuum extrapolation). Chiral extrapolations based on SU(2) staggered
{\Ch}PT at NNLO (with some coefficients fixed by Bayesian priors), and
also including one analytic NNNLO term. Residual error of 0.4-1.2\% depending on the bag parameter from
changing the Bayesian priors and fit method.}
\\[26.0ex] \hline \\[-1.0ex]

SWME 14C & \cite{Jang:2014aea} & 2+1 & 
\parbox[t]{1.8cm}{206/174,\, 195/222,\, 207/316}
 & \parbox[t]{6cm}{Valence/sea RMS $M_{\pi,\rm min}$ entries correspond to the
three lattice spacings. Chiral extrapolations performed via $B_i$-ratios that do not show SU(2)
NLO {\Ch}PT contribution and assuming various terms up to NNLO (with some coefficients fixed by Bayesian priors). 
}
\\[18.0ex] \hline \\[-1.0ex]

SWME 13A & \cite{Bae:2013tca} & 2+1 & 
\parbox[t]{1.8cm}{207/243,\, 196/262,\, 207/316}
 & \parbox[t]{6cm}{Valence/sea RMS $M_{\pi,\rm min}$ entries correspond to the
three lattice spacings. Chiral extrapolations performed via $B_i$-ratios that do not show SU(2)
NLO {\Ch}PT contribution and assuming various terms up to NNLO (with some coefficients fixed by Bayesian priors). 
Residual error in the valence of about 0.1\% from
doubling the widths of Bayesian priors. In the sea a combined error with the matching procedure of 4.4-5.6\% is reported.}
\\[29.0ex] \hline \\[-1.0ex]

RBC/UKQCD 12E & \cite{Boyle:2012qb} & 2+1 & 
\parbox[t]{1.8cm}{290/290}  &
\parbox[t]{6.0cm}{Chiral extrapolations based on polynomial and {\Ch}PT fits at NLO are carried out. 
  Central values are obtained from polynomial fits. Mild dependence on the quark mass. Systematic 
  uncertainties are estimated to about 4\% for all $B_i$'s.}
\\[15.0ex] 
\hline\hline%
\end{tabular*}
\caption{Chiral extrapolation/minimum pion mass in
  determinations of the BSM $B_i$ parameters with $\Nf=2+1$.}
}
\end{table}
\fi

\ifx\reducedapptables\undefined
\begin{table}[!ht]
{\footnotesize
\begin{tabular*}{\textwidth}{l @{\extracolsep{\fill}} c c c c l}
\hline\hline \\[-1.0ex]
{Collab.} & {Ref.} & {$\Nf$} & {$M_{\pi,\rm min}\,[\mev]$} & {Description}  
\\[1.0ex] \hline \hline \\[-1.0ex]
ETM 12D & \cite{Bertone:2012cu} & 2 & \parbox[t]{1.8cm}{400,
  270,\\ 300, 270} & 
  \parbox[t]{6cm}{Each $M_{\pi,\rm min}$ entry corresponds to a
  different lattice spacing. Simultaneous chiral \& continuum
  extrapolations, based on polynomial and {\Ch}PT at NLO, are carried out. 
 } 
\\ [10.0ex]
\hline \hline\\[-1.0ex]
\end{tabular*}
\caption{Chiral extrapolation/minimum pion mass in
  determinations of the BSM $B_i$ parameters with $\Nf=2$.}
}

\end{table}
\fi
\ifx\reducedapptables\undefined
\begin{table}[!ht]
{\footnotesize
\begin{tabular*}{\textwidth}{l @{\extracolsep{\fill}} c c c c l}
\hline\hline \\[-1.0ex]
Collab. & Ref. & $\Nf$ & $L$ [fm] & ${M_{\pi,\rm min}}L$ & Description 
\\[1.0ex] \hline \hline \\[-1.0ex]
%
ETM 15 & \cite{Carrasco:2015pra} & 2+1+1 & \parbox[t]{1.8cm}{2.1--2.8, 
  2.6, 3.0}& \parbox[t]{1.8cm}{3.5,\,3.2,\,3.2} &  
  \parbox[t]{5cm}{Each $L$ entry corresponds to a different
  lattice spacing, with two volumes at the coarsest lattice
  spacing; results from these two volumes at $M_{\pi} \sim 280$ MeV are
  compatible.} 
\\ [12.0ex]
\hline\hline%
\end{tabular*}
\caption{Finite-volume effects in determinations of the BSM $B_i$
  parameters with $\Nf=2+1+1$. If partially-quenched fits are used, the quoted
  $M_{\pi,\rm min}L$ is for lightest valence (RMS) pion.}  }
\end{table}
\fi

\ifx\reducedapptables\undefined
\begin{table}[!ht]
{\footnotesize
\begin{tabular*}{\textwidth}{l @{\extracolsep{\fill}} c c c c l}
\hline\hline \\[-1.0ex]
Collab. & Ref. & $\Nf$ & $L$ [fm] & ${M_{\pi,\rm min}}L$ & Description 
\\[1.0ex] \hline \hline \\[-1.0ex]
RBC/UKQCD 16 & \cite{Garron:2016mva} & 2+1
& \parbox[t]{1.8cm}{2.7, \, 2.7} & 
$4.5, \, 4.0$ 
& \parbox[t]{5cm}{Finite-volume effects are found to be negligible
  compared to systematic errors and are thus omitted in the final
  error budget.\\} 
\\
	\hline \\[-1.0ex]

SWME 15A & \cite{Jang:2015sla} & 2+1
& \parbox[t]{1.8cm}{2.4--3.4, 2.5--5.8, \,2.9--3.9,\,2.9} & 
$\gtrsim 3.8$ 
& \parbox[t]{5cm}{$L$ entries correspond to the four lattice spacings,
with several volumes in most cases. Finite-volume effects
estimated using NLO SU(2) S{\Ch}PT.}
\\[10.0ex] \hline \\[-1.0ex]

SWME 14C & \cite{Jang:2014aea} & 2+1
& \parbox[t]{1.8cm}{2.8--5.4,\, 2.8--3.8, 2.8} & 
5.6,\,3.7,\,2.9
& \parbox[t]{5cm}{$L$ entries correspond to the three lattice spacings,
with several volumes in most cases. Finite-volume effects
estimated using NLO {\Ch}PT.}
\\[10.0ex] \hline \\[-1.0ex]

SWME 13A & \cite{Bae:2013tca} & 2+1
& \parbox[t]{1.8cm}{2.4--3.4, 2.8--3.3, 2.8} & 
3.5,\,3.3,\,2.9
& \parbox[t]{5cm}{$L$ entries correspond to the three lattice spacings,
with several volumes in most cases. Finite-volume effects
estimated using NLO {\Ch}PT.}
\\[10.0ex] \hline \\[-1.0ex]

RBC/UKQCD 12E & \cite{Boyle:2012qb} 
& 2+1 & \parbox[t]{1.8cm}{2.8} &
$\gtrsim 4.0$ %
& \parbox[t]{5cm}{ 
The $L$ value corresponds to the unique lattice
spacing. Finite-volume effects, estimated using NLO {\Ch}PT are small, as 
it has also been found in the $B_K$ computation,    
and they have thus been  neglected in the final error budget analysis. 
}
\\[18.0ex] 
\hline\hline
\end{tabular*}
\caption{Finite-volume effects in determinations of the BSM $B_i$
  parameters with $\Nf=2+1$. If partially-quenched fits are used, the quoted
  $M_{\pi,\rm min}L$ is for lightest valence (RMS) pion.}  }
\end{table}
\fi

\ifx\reducedapptables\undefined
\begin{table}[!ht]
{\footnotesize
\begin{tabular*}{\textwidth}{l @{\extracolsep{\fill}} c c c c l}
\hline\hline \\[-1.0ex]
Collab. & Ref. & $\Nf$ & $L$ [fm] & ${M_{\pi,\rm min}}L$ & Description 
\\[1.0ex] \hline \hline \\[-1.0ex]
ETM 12D & \cite{Bertone:2012cu} & 2 & \parbox[t]{1.8cm}{2.1,\\
  2.2/2.9,\\ 2.2, 2.6}& \parbox[t]{1.8cm}{5,\\ 3.3/4.3,\\ 3.3, 3.5} &  
  \parbox[t]{5cm}{Each $L$ entry corresponds to a different
  lattice spacing, with two volumes at the second less coarse lattice
  spacing. Results from these two volumes at $M_{\pi} \sim 300$ MeV are
  compatible.} 
\\ [15.0ex] \hline\hline \\[-1.0ex]
\end{tabular*}
\caption{Finite-volume effects in determinations of the BSM $B_i$
  parameters with $\Nf=2$. If partially-quenched fits are used, the quoted
  $M_{\pi,\rm min}L$ is for lightest valence (RMS) pion.}  }
\end{table}
\fi

\ifx\reducedapptables\undefined
\begin{table}[!ht]
{\footnotesize
\begin{tabular*}{\textwidth}{l @{\extracolsep{\fill}} c c c c l}
\hline\hline \\[-1.0ex]
& & & & running & \\
\rb{Collab.} & \rb{Ref.} & \rb{$\Nf$} & \rb{Ren.} & match. & \rb{Description} 
\\[1.0ex] \hline \hline \\[-1.0ex]

ETM 15 & \cite{Carrasco:2015pra} & 2+1+1 & RI & PT1$\ell$ &
  \parbox[t]{5cm}{Uncertainty from RI renormalization combined with discretisation effects estimates are reported to be from about 4\% to 6\% . 
  Additional error from 1.8 to 3.9\% (depending on the bag parameter) for   the conversion to $\msbar$ at the scale of 3 GeV.}
\\ [18.0ex] 
\hline\hline%
\end{tabular*}
\caption{Running and matching in determinations of the BSM $B_i$ parameters with $\Nf=2+1+1$.}
}
\end{table}
\fi

\ifx\reducedapptables\undefined
\begin{table}[!ht]
{\footnotesize
\begin{tabular*}{\textwidth}{l @{\extracolsep{\fill}} c c c c l}
\hline\hline \\[-1.0ex]
& & & & running & \\
\rb{Collab.} & \rb{Ref.} & \rb{$\Nf$} & \rb{Ren.} & match. & \rb{Description} 
\\[1.0ex] \hline \hline \\[-1.0ex]
RBC/UKQCD 16 & \cite{Garron:2016mva} & 2+1 & RI & PT1$\ell$ &
  \parbox[t]{5cm}{Two different RI-SMOM schemes used to estimate
    systematic error in conversion to $\msbar$, which varies from
    1--4\%, depending on the four-quark operator.\\}
	\\
\hline\\[-1.0ex]

SWME 15A & \cite{Jang:2015sla} & 2+1 & PT1$\ell$ & PT1$\ell$ &
\parbox[t]{5cm}{Uncertainty from neglecting higher orders estimated from about 4.4\% to 9.6\% (depending on the bag parameter)
 by identifying the unknown 2-loop coefficient with result at the
  smallest lattice spacing.}
\\[15.0ex] \hline \\[-1.0ex]

SWME 14C & \cite{Jang:2014aea} & 2+1 & PT1$\ell$ & PT1$\ell$ &
\parbox[t]{5cm}{Uncertainty from neglecting higher orders estimated at
4.4\% by identifying the unknown 2-loop coefficient with result at the
  smallest lattice spacing.}
\\[12.0ex] \hline \\[-1.0ex]

SWME 13A & \cite{Bae:2013tca} & 2+1 & PT1$\ell$ & PT1$\ell$ &
\parbox[t]{5cm}{Uncertainty from neglecting higher orders estimated at
4.4\% (in combination with systematic uncertainty from 
CL and chiral extrapolation fit) by identifying the unknown 2-loop coefficient with result at the
  smallest lattice spacing.}
\\[18.0ex] \hline \\[-1.0ex]

RBC/UKQCD 12E & \cite{Boyle:2012qb} & 2+1 & RI & PT1$\ell$ &
\parbox[t]{5cm}{Computation in RI-MOM scheme. Systematic error from the conversion to $\msbar$
is estimated by taking the half of the difference between the LO and the NLO result.}
\\[12.0ex] 
\hline\hline%
\end{tabular*}
\caption{Running and matching in determinations of the BSM $B_i$ parameters with $\Nf=2+1$.}
}
\end{table}
\fi

\ifx\reducedapptables\undefined
\begin{table}[!ht]
{\footnotesize
\begin{tabular*}{\textwidth}{l @{\extracolsep{\fill}} c c c c l}
\hline\hline \\[-1.0ex]
& & & & running & \\
\rb{Collab.} & \rb{Ref.} & \rb{$\Nf$} & \rb{Ren.} & match. & \rb{Description} 
\\[1.0ex] \hline \hline \\[-1.0ex]
ETM 12D & \cite{Bertone:2012cu} & 2 & RI & PT1$\ell$ &
  \parbox[t]{5cm}{Uncertainty from RI renormalization estimated at 2.5\%.}
\\ [4.0ex] \hline \hline\\[-1.0ex]
\end{tabular*}
\caption{Running and matching in determinations of the BSM $B_i$ parameters with $\Nf=2$.}
}
\end{table}

\fi

\setcounter{subsection}{4}

\clearpage
\subsection{Notes to Sec.~\ref{sec:DDecays} on $D$-meson decay constants and form factors}
\label{app:DDecays}




\begin{table}[!htb]

{\footnotesize
\begin{tabular*}{\textwidth}{l @{\extracolsep{\fill}} c c c   l}
\hline\hline \\[-1.0ex]
{Collab.} & {Ref.} & {$\Nf$} & {$M_{\pi,\rm min}\,[\mev]$}  & {Description}  
\\[1.0ex] \hline \hline \\[-1.0ex]
  {$\chi$QCD 20A} & \cite{Chen:2020qma} & 2+1  & 114 & \parbox[t]{5.8cm}{Simulations are performed very close to the physical point
and linear interpolations/extrapolations are used to correct for mismatches.}\\[10.0ex]
\hline \\[-2.0ex]
\parbox[t]{2cm}{RBC/UKQCD~17, RBC/UKQCD 18A} & \cite{Boyle:2017jwu, Boyle:2018knm} & 2+1 & \parbox{2cm}{139, 139, 232} & \parbox[t]{5.8cm}{
The lattice spacing, pion-mass and charm-quark mass dependences are fit simultaneously through a Taylor expansion in $a^2$, $(m_\pi^2-
m_\pi^{2 phys})$ and $1/m_H-1/m_{D_{(s)}}$.}\\[12.0ex]
\ifx\reducedapptables\undefined
\hline \\[-2.0ex]
$\chi$QCD~14& \cite{Yang:2014sea} & 2+1 & \parbox[t]{2cm}{334, 296} & \parbox[t]{5.8cm}{Chiral and continuum extrapolations
 are peformed simultaneously using linear fits in $m_l$ (and quadratic or including partially quenched chiral logs, in order to assess the systematic error)
plus terms up to O$(a^2)$ and O$(a^4m_c^4)$.}
\\[15.0ex] \hline \\[-2.0ex]
HPQCD 12A& \cite{Na:2012iu} & 2+1 & \parbox[t]{2cm}{460, 329} & \parbox[t]{5.8cm}{Chiral and continuum extrapolations
 are peformed simultaneously using PQHM$\chi$PT augmented by $a$-dependent terms: $c_0(am_c)^2 + c_1(am_c)^4$.}
\\[10.0ex] \hline \\[-2.0ex]
FNAL/MILC 11 & \cite{Bazavov:2011aa} & 2+1 & \parbox[t]{2cm}{570, 440, 320} &
\parbox[t]{5.8cm}{Chiral and continuum extrapolations are peformed 
simultaneously using HM$\chi$PT for rooted staggered quarks.
Effects of hyperfine and flavor splittings are also included.}
\\[13.0ex] \hline \\[-2.0ex]
PACS-CS 11 & \cite{Namekawa:2011wt} & 2+1 & \parbox[t]{2cm}{152} & \parbox[t]{5.8cm}{Simulations are  
reweighted in the light- and strange-quark masses to the physical point.}
\\[7.0ex] \hline \\[-2.0ex]
HPQCD 10A & \cite{Davies:2010ip} & 2+1 & \parbox[t]{2cm}{542, 460, 329, 258, 334} & \parbox[t]{5.8cm}{Chiral and continuum 
extrapolations are performed simultaneously. Polynomials up to  $\left( \frac{m_{q,sea}-m_{q,phys}}{m_{q,phys}}\right)^2$ for $q=s,l$ and
up to  $(am_{c})^8$ are kept.}
\\[12.0ex] \hline \\[-2.0ex]
HPQCD/UKQCD 07 & \cite{Follana:2007uv} & 2+1 & \parbox[t]{2cm}{542, 460, 329} & \parbox[t]{5.8cm}{Combined chiral and continuum extrapolations using
HM$\chi$PT at NLO augmented by second and third-order polynomial terms in $m_q$ and terms up to $a^4$.}
\\[10.0ex] \hline \\[-2.0ex]
FNAL/MILC 05 & \cite{Aubin:2005ar} & 2+1 &
\parbox[t]{2cm}{$>~440\;, 440\;, 400$} & \parbox[t]{5.8cm}{Chiral extrapolations are first performed at each lattice spacing
uisng NLO HM$\chi$PT for rooted staggered quarks. Lattice artifacts are then extrapolated linearly in $a^2$. }
\\[13.0ex] 
%
\fi
\hline\hline
\end{tabular*}
\caption{Chiral extrapolation/minimum pion mass in $N_f=2+1$ determinations of the
  $D$- and $D_s$-meson decay constants.  For actions with multiple
 species of pions, masses quoted are the RMS pion masses.    The
 different $M_{\pi,\rm min}$ entries correspond to the different lattice
 spacings.} 
}
\end{table}
\begin{table}[!htb]
{\footnotesize
\begin{tabular*}{\textwidth}{l @{\extracolsep{\fill}} c c c   l}
\hline\hline \\[-1.0ex]
{Collab.} & {Ref.} & {$\Nf$} & {$M_{\pi,\rm min}\,[\mev]$}  & {Description}  
\\[1.0ex] \hline \hline \\[-1.0ex]
\parbox[t]{2cm}{Blossier~18, Balasubramanian 19} & \cite{Blossier:2018jol,Balasubramanian:2019net} & 2 &
\parbox[t]{2cm}{282, 194, 269} & \parbox[t]{5.8cm}{Linear fits (in $m_\pi^2$ and in $a^2$) are used in the 
  combined chiral/continuum extrapolation. NLO HM$\chi$PT expressions are
  used for a cross-check, concluding however that there are not enough data
  points to be sensitive to the NLO terms.\\}
\\[5.0ex]  \\[-2.0ex]
%
\hline\hline
\end{tabular*}
\caption{Chiral extrapolation/minimum pion mass in $N_f=2$ determinations of the
  $D$- and $D_s$-meson decay constants.  For actions with multiple
 species of pions, masses quoted are the RMS pion masses.    The
 different $M_{\pi,\rm min}$ entries correspond to the different lattice
 spacings.} 
}
\end{table}

\begin{table}[!htb]
{\footnotesize
\begin{tabular*}{\textwidth}{l @{\extracolsep{\fill}} c c c c l}
\hline\hline \\[-1.0ex]
Collab. & Ref. & $\Nf$ & $L$ [fm] & ${M_{\pi,\rm min}}L$ & Description 
\\[1.0ex] \hline \hline \\[-1.0ex]
       {$\chi$QCD 20A} & \cite{Chen:2020qma} & 2+1 &  \parbox[t]{1.75cm}{5.5} &  \parbox[t]{1.6cm}{3.2} & No explicit discussion of FSE.
\\[2.0ex] 
\hline \\[-2.0ex]
\parbox[t]{2.2cm}{RBC/UKQCD~17, RBC/UKQCD 18A} & \cite{Boyle:2017jwu, Boyle:2018knm} & 2+1 & 
 \parbox[t]{1.75cm}{5.5/2.7, 5.4/2.7, 3.4} & \parbox[t]{1.6cm}{$3.86, 3.78$, $4.11$}  &
\parbox[t]{4.9cm}{FV errors estimated to be at the permille level by either comparing values of $m_\pi L$
  to the study of FSE by MILC in~\cite{Bazavov:2014wgs} or by 1-loop HM$\chi$PT.} \\[12.0ex]
\ifx\reducedapptables\undefined
\hline \\[-1.0ex]
$\chi$QCD~14& \cite{Yang:2014sea} & 2+1 &
\parbox[t]{1.75cm}{2.7, 2.7}
& \parbox[t]{1.6cm}{4.6, 4.1}  &
\parbox[t]{4.7cm}{No explicit discussion of FV effects.}  
\\[5.0ex] \hline \\[-1.0ex]
HPQCD 12A& \cite{Na:2012iu} & 2+1 &
\parbox[t]{1.75cm}{2.4/2.8, 2.4/3.4}
& \parbox[t]{1.6cm}{3.8, 4.2}  &
\parbox[t]{4.7cm}{FV errors  estimated by comparing finite- and infinite-volume $\chi$PT.}  
\\[8.0ex] \hline \\[-1.0ex]
FNAL/MILC 11 & \cite{Bazavov:2011aa} & 2+1 & 
	     \parbox[t]{1.75cm}{2.4, 2.4/2.88, 2.52/3.6} & \parbox[t]{1.6cm}{3.9, 3.8, 4,2 }  &
\parbox[t]{4.7cm}{FV errors  estimated using  finite-volume $\chi$PT.}
\\[8.0ex] \hline \\[-1.0ex]
PACS-CS 11 & \cite{Namekawa:2011wt} & 2+1 &
	     \parbox[t]{1.75cm}{$2.88$} & \parbox[t]{1.6cm}{2.2 (before reweighting)}  &
\parbox[t]{4.7cm}{No discussion of FV effects.}
\\[8.0ex] \hline \\[-1.0ex]
HPQCD 10A & \cite{Davies:2010ip} & 2+1 &
             \parbox[t]{1.75cm}{2.4, 2.4/2.88/3.36, $2.52, 2.88$, 2.82} & \parbox[t]{1.6cm}{$3.9, 3.8$, $4.1, 4.5$, 4.6 }  &
\parbox[t]{4.7cm}{FV errors  estimated using finite- vs.~infinite-volume $\chi$PT.}
\\[10.0ex] \hline \\[-1.0ex]
HPQCD/UKQCD 07 & \cite{Follana:2007uv} & 2+1 & 
	     \parbox[t]{1.75cm}{2.4, 2.4/2.88, 2.52} & \parbox[t]{1.6cm}{3.9, 3.8, 4.1}  &
\parbox[t]{4.7cm}{FV errors  estimated using finite- vs.~infinite-volume $\chi$PT.}
\\[8.0ex] \hline \\[-1.0ex]
FNAL/MILC 05 & \cite{Aubin:2005ar} & 2+1 & \parbox[t]{1.75cm}{2.8, 2.9, 2.5} & \parbox[t]{1.6cm}{$3.8, 3.8,$ $4.1$ }  &
\parbox[t]{4.7cm}{FV errors estimated to be  $1.5\%$ or less from $\chi$PT.} 
\\[5.0ex] 
\fi
\hline\hline
\end{tabular*}
\caption{Finite-volume effects in $N_f=2+1$ determinations of the $D$- and $D_s$-meson
 decay constants.  Each $L$-entry corresponds to a different
 lattice spacing, with multiple spatial volumes at some lattice
 spacings. For actions with multiple species of pions, the lightest masses are
quoted.} 
}
\end{table}
\begin{table}[!htb]


{\footnotesize
\begin{tabular*}{\textwidth}{l @{\extracolsep{\fill}} c c c c l}
\hline\hline \\[-1.0ex]
Collab. & Ref. & $\Nf$ & $L$ [fm] & ${M_{\pi,\rm min}}L$ & Description 
\\[1.0ex] \hline \hline \\[-1.0ex]
\parbox[t]{2cm}{Blossier~18, Balasubramanian 19} & \cite{Blossier:2018jol,Balasubramanian:2019net} & 2 &
	     \parbox[t]{1.75cm}{2.4/3.6, 2.1/3.1/4.2, 2.3/3.1 } & \parbox[t]{1.6cm}{5.2, 4.1, 4.2}  &
\parbox[t]{4.7cm}{No explicit discussion of FV effects, but $m_\pi L>4$ always.}
\\[10.0ex]  
\ifx\reducedapptables\undefined
\hline \\[-1.0ex]
\parbox[t]{2.0cm}{TWQCD~14} & \cite{Chen:2014hva} & 2 & 
 \parbox[t]{1.75cm}{1.5} & \parbox[t]{1.6cm}{1.92}  &
\parbox[t]{4.7cm}{No explicit discussion of FV effects.}
\\[5.0ex]  \hline \\[-1.0ex]
\parbox[t]{2cm}{ALPHA~13B} & \cite{Heitger:2013oaa} & 2 &
	     \parbox[t]{1.75cm}{2.1/3.1/4.2, 2.3/3.1 } & \parbox[t]{1.6cm}{4, 4.2}  &
\parbox[t]{4.7cm}{No explicit discussion of FV effects, but $m_\pi L>4$ always.}
\\[5.0ex]  \hline \\[-1.0ex]
\parbox[t]{2cm}{ETM~09\\ ETM~11A\\ ETM~13} & \cite{Blossier:2009bx,Dimopoulos:2011gx,Carrasco:2013zta} & 2 &
	     \parbox[t]{1.75cm}{2.4, 2.0/2.7, 2.1, 2.6 } & \parbox[t]{1.6cm}{5, 3.3, 3.3, 3.5}  &
\parbox[t]{4.7cm}{FV errors are found to be negligible by comparing results at $m_\pi L = 3.3$ and $m_\pi L=4.3$ for $m_\pi \simeq 310$ MeV.}
\\[8.0ex]  \\[-1.0ex]
\fi
\hline\hline
\end{tabular*}
\caption{Finite-volume effects in $N_f=2$ determinations of the $D$- and $D_s$-meson
 decay constants.  Each $L$-entry corresponds to a different
 lattice spacing, with multiple spatial volumes at some lattice
 spacings. For actions with multiple species of pions, the lightest masses are
quoted.} 
}
\end{table}
\begin{table}[!htb]


{\footnotesize
\begin{tabular*}{\textwidth}{l @{\extracolsep{\fill}} c c l l l}
\hline\hline \\[-1.65ex]
Collab. & Ref. & $\Nf$ & $a$ [fm]  & Continuum extrapolation & Scale Setting
\\[.2ex] \hline \hline \\[-1.65ex]
  {$\chi$QCD 20A} & \cite{Chen:2020qma} & 2+1  & \parbox[t]{.9cm}{0.115} & \parbox[t]{3.8cm}{Cutoff effects are estimated
to be 2\% by comparing the result for $f_{D_s}$ with the one obtained in~\cite{Yang:2014sea}.} & \parbox[t]{3.4cm}{$m_{\Omega}$ used for scale
    setting by RBC/UKQCD, which produced the ensemble}\\[10.0ex]
  \hline \\[-2.0ex]
RBC/UKQCD~17, \\ RBC/UKQCD 18A & \cite{Boyle:2017jwu, Boyle:2018knm} & 2+1 &   \parbox[t]{.9cm}{0.11, 0.08, 0.07} &
\parbox[t]{3.8cm}{The lattice spacing, pion-mass and charm-quark mass dependences are fit simultaneously through a Taylor 
expansion in $a^2$, $(m_\pi^2-m_\pi^{2 phys})$ and $1/m_H-1/m_{D_{(s)}}$.} &
\parbox[t]{3.4cm}{The lattice scale and physical light-quark masses have been determined using $m_\pi$, $m_K$ and 
  $m_\Omega$ as inputs.}
\\[18.0ex]

\ifx\reducedapptables\undefined
\hline \\[-1.65ex]
$\chi$QCD~14 & \cite{Yang:2014sea} & 2+1 & \parbox[t]{.9cm}{0.113, 0.085} &
\parbox[t]{3.8cm}{
Chiral and continuum extrapolations  performed in global fits
including linear terms in $m_l$ and terms up to O$(a^2)$ and O$(a^4m_c^4)$.} 
& \parbox[t]{3.4cm}{Relative scale set through $r_0$, fixed together with the charm and strange quark masses
using $m_{D_s}$, $m_{D_s^*}$ amd $m_{J/\psi}$ as inputs.}
\\[15.0ex] \hline \\[-1.65ex]
HPQCD 12A& \cite{Na:2012iu} & 2+1 & \parbox[t]{.9cm}{0.12, 0.09} &
\parbox[t]{3.8cm}{
Chiral and continuum extrapolations  peformed 
simultaneously  using PQHM$\chi$PT augmented by $a$-dependent terms: $c_0(am_c)^2 + c_1(am_c)^4$.
} & \parbox[t]{3.4cm}{Relative scale set through $r_1$; absolute scale from $f_\pi$, $f_{\rm K}$ and the $\Upsilon$ splitting.
Uncertainties from both $r_1$ and $r_1/a$  propagated.}
\\[15.0ex] \hline \\[-1.65ex]
FNAL/MILC 11 & \cite{Bazavov:2011aa} & 2+1 &
             \parbox[t]{.9cm}{0.15, 0.12, 0.09} &
\parbox[t]{3.8cm}{
Chiral and continuum extrapolations  performed 
simultaneously using 1-loop HM$\chi$PT for rooted staggered quarks. Effects of hyperfine and flavor splittings
are also included. 
} & \parbox[t]{3.4cm}{Relative scale set through $r_1=0.3117(22)$. The error in $r_1$
comes from the spread of different absolute scale determinations using $f_\pi$, $f_{\rm K}$ and the $\Upsilon$ splitting.}
\\[21.0ex] \hline \\[-1.65ex]
PACS-CS 11 & \cite{Namekawa:2011wt} & 2+1 &
             \parbox[t]{.9cm}{$0.09$} & \parbox[t]{3.8cm}{
Cutoff effects from the heavy-quark action estimated by naive power counting to be at the percent level.} 
& \parbox[t]{3.4cm}{Scale set through $m_\Omega$.}
\\[12.0ex] \hline \\[-2.1ex]
HPQCD 10A & \cite{Davies:2010ip} & 2+1 &
\parbox[t]{.9cm}{0.15, 0.12, 0.09, 0.06, 0.044} &
\parbox[t]{3.8cm}{
Chiral and continuum 
extrapolations  performed simultaneously. Polynomials 
up to $am_{c}^8$ are kept (even powers only).} & \parbox[t]{3.4cm}{See the discussion for HPQCD 12A.}
\\[12.0ex] \hline \\[-1.65ex]
HPQCD/UKQCD 07 & \cite{Follana:2007uv} & 2+1 & 
\parbox[t]{.9cm}{0.15, 0.12, 0.09} &
\parbox[t]{3.8cm}{
Combined chiral and continuum extrapolations using
HM$\chi$PT at NLO augmented by second and third-order polynomial terms in $m_q$ and terms up to $a^4$.
} & \parbox[t]{3.4cm}{Scale set through $r_1$ obtained from the $\Upsilon$ spectrum using the 
nonrelativistic QCD action for $b$ quarks. Uncertainty propagated among the systematics.}
\\[18.0ex] \hline \\[-1.65ex]
FNAL/MILC 05 & \cite{Aubin:2005ar} & 2+1 &
\parbox[t]{.9cm}{0.175, 0.121, 0.086} &
\parbox[t]{3.8cm}{Most light-quark cutoff effects are removed through
 NLO HM$\chi$PT for rooted staggered quarks. Continuum values are then obtained 
by averaging the $a \approx 0.12$ and $a \approx 0.09$ fm results.}
 &  \parbox[t]{3.4cm}{Scale set through $r_1$ obtained from the $\Upsilon$ spectrum using the 
nonrelativistic QCD action for $b$ quarks.}
\\[21.0ex] 
%
%
\fi
\hline\hline
\end{tabular*}
\caption{Lattice spacings and description of actions used in $N_f=2+1$  determinations
 of the $D$- and $D_s$-meson decay constants.} 
}
\end{table}

\begin{table}[!htb]


{\footnotesize
\begin{tabular*}{\textwidth}{l @{\extracolsep{\fill}} c c l l l}
\hline\hline \\[-1.65ex]
Collab. & Ref. & $\Nf$ & $a$ [fm]  & Continuum extrapolation & Scale Setting
\\[.2ex] \hline \hline \\[-1.65ex]
\parbox[t]{2.0cm}{Blossier~18, Balasubramanian 19} & \cite{Blossier:2018jol,Balasubramanian:2019net} & 2 &
	     \parbox[t]{.9cm}{0.075, 0.065, 0.048 } &
\parbox[t]{3.8cm}{Linear fits (in $m_\pi^2$ and in $a^2$) are used in the 
combined chiral/continuum extrapolation.
\\} & \parbox[t]{3.4cm}{Scale set through $f_K$.}
\\[9.0ex] 
\ifx\reducedapptables\undefined
\hline \\[-2.1ex]
\parbox[t]{2.0cm}{TWQCD~14} & \cite{Chen:2014hva} & 2 & 
\parbox[t]{.9cm}{0.061} &
\parbox[t]{3.8cm}{Uncertainties associated to scale setting and discretization effects
estimated by performig the chiral fits once in physical and once in lattice units
($\approx 2 $ MeV on $f_{ D_{\rm s}}$).
\\} & \parbox[t]{3.4cm}{Scale set through the Wilson flow and $r_0$ set to 0.49 fm.}
\\[9.0ex] \hline \\[-2.1ex]
\parbox[t]{2.0cm}{ALPHA~13B} & \cite{Heitger:2013oaa} & 2 &
	     \parbox[t]{.9cm}{0.065, 0.048 } &
\parbox[t]{3.8cm}{Linear fits (in $m_\pi^2$ and in $a^2$) and
partially quenched HM$\chi$PT functional forms,
including terms linear in $a^2$,  are used in the 
combined chiral/continuum extrapolation.
\\} & \parbox[t]{3.4cm}{Scale set through $f_K$.}
\\[9.0ex] \hline \\[-2.1ex]
\parbox[t]{2.0cm}{ETM~09\\ ETM~11A\\ ETM~13} & \cite{Blossier:2009bx,Dimopoulos:2011gx,Carrasco:2013zta} & 2 &
	     \parbox[t]{.9cm}{0.10, 0.085, 0.065, 0.054 } &
\parbox[t]{3.8cm}{
NLO $SU(2)$ HM$\chi$PT 
supplemented by terms linear in $a^2$ and in $m_D a^2$ is used in the 
combined chiral/continuum extrapolation.
\\} & \parbox[t]{3.4cm}{Scale set through $f_\pi$.}
\\[9.0ex] \\[-2.1ex]
\fi
\hline\hline
\end{tabular*}
\caption{Lattice spacings and description of actions used in $N_f=2$ determinations of the $D$- and $D_s$-meson decay constants.} 
}
\end{table}



\begin{table}[!htb]
{\footnotesize
\begin{tabular*}{\textwidth}{l @{\extracolsep{\fill}} c c c l}
\hline\hline \\[-1.0ex]
Collab. & Ref. & $\Nf$ &  Ren. & Description
\\[1.0ex] \hline \hline \\[-1.0ex]
$\chi$QCD 20A & \cite{Chen:2020qma} & 2+1 & 
RI &
\parbox[t]{6.9cm}{The decay constants are extracted from an exact lattice Ward identity or from NP renormalized operators.}
\\[7.0ex]
\hline \\[-2.0ex]
RBC/UKQCD~17, \\ RBC/UKQCD 18A & \cite{Boyle:2017jwu, Boyle:2018knm} & 2+1 & mNPR &
\parbox[t]{6.9cm}{The local current is renormalized nonperturbatively for the case of the unmixed action; however
in the actual computation the domain wall height is chosen differently in the valence than in the sea and the effect
of that on the renormalization constant is estimated to be 0.4\% through a study in the RI/SMOM scheme.\\}
\\[3.0ex] 
\ifx\reducedapptables\undefined
\hline \\[-1.0ex]
$\chi$QCD~14 & \cite{Yang:2014sea} & 2+1 & 
RI &
\parbox[t]{6.9cm}{The decay constant is extracted from an exact lattice Ward identity and from the NP renormalized axial
current.\\}
\\[3.0ex] \hline \\[-1.0ex]
HPQCD 12A& \cite{Na:2012iu} & 2+1 & 
$-$ &
\parbox[t]{6.9cm}{The axial current is absolutely normalized.}
\\[3.0ex] \hline \\[-1.0ex]
FNAL/MILC 11 & \cite{Bazavov:2011aa} & 2+1 &
mNPR &
\parbox[t]{6.9cm}{2-loop and higher-order perturbative truncation errors estimated to be the full size of the 1-loop term.\\}
\\[3.0ex] \hline \\[-1.0ex]
PACS-CS 11 & \cite{Namekawa:2011wt} & 2+1 &
PT1$\ell$+NP &
\parbox[t]{6.9cm}{Mass dependent part of the renormalization constant of the axial current computed at 1-loop; the NP contribution
is added in the chiral limit.\\}
\\[3.0ex] \hline \\[-1.ex]
HPQCD 10A & \cite{Davies:2010ip} & 2+1 &
$-$ &
\parbox[t]{6.9cm}{The axial current is absolutely normalized.}
\\[3.0ex] \hline \\[-1.0ex]
HPQCD/UKQCD 07 & \cite{Follana:2007uv} & 2+1 & 
$-$&
\parbox[t]{6.9cm}{The axial current is absolutely normalized.}
\\[3.0ex] \hline \\[-1.0ex]
FNAL/MILC 05 & \cite{Aubin:2005ar} & 2+1 &
mNPR &
\parbox[t]{6.9cm}{Errors due to higher order corrections in the perturbative part are estimated to be $1.3\%$.}
\\[3.0ex] \\[-1.0ex]
\fi
\hline\hline
\end{tabular*}
\caption{Operator renormalization in $N_f=2+1$ determinations of the $D$- and $D_s$-meson decay constants.} 
}
\end{table}

\begin{table}[!htb]
{\footnotesize
\begin{tabular*}{\textwidth}{l @{\extracolsep{\fill}} c c c l}
\hline\hline \\[-1.0ex]
Collab. & Ref. & $\Nf$ &  Ren. & Description
\\[1.0ex] \hline \hline \\[-1.0ex]
\parbox[t]{2cm}{Blossier~18, Balasubramanian 19\\} & \cite{Blossier:2018jol,Balasubramanian:2019net} & 2 &
SF &
\parbox[t]{6.9cm}{NP renormalization and improvement of the axial current ($am$ terms included at 1-loop).}
\\[4.0ex]  
\ifx\reducedapptables\undefined
\hline \\[-1.0ex]
TWQCD~14 & \cite{Chen:2014hva} & 2 & 
$-$&
\parbox[t]{6.9cm}{The decay constant is extracted from an exact lattice Ward identity.}
\\[4.0ex] \hline \\[-1.0ex]
\parbox[t]{2cm}{ALPHA~13B } & \cite{Heitger:2013oaa} & 2 &
SF &
\parbox[t]{6.9cm}{NP renormalization and improvement of the axial current ($am$ terms included at 1-loop).}
\\[4.0ex]  
\hline \\[-1.0ex]
\parbox[t]{2cm}{ETM 09\\ ETM 11A\\ ETM~13} & \cite{Blossier:2009bx,Dimopoulos:2011gx,Carrasco:2013zta} & 2 &
$-$ &
\parbox[t]{6.9cm}{The axial current is absolutely normalized.}
\\[3.0ex]  \\[-1.0ex]
\fi
\hline\hline
\end{tabular*}
\caption{Operator renormalization in $N_f=2$ determinations of the $D$- and $D_s$-meson decay constants.} 
}
\end{table}

\begin{table}[!htb]
{\footnotesize
\begin{tabular*}{\textwidth}{l @{\extracolsep{\fill}} c c c l}
\hline\hline \\[-1.2ex]
Collab. & Ref. & $\Nf$ &  Action & Description
\\[1.0ex] \hline \hline \\[-1.2ex]
$\chi$QCD 20A & \cite{Chen:2020qma} & 2+1 & 
Overlap on DW &
\parbox[t]{3.9cm}{$am_c\approx 0.73$ set by using $m_{D_s}$.\\}
\\[4.0ex]
\hline \\[-1.2ex]
\parbox[t]{2cm}{RBC/UKQCD~17, RBC/UKQCD 18A} & \cite{Boyle:2017jwu,Boyle:2018knm} & 2+1 & 
\parbox[t]{2cm}{
M\"obius-DWF on  Shamir-DWF or M\"obius-DWF} &
\parbox[t]{3.9cm}{$0.18<am_h<0.4$. Charm discretization errors estimated using different ways to define the charm quark mass
(through $D$, $D_s$ or $\eta_c$) in the global fits.\\} 
\\[4.0ex]
\ifx\reducedapptables\undefined
\hline \\[-1.2ex]
$\chi$QCD~14 & \cite{Yang:2014sea} & 2+1 & 
Overlap on DW &
\parbox[t]{3.9cm}{$0.29<am_c<0.75$. Heavy-quark discretization errors estimated by including $(am_c)^2$ and $am_c)^4$ terms
in the chiral/continuum extrapolation.\\}
\\[4.0ex] \hline \\[-1.4ex]
HPQCD 12A& \cite{Na:2012iu} & 2+1 & 
HISQ &
\parbox[t]{3.9cm}{$0.41<am_c<0.62$. Heavy-quark discretization errors estimated using different fit ans\"atze to be $\approx~1.2\%$.\\}
\\[4.0ex] \hline \\[-1.4ex]
FNAL/MILC 11 & \cite{Bazavov:2011aa} & 2+1 &
Fermilab  &
\parbox[t]{3.9cm}{Discretization errors from charm quark estimated through a combination of Heavy-Quark and Symanzik Effective Theories to be around 3\% for $f_{D_{(s)}}$ and negligible for the ratio.\\}
\\[4.0ex] \hline \\[-1.4ex]
PACS-CS 11 & \cite{Namekawa:2011wt} & 2+1 &
Tsukuba &
\parbox[t]{3.9cm}{$am_c\approx 0.57$. Heavy-quark discretization errors estimated to be at the percent level by power counting.\\}
\\[4.0ex] \hline \\[-1.2ex]
HPQCD 10A & \cite{Davies:2010ip} & 2+1 &
HISQ &
\parbox[t]{3.9cm}{$0.193< am_c<0.825$. Heavy-quark discretization errors estimated by changing the fit-inputs to be $\approx~0.4\%$.\\}
\\[4.0ex] \hline \\[-1.2ex]
HPQCD/UKQCD 07 & \cite{Follana:2007uv} & 2+1 & 
HISQ &
\parbox[t]{3.9cm}{$0.43<am_c<0.85$.  Heavy-quark discretization errors estimated from the chiral/continuum fits to be $\approx~0.5\%$.\\
}
\\[4.0ex] \hline \\[-1.4ex]
FNAL/MILC 05 & \cite{Aubin:2005ar} & 2+1 &
Fermilab  &
\parbox[t]{3.9cm}{Discretization errors from charm quark estimated via heavy-quark power-counting at 4.2\% for $f_{D_{(s)}}$ and 0.5\% for the ratio.\\}
\\[4.0ex] 
%
\fi
\hline\hline
\end{tabular*}
\caption{Heavy-quark treatment in  $N_f=2+1$ determinations of the $D$- and $D_s$-meson decay constants.} 
}
\end{table}

\begin{table}[!htb]
{\footnotesize
\begin{tabular*}{\textwidth}{l @{\extracolsep{\fill}} c c c l}
\hline\hline \\[-1.2ex]
Collab. & Ref. & $\Nf$ &  Action & Description
\\[1.0ex] \hline \hline \\[-1.2ex]
\parbox[t]{2cm}{Blossier~18, Balasubramanian 19} & \cite{Blossier:2018jol,Balasubramanian:2019net} & 2 &
npSW &
\parbox[t]{3.9cm}{$am_c\leq 0.32$. Axial current nonperturbatively improved (O($am$) at 1-loop).\\}
\\[10.0ex] 
\ifx\reducedapptables\undefined
\hline  \\[-1.2ex]
\parbox[t]{2cm}{TWQCD~14} & \cite{Chen:2014hva} & 2 &
DW &
\parbox[t]{3.9cm}{$am_c\leq 0.55$. Optimal Domain Wall fermions~\cite{Chiu:2002ir} preserving chiral symmetry.\\}
\\[4.0ex] \hline  \\[-1.2ex]
\parbox[t]{2cm}{ALPHA~13B } & \cite{Heitger:2013oaa} & 2 &
npSW &
\parbox[t]{3.9cm}{$am_c\leq 0.28$. Axial current nonperturbatively improved (O($am$) at 1-loop).\\}
\\[4.0ex] 
\parbox[t]{2cm}{ETM~09\\ ETM~11A\\ ETM~13} & \cite{Blossier:2009bx,Dimopoulos:2011gx,Carrasco:2013zta} & 2 &
tmWil &
\parbox[t]{3.9cm}{$0.16<am_c<0.23$. $D(a_{\rm min})\approx 5\%$ in ETM~09.}
\\[4.0ex]  \\[-1.3ex]
\fi
\hline\hline
\end{tabular*}
\caption{Heavy-quark treatment in  $N_f=2$ determinations of the $D$- and $D_s$-meson decay constants.} 
}
\end{table}

\clearpage

\FloatBarrier
\subsubsection{Form factors for semileptonic decays of charmed hadrons}
\label{app:DtoPi_Notes}
\FloatBarrier

\begin{table}[!ht]

{\footnotesize
\begin{tabular*}{\textwidth}{l @{\extracolsep{\fill}} c c c l l}
\hline\hline \\[-1.0ex]
Collab. & Ref. & $\Nf$ & $a$ [fm] & Continuum extrapolation & Scale setting 
\\[1.0ex] \hline \hline \\[-1.0ex]
HPQCD 21A & \cite{Chakraborty:2021qav} & 2+1+1 & \parbox[t]{1.3cm}{0.042, 0.06, 0.09, 0.12, 0.15}  &
\parbox[t]{4.1cm}{Modified $z$-expansion fit combining the continuum and chiral extrapolations and the momentum-transfer dependence. Discretization effects assumed dominated by the charm scale. Discretization errors on form factors between 0.4\% and 1.2\% as a function of the momentum transfer.} & \parbox[t]{3.5cm}{Scale setting from $f_\pi$ via the flow quantity $w_0$~\cite{Dowdall:2013rya,Chakraborty:2014aca,Chakraborty:2016mwy}.}
\\[23.0ex]  \\ \hline
 \\[-2.0ex]
Zhang  21  & \cite{Zhang:2021oja} & 2+1 & \parbox[t]{1.3cm}{0.080, 0.11}   &
\parbox[t]{4.1cm}{Continuum extrapolation combined with fit to $q^2$-dependence of form factors in a ``modified'' $z$-expansion. Systematics estimated from difference between extrapolated results and results at smallest lattice spacing, and difference between two current renormalization methods.} & \parbox[t]{3.5cm}{Set from Wilson-flow quantity $w_0$.}
\\[25.0ex]  \\ \hline
 \\[-2.0ex]
HPQCD 20 & \cite{Cooper:2020wnj} & 2+1+1 & \parbox[t]{1.3cm}{0.06, 0.09, 0.12, 0.15}  &
\parbox[t]{4.1cm}{Modified $z$-expansion fit combining the continuum and chiral extrapolations and the momentum-transfer dependence, and, for the heavy-HISQ spectator $b$ quark, the dependence on $1/m_Q$. The analysis combines data with NRQCD $b$ quarks and data with HISQ heavy quarks.} & \parbox[t]{3.5cm}{Scale setting from $f_\pi$ via the flow quantity $w_0$~\cite{Dowdall:2013rya,Chakraborty:2014aca,Chakraborty:2016mwy}.}
\\[23.0ex]  \\ 
\hline\hline
\end{tabular*}
\caption{Continuum extrapolations/estimation of lattice artifacts in $N_f=2+1+1$ determinations of form factors for semileptonic decays of charmed hadrons.}
}
\end{table}

\ifx\reducedapptables\undefined

\begin{table}[!ht]

{\footnotesize
\begin{tabular*}{\textwidth}{l @{\extracolsep{\fill}} c c c l l}
\hline\hline \\[-1.0ex]
Collab. & Ref. & $\Nf$ & $a$ [fm] & Continuum extrapolation & Scale setting 
\\[1.0ex] \hline \hline \\[-1.0ex]
ETM 17D, 18 & \cite{Lubicz:2017syv,Lubicz:2018rfs} & 2+1+1 & \parbox[t]{1.3cm}{0.062, 0.082, 0.089}  &
\parbox[t]{4.1cm}{Modified $z$-expansion fit combining the continuum and chiral extrapolations and the momentum-transfer dependence. Lattice-spacing dependence through $\mathcal{O}(a^2)$, with systematic uncertainty estimated by adding $\mathcal{O}(a^4)$ terms constrained by priors. Additional terms included to fit artifacts due to the breaking of rotational invariance. Meson momenta tuned to be constant with changing lattice spacing and volume.\\} & \parbox[t]{3.5cm}{Relative scale set through $M_{c's'}$, the mass of a fictitious meson made of valence quarks of mass $r_0m_{s'} = 0.22$ and $r_0m_{c'} = 2.4$. Absolute scale from the
experimental value of $f_\pi$.
}
\\[17.0ex] \hline \\[-2.0ex]
JLQCD 17B & \cite{Kaneko:2017xgg} & 2+1 & \parbox[t]{1.3cm}{0.044, 0.055, 0.080}   &
\parbox[t]{4.1cm}{Joint chiral-continuum extrapolation, with mass dependence based on hard-pion HQ$\chi$PT.} & \parbox[t]{3.5cm}{Set from $t_0$ by using the value in physical units provided in~\cite{Borsanyi:2012zs}.}
\\[10.0ex] \hline \\[-2.0ex]
Meinel 16, 17  & \cite{Meinel:2016dqj,Meinel:2017ggx} & 2+1 & \parbox[t]{1.3cm}{0.085, 0.11}   &
\parbox[t]{4.1cm}{Joint chiral-continuum extrapolation, combined with fit to $q^2$-dependence of form factors in a ``modified'' $z$-expansion. Systematics estimated by varying fit form.} & \parbox[t]{3.5cm}{Set from $\Upsilon(2S)$--$\Upsilon(1S)$ splitting, cf.~\cite{Meinel:2010pv}.}
\\[16.0ex] \hline \\[-2.0ex]
HPQCD 10B, 11 & \cite{Na:2010uf,Na:2011mc} & 2+1 & \parbox[t]{1.3cm}{0.09, 0.12}  &
\parbox[t]{4.1cm}{Modified $z$-expansion fit combining the continuum and chiral extrapolations and the momentum-transfer dependence. Leading discretization errors from $(am_c)^n$ charm-mass effects (see Tab.~\ref{tab:DtoPiKHQ}). 
    Subleading $(aE)^n$ discretization corrections 
     estimated to be 1.0\% for both 
    $D\to\pi$ and $D\to K$.} & \parbox[t]{3.5cm}{Relative scale $r_1/a$ set from the static-quark potential.  Absolute scale $r_1$ set from several quantities including $f_\pi$, $f_K$, and $\Upsilon$ $2S-1S$ splitting c.f. HPQCD~09B~\cite{Davies:2009tsa}.  Scale uncertainty estimated to be 0.7\% in $D\to\pi$ and and 0.2\% in $D\to K$.}
\\[30.0ex] \hline \\[-2.0ex]
\parbox[t]{1cm}{FNAL/MILC~04} & \cite{Aubin:2004ej} & 2+1 & 0.12  &
\parbox[t]{4.1cm}{Discretization effects from light-quark sector estimated to be 4\% by power counting. Discretization effects from final-state pion and kaon energies estimated to be 5\%.} & \parbox[t]{3.5cm}{Scale set through $\Upsilon$ $2S-1S$ splitting c.f. HPQCD 03~\cite{Wingate:2003gm}.  Error in $a^{-1}$ estimated to be 1.2\%, but scale error in dimensionless form factor negligible compared to other uncertainties.}
\\[20.0ex]
\hline\hline
\end{tabular*}
\caption{Continuum extrapolations/estimation of lattice artifacts in $N_f=2+1$ determinations of form factors for semileptonic decays of charmed hadrons.}
}
\end{table}

\begin{table}[!ht]

{\footnotesize
\begin{tabular*}{\textwidth}{l @{\extracolsep{\fill}} c c c l l}
\hline\hline \\[-1.0ex]
Collab. & Ref. & $\Nf$ & $a$ [fm] & Continuum extrapolation & Scale setting 
\\[1.0ex] \hline \hline \\[-1.0ex]
ETM 11B & \cite{DiVita:2011py} & 2 & \parbox[t]{1.3cm}{0.068, 0.086, 0.102} &
\parbox[t]{4.1cm}{Discretization errors estimated to be 5\% for $D\to\pi$ and 3\% for $D\to K$ from comparison of results in the continuum limit to those at the finest lattice spacing.}
& \parbox[t]{3.5cm}{Scale set through $f_\pi$ c.f. ETM~07A~\cite{Boucaud:2007uk} and ETM~09C~\cite{Baron:2009wt}.}
\\[15.0ex]
\hline\hline
\end{tabular*}
\caption{Continuum extrapolations/estimation of lattice artifacts in $N_f=2$ determinations of form factors for semileptonic decays of charmed hadrons.}
}
\end{table}
\fi

\begin{table}[!ht]
{\footnotesize
\begin{tabular*}{\textwidth}{l @{\extracolsep{\fill}} c  c c  l}
\hline\hline \\[-1.0ex]
{Collab.} & {Ref.} & {$\Nf$} & {$M_{\pi,\rm min}\,[\mev]$}  & {Description}  
\\[1.0ex] \hline \hline \\[-1.0ex]
HPQCD 21A & \cite{Chakraborty:2021qav} & 2+1+1 & \parbox[t]{2cm}{315, 329, 129, 132, 131}
 & \parbox[t]{6cm}{Modified $z$-expansion fit combining the continuum and chiral extrapolations and the momentum-transfer dependence. Polynomial dependence on quark masses, supplemented by a pion chiral logarithm. Fit result compared with alternative approach based on cubic splines in $q^2$.}
\\[19.0ex] \hline \\[-2.0ex]

Zhang  21  & \cite{Zhang:2021oja} & 2+1 & \parbox[t]{2cm}{300, 290}   &
\parbox[t]{6cm}{Dependence on pion mass neglected. No estimate of resulting systematic uncertainty.}
\\[7.0ex] \hline \\[-2.0ex]
HPQCD 20 & \cite{Cooper:2020wnj} & 2+1+1 & \parbox[t]{2cm}{329, 316, 132/305, 131/305}  &
\parbox[t]{6cm}{Modified $z$-expansion fit combining the continuum and chiral extrapolations and the momentum-transfer dependence, and, for the heavy-HISQ spectator $b$ quark, the dependence on $1/m_Q$. The analysis combines data with NRQCD $b$ quarks and data with HISQ heavy quarks.}
\\[18.0ex] 
\ifx\reducedapptables\undefined
 \hline \\[-2.0ex]
ETM 17D, 18 & \cite{Lubicz:2017syv,Lubicz:2018rfs} & 2+1+1 & \parbox[t]{2cm}{220, 258, 275}
 & \parbox[t]{6cm}{Modified $z$-expansion fit combining the continuum and chiral extrapolations and the momentum-transfer dependence. Chiral log term in vector and scalar form factors set to hard-pion $\chi$PT prediction~\cite{Bijnens:2010ws}. Systematic uncertainty in tensor form factor estimated by comparing fits with and without chiral log terms.}
\\[21.0ex] \hline \\[-2.0ex]
JLQCD 17B & \cite{Kaneko:2017xgg} & 2+1 & \parbox[t]{2cm}{284, 296, 226}
 & \parbox[t]{6cm}{Joint chiral-continuum extrapolation, with mass dependence based on hard-pion Q$\chi$PT.}
\\[7.0ex] \hline \\[-2.0ex]
Meinel 17 & \cite{Meinel:2017ggx} & 2+1 & \parbox[t]{2.5cm}{227, 245 \\ (valence-valence) \\ 263, 294 \\ (valence-sea)}
& \parbox[t]{6cm}{Joint chiral-continuum extrapolation, combined with fit to $q^2$-dependence of form factors in a ``modified'' $z$-expansion. Only analytic NLO terms $\propto (m_\pi^2-m_{\pi,{\rm phys}}^2)$ included in light-mass dependence. Systematic uncertainty estimated by repeating fit with added higher-order terms.}
\\[21.0ex] \hline \\[-2.0ex]
Meinel 16 & \cite{Meinel:2016dqj} & 2+1 & \parbox[t]{2cm}{295, 139}
 & \parbox[t]{6cm}{Modified $z$-expansion fit combining the continuum and chiral extrapolations and the momentum-transfer dependence. Analytic function in $m_\pi, m_{\eta_s}$ used for mass dependence. ($\eta_s$ stands for a nonsinglet meson with two mass-degenerate valence quarks of mass $m_s$, used to set the strange scale.} 
\\[21ex] 
\hline \\[-2.0ex]
HPQCD 10B, 11 & \cite{Na:2010uf, Na:2011mc} & 2+1 & \parbox[t]{2cm}{390, 390}
& \parbox[t]{6cm}{Modified $z$-expansion fit combining the continuum and chiral extrapolations and the momentum-transfer dependence. Contributions to error budget from light valence and sea-quark mass dependence estimated to be 2.0\% for $D \to\pi$ and 1.0\% for $D \to K$. }
\\[18.0ex] \hline \\[-2.0ex]
FNAL/MILC~04 & \cite{Aubin:2004ej} & 2+1 & \parbox[t]{2cm}{510}
 & \parbox[t]{6cm}{Fit to S$\chi$PT, combined with the Becirevic-Kaidalov ansatz for the momentum-transfer dependence of form factors. Error estimated to be 3\% for $D \to\pi$ and 2\% for $D \to K$ by comparing fits with and without one extra analytic term.}
\\[15.0ex] \hline \\[-2.0ex]
ETM 11B & \cite{DiVita:2011py} & 2 & \parbox[t]{2cm}{270}
& \parbox[t]{6cm}{$SU(2)$ tmHM$\chi$PT plus Becirevic-Kaidalov ansatz for fits to the momentum-transfer dependence of form factors.  Fit uncertainty estimated to be 7\% for $D \to\pi$ and 5\% for $D \to K$
by considering fits with and without NNLO corrections of order $\cO(m_\pi^4)$ and/or higher-order terms through $E^5$, and by excluding data with $E \gtapprox 1$~GeV.}
\\[24.0ex]
\fi
\hline\hline
\end{tabular*}
\caption{Chiral extrapolation/minimum pion mass in
  determinations of form factors for semileptonic decays of charmed hadrons.  For actions with multiple species of pions, masses quoted are the RMS pion masses for $N_f=2+1$ and the Goldstone mode mass for $N_f=2+1+1$.  The different $M_{\pi,\rm min}$ entries correspond to the different lattice spacings.}
}
\end{table}

\begin{table}[!ht]
{\footnotesize
\begin{tabular*}{\textwidth}{l @{\extracolsep{\fill}} c c  c c l}
\hline\hline \\[-1.0ex]
Collab. & Ref. & $\Nf$ & $L$ [fm] & ${M_{\pi,\rm min}}L$ & Description 
\\[1.0ex] \hline \hline \\[-1.0ex]
HPQCD 21A & \cite{Chakraborty:2021qav} & 2+1+1 & \parbox[t]{1.8cm}{2.73, 2.72, 2.81/5.62, 2.93/5.87, 2.45/4.89} & $\gtrsim$ 3.7 & 
\parbox[t]{5cm}{Finite volume correction included in chiral fit, claimed to be a negligible effect.
Effect of frozen topology in finest ensemble not discussed.}
\\[10.0ex] \hline \\[-1.0ex]
Zhang  21  & \cite{Zhang:2021oja} & 2+1+1 & \parbox[t]{1.8cm}{2.6, 2.6} & $\gtrsim$ 3.8 & 
\parbox[t]{5cm}{No discussion of finite-volume effects.}
\\[5.0ex] \hline \\[-1.0ex]
HPQCD 20 & \cite{Cooper:2020wnj} & 2+1+1 & \parbox[t]{1.8cm}{2.72, 2.81, 2.93/5.87, 2.45/4.89} & $\gtrsim$ 3.8 & 
\parbox[t]{5cm}{Physical point ensemble at $a \simeq 0.15~{\rm fm}$ has $m_\pi L=3.3$; the statement $m_\pi L \gtrsim 3.8$ applies to the other five ensembles.}
\\[10.0ex]
\ifx\reducedapptables\undefined
\hline \\[-1.0ex]
ETM 17D, 18 & \cite{Lubicz:2017syv,Lubicz:2018rfs} & 2+1+1 & \parbox[t]{1.8cm}{2.97, 1.96/2.61, 2.13/2.84} & 3.31, 3.42, 3.49 & 
\parbox[t]{5cm}{Extrapolation to infinite volume performed by including term $\propto e^{-M_\pi L}/(M_\pi L)$ in global fit.}
\\[8.0ex] \hline \\[-1.0ex]
JLQCD 17B & \cite{Kaneko:2017xgg} & 2+1 & \parbox[t]{1.8cm}{2.8, 2.6, 2.6/3.9} & 4.0, 3.9, 4.4 & 
\parbox[t]{5cm}{No discussion of finite-volume effects.}
\\[5.0ex] \hline \\[-1.0ex]
Meinel 17 & \cite{Meinel:2017ggx} & 2+1 & \parbox[t]{1.8cm}{2.7, 2.7} & \parbox[t]{1.8cm}{$\gtrsim 3.6$ \\ (valence-sea pion)}  &
\parbox[t]{5cm}{FV effect estimated at 3\% from experience on $\chi$PT estimates of FV effects for heavy-baryon axial couplings.
}
\\[12.0ex] \hline \\[-1.0ex]
Meinel 16 & \cite{Meinel:2016dqj} & 2+1 & \parbox[t]{1.8cm}{2.7, 2.6/5.3} & 4.1, 3.7 & 
\parbox[t]{5cm}{Finite-volume effect estimated to be at 1.0\% level.}
\\[5.0ex] 
\hline \\[-1.0ex]
HPQCD 10B, 11 & \cite{Na:2010uf,Na:2011mc} & 2+1 & \parbox[t]{1.8cm}{2.4, 2.4/2.9} & $\gtrsim 3.8$ & 
\parbox[t]{5cm}{Finite-volume effects estimated to be 0.04\% for $D\to\pi$ and 0.01\% for $D\to K$ by comparing the ``$m_\pi^2 {\rm log} (m_\pi^2)$" term in infinite and finite volume.}
\\[12.0ex] \hline \\[-1.0ex]
FNAL/MILC 04 & \cite{Aubin:2004ej} & 2+1 & \parbox[t]{1.8cm}{2.4/2.9} & $\gtrsim 3.8$ & 
\parbox[t]{5cm}{No explicit estimate of FV error, but expected to be small for simulation masses and volumes.}
\\[7.0ex] \hline \\[-1.0ex]
ETM 11B & \cite{DiVita:2011py} & 2 & \parbox[t]{1.8cm}{2.2, 2.1/2.8, 2.4} & $\gtrsim 3.7$ &
\parbox[t]{5cm}{Finite-volume uncertainty estimated to be at most 2\% by considering
fits with and without the lightest pion mass point at $m_\pi L \approx 3.7$.}
\\[12.0ex]
\fi
\hline\hline
\end{tabular*}
\caption{Finite-volume effects in determinations of form factors for semileptonic decays of charmed hadrons.  Each $L$-entry corresponds to a different lattice
spacing, with multiple spatial volumes at some lattice spacings.  For actions with multiple species of pions, the lightest pion masses are quoted.}
}
\end{table}
\begin{table}[!ht]
{\footnotesize
\begin{tabular*}{\textwidth}{l @{\extracolsep{\fill}} c c c l}
\hline\hline \\[-1.0ex]
Collab. & Ref. & $\Nf$ & Ren. & Description 
\\[1.0ex] \hline \hline \\[-1.0ex]
HPQCD 21A & \cite{Chakraborty:2021qav} & 2+1+1 & NP &
\parbox[t]{6cm}{Vector current normalized by imposing Ward identity at zero recoil.}
\\[5.0ex] \hline \\[-1.0ex]
Zhang  21  & \cite{Zhang:2021oja} & 2+1 & NP & 
\parbox[t]{6cm}{Local vector current renormalized using ratio to conserved vector current. Axial current renormalized using ratio of off-shell quark matrix elements. }
\\[12.0ex] \hline \\[-1.0ex]
HPQCD 20 & \cite{Cooper:2020wnj} & 2+1+1 & NP &
\parbox[t]{6cm}{Vector current normalized by imposing Ward identity at zero recoil.}
\\[5.0ex]
\ifx\reducedapptables\undefined
\hline \\[-1.0ex]
ETM 17D, 18 & \cite{Lubicz:2017syv,Lubicz:2018rfs} & 2+1+1 & RI'-MOM &
\parbox[t]{6cm}{Vector current normalization obtained nonperturbatively by imposing
charge conservation in $D\to D$, $K \to K$, and $\pi \to \pi$ transitions. Tensor current renormalization factors computed in 
RI'-MOM. Scalar form factor is absolutely normalized from chiral symmetry. Renormalized tensor 
current matched to $\overline{\rm MS}$ using 2-loop perturbation theory.}
\\[24.0ex] \hline \\[-1.0ex]
JLQCD 17B & \cite{Kaneko:2017xgg} & 2+1 & mNPR &
\parbox[t]{6cm}{Nonperturbative renormalization of vector current.}
\\[5.0ex] \hline \\[-1.0ex]
Meinel 16, 17  & \cite{Meinel:2016dqj,Meinel:2017ggx} & 2+1 & mNPR &
\parbox[t]{6cm}{Nonperturbative renormalization of singlet currents, residual factor computed at one loop in tadpole-improved perturbation theory.}
\\[10.0ex] 
\hline \\[-1.0ex]
HPQCD 10B, 11 & \cite{Na:2010uf, Na:2011mc} & 2+1 & --- &
\parbox[t]{6cm}{Form factor extracted from absolutely normalized scalar-current matrix element then using kinematic constraint at zero momentum-transfer $f_+(0) = f_0(0)$.}
\\[10.0ex] \hline \\[-1.0ex]
FNAL/MILC~04 & \cite{Aubin:2004ej} & 2+1 & mNPR &
\parbox[t]{6cm}{Size of 2-loop correction to current renormalization factor assumed to be negligible.}
\\[5.0ex] \hline \\[-1.0ex]
ETM 11B & \cite{DiVita:2011py} & 2 & --- &
\parbox[t]{6cm}{Form factors extracted from double ratios insensitive to current normalization.}
\\[4.0ex]
\fi
\hline\hline
\end{tabular*}
\caption{Operator renormalization in determinations of form factors for semileptonic decays of charmed hadrons.}
}
\end{table}
\begin{table}[!ht]
{\footnotesize
\begin{tabular*}{\textwidth}{l @{\extracolsep{\fill}} c c c l}
\hline\hline \\[-1.0ex]
Collab. & Ref. & $\Nf$ & Action & Description 
\\[1.0ex] \hline \hline \\[-1.0ex]
HPQCD 21A & \cite{Chakraborty:2021qav} & 2+1+1 & HISQ &
\parbox[t]{6cm}{Bare charm-quark mass $0.194 \lesssim am_c \lesssim 0.8605$.}
\\[4.0ex] \hline \\[-1.0ex]
Zhang  21  & \cite{Zhang:2021oja} & 2+1+1 & SW &
\parbox[t]{6cm}{Bare charm-quark mass $0.235 \lesssim am_c \lesssim 0.485$. No $\mathcal{O}(a)$ improvement of currents.}
\\[6.0ex] \hline \\[-1.0ex]
HPQCD 20 & \cite{Cooper:2020wnj} & 2+1+1 & \parbox{2cm}{ Charm: HISQ \\ Bottom \\ (spectator): \\ HISQ and NRQCD} & 
\parbox[t]{6cm}{ Bare charm-quark HIQS mass $0.274 \lesssim am_c \lesssim 0.827$. \\  Bare bottom-quark HIQS mass $0.274 \lesssim am_b \lesssim 0.8$.}
\\[12.0ex]
\ifx\reducedapptables\undefined
\hline \\[-1.0ex]
ETM 17D, 18 & \cite{Lubicz:2017syv,Lubicz:2018rfs} & 2+1+1 & tmWil &
\parbox[t]{6cm}{Bare charm-quark mass $0.14 \lesssim am_c \lesssim 0.29$.}
\\[4.0ex] \hline \\[-1.0ex]
JLQCD 17B & \cite{Kaneko:2017xgg} & 2+1 & M\"obius DWF &
\parbox[t]{6cm}{Charm quark matched to its physical value.}
\\[4.0ex] \hline \\[-1.0ex]
Meinel 16, 17  & \cite{Meinel:2016dqj,Meinel:2017ggx} & 2+1 & Anisotropic SW. &
\parbox[t]{6cm}{Residual $\mathcal{O}(a)$ improvement coefficients in currents computed in 1-loop tadpole-improved perturbation theory.}
\\[7.0ex] 
\hline \\[-1.0ex]
HPQCD 10B, 11 & \cite{Na:2010uf,Na:2011mc} & 2+1 & HISQ &
\parbox[t]{6cm}{Bare charm-quark mass $am_c \sim 0.41$--0.63.  Errors of $(am_c)^n$ estimated within modified $z$-expansion to be 1.4\% for $D\to K$ and 2.0\% for $D \to\pi$.  Consistent with expected size of dominant 1-loop cutoff effects on the finest lattice spacing, $\cO(\alpha_S (am_c)^2 (v/c) )\sim 1.6$\%.}
\\[18.0ex] \hline \\[-1.0ex]
FNAL/MILC~04 & \cite{Aubin:2004ej} & 2+1 & Fermilab &
\parbox[t]{6cm}{Discretization errors from charm quark estimated via heavy-quark power-counting to be 7\%.}
\\[7.0ex] \hline \\[-1.0ex]
ETM 11B & \cite{DiVita:2011py} & 2 & tmWil &
\parbox[t]{6cm}{Bare charm-quark mass $am_c \sim 0.17$--0.30.   Expected size of $\cO((am_c)^2)$ cutoff effects on the finest lattice spacing consistent with quoted 5\% continuum-extrapolation uncertainty.}
\\[13.0ex]
\fi
\hline\hline
\end{tabular*}
\caption{Heavy-quark treatment in determinations of form factors for semileptonic decays of charmed hadrons. \label{tab:DtoPiKHQ} 
}}
\end{table}

\clearpage

\subsection{Notes to Sec.~\ref{sec:BDecays} on $B$-meson decay
  constants, mixing parameters and form factors}
\label{app:BDecays}



\subsubsection{$B_{(s)}$-meson decay constants}
\label{app:fB_Notes}
\ifx\reducedapptables\undefined
\begin{table}[!htb]
{\footnotesize
\begin{tabular*}{\textwidth}{l @{\extracolsep{\fill}} c c c   l}
\hline\hline \\[-1.0ex]
Collab. & \hspace{-0.4cm}Ref. & $\Nf$ & $M_{\pi,\rm min}\,[\mev]$  & Description 
\\[1.0ex] \hline \hline \\[-2.0ex]
\parbox[t]{2.8cm}{FNAL/MILC 17 }
& \parbox[t]{1.0cm}{\cite{Bazavov:2017lyh} } & 2+1+1
                                      & \parbox[t]{1.3cm}{130, 133,
                                        130, 135, 134, 309} &
\parbox[t]{7.0cm}{Multiple values of pion masses at each lattice
                                                            spacing,
                                                              except
                                                              for the
                                                              finest
                                                              lattice.
                                                        Chiral
                                                              extrapolation
                                                              performed
                                                              using
                                                              the
                                                              heavy-meson
                                                              rooted
                                                              all-staggered
                                                              $\chi$PT~\cite{Bernard:2013qwa}.
                                                             }
\\[10.0ex] \hline \\[-2.0ex]
\parbox[t]{2.8cm}{HPQCD 17A  \\ 
\ifx\reducedapptables\undefined
HPQCD 13
\fi
}
& \parbox[t]{1.0cm}{\cite{Hughes:2017spc} \\ 
\ifx\reducedapptables\undefined
\cite{Dowdall:2013tga} 
\fi
}
& 2+1+1 & \parbox[t]{1.3cm}{310, 294, 173} & 
\parbox[t]{7.0cm}{Two or three pion masses at each lattice spacing, one each 
with a physical mass GB pion. NLO (full QCD) HM$\chi$PT supplemented by 
generic $a^2$ and $a^4$ terms is used to
extrapolate to the physical pion mass.}
\\[12.0ex] \hline \\[-2.0ex]
\parbox[t]{2.8cm}{ETM 16B  \\ 
\ifx\reducedapptables\undefined
ETM 13E
\fi
}
& \parbox[t]{1.0cm}{\cite{Bussone:2016iua} \\ 
\ifx\reducedapptables\undefined
\cite{Carrasco:2013naa}
\fi
 }
& 2+1+1 & \parbox[t]{1.3cm}{245, 239, 211}
& \parbox[t]{7.0cm}{$M_{\pi,{\rm min}}$ refers to the
  charged pions.  Linear and NLO (full QCD) 
HM$\chi$PT formulae supplemented by an $a^2$ term are used for the
  chiral-continuum extrapolation.  In ETC 13, the chiral fit 
error is estimated from the difference between
the NLO HM$\chi$PT and linear fits with half the difference
used as estimate of the systematic error. The ratio $z_s$ is fit using just linear HM$\chi$PT 
supplemented by an $a^2$ term.  On the other hand, in ETC 16B, the systematic error is estimated by using data
  points with $M_{\pi} < 350$ MeV for the chiral-continuum fit. }
\\[29.0ex]
\hline\hline
\end{tabular*}
\caption{Chiral extrapolation/minimum pion mass in determinations of the
  $B$- and $B_s$-meson decay constants for $N_f=2+1+1$ simulations.  
  For actions with multiple species of pions, masses quoted are the RMS pion masses.    
  The different $M_{\pi,\rm min}$ entries correspond to the different lattice
 spacings.} 
}
\end{table}
\fi


\begin{table}[!htb]
{\footnotesize
\begin{tabular*}{\textwidth}{l @{\extracolsep{\fill}} c c c   l}
\hline\hline \\[-1.0ex]
Collab. & \hspace{-0.4cm}Ref. & $\Nf$ & $M_{\pi,\rm min}\,[\mev]$  & Description
\\[1.0ex] \hline \hline \\[-2.0ex]
\parbox[t]{2.8cm}{RBC/UKQCD 18A}
& \parbox[t]{1.0cm}{\cite{Boyle:2018knm} } & 2+1 & \parbox[t]{1.3cm}{139,139,232} &
\parbox[t]{7.0cm}{Three or four light-quark masses per lattice spacing except for the finest lattice spacing. Generic fits to $m_\pi^2-(m_\pi^{\mathrm{phys}})^2$ and $a^2$ in the combined chiral-continuum extrapolation, with systematic errors estimated to
be from $0.3\%$ to $0.5\%$ in $f_{B_{s}}/f_B$.}
\\[15.0ex] 
\ifx\reducedapptables\undefined
\hline \\[-2.0ex]
\parbox[t]{2.8cm}{RBC/UKQCD~14  \\ RBC/UKQCD 13}
& \parbox[t]{1.0cm}{\cite{Christ:2014uea} \\ \cite{Witzel:2013sla} } & 2+1 & \parbox[t]{1.3cm}{329, 289} &
\parbox[t]{7.0cm}{Two or three light-quark masses per lattice spacing.
  In RBC/UKQCD~14, three to four light valence-quark  masses that are heavier than
  the sea-quark masses are also employed to have partially-quenched points.
NLO $SU(2)$ HM$\chi$PT is used.  In RBC/UKQCD~14, the fit with only the
unitary points is the central analysis procedure, and the systematic
errors in the combined chiral-continuum extrapolation are estimated to
be from $3.1\%$ to $5.9\%$ in the decay constants and the $SU(3)$-breaking ratios.}
\\[29.0ex] \hline \\[-2.0ex]
\parbox[t]{2.8cm}{RBC/UKQCD~14A}
& \parbox[t]{1.0cm}{\cite{Aoki:2014nga} } & 2+1 & \parbox[t]{1.3cm}{327, 289} &
\parbox[t]{7.0cm}{Two or three light-quark masses per lattice spacing.
NLO $SU(2)$ HM$\chi$PT is used in the combined chiral-continuum
extrapolation.  The systematic
errors in this extrapolation are estimated to
be $3.54\%$ for $f_{B}$, $1.98\%$ for $f_{B_{s}}$, and $2.66\%$ for $f_{B_{s}}/f_{B}$.}
\\[15.0ex] \hline \\[-2.0ex]
\parbox[t]{2.8cm}{HPQCD 12} &  \parbox[t]{1.0cm}{\cite{Na:2012sp} }  & 2+1 & \parbox[t]{1.3cm}{390, 390} & 
\parbox[t]{7.0cm}{Two or three pion masses at each lattice spacing. NLO (full QCD) 
HM$\chi$PT supplemented by NNLO analytic terms and 
generic $a^2$ and $a^4$ terms is used. The systematic error is estimated by 
varying the fit Ansatz, in particular for the NNLO analytic terms and the $a^{2n}$ terms. }
\\[15.0ex] \hline \\[-2.0ex]
\parbox[t]{2.8cm}{HPQCD 11A} & \parbox[t]{1.0cm}{\cite{McNeile:2011ng}
} & 2+1 & \parbox[t]{1.3cm}{570, 450, 390, 330, 330}  &
\parbox[t]{7.0cm}{One light sea-quark mass only at each lattice spacing. The sea-quark mass 
dependence is assumed to be negligible, based on the calculation of $f_{D_s}$ in 
Ref.~\cite{Davies:2010ip}, where the sea-quark extrapolation error is estimated as $0.34\%$. } 
\\[13.0ex] \hline \\[-2.0ex]
\parbox[t]{2.8cm}{FNAL/MILC 11}
& \parbox[t]{1.0cm}{\cite{Bazavov:2011aa} } & 2+1 & \parbox[t]{1.3cm}{570, 440, 320} &
\parbox[t]{7.0cm}{Three to five sea-quark masses per lattice spacing, 
and  $9-12$ valence light quark masses per ensemble. NLO partially quenched HMrS$\chi$PT 
including $1/m$ terms and  supplemented by NNLO analytic and $\alpha_s^2a^2$ terms is used. 
The systematic error is estimated by varying the fit Ansatz,  in particular the NNLO analytic 
terms and the chiral scale. }
\\[18.0ex] \hline \\[-2.0ex]
\parbox[t]{2.8cm}{RBC/UKQCD 10C}
& \parbox[t]{1.0cm}{\cite{Albertus:2010nm} } & 2+1 & \parbox[t]{1.3cm}{430} &
\parbox[t]{7.0cm}{Three light-quark masses at one lattice spacing. 
NLO $SU(2)$ $\chi$PT is used. The systematic error is estimated
 from the difference between NLO $\chi$PT and linear fits as $\sim 7\%$.}
\\[10.0ex] \hline \\[-2.0ex]
\parbox[t]{2.8cm}{HPQCD 09} &  \parbox[t]{1.0cm}{\cite{Gamiz:2009ku} }
& 2+1 & \parbox[t]{1.3cm}{440, 400} & 
\parbox[t]{7.0cm}{Four or two pion masses per lattice spacing. NLO (full QCD) 
HMrS$\chi$PT supplemented by NNLO analytic terms and 
 $\alpha_s a^2, a^4$ terms is used. The chiral fit error is estimated by varying the fit Ansatz,
in particular, by adding or removing NNLO and discretization terms. }
\\[15.0ex]
\fi
\hline\hline
\end{tabular*}
\caption{Chiral extrapolation/minimum pion mass in determinations of the
  $B$- and $B_s$-meson decay constants for $N_f=2+1$ simulations.  
  For actions with multiple species of pions, masses quoted are the RMS pion masses.    
  The different $M_{\pi,\rm min}$ entries correspond to the different lattice
 spacings.} 
}
\end{table}


\begin{table}[!htb]
{\footnotesize
\begin{tabular*}{\textwidth}{l @{\extracolsep{\fill}} c c c   l}
\hline\hline \\[-1.0ex]
Collab. & \hspace{-0.4cm}Ref. & $\Nf$ & $M_{\pi,\rm min}\,[\mev]$  & Description 
\\[1.0ex] \hline \hline \\[-2.0ex]
\parbox[t]{2.8cm}{Balasubramamian 19} & \parbox[t]{1.0cm}{\cite{Balasubramanian:2019net}} 
& 2 & \parbox[t]{1.3cm}{282, 194, 269}
& \parbox[t]{7.0cm}{Two pion masses for the coarsest and finest lattice spacings, and four pion masses for the intermediate lattice spacing. Generic fits to $m_\pi^2$ and $a^2$ in the combined continuum-chiral extrapolation, with systematic effects obtained by including a generic NLO chiral term $m_\pi^2 \log m_\pi^2$. Detailed error budget not provided, but total systematic uncertainty estimated as $1.1$-$1.2\%$.}
\\[24.0ex]
\ifx\reducedapptables\undefined
\hline \\[-2.0ex]
\parbox[t]{2.8cm}{ALPHA 14  \\ ALPHA 13 \\ ALPHA 12A } & 
\parbox[t]{1.0cm}{\cite{Bernardoni:2014fva} \\
    \cite{Bernardoni:2013oda} \\ \cite{Bernardoni:2012ti} } & 
2 & \parbox[t]{1.3cm}{280, 190, 270}
& \parbox[t]{7.0cm}{LO and NLO HM$\chi$PT supplemented by a term linear in $a^2$ are 
used. In ALPHA 13 and ALPHA 12A, the final result is an average between LO and NLO with half the difference
used as estimate of the systematic error.  In ALPHA 14, the NLO fit is
used as the central analysis procedure, and the LO results are used to
estimate the systematic errors ($0.9\%$ for $f_{B_{s}}$, $1.1\%$ for
$f_{B}$, and $1.6\%$ for $f_{B_{s}}/f_{B}$).}
\\[24.0ex]\hline \\[-2.0ex]
\parbox[t]{2.8cm}{ETM 13B, 13C \\ ETM 12B  \\ ETM 11A}  & 
\parbox[t]{1.0cm}{\cite{Carrasco:2013zta,Carrasco:2013iba} \\
  \cite{Carrasco:2012de} \\ \cite{Dimopoulos:2011gx} } & 2 & 
\parbox[t]{1.3cm}{410, 275, 300, 270} &
 \parbox[t]{7.0cm}{$M_{\pi,{\rm min}}$ refers to the charged pions. Linear and NLO (full QCD) 
HM$\chi$PT supplemented by an $a^2$ term is used.The chiral fit error is estimated from the difference between
the NLO HM$\chi$PT and linear fits with half the difference
used as estimate of the systematic error. 
For the static limit calculation in ETM 11A, $\Phi_s^{\rm stat}$ is extrapolated 
assuming a constant in light-quark mass.
The ratio $\Phi_s^{\rm stat}/\Phi_{\ell}^{\rm stat}$ is fit using three 
different chiral fit forms (NLO HM$\chi$PT, linear, and quadratic) to estimate the chiral fir error. }
\\[29.0ex] \hline \\[-2.0ex]
\parbox[t]{2.8cm}{ALPHA 11}
& \parbox[t]{1.0cm}{\cite{Blossier:2011dk} }  & 
2 & \parbox[t]{1.3cm}{331, 268, 267}
& \parbox[t]{7.0cm}{Linear and NLO (full QCD) HM$\chi$PT supplemented by a term linear in $a^2$ are 
used. The final result is an average between linear and NLO fits with half the difference
used as estimate of the systematic error. }
\\[12.0ex]\hline \\[-2.0ex]
\parbox[t]{2.8cm}{ETM 09D} & \parbox[t]{1.0cm}{\cite{Blossier:2009hg} } & 2 & \parbox[t]{1.3cm}{410, 275, 300}
& \parbox[t]{7.0cm}{$M_{\pi,{\rm min}}$ refers to the charged pions.  
Linear and NLO (full QCD) 
HM$\chi$PT is used. The final result given by the average of NLO HM$\chi$PT and linear
{\it Ans\"atze} $\pm$ half the difference).}
\\[9.0ex] 
\fi
\hline\hline
\end{tabular*}
\caption{Chiral extrapolation/minimum pion mass in determinations of the
  $B$- and $B_s$-meson decay constants for $N_f=2$ simulations.  
  For actions with multiple species of pions, masses quoted are the RMS pion masses (where available).    
  The different $M_{\pi,\rm min}$ entries correspond to the different lattice
 spacings.} 
}
\end{table}


\ifx\reducedapptables\undefined

\begin{table}[!htb]
{\footnotesize
\begin{tabular*}{\textwidth}{l @{\extracolsep{\fill}} c c c c l}
\hline\hline \\[-1.0ex]
Collab. & \hspace{-0.4cm}Ref. & $\Nf$ & $L$ [fm] & ${M_{\pi,\rm min}}L$ & Description 
\\[1.0ex] \hline \hline \\[-2.0ex]
\parbox[t]{2.8cm}{FNAL/MILC 17 }
& \parbox[t]{1.0cm}{\cite{Bazavov:2017lyh}}
& 2+1+1 & 
\parbox[t]{1.8cm}{2.4$\sim$4.8
          /\\3.0$\sim$5.8/\\3.0$\sim$5.6/\\2.9$\sim$5.4/\\2.9, 6.1/\\3.3}  & 
\parbox[t]{1.5cm}{3.2, 3.9, 3.7, 3.7, 4.2, $\mbox{ }\mbox{ }\mbox{ }$4.8 } &
\parbox[t]{4.9cm}{ Finite-size effects estimated, using alternative EFT
                                                                             fits,
                                                                             to
                                                                             be
                                                                             $0.1\%$
                                                          for
                                                                             $f_{B^{0}}$,
                                                                             $0.07\%$
                                                                             for
                                                                             $f_{B_{s}}$,
                                                                             and
                                                                             $0.03\%$
                                                                             for $f_{B_{s}}/f_{B^{0}}$.
                                                                            }
\\[15.0ex] \hline \\[-2.0ex]
\parbox[t]{2.8cm}{HPQCD 17A  \\ 
\ifx\reducedapptables\undefined
HPQCD 13
\fi
}
& \parbox[t]{1.0cm}{\cite{Hughes:2017spc} \\ 
\ifx\reducedapptables\undefined
\cite{Dowdall:2013tga}
\fi
 }
& 2+1+1 & 
\parbox[t]{1.8cm}{2.4/3.5/4.7, 2.9/3.8/5.8, 2.8/5.6}  & 
\parbox[t]{1.5cm}{3.30, 3.88, 3.66} &
\parbox[t]{4.9cm}{ The analysis uses finite-volume
                                      $\chi$PT.  No explicit
                                      estimation of the systematic
                                      error arising from finite-size effects. }
\\[10.0ex] \hline \\[-2.0ex]
\parbox[t]{2.8cm}{ETM 16B  \\ 
\ifx\reducedapptables\undefined
ETM 13E
\fi
}
& \parbox[t]{1.0cm}{\cite{Bussone:2016iua} \\ 
\ifx\reducedapptables\undefined
\cite{Carrasco:2013naa}
\fi
 }
& 2+1+1 & 
\parbox[t]{1.8cm}{2.84/2.13, 2.61/1.96, 2.97} &
\parbox[t]{1.5cm}{3.53, 3.16, 3.19} &
\parbox[t]{4.9cm}{ The data show no statistically discernible
                                      finite-volume effects, and the
                                      related systematic error is not
                                      explicitly estimated }
\\[10.0ex] 
\hline\hline
\end{tabular*}
\caption{Finite-volume effects in determinations of the $B$- and $B_s$-meson
 decay constants for $N_f=2+1+1$ simulations.  Each $L$-entry corresponds to a different
 lattice spacing, with multiple spatial volumes at some lattice
 spacings.} 
}
\end{table}
\fi
%

\begin{table}[!htb]
{\footnotesize
\begin{tabular*}{\textwidth}{l @{\extracolsep{\fill}} c c c c l}
\hline\hline \\[-1.0ex]
Collab. & \hspace{-0.4cm}Ref. & $\Nf$ & $L$ [fm] & ${M_{\pi,\rm min}}L$ & Description 
\\[1.0ex] \hline \hline \\[-2.0ex]
\parbox[t]{2.8cm}{RBC/UKQCD 18A}
& \parbox[t]{1.0cm}{\cite{Boyle:2018knm} } & 2+1 & \parbox[t]{1.8cm}{2.65/5.47, 2.65/5.35, 3.49} &  \parbox[t]{1.5cm}{3.86, 3.78, 4.11} & 
\parbox[t]{4.9cm}{Finite-volume effects are estimated to be $0.18\%$ in $f_{B_{s}}/f_B$.}
\\[7.0ex] 
\ifx\reducedapptables\undefined
\hline \\[-2.0ex]
\parbox[t]{2.8cm}{RBC/UKQCD~14 \\ RBC/UKQCD 13} & 
\parbox[t]{1.0cm}{ \cite{Christ:2014uea}\\  \cite{Witzel:2013sla}} 
 & 2+1 & 
\parbox[t]{1.8cm}{2.64, 2.75} & \parbox[t]{1.5cm}{4.38, 4.03}& 
\parbox[t]{4.9cm}{In RBC/UKQCD~14, finite-volume effects are
  estimated to be negligible for $f_{B_{s}}$, $0.4\%$ for $f_{B^{0}}$,
$0.5\%$ for $f_{B^{+}}$ and the $SU(3)$-breaking ratios.}
\\[13.0ex] \hline \\[-2.0ex]
\parbox[t]{2.8cm}{RBC/UKQCD~14A} & \parbox[t]{1.0cm}{\cite{Aoki:2014nga}} & 2+1 & \parbox[t]{1.8cm}{2.74, 2.76}  & 
\parbox[t]{1.8cm}{4.53, 4.05} &
\parbox[t]{4.9cm}{Finite-volume effects are
  estimated to be negligible for $f_{B_{s}}$, $0.82\%$ for $f_{B}$, and
$1\%$ for $f_{B_{s}}/f_{B}$.}
\\[7.0ex] \hline \\[-2.0ex]
\parbox[t]{2.8cm}{HPQCD 12} & \parbox[t]{1.0cm}{\cite{Na:2012sp}} & 2+1 & \parbox[t]{1.8cm}{2.4/2.9, 2.5/3.6}  & 
\parbox[t]{1.5cm}{3.84, 4.21} &
\parbox[t]{4.9cm}{FV error is taken from Ref.~\cite{Follana:2007uv} for HPQCD's $D$ 
meson analysis, where it was estimated using finite-volume $\chi$PT .}
\\[9.0ex] \hline \\[-2.0ex]
\parbox[t]{2.8cm}{HPQCD 11A} & \parbox[t]{1.0cm}{\cite{McNeile:2011ng}} & 2+1 & \parbox[t]{1.8cm}{2.4, 2.4, 2.5, 2.9, 2.9} &
\parbox[t]{1.5cm}{3.93, 4.48, 4.14, 4.49, 4.54}  & 
\parbox[t]{4.9cm}{FV error is assumed to negligible.}
\\[6.0ex] \hline \\[-2.0ex]
\parbox[t]{2.8cm}{FNAL/MILC 11} & \parbox[t]{1.0cm}{\cite{Bazavov:2011aa}} & 2+1 & \parbox[t]{1.8cm}{2.4, 2.4/2.9, 2.5/3.6} &
\parbox[t]{1.5cm}{3.93, 3.78, 4.14}  & 
\parbox[t]{4.9cm}{FV error is estimated using  finite-volume $\chi$PT.}
\\[4.0ex] \hline \\[-2.0ex]
\parbox[t]{2.8cm}{RBC/UKQCD 10C} & \parbox[t]{1.0cm}{\cite{Albertus:2010nm}} & 2+1 & 
\parbox[t]{1.8cm}{1.8} & $3.9$ & 
\parbox[t]{4.9cm}{FV error estimated using  finite-volume $\chi$PT to be $1\%$
 for $SU(3)$-breaking ratios. }
\\[7.0ex] \hline \\[-2.0ex]
\parbox[t]{2.8cm}{HPQCD 09} & \parbox[t]{1.0cm}{\cite{Gamiz:2009ku}} & 2+1 & \parbox[t]{1.8cm}{2.4/2.9, 2.5} &
\parbox[t]{1.5cm}{3.78, 4.14}  & 
\parbox[t]{4.9cm}{FV error is assumed to negligible.}
\\[1.0ex] 
\fi
\hline\hline
\end{tabular*}
\caption{Finite-volume effects in determinations of the $B$- and $B_s$-meson
 decay constants for $N_f=2+1$ simulations.  Each $L$-entry corresponds to a different
 lattice spacing, with multiple spatial volumes at some lattice
 spacings.} 
}
\end{table}


\begin{table}[!htb]
{\footnotesize
\begin{tabular*}{\textwidth}{l @{\extracolsep{\fill}} c c c c l}
\hline\hline \\[-1.0ex]
Collab. & \hspace{-0.4cm}Ref. & $\Nf$ & $L$ [fm] & ${M_{\pi,\rm min}}L$ & Description 
\\[1.0ex] \hline \hline \\[-2.0ex]
\parbox[t]{3.0cm}{Balasubramamian 19} & \parbox[t]{1.0cm}{\cite{Balasubramanian:2019net}} 
& 2 & \parbox[t]{1.8cm}{2.4/3.6, 2.1/3.1/4.2, 2.3/3.1} & \parbox[t]{1.5cm}{4.1, 4.1, 4.0} 
& \parbox[t]{4.9cm}{No explicit estimate of FV errors, but expected to be much smaller than other uncertainties.}
\\[6.5ex]
\ifx\reducedapptables\undefined
\hline \\[-2.0ex]
\parbox[t]{2.8cm}{ALPHA 14 \\ALPHA 13 \\ALPHA 12A \\ALPHA 11} & 
\parbox[t]{1.0cm}{\cite{Bernardoni:2014fva}\\  \cite{Bernardoni:2013oda}\\ \cite{Bernardoni:2012ti}\\ \cite{Blossier:2011dk}} & 2 &
\parbox[t]{1.8cm}{2.4/3.6, 2.1/3.1/4.2, 2.3/3.1} & \parbox[t]{1.5cm}{5.2, 4.1, 4.2}  & 
\parbox[t]{4.9cm}{No explicit estimate of FV errors, but expected to be much smaller than other uncertainties.}
\\[9.0ex] \hline \\[-2.0ex]
\parbox[t]{2.8cm}{ETM 13B, 13C \\ETM 12B \\ETM 11A} & 
\parbox[t]{1.0cm}{ \cite{Carrasco:2013zta,Carrasco:2013iba}\\ \cite{Carrasco:2012de}\\ \cite{Dimopoulos:2011gx}} & 2 &
	     \parbox[t]{1.8cm}{2.4, 2.0/2.7, 2.1, 1.7/2.6} &
	      \parbox[t]{1.5cm}{5, 3.7, 3.3, 3.5}  &
\parbox[t]{4.9cm}{FV errors are found to be negligible by comparing results at $m_\pi L = 3.3$ and $m_\pi L=4.3$ for $m_\pi \simeq 310$ MeV.}
\\[9.0ex]
\fi
\hline\hline
\end{tabular*}
\caption{Finite-volume effects in determinations of the $B$- and $B_s$-meson
 decay constants for $N_f=2$ simulations.  Each $L$-entry corresponds to a different
 lattice spacing, with multiple spatial volumes at some lattice
 spacings.} 
}
\end{table}


\ifx\reducedapptables\undefined
\begin{table}[!htb]
{\footnotesize
\begin{tabular*}{\textwidth}{l @{\extracolsep{\fill}} c c c l l}
\hline\hline \\[-1.0ex]
Collab. & \hspace{-0.4cm}Ref. & $\Nf$ & $a$ [fm]  & Continuum extrapolation & Scale setting
\\[1.0ex] \hline \hline \\[-2.0ex]
\parbox[t]{2.8cm}{FNAL/MILC 17 }
& \parbox[t]{1.0cm}{\cite{Bazavov:2017lyh} }
& 2+1+1 & \parbox[t]{0.9cm}{0.15, 0.12, 0.09, 0.06, 0.042, 0.032}
& \parbox[t]{3.8cm}{Continuum extrapolation, linear in $a^2$, combined
  with the EFT study for the quark-mass dependence. }
& \parbox[t]{3.6cm}{ Scale set by $f_{\pi}$, with details described in
  Ref.~\cite{Bazavov:2014wgs}. }
\\[15.0ex] \hline \\[-1.0ex]
\parbox[t]{2.8cm}{HPQCD 17A  \\ 
\ifx\reducedapptables\undefined
HPQCD 13
\fi
}
& \parbox[t]{1.0cm}{\cite{Hughes:2017spc} \\ 
\ifx\reducedapptables\undefined
\cite{Dowdall:2013tga} 
\fi
}
& 2+1+1 & \parbox[t]{0.9cm}{0.15, 0.12, 0.09}
& \parbox[t]{3.8cm}{Combined  continuum and chiral extrapolation. 
Continuum extrapolation errors estimated to be $0.7\%$ in HPQCD 13,
  and $1.1\%$ in HPQCD 17A. 
}
& \parbox[t]{3.6cm}{Scale set from $\Upsilon$(2S-1S)
splitting, see Ref.~\cite{Dowdall:2011wh}.  Scale uncertainty included
  in statistical error. }
\\[15.0ex] \hline \\[-1.0ex]
\parbox[t]{2.8cm}{ETM 16B  \\ 
\ifx\reducedapptables\undefined
ETM 13E
\fi
}
& \parbox[t]{1.0cm}{\cite{Bussone:2016iua} \\ 
\ifx\reducedapptables\undefined
\cite{Carrasco:2013naa} 
\fi
}
& 2+1+1 & \parbox[t]{0.9cm}{0.089, 0.082, 0.062}
& \parbox[t]{3.8cm}{Combined  continuum and chiral extrapolation, linear in $a^2$.
}
& \parbox[t]{3.8cm}{Scale set from $f_\pi$. Scale setting uncertainty included in combined 
statistical and systematic error. Discretization effects are also estimated
  by removing data at the coarest lattice in the chiral-continuum fit.}
\\[23.0ex] 
\hline\hline
\end{tabular*}
\caption{Continuum extrapolations/estimation of lattice artifacts in determinations of 
the $B$- and $B_s$-meson decay constants for $N_f=2+1+1$ simulations.} 
}
\end{table}

\fi


\begin{table}[!htb]
{\footnotesize
\begin{tabular*}{\textwidth}{l @{\extracolsep{\fill}} c c c l l}
\hline\hline \\[-1.0ex]
Collab. & Ref. & $\Nf$ & $a$ [fm]  & Continuum extrapolation & Scale setting
\\[1.0ex] \hline \hline \\[-2.0ex]
\parbox[t]{2.8cm}{RBC/UKQCD 18A}
& \parbox[t]{0.5cm}{\cite{Boyle:2018knm} } & 2+1 & \parbox[t]{1.3cm}{0.11, 0.084, 0.073} &
\parbox[t]{4.3cm}{Combined  continuum and chiral extrapolation with linear in $a^2$ term. Systematic errors estimated to
be from $0.3\%$ to $0.5\%$ in $f_{B_{s}}/f_B$.}
& \parbox[t]{4.3cm}{Scale set by the $\Omega$ baryon mass. No estimate for scale
  uncertainty, but expected to be negligible in $f_{B_s}/f_B$.}
\\[12.0ex] 
\ifx\reducedapptables\undefined
\hline \\[-2.0ex]
\parbox[t]{2.8cm}{RBC/UKQCD~14  \\ RBC/UKQCD 13A}
& \parbox[t]{0.5cm}{\cite{Christ:2014uea} \\ \cite{Witzel:2013sla}} & 2+1 & \parbox[t]{0.9cm}{0.11, 0.086}
& \parbox[t]{4.3cm}{Combined  continuum and chiral extrapolation with linear in $a^2$ term. 
In RBC/UKQCD 14,  the systematic errors from this procedure are
estimated to be from $3.1\%$ to $5.9\%$ in the decay constants and the
$SU(3)$-breaking ratios.
}
& \parbox[t]{4.3cm}{Scale set by the $\Omega$ baryon mass.  In
  RBC/UKQCD 14, scale
  uncertainty estimated to be $1.5\%$ in the decay constants, and
  $0.1\%$ in the $SU(3)$-breaking ratios.}
\\[21.0ex] \hline \\[-2.0ex]
\parbox[t]{2.8cm}{RBC/UKQCD~14A}
& \parbox[t]{0.5cm}{\cite{Aoki:2014nga} } & 2+1 & \parbox[t]{0.9cm}{0.11, 0.086}
& \parbox[t]{4.3cm}{Chiral-continuum extrapolation with
  linear in $a^2$ term is employed, with the systematic errors 
  estimated to be from $1.98\%$ to $3.54\%$ in the decay constants and
  $f_{B_{s}}/f_{B}$.  Discretization errors at $\cO(\alpha_{s} a)$ in
  the static-light system are estimated to be $1\%$
  in the decay constants, and $0.2\%$ in $f_{B_{s}}/f_{B}$. 
}
& \parbox[t]{4.3cm}{Scale set by the $\Omega$ baryon mass.}
\\[29.0ex] \hline \\[-2.0ex]
\parbox[t]{2.8cm}{HPQCD 12} & \parbox[t]{0.5cm}{\cite{Na:2012sp} } & 2+1 & \parbox[t]{0.9cm}{0.12, 0.09}
& \parbox[t]{4.3cm}{Combined  continuum and chiral extrapolation. 
Continuum extrapolation errors estimated to be $0.9\%$. 
}
& \parbox[t]{4.3cm}{Relative scale $r_1/a$ from the static-quark potential.  
Absolute scale $r_1$ from $f_\pi$, $f_K$, and $\Upsilon$(2S-1S) 
splitting.  Scale uncertainty estimated to be $1.1\%$.}
\\[15.0ex] \hline \\[-2.0ex]
\parbox[t]{2.8cm}{HPQCD 11A} & \parbox[t]{0.5cm}{\cite{McNeile:2011ng}} & 2+1 & \parbox[t]{0.9cm}{0.15, 0.12, 0.09, 0.06, 0.045}
& \parbox[t]{4.3cm}{ $am_Q \approx 0.2 - 0.85$. Combined  continuum and HQET fit. 
 Continuum extrapolation error estimated by varying the fit ansatz and the included data
 points to be $0.63\%$. Discretization errors appear to decrease with increasing heavy-meson 
 mass. 
 } 
& \parbox[t]{4.3cm}{Relative scale $r_1/a$ from the static-quark potential.  
Absolute scale $r_1$ from $f_\pi$, $f_K$, and 
$\Upsilon$(2S-1S) splitting.  Scale uncertainty estimated to be $0.74\%$.}
\\[23.0ex] \hline \\[-2.0ex]
\parbox[t]{2.8cm}{FNAL/MILC 11} &\parbox[t]{0.5cm}{\cite{Bazavov:2011aa} } & 2+1 & \parbox[t]{0.9cm}{0.15, 0.12, 0.09}
& \parbox[t]{4.3cm}{Combined continuum and chiral extrapolation. Continuum extrapolation 
errors estimated to be 1.3\%.
}
& \parbox[t]{4.3cm}{Relative scale $r_1/a$ from the static-quark potential.  
Absolute scale $r_1$ from $f_\pi$, $f_K$, and $\Upsilon$(2S-1S) splitting.  
Scale uncertainty estimated to be 1~MeV.}
\\[15.0ex] \hline \\[-2.0ex]
\parbox[t]{2.8cm}{RBC/UKQCD 10C} &\parbox[t]{0.5cm}{\cite{Albertus:2010nm}}  & 2+1 & \parbox[t]{0.9cm}{0.11}
& \parbox[t]{4.3cm}{One lattice spacing with discretization errors estimated 
by power counting as  3\%.}
& \parbox[t]{4.3cm}{Scale set by the $\Omega$ baryon mass. Combined scale and mass tuning
uncertainties on $f_{B_s}/f_B$ estimated as 1\%}
\\[9.0ex] \hline \\[-2.0ex]
\parbox[t]{2.8cm}{HPQCD 09} & \parbox[t]{0.5cm}{\cite{Gamiz:2009ku} } & 2+1 & \parbox[t]{0.9cm}{0.12, 0.09}
& \parbox[t]{4.3cm}{Combined continuum and chiral extrapolation. Continuum 
extrapolation errors estimated to be 3\%. 
}
& \parbox[t]{4.3cm}{Relative scale $r_1/a$ from the static-quark potential.  
Absolute scale $r_1$  from the $\Upsilon$(2S-1S) splitting.  Scale uncertainty
 estimated to be $2.3\%$. }
\\[12.0ex] 
\fi
\hline\hline
\end{tabular*}
\caption{Continuum extrapolations/estimation of lattice artifacts in determinations of 
the $B$ and $B_s$ meson decay constants for $N_f=2+1$ simulations.} 
}
\end{table}


\ifx\reducedapptables\undefined

\begin{table}[!htb]
{\footnotesize
\begin{tabular*}{\textwidth}{l @{\extracolsep{\fill}} l c c l l}
\hline\hline \\[-1.0ex]
Collab. & Ref. & $\Nf$ & $a$ [fm]  & Continuum extrapolation & Scale setting
\\[1.0ex] \hline \hline \\[-2.0ex]
\parbox[t]{3.0cm}{Balasubramamian 19} & \parbox[t]{0.5cm}{\cite{Balasubramanian:2019net}} 
& 2 & \parbox[t]{0.9cm}{0.075, 0.065, 0.048}
& \parbox[t]{4.3cm}{Combined continuum and chiral extrapolation with linear in  
$a^2$ term. Detailed error budget not provided, but total systematic uncertainty estimated as $1.1$-$1.2\%$.}
& \parbox[t]{4.3cm}{Scale set from $f_\pi$. Scale setting uncertainty estimated as $0.9\%$.}
\\[24.0ex]\hline \\[-2.0ex]
\parbox[t]{2.0cm}{ALPHA 14  \\ ALPHA 13 \\ ALPHA 12A \\ ALPHA 11} & 
\parbox[t]{0.5cm}{\cite{Bernardoni:2014fva} \\
    \cite{Bernardoni:2013oda} \\ \cite{Bernardoni:2012ti} \\ \cite{Blossier:2011dk}}
  & 2 &
\parbox[t]{0.9cm}{0.075, 0.065, 0.048} & 
\parbox[t]{4.3cm}{Combined continuum and chiral extrapolation with linear in  
$a^2$ term. Continuum extrapolation 
errors estimated to be 5~MeV in ALPHA 11.  The
continuum extrapolation with a term linear in $a$ also investigated
in ALPHA 14, and within the statistical error no discernable difference was observed.
}
& \parbox[t]{4.3cm}{Relative scale set from $r_0$. Absolute scale set from $f_K$. Scale setting uncertainty included in combined statistical and extrapolation error.}
\\[29.0ex] \hline \\[-2.0ex]
\parbox[t]{2.4cm}{ETM 13B, 13C \\ ETM 12B  \\ ETM 11A}  & 
\parbox[t]{0.8cm}{\cite{Carrasco:2013zta,Carrasco:2013iba} \\
  \cite{Carrasco:2012de} \\ \cite{Dimopoulos:2011gx} }& 2 &
	     \parbox[t]{0.9cm}{0.098, 0.085, 0.067, 0.054} &
\parbox[t]{4.3cm}{Combined continuum and chiral extrapolation, with a term linear in $a^2$.
ETM 12 and 13 include heavier masses than ETM 11A. Discretization error included in combined
statistical and systematic error, estimated by dropping the data at the coarsest lattice spacing
as $\sim 0.5 - 1\%$. 
}
& \parbox[t]{4.3cm}{Scale set from $f_\pi$. Scale setting uncertainty included in combined 
statistical and systematic error. }
\\[26.0ex] \hline \\[-2.0ex]
\parbox[t]{2.0cm}{ETM 09D} & \parbox[t]{0.5cm}{\cite{Blossier:2009hg} } & 2 &
	     \parbox[t]{0.9cm}{0.098, 0.085, 0.067} &
\parbox[t]{4.3cm}{Combined continuum and chiral extrapolation with a term linear in 
$a^2$. 
}
& \parbox[t]{4.3cm}{Scale set from $f_\pi$. Scale setting uncertainty included in 
combined statistical and systematic error.}
\\[9.0ex] 
\hline\hline
\end{tabular*}
\caption{Continuum extrapolations/estimation of lattice artifacts in determinations of 
the $B$- and $B_s$- meson decay constants for $N_f=2$ simulations.} 
}
\end{table}

\fi

%
\ifx\reducedapptables\undefined
\begin{table}[!htb]
{\footnotesize
\begin{tabular*}{\textwidth}{l @{\extracolsep{\fill}} c c c l}
\hline\hline \\[-1.0ex]
Collab.  &  Ref. & $\Nf$ & Ren. & Description  
\\[1.0ex] \hline \hline \\[-2.0ex]
\parbox[t]{2.8cm}{FNAL/MILC 17 }
& \parbox[t]{0.5cm}{\cite{Bazavov:2017lyh} }
& 2+1+1 & -- &
\parbox[t]{7.5cm}{The current used for this work is absolutely normalized. }
\\[4.0ex] \hline \\[-2.0ex]
\parbox[t]{2.8cm}{HPQCD 17A  \\ 
\ifx\reducedapptables\undefined
HPQCD 13
\fi
}
& \parbox[t]{0.5cm}{\cite{Hughes:2017spc} \\ 
\ifx\reducedapptables\undefined
\cite{Dowdall:2013tga} 
\fi
}
& 2+1+1 & PT1$\ell$
& \parbox[t]{7.5cm}{ The NRQD effective current is matched through $O(1/m$) and renormalized using 1-loop PT.  Included are all terms though
$O(\alpha_s)$,  $O(\alpha_s \, a)$, $O(\Lambda_{\rm QCD}/M)$,
  $O(\alpha_s/a M)$ ,  $O(\alpha_s \, \Lambda_{\rm QCD}/M)$. The
  dominant error is due unknown $O(\alpha_s^2)$ contributions to the
  current renormalization. The error is estimated as  $\sim1.4$\% in
  HPQCD 13, and as $\sim2$\% in
  HPQCD 17A. }
\\[18.0ex]  \hline \\[-2.0ex]
\parbox[t]{2.8cm}{ETM 16B  \\ 
\ifx\reducedapptables\undefined
ETM 13E
\fi
}
& \parbox[t]{0.5cm}{\cite{Bussone:2016iua} \\ \ifx\reducedapptables\undefined
\cite{Carrasco:2013naa} 
\fi
}
& 2+1+1 & --, PT &
\parbox[t]{7.5cm}{The current used for the relativistic decay constants is absolutely normalized. 
 The ratio is constructed from the relativistic decay constant data
 and the heavy-quark pole masses. 
 Ratios of pole-to-$\msbar$ mass conversion factors are included 
at N$^{3}$LO in continuum perturbation theory, and the matching
                          between the continuum QCD and HQET currents is performed at
                          N$^{2}$LO.}
\\[20.0ex]  \\[-2.0ex]
\hline\hline
\end{tabular*}
\caption{Description of the renormalization/matching procedure adopted
  in the determinations of the $B$- and $B_s$-meson decay constants for
  $N_{f} = 2 + 1+1$ simulations.} 
}
\end{table}
\fi



\begin{table}[!htb]
{\footnotesize
\begin{tabular*}{\textwidth}{l @{\extracolsep{\fill}} c c c l}
\hline\hline \\[-1.0ex]
\rb{Collab.}  &  \hspace{-0.4cm}\rb{Ref.} & \rb{$\Nf$} & \rb{Ren.} & \rb{Description}  
\\[1.0ex] \hline \hline \\[-2.0ex]
\parbox[t]{2.8cm}{RBC/UKQCD 18A}
& \parbox[t]{1.3cm}{\cite{Boyle:2018knm} } & 2+1 & \parbox[t]{1.3cm}{$-$} &
\parbox[t]{7.5cm}{Mass-independent operator renormalization cancels in $f_{B_{s}}/f_B$.}
\\[3.0ex] 
\ifx\reducedapptables\undefined
\hline \\[-2.0ex]
\parbox[t]{2.8cm}{RBC/UKQCD~14 \\RBC/UKQCD 13}
& \parbox[t]{1.3cm}{ \cite{Christ:2014uea} \\ \cite{Witzel:2013sla}} & 2+1 & mNPR 
& \parbox[t]{7.5cm}{In RBC/UKQCD~14, the error is dominated by the
  perturbative aspect, and is estimated to be $1.7\%$ for the decay
  constants by taking the full size of the
  1-loop correction for the fine lattice. }
\\[9.0ex] \hline \\[-2.0ex]
\parbox[t]{2.8cm}{RBC/UKQCD~14A}
& \parbox[t]{1.3cm}{ \cite{Aoki:2014nga}} & 2+1 & PT1$\ell$ 
& \parbox[t]{7.5cm}{A two-step matching procedure is employed, first from
  QCD to HQET in the continuum at $m_{b}$, then to HQET on the lattice
  at $a^{-1}$ with $O(pa)$ and $O(m_{q} a)$ errors included.  Both
  matching steps are accurate to 1-loop, and the
  running between $m_{b}$ and $a^{-1}$ is performed at 2-loop
  accordingly.  The error is estimated using a power-counting argument
to be $6\%$ for the decay constants.}
\\[20.0ex] \hline \\[-2.0ex]
\parbox[t]{2.8cm}{HPQCD 12, 09} & \parbox[t]{1.3cm}{\cite{Na:2012sp,Gamiz:2009ku}} & 2+1 & PT1$\ell$
& \parbox[t]{7.5cm}{The NRQD effective current is matched through $O(1/m$) and renormalized using 1-loop PT.  Included are all terms though
$O(\alpha_s)$,  $O(\alpha_s \, a)$, $O(\Lambda_{\rm QCD}/M)$, $O(\alpha_s/a M)$ ,  $O(\alpha_s \, \Lambda_{\rm QCD}/M)$. The dominant error is due unknown $O(\alpha_s^2)$ contributions to the current renormalization. The authors take the perturbative error as $\sim 2  \rho_0 \, \alpha_s^2$, where $\rho_0$ is the coefficient of the 1-loop correction to the leading term, which yields an
error of $\sim4$\%. }
\\[23.0ex] \hline \\[-2.0ex]
\parbox[t]{2.8cm}{HPQCD 11A} & \parbox[t]{1.2cm}{\cite{McNeile:2011ng}} & 2+1 & -- 
& \parbox[t]{7.5cm}{This work uses PCAC together with an absolutely normalized current.}
\\[4.0ex] \hline \\[-2.0ex]
\parbox[t]{2.8cm}{FNAL/MILC 11} & \parbox[t]{1.3cm}{\cite{Bazavov:2011aa}}  & 2+1 & mNPR 
& \parbox[t]{7.5cm}{The authors' estimate of the perturbative errors is comparable in size to the actual 
1-loop corrections. }
\\[4.0ex] \hline \\[-2.0ex]
\parbox[t]{2.8cm}{RBC/UKQCD 10C} & \parbox[t]{1.2cm}{\cite{Albertus:2010nm}} & 2+1 & PT1$\ell$ & 
\parbox[t]{7.5cm}{The static-light current is matched through
 $O(\alpha_s a, \alpha_s)$ and renormalized using 1-loop
tadpole improved PT. For massless light quarks, the renormalization factors
cancel in the ratio of decay constants.}
\\[12.0ex] 
\fi
\hline\hline
\end{tabular*}
\caption{Description of the renormalization/matching procedure adopted
  in the determinations of the $B$- and $B_s$-meson decay constants for
  $N_{f} = 2+1$ simulations.} 
}
\end{table}

\begin{table}[!htb]
{\footnotesize
\begin{tabular*}{\textwidth}{l @{\extracolsep{\fill}} c c c l}
\hline\hline \\[-1.0ex]
\rb{Collab.}  &  \hspace{-0.4cm}\rb{Ref.} & \rb{$\Nf$} & \rb{Ren.} & \rb{Description}  
\\[1.0ex] \hline \hline \\[-2.0ex]
\parbox[t]{3.0cm}{Balasubramamian 19} & \parbox[t]{1.0cm}{\cite{Balasubramanian:2019net}} 
& 2 & PT2$\ell$
& \parbox[t]{7.50cm}{Perturbative coefficient relating HQET and QCD currents included at NNLO in continuum perturbation theory. Systematic uncertainties estimated using fits including only NLO coefficients.}
\\[12.0ex]
\ifx\reducedapptables\undefined
\hline \\[-2.0ex]
\parbox[t]{2.8cm}{ALPHA 14 \\ALPHA 13 \\ALPHA 12A \\ALPHA 11} & 
\parbox[t]{1.3cm}{ \cite{Bernardoni:2014fva}\\ \cite{Bernardoni:2013oda}\\ \cite{Bernardoni:2012ti}\\ \cite{Blossier:2011dk}} & 
2 & NPR & 
\parbox[t]{7.5cm}{The authors use the Schr\"odinger functional for the NP matching. }
\\[9.0ex] \hline \\[-2.0ex]
\parbox[t]{2.8cm}{ETM 13B, 13C \\ETM 12B \\ETM 11A} & 
\parbox[t]{1.3cm}{ \cite{Carrasco:2013zta,Carrasco:2013iba}\\ \cite{Carrasco:2012de}\\ \cite{Dimopoulos:2011gx}} & 
 2 & --, PT1$\ell$ & 
\parbox[t]{7.5cm}{The current used for the relativistic decay constants is absolutely normalized. 
{\bf Interpolation method:} The static limit current renormalization is calculated in 1-loop 
mean field improved perturbation theory, there half the correction is used to estimate the
 error.  {\bf Ratio method:} The ratio is constructed from the relativistic decay constant data
 and the heavy-quark pole masses. 
 Ratios of pole-to-$\msbar$ mass  conversion factors are included 
at NLO in continuum perturbation theory.}
\\[25.0ex]  \\[-2.0ex]
\fi
\hline\hline
\end{tabular*}
\caption{Description of the renormalization/matching procedure adopted
  in the determinations of the $B$- and $B_s$-meson decay constants for
  $N_{f} = 2$ simulations.} 
}
\end{table}



%
\ifx\reducedapptables\undefined
\begin{table}[!htb]
{\footnotesize
\begin{tabular*}{\textwidth}{l @{\extracolsep{\fill}} c c c l}
\hline\hline \\[-2.0ex]
Collab. & \hspace{-0.4cm}Ref. & $\Nf$ &  Action  &  Description
\\[1.0ex] \hline \hline \\[-1.0ex]
\parbox[t]{2.8cm}{FNAL/MILC 17 }
& \parbox[t]{1.0cm}{\cite{Bazavov:2017lyh} }
& 2+1+1 & HISQ 
& \parbox[t]{7.5cm}{ The discretization effects are estimated by
  repeating the fit by either adding the coarsest ($a \approx 0.15$ fm)  ensembles or omitting
  the finest ($a \approx 0.03$ fm) ensembles, as well as changing the fit ansatz in the
  combined EFT and continuum-limit analysis.}
\\[12.0ex] \hline  \\[-2.0ex]
\parbox[t]{2.8cm}{HPQCD 17A  \\ 
\ifx\reducedapptables\undefined
HPQCD 13
\fi
}
& \parbox[t]{1.0cm}{\cite{Hughes:2017spc} \\ 
\ifx\reducedapptables\undefined
\cite{Dowdall:2013tga} 
\fi
}
& 2+1+1 & NRQCD 
& \parbox[t]{7.5cm}{HQ truncation effects estimated as in HPQCD 09  to be $1.0\%$.}
\\[4.0ex] \hline  \\[-2.0ex]
\parbox[t]{2.8cm}{ETM 16B  \\ 
\ifx\reducedapptables\undefined
ETM 13E
\fi
}
& \parbox[t]{1.0cm}{\cite{Bussone:2016iua} \\ 
\ifx\reducedapptables\undefined
\cite{Carrasco:2013naa} 
\fi
}
& 2+1+1 & tmWil &
\parbox[t]{7.5cm}{The estimate of the discretization effects is described in the 
continuum table. Ratios of pole-to-$\msbar$ mass conversion factors are included 
at N$^{3}$LO in continuum perturbation theory, and the matching
                          between the continuum QCD and HQET currents is performed at
                          N$^{2}$LO.  The systematic error in this
                  procedure is very small compared to other systematic
                  effects.  }
\\[18.0ex] 
\hline\hline
\end{tabular*}
\caption{Heavy-quark treatment in determinations of the $B$- and $B_s$-meson decay constants for $N_{f} = 2+ 1 +1$ simulations.} 
}
\end{table}
\fi


\begin{table}[!htb]
{\footnotesize
\begin{tabular*}{\textwidth}{l @{\extracolsep{\fill}} c c c l}
\hline\hline \\[-2.0ex]
Collab. & \hspace{-0.4cm}Ref. & $\Nf$ &  Action  &  Description
\\[1.0ex] \hline \hline \\[-1.0ex]
\parbox[t]{2.8cm}{RBC/UKQCD 18A}
& \parbox[t]{1.0cm}{\cite{Boyle:2018knm} } & 2+1 & DWF &
\parbox[t]{7.5cm}{Linear extrapolation in inverse heavy quark mass $1/M_H$, with systematic errors estimated, by excluding heaviest and lightest valence quark masses, to
be $0.47\%$ in $f_{B_{s}}/f_B$. HQ discretization effects are  estimated to be $0.01\%$ in$f_{B_{s}}/f_B$.}
\\[12.0ex] 
\ifx\reducedapptables\undefined
\hline \\[-2.0ex]
\parbox[t]{2.8cm}{RBC/UKQCD~14 \\RBC/UKQCD 13A} & \parbox[t]{1.2cm}{
  \cite{Christ:2014uea} \\ \cite{Witzel:2013sla}} & 2+1 & RHQ & 
\parbox[t]{7.5cm}{In RBC/UKQCD 14, the heavy-quark discretization
  errors are estimated to be $1.7\%$ in the decay constants, and
  $0.3\%$ in the $SU(3)$-breaking ratios.}
\\[6.0ex] \hline \\[-2.0ex]
\parbox[t]{2.8cm}{RBC/UKQCD~14A} & \parbox[t]{1.2cm}{
  \cite{Aoki:2014nga}} & 2+1 & Static & 
\parbox[t]{7.5cm}{Static-limit computation, with $\cO(\Lambda_{{\rm
      QCD}}/m_{h})$ errors estimated to be $10\%$ for the decay
  constants, and $2.2\%$ for $f_{B_{s}}/f_{B}$.}
\\[6.0ex] \hline \\[-2.0ex]
\parbox[t]{2.8cm}{HPQCD 12} & \parbox[t]{1.2cm}{\cite{Na:2012sp}} & 2+1 & NRQCD 
& \parbox[t]{7.5cm}{HQ truncation effects estimated as in HPQCD 09  to be $1.0\%$}
\\[3.0ex] \hline \\[-2.0ex]
\parbox[t]{2.8cm}{HPQCD 11A} & \parbox[t]{1.2cm}{\cite{McNeile:2011ng}} & 2+1 & HISQ
& \parbox[t]{7.5cm}{ The analysis uses a combined continuum and $1/m$ extrapolation. }
\\[3.0ex] \hline \\[-2.0ex]
\parbox[t]{2.8cm}{FNAL/MILC 11} & \parbox[t]{1.2cm}{\cite{Bazavov:2011aa}} & 2+1 & Fermilab
& \parbox[t]{7.5cm}{HQ discretization effects are  
included in the combined chiral and continuum fits, and are estimated by varying the fit Ansatz
and excluding the data at the coarsest lattice spacing to be $\sim 2\%$, consistent with
simple power counting estimates but larger than the residual discretization errors observed
in the data.}
\\[15.0ex] \hline \\[-2.0ex]
\parbox[t]{2.8cm}{RBC/UKQCD 10C} & \parbox[t]{1.2cm}{\cite{Albertus:2010nm}} & 2+1 & Static & 
\parbox[t]{7.5cm}{Truncation effects of $\cO(1/m_h)$ on the $SU(3)$-breaking ratios are
estimated by power counting to be 2\%.}
\\[3.0ex] \hline \\[-2.0ex]
\parbox[t]{2.8cm}{HPQCD 09} & \parbox[t]{1.2cm}{\cite{Gamiz:2009ku}} & 2+1 & NRQCD
& \parbox[t]{7.5cm}{The leading HQ truncation effects are of 
$\cO(\alpha_s \Lambda_{\rm QCD}/m_h)$ due to the
tree-level coefficient of the ${\bm \sigma}\cdot{\rm\bf B}$ term. 
The error is estimated by calculating the $B^*-B$ hyperfine
splitting and comparing with experiment as $1\%$.}
\\[12.0ex] 
\fi
\hline\hline
\end{tabular*}
\caption{Heavy-quark treatment in $N_f=2+1$ determinations of the $B$-and $B_s$-meson decay constants.} 
}
\end{table}

\begin{table}[!htb]
{\footnotesize
\begin{tabular*}{\textwidth}{l @{\extracolsep{\fill}} c c c l}
\hline\hline \\[-2.0ex]
Collab. & \hspace{-0.4cm}Ref. & $\Nf$ &  Action  &  Description
\\[1.0ex] \hline \hline \\[-1.0ex]
\parbox[t]{3.0cm}{Balasubramamian 19} & \parbox[t]{1.0cm}{\cite{Balasubramanian:2019net}} 
& 2 & Wilson
& \parbox[t]{7.0cm}{Expansion in inverse of pseudoscalar heavy-strange meson mass, $1/M_{H_{s}}$. HQ discretization effects included in generic fit linear in $a^2(aM_{H_{s}})^2$. Detailed error budget not provided, but systematic uncertainty estimated as $1.1$-$2.3\%$ in $f_{B_s}$.}
\\[12.0ex]
\ifx\reducedapptables\undefined
\hline \\[-2.0ex]
\parbox[t]{2.8cm}{ALPHA 14 \\ALPHA 13 \\ALPHA 12A \\ALPHA 11} & 
\parbox[t]{1.2cm}{ \cite{Bernardoni:2014fva}\\ \cite{Bernardoni:2013oda}\\ \cite{Bernardoni:2012ti}\\ \cite{Blossier:2011dk}} & 2 & HQET &
\parbox[t]{7.5cm}{NP improved through $\cO(1/m_h)$. Truncation errors of 
$\cO\left [ (\Lambda_{\rm QCD}/m_h)^2 \right ]$ are not included. \\}
\\[9.0ex] \hline \\[-2.0ex]
\parbox[t]{2.8cm}{ETM 13B, 13C \\ETM 12B \\ETM 11A} & 
\parbox[t]{1.2cm}{ \cite{Carrasco:2013zta,Carrasco:2013iba}\\ \cite{Carrasco:2012de}\\ \cite{Dimopoulos:2011gx}} & 
2 & tmWil &
\parbox[t]{7.5cm}{The estimate of the discretization effects is described in the 
continuum table. In both methods the relativistic data are 
matched to HQET using NLO continuum PT in an intermediate step, and 
converted back to QCD at the end. 
The error due to HQET matching (estimated by replacing the NLO expressions  with LO)
 is a very small contribution to the systematic error due to the 
heavy-quark mass dependence. The variation observed from adding 
heavier masses to their data and/or including $1/m_h^3$ terms is 
$0.4-1.3\%$.}
\\[29.0ex]
\fi
\hline\hline
\end{tabular*}
\caption{Heavy-quark treatment in $N_f=2$ determinations of the $B$-and $B_s$-meson decay constants.} 
}
\end{table}


\clearpage

\subsubsection{$B_{(s)}$-meson mixing matrix elements}
\label{app:BMix_Notes}
\begin{table}[!ht]
{\footnotesize
\begin{tabular*}{\textwidth}{l @{\extracolsep{\fill}} c c c l l}
\hline\hline \\[-1.0ex]
Collab. & Ref. & $\Nf$ & $a$ [fm] & Continuum extrapolation & Scale setting 
\\[1.0ex] \hline \hline \\[-1.0ex]
HPQCD 19A & \cite{Dowdall:2019bea} & 2+1+1 & 
\parbox[t]{0.9cm}{0.15, 0.12, 0.09} & 
\parbox[t]{4.3cm}{Discretization errors start from $\alpha_sa^2$ and are included in the systematic error.
		 It is estimated as $1.8$\% for individual bag parameters.
		 Residual $\alpha_sa^2$ and $a^4$ errors from wrong-spin contributions are subtracted by including them in the chiral  fit.} &
\parbox[t]{4.2cm}{Scale setting done using $\Upsilon$ and $\Upsilon'$ mass splitting \cite{Dowdall:2011wh}.}
\\[24.0ex] 
\hline\hline
\end{tabular*}
\caption{Continuum extrapolations/estimation of lattice artifacts 
 in determinations of the neutral $B$-meson mixing matrix elements for
 $N_f=2+1+1$ simulations.}
\label{tab:BBcontinum4}
}
\end{table}

\begin{table}[!ht]
{\footnotesize
\begin{tabular*}{\textwidth}{l @{\extracolsep{\fill}} c c c l l}
\hline\hline \\[-1.0ex]
Collab. & Ref. & $\Nf$ & $a$ [fm] & Continuum extrapolation & Scale setting 
\\[1.0ex] \hline \hline \\[-1.0ex]
RBC/UKQCD 18A & \cite{Boyle:2018knm} & 2+1 & 
\parbox[t]{0.9cm}{0.11, 0.08, 0.07} & 
\parbox[t]{4.3cm}{Combined continuum ($a^2$) and heavy quark ($1/m_H$)
		 extrapolation with the LO pion mass dependence ($m_\pi^2$)
		 in the global fit.} &
\parbox[t]{4.2cm}{Lattice scale and target quark masses are set using
		  $\Omega$, $K$ and $\pi$ masses \cite{Aoki:2010dy,Blum:2014tka,Boyle:2017jwu}.}
\\[12.0ex] 
\ifx\reducedapptables\undefined
\hline \\[-2.0ex]
FNAL/MILC 16& \cite{Bazavov:2016nty} & 2+1 & 
\parbox[t]{0.9cm}{0.12, 0.09, 0.06, 0.045} & 
\parbox[t]{4.3cm}{Combined continuum and chiral extrapolation with
		 NLO HMrS$\chi$PT and NNLO analytic terms as well as the
		 terms for the heavy quark
		 discretization errors up to $a^3$, heavy quark mass
		 mismatch, and renormalization error of $\alpha_s^2$.} &
\parbox[t]{4.2cm}{Relative scale $r_1/a$ is set via static-quark
		  potential. Absolute scale is set as $r_1=0.3117(22)$ fm.
		  See the description of FNAL/MILC 12 below.
		  The scale uncertainty on $\xi$, e.g., is estimated as 0.6\%.} 
\\[20.0ex] 
\hline \\[-2.0ex]
RBC/UKQCD 14A& \cite{Aoki:2014nga} & 2+1 & \parbox[t]{0.9cm}{0.11, 0.086} & 
\parbox[t]{4.3cm}{Combined continuum and chiral extrapolation with
		 $SU(2)$ NLO HM$\chi$PT and linear in quark mass
		 both with $\cO(a^2)$ terms.
		 The combined continuum and chiral extrapolation
		 uncertainty is estimated as 2.55, 2.13 and 3.08\% for
		 $f_B\sqrt{B_B}$, $f_{B_s}\sqrt{B_{B_s}}$ and $\xi$
		 respectively.} &
\parbox[t]{4.2cm}{Scale is set using the $\Omega^-$ mass as input 
		 \cite{Aoki:2010dy}. The scale uncertainty is estimated
		 as 1.84, 1.86 and 0.05\% for $f_B\sqrt{B_B}$, 
		 $f_{B_s}\sqrt{B_{B_s}}$ and $\xi$ respectively.}
\\[26.0ex] \hline \\[-2.0ex]
FNAL/MILC 12 & \cite{Bazavov:2012zs} & 2+1 & \parbox[t]{0.9cm}{0.12, 0.09} & 
\parbox[t]{4.3cm}{Combined continuum and chiral extrapolation with
		 NLO rHMS$\chi$PT, NNLO analytic and generic 
		 $\cO(\alpha_s^2 a^2, a^4)$ terms. 
		 Combined statistical, chiral and light-quark discretization error 
		 is estimated, by examining the variation with different fit Ans\"atze
		 to be 3.7\% on $\xi$.} &
\parbox[t]{4.2cm}{Relative scale $r_1/a$ is set via static-quark
		  potential. Absolute scale $r_1=0.3117(22)$ fm is
		  determined \cite{Bazavov:2011aa} through averaging the
		  $f_\pi$ input and the estimate of HPQCD collaboration 
		  \cite{Davies:2009tsa}. The scale uncertainty on $\xi$ is estimated as 0.2\%.} 
\\[23.0ex] \hline \\[-2.0ex]
FNAL/MILC 11A & \cite{Bouchard:2011xj} & 2+1 & \parbox[t]{0.9cm}{0.12, 0.09, 0.06} & 
\parbox[t]{4.3cm}{Combined continuum and chiral extrapolation with
		 NLO rHMS$\chi$PT, NNLO analytic and and generic 
		 $\cO(\alpha_s^2 a^2, a^4)$ terms.} & 
\parbox[t]{4.2cm}{See above. The error in $r_1$ yields a 3\% 
		 uncertainty on $f^2_B B_B$.} 
\\[12.0ex] \hline \\[-2.0ex]
\parbox[t]{2.8cm}{RBC/UKQCD 10C} & \parbox[t]{0.5cm}{\cite{Albertus:2010nm}} & 2+1 & \parbox[t]{0.9cm}{0.11} &
\parbox[t]{4.3cm}{Only one lattice spacing is used. Discretization error
		 is estimated to be 4\% on $\xi$ by power counting.}&
\parbox[t]{4.2cm}{Scale is set using the $\Omega^-$ mass as input 
		 \cite{Allton:2008pn}.The error on $\xi$ due to the combined scale and 
		 light-quark mass uncertainties is estimated as 1\%.} 
\\[12.0ex] \hline \\[-2.0ex]
HPQCD 09 & \cite{Gamiz:2009ku} & 2+1 & \parbox[t]{1.1cm}{0.12, 0.09} &
\parbox[t]{4.3cm}{Combined continuum and chiral extrapolation with
		 NLO rHMS$\chi$PT and NNLO analytic terms. Light-quark
		 discretization error
		 is estimated as 3, 2 and 0.3\% for $f_B\sqrt{B_B}$,
		 $f_{B_s}\sqrt{B_{B_s}}$ and $\xi$ respectively.} &
\parbox[t]{4.2cm}{Relative scale $r_1/a$ is set via static-quark
		  potential. Absolute scale $r_1=0.321(5)$ fm is
		  determined through $\Upsilon$ mass \cite{Gray:2005ur}.
		  The error on $f_B\sqrt{B_B}$ due to the scale uncertainty
		  is estimated as 2.3\%.} 
\\[18.0ex] \hline \\[-2.0ex]
HPQCD 06A & \cite{Dalgic:2006gp} & 2+1 & \parbox[t]{1.1cm}{0.12} & 
\parbox[t]{4.3cm}{Only one lattice spacing is used. Light-quark discretization error
		 on $f_{B_s}^2B_{B_s}$ is estimated as 4\% 
		 by power counting.} &
\parbox[t]{4.2cm}{Scale is set using the $\Upsilon$ $2S-1S$ splitting as 
		 input \cite{Gray:2005ur}. The error on $f^2_B B_B$ due to the scale uncertainty
		  is estimated as 5\%.} 
\\[12.0ex]
\fi
\hline\hline
\end{tabular*}
\caption{Continuum extrapolations/estimation of lattice artifacts 
 in determinations of the neutral $B$-meson mixing matrix elements for
 $N_f=2+1$ simulations.}
\label{tab:BBcontinum}
}
\end{table}

\ifx\reducedapptables\undefined

\begin{table}[!ht]
{\footnotesize
\begin{tabular*}{\textwidth}{l @{\extracolsep{\fill}}c c c l l}
\hline\hline \\[-1.0ex]
Collab. & Ref. & $\Nf$ & $a$ [fm] & Continuum extrapolation & Scale setting 
\\[1.0ex] \hline \hline \\[-1.0ex]
ETM 13B  & \cite{Carrasco:2013zta} & 2 & \parbox[t]{1.1cm}{0.098,
	     0.085, 0.067, 0.054} & 
\parbox[t]{4.3cm}{
	     Combined chiral and continuum extrapolation, with a term linear in $a^2$.
	     Discretization error is estimated by omitting the
	     coarsest lattice as 0.5, 1.7, 1.3 and 1.0 \% for 
	     $B_{B_s}$, $B_B$, $B_{B_s}/B_B$ and $\xi$ respectively.
	     The heavy-quark masses vary in the range $0.13 \lsim  a m_h \lsim 0.85$.} &
\parbox[t]{4.0cm}{See below.} 
\\[26.0ex] 
\hline \\[-2.0ex]
\parbox[t]{2.4cm}{ETM 12A, 12B}  & \cite{Carrasco:2012dd,Carrasco:2012de} & 2 & \parbox[t]{1.1cm}{0.098,
	     0.085, 0.067} & 
\parbox[t]{4.3cm}{Combined chiral and continuum extrapolation, with a term linear in $a^2$.
		Discretization error included in combined statistical, chiral and continuum extrapolation error
		and estimated as $4.5$\%. 
		The heavy-quark masses vary in the range $0.25 \lsim  a m_h \lsim 0.6$.} &
\parbox[t]{4.0cm}{Relative scale $r_0/a$ set from the static-quark
		     potential. Absolute scale set from $f_\pi$. Scale setting uncertainty 
		     included in combined statistical and systematic error.} 
\\[23.0ex] 
\hline\hline
\end{tabular*}
\caption{Continuum extrapolations/estimation of lattice artifacts 
 in determinations of the neutral $B$-meson mixing matrix elements for
 $N_f=2$ simulations.}
\label{tab:BBcontinum2}
}
\end{table}

\fi

\begin{table}[!ht]
{\footnotesize
\begin{tabular*}{\textwidth}{l @{\extracolsep{\fill}} c c c  l}
\hline\hline \\[-1.0ex]
Collab. & Ref. & $\Nf$ & {$M_{\pi,\rm min}\,[\mev]$}  & {Description}
\\[1.0ex] \hline \hline \\[-1.0ex]
HPQCD 19A& \cite{Dowdall:2019bea} & 2+1+1 & \parbox[t]{2cm}{311, 241, --}
	     &  \parbox[t]{6cm}{Pion mass in the Goldston channel is as small as 130 MeV
		 for two coarser lattices. NLO HMrS$\chi$PT is used with NNLO analytic terms and other
		 discretization errors. Staggered wrong-spin contributions are included.
		 }
\\[7.0ex] 
\hline \\[-2.0ex]
RBC/UKQCD 18A& \cite{Boyle:2018knm} & 2+1 & \parbox[t]{2cm}{139, 139, 234}
	     &  \parbox[t]{6cm}{Combined continuum ($a^2$) and heavy quark ($1/m_H$)
		 extrapolation with the LO pion mass dependence ($m_\pi^2$)
		 in the global fit.}
\\[9.0ex] 
\ifx\reducedapptables\undefined
\hline \\[-2.0ex]
FNAL/MILC 16& \cite{Bazavov:2016nty} & 2+1 & \parbox[t]{2cm}{464,
	     280, 257, 332}
	     &  \parbox[t]{6cm}{Combined continuum and chiral extrapolation with
		 NLO HMrS$\chi$PT, NNLO analytic terms and other
		 discretization errors.
		 See the entry in Tab.~\ref{tab:BBcontinum}.
		 The breakdown of the chiral error on $\xi$ is 0.4\% and
		 not the dominant one.}
\\[15.0ex] 
\hline \\[-2.0ex]
RBC/UKQCD 14A& \cite{Aoki:2014nga} & 2+1 & \parbox[t]{2cm}{327, 289}
	     &  \parbox[t]{6cm}{Combined continuum and chiral extrapolation with
		 $SU(2)$ NLO HM$\chi$PT and linear in quark mass
		 both with $\cO(a^2)$ terms.
		 The chiral fit error is estimated from 
		 difference between the NLO HM$\chi$PT and linear fits,
		 and further from eliminating the heaviest $ud$ quark
		 mass point.}
\\[18.0ex] \hline \\[-2.0ex]
FNAL/MILC 12 & \cite{Bazavov:2012zs} & 2+1 & \parbox[t]{2cm}{440, 320}
	     &  \parbox[t]{6cm}{Combined continuum and chiral extrapolation with
		 NLO rHMS$\chi$PT and NNLO analytic terms.
		 See the entry in Tab.~\ref{tab:BBcontinum}.
		 The omission of wrong-spin contributions \cite{Bernard:2013dfa} 
		 in the HMrS$\chi$PT is treated as a systematic error and
		 estimated to be 3.2\% for $\xi$.}
\\[15.0ex] \hline \\[-2.0ex]
FNAL/MILC 11A & \cite{Bouchard:2011xj} & 2+1 & \parbox[t]{2cm}{440, 320, 250} & 
\parbox[t]{6cm}{Combined continuum and chiral extrapolation with
		NLO rHMS$\chi$PT and NNLO analytic terms.}
\\[6.0ex] \hline \\[-2.0ex]
RBC/UKQCD 10C & \cite{Albertus:2010nm} & 2+1 & \parbox[t]{2cm}{430} &
\parbox[t]{6cm}{Linear fit matched with $SU(2)$ NLO HM$\chi$PT at the lightest
		$ud$ mass point is used as the preferred fit. Many different fit Ans\"atze
		are considered. The systematic
		 error is estimated from the difference between the $SU(2)$ 
		 HM$\chi$PT fit described above and a linear fit. }
\\[18.0ex]\hline \\[-2.0ex] 
HPQCD 09 & \cite{Gamiz:2009ku} & 2+1 & \parbox[t]{2cm}{440, 400} &
\parbox[t]{6cm}{Combined continuum and chiral extrapolation with
		NLO rHMS$\chi$PT and NNLO analytic terms.} 
\\[6.0ex] \hline \\[-2.0ex]
HPQCD 06A & \cite{Dalgic:2006gp} & 2+1 & \parbox[t]{2cm}{510} & 
\parbox[t]{6cm}{Two sea $ud$ quark masses $m_{ud}/m_s=0.25$ and $0.5$
		 are used to calculate the matrix element for $B_s$
		 meson at the predetermined value of the strange-quark
		 mass. No significant sea-quark mass dependence is observed
		 and the value at the lighter sea $ud$ mass is taken as the result.} 
\\[18.0ex] \hline \\[-2.0ex]
\parbox[t]{2.2cm}{ETM 13B\\ ETM 12A, 12B}  & \parbox[t]{1.4cm}{\cite{Carrasco:2013zta}\\ \cite{Carrasco:2012dd,Carrasco:2012de}} & 2 & \parbox[t]{2cm}{410, 275, 300, 270} & 
\parbox[t]{6cm}{$M_{\pi,{\rm min}}$ refers to the charged pions, where 270 MeV
	on the finest lattice only included in ETM 13B. Linear and NLO (full
	QCD) HM$\chi$PT supplemented by an $a^2$ term is used. The chiral fit error is
	estimated from the difference between the NLO HM$\chi$PT and linear fits. }
\\[18.0ex] 
%
\fi
\hline\hline
\end{tabular*}
\caption{Chiral extrapolation/minimum pion mass
 in determinations of the neutral $B$-meson mixing matrix elements.
 For actions with multiple species of pions, masses quoted are the RMS
 pion masses (where available).  The different $M_{\pi,\rm min}$ entries correspond to the
 different lattice spacings.}
}
\end{table}

\begin{table}[!ht]
{\footnotesize
\begin{tabular*}{\textwidth}{l @{\extracolsep{\fill}} c c c c l}
\hline\hline \\[-1.0ex]
Collab. & Ref. & $\Nf$ & $L$ [fm] & ${M_{\pi,\rm min}}L$ & Description
\\[1.0ex] \hline \hline \\[-1.0ex]
HPQCD 19A & \cite{Dowdall:2019bea} & 2+1+1 & 
\parbox[t]{2.0cm}{2.4/3.5/4.6, 2.9/3.8/5.7, 2.8} & 
\parbox[t]{1.8cm}{7.3, 7.0, -} & 
\parbox[t]{5.1cm}{FV error is estimated to be negligible from FV HM$\chi$PT.} 
\\[7.0ex] 
\hline \\[-2.0ex]
RBC/UKQCD 18A & \cite{Boyle:2018knm} & 2+1 & 
\parbox[t]{2.0cm}{2.7/5.5, 2.6/5.3, 3.5} & 
\parbox[t]{1.8cm}{3.9, 3.8, 4.0} & 
\parbox[t]{5.1cm}{FV error is estimated to be less than $0.18\%$ for
		  $SU(3)$-breaking ratios from FV HM$\chi$PT.} 
\\[7.0ex] 
\ifx\reducedapptables\undefined
\hline \\[-2.0ex]
FNAL/MILC 16& \cite{Bazavov:2016nty} & 2+1 & 
\parbox[t]{2.0cm}{2.4/2.9, 2.5/2.9/3.6/5.8, 2.9/3.4/3.8, 2.9} & 
\parbox[t]{1.8cm}{6.8, 8.2, 5.0, 4.8} & 
\parbox[t]{5.1cm}{FV error is estimated to be less than $0.1\%$ for
		  $SU(3)$-breaking ratios from FV HMrS$\chi$PT.} 
\\[9.0ex] 
\hline \\[-2.0ex]
RBC/UKQCD 14A
& \cite{Aoki:2014nga} & 2+1 & 
\parbox[t]{1.8cm}{2.74, 2.76} & 4.5, 4.0 & 
\parbox[t]{5.1cm}{FV error is estimated from $SU(2)$ $\chi$PT to be 
		  0.76, 0, 1.07\% for $f_B\sqrt{B_B}$,
		   $f_{B_s}\sqrt{B_{B_s}}$ and $\xi$ respectively.} 
\\[6.0ex] \hline \\[-2.0ex]
FNAL/MILC 12 & \cite{Bazavov:2012zs} & 2+1 & 
\parbox[t]{1.8cm}{2.4/2.9, 2.5} & 6.4, 5.1 & 
\parbox[t]{5.1cm}{FV error is estimated to be less than $0.1\%$ for
		     $SU(3)$-breaking ratios from FV HMrS$\chi$PT.} 
\\[6.0ex] \hline \\[-2.0ex]
FNAL/MILC 11A & \cite{Bouchard:2011xj} & 2+1 & 
\parbox[t]{1.8cm}{2.4/2.9, 2.5/2.9/3.6, 3.8} & 6.4, 5.8, 4.9 &
\parbox[t]{5.1cm}{FV error on $f_B\sqrt{B_B}$ is estimated to be less
		  than $1\%$, 
		  which is inferred from the study of the $B$-meson decay
		  constant using FV HM$\chi$PT \cite{Bazavov:2011aa}.} 
\\[9.0ex] \hline \\[-2.0ex]
RBC/UKQCD 10C & \cite{Albertus:2010nm} & 2+1 & 
\parbox[t]{1.8cm}{1.8} & 3.9 & 
\parbox[t]{5.1cm}{FV error estimated through FV HM$\chi$PT as $1\%$ for $SU(3)$-breaking ratios. }
\\[6.0ex]\hline \\[-2.0ex] 
HPQCD 09 & \cite{Gamiz:2009ku} & 2+1 & 
\parbox[t]{1.8cm}{2.4/2.9, 2.5} & 6.4, 5.1 &
\parbox[t]{5.1cm}{No explicit estimate of FV error, but expected to be much smaller than other uncertainties.}
\\[6.0ex] \hline \\[-2.0ex]
HPQCD 06A & \cite{Dalgic:2006gp} & 2+1 & 
\parbox[t]{1.8cm}{2.4} & 6.2 & 
\parbox[t]{5.1cm}{No explicit estimate of FV error, but expected to be much smaller than other uncertainties.}
\\[6.0ex] \hline \\[-2.0ex]
\parbox[t]{2.2cm}{ETM 13B\\ ETM 12A, 12B}  &
     \parbox[t]{1.4cm}{\cite{Carrasco:2013zta}\\ \cite{Carrasco:2012dd,Carrasco:2012de}} & 2 & 
\parbox[t]{1.8cm}{2.4, 2.0/2.7, 2.1, 1.7/2.6} & 
\parbox[t]{1.8cm}{5.0, 3.7, 3.3, 3.5} & 
\parbox[t]{5.1cm}{$L=1.7/2.6\, {\rm fm}$ only included in ETM~13B. FV error is assumed to be negligible based on the study of $D$-meson decay constants in 
Ref.~\cite{Blossier:2009bx}.}
\\[9.0ex] 
\fi
\hline\hline
\end{tabular*}
\caption{Finite-volume effects 
 in determinations of the neutral $B$-meson mixing matrix elements.
 Each $L$-entry corresponds to a different
 lattice spacing, with multiple spatial volumes at some lattice
 spacings. 
 For actions with multiple species of pions, masses quoted are the RMS
 pion masses (where available).
}
}
\end{table}

\begin{table}[!ht]
{\footnotesize
\begin{tabular*}{\textwidth}{l @{\extracolsep{\fill}} c c c l}
\hline\hline \\[-1.0ex]
Collab. & Ref. & $\Nf$ & Ren. & Description 
\\[1.0ex] \hline \hline \\[-1.0ex]
HPQCD 19A & \cite{Dowdall:2019bea} & 2+1+1 & PT1$\ell$  & 
\parbox[t]{6cm}{HISQ-NRQCD 4-quark operators are matched through $O(1/M)$ and renormalized to 1-loop:
		 included are those of $O(\alpha_s)$, $O(\Lambda_{\rm QCD}/M)$, $O(\alpha_s/a M)$, 
		 $O(\alpha_s \, \Lambda_{\rm QCD}/M)$. 
		 Remnant error is dominated by $O(\alpha_s \Lambda_{\rm QCD}/M)~2.9$\% 
		 and $O(\alpha_s^2)~2.1$\% for individual bag parameters.
		 Associated error for their SU(3) breaking ratio are negligible.}
\\[24.0ex] 
\hline \\[-2.0ex]
RBC/UKQCD 18A & \cite{Boyle:2018knm} & 2+1 & - & 
\parbox[t]{6cm}{Operators are renormalized multiplicatively 
		 due to chiral symmetry of DWF. 
		 No need to calculate the renormalization factor
		 since only the $SU(3)$ breaking ratios are examined.}
\\[12.0ex] 
\ifx\reducedapptables\undefined
\hline \\[-2.0ex]
FNAL/MILC 16& \cite{Bazavov:2016nty}  & 2+1 & mNPR & 
\parbox[t]{6cm}{mNPR is used with 1-loop lattice perturbation theory
		to renormalize the four-quark operators
		with heavy quarks rotated to eliminate tree-level $\cO(a)$
		errors. The error from neglecting higher order corrections 
		is estimated to be $0.5$\% on $\xi$.} 
\\[15.0ex] 
\hline \\[-2.0ex]
RBC/UKQCD 14A & \cite{Aoki:2014nga} & 2+1 & PT$1l$ & 
\parbox[t]{6cm}{Static-light four-quark operators are 
		renormalized with 1-loop
		mean field improved PT.
		The errors due to neglected higher order effects 
		are estimated for purely $\cO\alpha_s^2$ to be
		$6$\% on the matrix elements 
		or $1.2$\% on $\xi$ and for $\cO\alpha_s^2 a^2$
		to be $1$\% or $0.2$\% respectively.}
\\[18.0ex] \hline \\[-2.0ex]
FNAL/MILC 12 & \cite{Bazavov:2012zs} & 2+1 & PT$1l$ & 
\parbox[t]{6cm}{1-loop mean-field improved PT
		 is used to renormalize the four-quark operators
		 with heavy quarks rotated to eliminate tree-level $\cO(a)$
		 errors. The error from neglecting higher order corrections 
		 is estimated to be $0.5$\% on $\xi$.} 
\\[15.0ex] \hline \\[-2.0ex]
FNAL/MILC 11A & \cite{Bouchard:2011xj} & 2+1 & PT$1l$ &
\parbox[t]{6cm}{1-loop mean-field improved PT is used to 
                renormalize the four-quark operators
		 with heavy quarks rotated to eliminate tree-level $\cO(a)$
		 errors. The error from neglected higher order corrections 
		 is estimated to be $4$\% on $f_B\sqrt{B_B}$.} 
\\[15.0ex] \hline \\[-2.0ex]
RBC/UKQCD 10C & \cite{Albertus:2010nm} & 2+1 & PT$1l$ & 
\parbox[t]{6cm}{Static-light four-quark operators are 
		 renormalized with 1-loop
		 mean field improved PT.
		 The error due to neglected higher order effects 
		 is estimated to be $2.2$\% on $\xi$.}
\\[9.0ex]\hline \\[-2.0ex]
HPQCD 09 & \cite{Gamiz:2009ku} & 2+1 & PT$1l$ &
\parbox[t]{6cm}{Four-quark operators in lattice NRQCD are matched to
		 QCD through order $\alpha_s$, $\Lambda_{\rm QCD}/M$ 
		 and  $\alpha_s/(aM)$ \cite{Gamiz:2008sk} using 1-loop PT. 
		 The error due to neglected higher order effects is estimated to 
		 be 4\% on $f_B\sqrt{B_B}$ and 0.7\% on $\xi$.}
\\[15.0ex] \hline \\[-2.0ex]
HPQCD 06A & \cite{Dalgic:2006gp} & 2+1 & PT$1l$ &
\parbox[t]{6cm}{Four-quark operators in lattice NRQCD are matched to
		 full QCD through order $\alpha_s$, $\Lambda_{\rm QCD}/M$ 
		 and  $\alpha_s/(aM)$ \cite{Gamiz:2008sk}. The error  
		 is estimated as $\sim 1\cdot \alpha_s^2$ to be 9\% on
		 $f_{B_s}^2 B_{B_s}$}
\\[9.0ex] \hline \\[-2.0ex]
ETM 13B, 12A, 12B & \cite{Carrasco:2013zta,Carrasco:2012dd,Carrasco:2012de} & 2 & NPR & 
\parbox[t]{6cm}{The bag parameters are nonperturbatively 
	         renormalized in the RI'-MOM scheme. They are calculated
	         as functions of the ($\msbar$) heavy-quark mass (renormalized
	         nonperturbatively in RI/MOM). }
\\[12.0ex] 
%
%
%
\fi
\hline\hline
\end{tabular*}
\caption{Operator renormalization 
 in determinations of the neutral $B$-meson mixing matrix elements.
}
\label{tab:BBmatching}
}
\end{table}

\begin{table}[!ht]
{\footnotesize
\begin{tabular*}{\textwidth}{l @{\extracolsep{\fill}} c c c l}
\hline\hline \\[-1.0ex]
Collab. & Ref. & $\Nf$ & Action & Description
\\[1.0ex] \hline \hline \\[-1.0ex]
HPQCD 19A & \cite{Dowdall:2019bea} & 2+1+1 & NRQCD &
\parbox[t]{8cm}{See the entry in Tab.~\ref{tab:BBmatching}.}
\\[4.0ex] 
\hline \\[-2.0ex]
RBC/UKQCD 18A & \cite{Boyle:2018knm} & 2+1 & DWF &
\parbox[t]{8cm}{Domain-wall fermion with 3 stout-smearing extends the reach to heavy mass,
		 allowing to simulate up to half of the b-quark mass.
		 Heavy mass errors on $\xi$ are estimated as 0.8\% from fitting range and 0.4\% 
		 from higher order ($1/M^2$) by power counting.}
\\[10.0ex] 
\ifx\reducedapptables\undefined
\hline \\[-2.0ex]
FNAL/MILC 16& \cite{Bazavov:2016nty} & 2+1 & Fermilab & 
\parbox[t]{8cm}{The heavy-quark discretization error is a dominant error 
		comparable to the statistical error. It reads 4.6\%,
		3.2\% or 0.7\% for the $B_d$, $B_s$ matrix element or $\xi$.
		}
\\[6.0ex] 
\hline \\[-2.0ex]
RBC/UKQCD 14A& \cite{Aoki:2014nga} & 2+1 & Static & 
\parbox[t]{8cm}{Two different static-quark actions with HYP1 and HYP2
		 smearings are used and the continuum extrapolation is
		 constrained so the two values converges in the limit.
		 The error due to the missing $1/m_b$ corrections is
		 estimated to be $12$\% for individual matrix
		 elements or $2.2$\% on $\xi$ using power-counting.}
\\[15.0ex] \hline \\[-2.0ex]
FNAL/MILC 12 & \cite{Bazavov:2012zs} & 2+1 & Fermilab & 
\parbox[t]{8cm}{The heavy-quark discretization error on $\xi$ is estimated
		to be 0.3 \%. The error on $\xi$ 
		 due to the uncertainty in the $b$-quark mass is
		 estimated to be 0.4 \%.} 
\\[6.0ex] \hline \\[-2.0ex]
FNAL/MILC 11A & \cite{Bouchard:2011xj} & 2+1 & Fermilab &
\parbox[t]{8cm}{The 
		 heavy-quark discretization error on $f_B\sqrt{B_B}$
		 is estimated as 4\% using power-counting.} 
\\[3.0ex] \hline \\[-2.0ex]
RBC/UKQCD 10C & \cite{Albertus:2010nm} & 2+1 & Static & 
\parbox[t]{8cm}{Two different static-quark actions with Ape and HYP
		 smearings are used. The discretization error
		 on $\xi$ is estimated as  $\sim 4$\% and  the error due to the missing
		 $1/m_b$ corrections  as
		 $\sim 2$\%, both using power-counting.}
\\[9.0ex]\hline \\[-2.0ex]
HPQCD 09 & \cite{Gamiz:2009ku} & 2+1 & NRQCD &
\parbox[t]{8cm}{Heavy-quark truncation errors due to relativistic corrections
		 are estimated to be 2.5, 2.5 and 0.4 \% for $f_B\sqrt{B_B}$,
		 $f_{B_s}\sqrt{B_{B_s}}$ and $\xi$ respectively.}
\\[6.0ex] \hline \\[-2.0ex]
HPQCD 06A & \cite{Dalgic:2006gp} & 2+1 & NRQCD &
\parbox[t]{8cm}{Heavy-quark truncation errors due to relativistic corrections
		 are estimated to be 3\% for $f_{B_s}^2{B_{B_s}}$.}
\\[3.0ex] \hline \\[-2.0ex]
\parbox[t]{2.2cm}{ETM 13B\\ ETM 12A,12B}  &
     \parbox[t]{1.4cm}{\cite{Carrasco:2013zta}\\ \cite{Carrasco:2012dd,Carrasco:2012de}} & 2 & 
	 tmWil & 
\parbox[t]{8cm}{The ratio method is used to perform an interpolation to
	the physical $b$ quark mass from the simulated
	heavy mass and the known static limit.
	In an intermediate step, the ratios include HQET matching
	factors calculated to tree-level, leading-log, 
	and next-to-leading-log (ETM 13B only)
	in continuum PT.
	The interpolation uses a polynomial up to
	quadratic in the inverse quark-mass. 
	The systematic errors added together with those of the chiral
	fit are estimated as $1.3-1.6$\% for bag parameters for ETM 13B,
	while they are estimated from
	changing the interpolating polynomial as $2$\% and from changing the order of HQET 
	matching factors as $3$\% for ETM 12A and 12B.}
\\[32.0ex]
\fi
\hline\hline
\end{tabular*}
\caption{Heavy-quark treatment
 in determinations of the neutral $B$-meson mixing matrix elements.
}}
\end{table}

\clearpage


\FloatBarrier
\subsubsection{Form factors entering determinations of $|V_{ub}|$ ($B \to \pi\ell\nu$, $B_s \to K\ell\nu$, $\Lambda_b\to p\ell\bar{\nu}$)}
\label{app:BtoPi_Notes}
\FloatBarrier



\begin{table}[!ht]

{\footnotesize
\begin{tabular*}{\textwidth}{l c c c l l}
\hline\hline \\[-1.0ex]
Collab. & Ref. & $\Nf$ & $a$ [fm] & Continuum extrapolation & Scale setting 
\\[1.0ex] \hline \hline \\[-1.0ex]
\SLfnalmilcBsK & \cite{Bazavov:2019aom} & 2+1 & \parbox[t]{20mm}{0.06, 0.09, 0.12}   &
\parbox[t]{4.2cm}{
HMrS$\chi$PT expansion used at next-to-leading
order in $SU(2)$ and leading order in $1/M_B$, including next-to-next-to-leading-order
(NNLO) analytic and generic discretization terms. Hard kaons assumed to decouple. Systematic uncertainties
estimated by varying fit ansatz and data range. The (stat + chiral extrap + HQ discretization + $g_{\pi}$) uncertainty
dominates the error budget, ranging from 2--3\% at $q^2 \gtrsim 21~{\rm GeV}^2$ to up to 8-10\%
in the lower end of the accessed $q^2$ interval.} & \parbox[t]{30mm}{Relative scale $r_1/a$ set from the static-quark potential.  Absolute scale $r_1$, including related uncertainty estimates, taken from \cite{Bazavov:2011aa}.}
\\[26.0ex]
\ifx\reducedapptables\undefined
\hline \\[-2.0ex]
\SLfnalmilcBpi & \cite{Lattice:2015tia} & 2+1 & \parbox[t]{20mm}{0.045, 0.06, 0.09, 0.12}   &
\parbox[t]{4.2cm}{Fit to HMrS$\chi$PT to remove light-quark discretization errors. Residual heavy-quark discretization errors estimated with power-counting. Total (stat + chiral extrap + HQ discretization + $g_{B^*B\pi}$) error estimated to be 3.1\% for $f_+$ and 3.8\% for $f_0$ at $q^2=20~{\rm GeV}^2$.} & \parbox[t]{30mm}{Relative scale $r_1/a$ set from the static-quark potential.  Absolute scale $r_1$, including related uncertainty estimates, taken from \cite{Bazavov:2011aa}.}
\\[26.0ex] \hline \\[-2.0ex]
\parbox[t]{20mm}{\SLLambdabp} & \cite{Detmold:2015aaa} & 2+1 & \parbox[t]{20mm}{0.0849(12), 0.1119(17)}   &
\parbox[t]{4.2cm}{Joint chiral-continuum extrapolation, combined with fit to $q^2$-dependence of form factors in a ``modified'' $z$-expansion. Systematics estimated by varying fit form and $\cO(a)$ improvement parameter values.} & \parbox[t]{30mm}{Set from $\Upsilon(2S)$--$\Upsilon(1S)$ splitting, cf.~\cite{Meinel:2010pv}.}
\\[21.0ex] \hline \\[-2.0ex]
\SLrbcukqcdBpi & \cite{Flynn:2015mha} & 2+1 & \parbox[t]{20mm}{0.086,0.11}   &
\parbox[t]{4.2cm}{Joint chiral-continuum extrapolation using $SU(2)$ hard-pion HM$\chi$PT.
Systematic uncertainty estimated by varying fit ansatz and form of coefficients, as well as implementing different
cuts on data; ranges from 5.0\% to 10.9\% for $B\to\pi$ form factors, and 2.5\% to 5.1\% for $B_s\to K$.
Light-quark and gluon discretization errors estimated at 1.1\% and 1.3\%, respectively.} & \parbox[t]{30mm}{Scale implicitly set in the light-quark sector using the $\Omega^-$ mass, cf.~\cite{Aoki:2010dy}.}
\\[35.0ex] \hline \\[-2.0ex]
\SLhpqcdBsK & \cite{Bouchard:2014ypa} & 2+1 & \parbox[t]{20mm}{0.09,0.12}   &
\parbox[t]{4.2cm}{Combined chiral-continuum extrapolation using hard-pion rHMS$\chi$PT. (No explicit estimate of discretization effects.)} & \parbox[t]{30mm}{Relative scale $r_1/a$ set from the static-quark potential. Absolute scale $r_1$
set to 0.3133(23)~fm.}
\\[12.0ex] \hline \\[-2.0ex]
HPQCD 06 & \cite{Dalgic:2006dt} & 2+1 & \parbox[t]{20mm}{0.09,0.12}  &
\parbox[t]{4.2cm}{Central values obtained from data at $a=0.12$~fm.  Discretization errors observed to be within the statistical error by comparison with data at $a=0.09$~fm.} &  \parbox[t]{30mm}{Relative scale $r_1/a$ set from the static-quark potential.  Absolute scale $r_1$ set through $\Upsilon$ $2S-1S$ splitting c.f.~HPQCD 05B~\cite{Gray:2005ur}.}
\\[18.0ex]
\fi
\hline\hline
\end{tabular*}
\caption{Continuum extrapolations/estimation of lattice artifacts in
  determinations of $B\to\pi\ell\nu$, $B_s\to K\ell\nu$, and $\Lambda_b\to p\ell\bar{\nu}$ form factors.}
}
\end{table}
\begin{table}[!ht]
{\footnotesize
\begin{tabular*}{\textwidth}{l @{\extracolsep{\fill}} c c c   l}
\hline\hline \\[-1.0ex]
{Collab.} & {Ref.} & {$\Nf$} & {$M_{\pi,\rm min}\,[\mev]$}  & {Description}  
\\[1.0ex] \hline \hline \\[-1.0ex]
\SLfnalmilcBsK & \cite{Bazavov:2019aom} & 2+1 & \parbox[t]{2cm}{255, 277, 456}
& \parbox[t]{5cm}{HMrS$\chi$PT expansion used at next-to-leading
order in $SU(2)$ and leading order in $1/M_B$, including next-to-next-to-leading-order
(NNLO) analytic and generic discretization terms. Hard kaons assumed to decouple. Systematic uncertainties
estimated by varying fit ansatz and data range. }
\\[18.0ex]
\ifx\reducedapptables\undefined
\hline \\[-2.0ex]
\SLfnalmilcBpi & \cite{Lattice:2015tia} & 2+1 & \parbox[t]{2cm}{330, 260, 280, 470}
& \parbox[t]{5cm}{Simultaneous chiral-continuum extrapolation and $q^2$ interpolation using NNLO $SU(2)$ hard-pion HMrS$\chi$PT.  Systematic error estimated by adding higher-order analytic terms and varying the $B^*$-$B$-$\pi$ coupling.}
\\[18.0ex] \hline \\[-2.0ex]
\parbox[t]{20mm}{\SLLambdabp} & \cite{Detmold:2015aaa} & 2+1 & \parbox[t]{2.5cm}{227, 245 \\ (valence-valence) \\ 263, 294 \\ (valence-sea)}
& \parbox[t]{5cm}{Joint chiral-continuum extrapolation, combined with fit to $q^2$-dependence of form factors in a ``modified'' $z$-expansion. Only analytic NLO terms $\propto (m_\pi^2-m_{\pi,{\rm phys}}^2)$ included in light-mass dependence. Systematic uncertainty estimated by repeating fit with added higher-order terms.}
\\[23.0ex] \hline \\[-2.0ex]
\SLrbcukqcdBpi & \cite{Flynn:2015mha} & 2+1 & \parbox[t]{2cm}{289, 329}
& \parbox[t]{5cm}{Joint chiral-continuum extrapolation using $SU(2)$ hard-pion HM$\chi$PT.
Systematic uncertainty estimated by varying fit ansatz and form of coefficients, as well as implementing different
cuts on data; ranges from 5.0\% to 10.9\% for $B\to\pi$ form factors, and 2.5\% to 5.1\% for $B_s\to K$.}
\\[23.0ex] \hline \\[-2.0ex]
\SLhpqcdBsK & \cite{Bouchard:2014ypa} & 2+1 & \parbox[t]{2cm}{295, 260}
& \parbox[t]{5cm}{Combined chiral-continuum extrapolation using hard-pion rHMS$\chi$PT. (No explicit estimate of extrapolation systematics.)}
\\[9.0ex] \hline \\[-2.0ex]
%
%
HPQCD 06 & \cite{Dalgic:2006dt} & 2+1 & \parbox[t]{2cm}{400, 440}
& \parbox[t]{5cm}{First interpolate data at fixed quark mass to fiducial values of $E_\pi$ using the Becirevic-Kaidalov and Ball-Zwicky ans\"atze, then extrapolate data at fixed $E_\pi$ to physical quark masses using $SU(3)$ rHMS$\chi$PT. Systematic error estimated by varying interpolation and extrapolation fit functions.}
\\[24.0ex]
%
\fi
\hline\hline
\end{tabular*}
\caption{Chiral extrapolation/minimum pion mass in
  determinations of $B\to\pi\ell\nu$, $B_s\to K\ell\nu$, and $\Lambda_b\to p\ell\bar{\nu}$ form factors.  For actions with multiple species of pions, masses quoted are the RMS pion masses.    The different $M_{\pi,\rm min}$ entries correspond to the different lattice spacings.}
}
\end{table}
\begin{table}[!ht]
{\footnotesize
\begin{tabular*}{\textwidth}{l @{\extracolsep{\fill}} c c c c l}
\hline\hline \\[-1.0ex]
Collab. & Ref. & $\Nf$ & $L$ [fm] & ${M_{\pi,\rm min}}L$ & Description 
\\[1.0ex] \hline \hline \\[-1.0ex]
\SLfnalmilcBsK & \cite{Bazavov:2019aom} & 2+1 & \parbox[t]{1.8cm}{3.8, 2.5/2.9/3.6/5.8, 2.9} & $\gtrsim 3.8$  &
\parbox[t]{5cm}{FV effects estimated by comparing infinite volume integrals with finite sums in HMrS$\chi$PT, found to be negligible.
}
\\[9.0ex]
\ifx\reducedapptables\undefined
\hline \\[-1.0ex]
\SLfnalmilcBpi & \cite{Lattice:2015tia} & 2+1 & \parbox[t]{1.8cm}{2.9, 2.9/3.4/3.8, 2.5/2.9/3.6/5.8, 2.4/2.9} & $\gtrsim 3.8$  &
\parbox[t]{5cm}{FV effects estimated by replacing infinite-volume chiral logs with sums over discrete momenta, found to be negligible.
}
\\[9.0ex] \hline \\[-1.0ex]
\parbox[t]{20mm}{\SLLambdabp} & \cite{Detmold:2015aaa} & 2+1 & \parbox[t]{1.8cm}{2.7, 2.7} & \parbox[t]{1.8cm}{$\gtrsim 3.6$ \\ (valence-sea pion)}  &
\parbox[t]{5cm}{FV effect estimated at 3\% from experience on $\chi$PT estimates of FV effects for heavy-baryon axial couplings.
}
\\[9.0ex] \hline \\[-1.0ex]
\SLrbcukqcdBpi & \cite{Flynn:2015mha} & 2+1 & \parbox[t]{1.8cm}{2.8, 2.6} & $4.0, 4.4$  &
\parbox[t]{5cm}{FV effects estimated by correction to chiral logs due to sums over discrete momenta;
quoted 0.3-0.5\% for $f_+$ and 0.4-0.7\% for $f_0$ for $B\to\pi$, and 0.2\% for $f_+$ and 0.1-0.2\% for $f_0$ for $B_s\to K$.
}
\\[15.0ex] \hline \\[-1.0ex]
\SLhpqcdBsK & \cite{Bouchard:2014ypa} & 2+1 & \parbox[t]{1.8cm}{2.5, 2.4/2.9} & $\gtrsim 3.8$  &
\parbox[t]{5cm}{FV effects estimated by shift of pion log, found to be negligible.
}
\\[4.0ex] \hline \\[-1.0ex]
%
%
HPQCD 06 & \cite{Dalgic:2006dt} & 2+1 & \parbox[t]{1.8cm}{2.4/2.9} & $\gtrsim 3.8$ & 
\parbox[t]{5cm}{No explicit estimate of FV error, but expected to be much smaller than other uncertainties.}
\\[7.0ex]
%
%
\fi
\hline\hline
\end{tabular*}
\caption{Finite-volume effects in determinations of $B\to\pi\ell\nu$, $B_s\to K\ell\nu$, and $\Lambda_b\to p\ell\bar{\nu}$ form factors.  Each $L$-entry corresponds to a different lattice
spacing, with multiple spatial volumes at some lattice spacings.  For actions with multiple species of pions, the lightest masses are quoted.}
}
\end{table}
\begin{table}[!ht]
{\footnotesize
\begin{tabular*}{\textwidth}{l @{\extracolsep{\fill}} c c c l}
\hline\hline \\[-1.0ex]
Collab. & Ref. & $\Nf$ & Ren. & Description 
\\[1.0ex] \hline \hline \\[-1.0ex]
\SLfnalmilcBsK & \cite{Bazavov:2019aom} & 2+1 & mNPR &
\parbox[t]{6cm}{Perturbative truncation error estimated at 1\% with size of 1-loop correction on next-to-finest ensemble.}
\\[7.0ex]
\ifx\reducedapptables\undefined
\hline \\[-1.0ex]
\SLfnalmilcBpi & \cite{Lattice:2015tia} & 2+1 & mNPR &
\parbox[t]{6cm}{Perturbative truncation error estimated at 1\% with size of 1-loop correction on next-to-finest ensemble.}
\\[7.0ex] \hline \\[-1.0ex]
\parbox[t]{20mm}{\SLLambdabp} & \cite{Detmold:2015aaa} & 2+1 & mNPR &
\parbox[t]{6cm}{Perturbative truncation error estimated at 1\% with size of 1-loop correction on next-to-finest ensemble.}
\\[7.0ex] \hline \\[-1.0ex]
\SLrbcukqcdBpi & \cite{Flynn:2015mha} & 2+1 & mNPR &
\parbox[t]{6cm}{Perturbative truncation error estimated as largest of power counting, effect from value of $\alpha_s$ used, numerical integration. Nonperturbative normalization of flavour-diagonal currents computed by fixing values of ratios of meson two-point functions to three-point functions with an extra current inversion, cf.~\cite{Christ:2014uea}}
\\[21.0ex] \hline \\[-1.0ex]
\SLhpqcdBsK & \cite{Bouchard:2014ypa} & 2+1 & mNPR &
\parbox[t]{6cm}{Currents matched using 1-loop HISQ lattice perturbation theory, omitting $\cO(\alpha_s\Lambda_{\rm QCD}/m_b$. Systematic uncertainty resulting from 1-loop matching and neglecting $\cO(\Lambda_{\rm QCD}^2/m_b^2$ terms estimated at 4\% from power counting.}
\\[16.0ex] \hline \\[-1.0ex]
%
%
HPQCD 06 & \cite{Dalgic:2006dt} & 2+1 & PT1$\ell$ &
\parbox[t]{6cm}{Currents included through $\cO(\alpha_S \Lambda_{\rm QCD}/M$, $\alpha_S/(aM)$, $\alpha_S\, a \Lambda_{\rm QCD})$. Perturbative truncation error estimated from power-counting.}
\\[10.0ex]
%
%
\fi
\hline\hline
\end{tabular*}
\caption{Operator renormalization in determinations of $B\to\pi\ell\nu$, $B_s\to K\ell\nu$, and $\Lambda_b\to p\ell\bar{\nu}$ form factors.}
}
\end{table}
\begin{table}[!ht]
{\footnotesize
\begin{tabular*}{\textwidth}{l @{\extracolsep{\fill}} c c c l}
\hline\hline \\[-1.0ex]
Collab. & Ref. & $\Nf$ & Action & Description 
\\[1.0ex] \hline \hline \\[-1.0ex]
\SLfnalmilcBsK & \cite{Bazavov:2019aom} & 2+1 & Fermilab &
\parbox[t]{8cm}{(See comments for continuum limit extrapolation.)}
\\[3.0ex]
\ifx\reducedapptables\undefined
\hline \\[-1.0ex]
\SLfnalmilcBpi & \cite{Lattice:2015tia} & 2+1 & Fermilab &
\parbox[t]{8cm}{Total statistical + chiral extrapolation + heavy-quark discretization + $g_{B^*B\pi}$ error estimated to be 3.1\% for $f_+$ and 3.8\% for $f_0$ at $q^2=20~{\rm GeV}^2$.}
\\[7.0ex] \hline \\[-1.0ex]
\parbox[t]{20mm}{\SLLambdabp} & \cite{Detmold:2015aaa} & 2+1 & Columbia RHQ &
\parbox[t]{8cm}{Discretization errors discussed as part of combined chiral-continuum-$q^2$ fit, stemming from $a^2|\mathbf{p}|^2$ terms.}
\\[4.0ex] \hline \\[-1.0ex]
\SLrbcukqcdBpi & \cite{Flynn:2015mha} & 2+1 & Columbia RHQ &
\parbox[t]{8cm}{Discretization errors estimated by power counting to be 1.8\% for $f_+$ and 1.7\% for $f_0$.}
\\[4.0ex] \hline \\[-1.0ex]
\SLhpqcdBsK & \cite{Bouchard:2014ypa} & 2+1 & NRQCD &
\parbox[t]{8cm}{Currents matched using 1-loop HISQ lattice perturbation theory, omitting $\cO(\alpha_s\Lambda_{\rm QCD}/m_b$. Systematic uncertainty resulting from 1-loop matching and neglecting $\cO(\Lambda_{\rm QCD}^2/m_b^2$ terms estimated at 4\% from power counting.}
\\[12.0ex] \hline \\[-1.0ex]
%
%
HPQCD 06 & \cite{Dalgic:2006dt} & 2+1 & NRQCD &
\parbox[t]{8cm}{Discretization errors in $f_+(q^2)$ estimated to be $\cO (\alpha_s (a \Lambda_{\rm QCD})^2) \sim  3\%$.  Relativistic errors estimated to be $\cO((\Lambda_{\rm QCD}/M)^2) \sim 1\%$.}
\\[7.0ex]
%
%
\fi
\hline\hline
\end{tabular*}
\caption{Heavy-quark treatment in determinations of $B\to\pi\ell\nu$, $B_s\to K\ell\nu$, and $\Lambda_b\to p\ell\bar{\nu}$ form factors.}
}
\end{table}

\FloatBarrier
\subsubsection{Form factors for rare decays of beauty hadrons}
\label{app:BtoK_Notes}
\FloatBarrier

\begin{table}[!ht]

{\footnotesize
\begin{tabular*}{\textwidth}{l c c c l l}
\hline\hline \\[-1.0ex]
Collab. & Ref. & $\Nf$ & $a$ [fm] & Continuum extrapolation & Scale setting 
\\[1.0ex] \hline \hline \\[-1.0ex]
\ifx\reducedapptables\undefined
Detmold 16 & \cite{Detmold:2016pkz} & 2+1 & \parbox[t]{20mm}{0.0849(12), 0.1119(17)}   &
\parbox[t]{4.2cm}{Joint chiral-continuum extrapolation, combined with fit to $q^2$-dependence of form factors in a ``modified'' $z$-expansion. Systematics estimated by varying fit form and $\cO(a)$ improvement parameter values. Contrary to Detmold 15 $\Lambda_b \rightarrow p$, no odd powers of spatial momenta are included in the fit ansatz, based on cubic symmetry.} & \parbox[t]{30mm}{Set from $\Upsilon(2S)$--$\Upsilon(1S)$ splitting, cf.~\cite{Meinel:2010pv}.}
\\[32.0ex] \hline \\[-2.0ex]
FNAL/MILC 15E & \cite{Bailey:2015nbd} & 2+1 & \parbox[t]{20mm}{0.045, 0.06, 0.09, 0.12}   &
\parbox[t]{4.2cm}{Fit to $SU(2)$ HMrS$\chi$PT for the combined chiral-continuum limit extrapolation.
Combined stat + chiral extrap + HQ discretization + $g_{B^*B\pi}$ error provided as a function of $q^2$ for each form factor; representative impact on $f_T$ quoted as 3.8\% at $q^2=20~{\rm GeV}^2$.} & \parbox[t]{30mm}{Relative scale $r_1/a$ set from the static-quark potential. Absolute scale $r_1$, including related uncertainty estimates, taken from \cite{Bazavov:2011aa}.}
\\[26.0ex] \hline \\[-2.0ex]
\fi
Meinel 20 & \cite{Meinel:2020owd} & 2+1 & \parbox[t]{20mm}{0.0828(3), 0.1106(3)}   & \parbox[t]{4.2cm}{Combined chiral-continuum extrapolation as part of the expansion of form factor shape in powers of $w-1$. Systematic uncertainty estimated by repeating fit with added higher-order terms.} & \parbox[t]{30mm}{Scale setting using $\Omega$ mass in Ref.~\cite{Blum:2014tka}.} 
\\[18.0ex]
\ifx\reducedapptables\undefined
 \hline \\[-2.0ex]
 FNAL/MILC 15D & \cite{Bailey:2015dka} & 2+1 & \parbox[t]{20mm}{0.045, 0.06, 0.09, 0.12}   &
\parbox[t]{4.2cm}{Fit to $SU(2)$ HMrS$\chi$PT for the combined chiral-continuum limit extrapolation.
Combined stat + chiral extrap + HQ discretization + $g_{B^*B\pi}$ error provided as a function of $q^2$ for each form factor, ranging between $\sim 1.4$\% and $\sim 2.8$\%.} & \parbox[t]{30mm}{Relative scale $r_1/a$ set from the static-quark potential. Absolute scale $r_1$, including related uncertainty estimates, taken from \cite{Bazavov:2011aa}.}
\\[23.0ex] \hline \\[-2.0ex]
HPQCD 13E & \cite{Bouchard:2013mia} & 2+1 & \parbox[t]{20mm}{0.09,0.12}   &
\parbox[t]{4.2cm}{Combined chiral-continuum extrapolation using rHMS$\chi$PT. Errors provided as a function of $q^2$, combined total ranging from $\sim 3$\% to $\sim 5$\% in data region.} & \parbox[t]{30mm}{Relative scale $r_1/a$ set from the static-quark potential. Absolute scale $r_1$
set to 0.3133(23)~fm.}
\\[15.0ex]
\fi
\hline\hline
\end{tabular*}
\caption{Continuum extrapolations/estimation of lattice artifacts in
  determinations of form factors for rare decays of beauty hadrons.}
}
\end{table}
\begin{table}[!ht]
{\footnotesize
\begin{tabular*}{\textwidth}{l @{\extracolsep{\fill}} c c c   l}
\hline\hline \\[-1.0ex]
{Collab.} & {Ref.} & {$\Nf$} & {$M_{\pi,\rm min}\,[\mev]$}  & {Description}  
\\[1.0ex] \hline \hline \\[-1.0ex]
Meinel 20 & \cite{Meinel:2020owd} & 2+1 &  \parbox[t]{2cm}{303, 340}
& \parbox[t]{5cm}{Combined chiral-continuum extrapolation as part of the expansion of form factor shape in powers of $w-1$. Systematic uncertainty estimated by repeating fit with added higher-order terms.}
\\[16.0ex]
\ifx\reducedapptables\undefined
\hline \\[-2.0ex]
FNAL/MILC 15D & \cite{Bailey:2015dka} & 2+1 & \parbox[t]{2cm}{330, 260, 280, 470}
& \parbox[t]{5cm}{Simultaneous chiral-continuum extrapolation and $q^2$ interpolation using $SU(2)$ HMrS$\chi$PT,
with a hard-kaon $\chi$PT treatment of high-energy kaons. Combined stat + chiral extrap + HQ discretization + $g_{B^*B\pi}$ error provided as a function of $q^2$ for each form factor, ranging between $\sim 1.4$\% and $\sim 2.8$\%.}
\\[23.0ex] \hline \\[-2.0ex]
HPQCD 13E & \cite{Bouchard:2013mia} & 2+1 & \parbox[t]{2cm}{295, 260}
& \parbox[t]{5cm}{Combined chiral-continuum extrapolation using rHMS$\chi$PT. Errors provided as a function of $q^2$, combined total ranging from $\sim 3$\% to $\sim 5$\% in data region.}
\\[12.0ex]
\hline \\[-2.0ex]
Detmold 16 & \cite{Detmold:2016pkz} & 2+1 &  \parbox[t]{2.5cm}{227, 245 \\ (valence-valence) \\ 263, 294 \\ (valence-sea)}
& \parbox[t]{5cm}{Joint chiral-continuum extrapolation, combined with fit to $q^2$-dependence of form factors in a ``modified'' $z$-expansion. Only analytic NLO terms $\propto (m_\pi^2-m_{\pi,{\rm phys}}^2)$ included in light-mass dependence. Systematic uncertainty estimated by repeating fit with added higher-order terms.}
\\[23.0ex] \hline \\[-2.0ex]
FNAL/MILC 15E & \cite{Bailey:2015nbd} & 2+1 & \parbox[t]{2cm}{330, 260, 280, 470}
& \parbox[t]{5cm}{Simultaneous chiral-continuum extrapolation and $q^2$ interpolation using $SU(2)$ HMrS$\chi$PT,
with a hard-pion $\chi$PT treatment of high-energy pions.}
\\[12.0ex] \hline \\[-2.0ex]
\fi
\hline\hline
\end{tabular*}
\caption{Chiral extrapolation/minimum pion mass in
  determinations of form factors for rare decays of beauty hadrons.  For actions with multiple species of pions, masses quoted are the RMS pion masses. The different $M_{\pi,\rm min}$ entries correspond to the different lattice spacings.}
}
\end{table}
\begin{table}[!ht]
{\footnotesize
\begin{tabular*}{\textwidth}{l @{\extracolsep{\fill}} c c c c l}
\hline\hline \\[-1.0ex]
Collab. & Ref. & $\Nf$ & $L$ [fm] & ${M_{\pi,\rm min}}L$ & Description 
\\[1.0ex] \hline \hline \\[-1.0ex]
Meinel 20 & \cite{Meinel:2020owd} & 2+1 & \parbox[t]{1.8cm}{2.7, 2.7}  & \parbox[t]{1.8cm}{4.1, 4.6} & \parbox[t]{4.8cm}{FV effects not quantified. Effects from unstable $\Lambda^*(1520)$ not quantified.}
\\[8.0ex]
\ifx\reducedapptables\undefined
\hline \\[-1.0ex]
FNAL/MILC 15D & \cite{Bailey:2015dka} & 2+1 & \parbox[t]{1.8cm}{2.9, 2.9/3.8, 2.5/2.9/3.6/5.8, 2.4/2.9} & $\gtrsim 3.8$  &
\parbox[t]{4.8cm}{FV effects estimated by replacing infinite-volume chiral logs with sums over discrete momenta, found to be negligible.
}
\\[9.0ex] \hline \\[-1.0ex]
HPQCD 13E & \cite{Bouchard:2013mia} & 2+1 & \parbox[t]{1.8cm}{2.5, 2.4/2.9} & $\gtrsim 3.8$  &
\parbox[t]{4.8cm}{FV effects included in combined chiral-continuum extrapolation.
}
\\[4.0ex] 
\hline \\[-2.0ex]
Detmold 16 & \cite{Detmold:2016pkz} & 2+1 & \parbox[t]{1.8cm}{2.7, 2.7} & \parbox[t]{1.8cm}{$\gtrsim 3.6$ \\ (valence-sea pion)}  &
\parbox[t]{4.8cm}{FV effect estimated at 3\% from experience on $\chi$PT estimates of FV effects for heavy-baryon axial couplings.
}
\\[9.0ex] \hline \\[-1.0ex]
\parbox[t]{2.7cm}{FNAL/MILC 15E} & \cite{Bailey:2015nbd} & 2+1 & \parbox[t]{1.8cm}{2.9, 2.9/3.8, 2.5/2.9/3.6/5.8, 2.4/2.9} & $\gtrsim 3.8$  &
\parbox[t]{4.8cm}{FV effects estimated by replacing infinite-volume chiral logs with sums over discrete momenta, found to be negligible.
}
\\[9.0ex] \hline \\[-1.0ex]
\fi
\hline\hline
\end{tabular*}
\caption{Finite-volume effects in determinations of form factors for rare decays of beauty hadrons.  Each $L$-entry corresponds to a different lattice
spacing, with multiple spatial volumes at some lattice spacings.  For actions with multiple species of pions, the lightest masses are quoted.}
}
\end{table}
\begin{table}[!ht]
{\footnotesize
\begin{tabular*}{\textwidth}{l @{\extracolsep{\fill}} c c c l}
\hline\hline \\[-1.0ex]
Collab. & Ref. & $\Nf$ & Ren. & Description 
\\[1.0ex] \hline \hline \\[-1.0ex]
Meinel 20 & \cite{Meinel:2020owd} & 2+1 & mNPR  & \parbox[t]{6cm}{Residual matching factors $\rho$ computed at one-loop for vector and axial-vector currents, but at tree-level only for tensor currents. A systematic uncertainty is assigned to $\rho_{T^{\mu\nu}}$ as the double of ${\rm max}(|\rho_{A^{\mu}}-1|,|\rho_{V^{\mu}}-1|)$.}
\\[16.0ex]
\ifx\reducedapptables\undefined
\hline \\[-1.0ex]
FNAL/MILC 15D & \cite{Bailey:2015dka} & 2+1 & mNPR &
\parbox[t]{6cm}{Perturbative truncation error estimated at 1\% for $f_+$ and $f_0$ and 2\% for $f_T$, using size of 1-loop correction on next-to-finest ensemble.}
\\[10.0ex] \hline \\[-1.0ex]
HPQCD 13E & \cite{Bouchard:2013mia} & 2+1 & mNPR &
\parbox[t]{6cm}{Currents matched using 1-loop massless-HISQ lattice perturbation theory. Associated systematic uncertainty dominates quoted 4\% uncertainty from matching, charm quenching, and electromagnetic and isospin-breaking effects.}
\\[16.0ex]
\hline \\[-2.0ex]
Detmold 16 & \cite{Detmold:2016pkz} & 2+1 & mNPR &
\parbox[t]{6cm}{Same procedure as in Detmold 15 $\Lambda_p \rightarrow p$ for vector and axial currents. For tensor currents, the residual renormalization factor is set to its tree-level value $\rho_{T^{\mu\nu}}=1$. A systematic uncertainty is assigned as the double of ${\rm max}[1-\rho_{A^{\mu}},1-\rho_{V^{\mu}}]$, using the known 1-loop values of the residual matchings for the vector and axial currents.}
\\[21.0ex] \hline \\[-1.0ex]
FNAL/MILC 15E & \cite{Bailey:2015nbd} & 2+1 & mNPR &
\parbox[t]{6cm}{Perturbative truncation error on $f_T$ estimated at 0.7\% from the ratio of singlet renormalization constants and 2.0\% from the residual renormalization.}
\\[10.0ex] \hline \\[-1.0ex]
\fi
\hline\hline
\end{tabular*}
\caption{Operator renormalization in determinations of form factors for rare decays of beauty hadrons.}
}
\end{table}
\begin{table}[!ht]
{\footnotesize
\begin{tabular*}{\textwidth}{l @{\extracolsep{\fill}} c c c l}
\hline\hline \\[-1.0ex]
Collab. & Ref. & $\Nf$ & Action & Description 
\\[1.0ex] \hline \hline \\[-1.0ex]
Meinel 20 & \cite{Meinel:2020owd} & 2+1 & Columbia RHQ & \parbox[t]{7.8cm}{Discretization errors discussed as part of combined chiral-continuum-$w$ fit. Higher-order fit also includes $\mathcal{O}(\alpha_s a |\mathbf{p}|)$ terms to account for missing radiative corrections to $\mathcal{O}(a)$ improvement of the currents.}
\\[10.0ex]
\ifx\reducedapptables\undefined
\hline \\[-1.0ex]
FNAL/MILC 15D & \cite{Bailey:2015dka} & 2+1 & Fermilab &
\parbox[t]{7.8cm}{Combined stat + chiral extrap + HQ discretization + $g_{B^*B\pi}$ error provided as a function of $q^2$ for each form factor, ranging between $\sim 1.4$\% and $\sim 2.8$\%.}
\\[7.0ex] \hline \\[-1.0ex]
HPQCD 13E & \cite{Bouchard:2013mia} & 2+1 & NRQCD &
\parbox[t]{7.8cm}{Currents matched using 1-loop massless-HISQ lattice perturbation theory. Associated systematic uncertainty dominates quoted 4\% uncertainty from matching, charm quenching, and electromagnetic and isospin-breaking effects.}
\\[12.0ex]
\hline \\[-2.0ex]
Detmold 16 & \cite{Detmold:2016pkz} & 2+1 & Columbia RHQ &
\parbox[t]{7.8cm}{Discretization errors discussed as part of combined chiral-continuum-$q^2$ fit, stemming from $a^2|\mathbf{p}|^2$ terms.}
\\[4.0ex] \hline \\[-1.0ex]
\parbox[t]{2.7cm}{FNAL/MILC 15E} & \cite{Bailey:2015nbd} & 2+1 & Fermilab &
\parbox[t]{7.8cm}{Combined stat + chiral extrap + HQ discretization + $g_{B^*B\pi}$ error provided as a function of $q^2$ for each form factor; representative impact on $f_T$ quoted as 3.8\% at $q^2=20~{\rm GeV}^2$.}
\\[10.0ex] \hline \\[-1.0ex]
\fi
\hline\hline
\end{tabular*}
\caption{Heavy-quark treatment in determinations of form factors for rare decays of beauty hadrons.}
}
\end{table}

\clearpage
\subsubsection{Form factors entering determinations of $|V_{cb}|$ ($B_{(s)} \to D_{(s)}^{(*)}\ell\nu$,  $\Lambda_b \to \Lambda_c^{(*)} \ell \bar{\nu}$) and $R(D_{(s)})$}
\label{app:BtoD_Notes}
\vspace{-0.48cm}
\begin{table}[!ht]
{\footnotesize
\begin{tabular*}{\textwidth}{l @{\extracolsep{\fill}} c c c l l}
\hline\hline \\[-1.0ex]
Collab. & Ref. & $\Nf$ & $a$ [fm] & Continuum extrapolation & Scale setting 
\\[1.0ex] \hline \hline \\[-1.0ex]
Meinel 21 & \cite{Meinel:2021rbm} & 2+1 & \parbox[t]{1.5cm}{0.0828(3), 0.1106(3)}   & \parbox[t]{4.2cm}{Combined chiral-continuum extrapolation as part of the expansion of form factor shape in powers of $w-1$. Systematics estimated by varying fit form.} & \parbox[t]{35mm}{Scale setting using $\Omega$ mass in Ref.~\cite{Blum:2014tka}.} 
\\[16.0ex] \hline \\[-2.0ex]
\parbox[t]{2.0cm}{HPQCD 19} & \cite{McLean:2019qcx} & 2+1+1 & \parbox[t]{1.5cm}{0.044, 0.059, $\approx$0.09} &
\parbox[t]{4.5cm}{Combined chiral-continuum extrapolation. Fractional error from continuum limit and
extrapolation to physical $b$ mass at zero recoil quoted as 1.20\%.}
& \parbox[t]{35mm}{Scale setting from $f_\pi$ via the flow quantity $w_0$~\cite{Dowdall:2013rya,Chakraborty:2014aca,Chakraborty:2016mwy}.} 
\\[14.0ex] \hline \\[-2.0ex]
\parbox[t]{2.0cm}{HPQCD 19B} & \cite{McLean:2019sds} & 2+1+1 & \parbox[t]{1.5cm}{0.044, 0.059, $\approx$0.09} &
\parbox[t]{4.5cm}{Combined chiral-continuum extrapolation. Fractional error from continuum limit and
extrapolation to physical $b$ mass at zero recoil quoted as 0.73\% and 0.69\%, respectively.}
& \parbox[t]{35mm}{Scale setting from $f_\pi$ via the flow quantity $w_0$~\cite{Dowdall:2013rya,Chakraborty:2014aca,Chakraborty:2016mwy}.} 
\\[16ex]
\ifx\reducedapptables\undefined
\hline
\parbox[t]{2.0cm}{HPQCD 17B} & \cite{Harrison:2017fmw} & 2+1+1 & \parbox[t]{1.5cm}{0.09, 0.12, 0.15} &
\parbox[t]{4.5cm}{Combined chiral-continuum extrapolation. $\mathcal{O}(a^2)$ uncertainty on the two relevant form factors at zero recoil estimated to be 0.7\% and 1.4\%.}
& \parbox[t]{35mm}{Determined from $\Upsilon(2S-1S)$ splitting in~\cite{Dowdall:2011wh}.} 
\\[12ex]  \hline
\parbox[t]{2.6cm}{HPQCD 17} & \cite{Monahan:2017uby}
 & 2+1 & \parbox[t]{1.5cm}{0.09, 0.12} &
\parbox[t]{4.2cm}{Combined chiral-continuum extrapolation as part of modified $z$-expansion of form factors, which also includes uncertainty related to matching of NRQCD and relativistic currents.}
& \parbox[t]{30mm}{Implicitly set from $r_1$.} 
\\[18ex] \hline \\[-2.0ex]
\parbox[t]{26mm}{Datta 17} & \cite{Datta:2017aue} & 2+1 & \parbox[t]{1.5cm}{0.0849(12), 0.1119(17)}   &
\parbox[t]{4.2cm}{Joint chiral-continuum extrapolation, combined with fit to $q^2$-dependence of form factors in a ``modified'' $z$-expansion. Systematics estimated by varying fit form and $\cO(a)$ improvement parameter values.} & \parbox[t]{30mm}{Set from $\Upsilon(2S)$--$\Upsilon(1S)$ splitting, cf.~\cite{Meinel:2010pv}.}
\\[21.0ex] 
\hline \\[-2.0ex]
\parbox[t]{26mm}{\SLfnalmilcBD} & \cite{Lattice:2015rga} & 2+1 & \parbox[t]{1.5cm}{ 0.045, 0.06, 0.09, 0.12} &
\parbox[t]{4.2cm}{Combined chiral-continuum extrapolation using HMrS$\chi$PT. Form factors fitted to NLO $\chi$PT, with chiral logs taken from staggered version of the Chow-Wise result, modified to include taste-breaking terms. $\cO(a^2)$ terms introduced based on power-counting arguments. Total uncertainty estimated at 0.6\% for $f_+$ and 0.5\% for $f_0$ for the largest recoil.}
& \parbox[t]{30mm}{Relative scale $r_1/a$ set from the static-quark potential.  Absolute scale $r_1$, including related uncertainty estimates, taken from \cite{Bazavov:2011aa}. Uncertainty related to scale setting estimated at 0.2\%.} 
\\[35ex] \hline \\[-2.0ex]
\parbox[t]{26mm}{\SLLambdabc} & \cite{Detmold:2015aaa} & 2+1 & \parbox[t]{1.5cm}{0.0849(12), 0.1119(17)}   &
\parbox[t]{4.2cm}{Joint chiral-continuum extrapolation, combined with fit to $q^2$-dependence of form factors in a ``modified'' $z$-expansion. Systematics estimated by varying fit form and $\cO(a)$ improvement parameter values.} & \parbox[t]{30mm}{Set from $\Upsilon(2S)$--$\Upsilon(1S)$ splitting, cf.~\cite{Meinel:2010pv}.}
\\[21.0ex] \hline \\[-2.0ex]
\parbox[t]{26mm}{\SLfnalmilcBDstar} & \cite{Bailey:2014tva} & 2+1 & \parbox[t]{1.5cm}{ 0.045, 0.06, 0.09, 0.12, 0.15} &
\parbox[t]{4.2cm}{Combined chiral-continuum extrapolation using HMrS$\chi$PT. Total uncertainty quoted at 0.5\%.}
& \parbox[t]{30mm}{Relative scale $r_1/a$ set from the static-quark potential.  Absolute scale $r_1$, including related uncertainty estimates, taken from \cite{Bazavov:2011aa}. Uncertainty related to scale setting estimated at 0.1\%.} 
\\[27.0ex] 
\fi
\hline\hline
\end{tabular*}
\caption{Continuum extrapolations/estimation of lattice artifacts in
    $N_f=2+1$ determinations of $B_{(s)} \to D_{(s)}^{(*)}\ell\nu$ and $\Lambda_b \to \Lambda_c^{(*)} \ell \bar{\nu}$ form factors, and of $R(D_{(s)})$.}
}
\end{table}

\ifx\reducedapptables\undefined

\begin{table}[!ht]
{\footnotesize
\begin{tabular*}{\textwidth}{l @{\extracolsep{\fill}} c c c l l}
\hline\hline \\[-1.0ex]
Collab. & Ref. & $\Nf$ & $a$ [fm] & Continuum extrapolation & Scale setting 
\\[1.0ex] \hline \hline \\[-1.0ex]
\parbox[t]{26mm}{Atoui  13} & \cite{Atoui:2013zza} & 2& \parbox[t]{20mm}{0.054, 0.067, 0.085, 0.098} &
\parbox[t]{4.2cm}{Combined continuum and chiral extrapolation, with linear terms in $a^2$ and $m_{\rm sea}$.
No dependence on $a$ or $m_{\rm sea}$ observed within errors. Stability of results vs.~fits with no $m_{\rm sea}$-dependence checked.}
& \parbox[t]{30mm}{Scale set through $F_\pi$.} 
\\[18ex]
\hline\hline
\end{tabular*}
\caption{Continuum extrapolations/estimation of lattice artifacts in
    $N_f=2$ determinations of $B_{(s)} \to D_{(s)}^{(*)}\ell\nu$ and $\Lambda_b \to \Lambda_c^{(*)} \ell \bar{\nu}$ form factors, and of $R(D_{(s)})$.}
}
\end{table}

\fi

\begin{table}[!ht]
{\footnotesize
\begin{tabular*}{\textwidth}{l @{\extracolsep{\fill}} c c c  l}
\hline\hline \\[-1.0ex]
{Collab.} & {Ref.} & {$\Nf$} & {$M_{\pi,\rm min}\,[\mev]$}  & {Description}  
\\[1.0ex] \hline \hline \\[-1.0ex]
Meinel 21 & \cite{Meinel:2021rbm} & 2+1 & \parbox[t]{2cm}{303, 340}
& \parbox[t]{5cm}{Combined chiral-continuum extrapolation as part of the expansion of form factor shape in powers of $w-1$. Systematic uncertainty estimated by repeating fit with added higher-order terms.}
\\[16.0ex]
\hline \\[-2.0ex]
\parbox[t]{26mm}{HPQCD 19} & \cite{McLean:2019qcx} & 2+1+1 & \parbox[t]{2cm}{315, 329, 129 \cite{Bazavov:2017lyh}}
 & \parbox[t]{5cm}{Combined chiral-continuum extrapolation using rS$\chi$PT. No specific uncertainty coming from chiral extrapolation quoted.}
\\[11.0ex]
\hline \\[-2.0ex]
\parbox[t]{26mm}{HPQCD 19B} & \cite{McLean:2019sds} & 2+1+1 & \parbox[t]{2cm}{315, 329, 129 \cite{Bazavov:2017lyh}}
 & \parbox[t]{5cm}{Combined chiral-continuum extrapolation using rS$\chi$PT. No specific uncertainty coming from chiral extrapolation quoted.}
\\[10.0ex]
\ifx\reducedapptables\undefined
\hline
\parbox[t]{26mm}{HPQCD 17B} & \cite{Harrison:2017fmw} & 2+1+1 & \parbox[t]{2cm}{130, 133, 130}
 & \parbox[t]{5cm}{Combined chiral-continuum extrapolation using rS$\chi$PT. No specific uncertainty coming from chiral extrapolation quoted.}
\\[10.0ex] \hline
\parbox[t]{26mm}{HPQCD 17} & \cite{Monahan:2017uby}  & 2+1 & \parbox[t]{2cm}{295, 260}
 & \parbox[t]{5cm}{Combined chiral-continuum extrapolation as part of modified $z$-expansion of form factors. Systematic uncertainties at $q^2=0$ estimated to 0.80\% ($B_s \to D_s$) and 1.14\% ($B \to D$, with input from hard-pion $\chi$PT).}
\\[18.0ex] \hline \\[-2.0ex]
\parbox[t]{26mm}{Datta 17} & \cite{Datta:2017aue} & 2+1 & \parbox[t]{2.5cm}{227, 245 \\ (valence-valence) \\ 263, 294 \\ (valence-sea)}
& \parbox[t]{5cm}{Joint chiral-continuum extrapolation, combined with fit to $q^2$-dependence of form factors in a ``modified'' $z$-expansion. Only analytic $NLO$ terms $\propto (m_\pi^2-m_{\pi,{\rm phys}}^2)$ included in light-mass dependence. Systematic uncertainty estimated by repeating fit with added higher-order terms.}
\\[23.0ex] 
\hline \\[-2.0ex]
\parbox[t]{26mm}{\SLfnalmilcBD} & \cite{Lattice:2015rga} & 2+1 & \parbox[t]{2cm}{330, 260, 280, 470}
 & \parbox[t]{5cm}{Combined chiral-continuum extrapolation using HMrS$\chi$PT. Form factors fitted to NLO $\chi$PT, with chiral logs taken from staggered version of the Chow-Wise result, modified to include taste-breaking terms. $\cO(a^2)$ terms introduced based on power-counting arguments. Total uncertainty estimated at 0.6\% for $f_+$ and 0.5\% for $f_0$ for the largest recoil.}
\\[29.0ex] \hline \\[-2.0ex]
\parbox[t]{26mm}{\SLLambdabc} & \cite{Detmold:2015aaa} & 2+1 & \parbox[t]{2.5cm}{227, 245 \\ (valence-valence) \\ 263, 294 \\ (valence-sea)}
& \parbox[t]{5cm}{Joint chiral-continuum extrapolation, combined with fit to $q^2$-dependence of form factors in a ``modified'' $z$-expansion. Only analytic $NLO$ terms $\propto (m_\pi^2-m_{\pi,{\rm phys}}^2)$ included in light-mass dependence. Systematic uncertainty estimated by repeating fit with added higher-order terms.}
\\[23.0ex] \hline \\[-2.0ex]
\parbox[t]{26mm}{\SLfnalmilcBDstar} & \cite{Bailey:2014tva} & 2+1 & \parbox[t]{2cm}{330, 260, 280, 470, 590}
 & \parbox[t]{5cm}{Combined chiral-continuum extrapolation using HMrS$\chi$PT.  Systematic errors estimated by adding higher-order analytic terms and varying the $D^*$-$D$-$\pi$ coupling. Total uncertainty quoted at 0.5\%.}
\\[15.0ex] 
\fi
\hline\hline
\end{tabular*}
\caption{Chiral extrapolation/minimum pion mass in $N_f=2+1$
  determinations of $B_{(s)} \to D_{(s)}^{(*)}\ell\nu$ and $\Lambda_b \to \Lambda_c^{(*)} \ell \bar{\nu}$ form factors, and of $R(D_{(s)})$.  For actions with multiple species of pions, masses quoted are the RMS pion masses for $N_f=2+1$ and the Goldstone mode mass for $N_f=2+1+1$.  The different $M_{\pi,\rm min}$ entries correspond to the different lattice spacings.}
}
\end{table}

\ifx\reducedapptables\undefined

\begin{table}[!ht]
{\footnotesize
\begin{tabular*}{\textwidth}{l @{\extracolsep{\fill}} c c c  l}
\hline\hline \\[-1.0ex]
{Collab.} & {Ref.} & {$\Nf$} & {$M_{\pi,\rm min}\,[\mev]$}  & {Description}  
\\[1.0ex] \hline \hline \\[-1.0ex]
Atoui  13 & \cite{Atoui:2013zza} & 2 & \parbox[t]{2cm}{270, 300, 270, 410}
 & \parbox[t]{5cm}{Combined continuum and chiral extrapolation, with linear terms in $a^2$ and $m_{\rm sea}$.
No dependence on $a$ or $m_{\rm sea}$ observed within errors. Stability of results vs.~fits with no $m_{\rm sea}$
dependence checked.}
\\[15.0ex]
\hline\hline
\end{tabular*}
\caption{Chiral extrapolation/minimum pion mass in $N_f=2$
  determinations of $B_{(s)} \to D_{(s)}^{(*)}\ell\nu$ and $\Lambda_b \to \Lambda_c^{(*)} \ell \bar{\nu}$ form factors, and of $R(D_{(s)})$.  For actions with multiple species of pions, masses quoted are the RMS pion masses.  The different $M_{\pi,\rm min}$ entries correspond to the different lattice spacings.}
}
\end{table}

\fi
\begin{table}[!ht]
{\footnotesize
\begin{tabular*}{\textwidth}{l @{\extracolsep{\fill}} c c c c l}
\hline\hline \\[-1.0ex]
Collab. & Ref. & $\Nf$ & $L$ [fm] & ${M_{\pi,\rm min}}L$ & Description 
\\[1.0ex] \hline \hline \\[-1.0ex]
Meinel 21 & \cite{Meinel:2021rbm} & 2+1 & \parbox[t]{1.8cm}{2.7, 2.7}  & \parbox[t]{1.8cm}{4.1, 4.6} & \parbox[t]{3.5cm}{FV effects not quantified. Effects from unstable $\Lambda_c^*$ not quantified.}
\\[10.0ex] \hline \\[-2.0ex]
\parbox[t]{26mm}{HPQCD 19} & \cite{McLean:2019qcx} & 2+1+1 & \parbox[t]{1.8cm}{2.73, 2.72, 2.81/5.62} & \parbox[t]{1.8cm}{4.3, 4.5, 3.7/4.5, 3.9/4.5, 3.3/3.8}  &
\parbox[t]{3.5cm}{FV effects neglected. This includes the effect of frozen topology in the finest ensemble.}
\\[10.0ex] \hline \\[-2.0ex]
\parbox[t]{26mm}{HPQCD 19B} & \cite{McLean:2019sds} & 2+1+1 & \parbox[t]{1.8cm}{2.73, 2.72, 2.81/5.62} & \parbox[t]{1.8cm}{4.3, 4.5, 3.7/4.5, 3.9/4.5, 3.3/3.8}  &
\parbox[t]{3.5cm}{FV effects neglected. This includes the effect of frozen topology in the finest ensemble.}
\\[10.0ex]
\ifx\reducedapptables\undefined
\hline \\[-1.0ex]
\parbox[t]{26mm}{HPQCD 17B} & \cite{Harrison:2017fmw} & 2+1+1 & \parbox[t]{1.8cm}{2.4/3.7/4.8, 3.0/3.9/5.8, 3.0/5.6} & $\gtrsim 3.2$  &
\parbox[t]{3.5cm}{FV effects estimated and subtracted using rs$\chi$PT formulae. No associated uncertainty quoted.}
\\[10.0ex] \hline \\[-1.0ex]
\parbox[t]{26mm}{HPQCD 17} & \cite{Monahan:2017uby} & 2+1 & \parbox[t]{1.8cm}{2.5, 2.4/2.9} & $\gtrsim 3.8$  &
\parbox[t]{3.5cm}{FV effects estimated to be below $0.01\%$.}
\\[4.0ex] \hline \\[-1.0ex]
\parbox[t]{26mm}{Datta 17} & \cite{Datta:2017aue} & 2+1 & \parbox[t]{1.8cm}{2.7, 2.7} & \parbox[t]{1.8cm}{$\gtrsim 3.6$ \\ (valence-sea pion)}  &
\parbox[t]{3.5cm}{No explicit discussion of finite-volume effects. Analysis identical to that of vector and axial form factors in \SLLambdabc.}
\\[15.0ex] 
\hline \\[-1.0ex]
\parbox[t]{26mm}{\SLfnalmilcBD} & \cite{Lattice:2015rga} & 2+1 & \parbox[t]{1.8cm}{2.9, 2.9--3.8, 2.5--5.8, 2.4/2.9} & $\gtrsim 3.8$ & 
\parbox[t]{3.5cm}{ FV error estimated to be negligible in \cite{Bailey:2012rr}.}
\\[7.0ex] \hline \\[-1.0ex]
\parbox[t]{26mm}{\SLLambdabc} & \cite{Detmold:2015aaa} & 2+1 & \parbox[t]{1.8cm}{2.7, 2.7} & \parbox[t]{1.8cm}{$\gtrsim 3.6$ \\ (valence-sea pion)}  &
\parbox[t]{3.5cm}{FV effect estimated at 1.5\% from experience on $\chi$PT estimates of FV effects for heavy-baryon axial couplings.
}
\\[13.0ex] \hline \\[-1.0ex]
\parbox[t]{26mm}{\SLfnalmilcBDstar} & \cite{Bailey:2014tva} & 2+1 & \parbox[t]{1.8cm}{2.9, 2.9--3.8, 2.4--5.5, 2.4/2.9, 2.4} & $\gtrsim 3.8$ & 
\parbox[t]{3.5cm}{ FV error estimated to be negligible.}
\\[7.0ex] \hline \\[-1.0ex]
Atoui  13 & \cite{Atoui:2013zza} & 2 & \parbox[t]{1.8cm}{1.7/2.6, 2.1, 2.0/2.7, 2.4} & $\gtrsim 3.6$ & 
\parbox[t]{3.5cm}{No volume dependence observed within errors.}
\\[4.0ex]
\fi
\hline\hline
\end{tabular*}
\caption{Finite-volume effects in determinations of $B_{(s)} \to D_{(s)}^{(*)}\ell\nu$ and $\Lambda_b \to \Lambda_c^{(*)} \ell \bar{\nu}$ form factors, and of $R(D_{(s)})$.  Each $L$-entry corresponds to a different lattice spacing, with multiple spatial volumes at some lattice spacings.  For actions with multiple species of pions, the lightest pion masses are quoted.}
}
\end{table}
\begin{table}[!ht]
{\footnotesize
\begin{tabular*}{\textwidth}{l @{\extracolsep{\fill}} c c c l}
\hline\hline \\[-1.0ex]
Collab. & Ref. & $\Nf$ & Ren. & Description 
\\[1.0ex] \hline \hline \\[-1.0ex]
Meinel 21 & \cite{Meinel:2021rbm} & 2+1 & mNPR  & \parbox[t]{6cm}{Residual matching factors $\rho$ computed at one-loop for vector and axial-vector currents, but at tree-level only for tensor currents. A systematic uncertainty is assigned to $\rho_{T^{\mu\nu}}$ as the double of ${\rm max}(|\rho_{A^{\mu}}-1|,|\rho_{V^{\mu}}-1|)$.}
\\[16.0ex] \hline \\[-2.0ex]
\parbox[t]{26mm}{HPQCD 19} & \cite{McLean:2019qcx} & 2+1+1 & NP &
\parbox[t]{6cm}{Currents normalized non-perturbatively by imposing that $f_+(0)=f_0(0)=1$
for transitions involving initial and final mesons with identical masses.}
\\[11.0ex] \hline \\[-2.0ex]
\parbox[t]{26mm}{HPQCD 19B} & \cite{McLean:2019sds} & 2+1+1 & NP &
\parbox[t]{6cm}{Currents renormalized nonperturbatively using PCAC relation.}
\\[5.0ex]
\ifx\reducedapptables\undefined
\hline \\[-1.0ex]
\parbox[t]{26mm}{HPQCD 17B} & \cite{Harrison:2017fmw} & 2+1+1 & 1-loop &
\parbox[t]{6cm}{1-loop matching of currents taken from~\cite{Monahan:2012dq}.}
\\[4.0ex] \hline \\[-1.0ex]
\parbox[t]{26mm}{HPQCD 17} & \cite{Monahan:2017uby} & 2+1 & 1-loop &
\parbox[t]{6cm}{1-loop matching of currents taken from~\cite{Monahan:2012dq}.}
\\[4.0ex] \hline \\[-1.0ex]
\parbox[t]{26mm}{Datta 17} & \cite{Datta:2017aue} & 2+1 & (incomplete) &
\parbox[t]{6cm}{mNPR procedure followed, fixing the residual renormalization factor to its tree-level value $\rho_{T^{\mu\nu}}=1$. A systematic uncertainty is assigned as the double of ${\rm max}[1-\rho_{A^{\mu}},1-\rho_{V^{\mu}}]$, using the known 1-loop values of the residual matchings for the vector and axial currents.}
\\[18.0ex] 
\hline \\[-1.0ex]
\parbox[t]{26mm}{\SLfnalmilcBD} & \cite{Lattice:2015rga} & 2+1 & mNPR &
\parbox[t]{6cm}{Form factors extracted from ratios of correlators that renormalize with
ratios of current normalizations, computed at 1-loop in perturbation theory.
Dependence of renormalization factor on recoil parameter $w$ neglected.
Systematic uncertainty due to perturbative truncation and $w$-dependence estimated
by power counting to 0.7\%.}
\\[21.0ex] \hline \\[-1.0ex]
\parbox[t]{26mm}{\SLLambdabc} & \cite{Detmold:2015aaa} & 2+1 & mNPR &
\parbox[t]{6cm}{Perturbative truncation error estimated at 1\% with size of 1-loop correction on next-to-finest ensemble.}
\\[7.0ex] \hline \\[-1.0ex]
\parbox[t]{26mm}{\SLfnalmilcBDstar} & \cite{Bailey:2014tva} & 2+1 & mNPR &
\parbox[t]{6cm}{Majority of current renormalization factor cancels in double ratio of lattice correlation functions.  Remaining correction calculated with 1-loop tadpole-improved lattice perturbation theory. Systematic uncertainty estimated at 0.4\%.}
\\[15.0ex] \hline \\[-1.0ex]
Atoui  13 & \cite{Atoui:2013zza} & 2 & --- &
\parbox[t]{6cm}{Observables obtained from ratios that do not require renormalization. Checks performed by comparing with results coming from currents that are renormalized separately with nonperturbative $Z_{\rm V}$.}
\\[12.0ex]
\fi
\hline\hline
\end{tabular*}
\caption{Operator renormalization in determinations of $B_{(s)} \to D_{(s)}^{(*)}\ell\nu$ and $\Lambda_b \to \Lambda_c^{(*)} \ell \bar{\nu}$ form factors, and of $R(D_{(s)})$.}
}
\end{table}
\begin{table}[!ht]
{\footnotesize
\begin{tabular*}{\textwidth}{l @{\extracolsep{\fill}} c c c l}
\hline\hline \\[-1.0ex]
Collab. & Ref. & $\Nf$ & Action & Description 
\\[1.0ex] \hline \hline \\[-1.0ex]
Meinel 21 & \cite{Meinel:2021rbm} & 2+1 & \parbox[t]{30mm}{Columbia RHQ for both the $b$ and $c$ quarks.} & \parbox[t]{6cm}{Discretization errors discussed as part of combined chiral-continuum-$w$ fit. Higher-order fit also includes $\mathcal{O}(\alpha_s a |\mathbf{p}|)$ terms to account for missing radiative corrections to $\mathcal{O}(a)$ improvement of the currents.}
\\[12.0ex] \hline \\[-2.0ex]
\parbox[t]{26mm}{HPQCD 19} & \cite{McLean:2019qcx} & 2+1+1 & \parbox[t]{30mm}{HISQ for both the $b$ and $c$ quarks.} &
\parbox[t]{6cm}{Values of bare heavy quark masses up to $am_h=0.8$. Fractional error from continuum limit and
extrapolation to physical $b$ mass at zero recoil quoted as 1.20\%.}
\\[10.0ex] \hline \\[-2.0ex]
\parbox[t]{26mm}{HPQCD 19B} & \cite{McLean:2019sds} & 2+1+1 & \parbox[t]{30mm}{HISQ for both the $b$ and $c$ quarks.} &
\parbox[t]{6cm}{Values of bare heavy-quark masses up to $am_h=0.8$. Fractional error from
extrapolation to physical $b$ mass quoted as 0.69\%.}
\\[7.0ex]
\ifx\reducedapptables\undefined
\hline \\[-1.0ex]
\parbox[t]{26mm}{HPQCD 17B} & \cite{Harrison:2017fmw} & 2+1+1 & \parbox[t]{30mm}{NRQCD for b quark,\\ HISQ for c quark} &
\parbox[t]{6cm}{$\mathcal{O}(\alpha_s^2)$ uncertainty quoted as 2.1\% and 2.5\% for the two
relevant form factors; $\mathcal{O}(\alpha_s\Lambda_{\rm QCD}/m_b,(\Lambda_{\rm QCD}/m_b)^2)$ errors quoted as 0.9\% and 0.8\% in both cases.}
\\[10.0ex] \hline \\[-1.0ex]
\parbox[t]{26mm}{HPQCD 17} & \cite{Monahan:2017uby} & 2+1 & \parbox[t]{30mm}{NRQCD for b quark,\\ HISQ for c quark} &
\parbox[t]{6cm}{Discretization errors estimated 
via power counting to be 2.47\% ($B_s \to D_s$) and 2.59\% ($B \to D$) at $q^2=0$.}
\\[7.0ex] \hline \\[-1.0ex]
\parbox[t]{26mm}{Datta 17} & \cite{Datta:2017aue} & 2+1 & Columbia RHQ &
\parbox[t]{6cm}{Discretization errors discussed as part of combined chiral-continuum-$q^2$ fit, stemming from $a^2|\mathbf{p}|^2$ terms.}
\\[7.0ex] 
\hline \\[-1.0ex]
\parbox[t]{26mm}{\SLfnalmilcBD} & \cite{Lattice:2015rga} & 2+1 & Fermilab &
\parbox[t]{6cm}{Discretization errors of form factors estimated 
via power counting to be 0.4\%.}
\\[4.0ex] \hline \\[-1.0ex]
\parbox[t]{26mm}{\SLLambdabc} & \cite{Detmold:2015aaa} & 2+1 & Columbia RHQ &
\parbox[t]{6cm}{Discretization errors discussed as part of combined chiral-continuum-$q^2$ fit, stemming from $a^2|\mathbf{p}|^2$ terms.}
\\[7.0ex] \hline \\[-1.0ex]
\parbox[t]{26mm}{\SLfnalmilcBDstar} & \cite{Bailey:2014tva} & 2+1 & Fermilab &
\parbox[t]{6cm}{Discretization errors estimated 
via power counting to be 1\%.}
\\[4.0ex] \hline \\[-1.0ex]
Atoui  13 & \cite{Atoui:2013zza} & 2 & tmWil &
\parbox[t]{6cm}{Results obtained from step-scaling in heavy quark mass via the ratio method. Separate continuum limit extrapolations with mild $a^2$-dependence carried out for each mass point separately. Result at physical value of $m_b$ obtained by interpolation between data region and known exact HQET limit.}
\\[21.0ex]
\fi
\hline\hline
\end{tabular*}
\caption{Heavy-quark treatment in determinations of $B_{(s)} \to D_{(s)}^{(*)}\ell\nu$ and $\Lambda_b \to \Lambda_c^{(*)} \ell \bar{\nu}$ form factors, and of $R(D_{(s)})$.}
}
\end{table}

\clearpage
%
%
%
%
%
%


\subsection{Notes to Sec.~\ref{sec:alpha_s} on the strong coupling $\alpha_{\rm s}$}
 In this section we provide more detailed information on the
 simulations used to calculate the strong coupling $\alpha_s$.
 \ifx\reducedapptables\undefined \else 
 We present this information only
 for results that have appeared since FLAG 19. 
 For information on previous 
 calculations not listed here we refer the previous reports FLAG 19 \cite{Aoki:2019cca}
 and FLAG 16~\cite{Aoki:2016frl}. 
\fi


\subsubsection{Renormalization scale and perturbative behaviour}



\begin{table}[!htb]
   \footnotesize
\begin{tabular*}{\textwidth}[l]{l@{\extracolsep{-0.005\textwidth}}rlllllllll}
\hline\hline \\[-1.0ex]
      Collab. & Ref. & $\Nf$ & $\alpha_{\rm eff}$ & $n_l$ & Description \\
      & & & &  \\
      \hline
      \hline
      & & & &  \\[-0.1cm]

Nada 20 & \cite{Nada:2020jay} & 0  &   
  \parbox[t]{0.15\textwidth}{SF: 0.08--0.15 \\GF: 0.08--0.24}   & 2
                  & \parbox[t]{0.555\textwidth}
                    {step-scaling with GF, non-pert matching to SF,\\
                    same gauge configurations as in \cite{DallaBrida:2019wur}} \\
\\[-0.1cm] \hline\\[-1.5ex] 

Husung 20 & \cite{Husung:2020pxg} & 0  &   
  \parbox[t]{0.15\textwidth}{0.17--0.36}   & 3
                  & \parbox[t]{0.555\textwidth}{estimated from $\alpha_{qq}$}
                     \\
\hline
   \hline
\end{tabular*}
\caption{Renormalization scale and perturbative behaviour
         of $\alpha_s$ determinations for $N_f=0$. }
\label{tab_Nf=0_renormalization_a}
\end{table}


\ifx\reducedapptables\undefined

\begin{table}[!htb]
\addtocounter{table}{-1}
   \footnotesize
   \begin{tabular*}{\textwidth}[l]{l@{\extracolsep{\fill}}rlllllllll}
\hline\hline \\[-1.0ex]
      Collab. & Ref. & $\Nf$ & $\alpha_{\rm eff}$ & $n_l$ & Description \\
      & & & &  \\
      \hline
      \hline
      & & & &  \\[-0.1cm]

\parbox[t]{0.14\textwidth}{Boucaud 00A} & \cite{Boucaud:2000ey}         &  0       &
  \parbox[t]{0.09\textwidth}{0.35-0.55, 0.25-0.45, 0.22-0.28, 0.18-0.22} & 2
& \parbox[t]{0.555\textwidth}{$\alpha_{\widetilde{\rm MOM}}$ with
  $p =$ 2--10\,GeV. Fitted to $n_l = 2$ loop perturbation theory
  with power correction.
 ($\beta$ = 6.0, 6.2, 6.4, 6.8.)} \\
\\[-0.1cm] \hline\\[-1.5ex] 

\parbox[t]{0.14\textwidth}{Boucaud 00B} & \cite{Boucaud:2000nd}         &  0       &
  \parbox[t]{0.09\textwidth}{0.35--0.55, 0.25--0.45, 0.22--0.28, 0.18--0.22} & 2
& \parbox[t]{0.555\textwidth}{$\alpha_{\rm MOM}$ with
  $2 \leq \mu \leq 10$\, GeV. Consistency check of $n_l = 2$ loop
  perturbation formula with gluon condensate.
  $\beta_2^{MOM}=1.5\times \beta_2^{\widetilde{MOM}}$ is needed.
  ($\beta$ = 6.0, 6.2, 6.4, 6.8.)} \\
\hline
   \hline
\end{tabular*}
\caption{(cntd.) Renormalization scale and perturbative behaviour
         of $\alpha_s$ determinations for $N_f=0$.}
\label{tab_Nf=0_renormalization_b}
\end{table}

\fi


\ifx\reducedapptables\undefined

\begin{table}[!htb]
   \footnotesize
   \begin{tabular*}{\textwidth}[l]{l@{\extracolsep{\fill}}rlllllllll}
\hline\hline \\[-1.0ex]
      Collab. & Ref. & $\Nf$ & $\alpha_{\rm eff}$ & $n_l$ & Description \\
      &&&&&&& & \\[-0.1cm]
      \hline
      \hline
      &&&&&&&&& \\[-0.1cm]

\parbox[t]{2.0cm}{Karbstein 18} & \cite{Karbstein:2018mzo} & 2 
    & 0.28 - 0.41 & 3
    & \parbox[t]{0.555\textwidth}{$\alpha_V(p)$ for momentum 
                                $1.5 < p < 3.0\,\mbox{GeV}$.
                                Values computed from the 
                                quoted $\Lambda$ parameter with the 2-loop
                                $\beta$ function; larger values (0.32--0.62)
                                are obtained with 3-loop running. 
                                As with ETM 11C
                                central values are taken from 
                                $a = 0.042\,\mbox{fm}$ lattice
                                with $L = 1.3\,\mbox{fm}$ and 
                                $m_\pi = 350\,\mbox{MeV}$.}
\\ \\[-0.1cm] 
\hline\\[-1.5ex] 
ALPHA 12 & \cite{Fritzsch:2012wq} & 2  & 
\parbox[t]{0.1\textwidth}{see \\ ALPHA 04} & 2
    & \parbox[t]{0.5550\textwidth}{Determination of $\Lambda_\msbar / f_K$ using ALPHA 04}
\\
\hline\hline
\end{tabular*}
\caption{Renormalization scale and perturbative behaviour 
of $\alpha_s$ determinations for $N_f=2$.}
\label{tab_Nf=2_renormalization}
\end{table}

\fi


\begin{table}[!htb]
   \footnotesize
   \begin{tabular*}{\textwidth}[l]{l@{\extracolsep{\fill}}rlllllllll}
\hline\hline \\[-1.0ex]
      Collab. & Ref. & $\Nf$ & $\alpha_{\rm eff}$ & $n_l$ & Description \\
      &&&&& \\[-0.1cm]
      \hline
      \hline
      &&&&& \\[-0.1cm]

Petreczky 20 & \cite{Petreczky:2020tky} & 2+1 &   
   0.22--0.38   & 2
                 & \parbox[t]{0.555\textwidth}
                    {$m_h=m_c-4m_c$} \\
\\[-0.1cm] \hline\\[-1.5ex] 

Ayala 20 & \cite{Ayala:2020odx} & 2+1 &   
   0.2--0.4   & 3
                 & \parbox[t]{0.555\textwidth}
                    {$1/r>2$ GeV} \\
\hline
      \hline
\end{tabular*}
\caption{Renormalization scale and perturbative behaviour of $\alpha_s$
         determinations for $N_f=3$.}
\label{tab_Nf=3_renormalization_a}
\end{table}


\ifx\reducedapptables\undefined

\begin{table}[!htb]
\addtocounter{table}{-1}
   \footnotesize
   \begin{tabular*}{\textwidth}[l]{l@{\extracolsep{\fill}}rlllllllll}
\hline\hline \\[-1.0ex]
      Collab. & Ref. & $\Nf$ & $\alpha_{\rm eff}$ & $n_l$ & Description \\
      &&&&& \\[-0.1cm]
      \hline
      \hline
      &&&&& \\[-0.1cm]

HPQCD 08B    & \cite{Allison:2008xk}  & 2+1 &  
0.38  & 2 &\parbox[t]{0.54\textwidth}{Fit of the ratios to PT
at the charm mass
including $(a m)^{2i}$ terms with $i\leq 2\ldots 4$; coefficients constrained by priors. } 
\\ \\[-0.1cm] \hline\\[-1.5ex]
 {HPQCD 08A}
   & \cite{Davies:2008sw}
   & 2+1
   & \parbox[t]{0.1\textwidth}
        {0.15-0.4}
   & 2
   & \parbox[t]{0.54\textwidth}
        {$\alphav(q^*)$ for a variety of short-distance quantities,
         using same method as in HPQCD 05A.}
\\ \\[-0.1cm] \hline\\[-1.5ex]
 {Maltman 08}
   & \cite{Maltman:2008bx}
   & 2+1
   &  & 2
   & \parbox[t]{0.54\textwidth}
        {Re-analysis of HPQCD 05A for a restricted set of
         short-distance quantities with similar results.}
\\ \\[-0.1cm] \hline\\[-1.5ex]
 {HPQCD 05A}
   & \cite{Mason:2005zx}
   & 2+1
   & \parbox[t]{0.1\textwidth}
        {0.2-0.4}
   & 2
   & \parbox[t]{0.54\textwidth}
        {$\alphav(q^*)$ for a variety of short-distance quantities.
        }
\\
      &&&&& \\[-0.1cm]
      \hline
      \hline
\end{tabular*}
\caption{(cntd.) Renormalization scale and perturbative behaviour of $\alpha_s$
         determinations for $N_f=3$.}
\label{tab_Nf=3_renormalization_b}
\end{table}

\fi


\ifx\reducedapptables\undefined

\begin{table}[!htb]
   \vspace{3.0cm}
   \footnotesize
   \begin{tabular*}{\textwidth}[l]{l@{\extracolsep{\fill}}rlllllllll}
\hline\hline \\[-1.0ex]
      Collab. & Ref. & $\Nf$ & $\alpha_{\rm eff}$ & $n_l$ & Description \\
      &&&&& \\[-0.1cm]
      \hline
      \hline
      &&&&& \\[-0.1cm]

 HPQCD 14A   &  \cite{Chakraborty:2014aca} & 2+1+1 &  
 0.11-0.33 & 2 
   &  \parbox[t]{0.54\textwidth}{
           Range given for $\alpha_\mathrm{eff}$ from $\tilde R_4$. 
           Fit of ratios $R_n\;n=4\ldots10 $ to perturbation theory
           including $(a m)^{2i}$ terms with $i\leq $10--20 and 
           higher-order perturbative terms;
           coefficients constrained by priors.}
 \\ \\[-0.1cm] \hline\\[-1.5ex]
 { {ETM 13D}}       &  \cite{Blossier:2013ioa} & 2+1+1 &  0.26-0.7    & 3 & 
	\parbox[t]{0.54\textwidth}{$\alpha_\mathrm{T}(p)$ for $p=$1.6--6.5\,GeV. 
	Update of \cite{Blossier:2012ef}
                                   with improved power law determination.}
 \\ \\[-0.1cm] \hline\\[-1.5ex]
 { ETM 12C}       & \cite{Blossier:2012ef}   & 2+1+1 &
0.24-0.38 & 3 &
	\parbox[t]{0.54\textwidth}{$\alpha_\mathrm{T}(p)$ for $p=$1.7--6.8\,GeV. 
	Fit to PT with gluon condensate correction or higher power.}
	 \\ \\[-0.1cm] \hline\\[-1.5ex]
      ALPHA 10A & \cite{Tekin:2010mm} & 4                  & 0.07-0.28 & 2
        & \parbox[t]{0.54\textwidth}{$\alpha_\mathrm{SF}(1/L)$.
        Comparison to PT with 2-, 3-loop $\beta$-function.}
 \\ \\[-0.1cm] \hline\\[-1.5ex]
 {ETM 11D}      & \cite{Blossier:2011tf}   & 2+1+1 & 
	0.24-0.4& 3 &
	\parbox[t]{0.54\textwidth}{$\alpha_\mathrm{T}(p)$ for $p=$3.8--7.1\,GeV
        with H(4)-procedure. Fit to PT with gluon condensate correction.}
 \\ \\[-0.1cm] \hline\\[-1.5ex]
      Perez 10 & \cite{PerezRubio:2010ke} & 4           & 0.06-0.28 & 2
        & \parbox[t]{0.54\textwidth}{$\alpha_\mathrm{SF}(1/L)$.
          Comparison with 1-, 2-, 3-loop $\beta$-function.}\\
      &&&&&&\\[-0.1cm]
      \hline
      \hline
\end{tabular*}
\caption{Renormalization scale and perturbative behaviour of $\alpha_s$
         determinations for $N_f=4$.}
\label{tab_Nf=4_renormalization}
\end{table}

\fi

\clearpage


\subsubsection{Continuum limit}


\begin{table}[!htb]
   \footnotesize
   \begin{tabular*}{\textwidth}[l]{l@{\extracolsep{\fill}}rlllllllll}
\hline\hline \\[-1.0ex]
      Collab. & Ref. & $\Nf$ & $a\,\mu$  & Description\\
      & & & &  \\
      \hline
      \hline
      & & & &  \\[-0.1cm]

Nada 20 & \cite{Nada:2020jay} & 0  &   
       \parbox[t]{0.25\textwidth}{GF: $0.125 < a\mu < 0.042$ \\
                                  SF: $0.17 <  a\mu < 0.063$ }                 &                 
      \parbox[t]{0.39\textwidth}{step scaling GF scheme, \\ non-pert matching to SF scheme} \\  
\\[-0.1cm] \hline\\[-1.5ex] 

Husung 20 & \cite{Husung:2020pxg} & 0  &   
      \parbox[t]{0.25\textwidth}{$a \mu<0.57$}                 &                 
      \parbox[t]{0.39\textwidth}{only $r/a>3/5$ are used in the analysis}     \\
\hline
      \hline
\end{tabular*}
\caption{Continuum limit for $\alpha_s$ determinations with $N_f=0$.}
\label{tab_Nf=0_continuumlimit_a}
\end{table}


\ifx\reducedapptables\undefined

\begin{table}[!htb]
\addtocounter{table}{-1}
   \footnotesize
   \begin{tabular*}{\textwidth}[l]{l@{\extracolsep{\fill}
   }rlllllllll}
\hline\hline \\[-1.0ex]
      Collaboration & Ref. & $\Nf$ & $a\,\mu$  & Description\\
      & & & &  \\
      \hline
      \hline
      & & & &  \\[-0.1cm]

    {Boucaud 00A}      &\cite{Boucaud:2000ey}           &  0    & 
    {$a =$ 0.1, 0.07, 0.05, 0.03\,fm}  &
    {At $\alpha_s \leq 0.3$ $ap = $1.9, 1.4, 1.0, 0.6.}\\
\\[-0.1cm] \hline\\[-1.5ex]
    {Boucaud 00B}   & \cite{Boucaud:2000nd}      & 0         & 
    {$a =$ 0.1, 0.07, 0.05, 0.03\,fm} & 
    {At $\alpha_s \leq 0.3$ $ap =$ 1.9, 1.4, 1.0, 0.6.}\\
\hline
      \hline
\end{tabular*}
\caption{(cntd.) Continuum limit for $\alpha_s$ determinations with $N_f=0$.}
\label{tab_Nf=0_continuumlimit_b}
\end{table}

\fi


\ifx\reducedapptables\undefined

\begin{table}[!htb]
   \vspace{0.0cm}
   \footnotesize
   \begin{tabular*}{\textwidth}[l]{l@{\extracolsep{\fill}}rlllllllll}
\hline\hline \\[-1.0ex]
      Collab. & Ref. & $\Nf$ &  $a\,\mu$ & Description   \\
      &&&&&&& & \\[-0.1cm]
      \hline
      \hline
      &&&&&&&&& \\[-0.1cm]
Karbstein 14 & \cite{Karbstein:2014bsa} & 2  & 
\parbox[t]{0.1\textwidth}{0.32 -- 1.19 \\ 0.63 -- 1.19 }
    & 
\parbox[t]{0.555\textwidth}{at $p=1.5$\,GeV \\
at $p=3$\,GeV, roughly coincides with $\alpha_s=0.3$.}
\\ \\[-0.1cm] \hline\\[-1.5ex] 
ALPHA 12 & \cite{Fritzsch:2012wq} & 2 &  
\parbox[t]{0.25\textwidth}{$a=$ 0.049, 0.066, 0.076 fm from $f_K$} & 
\parbox[t]{0.46\textwidth}{2-loop $\cO(a)$ boundary improvement, 
linear extrapolation of $L_\mathrm{max} f_\mathrm{K}$ in $a^2$.}
 \\
\hline
      \hline
\end{tabular*}
\caption{Continuum limit for $\alpha_s$ determinations with $N_f=2$.}
\label{tab_Nf=2_continuumlimit}
\end{table}

\fi


\begin{table}[!htb]
   \vspace{0.0cm}
   \footnotesize
   \begin{tabular*}{\textwidth}[l]{l@{\extracolsep{\fill}}rlllllllll}
\hline\hline \\[-1.0ex]
      Collab. & Ref. & $\Nf$ & $a\,\mu$ & Description \\
      &&&&&&& & \\[-0.1cm]
      \hline
      \hline
      &&&&&&& \\[-0.1cm]

Petreczky 20 & \cite{Petreczky:2020tky} & 2+1 &   
      \parbox[t]{0.25\textwidth}{$a\mu=0.32-5.40$}                 &                 
      \parbox[t]{0.39\textwidth}{$a=0.025-0.104$ fm, $m_h=m_c-4m_c$ }                                   \\
\\[-0.1cm] \hline\\[-1.5ex] 

Ayala 20 & \cite{Ayala:2020odx} & 2+1 &   
      \parbox[t]{0.25\textwidth}{$a \mu<0.71$}                 &                 
      \parbox[t]{0.39\textwidth}{only $r/a>\sqrt{8}$ are considered}                                   \\
\hline
      \hline
\end{tabular*}
\caption{Continuum limit for $\alpha_s$ determinations with $N_f=3$.}
\label{tab_Nf=3_continuumlimit_a}
\end{table}


\ifx\reducedapptables\undefined

\begin{table}[!htb]
   \vspace{0.0cm}
   \addtocounter{table}{-1}
   \footnotesize
   \begin{tabular*}{\textwidth}[l]{l@{\extracolsep{\fill}}rlllllllll}
\hline\hline \\[-1.0ex]
      Collab. & Ref. & $\Nf$ & $a\,\mu$ & Description \\
      &&&&&&& & \\[-0.1cm]
      \hline
      \hline
      &&&&&&& \\[-0.1cm]

PACS-CS 09A& \cite{Aoki:2009tf} & 2+1 & 
$a/L=1/8,1/6,1/4$ & 
\parbox[t]{0.42\textwidth}{Tree-level
$\cO(a)$ boundary improvement, which has been seen to behave better
than 1-loop in simulations \cite{Takeda:2004xha}; 
weighted average of $a/L=1/8,1/6$ for step-scaling function
which agrees with a
linear extrapolation in $a/L$ of all data points of the 
SSF. Linear extrapolation  in $a/L$ of $L_\mathrm{max}m_\rho$ with 
$a/L_\mathrm{max} = 1/8, 1/6, 1/4$.}
 \\ \\[-0.1cm] \hline\\[-1.5ex]

{HPQCD 08B}   & \cite{Allison:2008xk}  & 2+1 &  
\parbox[t]{0.25\textwidth}{ $a\mu=2 a\bar m_c$ = 0.8,\, 1.2,\, 1.7,\, 2.1} & 
\parbox[t]{0.42\textwidth}{4 lattice spacings with heavy-quark mass
approximately the charm mass, where $\alpha_{R_4}(\mu) = 0.38$.}
 \\ \\[-0.1cm] \hline\\[-1.5ex]

{HPQCD 08A}
   & \cite{Davies:2008sw}
   & 2+1
   & \parbox[t]{0.25\textwidth}
        {6 lattice spacings with $a = 0.18$ - $0.045\,\mbox{fm}$}
   & \parbox[t]{0.555\textwidth}
        {$r_1$ using $\Upsilon$ spectrum splitting.}
 \\ \\[-0.1cm] \hline\\[-1.5ex]

{Maltman 08}
   & \cite{Maltman:2008bx}
   & 2+1
   & \parbox[t]{0.25\textwidth}
        {5 lattice spacings with $a = 0.18$ - $0.06\,\mbox{fm}$}
   & \parbox[t]{0.42\textwidth}
        {Re-analysis of HPQCD 05A with additional lattice
         spacings $a = 0.06$, $0.15\,\mbox{fm}$.}
 \\ \\[-0.1cm] \hline\\[-1.5ex]

{HPQCD 05A}
   & \cite{Mason:2005zx}
   & 2+1
   & \parbox[t]{0.25\textwidth}
        {3 lattice spacings with $a = 0.18$ - $0.09\,\mbox{fm}$}
   & \parbox[t]{0.555\textwidth}
        {$r_1$ using $\Upsilon$ spectrum splitting.}
\\ 
        &&&&&&&&& \\[-0.1cm]
   \hline
      \hline
\end{tabular*}
\caption{(cntd.) Continuum limit for $\alpha_s$ determinations with $N_f=3$.}
\label{tab_Nf=3_continuumlimit_b}
\end{table}

\fi


\ifx\reducedapptables\undefined

\begin{table}[!htb]
   \vspace{3.0cm}
   \footnotesize
   \begin{tabular*}{\textwidth}[l]{l@{\extracolsep{\fill}}rlllllllll}
\hline\hline \\[-1.0ex]
      Collab. & Ref. & $\Nf$ & $a\,\mu$ & Description \\
      &&&& \\[-0.1cm]
      \hline
      \hline
      &&&& \\[-0.1cm]
HPQCD 14A   &  \cite{Chakraborty:2014aca} & 2+1+1 & 
\parbox[t]{0.25\textwidth}{$a\mu=2 a\bar m_c = 0.78 - 2.09$}& 
\parbox[t]{0.42\textwidth}{4 lattice spacings;
             2 lattice spacings with $ a\mu  \leq 1.5$ 
             and one more lattice spacing with $ a\mu  \lsim 1.6$ 
             when $\alpha_{R_4}(\mu) \leq 0.3$.}
 \\ \\[-0.1cm] \hline\\[-1.5ex]
ETM 13D         &  \cite{Blossier:2013ioa} & 2+1+1 &    
  $a=0.060, 0.068$ fm from $f_\pi$          &For $\alpha_s \leq 0.3$, $ap=1.5,1.7$.
	\parbox[t]{0.555\textwidth}{Update of \cite{Blossier:2012ef}.}
\\ \\[-0.1cm] \hline\\[-1.5ex]      
{ETM 12C}       & \cite{Blossier:2012ef}   & 2+1+1 &  
      $a=0.061, 0.078$ fm from $f_\pi$ & 
      \parbox[t]{0.42\textwidth}{Global fit with $(ap)^2$ discretization effects.
       For $\alpha_s \leq 0.3$, $ap = 1.5, 2.2$.}       
\\ \\[-0.1cm] \hline\\[-1.5ex]      
{ETM 11D}      & \cite{Blossier:2011tf}   & 2+1+1 & 
      $a = 0.061, 0,078$\,fm & {For $\alpha_s \leq 0.3$, $ap = 1.5, 2.0$.}
\\ \\[-0.1cm] \hline\\[-1.5ex]      
      ALPHA 10A & \cite{Tekin:2010mm} & 4                  & 
      $a/L = 1/4, 1/6, 1/8$ & Constant or global linear fit in $(a/L)^2$.
\\ \\[-0.1cm] \hline\\[-1.5ex]      
      Perez 10 & \cite{PerezRubio:2010ke} & 4 & $a/L=1/4, 1/6, 1/8$ &
      \parbox[t]{0.42\textwidth}{Linear extrapolation in $(a/L)^2$.
      1-loop improvement at the boundary.}
\\  &&&&&&&&& \\[-0.1cm]
   \hline
   \hline
\end{tabular*}
\caption{Continuum limit for $\alpha_s$ determinations with $N_f=4$.}
\label{tab_Nf=4_continuumlimit}
\end{table}

\fi

%
%

\clearpage
\newpage
\subsection{Notes to Sec.~\ref{sec:NME} on nucleon matrix elements}
\label{subsec:Notes to NME}

\begin{table}[!ht]
{\footnotesize
\begin{tabular*}{\textwidth}{l @{\extracolsep{\fill}} c c c l}
\hline\hline \\[-1.0ex]
Collab. & Ref. & $\Nf$ & $a$ [fm] & Description 
\\[.3ex]\\[-2.ex] \hline \hline \\[-1.0ex]
%
CalLat 19 & \cite{Walker-Loud:2019cif} & 2+1+1 & 0.15,0.12,0.09
& \parbox[t]{6.3cm}{Extrapolation to the physical point via
  simultaneous fit in the lattice spacing, $M_\pi$ and $M_\pi L$,
  including terms of order $a^2$ and $a^4$.}
\\[0.5ex]\\[0.5ex]  \hline\\[-0.5ex]
ETM 19 & \cite{Alexandrou:2019brg} & 2+1+1 & 0.08
& \parbox[t]{6.3cm}{Single lattice spacing.}
\\[0.5ex]\\[0.5ex]
\ifx\reducedapptables\undefined
\hline \\[-0.5ex]
PNDME 18 & \cite{Gupta:2018qil} & 2+1+1 & 0.15,0.12,0.09,0.06
& \parbox[t]{6.3cm}{Extrapolation performed including a linear term in
  $a$ as part of a simultaneous fit in $a$, $M_\pi$ and $M_\pi L$.} 
\\[0.5ex]\\[0.5ex]  \hline\\[-0.5ex]
CalLat 18 & \cite{Chang:2018uxx} & 2+1+1 & 0.15,0.12,0.09
& \parbox[t]{6.3cm}{Extrapolation to the physical point via
  simultaneous fit in the lattice spacing, $M_\pi$ and $M_\pi L$,
  including terms of order $a^2$ and $a^4$.}
\\[0.5ex]\\[0.5ex]  \hline\\[-0.5ex]
CalLat 17 & \cite{Berkowitz:2017gql} & 2+1+1 & 0.15,0.12,0.09
& \parbox[t]{6.3cm}{Extrapolation to the physical point via
  simultaneous fit in the lattice spacing, $M_\pi$ and $M_\pi L$,
  including terms of order $a^2$ and $a^4$.}
\\[0.5ex]\\[0.5ex]  \hline\\[-0.5ex]
PNDME 16 & \cite{Bhattacharya:2016zcn} & 2+1+1 & 0.12,0.09,0.06
& \parbox[t]{6.3cm}{ Extrapolation performed including a linear term
  in $a$ as part of a simultaneous fit in $a$, $M_\pi$ and $M_\pi L$.}
\\[0.5ex]\\[0.5ex]  \hline\\[-0.5ex]
PNDME 15 & \cite{Bhattacharya:2015wna,Bhattacharya:2015esa} & 2+1+1
& 0.12,0.09,0.06 & \parbox[t]{6.3cm}{Extrapolation performed
  including a linear term in $a$ as part of a simultaneous fit in $a$,
  $M_\pi$ and $M_\pi L$.}
\\[0.5ex]\\[0.5ex]  \hline\\[-0.5ex]
PNDME 13 & \cite{Bhattacharya:2013ehc} & 2+1+1 & 0.12
& \parbox[t]{6.3cm}{Single lattice spacing.}
\\[0.5ex]\\[0.5ex]
\fi
\hline \hline \\[-1.0ex]
\end{tabular*}
\caption{Continuum extrapolations/estimation of lattice artifacts in
  determinations of the isovector axial, scalar and tensor charges
  with $\Nf=2+1+1$ quark flavours.}
}
\end{table}

\begin{table}[!ht]
{\footnotesize
\begin{tabular*}{\textwidth}{l @{\extracolsep{\fill}} c c c l}
\hline\hline \\[-1.0ex]
Collab. & Ref. & $\Nf$ & $a$ [fm] & Description 
\\[.3ex]\\[-2.ex] \hline \hline \\[-1.0ex]
%
$\chi$QCD 21 & \cite{Liu:2021irg} & 2+1 & 0.06,0.08,0.11,0.14
& \parbox[t]{6.3cm}{Extrapolation performed using a quadratic term
  in $a$ as part of a simultaneous fit in $a$, $M_\pi$ and $M_\pi L$.}
\\[0.5ex]\\[0.5ex]  \hline\\[-0.5ex]
NME 21 & \cite{Park:2021ypf} & 2+1 & 0.07,0.09,0.13
& \parbox[t]{6.3cm}{Extrapolation performed using a linear term
  in $a$ as part of a simultaneous fit in $a$, $M_\pi$ and $M_\pi L$.}
\\[0.5ex]\\[0.5ex]  \hline\\[-0.5ex]
RBC/UKQCD 19 & \cite{Abramczyk:2019fnf} & 2+1 & 0.14
& \parbox[t]{6.3cm}{Single lattice spacing.}
\\[0.5ex]\\[0.5ex]  \hline\\[-0.5ex]
LHPC 19 & \cite{Hasan:2019noy} & 2+1 & 0.09,0.12
& \parbox[t]{6.3cm}{No extrapolation performed; final result taken from finer lattice with an enlarged error bar to account for result at coarser $a$.}
\\[0.5ex]\\[0.5ex]  \hline\\[-0.5ex]
Mainz 19 & \cite{Harris:2019bih} & 2+1 & 0.05,0.06,0.08,0.09
& \parbox[t]{6.3cm}{Extrapolation performed as part of a simultaneous
fit in $a$, $M_\pi$ and $M_\pi L$.}
\\[0.5ex]\\[0.5ex]  \hline\\[-0.5ex]
PACS 18A & \cite{Shintani:2018ozy} & 2+1 & 0.085 &
\parbox[t]{6.3cm}{Single lattice spacing.}
\\[0.5ex]\\[0.5ex]
\ifx\reducedapptables\undefined
\hline \\[-0.5ex]
PACS 18 & \cite{Ishikawa:2018rew} & 2+1 & 0.09 & \parbox[t]{6.3cm}{Single lattice spacing.}
\\[0.5ex]\\[0.5ex]  \hline\\[-0.5ex]
$\chi$QCD 18 & \cite{Liang:2018pis} & 2+1 & 0.14,0.11,0.08
& \parbox[t]{6.3cm}{Extrapolation to the physical point via
  simultaneous fit in the lattice spacing, $M_\pi$ and $M_\pi L$,
  including terms of order $a^2$.}
\\[0.5ex]\\[0.5ex]  \hline\\[-0.5ex]
JLQCD 18 & \cite{Yamanaka:2018uud} & 2+1 & 0.11 & \parbox[t]{6.3cm}{Single lattice spacing.}
\\[0.5ex]\\[0.5ex]  \hline\\[-0.5ex]
LHPC 12 & \cite{Green:2012ej} & 2+1 & 0.12,0.09
&  \parbox[t]{6.3cm}{No extrapolation or uncertainty
  estimated. Comparison between two lattice spacings at a single pion
  mass shows no deviation within statistical errors.} 
\\[0.5ex]\\[0.5ex]  \hline\\[-0.5ex]
LHPC 12A & \cite{Green:2012ud} & 2+1 & 0.12,0.09 
& \parbox[t]{6.3cm}{No statistically significant discretization effects
  observed. Results assumed to be constant in $a$.}
\\[0.5ex]\\[0.5ex] 
\fi
\hline \hline \\[-1.0ex]
\end{tabular*}
\caption{Continuum extrapolations/estimation of lattice artifacts in
  determinations of the isovector axial, scalar and tensor charges
  with $\Nf=2+1$ quark flavours.}
}
\end{table}

\ifx\reducedapptables\undefined
\begin{table}[!ht]
\addtocounter{table}{-1}
{\footnotesize
\begin{tabular*}{\textwidth}{l @{\extracolsep{\fill}} c c c l}
\hline\hline \\[-1.0ex]
Collab. & Ref. & $\Nf$ & $a$ [fm] & Description 
\\[.3ex]\\[-2.ex] \hline \hline \\[-1.0ex]
LHPC 10 & \cite{Bratt:2010jn} & 2+1 & 0.12 & \parbox[t]{6.3cm}{Single lattice spacing.}
\\[0.5ex]\\[0.5ex]  \hline\\[-0.5ex]
RBC/UKQCD 10D & \cite{Aoki:2010xg} & 2+1 & 0.11 &  \parbox[t]{6.3cm}{Single lattice spacing.}
\\[0.5ex]\\[0.5ex]  \hline\\[-0.5ex]
RBC/UKQCD 09B & \cite{Yamazaki:2009zq} & 2+1 & 0.11 & \parbox[t]{6.3cm}{Single lattice spacing.}
\\[0.5ex]\\[0.5ex]  \hline\\[-0.5ex]
RBC/UKQCD 08B & \cite{Yamazaki:2008py} & 2+1 & 0.11 & \parbox[t]{6.3cm}{Single lattice spacing.}
\\[0.5ex]\\[0.5ex]  \hline\\[-0.5ex]
LHPC 05 & \cite{Edwards:2005ym} & 2+1 & 0.12 & \parbox[t]{6.3cm}{Single lattice spacing.}
\\[0.5ex]\\[0.5ex]
\hline \hline \\[-1.0ex]
\end{tabular*}
\caption{(cntd.) Continuum extrapolations/estimation of lattice artifacts in
  determinations of the isovector axial, scalar and tensor charges
  with $\Nf=2+1$ quark flavours.}
}
\end{table}

\begin{table}[!ht]
{\footnotesize
\begin{tabular*}{\textwidth}{l @{\extracolsep{\fill}} c c c l}
\hline\hline \\[-1.0ex]
Collab. & Ref. & $\Nf$ & $a$ [fm] & Description 
\\[.3ex]\\[-2.ex] \hline \hline \\[-1.0ex]
%
Mainz 17 & \cite{Capitani:2017qpc} & 2 & 0.08,0.06,0.05
& \parbox[t]{6.3cm}{No statistically significant discretization effects
  observed. Results assumed to be constant in $a$.}
\\[0.5ex]\\[0.5ex]  \hline\\[-0.5ex]
ETM 17 & \cite{Alexandrou:2017qyt}  & 2 & 0.09  & \parbox[t]{6.3cm}{Single lattice spacing.}
\\[0.5ex]\\[0.5ex]  \hline\\[-0.5ex]
ETM 17B & \cite{Alexandrou:2017hac}  & 2 & 0.09 & \parbox[t]{6.3cm}{Single lattice spacing.}
\\[0.5ex]\\[0.5ex]  \hline\\[-0.5ex]
ETM 15D & \cite{Abdel-Rehim:2015owa} & 2 & 0.09 & \parbox[t]{6.3cm}{Single lattice spacing.}
\\[0.5ex]\\[0.5ex]  \hline\\[-0.5ex]
RQCD 14 & \cite{Bali:2014nma} & 2 & 0.08,0.07,0.06
& \parbox[t]{6.3cm}{No statistically significant discretization effects
  observed. No extrapolation performed, but residual systematic error
  due to discretization effects is estimated.}
\\[0.5ex]\\[0.5ex]  \hline\\[-0.5ex]
QCDSF 13 & \cite{Horsley:2013ayv} & 2 & 0.08,0.07,0.06
& \parbox[t]{6.3cm}{No statistically significant discretization
  effects observed. Results assumed to be constant in $a$.}
\\[0.5ex]\\[0.5ex]  \hline\\[-0.5ex]
Mainz 12 & \cite{Capitani:2012gj} & 2 & 0.08,0.06,0.05
& \parbox[t]{6.3cm}{No statistically significant discretization
  effects observed. Results assumed to be constant in $a$.} 
\\[0.5ex]\\[0.5ex]  \hline\\[-0.5ex]
RBC 08 & \cite{Lin:2008uz} & 2 & 0.11 & \parbox[t]{6.3cm}{Single lattice spacing.}
\\[0.5ex]\\[0.5ex]  \hline\\[-0.5ex]
QCDSF 06 & \cite{Khan:2006de} & 2 & 0.09,0.08,0.07,0.06
& \parbox[t]{6.3cm}{No statistically significant discretization
  effects observed. Results assumed to be constant in $a$.}
\\[0.5ex]\\[0.5ex]
\hline \hline \\[-1.0ex]
\end{tabular*}
\caption{Continuum extrapolations/estimation of lattice artifacts in
  determinations of the isovector axial, scalar and tensor charges
  with $\Nf=2$ quark flavours.}
}
\end{table}
\fi

\begin{table}[!ht]
{\footnotesize
\begin{tabular*}{\textwidth}{l @{\extracolsep{\fill}} c c c l}
\hline\hline \\[-1.0ex]
{Collab.} & {Ref.} & {$\Nf$} & {$M_{\pi,\rm min}\,[\mev]$} & {Description}  
\\[.3ex]\\[-2.ex] \hline \hline \\[-1.0ex]
%
CalLat 19 & \cite{Walker-Loud:2019cif} & 2+1+1 & 135,130,220
& \parbox[t]{6.3cm}{Fit performed including analytic and non-analytic
  terms in $M_\pi$ up to order $M_\pi^4$.}
\\[0.5ex]\\[0.5ex]  \hline\\[-0.5ex]
ETM 19 & \cite{Alexandrou:2019brg} & 2+1+1 & 139
& \parbox[t]{6.3cm}{Single pion mass within 3\% of the physical value.}
\\[0.5ex]\\[0.5ex]
\ifx\reducedapptables\undefined
\hline \\[-0.5ex]
PNDME 18 & \cite{Gupta:2018qil} & 2+1+1 & 321,225,138,136
& \parbox[t]{6.3cm}{Fit performed including a linear term in $M_\pi^2$
  and investigating a logarithmic term as part of a simultaneous fit
  in $a$, $M_\pi$ and $M_\pi L$.}
\\[0.5ex]\\[0.5ex]  \hline\\[-0.5ex]
CalLat 18 & \cite{Chang:2018uxx} & 2+1+1 & 130,130,220
& \parbox[t]{6.3cm}{Fit performed including analytic and non-analytic
  terms in $M_\pi$ up to order $M_\pi^4$.}
\\[0.5ex]\\[0.5ex]  \hline\\[-0.5ex]
CalLat 17 & \cite{Berkowitz:2017gql} & 2+1+1 &
130,220,310 & \parbox[t]{6.3cm}{Fit performed including analytic and
  non-analytic terms in $M_\pi$ up to order $M_\pi^4$.}
\\[0.5ex]\\[0.5ex]  \hline\\[-0.5ex]
PNDME 16 & \cite{Bhattacharya:2016zcn} & 2+1+1 & 225,138,235
& \parbox[t]{6.3cm}{Fit performed including linear and logarithmic
  terms in $M_\pi^2$ as part of a simultaneous fit in $a$, $M_\pi$ and
  $M_\pi L$.}
\\[0.5ex]\\[0.5ex]  \hline\\[-0.5ex]
PNDME 15 & \cite{Bhattacharya:2015wna,Bhattacharya:2015esa} & 2+1+1
& 225,138,235 & \parbox[t]{6.3cm}{Fit performed including a linear
  term in $M_\pi^2$ as part of a simultaneous fit in $a$, $M_\pi$ and
  $M_\pi L$.}
\\[0.5ex]\\[0.5ex]  \hline\\[-0.5ex]
PNDME 13 & \cite{Bhattacharya:2013ehc} & 2+1+1 & 228
&  \parbox[t]{6.3cm}{Fit performed with a linear term in $M_\pi^2$.}
\\[0.5ex]\\[0.5ex]
\fi
\hline \hline \\[-1.0ex]
\end{tabular*}
\caption{Chiral extrapolation/minimum pion mass in determinations of
  the isovector axial, scalar and tensor charges with $\Nf=2+1+1$
  quark flavours.} 
}
\end{table}

\begin{table}[!ht]
{\footnotesize
\begin{tabular*}{\textwidth}{l @{\extracolsep{\fill}} c c c l}
\hline\hline \\[-1.0ex]
{Collab.} & {Ref.} & {$\Nf$} & {$M_{\pi,\rm min}\,[\mev]$} & {Description}  
\\[.3ex]\\[-2.ex] \hline \hline \\[-1.0ex]
%
$\chi$QCD 21 & \cite{Liu:2021irg} & 2+1 & 139, 171, 302, 337, 371
& \parbox[t]{6.3cm}{Extrapolation performed using quadratic terms
  in $M_{\pi}$ from partially quenched chiral perturbation theory as part of a simultaneous fit in $a$, $M_\pi$ and $M_\pi L$.}
\\[0.5ex]\\[0.5ex]  \hline\\[-0.5ex]
NME 21 & \cite{Park:2021ypf} & 2+1 & 170,170,270,285
& \parbox[t]{6.3cm}{Extrapolation performed using a quadratic term
  in $M_{\pi}$ as part of a simultaneous fit in $a$, $M_\pi$ and $M_\pi L$.}
\\[0.5ex]\\[0.5ex]  \hline\\[-0.5ex]
RBC/UKQCD 19 & \cite{Abramczyk:2019fnf} & 2+1 & 170,250
& \parbox[t]{6.3cm}{Does not quote extrapolated value, but shows low mass dependence.}
\\[0.5ex]\\[0.5ex]  \hline\\[-0.5ex]
LHPC 19 & \cite{Hasan:2019noy} & 2+1 & 133,137
&  \parbox[t]{6.3cm}{Does not perform extrapolation; comparison of results at two near-physical pion masses.}
\\[0.5ex]\\[0.5ex]  \hline\\[-0.5ex]
Mainz 19 & \cite{Harris:2019bih} & 2+1 & 220, 290, 200, 260
& \parbox[t]{6.3cm}{Extrapolation performed using logarithmic and
quadratic terms in $M_\pi$ as part of a simultaneous fit in $a$,
$M_\pi$ and $M_\pi L$.}
\\[0.5ex]\\[0.5ex]  \hline\\[-0.5ex]
PACS 18A & \cite{Shintani:2018ozy} & 2+1 & 135
& \parbox[t]{6.3cm}{Single, physical pion mass.}
\\[0.5ex]\\[0.5ex]
\ifx\reducedapptables\undefined
\hline \\[-0.5ex]
PACS 18 & \cite{Ishikawa:2018rew} & 2+1 & 146
& \parbox[t]{6.3cm}{Single, nearly physical pion mass.}
\\[0.5ex]\\[0.5ex]  \hline\\[-0.5ex]
$\chi$QCD 18 & \cite{Liang:2018pis} & 2+1 & 171,337,302
& \parbox[t]{6.3cm}{Linear fit in $M_\pi^2$, including terms
  containing $M_\pi L$.}
\\[0.5ex]\\[0.5ex]  \hline\\[-0.5ex]
JLQCD 18 & \cite{Yamanaka:2018uud} & 2+1 & 293
& \parbox[t]{6.3cm}{Fit performed with linear and quadratic terms in
  $M_\pi^2$.}
\\[0.5ex]\\[0.5ex] 
\fi
\hline \hline \\[-1.0ex]
\end{tabular*}
\caption{Chiral extrapolation/minimum pion mass in determinations of
  the isovector axial, scalar and tensor charges with $\Nf=2+1$
  quark flavours.} 
}
\end{table}

\ifx\reducedapptables\undefined
\begin{table}[!ht]
\addtocounter{table}{-1}
{\footnotesize
\begin{tabular*}{\textwidth}{l @{\extracolsep{\fill}} c c c l}
\hline\hline \\[-1.0ex]
{Collab.} & {Ref.} & {$\Nf$} & {$M_{\pi,\rm min}\,[\mev]$} & {Description}  
\\[.3ex]\\[-2.ex] \hline \hline \\[-1.0ex]
%
LHPC 12 & \cite{Green:2012ej} & 2+1 & 149,317
&  \parbox[t]{6.3cm}{Chiral fit performed with $g_A$ and $F_{\pi}$
  held fixed to NNLO in $SU(2)$ HBChPT. Several other $M_\pi$
  investigated.} 
\\[0.5ex]\\[0.5ex]  \hline\\[-0.5ex]
LHPC 12A & \cite{Green:2012ud} & 2+1 & 149,317
& \parbox[t]{6.3cm}{Chiral fit formula based on the ``small scale
  expansion'' to order $\epsilon^3$ with some coefficients fixed.}
\\[0.5ex]\\[0.5ex]  \hline\\[-0.5ex]
LHPC 10 & \cite{Bratt:2010jn} & 2+1 & 293 & \parbox[t]{6.3cm}{Chiral
  fit formula based on the ``small scale expansion'' to order
  $\epsilon^3$ with some coefficients fixed.}
\\[0.5ex]\\[0.5ex]  \hline\\[-0.5ex]
RBC/UKQCD 10D & \cite{Aoki:2010xg} & 2+1 & 329,416,555,668
&  \parbox[t]{6.3cm}{Constant fit to heaviest three and linear fit to
  lightest two pion masses gives the quoted range.}
\\[0.5ex]\\[0.5ex]  \hline\\[-0.5ex]
RBC/UKQCD 09B & \cite{Yamazaki:2009zq} & 2+1 & 329
& \parbox[t]{6.3cm}{Linear fit in $M_\pi^2$, supplemented by a term in
  $e^{-M_\pi L}$ to describe finite-volume effects.}
\\[0.5ex]\\[0.5ex]  \hline\\[-0.5ex]
RBC/UKQCD 08B & \cite{Yamazaki:2008py} & 2+1 & 329
& \parbox[t]{6.3cm}{Linear fit in $M_\pi^2$, supplemented by a term in 
  $e^{-M_\pi L}$ to describe finite-volume effects.}
\\[0.5ex]\\[0.5ex]  \hline\\[-0.5ex]
LHPC 05 & \cite{Edwards:2005ym} & 2+1 & 353 & \parbox[t]{6.3cm}{Chiral
  fit formula based on the ``small scale expansion'' with some
  coefficients fixed.}
\\[0.5ex]\\[0.5ex]
\hline \hline \\[-1.0ex]
\end{tabular*}
\caption{(cntd.) Chiral extrapolation/minimum pion mass in determinations of
  the isovector axial, scalar and tensor charges with $\Nf=2+1$
  quark flavours.} 
}
\end{table}

\begin{table}[!ht]
{\footnotesize
\begin{tabular*}{\textwidth}{l @{\extracolsep{\fill}} c c c l}
\hline\hline \\[-1.0ex]
{Collab.} & {Ref.} & {$\Nf$} & {$M_{\pi,\rm min}\,[\mev]$} & {Description}  
\\[.3ex]\\[-2.ex] \hline \hline \\[-1.0ex]
%
Mainz 17 & \cite{Capitani:2017qpc} & 2 & 268,193,261
& \parbox[t]{6.3cm}{Main result from linear fit in $M_\pi^2$ with mass
  cut of $M_\pi\leq 300$\,MeV. Systematic error estimated from
  considering fits containing logarithmic terms.}
\\[0.5ex]\\[0.5ex]  \hline\\[-0.5ex]
ETM 17 & \cite{Alexandrou:2017qyt} & 2 & 130
& \parbox[t]{6.3cm}{Single, nearly physical, pion mass.}
\\[0.5ex]\\[0.5ex]  \hline\\[-0.5ex]
ETM 17B & \cite{Alexandrou:2017hac} & 2 & 130
& \parbox[t]{6.3cm}{Single, nearly physical, pion mass.}
\\[0.5ex]\\[0.5ex]  \hline\\[-0.5ex]
ETM 15D & \cite{Abdel-Rehim:2015owa} & 2 & 130
& \parbox[t]{6.3cm}{Single, nearly physical, pion mass.}
\\[0.5ex]\\[0.5ex]  \hline\\[-0.5ex]
RQCD 14 & \cite{Bali:2014nma} & 2 & 280,150,260
& \parbox[t]{6.3cm}{Linear fit in $M_\pi^2$ with pion mass cut
  ($M_\pi^2<0.1$~GeV$^2$).} 
\\[0.5ex]\\[0.5ex]  \hline\\[-0.5ex]
QCDSF 13 & \cite{Horsley:2013ayv} & 2 & 259,158,262
& \parbox[t]{6.3cm}{Chiral fit of $g_A/f_\pi$ including linear,
  quadratic and logarithmic terms in $M_\pi^2$.}
\\[0.5ex]\\[0.5ex]  \hline\\[-0.5ex]
Mainz 12 & \cite{Capitani:2012gj} & 2 & 312,277,430
& \parbox[t]{6.3cm}{Main result from linear fit in
  $M_\pi^2$. Systematic error estimated from considering additional
  terms containing $M_\pi^4$ and $M_\pi L$.}
\\[0.5ex]\\[0.5ex]  \hline\\[-0.5ex]
RBC 08 & \cite{Lin:2008uz} & 2 & 493 &  \parbox[t]{6.3cm}{Linear fit
  in $M_\pi^2$.}
\\[0.5ex]\\[0.5ex]  \hline\\[-0.5ex]
QCDSF 06 & \cite{Khan:2006de} & 2 & 617,664,666,726
& \parbox[t]{6.3cm}{Chiral fit formula based on the ``small scale 
  expansion'' with some fixed coefficients and finite-volume
  corrections subtracted.}
\\[0.5ex]\\[0.5ex] 
\hline \hline \\[-1.0ex]
\end{tabular*}
\caption{Chiral extrapolation/minimum pion mass in determinations of
  the isovector axial, scalar and tensor charges with $\Nf=2$
  quark flavours.} 
}
\end{table}
\fi

\begin{table}[!ht]
{\footnotesize
\begin{tabular*}{\textwidth}{l @{\extracolsep{\fill}} c c c c l}
\hline\hline \\[-1.0ex]
Collab. & Ref. & $\Nf$ & $L$ [fm] & ${M_{\pi,\rm min}}L$ & Description 
\\[.3ex]\\[-2.ex] \hline \hline \\[-1.0ex]
%
CalLat 19 & \cite{Walker-Loud:2019cif} & 2+1+1
& \parbox[t]{1.4cm}{2.4--7.2,\\2.9--5.8,\\2.9--4.2}
& \parbox[t]{1.cm}{4.9,\\3.8,\\4.7} & \parbox[t]{5.9cm}{Fit
  performed including a term of the form
  ${M_\pi^2}e^{-M_\pi L}/{\sqrt{M_{\pi}L}}$ as part of a
  simultaneous fit in $a^2$, $M_\pi$ and $M_\pi L$.}
\\[0.5ex]\\[0.5ex]  \hline\\[-0.5ex]
ETM 19 & \cite{Alexandrou:2019brg} & 2+1+1
& \parbox[t]{1.4cm}{5.12,\\4.5-6.0}
& \parbox[t]{1.cm}{3.62,\\2.98 } & \parbox[t]{5.9cm}{Final result quoted from the $M_\pi L = 3.62$ ensemble only.}
\\[0.5ex]\\[0.5ex]
\ifx\reducedapptables\undefined
\hline \\[-0.5ex]
PNDME 18 & \cite{Gupta:2018qil} & 2+1+1
& \parbox[t]{1.4cm}{2.4,\\2.9--4.8,\\2.8--5.6,\\2.8--5.5}
& \parbox[t]{1.cm}{3.9,\\5.5,\\3.9,\\3.7} & \parbox[t]{5.9cm}{Fit
  performed including a term of the form $M_\pi^2e^{-M_\pi L}$ as
  part of a simultaneous fit in $a$, $M_\pi$ and $M_\pi L$.}
\\[0.5ex]\\[0.5ex]  \hline\\[-0.5ex]
CalLat 18 & \cite{Chang:2018uxx} & 2+1+1
& \parbox[t]{1.4cm}{2.4--4.8,\\2.9--5.8,\\2.9--4.2}
& \parbox[t]{1.cm}{3.2,\\3.9,\\4.7} & \parbox[t]{5.9cm}{Fit
  performed including a term of the form
  ${M_\pi^2}e^{-M_\pi L}/{\sqrt{M_{\pi}L}}$ as part of a
  simultaneous fit in $a^2$, $M_\pi$ and $M_\pi L$.}
\\[0.5ex]\\[0.5ex]  \hline\\[-0.5ex]
CalLat 17 & \cite{Berkowitz:2017gql} & 2+1+1
& \parbox[t]{1.4cm}{2.4--4.8,\\2.9--4.8,\\2.9}
& \parbox[t]{1.cm}{3.2,\\5.4,\\4.5} & \parbox[t]{5.9cm}{Fit
  performed including a term of the form
  ${M_\pi^2}e^{-M_\pi L}/{\sqrt{M_{\pi}L}}$ as part of a
  simultaneous fit in $a^2$, $M_\pi$ and $M_\pi L$.}
\\[0.5ex]\\[0.5ex]  \hline\\[-0.5ex]
PNDME 16 & \cite{Bhattacharya:2016zcn} & 2+1+1
& \parbox[t]{1.4cm}{2.9--4.8,\\2.8--5.6,\\2.8--3.7}
& \parbox[t]{1.cm}{5.5,\\3.9,\\4.4} & \parbox[t]{5.9cm}{Fit performed
  including a term of the form $M_\pi^2e^{-M_\pi L}$ as part of a
  simultaneous fit in $a$, $M_\pi$ and $M_\pi L$.} 
\\[0.5ex]\\[0.5ex]  \hline\\[-0.5ex]
PNDME 15 & \cite{Bhattacharya:2015wna,Bhattacharya:2015esa} & 2+1+1
& \parbox[t]{1.4cm}{2.9--4.8,\\2.8--5.6,\\2.8--3.7}
& \parbox[t]{1.cm}{5.5,\\3.9,\\4.4} & \parbox[t]{5.9cm}{Fit performed
  including a term of the form $\exp\{-M_\pi L\}$ as part of a
  simultaneous fit in $a$, $M_\pi$ and $M_\pi L$.} 
\\[0.5ex]\\[0.5ex]  \hline\\[-0.5ex]
PNDME 13 & \cite{Bhattacharya:2013ehc} & 2+1+1
& \parbox[t]{1.4cm}{2.9--3.8} & \parbox[t]{1.cm}{4.4}
& \parbox[t]{6.3cm}{Finite-volume effects not estimated.}
\\[0.5ex]\\[0.5ex]
\fi
\hline\hline  \\[-1.0ex]
\end{tabular*}
\caption{Finite-volume effects in determinations of the isovector
  axial, scalar and tensor charges with $\Nf=2+1+1$ quark flavours.}
}
\end{table}

\begin{table}[!ht]
{\footnotesize
\begin{tabular*}{\textwidth}{l @{\extracolsep{\fill}} c c c c l}
\hline\hline \\[-1.0ex]
Collab. & Ref. & $\Nf$ & $L$ [fm] & ${M_{\pi,\rm min}}L$ & Description 
\\[.3ex]\\[-2.ex] \hline \hline \\[-1.0ex]
%
$\chi$QCD 21 & \cite{Liu:2021irg} & 2+1 & \parbox[t]{1.4cm}{4.5\\5.3,\\2.6,\\2.6,\\1.92} & \parbox[t]{1.4cm}{3.7,\\3.9,\\3.6,\\3.8,\\3.6}
& \parbox[t]{6.3cm}{Extrapolation performed including a term of the form
  $ e^{-M_\pi L}$ as part of a simultaneous fit in $a$, $M_\pi$ and $M_\pi L$.}
\\[0.5ex]\\[0.5ex]  \hline\\[-0.5ex]
NME 21 & \cite{Park:2021ypf} & 2+1 & \parbox[t]{1.4cm}{4.2,\\2.9-4.3,\\4.3-5.8,\\3.4-5.0} & \parbox[t]{1.4cm}{5.87,\\4.09,\\3.75,\\4.28}
& \parbox[t]{6.3cm}{Extrapolation performed including a term of the form
  ${M_\pi^2}e^{-M_\pi L}/{\sqrt{M_{\pi}L}}$ as part of a
  simultaneous fit in $a^2$, $M_\pi$ and $M_\pi L$.}
\\[0.5ex]\\[0.5ex]  \hline\\[-0.5ex]
RBC/UKQCD 19 & \cite{Abramczyk:2019fnf} & 2+1 & \parbox[t]{1.4cm}{4.5} & \parbox[t]{1.4cm}{4.0}
& \parbox[t]{6.3cm}{Does not quote extrapolated value.}
\\[0.5ex]\\[0.5ex]  \hline\\[-0.5ex]
LHPC 19 & \cite{Hasan:2019noy} & 2+1 &	\parbox[t]{1.4cm}{5.9,\\5.6} & \parbox[t]{1.4cm}{4.0,\\3.9}
& \parbox[t]{6.3cm}{Does not quote extrapolated value.}
\\[0.5ex]\\[0.5ex]  \hline\\[-0.5ex]
Mainz 19 & \cite{Harris:2019bih} & 2+1 &
\parbox[t]{1.4cm}{2.8-4.1,\\2.4-3.6,\\2.1-4.1,2.4-3.2} &
\parbox[t]{1.4cm}{4.7,\\5.3,\\4.2,\\4.3}
& \parbox[t]{6.3cm}{Extrapolation performed including a term of the
form ${M_\pi^2}e^{-M_\pi L}/{\sqrt{M_{\pi}L}}$ as part of a
  simultaneous fit in $a^2$, $M_\pi$ and $M_\pi L$.}
\\[0.5ex]\\[0.5ex]  \hline\\[-0.5ex]
PACS 18A & \cite{Shintani:2018ozy} & 2+1 & \parbox[t]{1.4cm}{10.8}
& \parbox[t]{1.cm}{7.4} & \parbox[t]{5.9cm}{Single spatial volume of
  10.8\,fm at physical pion mass $M_\pi L=7.4$.}
\\[0.5ex]\\[0.5ex] 
\ifx\reducedapptables\undefined
\hline \\[-0.5ex]
PACS 18 & \cite{Ishikawa:2018rew} & 2+1 & \parbox[t]{1.4cm}{8.2}
& \parbox[t]{1.cm}{6.0} & \parbox[t]{5.9cm}{Single spatial volume of
  8.2\,fm at near-physical pion mass $M_\pi L=6$.}
\\[0.5ex]\\[0.5ex] 
\fi
\hline\hline  \\[-1.0ex]
\end{tabular*}
\caption{Finite-volume effects in determinations of the isovector
  axial, scalar and tensor charges with $\Nf=2+1$ quark flavours.}
}
\end{table}

\ifx\reducedapptables\undefined
\begin{table}[!ht]
\addtocounter{table}{-1}
{\footnotesize
\begin{tabular*}{\textwidth}{l @{\extracolsep{\fill}} c c c c l}
\hline\hline \\[-1.0ex]
Collab. & Ref. & $\Nf$ & $L$ [fm] & ${M_{\pi,\rm min}}L$ & Description 
\\[.3ex]\\[-2.ex] \hline \hline \\[-1.0ex]
$\chi$QCD 18 & \cite{Liang:2018pis} & 2+1
& \parbox[t]{1.4cm}{4.6,\\2.7,\\2.6}
& \parbox[t]{1.cm}{4.0,\\4.5,\\4.1} & \parbox[t]{5.9cm}{Fit performed
  including a term of the form $e^{-M_\pi L}$ as part of a
  simultaneous fit in $a$, $M_\pi$ and $M_\pi L$.}
\\[0.5ex]\\[0.5ex]  \hline\\[-0.5ex]
JLQCD 18 & \cite{Yamanaka:2018uud} & 2+1
& \parbox[t]{1.4cm}{1.8--2.7} & \parbox[t]{1.cm}{4.0}
& \parbox[t]{5.9cm}{Finite-volume effects not estimated.}
\\[0.5ex]\\[0.5ex]  \hline\\[-0.5ex]
LHPC 12 & \cite{Green:2012ej} & 2+1
& \parbox[t]{1.4cm}{2.8--5.6,\\2.9} & \parbox[t]{1.cm}{4.2,\\4.6}
& \parbox[t]{6.3cm}{No uncertainty estimated. Comparison between two
  volumes at a single pion mass shows no deviation within statistical
  errors.}
\\[0.5ex]\\[0.5ex]  \hline\\[-0.5ex]
LHPC 12A & \cite{Green:2012ud} & 2+1
& \parbox[t]{1.4cm}{2.8--5.6,\\2.9} & \parbox[t]{1.cm}{4.2,\\4.6}
& \parbox[t]{5.9cm}{Finite-volume effects investigated and found to be
  negligible.}
\\[0.5ex]\\[0.5ex]  \hline\\[-0.5ex]
LHPC 10 & \cite{Bratt:2010jn} & 2+1 & \parbox[t]{1.4cm}{2.5--3.5}
& \parbox[t]{1.cm}{3.7} & \parbox[t]{5.9cm}{Finite-volume effects 
  included in chiral fit formula and found to be negligible.}
\\[0.5ex]\\[0.5ex]  \hline\\[-0.5ex]
RBC/UKQCD 10D & \cite{Aoki:2010xg} & 2+1 & \parbox[t]{1.4cm}{2.7}
& \parbox[t]{1.cm}{4.6} & \parbox[t]{5.9cm}{No uncertainty
  estimated. Comparison between two volumes for the three heaviest 
  pion masses shows no deviation within statistical
  errors.}
\\[0.5ex]\\[0.5ex]  \hline\\[-0.5ex]
RBC/UKQCD 09B & \cite{Yamazaki:2009zq} & 2+1 & \parbox[t]{1.4cm}{2.7}
& \parbox[t]{1.cm}{4.6} & \parbox[t]{5.9cm}{Chiral fit ansatz contains
  term in $e^{-M_\pi L}$ to describe finite-volume effects.}
\\[0.5ex]\\[0.5ex]  \hline\\[-0.5ex]
RBC/UKQCD 08B & \cite{Yamazaki:2008py} & 2+1 & \parbox[t]{1.4cm}{2.7}
& \parbox[t]{1.cm}{4.6} & \parbox[t]{5.9cm}{Chiral fit ansatz contains
  term in $e^{-M_\pi L}$ to describe finite-volume effects.}
\\[0.5ex]\\[0.5ex]  \hline\\[-0.5ex]
LHPC 05 & \cite{Edwards:2005ym} & 2+1 & \parbox[t]{1.4cm}{2.5--3.5}
& \parbox[t]{1.cm}{6.2} & \parbox[t]{5.9cm}{Finite-volume effects
  included in chiral fit formula and found to be negligible.}
\\[0.5ex]\\[0.5ex]
\hline\hline  \\[-1.0ex]
\end{tabular*}
\caption{(cntd.) Finite-volume effects in determinations of the isovector
  axial, scalar and tensor charges with $\Nf=2+1$ quark flavours.}
}
\end{table}

\begin{table}[!ht]
{\footnotesize
\begin{tabular*}{\textwidth}{l @{\extracolsep{\fill}} c c c c l}
\hline\hline \\[-1.0ex]
Collab. & Ref. & $\Nf$ & $L$ [fm] & ${M_{\pi,\rm min}}L$ & Description 
\\[.3ex]\\[-2.ex] \hline \hline \\[-1.0ex]
%
Mainz 17 & \cite{Capitani:2017qpc} & 2
& \parbox[t]{1.4cm}{2.5--3.8,\\2.0--4.0,\\2.4--3.2}
& \parbox[t]{1.cm}{5.0,\\4.0,\\4.4} & \parbox[t]{5.9cm}{Finite-volume
  effects not estimated.}
\\[0.5ex]\\[0.5ex]  \hline\\[-0.5ex]
ETM 17 & \cite{Alexandrou:2017qyt} & 2 & \parbox[t]{1.4cm}{4.5}
& \parbox[t]{1.cm}{3.0} & \parbox[t]{6.3cm}{Finite-volume effects not
  estimated.}
\\[0.5ex]\\[0.5ex]  \hline\\[-0.5ex]
ETM 17B & \cite{Alexandrou:2017hac} & 2 & \parbox[t]{1.4cm}{4.5}
& \parbox[t]{1.cm}{3.0} & \parbox[t]{5.9cm}{Finite-volume effects not
  estimated.}
\\[0.5ex]\\[0.5ex]  \hline\\[-0.5ex]
ETM 15D & \cite{Abdel-Rehim:2015owa} & 2 & \parbox[t]{1.4cm}{4.5}
& \parbox[t]{1.cm}{3.0} & \parbox[t]{5.9cm}{Finite-volume effects not
  estimated.}
\\[0.5ex]\\[0.5ex]  \hline\\[-0.5ex]
RQCD 14 & \cite{Bali:2014nma} & 2
& \parbox[t]{1.4cm}{2.6,\\1.7--4.5,\\1.9--2.9}
& \parbox[t]{1.cm}{3.7,\\3.5,\\3.8} & \parbox[t]{5.9cm}{No uncertainty
  estimated. Comparison between volumes for two pion masses, including
  the lightest, shows no deviation within statistical errors.} 
\\[0.5ex]\\[0.5ex]  \hline\\[-0.5ex]
QCDSF 13 & \cite{Horsley:2013ayv} & 2
& \parbox[t]{1.4cm}{1.2--2.4,\\0.9--3.4,\\1.4--2.9}
& \parbox[t]{1.cm}{3.2,\\2.7,\\3.8} & \parbox[t]{5.9cm}{Finite-volume
  effects assumed to be negligible in ratio $g_A/f_\pi$.}
\\[0.5ex]\\[0.5ex]  \hline\\[-0.5ex]
Mainz 12 & \cite{Capitani:2012gj} & 2
& \parbox[t]{1.4cm}{2.5,\\2.0--3.0,\\2.4}
& \parbox[t]{1.cm}{4.0,\\4.3,\\5.2} & \parbox[t]{5.9cm}{Finite-volume
  corrections estimated in $\chi$PT and subtracted before chiral fit.}
\\[0.5ex]\\[0.5ex]  \hline\\[-0.5ex]
RBC 08 & \cite{Lin:2008uz} & 2 & \parbox[t]{1.4cm}{1.8}
& \parbox[t]{1.cm}{4.6} & \parbox[t]{5.9cm}{Finite-volume effects
  investigated, but not included in error budget. All volumes below 2\,fm.}
\\[0.5ex]\\[0.5ex]  \hline\\[-0.5ex]
QCDSF 06 & \cite{Khan:2006de} & 2
& \parbox[t]{1.4cm}{1.5,\\1.2--1.8,\\0.9--1.7,\\1.4}
& \parbox[t]{1.cm}{4.7,\\6.1,\\5.7,\\5.3}
& \parbox[t]{5.9cm}{Finite-volume effects included in fit formula
  based on ``small scale expansion''. All spatial volumes below
  2\,fm.}
\\[0.5ex]\\[0.5ex]
\hline\hline  \\[-1.0ex]
\end{tabular*}
\caption{Finite-volume effects in determinations of the isovector
  axial, scalar and tensor charges with $\Nf=2$ quark flavours.}
}
\end{table}
\fi

\begin{table}[!ht]
{\footnotesize
\begin{tabular*}{\textwidth}{l @{\extracolsep{\fill}} c c c l}
\hline\hline \\[-1.0ex]
Collab. & Ref. & $\Nf$ & Ren.
\\[.3ex]\\[-2.ex] \hline \hline \\[-1.0ex]
%
CalLat 19 & \cite{Walker-Loud:2019cif} & 2+1+1 & RI-MOM  \\[0.5ex]\\[0.5ex]
\hline\\[-0.5ex]
ETM 19 & \cite{Alexandrou:2019brg} & 2+1+1 & RI-MOM  \\[0.5ex]\\[0.5ex]
\ifx\reducedapptables\undefined
\hline \\[-0.5ex]
PNDME 18 & \cite{Gupta:2018qil} & 2+1+1 & RI-SMOM  \\[0.5ex]\\[0.5ex]
\hline\\[-0.5ex]
CalLat 18 & \cite{Chang:2018uxx} & 2+1+1 & RI-MOM  \\[0.5ex]\\[0.5ex]
\hline\\[-0.5ex]
CalLat 17 & \cite{Berkowitz:2017gql} & 2+1+1 & RI-MOM  \\[0.5ex]\\[0.5ex]
\hline\\[-0.5ex]
PNDME 16 & \cite{Bhattacharya:2016zcn} & 2+1+1 & RI-SMOM  \\[0.5ex]\\[0.5ex]
\hline\\[-0.5ex]
PNDME 15 & \cite{Bhattacharya:2015wna,Bhattacharya:2015esa} & 2+1+1 & RI-SMOM    \\[0.5ex]\\[0.5ex]
\hline\\[-0.5ex]
PNDME 13 & \cite{Bhattacharya:2013ehc} & 2+1+1 & RI-SMOM    \\[0.5ex]\\[0.5ex]
\fi
 \hline \hline \\[-1.0ex]
\end{tabular*}
\caption{Renormalization in determinations of the isovector axial,
  scalar and tensor charges with $\Nf=2+1+1$ quark
  flavours.}
}
\end{table}

\begin{table}[!ht]
{\footnotesize
\begin{tabular*}{\textwidth}{l @{\extracolsep{\fill}} c c c l}
\hline\hline \\[-1.0ex]
Collab. & Ref. & $\Nf$ & Ren.
\\[.3ex]\\[-2.ex] \hline \hline \\[-1.0ex]
$\chi$QCD 21 & \cite{Liu:2021irg} & 2+1 & RI-MOM  \\[0.5ex]\\[0.5ex]
\hline\\[-0.5ex]
NME 21 & \cite{Park:2021ypf} & 2+1 & RI-SMOM  \\[0.5ex]\\[0.5ex]
\hline\\[-0.5ex]
RBC/UKQCD 19 & \cite{Abramczyk:2019fnf} & 2+1 & RI-SMOM  \\[0.5ex]\\[0.5ex]
\hline\\[-0.5ex]
LHPC 19 & \cite{Hasan:2019noy} & 2+1 & RI-MOM, RI-SMOM  \\[0.5ex]\\[0.5ex]
\hline\\[-0.5ex]
Mainz 19 & \cite{Harris:2019bih} & 2+1 & RI-MOM, SF
\\[0.5ex]\\[0.5ex]
\hline\\[-0.5ex]
PACS 18A & \cite{Shintani:2018ozy} & 2+1 & SF  \\[0.5ex]\\[0.5ex]
\ifx\reducedapptables\undefined
\hline \\[-0.5ex]
PACS 18 & \cite{Ishikawa:2018rew} & 2+1 & SF  \\[0.5ex]\\[0.5ex]
\hline\\[-0.5ex]
$\chi$QCD 18 & \cite{Liang:2018pis} & 2+1 & RI-MOM  \\[0.5ex]\\[0.5ex]
\hline\\[-0.5ex]
JLQCD 18 & \cite{Yamanaka:2018uud} & 2+1 & RI-MOM  \\[0.5ex]\\[0.5ex]
\hline\\[-0.5ex]
LHPC 12 & \cite{Green:2012ej} & 2+1 & RI-MOM  \\[0.5ex]\\[0.5ex]
\hline\\[-0.5ex]
LHPC 12A & \cite{Green:2012ud} & 2+1 & RI-MOM  \\[0.5ex]\\[0.5ex]
\hline\\[-0.5ex]
LHPC 10 & \cite{Bratt:2010jn} & 2+1 & RI-MOM  \\[0.5ex]\\[0.5ex]
\hline\\[-0.5ex]
RBC/UKQCD 10D & \cite{Aoki:2010xg} & 2+1 & RI-MOM    \\[0.5ex]\\[0.5ex]
\hline\\[-0.5ex]
RBC/UKQCD 09B & \cite{Yamazaki:2009zq} & 2+1 & $g_A/g_V$  \\[0.5ex]\\[0.5ex]
\hline\\[-0.5ex]
RBC/UKQCD 08B & \cite{Yamazaki:2008py} & 2+1 & $g_A/g_V$  \\[0.5ex]\\[0.5ex]
\hline\\[-0.5ex]
LHPC 05 & \cite{Edwards:2005ym} & 2+1 & $g_A/g_V$  \\[0.5ex]\\[0.5ex]
\fi
 \hline \hline \\[-1.0ex]
\end{tabular*}
\caption{Renormalization in determinations of the isovector axial,
  scalar and tensor charges with $2+1$ quark
  flavours.}
}
\end{table}

\ifx\reducedapptables\undefined
\begin{table}[!ht]
{\footnotesize
\begin{tabular*}{\textwidth}{l @{\extracolsep{\fill}} c c c l}
\hline\hline \\[-1.0ex]
Collab. & Ref. & $\Nf$ & Ren.
\\[.3ex]\\[-2.ex] \hline \hline \\[-1.0ex]
%
Mainz 17 & \cite{Capitani:2017qpc} & 2 & SF  \\[0.5ex]\\[0.5ex]
\hline\\[-0.5ex]
ETM 17B & \cite{Alexandrou:2017hac} & 2 & RI'-MOM  \\[0.5ex]\\[0.5ex]
\hline\\[-0.5ex]
ETM 15D & \cite{Abdel-Rehim:2015owa} & 2 & RI'-MOM  \\[0.5ex]\\[0.5ex]
\hline\\[-0.5ex]
RQCD 14 & \cite{Bali:2014nma} & 2 & RI'-MOM  \\[0.5ex]\\[0.5ex]
\hline\\[-0.5ex]
QCDSF 13 & \cite{Horsley:2013ayv} & 2 & $g_A/f_\pi$  \\[0.5ex]\\[0.5ex]
\hline\\[-0.5ex]
Mainz 12 & \cite{Capitani:2012gj} & 2 & SF  \\[0.5ex]\\[0.5ex]
\hline\\[-0.5ex]
RBC 08 & \cite{Lin:2008uz} & 2 & $g_A/g_V$  \\[0.5ex]\\[0.5ex]
\hline\\[-0.5ex]
QCDSF 06 & \cite{Khan:2006de} & 2 & RI'-MOM  \\[0.5ex]\\[0.5ex]
 \hline \hline \\[-1.0ex]
\end{tabular*}
\caption{Renormalization in determinations of the isovector axial,
  scalar and tensor charges with $\Nf=2$ quark
  flavours.}
}
\end{table}
\fi

\begin{table}[!ht]
{\footnotesize
\begin{tabular*}{\textwidth}{l @{\extracolsep{\fill}} c c c l}
\hline\hline \\[-1.0ex]
Collab. & Ref. & $\Nf$ & $\tau$ [fm] & Description 
\\[.3ex]\\[-2.ex]\hline \hline \\[-1.0ex]
%
CalLat 19 & \cite{Walker-Loud:2019cif} & 2+1+1 & all
& \parbox[t]{5.4cm}{Two-state fits to the $\tau$-dependence of summed
  operator insertion for $\tau\geq0.3$\,fm.}
\\[0.5ex]\\[0.5ex]  \hline\\[-0.5ex]
ETM 19 & \cite{Alexandrou:2019brg} & 2+1+1
& \parbox[t]{3.5cm}{[0.64--1.6]}
& \parbox[t]{5.4cm}{Fits to the $\tau$- and $t$-dependence of three-point
  correlators using multiple methods (plateau, two- and three-state fits, summation), with the two-state fits quoted for the final result.}
\\[0.5ex]\\[0.5ex]
\ifx\reducedapptables\undefined
\hline \\[-0.5ex]
PNDME 18 & \cite{Gupta:2018qil} & 2+1+1
& \parbox[t]{3.5cm}{[0.8--1.4]\par [1.0--1.4,1.0--1.4]\par
  [0.9--1.2,0.9--1.2,0.7--1.4]\par [0.9--1.4,0.9--1.4,0.9--1.2]}
& \parbox[t]{5.4cm}{Fits to the $\tau$- and $t$-dependence of three-point
  correlators using two or three lowest-lying states.}
\\[0.5ex]\\[0.5ex]  \hline\\[-0.5ex]
CalLat 18 & \cite{Chang:2018uxx} & 2+1+1 & all
& \parbox[t]{5.4cm}{Two-state fits to the $\tau$-dependence of summed
  operator insertion for $\tau\geq0.3$\,fm.}
\\[0.5ex]\\[0.5ex]  \hline\\[-0.5ex]
CalLat 17 & \cite{Berkowitz:2017gql} & 2+1+1 & all
& \parbox[t]{5.4cm}{Two-state fits to the $\tau$-dependence of summed
  operator insertion for $\tau\geq0.3$\,fm.}
\\[0.5ex]\\[0.5ex]  \hline\\[-0.5ex]
PNDME 16 & \cite{Bhattacharya:2016zcn} & 2+1+1
& \parbox[t]{3.5cm}{[1.0--1.4,1.0--1.4]\par
  [0.9--1.2,0.9--1.2,0.9--1.2]\par [0.9--1.4,0.9--1.4]}
& \parbox[t]{5.4cm}{Fits to the $\tau$ and $t$-dependence of three-point
  correlators using two or three lowest-lying states.}
\\[0.5ex]\\[0.5ex]  \hline\\[-0.5ex]
PNDME 15 & \cite{Bhattacharya:2015wna,Bhattacharya:2015esa} & 2+1+1
& \parbox[t]{3.5cm}{[1.0--1.4,1.0--1.4]\par
  [0.9--1.2,0.9--1.2,0.9--1.2]\par [0.9--1.4,0.9--1.4]}
& \parbox[t]{5.4cm}{Fits including up to two states are
  investigated.} \\[0.5ex]\\[0.5ex] 
\hline\\[-0.5ex]
PNDME 13 & \cite{Bhattacharya:2013ehc} & 2+1+1
& \parbox[t]{3.5cm}{[1.0--1.4,0.9--1.4]} &   \parbox[t]{5.4cm}{Fits
  including up to two states are investigated.}
\\[0.5ex]\\[0.5ex]
\fi
\hline \hline \\[-1.0ex]
\end{tabular*}
\caption{Control of excited state contamination in determinations of
  the isovector axial, scalar and tensor charges with $\Nf=2+1+1$
  quark flavours. The comma-separated list of numbers in square brackets denote the range of source-sink separations $\tau$ (in fermi) at each value of the bare coupling.}
}
\end{table}

\begin{table}[!ht]
{\footnotesize
\begin{tabular*}{\textwidth}{l @{\extracolsep{\fill}} c c c l}
\hline\hline \\[-1.0ex]
Collab. & Ref. & $\Nf$ & $\tau$ [fm] & Description 
\\[.3ex]\\[-2.ex]\hline \hline \\[-1.0ex]
%
$\chi$QCD 21 & \cite{Liu:2021irg} &  2+1
& \parbox[t]{3.5cm}{[0.9--1.1]\par [1--1.2]\par[0.9--1.3,0.9--1.3]\par[1--1.5]}
& \parbox[t]{5.4cm}{Fits to the $\tau$- and $t$-dependence of three-point
  correlators using up to three lowest-lying states.}
\\[0.5ex]\\[0.5ex]  \hline\\[-0.5ex]
NME 21 & \cite{Park:2021ypf} & 2+1
& \parbox[t]{3.5cm}{[1.0--1.8]\par [0.75--1.3,0.75--1.7]\par
  [0.72--1.46,0.72--1.46]\par [0.8--1.4,0.9--1.5]}
& \parbox[t]{5.4cm}{Fits to the $\tau$- and $t$-dependence of three-point
  correlators using three lowest-lying states.}
\\[0.5ex]\\[0.5ex]  \hline\\[-0.5ex]
RBC/UKQCD 19 & \cite{Abramczyk:2019fnf} & 2+1 &  \parbox[t]{3.5cm}{[1.0,1.3]}
& \parbox[t]{5.4cm}{Result quoted from $t$=1.3~fm point only; consistency in results from $t$=1.0~fm is demonstrated.}
\\[0.5ex]\\[0.5ex]  \hline\\[-0.5ex]
LHPC 19 & \cite{Hasan:2019noy} & 2+1 &\parbox[t]{3.5cm}{[0.4--1.4]\par [0.9-1.5]}
& \parbox[t]{5.4cm}{Multiple analysis methods used (plateau, summation, and two-state fits to each), and combined to produce final result.}
\\[0.5ex]\\[0.5ex]  \hline\\[-0.5ex]
Mainz 19 & \cite{Harris:2019bih} & 2+1
& \parbox[t]{3.5cm}{[1.0--1.4,1.0--1.4,1.0--1.4]\par
  [1.0--1.5,1.0--1.5]\par [1.0--1.4,1.0--1.4,1.0--1.4,1.0--1.4]\par
  [1.0--1.4,1.0--1.3]} & \parbox[t]{5.4cm}{Fits to the $\tau$- and $t$-dependence of correlator ratios using the two lowest-lying states.}
\\[0.5ex]\\[0.5ex]  \hline\\[-0.5ex]
PACS 18A & \cite{Shintani:2018ozy} & 2+1 &
\parbox[t]{3.5cm}{[0.9--1.4]}
& \parbox[t]{5.4cm}{Average of plateau values at the three largest source-sink separations.}
\\[0.5ex]\\[0.5ex]
\ifx\reducedapptables\undefined
\hline \\[-0.5ex]
PACS 18 & \cite{Ishikawa:2018rew} & 2+1 & \parbox[t]{3.5cm}{[1.3]}
& \parbox[t]{5.4cm}{Plateau fits of correlator ratio at $\tau=1.3$\,fm.}
\\[0.5ex]\\[0.5ex]  \hline\\[-0.5ex]
$\chi$QCD 18 & \cite{Liang:2018pis} & 2+1 & \parbox[t]{3.5cm}{[1.0--1.6]\par [0.9--1.3]\par [1.0--1.2]} &  \\[0.5ex]\\[0.5ex]
\hline\\[-0.5ex]
JLQCD 18 & \cite{Yamanaka:2018uud} & 2+1 & all
& \parbox[t]{5.4cm}{Fits to the leading (ground state) $\tau$- and $t$-dependence.}
\\[0.5ex]\\[0.5ex]
\fi
\hline\hline\\[-0.5ex]
\end{tabular*}
\caption{Control of excited state contamination in determinations of
  the isovector axial, scalar and tensor charges with $\Nf=2+1$
  quark flavours. The comma-separated list of numbers in square brackets denote the range of source-sink separations $\tau$ (in fermi) at each value of the bare coupling.}
}
\end{table}

\ifx\reducedapptables\undefined
\begin{table}[!ht]
\addtocounter{table}{-1}
{\footnotesize
\begin{tabular*}{\textwidth}{l @{\extracolsep{\fill}} c c c l}
\hline\hline \\[-1.0ex]
Collab. & Ref. & $\Nf$ & $\tau$ [fm] & Description 
\\[.3ex]\\[-2.ex]\hline \hline \\[-1.0ex]
%
LHPC 12 & \cite{Green:2012ej} & 2+1
& \parbox[t]{3.5cm}{[0.9--1.4,0.9--1.4,0.9--1.4,0.9--1.4,0.9--1.4]\par
  [0.9--1.4]} &  \parbox[t]{5.3cm}{Result is given by the average
  over central points in plateau, and a fitting uncertainty is
  estimated from the variation over two to three points.}
\\[0.5ex]\\[0.5ex]  \hline\\[-0.5ex]
LHPC 12A & \cite{Green:2012ud} & 2+1
& \parbox[t]{3.5cm}{[0.9--1.4,0.9--1.4,0.9--1.4,0.9--1.4,0.9--1.4]\par
  [0.9--1.4]}
& \parbox[t]{5.4cm}{Fits to the leading (ground state) $\tau$-dependence of summed correlator ratios.}
\\[0.5ex]\\[0.5ex]  \hline\\[-0.5ex]
LHPC 10 & \cite{Bratt:2010jn} & 2+1
& \parbox[t]{3.5cm}{[1.1,1.1,1.1,1.1,1.1--1.2,1.1]}
& \parbox[t]{5.4cm}{Plateau fits of correlator ratio at
  $\tau=1.1$\,fm. Larger source-sink separation on one ensemble as
  cross check.} 
\\[0.5ex]\\[0.5ex] \hline\\[-0.5ex]
RBC/UKQCD 10D & \cite{Aoki:2010xg} & 2+1
& \parbox[t]{3.5cm}{[1.4,1.4,1.4,1.4]} & \parbox[t]{5.4cm}{Single
  source-sink separation considered.}
\\[0.5ex]\\[0.5ex]  \hline\\[-0.5ex]
RBC/UKQCD 09B & \cite{Yamazaki:2009zq} & 2+1
& \parbox[t]{3.5cm}{[1.4,1.4,1.4,1.4]} & \parbox[t]{5.4cm}{Plateau
  fits to correlator ratios for $\tau=1.4$\,fm.}
\\[0.5ex]\\[0.5ex]  \hline\\[-0.5ex]
RBC/UKQCD 08B & \cite{Yamazaki:2008py} & 2+1
& \parbox[t]{3.5cm}{[1.4,1.4,1.4,1.4]} & \parbox[t]{5.4cm}{Plateau
  fits to correlator ratios for $\tau=1.4$\,fm.}
\\[0.5ex]\\[0.5ex]  \hline\\[-0.5ex]
LHPC 05 & \cite{Edwards:2005ym} & 2+1
& \parbox[t]{3.5cm}{[1.1]} & \parbox[t]{5.4cm}{Plateau
  fits of correlator ratio at single value of $\tau$.}
\\[0.5ex]\\[0.5ex]  \hline \hline \\[-1.0ex]
\end{tabular*}
\caption{(cntd.) Control of excited state contamination in determinations of
  the isovector axial, scalar and tensor charges with $\Nf=2+1$
  quark flavours. The comma-separated list of numbers in square brackets denote the range of source-sink separations $\tau$ (in fermi) at each value of the bare coupling.}
}
\end{table}

\begin{table}[!ht]
{\footnotesize
\begin{tabular*}{\textwidth}{l @{\extracolsep{\fill}} c c c l}
\hline\hline \\[-1.0ex]
Collab. & Ref. & $\Nf$ & $\tau$ [fm] & Description 
\\[.3ex]\\[-2.ex]\hline \hline \\[-1.0ex]
%
Mainz 17 & \cite{Capitani:2017qpc} & 2
& \parbox[t]{3.5cm}{[0.8--1.3,0.8--1.3,0.8--1.3,0.8--1.3]\par
  [0.7--1.1,0.7--1.1,0.7--1.1,0.7--1.1]\par
  [0.7--1.1,0.7--1.1,0.7--1.1]} & \parbox[t]{7.3cm}{Two-state fits of
  correlator ratio with fixed energy gaps. Summation method consistent
  within the total quoted error.}
\\[0.5ex]\\[0.5ex]  \hline\\[-0.5ex]
ETM 17 & \cite{Alexandrou:2017qyt} & 2
& \parbox[t]{3.5cm}{[0.9--1.7]} & \parbox[t]{7.3cm}{Fits including up
  to two states and summation method are considered. Result taken from
  the largest source-sink separation and error estimated from
  comparison with two-state fits.} 
\\[0.5ex]\\[0.5ex]  \hline\\[-0.5ex]
ETM 17B & \cite{Alexandrou:2017hac} & 2
& \parbox[t]{3.5cm}{[0.9--1.3]} & \parbox[t]{7.3cm}{Fits including up
  to two states and the summation method are considered. No excited
  state contamination seen within statistical errors. Result taken
  from the smallest source-sink separation at which plateau fits and
  two-state fits agree. Systematic error estimated as the difference
  with the summation method.}
\\[0.5ex]\\[0.5ex]  \hline\\[-0.5ex]
ETM 15D & \cite{Abdel-Rehim:2015owa} & 2
& \parbox[t]{3.5cm}{[0.9--1.3]} & \parbox[t]{7.3cm}{Fits including up
  to two states and the summation method are considered. No excited
  state contamination seen within statistical errors. Result taken
  from the largest source-sink separation and no error estimated.}
\\[0.5ex]\\[0.5ex]  \hline\\[-0.5ex]
RQCD 14 & \cite{Bali:2014nma} & 2 & \parbox[t]{3.5cm}{[1.1]\par
  [1.1--1.2,0.5--1.2,0.6--1.1]\par [1.1,1.1,1.1]}
& \parbox[t]{7.3cm}{Multiple source-sink separations produced on two
  ensembles to determine a single source-sink separation to be used
  for all other ensembles.}
\\[0.5ex]\\[0.5ex]  \hline\\[-0.5ex]
QCDSF 13 & \cite{Horsley:2013ayv} & 2
& \parbox[t]{3.5cm}{[1.0,1.0,1.0,1.0,1.0]\par
  [1.0,1.0,1.0,0.8--1.3,1.0,1.0,1.0]\par [1.0,1.0,1.0,1.0,1.0,1.0]}
& \parbox[t]{7.3cm}{Plateau fits to $g_A/f_\pi$ for
  $\tau=1.0$\,fm. Multiple source-sink separations investigated on one
  ensemble as cross check.}
\\[0.5ex]\\[0.5ex]  \hline \hline \\[-1.0ex]
\end{tabular*}
\caption{Control of excited state contamination in determinations of
  the isovector axial, scalar and tensor charges with $\Nf=2$
  quark flavours. The comma-separated list of numbers in square brackets denote the range of source-sink separations $\tau$ (in fermi) at each value of the bare coupling.}
}
\end{table}

\begin{table}[!ht]
\addtocounter{table}{-1}
{\footnotesize
\begin{tabular*}{\textwidth}{l @{\extracolsep{\fill}} c c c l}
\hline\hline \\[-1.0ex]
Collab. & Ref. & $\Nf$ & $\tau$ [fm] & Description 
\\[.3ex]\\[-2.ex]\hline \hline \\[-1.0ex]
%
Mainz 12 & \cite{Capitani:2012gj} & 2
& \parbox[t]{3.5cm}{[0.8--1.3,0.8--1.3,0.8--1.3,0.8--1.3]\par
  [0.7--1.1,0.7--1.1,0.7--1.1,0.7--1.1,0.7--1.1]\par
  [0.7--1.1,0.7--1.1]} & \parbox[t]{7.3cm}{Fits to the leading (ground
  state) $\tau$-dependence of summed correlator ratios.}
\\[0.5ex]\\[0.5ex]  \hline\\[-0.5ex]
RBC 08 & \cite{Lin:2008uz} & 2
& \parbox[t]{3.5cm}{[1.1,1.1,1.1--1.4]} & \parbox[t]{7.3cm}{Two
  source-sink separations produced for lightest pion mass to determine
  a single source-sink separation to be used for all other ensembles.}
\\[0.5ex]\\[0.5ex]  \hline\\[-0.5ex]
QCDSF 06 & \cite{Khan:2006de} & 2
& \parbox[t]{3.5cm}{[1.2,1.2,1.2]\par [1.2,1.2,1.2]\par
  [1.2,1.2,1.2,1.2]\par [1.2,1.2,1.2]} & \parbox[t]{7.3cm}{Ground
  state fits performed. No investigation of excited-state effects.}
\\[0.5ex]\\[0.5ex]  \hline \hline \\[-1.0ex]
\end{tabular*}
\caption{(cntd.) Control of excited state contamination in determinations of
  the isovector axial, scalar and tensor charges with $\Nf=2$
  quark flavours. The comma-separated list of numbers in square brackets denote the range of source-sink separations $\tau$ (in fermi) at each value of the bare coupling.}
}
\end{table}
\fi

\clearpage
\clearpage
\begin{table}[!ht]
{\footnotesize
\begin{tabular*}{\textwidth}{l @{\extracolsep{\fill}} c c c l}
\hline\hline \\[-1.0ex]
Collab. & Ref. & $\Nf$ & $a$ [fm] & Description 
\\[.3ex]\\[-2.ex] \hline \hline \\[-1.0ex]
%
ETM 19 & \cite{Alexandrou:2019brg} & 2+1+1 & 0.08 & \parbox[t]{6.3cm}{Not estimated. }
\\[0.5ex]\\[0.5ex]
\ifx\reducedapptables\undefined
\hline \\[-0.5ex]
PNDME 18A & \cite{Lin:2018obj} & 2+1+1 & 0.15,0.12,0.09,0.06 & \parbox[t]{4.7cm}{  Connected: Joint continuum, chiral and volume fit with a term linear in $a$. \\ Disconnected: Joint continuum and chiral fit with a term linear in $a$. }   \\[0.5ex]\\[0.5ex]
\hline\\[-0.5ex]
$\chi$QCD 18 & \cite{Liang:2018pis} & 2+1 & 0.14,0.11,0.08 & \parbox[t]{4.7cm}{Joint continuum and chiral fit with a term linear in $a^2$.}   \\[0.5ex]\\[0.5ex]
\hline\\[-0.5ex]
JLQCD 18 & \cite{Yamanaka:2018uud} & 2+1 & 0.11 &  \parbox[t]{4.7cm}{  Discretization effects are estimated to be $8\%$ using $\cO(\Lambda^2_{QCD}a^2)$ with $\Lambda_{QCD}\approx 500$~MeV.}   \\[0.5ex]\\[0.5ex]
\hline\\[-0.5ex]
$\chi$QCD 15 & \cite{Gong:2015iir} & 2+1 & 0.11 & \parbox[t]{4.7cm}{ Single lattice spacing.}   \\[0.5ex]\\[0.5ex]
\hline\\[-0.5ex]
Engelhardt 12 & \cite{Engelhardt:2012gd} & 2+1 & 0.12 &  \parbox[t]{4.7cm}{ Single lattice spacing.}  \\[0.5ex]\\[0.5ex]
\hline\\[-0.5ex]
ETM 17C & \cite{Alexandrou:2017oeh} & 2 & 0.09 & \parbox[t]{4.7cm}{ Single lattice spacing.}  \\[0.5ex]\\[0.5ex]
\fi
\hline \hline \\[-1.0ex]
\end{tabular*}
\caption{
\ifx\reducedapptables\undefined
Continuum extrapolations/estimation of lattice artifacts in
  determinations of $g_A^q$.
\else
Continuum extrapolations/estimation of lattice artifacts in
  determinations of $g_A^q$ and $g_T^q$.
\fi
}
}
\end{table}

\begin{table}[!ht]
{\footnotesize
\begin{tabular*}{\textwidth}{l @{\extracolsep{\fill}} c c c l}
\hline\hline \\[-1.0ex]
{Collab.} & {Ref.} & {$\Nf$} & {$M_{\pi,\rm min}\,[\mev]$} & {Description}  
\\[.3ex]\\[-2.ex] \hline \hline \\[-1.0ex]
%
ETM 19 & \cite{Alexandrou:2019brg} & 2+1+1 & 139  & \parbox[t]{6.3cm}{ Simulate close to $M_\pi^{phys}$. }
\\[0.5ex]\\[0.5ex]
\ifx\reducedapptables\undefined
\hline \\[-0.5ex]
PNDME 18A & \cite{Lin:2018obj} & 2+1+1 &  \parbox[t]{2.5cm}{ $C$: 320,225,135 \\ $D_l$: 320, 235 \\ $D_s$: 320, 225, 138} &  \parbox[t]{4.7cm}{  Connected: Joint continuum, chiral and volume fit with a term linear in $M_\pi^2$. \par Disconnected: Joint continuum and chiral fit with a term linear in $M_\pi^2$. }  \\[0.5ex]\\[0.5ex]
\hline\\[-0.5ex]
$\chi$QCD 18 & \cite{Liang:2018pis} & 2+1 & 171,337,302 & \parbox[t]{4.7cm}{Joint continuum and chiral fit with a term linear in $M_{\pi,{\rm val}}^2$.}   \\[0.5ex]\\[0.5ex]
\hline\\[-0.5ex]
JLQCD 18 & \cite{Yamanaka:2018uud} & 2+1 & 293 & \parbox[t]{4.7cm}{ Chiral fit with a term linear in $M_\pi^2$. } \\[0.5ex]\\[0.5ex]
\hline\\[-0.5ex]
$\chi$QCD 15 & \cite{Gong:2015iir} & 2+1 & 330 & \parbox[t]{4.7cm}{ Chiral fit with a term linear in $M_{\pi^2, {\rm val}}$. }   \\[0.5ex]\\[0.5ex]
\hline\\[-0.5ex]
Engelhardt 12 & \cite{Engelhardt:2012gd} & 2+1 & 495, 356, 293 & \parbox[t]{4.7cm}{Value from chiral log fit. Difference from constant fit added as a 
 systematic uncertainty.}   \\[0.5ex]\\[0.5ex]
\hline\\[-0.5ex]
ETM 17C & \cite{Alexandrou:2017oeh} & 2 & 130 & \parbox[t]{4.7cm}{Single simulation at $M_\pi = 130$ MeV. }  \\[0.5ex]\\[0.5ex]
\fi
\hline\hline
\end{tabular*}
\caption{
\ifx\reducedapptables\undefined
Chiral extrapolation/minimum pion mass in determinations of $g_A^q$.
\else
Chiral extrapolation/minimum pion mass in determinations of $g_A^q$ and $g_T^q$.
\fi
}
}
\end{table}

\begin{table}[!ht]
{\footnotesize
\begin{tabular*}{\textwidth}{l @{\extracolsep{\fill}} c c c c l}
\hline\hline \\[-1.0ex]
Collab. & Ref. & $\Nf$ & $L$ [fm] & ${M_{\pi,\rm min}}L$ & Description 
\\[.3ex]\\[-2.ex] \hline \hline \\[-1.0ex]
%
ETM 19 & \cite{Alexandrou:2019brg} & 2+1+1 & \parbox[t]{1.4cm}{5.1} &
\parbox[t]{1.cm}{3.6} & \parbox[t]{6.3cm}{FVE are anticipated to be small based on an investigation using two
near physical point $N_f=2$ ensembles with $M_\pi L=3.0$ and $4.0$.}
\\[0.5ex]\\[0.5ex]
\ifx\reducedapptables\undefined
\hline \\[-0.5ex]
PNDME 18A & \cite{Lin:2018obj} & 2+1+1 & \parbox[t]{1.4cm}{2.4,\\2.9--3.8,\\2.8--5.6,\\2.8} & \parbox[t]{1.cm}{3.9,\\4.4,\\3.9,\\4.5} & \parbox[t]{4.7cm}{ Connected: Joint continuum, chiral and volume fit with the term  $M_\pi^2 e^{-M_\pi L}$. \\ Disconnected: Neglect volume dependence. }  \\[0.5ex]\\[0.5ex]
\hline\\[-0.5ex]
$\chi$QCD 18 & \cite{Liang:2018pis} & 2+1 & \parbox[t]{1.4cm}{4.6,\\2.7,\\2.6} & \parbox[t]{1.cm}{4.0,\\4.5,\\4.1} &  \parbox[t]{4.7cm}{ FV correction expected to be small. }\\[0.5ex]\\[0.5ex]
\hline\\[-0.5ex]
JLQCD 18 & \cite{Yamanaka:2018uud} & 2+1 & \parbox[t]{1.4cm}{1.8--2.7} & \parbox[t]{1.cm}{4.0} & \parbox[t]{4.7cm}{ FV correction expected to be small. }  \\[0.5ex]\\[0.5ex]
\hline\\[-0.5ex]
$\chi$QCD 15 & \cite{Gong:2015iir} & 2+1 & \parbox[t]{1.4cm}{2.6} & \parbox[t]{1.cm}{3.3} &  \parbox[t]{4.7cm}{ FV correction neglected. }   \\[0.5ex]\\[0.5ex]
\hline\\[-0.5ex]
Engelhardt 12 & \cite{Engelhardt:2012gd} & 2+1 & \parbox[t]{1.4cm}{2.5} & \parbox[t]{1.cm}{3.7} & \parbox[t]{4.7cm}{ A generic 10\% error is added based on possible FVE in $g_A$.}   \\[0.5ex]\\[0.5ex]
\hline\\[-0.5ex]
ETM 17C & \cite{Alexandrou:2017oeh} & 2 & \parbox[t]{1.4cm}{3.7} & \parbox[t]{1.cm}{3.0} &  \parbox[t]{5.2cm}{Single simulation with $M_\pi L = 3.0$. }  \\[0.5ex]\\[0.5ex]
\fi
\hline\hline
\end{tabular*}
\caption{
\ifx\reducedapptables\undefined
Finite-volume effects in determinations of $g_A^q$.
\else
Finite-volume effects in determinations of $g_A^q$ and $g_T^q$.
\fi
}
}
\end{table}

\begin{table}[!ht]
{\footnotesize
\begin{tabular*}{\textwidth}{l @{\extracolsep{\fill}} c c c }
\hline\hline \\[-1.0ex]
Collab. & Ref. & $\Nf$ & Ren. 
\\[.3ex]\\[-2.ex] \hline \hline \\[-1.0ex]
%
ETM 19 & \cite{Alexandrou:2019brg} & 2+1+1 & RI-SMOM   \\[0.5ex]\\[0.5ex]
\ifx\reducedapptables\undefined
\hline \\[-0.5ex]
PNDME 18A & \cite{Lin:2018obj} & 2+1+1 & RI-SMOM    \\[0.5ex]\\[0.5ex]
\hline\\[-0.5ex]
$\chi$QCD 18 & \cite{Liang:2018pis} & 2+1 & NP    \\[0.5ex]\\[0.5ex]
\hline\\[-0.5ex]
JLQCD 18 & \cite{Yamanaka:2018uud} & 2+1 & RI-MOM    \\[0.5ex]\\[0.5ex]
\hline\\[-0.5ex]
$\chi$QCD 15 & \cite{Gong:2015iir} & 2+1 & NP    \\[0.5ex]\\[0.5ex]
\hline\\[-0.5ex]
Engelhardt 12 & \cite{Engelhardt:2012gd} & 2+1 & NP    \\[0.5ex]\\[0.5ex]
\hline\\[-0.5ex]
ETM 17C & \cite{Alexandrou:2017oeh} & 2 & RI'-MOM    \\[0.5ex]\\[0.5ex]
\fi
 \hline \hline \\[-1.0ex]
\end{tabular*}
\caption{
\ifx\reducedapptables\undefined
Renormalization in determinations of $g_A^q$.
\else
Renormalization in determinations of $g_A^q$ and $g_T^q$.
\fi
}
}
\end{table}
\begin{table}[!ht]

{\footnotesize
\begin{tabular*}{\textwidth}{l @{\extracolsep{\fill}} c c c l}
\hline\hline \\[-1.0ex]
Collab. & Ref. & $\Nf$ & $\tau$ [fm] & Description 
\\[.3ex]\\[-2.ex]\hline \hline \\[-1.0ex]
%
ETM 19 & \cite{Alexandrou:2019brg} & 2+1+1 &  \parbox[t]{3.5cm}{[0.6--1.6]/all}  &  \parbox[t]{6.3cm}{Two-state fit to all $\tau$. A comparison is made with three-state fits, plateau fits and the summation method.} \\[0.5ex]\\[0.5ex]
\ifx\reducedapptables\undefined
\hline \\[-0.5ex]
PNDME 18A & \cite{Lin:2018obj} & 2+1+1 & \parbox[t]{3.5cm}{[0.8--1.4]\par [1.0--1.4,0.9--1.4]\par [0.9--1.2,0.9--1.2,0.7--1.4]\par [0.9--1.4]} &  \parbox[t]{4.7cm}{ Connected: three-state fit. \\ Disconnected: two-state fit.} \\[0.5ex]\\[0.5ex]
\hline\\[-0.5ex]
$\chi$QCD 18 & \cite{Liang:2018pis} & 2+1 & \parbox[t]{3.5cm}{[1.0--1.6]\par [0.9--1.3]\par [1.0--1.2]} & \parbox[t]{4.7cm}{ Two-state fit. }  \\[0.5ex]\\[0.5ex]
\hline\\[-0.5ex]
JLQCD 18 & \cite{Yamanaka:2018uud} & 2+1 & \parbox[t]{3.5cm}{[1.0--1.54]} &   \parbox[t]{4.7cm}{ Constant fit to all $\tau$ \\ and $t \in [4 - (\tau-4)] $ data. } \\[0.5ex]\\[0.5ex]
\hline\\[-0.5ex]
$\chi$QCD 15 & \cite{Gong:2015iir} & 2+1 & \parbox[t]{3.5cm}{[0.68--1.14]} & \parbox[t]{4.7cm}{ Two-state fit. }  \\[0.5ex]\\[0.5ex]
\hline\\[-0.5ex]
Engelhardt 12 & \cite{Engelhardt:2012gd} & 2+1 & \parbox[t]{3.5cm}{[1.24--1.5]} & \parbox[t]{4.7cm}{Not quantified.}  \\[0.5ex]\\[0.5ex]
\hline\\[-0.5ex]
ETM 17C & \cite{Alexandrou:2017oeh} & 2 & \parbox[t]{3.5cm}{[0.9--1.3]} & \parbox[t]{4.7cm}{ Plateau value at 1.31 fm. Included systematic error as the difference from the two-state fit.} \\[0.5ex]\\[0.5ex]
\fi
\hline \hline \\[-1.0ex]
\end{tabular*}
\caption{
\ifx\reducedapptables\undefined
Control of excited state contamination in determinations of $g_A^q$.
\else
Control of excited state contamination in determinations of $g_A^q$ and $g_T^q$.
\fi
The comma-separated list of numbers in square brackets denote the range of source-sink separations $\tau$ (in fermi) at each value of the bare coupling. } }
\end{table}

\clearpage
\begin{table}[!ht]

{\footnotesize
\begin{tabular*}{\textwidth}{l @{\extracolsep{\fill}} c c c l}
\hline\hline \\[-1.0ex]
Collab. & Ref. & $\Nf$ & $a$ [fm] & Description 
\\[.3ex]\\[-2.ex] \hline \hline \\[-1.0ex]
%
ETM 19 & \cite{Alexandrou:2019brg} & 2+1+1 & 0.08 & \parbox[t]{6.3cm}{Not estimated. }
\\[0.5ex]\\[0.5ex]
\ifx\reducedapptables\undefined
\hline \\[-0.5ex]
JLQCD 18 & \cite{Yamanaka:2018uud} & 2+1 & 0.11 &  \parbox[t]{6.3cm}{  Discretization effects are estimated to be $\cO(\Lambda^2_{QCD}a^2)\sim 8\%$ with $\Lambda_{QCD}\sim 500$~MeV. This error is neglected for $\sigma_s$ as it is much smaller than the overall uncertainty.} \\[0.5ex]\\[0.5ex]
\hline\\[-0.5ex]
$\chi$QCD 15A & \cite{Yang:2015uis} & 2+1 & 0.11,0.08 &  \parbox[t]{6.3cm}{  Joint continuum, chiral and volume fit with a linear term in $a^2$.} \\[0.5ex]\\[0.5ex]
\hline\\[-0.5ex]
$\chi$QCD 13A & \cite{Gong:2013vja} & 2+1 & 0.11 &  \parbox[t]{6.3cm}{Not estimated. } \\[0.5ex]\\[0.5ex]
\hline\\[-0.5ex]
JLQCD 12A & \cite{Oksuzian:2012rzb} & 2+1 & 0.11 &  \parbox[t]{6.3cm}{  Discretization effects are estimated at $9\%$ taking $\cO(\Lambda^2_{QCD}a^2)$ with $\Lambda_{QCD}\approx 500$~MeV. This error is neglected as this is much smaller than the statistical accuracy.} \\[0.5ex]\\[0.5ex]
\hline\\[-0.5ex]
Engelhardt 12 & \cite{Engelhardt:2012gd} & 2+1 & 0.12 &  \parbox[t]{6.3cm}{  Estimate $3\%$ discretization error based on a study of the nucleon mass in Ref.~\cite{WalkerLoud:2008bp}.} \\[0.5ex]\\[0.5ex]
\hline\\[-0.5ex]
ETM 16A & \cite{Abdel-Rehim:2016won} & 2 & 0.09 &  \parbox[t]{6.3cm}{ Uncertainty in the fixing the lattice spacing is given based on the analysis of Ref.~\cite{Abdel-Rehim:2015pwa}.  } \\[0.5ex]\\[0.5ex]
\hline\\[-0.5ex]
RQCD 16 & \cite{Bali:2016lvx} & 2 & 0.08,0.07,0.06 &  \parbox[t]{6.3cm}{  No significant discretization effects observed.} \\[0.5ex]\\[0.5ex]
\hline\hline\\[-0.5ex]
MILC 12C & \cite{Freeman:2012ry} & 2+1+1 & 0.15,0.12,0.09,0.06 &  \parbox[t]{6.3cm}{  Combined chiral and continuum fit with an $a^2$ term. Coefficient constrained with a Bayesian prior with a 1 standard deviation width corresponding to a $10\%$ effect at $a=0.12$~fm.} \\[0.5ex]\\[0.5ex]
\hline\\[-0.5ex]
MILC 12C & \cite{Freeman:2012ry} & 2+1 & 0.12,0.09,0.06 &  \parbox[t]{6.3cm}{Analysis and estimation
of discretisation effects as for the $2+1+1$ result in the same publication~(see above).} \\[0.5ex]\\[0.5ex]
\hline\\[-0.5ex]
MILC 09D & \cite{Toussaint:2009pz} & 2+1 & 0.12,0.09,0.06 &  \parbox[t]{6.3cm}{Analysis and estimation         
of discretisation effects as in Ref.~\cite{Freeman:2012ry}~(see above).} \\[0.5ex]\\[0.5ex]
\fi
\hline \hline \\[-1.0ex]
\end{tabular*}
\caption{
\ifx\reducedapptables\undefined
Continuum extrapolation/estimation of lattice artifacts in the determinations of $\sigma_{\pi N}$ and $\sigma_s$. 
The calculations of $\sigma_s$ by MILC 12C and MILC 09D employ a hybrid approach, while all other results were obtained 
from a direct calculation.
\else
  Continuum extrapolation/estimation of lattice artifacts in direct determinations of $\sigma_{\pi N}$ and $\sigma_s$. 
\fi
}
}
\end{table}

\begin{table}[!ht]
{\footnotesize
\begin{tabular*}{\textwidth}{l @{\extracolsep{\fill}} c c c l}
\hline\hline \\[-1.0ex]
{Collab.} & {Ref.} & {$\Nf$} & {$M_{\pi,\rm min}\,[\mev]$} & {Description}  
\\[.3ex]\\[-2.ex] \hline \hline \\[-1.0ex]
%
ETM 19 & \cite{Alexandrou:2019brg} & 2+1+1 & 139  & \parbox[t]{6.3cm}{ Simulate close to $M_\pi^{phys}$. }
\\[0.5ex]\\[0.5ex]
\ifx\reducedapptables\undefined
\hline \\[-0.5ex]
JLQCD 18 & \cite{Yamanaka:2018uud} & 2+1 & 293 & \parbox[t]{7.1cm}{$\sigma_{\pi N}$:  Fit including linear and quadratic terms in $M_\pi^2$. Difference with a linear fit taken as the systematic error. $\sigma_s$:  Linear fit in $M_\pi^2$. Difference with a constant fit taken as the systematic error.}  \\[0.5ex]\\[0.5ex]
\hline\\[-0.5ex]
$\chi$QCD 15A & \cite{Yang:2015uis} & 2+1 & 139,300 & \parbox[t]{7.1cm}{  Joint continuum, chiral and volume fit. Partially quenched analysis with $M_\pi \in [114,400]$~MeV. $\sigma_{\pi N}$: employ partially quenched $SU(2)$ $\chi$PT~\cite{Beane:2002vq,Tiburzi:2005is,WalkerLoud:2008bp}. $\sigma_s$: linear terms in the sea and valence $M_\pi^2$ are included. Unitary $M_{\pi}$ given.}  \\[0.5ex]\\[0.5ex]
\hline\\[-0.5ex]
$\chi$QCD 13A & \cite{Gong:2013vja} & 2+1 & 331 & \parbox[t]{7.1cm}{  Partially quenched study with a valence $M_{\pi,min}=250$~MeV. Combined chiral and excited state fit. Linear fit in
the valence light-quark mass~(and $m_s$ to interpolate to the physical strange-quark mass). Unitary $M_{\pi}$ given.}  \\[0.5ex]\\[0.5ex]
\hline\\[-0.5ex]
JLQCD 12A & \cite{Oksuzian:2012rzb} & 2+1 & 300 & \parbox[t]{7.1cm}{  Linear fit in $m_{ud}$~(and $m_s$ to interpolate to the physical strange quark mass). Difference with $SU(3)$ HB$\chi$PT~\cite{WalkerLoud:2004hf} fit taken as the systematic error.}  \\[0.5ex]\\[0.5ex]
\hline\\[-0.5ex]
Engelhardt 12 & \cite{Engelhardt:2012gd} & 2+1 & 293 & \parbox[t]{7.1cm}{  Linear fit in $M_\pi^2$. Systematic error of $6\%$ estimated from comparing with a constant fit.}  \\[0.5ex]\\[0.5ex]
\hline\\[-0.5ex]
ETM 16A & \cite{Abdel-Rehim:2016won} & 2 & 130 & \parbox[t]{7.1cm}{  Simulate close to $M_\pi^{phys}$.}  \\[0.5ex]\\[0.5ex]
\hline\\[-0.5ex]
RQCD 16 & \cite{Bali:2016lvx} & 2 & 280,150,260 & \parbox[t]{7.1cm}{$\sigma_{\pi N}$:  Rescaling with $M_\pi^2$ of the result at lightest $M_\pi$. Result is consistent with~(but the error more conservative than) performing a baryon $\chi$PT~\cite{Steininger:1998ya,Becher:1999he} fit to $M_\pi\le 420$~MeV. $\sigma_s$: Linear fit in $M_\pi^2$.}  \\[0.5ex]\\[0.5ex]
\fi
\hline\hline\\[-0.5ex]
\end{tabular*}
\caption{
Chiral extrapolation/minimum pion mass in
direct determinations of $\sigma_{\pi N}$ and $\sigma_s$.
}
}
\end{table}

\ifx\reducedapptables\undefined
\begin{table}[!ht]
{\footnotesize
\begin{tabular*}{\textwidth}{l @{\extracolsep{\fill}} c c c l}
\hline\hline \\[-1.0ex]
{Collab.} & {Ref.} & {$\Nf$} & {$M_{\pi,\rm min}\,[\mev]$} & {Description}  
\\[.3ex]\\[-2.ex] \hline \hline \\[-1.0ex]
MILC 12C & \cite{Freeman:2012ry} & 2+1+1 & 131,216,221,329 & \parbox[t]{7.1cm}{  Combined chiral and continuum fit with a linear term in $m_{ud}$. Bayesian prior used to constrain the coefficient with the central value taken from the $N_f=2+1$ analysis and a width equal to the error of the slope. A $7\%$ error is added to account for omitted higher order terms in $\chi$PT.}  \\[0.5ex]\\[0.5ex]
\hline\\[-0.5ex]
MILC 12C & \cite{Freeman:2012ry} & 2+1 & 269,247,224 & \parbox[t]{7.1cm}{  Combined chiral and continuum fit with a linear term in $m_{ud}$. A $7\%$ error is added to account for omitted higher order terms in $\chi$PT.}  \\[0.5ex]\\[0.5ex]
\hline\\[-0.5ex]
MILC 09D & \cite{Toussaint:2009pz} & 2+1 & 269,247,224 & \parbox[t]{7.1cm}{  Combined chiral and continuum fit with a linear term in $m_{ud}$. A $7\%$ error is added to account for omitted higher order terms in $\chi$PT.}  \\[0.5ex]\\[0.5ex]
\hline\hline
\end{tabular*}
\caption{Chiral extrapolation/minimum pion mass in
hybrid calculations of $\sigma_s$.}
}
\end{table}
\fi

\begin{table}[!ht]
{\footnotesize
\begin{tabular*}{\textwidth}{l @{\extracolsep{\fill}} c c c c l}
\hline\hline \\[-1.0ex]
Collab. & Ref. & $\Nf$ & $L$ [fm] & ${M_{\pi,\rm min}}L$ & Description 
\\[.3ex]\\[-2.ex] \hline \hline \\[-1.0ex]
%
ETM 19 & \cite{Alexandrou:2019brg} & 2+1+1 & \parbox[t]{1.4cm}{5.1} &
\parbox[t]{1.cm}{3.6} & \parbox[t]{6.3cm}{FVE are found to be small from an investigation using two
near physical point $N_f=2$ ensembles with $M_\pi L=3.0$ and $4.0$.}
\\[0.5ex]\\[0.5ex]
\ifx\reducedapptables\undefined
\hline \\[-0.5ex]
JLQCD 18 & \cite{Yamanaka:2018uud} & 2+1 & \parbox[t]{1.4cm}{1.8--2.7} & \parbox[t]{1.cm}{4.0} & \parbox[t]{5.7cm}{  FVE expected to be small for $M_\pi L\gtrsim 4$.}  \\[0.5ex]\\[0.5ex]
\hline\\[-0.5ex]
$\chi$QCD 15A & \cite{Yang:2015uis} & 2+1 & \parbox[t]{1.4cm}{2.7--5.5,\\2.6} & \parbox[t]{1.cm}{3.9,\\4.0} & \parbox[t]{5.7cm}{  Joint continuum, chiral and volume fit with 
leading order FV terms included for $\sigma_{\pi N}$ and a $e^{-M_\pi L}$ term included for $\sigma_s$. Unitary $M_{\pi,min}L$ given.}  \\[0.5ex]\\[0.5ex]
\hline\\[-0.5ex]
$\chi$QCD 13A & \cite{Gong:2013vja} & 2+1 & \parbox[t]{1.4cm}{2.7} & \parbox[t]{1.cm}{4.6} & \parbox[t]{5.7cm}{Not estimated.  Unitary $M_{\pi,min}L$ given.}  \\[0.5ex]\\[0.5ex]
\hline\\[-0.5ex]
JLQCD 12A & \cite{Oksuzian:2012rzb} & 2+1 & \parbox[t]{1.4cm}{1.8--2.7} & \parbox[t]{1.cm}{4.1} & \parbox[t]{5.7cm}{  No significant effects observed when comparing $L/a=16$ and 24 for the two lightest $M_\pi$.}  \\[0.5ex]\\[0.5ex]
\hline\\[-0.5ex]
Engelhardt 12 & \cite{Engelhardt:2012gd} & 2+1 & \parbox[t]{1.4cm}{2.5} & \parbox[t]{1.cm}{3.7} & \parbox[t]{5.7cm}{  A generic $10\%$ FVE is added based on possible FVE in $g_A$.}  \\[0.5ex]\\[0.5ex]
\hline\\[-0.5ex]
ETM 16A & \cite{Abdel-Rehim:2016won} & 2 & \parbox[t]{1.4cm}{4.5} & \parbox[t]{1.cm}{3.0} & \parbox[t]{5.7cm}{ FVE from baryon $\chi$PT following Ref.~\cite{AliKhan:2003ack} estimated to lead to a $5\%$ increase in $\sigma_{\pi N}$, included as a systematic. This increase is taken as an upper bound on FVE for $\sigma_s$.}  \\[0.5ex]\\[0.5ex]
\hline\\[-0.5ex]
RQCD 16 & \cite{Bali:2016lvx} & 2 & \parbox[t]{1.4cm}{2.6,\\1.7--4.5,\\1.9--2.9} & \parbox[t]{1.cm}{3.7,\\3.5,\\3.8} & \parbox[t]{5.7cm}{  No significant FVE seen for $LM_{\pi}=3.4-6.7$ at $M_\pi=290$~MeV.}  \\[0.5ex]\\[0.5ex]
\hline\hline\\[-0.5ex]
MILC 12C & \cite{Freeman:2012ry} & 2+1+1 & \parbox[t]{1.4cm}{2.5--4.9,\\2.9--4.9,\\2.8--4.2,\\2.7} & \parbox[t]{1.cm}{3.3,\\5.4,\\4.7,\\4.5} & \parbox[t]{5.7cm}{  Estimated to be $2\%$, less than for the $\Nf=2+1$ analysis, since the volumes are larger.}  \\[0.5ex]\\[0.5ex]
\hline\\[-0.5ex]
MILC 12C & \cite{Freeman:2012ry} & 2+1 & \parbox[t]{1.4cm}{2.3--2.8,\\2.3--3.4,\\2.8--3.8} & \parbox[t]{1.cm}{3.8,\\4.2,\\4.3} & \parbox[t]{5.7cm}{  Estimated to be $3\%$, based on FVE observed in the nucleon mass.}  \\[0.5ex]\\[0.5ex]
\hline\\[-0.5ex]
MILC 09D & \cite{Toussaint:2009pz} & 2+1 & \parbox[t]{1.4cm}{2.3--2.8,\\2.3--3.4,\\2.8--3.8} & \parbox[t]{1.cm}{3.8,\\4.2,\\4.3} & \parbox[t]{5.7cm}{  Estimated to be $3\%$, based on FVE observed in the nucleon mass.}  \\[0.5ex]\\[0.5ex]
\fi
\hline\hline
\end{tabular*}
\caption{
\ifx\reducedapptables\undefined
Finite-volume effects in the determinations of $\sigma_{\pi N}$ and  $\sigma_s$. 
The calculations of $\sigma_s$ by MILC 12C and MILC 09D employ a hybrid approach, while all other results were obtained 
from a direct calculation.
\else
Finite-volume effects in direct determinations of $\sigma_{\pi N}$ and  $\sigma_s$. 
\fi
}
}
\end{table}

\begin{table}[!ht]
{\footnotesize
\begin{tabular*}{\textwidth}{l @{\extracolsep{\fill}} c c c l}
\hline\hline \\[-1.0ex]
Collab. & Ref. & $\Nf$ & Ren. &  Description
\\[.3ex]\\[-2.ex] \hline \hline \\[-1.0ex]
%
ETM 19 & \cite{Alexandrou:2019brg} & 2+1+1 & na/na   &  \parbox[t]{6.3cm}{ } \\[0.5ex]\\[0.5ex]
\ifx\reducedapptables\undefined
\hline \\[-0.5ex]
JLQCD 18 & \cite{Yamanaka:2018uud} & 2+1 & na/na  &  \parbox[t]{6.3cm}{ } \\[0.5ex]\\[0.5ex]
\hline\\[-0.5ex]
$\chi$QCD 15A & \cite{Yang:2015uis} & 2+1 & na/na   &  \parbox[t]{6.3cm}{ } \\[0.5ex]\\[0.5ex]
\hline\\[-0.5ex]
$\chi$QCD 13A & \cite{Gong:2013vja} & 2+1 & $-$/na   &  \parbox[t]{6.3cm}{ } \\[0.5ex]\\[0.5ex]
\hline\\[-0.5ex]
JLQCD 12A & \cite{Oksuzian:2012rzb} & 2+1 & $-$/na   &  \parbox[t]{6.3cm}{ } \\[0.5ex]\\[0.5ex]
\hline\\[-0.5ex]
Engelhardt 12 & \cite{Engelhardt:2012gd} & 2+1 &$-$/na   &  \parbox[t]{6.3cm}{  Flavour mixing due to residual chiral symmetry breaking for a DW action estimated to be $1\%$ and neglected.} \\[0.5ex]\\[0.5ex]
\hline\\[-0.5ex]
ETM 16A & \cite{Abdel-Rehim:2016won} & 2 & na/na   &  \parbox[t]{6.3cm}{ } \\[0.5ex]\\[0.5ex]
\hline\\[-0.5ex]
RQCD 16 & \cite{Bali:2016lvx} & 2 & na/NP   &  \parbox[t]{6.3cm}{  Flavour mixing occurs due to breaking of chiral symmetry when evaluating $\sigma_s$. The ratio of $Z^{ns}/Z^{s}$ is computed nonperturbatively.} \\[0.5ex]\\[0.5ex]
\hline\hline\\[-0.5ex]
MILC 12C & \cite{Freeman:2012ry} & 2+1+1 & 2-loop   &  \parbox[t]{6.3cm}{  $\langle N|s\bar{s}|N\rangle$ computed. $Z_m$ for Asqtad action applied. The 1-loop factors for Asqtad and HISQ are very similar, cf Refs.~\cite{McNeile:2012xh} and~\cite{Mason:2005bj}.} \\[0.5ex]\\[0.5ex]
\hline\\[-0.5ex]
MILC 12C & \cite{Freeman:2012ry} & 2+1 & 2-loop   &  \parbox[t]{6.3cm}{  $\langle N|s\bar{s}|N\rangle$ computed. $Z_m$ from Ref.~\cite{Mason:2005bj}.} \\[0.5ex]\\[0.5ex]
\hline\\[-0.5ex]
MILC 09D & \cite{Toussaint:2009pz} & 2+1 & na   &  \parbox[t]{6.3cm}{ } \\[0.5ex]\\[0.5ex]
\fi
 \hline \hline \\[-1.0ex]
\end{tabular*}
\caption{
\ifx\reducedapptables\undefined
Renormalization for determinations of $\sigma_{\pi N}$ and $\sigma_s$.
 The calculations of $\sigma_s$ by MILC 12C and MILC 09D employ a hybrid approach.
 For the remaining direct determinations, the 
\else
  Renormalization for direct determinations of $\sigma_{\pi N}$ and $\sigma_s$.
The 
  \fi
type of
  renormalization~(Ren.) is given for $\sigma_{\pi N}$ first and
  $\sigma_s$ second. The label 'na' indicates that no renormalization is required.  
}

}
\end{table}
\begin{table}[ht!]
  {\footnotesize
\begin{tabular*}{\textwidth}{l @{\extracolsep{\fill}} c c c l}
\hline\hline \\[-1.0ex]
Collab. & Ref. & $\Nf$ & $\tau$ [fm] & Description 
\\[.3ex]\\[-2.ex]\hline \hline \\[-1.0ex]
%
ETM 19 & \cite{Alexandrou:2019brg} & 2+1+1 &  \parbox[t]{3.5cm}{[0.6--1.6]/all}  &  \parbox[t]{6.3cm}{Two-state fit to all $\tau$. A comparison is made with three-state fits, plateau fits and the summation method.} \\[0.5ex]\\[0.5ex]
\ifx\reducedapptables\undefined
\hline \\[-0.5ex]
JLQCD 18 & \cite{Yamanaka:2018uud} & 2+1 & all/all &  \parbox[t]{6.3cm}{  Simultaneous plateau fit for $\tau/a\gtrsim 11$.} \\[0.5ex]\\[0.5ex]
\hline\\[-0.5ex]
$\chi$QCD 15A & \cite{Yang:2015uis} & 2+1 & \parbox[t]{3.5cm}{[0.9--1.3,0.9--1.4]\par [1.0--1.2]/all} &  \parbox[t]{6.3cm}{  Two-state fit with $t\in[0.2,\tau-0.2]$~fm.} \\[0.5ex]\\[0.5ex]
\hline\\[-0.5ex]
$\chi$QCD 13A & \cite{Gong:2013vja} & 2+1 & $-$/all &  \parbox[t]{6.3cm}{  Combined chiral and excited state fit using the summation method with $\tau/a=7-14$.} \\[0.5ex]\\[0.5ex]
\hline\\[-0.5ex]
JLQCD 12A & \cite{Oksuzian:2012rzb} & 2+1 & $-$/all &  \parbox[t]{6.3cm}{  Simultaneous plateau fit to $\tau/a\in [12,23]$ with $t/a\in [5,\tau-5]$.} \\[0.5ex]\\[0.5ex]
\hline\\[-0.5ex]
Engelhardt 12 & \cite{Engelhardt:2012gd} & 2+1 & $-$/all &  \parbox[t]{6.3cm}{  Ratio of three-point to two-point function averaged over $t/a\in [3,7]$ for $\tau/a=10$.} \\[0.5ex]\\[0.5ex]
\hline\\[-0.5ex]
ETM 16A & \cite{Abdel-Rehim:2016won} & 2 & \parbox[t]{3.5cm}{[0.9--1.7]/all} &  \parbox[t]{6.3cm}{  Plateau fit to $\tau=1.7$~fm. Systematic error estimated from plateau fits for $\tau \ge 1.5$~fm. Comparison made with 2-state fits and the summation method.} \\[0.5ex]\\[0.5ex]
\hline\\[-0.5ex]
RQCD 16 & \cite{Bali:2016lvx} & 2 & \parbox[t]{3.5cm}{[1.1]\par [1.1--1.2,0.5--1.2,0.6--1.1]\par [1.1,1.1,1.1]/all} &  \parbox[t]{6.3cm}{  Two state~(simultaneous plateau) fit for the connected~(disconnected) terms.} \\[0.5ex]\\[0.5ex]
\fi
\hline \hline \\[-1.0ex]
\end{tabular*}
\caption{
  Control of excited state contamination in
  direct determinations of $\sigma_{\pi N}$ and $\sigma_s$. 
  The comma-separated list of numbers in
  square brackets denote the range of source-sink separations $\tau$
  (in fermi) at each value of the bare coupling. The range of $\tau$ for the connected~(disconnected)
  contributions to the three-point correlation functions is given
  first~(second). If a wide range of $\tau$ values is available this
  is indicated by ``all'' in the table. } }
\end{table}

\ifx\reducedapptables\undefined
\begin{table}[ht!]
{\footnotesize
\begin{tabular*}{\textwidth}{l @{\extracolsep{\fill}} c c c l}
\hline\hline \\[-1.0ex]
Collab. & Ref. & $\Nf$ & $\tau$ [fm] & Description 
\\[.3ex]\\[-2.ex]\hline \hline \\[-1.0ex]
%
MILC 12C & \cite{Freeman:2012ry} & 2+1+1 & all &  \parbox[t]{6.3cm}{A modified nucleon correlator is fitted including a positive and negative parity state starting from $\approx 0.6$~fm. A $5\%$ error for possible excited state contamination is included.} \\[0.5ex]\\[0.5ex]
\hline\\[-0.5ex]
MILC 12C & \cite{Freeman:2012ry} & 2+1 & all &  \parbox[t]{6.3cm}{Analysis and estimation
of excited state contamination as for the $2+1+1$ result in the same publication~(see above).} \\[0.5ex]\\[0.5ex]
\hline\\[-0.5ex]
MILC 09D & \cite{Toussaint:2009pz} & 2+1 & all &  \parbox[t]{6.3cm}{A modified nucleon correlator is fitted including a positive and negative parity state starting from $\approx 0.6$~fm. A $10\%$ error for possible excited state contamination is included.} \\[0.5ex]\\[0.5ex]
\hline \hline \\[-1.0ex]
\end{tabular*}
\caption{Control of excited state contamination in
 determinations of $\sigma_s$ using a hybrid approach. 
If a wide range of $\tau$ values is available this
  is indicated by ``all'' in the table. } }
\end{table}
\fi

\clearpage
\begin{table}[!ht]

{\footnotesize
\begin{tabular*}{\textwidth}{l @{\extracolsep{\fill}} c c c l}
\hline\hline \\[-1.0ex]
Collab. & Ref. & $\Nf$ & $a$ [fm] & Description 
\\[.3ex]\\[-2.ex] \hline \hline \\[-1.0ex]
%
BMW 20A & \cite{Borsanyi:2020bpd} & 1+1+1+1 & 0.10,0.09,0.08,0.06 &  \parbox[t]{6.8cm}{ Combined continuum, chiral and volume fit within an extended frequentist method. $\cO(\alpha a)$ or $\cO(a^2)$ terms are included. $M_N$ used to fix the lattice spacing. } \\[0.5ex]\\[0.5ex]
\hline \\[-0.5ex]
\ifx\reducedapptables\undefined
ETM 14A & \cite{Alexandrou:2014sha} & 2+1+1 & 0.09,0.08,0.06 &  \parbox[t]{6.8cm}{  No significant discretization effects are observed. $M_N$ used to fix the lattice spacing.} \\[0.5ex]\\[0.5ex]
\hline\\[-0.5ex]
BMW 15 & \cite{Durr:2015dna} & 2+1 & \parbox[t]{2.5cm}{0.12,0.09,0.08,\\0.07,0.05} &  \parbox[t]{6.8cm}{  Combined continuum, chiral and volume fit to $M_{N,\Omega,\pi,K}$ within an extended frequentist method. For $M_N$, $\cO(\alpha a)$ or $\cO(a^2)$ errors on the physical limit are included. Systematics include shift when adding discretization terms to the slope of $M_N$ with $m_{ud,s}$, fixing $M_N$ to expt..} \\[0.5ex]\\[0.5ex]
\hline\\[-0.5ex]
Junnarkar 13 & \cite{Junnarkar:2013ac} & 2+1 & 0.12,0.09 &  \parbox[t]{6.8cm}{Discretization effects are not resolved.} \\[0.5ex]\\[0.5ex]
\hline\\[-0.5ex]
Shanahan 12 & \cite{Shanahan:2012wh} & 2+1 & 0.09 &  \parbox[t]{6.8cm}{  Two results quoted, differing on setting the lattice spacing per ensemble or in the chiral limit.} \\[0.5ex]\\[0.5ex]
\hline\\[-0.5ex]
JLQCD 12A & \cite{Oksuzian:2012rzb} & 2+1 & 0.11 &  \parbox[t]{6.8cm}{  Discretization effects are estimated to be $\cO(\Lambda^2_{QCD}a^2)\sim 9\%$ with $\Lambda_{QCD}\sim 500$~MeV and neglected as the statistical error is much larger.} \\[0.5ex]\\[0.5ex]
\hline\\[-0.5ex]
QCDSF 11 & \cite{Horsley:2011wr} & 2+1 & 0.08 &  \parbox[t]{6.8cm}{  Not estimated.} \\[0.5ex]\\[0.5ex]
\hline\\[-0.5ex]
BMW 11A & \cite{Durr:2011mp} & 2+1 & 0.12,0.08,0.06 &  \parbox[t]{6.8cm}{  Joint continuum and chiral extrapolation of octet baryons masses with $\cO(a)$, $\cO(a^2)$ terms or no discretization terms. Final result from consideration of all fit combinations weighted with the fit quality.} \\[0.5ex]\\[0.5ex]
\hline\\[-0.5ex]
Martin~Camalich 10 & \cite{MartinCamalich:2010fp} & 2+1 & 0.09 &  \parbox[t]{6.8cm}{  Not estimated.} \\[0.5ex]\\[0.5ex]
\hline\\[-0.5ex]
PACS-CS 09 & \cite{Ishikawa:2009vc} & 2+1 & 0.09 &  \parbox[t]{6.3cm}{  Not estimated.} \\[0.5ex]\\[0.5ex]
\hline\hline
\end{tabular*}
\caption{Finite-volume effects in determinations of $\sigma_{\pi N}$  and $\sigma_s$  from the Feynman-Hellmann method.}
}
\end{table}
\fi
\ifx\reducedapptables\undefined
\begin{table}[!ht]
\addtocounter{table}{-1}
{\footnotesize
\begin{tabular*}{\textwidth}{l @{\extracolsep{\fill}} c c c l}
\hline\hline \\[-1.0ex]
Collab. & Ref. & $\Nf$ & $a$ [fm] & Description 
\\[.3ex]\\[-2.ex] \hline \hline \\[-1.0ex]
%
\fi
Walker-Loud 08 & \cite{Walker-Loud:2008rui} & 2+1 & 0.12 &  \parbox[t]{6.3cm}{Not estimated. } \\[0.5ex]\\[0.5ex]
\ifx\reducedapptables\undefined
\hline\\[-0.5ex]
QCDSF 12 & \cite{Bali:2012qs} & 2 & 0.08,0.07,0.06 &  \parbox[t]{6.3cm}{  No significant discretization effects observed. $M_N$ is used to fix the lattice spacing.} \\[0.5ex]\\[0.5ex]
\hline\\[-0.5ex]
JLQCD 08B & \cite{Ohki:2008ff} & 2 & 0.12 &  \parbox[t]{6.3cm}{  Not estimated.} \\[0.5ex]\\[0.5ex]
\fi
\hline \hline \\[-1.0ex]
\end{tabular*}
\ifx\reducedapptables\undefined
\caption{(cntd) Continuum extrapolations/estimation of lattice artifacts in
  determinations of $\sigma_{\pi N}$ and $\sigma_s$ from the Feynman-Hellmann method.}
\else
\caption{Continuum extrapolations/estimation of lattice artifacts in
  determinations of $\sigma_{\pi N}$ and $\sigma_s$ from the Feynman-Hellmann method.}
\fi
}
\end{table}

\begin{table}[!ht]
{\footnotesize
\begin{tabular*}{\textwidth}{l @{\extracolsep{\fill}} c c c l}
\hline\hline \\[-1.0ex]
{Collab.} & {Ref.} & {$\Nf$} & {$M_{\pi,\rm min}\,[\mev]$} & {Description}  
\\[.3ex]\\[-2.ex] \hline \hline \\[-1.0ex]
%
BMW 20A & \cite{Borsanyi:2020bpd} & 1+1+1+1 & 238,200,219,199 &  \parbox[t]{6.8cm}{ Combined continuum, chiral and volume fit within an extended frequentist method. Terms with $M_\pi^{2,3,4}$ and $M_K^2$ are included and cuts of $M_\pi < 420$~MeV and $M_\pi < 360$~MeV are made.} \\[0.5ex]\\[0.5ex]
\hline \\[-0.5ex]
\ifx\reducedapptables\undefined
\else
Walker-Loud 08 & \cite{Walker-Loud:2008rui} & 2+1 & 294 &  \parbox[t]{6.3cm}{Fit using NNLO $SU(2)$ HB$\chi$PT~\cite{Tiburzi:2005na} with a zero and non-zero value for $g_{\Delta N}$. Errors on the LECs that are not free parameters determine the systematic uncertainty. } \\[0.5ex]\\[0.5ex]
\fi
\ifx\reducedapptables\undefined
ETM 14A & \cite{Alexandrou:2014sha} & 2+1+1 & 261,256,213 & \parbox[t]{6.3cm}{  Fit using $SU(2)$ HB$\chi$PT to $\cO(p^3)$~\cite{Gasser:1987rb}. Error assigned to difference with fit using HB$\chi$PT in SSE to $\cO(p^4)$~\cite{Procura:2006bj}.}  \\[0.3ex]\\
\hline\\[-0.5ex]
BMW 15 & \cite{Durr:2015dna} & 2+1 & 136,131,120,182,219 & \parbox[t]{6.3cm}{  Combined continuum, chiral and volume fit to $M_{N,\Omega,\pi,K}$ within an extended frequentist method~\cite{Durr:2008zz}. Terms linear in $m_{ud,s}$ are included and cuts of $M_\pi\le 480$~MeV and $M_\pi \le 320$~MeV are made. Higher order terms in the fit are also considered.}  \\[0.3ex]\\
\hline\\[-0.5ex]
Junnarkar 13 & \cite{Junnarkar:2013ac} & 2+1 & 380,238 & \parbox[t]{6.3cm}{  $\sigma_s$ from $M_N$ with one $m_s$ value above and one below the physical $m_s$. Weighted average is made of fits to $\sigma_s$~(including a $M_\pi^2$ term) and $f_{T_s}$~(including a $M_\pi$ or $M_\pi^2$ term). Correlations are taken into account.}  \\[0.3ex]\\
\hline\\[-0.5ex]
Shanahan 12 & \cite{Shanahan:2012wh} & 2+1 & 296 & \parbox[t]{6.3cm}{  Fit to baryon octet masses using finite-range regularization of baryon $\chi$PT~\cite{Young:2002cj}. Coefficients held fixed in the fit are varied by $10\%$.}  \\[0.3ex]\\
\hline\\[-0.5ex]
JLQCD 12A & \cite{Oksuzian:2012rzb} & 2+1 & 300 & \parbox[t]{6.3cm}{  Only $\sigma_s$ computed. Reweighting used to vary $M_N$ in a region $m_s^\prime\in[m_s-25\text{\,MeV},m_s+25\text{\,MeV}]$. Linear fit to extract the slope at physical $m_s$ for two different sea $m_s$. Linear fit in $m_{ud}$ and $m_s$. 1-loop $SU(3)$ HB$\chi$PT~\cite{WalkerLoud:2004hf} used to estimate the systematic error.}  \\[0.3ex]\\
\hline\\[-0.5ex]
QCDSF 11 & \cite{Horsley:2011wr} & 2+1 & 328 & \parbox[t]{6.3cm}{$SU(3)$ flavour expansion along simulation trajectory with average quark mass held fixed~\cite{Bietenholz:2011qq}. Fit with linear flavour breaking terms, with systematics from next order terms included in the error.}  \\[0.3ex]\\
\hline\\[-0.5ex]
BMW 11A & \cite{Durr:2011mp} & 2+1 & 273,197,321 & \parbox[t]{6.3cm}{  Joint continuum and chiral extrapolation of octet baryon masses. Fits involving Taylor and Pad\'{e} expansions and $SU(3)$ baryon $\chi$PT~\cite{Oller:2006yh,Frink:2006hx} and cuts of $M_\pi<410$~MeV and $M_\pi<550$~MeV. Final result from consideration of all fit combinations weighted with the fit quality.}  \\[0.3ex]\\
\fi
\hline\hline
\end{tabular*}
\caption{Chiral extrapolation/minimum pion mass in
  determinations of $\sigma_{\pi N}$  and $\sigma_s$ from the Feynman-Hellmann method.}
}
\end{table}

\ifx\reducedapptables\undefined
\begin{table}[!ht]
\addtocounter{table}{-1}
{\footnotesize
\begin{tabular*}{\textwidth}{l @{\extracolsep{\fill}} c c c l}
\hline\hline \\[-1.0ex]
{Collab.} & {Ref.} & {$\Nf$} & {$M_{\pi,\rm min}\,[\mev]$} & {Description}  
\\[.3ex]\\[-2.ex] \hline \hline \\[-1.0ex]
%
Martin~Camalich 10 & \cite{MartinCamalich:2010fp} & 2+1 & 156 & \parbox[t]{6.3cm}{  Fit to baryon octet masses using NLO covariant baryon $\chi$PT in the EOMS scheme~\cite{Gegelia:1999gf,Fuchs:2003qc}. Uncertainty from omitted higher order terms estimated as half the difference between LO and NLO.}  \\[0.3ex]\\
\hline\\[-0.5ex]
PACS-CS 09 & \cite{Ishikawa:2009vc} & 2+1 & 156 & \parbox[t]{6.3cm}{  Fit $M_N$ with $\cO(p^3)$ $SU(2)$ HB$\chi$PT~\cite{Gasser:1987rb}.}  \\[0.5ex]\\[0.5ex]
\hline\\[-0.5ex]
Walker-Loud 08 & \cite{Walker-Loud:2008rui} & 2+1 & 294 &  \parbox[t]{6.3cm}{Fit using NNLO $SU(2)$ HB$\chi$PT~\cite{Tiburzi:2005na} with a zero and non-zero value for $g_{\Delta N}$. Errors on the LECs that are not free parameters determine the systematic uncertainty. } \\[0.5ex]\\[0.5ex]
\hline\\[-0.5ex]
QCDSF 12 & \cite{Bali:2012qs} & 2 & 478,158,262 & \parbox[t]{6.3cm}{  Fit $M_N$ to $\cO(p^4)$ baryon $\chi$PT~\cite{Procura:2003ig}. Slope at $M_\pi=290$~MeV is fixed from a direct determination of $\sigma_{\pi N}$. Results consistent with $\cO(p^2)$ and $\cO(p^3)$ fits.}  \\[0.5ex]\\[0.5ex]
\hline\\[-0.5ex]
JLQCD 08B & \cite{Ohki:2008ff} & 2 & 288 & \parbox[t]{6.3cm}{  Fit $M_N$ using a reduced form of covariant baryon $\chi$PT~\cite{Procura:2003ig} with the systematic error determined from $\cO(p^3)$ and $\cO(p^4)$ fits.}  \\[0.5ex]\\[0.5ex]
\hline\hline
\end{tabular*}
\caption{(cntd.) Chiral extrapolation/minimum pion mass in
  determinations of $\sigma_{\pi N}$  and $\sigma_s$ from the Feynman-Hellmann method.}
}
\end{table}
\fi

\begin{table}[!ht]
{\footnotesize
\begin{tabular*}{\textwidth}{l @{\extracolsep{\fill}} c c c c l}
\hline\hline \\[-1.0ex]
Collab. & Ref. & $\Nf$ & $L$ [fm] & ${M_{\pi,\rm min}}L$ & Description 
\\[.3ex]\\[-2.ex] \hline \hline \\[-1.0ex]
%
BMW 20A & \cite{Borsanyi:2020bpd} & 1+1+1+1 & \parbox[t]{1.4cm}{3.2,\\2.9--4.3,\\2.6--3.8,\\1.9--3.8} & \parbox[t]{1.cm}{4.0,\\4.3,\\4.1,\\4.1}  &  \parbox[t]{5.3cm}{ Combined continuum, chiral and volume fit within an extended frequentist method. FV term $M_\pi^{1/2}L^{-3/2}e^{-M_\pi L}$ added.} \\[0.5ex]\\[0.5ex]
\hline \\[-0.5ex]
\ifx\reducedapptables\undefined
ETM 14A & \cite{Alexandrou:2014sha} & 2+1+1 & \parbox[t]{1.4cm}{1.9--3.0,\\2.0--2.6,\\2.1--3.1} & \parbox[t]{1.cm}{4.0,\\3.4,\\3.4} & \parbox[t]{5.3cm}{  No significant finite-volume effects are observed.}  \\[0.5ex]\\[0.5ex]
\hline\\[-0.5ex]
BMW 15 & \cite{Durr:2015dna} & 2+1 & \parbox[t]{1.4cm}{1.9--5.6,\\1.5--5.9,\\2.5--4.9,\\2.1--4.2,\\1.7--3.4} & \parbox[t]{1.cm}{3.9,\\3.9,\\3.0,\\3.9,\\3.8} & \parbox[t]{5.3cm}{  Combined continuum, chiral and volume fit to $M_{N,\Omega,\pi,K}$ within an extended frequentist method with finite-volume corrections following Ref.~\cite{Colangelo:2005cg}}  \\[0.5ex]\\[0.5ex]
\hline\\[-0.5ex]
Junnarkar 13 & \cite{Junnarkar:2013ac} & 2+1 & \parbox[t]{1.4cm}{2.4,\\3.6} & \parbox[t]{1.cm}{4.4,\\4.1} & \parbox[t]{5.3cm}{  Not estimated.}  \\[0.5ex]\\[0.5ex]
\hline\\[-0.5ex]
Shanahan 12 & \cite{Shanahan:2012wh} & 2+1 & \parbox[t]{1.4cm}{2.9} & \parbox[t]{1.cm}{4.3} & \parbox[t]{5.3cm}{  Baryons masses are finite-volume corrected~\cite{AliKhan:2003ack}.}  \\[0.5ex]\\[0.5ex]
\hline\\[-0.5ex]
JLQCD 12A & \cite{Oksuzian:2012rzb} & 2+1 & \parbox[t]{1.4cm}{1.8--2.7} & \parbox[t]{1.cm}{4.1} & \parbox[t]{5.3cm}{  No significant effects observed when comparing $L/a=16$ and 24 for the two lightest $M_\pi$.}  \\[0.5ex]\\[0.5ex]
\hline\\[-0.5ex]
QCDSF 11 & \cite{Horsley:2011wr} & 2+1 & \parbox[t]{1.4cm}{1.8--2.5} & \parbox[t]{1.cm}{4.1} & \parbox[t]{5.3cm}{  Not estimated.}  \\[0.5ex]\\[0.5ex]
\hline\\[-0.5ex]
BMW 11A & \cite{Durr:2011mp} & 2+1 & \parbox[t]{1.4cm}{2.0--3.9,\\2.0--3.9,\\2.0--2.5} & \parbox[t]{1.cm}{4.1,\\3.9,\\4.1} & \parbox[t]{5.3cm}{  FVE found to be small in Ref.~\cite{Durr:2008zz}.}  \\[0.5ex]\\[0.5ex]
\hline\\[-0.5ex]
Martin~Camalich 10 & \cite{MartinCamalich:2010fp} & 2+1 & \parbox[t]{1.4cm}{2.9} & \parbox[t]{1.cm}{2.3} & \parbox[t]{5.3cm}{  Baryons masses are finite-volume corrected using Ref.~\cite{Beane:2004tw}.}  \\[0.5ex]\\[0.5ex]
\hline\\[-0.5ex]
PACS-CS 09 & \cite{Ishikawa:2009vc} & 2+1 & \parbox[t]{1.4cm}{2.9} & \parbox[t]{1.cm}{2.3} & \parbox[t]{5.3cm}{  Estimated to be less than $1\%$ in $M_N$ using Ref.~\cite{Beane:2004tw}.}  \\[0.5ex]\\[0.5ex]
\hline\hline
\end{tabular*}
\caption{Finite-volume effects in determinations of $\sigma_{\pi N}$  and $\sigma_s$  from the Feynman-Hellmann method.}
}
\end{table}
\fi
\ifx\reducedapptables\undefined
\begin{table}[!ht]
\addtocounter{table}{-1}
{\footnotesize
\begin{tabular*}{\textwidth}{l @{\extracolsep{\fill}} c c c c l}
\hline\hline \\[-1.0ex]
Collab. & Ref. & $\Nf$ & $L$ [fm] & ${M_{\pi,\rm min}}L$ & Description 
\\[.3ex]\\[-2.ex] \hline \hline \\[-1.0ex]
%
\fi
Walker-Loud 08 & \cite{Walker-Loud:2008rui} & 2+1 & \parbox[t]{1.4cm}{2.5} & \parbox[t\
]{1.cm}{3.7} &\parbox[t]{5.3cm}{FV effects expected to be $1\%$ or less for the lightest pion mass based on~\cite{Beane:2004tw}. } \\[0.5ex]\\[0.5ex]
\ifx\reducedapptables\undefined
\hline\\[-0.5ex]
QCDSF 12 & \cite{Bali:2012qs} & 2 & \parbox[t]{1.4cm}{1.8,\\1.7--3.4,\\1.9--2.9} & \parbox[t]{1.cm}{4.4,\\2.7,\\3.8} & \parbox[t]{5.3cm}{  Finite-volume corrections are applied~\cite{AliKhan:2003ack}.}  \\[0.5ex]\\[0.5ex]
\hline\\[-0.5ex]
JLQCD 08B & \cite{Ohki:2008ff} & 2 & \parbox[t]{1.4cm}{1.9} & \parbox[t]{1.cm}{2.8} & \parbox[t]{5.3cm}{ Systematic error based on fits with and without FVE of Ref.~\cite{AliKhan:2003ack}.}  \\[0.5ex]\\[0.5ex]
\fi
\hline\hline
\end{tabular*}
\ifx\reducedapptables\undefined
\caption{(cntd.) Finite-volume effects in determinations of $\sigma_{\pi N}$  and $\sigma_s$  from the Feynman-Hellmann method.}
}
\else
\caption{Finite-volume effects in determinations of $\sigma_{\pi N}$  and $\sigma_s$  from the Feynman-Hellmann method.}
}
\fi
\end{table}

 \clearpage
\ifx\reducedapptables\undefined
\begin{table}[!ht]

{\footnotesize
\begin{tabular*}{\textwidth}{l @{\extracolsep{\fill}} c c c l}
\hline\hline \\[-1.0ex]
Collab. & Ref. & $\Nf$ & $a$ [fm] & Description 
\\[.3ex]\\[-2.ex] \hline \hline \\[-1.0ex]
%
ETM 19 & \cite{Alexandrou:2019brg} & 2+1+1 & 0.08 & \parbox[t]{6.3cm}{Not estimated. }
\\[0.5ex]\\[0.5ex]
\hline\\[-0.5ex]
PNDME 18B & \cite{Gupta:2018lvp} & 2+1+1 & 0.15,0.12,0.09,0.06 & \parbox[t]{4.7cm}{  Connected: Joint continuum, chiral and volume fit with a term linear in $a$. \\ Disconnected: Joint continuum and chiral fit with a term linear in $a$. } \\[0.5ex]\\[0.5ex]
\hline\\[-0.5ex]
PNDME 16 & \cite{Bhattacharya:2016zcn} & 2+1+1 & 0.12,0.09,0.06 &\parbox[t]{4.7cm}{  Connected: Joint continuum, chiral and volume fit with a term linear in $a$. \\ Disconnected contribution neglected. } \\[0.5ex]\\[0.5ex]
\hline\\[-0.5ex]
PNDME 15 & \cite{Bhattacharya:2015wna,Bhattacharya:2015esa} & 2+1+1 & 0.12,0.09,0.06 &  \parbox[t]{4.7cm}{ Connected: Joint continuum, chiral and volume fit with a term linear in $a$. \\ Disconnected strange: Joint continuum and chiral fit with a term linear in $a$. } \\[0.5ex]\\[0.5ex]
\hline\\[-0.5ex]
JLQCD 18 & \cite{Yamanaka:2018uud} & 2+1 & 0.11 &   \parbox[t]{4.7cm}{  Discretization effects are estimated to be $8\%$ using $\cO(\Lambda^2_{QCD}a^2)$ with $\Lambda_{QCD}\approx 500$~MeV.} \\[0.5ex]\\[0.5ex]
\hline\\[-0.5ex]
ETM 17 & \cite{Alexandrou:2017qyt} & 2 & 0.09 &  \parbox[t]{4.7cm}{ Single lattice spacing.}  \\[0.5ex]\\[0.5ex]
\hline \hline \\[-1.0ex]
\end{tabular*}
\caption{Continuum extrapolations/estimation of lattice artifacts in determinations of $g_T^q$.}
}
\end{table}

\begin{table}[!ht]
{\footnotesize
\begin{tabular*}{\textwidth}{l @{\extracolsep{\fill}} c c c l}
\hline\hline \\[-1.0ex]
{Collab.} & {Ref.} & {$\Nf$} & {$M_{\pi,\rm min}\,[\mev]$} & {Description}  
\\[.3ex]\\[-2.ex] \hline \hline \\[-1.0ex]
%
ETM 19 & \cite{Alexandrou:2019brg} & 2+1+1 & 139  & \parbox[t]{6.3cm}{ Simulate close to $M_\pi^{phys}$. }
\\[0.5ex]\\[0.5ex]
\hline\\[-0.5ex]
PNDME 18B & \cite{Gupta:2018lvp} & 2+1+1 &  \parbox[t]{2.5cm}{ $C$: 320,225,135 \\ $D_l$: 320, 235 \\ $D_s$: 320, 225, 138}&  \parbox[t]{4.7cm}{  Connected: Joint continuum, chiral and volume fit with a term linear in $M_\pi^2$. \par Disconnected: Joint continuum and chiral fit with a term linear in $M_\pi^2$. }  \\[0.5ex]\\[0.5ex]
\hline\\[-0.5ex]
PNDME 16 & \cite{Bhattacharya:2016zcn} & 2+1+1 & \parbox[t]{2.5cm}{ $C$: 310,225,138 } &   \parbox[t]{4.7cm}{  Connected: Joint continuum, chiral and volume fit with a term linear in $M_\pi^2$. } \\[0.5ex]\\[0.5ex]
\hline\\[-0.5ex]
PNDME 15 & \cite{Bhattacharya:2015wna,Bhattacharya:2015esa} & 2+1+1 & \parbox[t]{2.5cm}{ $C$: 310,225,138 \\ $D_s$: 310, 228} & \parbox[t]{4.7cm}{  Connected: Joint continuum, chiral and volume fit with a term linear in $M_\pi^2$. \par Disconnected: Joint continuum and chiral fit with a term linear in $M_\pi^2$. }  \\[0.5ex]\\[0.5ex]
\hline\\[-0.5ex]
JLQCD 18 & \cite{Yamanaka:2018uud} & 2+1 & 293 &  Fit linear in $M_\pi^2$. \\[0.5ex]\\[0.5ex]
\hline\\[-0.5ex]
ETM 17 & \cite{Alexandrou:2017qyt} & 2 & 130 & \parbox[t]{4.7cm}{Single simulation at $M_\pi = 130$ MeV. } \\[0.5ex]\\[0.5ex]
\hline\hline
\end{tabular*}
\caption{Chiral extrapolation/minimum pion mass in determinations of $g_T^q$.}
}
\end{table}

\begin{table}[!ht]
{\footnotesize
\begin{tabular*}{\textwidth}{l @{\extracolsep{\fill}} c c c c l}
\hline\hline \\[-1.0ex]
Collab. & Ref. & $\Nf$ & $L$ [fm] & ${M_{\pi,\rm min}}L$ & Description 
\\[.3ex]\\[-2.ex] \hline \hline \\[-1.0ex]
%
ETM 19 & \cite{Alexandrou:2019brg} & 2+1+1 & \parbox[t]{1.4cm}{5.1} &
\parbox[t]{1.cm}{3.6} & \parbox[t]{6.3cm}{FVE are anticipated to be small based on an investigation using two
near physical point $N_f=2$ ensembles with $M_\pi L=3.0$ and $4.0$.}
\\[0.5ex]\\[0.5ex]
\hline\\[-0.5ex]
PNDME 18B & \cite{Gupta:2018lvp} & 2+1+1 & \parbox[t]{1.4cm}{2.4,\\2.9--3.8,\\2.8--5.6,\\2.8} & \parbox[t]{1.cm}{3.9,\\4.4,\\3.9,\\4.5} &  \parbox[t]{4.7cm}{ Connected: Joint continuum, chiral and volume fit with the term  $M_\pi^2 e^{-M_\pi L}$. \\ Disconnected: Neglect volume dependence. }  \\[0.5ex]\\[0.5ex]
\hline\\[-0.5ex]
PNDME 16 & \cite{Bhattacharya:2016zcn} & 2+1+1 & \parbox[t]{1.4cm}{2.9--4.8,\\2.8--5.6,\\2.8--3.7} & \parbox[t]{1.cm}{5.5,\\3.9,\\4.4} & \parbox[t]{4.7cm}{ Connected: Joint continuum, chiral and volume fit with the term  $M_\pi^2 e^{-M_\pi L}$.  }   \\[0.5ex]\\[0.5ex]
\hline\\[-0.5ex]
PNDME 15 & \cite{Bhattacharya:2015wna,Bhattacharya:2015esa} & 2+1+1 & \parbox[t]{1.4cm}{2.9--4.8,\\2.8--5.6,\\2.8--3.7} & \parbox[t]{1.cm}{5.5,\\3.9,\\4.4} & \parbox[t]{4.7cm}{ Connected: Joint continuum, chiral and volume fit with the term  $M_\pi^2 e^{-M_\pi L}$. \\ Disconnected strange: Neglect volume dependence. }   \\[0.5ex]\\[0.5ex]
\hline\\[-0.5ex]
JLQCD 18 & \cite{Yamanaka:2018uud} & 2+1 & \parbox[t]{1.4cm}{1.8--2.7} & \parbox[t]{1.cm}{4.0} &  \parbox[t]{4.7cm}{FV correction expected to be small for $M_\pi L \gtrsim 4$.} \\[0.5ex]\\[0.5ex]
\hline\\[-0.5ex]
ETM 17 & \cite{Alexandrou:2017qyt} & 2 & \parbox[t]{1.4cm}{4.5} & \parbox[t]{1.cm}{3.0} & \parbox[t]{4.7cm}{Single simulation with $M_\pi L = 3.0$. }  \\[0.5ex]\\[0.5ex]
\hline\hline
\end{tabular*}
\caption{Finite-volume effects in determinations of $g_T^q$.}
}
\end{table}

\begin{table}[!ht]
{\footnotesize
\begin{tabular*}{\textwidth}{l @{\extracolsep{\fill}} c c c }
\hline\hline \\[-1.0ex]
Collab. & Ref. & $\Nf$ & Ren. 
\\[.3ex]\\[-2.ex] \hline \hline \\[-1.0ex]
%
ETM 19 & \cite{Alexandrou:2019brg} & 2+1+1 & RI-SMOM   \\[0.5ex]\\[0.5ex]
\hline\\[-0.5ex]
PNDME 18B & \cite{Gupta:2018lvp} & 2+1+1 & RI-SMOM    \\[0.5ex]\\[0.5ex]
\hline\\[-0.5ex]
PNDME 16 & \cite{Bhattacharya:2016zcn} & 2+1+1 & RI-SMOM    \\[0.5ex]\\[0.5ex]
\hline\\[-0.5ex]
PNDME 15 & \cite{Bhattacharya:2015wna,Bhattacharya:2015esa} & 2+1+1 & RI-SMOM    \\[0.5ex]\\[0.5ex]
\hline\\[-0.5ex]
JLQCD 18 & \cite{Yamanaka:2018uud} & 2+1 & NP    \\[0.5ex]\\[0.5ex]
\hline\\[-0.5ex]
ETM 17 & \cite{Alexandrou:2017qyt} & 2 & RI'-MOM    \\[0.5ex]\\[0.5ex]
 \hline \hline \\[-1.0ex]
\end{tabular*}
\caption{Renormalization in determinations of $g_T^q$.}
}
\end{table}
\begin{table}[!ht]

{\footnotesize
\begin{tabular*}{\textwidth}{l @{\extracolsep{\fill}} c c c l}
\hline\hline \\[-1.0ex]
Collab. & Ref. & $\Nf$ & $\tau$ [fm] & Description 
\\[.3ex]\\[-2.ex]\hline \hline \\[-1.0ex]
%
ETM 19 & \cite{Alexandrou:2019brg} & 2+1+1 &  \parbox[t]{3.5cm}{[0.6--1.6]/all}  &  \parbox[t]{6.3cm}{Two-state fit to all $\tau$. A comparison is made with three-state fits, plateau fits and the summation method.} \\[0.5ex]\\[0.5ex]
\hline\\[-0.5ex]
PNDME 18B & \cite{Gupta:2018lvp} & 2+1+1 & \parbox[t]{3.5cm}{[0.8--1.4]\par [1.0--1.4,0.9--1.4]\par [0.9--1.2,0.9--1.2,0.7--1.4]\par [0.9--1.4]} &  \parbox[t]{4.7cm}{ Connected: three-state fit. \\ Disconnected: constant fit.} \\[0.5ex]\\[0.5ex]
\hline\\[-0.5ex]
PNDME 16 & \cite{Bhattacharya:2016zcn} & 2+1+1 & \parbox[t]{3.5cm}{[1.0--1.4,1.0--1.4]\par [0.9--1.2,0.9--1.2,0.9--1.2]\par [0.9--1.4,0.9--1.4]} & \parbox[t]{4.7cm}{ Connected: two-state fit. } \\[0.5ex]\\[0.5ex]
\hline\\[-0.5ex]
PNDME 15 & \cite{Bhattacharya:2015wna,Bhattacharya:2015esa} & 2+1+1 & \parbox[t]{3.5cm}{[1.0--1.4,1.0--1.4]\par [0.9--1.2,0.9--1.2,0.9--1.2]\par [0.9--1.4,0.9--1.4]} & \parbox[t]{4.7cm}{ Connected: two-state fit. \\ Disconnected: two-state fit. } \\[0.5ex]\\[0.5ex]
\hline\\[-0.5ex]
JLQCD 18 & \cite{Yamanaka:2018uud} & 2+1 & \parbox[t]{3.5cm}{[1.0--1.54]} & Constant fit to all data \\[0.5ex]\\[0.5ex]
\hline\\[-0.5ex]
ETM 17 & \cite{Alexandrou:2017qyt} & 2 & \parbox[t]{3.5cm}{[0.9--1.3]} & \parbox[t]{4.5cm}{ $C$:  Plateau fit to 1.31 fm data \\ $D$: Plateau fit to 0.75 fm data}   \\[0.5ex]\\[0.5ex]
\hline \hline \\[-1.0ex]
\end{tabular*}
\caption{Control of excited state contamination in determinations of $g_T^q$. The comma-separated list of numbers in square brackets denote the range of source-sink separations $\tau$ (in fermi) at each value of the bare coupling. } }
\end{table}

\fi

\clearpage
\subsection{Notes to Sec.~\ref{sec:scalesetting} on scale setting}

\begin{table}[!ht]
{\footnotesize
\begin{tabular*}{\textwidth}[l]{l @{\extracolsep{\fill}} c c c l}
\hline\hline \\[-1.0ex]
Collab. & Ref. & $\Nf$ & $a$ [fm] & Description
\\[1.0ex] \hline \hline \\[-1.0ex]

ETM 21 & \cite{Alexandrou:2021bfr} & 2+1+1 & 0.069, 0.079, 0.097 & \\
[1.0ex] \hline \\[-1.0ex]
CalLat 20A & \cite{Miller:2020evg} & 2+1+1 & \parbox[t]{3.5cm}{$[0.056, 0.082, 0.11, 0.13]$, $[0.057, 0.087, 0.12, 0.15]$}   & \parbox[t]{5.5cm}{M\"obius Domain-wall valence quarks on HISQ /  MILC sea; lattice spacings depending on scheme.} \\
[7.0ex] \hline \\[-1.0ex]
 BMW 20   &  \cite{Borsanyi:2020mff} & 1+1+1+1 & \parbox[t]{3.5cm}{\centering 0.0640, 0.0787, 0.0952, 0.1116, 0.1191, 0.1315} &  \parbox[t]{5.5cm}{Staggered fermion computation with isospin breaking and QED.}  \\[4.0ex] 
\hline\\[-1.0ex]
  ETM 20 & \cite{Dimopoulos:2020eqd} & 2+1+1 & 0.069, 0.079, 0.097 & \parbox[t]{5.5cm}{Wilson TM
  fermions at maximal twist including clover term. Note that  lattice spacings are not explicitely given in \cite{Dimopoulos:2020eqd}.}\\
  [10.0ex] \hline \\[-1.0ex]
ETM 18A & \cite{Alexandrou:2018egz} & 2+1+1 & 0.08 & \parbox[t]{5.5cm}{Wilson TM fermions at maximal twist including clover term.}\\
  [4.0ex] \hline \\[-1.0ex]
FNAL/MILC 17 &  \cite{Bazavov:2017lyh} & 2+1+1 &  \parbox[t]{3.5cm}{\centering 0.03, 0.042, 0.06, 0.09, 0.12, 0.15} & Determination of $f_{p4s}$.\\
[4.0ex]
\hline \\[-1.0ex]
MILC 15   & \cite{Bazavov:2015yea}  & 2+1+1 & \parbox[t]{3.5cm}{\centering 0.06, 0.09, 0.12, 0.15\\}& \parbox[t]{5.5cm}{Highly improved staggered quarks.} \\
[2.0ex] \hline \\[-1.0ex]
ETM 14 & \cite{Carrasco:2014cwa} & 2+1+1 & 0.089, 0.082, 0.062 & \parbox[t]{5.5cm}{Wilson TM fermions at maximal twist.}\\
[2.0ex] \hline \\[-1.0ex]
FNAL/MILC 14A &  \cite{Bazavov:2014wgs} & 2+1+1 & 0.06, 0.09, 0.12, 0.15  & \\
[2.0ex]
\hline \\[-1.0ex]
HPQCD 13A &  \cite{Dowdall:2013rya} & 2+1+1 &  0.15, 0.12, 0.09 & \\
[2.0ex] \hline \\[-1.0ex]
HPQCD 11B &  \cite{Dowdall:2011wh} & 2+1+1 & 0.09, 0.12, 0.15 &
                                                                \parbox[t]{5.5cm}{NRQCD for $b$ quark and HISQ for valence light quarks and MILC sea.} \\
[4.0ex]
\hline \hline\\[-1.0ex]
\end{tabular*}
\caption{Continuum extrapolations/estimation of lattice artifacts in scale
  determinations with $N_f=2+1+1$ quark flavours.}
}
\end{table}

\begin{table}[!ht]
{\footnotesize
\begin{tabular*}{\textwidth}[l]{l @{\extracolsep{\fill}} c c c l}
\hline\hline \\[-1.0ex]
Collab. & Ref. & $\Nf$ & $a$ [fm] & Description
\\[1.0ex] \hline \hline \\[-1.0ex]
CLS 16 & \cite{Bruno:2016plf} & 2+1 & 0.085, 0.065, 0.05 &
                                                              \parbox[t]{5.5cm}{NP ${\cal O}(a)$-improved Wilson fermions with LW gauge action.} \\
[4.0ex] \hline \\[-1.0ex]
QCDSF/UKQCD 15B & \cite{Bornyakov:2015eaa} & 2+1 &  \parbox[t]{3.5cm}{\centering 0.059, 0.068, 0.074, 0.082} &  $(a_\text{max}/a_\text{min})^2=1.94$ \\
[4.0ex]\hline \\[-1.0ex]
$\chi$QCD 14& \cite{Yang:2014sea} & 2+1 & 0.084, 0.112 & \parbox[t]{5.5cm}{Valence overlap fermions on domain wall fermion gauge configurations.}\\
[4.0ex] \hline \\[-1.0ex]
HotQCD 14 & \cite{Bazavov:2014pvz} & 2+1 & $[0.04,0.25]$ & \parbox[t]{5.5cm}{HISQ and tree-level improved Symanzik gauge action. Several ensembles with $a$ in the quoted range and with a single $M_\pi\approx 160\MeV$.} \\[10.0ex]
\hline\\ [-1.0ex]
RBC/UKQCD 14  & \cite{Blum:2014tka}  & 2+1 & \parbox[t]{3.5cm}{\centering 0.06, 0.08, 0.11, 0.14\\}& \parbox[t]{5.5cm}{Two versions of Domain Wall Fermions combined each with the Iwasaki gauge action.} \\[7.0ex]
\hline\\ [-1.0ex]
BMW 12A &  \cite{Borsanyi:2012zs} & 2+1 & \parbox[t]{3.5cm}{\centering 0.053, 0.065, 0.077, 0.093} & \parbox[t]{5.5cm}{ Wilson fermion computation.} \\
[4.0ex]\hline \\[-1.0ex]
HotQCD 11 & \cite{Bazavov:2011nk} & 2+1 & \parbox[t]{3.5cm}{\centering $[0.066,0.25]$\\}& \parbox[t]{5.5cm}{HISQ and tree-level improved Symanzik gauge action. Several ensembles with $a$ in the quoted range.} \\[7.0ex]
\hline \\[-1.0ex]
RBC/UKQCD 10A &  \cite{Aoki:2010dy} & 2+1 & 0.087, 0.114 & \\
[2.0ex]
\hline \\[-1.0ex]
MILC 10 & \cite{Bazavov:2010hj} & 2+1 & 0.045, 0.06, 0.09 & Asqtad staggered quarks.
\\[2.0ex] \hline \\[-1.0ex]
HPQCD 09B &  \cite{Davies:2009tsa} & 2+1 &
                                           \parbox[t]{3.5cm}{\centering 0.044, 0.059, 0.085, 0.12, 0.15} & \parbox[t]{5.5cm}{NRQCD for $b$ quark and HISQ for valence light quarks on MILC asqtad sea.}
 \\
[4.0ex]\hline \\[-1.0ex]
MILC 09 & \parbox[t]{.5cm}{\cite{Bazavov:2009bb,Bazavov:2009fk,Bazavov:2009tw} }& 2+1 & \parbox[t]{3.5cm}{\centering $[0.045,0.18]$\\}& \parbox[t]{5.5cm}{Asqtad staggered quarks, 6 different lattice spacings in the quoted range.} \\
[7.0ex]\hline \\[-1.0ex]
PACS-CS 08 & \cite{Aoki:2008sm} & 2+1 & \parbox[t]{3.5cm}{\centering $0.09$\\}& \parbox[t]{5.5cm}{Several ensembles of Wilson clover quarks and Iwasaki gauge action at a single lattice spacing.} \\
[7.0ex] \hline \\[-1.0ex]
HPQCD 05B &  \cite{Gray:2005ur} & 2+1 & 0.09, 0.12, 0.17  & \parbox[t]{5.5cm}{NRQCD with tree level tadpole improved couplings on MILC asqtad sea.} \\
[4.0ex]\hline \\[-1.0ex]
Aubin 04 & \cite{Aubin:2004wf} & 2+1 & \parbox[t]{3.5cm}{\centering 0.09, 0.12\\}& \parbox[t]{5.5cm}{Asqtad staggered quarks.}\\ [2.0ex]
\hline \hline \\[-1.0ex]
\end{tabular*}
\caption{Continuum extrapolations/estimation of lattice artifacts in scale
  determinations with $N_f=2+1$ quark flavours.}
}
\end{table}

\begin{table}
{\footnotesize
\begin{tabular*}{\textwidth}[l]{l @{\extracolsep{\fill}} c c c c l}
\hline \hline  \\[-1.0ex]
Collab. & Ref. & $\Nf$ & $M_{\pi,\text{min}}$ [MeV] &  $M_{\pi} L$  & Description
\\[1.0ex] \hline \hline \\[-1.0ex]

  ETM 21 & \cite{Alexandrou:2021bfr} & 2+1+1 & & & \\
[1.0ex] \hline \\[-1.0ex]
CalLat 20A & \cite{Miller:2020evg}    & 2+1+1 &  130 & 3.9 &  \\
[2.0ex] \hline \\[-1.0ex]
BMW 20   &  \cite{Borsanyi:2020mff} & 1+1+1+1 &   \parbox[t]{2.75cm}{Several pion masses within $\pm$3\% of physical value.}  &  3.0   &       \\
[7.0ex] \hline\\[-1.0ex]
ETM 20 & \cite{Dimopoulos:2020eqd} & 2+1+1 &  135 & 3.5 & \parbox[t]{5.5cm}{Only  indirectly inferred.}\\
[2.0ex] \hline \\[-1.0ex]
  ETM 18A & \cite{Alexandrou:2018egz} & 2+1+1 &  139.8 & 3.6 & One ensemble only.\\
[2.0ex] \hline \\[-1.0ex]
FNAL/MILC 17 &  \cite{Bazavov:2014wgs} & 2+1+1 & 129 & 3.7  & Determination of $f_{4ps}, M_{4ps}$.\\
[2.0ex]
\hline \\[-1.0ex]
MILC 15  & \cite{Bazavov:2015yea} & 2+1+1 & 125 & 3.7
&  \parbox[t]{5.5cm}{Four ensembles at physical point.}\\
[2.0ex] \hline \\[-1.0ex]
ETM 14 & \cite{Carrasco:2014cwa} & 2+1+1 & 211 & 3.19 &\\
[2.0ex] \hline \\[-1.0ex]
FNAL/MILC 14A &  \cite{Bazavov:2014wgs} & 2+1+1 & 130 & 3.7 & Determination of $f_{4ps}, M_{4ps}$.\\
[2.0ex]
\hline \\[-1.0ex]
HPQCD 13A &  \cite{Dowdall:2013rya} & 2+1+1 & 128 & 3.7 & \\
[2.0ex] \hline \\[-1.0ex]
HPQCD 11B & \cite{Dowdall:2011wh} & 2+1+1 & 211 & 4.0 & \parbox[t]{5.5cm}{NRQCD $\Upsilon$ 2s-1s splitting and $\eta_s$ as input.}
\\[4.0ex] \hline\hline \\[-1.0ex]
%
%
\end{tabular*}
\caption{Chiral extrapolation and finite-volume effects in scale determinations 
  with $N_f=2+1+1$ quark flavours. We list the minimum pion mass $M_{\pi,\text{min}}$ and $M_{\pi} L \equiv M_{\pi,\text{min}} [L(M_{\pi,\text{min}})]_\text{max}$ is evaluated at the maximum value of $L$ available at $M_{\pi} =M_{\pi,\text{min}}$.  }
}
\end{table}

\begin{table}
{\footnotesize
\begin{tabular*}{\textwidth}[l]{l @{\extracolsep{\fill}} c c c c l}
\hline \hline  \\[-1.0ex]
Collab. & Ref. & $\Nf$ & $M_{\pi,\text{min}}$ [MeV] &  $M_{\pi} L$  & Description
\\[1.0ex] \hline \hline \\[-1.0ex]
CLS 16 & \cite{Bruno:2016plf} & 2+1 & 200 & 4.2 & \parbox[t]{5.0cm}{Extrapolation along line $m_u+m_d+m_s=\mathrm{const.}$} \\
[4.0ex] 
\hline\\[-1.0ex]
QCDSF/UKQCD 15B & \cite{Bornyakov:2015eaa} & 2+1 & 228 & 4.1 & \\
[2.0ex]
\hline\\[-1.0ex]
$\chi$QCD 14& \cite{Yang:2014sea} & 2+1 & 290 & 4.0 & \parbox[t]{5.0cm}{}\\
[2.0ex] \hline \\[-1.0ex]
HotQCD 14 & \cite{Bazavov:2014pvz} & 2+1 & 160 & 4.8 & \parbox[t]{5.0cm}{HISQ staggered quarks, the pion mass quoted is $M_\pi$. Results on 20 ensembles with volumes in the range $[2.6,6.1]$~fm, we quote $M_\pi^{min} L^{max}$.}\\ 
[12.0ex]\hline \\[-1.0ex]
RBC/UKQCD 14  & \cite{Blum:2014tka}  & 2+1 &  $139$ & $3.9$ &  \\
[4.0ex] 
\hline\\[-1.0ex]
BMW 12A &  \cite{Borsanyi:2012zs} & 2+1 &  131 and 120 & 3.92 and 3.0 & \\
[2.0ex]\hline \\[-1.0ex]
HotQCD 11 & \cite{Bazavov:2011nk}    & 2+1 & $160$  & $4.8$ &  \parbox[t]{5.0cm}{HISQ staggered quarks, the pion mass quoted is $M_\pi$. Results on 20 ensembles with volumes in the range $[3.2,6.1]$~fm, we quote $M_\pi^{min}L^{max}$. }\\
[12.0ex]
\hline \\[-1.0ex]
RBC/UKQCD 10A &  \cite{Aoki:2010dy} & 2+1 & 290 & 4.0 & \\
[2.0ex]
\hline \\[-1.0ex]
MILC 10 & \cite{Bazavov:2010hj} & 2+1 & 255 & 4.8 & \parbox[t]{5.0cm}{Determination from global fit, scale from $f_\pi$.}
\\[4.0ex] \hline \\[-1.0ex]
HPQCD 09B&  \cite{Davies:2009tsa} & 2+1 & 211 & 4.0 & Extension of HPQCD 05B  \cite{Gray:2005ur}.\\
[2.0ex]\hline \\[-1.0ex]
MILC 09 & \parbox[t]{.5cm}{\cite{Bazavov:2009bb,Bazavov:2009fk,Bazavov:2009tw}}    & 2+1 & 258  & $4.3$ &  \parbox[t]{5.0cm}{Asqtad staggered quarks, the pion mass quoted is $M_\pi^{RMS}$.}\\
[7.0ex]\hline \\[-1.0ex]
PACS-CS 08 & \cite{Aoki:2008sm} & 2+1 & 156  & $2.3$ &  \parbox[t]{5.0cm}{Clover quarks, several pion masses in the range $[156,702]$~MeV, single lattice spacing.}\\
[7.0ex]\hline \\[-1.0ex]
HPQCD 05B & \cite{Gray:2005ur} & 2+1 & 270 & 3.7 & NRQCD $\Upsilon$ 2s-1s splitting as input.\\
[2.0ex]\hline \\[-1.0ex]
Aubin 04 & \cite{Aubin:2004wf}    & 2+1 & 253  & $3.8$ &  \parbox[t]{5.0cm}{Asqtad staggered quarks, the pion mass quoted is the Goldstone mass.}\\
[4.0ex]
\hline \hline \\[-1.0ex]
\end{tabular*}
\caption{Chiral extrapolation and finite-volume effects in scale determinations 
  with $N_f=2+1$ quark flavours. We list the minimum pion mass $M_{\pi,\text{min}}$ and $M_{\pi} L \equiv M_{\pi,\text{min}} [L(M_{\pi,\text{min}})]_\text{max}$ is evaluated at the maximum value of $L$ available at $M_{\pi} =M_{\pi,\text{min}}$.  }
}
\end{table}

\end{appendix}


\clearpage
\settocbibname{References}
\bibliography{FLAG}
\bibliographystyle{JHEP_FLAG}

\end{document}